# Research on the Work-Energy Principle Based Characteristic Mode Theory for Scattering Systems

(English Version of a Doctoral Dissertation Submitted to University of Electronic Science and Technology of China)

Discipline: **Electromagnetic Field and Microwave Technology**

Author: **Renzun Lian**

**rzlian@vip.163.com**

Supervisor: **Prof. Jin Pan**

University: **University of Electronic Science and Technology of China**

School: **School of Electronic Science and Engineering**

# ABSTRACT

Electromagnetic (EM) scattering systems widely exist in EM engineering domain. For a certain objective scattering system, all of its working modes constitute a linear space —— modal space. Characteristic mode theory (CMT) can effectively construct a basis of the space —— characteristic modes (CMs), and the CMs only depend on the inherent physical properties of the objective system, such as the topological structure and the material parameter of the objective system. Thus, CMT is very valuable for analyzing and designing the inherent EM scattering characters of the objective system.

Since Prof. Harrington et al. transformed the framework for carrying CMT from Prof. Garbacz's scattering matrix (SM) framework to integral equation (IE) framework, the theoretical research and engineering application of CMT have achieved quick developments. But, IE-based CMT (IE-CMT) still has some important problems which have not been completely solved, for example:

**Old Problem 1 (OP1).** The IE-based CM formulations are usually derived from various EM field boundary conditions, and the boundary conditions are mathematical equations. What physical properties do the CMs constructed by the mathematical equations have? In other words, what physical picture does IE-CMT have? This is one of the important questions in IE-CMT researching domain in recent years, and it has not been completely answered, and then some phenomena which have not been fully explained have appeared. **Unexplained Phenomenon A.** The CMs derived from the electric field integral equation based CMT (EFIE-CMT) for metallic scattering systems can guarantee the orthogonality among modal far fields, but the CMs derived from the volume integral equation based CMT (VIE-CMT) for lossy material scattering systems cannot guarantee the orthogonality among modal far fields; **Unexplained Phenomenon B.** The physical meanings of the characteristic values calculated from VIE-CMT and EFIE-CMT are not the same; **Unexplained Phenomenon C.** For a certain objective metallic scattering system, the CMs derived from EFIE-CMT, magnetic field integral equation based CMT (MFIE-CMT), and combined field integral equation based CMT (CFIE-CMT) are not the same.

**Old Problem 2 (OP2).** For a metallic scattering system, non-radiative mode set (such as internal resonant mode set) is an important part of whole modal space. But, IE-CMT cannot construct non-radiative modes, so cannot provide a complete CM set to the objective metallic scattering system.

**Old Problem 3 (OP3).** Existing IE-CMT cannot provide effective CM formulations to some material scattering systems which have complicated topological structures (such as stacked dielectric resonator antennas) or complicated material parameters (such as Lüneburg lens antenna).

**Old Problem 4 (OP4).** Existing IE-CMT cannot provide effective CM formulations to some metal-material composite scattering systems which have complicated topological structures (such as printed microstrip antennas) or complicated material parameters.

**Old Problem 5 (OP5).** For some complicated scattering systems, it is not completely solved how to effectively suppress the spurious CMs derived from IE-CMT.

This dissertation is committed to further sovling the above-mentioned old problems OP1~OP5. In the process of solving old problems OP1, OP3, and OP4, and after having solved old problems OP2 and OP5, this dissertation also raises a series of new questions as follows:

**New Problem 1 (NP1).** In the process of solving old problem OP1, this dissertation raises a new question on "Whether or not IE is the best framework for carrying CMT?".

**New Problem 2 (NP2).** After having solved old problem OP2, this dissertation raises a new question on "the detailed classification for working modes and the orthogonal decomposition for modal space".

**New Problem 3 (NP3).** In the process of solving old problem OP3, this dissertation raises a new question on "How to derive the mathematical expressions of the surface equivalence principle for complicated material scattering systems?".

**New Problem 4 (NP4).** In the process of solving old problem OP4, this dissertation raises a new question on "How to establish the line-surface equivalence principle of complicated metal-material composite scattering systems, and how to derive the corresponding mathematical expressions?".

**New Problem 5 (NP5).** After having solved old problem OP5, this dissertation raises a new question on "the physical meaning and numerical influence of the singular EM current term (SCT) contained in the CM formulations of complicated scattering systems".

Focusing on above {OP1, NP1}, {OP2, NP2}, …, and {OP5, NP5}, this dissertation does some innovative works. The main contributions of this dissertation are as follows:

**Main Contribution 1. Contribution in the Aspect of Solving {OP1, NP1}**

After some studies, this dissertation finds out that IE is not the best framework for carrying CMT. This dissertation proposes a completely new framework for carrying CMT —— work-energy principle (WEP) framework, and at the same time proposes a completely new method for constructing CMs —— orthogonalizing driving power operator (DPO) method. In new WEP framework and based on new orthogonalizing DPO method, this dissertation draws a clear physical picture for the IE-CMT established by Prof. Harrington et al. —— constructing a series of steadily working modes which have not net energy exchange in any integral period. Employing the physical picture, this dissertation reveals the causes of the unexplained phenomena A, B, and C mentioned above.

**Main Contribution 2. Contribution in the Aspect of Solving {OP2, NP2}**

This dissertation derives the mathematical expression of the WEP corresponding to metallic scattering systems, and then obtains the operator expression of the driving power (DP) corresponding to metallic scattering systems. In new WEP framework and based on new orthogonalizing DPO method, this dissertation proves that the CM set constructed by the traditional IE-CMT for metallic scattering systems is not complete, and provides a practical method for completing the CM set. Based on the complete CM set, this dissertation orthogonally decomposes whole modal space and detailedly classifies all working modes, and then orthogonally decomposes various working modes in the simplest way. The obtained modal simplest orthogonal decomposition is very valuable for revealing the working mechanism of various modes.

**Main Contribution 3. Contribution in the Aspect of Solving {OP3, NP3}**

This dissertation generalizes the traditional surface equivalence principle (SEP) of material scattering systems to generalized surface equivalence principle (GSEP) from the aspects of external EM environment, system topological structure, and system material parameter, and derives the corresponding mathematical expressions. This dissertation derives the mathematical expression of the WEP corresponding to material scattering systems, and then obtains the operator expression of the DP corresponding to material scattering systems. In new WEP framework and based on new orthogonalizing DPO method and employing newly obtained GSEP, this dissertation generalizes the traditional IE-CMT for material scattering systems from the aspects of external environment, topological structure, and material parameter, and the CM formulations built in this dissertation are completely new. The new formulations have advantages over the traditional formulations in the aspects of applicable range, expression form, physical picture, and computational burden. In addition, the new formulations have not been able to be effectively established in traditional IE framework.

**Main Contribution 4. Contribution in the Aspect of Solving {OP4, NP4}**

This dissertation establishes the line-surface equivalence principle (LSEP) for complicated metal-material composite scattering systems, and derives the corresponding mathematical expressions. This dissertation derives the mathematical expression of the WEP corresponding to metal-material composite scattering systems, and then obtains the operator expression of the DP corresponding to metal-material composite scattering systems. In new WEP framework and based on new orthogonalizing DPO method and employing newly obtained LSEP, this dissertation derives the CM formulations for complicated metal-material composite scattering systems, and the obtained CM formulations have a very wide applicable range from the aspects of external environment, topological structure, and material parameter, and the CM formulations derived in this dissertation are completely new. The new CM formulations have advantages over the traditional formulations, and the new formulations have not been able to be effectively established in traditional IE framework.

**Main Contribution 5. Contribution in the Aspect of Solving {OP5, NP5}**

In new WEP framework and based on new orthogonalizing DPO method and employing newly obtained GSEP and LSEP, this dissertation deeply studies the forming reasons and the suppressing methods for spurious CMs, which widely exist in CMT, and systematically summarizes the operating process for suppressing the spurious CMs for arbitrary scattering systems. This dissertation reveals the physical meaning of the SCTs in the DPOs of material scattering systems and metal-material composite scattering systems, and also researches the influence of the SCTs on the numerical performance of CM formulations.

To verify the correctness and validity of the above results and conclusions, this dissertation provides general mathematical proofs or special numerical evidence to the results and conclusions.

# Main Symbol Table

For the convenience of expressions, this dissertation utilizes some abbreviations frequently. Now, we list the abbreviations and their full names in the following table:

| Abbreviations | Full Names |
|---|---|
| MetSca | metallic scattering system |
| MatSca | material scattering system |
| ComSca | metal-material composite scattering system |
| EFIE | electric field integral equation |
| MFIE | magnetic field integral equation |
| CFIE | combined field integral equation |
| VIE | volume integral equation |
| SIE | surface integral equation |
| SEM | singularity expansion method |
| EMT | eigen-mode theory |
| HWE-CloMetSca-EMT | homogeneous wave equation based EMT for closed metallic scattering systems |
| CM | characteristic mode |
| CMT | characteristic mode theory |
| SM | scattering matrix |
| PM | perturbation matrix |
| PMO | perturbation matrix operator |
| Garbacz’s CMT | CMT established by Prof. Garbacz |
| SM-MetSca-CMT | SM-based CMT for metallic scattering systems |
| IE | integral equation |
| IM | impedance matrix |
| IMO | impedance matrix operator |
| Harrington’s CMT | CMT established by Prof. Harrington et al. |
| IE-ScaSys-CMT | IE-based CMT for scattering systems |
| EFIE-OpeMetSca-CMT | EFIE-based CMT for open metallic scattering systems |
| VIE-MatSca-CMT | VIE-based CMT for material scattering systems |
| SIE-MatSca-CMT | SIE-based CMT for material scattering systems |
| WEP | work-energy principle |
| DP | driving power |
| DPO | driving power operator |
| DP-CM | CM derived from orthogonalizing DPO |
| WEP-ScaSys-CMT | WEP-based CMT for scattering systems |
| WEP-MetSca-CMT | WEP-based CMT for metallic scattering systems |
| WEP-MatSca-CMT | WEP-based CMT for material scattering systems |
| Vol-WEP-MatSca-CMT | volume formulation of WEP-MatSca-CMT |
| Surf-WEP-MatSca-CMT | surface formulation of WEP-MatSca-CMT |
| WEP-ComSca-CMT | WEP-based CMT for metal-material composite scattering systems |
| LS-WEP-ComSca-CMT | line-surface formulation of WEP-ComSca-CMT |
| MS | modal significance |
| SEP | surface equivalence principle |
| HFP | Huygens-Fresnel principle |
| BET | backward extinction theorem |
| FHF | Franz-Harrington formulation |
| GSEP | generalized SEP |
| LSEP | line-surface equivalence principle |
| GHFP | generalized HFP |
| GBET | generalized BET |
| GFHF | generalized FHF |
| SL | scattering line source |
| SS | scattering surface source |
| SV | scattering volume source |
| CV | conduction volume electric current |
| PV | polarization volume electric current |
| MV | magnetization volume magnetic current |
| EL | equivalent line source |
| ES | equivalent surface source |
| P.V. $\mathcal{K}$ | principal value of operator $\mathcal{K}$ |
| PVT | principal value term |
| SCT | singular current term |
| BV | basic variable |
| OP | old problem existing in Harrington's CMT |
| NP | new problem closely related to CMT |
| MC | main contribution of this dissertation |
| LHS | left-hand side |
| RHS | right-hand side |

# Contents

# Chapter 1 Introduction

There are to ways to do calculations.
The first way, which I prefer, is to have a clear physical picture. The second way is to have a rigorous mathematical formalism. [103]
—— Enrico Fermi (Nobel Prize in Physics, 1938)

In this chapter, we introduce the research background and significance (of this dissertation), the research history and state (related to this dissertation), the important problems and challenges (focused on by this dissertation), the contributions and innovations of this dissertation, and the outline of this dissertation.

## 1.1 Research Background and Significance

Electromagnetic (EM) scattering structures widely exist in EM engineering, and they have been the main research objects in wireless communication[1,2], microwave remote sensing[3-5], radar technology[6,7], target recognition[8], antenna engineering[9,10], stealth and anti-stealth technology[11,12], biomedical EM simulation[13], and EM compatibility analysis and design[14], and one of the main research topics in the above-mentioned research domains is the method to extract the inherent EM scattering characters of objective structures. In general, the inherent EM scattering characters of an objective structure need to be extracted from a large number of working modes① of the objective structure.

For a certain linear scattering structure, all of its possible modes constitute a linear space —— modal space[15,16]. Any set of independent modes which can span the space is a basis of the space. In some sense, the scattering structure and its modal space are one-to-one correspondence, but the basis of the modal space is not unique[15,16]. In mathematical physics, the modes which can reflect the inherent physical properties of the scattering structure are usually called as fundamental modes, and the theory to research the fundamental modes and their constructing methods is usually called as modal theory or modal analysis[15,16]. Thus, modal theory plays an important and positive role in the

① In the following parts of this dissertation, the terminology "working mode" is simply called as "mode".

process of extracting the inherent EM scattering characters of the objective structure.

In EM engineering, there have been some modal theories, such as cavity model theory[17-19], dielectric waveguide model theory[20-22], singularity expansion method (SEM)[23-25], eigen-mode theory (EMT)[26-28], and characteristic mode theory (CMT)[29-35], etc. Cavity model theory and dielectric waveguide model theory are usually used to calculate the fundamental radiated modes of printed microstrip antennas and dielectric resonator antennas respectively, but they are the model-based approximate calculation theories based on metallic cavity model and dielectric waveguide model respectively, so their calculation results are much different from measured results when the approximation conditions are not satisfied[17-22], and then their applicable ranges are very narrow; SEM is usually used to construct the fundamental modes of the open EM structures which work at naturally evanescent states, and the obtained fundamental modes are usually called as natural modes, but SEM usually outputs some spurious poles[36], and the spurious poles have not been able to be suppressed effectively, so SEM has not been applied extensively; EMT is usually used to construct the fundamental modes of the closed EM structures which work at steadily self-oscillation states, and the obtained fundamental modes are usually called as eigen-modes, but EMT is not suitable for being directly applied to open EM structures; CMT is an important modal theory which has been extensively applied to construct the fundamental modes of open EM structures, especially antenna structures[35], in recent years, and the obtained fundamental modes are usually called as characteristic modes (CMs).

Recently, some researchers[37] uniquely decomposed whole modal space into the direct sum of four weightedly orthogonal subspaces (pure capacitance space, nonradiation space, high-quality radiation space, and pure inductance space①) by employing the CM-based modal expansion, and found out that: the modes in pure capacitance space have a stronger attribute to reactively store electric energy, and all of the capacitive CMs constitute the basis of pure capacitance space; nonradiation space is just the traditional internal resonance space, and the modes in nonradiation space are just the traditional internally resonant modes, and all of the nonradiative CMs constitute the basis of nonradiation space②; the modes in high-quality radiation space have a stronger attribute

① The rigorous definitions for these subspaces can be found in literature [37] and the Section 3.3 of this dissertation.
② Based on this, we can effectively establish the corresponding relationship between nonradiative CMs and internally resonant eigen-modes and the connection between CMT and EMT. The detailed discussions for the corresponding relationship and connection can be found in literatures [37,40,41] and the Section 3.3 of this dissertation.

to scatter EM energy to infinity, and all of the radiative resonant CMs constitute the basis of high-quality radiation space; the modes in pure inductance space have a stronger attribute to reactively store magnetic energy, and all of the inductive CMs constitute the basis of pure inductance space.

After orthogonally decomposing whole modal space, the orthogonal decomposition for any working mode can be achieved by projecting the working mode onto the four subspaces, i.e., any one working mode can be expressed in terms of the superposition of four weightedly orthogonal fundamental components: working mode = purely capacitive component + nonradiative component + high-quality radiative component + purely inductive component. Obviously, the above-mentioned decomposition has important guidance significance for revealing the working mechanism of the working mode of the objective structure.

In summary, CMT has great application significance for extracting the inherent EM scattering characters of the objective structure. This is just one of the main reasons to research CMT, i.e. one of the main foundations to select the researching topic of this dissertation.

## 1.2 Research History and State

The concept of CM was introduced by Prof. R. Garbacz in literature [29] in 1965 for the first time. This dissertation calls the theory established in literature [29] as Garbacz's CMT to be distinguished from the other CMTs established by other scholars in other frameworks. Based on the theories related to scattering matrix (SM), Prof. Garbacz[29,30] proved that: for any selected lossless objective metallic scattering system① (MetSca), in its modal space there exists a set of CMs whose modal far fields are orthogonal to each others, and the external EM field of the lossless objective MetSca can be expressed in terms of the linear superposition of the CMs. Literatures [29,30] proved the existence of the CMs by employing a constructive method, i.e. orthogonalizing perturbation matrix operator (PMO) method, but the PMO-based orthogonalization method is very complicated, so the method has not been widely applied so far.

Inspired by Prof. Garbacz, Prof. R. Harrington et al.[31-34] established a new kind of CMT in integral equation (IE) framework in the 1970s, and this dissertation calls the CMT

① In the following parts of this dissertation, the terminology "metallic scattering system" is simply called as "metallic system", and simply denoted as "MetSca".

established by Prof. Harrington et al. as Harrington's CMT to be distinguished from the other CMTs established by other scholars in other frameworks (especially from the Garbacz's CMT established by Prof. Garbacz in SM framework). In literatures [31,32], Prof. Harrington and Dr. J. Mautz established the electric field integral equation (EFIE) based CMT for open metallic systems (EFIE-OpeMetSca-CMT), and the EFIE-OpeMetSca-CMT can construct a set of fundamental modes for open metallic systems (OpeMetSca) such that the modal far fields are orthogonal to each others, and the constructing method is very simple; in literature [33], Prof. Harrington et al. established the volume integral equation (VIE) based CMT for material scattering systems① (VIE-MatSca-CMT); in literature [34], Prof. Harrington et al. established the surface integral equation (SIE) based CMT for material systems (SIE-MatSca-CMT).

Because of its simplicity, Harrington's CMT (especially EFIE-OpeMetSca-CMT[31,32]) has been widely applied in antenna engineering domain. In the following parts of this section, we will briefly review the research state of Harrington's CMT from the aspects of theory and application.

### 1.2.1 Theoretical Aspects

In recent years, the theoretical researches on Harrington's CMT mainly focus on the following aspects:

**Theoretical Aspect I. On the Energy Relationships Which Harrington's CMs[31-34] Satisfy**

Literatures [35,37,40,41] did some in-depth studies on the energy relationship satisfied by the CMs derived from EFIE-OpeMetSca-CMT[31,32]. Professor Y. Chen[35,42], Prof. C.-F. Wang[35], Dr. Z. Miers[43], and Prof. B. Lau[43], et al.[44,45] did some systematical studies on the energy relationships satisfied by the CMs derived from VIE-MatSca-CMT[33] and SIE-MatSca-CMT[34]. Literatures [37,40,41,44,45] found out that the modal powers of Harrington's CMs are just the powers done by the corresponding modal incident fields on the corresponding modal scattered sources, and this conclusion will be further sublimated and generalized in the Chapters 2, 3, 4, and 5 of this dissertation.

**Theoretical Aspect II. On the Spurious Modes Which SIE-MatSca-CMT[34] Outputs**

In 2014, H. Alroughani et al.[46] first found out that the CM set constructed by SIE-

① In the following parts of this dissertation, the terminology "material scattering system" is simply called as "material system", and simply denoted as "MatSca".

MatSca-CMT[34] contains many spurious modes (the modes corresponding to the black curves shown in Figure 1-1(a) are just spurious modes), but literature [46] didn't explain the cause of the spurious modes and also didn't provide the suppression method for the spurious modes. Professor Chen[35,42] and Prof. Wang[35] pointed out that the cause of the spurious modes is that literature [34] overlooked the dependent relationship between the equivalent surface electric and magnetic currents on the boundary of material body, and Prof. Chen and Prof. Wang proposed a scheme to establish the transformation between the equivalent surface electric and magnetic currents. In addition, Dr. Miers and Prof. Lau[47] proposed a postprocessing scheme for filtering the spurious modes based on the energy relationships satisfied by CMs. Recently, Dr. F.-G. Hu[48], Prof. Wang[48], and Prof. P. Ylä-Oijala[49], et al.[44,45] proposed a series of new preprocessing schemes for suppressing the spurious modes. The preprocessing scheme in literature [44] is to establish the transformation between the equivalent surface electric and magnetic currents based on the continuation conditions of the tangential components of scattered EM fields, and the preprocessing scheme in literatures [45,49] is to establish the transformation between the equivalent surface electric and magnetic currents based on the definitions of the currents, and the results obtained in literature [45] is illustrated in Figure 1-1(b). In the Chapters 4, 5, and 6 of this dissertation, we will further discuss and extend the schemes proposed in literatures [44,45].

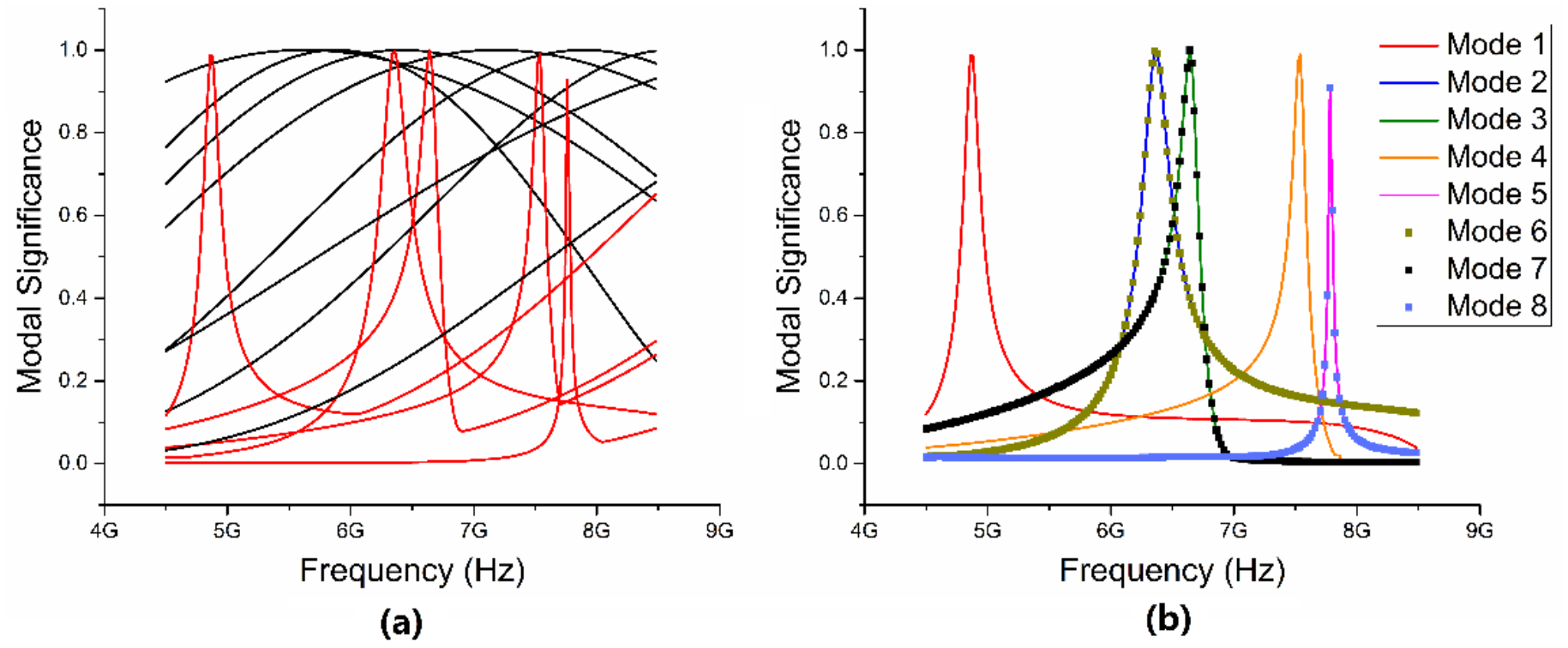

Figure 1-1 Modal Significance (MS) curves corresponding to the Harrington's CMs of a lossless dielectric cylinder, whose radius and hight are 5.25mm and 4.60mm respectively[45]. (a) results derived from the formulations established in literature [34], where the black lines correspond to spurious modes[44,45]; (b) results derived from the formulations proposed in literature [45], where the formulations employed the relationships between the equivalent surface electric and magnetic currents, and it is obvious that the spurious modes are successfully suppressed

**Theoretical Aspect III. On Establishing the CM Formulations for Metal-Material Composite Scattering Systems①**

Based on mixed potential integral equation (MPIE), Prof. Chen, Dr. L. Guo, and Prof. S. Yang[50] derived the formulation for constructing the CMs of the metallic structures in layered materials. By generalizing the concept of substructure CM, H. Alroughani et al.[51] derived some formulations for constructing the CMs of composite systems. Based on E-PMCHWT equation, R. Maximidis et al.[52] did some studies on constructing the CMs of composite systems. Doctor Guo, Prof. Chen, and Prof. Yang[53,54], et al.[55,56] generalized Harrington's EFIE-OpeMetSca-CMT[31,32] and SIE-MatSca-CMT[33,34] to composite systems, and it makes the applicable range of Harrington's CMT be enormously extended. In the Chapter 5 of this dissertation, we will establish some new CM calculation formulations for composite systems from a new point of view, and the new formulations are different from the traditional ones in IE framework, and the new formulations have advantages over the traditional ones in the aspects of extending applicable range, simplifying expressive form, and reducing computational burden.

**Theoretical Aspect IV. On Revealing and Clarifying the Imperfections of the Foundation of Harrington's CMT[31-34]**

Professor T. Sarkar[57] et al.[44] found out that there had been some imperfections in Harrington's CMT since the establishment of the theory. In the Subsections 3.4.1 and 4.2.2 of this dissertation, we will further analyze the causes of the imperfections, and then practically improve Harrington's CMT.

**Theoretical Aspect V. On the IE-Based Variants of Harrington's CMT[31-34], Which Was Established in IE Framework**

In recent years, some scholars provided some variants by imitating traditional Harrington's metallic CM calculation formulation[31,32], such as the magnetic field integral equation (MFIE) based[58], combined field integral equation (CFIE) based[59-61], and complex background Green's function based[62,63] CM calculation formulations for metallic systems.

**Theoretical Aspect VI. On the Modal Tracking Algorithms in CMT**

The calculations for CMs are done frequency by frequency. After obtaining the CMs at all frequencies, the one-to-one correspondences among the CMs at adjacent two frequencies are needed to be established, and the establishment process is usually called

① In the following parts of this dissertation, the terminology "metal-material composite scattering system" is simply called as "composite system", and simply denoted as "ComSca".

as modal tracking. The most commonly used modal tracking algorithm is based on the correlation between two characteristic vectors[64,65], and a systematical summary for this algorithm can be found in literature [35], and a pair of comparison results between pre-tracking and post-tracking is illustrated in Figure 1-2[35]. Besides the characteristic-vector-based tracking algorithm mentioned above, there also exist characteristic-current-based[66] and characteristic-far-field-based[67] tracking algorithms. In addition, some scholars[68,69] improved the characteristic-vector-based tracking algorithm by using approximate characteristic values. Recently, some scholars[70] also proposed the modal tracking algorithm based on enhancedly tracking characteristic values.

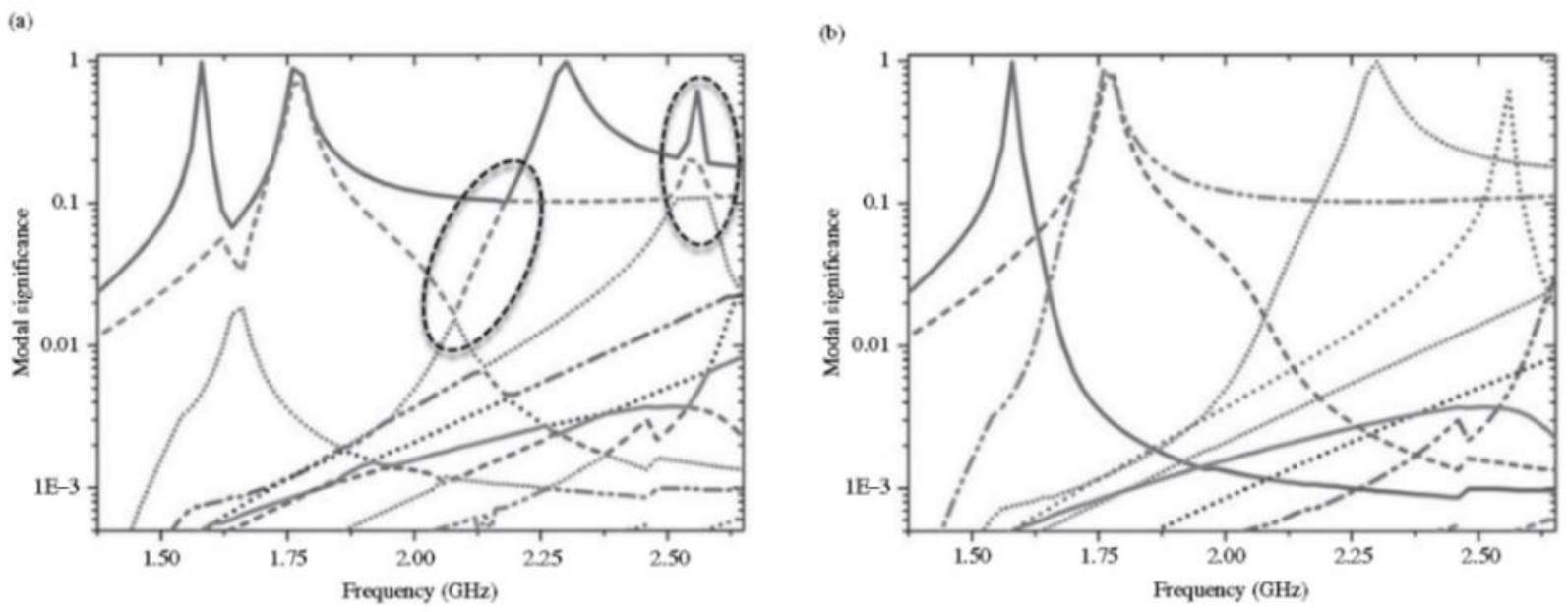

Figure 1-2 Modal Significance (MS) curves corresponding to the stacked printed microstrip antenna shown in the Figure 2.9 in classical literature [35], and this Figure 1-2 is just the Figure 2.10 given in classical literature [35]. Here, we have appropriately adjusted the arrangement of the subgraphs in the original Figure 2.10 in classical literature [35], to facilitate the setting of this dissertation. (a) results before tracking modes; (b) results after tracking modes

**Theoretical Aspect VII. On Exploring the New Carrying Framework for Harrington's CMT**

Harrington's CMs are a series of fundamental modes of an objective scattering system, and they are derived from orthogonalizing impedance matrix (IM) operator in IE framework. Recently, literatures [37,40,44,45,55,56] constructed the Harrington's CMs for metallic systems[37,40], material systems[44,45], and composite systems[55,56] by orthogonalizing a power operator, which corresponds to the power done by the incident EM fields on the scattered sources. In essence, the works of literatures [37,40,44,45,55,56] are to explore the new carrying framework for Harrington's CMT (for details see the Chapter 2 of this dissertation). In the Chapters 3, 4, 5, and 6 of this dissertation, we will further sublimate and generalize the works of literatures [37,40,44,45,55,56], and then prompt the works to develop into a complete theoretical formalism.

### 1.2.2 Application Aspects

In recent years, the application researches on Harrington's CMT mainly focus on the following aspects:

**Application Aspect I. Antenna Synthesis Based on Harrington's CMT**

In literature [71], Prof. Harrington and Dr. Mautz proposed a synthesis method for antenna radiation pattern by loading some appropriate CMs, and the method has already developed into a relatively complete set of design philosophy for mobile terminal antennas[72-74], and a typical design example is illustrated in Figure 1-3[74]. In literature [75], E. Newman proposed a synthesis method for the position of the antenna on platform, and the method has already developed into a relatively complete set of analysis and design philosophy for the platform-integration antennas[35,76-80], and a typical analysis example is illustrated in Figure 1-4[78]. In literatures [81,82], Prof. Garbacz, Prof. Pozar, and Dr. D. Liu proposed a synthesis method for antenna shape, and the method has already developed into a relatively complete set of design philosophy for synthesizing antenna shape[83-85], and a typical design example is illustrated in Figure 1-5[84].

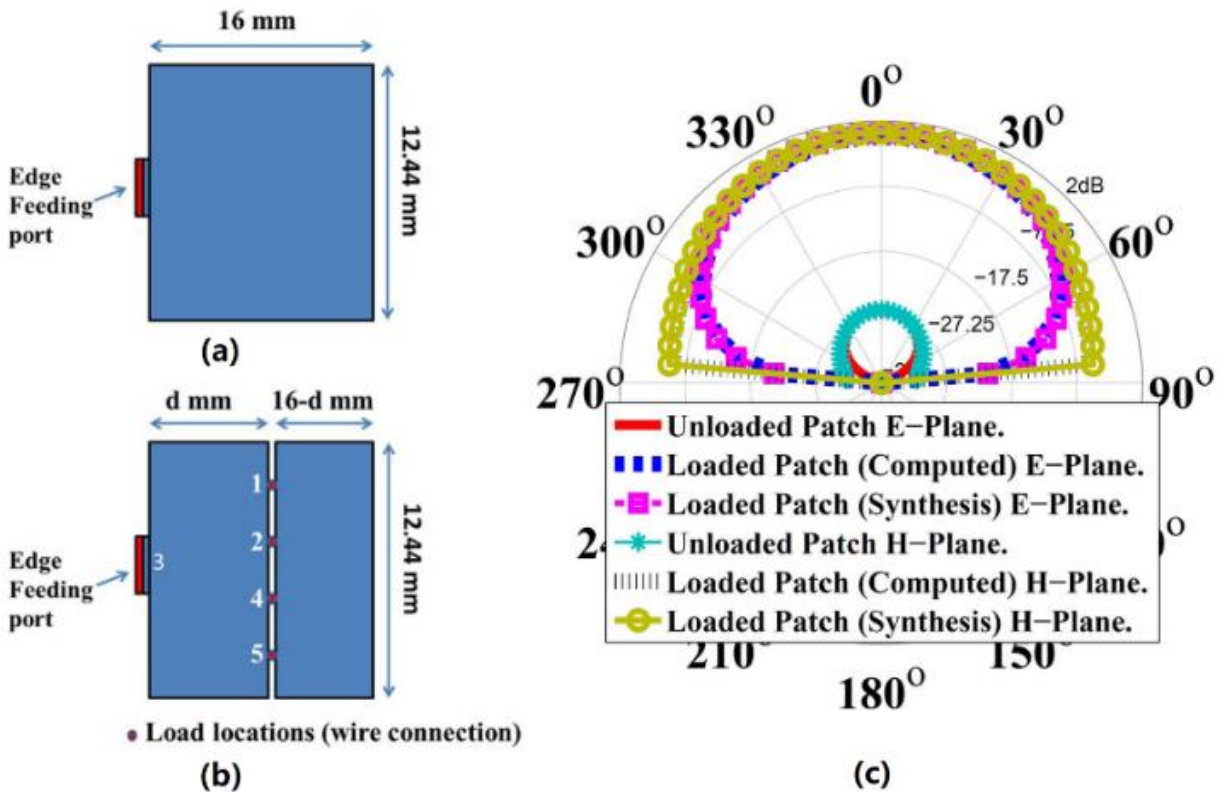

Figure 1-3 Reactance-load-based antenna design examples provided in literature [74]. (a) printed microstrip antenna before loading reactance; (b) printed microstrip antenna after loading reactance; (c) antenna gain curves corresponding to the antennas shown in Figures 1-3(a) and 1-3(b)

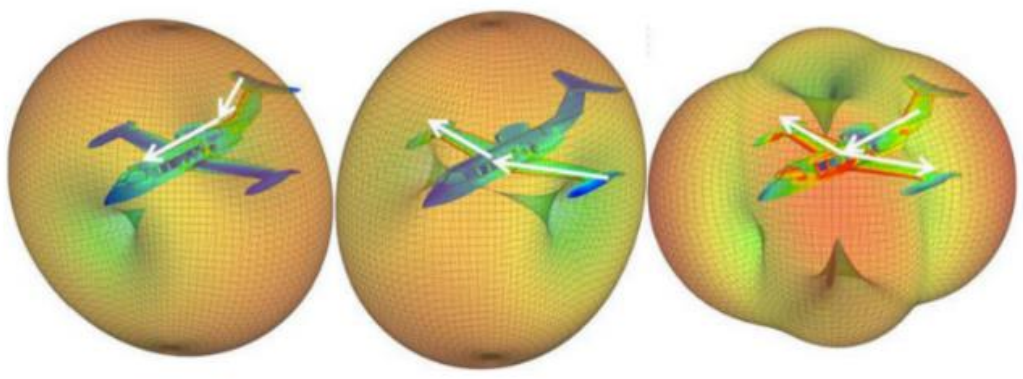

Figure 1-4 Modal electric currents and modal radiation patterns corresponding to three typical CMs of the airplane considered in literature [78]

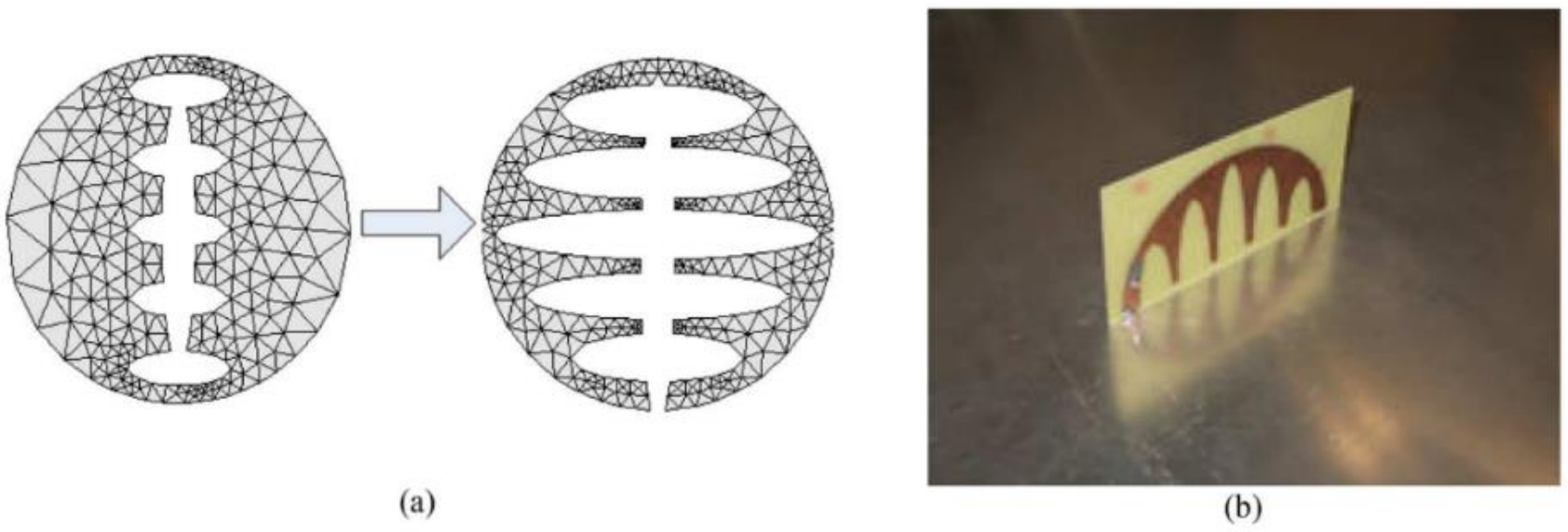

Figure 1-5 CM-based antenna shape synthesis example provided in literature [84]. (a) process of antenna shape synthesis; (b) antenna prototype designed by employing the process shown in Figure 1-5(a)

**Application Aspect II. Antenna Design Based on Harrington's CMT**

CM-based antenna design philosophy was applied to design multi-input multi-output systems (MIMO systems) in literatures [86-91], and was applied to design plasmonic nanoantennas in literatures [92,93], and was applied to design metasurface antennas in literatures [94-99], and was applied to analyze the EM scattering characters of impedance surfaces in literature [100], and was applied to design carbon nanotubes in literature [101].

## 1.3 Important Problems and Challenges

Throughout the development history of CMT, it is easy to find out that: CMT has achieved rapid developments in both theoretical studies and engineering applications since that Prof. Harrington et al.[31-34] established their CMT in IE framework in the 1970s.

### 1.3.1 Old Problems

However, there still exist some important problems which have not been completely solved (in what follows, these problems are simply called as old problems, and simply denoted as OPs), for example:

**Old Problem 1 (OP1).** After the establishment of Harrington's CMT, the theory is mainly treated as a numerical calculation method for constructing the fundamental modes of open EM systems, so the studies on Harrington's CMT mainly focused on the relatively mathematical aspect of generalizing formulations and on the relatively engineering aspect of designing antennas, but the study on the physical picture (i.e. the fundamental formalism in the aspect of physics) of Harrington's CMT is scarce. However, a clear physical picture is of great importance to a numerical calculation method①. In recent years,

① As Prof. Enrico Fermi (Nobel Prize in Physics, 1938) said: "There are two ways to do calculations. The first way,

only a few scholars[35,37,40,41,44,45] did some studies for drawing the physical picture of Harrington's CMT, but the obtained physical picture remains unclear. Just due to the lack of clear physical picture, some academic debates on Harrington's CMs have already emerged, for example:

**Debate A.** Around the question "Whether or not the Harrington's CMs of material systems[33,34] should have orthogonal modal far fields just like the Harrington's CMs of metallic systems[31,32]?", some debates have appeared recently (for details see literature [57]), and there has not been final conclusion now.

**Debate B.** Around the question "Whether or not the Harrington's characteristic values of material systems[33,34] should have the same physical meaning as the Harrington's characteristic values of metallic systems[31,32]?", some debates have appeared recently (for details see literatures [42-44]), and there has not been final conclusion now.

**Debate C.** Recently, some IE-based variants of the traditional Harrington's CMT[31,32] have appeared, such as the MFIE-based[58], CFIE-based[59-61], and complex background Green's function based[62] CMTs for metallic systems. Some studies have presented that the CMs derived from these variants are not identical to the ones derived from the traditional Harrington's CMT, as illustrated in Figure 1-6[60] (the CFIE-based CMs are not identical to the traditional Harrington's EFIE-based CMs) and Figure 1-7[63] (the complex background Green's function based CMs are not identical to the traditional Harrington's free space Green's function based CMs). However, as the embodiment of the inherent physical properties of the objective scattering system, the CMs should only depend on the objective scattering system itself[31,32,35], and should not depend on the numerical calculation method.

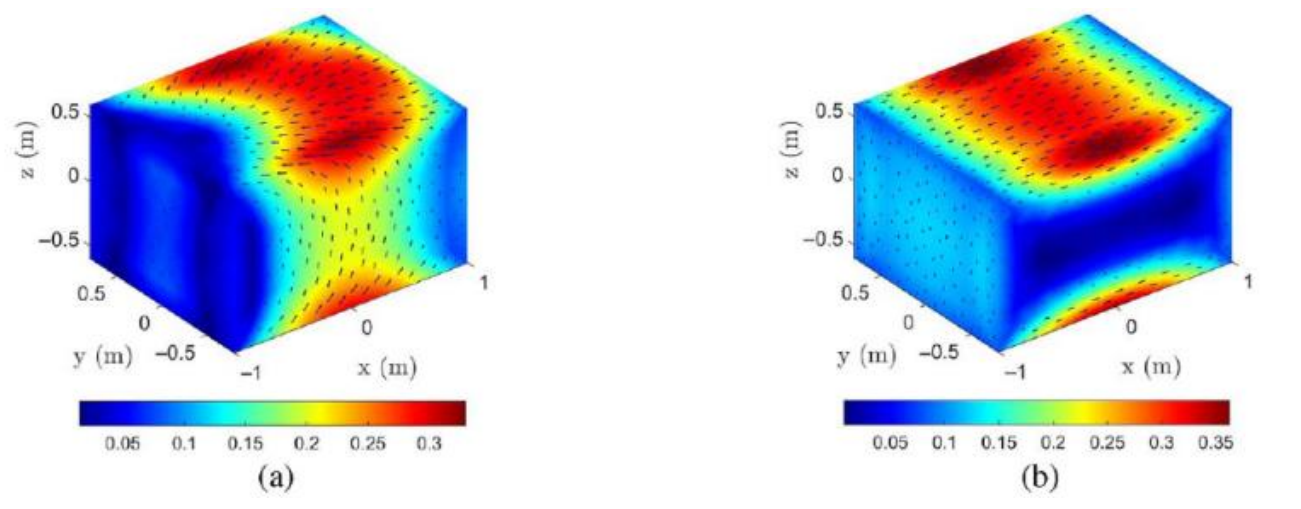

Figure 1-6 Modal electric currents corresponding to the CM3 of a metallic cube working at 119.88MHz given in literature [60]. (a) results derived from EFIE-based formulation; (b) results derived from CFIE-based formulation

which I prefer, is to have a clear physical picture. The second way is to have a rigorous mathematical formalism.[103]".[102]

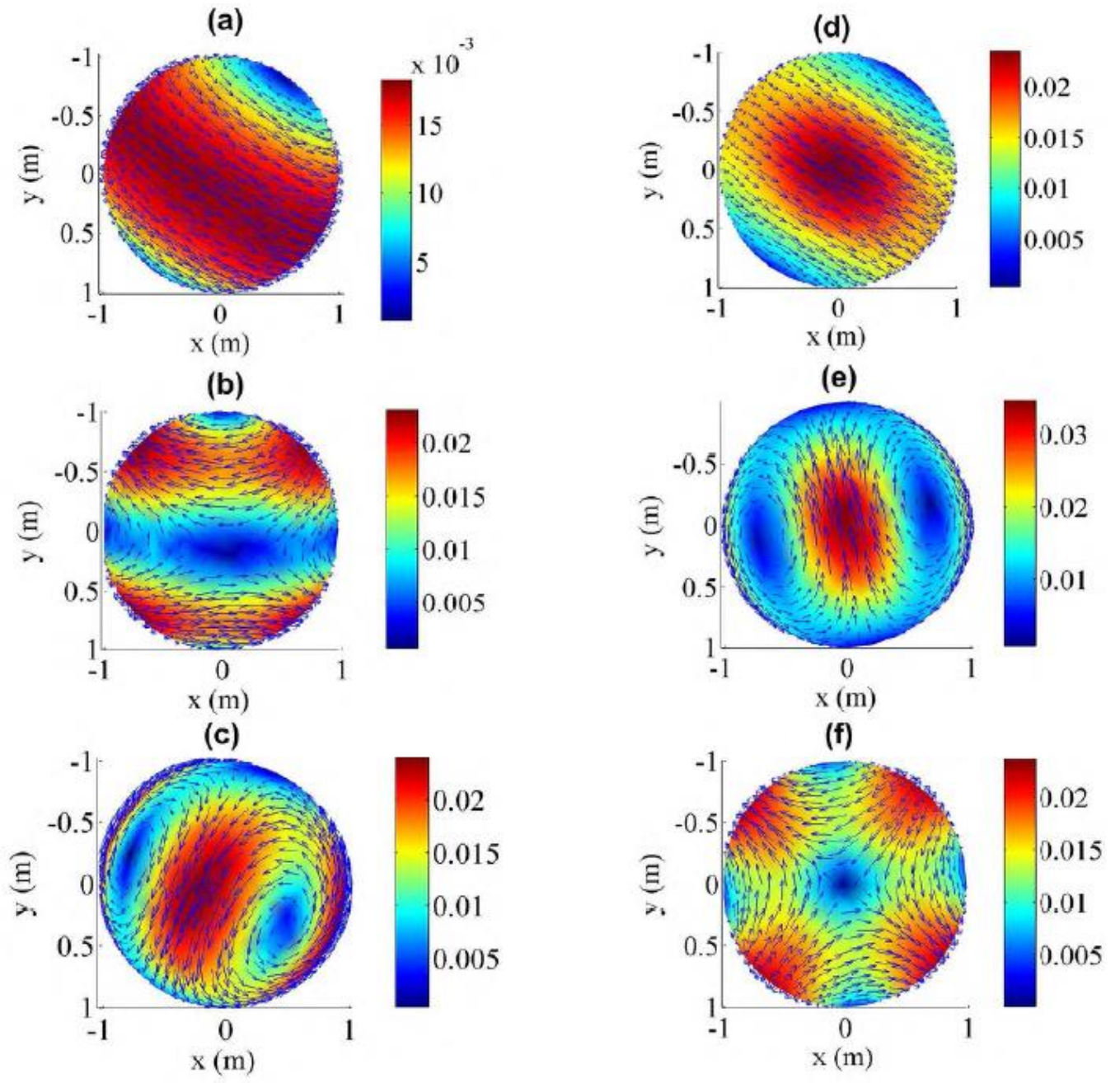

Figure 1-7 Modal electric currents corresponding to some typical CMs of a metallic sphere considered in literature [63] based on free-space Green's function (FS-GF) and layer-medium Green's function (LM-GF) respectively. (a) modal electric current of the CM1 based on FS-GF; (b) modal electric current of the CM1 based on LM-GF; (c) modal electric current of the CM4 based on FS-GF; (d) modal electric current of the CM4 based on LM-GF; (e) modal electric current of the CM6 based on FS-GF; (f) modal electric current of the CM6 based on LM-GF

**Old Problem 2 (OP2).** The Harrington's CMT for metallic systems (EFIE-OpeMetSca-CMT[31,32]) has not been able to construct the nonradiative CMs of metallic systems, so it has not been able to provide complete CM sets to metallic systems (for details see literatures [37,40,41]).

**Old Problem 3 (OP3).** For some material systems which have complicated topological structures (such as stacked dielectric resonator antennas[20,21]) or complicated material parameters (such as Lüneburg lens antenna[9,10]), SIE-MatSca-CMT[34] has not been able to provide effective CM calculation formulation.

**Old Problem 4 (OP4).** For some composite systems which have complicated topological structures (such as printed microstrip antennas[17-19]) or complicated material parameters, existing Harrington's CMT[53,54] has not been able to provide effective CM calculation formulation.

**Old Problem 5 (OP5).** The existing Harrington's CMT for complicated scattering systems often outputs some spurious modes, and it has not been thoroughly solved how to effectively suppress the spurious modes.

Because of the existence of the old problems OP1~OP5 mentioned above, it has great study value for the fundamentals of Harrington's CMT, and this is just the second main reason why this dissertation selects to research CMT, i.e. the second main foundation to select the researching topic of this dissertation (the first foundation is given in the last paragraph of Section 1.1).

### 1.3.2 New Problems

This dissertation is committed to further solving the old problems OP1~OP5 mentioned above. During solving OP1, OP3, and OP4, and after solving OP2 and OP5, this dissertation further extends a series of important problems which have not been widely concerned (in what follows, these problems are simply called as new problems, and simply denoted as NPs), as follows:

**New Problem 1 (NP1).** During solving OP1, this dissertation, for the first time, puts forward the new problem "Whether or not IE is the best framework for carrying CMT?".

**New Problem 2 (NP2).** After solving OP2, this dissertation, for the first time, puts forward the new problem "How to finely classify all of the working modes of an objective metallic system, and how to orthogonally decompose the modal space of the objective metallic system?".

**New Problem 3 (NP3).** During solving OP3, this dissertation puts forward the new problem "How to generalize the traditional surface equivalence principle for simple material bodies to the one for complicated material systems, and how to derive the corresponding mathematical expressions?".

**New Problem 4 (NP4).** During solving OP4, this dissertation, for the first time, puts forward the new problem "How to establish the line-surface equivalence principle for complicated metal-material composite systems, and how to derive the corresponding mathematical expressions?".

**New Problem 5 (NP5).** After solving OP5, this dissertation, for the first time, puts forward the new problem "What is the physical meaning of the singular electromagnetic current term in the CM calculation formulations for complicated scattering systems?".

The above-mentioned new problems NP1~NP5 imply that there exists a large area of uncultivated land, so the related domain has a large research space, and this is just the third main reason why this dissertation selects to research CMT, i.e. the third main foundation to select the researching topic of this dissertation.

## 1.4 Contributions and Innovations of This Dissertation

In fact, the changement from Garbacz's CMT[29,30] to Harrington's CMT[31-34] is a transformation for the framework carrying CMT —— from scattering matrix (SM) framework to integral equation (IE) framework. Accompanying the framework transformation is a transformation for the method constructing CMs —— from orthogonalizing perturbation matrix operator (PMO) method to orthogonalizing impedance matrix operator (IMO) method. The above-mentioned transformations have played positive roles in accelerating the development of CMT, because IE framework provides a set of efficient calculation methods to CMs.

However, the CMT under IE framework still has a series of important problems which have not been completely solved or widely concerned, such as {OP1, NP1}, {OP2, NP2}, …, and {OP5, NP5}. To promote further solving the above-mentioned 5 pairs of important problems, this dissertation does a series of innovative studies (in what follows, these innovative studies are simply called as main contributions, and simply denoted as MCs) as follows:

**Main Contribution 1 (MC1). Contribution in the Aspect of Solving {OP1, NP1}**

Based on the works of literatures [37,40,41,44,45,56], this dissertation does the second transformation for the carrying framework of CMT —— from integral equation (IE) framework to work-energy principle (WEP)① framework, and then triggers the second transformation for the constructing method of CMs —— from orthogonalizing impedance matrix operator (IMO) method to orthogonalizing driving power operator (DPO) method②, and the above transformations are as illustrated in Figure 1-8. Based on the second transformation, this dissertation, for the first time, clearly reveals the physical picture of Harrington's CMT[31-34], and we, taking the metallic system case as an example, provide a brief presentation as follows:

① Work-energy principle (WEP)[104,105] is an important manifestation of the conservation law of energy. This dissertation, in classical electromagnetics framework, derives the counterpart of the WEP in classical mechanics. The WEP in classical mechanics is usually stated as that: "… the work done by all forces acting on a particle (the work of the resultant force) equals the change in the kinetic energy of the particle. [104]". Deriving the mathematical expression of the WEP in classical electromagnetics from Maxwell's equations is one of the main contributions of this dissertation, and the detailed processs can be found in the Sections 3.2, 4.2, and 5.5 of this dissertation.

② The WEP in classical mechanics points out that: the work done by the resultant force acting on a particle is the energy source to sustain the variation of the kinetic energy of the particle. After some studies, this dissertation finds out that: the work done by the resultant field acting on the scattered source distributing on a scattering system is just the energy source to sustain outputting EM energy from the scattering system. Based on this, this dissertation calls the power done by the resultant field acting on scattered source as driving power (DP). In fact, frequency-domain DPO based orthogonalization has a profound physical meaning: in any integral period, there is not net energy exchange between any two different CMs.

When resultant fields $\{\vec{E}^{\text{inc}},\vec{H}^{\text{inc}}\}$ are incident on a metallic system $D_{\text{met sys}}$, an scattered electric current $\vec{J}^{\text{sca}}$ will be induced on the boundary $\partial D_{\text{met sys}}$ of $D_{\text{met sys}}$. The WEP corresponding to this scattering problem can be mathematically expressed as that $(1/2) < \vec{J}^{\text{sca}},\vec{E}^{\text{inc}} >_{\partial D_{\text{met sys}}} = P^{\text{rad}} + j2\omega(W_{\text{vac}}^{\text{mag}} - W_{\text{vac}}^{\text{ele}})$. Here, $(1/2) < \vec{J}^{\text{sca}},\vec{E}^{\text{inc}} >_{\partial D_{\text{met sys}}}$ is the frequency-domain version of the power done by $\vec{E}^{\text{inc}}$ on $\vec{J}^{\text{sca}}$ ①, and it is called as driving power (DP) in this dissertation. Literatures [37,40,41] pointed out that: the essential objective of EFIE-OpeMetSca-CMT[31,32] is to construct a series of CMs having ability to orthogonalize DPO $(1/2) < \vec{J}^{\text{sca}},\vec{E}^{\text{inc}} >_{\partial D_{\text{met sys}}}$, and the obtained CMs satisfy orthogonality $(1/2) < \vec{J}_{\xi}^{\text{sca}},\vec{E}_{\zeta}^{\text{inc}} >_{\partial D_{\text{met sys}}} = 0$ where $\xi \neq \zeta$. If $T$ is the time period of $\vec{E}_{\zeta}^{\text{inc}}$ and $\vec{J}_{\xi}^{\text{sca}}$, then $\text{Re}\{(1/2) < \vec{J}_{\xi}^{\text{sca}},\vec{E}_{\zeta}^{\text{inc}} >_{\partial D_{\text{met sys}}}\} = (1/T)\int_0^T < \vec{J}_{\xi}^{\text{sca}}(t),\vec{E}_{\zeta}^{\text{inc}}(t) >_{\partial D_{\text{met sys}}} dt$ [9], where $\text{Re}\{z\}$ represents the real part of complex number $z$. Thus, the CMs derived from EFIE-OpeMetSca-CMT[31,32] satisfy that: in any integral period, the net work done by the modal incident electric field $\vec{E}_{\zeta}^{\text{inc}}$ corresponding to the $\zeta$-th CM on the modal scattered electric current $\vec{J}_{\xi}^{\text{sca}}$ corresponding to the $\xi$-th CM is zero, if $\xi \neq \zeta$. Obviously, new WEP framework and new orthogonaling DPO method draw a clear physical picture —— constructing a series of steadily working modes which don't have net energy exchange in any integral period —— for the EFIE-OpeMetSca-CMT[31,32] in IE framework.

The above-mentioned physical picture reveals the following facts for the first time: **A.** The essential objective of Harrington's CMT is to construct the modes which don't have net energy exchange in any integral period, rather than the modes which have orthogonal modal far fields, and then it becomes clear why the VIE-MatSca-CMT[33] for lossy material systems cannot guarantee orthogonal modal far fields; **B.** The consistency between the Harrington's CMTs for metallic and material systems is embodied in the aspect of physical picture (i.e., all of the Harrington's CMTs for various scattering systems are committed to constructing a series of steadily working modes not having net energy exchange in any integral period) rather than in the aspect of the physical meaning of characteristic values; **C.** The original Harrington's CMT (such as the CMT for metallic systems based on EFIE and vacuum Green's function[31,32]) has some IE-based variants (such as the MFIE-based[58], CFIE-based[59-61], and complex background Green's function based[62] CMTs for metallic systems, etc.), and the essential objectives of these IE-based

① Here, power $P^{\text{rad}}$ is the radiated power carried by scattered field, and energies $W_{\text{vac}}^{\text{mag}}$ and $W_{\text{vac}}^{\text{ele}}$ are respectively the magnetic energy and electric energy corresponding to scattered field. For details see the Section 3.2 of this dissertation.

variants are not to orthogonalize DPO, and then it cannot be guaranteed that the CMs derived from these IE-based variants don't have net energy exchange in any integral period, and then it becomes clear why the CMs derived from these IE-based variants are not completely identical to the ones derived from the original Harrington's CMTs.

**Main Contribution 2 (MC2). Contribution in the Aspect of Solving {OP2, NP2}**

In new WEP framework and based on new orthogonalizing DPO method, this dissertation proves that the CM set derived from EFIE-OpeMetSca-CMT[31,32] is not complete, and provides a practical scheme to complete the CM set, and then, for the first time, integrates EFIE-OpeMetSca-CMT[31,32] and the homogeneous wave equation based EMT for closed metallic scattering systems (HWE-CloMetSca-EMT)[26-28] into a whole modal theory —— WEP-based CMT for arbitrary metallic scattering systems (WEP-MetSca-CMT).

In addition, based on the complete CM set derived from WEP-MetSca-CMT, this dissertation, for the first time, realizes the detailed classification for all working modes, the orthogonal decomposition for whole modal space, and the simplest orthogonal decompositions for the modes in various modal subclasses. The above-mentioned modal classification and modal decomposition have important instruction significance for understanding the working mechanisms of various modes, so have great application significance for extracting the inherent EM scattering characters of the objective EM structure.

**Main Contribution 3 (MC3). Contribution in the Aspect of Solving {OP3, NP3}**

From the aspects of external EM environment, system topological structure, and system material parameter, this dissertation generalizes the traditional surface equivalence principle (SEP) for the simple material bodies placed in vacuum to the generalized surface equivalence principle (GSEP) for the complicated material systems placed in complex environment, and also derives the mathematical expressions related to the GSEP.

In new WEP framework, based on new orthogonalizing DPO method, and employing newly obtained GSEP, this dissertation generalizes traditional VIE-MatSca-CMT[33] and SIE-MatSca-CMT[34] from the aspects of external EM environment, system topological structure, and system material parameter. The generalized version is independent of the external EM environment surrounding the objective system, and only depends on the inherent physical properties of the objective system itself, and this feature of the generalized version is significant for extracting the inherent EM scattering characters of the objective system; the generalized version is not only applicable to simply connected material systems, but also suitable for multiply connected material systems, even valid for non-connected material systems; the generalized version is not only applicable to homogeneous isotropic material systems, but also suitable for inhomogeneous anisotropic material systems, even valid for piecewise inhomogeneous anisotropic material systems.

**Main Contribution 4 (MC4). Contribution in the Aspect of Solving {OP4, NP4}**

This dissertation, for the first time, establishes the line-surface equivalence principle (LSEP) for the complicated metal-material composite systems placed in complex EM environment, and also derives the the mathematical expressions related to the LSEP.

In new WEP framework, based on new orthogonalizing DPO method, and employing newly obtained LSEP, this dissertation establishes the CMT for metal-material composite systems, and the CMT has a wide applicability in the aspects of external EM environment, system topological structure, and system material parameter. In the aspect of EM environment, the CMT is independent of the EM environment surrounding the objective composite system; in the aspect of topological structure, the metallic part of the objective composite system can be metallic line, metallic surface, metallic body, or line-surface-body composite structure, and the metallic and material parts of the objective composite system can be in contact or separate, and the metallic part of the objective composite system can be completely of partially submerged into the material part of the objective composite system; in the aspect of material parameter, the material part of the objective composite system can be either homogeneous isotropic or inhomogeneous anisotropic.

**Main Contribution 5 (MC5). Contribution in the Aspect of Solving {OP5, NP5}**

In new WEP framework, based on new orthogonalizing DPO method, and employing newly obtained GSEP and LSEP, this dissertation does an in-depth study on the cause of the spurious modes derived from CMT, and, for the first time, obtains the systematical scheme for suppressing the spurious modes of complicated scattering systems.

In addition, this dissertation, for the first time, reveals the physical meaning of the singular EM current terms (SCTs) contained in the DPOs of material systems and composite systems.

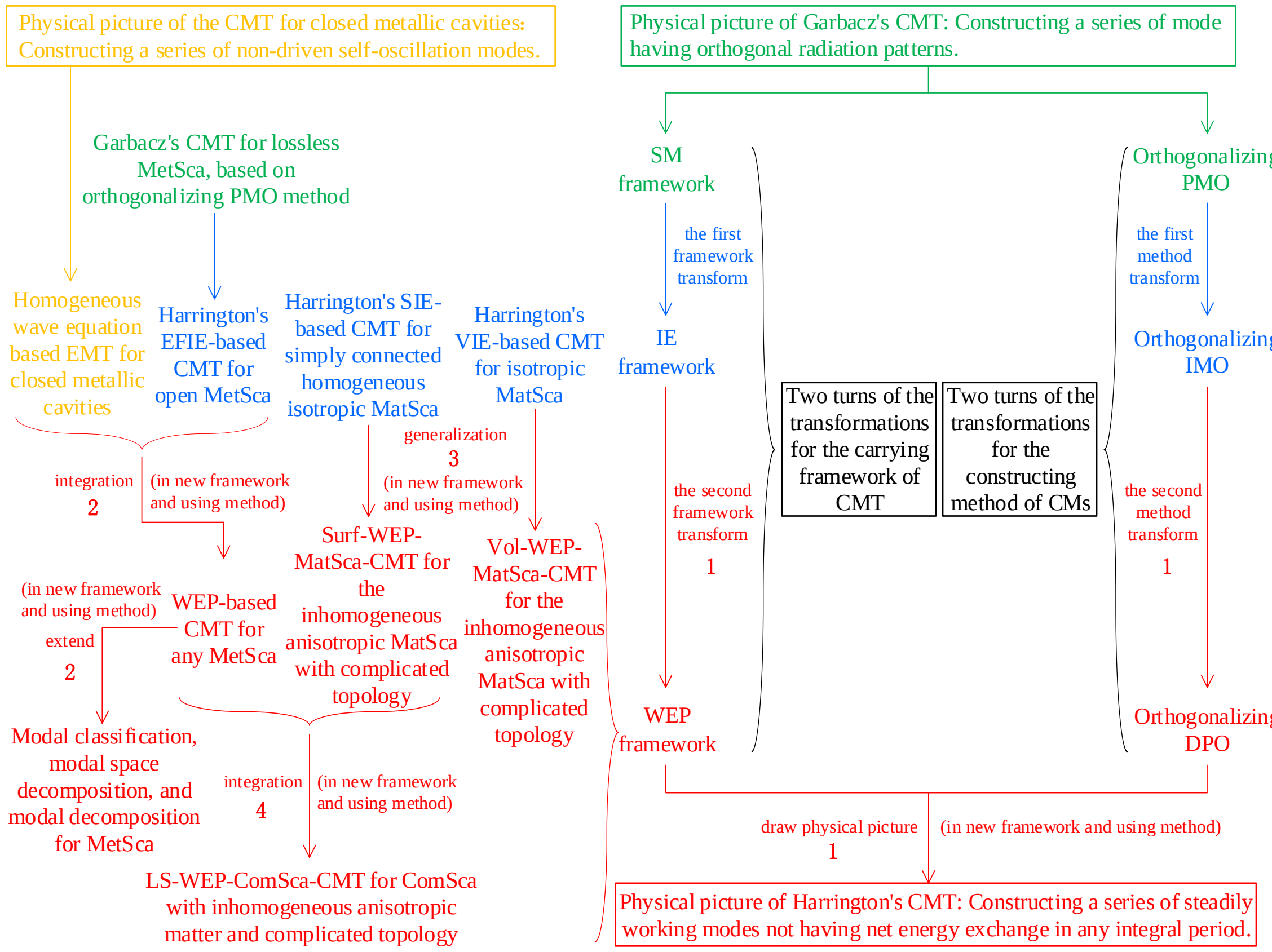

Figure 1-8 Logical relationships among the main contributions of this dissertation

The logical relationships among the above-mentioned main contributions MC1~CM4 are illustrated in Figure 1-8. The main contribution MC5 is simultaneously related to the CMTs for various scattering systems. The numbers marked in Figure 1-8 correspond to the main contributions MC1~MC4 mentioned above. The red parts in Figure 1-8 are the innovative works of this dissertation, and the green and blue parts in Figure 1-8 are respectively Prof. Garbacz's works (1965) and the works done by Prof. Harrington et al. (1970s).

In addition, besides new framework for carrying CMT and new method for constructing CMs, this dissertation also derives a series of new formulations for calculating CMs, and the new formulations have not been able to be effectively established in IE framework. It is verified in this dissertation that: for the simple scattering systems to which both the new formulations and the traditional formulations are applicable, the CMs derived from the new formulations and the traditional formulations are the same. However, it should be emphasized that: there exist some complicated scattering systems to which the traditional formulations are not applicable but the new formulations are applicable.

## 1.5 Outline of This Dissertation

In above Sections 1.1 and 1.3, we have pointed out three main foundations to select the researching topic of this dissertation, and we simply review them as below.

**The First Main Reason Why This Dissertation Selects to Research CMT:** CMT has important instruction significance for understanding the working mechanisms of various modes, and then has great application significance for extracting the inherent EM scattering characters of the objective structure, so this dissertation is committed to doing some relatively fundamental theoretical studies for CMT.

**The Second Main Reason Why This Dissertation Selects to Research CMT:** In Harrington's CMT, there exist a series of important problems OP1~OP5 which have not been completely solved, and this dissertation is committed to promoting the further solving for OP1~OP5.

**The Third Main Reason Why This Dissertation Selects to Research CMT:** During and after solving old problems OP1~OP5, this dissertation further extends a series of important new problems NP1~NP5 which are closely related to CMT and have not been widely concerned, and this dissertation is committed to studying these new problems, and then to further improving and expanding the whole theoretical formalism of CMT.

Taking the above-mentioned 3 main reasons for researching CMT and 5 pairs of important problems {OP1, NP1}, {OP2, NP2}, …, and {OP5, NP5} existing in CMT as an outline, we organize the chapters and sections of this dissertation as follows:

**Chapter 1** demonstrates, from the aspects of EM theory and EM engineering, the important status and study value of CMT (Section 1.1), and summarizes the research history and state of CMT (Section 1.2), and points out the unsolved important problems and challenges in CMT (Section 1.3). To further solve the unsolved important problems, this dissertation does a series of innovative studies, and the logical relationships among the studies are summarized in Section 1.4. In Section 1.5, we simply retrospects the 3 main reasons for researching CMT and the 5 pairs of important problems existing in CMT, and we, based on them, draw the research route map of this dissertation as illustrated in Figure 1-9. In the research route map (Figure 1-9), we have marked the corresponding relationships among various plates, various chapters (Chapter 1 ~ Chapter 7), various old problems (OP1~OP5) which have not been completely solved, various new problems (NP1~NP5) which have not been widely concerned, and the main contributions (MC1~MC5) made by this dissertation.

**Chapter 2** is mainly committed to solving the first pair of important problems {OP1, NP1}. Firstly (in Section 2.2), Garbacz's CM construction method in SM framework —— orthogonalizing PMO method —— is simply retrospected, and it is emphasized that the physical picture of Garbacz's CMT is to construct a series of fundamental modes whose modal far fields are orthogonal to each others; secondly (in Section 2.3), Harrington's CM construction method in IE framework —— orthogonalizing IMO method —— is simply retrospected, and it is pointed out that the physical pictures of Harrington's CMT and Garbacz's CMT are not the same; thirdly (in Section 2.4), based on the works of literatures [37,40,41,44,45], the carrying framework of Harrington's CMT is transformed —— from IE framework to WEP framework, and at the same time the constructing method for Harrington's CMs is also transformed —— from orthogonalizing IMO method to orthogonalizing DPO method; finally (in Section 2.4), based on the new framework and the new method, the physical picture of Harrington's CMT —— constructing a series of steadily working modes which don't have net energy exchange in any integral period —— is clearly revealed.

**Chapter 3** is mainly committed to solving the second pair of important problems {OP2, NP2}. Firstly (in Section 3.2), the mathematical expression of the WEP corresponding to metallic systems is derived, and then the operator expression of the DP of metallic systems is obtained, and traditional EFIE-OpeMetSca-CMT[31,32] is rebuilt by orthogonalizing the DPO; afterwards (in Section 3.2), based on normalized modal DPs, the homogeneous wave equation based EMT for closed metallic scattering systems (HWE-CloMetSca-EMT)[26-28] is equivalently rebuilt in the formalism of CMT; then (in Section 3.2), EFIE-OpeMetSca-CMT and HWE-CloMetSca-EMT are integrated into a whole modal theory —— WEP-based CMT for arbitrary metallic scattering systems (WEP-MetSca-CMT), and at the same time a complete CM set for any objective metallic system is obtained, and it is also proven that any two different CMs don't have net energy exchange in any integral period; after that (in Section 3.3), by employing CM-based modal expansion, all of working modes are finely classified, and whole modal space is orthogonally decomposed, and then the simplest orthogonal decompositions for the modes in various modal classes are obtained; finally (in Section 3.4), the imperfections (which have existed in EFIE-OpeMetSca-CMT[31,32] since EFIE-OpeMetSca-CMT was established in 1971) are carefully analyzed and improved, and the comparisons and discussions for some kinds of typical fundamental modes of metallic systems are done.

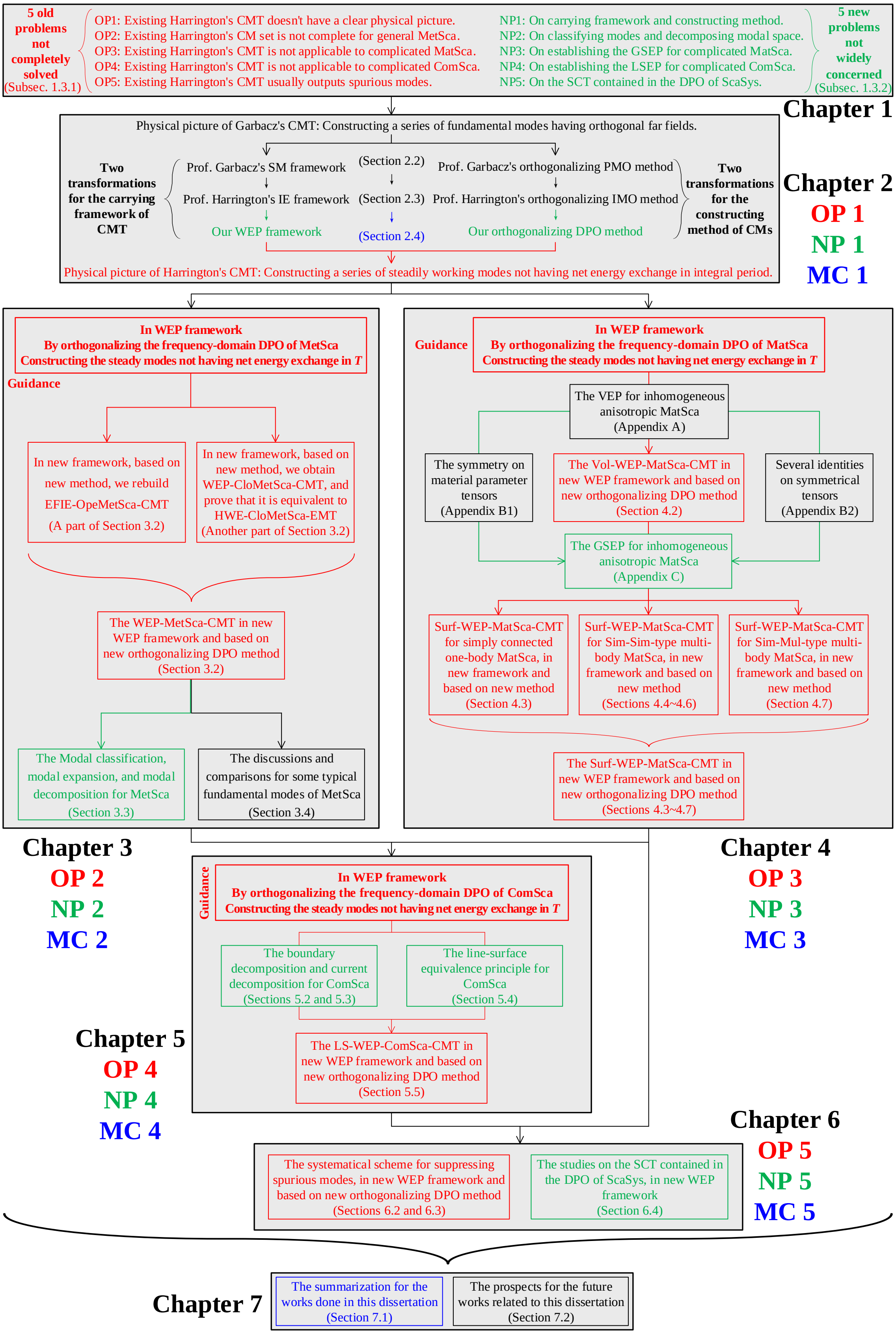

Figure 1-9 The research route map of this dissertation and the logical relationships among the chapters and sections of this dissertation (the full names of all of the abbreviations here can be found in Sections 1.1~1.5)

**Chapter 4** is mainly committed to solving the third pair of important problems {OP3, NP3}. Firstly (in Section 4.2), the mathematical expression of the WEP corresponding to material systems is derived, and then the operator expression of the DP of material systems is obtained by employing the results given in Appendixes A~C; afterwards (in Sections 4.3~4.7), the volume formulation of the WEP-based CMT for material scattering systems (Vol-WEP-MatSca-CMT) and the surface formulation of the WEP-based CMT for material scattering systems (Surf-WEP-MatSca-CMT) are established in WEP framework by orthogonalizing DPO, and then a series of steadily working modes which don't have net energy exchange in any integral period are derived from Vol-WEP-MatSca-CMT and Surf-WEP-MatSca-CMT.

**Chapter 5** is mainly committed to solving the fourth pair of important problems {OP4, NP4}. Firstly (in Sections 5.2 and 5.3), the decompositions for the metallic boundaries and material boundaries in composite systems are done, and then the EM currents distributing on the boundaries are decomposed correspondingly; secondly (in Section 5.4), the line-surface equivalence principle (LSEP) for composite systems is established and the corresponding mathematical expressions are also derived, by employing the boundary decompositions, the current decompositions, and the generalized surface equivalence principle (GSEP) given in Appendix C; finally (in Section 5.5), in WEP framework, the operator expression of the DP for composite systems is derived by utilizing LSEP, and then the line-surface formulation of the WEP-based CMT for metal-material composite scattering systems (LS-WEP-ComSca-CMT) is established, and then a series of steadily working modes which don't have net energy exchange in any integral period are derived from LS-WEP-ComSca-CMT.

**Chapter 6** is mainly committed to solving the fifth pair of important problems {OP5, NP5}. Firstly (in Section 6.2), the linear correlations among the variables contained in the DPO of scattering systems are researched, and it is studied how to select a series of complete and independent variables from all of the variables contained in DPO, and the linear transformation from the complete and independent variables to the dependent variables is established, and then a systematical scheme for suppressing spurious modes is obtained for arbitrary scattering systems; afterwards (in Section 6.4), the physical meaning of the singular EM current terms (SCTs) contained in the DPOs of material systems and composite systems is revealed, and then the SCT's influence on the numerical performance of the related CM calculation formulations is investigated.

**Chapter 7** systematically summarizes the works of this dissertation (in Section 7.1). The prospect of some future works related to this dissertation is simply discussed too (in Section 7.2).

The above-mentioned {WEP-MetSca-CMT, Vol-WEP-MatSca-CMT, Surf-WEP-MatSca-CMT, LS-WEP-ComSca-CMT} together constitute the basic skeleton of the WEP-based CMT for scattering systems (WEP-ScaSys-CMT). To emphasize the most complete physical picture of WEP-ScaSys-CMT and the most fundamental physical feature of the CMs derived from WEP-ScaSys-CMT —— orthogonalizing the DPO in frequency domain (this is just the mathematical denotation of the physical connotation "there is not net energy exchange in any integral period"), this dissertation specially calls the CMs derived from WEP-ScaSys-CMT as DPO-based CMs (DP-CMs).

In addition, the $e^{j\omega t}$ convention is used in this dissertation. For the time-domain physical quantities, this dissertation explicitly exhibits their time variable $t$, for example: time-domain quantity $Q(t)$. The quantities not explicitly containing time variable $t$ are in frequency domain, for example: frequency-domain quantity $Q$. If $\{Q(t), Q\}$ correspond to a linear physical quantity, then $Q(t) = \text{Re}\{Qe^{j\omega t}\}$. To assist readers to understand the symbols with superscripts or subscripts, this dissertation provides the meanings of the superscripts or subscripts when the symbols first appear, for example: the driving power $P_{\text{met sys}}^{\text{Driving}}$ of metallic systems. For all of the abbreviations used in this dissertation, their full names are provided when they first appear, for example: characteristic mode (CM). We also list all of the abbreviations in the Main Symbol Table after the Abstract of this dissertation.

# Chapter 2 Carrying Framework and Physical Picture of the CMT for Scattering Systems

横看成岭侧成峰，远近高低各不同；
不识庐山真面目，只缘身在此山中。[150]
—— 苏轼（北宋文学家）

In this chapter, we simply retrospect the Garbacz's CMT in SM framework, and focus on drawing a clear physical picture for the Harrington's CMT in IE framework, and then propose the second transformation for the carrying framework of CMT and the second transformation for the constructing method of CMs.

## 2.1 Chapter Introduction

In 1965, Prof. Garbacz[29], for the first time, established his CMT for metallic systems by orthogonalizing the PMO in SM framework (SM-MetSca-CMT), and the CMs derived from SM-MetSca-CMT have orthogonal modal far fields. However, it is very difficult to obtain the PMO for general metallic systems except some topologically simple metallic systems, so SM-MetSca-CMT has not been widely used in EM engineering domain.

In 1971, Prof. Harrington and Dr. Mautz[31,32], for the first time, established their CMT for open metallic systems by orthogonalizing EFIE-based IMO in IE framework (EFIE-OpeMetSca-CMT), and the CMs derived from EFIE-OpeMetSca-CMT have orthogonal modal far fields. In 1972 and 1977, Prof. Harrington et al. established their CMT for material systems by orthogonalizing VIE-based IMO (VIE-MatSca-CMT)[33] and SIE-based IMO (SIE-MatSca-CMT)[34] in IE framework, and the CMs derived from VIE-MatSca-CMT and SIE-MatSca-CMT satisfy some certain orthogonalities.

For the Garbacz's theory, the carrying framework of CMT, the constructing method of CMs, and the physical picture of CMT are very explicit, as listed in the second column of Table 2-1. For the Harrington's theory, the carrying framework of CMT and the constructing method of CMs are very explicit too, as listed in the third column of Table 2-1. But, the physical picture of Harrington's CMT has not been thoroughly clarified until now.

Professor Harrington and Dr. Mautz pointed out in literatures [31,32] that: just like the CMs derived from SM-MetSca-CMT[29,30], the CMs derived from EFIE-OpeMetSca-CMT[31,32] also satisfy the orthogonality among modal far fields①. But, this doesn't imply that the physical pictures of Harrington's CMT and Garbacz's CMT are the same, because Prof. Harrington et al. never said that their CMT is for orthogonalizing modal far fields②.

Table 2-1 Comparisons between Garbacz's CMT and Harrington's CMT from the aspects of carrying framework, constructing method, and physical picture

| Two Theories / Three Aspects | **Garbacz's CMT[29,30]** | **Harrington's CMT[31-34]** |
|---|---|---|
| **Carrying Framework of CMT** | SM framework | IE framework |
| **Constructing Method of CMs** | Orthogonalizing PMO method | Orthogonalizing IMO method |
| **Physical Picture of CMT** | Constructing a series of CMs whose modal far fields are orthogonal to each others | |

In fact, how to draw a clear physical picture for Harrington's CMT is just one of the unsolved important problems in CMT domain (i.e. the old problem OP1 discussed in Subsection 1.3.1). Just because of the lack of clear physical picture, some perplexities have appeared, for example:

**Perplex A.** As everyone knows, the CMs derived from EFIE-OpeMetSca-CMT[31,32] satisfy far-field orthogonality[31,32,35,38,39,57]. But, Prof. Sarkar[57] et al.[55] recently found out that the CMs derived from the VIE-MatSca-CMT[33] for lossy material systems (i.e. the material systems whose conductivity is not 0) don't always satisfy far-field orthogonality. Then, whether or not Harrington's CMs should satisfy far-field orthogonality?

**Perplex B.** As everyone knows, the characteristic value calculated from EFIE-OpeMetSca-CMT[31,32] equals the ratio of modal imaginary power to modal real

① A part of the abstract of classical literature [31] is that: "A theory of characteristic modes for conducting bodies is developed starting from the operator formulation for current. The mode currents form a weighted orthogonal set over the conductor surface, and the mode fields form an orthogonal set over the sphere at infinity. It is shown that the modes are the same ones introduced by Garbacz to diagonalize the scattering matrix of the body. [31]".

② Literature [57] pointed out that: it cannot be guaranteed that the Harrington's CMs of lossy material systems have orthogonal modal far fields. In addition, based on the studies of literatures [37,40,41,44,45,55,56], this dissertation finds out that: Harrington's CMT indeed has a different physical picture from Garbacz's CMT.

power[31,32,35,40,41]. But, Dr. Miers and Prof. Lau[43] et al.[44,45] recently found out that the characteristic values calculated from the VIE-MatSca-CMT[33] and SIE-MatSca-CMT[34] for magnetic material systems (i.e. the material systems whose relative magnetic permeabilities are not 1) don't always equal the ratio of modal imaginary power to modal real power. Then, whether or not Harrington's characteristic value should equal the ratio of modal imaginary power to modal real power?

**Perplex C.** As everyone knows, CMs are the reflection for the inherent physical properties of the objective system[29-35,38,39,57], so CMs should depend only on the objective system itself and be completely independent of the process of numerical calculations. But, M. Meng and Prof. Z. Nie[63] recently found out that the original Harrington's EFIE-OpeMetSca-CMT[31,32] (using free space Green's function, i.e. vacuum Green's function) and its IE-based variant (using complex background Green's function[62]) cannot provide the same CM set for a certain objective metallic system; Dr. Dai and Prof. W. Chew et al.[59,60] recently found out that the CMs derived from the CFIE-based CMT for metallic systems depend on the combination coefficients $\{a,b\}$ ① of EFIE and MFIE. Then, why the CMs derived from original Harrington's EFIE-OpeMetSca-CMT[31,32] and its IE-based variants[59,60,62] are not the same for a certain objective metallic system?

After some studies, this dissertation finds out that the cause of the above mentioned Perplexes A~C is the lack of the clear physical picture of Harrington's CMT, so one of the main destinations of this chapter is to draw a clear physical picture for Harrington's CMT. After obtaining the physical picture, we will eliminate the perplexes by employing the physical picture.

This chapter is organized as follows: in Section 2.2, we simply retrospect the carrying framework of Garbacz's CMT, the constructing method of Garbacz's CMs, and the physical picture of Garbacz's CMT; in Section 2.3, we will simply retrospect the carrying framework of Harrington's CMT and the constructing method of Harrington's CMs; in Section 2.4, we will, based on the works of literatures [37,40,41,44,45], draw a clear physical picture for Harrington's CMT, and then we will trigger a new transformation for the carrying framework of Harrington's CMT and a new transformation for the constructing method of Harrington's CMs.

① CFIE is the linear combination of EFIE and MFIE, i.e., $\text{CFIE} = a\,\text{EFIE} + b\,\text{MFIE}$ [106-109], where coefficients $\{a,b\}$ can be arbitrarily adjusted, and the spectrum of CFIE-based IMO depends on the coefficients[106,108].

## 2.2 CMT in SM Framework and Its Physical Picture

The EM scattering problem shown in Fgure 2-1 is considered in this section, where the metallic system is denoted as $D_{\text{met sys}}$. Under the excitation of incident field $\vec{F}^{\text{inc}}$, a scattered electric current $\vec{J}^{\text{sca}}$ will be induced on $\partial D_{\text{met sys}}$ [110-115], where $\partial D_{\text{met sys}}$ is the boundary of $D_{\text{met sys}}$. Current $\vec{J}^{\text{sca}}$ will generate scattered field $\vec{F}^{\text{sca}}$ on whole three-dimensional Euclidean space $\mathbb{R}^3$, where $F = E, H$.

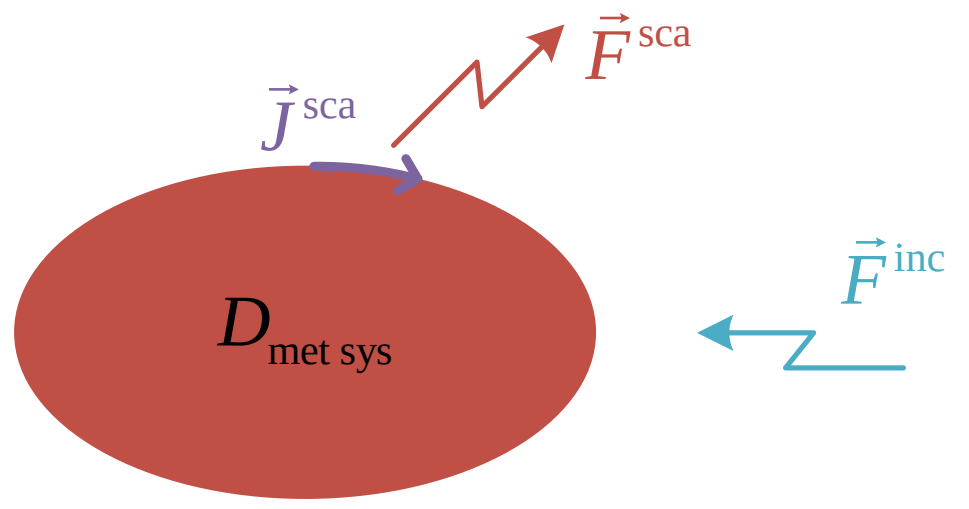

Figure 2-1 Metal scattering problem considered in this chapter

### 1) Perturbation Matrix and Radiated Power

In a source-free region, incident electric field $\vec{E}^{\text{inc}}$ and scattered electric field $\vec{E}^{\text{sca}}$ can be expressed in terms of the superpositions of multipoles, as follows[112]:

$$\vec{E}^{\text{inc}}(\vec{r}) = \sum\nolimits_{l,m} c_E^{\text{inc}}(l,m)\vec{N}_{lm}^E + c_M^{\text{inc}}(l,m)\vec{N}_{lm}^M = \bar{\boldsymbol{X}} \cdot \bar{c}^{\text{inc}} \quad , \quad \vec{r} \in \text{source-free region} \tag{2-1a}$$

$$\vec{E}^{\text{sca}}(\vec{r}) = \sum\nolimits_{l,m} c_E^{\text{sca}}(l,m)\vec{N}_{lm}^E + c_M^{\text{sca}}(l,m)\vec{N}_{lm}^M = \bar{\boldsymbol{X}} \cdot \bar{c}^{\text{sca}} \quad , \quad \vec{r} \in \text{source-free region} \tag{2-1b}$$

where $l = 0,1,2,\cdots$, and $m = 0,\pm1,\pm2,\cdots,\pm l$, and

$$\bar{\boldsymbol{X}} = \begin{bmatrix} \cdots & \vec{N}_{lm}^E & \vec{N}_{lm}^M & \cdots \end{bmatrix} \tag{2-2}$$

$$\bar{c}^{\text{inc}} = \begin{bmatrix} \cdots & c_E^{\text{inc}}(l,m) & c_M^{\text{inc}}(l,m) & \cdots \end{bmatrix}^T \tag{2-3a}$$

$$\bar{c}^{\text{sca}} = \begin{bmatrix} \cdots & c_E^{\text{sca}}(l,m) & c_M^{\text{sca}}(l,m) & \cdots \end{bmatrix}^T \tag{2-3b}$$

$\bar{c}^{\text{inc}}$ and $\bar{c}^{\text{sca}}$ are the vectors constituted by expansion coefficients, and then this dissertation simply calls them as expansion vectors. In formulation (2-3), the superscript "$T$" represents the transpose operation for a matrix or vector. In formulation (2-2), the $\vec{N}_{lm}^E$ and $\vec{N}_{lm}^M$ are defined as follows[112]:

$$\vec{N}_{lm}^E = j\left(\eta_0^{3/2}/\sqrt{2}k_0^2\right)\nabla\times\left[f_l(k_0 r)\vec{X}_{lm}\right] \tag{2-4a}$$

$$\vec{N}_{lm}^M = \left(\eta_0^{3/2}/\sqrt{2}k_0\right)g_l(k_0 r)\vec{X}_{lm} \tag{2-4b}$$

In formulation (2-4), $j$ is imaginary unit; $\eta_0 = \sqrt{\mu_0/\varepsilon_0}$ and $k_0 = \omega\sqrt{\mu_0\varepsilon_0}$ are the wave impedance in vacuum and the wave number in vacuum respectively, where $\mu_0$ and $\varepsilon_0$ are the permeability of vacuum and the permittivity of vacuum respectively, and $\omega$ is the angular frequency of EM fields; the definitions of wave functions $f_l(k_0 r)$, $g_l(k_0 r)$, and $\vec{X}_{lm}$ are the same as the ones given in classical literature [112], and they will not be repeated here.

Expansion vectors $\bar{c}^{\text{inc}}$ and $\bar{c}^{\text{sca}}$ satisfy the following linear transformation relationship[29,30]:

$$\bar{c}^{\text{sca}} = \bar{\bar{\mathcal{P}}} \cdot \bar{c}^{\text{inc}} \tag{2-5}$$

where $\bar{\bar{\mathcal{P}}}$ is perturbation matrix, and $\bar{\bar{\mathcal{P}}}$ and scattering matrix $\bar{\bar{S}}$ satisfy the relationship that $\bar{\bar{\mathcal{P}}} = \bar{\bar{S}} - \bar{\bar{I}}$ [30].

The radiated power $P^{\text{rad}}$ carried by scattered field $\vec{F}^{\text{sca}}$ satisfies the following relationship[112]:

$$\begin{aligned} P^{\text{rad}} &= (1/2) \oiint_{S_\infty} \left[ \vec{E}^{\text{sca}} \times \left( \vec{H}^{\text{sca}} \right)^* \right] \cdot \hat{n}_\infty^+ dS \\ &= \sum_{l,m} \left[ \left| c_E^{\text{sca}}(l,m) \right|^2 + \left| c_M^{\text{sca}}(l,m) \right|^2 \right] \\ &= \left( \bar{c}^{\text{sca}} \right)^H \cdot \bar{c}^{\text{sca}} \\ &= \left( \bar{c}^{\text{inc}} \right)^H \cdot \bar{\bar{\mathcal{P}}}^H \cdot \bar{\bar{\mathcal{P}}} \cdot \bar{c}^{\text{inc}} \end{aligned} \tag{2-6}$$

In formulation (2-6), integral domain $S_\infty$ is a spherical surface at infinity; unit vector $\hat{n}_\infty^+$ is the normal vector of $S_\infty$, and it points to infinity; superscript "$*$" represents the conjugate operation for complex number; superscript "$H$" represents the conjugate transpose for a matrix or vector. In formulation (2-6), the first equality is based on the definition of radiated power; the second equality is based on the results given in classical literature [112]; the third equality is based on formulation (2-3b); the fourth equality is based on formulation (2-5).

**2) Calculation Formulation for Garbacz's CM**

In literature [30], Prof. Garbacz proved that: if metallic system $D_{\text{met sys}}$ is lossless, then perturbation matrix $\bar{\bar{\mathcal{P}}}$ can be a normal matrix. When $\bar{\bar{\mathcal{P}}}$ is a normal matrix, there must be a unitary matrix $\bar{\bar{U}}^{\text{inc}}$ such that[116]

$$\left( \bar{\bar{U}}^{\text{inc}} \right)^H \cdot \bar{\bar{\mathcal{P}}} \cdot \bar{\bar{U}}^{\text{inc}} = \text{diag}\left\{ \lambda_1^{\text{SM}}, \lambda_2^{\text{SM}}, \cdots, \lambda_\xi^{\text{SM}}, \cdots \right\} \tag{2-7}$$

where $\text{diag}\{\lambda_1^{\text{SM}}, \lambda_2^{\text{SM}}, \cdots, \lambda_\xi^{\text{SM}}, \cdots\}$ is a diagonal matrix, whose diagonal elements are listed in the bracket. If $\bar{\bar{U}}^{\text{inc}}$ is partitioned as $\bar{\bar{U}}^{\text{inc}} = [\bar{u}_1^{\text{inc}} \quad \bar{u}_2^{\text{inc}} \quad \cdots \quad \bar{u}_\xi^{\text{inc}} \quad \cdots]$ based on its columns, the following orthogonality exists:

$$\lambda_{\xi}^{\rm SM}\delta_{\xi\zeta} = \left(\bar{u}_{\xi}^{\rm inc}\right)^{H}\cdot\bar{\bar{\mathcal{P}}}\cdot\bar{u}_{\zeta}^{\rm inc} \quad , \quad \left(\xi,\zeta=1,2,\cdots\right) \tag{2-8}$$

where $\delta_{\xi\zeta}$ is Kronecker's symbol.

As everyone knows, the vector group $\{\bar{u}_{1}^{\rm inc},\bar{u}_{2}^{\rm inc},\cdots,\bar{u}_{\xi}^{\rm inc},\cdots\}$, which satisfy orthogonality (2-8), can be obtained by solving the following characteristic equation[116]:

$$\bar{\bar{\mathcal{P}}}\cdot\bar{u}_{\xi}^{\rm inc} = \lambda_{\xi}^{\rm SM}\bar{u}_{\xi}^{\rm inc} \quad , \quad \xi=1,2,\cdots \tag{2-9a}$$

Obviously, in source-free region the characteristic fields corresponding to the above characteristic vectors are as follows:

$$\vec{E}_{\xi}^{\rm sca}\left(\vec{r}\right) = \bar{\boldsymbol{X}}\cdot\bar{u}_{\xi}^{\rm sca} = \bar{\boldsymbol{X}}\cdot\bar{\bar{\mathcal{P}}}\cdot\bar{u}_{\xi}^{\rm inc} \quad , \quad \vec{r}\in\text{source-free region} \tag{2-10a}$$

$$\vec{H}_{\xi}^{\rm sca}\left(\vec{r}\right) = -\left(1/j\omega\mu_0\right)\nabla\times\vec{E}_{\xi}^{\rm sca}\left(\vec{r}\right) \quad , \quad \vec{r}\in\text{source-free region} \tag{2-10b}$$

where $\xi=1,2,\cdots$. Above $\{\lambda_{\xi}^{\rm SM},\bar{u}_{\xi}^{\rm inc}\}_{\xi=1}^{\infty}$ are called as Garbacz's characteristic pairs, and above $\{\bar{u}_{\xi}^{\rm inc},\bar{u}_{\xi}^{\rm sca},\vec{E}_{\xi}^{\rm sca},\vec{H}_{\xi}^{\rm sca}\}_{\xi=1}^{\infty}$ are collectively referred to as Garbacz's CMs.

**3) Physical Picture of Garbacz's CMT**

Above matrix $\bar{\bar{U}}^{\rm inc}$ is an unitary matrix, so $\bar{\bar{U}}^{\rm inc}\cdot(\bar{\bar{U}}^{\rm inc})^{H}=\bar{\bar{I}}$ [116], and then $(\bar{\bar{U}}^{\rm inc})^{H}\cdot\bar{\bar{\mathcal{P}}}^{H}\cdot\bar{\bar{\mathcal{P}}}\cdot\bar{\bar{U}}^{\rm inc}=(\bar{\bar{U}}^{\rm inc})^{H}\cdot\bar{\bar{\mathcal{P}}}^{H}\cdot\bar{\bar{U}}^{\rm inc}\cdot(\bar{\bar{U}}^{\rm inc})^{H}\cdot\bar{\bar{\mathcal{P}}}\cdot\bar{\bar{U}}^{\rm inc}={\rm diag}\left\{|\lambda_{1}^{\rm SM}|^{2},\cdots,|\lambda_{\xi}^{\rm SM}|^{2},\cdots\right\}$. This implies that

$$\left|\lambda_{\xi}^{\rm SM}\right|^{2}\delta_{\xi\zeta} = \left(\bar{u}_{\xi}^{\rm inc}\right)^{H}\cdot\bar{\bar{\mathcal{P}}}^{H}\cdot\bar{\bar{\mathcal{P}}}\cdot\bar{u}_{\zeta}^{\rm inc} = \left(\bar{u}_{\xi}^{\rm sca}\right)^{H}\cdot\bar{u}_{\zeta}^{\rm sca} \quad , \quad \left(\xi,\zeta=1,2,\cdots\right) \tag{2-11}$$

where the second equality is based on transformation (2-5).

Based on relationship (2-6) and orthogonality (2-11), we can easily conclude that: the physical picture of Garbacz's CMT[29] is to construct a series of fundamental modes whose modal far fields satisfy orthogonality (2-11). In addition, it is easy to find out that: the physical meaning of $|\lambda_{\xi}^{\rm SM}|^{2}$ is just the modal radiated power of the $\xi$-th Garbacz's CM, i.e., $|\lambda_{\xi}^{\rm SM}|^{2}=P_{\xi}^{\rm rad}$.

In fact, matrix $\bar{\bar{\mathcal{P}}}^{H}\cdot\bar{\bar{\mathcal{P}}}$ is always Hermitian obviously, so the CMs satisfying orthogonality (2-11) can also be derived from solving the following generalized characteristic equation[116]:

$$\left(\bar{\bar{\mathcal{P}}}^{H}\cdot\bar{\bar{\mathcal{P}}}\right)\cdot\bar{u}_{\xi}^{\rm inc} = P_{\xi}^{\rm rad}\bar{u}_{\xi}^{\rm inc} \quad , \quad \xi=1,2,\cdots \tag{2-9b}$$

To obtain orthogonality (2-11) from orthogonality (2-7), it is necessary to guarantee that perturbation matrix $\bar{\bar{\mathcal{P}}}$ is a normal matrix, so it is necessary to require that the scattering system is lossless[30]. But, the derivation of orthogonality (2-11) from equation (2-9b) depends only on the Hermitian property of matrix $\bar{\bar{\mathcal{P}}}^H \cdot \bar{\bar{\mathcal{P}}}$, so equation (2-9b) is also applicable to lossy scattering systems.

## 2.3 CMT in IE Framework and Its Physical Picture

In this section, we, using three subsections (Subsections 2.3.1~2.3.3), simply retrospect the fundamental principles of the Harrington's EFIE-OpeMetSca-CMT[31,32], VIE-MatSca-CMT[33], and SIE-MatSca-CMT[34] in IE framework.

### 2.3.1 CM Calculation Formulations Corresponding to Harrington's EFIE-OpeMetSca-CMT

In this subsection, we will still consider the metallic scattering system shown in Figure 2-1.

**1) EFIE of Metallic Systems and the Corresponding IMO**

On the boundary $\partial D_{\text{met sys}}$ of metallic system $D_{\text{met sys}}$, the following EFIE exists[106-108,115]:

$$\begin{aligned} \vec{E}_{\tan}^{\text{inc}}(\vec{r}) &= -\vec{E}_{\tan}^{\text{sca}}(\vec{r}) \\ &= -\left[-j\omega\mu_0\mathcal{L}_0\left(\vec{J}^{\text{sca}}\right)\right]_{\tan} \quad , \quad \vec{r} \in \partial D_{\text{met sys}} \end{aligned} \tag{2-12}$$

where subscript "tan" is to emphasize that the EFIE is valid for the tangential components of the related EM fields; the operator $\mathcal{L}_0$ is defined as that $\mathcal{L}_0(\vec{X}) = [1+(1/k_0^2)\nabla\nabla\cdot]\int_\Omega G_0(\vec{r},\vec{r}\,')\vec{X}(\vec{r}\,')d\Omega'$ [107], and $G_0(\vec{r},\vec{r}\,') = e^{-jk_0|\vec{r}-\vec{r}\,'|}/4\pi|\vec{r}-\vec{r}\,'|$ is free space scalar Green's function[106-108].

If the $\vec{J}^{\text{sca}}$ in EFIE (2-12) is expanded in terms of basis functions $\{\vec{b}_\xi^J\}_{\xi=1}^{\Xi}$ (i.e., $\vec{J}^{\text{sca}} = \sum_{\xi=1}^{\Xi} a_\xi^J \vec{b}_\xi^J$ ) and EFIE (2-12) is tested by testing functions $\{\vec{b}_\zeta^J\}_{\zeta=1}^{\Xi}$, then EFIE (2-12) can be discretized into the following matrix equation[106-108,115]:

$$\bar{b} = \bar{\bar{Z}}^{\text{EFIE}} \cdot \bar{a}^J \tag{2-13}$$

where

$$\bar{b} = \begin{bmatrix} b_1 & b_2 & \cdots & b_\Xi \end{bmatrix}^T \tag{2-14a}$$

$$\bar{\bar{Z}}^{\text{EFIE}} = \begin{bmatrix} z_{\xi\zeta}^{\text{EFIE}} \end{bmatrix}_{\Xi\times\Xi} \tag{2-14b}$$

$$\bar{a}^J = \begin{bmatrix} a_1^J & a_2^J & \cdots & a_\Xi^J \end{bmatrix}^T \tag{2-14c}$$

and the elements in excitation vector $\bar{b}$ and impedance matrix $\bar{\bar{Z}}^{\text{EFIE}}$ can be calculated as follows:

$$b_\xi = \left\langle \vec{b}_\xi^J, \vec{E}^{\text{inc}} \right\rangle_{\partial D_{\text{met sys}}} \tag{2-15}$$

$$z_{\xi\zeta}^{\text{EFIE}} = \left\langle \vec{b}_\xi^J, -\left[ -j\omega\mu_0 \mathcal{L}_0\left(\vec{b}_\zeta^J\right)\right]\right\rangle_{\partial D_{\text{met sys}}} \tag{2-16}$$

where the inner product is defined as that $<\vec{f},\vec{g}>_\Omega = \int_\Omega \vec{f}^* \cdot \vec{g}\, d\Omega$.

**2) Calculation Formulation for Harrington's Metallic CMs**

Above impedance matrix $\bar{\bar{Z}}^{\text{EFIE}}$ is a complex matrix, and it can be decomposed as follows[31,32]:

$$\bar{\bar{Z}}^{\text{EFIE}} = \bar{\bar{R}}^{\text{EFIE}} + j\,\bar{\bar{X}}^{\text{EFIE}} \tag{2-17}$$

where $\bar{\bar{R}}^{\text{EFIE}}$ and $\bar{\bar{Z}}^{\text{EFIE}}$ are the real part of $\bar{\bar{Z}}^{\text{EFIE}}$ and the imaginary part of $\bar{\bar{Z}}^{\text{EFIE}}$ respectively, i.e., $\bar{\bar{R}}^{\text{EFIE}} = \text{Re}\{\bar{\bar{Z}}^{\text{EFIE}}\}$ and $\bar{\bar{X}}^{\text{EFIE}} = \text{Im}\{\bar{\bar{Z}}^{\text{EFIE}}\}$.

The Harrington's characteristic vectors of metallic systems can be derived from solving the following generalized characteristic equation[31,32]:

$$\bar{\bar{X}}^{\text{EFIE}} \cdot \bar{a}_\xi^J = \lambda_\xi^{\text{EFIE}} \bar{\bar{R}}^{\text{EFIE}} \cdot \bar{a}_\xi^J \tag{2-18}$$

Using the characteristic vectors, the corresponding characteristic currents can be obtained by employing the basis-function-based expansion formulation of $\vec{J}_\xi^{\text{sca}}$. The characteristic fields generated by $\vec{J}_\xi^{\text{sca}}$ can be expressed as follows:

$$\vec{E}_\xi^{\text{sca}}(\vec{r}) = -j\omega\mu_0 \mathcal{L}_0\left(\vec{J}_\xi^{\text{sca}}\right) \quad , \quad \vec{r} \in \mathbb{R}^3 \tag{2-19a}$$

$$\vec{H}_\xi^{\text{sca}}(\vec{r}) = \mathcal{K}_0\left(\vec{J}_\xi^{\text{sca}}\right) \quad , \quad \vec{r} \in \mathbb{R}^3 \tag{2-19b}$$

where operator $\mathcal{K}_0$ is defined as that $\mathcal{K}_0(\vec{X}) = \nabla \times \int_\Omega G_0(\vec{r},\vec{r}\,')\vec{X}(\vec{r}\,')d\Omega'$ [107]. Above-mentioned $\{\lambda_\xi^{\text{EFIE}}, \bar{a}_\xi^J\}_{\xi=1}^{\Xi}$ are called as the Harrington's characteristic pairs of metallic systems, and above-mentioned $\{\bar{a}_\xi^J, \vec{J}_\xi^{\text{sca}}, \vec{E}_\xi^{\text{sca}}, \vec{H}_\xi^{\text{sca}}\}_{\xi=1}^{\Xi}$ are collectively referred to as the Harrington's CMs of metallic systems.

## 2.3.2 CM Calculation Formulations Corresponding to Harrington's VIE-MatSca-CMT

In this subsection, we consider the EM scattering problem shown in Figure 2-2. In the figure, $V_{\text{mat sys}}$ is a material system, which is constituted by a single simply connected material body; $\vec{F}^{\text{inc}}$ and $\vec{F}^{\text{sca}}$ represent incident field and scattered field respectively, where $F = E, H$ ; total field $\vec{F}^{\text{tot}}$ is the summation of $\vec{F}^{\text{inc}}$ and $\vec{F}^{\text{sca}}$ , i.e., $\vec{F}^{\text{tot}} = \vec{F}^{\text{inc}} + \vec{F}^{\text{sca}}$; $\{\vec{J}^{\text{SV}}, \vec{M}^{\text{SV}}\}$ are the scattered volume currents distributing on the interior $\text{int}V_{\text{mat sys}}$ of $V_{\text{mat sys}}$ ; $\{\vec{J}^{\text{ES}}, \vec{M}^{\text{ES}}\}$ are the equivalent surface currents distributing on the boundary $\partial V_{\text{mat sys}}$ of $V_{\text{mat sys}}$. Here, $\vec{J}^{\text{ES}}$ and $\vec{M}^{\text{ES}}$ are respectively defined as that $\vec{J}^{\text{ES}} = \hat{n}_{\text{mat}}^{-} \times \vec{H}_{-;\tan}^{\text{tot}}$ and that $\vec{M}^{\text{ES}} = \vec{E}_{-;\tan}^{\text{tot}} \times \hat{n}_{\text{mat}}^{-}$ , where $\hat{n}_{\text{mat}}^{-}$ is the inner normal direction of $\partial V_{\text{mat sys}}$ .

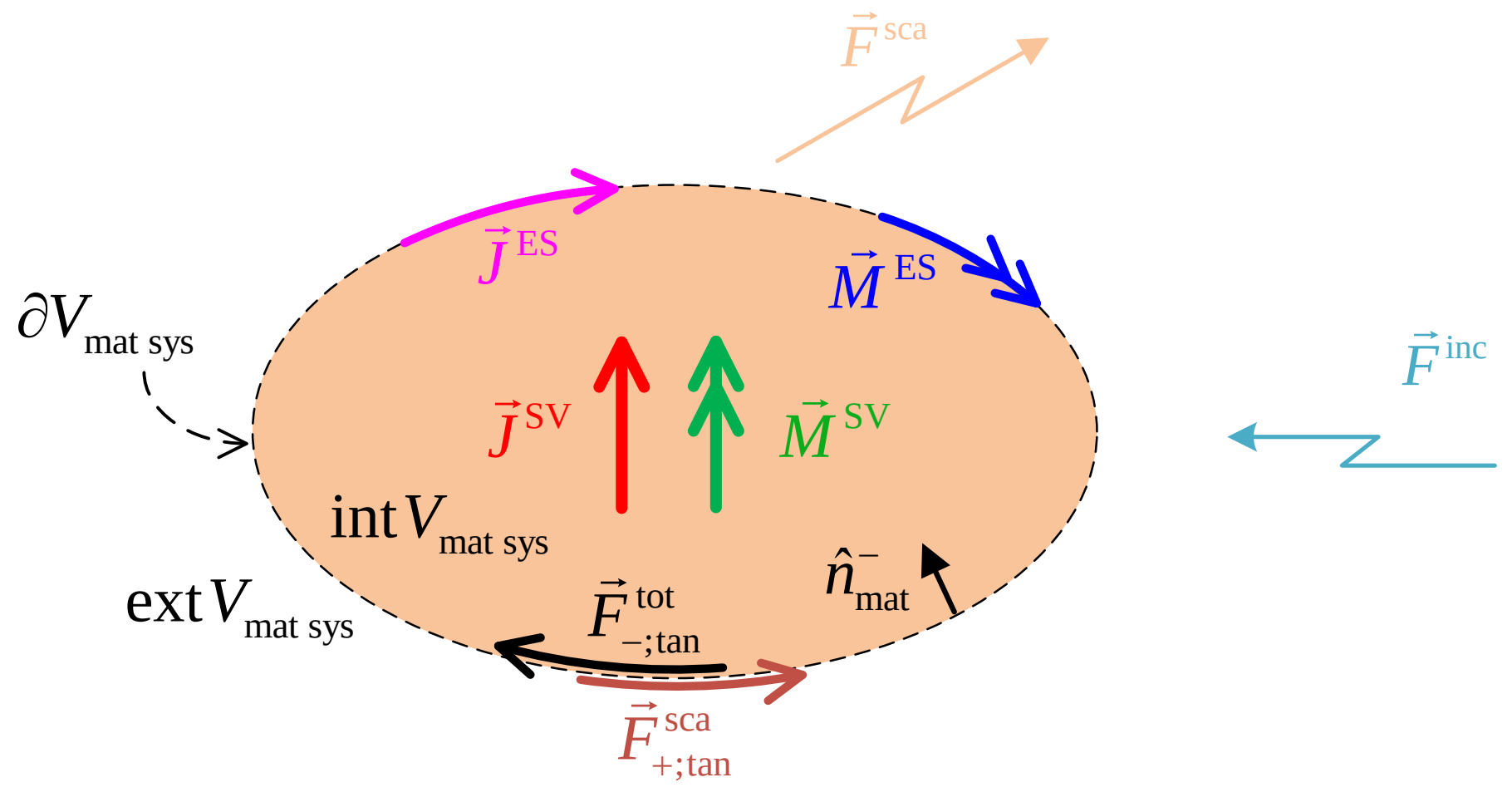

Figure 2-2 Material scattering problem considered in this chapter

**1) VIE of Material Systems and the Corresponding IMO**

On material system $V_{\text{mat sys}}$, various EM fields satisfy the following VIEs[33,106-108]:

$$\vec{E}^{\text{inc}}(\vec{r}) = \vec{E}^{\text{tot}}(\vec{r}) - \vec{E}^{\text{sca}}(\vec{r})$$
$$= \left(1/j\omega\Delta\varepsilon_{\text{mat}}^{\text{c}}\right)\vec{J}^{\text{SV}} - \left[-j\omega\mu_0\mathcal{L}_0\left(\vec{J}^{\text{SV}}\right) - \mathcal{K}_0\left(\vec{M}^{\text{SV}}\right)\right] \quad , \quad \vec{r} \in V_{\text{mat sys}} \qquad (2\text{-}20\text{a})$$

$$\vec{H}^{\text{inc}}(\vec{r}) = \vec{H}^{\text{tot}}(\vec{r}) - \vec{H}^{\text{sca}}(\vec{r})$$
$$= \left(1/j\omega\Delta\mu_{\text{mat}}\right)\vec{M}^{\text{SV}} - \left[-j\omega\varepsilon_0\mathcal{L}_0\left(\vec{M}^{\text{SV}}\right) + \mathcal{K}_0\left(\vec{J}^{\text{SV}}\right)\right] \quad , \quad \vec{r} \in V_{\text{mat sys}} \qquad (2\text{-}20\text{b})$$

where the first equality is based on superposition principle $\vec{F}^{\text{tot}} = \vec{F}^{\text{inc}} + \vec{F}^{\text{sca}}$; the second equality is based on volume equivalence principle $\{\vec{J}^{\text{SV}} = j\omega\Delta\varepsilon_{\text{mat}}^{\text{c}}\vec{E}^{\text{tot}}$, $\vec{M}^{\text{SV}} = j\omega\Delta\mu_{\text{mat}}\vec{H}^{\text{tot}}\}$ [33,106], and $\Delta\varepsilon_{\text{mat}}^{\text{c}} = (\sigma_{\text{mat}}/j\omega) + \varepsilon_{\text{mat}} - \varepsilon_0$ and $\Delta\mu_{\text{mat}} = \mu_{\text{mat}} - \mu_0$ , and $\varepsilon_{\text{mat}}$ , $\mu_{\text{mat}}$ , and $\sigma_{\text{mat}}$ are respectively the permittivity, permeability, and conductivity of material system $V_{\text{mat sys}}$ .

If $\vec{J}^{\text{SV}}$ and $\vec{M}^{\text{SV}}$ are respectively expanded in terms of basis functions $\{\vec{b}_{\xi}^{J}\}_{\xi=1}^{\Xi^J}$ and $\{\vec{b}_{\xi}^{M}\}_{\xi=1}^{\Xi^M}$ as that $\vec{J}^{\text{SV}} = \sum_{\xi=1}^{\Xi^J} a_{\xi}^{J}\vec{b}_{\xi}^{J}$ and that $\vec{M}^{\text{SV}} = \sum_{\xi=1}^{\Xi^M} a_{\xi}^{M}\vec{b}_{\xi}^{M}$, and VIEs (2-20a) and (2-20b) are tested by testing functions $\{\vec{b}_{\xi}^{J}\}_{\xi=1}^{\Xi^J}$ and $\{\vec{b}_{\xi}^{M}\}_{\xi=1}^{\Xi^M}$ respectively, then the integral equations can be discretized into the following matrix equation[33,106-108]:

$$\begin{bmatrix} \bar{b}^{JE} \\ \bar{b}^{MH} \end{bmatrix} = \underbrace{\begin{bmatrix} \bar{\bar{Z}}^{JEJ} & \bar{\bar{Z}}^{JEM} \\ \bar{\bar{Z}}^{MHJ} & \bar{\bar{Z}}^{MHM} \end{bmatrix}}_{\bar{\bar{Z}}^{\text{VIE}}} \cdot \underbrace{\begin{bmatrix} \bar{a}^{J} \\ \bar{a}^{M} \end{bmatrix}}_{\bar{a}^{JM}} \tag{2-21}$$

where

$$\bar{b}^{JE/MH} = \begin{bmatrix} b_1^{JE/MH} & b_2^{JE/MH} & \cdots & b_{\Xi^{J/M}}^{JE/MH} \end{bmatrix}^T \tag{2-22a}$$

$$\bar{\bar{Z}}^{JEJ/JEM/MHJ/MHM} = \begin{bmatrix} z_{\xi\zeta}^{JEJ/JEM/MHJ/MHM} \end{bmatrix}_{\Xi^{J/J/M/M} \times \Xi^{J/M/J/M}} \tag{2-22b}$$

$$\bar{a}^{J/M} = \begin{bmatrix} a_1^{J/M} & a_2^{J/M} & \cdots & a_{\Xi^{J/M}}^{J/M} \end{bmatrix}^T \tag{2-22c}$$

and the elements in excitation sub-vectors $\bar{b}^{JE}$ and $\bar{b}^{MH}$ and impedance sub-matrices $\bar{\bar{Z}}^{JEJ}$, $\bar{\bar{Z}}^{JEM}$, $\bar{\bar{Z}}^{MHJ}$, and $\bar{\bar{Z}}^{MHM}$ are calculated as follows:

$$b_\xi^{JE} = \left\langle \vec{b}_\xi^{J}, \vec{E}^{\text{inc}} \right\rangle_{V_{\text{mat sys}}} \tag{2-23a}$$

$$b_\xi^{MH} = \left\langle \vec{b}_\xi^{M}, \vec{H}^{\text{inc}} \right\rangle_{V_{\text{mat sys}}} \tag{2-23b}$$

$$z_{\xi\zeta}^{JEJ} = \left\langle \vec{b}_\xi^{J}, \left(1/j\omega\Delta\varepsilon_{\text{mat}}^{\text{c}}\right)\vec{b}_\zeta^{J} - \left[-j\omega\mu_0\mathcal{L}_0\left(\vec{b}_\zeta^{J}\right)\right] \right\rangle_{V_{\text{mat sys}}} \tag{2-24a}$$

$$z_{\xi\zeta}^{JEM} = \left\langle \vec{b}_\xi^{J}, -\left[-\mathcal{K}_0\left(\vec{b}_\zeta^{M}\right)\right] \right\rangle_{V_{\text{mat sys}}} \tag{2-24b}$$

$$z_{\xi\zeta}^{MHJ} = \left\langle \vec{b}_\xi^{M}, -\mathcal{K}_0\left(\vec{b}_\zeta^{J}\right) \right\rangle_{V_{\text{mat sys}}} \tag{2-24c}$$

$$z_{\xi\zeta}^{MHM} = \left\langle \vec{b}_\xi^{M}, \left(1/j\omega\Delta\mu_{\text{mat}}\right)\vec{b}_\zeta^{M} - \left[-j\omega\varepsilon_0\mathcal{L}_0\left(\vec{b}_\zeta^{M}\right)\right] \right\rangle_{V_{\text{mat sys}}} \tag{2-24d}$$

**2) Volume Version of the Calculation Formulation for Harrington's Material CMs**

Above impedance matrix $\bar{\bar{Z}}^{\text{VIE}}$ is a complex matrix, and it can be decomposed as follows[33]:

$$\bar{\bar{Z}}^{\text{VIE}} = \bar{\bar{R}}^{\text{VIE}} + j\,\bar{\bar{X}}^{\text{VIE}} \tag{2-25}$$

where $\bar{\bar{R}}^{\text{VIE}} = \text{Re}\{\bar{\bar{Z}}^{\text{VIE}}\}$ and $\bar{\bar{X}}^{\text{VIE}} = \text{Im}\{\bar{\bar{Z}}^{\text{VIE}}\}$.

The Harrington's characteristic vectors for material systems can be derived from solving the following generalized characteristic equation[33]:

$$\bar{\bar{X}}^{\text{VIE}} \cdot \bar{a}_\xi^{JM} = \lambda_\xi^{\text{VIE}} \bar{\bar{R}}^{\text{VIE}} \cdot \bar{a}_\xi^{JM} \quad , \quad \xi = 1, 2, \cdots, \Xi^{JM} \tag{2-26}$$

where $\Xi^{JM} = \Xi^{J} + \Xi^{M}$. Using the characteristic vectors, the corresponding characteristic currents can be obtained by employing the basis-function-based expansion formulations of $\vec{J}^{\text{SV}}$ and $\vec{M}^{\text{SV}}$. The characteristic fields corresponding to $\{\vec{J}_\xi^{\text{SV}}, \vec{M}_\xi^{\text{SV}}\}$ can be

expressed as follows:

$$\vec{E}_\xi^{\text{sca}}(\vec{r}) = -j\omega\mu_0\mathcal{L}_0(\vec{J}_\xi^{\text{SV}}) - \mathcal{K}_0(\vec{M}_\xi^{\text{SV}}) \quad , \quad \vec{r}\in\mathbb{R}^3 \tag{2-27a}$$

$$\vec{H}_\xi^{\text{sca}}(\vec{r}) = -j\omega\varepsilon_0\mathcal{L}_0(\vec{M}_\xi^{\text{SV}}) + \mathcal{K}_0(\vec{J}_\xi^{\text{SV}}) \quad , \quad \vec{r}\in\mathbb{R}^3 \tag{2-27b}$$

$$\vec{E}_\xi^{\text{tot}}(\vec{r}) = \left(1/j\omega\Delta\varepsilon_{\text{mat}}^{\text{c}}\right)\vec{J}_\xi^{\text{SV}} \quad , \quad \vec{r}\in V_{\text{mat sys}} \tag{2-28a}$$

$$\vec{H}_\xi^{\text{tot}}(\vec{r}) = \left(1/j\omega\Delta\mu_{\text{mat}}\right)\vec{M}_\xi^{\text{SV}} \quad , \quad \vec{r}\in V_{\text{mat sys}} \tag{2-28b}$$

$$\vec{E}_\xi^{\text{inc}}(\vec{r}) = \vec{E}_\xi^{\text{tot}}(\vec{r}) - \vec{E}_\xi^{\text{sca}}(\vec{r}) \quad , \quad \vec{r}\in V_{\text{mat sys}} \tag{2-29a}$$

$$\vec{H}_\xi^{\text{inc}}(\vec{r}) = \vec{H}_\xi^{\text{tot}}(\vec{r}) - \vec{H}_\xi^{\text{sca}}(\vec{r}) \quad , \quad \vec{r}\in V_{\text{mat sys}} \tag{2-29b}$$

Above-mentioned $\{\lambda_\xi^{\text{VIE}}, \overline{a}_\xi^{JM}\}_{\xi=1}^{\Xi^{JM}}$ are called as the Harrington's characteristic pairs of material systems, and above-mentioned $\{\overline{a}_\xi^{JM}, \vec{J}_\xi^{\text{SV}}, \vec{M}_\xi^{\text{SV}}, \vec{E}_\xi^{\text{sca}}, \vec{H}_\xi^{\text{sca}}, \vec{E}_\xi^{\text{tot}}, \vec{H}_\xi^{\text{tot}}, \vec{E}_\xi^{\text{inc}}, \vec{H}_\xi^{\text{inc}}\}_{\xi=1}^{\Xi^{JM}}$ are collectively referred to as the Harrington's CMs of material systems.

### 2.3.3 CM Calculation Formulations Corresponding to Harrington's SIE-MatSca-CMT

In this subsection, we still consider the EM scattering problem of the material system shown in Figure 2-2.

**1) SIE of Material Systems and the Corresponding IMO**

On the boundary $\partial V_{\text{mat sys}}$ of material system $V_{\text{mat sys}}$, the tangential components of various EM fields satisfy the following SIEs[34,106-108]:

$$\begin{aligned}\vec{E}_{\text{tan}}^{\text{inc}}(\vec{r}) &= \vec{E}_{-;\text{tan}}^{\text{tot}}(\vec{r}) - \vec{E}_{+;\text{tan}}^{\text{sca}}(\vec{r}) \\ &= \left[-j\omega\mu_{\text{mat}}\mathcal{L}(\vec{J}^{\text{ES}}) - \mathcal{K}(\vec{M}^{\text{ES}})\right] \\ &\quad -\left[+j\omega\mu_0\mathcal{L}_0(\vec{J}^{\text{ES}}) + \mathcal{K}_0(\vec{M}^{\text{ES}})\right] \quad , \quad \vec{r}\in\partial V_{\text{mat sys}}\end{aligned} \tag{2-30a}$$

$$\begin{aligned}\vec{H}_{\text{tan}}^{\text{inc}}(\vec{r}) &= \vec{H}_{-;\text{tan}}^{\text{tot}}(\vec{r}) - \vec{H}_{+;\text{tan}}^{\text{sca}}(\vec{r}) \\ &= \left[-j\omega\varepsilon_{\text{mat}}^{\text{c}}\mathcal{L}(\vec{M}^{\text{ES}}) + \mathcal{K}(\vec{J}^{\text{ES}})\right] \\ &\quad -\left[+j\omega\varepsilon_0\mathcal{L}_0(\vec{M}^{\text{ES}}) - \mathcal{K}_0(\vec{J}^{\text{ES}})\right] \quad , \quad \vec{r}\in\partial V_{\text{mat sys}}\end{aligned} \tag{2-30b}$$

In equations (2-30a) and (2-30b), the first equality is based on superposition principle $\vec{F}^{\text{tot}} = \vec{F}^{\text{inc}} + \vec{F}^{\text{sca}}$ and the tangential continuation condition of $\vec{F}^{\text{tot}}$, and $\vec{F}_-^{\text{tot}}$ and $\vec{F}_+^{\text{sca}}$ are respectively the total field on the internal surface of $\partial V_{\text{mat sys}}$ and the scattered field on the external surface of $\partial V_{\text{mat sys}}$; the second equality is based on the surface equivalence principle for a single simply connected homogeneous isotropic material body[34,44,45,106-108]; operators $\mathcal{L}$ and $\mathcal{K}$ are respectively defined as that $\mathcal{L}(\vec{X}) = [1+(1/k_{\text{mat}}^2)\nabla\nabla\cdot]\int_{\Omega} G(\vec{r},\vec{r}')\vec{X}(\vec{r}')d\Omega'$ and that $\mathcal{K}(\vec{X}) = \nabla\times\int_{\Omega} G(\vec{r},\vec{r}')\vec{X}(\vec{r}')d\Omega'$ [107], where $G(\vec{r},\vec{r}') = e^{-jk_{\text{mat}}|\vec{r}-\vec{r}'|}/4\pi|\vec{r}-\vec{r}'|$ is the scalar Green's function in homogeneous isotropic medium[106-108], and $k_{\text{mat}} = \omega\sqrt{\mu_{\text{mat}}\varepsilon_{\text{mat}}^{\text{c}}}$.

If $\vec{J}^{\text{ES}}$ and $\vec{M}^{\text{ES}}$ are respectively expanded in terms of the linear superpositions of basis functions $\{\vec{b}_{\xi}^{J}\}_{\xi=1}^{\Xi^J}$ and $\{\vec{b}_{\xi}^{M}\}_{\xi=1}^{\Xi^M}$, i.e., $\vec{J}^{\text{ES}} = \sum_{\xi=1}^{\Xi^J} a_{\xi}^{J}\vec{b}_{\xi}^{J}$ and $\vec{M}^{\text{ES}} = \sum_{\xi=1}^{\Xi^M} a_{\xi}^{M}\vec{b}_{\xi}^{M}$, and SIEs (2-30a) and (2-30b) are tested by testing functions $\{\vec{b}_{\xi}^{J}\}_{\xi=1}^{\Xi^J}$ and $\{\vec{b}_{\xi}^{M}\}_{\xi=1}^{\Xi^M}$ respectively, then integral equations (2-30a) and (2-30b) can be discretized into the following matrix equation[34,106-108]:

$$\begin{bmatrix} \bar{b}^{JE} \\ \bar{b}^{MH} \end{bmatrix} = \underbrace{\begin{bmatrix} \bar{\bar{Z}}^{JEJ} & \bar{\bar{Z}}^{JEM} \\ \bar{\bar{Z}}^{MHJ} & \bar{\bar{Z}}^{MHM} \end{bmatrix}}_{\bar{\bar{Z}}^{\text{SIE}}} \cdot \underbrace{\begin{bmatrix} \bar{a}^{J} \\ \bar{a}^{M} \end{bmatrix}}_{\bar{a}^{JM}} \tag{2-31}$$

where

$$\bar{b}^{JE/MH} = \begin{bmatrix} b_1^{JE/MH} & b_2^{JE/MH} & \cdots & b_{\Xi^{J/M}}^{JE/MH} \end{bmatrix}^T \tag{2-32a}$$

$$\bar{\bar{Z}}^{JEJ/JEM/MHJ/MHM} = \begin{bmatrix} z_{\xi\zeta}^{JEJ/JEM/MHJ/MHM} \end{bmatrix}_{\Xi^{J/J/M/M}\times\Xi^{J/M/J/M}} \tag{2-32b}$$

$$\bar{a}^{J/M} = \begin{bmatrix} a_1^{J/M} & a_2^{J/M} & \cdots & a_{\Xi^{J/M}}^{J/M} \end{bmatrix}^T \tag{2-32c}$$

and the elements of sub-vectors $\bar{b}^{JE}$ and $\bar{b}^{MH}$ and the elements of sub-matrices $\bar{\bar{Z}}^{JEJ}$, $\bar{\bar{Z}}^{JEM}$, $\bar{\bar{Z}}^{MHJ}$, and $\bar{\bar{Z}}^{MHM}$ are calculated as follows:

$$b_{\xi}^{JE} = \left\langle \vec{b}_{\xi}^{J}, \vec{E}^{\text{inc}} \right\rangle_{\partial V_{\text{mat sys}}} \tag{2-33a}$$

$$b_{\xi}^{MH} = \left\langle \vec{b}_{\xi}^{M}, \vec{H}^{\text{inc}} \right\rangle_{\partial V_{\text{mat sys}}} \tag{2-33b}$$

$$z_{\xi\zeta}^{JEJ} = \left\langle \vec{b}_{\xi}^{J}, -j\omega\mu_{\text{mat}}\mathcal{L}\left(\vec{b}_{\zeta}^{J}\right) - j\omega\mu_0\mathcal{L}_0\left(\vec{b}_{\zeta}^{J}\right) \right\rangle_{\partial V_{\text{mat sys}}} \tag{2-34a}$$

$$z_{\xi\zeta}^{JEM} = \left\langle \vec{b}_{\xi}^{J}, -\text{P.V.}\mathcal{K}\left(\vec{b}_{\zeta}^{M}\right) - \text{P.V.}\mathcal{K}_0\left(\vec{b}_{\zeta}^{M}\right) \right\rangle_{\partial V_{\text{mat sys}}} \tag{2-34b}$$

$$z_{\xi\zeta}^{MHJ} = \left\langle \vec{b}_{\xi}^{M}, +\text{P.V.}\mathcal{K}\left(\vec{b}_{\zeta}^{J}\right) + \text{P.V.}\mathcal{K}_0\left(\vec{b}_{\zeta}^{J}\right) \right\rangle_{\partial V_{\text{mat sys}}} \tag{2-34c}$$

$$z_{\xi\zeta}^{MHM} = \left\langle \vec{b}_{\xi}^{M}, -j\omega\varepsilon_{\text{mat}}^{\text{c}}\mathcal{L}\left(\vec{b}_{\zeta}^{M}\right) - j\omega\varepsilon_0\mathcal{L}_0\left(\vec{b}_{\zeta}^{M}\right) \right\rangle_{\partial V_{\text{mat sys}}} \tag{2-34d}$$

In above formulations (2-34b) and (2-34c), symbol "P.V." represents the principal value of the related integral[107].

**2) Surface Version of the Calculation Formulation for Harrington's Material CMs**

Above impedance matrix $\bar{\bar{Z}}^{\text{SIE}}$ is a complex matrix, and the matrix can be decomposed as follows[34]:

$$\overline{\overline{Z}}^{\text{SIE}} = \overline{\overline{R}}^{\text{SIE}} + j\,\overline{\overline{X}}^{\text{SIE}} \tag{2-35}$$

where $\overline{\overline{R}}^{\text{SIE}} = \text{Re}\{\overline{\overline{Z}}^{\text{SIE}}\}$ and $\overline{\overline{X}}^{\text{SIE}} = \text{Im}\{\overline{\overline{Z}}^{\text{SIE}}\}$.

The Harrington's characteristic vectors of material systems can be derived from the following generalized characteristic equation[34]:

$$\overline{\overline{X}}^{\text{SIE}} \cdot \bar{a}_{\xi}^{JM} = \lambda_{\xi}^{\text{SIE}} \overline{\overline{R}}^{\text{SIE}} \cdot \bar{a}_{\xi}^{JM} \quad , \quad \xi = 1, 2, \cdots, \Xi^{JM} \tag{2-36}$$

where $\Xi^{JM} = \Xi^{J} + \Xi^{M}$. Using the characteristic vectors, the corresponding characteristic currents can be obtained by employing the basis-function-based expansion formulations of $\vec{J}^{\text{ES}}$ and $\vec{M}^{\text{ES}}$. The characteristic fields corresponding to $\{\vec{J}_{\xi}^{\text{ES}}, \vec{M}_{\xi}^{\text{ES}}\}$ can be expressed as follows:

$$\vec{E}_{\xi}^{\text{sca}}(\vec{r}) = +j\omega\mu_0\mathcal{L}_0(\vec{J}_{\xi}^{\text{ES}}) + \mathcal{K}_0(\vec{M}_{\xi}^{\text{ES}}) \quad , \quad \vec{r} \in \text{ext}V_{\text{mat sys}} \tag{2-37a}$$

$$\vec{H}_{\xi}^{\text{sca}}(\vec{r}) = +j\omega\varepsilon_0\mathcal{L}_0(\vec{M}_{\xi}^{\text{ES}}) - \mathcal{K}_0(\vec{J}_{\xi}^{\text{ES}}) \quad , \quad \vec{r} \in \text{ext}V_{\text{mat sys}} \tag{2-37b}$$

$$\vec{E}_{\xi}^{\text{inc}}(\vec{r}) = -j\omega\mu_0\mathcal{L}_0(\vec{J}_{\xi}^{\text{ES}}) - \mathcal{K}_0(\vec{M}_{\xi}^{\text{ES}}) \quad , \quad \vec{r} \in \text{int}V_{\text{mat sys}} \tag{2-38a}$$

$$\vec{H}_{\xi}^{\text{inc}}(\vec{r}) = -j\omega\varepsilon_0\mathcal{L}_0(\vec{M}_{\xi}^{\text{ES}}) + \mathcal{K}_0(\vec{J}_{\xi}^{\text{ES}}) \quad , \quad \vec{r} \in \text{int}V_{\text{mat sys}} \tag{2-38b}$$

$$\vec{E}_{\xi}^{\text{tot}}(\vec{r}) = -j\omega\mu_{\text{mat}}\mathcal{L}(\vec{J}_{\xi}^{\text{ES}}) - \mathcal{K}(\vec{M}_{\xi}^{\text{ES}}) \quad , \quad \vec{r} \in \text{int}V_{\text{mat sys}} \tag{2-39a}$$

$$\vec{H}_{\xi}^{\text{tot}}(\vec{r}) = -j\omega\varepsilon_{\text{mat}}^{\text{c}}\mathcal{L}(\vec{M}_{\xi}^{\text{ES}}) + \mathcal{K}(\vec{J}_{\xi}^{\text{ES}}) \quad , \quad \vec{r} \in \text{int}V_{\text{mat sys}} \tag{2-39b}$$

Above-mentioned $\{\lambda_{\xi}^{\text{SIE}}, \bar{a}_{\xi}^{JM}\}_{\xi=1}^{\Xi^{JM}}$ are called as the Harrington's characteristic pairs of material systems, and above-mentioned $\{\bar{a}_{\xi}^{JM}, \vec{J}_{\xi}^{\text{ES}}, \vec{M}_{\xi}^{\text{ES}}, \vec{E}_{\xi}^{\text{sca}}, \vec{H}_{\xi}^{\text{sca}}, \vec{E}_{\xi}^{\text{inc}}, \vec{H}_{\xi}^{\text{inc}}, \vec{E}_{\xi}^{\text{tot}}, \vec{H}_{\xi}^{\text{tot}}\}_{\xi=1}^{\Xi^{JM}}$ are collectively referred to as the Harrington's CMs of material systems.

### 2.3.4 On the Physical Pictures of the Harrington's CMTs for Various Scattering Systems

Based on Poynting's theorem and some simple vector operations, it is easy to conclude that the Harrington's metallic CMs derived from characteristic equation (2-18) satisfy the following orthogonality[31,32,35]:

$$P_{\xi}^{\text{rad}}\delta_{\xi\zeta} = (1/2)\oint\!\!\!\!\oint_{S_\infty}\left[\vec{E}_{\zeta}^{\text{sca}} \times \left(\vec{H}_{\xi}^{\text{sca}}\right)^{*}\right]\cdot \hat{n}_{\infty}^{+} dS \quad , \quad (\xi, \zeta = 1, 2, \cdots, \Xi) \tag{2-40}$$

Obviously, Harrington's metallic CMs are the same as Garbacz's CMs in the aspect of satisfying the modal far-field orthogonality[31,32]. Due to this, it, for a very long time, has been considered that Harrington's material CMs should also have orthogonal modal far fields. However, it was recently found out that the Harrington's material CMs derived from equation (2-26) satisfy the following orthogonality[44]:

$$\left(P_{\xi}^{\text{los}} + P_{\xi}^{\text{rad}}\right)\delta_{\xi\zeta} = (1/2)\left\langle \sigma_{\text{mat}}\vec{E}_{\xi}^{\text{tot}}, \vec{E}_{\zeta}^{\text{tot}}\right\rangle_{V_{\text{mat sys}}} + (1/2)\oiint_{S_\infty}\left[\vec{E}_{\zeta}^{\text{sca}} \times \left(\vec{H}_{\xi}^{\text{sca}}\right)^*\right]\cdot \hat{n}_\infty^+ dS \quad (2\text{-}41)$$

where $\xi, \zeta = 1, 2, \cdots, \Xi^{JM}$, and $P_{\xi}^{\text{los}}$ is the lossy power corresponding to the $\xi$-th CM. Obviously, for lossy material systems, the existence of term $(1/2) < \sigma_{\text{mat}}\vec{E}_{\xi}^{\text{tot}}, \vec{E}_{\zeta}^{\text{tot}} >_{V_{\text{mat sys}}}$ will cause that Harrington's material CMs don't have orthogonal modal far fields[57]. Then, it leads to the Perplex A given in Section 2.1 —— whether or not Harrington's CMs should satisfy far-field orthogonality?

In addition, based on Poynting's theorem and some simple vector operations, it is easy to conclude that the Harrington's metallic characteristic values $\lambda_{\xi}^{\text{EFIE}}$ calculated from characteristic equation (2-18) satisfy the following relationship[31,32,35,40,41]:

$$\lambda_{\xi}^{\text{EFIE}} = \frac{2\omega\left[(1/4)\left\langle \vec{H}_{\xi}^{\text{sca}}, \mu_0\vec{H}_{\xi}^{\text{sca}}\right\rangle_{\mathbb{R}^3} - (1/4)\left\langle \varepsilon_0\vec{E}_{\xi}^{\text{sca}}, \vec{E}_{\xi}^{\text{sca}}\right\rangle_{\mathbb{R}^3}\right]}{(1/2)\oiint_{S_\infty}\left[\vec{E}_{\xi}^{\text{sca}} \times \left(\vec{H}_{\xi}^{\text{sca}}\right)^*\right]\cdot \hat{n}_\infty^+ dS} \quad (2\text{-}42)$$

Due to this, it, for a very long time, has been considered that Harrington's material characteristic values should also satisfy a kind of relationship being similar to relationship (2-42). However, it was recently found out that the Harrington's material characteristic values $\lambda_{\xi}^{\text{VIE}}$ and $\lambda_{\xi}^{\text{SIE}}$ derived from equations (2-26) and (2-36) satisfy the following relationship[44,45]:

$$\lambda_{\xi}^{\text{SIE}} = \lambda_{\xi}^{\text{VIE}} = \frac{P_1}{P_2} \quad (2\text{-}43)$$

where

$$\begin{aligned} P_1 = {} & 2\omega\left[(1/4)\left\langle \vec{H}_{\xi}^{\text{sca}}, \mu_0\vec{H}_{\xi}^{\text{sca}}\right\rangle_{\mathbb{R}^3} - (1/4)\left\langle \varepsilon_0\vec{E}_{\xi}^{\text{sca}}, \vec{E}_{\xi}^{\text{sca}}\right\rangle_{\mathbb{R}^3}\right] \\ & + 2\omega\left[(1/4)\left\langle \vec{H}_{\xi}^{\text{tot}}, \Delta\mu_{\text{mat}}\vec{H}_{\xi}^{\text{tot}}\right\rangle_{V_{\text{mat sys}}} - (1/4)\left\langle \Delta\varepsilon_{\text{mat}}\vec{E}_{\xi}^{\text{tot}}, \vec{E}_{\xi}^{\text{tot}}\right\rangle_{V_{\text{mat sys}}}\right] \\ & - 4\omega\,\text{Re}\left\{(1/4)\left\langle \vec{H}_{\xi}^{\text{inc}}, \Delta\mu_{\text{mat}}\vec{H}_{\xi}^{\text{tot}}\right\rangle_{V_{\text{mat sys}}}\right\} \quad (2\text{-}44\text{a}) \end{aligned}$$

$$P_2 = (1/2)\left\langle \sigma_{\text{mat}}\vec{E}_{\xi}^{\text{tot}}, \vec{E}_{\xi}^{\text{tot}}\right\rangle_{V_{\text{mat sys}}} + (1/2)\oiint_{S_\infty}\left[\vec{E}_{\xi}^{\text{sca}} \times \left(\vec{H}_{\xi}^{\text{sca}}\right)^*\right]\cdot \hat{n}_\infty^+ dS \quad (2\text{-}44\text{b})$$

Obviously, for the magnetic material systems whose permeabilities $\mu_{\text{mat}}$ are different from $\mu_0$, the existence of term $4\omega\text{Re}\{(1/4) < \vec{H}_{\xi}^{\text{inc}}, \Delta\mu_{\text{mat}}\vec{H}_{\xi}^{\text{tot}} >_{V_{\text{mat sys}}}\}$ will cause that the physical meaning of Harrington's material characteristic values is essentially different from the physical meaning of Harrington's metallic characteristic values[44,45]. Then, it leads to the Perplex B given in Section 2.1 —— whether or not Harrington's characteristic value should equal the ratio of modal imaginary power to modal real power?

On the foundations of the works of literatures [37,40,41,44,45], this dissertation, by some studies, finds out that: the reason leading to above Perplexes A and B and the Perplex C given in Section 2.1 is the lack of a clear physical picture of Harrington's CMT (which was established in IE framework). In the following Section 2.4, we will, based on the works of literatures [37,40,41,44,45], transform Harrington's CMT from its traditional IE framework to a completely new framework —— WEP framework, and develop a completely new constructing method for Harrington's CMs. In the new framework and based on the new method, we will draw a clear physical picture for Harrington's CMT, and then eliminate Perplexes A~C by employing the physical picture.

## 2.4 CMT in WEP Framework and Its Physical Picture

Literatures [37,40,41] pointed out that the essence of Harrington's EFIE-OpeMetSca-CMT[31,32] is to construct a series of CMs which can orthogonalize frequency-domain power operator $P_{\text{met sys}}^{\text{Driving}} = (1/2) < \vec{J}^{\text{sca}}, \vec{E}^{\text{inc}} >_{\partial D_{\text{met sys}}}$, and the obtained CMs satisfy the following orthogonality[37,40,41]:

$$P_{\text{met sys};\xi}^{\text{Driving}} \delta_{\xi\zeta} = (1/2) \left\langle \vec{J}_{\xi}^{\text{sca}}, \vec{E}_{\zeta}^{\text{inc}} \right\rangle_{\partial D_{\text{met sys}}} \quad , \quad (\xi, \zeta = 1, 2, \cdots, \Xi) \tag{2-45}$$

Literatures [44,45] pointed out that the essence of Harrington's VIE-MatSca-CMT[33] is to construct a series of CMs which can orthogonalize frequency-domain power operator $P_{\text{mat sys}}^{\text{driving}} = (1/2) < \vec{J}^{\text{SV}}, \vec{E}^{\text{inc}} >_{V_{\text{mat sys}}} + (1/2) < \vec{M}^{\text{SV}}, \vec{H}^{\text{inc}} >_{V_{\text{mat sys}}}$, and that the essence of Harrington's SIE-MatSca-CMT[34] is to construct a series of CMs which can orthogonalize frequency-domain power operator $-(1/2) < \vec{J}^{\text{ES}}, \vec{E}^{\text{inc}} >_{\partial V_{\text{mat sys}}}$ $-(1/2) < \vec{M}^{\text{ES}}, \vec{H}^{\text{inc}} >_{\partial V_{\text{mat sys}}}$, and the above two power operators equal to each other, i.e.,

$$\begin{aligned} P_{\text{mat sys}}^{\text{driving}} &= (1/2) \left\langle \vec{J}^{\text{SV}}, \vec{E}^{\text{inc}} \right\rangle_{V_{\text{mat sys}}} + (1/2) \left\langle \vec{M}^{\text{SV}}, \vec{H}^{\text{inc}} \right\rangle_{V_{\text{mat sys}}} \\ &= -(1/2) \left\langle \vec{J}^{\text{ES}}, \vec{E}^{\text{inc}} \right\rangle_{\partial V_{\text{mat sys}}} - (1/2) \left\langle \vec{M}^{\text{ES}}, \vec{H}^{\text{inc}} \right\rangle_{\partial V_{\text{mat sys}}} \end{aligned} \tag{2-46}$$

Thus, the CMs derived from Harrington's VIE-MatSca-CMT[33] and SIE-MatSca-CMT[34] satisfy the following orthogonality[44,45]:

$$P_{\text{mat sys};\xi}^{\text{driving}} \delta_{\xi\zeta} = (1/2) \left\langle \vec{J}_{\xi}^{\text{SV}}, \vec{E}_{\zeta}^{\text{inc}} \right\rangle_{D_{\text{mat sys}}} + (1/2) \left\langle \vec{M}_{\xi}^{\text{SV}}, \vec{H}_{\zeta}^{\text{inc}} \right\rangle_{D_{\text{mat sys}}} \ , \ \left(\xi, \zeta = 1, 2, \cdots, \Xi^{JM}\right) \tag{2-47}$$

Evidently, to clarify the physical meanings of power operators $P_{\text{met sys}}^{\text{Driving}}$ and $P_{\text{mat sys}}^{\text{driving}}$ is a necessary condition to effectively reveal the physical picture of Harrington's CMT. Thus, in the following Subsections 2.4.1 and 2.4.2, we will study the physical meanings of $P_{\text{met sys}}^{\text{Driving}}$ and $P_{\text{mat sys}}^{\text{driving}}$ firstly, and then draw a clear physical picture for Harrington's CMT.

### 2.4.1 Drawing the Physical Picture of Harrington's EFIE-OpeMetSca-CMT in WEP Framework

Literature [44] pointed out that: frequency-domain power operator $P_{\text{met sys}}^{\text{Driving}}$ is just the only frequency-domain form of time-domain power operator $P_{\text{met sys}}^{\text{Driving}}(t) =< \vec{J}^{\text{sca}}(t), \vec{E}^{\text{inc}}(t) >_{\partial D_{\text{met sys}}}$. The Chapter 3 of this dissertation will further derive the following time-domain power equation①:

$$P_{\text{met sys}}^{\text{Driving}}(t) = \left\langle \vec{J}^{\text{sca}}(t), \vec{E}^{\text{inc}}(t) \right\rangle_{\partial D_{\text{met sys}}} = P^{\text{rad}}(t) + \frac{d}{dt}\left[ W_{\text{vac}}^{\text{mag}}(t) + W_{\text{vac}}^{\text{ele}}(t) \right] \quad (2\text{-}48)$$

In equation (2-48), $P^{\text{rad}}(t)$ is the transient radiated power carried by scattered fields; $W_{\text{vac}}^{\text{mag}}(t)$ and $W_{\text{vac}}^{\text{ele}}(t)$ are the transient magnetic and electric energies stored in scattered fields. Integrating the time variable in equation (2-48) from time point $t_0$ to time point $t_0 + \Delta t$ (where $\Delta t > 0$), the following energy equation can be obtained[117]:

$$\begin{aligned} \int_{t_0}^{t_0+\Delta t} P_{\text{met sys}}^{\text{Driving}}(\tau) d\tau &= \int_{t_0}^{t_0+\Delta t} \left\langle \vec{J}^{\text{sca}}(\tau), \vec{E}^{\text{inc}}(\tau) \right\rangle_{\partial D_{\text{met sys}}} d\tau \\ &= \int_{t_0}^{t_0+\Delta t} P^{\text{rad}}(\tau) d\tau \\ &\quad + \left[ W_{\text{vac}}^{\text{mag}}(t_0+\Delta t) + W_{\text{vac}}^{\text{ele}}(t_0+\Delta t) \right] - \left[ W_{\text{vac}}^{\text{mag}}(t_0) + W_{\text{vac}}^{\text{ele}}(t_0) \right] \quad (2\text{-}49) \end{aligned}$$

Obviously, the above energy equation has a very clear physical meaning: in time interval $[t_0, t_0 + \Delta t]$, the work done by the incident field $\vec{E}^{\text{inc}}(t)$ acting on scattered source $\vec{J}^{\text{sca}}(t)$ is transverted (transfer & convert) to two parts —— a part is transferred to infinity by the radiative scattered field & the other part is converted to the energy stored in the reactive scattered field.

In fact, the energy equation (2-49) for the metallic systems in classical electromagnetics is just the counterpart of the work-energy principle (WEP) for the particles in classical mechanics. The WEP for the particles in classical mechanics is usually stated as follows[104]:

---

① To facilitate the systematical discussions for the related contents in Chapter 3, we will provide the detailed derivations for this equation in Section 3.2 instead of providing them here.

"… the work done by all forces acting on a particle (the work of the resultant force) equals the change in the kinetic energy of the particle. [104]"

Thus, this dissertation calls energy equation (2-49) as the work-energy principle (i.e. the principle on the transversion between work and energy) for the metallic systems in classical electromagnetics. In addition, this dissertation calls the energy origin $P_{\text{met sys}}^{\text{Driving}}(t) = < \vec{J}^{\text{sca}}(t), \vec{E}^{\text{inc}}(t) >_{\partial D_{\text{met sys}}}$, which sustains the energy transfer and conversion, as time-domain driving power (DP), and this is just the reason why this dissertation uses superscript "Driving" in $P_{\text{met sys}}^{\text{Driving}}(t)$ and its frequency-domain version $P_{\text{met sys}}^{\text{Driving}}$.

As everyone knows, for a quadratic quantity corresponding to time-harmonic EM fields, its time-domain version $Q(t)$ and frequency-domain version $Q$ satisfy relationship $\text{Re}\{Q\} = (1/T)\int_{t_0}^{t_0+T} Q(\tau) d\tau$ [9,114], where $T$ is the time period of time-harmonic EM fields. Thus, frequency-domain orthogonality (2-45) implies the following time-domain orthogonality:

$$\left[T \cdot \text{Re}\left\{P_{\text{met sys};\xi}^{\text{Driving}}\right\}\right]\delta_{\xi\zeta} = \int_{t_0}^{t_0+T} \left\langle \vec{J}_{\xi}^{\text{sca}}(\tau), \vec{E}_{\zeta}^{\text{inc}}(\tau) \right\rangle_{\partial D_{\text{met sys}}} d\tau \quad , \quad (\xi, \zeta = 1, 2, \cdots, \Xi) \tag{2-50}$$

Then, the essential physical destination of Harrington's EFIE-OpeMetSca-CMT[31,32] is to construct a series of orthogonal modes which don't have net energy exchange in any integral period. This is just the physical picture which this dissertation draws for Harrington's EFIE-OpeMetSca-CMT[31,32].

Based on the physical picture, it is easy to reveal that: the MFIE-based[58] and CFIE-based[59,60] CMTs for metallic systems are essentially different from Harrington's EFIE-OpeMetSca-CMT[31,32]. In fact, the essential difference between Harrington's EFIE-OpeMetSca-CMT[31,32] and the metallic CMT based on compelx background Green's functions can also be revealed by employing the physical picture, but we want to defer the related discussions until the Chapter 3 of this dissertation. So far, the Perplex C mentioned in Section 2.1 has been eliminated.

## 2.4.2 Drawing the Physical Pictures of Harrington's VIE-MatSca-CMT and SIE-MatSca-CMT in WEP Framework

Literature [44] pointed out that: frequency-domain power operator $P_{\text{mat sys}}^{\text{driving}} = (1/2) < \vec{J}^{\text{SV}}, \vec{E}^{\text{inc}} >_{V_{\text{mat sys}}} + (1/2) < \vec{M}^{\text{SV}}, \vec{H}^{\text{inc}} >_{V_{\text{mat sys}}}$ is the one of two frequency-domain forms of time-domain power operator $P_{\text{mat sys}}^{\text{Driving}}(t) = < \vec{J}^{\text{SV}}(t), \vec{E}^{\text{inc}}(t) >_{V_{\text{mat sys}}} + < \vec{H}^{\text{inc}}(t), \vec{M}^{\text{SV}}(t) >_{V_{\text{mat sys}}}$ , and the other frequency-domain form is $P_{\text{mat sys}}^{\text{DRIVING}} = (1/2) < \vec{J}^{\text{SV}}, \vec{E}^{\text{inc}} >_{V_{\text{mat sys}}} + (1/2) < \vec{H}^{\text{inc}}, \vec{M}^{\text{SV}} >_{V_{\text{mat sys}}}$ . The Chapter 4 of this dissertation will further derive the following time-domain power equation①:

$$P_{\text{mat sys}}^{\text{Driving}}(t) = \left\langle \vec{J}^{\text{SV}}(t), \vec{E}^{\text{inc}}(t) \right\rangle_{V_{\text{mat sys}}} + \left\langle \vec{H}^{\text{inc}}(t), \vec{M}^{\text{SV}}(t) \right\rangle_{V_{\text{mat sys}}}$$

$$= P^{\text{rad}}(t) + P^{\text{los}}(t) + \frac{d}{dt}\left[W_{\text{vac}}^{\text{mag}}(t) + W_{\text{vac}}^{\text{ele}}(t)\right] + \frac{d}{dt}\left[W_{\text{mat}}^{\text{mag}}(t) + W_{\text{mat}}^{\text{pol}}(t)\right] \quad (2\text{-}51)$$

In the above power equation, $P^{\text{los}}(t)$ is the transient lossy power of the objective material system; $P^{\text{rad}}(t)$ is the transient radiated power carried by the scattered EM fields; $W_{\text{vac}}^{\text{mag}}(t)$ and $W_{\text{vac}}^{\text{ele}}(t)$ are respectively the transient magnetic and electric energies stored in the scattered magnetic and electric fields; $W_{\text{mat}}^{\text{mag}}(t)$ and $W_{\text{mat}}^{\text{pol}}(t)$ are respectively the transient magnetization and polarization energies stored in the matters constituting the objective material system. Integrating power equation (2-51) in time interval $[t_0, t_0 + \Delta t]$, the following energy equation can be obtained[117]:

$$\begin{aligned}
\int_{t_0}^{t_0+\Delta t} P_{\text{mat sys}}^{\text{Driving}}(\tau)\, d\tau &= \int_{t_0}^{t_0+\Delta t} \left[ \left\langle \vec{J}^{\text{SV}}(\tau), \vec{E}^{\text{inc}}(\tau) \right\rangle_{V_{\text{mat sys}}} + \left\langle \vec{H}^{\text{inc}}(\tau), \vec{M}^{\text{SV}}(\tau) \right\rangle_{V_{\text{mat sys}}} \right] d\tau \\
&= \int_{t_0}^{t_0+\Delta t} P^{\text{rad}}(\tau)\, d\tau + \int_{t_0}^{t_0+\Delta t} P^{\text{los}}(\tau)\, d\tau \\
&\quad + \left[W_{\text{vac}}^{\text{mag}}(t_0 + \Delta t) + W_{\text{vac}}^{\text{ele}}(t_0 + \Delta t)\right] - \left[W_{\text{vac}}^{\text{mag}}(t_0) + W_{\text{vac}}^{\text{ele}}(t_0)\right] \\
&\quad + \left[W_{\text{mat}}^{\text{mag}}(t_0 + \Delta t) + W_{\text{mat}}^{\text{pol}}(t_0 + \Delta t)\right] - \left[W_{\text{mat}}^{\text{mag}}(t_0) + W_{\text{mat}}^{\text{pol}}(t_0)\right] \quad (2\text{-}52)
\end{aligned}$$

Obviously, the above energy equation has a very clear physical meaning: in time interval $[t_0, t_0 + \Delta t]$, the work done by the incident fields $\{\vec{E}^{\text{inc}}(t), \vec{H}^{\text{inc}}(t)\}$ acting on scattered sources $\{\vec{J}^{\text{SV}}(t), \vec{M}^{\text{SV}}(t)\}$ is transverted (transfer & convert) to four parts —— the part transferred to infinity by the radiative scattered field & the part converted to Joule heat & the part converted to the energy stored in the reactive scattered field & the part converted to the energy stored in the magnetized and polarized matter.

In fact, the energy equation (2-52) for the material systems in classical electromagnetics is just the counterpart of the work-energy principle (WEP) for the particles in classical mechanics, so this dissertation calls energy equation (2-52) as the work-energy principle for the material systems in classical electromagnetics. In addition, this dissertation calls the energy origin $P_{\text{mat sys}}^{\text{Driving}}(t) = < \vec{J}^{\text{SV}}(t), \vec{E}^{\text{inc}}(t) >_{V_{\text{mat sys}}} + < \vec{H}^{\text{inc}}(t), \vec{M}^{\text{SV}}(t) >_{V_{\text{mat sys}}}$, which sustains the energy transfer and the energy conversion, as time-domain driving power, and this is just the reason why this dissertation uses

① To facilitate the systematical discussions for the related contents in Chapter 4, we will provide the detailed derivations for this equation in Section 4.2 instead of providing them here.

superscripts "Driving", "driving", and "DRIVING" in $P_{\text{mat sys}}^{\text{Driving}}(t)$ and its frequency-domain versions $P_{\text{mat sys}}^{\text{driving}}$ and $P_{\text{mat sys}}^{\text{DRIVING}}$. Here, two different frequency-domain superscripts "driving" and "DRIVING" are to distinguish two different frequency-domain versions $P_{\text{mat sys}}^{\text{driving}}$ and $P_{\text{mat sys}}^{\text{DRIVING}}$ from each others.

Similarly to the orthogonality (2-50) for Harrington's metallic CMs, the frequency-domain orthogonality (2-47) for Harrington's material CMs implies the following time-domain orthogonality:

$$\left[T \cdot \text{Re}\left\{P_{\text{mat sys};\xi}^{\text{driving}}\right\}\right]\delta_{\xi\zeta} = \int_{t_0}^{t_0+T}\left[\left\langle \vec{J}_{\xi}^{\text{SV}}(\tau), \vec{E}_{\zeta}^{\text{inc}}(\tau)\right\rangle_{V_{\text{mat sys}}} + \left\langle \vec{M}_{\xi}^{\text{SV}}(\tau), \vec{H}_{\zeta}^{\text{inc}}(\tau)\right\rangle_{V_{\text{mat sys}}}\right]d\tau \quad (2\text{-}53)$$

where $\xi, \zeta = 1, 2, \cdots, \Xi^{JM}$.

It is thus evident that the common physical destination of VIE-MatSca-CMT[33] and SIE-MatSca-CMT[34] is to construct a series of orthogonal modes not having net energy exchange in any integral period, and this is just the physical picture which this dissertation draws for the Harrington's CMT for material systems. In the Section 4.2 of this dissertation, we will further prove that: only to orthogonalize frequency-domain DP operator (DPO) $P_{\text{mat sys}}^{\text{driving}}$, the above physical destination can be effectively achieved; the above physical destination cannot be achieved by orthogonalizing frequency-domain DPO $P_{\text{mat sys}}^{\text{DRIVING}}$.

Based on the physical picture, the Perplexes A and B given in Section 2.1 can be easily eliminated: **A.** The physical destination of Harrington's VIE-MatSca-CMT[33] and SIE-MatSca-CMT[34] is not to orthogonalize modal far fields, so the Harrington's CMs derived from VIE-MatSca-CMT[33] and SIE-MatSca-CMT[34] can have non-orthogonal modal far fields; **B.** the uniformity between the {VIE-MatSca-CMT[33], SIE-MatSca-CMT[34]} for material systems and the EFIE-OpeMetSca-CMT[31,32] for metallic systems is reflected in the aspect of their physical pictures (i.e., they have a common physical destination —— to construct a series of orthogonal modes not having net energy exchange in any integral period) rather than their characteristic values, so their characteristic values can have different forms. At this point, the Perplexes A and B mentioned in Section 2.1 have been eliminated successfully.

### 2.4.3 Transforming the Carrying Framework of Harrington's CMT From IE to WEP

In fact, on the foundations of the works done by literatures [37,40,41,44,45,56],

above Sections 2.4.1 and 2.4.2 accomplish the transformation for the carrying framework of Harrington's CMT (EFIE-OpeMetSca-CMT[31,32], VIE-MatSca-CMT[33], and SIE-MatSca-CMT[34]) —— from IE framework to WEP framework, as shown in Figure 2-3. Following the framework transformation, we obtain the transformation for the constructing method of Harrington's CMs —— from orthogonalizing IMO method to orthogonalizing DPO method, as shown in Figure 2-3.

| | | | | |
|---|---|---|---|---|
| SM framework ↓ IE framework ↓ WEP framework | Two turns of the transformations for carrying framework | & | Two turns of the transformations for constructing method | Orthogonalizing PMO method ↓ Orthogonalizing IMO method ↓ Orthogonalizing DPO method |

Figure 2-3 Two turns of transformations for the carrying framework of CMT and the constructing method of CMs

The above-mentioned second transformation gives a clear physical picture —— constructing a series of steadily working modes not having net energy exchange in any integral period —— to Harrington's CMT.

Based on the above several conclusions, we can further complete the Table 2-1 given in Section 2.1 as the following Table 2-2.

Table 2-2 Comparisons among Garbacz's CMT, Harrington's CMT, and driving power CMT from the aspects of carrying framework, constructing method, and physical picture

| **Two Theories** / **Three Aspects** | **Garbacz's CMT[29,30]** | **Harrington's CMT (Driving Power CMT)** | |
|---|---|---|---|
| | | **Scheme Used in Literatures [31-34]** | **Scheme Used in This Dissertation** |
| **Carrying Framework** | SM framework | IE framework | WEP framework |
| **Constructing Method** | Orthogonalizing PMO method | Orthogonalizing IMO method | Orthogonalizing DPO method |
| **Physical Picture** | Constructing a series of modes having orthogonal modal far fields | | Constructing a series of steadily working modes not having net energy exchange in any integral period |

The Sections 2.4.1 and 2.4.2 of this chapter have exhibited that: the above second transformation brings a clear physical picture such that some perplexes existing in IE framework can be successfully eliminated. However, the effect of the above second transformation is not restricted to this. In fact, the following chapters will exhibit the facts that: the new carrying framework and new constructing method change the whole face of Harrington's CMT; the old problems OP2~OP5 mentioned in Subsection 1.3.1 can be successfully solved in the new carrying framework and based on the new constructing method (just like that this chapter successfully solves the old problem OP1 mentioned in Subsection 1.3.1); the new problems NP2 and NP5 mentioned in Subsection 1.3.2 can be successfully solved in the new carrying framework and based on the new constructing method (just like that this chapter successfully solves the new problem NP1 mentioned in Subsection 1.3.2).

To emphasize the physical picture of Harrington's CMT, and also to emphasize that the following parts of this dissertation are discussed in WEP framework and based on orthogonalizing DPO method, we will alternatively call "Harrington's CMT" and "Harrington's CMs" as "driving power CMT" and "driving power CMs (DP-CMs)" respectively. In fact, this is just the reason why we label "driving power CMT" after the "Harrington's CMT" in Table 2-2.

## 2.5 Chapter Summary

Firstly, this chapter simply retrospects the carrying framework of Garbacz's CMT —— SM framework, and the constructing method for Garbacz's CMs —— orthogonalizing PMO method, and the physical picture of Garbacz's CMT —— constructing a series of fundamental modes having orthogonal modal far fields.

Afterwards, this chapter simply retrospects the carrying framework of Harrington's CMT —— IE framework, and the constructing method for Harrington's CMs —— orthogonalizing IMO method. At the same time, this chapter explicitly points out that the physical picture of Harrington's CMT is different from the physical picture of Garbacz's CMT, and a typical exemplification of this conclusion is that: the Harrington's CMs of lossy material systems don't have orthogonal modal far fields.

After that, based on the works of literatures [37,40,41,44,45], this chapter, for the first time, gives Harrington's CMT a clear physical picture —— constructing a series of steadily working modes not having net energy exchange in any integral period. The physical picture, for the first time, clarifies the ultimate cause leading to the different physical meanings between Harrington's metallic characteristic value and Harrington's material characteristic value. The physical picture, for the first time, reveals the essential differences between the original Harrington's CMT and its some variants in IE framework (such as the MFIE-based and CFIE-based CMTs for metallic systems).

Finally, during the process to reveal the physical picture of Harrington's CMT, this chapter establishes a completely new carrying framework for Harrington's CMT —— WEP framework, and at the same time develops a completely new constructing method for Harrington's CMs —— orthogonalizing DPO method. Compared with the traditional IE framework and orthogonalizing IMO method, the new carrying framework and constructing method can clearly reveal the physical picture of Harrington's CMT, so it is difficult to derive the inappropriate variants (which have the different physical pictures from the original Harrington's CMT) in the new framework and based on the new method. However, the effect of the new framework and the new method is not restricted to this. In fact, the following chapters will exhibit the facts that: the new carrying framework and the new constructing method change the whole face of Harrington's CMT; the old problems OP2~OP5 given in Subsection 1.3.1 can be successfully solved in the new carrying framework and based on the new constructing method (just like that this chapter successfully solves the old problem OP1 given in Subsection 1.3.1); the new problems NP2 and NP5 given in Subsection 1.3.2 can be successfully solved in the new carrying framework and based on the new constructing method (just like that this chapter successfully solves the new problem NP1 given in Subsection 1.3.2).

The above-mentioned various carrying frameworks, constructing methods, and physical pictures are listed in Table 2-2, and they will not be repeated here. In new WEP framework and based on new orthogonalizing DPO method, the following chapters will, for various scattering systems, construct a series of steadily working CMs not having net energy exchange in any integral period. To emphasize the CM construction method used in the following chapters —— orthogonalizing DPO method, we will call the CMs constructed in the following chapters as driving power CMs (DP-CMs).

# Chapter 3 WEP-Based DP-CMs of Metallic Scattering Systems

> The objective of this chapter is to introduce the energy point of view into the study of electromagnetic fields. The point of view is of great importance to us …
>
> …
>
> Our faith in the validity of the law of conservation of energy suggests that it should be possible to define energy functions for electromagnetic fields expressing the energy supplied to a field or delivered by it.[110]
>
> —— R. M. Fano, L. J. Chu, and R. B. Adler

This chapter, in WEP framework, focuses on constructing the DP-CMs of metallic systems (which don't have net energy exchange in any integral period) by orthogonalizing frequency-domain DPO, and doing some necessary analysis and discussions for the related topics.

## 3.1 Chapter Introduction

Metallic scattering systems widely exist in microwave remote sensing[3-5], target recognition[8], and antenna engineering[35], etc. domains. Analyzing and designing the inherent EM scattering characters of objective metallic systems plays a significant role in the above domains. Harrington's EFIE-OpeMetSca-CMT[31,32] has been widely applied to analyzing and designing the inherent EM scattering characters of open objective metallic systems, but there still exist some imperfections in the fundamental theoretical aspect of EFIE-OpeMetSca-CMT[40,57].

In WEP framework and based on orthogonalizing DPO method, this chapter points out that the CM set derived from the original EFIE-OpeMetSca-CMT[31,32] is incomplete, and develops a practical method to complete the CM set. In WEP framework and based on the complete CM set, this chapter, for an arbitrary metallic system, realizes the orthogonal decomposition for the modal space and the detailed modal classification for the working modes, and then accomplishes the simplest orthogonal decompositions for all types of working modes. The simplest orthogonal decompositions clearly reveal the working mechanisms of all types of working modes, so it has instructional significance to analyzing and designing the inherent EM scattering characters of metallic systems.

In addition, this chapter will review and compare the traditional Harrington's CMs, eigen-modes, natural modes, and resonant modes of metallic systems from the perspectives of WEP and DP.

## 3.2 DP-CMs of Metallic Systems

We now consider the EM scattering problem shown in Figure 3-1, where $D_{\text{met sys}}$, $D_{\text{env}}$, and $D_{\text{imp}}$ are the domains occupied by metallic system, external EM environment, and external impressed source respectively①. We denote the EM field generated by $D_{\text{imp}}$ as $\vec{F}_{\text{imp}}(t)$ ②. Under the excitation of $\vec{F}_{\text{imp}}(t)$, some electric currents $\vec{J}^{\text{sca}}(t)$ will be induced on the boundary $\partial D_{\text{met sys}}$ of $D_{\text{met sys}}$, and some electric currents $\vec{J}_{\text{env}}(t)$ and magnetic current $\vec{M}_{\text{env}}(t)$ will be induced on $D_{\text{env}}$ ③. If the fields generated by $\vec{J}^{\text{sca}}(t)$ and $\{\vec{J}_{\text{env}}(t), \vec{M}_{\text{env}}(t)\}$ are denoted as $\vec{F}^{\text{sca}}(t)$ and $\vec{F}_{\text{env}}(t)$ respectively, then it can be concluded, based on superposition principle[110-114], that: field $\vec{F}_{\text{imp}}(t) + \vec{F}_{\text{env}}(t)$ is just the resultant field used to drive the working of $D_{\text{met sys}}$. To keep the consistency of the terminologies in classical EM scattering theory[106-108], this dissertation equivalently calls the resultant field as incident field, and correspondingly denotes it as $\vec{F}^{\text{inc}}(t)$, i.e., $\vec{F}^{\text{inc}}(t) = \vec{F}_{\text{imp}}(t) + \vec{F}_{\text{env}}(t)$ ④. Correspondingly, this dissertation calls $\vec{J}^{\text{sca}}(t)$ and $\vec{F}^{\text{sca}}(t)$ as scattered electric current and scattered field respectively.

If $D_{\text{met sys}}$, $D_{\text{env}}$, and $D_{\text{imp}}$ are the linear structures which satisfy the law of causality, then: for a certain $D_{\text{met sys}}$, any given $\vec{F}^{\text{inc}}(t)$ will uniquely result in a $\vec{J}^{\text{sca}}(t)$

① In this dissertation, metallic scattering systems are just the scattering systems constructed by perfectly electric conductors (PECs). In the following parts of this dissertation, metallic scattering system, external EM environment, and external impressed excitation are simply called as metallic system, external environment, and external excitation respectively. In addition, symbols $D_{\text{met sys}}$, $D_{\text{env}}$, and $D_{\text{imp}}$ are used to represent not only the corresponding domains but also the metallic system, external environment, and external excitation themselves.

② In the following parts of this dissertation, EM field is simply called as field, and $F = E, H$. For vectorial fields and vectorial sources, we especially mark them with an over single arrow, such as $\vec{F}_{\text{imp}}$ and $\vec{J}^{\text{sca}}$, to distinguish them from the dyadic fields, dyadic sources, algebraic vectors, and algebraic matrices to appear later. In the aspect of designing symbols, this dissertation follows the philosophies that: **1.** we never use similar symbols to represent different kinds of quantities (physical quantity, geometrical quantity, and algebraical quantity, etc.) to avoid confusions; **2.** to simplify the symbols appearing in this dissertation, we try to avoid using unnecessary distinction methods for different kinds of quantities (for example, the difference between field vector $\vec{F}_{\text{imp}}$ and a scalar has been reflected in the over single arrow, so we will not further write $\vec{F}_{\text{imp}}$ as bold type $\boldsymbol{\vec{F}}_{\text{imp}}$).

③ On the induced electric and magnetic currents distributing on $D_{\text{env}}$, we will provide some detailed discussions in Appendix A, and we will not discuss them here.

④ The principle foundations for treating $\vec{F}_{\text{imp}}$ and $\vec{F}_{\text{env}}$ as a whole will be given in Subsection 3.2.1, Subsection 3.4.1, Subsection 4.2.1, and Appendix C.

distributing on $\partial D_{\text{met sys}}$ [106,108,110-114]; every possible $\vec{J}^{\text{sca}}(t)$ corresponds to a working mode of $D_{\text{met sys}}$; all possible working modes constitute a linear space[106,108,115], and this dissertation calls the space as modal space[37].

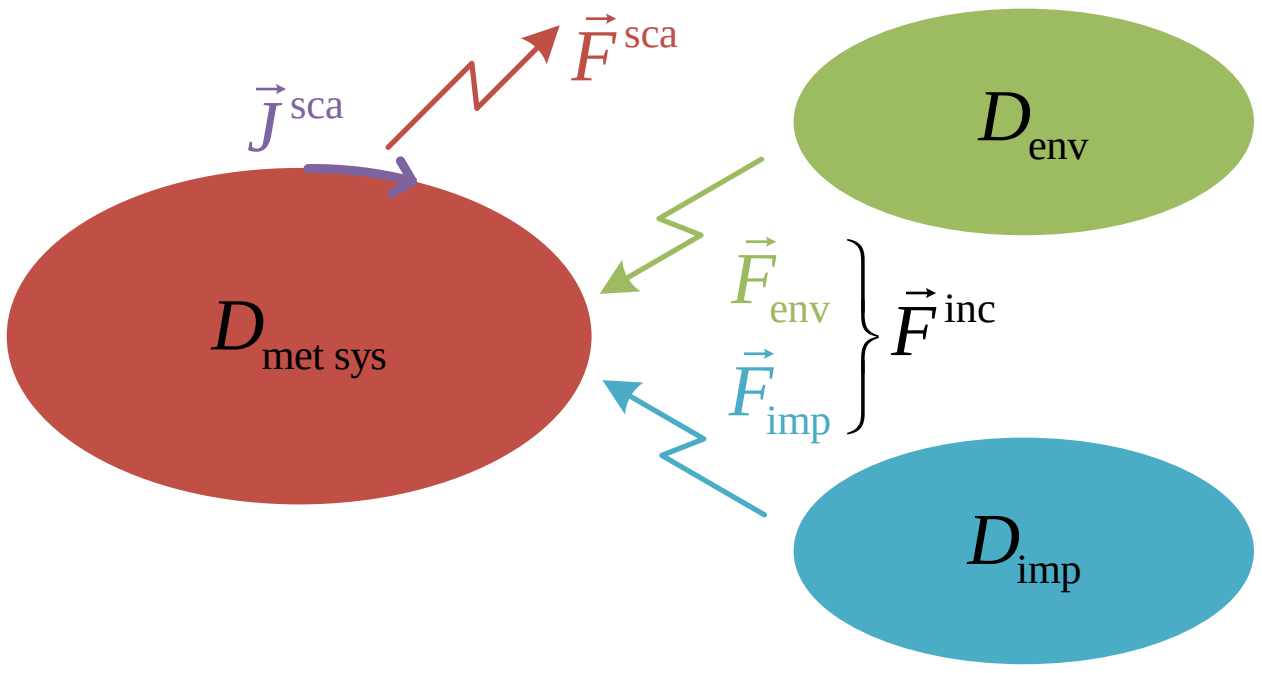

Figure 3-1 Metal scattering problem considered in Chapter 3

### 3.2.1 WEP and DP

In time domain, the instantaneous power $P_{\text{met sys}}^{\text{Driving}}(t)$ done by incident fields $\{\vec{E}^{\text{inc}}(t), \vec{H}^{\text{inc}}(t)\}$ on scattered source $\vec{J}^{\text{sca}}(t)$ is as follows:

$$\begin{aligned} P_{\text{met sys}}^{\text{Driving}}(t) &= \left\langle \vec{J}^{\text{sca}}(t), \vec{E}^{\text{inc}}(t) \right\rangle_{\partial D_{\text{met sys}}} \\ &= \left\langle \vec{J}^{\text{sca}}(t), \vec{E}^{\text{tot}}(t) - \vec{E}^{\text{sca}}(t) \right\rangle_{\partial D_{\text{met sys}}} \\ &= -\left\langle \vec{J}^{\text{sca}}(t), \vec{E}^{\text{sca}}(t) \right\rangle_{\partial D_{\text{met sys}}} \\ &= P^{\text{rad}}(t) + \frac{d}{dt}\left[ W_{\text{vac}}^{\text{mag}}(t) + W_{\text{vac}}^{\text{ele}}(t) \right] \end{aligned} \tag{3-1}$$

where the inner product is defined as that $< \vec{f}, \vec{g} >_{\Omega} = \int_{\Omega} \vec{f}^{*} \cdot \vec{g} \, d\Omega$, where superscript "$*$" represents the conjugate operation for a complex number, and symbol "$\cdot$" represents the dot product operation between two field vectors. In formulation (3-1), the first equality is based on the fact that the Lorentz's force acting on $\vec{J}^{\text{sca}}(t)$ by $\vec{H}^{\text{inc}}(t)$ doesn't do any work[110-114]; the second equality is based on superposition principle $\vec{F}^{\text{tot}}(t) = \vec{F}^{\text{inc}}(t) + \vec{F}^{\text{sca}}(t)$ [110-114]; the third equality is based on the tangential electric field boundary condition $\vec{E}_{\text{tan}}^{\text{tot}}(t) = 0$ on metallic boundary $\partial D_{\text{met sys}}$ [106-108]; the fourth equality is based on time-domain Poynting's theorem[110-113]. On the right-hand side (RHS) of the fourth equality in formulation (3-1), $P^{\text{rad}}(t)$ is the instantaneous radiated power carried by scattered field, and $W_{\text{vac}}^{\text{mag}}(t)$ and $W_{\text{vac}}^{\text{ele}}(t)$ are respectively the instantaneous magnetic and electric energies in vacuum①, and $P^{\text{rad}}(t) = \oiint_{S_{\infty}} [\vec{E}^{\text{sca}}(t) \times \vec{H}^{\text{sca}}(t)] \cdot \hat{n}_{\infty}^{+} dS$

① i.e. the magnetic and electric energies carried by scattered field.

and $W_{\text{vac}}^{\text{mag}}(t) = (1/2) < \vec{H}^{\text{sca}}(t), \mu_0 \vec{H}^{\text{sca}}(t) >_{\mathbb{R}^3}$ and $W_{\text{vac}}^{\text{ele}}(t) = (1/2) < \varepsilon_0 \vec{E}^{\text{sca}}(t), \vec{E}^{\text{sca}}(t) >_{\mathbb{R}^3}$, where integral domain $S_\infty$ is a spherical surface at infinity, and integral domain $\mathbb{R}^3$ is whole 3-D Euclidean space, and unit vector $\hat{n}_\infty^+$ is the normal direction of $S_\infty$ and points to infinity, and $\mu_0$ and $\varepsilon_0$ are respectively the permeability and permittivity in vacuum, and symbol "$\times$" is the cross product operation between two field vectors.

The physical meaning of formulation (3-1) is that: the power done by resultant field on scattered source is transverted to two parts —— a part is transferred to $S_\infty$ & the other part is converted to the EM energy stored in $\mathbb{R}^3$. In fact, formulation (3-1) is just the counterpart of the work-energy principle (WEP) for the particle in classical mechanics①, so this dissertation calls the formulation as the time-domain form of the work-energy principle for the metallic systems in classical electromagnetics②. In formulation (3-1), power term $P_{\text{met sys}}^{\text{Driving}}(t) =< \vec{J}^{\text{sca}}(t), \vec{E}^{\text{inc}}(t) >_{\partial D_{\text{met sys}}}$ is just the power source for driving $D_{\text{met sys}}$, so this dissertation calls it as the time-domain form of the driving power for metallic systems③.

In addition, the frequency-domain version of time-domain WEP (3-1) is as follows[41]:

$$\begin{aligned} P_{\text{met sys}}^{\text{Driving}} &= (1/2)\left\langle \vec{J}^{\text{sca}}, \vec{E}^{\text{inc}} \right\rangle_{\partial D_{\text{met sys}}} = (1/2)\left\langle \vec{J}^{\text{sca}}, \vec{E}^{\text{tot}} - \vec{E}^{\text{sca}} \right\rangle_{\partial D_{\text{met sys}}} = -(1/2)\left\langle \vec{J}^{\text{sca}}, \vec{E}^{\text{sca}} \right\rangle_{\partial D_{\text{met sys}}} \\ &= \underbrace{P^{\text{rad}}}_{P_{\text{met sys}}^{\text{act}} = \text{Re}\left\{P_{\text{met sys}}^{\text{Driving}}\right\}} + j \underbrace{\overbrace{2\omega\left(W_{\text{vac}}^{\text{mag}} - W_{\text{vac}}^{\text{ele}}\right)}^{P_{\text{vac}}^{\text{imag}}}}_{P_{\text{met sys}}^{\text{react}} = \text{Im}\left\{P_{\text{met sys}}^{\text{Driving}}\right\}} \end{aligned} \tag{3-2}$$

In above frequency-domain WEP (3-2), the $P_{\text{met sys}}^{\text{Driving}}$ on the left-hand side (LHS) of the first equality is the frequency-domain form of the DP for metallic system④; the coefficient $1/2$ on the RHS of the first equality is due to time average[114]; the $j$ on the RHS of the fourth equality is the unit of imaginary number; the $P_{\text{met sys}}^{\text{act}}$ & $P_{\text{met sys}}^{\text{react}}$ and $P^{\text{rad}}$ & $W_{\text{vac}}^{\text{mag}}$ & $W_{\text{vac}}^{\text{ele}}$ on the RHS of the fourth equality are respectively the active & reactive

① The WEP in classical mechanics is usually stated as that: "… the work done by all forces acting on a particle (the work of the resultant force) equals the change in the kinetic energy of the particle. [104]".

② In what follows, we will simply call it as the time-domain WEP of metallic systems. In Chapters 4 and 5, we will further provide the time-domain forms of the WEPs corresponding to the material scattering systems and metal-material composite scattering systems in classical electromagnetics.

③ In what follows, we will simply call it as the time-domain DP of metallic systems. In Chapters 4 and 5, we will further provide the time-domain forms of the DPs corresponding to the material scattering systems and metal-material composite scattering systems in classical electromagnetics.

④ In what follows, we will simply call it as the frequency-domain DP of metallic systems. In Chapters 4 and 5, we will further provide the frequency-domain forms of the DPs corresponding to the material scattering systems and metal-material composite scattering systems in classical electromagnetics.

powers acting $D_{\text{met sys}}$ by $\vec{F}^{\text{inc}}$ and the average radiated power & average magnetic energy in vacuum & average electric energy in vacuum, and $P^{\text{rad}} = P^{\text{act}}_{\text{met sys}} = \text{Re}\{P^{\text{Driving}}_{\text{met sys}}\}$ and $2\omega(W^{\text{mag}}_{\text{vac}} - W^{\text{ele}}_{\text{vac}}) = P^{\text{react}}_{\text{met sys}} = \text{Im}\{P^{\text{Driving}}_{\text{met sys}}\}$ and $P^{\text{rad}} = (1/2)\oiint_{S_\infty}[\vec{E}^{\text{sca}} \times (\vec{H}^{\text{sca}})^*] \cdot \hat{n}^+_\infty dS$ and $W^{\text{mag}}_{\text{vac}} = (1/4) < \vec{H}^{\text{sca}}, \mu_0\vec{H}^{\text{sca}} >_{\mathbb{R}^3}$ and $W^{\text{ele}}_{\text{vac}} = (1/4) < \varepsilon_0\vec{E}^{\text{sca}}, \vec{E}^{\text{sca}} >_{\mathbb{R}^3}$ [114], where $\omega$ is working angular frequency; the $2\omega(W^{\text{mag}}_{\text{vac}} - W^{\text{ele}}_{\text{vac}})$ on the RHS of the fourth equality is sometimes called as the imaginary power of metallic systems[37], so this dissertation correspondingly denotes it as $P^{\text{imag}}_{\text{vac}}$, i.e., $P^{\text{imag}}_{\text{vac}} = 2\omega(W^{\text{mag}}_{\text{vac}} - W^{\text{ele}}_{\text{vac}})$. In this dissertation, formulation (3-2) is called as the frequency-domain form of the work-energy principle for the metallic systems in classical electromagnetics①.

In frequency-domain WEP (3-2), the RHS of the first equality exhibits as the interaction between incident electric field and scattered electric current, so this dissertation calls it as the interaction form of frequency-domain DP; the RHS of the third equality exhibits as the self-interaction between scattered electric field and scattered electric current, so this dissertation calls it as the self-interaction form of frequency-domain DP; the RHS of the fourth equality only includes the various integrals for scattered field, so this dissertation calls it as the field form of frequency-domain DP. In addition, the self-interaction form of frequency-domain DP can also be equivalently rewritten as follows:

$$P^{\text{Driving}}_{\text{met sys}} = -(1/2)\left\langle \vec{J}^{\text{sca}}, -j\omega\mu_0\mathcal{L}_0\left(\vec{J}^{\text{sca}}\right)\right\rangle_{\partial D_{\text{met sys}}} \tag{3-3}$$

based on that $\vec{E}^{\text{sca}} = -j\omega\mu_0\mathcal{L}_0(\vec{J}^{\text{sca}})$ [107]. In formulation (3-3), operator $\mathcal{L}_0$ is defined as that $\mathcal{L}_0(\vec{X}) = [1 + (1/k_0^2)\nabla\nabla\cdot]\int_\Omega G_0(\vec{r},\vec{r}')\vec{X}(\vec{r}')d\Omega'$ [107], where $k_0 = \omega\sqrt{\mu_0\varepsilon_0}$ is the wave number in vacuum, and $G_0(\vec{r},\vec{r}') = e^{-jk_0|\vec{r}-\vec{r}'|}/4\pi|\vec{r}-\vec{r}'|$ is the scalar Green's function in vacuum[106-108]. The RHS of formulation (3-3) only includes current and the function of current, so this dissertation calls it as the current form of frequency-domain DP.

For a certain $D_{\text{met sys}}$, all possible $\vec{J}^{\text{sca}}$ constitute a linear space called as electric current space②. If $\{\vec{b}_\xi\}_{\xi=1}^{\Xi}$ is a series of independent and complete current basis functions defined on $\partial D_{\text{met sys}}$, then any $\vec{J}^{\text{sca}}$ can be expanded in terms of the basis functions as follows[106-108,115]:

---

① In what follows, we will simply call it as the frequency-domain WEP of metallic systems. In Chapters 4 and 5, we will further provide the frequency-domain forms of the WEPs corresponding to the material scattering systems and metal-material composite scattering systems in classical electromagnetics.

② In what follows, we will simply call it as current space.

$$\vec{J}^{\text{sca}}(\vec{r}) = \sum_{\xi=1}^{\Xi} a_\xi \vec{b}_\xi(\vec{r}) = \bar{\boldsymbol{B}} \cdot \bar{a} \quad , \quad \vec{r} \in \partial D_{\text{met sys}} \tag{3-4}$$

where $\bar{\boldsymbol{B}} = [\vec{b}_1, \vec{b}_2, \cdots, \vec{b}_\Xi]$ and $\bar{a} = [a_1, a_2, \cdots, a_\Xi]^T$, and superscript "$T$" represents the transpose operation for a matrix or vector①; symbol "$\cdot$" represents the traditional matrix multiplication②. $\bar{a}$ is the vector constituted by expansion coefficients, so it is called as expansion vector in this dissertation. In fact, expansion formulation (3-4) builds a one-to-one correspondence between $\vec{J}^{\text{sca}}$ and $\bar{a}$, so all of $\bar{a}$ also constitute a linear space called as expansion vector space[37]. Substituting expansion formulation (3-4) into the current form (3-3) of $P_{\text{met sys}}^{\text{Driving}}$, the current form will be discretized into the following matrix form[37,40,41]:

$$P_{\text{met sys}}^{\text{Driving}} = \bar{a}^H \cdot \bar{\bar{P}}_{\text{met sys}}^{\text{Driving}} \cdot \bar{a} \tag{3-5}$$

In formulation (3-5), superscript "$H$" represents the conjugate transpose operation for a matrix or vector; the elements of quadratic matrix $\bar{\bar{P}}_{\text{met sys}}^{\text{Driving}}$ are determined as that $p_{\text{met sys};\xi\zeta}^{\text{Driving}} = -(1/2) < \vec{b}_\xi, -j\omega\mu_0 \mathcal{L}_0(\vec{b}_\zeta) >_{\partial D_{\text{met sys}}}$.

The following parts of this dissertation are discussed in frequency domain, so we will not use modifier "frequency-domain" explicitly.

### 3.2.2 DP-CMs

Because $\bar{\bar{P}}_{\text{met sys}}^{\text{Driving}}$ is a square matrix, then the following Toeplitz's decomposition uniquely exists[116]:

$$\bar{\bar{P}}_{\text{met sys}}^{\text{Driving}} = \bar{\bar{P}}_{+}^{\text{Driving}} + j\, \bar{\bar{P}}_{-}^{\text{Driving}} \tag{3-6}$$

In above decomposition formulation (3-6), $\bar{\bar{P}}_{+}^{\text{Driving}} = [\bar{\bar{P}}_{\text{met sys}}^{\text{Driving}} + (\bar{\bar{P}}_{\text{met sys}}^{\text{Driving}})^H]/2$ and $\bar{\bar{P}}_{-}^{\text{Driving}} = [\bar{\bar{P}}_{\text{met sys}}^{\text{Driving}} - (\bar{\bar{P}}_{\text{met sys}}^{\text{Driving}})^H]/2j$. Obviously, $\bar{\bar{P}}_{+}^{\text{Driving}}$ and $\bar{\bar{P}}_{-}^{\text{Driving}}$ are Hermitian, so $\bar{a}^H \cdot \bar{\bar{P}}_{+}^{\text{Driving}} \cdot \bar{a}$ and $\bar{a}^H \cdot \bar{\bar{P}}_{-}^{\text{Driving}} \cdot \bar{a}$ must be real numbers for any complex vector $\bar{a}$ [116], and then previous formulations (3-2), (3-5), and (3-6) imply that $P^{\text{rad}} = \text{Re}\{P_{\text{met sys}}^{\text{Driving}}\} = \bar{a}^H \cdot \bar{\bar{P}}_{+}^{\text{Driving}} \cdot \bar{a}$ and $P_{\text{vac}}^{\text{imag}} = \text{Im}\{P_{\text{met sys}}^{\text{Driving}}\} = \bar{a}^H \cdot \bar{\bar{P}}_{-}^{\text{Driving}} \cdot \bar{a}$ [37,40,41]. Because $P^{\text{rad}} \ge 0$ for any $\bar{a}$, then $\bar{\bar{P}}_{+}^{\text{Driving}}$ is positive semi-definite[37,40,41]. In addition,

---

① In what follows, the algebraic vectors whose elements are field vectors will be denoted as the bold typefaces marked with single overlines, such as the $\bar{\boldsymbol{B}}$ in formulation (3-4); the algebraic vectors whose elements are scalars will be directly marked with single overlines, such as the $\bar{a}$ in formulation (3-4); the algebraic matrices whose elements are scalars will be directly marked with double overlines, such as the $\bar{\bar{P}}_{\text{met sys}}^{\text{Driving}}$ in formulation (3-4).

② In what follows, we will not especially distinguish "the dot product between two field vectors" and "the matrix multiplication between matrices or vectors" from the aspect of symbolic form, and the reasons are that: **1.** to avoid using many unnecessary symbols in this dissertation; **2.** the above manner will not lead to any confusion, because we have distinguished field vectors and algebraic quantities from the aspect of their symbols.

the elements of $\overline{\overline{P}}_{+}^{\text{Driving}}$ are the functions of working frequency $f$, so we will completely denote the matrix as $\overline{\overline{P}}_{+}^{\text{Driving}}(f)$.

If $\overline{\overline{P}}_{+}^{\text{Driving}}(f)$ is rigorously positive definite at frequency $f$, then there must be a non-singular matrix $\overline{\overline{A}}(f)$ such that both $\overline{\overline{A}}^{H}(f)\cdot\overline{\overline{P}}_{+}^{\text{Driving}}(f)\cdot\overline{\overline{A}}(f)$ and $\overline{\overline{A}}^{H}(f)\cdot\overline{\overline{P}}_{-}^{\text{Driving}}(f)\cdot\overline{\overline{A}}(f)$ are real diagonal matrices[116], and the column vectors of $\overline{\overline{A}}(f)$ can be derived from solving the following generalized characteristic equation①:

$$\overline{\overline{P}}_{-}^{\text{Driving}}(f)\cdot\bar{\alpha}_{\xi}(f) = \lambda_{\xi}(f)\,\overline{\overline{P}}_{+}^{\text{Driving}}(f)\cdot\bar{\alpha}_{\xi}(f) \tag{3-7}$$

where $\lambda_{\xi}(f)$ and $\bar{\alpha}_{\xi}(f)$ are called as characteristic value and characteristic vector② respectively. The corresponding characteristic electric current, characteristic electric field, and characteristic magnetic field are $\vec{J}_{\xi}^{\text{sca}}(f)=\bar{\boldsymbol{B}}\cdot\bar{\alpha}_{\xi}(f)$, $\vec{E}_{\xi}^{\text{sca}}(f)=-j\omega\mu_0\mathcal{L}_0(\vec{J}_{\xi}^{\text{sca}}(f))$, and $\vec{H}_{\xi}^{\text{sca}}(f)=\mathcal{K}_0(\vec{J}_{\xi}^{\text{sca}}(f))$ respectively, where operator $\mathcal{K}_0$ is defined as that $\mathcal{K}_0(\vec{X})=\nabla\times\int_{\Omega}G_0(\vec{r},\vec{r}\,')\vec{X}(\vec{r}\,')d\Omega'$[107]. In addition, the corresponding characteristic DP, characteristic radiated power, and characteristic imaginary power are $P_{\text{met sys};\xi}^{\text{Driving}}(f)=\bar{\alpha}_{\xi}^{H}(f)\cdot\overline{\overline{P}}_{\text{met sys}}^{\text{Driving}}(f)\cdot\bar{\alpha}_{\xi}(f)$, $P_{\xi}^{\text{rad}}(f)=\bar{\alpha}_{\xi}^{H}(f)\cdot\overline{\overline{P}}_{+}^{\text{Driving}}(f)\cdot\bar{\alpha}_{\xi}(f)$, and $P_{\text{vac};\xi}^{\text{imag}}(f)=\bar{\alpha}_{\xi}^{H}(f)\cdot\overline{\overline{P}}_{-}^{\text{Driving}}(f)\cdot\bar{\alpha}_{\xi}(f)$ respectively. Obviously, the above characteristic powers and characteristic value satisfy the relationships that $P_{\text{met sys};\xi}^{\text{Driving}}(f)=P_{\xi}^{\text{rad}}(f)+jP_{\text{vac};\xi}^{\text{imag}}(f)$ and $\lambda_{\xi}(f)=P_{\text{vac};\xi}^{\text{imag}}(f)/P_{\xi}^{\text{rad}}(f)$. Thus, characteristic value $\lambda_{\xi}(f)$ must be real number, when $\overline{\overline{P}}_{+}^{\text{Driving}}(f)$ is rigorous positive definite.

If $\overline{\overline{P}}_{+}^{\text{Driving}}(f)$ is positive semi-definite at frequency $f_0$ (i.e., $\overline{\overline{P}}_{+}^{\text{Driving}}(f_0)$ is positive semi-definite), then the characteristic pairs $\{\lambda_{\xi}(f_0),\bar{\alpha}_{\xi}(f_0)\}_{\xi=1}^{\Xi}$ at frequency $f_0$ can be obtained by using the following limitations[40]:

$$\lambda_{\xi}(f_0) = \lim\nolimits_{f\to f_0}\lambda_{\xi}(f) \tag{3-8}$$

$$\bar{\alpha}_{\xi}(f_0) = \lim\nolimits_{f\to f_0}\bar{\alpha}_{\xi}(f) \tag{3-9}$$

When $\overline{\overline{P}}_{+}^{\text{Driving}}(f_0)$ is positive semi-definite, the corresponding characteristic electric current, characteristic fields, and characteristic powers can be obtained similarly to the case that $\overline{\overline{P}}_{+}^{\text{Driving}}(f_0)$ is rigorous positive definite, and they will not be explicitly given here. For the modes satisfying that $P_{\xi}^{\text{rad}}(f_0)=0$, their radiated and imaginary powers

① To obtain the column vectors of matrix $\overline{\overline{A}}(f)$, it is sometimes necessary to do some orthogonalizations for degenerate characteristic vectors, besides solving equation (3-7). The contents related to the orthogonalization will be discussed in detail in the following Subsection 3.2.3. In 1971, Prof. Harrington et al.[31], for the first time, proposed a characteristic equation which is similar to equation (3-7) in form. The differences between Harrington's equation[31] and equation (3-7) are mainly manifested in the aspects of carrying framework and the related matrices, and the detailed analysis for the differences will be given in the Subsection 3.4.1 of this dissertation.

② In this dissertation, characteristic vectors are especially denoted as $\bar{\alpha}_{\xi}$ to be distinguished from the expansion vectors $\bar{a}$ corresponding to general working modes.

satisfy that $\lim_{f \to f_0} P_\xi^{\text{rad}}(f) = 0 = \lim_{f \to f_0} P_{\text{vac};\xi}^{\text{imag}}(f)$ [37,40,41], so: if $P_\xi^{\text{rad}}(f_0) = 0$, then limitation $\lim_{f \to f_0} \lambda_\xi(f) = \lim_{f \to f_0} [P_{\text{vac};\xi}^{\text{imag}}(f) / P_\xi^{\text{rad}}(f)]$ is a $0/0$-type indeterminate form[117]. Generally speaking, for the modes satisfying that $P_\xi^{\text{rad}}(f_0) = 0$, their radiated and imaginary powers satisfy that $\lim_{f \to f_0} (dP_{\text{vac};\xi}^{\text{imag}}(f)/df) \neq 0$ and $\lim_{f \to f_0} (dP_\xi^{\text{rad}}(f)/df) = 0$, so the L'Hôpital's rule[117] in Calculus gives the conclusion that: if $P_\xi^{\text{rad}}(f_0) = 0$, then $\lim_{f \to f_0} \lambda_\xi(f) = +\infty$ or $\lim_{f \to f_0} \lambda_\xi(f) = -\infty$. In the last section of this chapter, we will provide some typical examples to verify this conclusion.

Every group of $\{\lambda_\xi, \bar{\alpha}_\xi, \vec{J}_\xi^{\text{sca}}, \vec{F}_\xi^{\text{sca}}, P_{\text{met sys};\xi}^{\text{Driving}}\}$ corresponds to a characteristic working state of $D_{\text{met sys}}$, and the state is usually called as characteristic mode (CM)①. In the following Subsection 3.2.3, we will exhibit that the above CMs can orthogonalize DPO. Based on this, the CMs are specifically called as driving power CMs, and simply denoted as DP-CMs.

### 3.2.3 Orthogonality Among DP-CMs

The completeness of the DP-CMs is obvious. In this subsection, we will discuss the orthogonality among DP-CMs, and for the first time discuss the modal orthogonality related to nonradiative modes and the orthogonalization for degenerate CMs.

**1) $\lambda_\xi \neq \lambda_\zeta$ and $\lambda_\xi, \lambda_\zeta \neq \pm\infty$ Case.**

Because both $\bar{\bar{P}}_+^{\text{Driving}}$ and $\bar{\bar{P}}_-^{\text{Driving}}$ are Hermitian matrices and both $\lambda_\xi$ and $\lambda_\zeta$ are real numbers, then characteristic equations $\bar{\bar{P}}_-^{\text{Driving}} \cdot \bar{\alpha}_\xi = \lambda_\xi \bar{\bar{P}}_+^{\text{Driving}} \cdot \bar{\alpha}_\xi$ and $\bar{\bar{P}}_-^{\text{Driving}} \cdot \bar{\alpha}_\zeta = \lambda_\zeta \bar{\bar{P}}_+^{\text{Driving}} \cdot \bar{\alpha}_\zeta$ imply relationships $\bar{\alpha}_\xi^H \cdot \bar{\bar{P}}_-^{\text{Driving}} \cdot \bar{\alpha}_\zeta = \lambda_\xi \bar{\alpha}_\xi^H \cdot \bar{\bar{P}}_+^{\text{Driving}} \cdot \bar{\alpha}_\zeta$ and $\bar{\alpha}_\xi^H \cdot \bar{\bar{P}}_-^{\text{Driving}} \cdot \bar{\alpha}_\zeta = \lambda_\zeta \bar{\alpha}_\xi^H \cdot \bar{\bar{P}}_+^{\text{Driving}} \cdot \bar{\alpha}_\zeta$. The difference between the above two relationships gives that $0 = (\lambda_\xi - \lambda_\zeta) \bar{\alpha}_\xi^H \cdot \bar{\bar{P}}_+^{\text{Driving}} \cdot \bar{\alpha}_\zeta$. Because $\lambda_\xi \neq \lambda_\zeta$, then $\lambda_\xi - \lambda_\zeta \neq 0$, and this implies that: $\bar{\alpha}_\xi$ and $\bar{\alpha}_\zeta$ satisfy orthogonality $\bar{\alpha}_\xi^H \cdot \bar{\bar{P}}_+^{\text{Driving}} \cdot \bar{\alpha}_\zeta = P_\xi^{\text{rad}} \delta_{\xi\zeta}$, where $\delta_{\xi\zeta}$ is Kronecker's symbol. Then, relationships $\bar{\alpha}_\xi^H \cdot \bar{\bar{P}}_-^{\text{Driving}} \cdot \bar{\alpha}_\zeta = \lambda_\xi \bar{\alpha}_\xi^H \cdot \bar{\bar{P}}_+^{\text{Driving}} \cdot \bar{\alpha}_\zeta$ and $\lambda_\xi = P_{\text{vac};\xi}^{\text{imag}} / P_\xi^{\text{rad}}$ imply that②: $\bar{\alpha}_\xi$ and $\bar{\alpha}_\zeta$ also satisfy orthogonality $\bar{\alpha}_\xi^H \cdot \bar{\bar{P}}_-^{\text{Driving}} \cdot \bar{\alpha}_\zeta = P_{\text{vac};\xi}^{\text{imag}} \delta_{\xi\zeta}$.

**2) $\lambda_{\xi_1} = \lambda_{\xi_2} = \cdots = \lambda_{\xi_M} = \lambda_\xi \neq \pm\infty$ Case③.**

In this case, characteristic vectors $\{\bar{\alpha}_{\xi_1}, \bar{\alpha}_{\xi_2}, \cdots, \bar{\alpha}_{\xi_M}\}$ can be transformed into a set of new vectors $\{\bar{\alpha}'_{\xi_1}, \bar{\alpha}'_{\xi_2}, \cdots, \bar{\alpha}'_{\xi_M}\}$ by using the following Cramer-Schmidt

① The concept of CM was introduced by Prof. Garbacz in classical literature [29] for the first time.

② Based on the conclusions obtained in the Subsection 3.2.2 of this dissertation, we know that $P_\xi^{\text{rad}} \neq 0$, because $\lambda_\xi \neq \pm\infty$.

③ In fact, this case corresponds to the radiative CM whose characteristic value and degeneracy order are $\lambda_\xi$ and $M$ respectively, where $M$ is an integer being larger than 1.

orthogonalization method[116]:

$$\left.\begin{array}{rcl} \bar{\alpha}_{\xi_1} &=& \bar{\alpha}'_{\xi_1} \\ \bar{\alpha}_{\xi_2} - \chi_{12}\bar{\alpha}'_{\xi_1} &=& \bar{\alpha}'_{\xi_2} \\ \bar{\alpha}_{\xi_3} - \chi_{23}\bar{\alpha}'_{\xi_2} - \chi_{13}\bar{\alpha}'_{\xi_1} &=& \bar{\alpha}'_{\xi_3} \\ \bar{\alpha}_{\xi_4} - \chi_{34}\bar{\alpha}'_{\xi_3} - \chi_{24}\bar{\alpha}'_{\xi_2} - \chi_{14}\bar{\alpha}'_{\xi_1} &=& \bar{\alpha}'_{\xi_4} \\ && \cdots \\ \bar{\alpha}_{\xi_M} - \cdots - \chi_{4M}\bar{\alpha}'_{\xi_4} - \chi_{3M}\bar{\alpha}'_{\xi_3} - \chi_{2M}\bar{\alpha}'_{\xi_2} - \chi_{1M}\bar{\alpha}'_{\xi_1} &=& \bar{\alpha}'_{\xi_M} \end{array}\right\} \tag{3-10}$$

where the combination coefficients are determined as follows:

$$\left.\begin{array}{r} \chi_{12} = \dfrac{\left(\bar{\alpha}'_{\xi_1}\right)^H \cdot \bar{\bar{P}}_+^{\text{Driving}} \cdot \bar{\alpha}_{\xi_2}}{\left(\bar{\alpha}'_{\xi_1}\right)^H \cdot \bar{\bar{P}}_+^{\text{Driving}} \cdot \bar{\alpha}'_{\xi_1}} \\ \chi_{23} = \dfrac{\left(\bar{\alpha}'_{\xi_2}\right)^H \cdot \bar{\bar{P}}_+^{\text{Driving}} \cdot \bar{\alpha}_{\xi_3}}{\left(\bar{\alpha}'_{\xi_2}\right)^H \cdot \bar{\bar{P}}_+^{\text{Driving}} \cdot \bar{\alpha}'_{\xi_2}} \ , \ \chi_{13} = \dfrac{\left(\bar{\alpha}'_{\xi_1}\right)^H \cdot \bar{\bar{P}}_+^{\text{Driving}} \cdot \bar{\alpha}_{\xi_3}}{\left(\bar{\alpha}'_{\xi_1}\right)^H \cdot \bar{\bar{P}}_+^{\text{Driving}} \cdot \bar{\alpha}'_{\xi_1}} \\ \chi_{34} = \dfrac{\left(\bar{\alpha}'_{\xi_3}\right)^H \cdot \bar{\bar{P}}_+^{\text{Driving}} \cdot \bar{\alpha}_{\xi_4}}{\left(\bar{\alpha}'_{\xi_3}\right)^H \cdot \bar{\bar{P}}_+^{\text{Driving}} \cdot \bar{\alpha}'_{\xi_3}} \ , \ \chi_{24} = \dfrac{\left(\bar{\alpha}'_{\xi_2}\right)^H \cdot \bar{\bar{P}}_+^{\text{Driving}} \cdot \bar{\alpha}_{\xi_4}}{\left(\bar{\alpha}'_{\xi_2}\right)^H \cdot \bar{\bar{P}}_+^{\text{Driving}} \cdot \bar{\alpha}'_{\xi_2}} \ , \ \chi_{14} = \dfrac{\left(\bar{\alpha}'_{\xi_1}\right)^H \cdot \bar{\bar{P}}_+^{\text{Driving}} \cdot \bar{\alpha}_{\xi_4}}{\left(\bar{\alpha}'_{\xi_1}\right)^H \cdot \bar{\bar{P}}_+^{\text{Driving}} \cdot \bar{\alpha}'_{\xi_1}} \\ \cdots \end{array}\right\} \tag{3-11}$$

It is easy to prove that: new characteristic vectors $\{\bar{\alpha}'_{\xi_1}, \bar{\alpha}'_{\xi_2}, \cdots, \bar{\alpha}'_{\xi_M}\}$ satisfy orthogonality $(\bar{\alpha}'_{\xi_m})^H \cdot \bar{\bar{P}}_+^{\text{Driving}} \cdot \bar{\alpha}'_{\xi_n} = P_{\xi_m}^{\text{rad}\prime}\delta_{\xi_m\xi_n}$, where $\xi_m, \xi_n = \xi_1, \xi_2, \cdots, \xi_M$. Then, $\{\bar{\alpha}'_{\xi_1}, \bar{\alpha}'_{\xi_2}, \cdots, \bar{\alpha}'_{\xi_M}\}$ naturally satisfy orthogonality $(\bar{\alpha}'_{\xi_m})^H \cdot \bar{\bar{P}}_-^{\text{Driving}} \cdot \bar{\alpha}'_{\xi_n} = P_{\text{vac};\xi_m}^{\text{imag}\ \prime}\delta_{\xi_m\xi_n}$, because $(\bar{\alpha}'_{\xi_m})^H \cdot \bar{\bar{P}}_-^{\text{Driving}} \cdot \bar{\alpha}'_{\xi_n} = \lambda_{\xi}(\bar{\alpha}'_{\xi_m})^H \cdot \bar{\bar{P}}_+^{\text{Driving}} \cdot \bar{\alpha}'_{\xi_n}$ and $\lambda_{\xi} = P_{\text{vac};\xi_m}^{\text{imag}\ \prime} / P_{\xi_m}^{\text{rad}\prime}$. In addition, both $P_{\xi_m}^{\text{rad}\prime}$ and $P_{\text{vac};\xi_m}^{\text{imag}\ \prime}$ are real numbers, and both $\bar{\bar{P}}_+^{\text{Driving}}$ and $\bar{\bar{P}}_-^{\text{Driving}}$ are Hermitian matrices, and $\delta_{\xi_m\xi_n} = \delta_{\xi_n\xi_m}$, so $\{\bar{\alpha}'_{\xi_1}, \bar{\alpha}'_{\xi_2}, \cdots, \bar{\alpha}'_{\xi_M}\}$ satisfy orthogonality $(\bar{\alpha}'_{\xi_n})^H \cdot \bar{\bar{P}}_+^{\text{Driving}} \cdot \bar{\alpha}'_{\xi_m} = P_{\xi_n}^{\text{rad}\prime}\delta_{\xi_n\xi_m}$ and $(\bar{\alpha}'_{\xi_n})^H \cdot \bar{\bar{P}}_-^{\text{Driving}} \cdot \bar{\alpha}'_{\xi_m} = P_{\text{vac};\xi_n}^{\text{imag}\ \prime}\delta_{\xi_n\xi_m}$ simultaneously.

**3) $\lambda_{\xi} = +\infty$ or $-\infty$ Case.**

In this case, it must be that $\bar{\alpha}_{\xi}^H \cdot \bar{\bar{P}}_+^{\text{Driving}} \cdot \bar{\alpha}_{\xi} = P_{\xi}^{\text{rad}} = 0$. Based on this and the conclusions given in literatures [118,119] and the method used in literatures [40,41], it is easy to prove that: $(\bar{\bar{P}}_+^{\text{Driving}} + j\bar{\bar{P}}_-^{\text{Driving}}) \cdot \bar{\alpha}_{\xi} = \bar{\bar{P}}^{\text{Driving}} \cdot \bar{\alpha}_{\xi} = 0$. Because $\bar{\bar{P}}_+^{\text{Driving}}$ is positive definite or semi-definite, then $\bar{\alpha}_{\xi}^H \cdot \bar{\bar{P}}_+^{\text{Driving}} \cdot \bar{\alpha}_{\xi} = 0$ if and only if $\bar{\bar{P}}_+^{\text{Driving}} \cdot \bar{\alpha}_{\xi} = 0$ [116]. These above imply that $\bar{\bar{P}}_-^{\text{Driving}} \cdot \bar{\alpha}_{\xi} = 0$ and $P_{\text{vac};\xi}^{\text{imag}} = 0$. Thus $\bar{\alpha}_{\xi}^H \cdot \bar{\bar{P}}_{\pm}^{\text{Driving}} = 0$, because $\bar{\bar{P}}_{\pm}^{\text{Driving}}$ is Hermitian. Then, for any $\bar{\alpha}_{\zeta}$, there exist the relationships that $\bar{\alpha}_{\zeta}^H \cdot \bar{\bar{P}}_+^{\text{Driving}} \cdot \bar{\alpha}_{\xi} = P_{\zeta}^{\text{rad}}\delta_{\zeta\xi} = P_{\xi}^{\text{rad}}\delta_{\zeta\xi} = 0$ and $\bar{\alpha}_{\zeta}^H \cdot \bar{\bar{P}}_-^{\text{Driving}} \cdot \bar{\alpha}_{\xi} = P_{\text{vac};\zeta}^{\text{imag}}\delta_{\zeta\xi} = P_{\text{vac};\xi}^{\text{imag}}\delta_{\zeta\xi} = 0$ and $\bar{\alpha}_{\xi}^H \cdot \bar{\bar{P}}_+^{\text{Driving}} \cdot \bar{\alpha}_{\zeta} = P_{\xi}^{\text{rad}}\delta_{\xi\zeta} = 0$ and $\bar{\alpha}_{\xi}^H \cdot \bar{\bar{P}}_-^{\text{Driving}} \cdot \bar{\alpha}_{\zeta} = P_{\text{vac};\xi}^{\text{imag}}\delta_{\xi\zeta} = 0$.

**4) $\lambda_{\zeta_1} = \lambda_{\zeta_2} = \cdots = \lambda_{\zeta_N} = \lambda_{\zeta} = +\infty$ or $-\infty$ Case①.**

In this case, the original characteristic vectors $\{\bar{\alpha}_{\zeta_1}, \bar{\alpha}_{\zeta_2}, \cdots, \bar{\alpha}_{\zeta_N}\}$ can be transformed into a series of new characteristic vectors $\{\bar{\alpha}'_{\zeta_1}, \bar{\alpha}'_{\zeta_2}, \cdots, \bar{\alpha}'_{\zeta_N}\}$ by employing the following Cramer-Schmidt orthogonalization method[116]:

$$\left.\begin{aligned}
\bar{\alpha}_{\zeta_1} &= \bar{\alpha}'_{\zeta_1} \\
\bar{\alpha}_{\zeta_2} - \gamma_{12}\bar{\alpha}'_{\zeta_1} &= \bar{\alpha}'_{\zeta_2} \\
\bar{\alpha}_{\zeta_3} - \gamma_{23}\bar{\alpha}'_{\zeta_2} - \gamma_{13}\bar{\alpha}'_{\zeta_1} &= \bar{\alpha}'_{\zeta_3} \\
\bar{\alpha}_{\zeta_4} - \gamma_{34}\bar{\alpha}'_{\zeta_3} - \gamma_{24}\bar{\alpha}'_{\zeta_2} - \gamma_{14}\bar{\alpha}'_{\zeta_1} &= \bar{\alpha}'_{\zeta_4} \\
&\cdots \\
\bar{\alpha}_{\zeta_N} - \cdots - \gamma_{4N}\bar{\alpha}'_{\zeta_4} - \gamma_{3N}\bar{\alpha}'_{\zeta_3} - \gamma_{2N}\bar{\alpha}'_{\zeta_2} - \gamma_{1N}\bar{\alpha}'_{\zeta_1} &= \bar{\alpha}'_{\zeta_N}
\end{aligned}\right\} \quad (3\text{-}12)$$

The above combination coefficients are determined as follows:

$$\left.\begin{aligned}
&\gamma_{12} = \frac{\left(\bar{\alpha}'_{\zeta_1}\right)^H \cdot \bar{\bar{J}}^{\text{sca}} \cdot \bar{\alpha}_{\zeta_2}}{\left(\bar{\alpha}'_{\zeta_1}\right)^H \cdot \bar{\bar{J}}^{\text{sca}} \cdot \bar{\alpha}'_{\zeta_1}} \\
\gamma_{23} = \frac{\left(\bar{\alpha}'_{\zeta_2}\right)^H \cdot \bar{\bar{J}}^{\text{sca}} \cdot \bar{\alpha}_{\zeta_3}}{\left(\bar{\alpha}'_{\zeta_2}\right)^H \cdot \bar{\bar{J}}^{\text{sca}} \cdot \bar{\alpha}'_{\zeta_2}} \quad , \quad &\gamma_{13} = \frac{\left(\bar{\alpha}'_{\zeta_1}\right)^H \cdot \bar{\bar{J}}^{\text{sca}} \cdot \bar{\alpha}_{\zeta_3}}{\left(\bar{\alpha}'_{\zeta_1}\right)^H \cdot \bar{\bar{J}}^{\text{sca}} \cdot \bar{\alpha}'_{\zeta_1}} \\
\gamma_{34} = \frac{\left(\bar{\alpha}'_{\zeta_3}\right)^H \cdot \bar{\bar{J}}^{\text{sca}} \cdot \bar{\alpha}_{\zeta_4}}{\left(\bar{\alpha}'_{\zeta_3}\right)^H \cdot \bar{\bar{J}}^{\text{sca}} \cdot \bar{\alpha}'_{\zeta_3}} \quad , \quad \gamma_{24} = \frac{\left(\bar{\alpha}'_{\zeta_2}\right)^H \cdot \bar{\bar{J}}^{\text{sca}} \cdot \bar{\alpha}_{\zeta_4}}{\left(\bar{\alpha}'_{\zeta_2}\right)^H \cdot \bar{\bar{J}}^{\text{sca}} \cdot \bar{\alpha}'_{\zeta_2}} \quad , \quad &\gamma_{14} = \frac{\left(\bar{\alpha}'_{\zeta_1}\right)^H \cdot \bar{\bar{J}}^{\text{sca}} \cdot \bar{\alpha}_{\zeta_4}}{\left(\bar{\alpha}'_{\zeta_1}\right)^H \cdot \bar{\bar{J}}^{\text{sca}} \cdot \bar{\alpha}'_{\zeta_1}} \\
&\cdots
\end{aligned}\right\} \quad (3\text{-}13)$$

It is easy to prove that: above new characteristic vectors $\{\bar{\alpha}'_{\zeta_1}, \bar{\alpha}'_{\zeta_2}, \cdots, \bar{\alpha}'_{\zeta_N}\}$ satisfy orthogonality $(\bar{\alpha}'_{\zeta_m})^H \cdot \bar{\bar{J}}^{\text{sca}} \cdot \bar{\alpha}'_{\zeta_n} = J^{\text{sca}\prime}_{\zeta_m} \delta_{\zeta_m \zeta_n}$, where $\zeta_m, \zeta_n = \zeta_1, \zeta_2, \cdots, \zeta_N$. The elements $j^{\text{sca}}_{\xi\zeta}$ in the matrix $\bar{\bar{J}}^{\text{sca}}$ used in formulation (3-13) are calculated as $j^{\text{sca}}_{\xi\zeta} = (1/2) < \vec{b}_{\xi}, \vec{b}_{\zeta} >_{\partial D_{\text{met sys}}}$.

In summary, we can obtain a series of DP-CMs which satisfy the following orthogonality:

$$P^{\text{rad}}_{\xi} \delta_{\xi\zeta} = \bar{\alpha}^H_{\xi} \cdot \bar{\bar{P}}^{\text{Driving}}_{+} \cdot \bar{\alpha}_{\zeta} \quad (3\text{-}14)$$

$$P^{\text{imag}}_{\text{vac};\xi} \delta_{\xi\zeta} = \bar{\alpha}^H_{\xi} \cdot \bar{\bar{P}}^{\text{Driving}}_{-} \cdot \bar{\alpha}_{\zeta} \quad (3\text{-}15)$$

and then the following orthogonality:

$$\underbrace{\left(P^{\text{rad}}_{\xi} + jP^{\text{imag}}_{\text{vac};\xi}\right)}_{P^{\text{Driving}}_{\text{met sys};\xi}} \delta_{\xi\zeta} = \bar{\alpha}^H_{\xi} \cdot \underbrace{\left(\bar{\bar{P}}^{\text{Driving}}_{+} + j\,\bar{\bar{P}}^{\text{Driving}}_{-}\right)}_{\bar{\bar{P}}^{\text{Driving}}_{\text{met sys}}} \cdot \bar{\alpha}_{\zeta} \quad (3\text{-}16)$$

① In fact, this case corresponds to the nonradiative CM whose degeneracy order is $N$, where $N$ is an integer being larger than 1.

To simplify the symbolic system of this dissertation, the DP-CMs $\{\bar{\alpha}'_{\xi_1},\bar{\alpha}'_{\xi_2},\cdots,\bar{\alpha}'_{\xi_M}\}$ in case $\lambda_{\xi_1}=\lambda_{\xi_2}=\cdots=\lambda_{\xi_M}=\lambda_{\xi}\neq\pm\infty$ have been simply denoted as $\{\bar{\alpha}_{\xi_1},\bar{\alpha}_{\xi_2},\cdots,\bar{\alpha}_{\xi_M}\}$; the DP-CMs $\{\bar{\alpha}'_{\zeta_1},\bar{\alpha}'_{\zeta_2},\cdots,\bar{\alpha}'_{\zeta_N}\}$ in case $\lambda_{\zeta_1}=\lambda_{\zeta_2}=\cdots=\lambda_{\zeta_N}=\lambda_{\zeta}=+\infty$ or $-\infty$ have been simply denoted as $\{\bar{\alpha}_{\zeta_1},\bar{\alpha}_{\zeta_2},\cdots,\bar{\alpha}_{\zeta_N}\}$. Above formulations (3-14)~(3-16) are collectively referred to as the matrix-vector multiplication form of the orthogonality among DP-CMs. According to the physical meanings of the elements in $\bar{\bar{P}}^{\text{Driving}}_{\text{met sys}}$, orthogonality (3-16) can be equivalently rewritten as the following current form:

$$P^{\text{Driving}}_{\text{met sys};\xi}\delta_{\xi\zeta} = -(1/2)\left\langle \vec{J}^{\text{sca}}_{\xi}, -j\omega\mu_0\mathcal{L}_0\left(\vec{J}^{\text{sca}}_{\zeta}\right)\right\rangle_{\partial D_{\text{met sys}}} \tag{3-17}$$

and the following self-interaction form:

$$P^{\text{Driving}}_{\text{met sys};\xi}\delta_{\xi\zeta} = -(1/2)\left\langle \vec{J}^{\text{sca}}_{\xi}, \vec{E}^{\text{sca}}_{\zeta}\right\rangle_{\partial D_{\text{met sys}}} \tag{3-18}$$

In addition, based on that $\bar{\bar{P}}^{\text{Driving}}_{+}=[\bar{\bar{P}}^{\text{Driving}}_{\text{met sys}}+(\bar{\bar{P}}^{\text{Driving}}_{\text{met sys}})^H]/2$ and that $\bar{\bar{P}}^{\text{Driving}}_{-}=[\bar{\bar{P}}^{\text{Driving}}_{\text{met sys}}-(\bar{\bar{P}}^{\text{Driving}}_{\text{met sys}})^H]/2j$ and the Maxwell's equations corresponding to DP-CMs, orthogonality (3-14) and (3-15) can also be alternatively rewritten as the following field form:

$$\begin{aligned} P^{\text{rad}}_{\xi}\delta_{\xi\zeta} &= (1/2)\oint\!\!\oint_{S_\infty}\left[\vec{E}^{\text{sca}}_{\zeta}\times\left(\vec{H}^{\text{sca}}_{\xi}\right)^*\right]\cdot d\vec{S} = (1/2\eta_0)\left\langle \vec{E}^{\text{sca}}_{\xi},\vec{E}^{\text{sca}}_{\zeta}\right\rangle_{S_\infty} \\ &= (\eta_0/2)\left\langle \vec{H}^{\text{sca}}_{\xi},\vec{H}^{\text{sca}}_{\zeta}\right\rangle_{S_\infty} \end{aligned} \tag{3-19}$$

$$P^{\text{imag}}_{\text{vac};\xi}\delta_{\xi\zeta} = 2\omega\left[(1/4)\left\langle \vec{H}^{\text{sca}}_{\xi},\mu_0\vec{H}^{\text{sca}}_{\zeta}\right\rangle_{\mathbb{R}^3} - (1/4)\left\langle \varepsilon_0\vec{E}^{\text{sca}}_{\xi},\vec{E}^{\text{sca}}_{\zeta}\right\rangle_{\mathbb{R}^3}\right] \tag{3-20}$$

In orthogonality (3-19), $\eta_0=\sqrt{\mu_0/\varepsilon_0}$ is the wave impedance in vacuum; the second and third equalities are based on the relationships $\vec{E}^{\text{sca}}_{\xi}=\eta_0\vec{H}^{\text{sca}}_{\xi}\times\hat{n}^+_\infty$ and $\vec{H}^{\text{sca}}_{\zeta}=\hat{n}^+_\infty\times\vec{E}^{\text{sca}}_{\zeta}/\eta_0$ on $S_\infty$, and these two relationships can be derived from Sommerfeld's radiation condition $\lim_{|\vec{r}|\to+\infty}(\nabla\times\vec{F}+jk_0\hat{n}^+_\infty\times\vec{F})=0$ [120] and the homogeneous Maxwell's equations $\{\nabla\times\vec{H}=j\omega\varepsilon_0\vec{E}\,;\nabla\times\vec{E}=-j\omega\mu_0\vec{H}\}$ [114] at infinity. Obviously, orthogonality (3-19) implies that the modal far fields of DP-CMs are orthogonal to each others. In addition, for the nonradiative DP-CMs $\{\bar{\alpha}_{\zeta_1},\bar{\alpha}_{\zeta_2},\cdots,\bar{\alpha}_{\zeta_N}\}$, the following orthogonality among modal currents also exists:

$$J^{\text{sca}}_{\zeta_m}\delta_{\zeta_m\zeta_n} = (1/2)\left\langle \vec{J}^{\text{sca}}_{\zeta_m},\vec{J}^{\text{sca}}_{\zeta_n}\right\rangle_{\partial D_{\text{met sys}}} \quad , \quad \left(\zeta_m,\zeta_n=\zeta_1,\zeta_2,\cdots,\zeta_N\right) \tag{3-21}$$

Evidently, $J^{\text{sca}}_{\zeta_m}=(1/2)<\vec{J}^{\text{sca}}_{\zeta_m},\vec{J}^{\text{sca}}_{\zeta_m}>_{\partial D_{\text{met sys}}}\in\mathbb{R}^+$.

In fact, orthogonality (3-18) is equivalent to that $(-T\cdot P^{\text{rad}}_{\xi})\delta_{\xi\zeta}=\int_0^T<\vec{J}^{\text{sca}}_{\xi}(t),\vec{E}^{\text{sca}}_{\zeta}(t)>_{\partial D_{\text{met sys}}}dt$, because $P^{\text{rad}}_{\xi}=\text{Re}\{P^{\text{Driving}}_{\text{met sys};\xi}\}$ and $\text{Re}\{-(1/2)\langle \vec{J}_{\xi}^{\text{sca}}, \vec{E}_{\zeta}^{\text{sca}} \rangle_{\partial D_{\text{met sys}}}\} = -(1/T)\int_0^T \langle \vec{J}_{\xi}^{\text{sca}}(t), \vec{E}_{\zeta}^{\text{sca}}(t) \rangle_{\partial D_{\text{met sys}}} dt$ [9,114]. This implies that any two different DP-CMs don't exchange net energy in any integral period.

## 3.2.4 Numerical Examples Corresponding to Typical Structures

In this subsection, we provide some typical numerical examples to verify the above conclusions.

### 1) DP-CMs of Metallic Cylinder

Now we consider a metallic cylinder whose radius is 20mm and height is also 20mm. The topological structure and surface triangular meshes of the cylinder are illustrated in Figure 3-2. In frequency band 5~11GHz, some characteristic quantity curves corresponding to the first 10 typical DP-CMs are illustrated in Figure 3-3, Figure 3-4, and Figure 3-5.

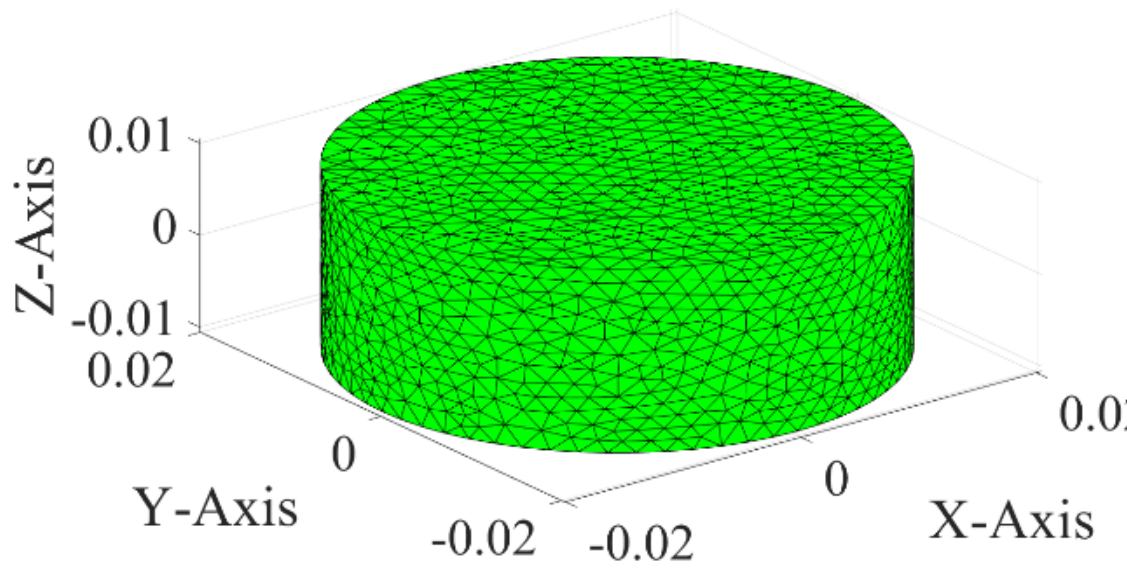

Figure 3-2 The topological structure and surface triangular meshes of a metallic cylinder whose radius and height are both 20mm

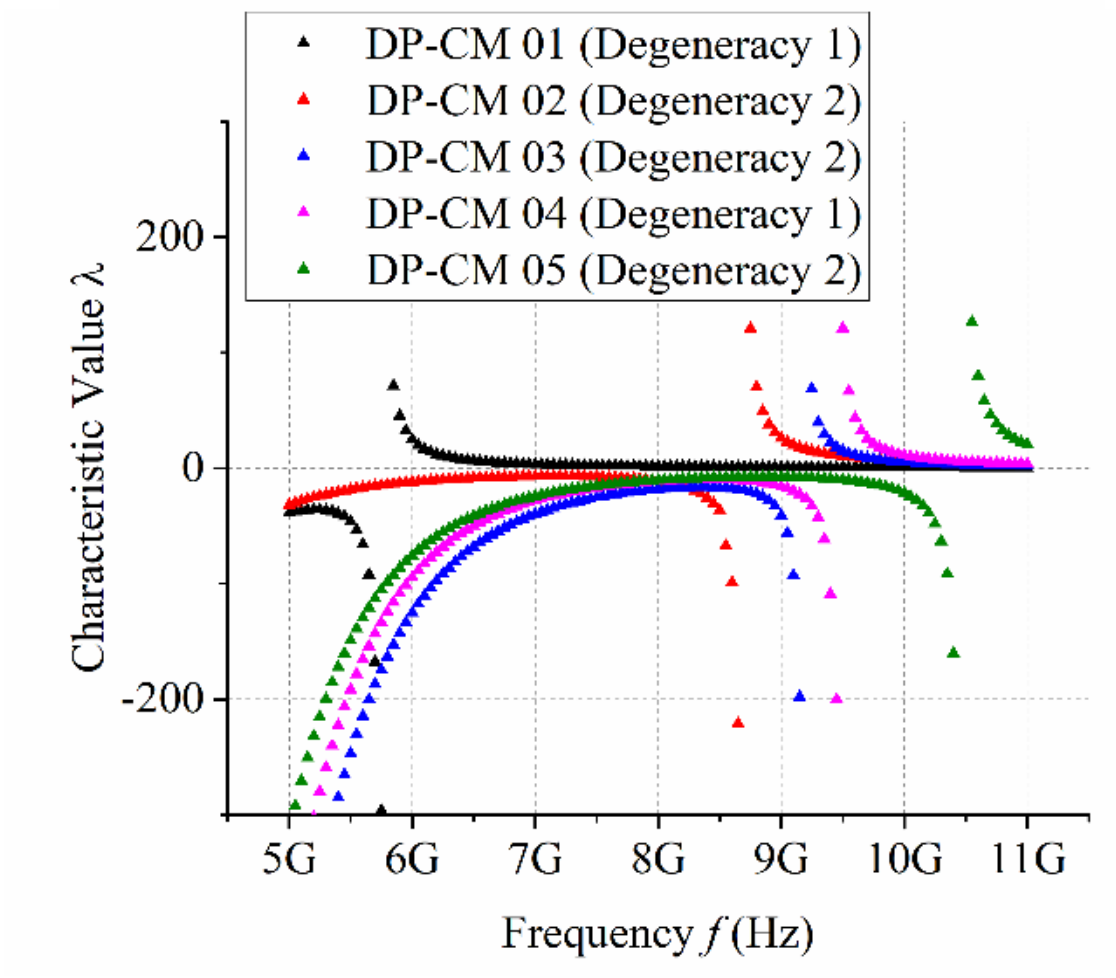

(a)

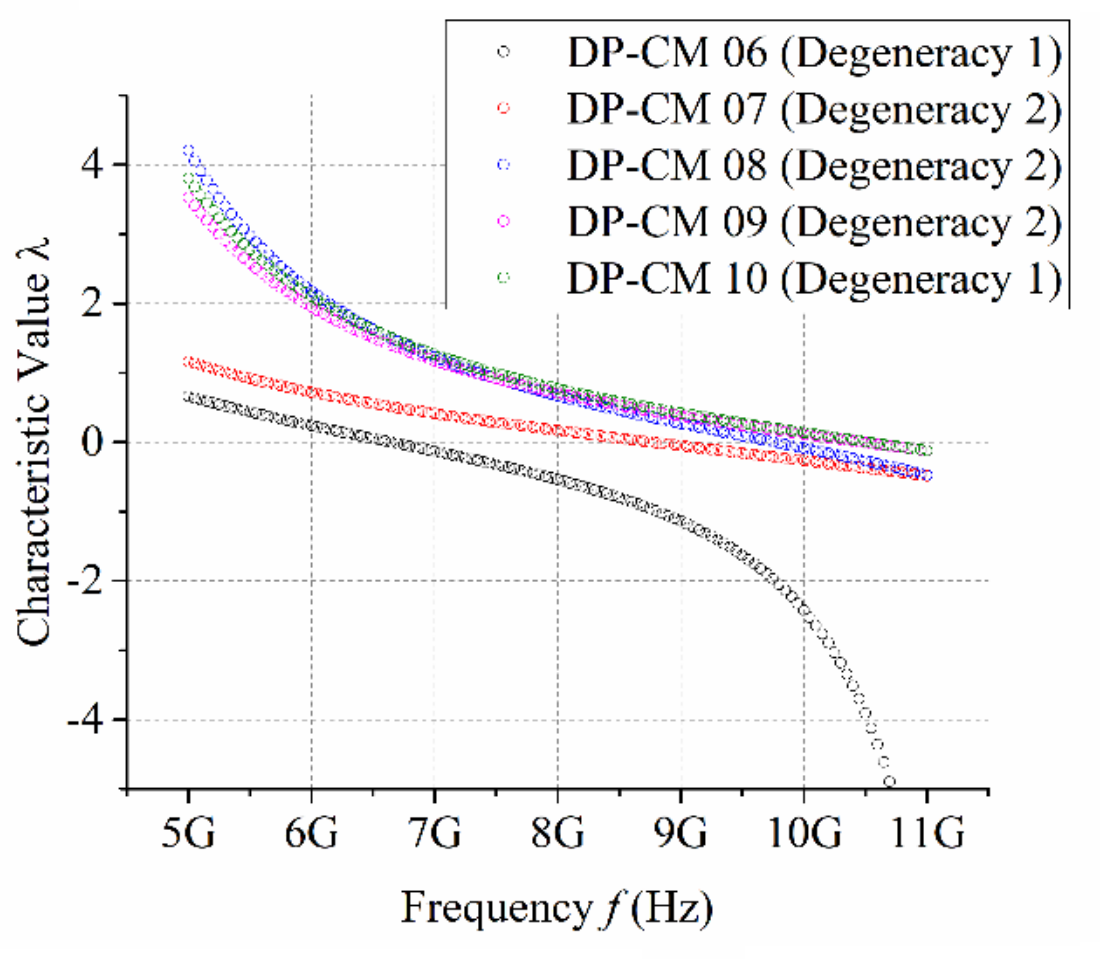

(b)

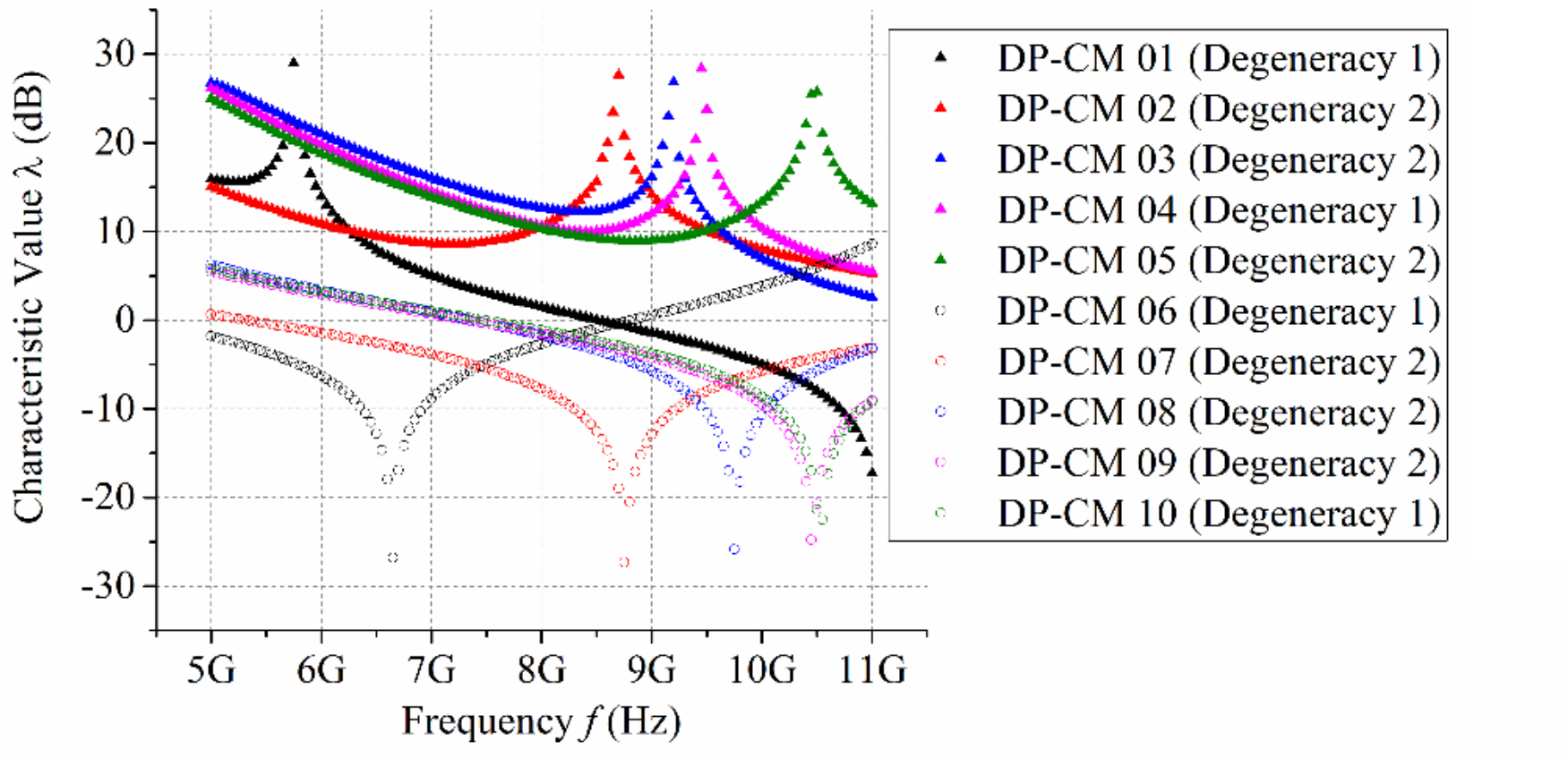

(c)

Figure 3-3 The characteristic value curves corresponding to 10 typical DP-CMs of the metallic cylinder shown in Figure 3-2. (a) the curves corresponding to DP-CM01 ~ DP-CM05; (b) the curves corresponding to DP-CM06 ~ DP-CM10; (c) the dB curves corresponding to DP-CM01 ~ DP-CM10

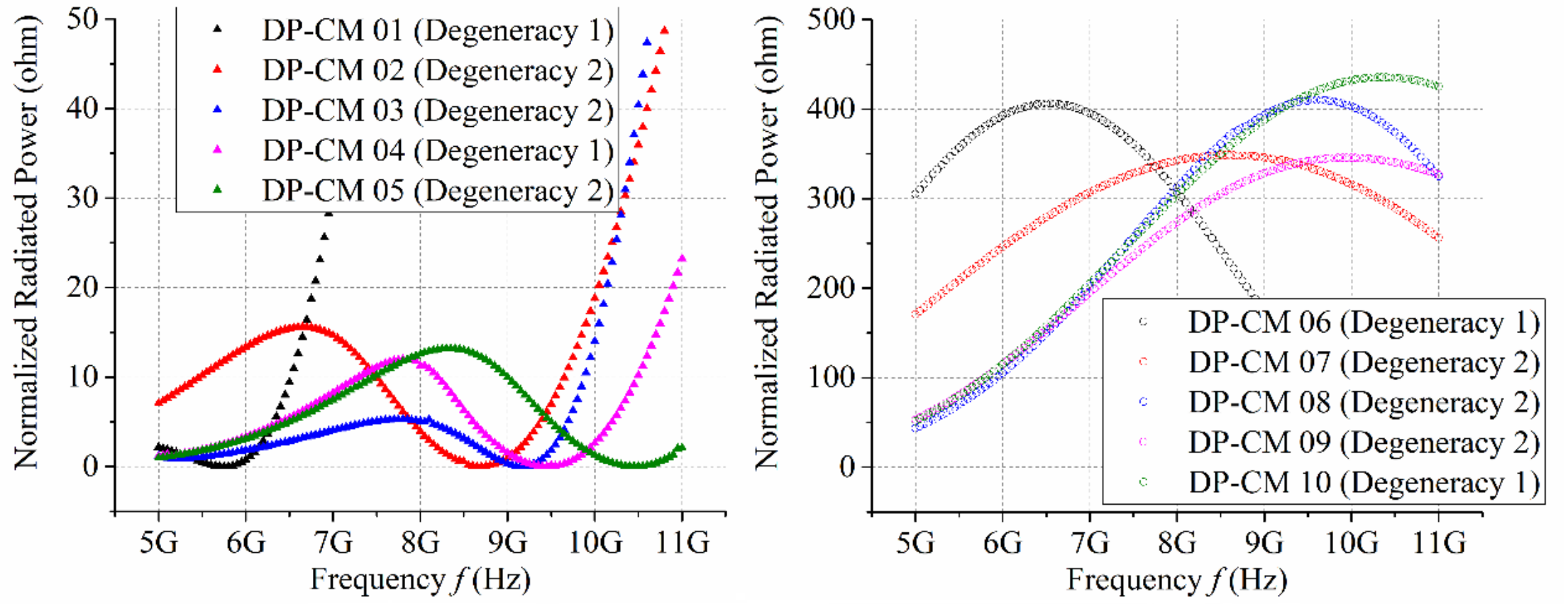

(a) (b)

Figure 3-4 The normalized modal radiated power curves corresponding to some typical DP-CMs of the metallic cylinder shown in Figure 3-2. (a) the curves corresponding to the 5 DP-CMs shown in Figure 3-3(a); (b) the curves corresponding to the 5 DP-CMs shown in Figure 3-3(b)

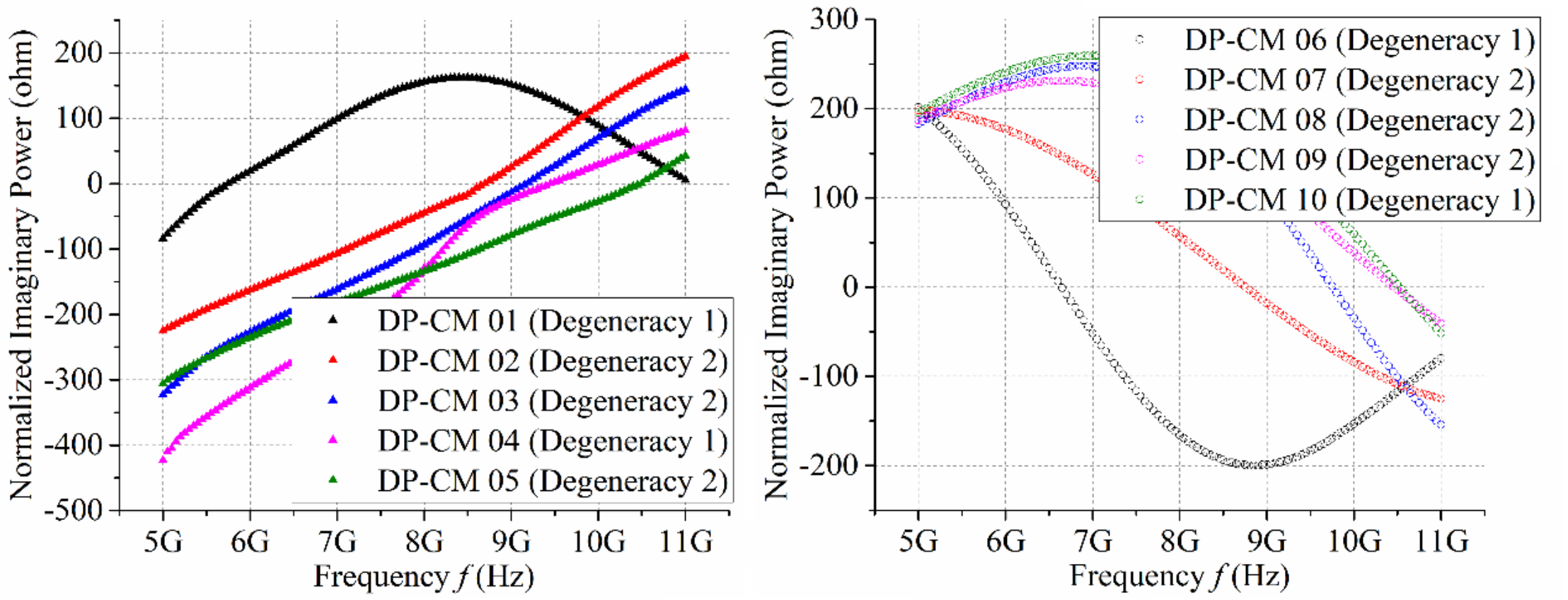

(a) (b)

Figure 3-5 The normalized modal imaginary power curves corresponding to some typical DP-CMs of the metallic cylinder shown in Figure 3-2. (a) the curves corresponding to the 5 DP-CMs shown in Figure 3-3(a); (b) the curves corresponding to the 5 DP-CMs shown in Figure 3-3(b)

From Figures 3-3(a), 3-4(a), and 3-5(a), it is easy to find out that: DP-CM1 is nonradiative at 5.75GHz, and it satisfies that $\lim_{f\to 5.75\text{G}}(dP_{\text{vac};1}^{\text{imag}}(f)/df)\neq 0$ and $\lim_{f\to 5.75\text{G}}(dP_1^{\text{rad}}(f)/df)=0$ simultaneously, so the characteristic value of DP-CM01 satisfies that $\lim_{f\to 5.75\text{G}^{\pm}}\lambda_1(f)=\pm\infty$ ①; DP-CM02 is nonradiative at 8.69GHz, and it satisfies that $\lim_{f\to 8.69\text{G}}(dP_{\text{vac};2}^{\text{imag}}(f)/df)\neq 0$ and $\lim_{f\to 8.69\text{G}}(dP_2^{\text{rad}}(f)/df)=0$ simultaneously, so the characteristic value of DP-CM02 satisfies that $\lim_{f\to 8.69\text{G}^{\pm}}\lambda_2(f)=\pm\infty$; DP-CM03 is nonradiative at 9.15GHz, and it satisfies that $\lim_{f\to 9.15\text{G}}(dP_{\text{vac};3}^{\text{imag}}(f)/df)\neq 0$ and $\lim_{f\to 9.15\text{G}}(dP_3^{\text{rad}}(f)/df)=0$ simultaneously, so the characteristic value of DP-CM03 satisfies that $\lim_{f\to 9.15\text{G}^{\pm}}\lambda_3(f)=\pm\infty$; DP-CM04 is nonradiative at 9.45GHz, and it satisfies that $\lim_{f\to 9.45\text{G}}(dP_{\text{vac};4}^{\text{imag}}(f)/df)\neq 0$ and $\lim_{f\to 9.45\text{G}}(dP_4^{\text{rad}}(f)/df)=0$ simultaneously, so the characteristic value of DP-CM04 satisfies that $\lim_{f\to 9.45\text{G}^{\pm}}\lambda_4(f)=\pm\infty$; DP-CM05 is nonradiative at 10.50GHz, and it satisfies that $\lim_{f\to 10.50\text{G}}(dP_{\text{vac};5}^{\text{imag}}(f)/df)\neq 0$ and $\lim_{f\to 10.50\text{G}}(dP_5^{\text{rad}}(f)/df)=0$ simultaneously, so the characteristic value of DP-CM05 satisfies that $\lim_{f\to 10.50\text{G}^{\pm}}\lambda_5(f)=\pm\infty$.

To compare the above nonradiative DP-CMs with the traditional internally resonant eigen-modes of closed metallic cylindrical cavity, we list their working frequencies and degeneracy orders in Table 3-1, and it is easy to find out that there exist some corresponding relationships between nonradiative DP-CMs and internally resonant eigen-modes.

Table 3-1 The working frequencies (GHz) and the degeneracy orders corresponding to the nonradiative DP-CMs shown in Figure 3-3 and the internally resonant eigen-modes of the metallic cylinder shown in Figure 3-2

| Modal Type / Modal Index | **Nonradiative DP-CM[37]** | **Internally Resonant Eigen-mode[26,28]** |
|---|---|---|
| **Mode 01** | 05.75 (1) | 05.74 (1) |
| **Mode 02** | 08.69 (2) | 08.69 (2) |
| **Mode 03** | 09.15 (2) | 09.15 (2) |
| **Mode 04** | 09.45 (1) | 09.45 (1) |
| **Mode 05** | 10.50 (2) | 10.49 (2) |

① Here, $\lim_{f\to 5.75\text{G}^{-}}\lambda_1(f)$ and $\lim_{f\to 5.75\text{G}^{+}}\lambda_1(f)$ respectively represent the left and right limits[117] at 5.75GHz.

To reveal the corresponding relationships, we provide the electric current and electric field distributions of the DP-CM01 at 5.75GHz as shown in Figures 3-6(a) and 3-6(b), and we also provide the electric current and electric field distributions of the two degenerate states of DP-CM02 at 5.75GHz as shown in Figures 3-7 and 3-8. From Figure 3-6, it is easy to find out that: nonradiative DP-CM01 satisfies homogeneous tangential electric field boundary condition, and it is just TE111 eigen-mode. From Figures 3-7 and 3-8, it is easy to find out that: the two degenerate nonradiative DP-CMs are just TM111 eigen-modes. Whether or not all nonradiative DP-CMs and all internally resonant eigen-modes are one-to-one correspondence? We will provide a systematical answer for this question in the Section 3.3 of this dissertation.

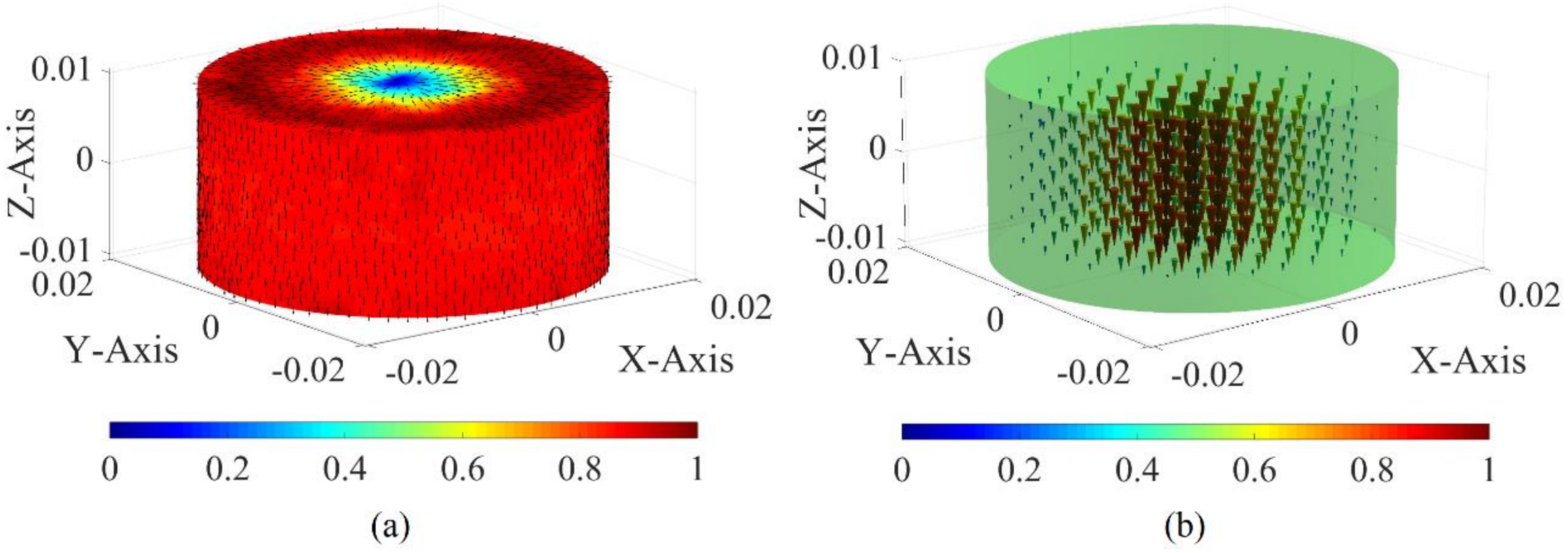

Figure 3-6 Some typical physical quantity distributions corresponding to the nonradiative DP-CM01 listed in Table 3-1. (a) modal electric current distribution; (b) modal electric field distributing in internal cavity

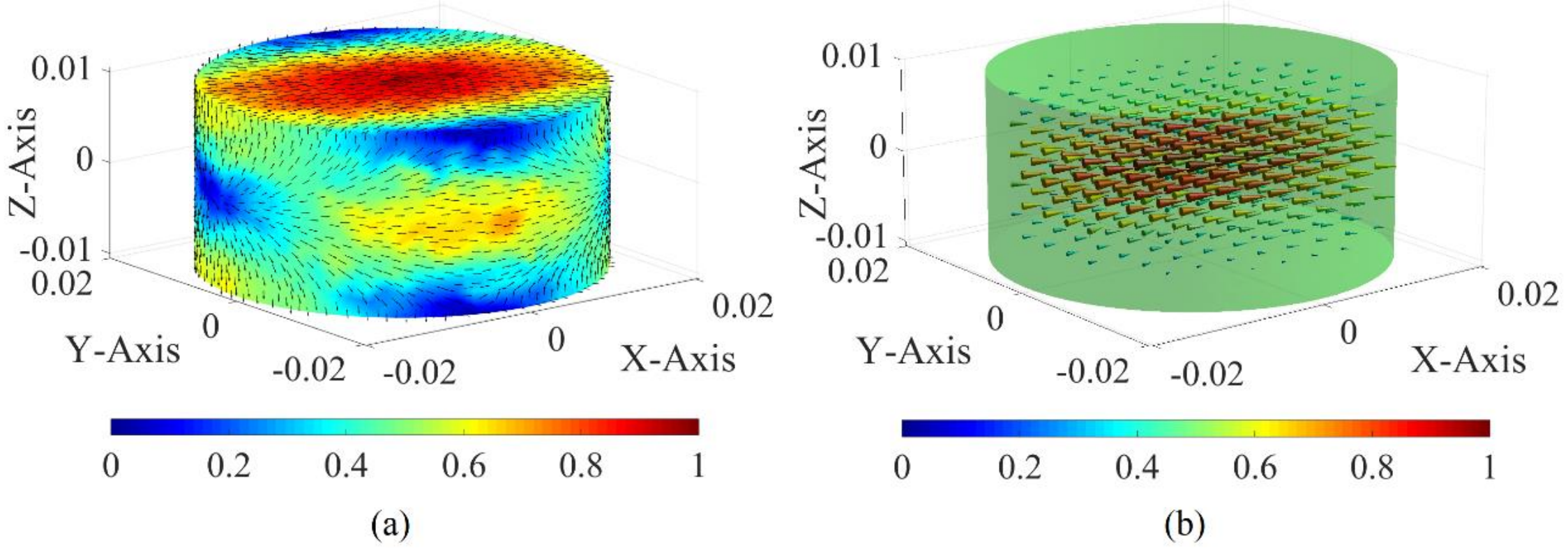

Figure 3-7 Some typical physical quantity distributions corresponding to the first degenerate state of the nonradiative DP-CM02 listed in Table 3-1. (a) modal electric current distribution; (b) modal electric field distributing in internal cavity

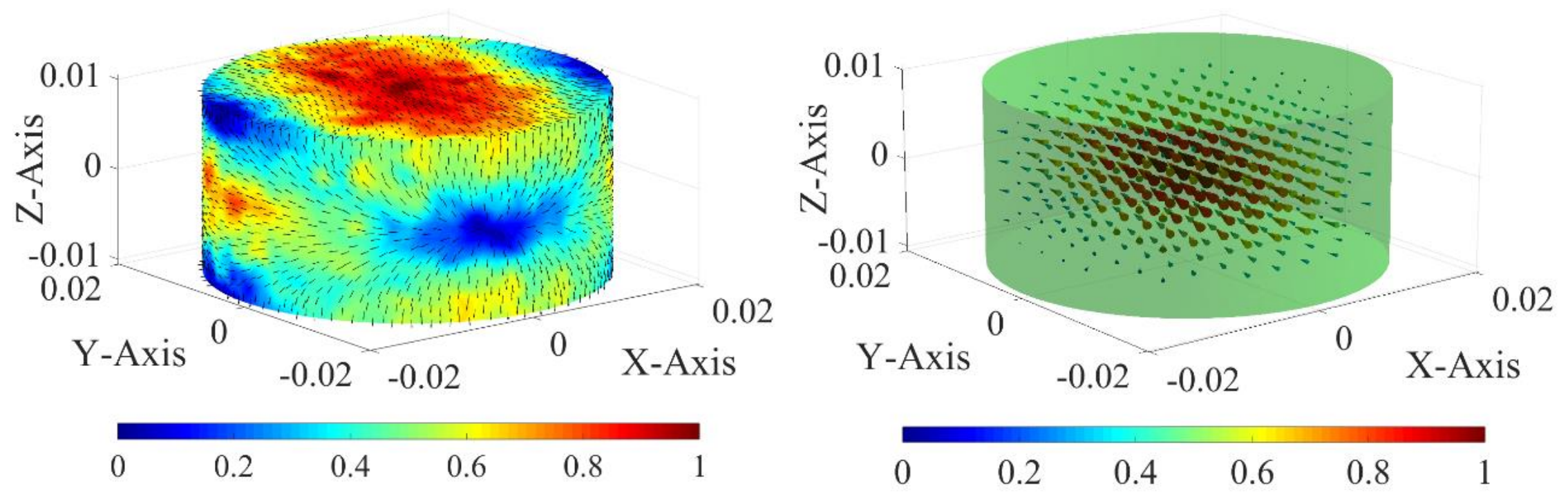

(a) (b)

Figure 3-8 Some typical physical quantity distributions corresponding to the second degenerate state of the nonradiative DP-CM02 listed in Table 3-1. (a) modal electric current distribution; (b) modal electric field distributing in internal cavity

In addition, from Figure 3-4(b) it is easy to find out that: DP-CM06, DP-CM07, DP-CM08, DP-CM09, and DP-CM10 are radiative in whole frequency band, and they are resonant (i.e., their imaginary power are zero) at 6.65GHz, 8.75GHz, 9.75GHz, 10.45GHz, and 10.55GHz respectively. Thus, the characteristic value curves of the above 5 DP-CMs pass through the lateral axis in Figure 3-3(b). Taking the radiative resonant DP-CM06 at 6.65GHz as a typical example, we provide its electric current distribution and radiation pattern as shown in Figures 3-9(a) and 3-9(b), and it is easy to find out that the working state of this DP-CM is similar to magnetic dipole. In addition, we also provide the electric current distributions and radiation patterns of the two degenerate states of the DP-CM07 at 8.75GHz as shown in Figures 3-10 and 3-11.

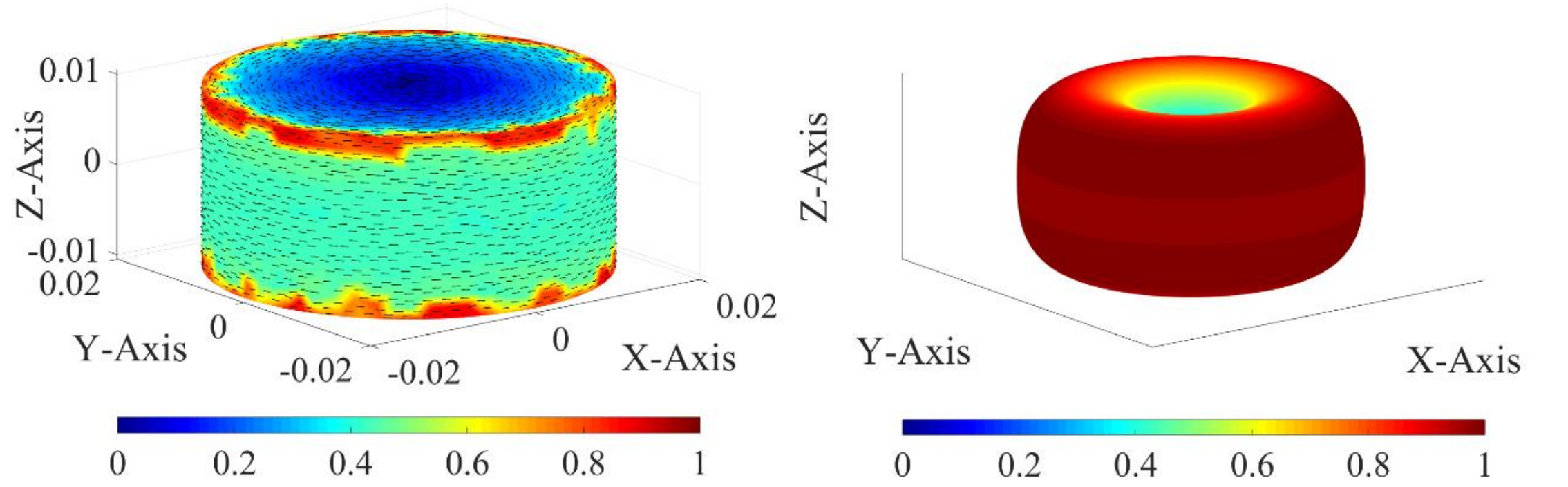

(a) (b)

Figure 3-9 The modal electric current and radiation pattern corresponding to the radiative resonant DP-CM06 working at 6.65GHz. (a) modal electric current; (b) modal radiation pattern

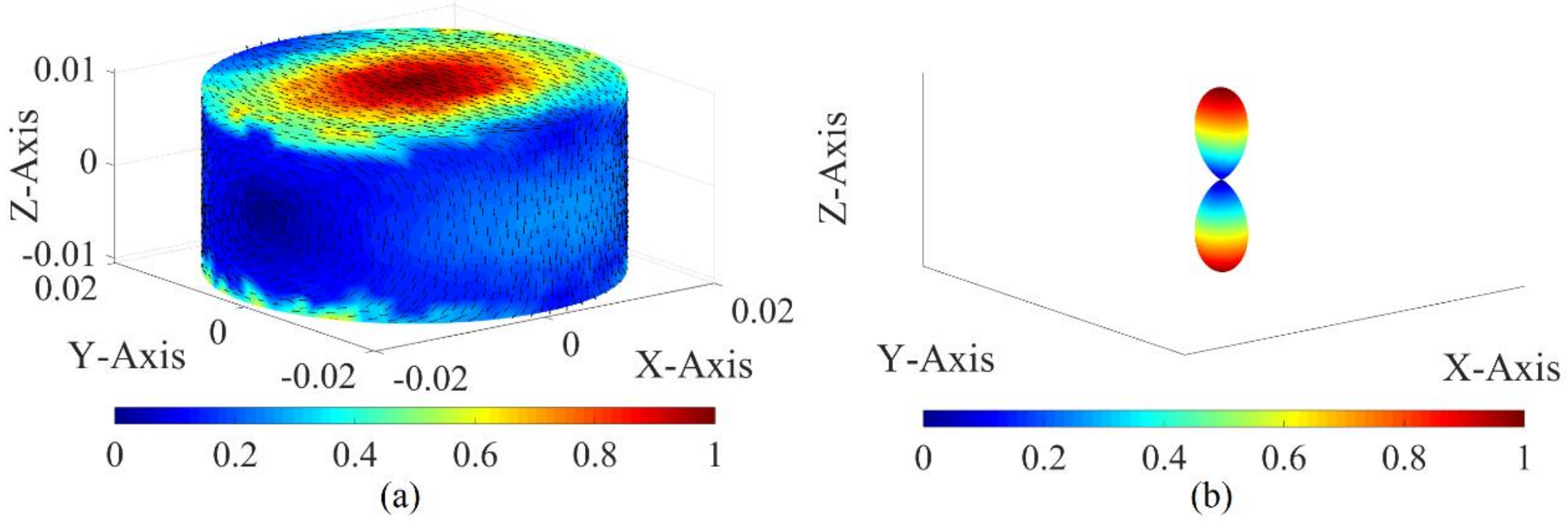

Figure 3-10 The modal electric current and radiation pattern corresponding to the first degenerate state of the radiative resonant DP-CM07 working at 8.75GHz. (a) modal electric current; (b) modal radiation pattern

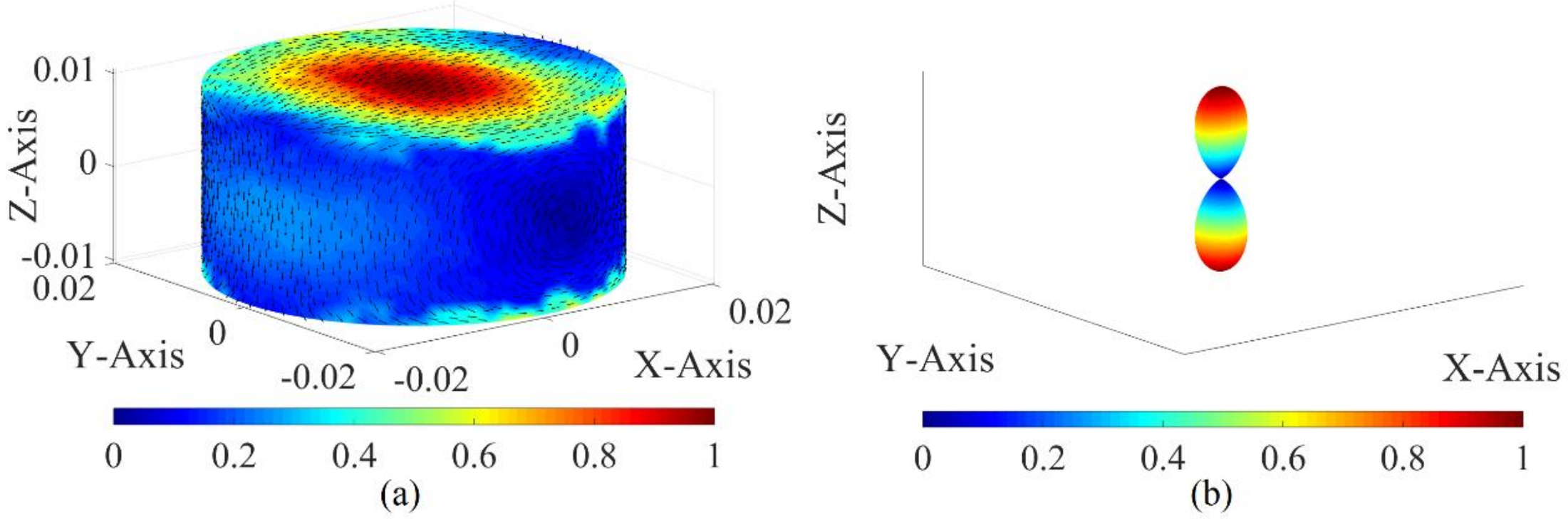

Figure 3-11 The modal electric current and radiation pattern corresponding to the second degenerate state of the radiative resonant DP-CM07 working at 8.75GHz. (a) modal electric current; (b) modal radiation pattern

To verify the various orthogonalities given in Subsection 3.2.3, we, taking the two degenerate states of the radiative resonant DP-CM07 at 8.75GHz as examples, list the orthogonality (3-16) for the radiative degenerate DP-CMs in Table 3-2; we, taking the two degenerate states of the nonradiative resonant DP-CM02 at 8.69GHz as examples, list the orthogonality (3-16) for the nonradiative degenerate DP-CMs in Table 3-3; we, taking the DP-CMs at 7.50GHz as examples, list the orthogonality (3-16) for the DP-CMs with different characteristic values in Table 3-4.

Table 3-2 The (3-16)-based orthogonality between two degenerate states of the radiative resonant DP-CM07 working at 8.75GHz

| $\bar{\alpha}_\zeta$ in (3-16) / $\bar{\alpha}_\xi$ in (3-16) | **The First Degenerate State** | **The Second Degenerate State** |
|---|---|---|
| **The First Degenerate State** | 348.3074 + $j$0.6461 | 1.0925e-10 - $j$3.6087e-10 |
| **The Second Degenerate State** | 1.0788e-10 + $j$3.6127e-10 | 348.3866 + $j$0.6610 |

Table 3-3 The (3-16)-based orthogonality between two degenerate states of the nonradiative DP-CM02 working at 8.69GHz

| $\bar{\alpha}_\zeta$ in (3-16) / $\bar{\alpha}_\xi$ in (3-16) | **The First Degenerate State** | **The Second Degenerate State** |
|---|---|---|
| **The First Degenerate State** | -0.2018 - $j$0.0981 | 1.2960e-09 + $j$1.0164e-10 |
| **The Second Degenerate State** | -1.2957e-09 + $j$1.0688e-10 | -0.1417 - $j$0.0654 |

Table 3-4 The (3-16)-based orthogonality among some DP-CMs which work at 7.50GHz and have different characteristic values

| $\bar{\alpha}_\zeta$ / $\bar{\alpha}_\xi$ | **CM 1** | **CM 2** | **CM 3** | **CM 4** | **CM 5** | **CM 6** | **CM 7** | **CM 8** |
|---|---|---|---|---|---|---|---|---|
| **CM 1** | 65.91 + $j$133.16 | -1.53e-12 + -$j$1.91e-13 | -2.81e-13 + $j$2.75e-13 | 3.11e-13 + -$j$4.63e-13 | -1.23e-13 + $j$1.09e-13 | -2.94e-16 + $j$2.39e-13 | -3.85e-13 + $j$1.93e-12 | -1.60e-14 + $j$8.44e-14 |
| **CM 2** | 1.56e-12 + $j$1.54e-13 | 9.88 + -$j$75.32 | 1.72e-12 + $j$5.37e-12 | -4.98e-12 + -$j$4.58e-12 | 2.24e-12 + -$j$6.87e-13 | 1.21e-12 + -$j$2.75e-13 | -1.46e-12 + -$j$1.65e-12 | -4.98e-13 + -$j$7.27e-13 |
| **CM 3** | 5.35e-13 + $j$8.63e-14 | -2.90e-12 + $j$4.66e-12 | 505.90 + -$j$128.87 | 1.00e-12 + -$j$8.48e-12 | 4.39e-12 + $j$4.37e-13 | 4.09e-13 + -$j$7.30e-14 | 1.26e-12 + -$j$1.89e-12 | 5.14e-13 + -$j$1.27e-12 |
| **CM 4** | 6.21e-15 + -$j$2.26e-13 | 6.08e-12 + -$j$3.40e-12 | -2.22e-13 + -$j$8.61e-12 | 11.10 + -$j$190.06 | -1.74e-11 + $j$2.46e-11 | 3.76e-15 + $j$1.47e-13 | -8.41e-13 + -$j$4.64e-13 | -2.32e-13 + -$j$2.44e-13 |
| **CM 5** | 1.00e-13 + $j$9.02e-14 | -1.92e-12 + -$j$9.45e-13 | -4.32e-12 + -$j$2.13e-13 | 1.41e-11 + $j$2.66e-11 | 10.22 + -$j$158.26 | 6.02e-14 + -$j$1.12e-13 | 8.80e-14 + -$j$7.69e-14 | -5.04e-14 + $j$1.92e-13 |
| **CM 6** | 3.75e-13 + $j$5.09e-13 | -9.30e-13 + -$j$8.86e-13 | -1.27e-13 + $j$3.30e-14 | -1.48e-13 + -$j$2.48e-13 | 1.35e-13 + -$j$1.24e-13 | 360.91 + -$j$116.46 | -5.62e-13 + -$j$1.39e-12 | -1.95e-13 + -$j$1.77e-13 |
| **CM 7** | 8.52e-13 + -$j$1.85e-12 | -9.43e-13 + -$j$6.36e-13 | 2.98e-13 + $j$1.87e-12 | -9.43e-13 + -$j$1.33e-15 | -5.77e-13 + -$j$1.68e-13 | -1.09e-12 + $j$1.12e-12 | 328.36 + $j$93.39 | 3.61e-12 + -$j$1.99e-12 |
| **CM 8** | 6.00e-14 + $j$2.30e-13 | 4.50e-13 + -$j$5.35e-13 | -8.53e-13 + $j$8.15e-13 | -4.46e-13 + -$j$1.22e-13 | 1.96e-14 + $j$1.10e-13 | -3.30e-13 + -$j$8.49e-14 | 1.87e-12 + $j$6.61e-12 | 257.18 + $j$236.30 |

### 2) DP-CMs of Metallic Sphere

We now consider a metallic sphere whose radius is 32mm, and its topological structure and surface triangular meshes are illustrated in Figure 3-12. In frequency band 3~10GHz, some characteristic quantity curves of 8 typical DP-CMs are illustrated in Figures 3-13, 3-14, and 3-15.

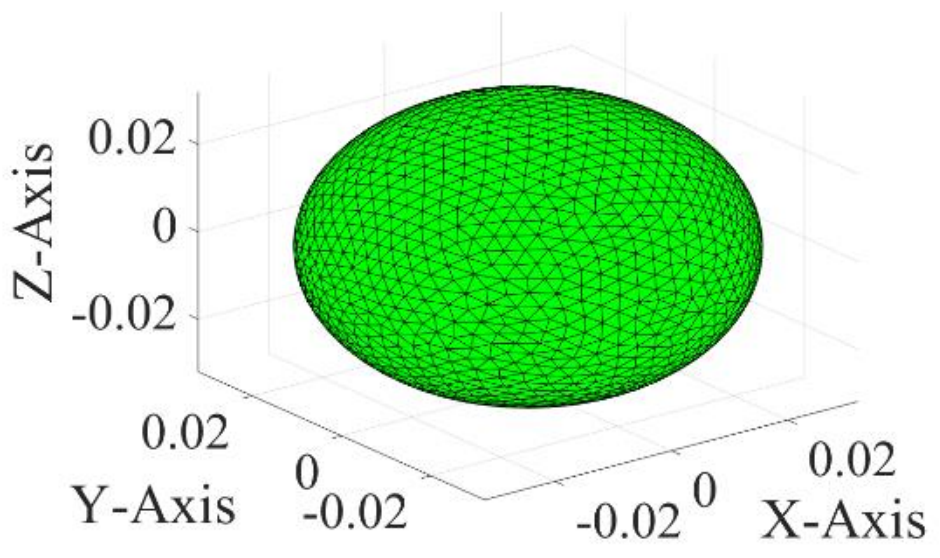

Figure 3-12 The topological structure and surface triangular meshes of a metallic sphere whose radius is 32mm

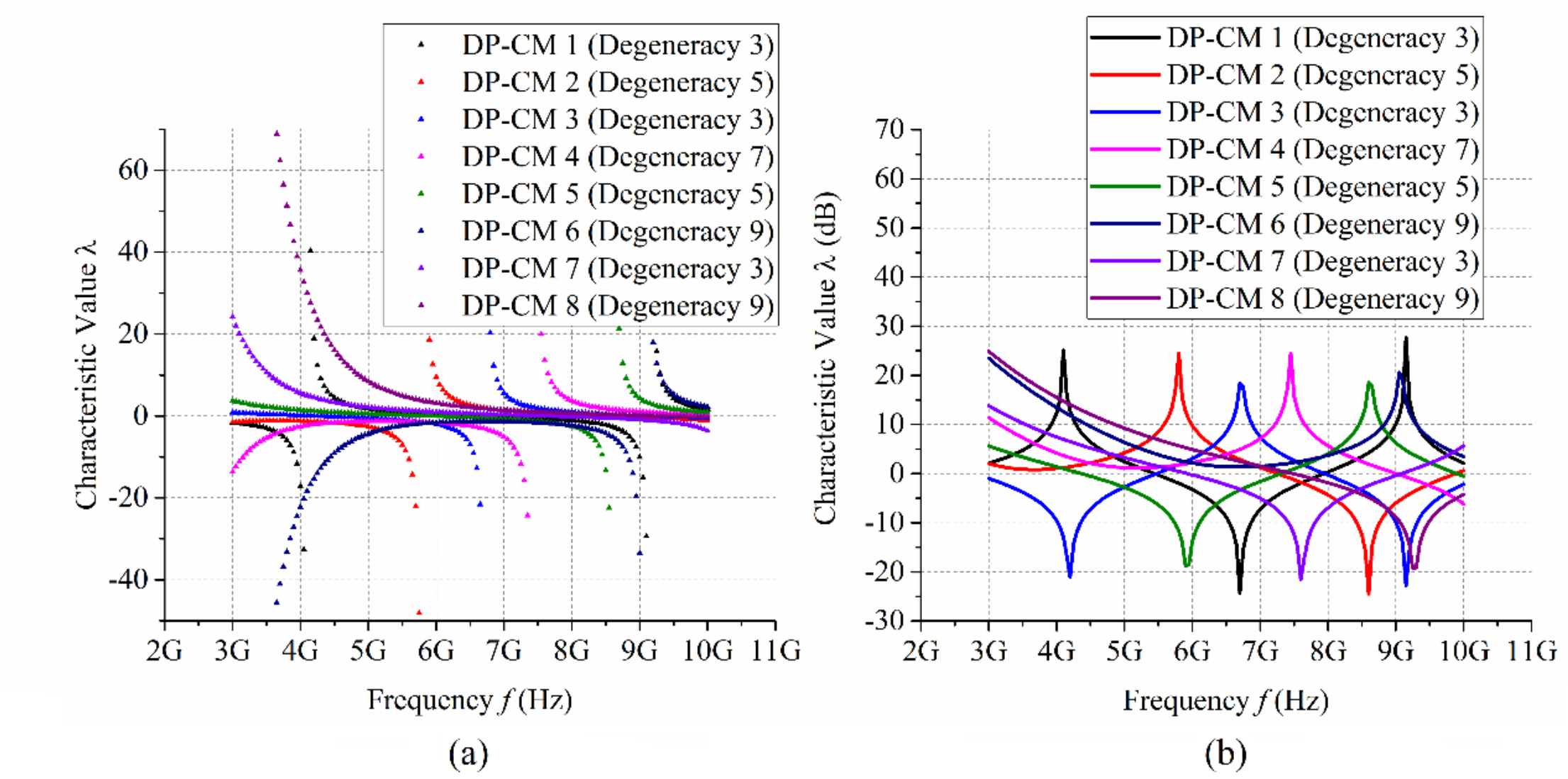

Figure 3-13 The characteristic quantity curves corresponding to 8 typical DP-CMs of the metallic sphere shown in Figure 3-12. (a) characteristic value curves; (b) characteristic value (dB) curves

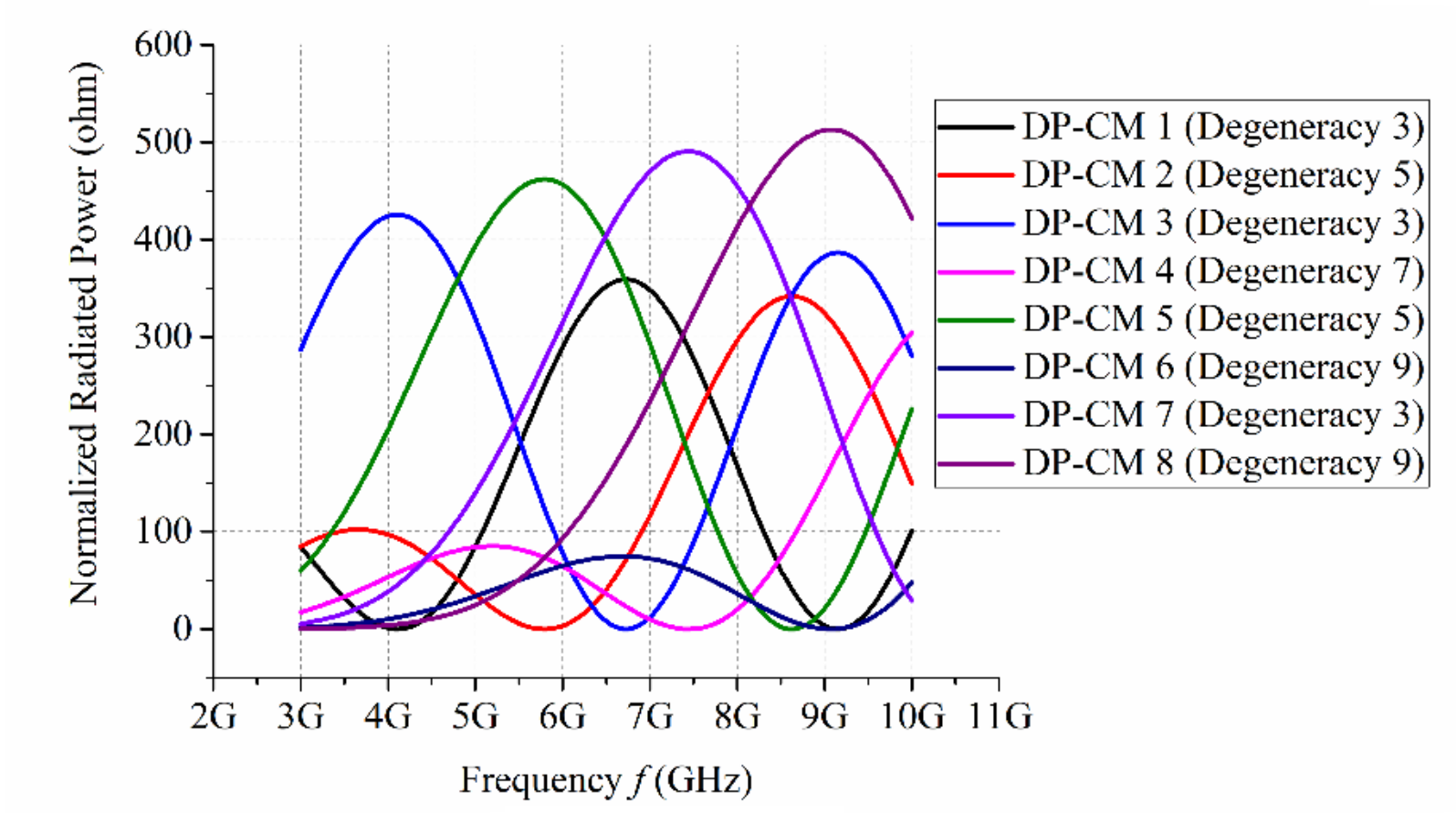

Figure 3-14 The normalized radiated power curves corresponding to the DP-CMs shown in Figure 3-13

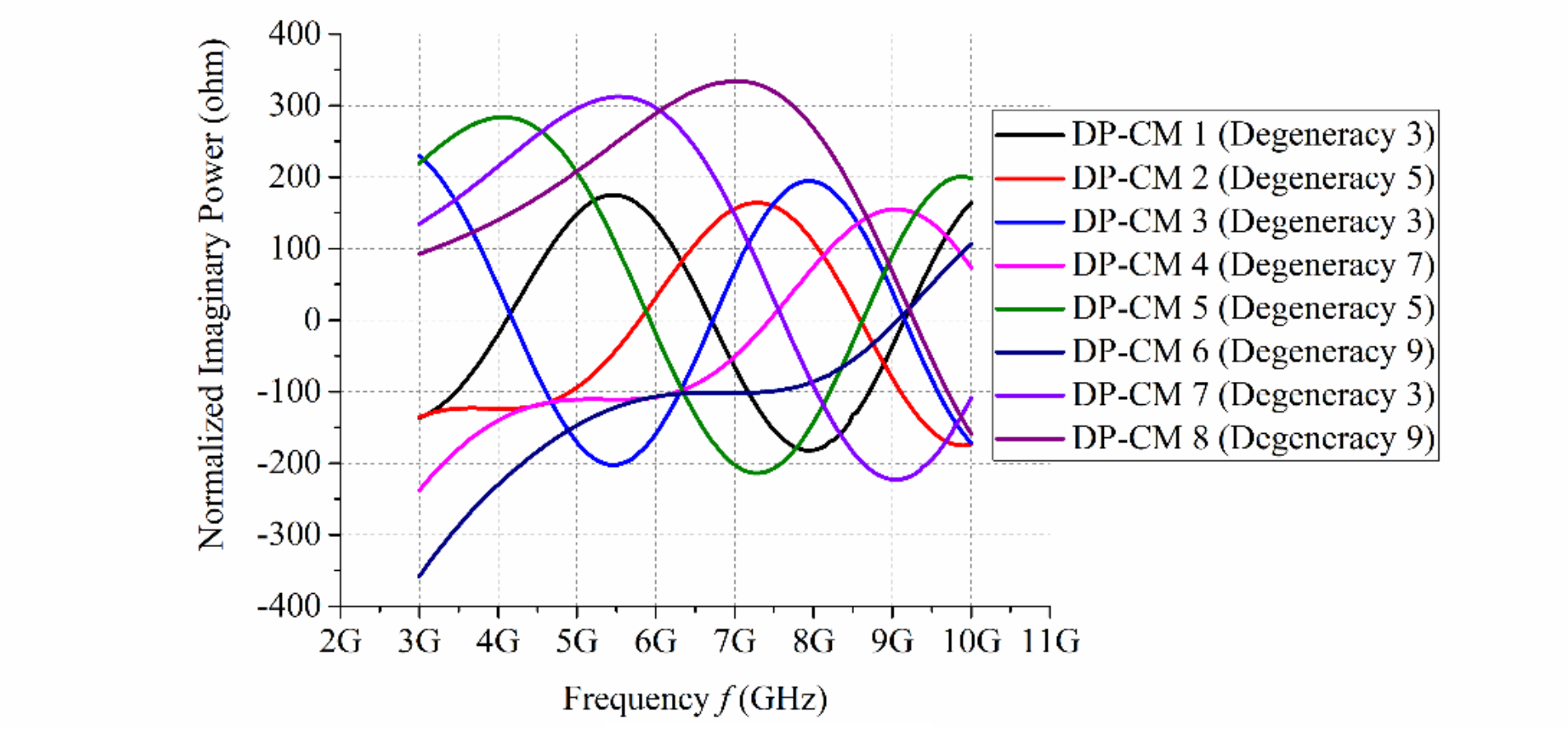

Figure 3-15 The normalized imaginary power curves corresponding to the DP-CMs shown in Figure 3-13

From Figures 3-13, 3-14, and 3-15, it is easy to find out that: DP-CM1 is nonradiative at 4.10GHz, and it satisfies that $\lim_{f \to 4.10\text{G}} (dP_{\text{vac};1}^{\text{imag}}(f) / df) \neq 0$ and $\lim_{f \to 4.10\text{G}} (dP_1^{\text{rad}}(f) / df) = 0$ simultaneously, so the characteristic value of DP-CM1 satisfies that $\lim_{f \to 4.10\text{G}^{\pm}} \lambda_1(f) = \pm\infty$; DP-CM2 is nonradiative at 5.80GHz, and it satisfies that $\lim_{f \to 5.80\text{G}} (dP_{\text{vac};2}^{\text{imag}}(f) / df) \neq 0$ and $\lim_{f \to 5.80\text{G}} (dP_2^{\text{rad}}(f) / df) = 0$ simultaneously, so the characteristic value of DP-CM2 satisfies that $\lim_{f \to 5.80\text{G}^{\pm}} \lambda_2(f) = \pm\infty$; DP-CM3 is nonradiative at 6.75GHz, and it satisfies that $\lim_{f \to 6.75\text{G}} (dP_{\text{vac};3}^{\text{imag}}(f) / df) \neq 0$ and $\lim_{f \to 6.75\text{G}} (dP_3^{\text{rad}}(f) / df) = 0$ simultaneously, so the characteristic value of DP-CM3 satisfies that $\lim_{f \to 6.75\text{G}^{\pm}} \lambda_3(f) = \pm\infty$; DP-CM4 is nonradiative at 7.45GHz, and it satisfies that $\lim_{f \to 7.75\text{G}} (dP_{\text{vac};4}^{\text{imag}}(f) / df) \neq 0$ and $\lim_{f \to 7.45\text{G}} (dP_4^{\text{rad}}(f) / df) = 0$ simultaneously, so the characteristic value of DP-CM4 satisfies that $\lim_{f \to 7.45\text{G}^{\pm}} \lambda_4(f) = \pm\infty$; DP-CM5 is nonradiative at 8.65GHz, and it satisfies that $\lim_{f \to 8.65\text{G}} (dP_{\text{vac};5}^{\text{imag}}(f) / df) \neq 0$ and $\lim_{f \to 8.65\text{G}} (dP_5^{\text{rad}}(f) / df) = 0$ simultaneously, so the characteristic value of DP-CM5 satisfies that $\lim_{f \to 8.65\text{G}^{\pm}} \lambda_5(f) = \pm\infty$; DP-CM6 is nonradiative at 9.05GHz, and it satisfies that $\lim_{f \to 9.05\text{G}} (dP_{\text{vac};6}^{\text{imag}}(f) / df) \neq 0$ and $\lim_{f \to 9.05\text{G}} (dP_6^{\text{rad}}(f) / df) = 0$ simultaneously, so the characteristic value of DP-CM6 satisfies that $\lim_{f \to 9.05\text{G}^{\pm}} \lambda_6(f) = \pm\infty$. In addition, DP-CM1 is also nonradiative at 9.15GHz, and it satisfies that $\lim_{f \to 9.15\text{G}} (dP_{\text{vac};1}^{\text{imag}}(f) / df) \neq 0$ and $\lim_{f \to 9.15\text{G}} (dP_1^{\text{rad}}(f) / df) = 0$ simultaneously, so the characteristic value of DP-CM1 satisfies that $\lim_{f \to 9.15\text{G}^{\pm}} \lambda_1(f) = \pm\infty$.

To compare the above nonradiative DP-CMs with the traditional internally resonant eigen-modes of the closed metallic spherical cavity, we list their working frequencies and degeneracy orders in Table 3-5, and it is easy to find out that there exist some corresponding relationships between the nonradiative DP-CMs and the internally resonant eigen-modes. To reveal the corresponding relationships, we provide the electric current and electric field distributions of the three degenerate states of the DP-CM1 at 4.10GHz as shown in Figures 3-16, 3-17, and 3-18, and it is easy to find out that: these nonradiative DP-CMs satisfy homogeneous tangential electric field boundary conditions. We will further discuss the corresponding relationships between nonradiative DP-CMs and internally resonant eigen-modes in the Section 3.3 of this dissertation.

Table 3-5 The working frequencies (GHz) and the degeneracy orders corresponding to the nonradiative DP-CMs shown in Figure 3-13 and the internally resonant eigen-modes of the metallic spherical cavity shown in Figure 3-12

| **Modal Type** / **Modal Index** | **Nonradiative DP-CM[37,40,41]** | **Internally Resonant Eigen-mode[26,28]** |
|---|---|---|
| **Mode 1** | 4.10 (3) | 4.09 (3) |
| **Mode 2** | 5.80 (5) | 5.77 (5) |
| **Mode 3** | 6.75 (3) | 6.70 (3) |
| **Mode 4** | 7.45 (7) | 7.42 (7) |
| **Mode 5** | 8.65 (5) | 8.60 (5) |
| **Mode 6** | 9.05 (9) | 9.05 (9) |
| **Mode 7** | 9.15 (3) | 9.13 (3) |

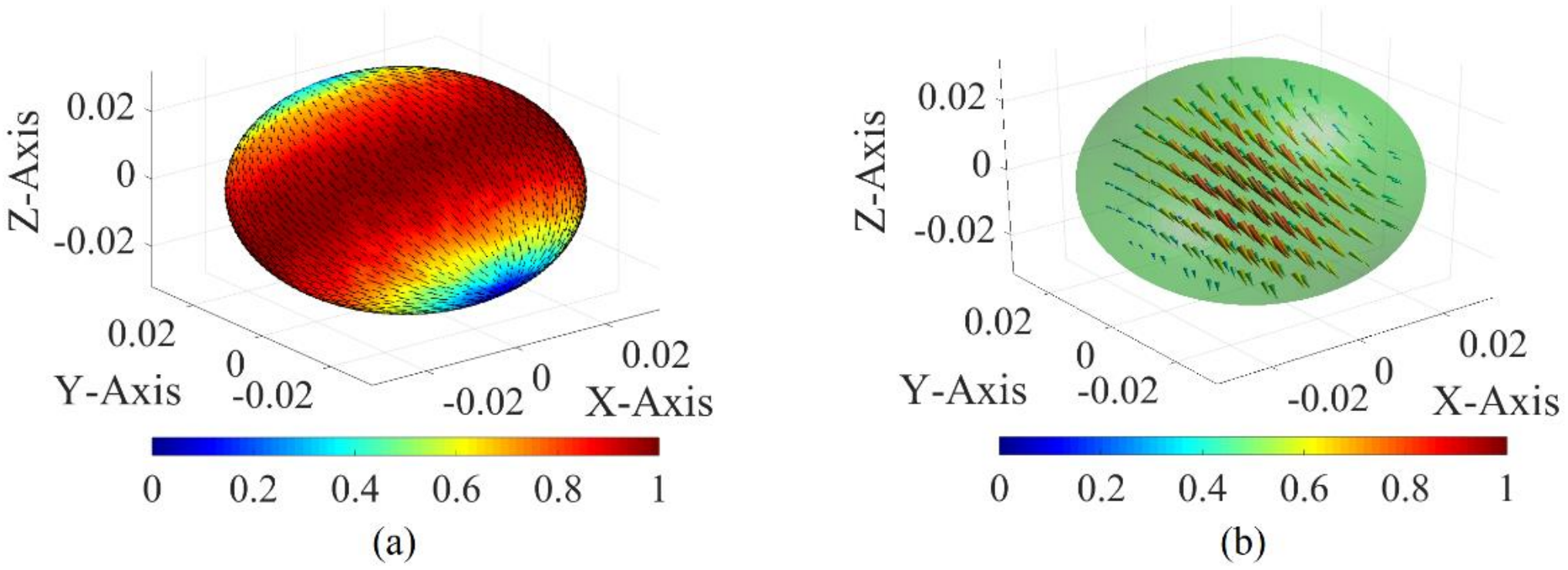

Figure 3-16 Some typical physical quantity distributions corresponding to the first degenerate state of the nonradiative DP-CM1 listed in Table 3-5. (a) modal electric current distribution; (b) modal electric field distributing in internal cavity

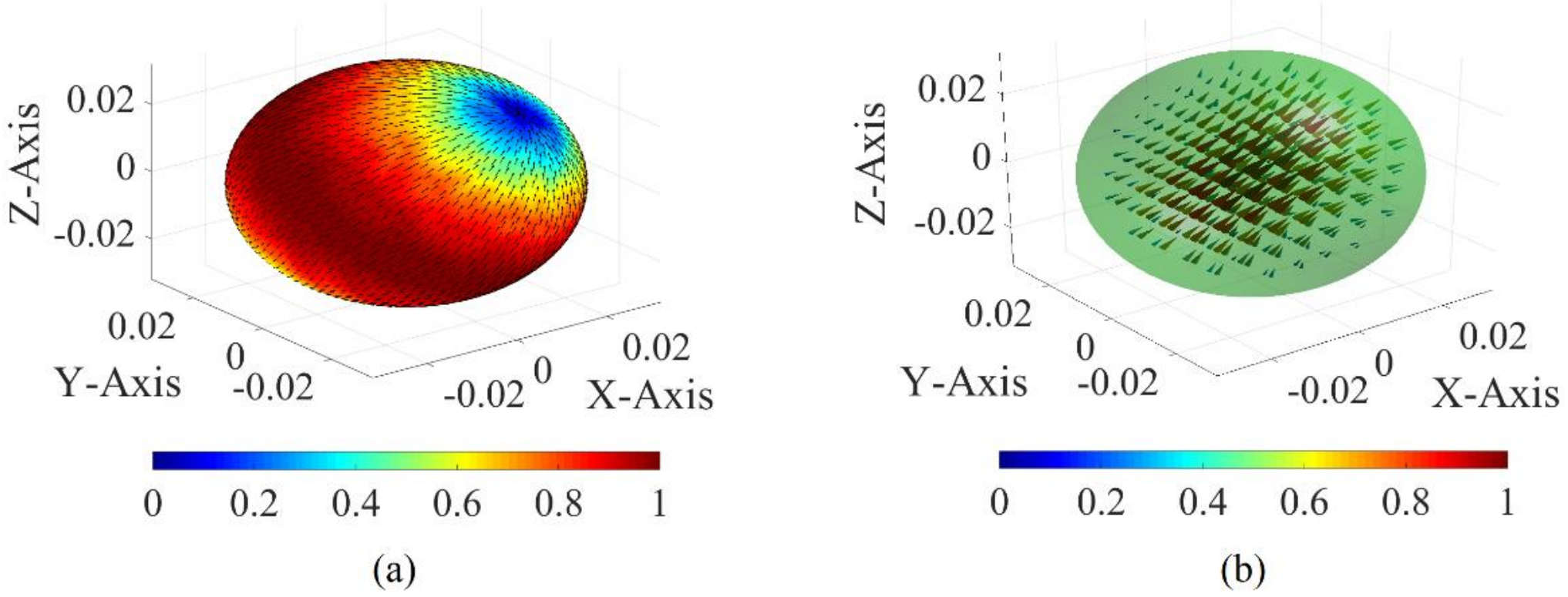

Figure 3-17 Some typical physical quantity distributions corresponding to the second degenerate state of the nonradiative DP-CM1 listed in Table 3-5. (a) modal electric current distribution; (b) modal electric field distributing in internal cavity

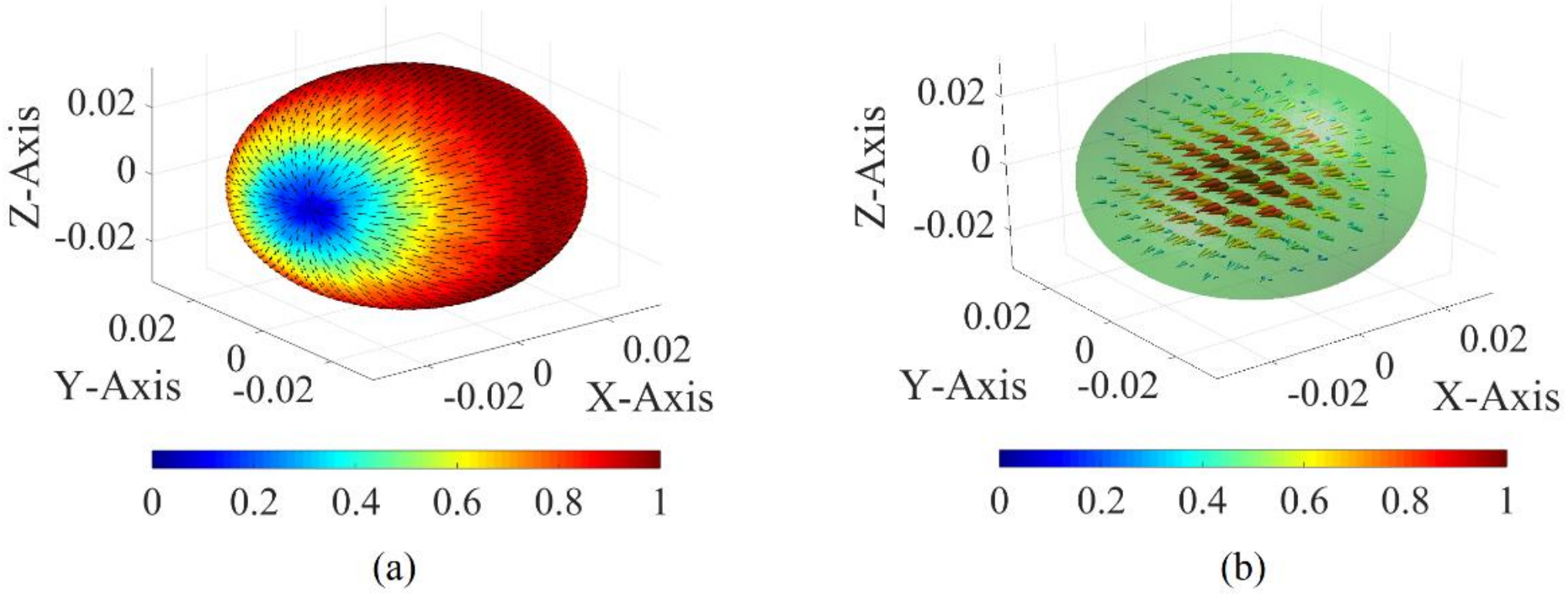

Figure 3-18 Some typical physical quantity distributions corresponding to the third degenerate state of the nonradiative DP-CM1 listed in Table 3-5. (a) modal electric current distribution; (b) modal electric field distributing in internal cavity

In addition, from Figure 3-14 it is easy to find out that: DP-CM7 and DP-CM8 are radiative in whole frequency band, and they are resonant (i.e., their imaginary powers are zero) at 7.60GHz and 9.25GHz respectively. Thus, the characteristic value curves of the above two DP-CMs pass through the lateral axis in Figure 3-15; DP-CM3 is resonant and radiative at 4.10GHz and 9.15GHz; DP-CM5 is resonant and radiative at 5.90GHz; DP-CM1 is resonant and radiative at 6.70GHz; CM2 is resonant and radiative at 8.60GHz. Taking the radiative resonant DP-CM3 at 4.10GHz as a typical example, we provide the electric current distribution and radiation pattern of its three degenerate states as shown Figures 3-19, 3-20, and 3-21.

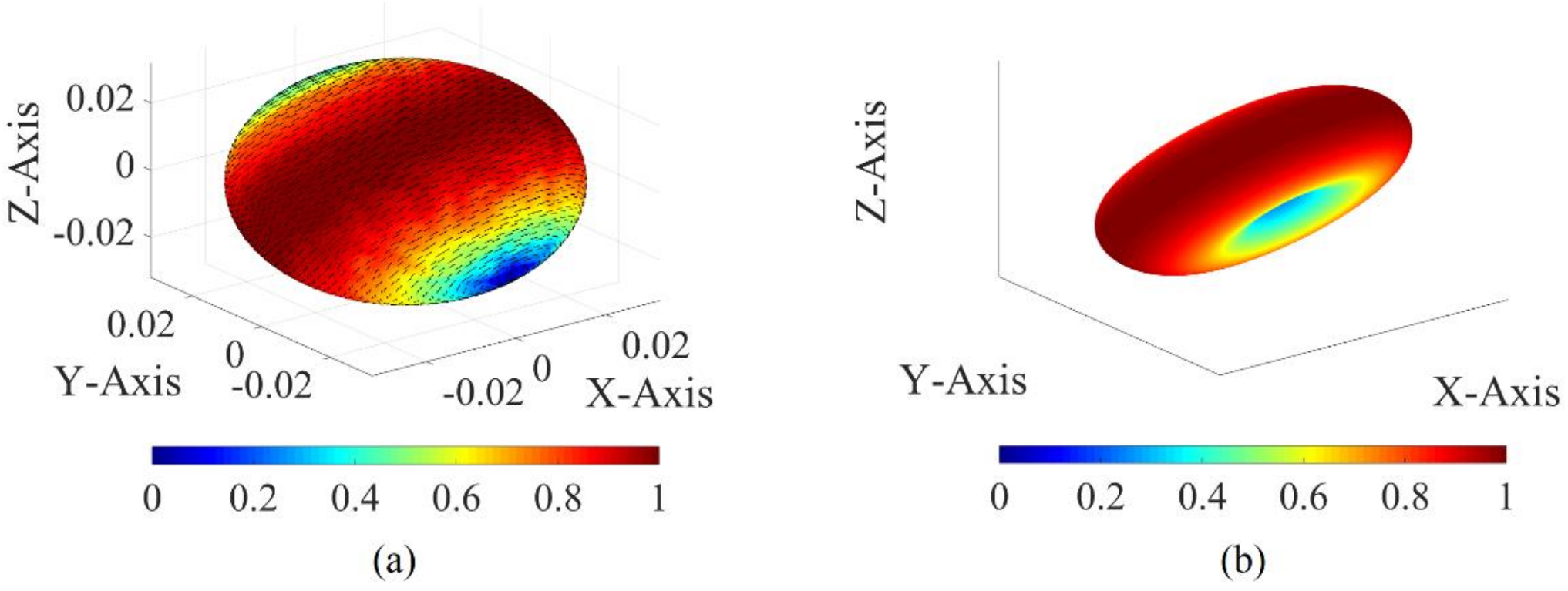

Figure 3-19 The modal electric current and radiation pattern corresponding to the first degenerate state of the radiative resonant DP-CM3 working at 4.10GHz. (a) modal electric current; (b) modal radiation pattern

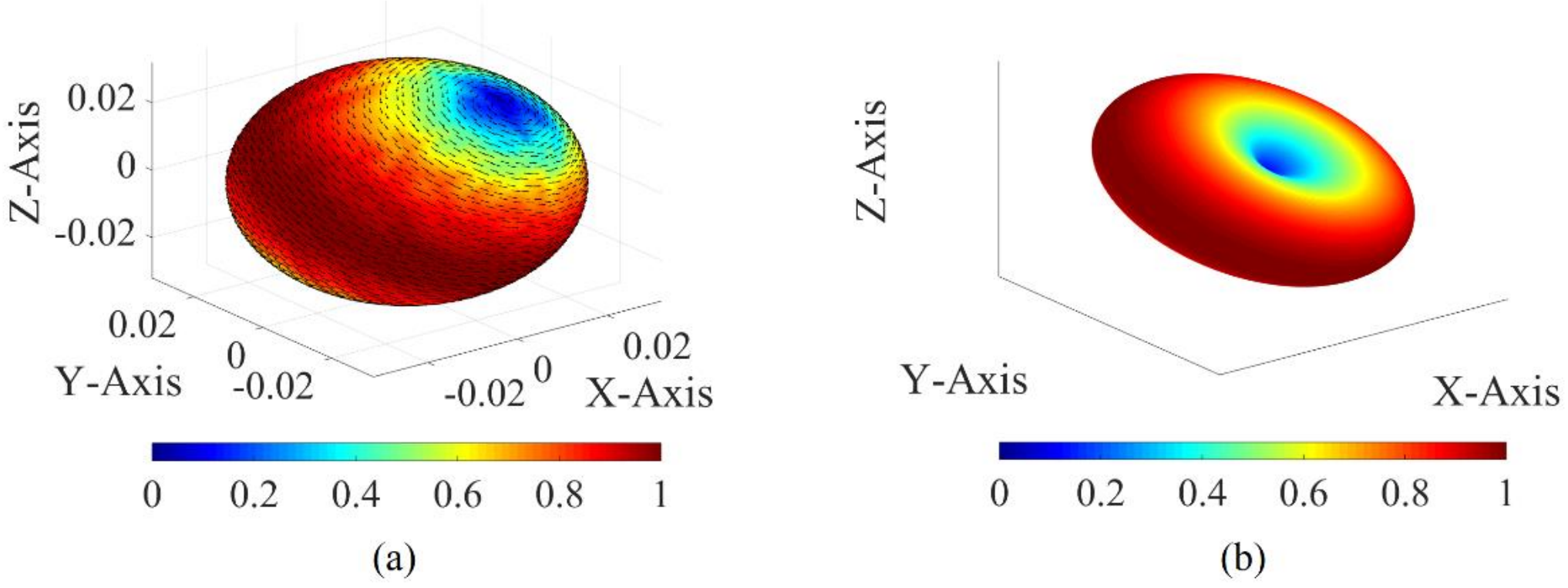

Figure 3-20 The modal electric current and radiation pattern corresponding to the second degenerate state of the radiative resonant DP-CM3 working at 4.10GHz. (a) modal electric current; (b) modal radiation pattern

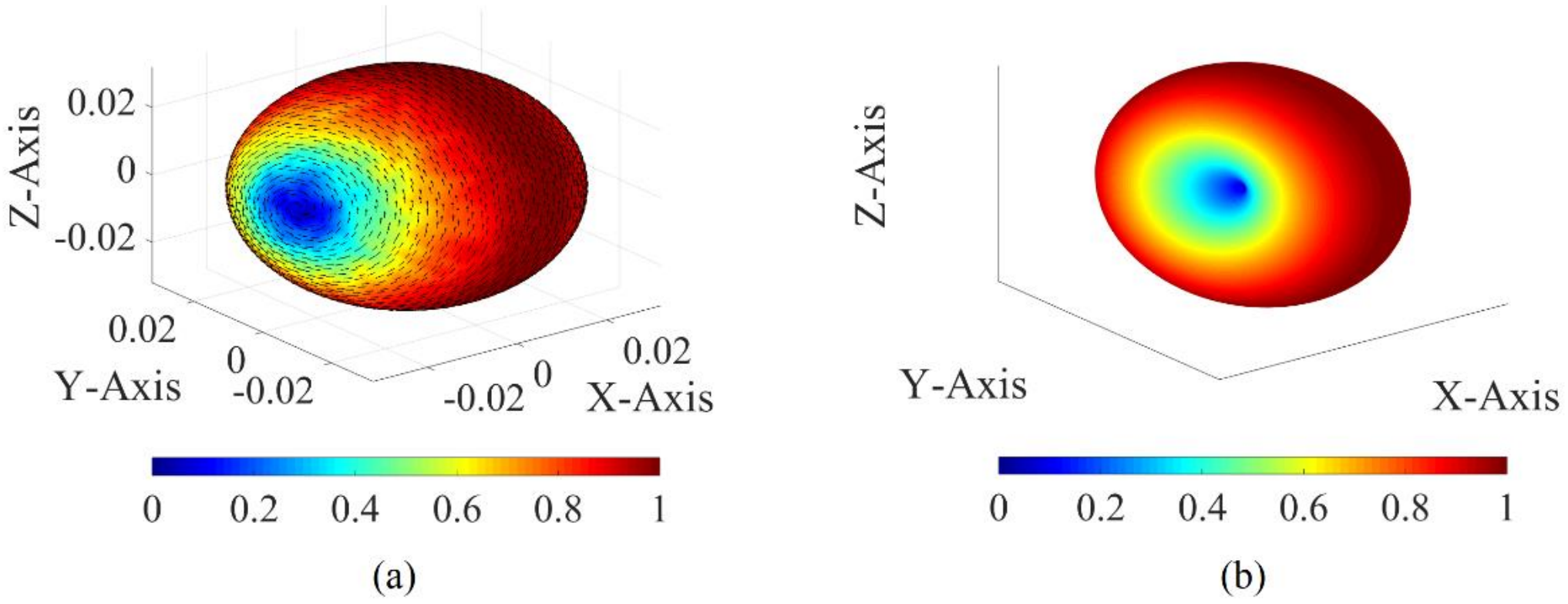

Figure 3-21 The modal electric current and radiation pattern corresponding to the third degenerate state of the radiative resonant DP-CM3 working at 4.10GHz. (a) modal electric current; (b) modal radiation pattern

In fact, the DP-CMs of the metallic sphere shown in Figure 3-12 also satisfy the orthogonality similarly to the DP-CMs of the metallic cylinder shown in Figure 3-2, and it will not be repeated here.

## 3.3 Modal Classification, Modal Expansion, and Modal Decomposition for Metallic Systems

In this section, we firstly reclassify all of the working modes of any objective metallic system based on DP; secondly, based on the new modal classification and the completeness and orthogonality of the DP-CMs of the metallic system, we expands an arbitrary working mode of the metallic system in terms of the linear combinations of the DP-CMs; finally, we accomplish the orthogonal decompositions for the working modes and whole modal space of the metallic system by employing DP-CM-based modal expansion. For the convenience of the following discussions, the following parts of this section are discussed in expansion vector space.

### 3.3.1 DP-Based Modal Classification

In this subsection, we simply retrospect the traditional classification for all working modes, and then propose a new modal classification. Afterwards, we classify all DP-CMs based on the new classification method.

**1) Traditional Classification for Working Modes**

The power matrix $\overline{\overline{P}}_{+}^{\text{Driving}}$ given in Section 3.2 is positive definite or semi-definite, so $\overline{a}^H \cdot \overline{\overline{P}}_{+}^{\text{Driving}} \cdot \overline{a} \geq 0$ for any working mode $\overline{a}$ [116]. Traditionally, the modes corresponding to $P^{\text{rad}} = \overline{a}^H \cdot \overline{\overline{P}}_{+}^{\text{Driving}} \cdot \overline{a} = 0$ and $P^{\text{rad}} = \overline{a}^H \cdot \overline{\overline{P}}_{+}^{\text{Driving}} \cdot \overline{a} > 0$ are called as non-radiative and radiative modes respectively, and this dissertation denotes them as $\overline{a}_{\text{non-rad}}$ and $\overline{a}_{\text{rad}}$ respectively. In addition, the positive definiteness and semi-definiteness of power matrix $\overline{\overline{P}}_{+}^{\text{Driving}}$ guarantee that: $\overline{a}^H \cdot \overline{\overline{P}}_{+}^{\text{Driving}} \cdot \overline{a} = 0$ if and only if $\overline{\overline{P}}_{+}^{\text{Driving}} \cdot \overline{a} = 0$ [116]. In fact, this implies the following equivalence relationships[37]:

$$\overline{\overline{P}}_{+}^{\text{Driving}} \cdot \overline{a} = 0 \quad \Leftrightarrow \quad \overline{a}^H \cdot \overline{\overline{P}}_{+}^{\text{Driving}} \cdot \overline{a} = 0 \quad \Leftrightarrow \quad \overline{a} \text{ is a nonradiative mode}. \tag{3-22}$$

Thus, all $\overline{a}_{\text{non-rad}}$ constitute a linear space (i.e. the null space of matrix $\overline{\overline{P}}_{+}^{\text{Driving}}$ denoted as $\text{nullspace}\,\overline{\overline{P}}_{+}^{\text{Driving}}$ [37,116]) ①, and this dissertation calls the space as nonradiation space. In addition, it is obvious that any $\overline{a}_{\text{non-rad}}$ and any $\overline{a}$ satisfy the following orthogonality[14]:

① This space is also the same as the internal resonance space of the metallic system[37,40,41].

$$\bar{a}^H \cdot \bar{\bar{P}}_+^{\text{Driving}} \cdot \bar{a}_{\text{non-rad}} = 0 = \bar{a}_{\text{non-rad}}^H \cdot \bar{\bar{P}}_+^{\text{Driving}} \cdot \bar{a} \tag{3-23}$$

due to that $\bar{\bar{P}}_+^{\text{Driving}}$ is Hermitian.

The power matrix $\bar{\bar{P}}_-^{\text{Driving}}$ given in Section 3.2 is indefinite, so real number $\bar{a}^H \cdot \bar{\bar{P}}_-^{\text{Driving}} \cdot \bar{a}$ can be negative or zero or positive. Traditionally, the modes corresponding to $\bar{a}^H \cdot \bar{\bar{P}}_-^{\text{Driving}} \cdot \bar{a} < 0$, $\bar{a}^H \cdot \bar{\bar{P}}_-^{\text{Driving}} \cdot \bar{a} = 0$, and $\bar{a}^H \cdot \bar{\bar{P}}_-^{\text{Driving}} \cdot \bar{a} > 0$ are called as capacitive modes, resonant modes, and inductive modes respectively, and this dissertation denotes them as $\bar{a}^{\text{cap}}$, $\bar{a}^{\text{res}}$, and $\bar{a}^{\text{ind}}$ respectively. In fact, according to whether or not the resonant modes radiate EM energy, the resonant modes can be further classified into externally resonant modes and internally resonant modes①. Based on the conclusions given in literatures [118,119] and the methods used in literatures [40,41], it can be proven that all nonradiative modes are resonant, and then all capacitive and inductive modes are radiative. Thus, this dissertation more completely calls capacitive modes and inductive modes as radiative capacitive modes and radiative inductive modes respectively, and correspondingly denotes them as $\bar{a}_{\text{rad}}^{\text{cap}}$ and $\bar{a}_{\text{rad}}^{\text{ind}}$ respectively.

**2) New Classification for Working Modes**

Matrix $\bar{\bar{P}}_-^{\text{Driving}}$ is indefinite, so $\bar{a}^H \cdot \bar{\bar{P}}_-^{\text{Driving}} \cdot \bar{a} = 0$ doesn't imply that $\bar{\bar{P}}_-^{\text{Driving}} \cdot \bar{a} = 0$[116], though $\bar{\bar{P}}_-^{\text{Driving}} \cdot \bar{a} = 0$ always implies that $\bar{a}^H \cdot \bar{\bar{P}}_-^{\text{Driving}} \cdot \bar{a} = 0$. Thus, there exists the following relationships[37]:

$$\bar{\bar{P}}_-^{\text{Driving}} \cdot \bar{a} = 0 \;\; \underset{\not\Leftarrow}{\Rightarrow} \;\; \bar{a}^H \cdot \bar{\bar{P}}_-^{\text{Driving}} \cdot \bar{a} = 0 \;\; \Leftrightarrow \;\; \bar{a} \text{ is a resonant mode}. \tag{3-24}$$

In other words, $\bar{\bar{P}}_-^{\text{Driving}} \cdot \bar{a} = 0$ is a stronger condition than $\bar{a}^H \cdot \bar{\bar{P}}_-^{\text{Driving}} \cdot \bar{a} = 0$ in the aspect of guaranteeing resonance. Based on this, this dissertation calls $\bar{a}^H \cdot \bar{\bar{P}}_-^{\text{Driving}} \cdot \bar{a} = 0$ as resonance condition, and calls $\bar{\bar{P}}_-^{\text{Driving}} \cdot \bar{a} = 0$ as pure resonance condition②.

Correspondingly, this dissertation calls the modes satisfying $\bar{\bar{P}}_-^{\text{Driving}} \cdot \bar{a} = 0$ as purely resonant modes, and denotes the modes as $\bar{a}^{\text{pur res}}$; this dissertation calls the modes which satisfy $\bar{a}^H \cdot \bar{\bar{P}}_-^{\text{Driving}} \cdot \bar{a} = 0$ but don't satisfy $\bar{\bar{P}}_-^{\text{Driving}} \cdot \bar{a} = 0$ as impurely resonant modes, and denotes the modes as $\bar{a}^{\text{impur res}}$③. Evidently, all purely resonant modes constitute a linear space, i.e. the null space of $\bar{\bar{P}}_-^{\text{Driving}}$ —— nullspace $\bar{\bar{P}}_-^{\text{Driving}}$ [37,116], and

① So-called externally resonant modes are the modes which are both radiative and resonant. So-called internally resonant modes are the modes which are both nonradiative and resonant.[57,61]

② The reason to utilize modifier "pure" here will be given in Subsection 3.3.3.

③ Here, it needs to be emphasized that: both purely resonant modes and impurely resonant modes are resonant modes. The formers satisfy the stronger condition —— pure resonance condition $\bar{\bar{P}}_-^{\text{Driving}} \cdot \bar{a} = 0$, so they naturally satisfy the weaker condition —— resonance condition $\bar{a}^H \cdot \bar{\bar{P}}_-^{\text{Driving}} \cdot \bar{a} = 0$. However, the latters only satisfy the weaker condition $\bar{a}^H \cdot \bar{\bar{P}}_-^{\text{Driving}} \cdot \bar{a} = 0$, but don't satisfy the stronger condition $\bar{\bar{P}}_-^{\text{Driving}} \cdot \bar{a} = 0$.

the space is called as pure resonance space in this dissertation. Similarly to orthogonality (3-23), any $\overline{a}^{\text{pur res}}$ and any $\overline{a}$ satisfy the following relationship[37]:

$$\overline{a}^H \cdot \overline{\overline{P}}_-^{\text{Driving}} \cdot \overline{a}^{\text{pur res}} = 0 = \left(\overline{a}^{\text{pur res}}\right)^H \cdot \overline{\overline{P}}_-^{\text{Driving}} \cdot \overline{a} \tag{3-25}$$

This dissertation calls the $\overline{a}^{\text{pur res}}$ which satisfy condition $(\overline{a}^{\text{pur res}}) \cdot \overline{\overline{P}}_+^{\text{Driving}} \cdot \overline{a}^{\text{pur res}} = 0$ as nonradiative purely resonant modes, and denotes them as $\overline{a}_{\text{non-rad}}^{\text{pur res}}$. This dissertation calls the $\overline{a}^{\text{pur res}}$ which satisfy condition $(\overline{a}^{\text{pur res}}) \cdot \overline{\overline{P}}_+^{\text{Driving}} \cdot \overline{a}^{\text{pur res}} > 0$ as radiative purely resonant modes, and denotes them as $\overline{a}_{\text{rad}}^{\text{pur res}}$. Because $\overline{\overline{P}}_+^{\text{Driving}} \cdot \overline{a} = 0$ is a sufficient condition of $\overline{\overline{P}}_-^{\text{Driving}} \cdot \overline{a} = 0$①, then: pure resonance space $\text{nullspace}\,\overline{\overline{P}}_-^{\text{Driving}}$ contains whole nonradiation space $\text{nullspace}\,\overline{\overline{P}}_+^{\text{Driving}}$. This implies that: all nonradiative purely resonant modes constitute a linear space, and this space is just nonradiation space $\text{nullspace}\,\overline{\overline{P}}_+^{\text{Driving}}$; all impurely resonant modes must be radiative modes, so this dissertation further denotes them as $\overline{a}_{\text{rad}}^{\text{impur res}}$; any $\overline{a}_{\text{non-rad}}^{\text{pur res}}$ and any $\overline{a}$ satisfy the following orthogonality[37]:

$$\overline{a}^H \cdot \overline{\overline{P}}_\pm^{\text{Driving}} \cdot \overline{a}_{\text{non-rad}}^{\text{pur res}} = 0 = \left(\overline{a}_{\text{non-rad}}^{\text{pur res}}\right)^H \cdot \overline{\overline{P}}_\pm^{\text{Driving}} \cdot \overline{a} \tag{3-26}$$

By introducing the concepts of pure resonance and impure resonance, we have decomposed whole resonant mode set $\{\overline{a}^{\text{res}}\}$ into three subsets —— nonradiative purely resonant mode set $\{\overline{a}_{\text{non-rad}}^{\text{pur res}}\}$, radiative purely resonant mode set $\{\overline{a}_{\text{rad}}^{\text{pur res}}\}$, and impurely resonant mode set $\{\overline{a}_{\text{rad}}^{\text{impur res}}\}$. Set $\{\overline{a}_{\text{non-rad}}^{\text{pur res}}\}$ is just the internally resonant mode set[37], so the introductions of sets $\{\overline{a}_{\text{rad}}^{\text{pur res}}\}$ and $\{\overline{a}_{\text{rad}}^{\text{impur res}}\}$ are essentially a fine classification for traditional externally resonant modes. In addition, traditional capacitive modes and traditional inductive modes can also be finely classified similarly, and the related contents will be detailedly discussed in the Subsection 3.3.3 of this dissertation.

**3) Classification for DP-CMs**

The above traditional modal classification and new modal classification are applicable to whole modal space $\{\overline{a}\}$, so they can also be applied to DP-CM set $\{\overline{\alpha}_\xi\}$, because $\{\overline{\alpha}_\xi\}$ is a proper subset of $\{\overline{a}\}$ (i.e., $\{\overline{\alpha}_\xi\} \subset \{\overline{a}\}$)②.

Specifically, this dissertation decomposes whole DP-CM set $\{\overline{\alpha}_\xi\}$ into four subsets: the set $\{\overline{\alpha}_{\text{rad};\xi}^{\text{cap}}\}$ constituted by all radiative capacitive DP-CMs, the set $\{\overline{\alpha}_{\text{non-rad};\xi}^{\text{res}}\}$

① According to the conclusions given in literatures [118,119] and the methods developed in literatures [40,41], this conclusion can be easily proven. In addition, literature [37] also provided some typical examples corresponding to this conclusion.

② Please notice that: the vector $\overline{a}$ in this dissertation represents the expansion vector of a general working mode; the vector $\overline{\alpha}_\xi$ in this dissertation represents the expansion vector of the $\xi$-th DP-CM.

constituted by all nonradiative resonant DP-CMs, the set $\{\bar{\alpha}_{\text{rad};\xi}^{\text{res}}\}$ constituted by all radiative resonant DP-CMs, and the set $\{\bar{\alpha}_{\text{rad};\xi}^{\text{ind}}\}$ constituted by all radiative inductive DP-CMs①. In addition, the union of $\{\bar{\alpha}_{\text{non-rad};\xi}^{\text{res}}\}$ and $\{\bar{\alpha}_{\text{rad};\xi}^{\text{res}}\}$ is denoted as $\{\bar{\alpha}_{\xi}^{\text{res}}\}$, and it is obvious that $\{\bar{\alpha}_{\xi}^{\text{res}}\}$ is just the set constituted by all resonant DP-CMs. As mentioned in Subsection 3.2.2, all $\bar{\alpha}_{\xi}^{\text{res}}$ satisfy equation $\bar{\bar{P}}_{-}^{\text{Driving}} \cdot \bar{\alpha}_{\xi}^{\text{res}} = 0$. This implies that: all resonant CMs are purely resonant. Thus, all nonradiative resonant CMs and all radiative resonant CMs are purely resonant. Based on these above, this dissertation denotes $\bar{\alpha}_{\xi}^{\text{res}}$ as $\bar{\alpha}_{\xi}^{\text{pur res}}$, and denotes $\bar{\alpha}_{\text{non-rad};\xi}^{\text{res}}$ and $\bar{\alpha}_{\text{rad};\xi}^{\text{res}}$ as $\bar{\alpha}_{\text{non-rad};\xi}^{\text{pur res}}$ and $\bar{\alpha}_{\text{rad};\xi}^{\text{pur res}}$ respectively.

Based on the related conclusions in matrix theory[116], it is easy to conclude that: the number of the elements in set $\{\bar{\alpha}_{\xi}^{\text{pur res}}\}$ equals the rank of space $\text{nullspace}\,\bar{\bar{P}}_{-}^{\text{Driving}}$, and the number of the elements in set $\{\bar{\alpha}_{\text{non-rad};\xi}^{\text{pur res}}\}$ equals the rank of space $\text{nullspace}\,\bar{\bar{P}}_{+}^{\text{Driving}}$. In addition, all $\bar{\alpha}_{\xi}$ are independent of each others, so all $\bar{\alpha}_{\xi}^{\text{pur res}}$ and all $\bar{\alpha}_{\text{non-rad};\xi}^{\text{pur res}}$ are independent respectively[116]. These above imply that: $\{\bar{\alpha}_{\xi}^{\text{pur res}}\}$ constitute the basis of $\text{nullspace}\,\bar{\bar{P}}_{-}^{\text{Driving}}$, so any purely resonant mode can be expanded in terms of the linear combination of $\{\bar{\alpha}_{\xi}^{\text{pur res}}\}$; $\{\bar{\alpha}_{\text{non-rad};\xi}^{\text{pur res}}\}$ constitute the basis of $\text{nullspace}\,\bar{\bar{P}}_{+}^{\text{Driving}}$, so any nonradiative mode can be expanded in terms of the linear combination of $\{\bar{\alpha}_{\text{non-rad};\xi}^{\text{pur res}}\}$.

In addition, based on the conclusions given in Section 3.2 and this section, it can be concluded that: $\bar{\alpha}_{\zeta} \in \{\bar{\alpha}_{\text{rad};\xi}^{\text{cap}}\}$ if and only if $\lambda_{\zeta} < 0$; $\bar{\alpha}_{\zeta} \in \{\bar{\alpha}_{\text{non-rad};\xi}^{\text{pur res}}\}$ if and only if $\lambda_{\zeta} = +\infty$ or $-\infty$; $\bar{\alpha}_{\zeta} \in \{\bar{\alpha}_{\text{rad};\xi}^{\text{pur res}}\}$ if and only if $\lambda_{\zeta} = 0$; $\bar{\alpha}_{\zeta} \in \{\bar{\alpha}_{\text{rad};\xi}^{\text{ind}}\}$ if and only if $\lambda_{\zeta} > 0$. In Subsection 3.2.4, we have provided some typical examples corresponding to the above conclusions, and we will not further provide other examples here.

The above contents discussed in this section are collectively referred to as DP-based modal classification②. To further verify the important conclusion that nonradiative modes are just internally resonant modes, we provide the distributions of the currents and internal surface tangential magnetic fields corresponding to the nonradiative DP-CM at 5.75GHz listed in Table 3-1, the two degenerate states of the nonradiative DP-CM at 8.69GHz listed in Table 3-1, the two degenerate states of the nonradiative DP-CM at 9.15GHz listed in Table 3-1, and the three degenerate states of the nonradiative DP-CM at 4.10GHz as shown in the following Figure 3-22 ~ Figure 3-29.

① In what follows, $\bar{\alpha}_{\text{rad};\xi}^{\text{cap}}$, $\bar{\alpha}_{\text{non-rad};\xi}^{\text{res}}$, $\bar{\alpha}_{\text{rad};\xi}^{\text{res}}$, and $\bar{\alpha}_{\text{rad};\xi}^{\text{ind}}$ will be simply called as capacitive DP-CM, nonradiative DP-CM, radiative resonant DP-CM, and inductive DP-CM respectively. Obviously, this manner will not lead to any confusion, because: all capacitive modes (including capacitive DP-CMs) are radiative; all nonradiative modes (including nonradiative DP-CMs) are resonant; all inductive modes (including inductive DP-CMs) are radiative.

② In what follows, it will be simply called as modal classification.

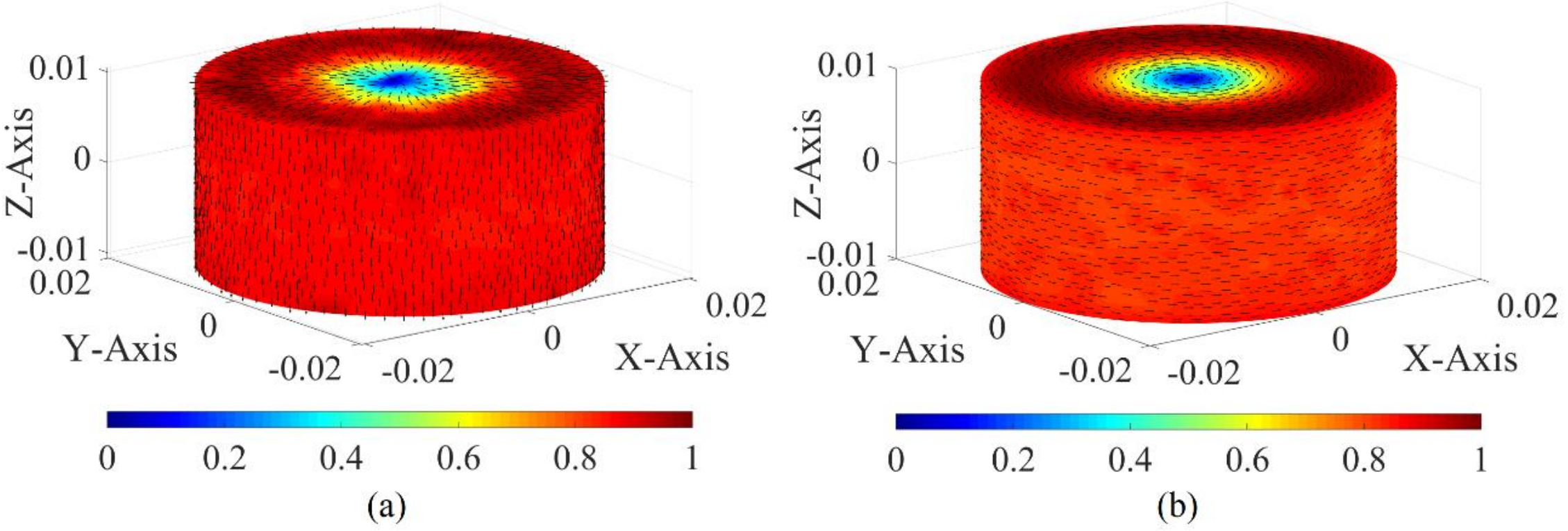

Figure 3-22 Some typical physical quantity distributions corresponding to the nonradiative DP-CM01 listed in Table 3-1. (a) modal electric current distribution; (b) tangential modal magnetic field distributing on the internal boundary of the metallic cylindrical cavity

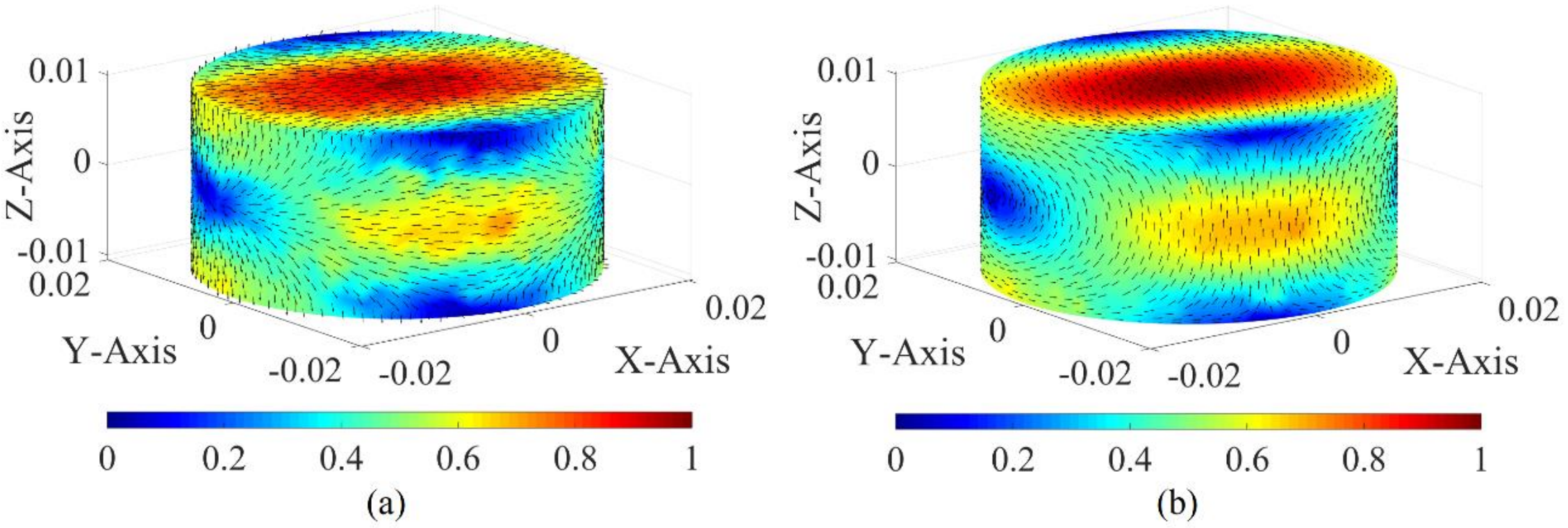

Figure 3-23 Some typical physical quantity distributions corresponding to the first degenerate state of the nonradiative DP-CM02 listed in Table 3-1. (a) modal electric current distribution; (b) tangential modal magnetic field distributing on the internal boundary of the metallic cylindrical cavity

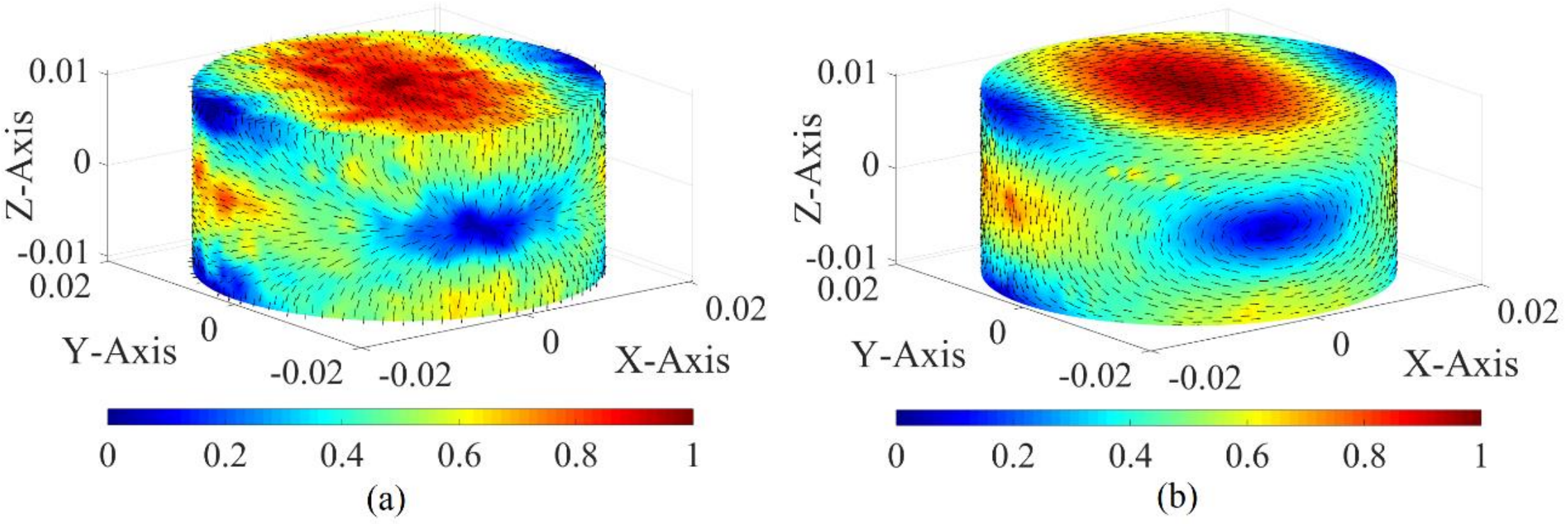

Figure 3-24 Some typical physical quantity distributions corresponding to the second degenerate state of the nonradiative DP-CM02 listed in Table 3-1. (a) modal electric current distribution; (b) tangential modal magnetic field distributing on the internal boundary of the metallic cylindrical cavity

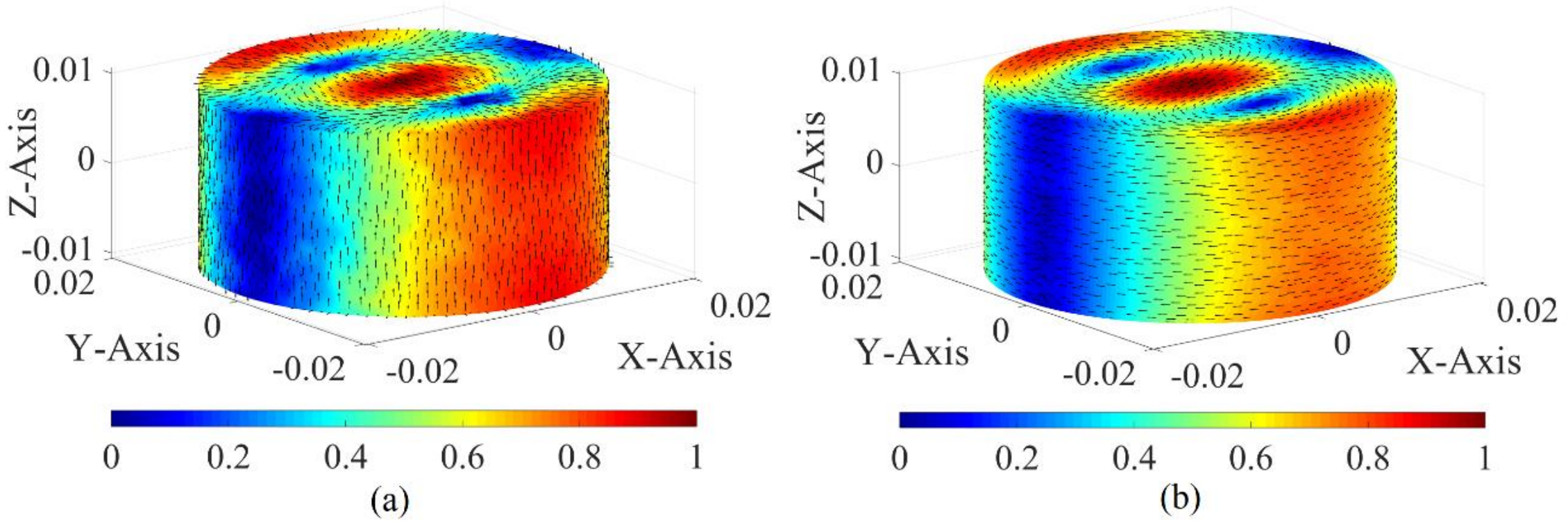

Figure 3-25 Some typical physical quantity distributions corresponding to the first degenerate state of the nonradiative DP-CM03 listed in Table 3-1. (a) modal electric current distribution; (b) tangential modal magnetic field distributing on the internal boundary of the metallic cylindrical cavity

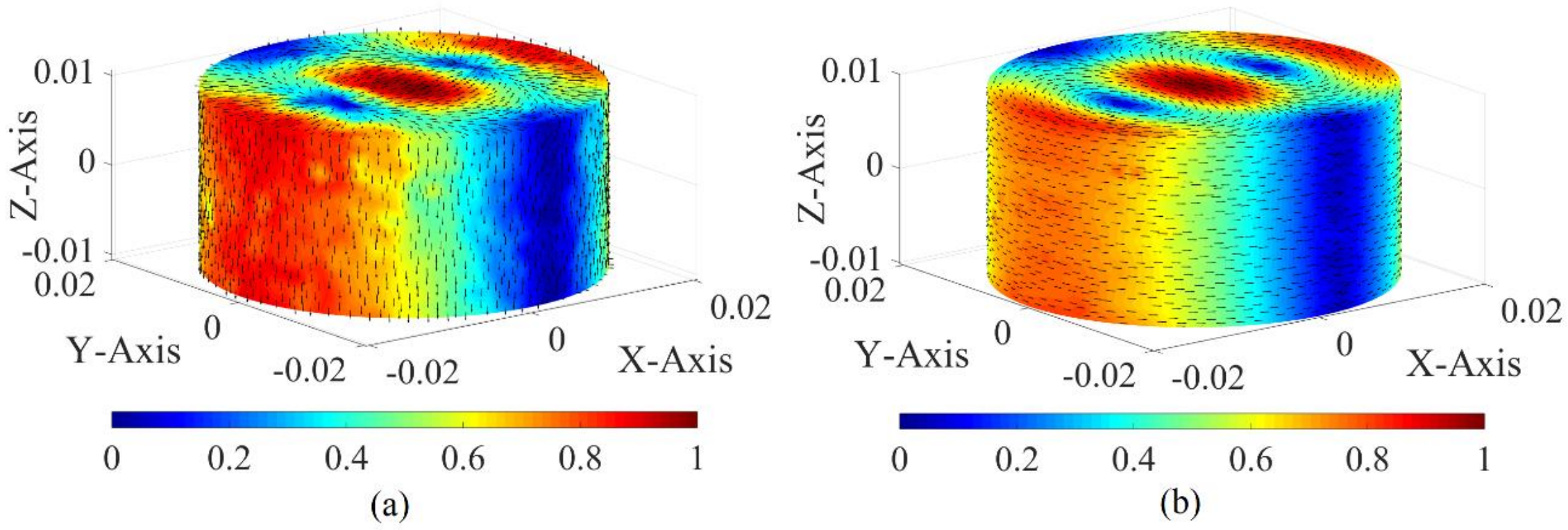

Figure 3-26 Some typical physical quantity distributions corresponding to the second degenerate state of the nonradiative DP-CM03 listed in Table 3-1. (a) modal electric current distribution; (b) tangential modal magnetic field distributing on the internal boundary of the metallic cylindrical cavity

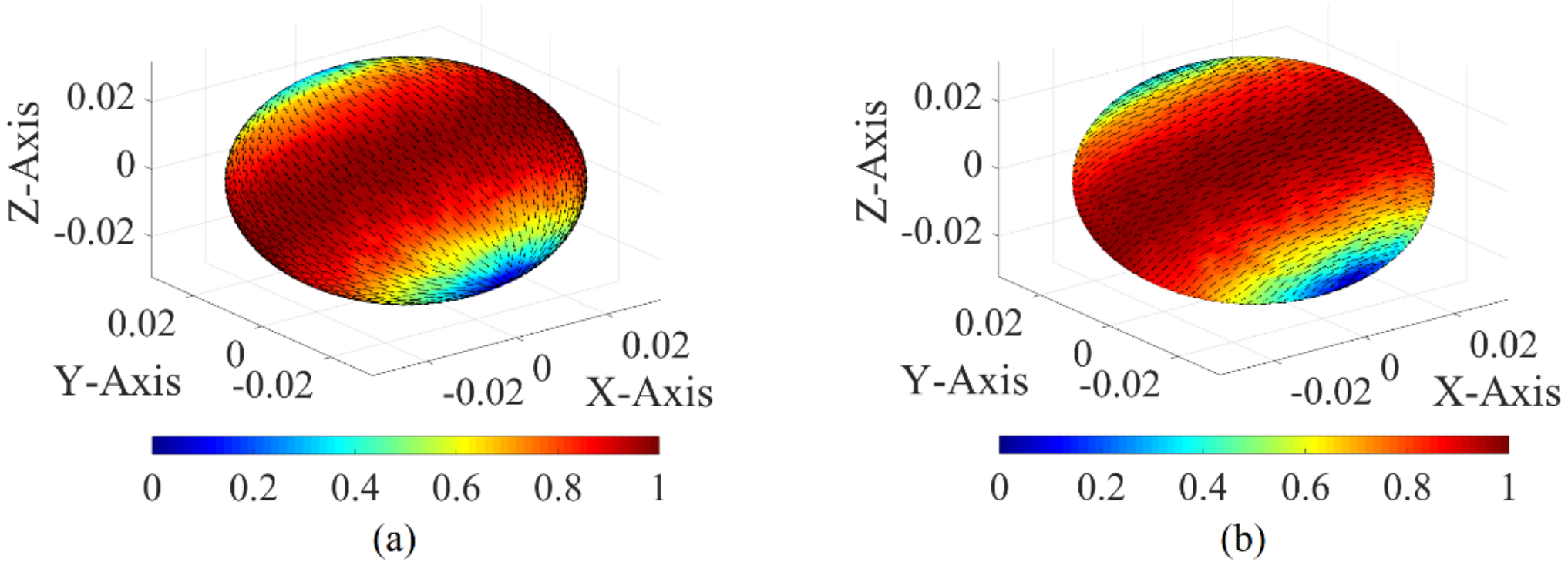

Figure 3-27 Some typical physical quantity distributions corresponding to the first degenerate state of the nonradiative DP-CM1 listed in Table 3-5. (a) modal electric current distribution; (b) tangential modal magnetic field distributing on the internal boundary of the metallic spherical cavity

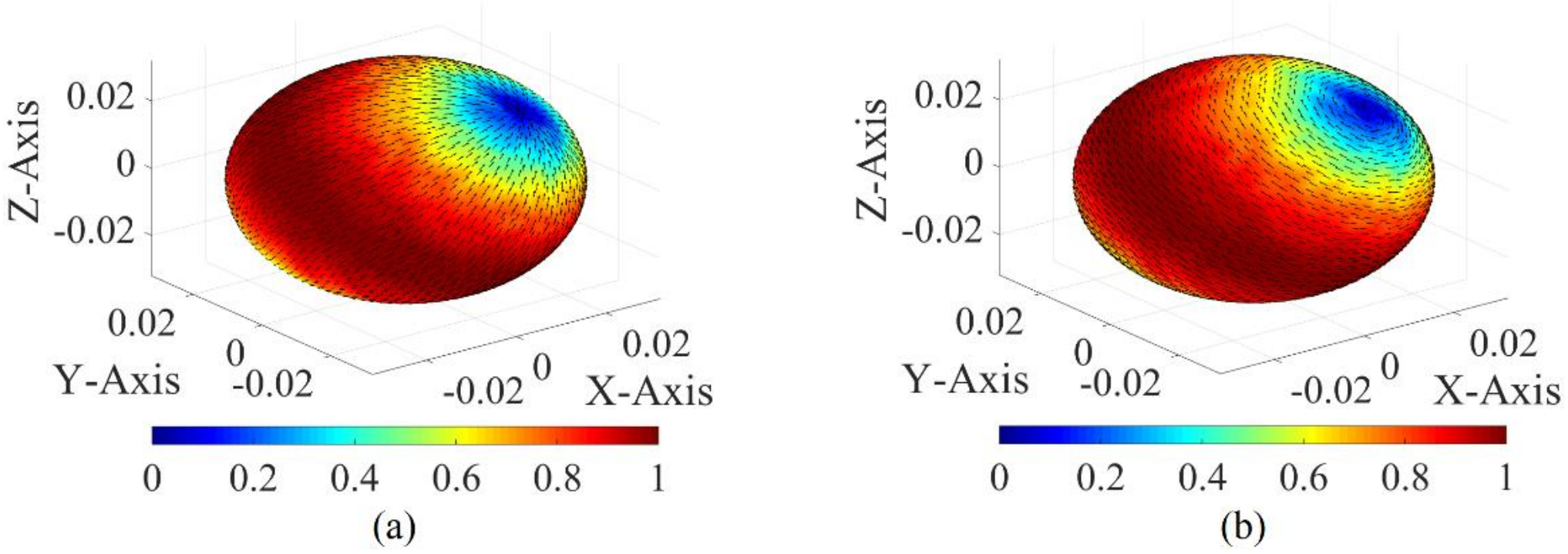

Figure 3-28 Some typical physical quantity distributions corresponding to the second degenerate state of the nonradiative DP-CM1 listed in Table 3-5. (a) modal electric current distribution; (b) tangential modal magnetic field distributing on the internal boundary of the metallic spherical cavity

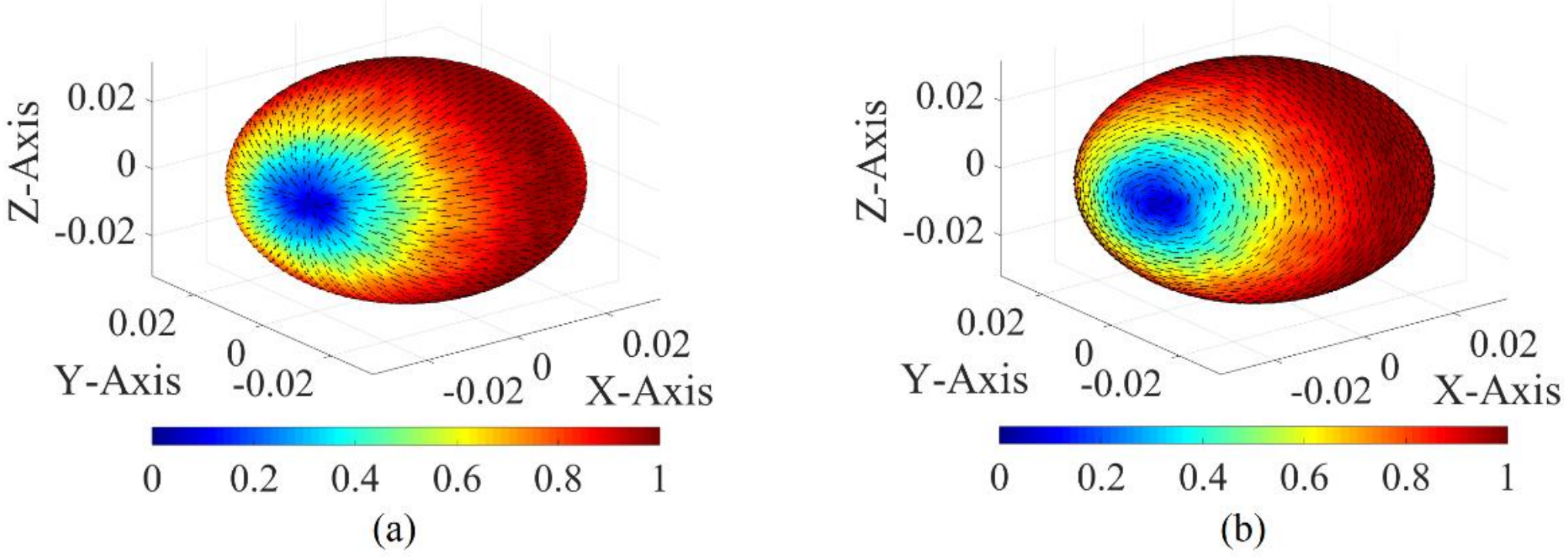

Figure 3-29 Some typical physical quantity distributions corresponding to the third degenerate state of the nonradiative DP-CM1 listed in Table 3-5. (a) modal electric current distribution; (b) tangential modal magnetic field distributing on the internal boundary of the metallic spherical cavity

It is thus clear that: the nonradiative modes not only satisfy the homogeneous tangential electric field boundary condition $0 = \hat{n}_{\text{inner normal}} \times \vec{E}_{\text{internal resonance}}$ of internally resonant modes[26-28,114] (as shown in Figures 3-6, 3-7, 3-8, 3-16, 3-17, and 3-18), but also satisfy the internal surface tangential magnetic field boundary condition $\vec{J}_{\text{internal resonance}} = \hat{n}_{\text{inner normal}} \times \vec{H}_{\text{internal resonance}}$ of internally resonant modes[26-28,114] (as shown in Figures 3-22, 3-23, 3-24, 3-25, 3-26, 3-27, 3-28, and 3-29).

## 3.3.2 DP-CM-Based Modal Expansion

In this subsection, we expand general working modes, general resonant modes, purely resonant modes, and impurely resonant modes in terms of the linear combinations of DP-CMs. At the same time, we also give some important conclusions related to these modal expansions.

**1) DP-CM-Based Linear Expansion for General Working Modes**

Because $\{\bar{\alpha}_\xi\}$ are independent and complete, then any $\bar{a}$ can be uniquely expanded in terms of the linear combination of some capacitive DP-CMs $\bar{\alpha}^{\text{cap}}_{\text{rad};\xi}$, some nonradiative DP-CMs $\bar{\alpha}^{\text{pur res}}_{\text{non-rad};\xi}$, some radiative resonant DP-CMs $\bar{\alpha}^{\text{pur res}}_{\text{rad};\xi}$, and some inductive DP-CMs $\bar{\alpha}^{\text{ind}}_{\text{rad};\xi}$ as follows[37]:

$$\bar{a} \sim \sum c^{\text{cap}}_{\text{rad};\xi}\bar{\alpha}^{\text{cap}}_{\text{rad};\xi} + \sum c^{\text{pur res}}_{\text{non-rad};\xi}\bar{\alpha}^{\text{pur res}}_{\text{non-rad};\xi} + \sum c^{\text{pur res}}_{\text{rad};\xi}\bar{\alpha}^{\text{pur res}}_{\text{rad};\xi} + \sum c^{\text{ind}}_{\text{rad};\xi}\bar{\alpha}^{\text{ind}}_{\text{rad};\xi} \tag{3-27}$$

Here, the reason to utilize symbol " ~ " instead of " = " will be provided in Subsection 3.3.3. Based on expansion formulation (3-27), a series of important conclusions shown in Figure 3-30[37] can be derived.

$$\left\{c^{\text{cap}}_{\text{rad};\xi}\right\} \;,\; \left\{c^{\text{ind}}_{\text{rad};\xi}\right\} = \{0\}\ .$$

$$1\Downarrow\ \Uparrow 11$$

$$\sum c^{\text{cap}}_{\text{rad};\xi}\bar{\alpha}^{\text{cap}}_{\text{rad};\xi}\ ,\ \sum c^{\text{ind}}_{\text{rad};\xi}\bar{\alpha}^{\text{ind}}_{\text{rad};\xi} = 0\ .$$

$$2\Downarrow\ \Uparrow 10$$

$$\sum c^{\text{cap}}_{\text{rad};\xi}\bar{\alpha}^{\text{cap}}_{\text{rad};\xi} + \sum c^{\text{ind}}_{\text{rad};\xi}\bar{\alpha}^{\text{ind}}_{\text{rad};\xi} = 0\ .$$

$$3\Downarrow\ \Uparrow 9$$

$$\sum c^{\text{cap}}_{\text{rad};\xi}\bar{\alpha}^{\text{cap}}_{\text{rad};\xi} + \sum c^{\text{ind}}_{\text{rad};\xi}\bar{\alpha}^{\text{ind}}_{\text{rad};\xi} \text{ is purely resonant.}$$

$$4\Downarrow\ \Uparrow 8$$

Working mode $\bar{a}$ is purely resonant.

$$5\Downarrow\ \not\Uparrow$$

Working mode $\bar{a}$ is resonant.

$$6\Downarrow\ \Uparrow 7$$

$$\sum c^{\text{cap}}_{\text{rad};\xi}\bar{\alpha}^{\text{cap}}_{\text{rad};\xi} + \sum c^{\text{ind}}_{\text{rad};\xi}\bar{\alpha}^{\text{ind}}_{\text{rad};\xi} \text{ is resonant.}$$

Figure 3-30 Some "equivalence relationships" related to resonance

Now, we prove the conclusions given in Figure 3-30 as follows[37]:

- **Proof for " 1$\Downarrow$ ":** $\{c^{\text{cap}}_{\text{rad};\xi}\},\{c^{\text{ind}}_{\text{rad};\xi}\} = \{0\}$ represents that all coefficients corresponding to capacitive and inductive terms are zero, so both $\sum c^{\text{cap}}_{\text{rad};\xi}\bar{\alpha}^{\text{cap}}_{\text{rad};\xi}$ and $\sum c^{\text{ind}}_{\text{rad};\xi}\bar{\alpha}^{\text{ind}}_{\text{rad};\xi}$ must be zero.
- **Proof for " 2$\Downarrow$ ":** The correctness of " 2$\Downarrow$ " is obvious.
- **Proof for " 3$\Downarrow$ ":** It is obvious that $\underline{\bar{\bar{P}}}^{\text{Driving}} \cdot 0 = 0$, so mode 0 is purely resonant. Thus, if $\sum c^{\text{cap}}_{\text{rad};\xi}\bar{\alpha}^{\text{cap}}_{\text{rad};\xi} + \sum c^{\text{ind}}_{\text{rad};\xi}\bar{\alpha}^{\text{ind}}_{\text{rad};\xi} = 0$, then mode $\sum c^{\text{cap}}_{\text{rad};\xi}\bar{\alpha}^{\text{cap}}_{\text{rad};\xi} + \sum c^{\text{ind}}_{\text{rad};\xi}\bar{\alpha}^{\text{ind}}_{\text{rad};\xi}$ must be purely resonant.
- **Proofs for " 4$\Downarrow$ " and " $\Uparrow$8 ":** Obviously, $\sum c^{\text{pur res}}_{\text{non-rad};\xi}\bar{\alpha}^{\text{pur res}}_{\text{non-rad};\xi} + \sum c^{\text{pur res}}_{\text{rad};\xi}\bar{\alpha}^{\text{pur res}}_{\text{rad};\xi}$ is purely resonant. Based on this and the pure resonance condition introduced in Subsection 3.3.1, it is easy to find out that $\bar{a}$ is a purely resonant mode if and only if $\sum c_{\text{rad};\xi}^{\text{cap}} \bar{\alpha}_{\text{rad};\xi}^{\text{cap}} + \sum c_{\text{rad};\xi}^{\text{ind}} \bar{\alpha}_{\text{rad};\xi}^{\text{ind}}$ is a purely resonant mode.

• **Proof for " 5 $\Downarrow$":** Based on relationship (3-24) and the definition of purely resonant modes, the correctness of " 5 $\Downarrow$ " can be proven. In addition, formulation (3-24) also provides the reason leading to the " $\not\Updownarrow$ " on the right side of " 5 $\Downarrow$ ".

• **Proofs for " 6 $\Downarrow$ " and " $\Uparrow$ 7 ":** $\sum c_{\text{non-rad};\xi}^{\text{pur res}} \bar{\alpha}_{\text{non-rad};\xi}^{\text{pur res}} + \sum c_{\text{rad};\xi}^{\text{pur res}} \bar{\alpha}_{\text{rad};\xi}^{\text{pur res}}$ is purely resonant. Based on this and orthogonality (3-25), it can be concluded that $\bar{a}$ and $\sum c_{\text{rad};\xi}^{\text{cap}} \bar{\alpha}_{\text{rad};\xi}^{\text{cap}} + \sum c_{\text{rad};\xi}^{\text{ind}} \bar{\alpha}_{\text{rad};\xi}^{\text{ind}}$ have the same imaginary power. Then, " 6 $\Downarrow$ " and " $\Uparrow$ 7 " are proven simultaneously.

• **Proofs for " $\Uparrow$ 9 ", " $\Uparrow$ 10 ", and " $\Uparrow$ 11 ":** If $\sum c_{\text{rad};\xi}^{\text{cap}} \bar{\alpha}_{\text{rad};\xi}^{\text{cap}} + \sum c_{\text{rad};\xi}^{\text{ind}} \bar{\alpha}_{\text{rad};\xi}^{\text{ind}}$ is purely resonant, then $\bar{a}$ is purely resonant (based on " 4 $\Downarrow$ "). This implies that $\bar{a}$ can be linearly expanded by $\{\bar{\alpha}_{\xi}^{\text{pur res}}\}$ (based on the conclusions obtained in Subsection 3.3.1). Then, $\bar{a}$ can be linearly expanded by $\{\bar{\alpha}_{\text{non-rad};\xi}^{\text{pur res}}\} \bigcup \{\bar{\alpha}_{\text{rad};\xi}^{\text{pur res}}\}$, because $\{\bar{\alpha}_{\xi}^{\text{pur res}}\} = \{\bar{\alpha}_{\text{non-rad};\xi}^{\text{pur res}}\} \bigcup \{\bar{\alpha}_{\text{rad};\xi}^{\text{pur res}}\}$. Due to the independence of whole DP-CM set $\{\bar{\alpha}_{\xi}\}$, it can be known that expansion formulation (3-27) is unique[116]. Thus, when $\sum c_{\text{rad};\xi}^{\text{cap}} \bar{\alpha}_{\text{rad};\xi}^{\text{cap}} + \sum c_{\text{rad};\xi}^{\text{ind}} \bar{\alpha}_{\text{rad};\xi}^{\text{ind}}$ is purely resonant, the coefficients $\{c_{\text{rad};\xi}^{\text{cap}}\}$ and $\{c_{\text{rad};\xi}^{\text{ind}}\}$ in expansion formulation (3-27) must be zero. Based on this and " 1 $\Downarrow$ " and " 2 $\Downarrow$ ", it can be proven that " $\Uparrow$ 9 ", " $\Uparrow$ 10 ", and " $\Uparrow$ 11 " hold.

**2) DP-CM-Based Linear Expansion for Resonant Modes**

Obviously, any resonant mode $\bar{a}^{\text{res}}$ can be expanded in terms of DP-CMs as follows[37]:

$$\bar{a}^{\text{res}} \simeq \underline{\sum c_{\text{rad};\xi}^{\text{cap}} \bar{\alpha}_{\text{rad};\xi}^{\text{cap}}} + \sum c_{\text{non-rad};\xi}^{\text{pur res}} \bar{\alpha}_{\text{non-rad};\xi}^{\text{pur res}} + \sum c_{\text{rad};\xi}^{\text{pur res}} \bar{\alpha}_{\text{rad};\xi}^{\text{pur res}} + \underline{\sum c_{\text{rad};\xi}^{\text{ind}} \bar{\alpha}_{\text{rad};\xi}^{\text{ind}}} \tag{3-28}$$

Here, the reason to use symbol " $\simeq$ " instead of " $=$ " and the reason to mark terms $\sum c_{\text{rad};\xi}^{\text{cap}} \bar{\alpha}_{\text{rad};\xi}^{\text{cap}}$ and $\sum c_{\text{rad};\xi}^{\text{ind}} \bar{\alpha}_{\text{rad};\xi}^{\text{ind}}$ with single underlines will be given in Subsection 3.3.3.

**3) DP-CM-Based Linear Expansion for Purely Resonant Modes**

Based on the conclusions given in Subsection 3.3.1 and Figure 3-30, any purely resonant mode $\bar{a}^{\text{pur res}}$ can be expressed in terms of the linear combination of $\{\bar{\alpha}_{\text{non-rad};\xi}^{\text{pur res}}\} \bigcup \{\bar{\alpha}_{\text{rad};\xi}^{\text{pur res}}\}$ as follows[37]:

$$\bar{a}^{\text{pur res}} \sim \sum c_{\text{non-rad};\xi}^{\text{pur res}} \bar{\alpha}_{\text{non-rad};\xi}^{\text{pur res}} + \sum c_{\text{rad};\xi}^{\text{pur res}} \bar{\alpha}_{\text{rad};\xi}^{\text{pur res}} \tag{3-29}$$

Expansion formulation (3-29) implies that: in the expansion formulation of purely resonant modes, there doesn't exist any capacitive DP-CM term, and there also doesn't exist any inductive DP-CM term.

As pointed out in Subsection 3.3.1, any nonradiative mode $\bar{a}_{\text{non-rad}}^{\text{pur res}}$ can be expanded in terms of $\{\bar{\alpha}_{\text{non-rad};\xi}^{\text{pur res}}\}$ as follows[37]:

$$\bar{a}_{\text{non-rad}}^{\text{pur res}} \sim \sum c_{\text{non-rad};\xi}^{\text{pur res}} \bar{\alpha}_{\text{non-rad};\xi}^{\text{pur res}} \tag{3-30}$$

Expansion formulation (3-30) implies that: in the expansion formulation of nonradiative mode, there only exists nonradiative DP-CM term. But, due to the existence of orthogonality (3-26), we cannot guarantee that the expansion formulation of a radiative purely resonant mode $\bar{a}_{\text{rad}}^{\text{pur res}}$ only contains radiative purely resonant DP-CM term. Thus, the DP-CM-based expansion formulation for any radiative purely resonant mode $\bar{a}_{\text{rad}}^{\text{pur res}}$ is as follows[37]:

$$\bar{a}_{\text{rad}}^{\text{pur res}} \cong \sum c_{\text{non-rad};\xi}^{\text{pur res}} \bar{\alpha}_{\text{non-rad};\xi}^{\text{pur res}} + \underline{\underline{\sum c_{\text{rad};\xi}^{\text{pur res}} \bar{\alpha}_{\text{rad};\xi}^{\text{pur res}}}} \tag{3-31}$$

Here, the reason to utilize symbol "$\cong$" instead of "$=$" and the reason to mark term $\sum c_{\text{rad};\xi}^{\text{ind}} \bar{\alpha}_{\text{rad};\xi}^{\text{ind}}$ with a double underline will be given in Subsection 3.3.3.

#### 4) DP-CM-Based Linear Expansion for Impurely Resonant Modes

Obviously, any impurely resonant mode $\bar{a}_{\text{rad}}^{\text{impur res}}$ can be expanded in terms of DP-CMs as follows[37]:

$$\bar{a}_{\text{rad}}^{\text{impur res}} \cong \underline{\underline{\sum c_{\text{rad};\xi}^{\text{cap}} \bar{\alpha}_{\text{rad};\xi}^{\text{cap}}}} + \sum c_{\text{non-rad};\xi}^{\text{pur res}} \bar{\alpha}_{\text{non-rad};\xi}^{\text{pur res}} + \sum c_{\text{rad};\xi}^{\text{pur res}} \bar{\alpha}_{\text{rad};\xi}^{\text{pur res}} + \underline{\underline{\sum c_{\text{rad};\xi}^{\text{ind}} \bar{\alpha}_{\text{rad};\xi}^{\text{ind}}}} \tag{3-32}$$

Based on the "3$\Downarrow$" and "$\Uparrow$9" in Figure 3-30, it is easy to conclude that: in expansion formulation (3-32), $\sum c_{\text{rad};\xi}^{\text{cap}} \bar{\alpha}_{\text{rad};\xi}^{\text{cap}} + \sum c_{\text{rad};\xi}^{\text{ind}} \bar{\alpha}_{\text{rad};\xi}^{\text{ind}} \neq 0$. In fact, we can further conclude that $\sum c_{\text{rad};\xi}^{\text{cap}} \bar{\alpha}_{\text{rad};\xi}^{\text{cap}} \neq 0$ and at the same time $\sum c_{\text{rad};\xi}^{\text{ind}} \bar{\alpha}_{\text{rad};\xi}^{\text{ind}} \neq 0$, because: if $\sum c_{\text{rad};\xi}^{\text{cap}} \bar{\alpha}_{\text{rad};\xi}^{\text{cap}} \neq 0$ and $\sum c_{\text{rad};\xi}^{\text{ind}} \bar{\alpha}_{\text{rad};\xi}^{\text{ind}} = 0$, then orthogonality (3-25) implies that the imaginary power $\bar{a}_{\text{rad}}^{\text{impur res}}$ is smaller than 0, and this leads to the contradiction of that $\bar{a}_{\text{rad}}^{\text{impur res}}$ is resonant; if $\sum c_{\text{rad};\xi}^{\text{cap}} \bar{\alpha}_{\text{rad};\xi}^{\text{cap}} = 0$ and $\sum c_{\text{rad};\xi}^{\text{ind}} \bar{\alpha}_{\text{rad};\xi}^{\text{ind}} \neq 0$, then orthogonality (3-25) implies that the imaginary power $\bar{a}_{\text{rad}}^{\text{impur res}}$ is larger than 0, and this leads to the contradiction of that $\bar{a}_{\text{rad}}^{\text{impur res}}$ is resonant; if $\sum c_{\text{rad};\xi}^{\text{cap}} \bar{\alpha}_{\text{rad};\xi}^{\text{cap}} = 0$ and $\sum c_{\text{rad};\xi}^{\text{ind}} \bar{\alpha}_{\text{rad};\xi}^{\text{ind}} = 0$, then $\sum c_{\text{rad};\xi}^{\text{cap}} \bar{\alpha}_{\text{rad};\xi}^{\text{cap}} + \sum c_{\text{rad};\xi}^{\text{ind}} \bar{\alpha}_{\text{rad};\xi}^{\text{ind}} = 0$, and this leads to the contradiction of $\sum c_{\text{rad};\xi}^{\text{cap}} \bar{\alpha}_{\text{rad};\xi}^{\text{cap}} + \sum c_{\text{rad};\xi}^{\text{ind}} \bar{\alpha}_{\text{rad};\xi}^{\text{ind}} \neq 0$.

The contents discussed in this subsection are collectively referred to as DP-CM-based modal expansion①.

① In what follows, it is simply called as modal expansion.

### 3.3.3 Modal Decomposition Based on Modal Classification and Modal Expansion

To simplify the symbolic system of the following discussions, we simply denote the building block terms $\sum c_{\text{rad};\xi}^{\text{cap}} \bar{\alpha}_{\text{rad};\xi}^{\text{cap}}$, $\sum c_{\text{non-rad};\xi}^{\text{pur res}} \bar{\alpha}_{\text{non-rad};\xi}^{\text{pur res}}$, $\sum c_{\text{rad};\xi}^{\text{pur res}} \bar{\alpha}_{\text{rad};\xi}^{\text{pur res}}$, and $\sum c_{\text{rad};\xi}^{\text{ind}} \bar{\alpha}_{\text{rad};\xi}^{\text{ind}}$ used in Subsection 3.3.2 as $\bar{\beta}_{\text{rad}}^{\text{cap}}$, $\bar{\beta}_{\text{non-rad}}^{\text{pur res}}$, $\bar{\beta}_{\text{rad}}^{\text{pur res}}$, and $\bar{\beta}_{\text{rad}}^{\text{ind}}$ respectively. Then, expansion formulations (3-27)~(3-32) can be correspondingly rewritten as follows[37]:

$$\bar{a} \quad \sim \quad \bar{\beta}_{\text{rad}}^{\text{cap}} + \bar{\beta}_{\text{non-rad}}^{\text{pur res}} + \bar{\beta}_{\text{rad}}^{\text{pur res}} + \bar{\beta}_{\text{rad}}^{\text{ind}} \tag{3-33}$$

and

$$\bar{a}^{\text{res}} \quad \simeq \quad \underline{\bar{\beta}_{\text{rad}}^{\text{cap}}} + \bar{\beta}_{\text{non-rad}}^{\text{pur res}} + \bar{\beta}_{\text{rad}}^{\text{pur res}} + \underline{\bar{\beta}_{\text{rad}}^{\text{ind}}} \tag{3-34}$$

$$\bar{a}^{\text{pur res}} \quad \sim \quad \bar{\beta}_{\text{non-rad}}^{\text{pur res}} + \bar{\beta}_{\text{rad}}^{\text{pur res}} \tag{3-35}$$

$$\bar{a}_{\text{non-rad}}^{\text{pur res}} \quad \sim \quad \bar{\beta}_{\text{non-rad}}^{\text{pur res}} \tag{3-36}$$

$$\bar{a}_{\text{rad}}^{\text{pur res}} \quad \cong \quad \bar{\beta}_{\text{non-rad}}^{\text{pur res}} + \underline{\underline{\bar{\beta}_{\text{rad}}^{\text{pur res}}}} \tag{3-37}$$

$$\bar{a}_{\text{rad}}^{\text{impur res}} \quad \cong \quad \underline{\underline{\bar{\beta}_{\text{rad}}^{\text{cap}}}} + \bar{\beta}_{\text{non-rad}}^{\text{pur res}} + \bar{\beta}_{\text{rad}}^{\text{pur res}} + \underline{\underline{\bar{\beta}_{\text{rad}}^{\text{ind}}}} \tag{3-38}$$

We call above formulations (3-33)~(3-38) as the modal decompositions based on the DP and the corresponding DP-CMs①. In fact, the above modal decompositions can also be similarly generalized to capacitive modes and inductive modes as follows[37]:

$$\bar{a}_{\text{rad}}^{\text{cap}} \quad \cong \quad \underline{\underline{\bar{\beta}_{\text{rad}}^{\text{cap}}}} + \bar{\beta}_{\text{non-rad}}^{\text{pur res}} + \bar{\beta}_{\text{rad}}^{\text{pur res}} + \bar{\beta}_{\text{rad}}^{\text{ind}} \tag{3-39}$$

$$\bar{a}_{\text{rad}}^{\text{ind}} \quad \cong \quad \bar{\beta}_{\text{rad}}^{\text{cap}} + \bar{\beta}_{\text{non-rad}}^{\text{pur res}} + \bar{\beta}_{\text{rad}}^{\text{pur res}} + \underline{\underline{\bar{\beta}_{\text{rad}}^{\text{ind}}}} \tag{3-40}$$

As the continuations of the conclusions obtained in Subsections 3.3.1 and 3.3.2, we will, based on the above modal decompositions, obtain some further conclusions as follows[37]:

**Further Conclusion 1.** In formulations (3-33), (3-35), and (3-36), all terms in RHS can be either zero or nonzero. In formulation (3-34), terms $\bar{\beta}_{\text{non-rad}}^{\text{pur res}}$ and $\bar{\beta}_{\text{rad}}^{\text{pur res}}$ can be either zero or nonzero, and the terms $\bar{\beta}_{\text{rad}}^{\text{cap}}$ and $\bar{\beta}_{\text{rad}}^{\text{ind}}$ with single underlines can be simultaneously zero or simultaneously nonzero. In formulation (3-38), terms $\bar{\beta}_{\text{non-rad}}^{\text{pur res}}$ and $\bar{\beta}_{\text{rad}}^{\text{pur res}}$ can be either zero or nonzero, and the terms $\bar{\beta}_{\text{rad}}^{\text{cap}}$ and $\bar{\beta}_{\text{rad}}^{\text{ind}}$ with double underlines must be simultaneously nonzero. In formulations (3-37), (3-39), and (3-40), the terms with double underlines must be nonzero, and the terms without underlines can be either zero or nonzero. The above these are just the reasons to use "$\sim$", "$\simeq$", and "$\cong$"

① In what follows, it is simply called as modal decomposition.

in modal expansions (3-27)~(3-32) and modal decompositions (3-33)~(3-40) rather than simply using "$=$", and also the reasons why some terms are marked by single or double underlines.

**Further Conclusion 2.** The term $\bar{\beta}_{\text{non-rad}}^{\text{pur res}}$ in the RHS of formulation (3-37) can be nonzero, so the set constituted by all $\bar{a}_{\text{rad}}^{\text{pur res}}$ is not closed for addition①, and then all $\bar{a}_{\text{rad}}^{\text{pur res}}$ cannot constitute a linear space[116]. Based on modal decompositions (3-38)~(3-40), we can similarly conclude that: all $\bar{a}_{\text{rad}}^{\text{impur res}}$ cannot constitute a linear space; all $\bar{a}_{\text{rad}}^{\text{cap}}$ cannot constitute a linear space; all $\bar{a}_{\text{rad}}^{\text{ind}}$ cannot constitute a linear space.

In addition, all resonant modes also cannot constitute a linear space. For example, if the imaginary powers of $\bar{\alpha}_{\text{rad};1}^{\text{cap}}$ and $\bar{\alpha}_{\text{rad};1}^{\text{ind}}$ are normalized to $-1$ and $+1$ respectively, then both of modes $\bar{a}^{\text{res}} = \bar{\alpha}_{\text{rad};1}^{\text{cap}} + \bar{\alpha}_{\text{rad};1}^{\text{ind}}$ and $\bar{a}_{\varphi}^{\text{res}} = \bar{\alpha}_{\text{rad};1}^{\text{cap}} + e^{j\varphi}\bar{\alpha}_{\text{rad};1}^{\text{ind}}$ must be resonant for any real number $\varphi$②. However, working mode $\bar{a}^{\text{res}} + \bar{a}_{\varphi}^{\text{res}}$ may be non-resonant, because $\varphi$ is arbitrary③. These above imply that the set constituted by all resonant modes is not closed for addition.

**Further Conclusion 3.** Modal decomposition (3-34) implies that $\bar{a}^{\text{res}}$ may include components $\bar{\beta}_{\text{rad}}^{\text{cap}}$, $\bar{\beta}_{\text{non-rad}}^{\text{pur res}}$, and $\bar{\beta}_{\text{rad}}^{\text{ind}}$. Modal decompositions (3-35) and (3-37) imply that $\bar{a}^{\text{pur res}}$ and $\bar{a}_{\text{rad}}^{\text{pur res}}$ may include component $\bar{\beta}_{\text{non-rad}}^{\text{pur res}}$. Modal decomposition (3-38) implies that $\bar{a}_{\text{rad}}^{\text{impur res}}$ must include components $\bar{\beta}_{\text{rad}}^{\text{cap}}$ and $\bar{\beta}_{\text{rad}}^{\text{ind}}$, and may include component $\bar{\beta}_{\text{non-rad}}^{\text{pur res}}$. In fact, this is just the reason why literature [41] pointed out that the above kinds of resonant modes cannot guarantee the most efficient radiation. So, which kind of resonant modes includes neither component $\bar{\beta}_{\text{non-rad}}^{\text{pur res}}$ nor components $\bar{\beta}_{\text{rad}}^{\text{cap}}$ & $\bar{\beta}_{\text{rad}}^{\text{ind}}$? To answer this question, we introduce so-called "high-quality purely resonant mode" and some related concepts as below.

We denote the space spanned by $\{\bar{\alpha}_{\text{rad};\xi}^{\text{pur res}}\}$ as $\text{span}\{\bar{\alpha}_{\text{rad};\xi}^{\text{pur res}}\}$. Any element in space $\text{span}\{\bar{\alpha}_{\text{rad};\xi}^{\text{pur res}}\}$ only includes component $\bar{\beta}_{\text{rad}}^{\text{pur res}}$, and this dissertation calls the space as high-quality radiative pure resonance space or simply as high-quality radiation space. Obviously, $\text{span}\{\bar{\alpha}_{\xi}^{\text{pur res}}\} = \text{span}\{\bar{\alpha}_{\text{non-rad};\xi}^{\text{pur res}}\} \oplus \text{span}\{\bar{\alpha}_{\text{rad};\xi}^{\text{pur res}}\}$, where symbol "$\oplus$" represents the direct sum of two spaces. To effectively distinguish the radiative modes in $\text{span}\{\bar{\alpha}_{\xi}^{\text{pur res}}\}$ from the radiative modes in $\text{span}\{\bar{\alpha}_{\text{rad};\xi}^{\text{pur res}}\}$, this dissertation specifically

① For example: the summation of radiative purely resonant mode $\bar{\beta}_{\text{non-rad}}^{\text{pur res}} + \bar{\beta}_{\text{rad}}^{\text{pur res}}$ and radiative purely resonant mode $-\bar{\beta}_{\text{rad}}^{\text{pur res}}$ is a nonradiative mode $\bar{\beta}_{\text{non-rad}}^{\text{pur res}}$.

② This conclusion is based on the orthogonality (3-15) among DP-CMs.

③ Based on orthogonality (3-15) and some simple operations, it can be known that: the imaginary power of mode $\bar{a}^{\text{res}} + \bar{a}_{\varphi}^{\text{res}}$ is $2(\cos\varphi - 1)$.

calls the latters as high-quality radiative purely resonant modes, and correspondingly denotes them as $\bar{a}_{\text{Rad};\xi}^{\text{pur res}}$. Obviously, $\bar{\beta}_{\text{rad}}^{\text{pur res}}$ and $\bar{\alpha}_{\text{rad};\xi}^{\text{pur res}}$ themselves are just high-quality radiative purely resonant modes, so the following parts of this dissertation denote them as $\bar{\beta}_{\text{Rad}}^{\text{pur res}}$ and $\bar{\alpha}_{\text{Rad};\xi}^{\text{pur res}}$ respectively. At the same time, coefficient $c_{\text{rad};\xi}^{\text{pur res}}$ is also denoted as $c_{\text{Rad};\xi}^{\text{pur res}}$ correspondingly. Then,

$$\bar{a}_{\text{Rad}}^{\text{pur res}} = \underline{\underline{\bar{\beta}_{\text{Rad}}^{\text{pur res}}}} = \underline{\underline{\sum c_{\text{Rad};\xi}^{\text{pur res}} \bar{\alpha}_{\text{Rad};\xi}^{\text{pur res}}}} \tag{3-41}$$

However, we must explicitly emphasize that: mode 0 is an element of high-quality radiative pure resonance space①, though it doesn't radiate any EM energy. In addition, this dissertation calls the working modes in set $\{\bar{a}^{\text{pur res}}\}\setminus(\{\bar{a}_{\text{non-rad}}^{\text{pur res}}\}\bigcup\{\bar{a}_{\text{Rad}}^{\text{pur res}}\})$ as non-high-quality radiative purely resonant modes.

**Further Conclusion 4.** Modal decompositions (3-35)~(3-37) imply that various purely resonant modes only include purely resonant components. In other words, the components of purely resonant modes are purer than the components of general resonant modes and impurely resonant modes. In fact, this is just the reason why this dissertation calls equation $\bar{\bar{P}}_{-}^{\text{Driving}} \cdot \bar{a} = 0$ as pure resonance condition, and why this dissertation calls the modes satisfying $\bar{\bar{P}}_{-}^{\text{Driving}} \cdot \bar{a} = 0$ as purely resonant modes.

Similarly, this dissertation calls the capacitive modes which don't include components $\bar{\beta}_{\text{non-rad}}^{\text{pur res}}$, $\bar{\beta}_{\text{Rad}}^{\text{pur res}}$, and $\bar{\beta}_{\text{rad}}^{\text{ind}}$ as purely capacitive modes, and correspondingly denotes them as $\bar{a}_{\text{rad}}^{\text{pur cap}}$; this dissertation calls the inductive modes which don't include components $\bar{\beta}_{\text{rad}}^{\text{cap}}$, $\bar{\beta}_{\text{non-rad}}^{\text{pur res}}$, and $\bar{\beta}_{\text{Rad}}^{\text{pur res}}$ as purely inductive modes, and correspondingly denotes them as $\bar{a}_{\text{rad}}^{\text{pur ind}}$. Obviously, building block terms $\bar{\beta}_{\text{rad}}^{\text{cap}}$ and $\bar{\beta}_{\text{rad}}^{\text{ind}}$ themselves are just purely capacitive and purely inductive respectively, so they are denoted as $\bar{\beta}_{\text{rad}}^{\text{pur cap}}$ and $\bar{\beta}_{\text{rad}}^{\text{pur ind}}$ respectively; DP-CMs $\bar{\alpha}_{\text{rad};\xi}^{\text{cap}}$ and $\bar{\alpha}_{\text{rad};\xi}^{\text{ind}}$ themselves are just purely capacitive and purely inductive respectively, so they are denoted as $\bar{\alpha}_{\text{rad};\xi}^{\text{pur cap}}$ and $\bar{\alpha}_{\text{rad};\xi}^{\text{pur ind}}$ respectively. In addition, this dissertation calls the capacitive modes which are not purely capacitive as impurely capacitive modes, and correspondingly denotes them as $\bar{a}_{\text{rad}}^{\text{impur cap}}$; this dissertation calls the inductive modes which are not purely inductive as impurely inductive modes, and correspondingly denotes them as $\bar{a}_{\text{rad}}^{\text{impur ind}}$.

Based on the above these, we can further rewrite modal decompositions (3-33)~(3-41) as follows[37]:

① The reason to include mode 0 into high-quality radiation space is to make the set constituted by all of the modes having form $\bar{\beta}_{\text{Rad}}^{\text{pur res}}$ be closed and then be a linear space.

$$\overline{a} \quad \sim \quad \overline{\beta}_{\text{rad}}^{\text{pur cap}} + \overline{\beta}_{\text{non-rad}}^{\text{pur res}} + \overline{\beta}_{\text{Rad}}^{\text{pur res}} + \overline{\beta}_{\text{rad}}^{\text{pur ind}} \tag{3-42}$$

and

$$\overline{a}_{\text{rad}}^{\text{cap}} \quad \cong \quad \underline{\underline{\overline{\beta}_{\text{rad}}^{\text{pur cap}}}} + \overline{\beta}_{\text{non-rad}}^{\text{pur res}} + \overline{\beta}_{\text{Rad}}^{\text{pur res}} + \overline{\beta}_{\text{rad}}^{\text{pur ind}} \tag{3-43}$$

$$\overline{a}_{\text{rad}}^{\text{pur cap}} \quad = \quad \underline{\underline{\overline{\beta}_{\text{rad}}^{\text{pur cap}}}} \tag{3-44}$$

$$\overline{a}_{\text{rad}}^{\text{impur cap}} \quad = \quad \underline{\underline{\overline{\beta}_{\text{rad}}^{\text{pur cap}}}} + \underline{\underline{\overline{\beta}_{\text{non-rad}}^{\text{pur res}} + \overline{\beta}_{\text{Rad}}^{\text{pur res}} + \overline{\beta}_{\text{rad}}^{\text{pur ind}}}} \tag{3-45}$$

and

$$\overline{a}^{\text{res}} \quad \simeq \quad \underline{\overline{\beta}_{\text{rad}}^{\text{pur cap}}} + \overline{\beta}_{\text{non-rad}}^{\text{pur res}} + \overline{\beta}_{\text{Rad}}^{\text{pur res}} + \underline{\overline{\beta}_{\text{rad}}^{\text{pur ind}}} \tag{3-46}$$

$$\overline{a}^{\text{pur res}} \quad \sim \quad \overline{\beta}_{\text{non-rad}}^{\text{pur res}} + \overline{\beta}_{\text{Rad}}^{\text{pur res}} \tag{3-47}$$

$$\overline{a}_{\text{non-rad}}^{\text{pur res}} \quad = \quad \underline{\underline{\overline{\beta}_{\text{non-rad}}^{\text{pur res}}}} \tag{3-48}$$

$$\overline{a}_{\text{Rad}}^{\text{pur res}} \quad = \quad \underline{\underline{\overline{\beta}_{\text{Rad}}^{\text{pur res}}}} \tag{3-49}$$

$$\overline{a}_{\text{rad}}^{\text{pur res}} \quad \cong \quad \overline{\beta}_{\text{non-rad}}^{\text{pur res}} + \underline{\underline{\overline{\beta}_{\text{Rad}}^{\text{pur res}}}} \tag{3-50}$$

$$\overline{a}_{\text{rad}}^{\text{impur res}} \quad \cong \quad \underline{\underline{\overline{\beta}_{\text{rad}}^{\text{pur cap}}}} + \overline{\beta}_{\text{non-rad}}^{\text{pur res}} + \overline{\beta}_{\text{Rad}}^{\text{pur res}} + \underline{\underline{\overline{\beta}_{\text{rad}}^{\text{pur ind}}}} \tag{3-51}$$

and

$$\overline{a}_{\text{rad}}^{\text{ind}} \quad \cong \quad \overline{\beta}_{\text{rad}}^{\text{pur cap}} + \overline{\beta}_{\text{non-rad}}^{\text{pur res}} + \overline{\beta}_{\text{Rad}}^{\text{pur res}} + \underline{\underline{\overline{\beta}_{\text{rad}}^{\text{pur ind}}}} \tag{3-52}$$

$$\overline{a}_{\text{rad}}^{\text{pur ind}} \quad = \quad \underline{\underline{\overline{\beta}_{\text{rad}}^{\text{pur ind}}}} \tag{3-53}$$

$$\overline{a}_{\text{rad}}^{\text{impur ind}} \quad = \quad \underline{\underline{\overline{\beta}_{\text{rad}}^{\text{pur cap}} + \overline{\beta}_{\text{non-rad}}^{\text{pur res}} + \overline{\beta}_{\text{Rad}}^{\text{pur res}}}} + \underline{\underline{\overline{\beta}_{\text{rad}}^{\text{pur ind}}}} \tag{3-54}$$

In modal decompositions (3-45) and (3-54), the terms $\overline{\beta}_{\text{non-rad}}^{\text{pur res}} + \overline{\beta}_{\text{Rad}}^{\text{pur res}} + \overline{\beta}_{\text{rad}}^{\text{pur ind}}$ and $\overline{\beta}_{\text{rad}}^{\text{pur cap}} + \overline{\beta}_{\text{non-rad}}^{\text{pur res}} + \overline{\beta}_{\text{Rad}}^{\text{pur res}}$ with double underlines must be non-zero. Actually, this is just the most essential property of impurely capacitive and inductive modes. In addition, the reason why we use "$=$" in formulation (3-48) instead of "$\sim$" like formulation (3-36) is that mode 0 can be written as the form of $\overline{\beta}_{\text{non-rad}}^{\text{pur res}}$, and it is also the reason why we mark the term $\overline{\beta}_{\text{non-rad}}^{\text{pur res}}$ in formulation (3-48) by a double underline.

**Further Conclusion 5.** As pointed out in previous Subsection 3.3.1, whole modal space $\{\overline{a}\}$ can be spanned by $\{\overline{\alpha}_{\xi}\}$; whole pure resonance space $\text{nullspace}\, \overline{\overline{P}}_{-}^{\text{Driving}}$ can be spanned by $\{\overline{\alpha}_{\xi}^{\text{pur res}}\}$; whole nonradiation space $\text{nullspace}\, \overline{\overline{P}}_{+}^{\text{Driving}}$ can be spanned by $\{\overline{\alpha}_{\text{non-rad};\xi}^{\text{pur res}}\}$. As pointed out in the previous parts of this subsection, high-quality radiation space can be spanned by $\{\overline{\alpha}_{\text{Rad};\xi}^{\text{pur res}}\}$.

In addition, it is easy to find out that: mode 0 and all $\overline{a}_{\text{rad}}^{\text{pur cap}}$ constitute a space $\{0\} \bigcup \{\overline{a}_{\text{rad}}^{\text{pur cap}}\}$, and the space is called as pure capacitance space, and denoted as $\text{span}\{\overline{\alpha}_{\text{rad};\xi}^{\text{pur cap}}\}$ because it is spanned by $\{\overline{\alpha}_{\text{rad};\xi}^{\text{pur cap}}\}$; mode 0 and all $\overline{a}_{\text{rad}}^{\text{pur ind}}$ constitute a space $\{0\} \bigcup \{\overline{a}_{\text{rad}}^{\text{pur ind}}\}$, and the space is called as pure inductance space, and denoted as $\text{span}\{\overline{\alpha}_{\text{rad};\xi}^{\text{pur ind}}\}$ because it is spanned by $\{\overline{\alpha}_{\text{rad};\xi}^{\text{pur ind}}\}$.

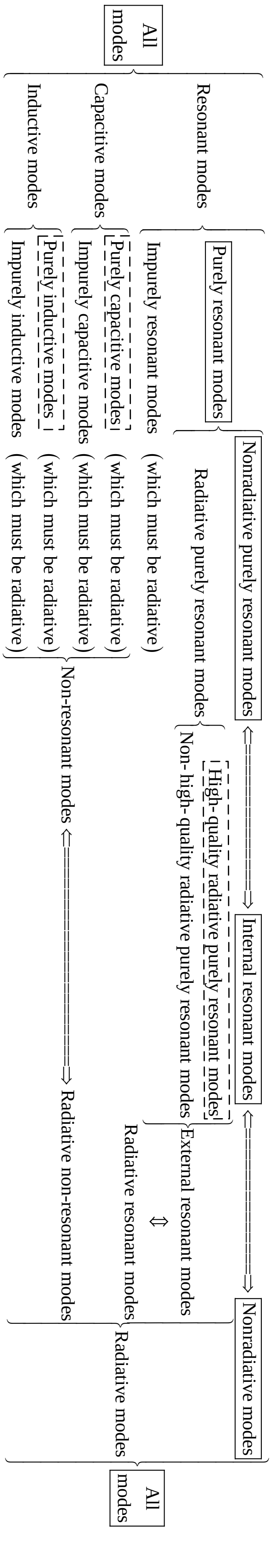

Figure 3-31 The modal classification proposed in this section and the relationships among various modal sub-classes

**Further Conclusion 6.** To macroscopically grasp the modal classification obtained above, the relationships of various modes are illustrated in Figure 3-31[37]. In Figure 3-31, the modal classes in the solid-line boxes are linear spaces; the modal classes in dotted-line boxes will become linear spaces, if mode 0 is added to the classes; the other modal classes are usually not linear spaces.

## 3.3.4 Uniqueness of Modal Decomposition

Obviously, for a certain $\overline{\overline{P}}{}_{\text{met sys}}^{\text{Driving}}$, the $\overline{\overline{P}}{}_{+}^{\text{Driving}}$ and $\overline{\overline{P}}{}_{-}^{\text{Driving}}$ derived from Toeplitz's decomposition (3-6) are uniquely determined[116], so the characteristic values calculated from characteristic equation (3-7) are also uniquely determined[116]. In addition, if $\lambda_\xi$ is a characteristic value, then vector 0 and all of the vectors satisfying equation (3-7) constitute a linear space, i.e. the characteristic subspace corresponding to $\lambda_\xi$ [116]; the characteristic subspace corresponding to $\lambda_\xi$ is unique[116]; all characteristic vectors corresponding to $\lambda_\xi$ constitute a basis of the characteristic subspace of $\lambda_\xi$, but the basis is not unique[116].

Thus, for a certain $\overline{\overline{P}}{}_{\text{met sys}}^{\text{Driving}}$, we have conclusions: characteristic values $\{\lambda_{\text{rad};\xi}^{\text{pur cap}}\}$ are unique, and the characteristic subspace corresponding to any $\lambda_{\text{rad};\xi}^{\text{pur cap}}$ is also unique, but characteristic vectors $\{\overline{\alpha}_{\text{rad};\xi}^{\text{pur cap}}\}$ are not unique; the characteristic subspace corresponding to $\lambda_{\text{non-rad};\xi}^{\text{pur res}} = +\infty / -\infty$ (i.e. nonradiation space) is unique, but characteristic vectors $\{\overline{\alpha}_{\text{non-rad};\xi}^{\text{pur res}}\}$ are not unique; the characteristic subspace corresponding to $\lambda_{\text{Rad};\xi}^{\text{pur res}} = 0$ ① is unique, but characteristic vectors $\{\overline{\alpha}_{\text{Rad};\xi}^{\text{pur res}}\}$ are not unique; characteristic values $\{\lambda_{\text{rad};\xi}^{\text{pur ind}}\}$ are unique, and the characteristic subspace corresponding to any $\lambda_{\text{rad};\xi}^{\text{pur ind}}$ is also unique, but characteristic vectors $\{\overline{\alpha}_{\text{rad};\xi}^{\text{pur ind}}\}$ are not unique. Considering that characteristic vector sets $\{\overline{\alpha}_{\text{rad};\xi}^{\text{pur cap}}\}$, $\{\overline{\alpha}_{\text{non-rad};\xi}^{\text{pur res}}\}$, $\{\overline{\alpha}_{\text{Rad};\xi}^{\text{pur res}}\}$, and $\{\overline{\alpha}_{\text{rad};\xi}^{\text{pur ind}}\}$ are not unique, now we want to know whether or not the building block components $\overline{\beta}_{\text{rad}}^{\text{pur cap}}$, $\overline{\beta}_{\text{non-rad}}^{\text{pur res}}$, $\overline{\beta}_{\text{Rad}}^{\text{pur res}}$, and $\overline{\beta}_{\text{rad}}^{\text{pur ind}}$ constructed by the sets are unique? In other words, for a certain working mode $\overline{a}$, whether or not the modal decompositions obtained in Subsection 3.3.3 are unique?

Our answer to the above question is YES, and we give the explanations as follows:

**1)** For a certain $\overline{\overline{P}}{}_{\text{met sys}}^{\text{Driving}}$, matrices $\overline{\overline{P}}{}_{+}^{\text{Driving}}$ and $\overline{\overline{P}}{}_{-}^{\text{Driving}}$ are unique[116]. Thus, for a certain $\overline{\overline{P}}{}_{\text{met sys}}^{\text{Driving}}$, nonradiation space $\text{span}\{\overline{\alpha}_{\text{non-rad};\xi}^{\text{pur res}}\}$, i.e. null space $\text{nullspace}\,\overline{\overline{P}}{}_{+}^{\text{Driving}}$, is unique; pure resonance space $\text{span}\{\overline{\alpha}_{\xi}^{\text{pur res}}\}$, i.e. null space $\text{nullspace}\,\overline{\overline{P}}{}_{-}^{\text{Driving}}$, is unique.

① i.e. high-quality radiation space $\text{span}\{\overline{\alpha}_{\text{Rad};\xi}^{\text{pur res}}\}$.

Thus, high-quality radiation space $\text{span}\{\bar{\alpha}_{\text{Rad};\xi}^{\text{pur res}}\}$ is also unique, because of that $\text{span}\{\bar{\alpha}_{\text{non-rad};\xi}^{\text{pur res}}\} \oplus \text{span}\{\bar{\alpha}_{\text{Rad};\xi}^{\text{pur res}}\} = \text{span}\{\bar{\alpha}_{\xi}^{\text{pur res}}\}$ and $\text{span}\{\bar{\alpha}_{\text{non-rad};\xi}^{\text{pur res}}\} \cap \text{span}\{\bar{\alpha}_{\text{Rad};\xi}^{\text{pur res}}\} = \{0\}$.

**2)** For a certain $\bar{\bar{P}}_{\text{met sys}}^{\text{Driving}}$, mode 0 is the only common element of two characteristic subspaces corresponding to two different characteristic values. Thus, pure capacitance space is actually the direct sum of the characteristic subspaces corresponding to all negative characteristic values[116]; pure inductance space is actually the direct sum of the characteristic subspaces corresponding to all positive characteristic values[116]. Due to the uniqueness of the characteristic subspaces, the direct sums are also unique, i.e., both of pure capacitance space and pure inductance space are unique.

**3)** Obviously, mode 0 is the only element of the intersections between pairs of pure capacitance space $\text{span}\{\bar{\alpha}_{\text{rad};\xi}^{\text{pur cap}}\}$, nonradiation space $\text{span}\{\bar{\alpha}_{\text{non-rad};\xi}^{\text{pur res}}\}$, high-quality radiation space $\text{span}\{\bar{\alpha}_{\text{Rad};\xi}^{\text{pur res}}\}$, and pure inductance space $\text{span}\{\bar{\alpha}_{\text{rad};\xi}^{\text{pur ind}}\}$. Then, whole modal space $\{\bar{a}\}$ is actually the direct sum of $\text{span}\{\bar{\alpha}_{\text{rad};\xi}^{\text{pur cap}}\}$, $\text{span}\{\bar{\alpha}_{\text{non-rad};\xi}^{\text{pur res}}\}$, $\text{span}\{\bar{\alpha}_{\text{Rad};\xi}^{\text{pur res}}\}$, and $\text{span}\{\bar{\alpha}_{\text{rad};\xi}^{\text{pur ind}}\}$ [116], i.e.,

$$\underbrace{\overbrace{\text{span}\left\{\bar{\alpha}_{\xi}\right\}}^{\text{Modal space}}}_{\{\bar{a}\}} = \underbrace{\overbrace{\text{span}\left\{\bar{\alpha}_{\text{rad};\xi}^{\text{pur cap}}\right\}}^{\text{Pure capacitance space}}}_{\{\bar{a}_{\text{rad}}^{\text{pur cap}}\}\cup\{0\}} \oplus \underbrace{\overbrace{\underbrace{\overbrace{\text{span}\left\{\bar{\alpha}_{\text{non-rad};\xi}^{\text{pur res}}\right\}}^{\text{Nonradiation space}}}_{\{\bar{a}_{\text{non-rad}}^{\text{pur res}}\}} \oplus \underbrace{\overbrace{\text{span}\left\{\bar{\alpha}_{\text{Rad};\xi}^{\text{pur res}}\right\}}^{\text{High-quality radiation space}}}_{\{0\}\cup\{\bar{a}_{\text{Rad}}^{\text{pur res}}\}}}^{\text{Pure resonance space}}}_{\{\bar{a}^{\text{pur res}}\} = \text{span}\{\bar{\alpha}_{\xi}^{\text{pur res}}\}} \oplus \underbrace{\overbrace{\text{span}\left\{\bar{\alpha}_{\text{rad};\xi}^{\text{pur ind}}\right\}}^{\text{Pure inductance space}}}_{\{0\}\cup\{\bar{a}_{\text{rad}}^{\text{pur ind}}\}} \tag{3-55}$$

In addition, orthogonalities (3-14) and (3-15) imply that: subspaces $\text{span}\{\bar{\alpha}_{\text{rad};\xi}^{\text{pur cap}}\}$, $\text{span}\{\bar{\alpha}_{\text{non-rad};\xi}^{\text{pur res}}\}$, $\text{span}\{\bar{\alpha}_{\text{Rad};\xi}^{\text{pur res}}\}$, and $\text{span}\{\bar{\alpha}_{\text{rad};\xi}^{\text{pur ind}}\}$ are pairwisely weighted orthogonal, and the weighted matrices are $\bar{\bar{P}}_{+}^{\text{Driving}}$ and $\bar{\bar{P}}_{-}^{\text{Driving}}$. Thus, formulation (3-55) can also be called as the orthogonal decomposition for whole modal space $\{\bar{a}\}$.

**4)** The above-mentioned 1), 2), and 3) together lead to that: any mode $\bar{a}$ can be uniquely projected on subspaces $\text{span}\{\bar{\alpha}_{\text{rad};\xi}^{\text{pur cap}}\}$, $\text{span}\{\bar{\alpha}_{\text{non-rad};\xi}^{\text{pur res}}\}$, $\text{span}\{\bar{\alpha}_{\text{Rad};\xi}^{\text{pur res}}\}$, and $\text{span}\{\bar{\alpha}_{\text{rad};\xi}^{\text{pur ind}}\}$ [116], i.e., the terms $\bar{\beta}_{\text{rad}}^{\text{pur cap}}$, $\bar{\beta}_{\text{non-rad}}^{\text{pur res}}$, $\bar{\beta}_{\text{Rad}}^{\text{pur res}}$, and $\bar{\beta}_{\text{rad}}^{\text{pur ind}}$ in the modal decompositions are unique. This dissertation calls modal decompositions (3-42)~(3-54) as the orthogonal decompositions for various working modes.

## 3.4 Discussions and Comparisons for Several Kinds of Typical Fundamental Modes of Metallic Systems

In theoretical and engineering electromagnetics, the studies on the fundamental modes of metallic systems have had a very long history, and there have been some relatively systematical theoretical formalisms. Under the theoretical formalisms, scholars introduced some different kinds of fundamental modes, such as Harrington's CMs[31,32], eigen-modes[26-28], and natural modes[23-25], etc. This dissertation will discuss and compare these fundamental modes from the perspectives of WEP and DP.

### 3.4.1 Harrington's CMs of Metallic Systems

In 1971, Prof. Harrington and Dr. Mautz[31,32] established the EFIE-based CMT for open metallic systems (EFIE-OpeMetSca-CMT). EFIE-OpeMetSca-CMT has had many successful engineering applications, and literature [35] provides a very comprehensive review for the various antenna applications of EFIE-OpeMetSca-CMT.

In what follows, we simply discuss the essential principles of EFIE-OpeMetSca-CMT.

**1)** The essential destination of EFIE-OpeMetSca-CMT is to establish a characteristic transformation, such that a series of preselected basis functions are transformed into a series of new basis functions —— Harrington's CMs, and the Harrington's CMs satisfy the orthogonality (17), (20), and (22)~(24) given in literature [31]. In fact, the orthogonality (17), (20), and (22)~(24) given in literature [31] are equivalent to the orthogonality (3-14)~(3-20) given in this dissertation, so EFIE-OpeMetSca-CMT and WEP-MetSca-CMT (the WEP-based CMT for metallic scattering systems established in this dissertation) have the same physical destination —— constructing a series of steadily working modes not having net energy exchange in any integral period. However, the carrying frameworks of the two theories are very different: EFIE-OpeMetSca-CMT was established in IE framework, but WEP-MetSca-CMT is established in WEP framework. This difference is not especially noticeable in the process to construct the CMs of metallic systems, but it will be very noticeable in the processes to construct the CMs of material systems (Chapter 4) and the CMs of metal-material composite systems (Chapter 5).

**2)** Literatures [31,32] claimed that: by decomposing EFIE-based IMO $\overline{\overline{Z}}^{\text{EFIE}}$ into its real part $\overline{\overline{R}}^{\text{EFIE}} = \text{Re}\{\overline{\overline{Z}}^{\text{EFIE}}\}$ and imaginary part $\overline{\overline{X}}^{\text{EFIE}} = \text{Im}\{\overline{\overline{Z}}^{\text{EFIE}}\}$, and solving characteristic equation $\overline{\overline{X}}^{\text{EFIE}} \cdot \overline{\alpha} = \lambda \overline{\overline{R}}^{\text{EFIE}} \cdot \overline{\alpha}$, the destination in 1) can be achieved.

**3)** In fact, to achieve the destination in 1) by using the method in 2), it is necessary to guarantee three additional conditions as follows:

**(3.1)** the relationship $P_{\text{met sys}}^{\text{Driving}} = \overline{a}^H \cdot \overline{\overline{Z}}^{\text{EFIE}} \cdot \overline{a}$ must be satisfied, to ensure that the power-based orthogonality (17), (20), and (22)~(24) given in literature [31] can be realized;

**(3.2)** matrices $\overline{\overline{R}}^{\text{EFIE}}$ and $\overline{\overline{X}}^{\text{EFIE}}$ must be symmetrical, to ensure that $\bar{a}^H \cdot \overline{\overline{R}}^{\text{EFIE}} \cdot \bar{a}$ and $\bar{a}^H \cdot \overline{\overline{X}}^{\text{EFIE}} \cdot \bar{a}$ correspond to $P^{\text{rad}}$ and $P_{\text{vac}}^{\text{imag}}$ respectively, i.e., $P^{\text{rad}} = \bar{a}^H \cdot \overline{\overline{R}}^{\text{EFIE}} \cdot \bar{a}$ and $P_{\text{vac}}^{\text{imag}} = \bar{a}^H \cdot \overline{\overline{X}}^{\text{EFIE}} \cdot \bar{a}$ ;

**(3.3)** matrix $\overline{\overline{R}}^{\text{EFIE}}$ must be positive definite, to ensure that equation $\overline{\overline{X}}^{\text{EFIE}} \cdot \bar{\alpha} = \lambda\, \overline{\overline{R}}^{\text{EFIE}} \cdot \bar{\alpha}$ can provide enough independent characteristic vectors.

**4)** To realize the three conditions listed in above 3), EFIE-OpeMetSca-CMT need to be established as follows:

**(4.1)** to realize condition (3.1), matrix $\overline{\overline{Z}}^{\text{EFIE}}$ must be derived from using inner product, and the testing functions must be the same as the basis functions (i.e. Galerkin's method), and a coefficient 1/2 should be included;

**(4.2)** to realize condition (3.2), matrix $\overline{\overline{Z}}^{\text{EFIE}}$ must be derived from using symmetrical product, and the testing functions must be the same as the basis functions, and the line charges distributing on the boundary lines of open metallic surfaces must be ignored;

**(4.3)** to realize condition (3.3), the objective metallic system must radiate some EM energies.

To realize above (4.1) and (4.2) simultaneously, EFIE-OpeMetSca-CMT requires the basis functions used to expand the scattered electric current to be real, so the characteristic currents corresponding to Harrington's metallic CMs must be real[31,32], and the real characteristic electric currents correspond to standing wave currents. This implies that: EFIE-OpeMetSca-CMT has only ability to construct standing-wave-type real characteristic electric currents, but has no ability to construct traveling-wave-type or traveling-standing-wave-type complex characteristic electric currents.

To realize above (4.3), EFIE-OpeMetSca-CMT requires that the objective metallic system cannot work at internal resonance frequencies, so EFIE-OpeMetSca-CMT has no ability to research the internally resonant modes of metallic systems and the working states of the metallic systems at internal resonance frequencies[40].

The decomposition method (for quadratic matrix $\overline{\overline{P}}_{\text{met sys}}^{\text{Driving}}$) used in this dissertation is Toeplitz's decomposition (3-6). The Toeplitz's decomposition can guarantee that the obtained matrices $\overline{\overline{P}}_{+}^{\text{Driving}}$ and $\overline{\overline{P}}_{-}^{\text{Driving}}$ must be Hermitian, and then guarantee that $P^{\text{rad}} = \bar{a}^H \cdot \overline{\overline{P}}_{+}^{\text{Driving}} \cdot \bar{a}$ and $P_{\text{vac}}^{\text{imag}} = \bar{a}^H \cdot \overline{\overline{P}}_{-}^{\text{Driving}} \cdot \bar{a}$ always hold. Thus, there is no need to do any restriction for basis functions except independence and completeness, in this dissertation. Then, the WEP-MetSca-CMT established in this dissertation has ability to construct both real characteristic electric currents and complex characteristic electric currents.

Based on the limitation thought in Calculus, this dissertation proposes a method to construct the DP-CMs at internal resonance frequencies, such that WEP-MetSca-CMT is applicable to all working frequencies[37,40].

After establishing EFIE-OpeMetSca-CMT, Prof. Harrington and Dr. Mautz didn't develop any EFIE-OpeMetSca-CMT based variants in IE framework. But, some followers developed some EFIE-OpeMetSca-CMT based variants in IE framework①, such as MFIE-based metallic CMT[58], CFIE-based metallic CMT[59,60], and complex background Green's function based metallic CMT[62], etc.

Obviously, only the EFIE-based IMO derived from Galerkin's method corresponds to the matrix form of DPO, but the IMO derived from non-Galerkin's method and MFIE-based and CFIE-based IMOs don't correspond to the matrix form of DPO.

In addition, in the process of constructing DP-CMs, the vacuum Green's function should be selected (rather than non-vacuum Green's functions), because: the DP introduced from WEP is the power done by resultant fields on scattered sources, and the fields generated by environment have been contained in the resultant fields. This is also one of the principle foundations of that this dissertation treats the summation of excitation field $\vec{F}_{\text{imp}}$ and environment field $\vec{F}_{\text{env}}$ as a whole (i.e. incident field $\vec{F}^{\text{inc}}$), and another principle foundation is Huygens-Fresnel principle, which will be detailedly discussed in the Appendix C of this dissertation. In fact, this is just the basic reason why some scholars[63] recently discovered that for a certain objective metallic system the CMs calculated from the formulations based on vacuum Green's function are different from the CMs derived from original EFIE-OpeMetSca-CMT.

In this dissertation, the WEP-MetSca-CMT for the metallic systems placed in vacuum environment and the WEP-MetSca-CMT for the metallic systems placed in complex environment don't have any difference in mathematical form. But, it is very important to clearly point out that WEP-MetSca-CMT is directly applicable to the complex environment case, because it clearly points out that: when metallic systems are selected as objective systems, the non-vacuum Green's functions should never be used. If we want to take the influences of environment into account, then the modal problem

① This dissertation calls them as the IE-based variants of Harrington's EFIE-OpeMetSca-CMT.

we are facing has not been “constructing the CMs of the metallic scattering structures” but “constructing the CMs of the EM systems constructed by the metallic scattering structures and the environments”, and the latter modal problem can be solved by using the method developed in the Chapter 5 of this dissertation.

In summary, the EFIE-OpeMetSca-CMT established in literatures [31,32] and the WEP-MetSca-CMT established in this dissertation are equivalent to each other in the aspect of physical destination —— constructing a series of steadily working modes not having net energy exchange in any integral period, so the following parts of this dissertation will not especially distinguish them. Now that EFIE-OpeMetSca-CMT and WEP-MetSca-CMT have equivalent physical destinations, why does Section 3.2 establish WEP-MetSca-CMT? The reasons are that: compared with traditional EFIE-OpeMetSca-CMT, new WEP-MetSca-CMT has a clearer physical picture; based on the clear physical picture, it is easy to recognize and then to avoid the IE-based variants whose physical destinations are not equivalent to traditional EFIE-OpeMetSca-CMT; the clear physical picture is indispensable for properly establishing the Harrington’s CMTs for material systems (for details see Chapter 4) and composite systems (for details see Chapter 5). In addition, we want to emphasize that: it should never confuse EFIE-OpeMetSca-CMT[31,32] and its variants in IE framework.

Before finishing this subsection, we will give some necessary discussions to the physical meaning of an important characteristic quantity Modal Significance (MS) in EFIE-OpeMetSca-CMT. For radiative CMs, the definition[35] and some equivalent forms for $\text{MS}_{\text{rad};\xi}$ ① are as follows:

$$\text{MS}_{\text{rad};\xi} = \frac{1}{\left|1+j\lambda_\xi\right|} = \frac{\left|P_\xi^{\text{rad}}\right|}{\left|P_\xi^{\text{rad}} + jP_{\text{vac};\xi}^{\text{imag}}\right|} = \frac{P_\xi^{\text{rad}}}{\left|P_{\text{met sys};\xi}^{\text{Driving}}\right|} \tag{3-56}$$

where the second equality is based on that $\lambda_\xi = P_{\text{vac};\xi}^{\text{imag}} / P_\xi^{\text{rad}}$, and the third equality is based on that $P_\xi^{\text{rad}} > 0$. Obviously, the physical meaning of $\text{MS}_{\text{rad};\xi}$ is “the weight of $P_\xi^{\text{rad}}$ on whole $P_{\text{met sys};\xi}^{\text{Driving}}$”. In fact, the above definition is also suitable for the $\text{MS}_{\text{non-rad};\xi}$ ② corresponding to nonradiative CMs, and

$$\text{MS}_{\text{non-rad};\xi} = \frac{1}{\left|1+j\lambda_\xi\right|} = \frac{1}{\left|1\pm j\infty\right|} = 0 \tag{3-57}$$

① The subscript “ rad ” used here is to emphasize that $\text{MS}_{\text{rad};\xi}$ corresponds to a radiative CM.
② The subscript “ non-rad ” used here is to emphasize that $\text{MS}_{\text{non-rad};\xi}$ corresponds to a nonradiative CM.

where the second equality is based on the conclusions given in Subsection 3.2.2. Based on these, this dissertation collectively defines the MS for all CMs as $\text{MS}_\xi = 1/|1 + j\lambda_\xi|$.

As the verifications for above relationship (3-57), we provide two groups of typical examples —— the MSs corresponding to the nonradiative DP-CMs of a metallic cylinder and a metallic sphere. The MS curves of the DP-CMs shown in Figure 3-3 (metallic cylinder case) are illustrated in Figure 3-32. From Figure 3-32, it is easy to find out that: when the DP-CMs work at nonradiation states, the corresponding MS is indeed zero, i.e., relationship (3-57) surely hold for the DP-CMs.

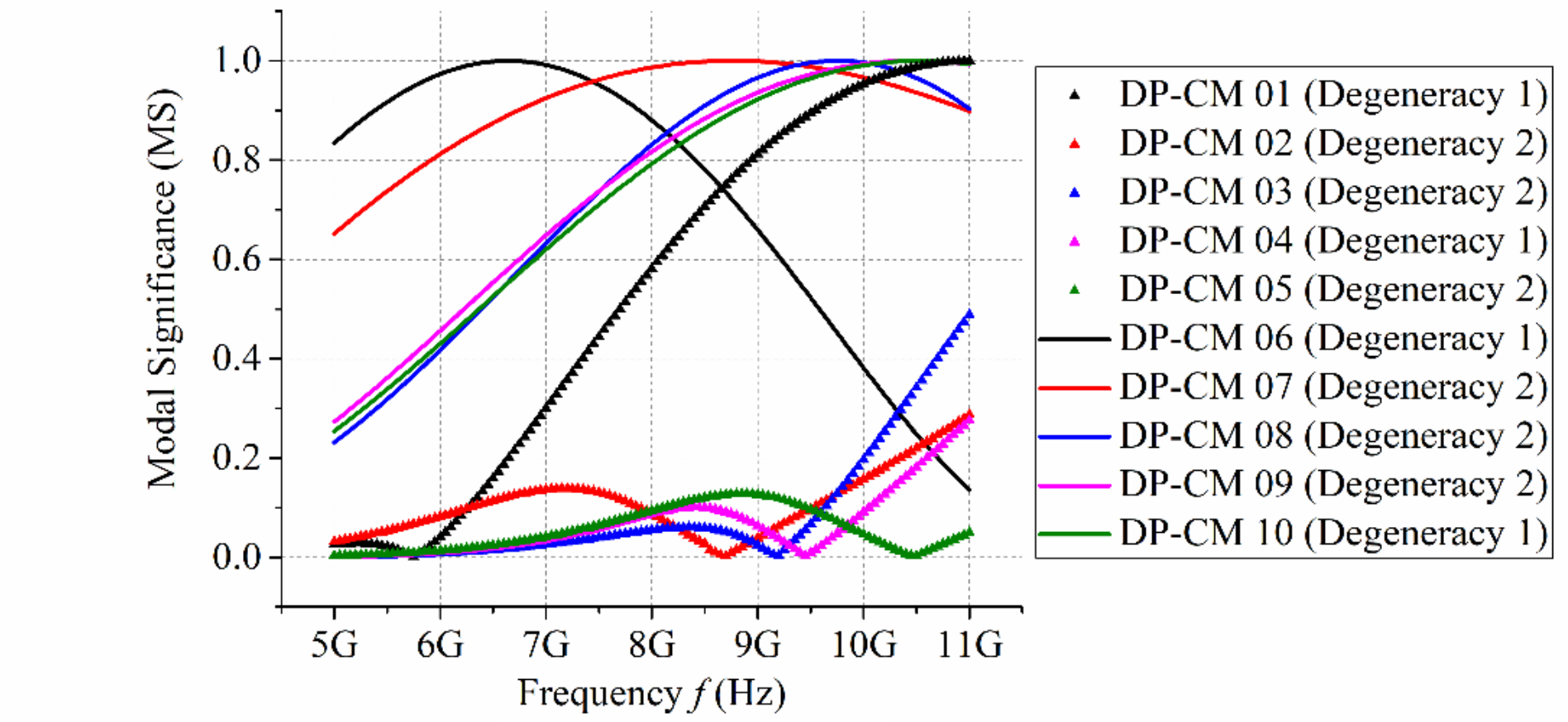

Figure 3-32 The MS curves corresponding to the first 10 typical DP-CMs of the metallic cylinder shown in Figure 3-2

The MS curves of the DP-CMs shown in Figure 3-13 (metallic sphere case) are illustrated in Figure 3-33. From Figure 3-33, it is easy to find out that: when the DP-CMs work at nonradiation states, the corresponding MS is indeed zero, i.e., relationship (3-57) surely hold for the DP-CMs.

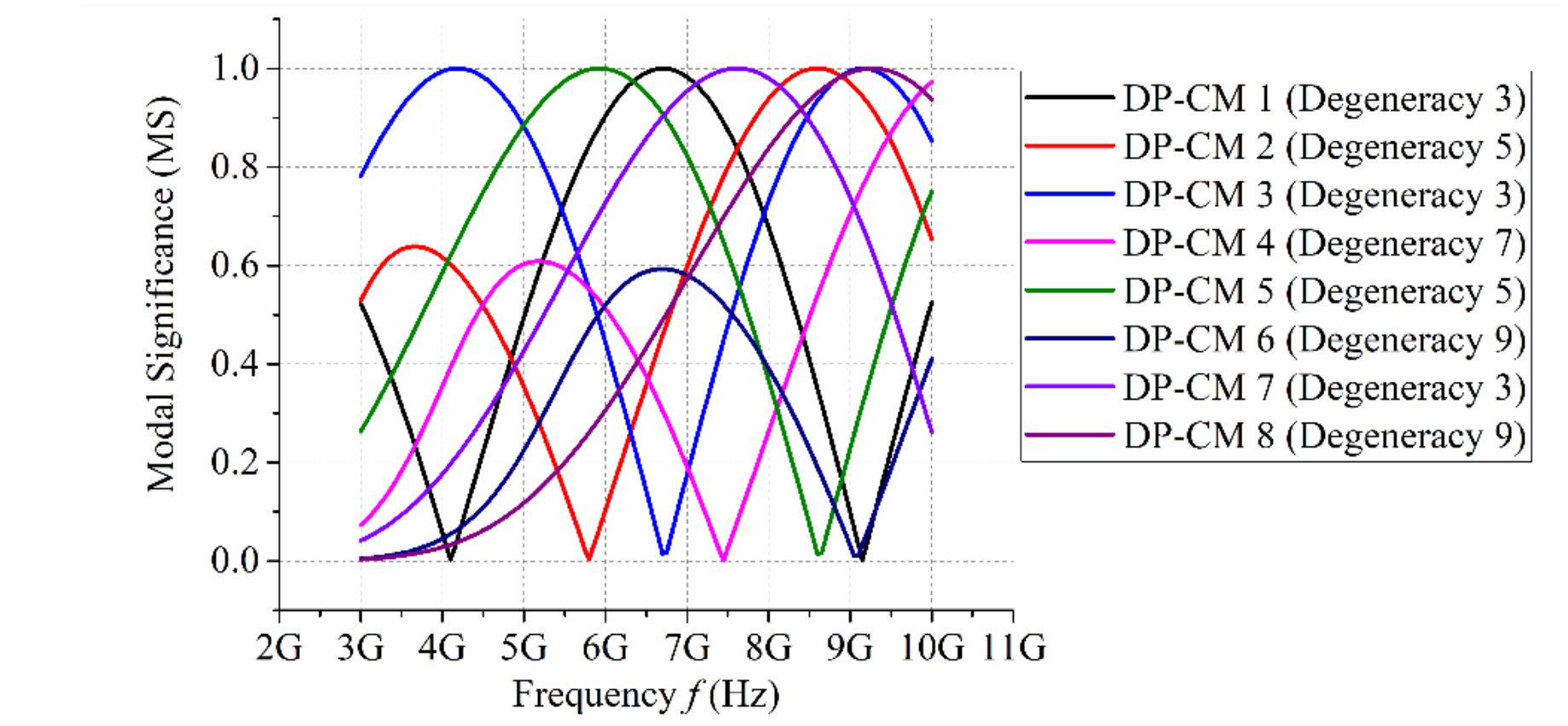

Figure 3-33 The MS curves corresponding to the first 8 typical DP-CMs of the metallic sphere shown in Figure 3-12

### 3.4.2 Internally Resonant Eigen-modes of Metallic Systems

If $f_0$ is an internal resonance frequency of metallic system, then matrix $\overline{\overline{P}}{}_{+}^{\text{Driving}}(f_0)$ is positive semi-definite. If the degeneracy order of the nonradiative DP-CMs at frequency $f_0$ is $N$, then there exists the following orthogonality[37,116]:

$$\overline{\overline{A}}{}^{H} \cdot \overline{\overline{P}}{}_{\text{met sys}}^{\text{Driving}} \cdot \overline{\overline{A}} = \overline{\overline{\Lambda}} \tag{3-58}$$

In orthogonality (3-58), $\overline{\overline{A}} = [\bar{\alpha}_{\text{rad;1}}, \bar{\alpha}_{\text{rad;2}}, \cdots, \bar{\alpha}_{\text{rad;}\Xi-N}, \bar{\alpha}_{\text{non-rad;1}}^{\text{pur res}}, \bar{\alpha}_{\text{non-rad;2}}^{\text{pur res}}, \cdots, \bar{\alpha}_{\text{non-rad;}N}^{\text{pur res}}]$ is the square matrix constructed by all characteristic vectors; $\overline{\overline{\Lambda}} = \text{diag}\{P_{\text{rad;1}}^{\text{Driving}}, P_{\text{rad;2}}^{\text{Driving}}, \cdots, P_{\text{rad;}\Xi-N}^{\text{Driving}}, P_{\text{non-rad;1}}^{\text{Driving}}, P_{\text{non-rad;2}}^{\text{Driving}}, \cdots, P_{\text{non-rad;}N}^{\text{Driving}}\}$ is the diagonal matrix whose diagonal elements are the ones listed in the bracket, where diagonal elements $P_{\text{rad;1}}^{\text{Driving}}$ and $P_{\text{non-rad;1}}^{\text{Driving}}$ are respectively the DPs of the first radiative DP-CM and the first nonradiative DP-CM, and the other diagonal elements can be explained similarly.

Because all characteristic vectors are linearly independent, then $\overline{\overline{A}}$ is invertible. Thus, the ranks of $\overline{\overline{P}}{}_{\text{met sys}}^{\text{Driving}}$ and $\overline{\overline{\Lambda}}$ are the same, i.e., $\text{rank}\,\overline{\overline{P}}{}_{\text{met sys}}^{\text{Driving}} = \text{rank}\,\overline{\overline{\Lambda}}$ [116]. Because $\text{rank}\,\overline{\overline{P}}{}_{\text{met sys}}^{\text{Driving}} = \Xi - \text{rank nullspace}\,\overline{\overline{P}}{}_{\text{met sys}}^{\text{Driving}}$ [116] and $\text{rank}\,\overline{\overline{\Lambda}} = \Xi - N$ ①, then $\text{rank nullspace}\,\overline{\overline{P}}{}_{\text{met sys}}^{\text{Driving}} = N$. Because $\overline{\overline{P}}{}_{\text{met sys}}^{\text{Driving}} = (1/2)\overline{\overline{Z}}{}^{\text{EFIE}}$, then $\text{nullspace}\,\overline{\overline{P}}{}_{\text{met sys}}^{\text{Driving}}$ is just the internal resonance space of metallic system[37,40,41]. In addition, $\bar{\alpha}_{\text{non-rad;1}}^{\text{pur res}}, \bar{\alpha}_{\text{non-rad;2}}^{\text{pur res}}, \cdots, \bar{\alpha}_{\text{non-rad;}N}^{\text{pur res}} \in \text{nullspace}\,\overline{\overline{P}}{}_{\text{met sys}}^{\text{Driving}}$, and $\bar{\alpha}_{\text{non-rad;1}}^{\text{pur res}}, \bar{\alpha}_{\text{non-rad;2}}^{\text{pur res}}, \cdots, \bar{\alpha}_{\text{non-rad;}N}^{\text{pur res}}$ are independent, so $\bar{\alpha}_{\text{non-rad;1}}^{\text{pur res}}, \bar{\alpha}_{\text{non-rad;2}}^{\text{pur res}}, \cdots, \bar{\alpha}_{\text{non-rad;}N}^{\text{pur res}}$ constitute the basis of internal resonance space.

Thus, in the aspect of spanning whole internal resonance space, nonradiative DP-CMs are equivalent to traditional internally resonant eigen-modes②. Then, the traditional EMT for closed metallic systems is successfully classified into the formalism of WEP-MetSca-CMT, and then it is realized that: the traditional EMT for closed metallic systems and the traditional EFIE-OpeMetSca-CMT[31,32] for open metallic systems are integrated into a whole modal theory —— WEP-MetSca-CMT for any metallic systems.

The above integration is very valuable for the modal analysis focusing on metallic systems. Specifically, if the term $\bar{\beta}_{\text{rad}}^{\text{pur cap}} + \bar{\beta}_{\text{rad}}^{\text{pur res}} + \bar{\beta}_{\text{rad}}^{\text{pur ind}}$ in decomposition (3-42) is nonzero, then the corresponding $\bar{a}$ cannot be expanded in terms of eigen-modes; if the term $\bar{\beta}_{\text{non-rad}}^{\text{pur res}}$ in decomposition (3-42) is nonzero, then the corresponding $\bar{a}$ cannot be expanded in terms of traditional Harrington's CMs[31,32]; if both the terms $\bar{\beta}_{\text{rad}}^{\text{pur cap}} + \bar{\beta}_{\text{rad}}^{\text{pur res}} + \bar{\beta}_{\text{rad}}^{\text{pur ind}}$ and $\bar{\beta}_{\text{non-rad}}^{\text{pur res}}$ in decomposition (3-42) are nonzero, then the corresponding $\bar{a}$ cannot be expanded in terms of not only eigen-modes but also

① Because $P_{\text{non-rad;1}}^{\text{Driving}}, P_{\text{non-rad;2}}^{\text{Driving}}, \cdots, P_{\text{non-rad;}N}^{\text{Driving}} = 0$.

② Such as the examples given in Subsection 3.2.4.

traditional Harrington's CMs, but it can be expanded in terms of the DP-CMs constructed in this dissertation.

### 3.4.3 Natural Modes of Metallic Systems

In essence, singularity expansion method (SEM)[23-25] is, in complex frequency domain①, to construct the basis of the null space of the EFIE-IMO $\overline{\overline{Z}}^{\text{EFIE}}$ of the objective metallic system, under the condition that resultant field is zero, and the basis are usually called as natural modes. Because the resultant field is zero, then the DPs corresponding to natural modes are zero, so all natural modes are resonant[37] ②.

Obviously, at internal resonance frequencies, the natural modes contain the basis of traditional internal resonance space, so also contain the basis of the nonradiation space $\text{nullspace}\,\overline{\overline{P}}_{+}^{\text{Driving}}$ defined in this dissertation. Thus, at internal resonance frequencies, natural mode set equivalently contains internally resonant eigen-mode set and nonradiative DP-CM set.

At the frequencies being different from internal resonance frequencies, the natural modes calculated from SEM are externally resonant evanescent modes, and the evanescent modes cannot continuously work at steady states. From the mathematical point of view, it is because of that: at the frequencies being different from internal resonance frequencies, equation $\overline{\overline{P}}_{\text{met sys}}^{\text{Driving}} \cdot \overline{a} = 0$ doesn't have nonzero solution in real frequency domain; in complex frequency domain, the so-called resonance frequencies making $\det \overline{\overline{P}}_{\text{met sys}}^{\text{Driving}} = 0$ have nonzero imaginary parts, and the imaginary parts lead to a radial evanescence of natural modes. From the physical point of view, it is because of that: to sustain the steady works of the modes of open systems, nonzero resultant fields are necessary[110], but there doesn't exist the resultant fields in SEM formalism.

It is thus clear that: except the nonradiative modes (i.e. the internally resonant modes), the modes focused on by this dissertation don't have any intersection with the natural modes. All of the modes focused on by this dissertation are the steady modes working in frequency domain. Except the nonradiative modes, all of the natural modes are the evanescent modes working in complex frequency domain.

### 3.4.4 Resonant Modes of Metallic Systems

According to the conclusions obtained in Subsection 3.3.3, we find out that:

① Complex frequency domain is not the same as frequency domain. It should never confuse them.
② But, it cannot be guaranteed that they are steadily resonant.

generally speaking, there exist some resonant modes at any frequency, such as the example $\bar{a}^{\text{res}} = \bar{\alpha}_{\text{rad};1}^{\text{pur cap}} + \bar{\alpha}_{\text{rad};1}^{\text{pur ind}}$ given in Subsection 3.3.3. In addition, according to the conclusions obtained in Subsection 3.3.3, we also find out that: a general resonant mode may include many components which cannot efficiently radiate EM energies; the modes having relatively strong radiation ability are contained in high-quality radiation space, and radiative purely resonant DP-CMs $\{\bar{\alpha}_{\text{Rad};\xi}^{\text{pur res}}\}$ constitute the basis of the space.

## 3.5 Chapter Summary

This chapter mainly focuses on establishing the WEP-based CMT for metallic systems (WEP-MetSca-CMT) by orthogonalizing DPO, and is also committed to deriving some important conclusions by employing WEP-MetSca-CMT.

Firstly, this chapter gives the mathematical expression for the WEP in classical electromagnetics, and then introduces the concept of DP for metallic systems —— power done by the resultant fields acting on scattered sources. By orthogonalizing frequency-domain DPO, this chapter, for metallic systems, constructs a series of steadily working modes not having net energy exchange in any integral period —— DP-CMs of metallic systems.

Afterwards, this chapter proves that all radiative DP-CMs are just traditional Harrington's metallic CMs, and that all nonradiative DP-CMs are equivalent to traditional internally resonant eigen-modes in the aspect of spanning whole nonradiation space (i.e. internal resonance space). Then, this chapter integrates the traditional EFIE-based CMT for open metallic systems (EFIE-OpeMetSca-CMT) and the traditional EMT for closed metallic systems into a complete set of modal theory —— WEP-based CMT for arbitrary metallic systems (WEP-MetSca-CMT). Based on the above integration, this chapter completes the fundamental mode set of metallic systems.

After that, based on the DP of metallic systems, this chapter finely classifies the working modes of metallic systems as shown in Figure 3-31. Based on the completeness and orthogonality of DP-CMs, this chapter realizes the orthogonal decomposition for whole modal space as illustrated in formulation (3-55) and the simplest orthogonal decompositions for various working modes as illustrated in formulations (3-42)~(3-54), and proves the uniqueness of the above decompositions. The simplest orthogonal decompositions for various working modes have important instruction significance for understanding the working mechanisms of various modes, so have great application significance for extracting the inherent EM scattering characters of objective structure.

Finally, based on the physical picture of EFIE-OpeMetSca-CMT (constructing a series of steadily working modes not having net energy exchange in any integral period), this chapter clarifies and improves the imperfections of original EFIE-OpeMetSca-CMT, and at the same time reveals the essential differences between Harrington's EFIE-OpeMetSca-CMT and its some IE-based variants. In WEP framework, this chapter reveals the essential differences between CMT and SEM (the former is a modal theory in frequency domain, but the latter is a modal theory in complex frequency domain) and the essential differences between CMs and natural modes (the formers are some steadily working modes in frequency domain, but the latters are some evanescent modes in complex frequency domain), and at the same time provides the physical reason leading to the above differences —— SEM is lack of the nonzero resultant fields to sustain the steady works of modes.

# Chapter 4 WEP-Based DP-CMs of Material Scattering Systems

To see a world in a grain of sand;

And a heaven in a wild flower. [151]

—— William Blake (English poet)

This chapter, in WEP framework, focuses on constructing the DP-CMs of material systems (which don't have net energy exchange in any integral period) by orthogonalizing frequency-domain DPO, and doing some necessary analysis and discussions for the related topics.

## 4.1 Chapter Introduction

Due to their small size, wide band, and high efficiency, etc., dielectric resonator antennas (DRAs) have been widely applied in antenna engineering domain[20-22]. In fact, the DRAs excluding their feeding structures are just material scattering systems. It is thus clear that: to effectively construct the inherent working modes of material scattering systems has instructional significance to analyzing and designing DRAs.

In theoretical and engineering electromagnetics domains, there has been a long history of studying the inherent working modes of material scattering systems, and has appeared some modal theories, such as dielectric waveguide model (DWM) theory[20-22], volume integral equation based CMT for material scattering systems (VIE-MatSca-CMT)[33], and surface integral equation based CMT for material scattering systems (SIE-MatSca-CMT)[34,35,42-49] ①. DWM theory is an approximate calculation theory based on the DWMs having regular topological structures, and the obtained results will have great errors if the approximation conditions are not satisfied[20-22], so it is very difficult to widely apply the theory to various engineering applications. Harrington's IE-MatSca-CMT is widely applicable, and the modes derived from it can effectively reflect the inherent EM scattering characters of material scattering systems, so it has attracted much attention[35].

Compared with the original IE-MatSca-CMT, which was established by Prof. Harrington et al.[33,34] in the 1970s, IE-MatSca-CMT has achieved certain development,

① In what follows, VIE-MatSca-CMT and SIE-MatSca-CMT are collectively referred to as Harrington's IE-based CMT for material scattering systems (IE-MatSca-CMT).

but it still has some imperfections. For example, in the aspect of physical picture, IE-MatSca-CMT, which is usually treated as a calculation method, is still lack of a clear physical picture, so the progress of further developing the fundamental theory of IE-MatSca-CMT is impeded tremendously①; in the aspect of applicable range, existing SIE-MatSca-CMT has not been able to be applied to some complicated material systems, such as inhomogeneous anisotropic material systems, multiply connected material systems, and multi-body material systems; in the aspect of some technical operations, the method to suppress the spurious modes outputted from SIE-MatSca-CMT has not been systematically established.

Following the direction proposed in Chapter 3②, this chapter will further solve the above-mentioned imperfections existing in Harrington's CMT. In Section 4.2, we focus on studying the volume formulation of the WEP-based CMT for material scattering systems; in Sections 4.3~4.7, we focus on studying the surface formulation of the WEP-based CMT for material scattering systems. It is necessary to emphasize that: in Sections 4.3~4.7, the CM calculation formulations developed in WEP framework are completely new, and the new formulations have wider applicable range, more concise mathematical expressions, and smaller requirements for computational resources.

## 4.2 Volume Formulation for Calculating the DP-CMs of Material Systems

In this section, we consider the scattering problem corresponding to the inhomogeneous anisotropic material scattering system③ shown in Figure 4-1. In Figure 4-1, $V_{\text{mat sys}}$, $D_{\text{imp}}$, and $D_{\text{env}}$ represent material system, external excitation, and EM environment respectively. $\vec{J}_{\text{imp}}$ is the excitation source distributing on $D_{\text{imp}}$, and $\vec{F}_{\text{imp}}$ is the EM field generated by $\vec{J}_{\text{imp}}$, where $F = E,H$. Under the excitation of $\vec{F}_{\text{imp}}$, conduction phenomenon, polarization phenomenon, and magnetization phenomenon will occur on $V_{\text{mat sys}}$ and $D_{\text{env}}$, and the phenomena will lead to some induced sources

① As Prof. Enrico Fermi (Nobel Prize in Physics, 1938) said: "There are two ways to do calculations. The first way, which I prefer, is to have a clear physical picture. The second way is to have a rigorous mathematical formalism.[103]". In fact, this chapter will, based on some facts, demonstrate that: clear physical picture plays an irreplaceable active role in the aspect of further developing the CMT for material scattering systems.

② Using WEP as the carrying framework of CMT & aiming at constructing a series of steadily working CMs which are complete and don't have net energy exchange in any integral period & basing on orthogonalizing frequency-domain DPO method.

③ In what follows, the terminology "inhomogeneous anisotropic material scattering system" will be simply called as "material system".

distributing on $V_{\text{mat sys}}$ and $D_{\text{env}}$ [110,121].

The induced source on $V_{\text{mat sys}}$ generates scattered field $\vec{F}^{\text{sca}}$, and the induced source on $D_{\text{env}}$ generates environment field $\vec{F}_{\text{env}}$. The summation of $\vec{F}_{\text{imp}}$ and $\vec{F}_{\text{env}}$ is just the resultant field used to drive $V_{\text{mat sys}}$, and this dissertation also calls it as incident field, and correspondingly denotes it as $\vec{F}^{\text{inc}}$, i.e., $\vec{F}^{\text{inc}} = \vec{F}_{\text{imp}} + \vec{F}_{\text{env}}$ ①. In addition, the summation of $\vec{F}^{\text{inc}}$ and $\vec{F}^{\text{sca}}$ is just the total field corresponding to the scattering problem shown in Figure 4-1, and it is denoted as $\vec{F}^{\text{tot}}$, i.e., $\vec{F}^{\text{tot}} = \vec{F}^{\text{inc}} + \vec{F}^{\text{sca}}$.

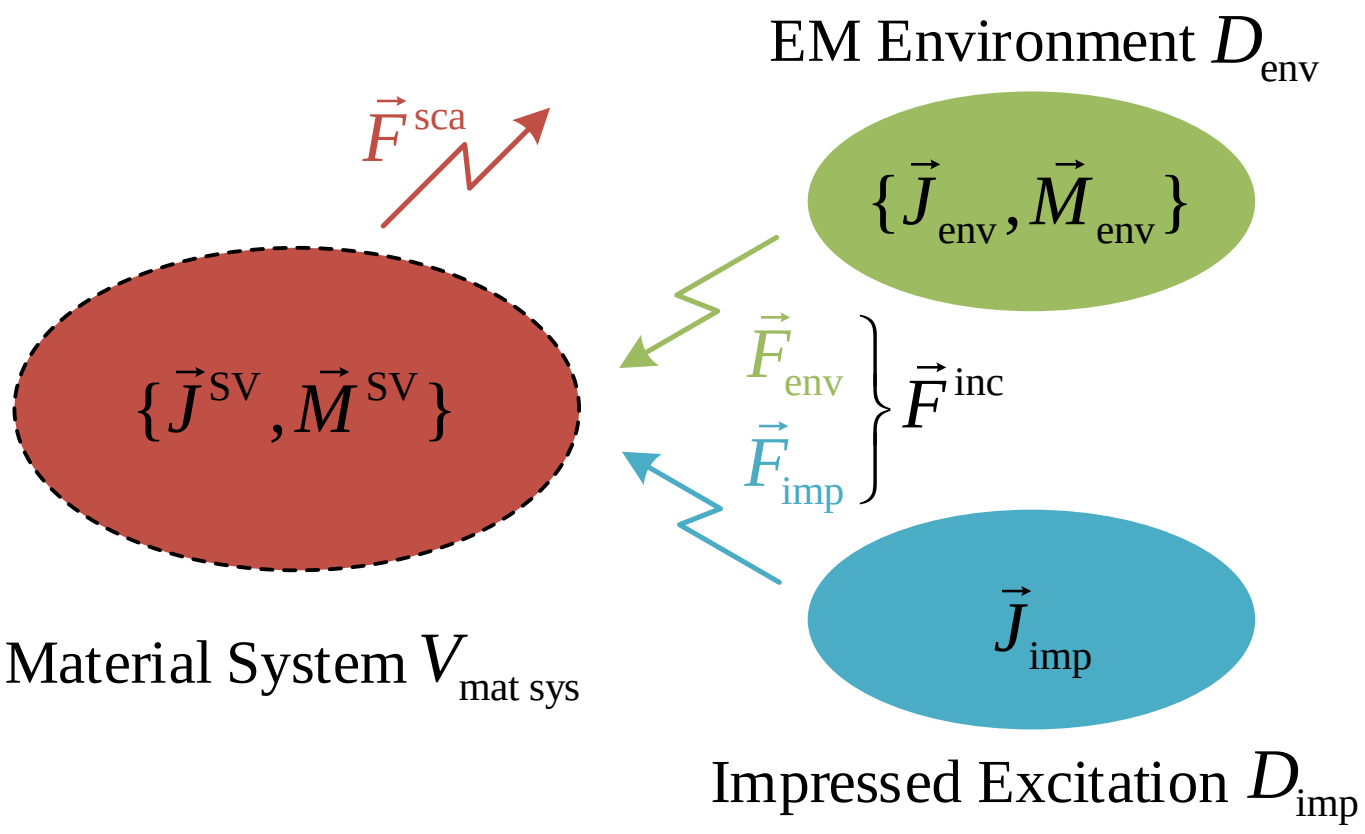

Figure 4-1 Material scattering problem considered in this section

Literature [110] and literature [121] pointed out that: if the conduction phenomenon, polarization phenomenon, and magnetization phenomenon are described by conduction electric current model, polarization electric current model, and magnetization electric current model respectively, then there will exist the conduction volume electric current, polarization volume electric current, and magnetization volume electric current distributing in the interior of $V_{\text{mat sys}}$, and at the same time there also will exist the magnetization surface electric current distributing on the boundary of $V_{\text{mat sys}}$. Based on a rigorous mathematical method, literature [121] proved the above conclusion, and also derived the quantitive mathematical relationships between various induced currents and total fields $\vec{E}^{\text{tot}} \,\&\, \vec{H}^{\text{tot}}$.

Literature [110] also pointed out that: if the conduction phenomenon, polarization phenomenon, and magnetization phenomenon are described by conduction electric current model, polarization electric current model, and magnetization magnetic current model respectively, then there will exist the conduction volume electric current, polarization volume electric current, and magnetization volume magnetic current

① The reason to treat $\vec{F}_{\text{imp}}$ and $\vec{F}_{\text{env}}$ as a whole has been given in Chapter 3, and it will not be repeated here.

distributing in the interior of $V_{\text{mat sys}}$, but there will not exist any induced surface current distributing on the boundary of $V_{\text{mat sys}}$. Similarly to literature [121], the Appendix A of this dissertation proves the above conclusion based on a rigorous mathematical method, and also derives the quantitive mathematical relationships between various induced currents and total fields $\vec{E}^{\text{tot}}$ & $\vec{H}^{\text{tot}}$.

In fact, literature [110] also pointed out that: the second model is more advantageous than the first model, and the second model is indeed applied in computational electromagnetics[106] widely, so this dissertation selects to use the second model. Based on the second model, this section denotes the conduction volume electric current, polarization volume electric current, and magnetization volume magnetic current on $V_{\text{mat sys}}$ as $\vec{J}^{\text{CV}}$, $\vec{J}^{\text{PV}}$, and $\vec{M}^{\text{MV}}$ respectively, and the summation of $\vec{J}^{\text{CV}}$ and $\vec{J}^{\text{PV}}$ is denoted as $\vec{J}^{\text{SV}}$ (i.e., $\vec{J}^{\text{SV}} = \vec{J}^{\text{CV}} + \vec{J}^{\text{PV}}$), and $\vec{M}^{\text{MV}}$ is similarly denoted as $\vec{M}^{\text{SV}}$ (i.e., $\vec{M}^{\text{SV}} = \vec{M}^{\text{MV}}$); in this dissertation, the induced electric and magnetic currents on $D_{\text{env}}$ are denoted as $\vec{J}_{\text{env}}$ and $\vec{M}_{\text{env}}$ respectively.

In the following parts of this section, we will, based on WEP, introduce the concept of DP for material systems, and provide the volume formulation for constructing the DP-CMs of $V_{\text{mat sys}}$.

## 4.2.1 WEP and DP

In time domain, the instantaneous power $P^{\text{Driving}}_{\text{mat sys}}(t)$ done by the incident fields $\{\vec{E}^{\text{inc}}(t), \vec{H}^{\text{inc}}(t)\}$ acting on scattered sources $\{\vec{J}^{\text{SV}}(t), \vec{M}^{\text{SV}}(t)\}$ is as follows[44]:

$$\begin{aligned}
&P^{\text{Driving}}_{\text{mat sys}}(t)\\
=\;& \left\langle \vec{J}^{\text{SV}}(t), \vec{E}^{\text{inc}}(t)\right\rangle_{V_{\text{mat sys}}} + \left\langle \vec{H}^{\text{inc}}(t), \vec{M}^{\text{SV}}(t)\right\rangle_{V_{\text{mat sys}}}\\
=\;& \left\langle \vec{J}^{\text{SV}}(t), \vec{E}^{\text{tot}}(t) - \vec{E}^{\text{sca}}(t)\right\rangle_{V_{\text{mat sys}}} + \left\langle \vec{H}^{\text{tot}}(t) - \vec{H}^{\text{sca}}(t), \vec{M}^{\text{SV}}(t)\right\rangle_{V_{\text{mat sys}}}\\
=\;& \left\langle \vec{\vec{\sigma}}_{\text{mat}} \cdot \vec{E}^{\text{tot}}(t) + \frac{\partial}{\partial t}\left[\Delta\vec{\vec{\varepsilon}}_{\text{mat}} \cdot \vec{E}^{\text{tot}}(t)\right], \vec{E}^{\text{tot}}(t)\right\rangle_{V_{\text{mat sys}}} + \left\langle \vec{H}^{\text{tot}}(t), \frac{\partial}{\partial t}\left[\Delta\vec{\vec{\mu}}_{\text{mat}} \cdot \vec{H}^{\text{tot}}(t)\right]\right\rangle_{V_{\text{mat sys}}}\\
&- \left\langle \vec{J}^{\text{SV}}(t), \vec{E}^{\text{sca}}(t)\right\rangle_{V_{\text{mat sys}}} - \left\langle \vec{H}^{\text{sca}}(t), \vec{M}^{\text{SV}}(t)\right\rangle_{V_{\text{mat sys}}}\\
=\;& P^{\text{los}}(t) + P^{\text{rad}}(t) + \frac{d}{dt}\left[W^{\text{mag}}_{\text{vac}}(t) + W^{\text{ele}}_{\text{vac}}(t)\right] + \frac{d}{dt}\left[W^{\text{mag}}_{\text{mat}}(t) + W^{\text{pol}}_{\text{mat}}(t)\right]
\end{aligned} \tag{4-1}$$

In above formulation (4-1), the inner product is defined as the one used in Chapters 2 and 3; the first equality is based on the fact that incident electric and magnetic fields don't do any work on scattered magnetic and electric currents respectively[110-114]; the second equality is based on linear superposition principle $\vec{F}^{\text{tot}}(t)=\vec{F}^{\text{inc}}(t)+\vec{F}^{\text{sca}}(t)$ [110-114]; the third equality is based on that $\vec{J}^{\text{SV}}(t)=\vec{J}_{\text{mat}}^{\text{CV}}(t)+\vec{J}_{\text{mat}}^{\text{PV}}(t)$ & $\vec{M}^{\text{SV}}(t)=\vec{M}_{\text{mat}}^{\text{MV}}(t)$ and $\vec{J}_{\text{mat}}^{\text{CV}}(t)=\vec{\vec{\sigma}}_{\text{mat}}\cdot\vec{E}^{\text{tot}}(t)$ & $\vec{J}_{\text{mat}}^{\text{PV}}(t)=\partial[\Delta\vec{\vec{\varepsilon}}_{\text{mat}}\cdot\vec{E}^{\text{tot}}(t)]/\partial t$ & $\vec{M}_{\text{mat}}^{\text{MV}}(t)=\partial[\Delta\vec{\vec{\mu}}_{\text{mat}}\cdot\vec{H}^{\text{tot}}(t)]/\partial t$ ①; the fourth equality is based on time-domain Poynting's theorem[110-113]; on the RHS of the fourth equality, $P^{\text{los}}(t)$ and $P^{\text{rad}}(t)$ are respectively the instantaneous ohmic loss power corresponding to total electric field and the instantaneous radiated power carried by scattered field, and $W_{\text{vac}}^{\text{mag}}(t)$ and $W_{\text{vac}}^{\text{ele}}(t)$ are respectively the instantaneous magnetic and electric energies in vacuum, and $W_{\text{mat}}^{\text{mag}}(t)$ and $W_{\text{mat}}^{\text{pol}}(t)$ are respectively the instantaneous magnetization and polarization energies in matter, and $P^{\text{los}}(t)=<\vec{\vec{\sigma}}_{\text{mat}}\cdot\vec{E}^{\text{tot}}(t),\vec{E}^{\text{tot}}(t)>_{V_{\text{mat sys}}}$ and $P^{\text{rad}}(t)=\oint\!\!\oint_{S_\infty}[\vec{E}^{\text{sca}}(t)\times\vec{H}^{\text{sca}}(t)]\cdot\hat{n}_\infty^+ dS$ and $W_{\text{vac}}^{\text{mag}}(t)=(1/2)<\vec{H}^{\text{sca}}(t),\mu_0\vec{H}^{\text{sca}}(t)>_{\mathbb{R}^3}$ and $W_{\text{vac}}^{\text{ele}}(t)=(1/2)<\varepsilon_0\vec{E}^{\text{sca}}(t),\vec{E}^{\text{sca}}(t)>_{\mathbb{R}^3}$ and $W_{\text{mat}}^{\text{mag}}(t)=(1/2)<\vec{H}^{\text{tot}}(t),\Delta\vec{\vec{\mu}}_{\text{mat}}\cdot\vec{H}^{\text{tot}}(t)>_{V_{\text{mat sys}}}$ and $W_{\text{mat}}^{\text{pol}}(t)=(1/2)<\Delta\vec{\vec{\varepsilon}}_{\text{mat}}\cdot\vec{E}^{\text{tot}}(t),\vec{E}^{\text{tot}}(t)>_{V_{\text{mat sys}}}$.

The physical meaning of formulation (4-1) is that: the instantaneous power done by resultant fields on scattered currents are transverted to four parts —— the radiated power transferred to infinity by scattered fields, the lossy power converted to Joule's heat, the part converted to the reactive power corresponding to the electric and magnetic energies in vacuum, and the part converted to the reactive power corresponding to the polarization electric and magnetization magnetic energies in matter. Similarly to Chapter 3, this chapter calls power term $P_{\text{mat sys}}^{\text{Driving}}(t)=<\vec{J}^{\text{SV}}(t),\vec{E}^{\text{inc}}(t)>_{V_{\text{mat sys}}}+<\vec{H}^{\text{inc}}(t),\vec{M}^{\text{SV}}(t)>_{V_{\text{mat sys}}}$ as the time-domain form of the DP for material scattering systems.

### 1) Frequency-Domain Version I of the Time-Domain DP $P_{\text{mat sys}}^{\text{Driving}}(t)$

The frequency-domain version I of time-domain DP $P_{\text{mat sys}}^{\text{Driving}}(t)$ is as follows[44]:

$$\begin{aligned}
P_{\text{mat sys}}^{\text{DRIVING}} &= (1/2)\left\langle\vec{J}^{\text{SV}},\vec{E}^{\text{inc}}\right\rangle_{V_{\text{mat sys}}}+(1/2)\left\langle\vec{H}^{\text{inc}},\vec{M}^{\text{SV}}\right\rangle_{V_{\text{mat sys}}}\\
&= (1/2)\left\langle\vec{J}^{\text{SV}},\vec{E}^{\text{tot}}-\vec{E}^{\text{sca}}\right\rangle_{V_{\text{mat sys}}}+(1/2)\left\langle\vec{H}^{\text{tot}}-\vec{H}^{\text{sca}},\vec{M}^{\text{SV}}\right\rangle_{V_{\text{mat sys}}}\\
&= (1/2)\left\langle\vec{J}^{\text{SV}},\vec{E}^{\text{tot}}\right\rangle_{V_{\text{mat sys}}}+(1/2)\left\langle\vec{H}^{\text{tot}},\vec{M}^{\text{SV}}\right\rangle_{V_{\text{mat sys}}}\\
&\quad -(1/2)\left\langle\vec{J}^{\text{SV}},\vec{E}^{\text{sca}}\right\rangle_{V_{\text{mat sys}}}-(1/2)\left\langle\vec{H}^{\text{sca}},\vec{M}^{\text{SV}}\right\rangle_{V_{\text{mat sys}}}
\end{aligned}\tag{4-2}$$

In above formulation (4-2), the $P_{\text{mat sys}}^{\text{DRIVING}}$ on the LHS of the first equality is the frequency-domain version I of the DP of material systems, where using superscript "DRIVING" instead of "Driving" is to be distinguished from the frequency-domain

① The detailed process to derive the relationships between various induced currents and total fields can be found in the Appendix A of this dissertation.

version II to be introduced subsequently; the coefficient $1/2$ used in the RHS of the first equality originates from averaging the quadratic quantity of time-harmonic EM fields[114]; the second equality is based on superposition principle $\vec{F}^{\text{tot}} = \vec{F}^{\text{inc}} + \vec{F}^{\text{sca}}$; the third equality is based on the linear property of inner product.

Similarly to Chapter 3, it is easy to rewrite the RHS of the third equality in formulation (4-2) as follows:

$$\begin{aligned} P_{\text{mat sys}}^{\text{DRIVING}} = & \ (1/2)\left\langle \vec{J}^{\text{SV}}, (j\omega\Delta\overleftrightarrow{\varepsilon}_{\text{mat}}^{\text{c}})^{-1}\cdot\vec{J}^{\text{SV}}\right\rangle_{V_{\text{mat sys}}} + (1/2)\left\langle (j\omega\Delta\overleftrightarrow{\mu}_{\text{mat}})^{-1}\cdot\vec{M}^{\text{SV}}, \vec{M}^{\text{SV}}\right\rangle_{V_{\text{mat sys}}} \\ & - \frac{1}{2}\left\langle \vec{J}^{\text{SV}}, -j\omega\mu_0\mathcal{L}_0(\vec{J}^{\text{SV}}) - \mathcal{K}_0(\vec{M}^{\text{SV}})\right\rangle_{V_{\text{mat sys}}} - \frac{1}{2}\left\langle -j\omega\varepsilon_0\mathcal{L}_0(\vec{M}^{\text{SV}}) + \mathcal{K}_0(\vec{J}^{\text{SV}}), \vec{M}^{\text{SV}}\right\rangle_{V_{\text{mat sys}}} \\ = & \ (1/2)\left\langle \vec{J}^{\text{SV}}, (j\omega\Delta\overleftrightarrow{\varepsilon}_{\text{mat}}^{\text{c}})^{-1}\cdot\vec{J}^{\text{SV}}\right\rangle_{V_{\text{mat sys}}} + (1/2)\left\langle (j\omega\Delta\overleftrightarrow{\mu}_{\text{mat}})^{-1}\cdot\vec{M}^{\text{SV}}, \vec{M}^{\text{SV}}\right\rangle_{V_{\text{mat sys}}} \\ & - (1/2)\left\langle \vec{J}^{\text{SV}}, -j\omega\mu_0\mathcal{L}_0(\vec{J}^{\text{SV}})\right\rangle_{V_{\text{mat sys}}} \\ & - (1/2)\left\langle \vec{J}^{\text{SV}}, -\mathcal{K}_0(\vec{M}^{\text{SV}})\right\rangle_{V_{\text{mat sys}}} - (1/2)\left\langle \mathcal{K}_0(\vec{J}^{\text{SV}}), \vec{M}^{\text{SV}}\right\rangle_{V_{\text{mat sys}}} \\ & - (1/2)\left\langle -j\omega\varepsilon_0\mathcal{L}_0(\vec{M}^{\text{SV}}), \vec{M}^{\text{SV}}\right\rangle_{V_{\text{mat sys}}} \end{aligned} \tag{4-3}$$

where the superscript "$-1$" represents the inverse of two-order tensor. In addition, the definitions of operators $\mathcal{L}_0(\vec{X})$ and $\mathcal{K}_0(\vec{X})$ are the same as the ones given in Chapter 3, and they will not be repeated here.

If $\{\vec{b}_\xi^J\}_{\xi=1}^{\Xi^J}$ and $\{\vec{b}_\xi^M\}_{\xi=1}^{\Xi^M}$ are respectively the complete basis functions for the $\vec{J}^{\text{SV}}$ and $\vec{M}^{\text{SV}}$ on $V_{\text{mat sys}}$, then $\vec{J}^{\text{SV}}$ and $\vec{M}^{\text{SV}}$ can be expanded as follows[106-108,115]:

$$\vec{C}^{\text{SV}}(\vec{r}) = \sum_{\xi=1}^{\Xi^C} a_\xi^C \vec{b}_\xi^C(\vec{r}) = \overline{\boldsymbol{B}}^C \cdot \bar{a}^C \quad , \quad \vec{r} \in V_{\text{mat sys}} \tag{4-4}$$

In formulation (4-4), $C = J, M$; $\overline{\boldsymbol{B}}^C = [\vec{b}_1^C, \vec{b}_2^C, \cdots, \vec{b}_{\Xi^C}^C]$; $\bar{a}^C = [a_1^C, a_2^C, \cdots, a_{\Xi^C}^C]^T$. Similarly to the manner to discretize the DPO $P_{\text{met sys}}^{\text{Driving}}$ of metallic systems into matrix form used in Chapter 3, the DPO $P_{\text{mat sys}}^{\text{DRIVING}}$ of material systems can be easily discretized by inserting expansion formulation (4-4) into operator formulation (4-3) as follows:

$$\begin{aligned} P_{\text{mat sys}}^{\text{DRIVING}} &= \begin{bmatrix} \bar{a}^J \\ \bar{a}^M \end{bmatrix}^H \cdot \begin{bmatrix} \overline{\overline{P}}^{\text{ele}} & 0 \\ 0 & \overline{\overline{P}}^{\text{mag}} \end{bmatrix} \cdot \begin{bmatrix} \bar{a}^J \\ \bar{a}^M \end{bmatrix} + \begin{bmatrix} \bar{a}^J \\ \bar{a}^M \end{bmatrix}^H \cdot \begin{bmatrix} \overline{\overline{P}}_0^{JJ} & \overline{\overline{P}}_0^{JM} \\ 0 & \overline{\overline{P}}_0^{MM} \end{bmatrix} \cdot \begin{bmatrix} \bar{a}^J \\ \bar{a}^M \end{bmatrix} \\ &= \begin{bmatrix} \bar{a}^J \\ \bar{a}^M \end{bmatrix}^H \cdot \underbrace{\begin{bmatrix} \overline{\overline{P}}^{\text{ele}} + \overline{\overline{P}}_0^{JJ} & \overline{\overline{P}}_0^{JM} \\ 0 & \overline{\overline{P}}^{\text{mag}} + \overline{\overline{P}}_0^{MM} \end{bmatrix}}_{\overline{\overline{P}}_{\text{mat sys}}^{\text{DRIVING}}} \cdot \begin{bmatrix} \bar{a}^J \\ \bar{a}^M \end{bmatrix} \end{aligned} \tag{4-5}$$

where the 0s are some zero matrices with proper line and column numbers. In addition, the elements of the submatrices in formulation (4-5) are calculated as follows:

$$p_{\xi\zeta}^{\text{ele}} = (1/2)\left\langle \vec{b}_{\xi}^{J}, \left(j\omega\Delta\overset{\leftrightarrow}{\varepsilon}_{\text{mat}}^{\text{c}}\right)^{-1}\cdot\vec{b}_{\zeta}^{J}\right\rangle_{V_{\text{mat sys}}} \tag{4-6a}$$

$$p_{\xi\zeta}^{\text{mag}} = (1/2)\left\langle \left(j\omega\Delta\overset{\leftrightarrow}{\mu}_{\text{mat}}\right)^{-1}\cdot\vec{b}_{\xi}^{M}, \vec{b}_{\zeta}^{M}\right\rangle_{V_{\text{mat sys}}} \tag{4-6b}$$

$$p_{0;\xi\zeta}^{JJ} = -(1/2)\left\langle \vec{b}_{\xi}^{J}, -j\omega\mu_0\mathcal{L}_0\left(\vec{b}_{\zeta}^{J}\right)\right\rangle_{V_{\text{mat sys}}} \tag{4-6c}$$

$$p_{0;\xi\zeta}^{JM} = -(1/2)\left\langle \vec{b}_{\xi}^{J}, -\mathcal{K}_0\left(\vec{b}_{\zeta}^{M}\right)\right\rangle_{V_{\text{mat sys}}} - (1/2)\left\langle \mathcal{K}_0\left(\vec{b}_{\xi}^{J}\right), \vec{b}_{\zeta}^{M}\right\rangle_{V_{\text{mat sys}}} \tag{4-6d}$$

$$p_{0;\xi\zeta}^{MM} = -(1/2)\left\langle -j\omega\varepsilon_0\mathcal{L}_0\left(\vec{b}_{\xi}^{M}\right), \vec{b}_{\zeta}^{M}\right\rangle_{V_{\text{mat sys}}} \tag{4-6e}$$

where the subscripts "0" in various elements represent that the corresponding operators $\mathcal{L}_0$ and $\mathcal{K}_0$ are the vacuum versions.

### 2) Frequency-Domain Version II of the Time-Domain DP $P_{\text{mat sys}}^{\text{Driving}}(t)$

In fact, time-domain DP $P_{\text{mat sys}}^{\text{Driving}}(t)$ also exists the following frequency-domain version II[44,45]:

$$\begin{aligned} P_{\text{mat sys}}^{\text{driving}} &= (1/2)\left\langle \vec{J}^{\text{SV}}, \vec{E}^{\text{inc}}\right\rangle_{V_{\text{mat sys}}} + (1/2)\left\langle \vec{M}^{\text{SV}}, \vec{H}^{\text{inc}}\right\rangle_{V_{\text{mat sys}}} \\ &= (1/2)\left\langle \vec{J}^{\text{SV}}, \vec{E}^{\text{tot}} - \vec{E}^{\text{sca}}\right\rangle_{V_{\text{mat sys}}} + (1/2)\left\langle \vec{M}^{\text{SV}}, \vec{H}^{\text{tot}} - \vec{H}^{\text{sca}}\right\rangle_{V_{\text{mat sys}}} \\ &= (1/2)\left\langle \vec{J}^{\text{SV}}, \vec{E}^{\text{tot}}\right\rangle_{V_{\text{mat sys}}} + (1/2)\left\langle \vec{M}^{\text{SV}}, \vec{H}^{\text{tot}}\right\rangle_{V_{\text{mat sys}}} \\ &\quad -(1/2)\left\langle \vec{J}^{\text{SV}}, \vec{E}^{\text{sca}}\right\rangle_{V_{\text{mat sys}}} - (1/2)\left\langle \vec{M}^{\text{SV}}, \vec{H}^{\text{sca}}\right\rangle_{V_{\text{mat sys}}} \end{aligned} \tag{4-7}$$

Similarly to deriving formulation (4-3) from formulation (4-2), we can, from above formulation (4-7), derive the following EM current form (4-8):

$$\begin{aligned} P_{\text{mat sys}}^{\text{driving}} &= (1/2)\left\langle \vec{J}^{\text{SV}}, \left(j\omega\Delta\overset{\leftrightarrow}{\varepsilon}_{\text{mat}}^{\text{c}}\right)^{-1}\cdot\vec{J}^{\text{SV}}\right\rangle_{V_{\text{mat sys}}} + (1/2)\left\langle \vec{M}^{\text{SV}}, \left(j\omega\Delta\overset{\leftrightarrow}{\mu}_{\text{mat}}\right)^{-1}\cdot\vec{M}^{\text{SV}}\right\rangle_{V_{\text{mat sys}}} \\ &\quad -(1/2)\left\langle \vec{J}^{\text{SV}}, -j\omega\mu_0\mathcal{L}_0\left(\vec{J}^{\text{SV}}\right) - \mathcal{K}_0\left(\vec{M}^{\text{SV}}\right)\right\rangle_{V_{\text{mat sys}}} \\ &\quad -(1/2)\left\langle \vec{M}^{\text{SV}}, -j\omega\varepsilon_0\mathcal{L}_0\left(\vec{M}^{\text{SV}}\right) + \mathcal{K}_0\left(\vec{J}^{\text{SV}}\right)\right\rangle_{V_{\text{mat sys}}} \end{aligned} \tag{4-8}$$

Inserting expansion formulation (4-4) into above EM current form (4-8), $P_{\text{mat sys}}^{\text{driving}}$ will be discretized into the following matrix form:

$$\begin{aligned} P_{\text{mat sys}}^{\text{driving}} &= \begin{bmatrix} \bar{a}^{J} \\ \bar{a}^{M} \end{bmatrix}^{H} \cdot \begin{bmatrix} \overline{\overline{P}}^{\text{ele}} & 0 \\ 0 & \left(\overline{\overline{P}}^{\text{mag}}\right)^{H} \end{bmatrix} \cdot \begin{bmatrix} \bar{a}^{J} \\ \bar{a}^{M} \end{bmatrix} + \begin{bmatrix} \bar{a}^{J} \\ \bar{a}^{M} \end{bmatrix}^{H} \cdot \begin{bmatrix} \overline{\overline{P}}_0^{JJ} & \overline{\overline{Q}}_0^{JM} \\ \overline{\overline{Q}}_0^{MJ} & \left(\overline{\overline{P}}_0^{MM}\right)^{H} \end{bmatrix} \cdot \begin{bmatrix} \bar{a}^{J} \\ \bar{a}^{M} \end{bmatrix} \\ &= \begin{bmatrix} \bar{a}^{J} \\ \bar{a}^{M} \end{bmatrix}^{H} \cdot \underbrace{\begin{bmatrix} \overline{\overline{P}}^{\text{ele}} + \overline{\overline{P}}_0^{JJ} & \overline{\overline{Q}}_0^{JM} \\ \overline{\overline{Q}}_0^{MJ} & \left(\overline{\overline{P}}^{\text{mag}} + \overline{\overline{P}}_0^{MM}\right)^{H} \end{bmatrix}}_{\overline{\overline{P}}_{\text{mat sys}}^{\text{driving}}} \cdot \begin{bmatrix} \bar{a}^{J} \\ \bar{a}^{M} \end{bmatrix} \end{aligned} \tag{4-9}$$

In formulation (4-9), matrices $\overline{\overline{P}}^{\text{ele}}$, $\overline{\overline{P}}^{\text{mag}}$, $\overline{\overline{P}}_0^{JJ}$, and $\overline{\overline{P}}_0^{MM}$ are just the ones used in formulation (4-5); the elements of matrices $\overline{\overline{Q}}_0^{JM}$ and $\overline{\overline{Q}}_0^{MJ}$ are calculated as follows:

$$q_{0;\xi\zeta}^{JM} = -(1/2)\left\langle \vec{b}_\xi^J, -\mathcal{K}_0\left(\vec{b}_\zeta^M\right)\right\rangle_{V_{\text{mat sys}}} \tag{4-10a}$$

$$q_{0;\xi\zeta}^{MJ} = -(1/2)\left\langle \vec{b}_\xi^M, \mathcal{K}_0\left(\vec{b}_\zeta^J\right)\right\rangle_{V_{\text{mat sys}}} \tag{4-10b}$$

where the subscripts "0" in various elements represent that the corresponding operator $\mathcal{K}_0$ is the vacuum version.

In addition, the following parts of this chapter are discussed in frequency domain, so we will not especially use the modifier "frequency-domain".

## 4.2.2 DP-CMs and Their Orthogonality

In this section, we will construct two sets of CMs for material systems, respectively based on DPO version I $P_{\text{mat sys}}^{\text{DRIVING}}$ and DPO version II $P_{\text{mat sys}}^{\text{driving}}$. By comparing the orthogonality satisfied by the two CM sets, we find out that: the CMs derived from $P_{\text{mat sys}}^{\text{driving}}$ are completely decoupled, but the CMs derived from $P_{\text{mat sys}}^{\text{DRIVING}}$ are not completely decoupled. Based on this, this dissertation will always use the $P_{\text{mat sys}}^{\text{driving}}$-based constructing method for CMs, and will compare the CMs with the ones derived from Harrington's VIE-MatSca-CMT[33].

### 1) CMs Derived From DPO Version I $P_{\text{mat sys}}^{\text{DRIVING}}$

Square matrix $\overline{\overline{P}}_{\text{mat sys}}^{\text{DRIVING}}$ exists the following Toeplitz's decomposition[116]:

$$\overline{\overline{P}}_{\text{mat sys}}^{\text{DRIVING}} = \overline{\overline{P}}_{+}^{\text{DRIVING}} + j\,\overline{\overline{P}}_{-}^{\text{DRIVING}} \tag{4-11}$$

where $\overline{\overline{P}}_{+}^{\text{DRIVING}} = [\overline{\overline{P}}_{\text{mat sys}}^{\text{DRIVING}} + (\overline{\overline{P}}_{\text{mat sys}}^{\text{DRIVING}})^H]/2$ and $\overline{\overline{P}}_{-}^{\text{DRIVING}} = [\overline{\overline{P}}_{\text{mat sys}}^{\text{DRIVING}} - (\overline{\overline{P}}_{\text{mat sys}}^{\text{DRIVING}})^H]/2j$. Obviously, both $\overline{\overline{P}}_{+}^{\text{DRIVING}}$ and $\overline{\overline{P}}_{-}^{\text{DRIVING}}$ are Hermitian. In general, $\overline{\overline{P}}_{+}^{\text{DRIVING}}$ is positive definite, because $V_{\text{mat sys}}$ is an open EM system and there doesn't exist any scattered surface current distributing on its boundary.

Based on operator $P_{\text{mat sys}}^{\text{DRIVING}}$, the characteristic vectors of material systems can be derived from solving the following generalized characteristic equation:

$$\overline{\overline{P}}_{-}^{\text{DRIVING}} \cdot \bar{\alpha}_\xi^{\text{DRIVING}} = \lambda_{\text{mat sys};\xi}^{\text{DRIVING}}\, \overline{\overline{P}}_{+}^{\text{DRIVING}} \cdot \bar{\alpha}_\xi^{\text{DRIVING}} \text{ ①} \tag{4-12}$$

① In Chapter 3, we didn't add the superscripts and subscripts (such as " Driving ", " met sys ", and " $J$ ", etc.) to the characteristic values and characteristic vectors to simplify the symbolic system of Chapter 3, because Chapter 3 discussed {modal classification, modal expansion, modal decomposition} and the related contents need to add some superscripts and subscripts (such as " cap ", " res ", " ind ", " rad ", and " non- rad ", etc.) to the characteristic vectors.

In characteristic equation (4-12), $(\bar{\alpha}_{\xi}^{\text{DRIVING}})^T = [(\bar{\alpha}_{J;\xi}^{\text{DRIVING}})^T \quad (\bar{\alpha}_{M;\xi}^{\text{DRIVING}})^T]$ ; the corresponding characteristic electric and magnetic currents are $\vec{C}^{\text{SV}} = \bar{\boldsymbol{B}}^C \cdot \bar{\alpha}_{C;\xi}^{\text{DRIVING}}$ , where $C = J, M$ ; in the interior of $V_{\text{mat sys}}$, the corresponding characteristic total fields are $\vec{E}_{\xi}^{\text{tot}} = (j\omega\Delta\vec{\vec{\varepsilon}}_{\text{mat}}^{\text{c}})^{-1} \cdot \vec{J}_{\xi}^{\text{SV}}$ and $\vec{H}_{\xi}^{\text{tot}} = (j\omega\Delta\vec{\vec{\mu}}_{\text{mat}})^{-1} \cdot \vec{M}_{\xi}^{\text{SV}}$ ; in whole $\mathbb{R}^3$ , the corresponding characteristic scattered fields are $\vec{E}_{\xi}^{\text{sca}} = -j\omega\mu_0\mathcal{L}_0(\vec{J}_{\xi}^{\text{SV}}) - \mathcal{K}_0(\vec{M}_{\xi}^{\text{SV}})$ and $\vec{H}_{\xi}^{\text{sca}} = -j\omega\varepsilon_0\mathcal{L}_0(\vec{M}_{\xi}^{\text{SV}}) + \mathcal{K}_0(\vec{J}_{\xi}^{\text{SV}})$ ; in the interior of $V_{\text{mat sys}}$ , the corresponding characteristic incident fields are $\vec{F}_{\xi}^{\text{inc}} = \vec{F}_{\xi}^{\text{tot}} - \vec{F}_{\xi}^{\text{sca}}$, where $F = E, H$ .

Based on the method similar to the one used in Chapter 3, it is easy to prove that the above CMs satisfy the following orthogonality:

$$P_{\text{mat sys};\xi}^{\text{DRIVING}}\delta_{\xi\zeta} = (1/2)\langle \vec{J}_{\xi}^{\text{SV}}, \vec{E}_{\zeta}^{\text{inc}} \rangle_{V_{\text{mat sys}}} + (1/2)\langle \vec{H}_{\xi}^{\text{inc}}, \vec{M}_{\zeta}^{\text{SV}} \rangle_{V_{\text{mat sys}}} \tag{4-13}$$

where $\delta_{\xi\zeta}$ is Kronecker's delta symbol.

### 2) CMs Derived From DPO Version II $P_{\text{mat sys}}^{\text{driving}}$

Based on operator $P_{\text{mat sys}}^{\text{driving}}$, the characteristic vectors of material systems can be derived from solving the following generalized characteristic equation:

$$\bar{\bar{P}}_{-}^{\text{driving}} \cdot \bar{\alpha}_{\xi}^{\text{driving}} = \lambda_{\text{mat sys};\xi}^{\text{driving}} \bar{\bar{P}}_{+}^{\text{driving}} \cdot \bar{\alpha}_{\xi}^{\text{driving}} \tag{4-14}$$

In the above characteristic equation, $\bar{\bar{P}}_{+}^{\text{driving}} = [\bar{\bar{P}}_{\text{mat sys}}^{\text{driving}} + (\bar{\bar{P}}_{\text{mat sys}}^{\text{driving}})^H]/2$ and $\bar{\bar{P}}_{-}^{\text{driving}} = [\bar{\bar{P}}_{\text{mat sys}}^{\text{driving}} - (\bar{\bar{P}}_{\text{mat sys}}^{\text{driving}})^H]/2j$ ; $(\bar{\alpha}_{\xi}^{\text{driving}})^T = [(\bar{\alpha}_{J;\xi}^{\text{driving}})^T \quad (\bar{\alpha}_{M;\xi}^{\text{driving}})^T]$ ; the corresponding characteristic currents and characteristic fields can be obtained by employing the formulations used in previous 1), and they will not be explicitly given here. Based on the method similar to the one used in Chapter 3, it is easy to prove that the above CMs satisfy the following orthogonality:

$$P_{\text{mat sys};\xi}^{\text{driving}}\delta_{\xi\zeta} = (1/2)\langle \vec{J}_{\xi}^{\text{SV}}, \vec{E}_{\zeta}^{\text{inc}} \rangle_{V_{\text{mat sys}}} + (1/2)\langle \vec{M}_{\xi}^{\text{SV}}, \vec{H}_{\zeta}^{\text{inc}} \rangle_{V_{\text{mat sys}}} \tag{4-15}$$

where we don't explicitly distinguish the kinds of CMs from the aspect of symbolic system (except characteristic values and characteristic vectors) to simplify the symbolic system of this chapter.

### 3) Comparisons and Discussions for the Orthogonality Satisfied by $\bar{\bar{P}}_{\text{mat sys}}^{\text{DRIVING}}$-Based CMs and the Orthogonality Satisfied by $P_{\text{mat sys}}^{\text{driving}}$-Based CMs

By comparing orthogonality (4-13) and orthogonality (4-15), it is easy to find out

---

But, this chapter and the following chapters will specially add some superscripts and subscripts (such as "DRIVING" and "mat sys" etc.) to the characteristic values and characteristic vectors, to effectively distinguish the different kinds of CMs.

that: the CMs derived from the $P_{\text{mat sys}}^{\text{DRIVING}}$-based method are not completely decoupled, but the CMs derived from the $P_{\text{mat sys}}^{\text{driving}}$-based method are completely decoupled —— the frequency-domain power done by modal field $\{\vec{E}_\zeta^{\text{inc}},\vec{H}_\zeta^{\text{inc}}\}$ on modal sources $\{\vec{J}_\xi^{\text{SV}},\vec{M}_\xi^{\text{SV}}\}$ is zero (this implies that in any integral period there doesn't exist net energy exchange between mode $\zeta$ and mode $\xi$). Based on this conclusion, the following parts of this dissertation will always focus on the completely decoupled CMs derived from the $P_{\text{mat sys}}^{\text{driving}}$-based method rather than the CMs derived from the $P_{\text{mat sys}}^{\text{DRIVING}}$-based method, and calls the $P_{\text{mat sys}}^{\text{driving}}$-based CMs as the DP-CMs of material systems, and calls the $P_{\text{mat sys}}^{\text{driving}}$-based method as the volume formulation of the WEP-based CMT for material systems (Vol-WEP-MatSca-CMT)

**4) Comparisons and Discussions for DP-CMs and Harrington's CMs**

In 1972, Prof. Harrington et al.[33] established VIE-MatSca-CMT. Literature [45] pointed out that: the physical destination of VIE-MatSca-CMT is the same as Vol-WEP-MatSca-CMT. Based on this, Vol-WEP-MatSca-CMT can be viewed as the generalization for VIE-MatSca-CMT in the aspect of the material parameters of material systems —— from isotropic matter to anisotropic matter. However, Vol-WEP-MatSca-CMT is more than generalization, because Vol-WEP-MatSca-CMT is also the transformation for the carrying framework of VIE-MatSca-CMT —— from IE framework to WEP framework. Just based on the framework transformation, the following Sections 4.3~4.7 will derive a series of completely new CM calculation formulations, and the new formulations have advantages in many aspects.

In addition, just due to the inappropriate original carrying framework, VIE-MatSca-CMT has had some imperfections since it was established in 1972 (for details see literatures [44,57]). Similarly to the discussions for EFIE-MetSca-CMT[31,32] in Subsection 3.4.1, we can obtain the conclusions related to Harrington's VIE-MatSca-CMT that: by decomposing VIE-based IMO $\overline{\overline{Z}}^{\text{VIE}}$ into real part $\overline{\overline{R}}^{\text{VIE}}=\text{Re}\{\overline{\overline{Z}}^{\text{VIE}}\}$ and imaginary part $\overline{\overline{X}}^{\text{VIE}}=\text{Im}\{\overline{\overline{Z}}^{\text{VIE}}\}$, it cannot be guaranteed that $\overline{\overline{R}}^{\text{VIE}}$ and $\overline{\overline{X}}^{\text{VIE}}$ are Hermitian[44]; due to the existence of the polarization electric charges and magnetization magnetic charges① distributing on the boundaries of material systems, it cannot be guaranteed that $\overline{\overline{R}}^{\text{VIE}}$ and $\overline{\overline{X}}^{\text{VIE}}$ are symmetrical[44,57]; $\bar{a}^H\cdot\overline{\overline{R}}^{\text{VIE}}\cdot\bar{a}$ and

① The magnetization magnetic charges mentioned here originate from the cumulation of the magnetic dipoles on the material boundary, and they don't correspond the magnetic monopoles, as stated by Prof. Lan Jen Chu et al. in classical literature [110] that: "… we know that surface distributions of magnetic charges as well as volume distributions can arise from magnetized matter.". The detailed discussions on this topic can be found in the Chapters 5 and 7 of classical literature [110], and they will not be repeated here.

$\bar{a}^H \cdot \bar{\bar{X}}^{\text{VIE}} \cdot \bar{a}$ don't correspond to $\text{Re}\{P^{\text{driving}}_{\text{mat sys}}\}$ and $\text{Im}\{P^{\text{driving}}_{\text{mat sys}}\}$ respectively[44].

### 4.2.3 DP-CM-Based Modal Expansion

Based on the completeness of driving power CMs (DP-CMs), we can expand any working mode $\vec{F}^{\text{inc}}$ of $V_{\text{mat sys}}$ as follows:

$$\vec{E}^{\text{inc}}(\vec{r}) = \sum\nolimits_{\zeta=1}^{\Xi} c_\zeta \vec{E}_\zeta^{\text{inc}}(\vec{r}) \quad , \quad \vec{r} \in V_{\text{mat sys}} \tag{4-16a}$$

$$\vec{H}^{\text{inc}}(\vec{r}) = \sum\nolimits_{\zeta=1}^{\Xi} c_\zeta \vec{H}_\zeta^{\text{inc}}(\vec{r}) \quad , \quad \vec{r} \in V_{\text{mat sys}} \tag{4-16b}$$

where $\Xi = \Xi^J + \Xi^M$. If equations (4-16a) and (4-16b) are tested by $\{\vec{J}_\xi^{\text{SV}}/2\}_{\xi=1}^{\Xi}$ and $\{\vec{M}_\xi^{\text{SV}}/2\}_{\xi=1}^{\Xi}$ respectively, then we obtain the simultaneous equations of expansion coefficients $\{c_\zeta\}_{\xi=1}^{\Xi}$ as follows:

$$(1/2)\left\langle \vec{J}_\xi^{\text{SV}}, \vec{E}^{\text{inc}} \right\rangle_{V_{\text{mat sys}}} = \sum\nolimits_{\zeta=1}^{\Xi} c_\zeta (1/2)\left\langle \vec{J}_\xi^{\text{SV}}, \vec{E}_\zeta^{\text{inc}} \right\rangle_{V_{\text{mat sys}}} \quad , \quad \xi = 1,2,\cdots,\Xi \tag{4-17a}$$

$$(1/2)\left\langle \vec{M}_\xi^{\text{SV}}, \vec{H}^{\text{inc}} \right\rangle_{V_{\text{mat sys}}} = \sum\nolimits_{\zeta=1}^{\Xi} c_\zeta (1/2)\left\langle \vec{M}_\xi^{\text{SV}}, \vec{H}_\zeta^{\text{inc}} \right\rangle_{V_{\text{mat sys}}} , \quad \xi = 1,2,\cdots,\Xi \tag{4-17b}$$

Solving the simultaneous equations and employing orthogonality (4-15), it is easy to obtain the explicit expression of the expansion coefficients as follows:

$$c_\xi = \frac{1}{P^{\text{driving}}_{\text{mat sys};\xi}} \left[ (1/2)\left\langle \vec{J}_\xi^{\text{SV}}, \vec{E}^{\text{inc}} \right\rangle_{V_{\text{mat sys}}} + (1/2)\left\langle \vec{M}_\xi^{\text{SV}}, \vec{H}^{\text{inc}} \right\rangle_{V_{\text{mat sys}}} \right] \quad , \quad \xi = 1,2,\cdots,\Xi \text{ ① } \tag{4-18}$$

where incident fields $\{\vec{E}^{\text{inc}}, \vec{H}^{\text{inc}}\}$ are known.

### 4.2.4 Numerical Examples Corresponding to Typical Structures

In this section, we provide some typical numerical examples to verify the validities of the above conclusions. All material systems considered in this section have the same topological structure as the one shown in Figure 4-2 —— a material cylinder whose radius and height are 5.25mm and 4.60mm respectively.

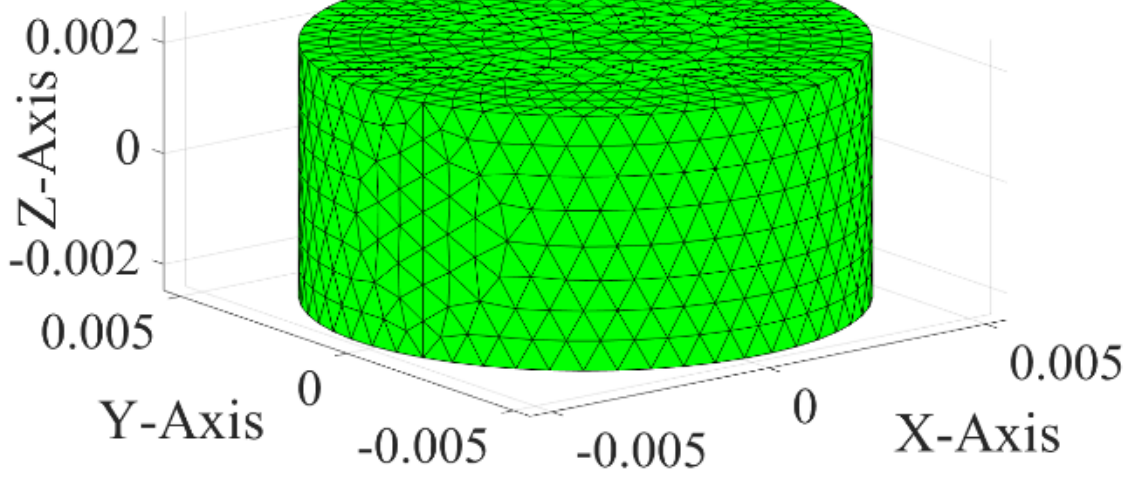

Figure 4-2 The topological structure and volume tetrahedron meshes of a material cylinder whose radius and height are 5.25mm and 4.60mm respectively

① The above expansion coefficient formulation has the same manifestation form as the one given in literature [33].

The differences among the various material cylinders considered in this section are reflected in their material parameters. In following 1), the material cylinder is lossless and homogeneous isotropic; in following 2), the material cylinder considered is lossless and homogeneous and electricaly uniaxial; in following 3), the material cylinder considered is lossless and homogeneous and electricaly biaxial; in following 4), the material cylinder considered is lossless and homogeneous and magnetically uniaxial; in following 5), the material cylinder considered is lossless and homogeneous and magnetically biaxial.

**1) DP-CMs of a Lossless Material Cylinder Whose Relative Permeability Is $\overleftrightarrow{\mu}^{\rm r}_{\rm mat} = \overleftrightarrow{I}2$ and Relative Permittivity Is $\overleftrightarrow{\varepsilon}^{\rm r}_{\rm mat} = \overleftrightarrow{I}18$**

If the material cylinder shown in Figure 4-2 is with the material parameters $\overleftrightarrow{\mu}^{\rm r}_{\rm mat} = \overleftrightarrow{I}2$ and $\overleftrightarrow{\varepsilon}^{\rm r}_{\rm mat} = \overleftrightarrow{I}18$, then some characteristic quantity curves corresponding to 6 typical DP-CMs derived from characteristic equation (4-14) are illustrated in Figure 4-3.

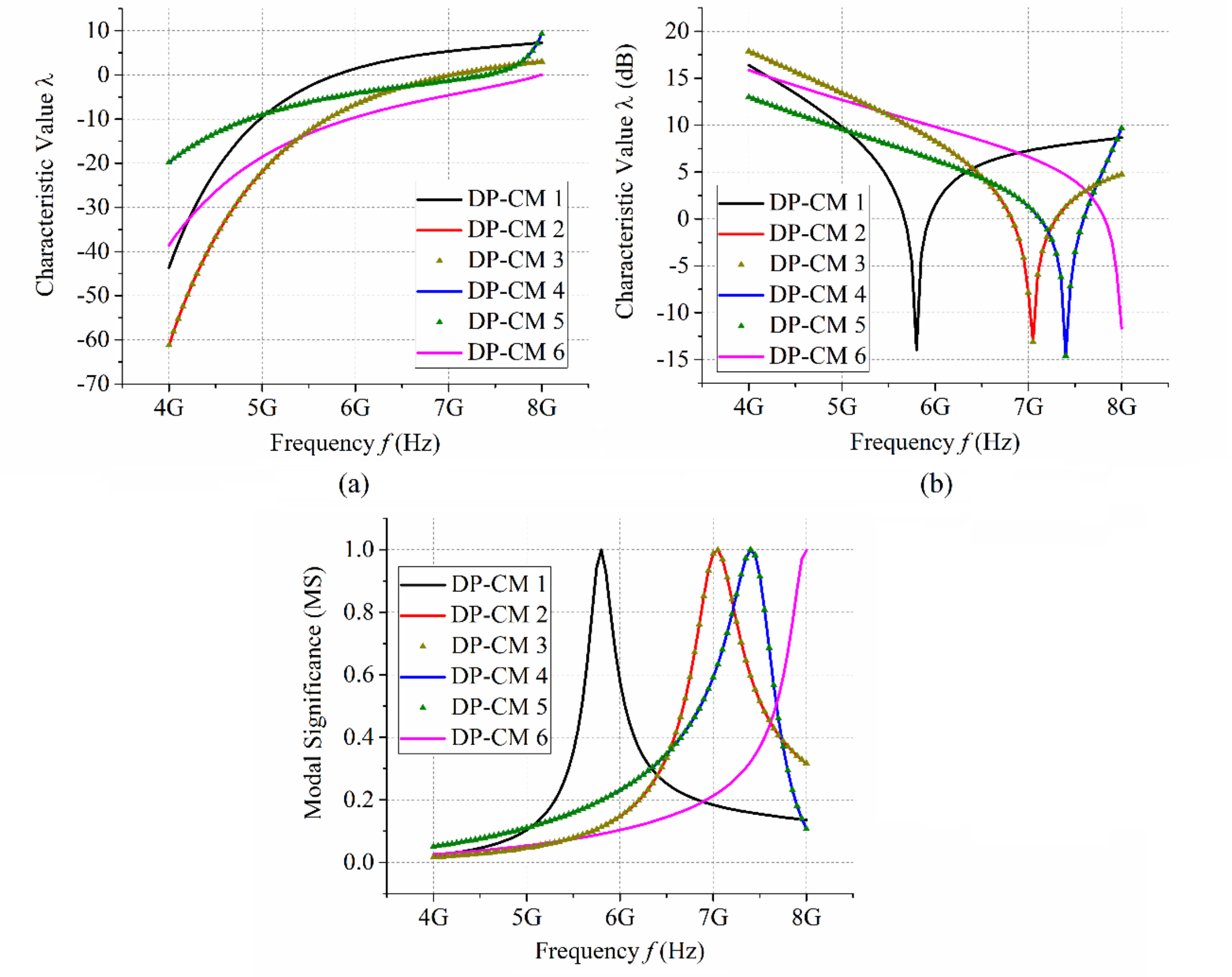

(c)

Figure 4-3 The characteristic quantity curves corresponding to the first 6 typical DP-CMs of the lossless material cylinder whose topological structure is shown in Figure 4-2 and relative permeability and relative permittivity are $\overleftrightarrow{\mu}^{\rm r}_{\rm mat} = \overleftrightarrow{I}2$ and $\overleftrightarrow{\varepsilon}^{\rm r}_{\rm mat} = \overleftrightarrow{I}18$ respectively. (a) characteristic valuve curves; (b) characteristic value dB curves; (c) MS curves

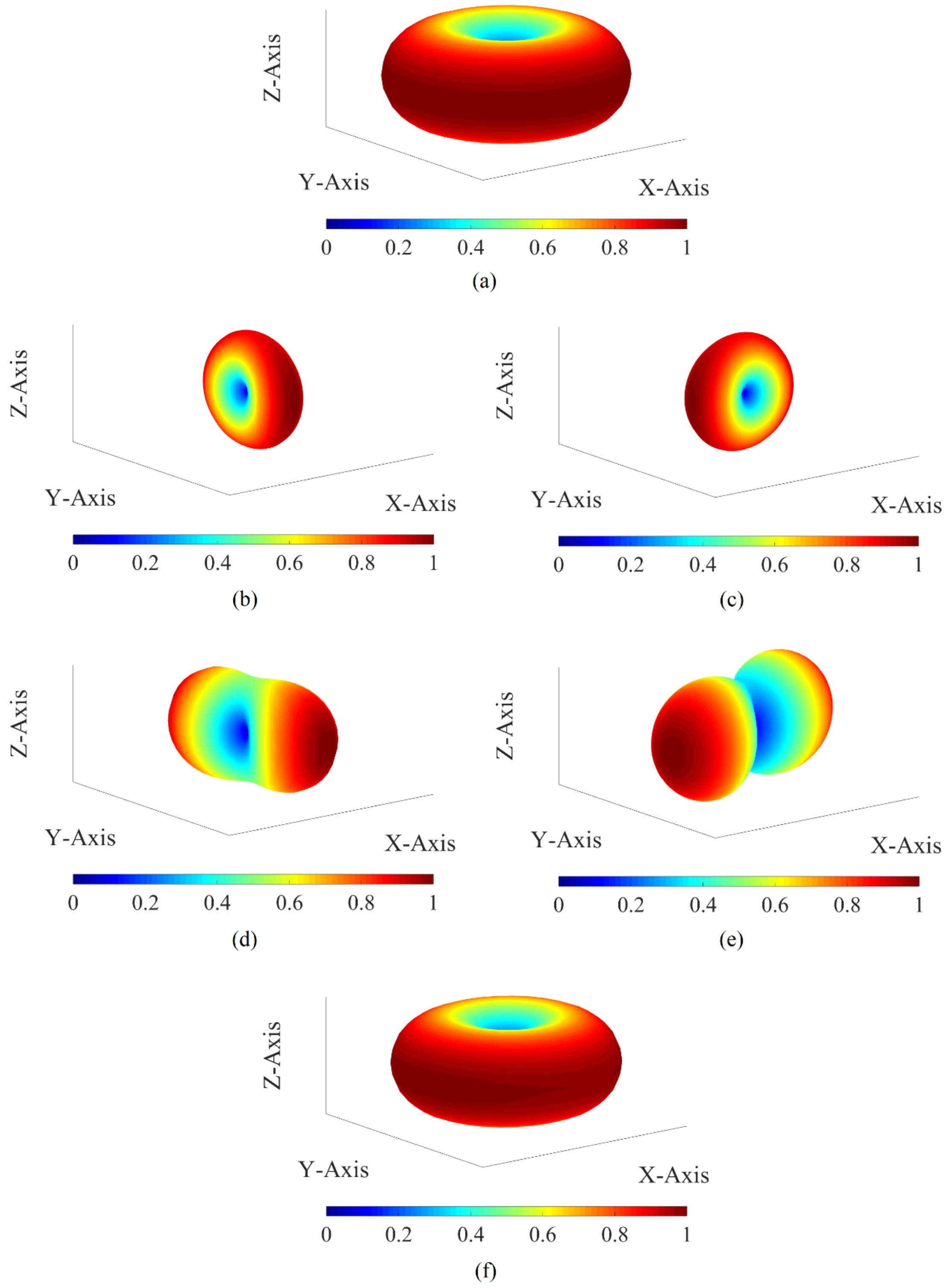

Figure 4-4 The radiation patterns corresponding to the typical DP-CMs shown in Figure 4-3. (a) the radiation pattern of the “resonant” DP-CM1 at 5.80GHz; (b) the radiation pattern of the “resonant” DP-CM2 at 7.05GHz; (c) the radiation pattern of the “resonant” DP-CM3 at 7.05GHz; (d) the radiation pattern of the “resonant” DP-CM4 at 7.40GHz; (e) the radiation pattern of the “resonant” DP-CM5 at 7.40GHz; (f) the radiation pattern of the “resonant” DP-CM6 at 8.00GHz

From above Figure 4-3, it is easy to find out that: DP-CM1 is “resonant” at 5.80GHz, and DP-CM2 and DP-CM3 are simultaneously “resonant” at 7.05GHz, and DP-CM4 and DP-CM5 are simultaneously “resonant” at 7.40GHz, and DP-CM6 is “resonant” at 8.00GHz. In addition, DP-CM2 and DP-CM3 have the same characteristic value at whole frequency band 4~8GHz, and DP-CM4 and DP-CM5 also have the same characteristic value at whole frequency band 4~8GHz. We show the radiation patterns of the “resonant” DP-CMs in Figure 4-4. Based on Figures 4-3 and 4-4, it is easy to find out that: DP-CM2 and DP-CM3 constitute a pair of degenerate modes, and DP-CM4 and DP-CM5 constitute another pair of degenerate modes. The reason leading to the degenerate modes is that: the material cylinder shown in Figure 4-2 is rotationally symmetric on z-axial.

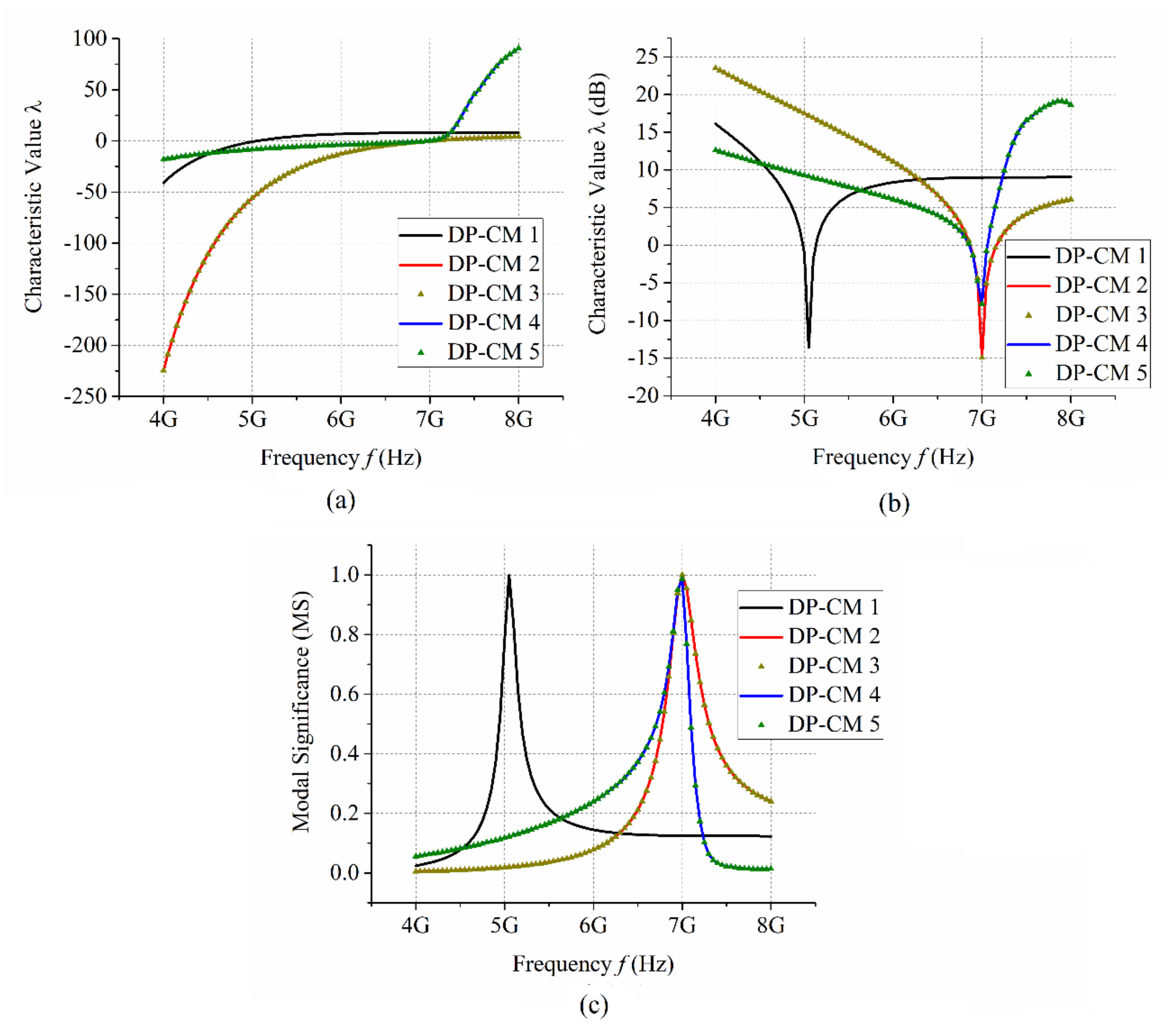

Figure 4-5 The characteristic quantity curves corresponding to the first 5 typical DP-CMs of the lossless electrically uniaxial cylinder whose topological structure is shown in Figure 4-2 and relative permeability and relative permittivity are $\ddot{\mu}^{\rm r}_{\rm mat} = \ddot{I}1$ and $\ddot{\varepsilon}^{\rm r}_{\rm mat} = \hat{x}\hat{x}36 + \hat{y}\hat{y}36 + \hat{z}\hat{z}26$ respectively. (a) characteristic valuve curves; (b) characteristic value dB curves; (c) MS curves

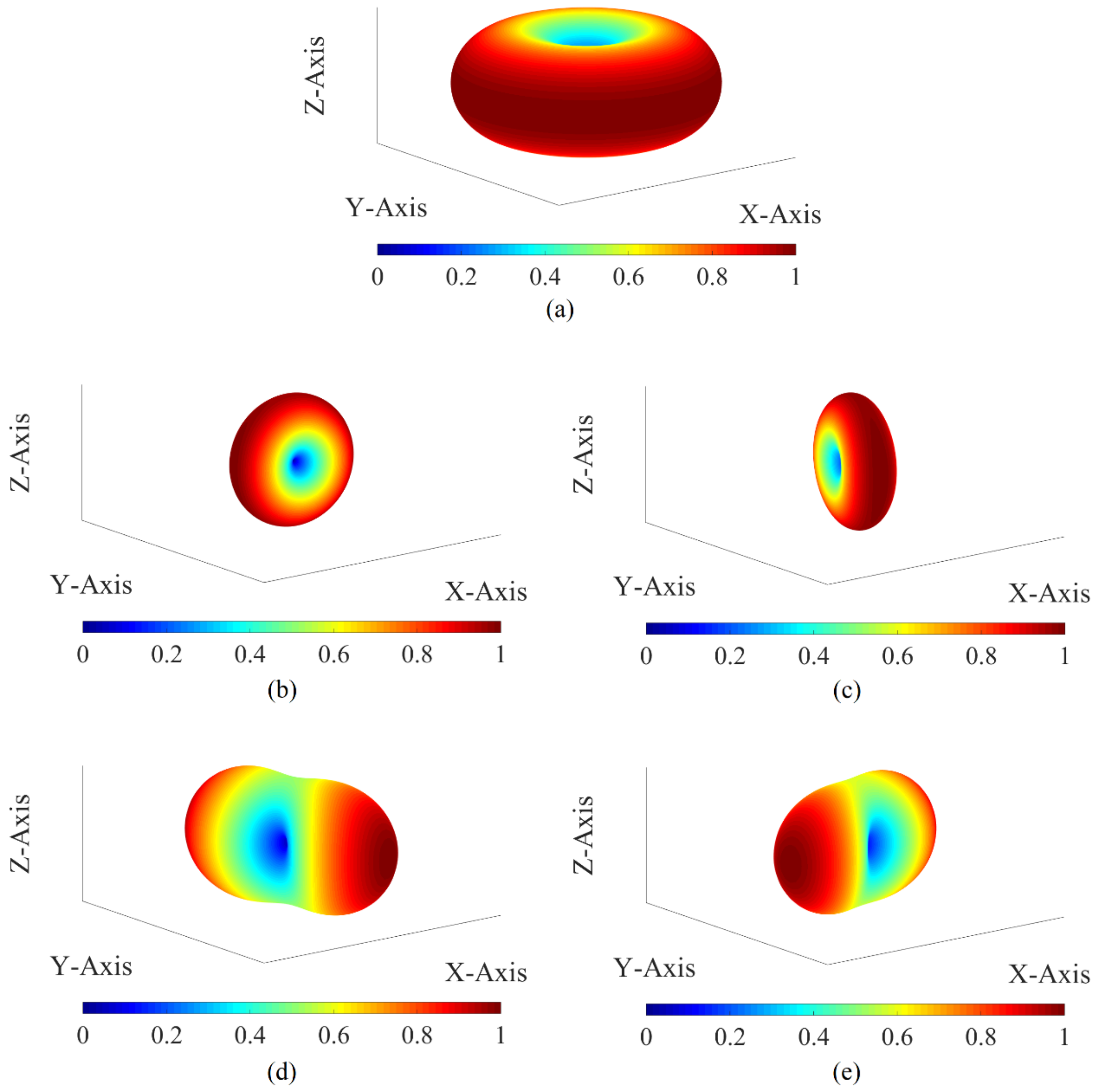

Figure 4-6 The radiation patterns corresponding to the typical DP-CMs shown in Figure 4-5. (a) the radiation pattern of the resonant DP-CM1 at 5.05GHz; (b) the radiation pattern of the resonant DP-CM2 at 6.09GHz; (c) the radiation pattern of the resonant DP-CM3 at 6.09GHz; (d) the radiation pattern of the resonant DP-CM4 at 6.09GHz; (e) the radiation pattern of the resonant DP-CM5 at 6.09GHz

**2) DP-CMs of a Lossless Electrically Uniaxial Cylinder Whose Relative Permeability Is $\overleftrightarrow{\mu}^{r}_{mat} = \overleftrightarrow{I}1$ and Relative Permittivity Is $\overleftrightarrow{\varepsilon}^{r}_{mat} = \hat{x}\hat{x}36 + \hat{y}\hat{y}36 + \hat{z}\hat{z}26$**

If the material cylinder shown in Figure 4-2 is with material parameters $\overleftrightarrow{\mu}^{r}_{mat} = \overleftrightarrow{I}1$ and $\overleftrightarrow{\varepsilon}^{r}_{mat} = \hat{x}\hat{x}36 + \hat{y}\hat{y}36 + \hat{z}\hat{z}26$ (electrically uniaxial), then some characteristic quantity curves corresponding to 5 typical DP-CMs derived from characteristic equation (4-14) are illustrated in Figure 4-5.

From above Figure 4-5, it is easy to find out that: DP-CM1 is resonant at 5.05GHz, and DP-CM2 and DP-CM3 are simultaneously resonant at 6.09GHz, and DP-CM4 and DP-CM5 are simultaneously resonant at 6.09GHz. In addition, DP-CM2 and DP-CM3 have the same characteristic value (i.e., they are degenerate) at whole frequency band 4~8GHz, and DP-CM4 and DP-CM5 also have the same characteristic value (i.e., they are degenerate) at whole frequency band 4~8GHz. We show the radiation patterns of the resonant DP-CMs in Figure 4-6.

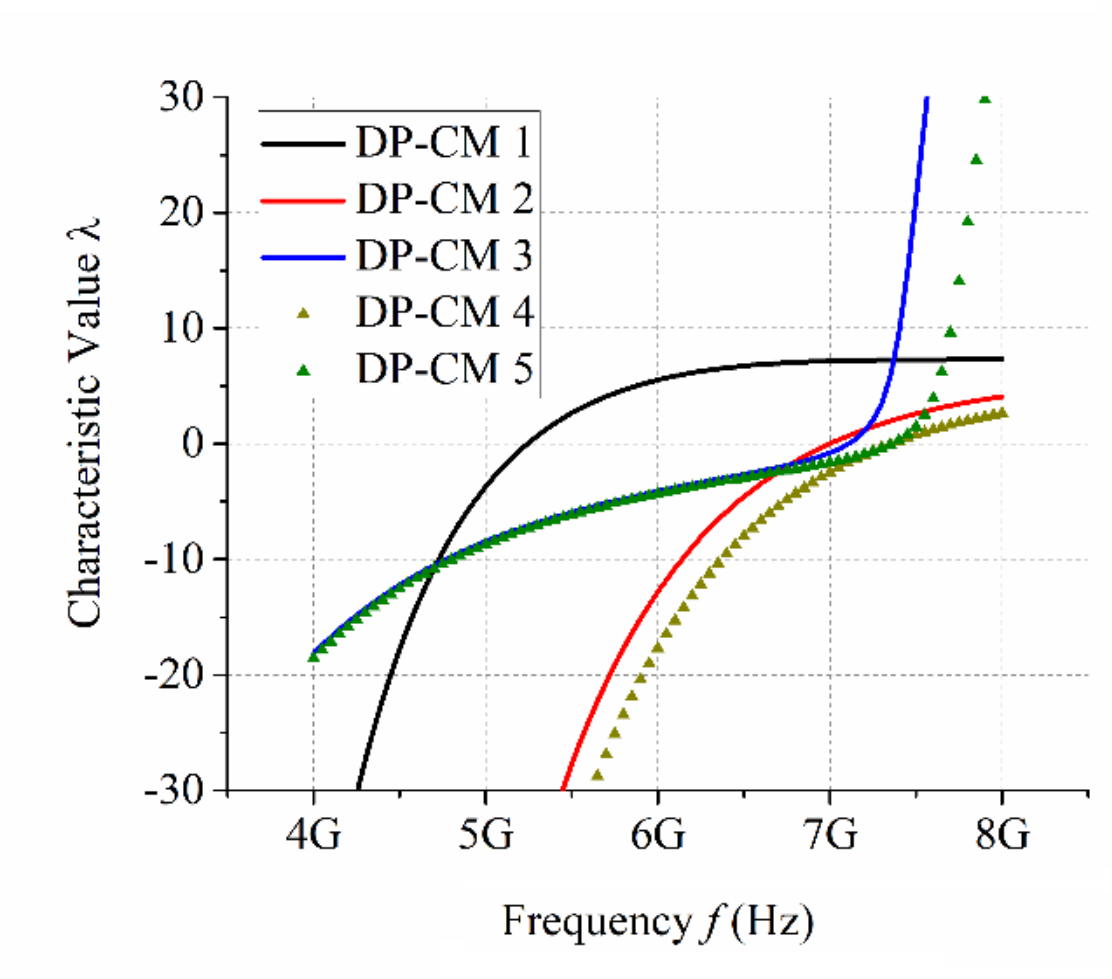

(a)

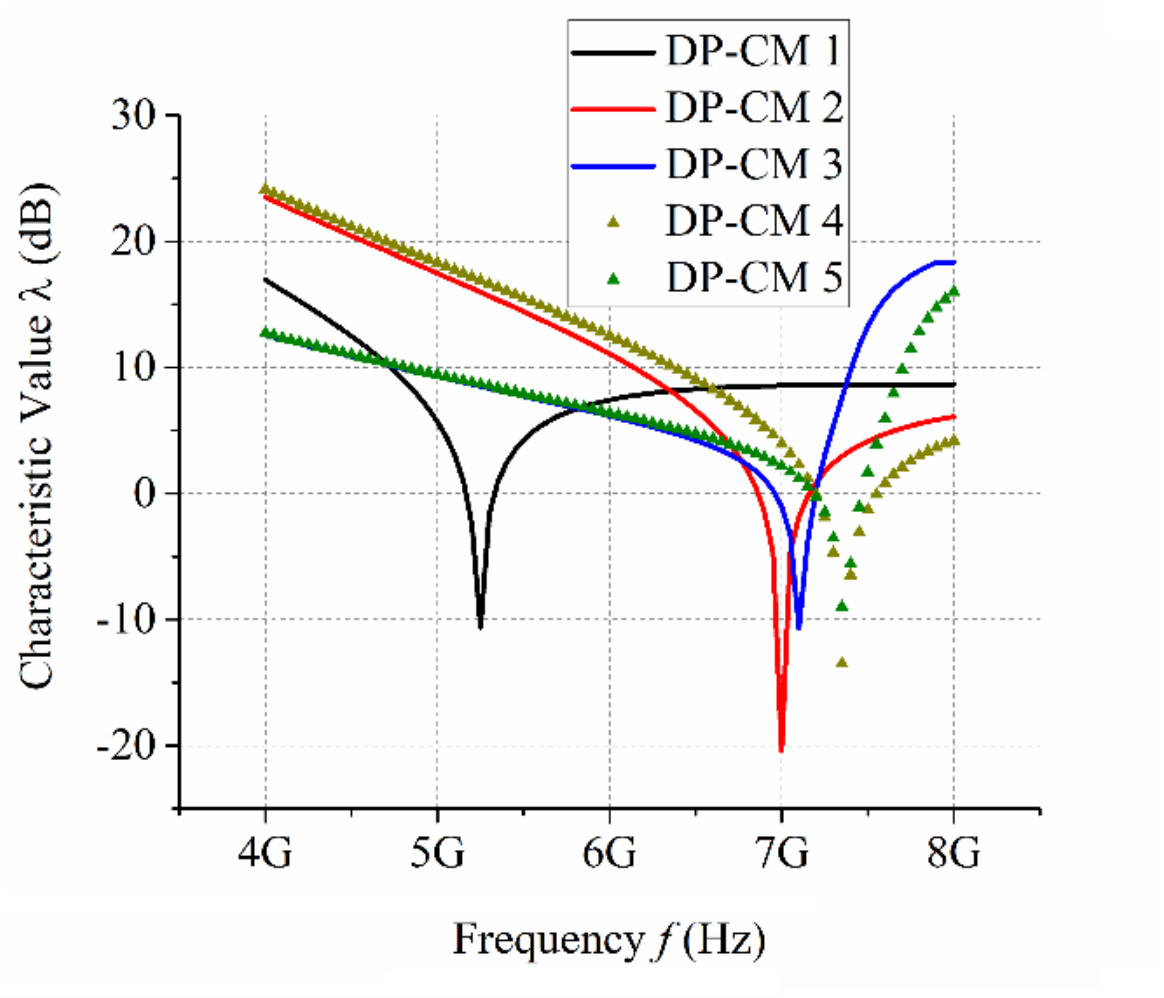

(b)

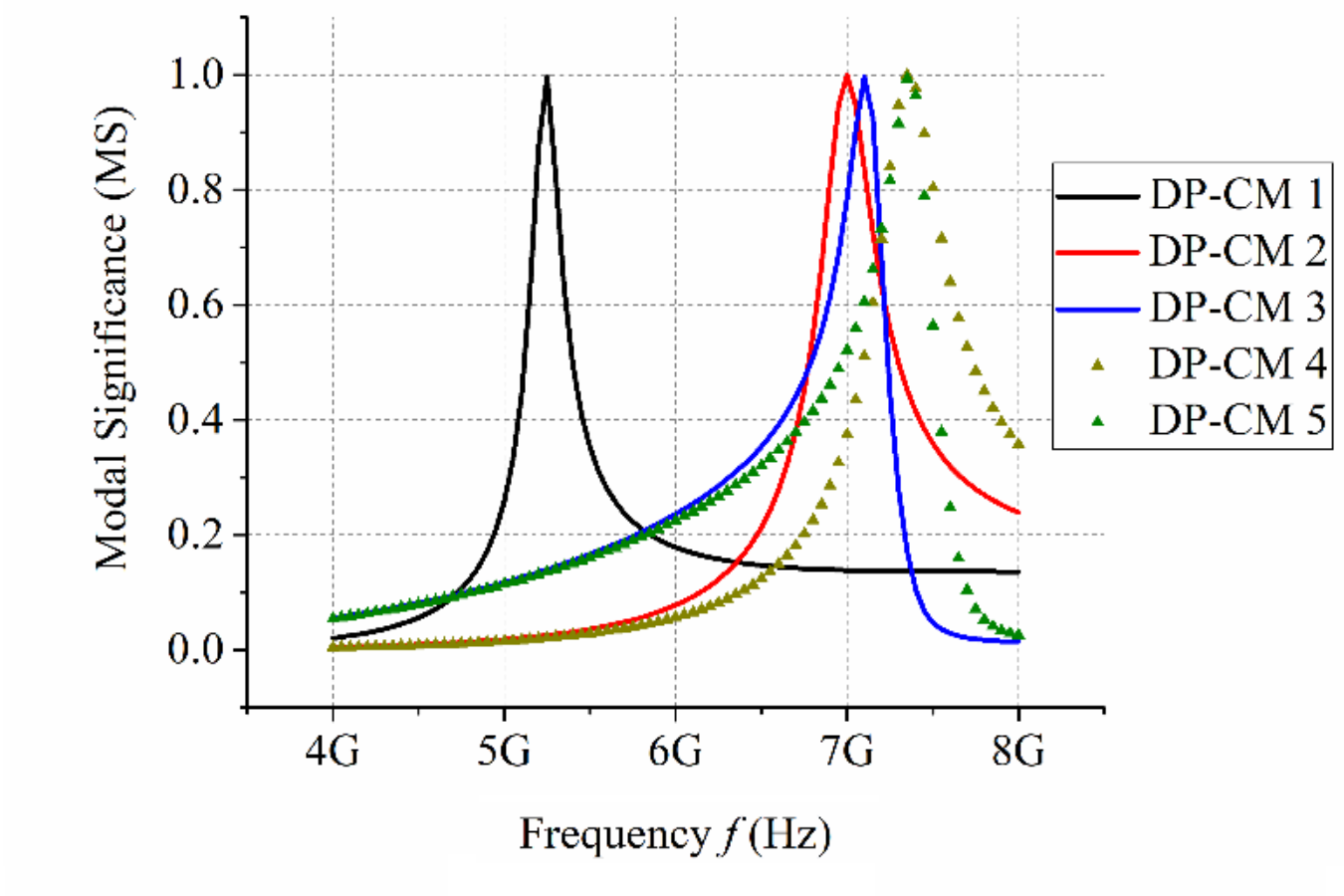

(c)

Figure 4-7 The characteristic quantity curves corresponding to the first 5 typical DP-CMs of the lossless electrically biaxial cylinder whose topological structure is shown in Figure 4-2 and relative permeability and relative permittivity are $\vec{\vec{\mu}}^{\rm r}_{\rm mat} = \vec{\vec{I}}1$ and $\vec{\vec{\varepsilon}}^{\rm r}_{\rm mat} = \hat{x}\hat{x}36 + \hat{y}\hat{y}31 + \hat{z}\hat{z}26$ respectively. (a) characteristic valuve curves; (b) characteristic value dB curves; (c) MS curves

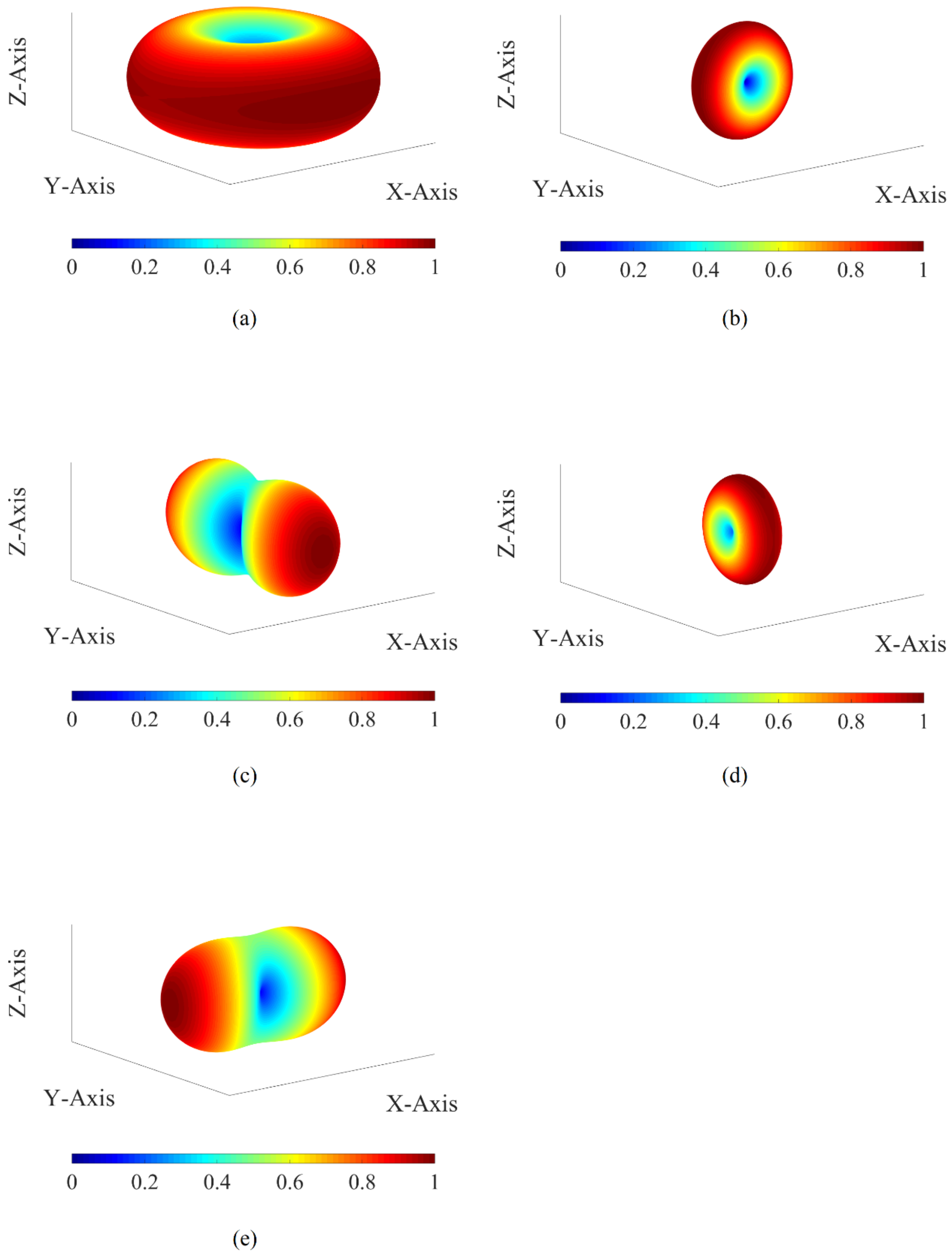

Figure 4-8 The radiation patterns corresponding to the typical DP-CMs shown in Figure 4-7. (a) the radiation pattern of the resonant DP-CM1 at 5.25GHz; (b) the radiation pattern of the resonant DP-CM2 at 7.00GHz; (c) the radiation pattern of the resonant DP-CM3 at 7.00GHz; (d) the radiation pattern of the resonant DP-CM4 at 7.35GHz; (e) the radiation pattern of the resonant DP-CM5 at 7.35GHz

**3) DP-CMs of a Lossless Electrically Biaxial Cylinder Whose Relative Permeability Is $\overleftrightarrow{\mu}_{\text{mat}}^{\text{r}} = \overleftrightarrow{I}1$ and Relative Permittivity Is $\overleftrightarrow{\varepsilon}_{\text{mat}}^{\text{r}} = \hat{x}\hat{x}36 + \hat{y}\hat{y}31 + \hat{z}\hat{z}26$**

If the material cylinder shown in Figure 4-2 is with material parameters $\overleftrightarrow{\mu}_{\text{mat}}^{\text{r}} = \overleftrightarrow{I}1$ and $\overleftrightarrow{\varepsilon}_{\text{mat}}^{\text{r}} = \hat{x}\hat{x}36 + \hat{y}\hat{y}31 + \hat{z}\hat{z}26$ (electrically biaxial), then some characteristic quantity curves corresponding to 5 typical DP-CMs derived from characteristic equation (4-14) are shown in Figure 4-7. Comparing Figure 4-5 with Figure 4-7, it is easy to find out that: original degenerate modes (the “DP-CM2 & DP-CM3” and “DP-CM4 & DP-CM5” shown in Figure 4-5) transform to some nondegenerate modes (the “DP-CM2 & DP-CM4” and “DP-CM3 & DP-CM5” shown in Figure 4-7), because the material parameters of the electrically biaxial material cylinder are not rotationally symmetrical on z-axial.

From above Figure 4-7, it is easy to find out that: DP-CM1 is resonant at 5.25GHz, and DP-CM2 is resonant at 7.00GHz, and DP-CM3 is resonant at 7.10GHz, and DP-CM4 and DP-CM5 are simultaneously resonant at 7.35GHz. We show the radiation patterns of the resonant DP-CMs in Figure 4-8. It is obvious to conclude from the radiation patterns that: DP-CM4 and DP-CM5 are not degenerate, though they are simultaneously resonant at 7.35GHz. By comparing Figures 4-6 and 4-8, the previous conclusion obtained from comparing Figures 4-5 and 4-7 can be further verified: because the permittivity tensor of the electrically biaxial material cylinder is not symmetrical on z-axial, then the original degenerate modes (the CM2&CM3 and CM4&CM5 shown in Figure 4-5) transform to some nondegenerate modes (the CM2&CM4 and CM3&CM5 shown in Figure 4-7).

**4) DP-CMs of a Lossless Magnetically Uniaxial Cylinder Whose Relative Permeability Is $\overleftrightarrow{\mu}_{\text{mat}}^{\text{r}} = \hat{x}\hat{x}36 + \hat{y}\hat{y}36 + \hat{z}\hat{z}26$ and Relative Permittivity Is $\overleftrightarrow{\varepsilon}_{\text{mat}}^{\text{r}} = \overleftrightarrow{I}1$**

If the material cylinder shown in Figure 4-2 is with material parameters $\overleftrightarrow{\mu}_{\text{mat}}^{\text{r}} = \hat{x}\hat{x}36 + \hat{y}\hat{y}36 + \hat{z}\hat{z}26$ and $\overleftrightarrow{\varepsilon}_{\text{mat}}^{\text{r}} = \overleftrightarrow{I}1$ (magnetically uniaxial), then some characteristic quantity curves corresponding to 5 typical DP-CMs derived from characteristic equation (4-14) are shown in Figure 4-9.

From above Figure 4-9, it is easy to find out that: DP-CM1 is resonant at 5.05GHz, and DP-CM2 and DP-CM3 are simultaneously resonant at 7.00GHz, and DP-CM4 and DP-CM5 are simultaneously resonant at 7.00GHz. In addition, DP-CM2 and DP-CM3 have the same characteristic value (i.e., they are degenerate) at whole frequency band 4~8GHz, and DP-CM4 and DP-CM5 also have the same characteristic value (i.e., they are degenerate) at whole frequency band 4~8GHz. We show the radiation patterns of the resonant DP-CMs in Figure 4-10.

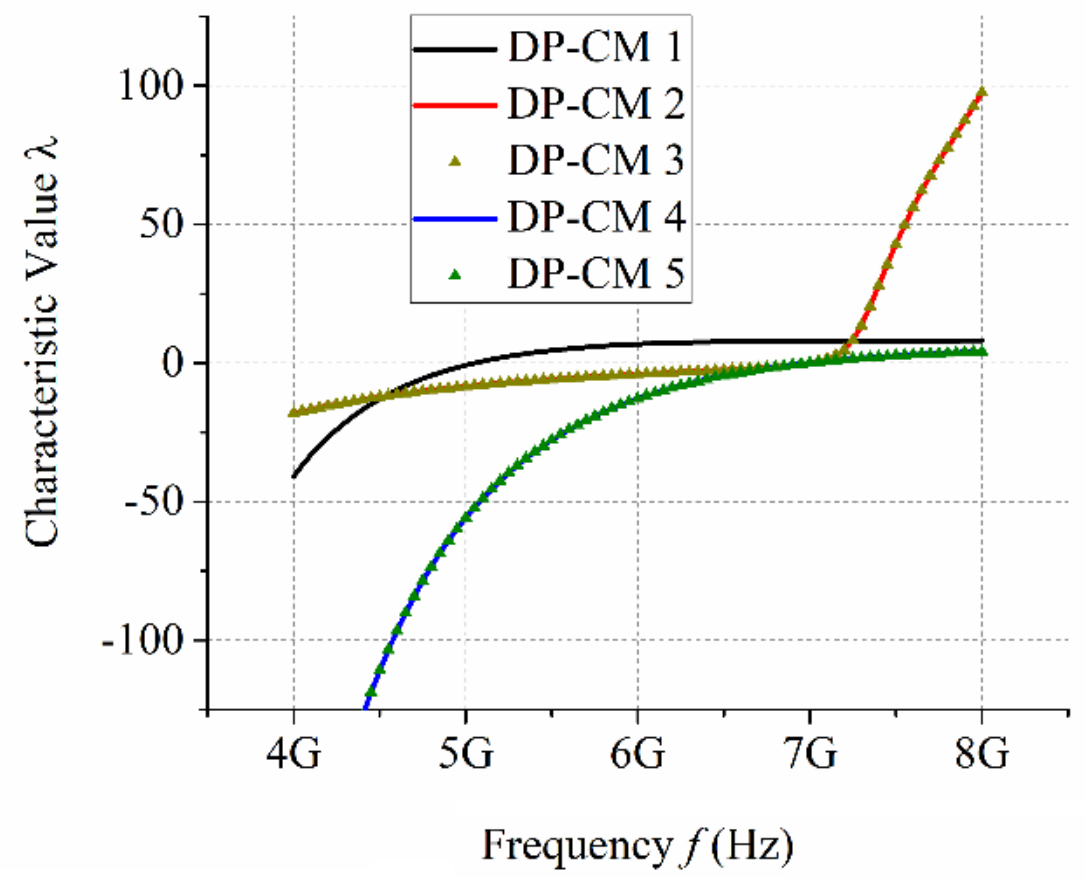

(a)

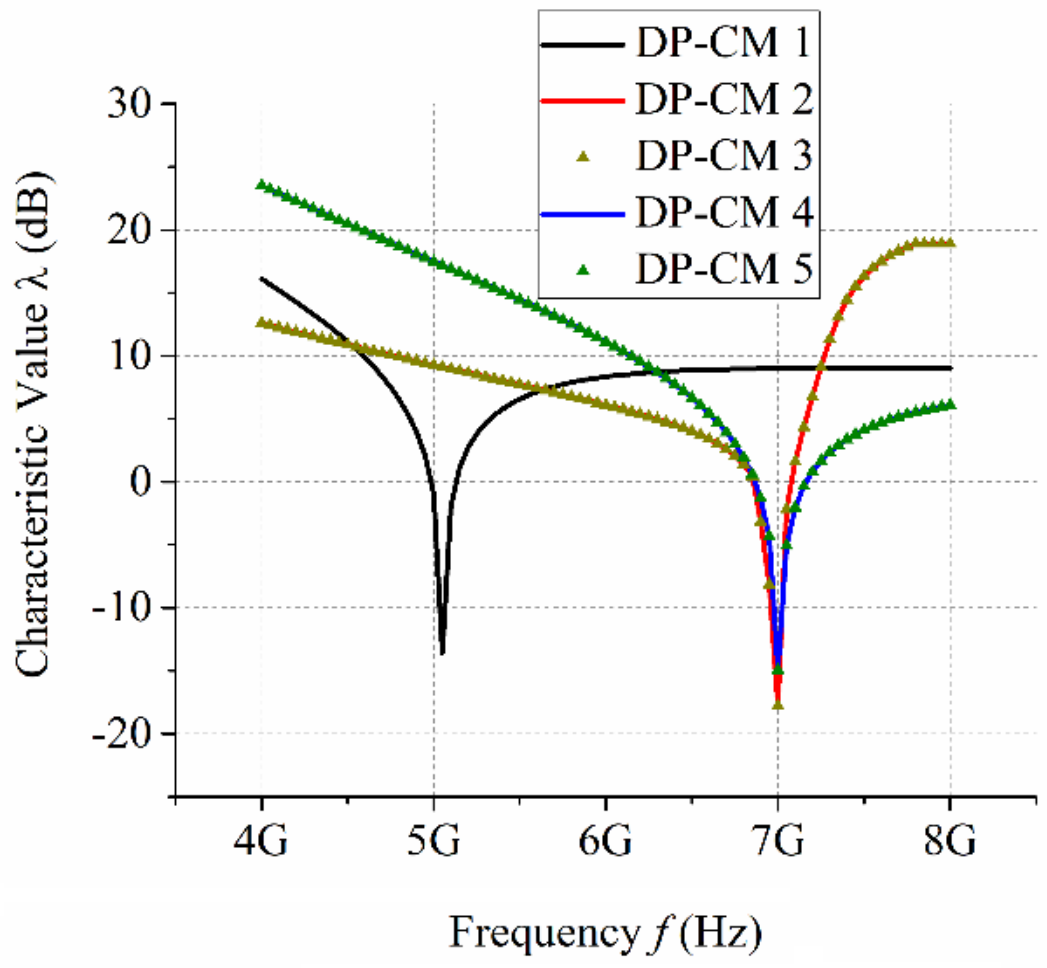

(b)

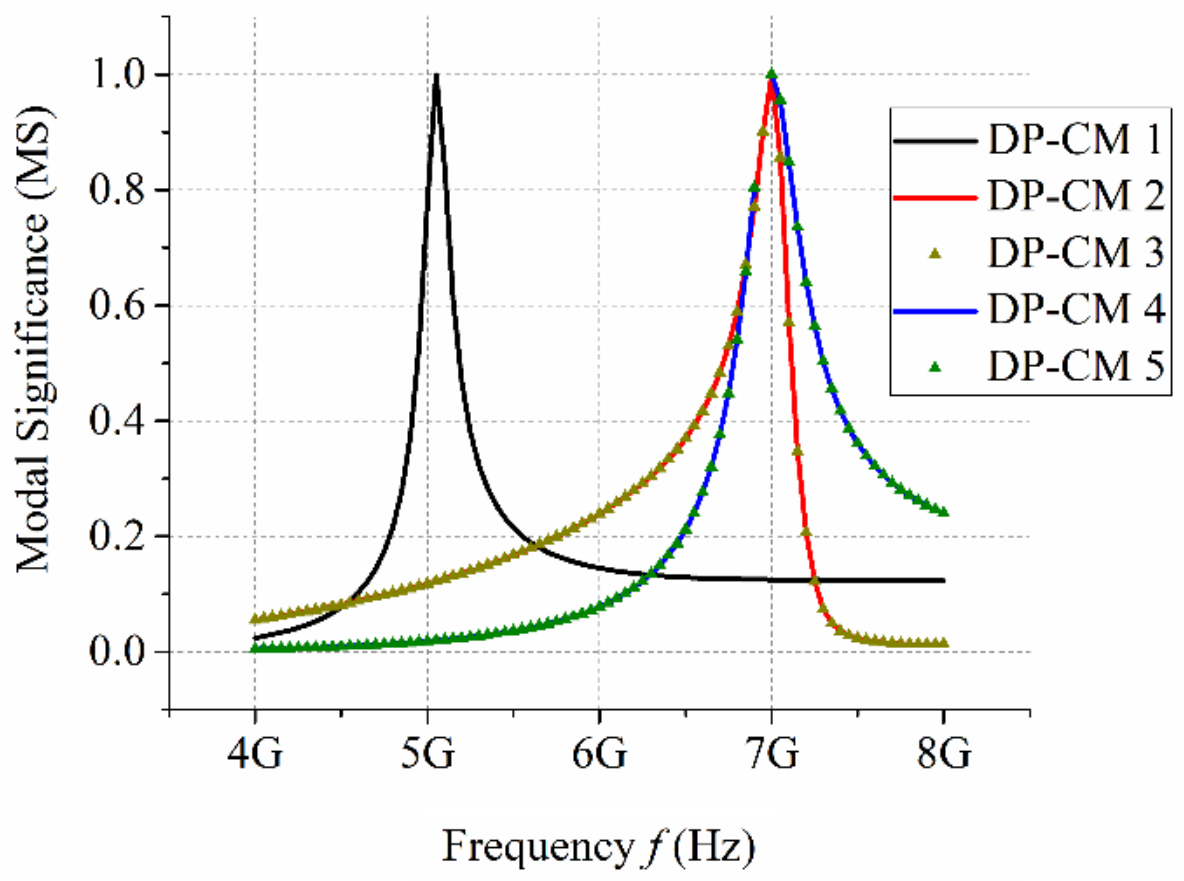

(c)

Figure 4-9 The characteristic quantity curves corresponding to the first 5 typical DP-CMs of the lossless magnetically uniaxial cylinder whose topological structure is shown in Figure 4-2 and relative permeability and relative permittivity are $\overleftrightarrow{\mu}_{\text{mat}}^{\text{r}} = \hat{x}\hat{x}36 + \hat{y}\hat{y}36 + \hat{z}\hat{z}26$ and $\overleftrightarrow{\varepsilon}_{\text{mat}}^{\text{r}} = \overleftrightarrow{I}1$ respectively. (a) characteristic valuve curves; (b) characteristic value dB curves; (c) MS curves

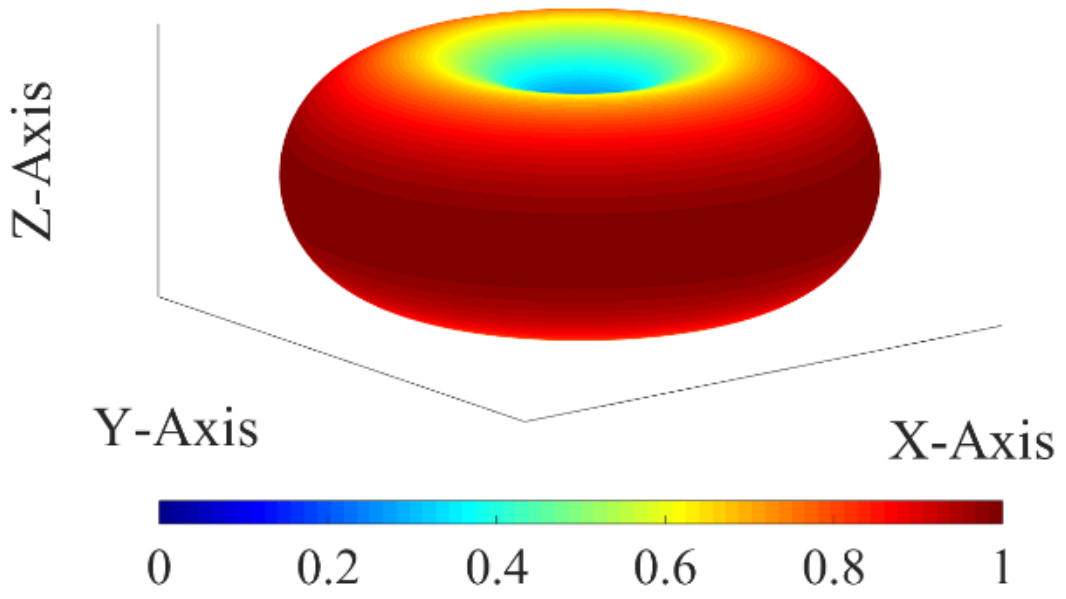

(a)

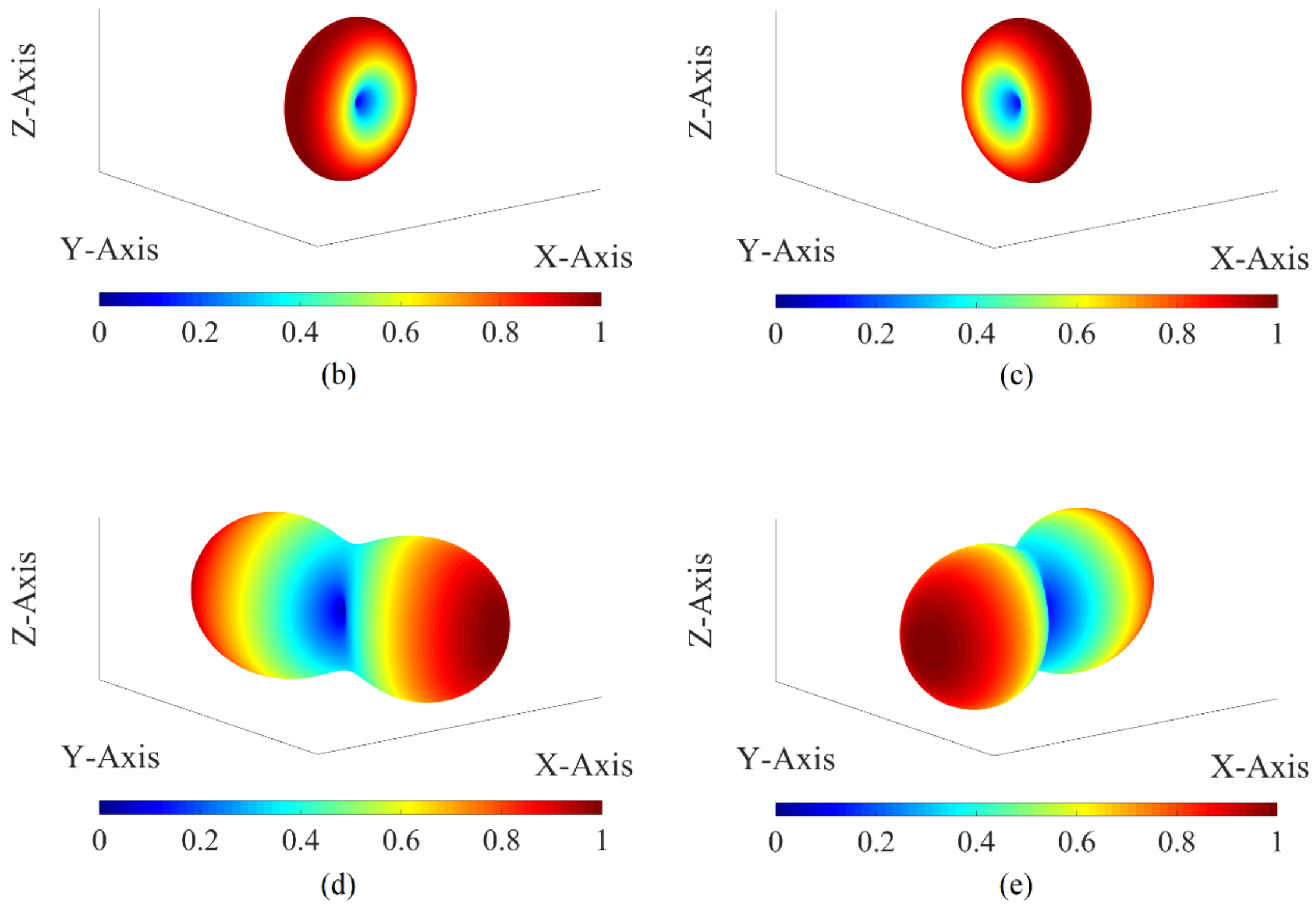

Figure 4-10 The radiation patterns corresponding to the typical DP-CMs shown in Figure 4-9. (a) the radiation pattern of the "resonant" DP-CM1 at 5.05GHz; (b) the radiation pattern of the "resonant" DP-CM2 at 7.00GHz; (c) the radiation pattern of the "resonant" DP-CM3 at 7.00GHz; (d) the radiation pattern of the "resonant" DP-CM4 at 7.00GHz; (e) the radiation pattern of the "resonant" DP-CM5 at 7.00GHz

**5) DP-CMs of a Lossless Magnetically Biaxial Cylinder Whose Relative Permeability Is $\overleftrightarrow{\mu}^{r}_{mat} = \hat{x}\hat{x}36 + \hat{y}\hat{y}31 + \hat{z}\hat{z}26$ and Relative Permittivity Is $\overleftrightarrow{\varepsilon}^{r}_{mat} = \overleftrightarrow{I}1$**

If the material cylinder shown in Figure 4-2 is with material parameters $\overleftrightarrow{\mu}^{r}_{mat} = \hat{x}\hat{x}36 + \hat{y}\hat{y}31 + \hat{z}\hat{z}26$ and $\overleftrightarrow{\varepsilon}^{r}_{mat} = \overleftrightarrow{I}1$ (magnetically biaxial), then some characteristic quantity curves corresponding to 5 typical DP-CMs derived from characteristic equation (4-14) are shown in Figure 4-11.

From above Figure 4-11, it is easy to find out that: DP-CM1 is resonant at 5.25GHz, and DP-CM2 is resonant at 7.00GHz, and DP-CM3 is resonant at 7.10GHz, and DP-CM4 is resonant at 7.35GHz, and DP-CM5 is also resonant at 7.35GHz. We show the radiation patterns of the resonant DP-CMs in Figure 4-12. It is obvious to conclude from the radiation patterns that: DP-CM4 and DP-CM5 are not degenerate, though they are simultaneously resonant at 7.35GHz. By comparing Figures 4-10 and 4-12, the previous conclusion obtained from comparing Figures 4-9 and 4-11 can be further verified: because the permeability tensor of the magnetically biaxial material cylinder is not symmetrical on z-axial, then the original degenerate modes (the CM2&CM3 and CM4&CM5 shown in Figure 4-9) transform to some nondegenerate modes (the CM2&CM4 and CM3&CM5 shown in Figure 4-11).

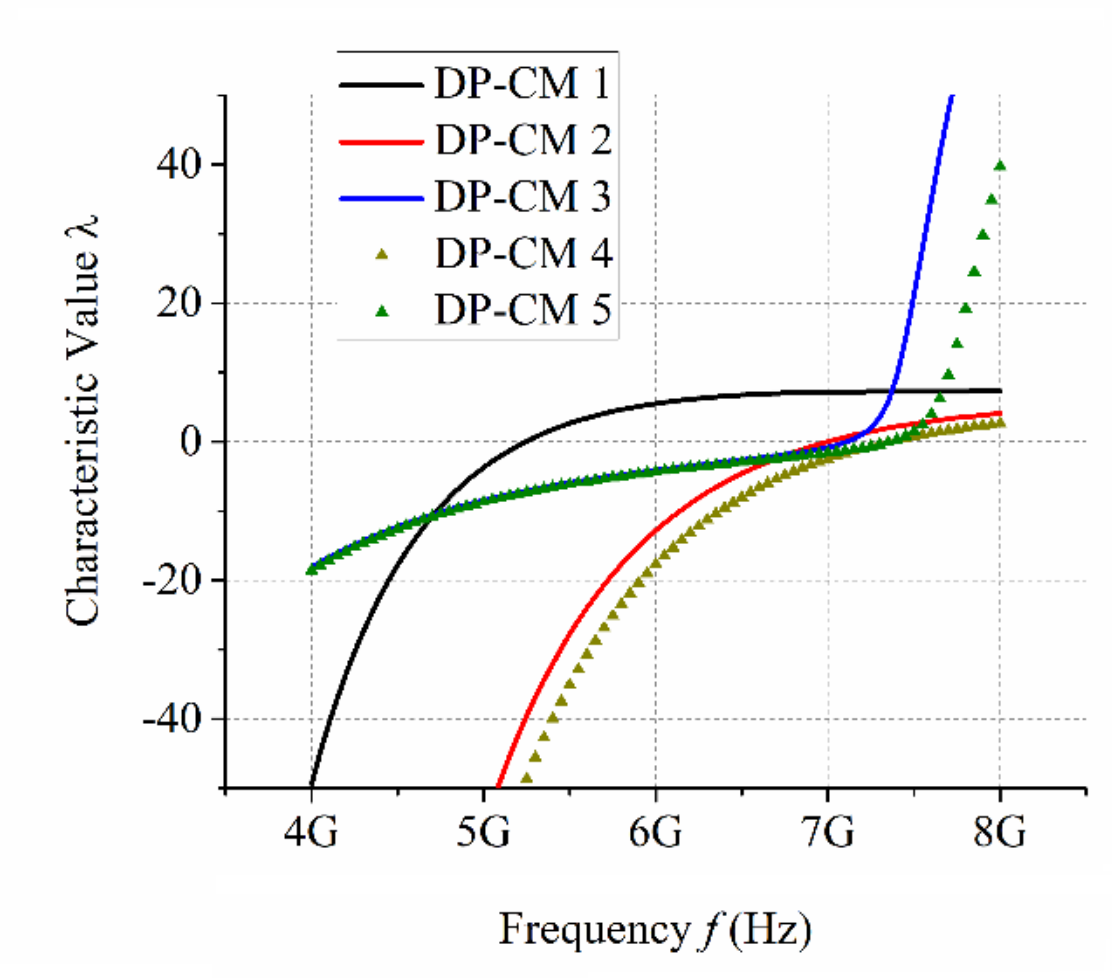

(a)

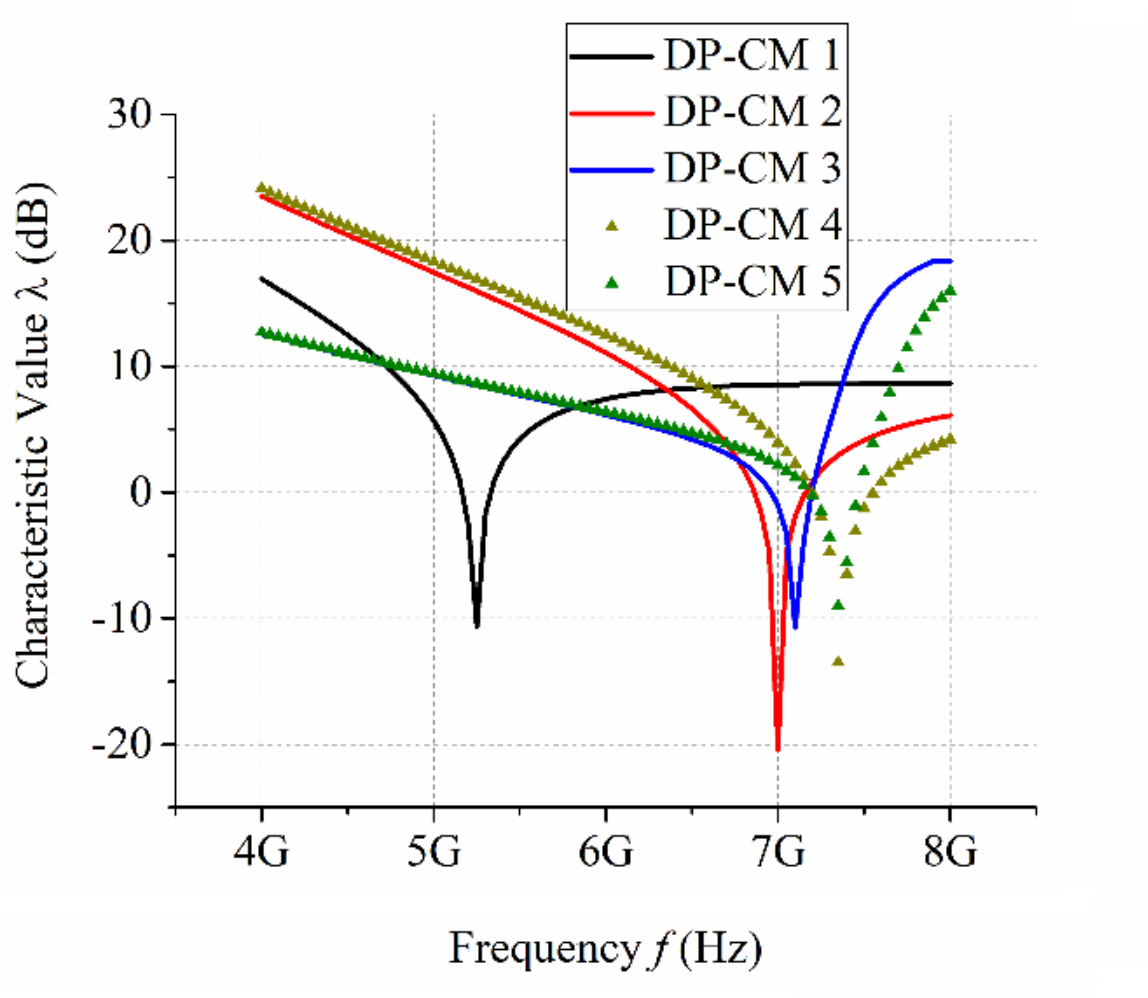

(b)

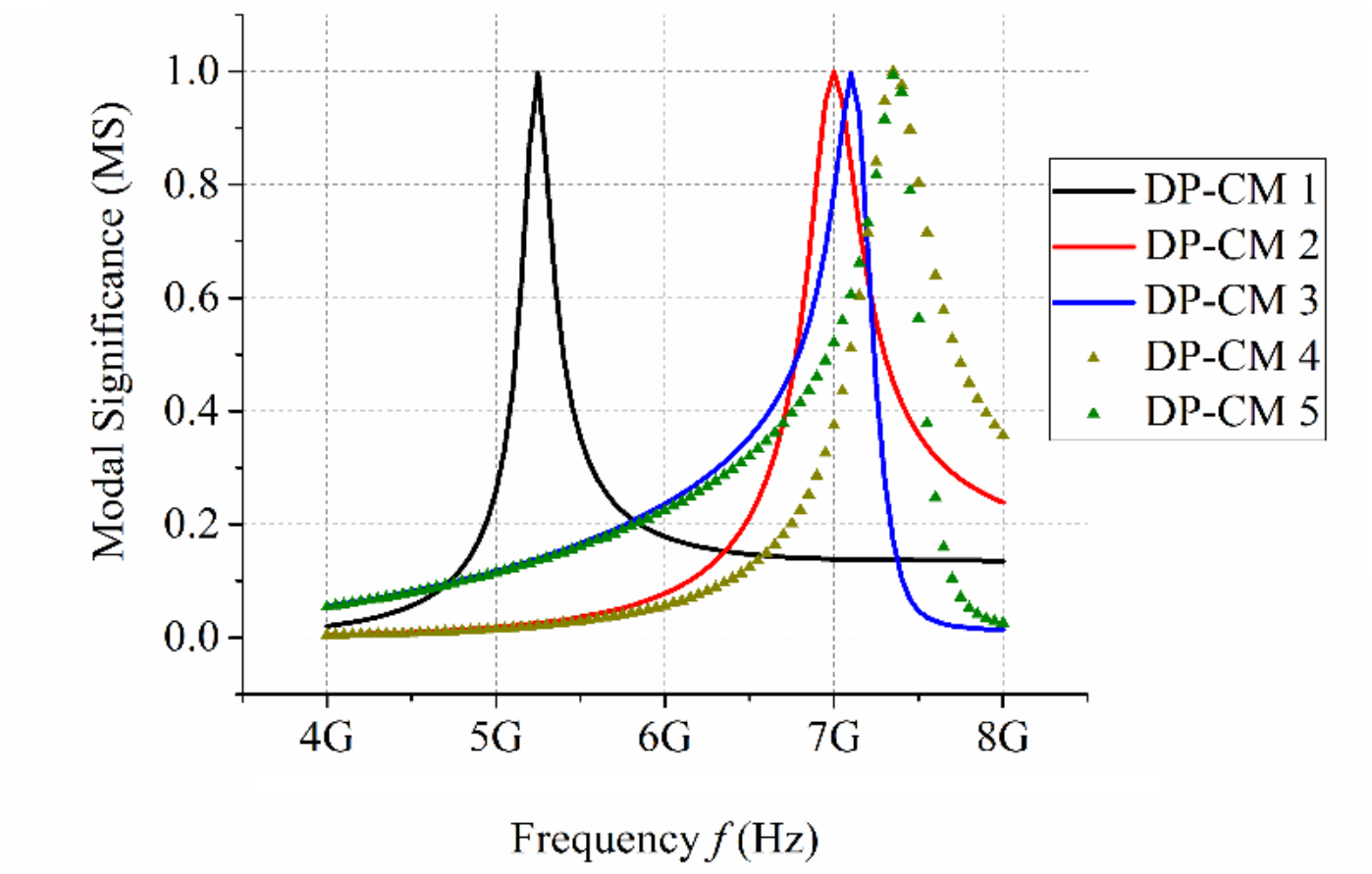

(c)

Figure 4-11 The characteristic quantity curves corresponding to the first 5 typical DP-CMs of the lossless magnetically biaxial cylinder whose topological structure is shown in Figure 4-2 and relative permeability and relative permittivity are $\ddot{\mu}^{\rm r}_{\rm mat} = \hat{x}\hat{x}36 + \hat{y}\hat{y}31 + \hat{z}\hat{z}26$ and $\ddot{\varepsilon}^{\rm r}_{\rm mat} = \ddot{I}1$ respectively. (a) characteristic valuve curves; (b) characteristic value dB curves; (c) MS curves

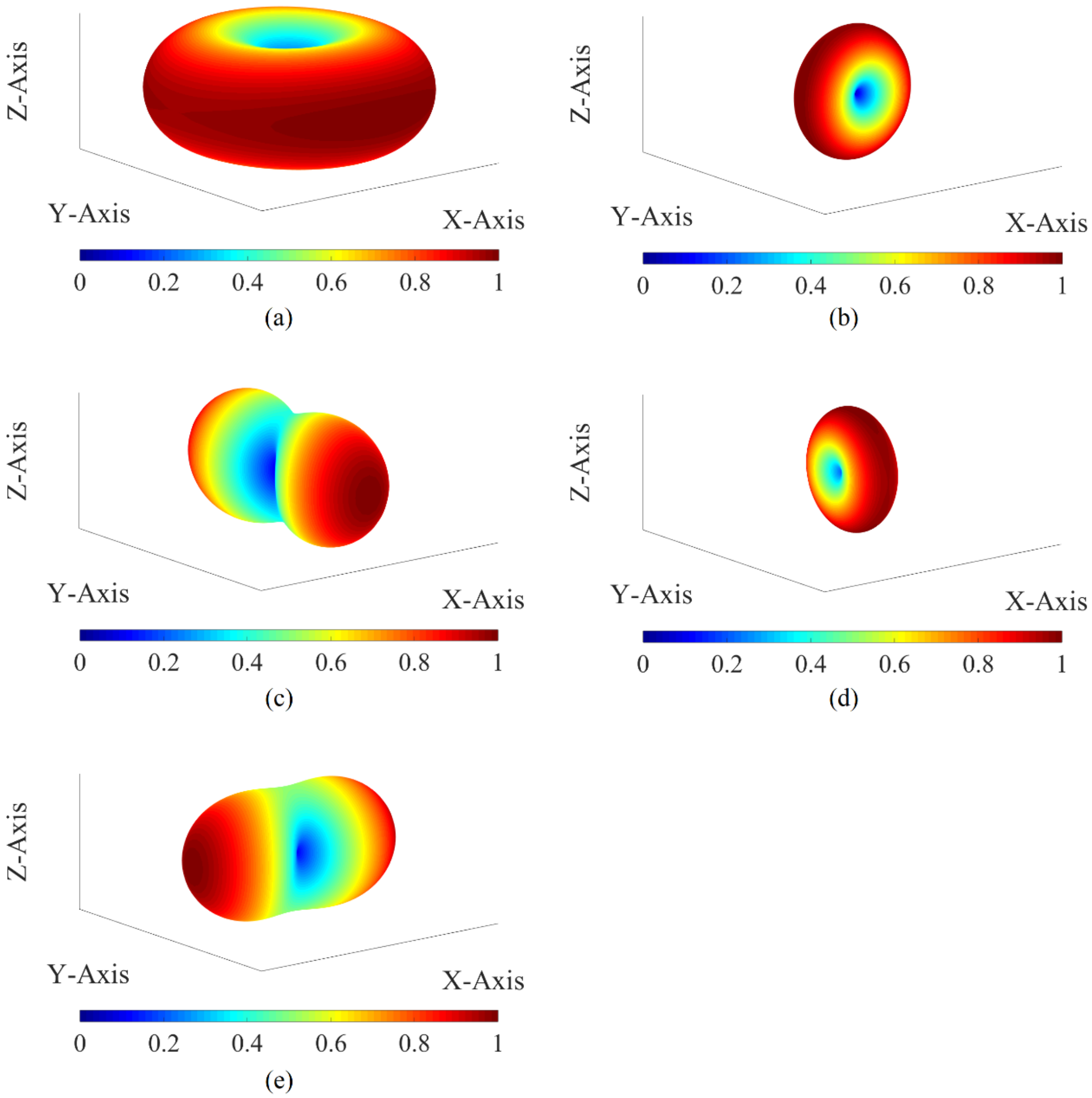

Figure 4-12 The radiation patterns corresponding to the typical DP-CMs shown in Figure 4-11. (a) the radiation pattern of the "resonant" DP-CM1 at 5.25GHz; (b) the radiation pattern of the "resonant" DP-CM2 at 7.00GHz; (c) the radiation pattern of the "resonant" DP-CM3 at 7.10GHz; (d) the radiation pattern of the "resonant" DP-CM4 at 7.35GHz; (e) the radiation pattern of the "resonant" DP-CM5 at 7.35GHz

In the following sections of this chapter, we continue focusing on, in WEP framework, establishing the CMT for inhomogeneous anisotropic material systems. The differences between the following sections and previous Section 4.2 are that: the variables in the DPO in the following sections are the equivalent surface sources distributing on the material boundaries of scattering systems, rather than the scattered volume sources distributing on the interior of scattering systems.

To make the results obtained in the following sections have a wider applicable range, we, firstly in the Appendix C of this dissertation, generalize the traditional surface equivalence principle (SEP) for a simply connected homogeneous isotropic material body to the inhomogeneous anisotropic material system whose topological structure is arbitrary; we secondly establish the surface formulations of the WEP-based CMT for various material systems in Section 4.3 (for a single simply connected material body), Sections 4.4~4.6 (for a two-body system constructed by two simply connected material bodies), and Section 4.7 (for a two-body system constructed by a simply connected material body and a multiply connected material body). Although the studies on establishing the surface CM calculation formulations of material systems in IE framework have had a long history, the following surface CM calculation formulations established in WEP framework are completely new.

## 4.3 Surface Formulation for Calculating the DP-CMs of Single Simply Connected Material Body

In this section, we will establish the surface DP-CM formulation of the simply connected inhomogeneous anisotropic material body shown in Figure 4-13, based on the generalized surface equivalence principle (GSEP) obtained in Appendix C5. Before establishing the general theory, we, firstly in Subsection 4.3.1, simply review the traditional Harrington's CMT for a single simply connected homogeneous isotropic material body.

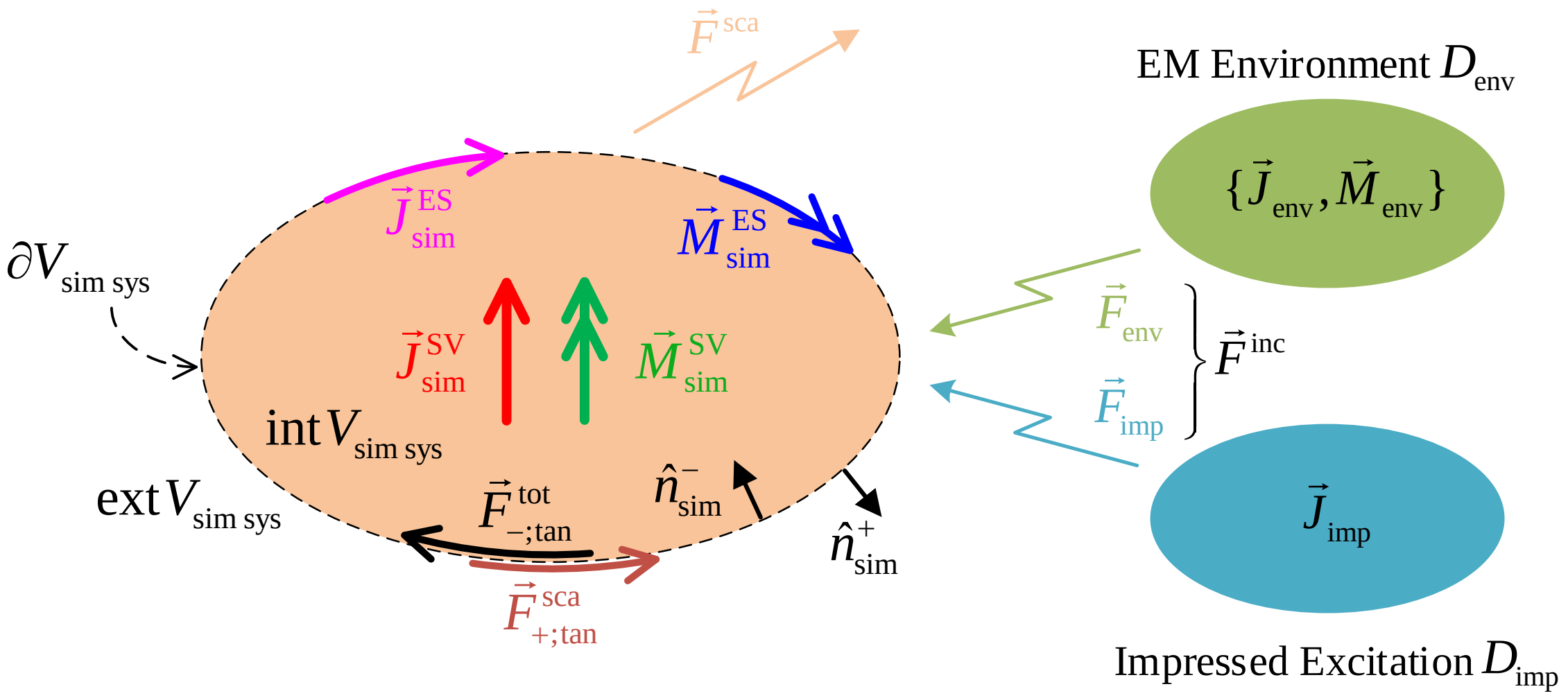

Figure 4-13 The scattering problem corresponding to a simply connected inhomogeneous anisotropic material body placed in a complex environment

## 4.3.1 Physical Picture of Harrington's SIE-MatSca-CMT

Focusing on a single simply connected homogeneous isotropic material body, Dr. Chang and Prof. Harrington[34], in IE framework, established SIE-MatSca-CMT in 1977, and the CMs derived from SIE-MatSca-CMT have ability to orthogonalize the following operator[44,45]:

$$(1/2)\left\langle -\vec{J}_{\text{sim}}^{\text{ES}}, \vec{E}_{-}^{\text{tot}} - \vec{E}_{+}^{\text{sca}} \right\rangle_{\partial V_{\text{sim sys}}} + (1/2)\left\langle -\vec{M}_{\text{sim}}^{\text{ES}}, \vec{H}_{-}^{\text{tot}} - \vec{H}_{+}^{\text{sca}} \right\rangle_{\partial V_{\text{sim sys}}} \tag{4-19}$$

In operator (4-19), the minus sign in the front of $\vec{J}_{\text{sim}}^{\text{ES}}$ originates from that literature [34] defined equivalent surface electric current as $\hat{n}_{\text{sim}}^{+} \times \vec{H}_{-}^{\text{tot}}$ (in this dissertation, $\vec{J}_{\text{sim}}^{\text{ES}}$ is defined as $\vec{J}_{\text{sim}}^{\text{ES}} = \hat{n}_{\text{sim}}^{-} \times \vec{H}_{-}^{\text{tot}}$), and the explanation for the minus sign in the front of $\vec{M}_{\text{sim}}^{\text{ES}}$ is similar. $\vec{F}_{-}^{\text{tot}}$ and $\vec{F}_{+}^{\text{sca}}$ are respectively the internal total field and external scattered field corresponding to $V_{\text{sim sys}}$. Obviously, to know the physical meaning of operator (4-19) is crucial to correctly understanding and applying the CMs constructed in literature [34]. In the following parts of this subsection, we focus on clarifying the physical meaning of operator (4-19), and then draw a clear physical picture for SIE-MatSca-CMT [34].

Operator (4-19) can be alternatively written as follows[44,45]:

$$\begin{aligned}&(1/2)\left\langle -\vec{J}_{\text{sim}}^{\text{ES}}, \vec{E}_{-}^{\text{tot}} - \vec{E}_{+}^{\text{sca}} \right\rangle_{\partial V_{\text{sim sys}}} + (1/2)\left\langle -\vec{M}_{\text{sim}}^{\text{ES}}, \vec{H}_{-}^{\text{tot}} - \vec{H}_{+}^{\text{sca}} \right\rangle_{\partial V_{\text{sim sys}}} \\ =\ &(1/2)\left\langle -\vec{J}_{\text{sim}}^{\text{ES}}, \vec{E}^{\text{inc}} \right\rangle_{\partial V_{\text{sim sys}}} + (1/2)\left\langle -\vec{M}_{\text{sim}}^{\text{ES}}, \vec{H}^{\text{inc}} \right\rangle_{\partial V_{\text{sim sys}}}\end{aligned} \tag{4-20}$$

because of linear superposition principle $\vec{F}^{\text{tot}} = \vec{F}^{\text{inc}} + \vec{F}^{\text{sca}}$ and the tangential continuity of the $\vec{F}^{\text{tot}}$ and $\vec{F}^{\text{sca}}$ on material boundary $\partial V_{\text{sim sys}}$.

On $V_{\text{sim sys}}$, there exist relationships $\vec{J}_{\text{sim}}^{\text{SV}} = j\omega\Delta\varepsilon_{\text{sim}}^{\text{c}}\vec{E}^{\text{tot}}$ and $\nabla\times\vec{H}^{\text{tot}} = j\omega\varepsilon_{\text{sim}}^{\text{c}}\vec{E}^{\text{tot}}$ ①. Based on this, it can be derived that

$$(1/2)\left\langle \vec{J}_{\text{sim}}^{\text{SV}}, \vec{E}^{\text{inc}} \right\rangle_{V_{\text{sim sys}}} = \left(\frac{\Delta\varepsilon_{\text{sim}}^{\text{c}}}{\varepsilon_{\text{sim}}^{\text{c}}}\right)^{*} (1/2)\left\langle \nabla\times\vec{H}^{\text{tot}}, \vec{E}^{\text{inc}} \right\rangle_{V_{\text{sim sys}}} \tag{4-21}$$

The $(1/2) < \nabla\times\vec{H}^{\text{tot}}, \vec{E}^{\text{inc}} >_{V_{\text{sim sys}}}$ on the RHS of above formulation (4-21) can be alternatively written as follows:

$$\begin{aligned}(1/2)\left\langle \nabla\times\vec{H}^{\text{tot}}, \vec{E}^{\text{inc}} \right\rangle_{V_{\text{sim sys}}} &= (1/2)\iiint_{V_{\text{sim sys}}} \nabla\cdot\left[\left(\vec{H}^{\text{tot}}\right)^{*}\times\vec{E}^{\text{inc}}\right]dV + (1/2)\left\langle \vec{H}^{\text{tot}}, \nabla\times\vec{E}^{\text{inc}} \right\rangle_{V_{\text{sim sys}}} \\ &= -(1/2)\left\langle \vec{J}_{\text{sim}}^{\text{ES}}, \vec{E}^{\text{inc}} \right\rangle_{\partial V_{\text{sim sys}}} + (1/2)\left\langle \vec{H}^{\text{tot}}, \nabla\times\vec{E}^{\text{inc}} \right\rangle_{V_{\text{sim sys}}}\end{aligned}$$

---

① Relationship $\vec{J}_{\text{sim}}^{\text{SV}} = j\omega\Delta\varepsilon_{\text{sim}}^{\text{c}}\vec{E}^{\text{tot}}$ is based on the volume equivalence principle given in Appendix A. Here, $\Delta\varepsilon_{\text{sim}}^{\text{c}} = \varepsilon_{\text{sim}}^{\text{c}} - \varepsilon_{0}$ is a constant scalar, because literature [34] focuses on homogeneous isotropic material bodies.

$$=-(1/2)\left\langle \vec{J}_{\text{sim}}^{\text{ES}},\vec{E}^{\text{inc}}\right\rangle_{\partial V_{\text{sim sys}}} + (1/2)\left\langle \vec{M}_{\text{sim}}^{\text{SV}} / j\omega\Delta\mu_{\text{sim}}, -j\omega\mu_0\vec{H}^{\text{inc}}\right\rangle_{V_{\text{sim sys}}} \tag{4-22}$$

In formulation (4-22), the first equality is based on vectorial identity $\nabla\cdot(\vec{a}\times\vec{b})=(\nabla\times\vec{a})\cdot\vec{b}-\vec{a}\cdot(\nabla\times\vec{b})$[122]; the second equality is based on Gauss' divergence theorem[122] and formulation (C-41a); the third equality is based on the relationship $\vec{M}_{\text{sim}}^{\text{SV}} = j\omega\Delta\mu_{\text{sim}}\vec{H}^{\text{tot}}$ ① on $V_{\text{sim sys}}$ and the Faraday's law $\nabla\times\vec{E}^{\text{inc}}=-j\omega\mu_0\vec{H}^{\text{inc}}$ of EM induction on $V_{\text{sim sys}}$. Inserting formulation (4-22) into formulation (4-21), we immediately obtain that

$$\Delta\mu_{\text{sim}}\Delta\varepsilon_{\text{sim}}^{\text{c}}(1/2)\left\langle \vec{J}_{\text{sim}}^{\text{ES}},\vec{E}^{\text{inc}}\right\rangle_{\partial V_{\text{sim sys}}}^{*} = -\varepsilon_{\text{sim}}^{\text{c}}\Delta\mu_{\text{sim}}(1/2)\left\langle \vec{J}_{\text{sim}}^{\text{SV}},\vec{E}^{\text{inc}}\right\rangle_{\partial V_{\text{sim sys}}}^{*} + \mu_0\Delta\varepsilon_{\text{sim}}^{\text{c}}(1/2)\left\langle \vec{M}_{\text{sim}}^{\text{SV}},\vec{H}^{\text{inc}}\right\rangle_{V_{\text{sim sys}}}^{*} \tag{4-23}$$

Similarly, we can also obtain that

$$\Delta\mu_{\text{sim}}\Delta\varepsilon_{\text{sim}}^{\text{c}}(1/2)\left\langle \vec{M}_{\text{sim}}^{\text{ES}},\vec{H}^{\text{inc}}\right\rangle_{\partial V_{\text{sim sys}}}^{*} = +\varepsilon_0\Delta\mu_{\text{sim}}(1/2)\left\langle \vec{J}_{\text{sim}}^{\text{SV}},\vec{E}^{\text{inc}}\right\rangle_{V_{\text{sim sys}}}^{*} - \mu_{\text{sim}}\Delta\varepsilon_{\text{sim}}^{\text{c}}(1/2)\left\langle \vec{M}_{\text{sim}}^{\text{SV}},\vec{H}^{\text{inc}}\right\rangle_{V_{\text{sim sys}}}^{*} \tag{4-24}$$

If above formulations (4-23) and (4-24) are treated as the simultaneous equations about $(1/2)<\vec{J}_{\text{sim}}^{\text{SV}},\vec{E}^{\text{inc}}>_{V_{\text{sim sys}}}$ and $(1/2)<\vec{M}_{\text{sim}}^{\text{SV}},\vec{H}^{\text{inc}}>_{V_{\text{sim sys}}}$, then it can be solved that[44]

$$(1/2)\left\langle \vec{J}_{\text{sim}}^{\text{SV}},\vec{E}^{\text{inc}}\right\rangle_{V_{\text{sim sys}}} = \gamma_{\text{sim}}\left[(1/2)\left\langle \vec{J}_{\text{sim}}^{\text{ES}},\vec{E}^{\text{inc}}\right\rangle_{\partial V_{\text{sim sys}}} + \frac{\mu_0}{\mu_{\text{sim}}}(1/2)\left\langle \vec{M}_{\text{sim}}^{\text{ES}},\vec{H}^{\text{inc}}\right\rangle_{\partial V_{\text{sim sys}}}\right] \tag{4-25}$$

$$(1/2)\left\langle \vec{M}_{\text{sim}}^{\text{SV}},\vec{H}^{\text{inc}}\right\rangle_{V_{\text{sim sys}}} = \gamma_{\text{sim}}'\left[\frac{\varepsilon_0}{\left(\varepsilon_{\text{sim}}^{\text{c}}\right)^*}(1/2)\left\langle \vec{J}_{\text{sim}}^{\text{ES}},\vec{E}^{\text{inc}}\right\rangle_{\partial V_{\text{sim sys}}} + (1/2)\left\langle \vec{M}_{\text{sim}}^{\text{ES}},\vec{H}^{\text{inc}}\right\rangle_{\partial V_{\text{sim sys}}}\right] \tag{4-26}$$

where $\gamma_{\text{sim}}=\mu_{\text{sim}}(\Delta\varepsilon_{\text{sim}}^{\text{c}})^*/[\mu_0\varepsilon_0-\mu_{\text{sim}}(\varepsilon_{\text{sim}}^{\text{c}})^*]$ and $\gamma_{\text{sim}}'=(\varepsilon_{\text{sim}}^{\text{c}})^*\Delta\mu_{\text{sim}}/[\varepsilon_0\mu_0-(\varepsilon_{\text{sim}}^{\text{c}})^*\mu_{\text{sim}}]$, and $\mu_0\varepsilon_0-\mu_{\text{sim}}(\varepsilon_{\text{sim}}^{\text{c}})^*\neq 0$. From formulations (4-25) and (4-26), it is immediately obtained that[44]

$$\frac{1}{2}\left\langle -\vec{J}_{\text{sim}}^{\text{ES}},\vec{E}^{\text{inc}}\right\rangle_{\partial V_{\text{sim sys}}} + \frac{1}{2}\left\langle -\vec{M}_{\text{sim}}^{\text{ES}},\vec{H}^{\text{inc}}\right\rangle_{\partial V_{\text{sim sys}}} = \frac{1}{2}\left\langle \vec{J}_{\text{sim}}^{\text{SV}},\vec{E}^{\text{inc}}\right\rangle_{V_{\text{sim sys}}} + \frac{1}{2}\left\langle \vec{M}_{\text{sim}}^{\text{SV}},\vec{H}^{\text{inc}}\right\rangle_{V_{\text{sim sys}}} = P_{\text{sim sys}}^{\text{driving}} \tag{4-27}$$

where the second equality is based on the formulation (4-7) obtained in Section 4.2.

① The derivation for this relationship can be found in Appendix A. Here, $\Delta\mu_{\text{sim}}=\mu_{\text{sim}}-\mu_0$ is a constant scalar, because literature [34] focuses on homogeneous isotropic material bodies.

Synthetically considering formulations (4-20) and (4-27), we have that

$$(1/2)\left\langle -\vec{J}_{\text{sim}}^{\text{ES}}, \vec{E}_{-}^{\text{tot}} - \vec{E}_{+}^{\text{sca}} \right\rangle_{\partial V_{\text{sim sys}}} + (1/2)\left\langle -\vec{M}_{\text{sim}}^{\text{ES}}, \vec{H}_{-}^{\text{tot}} - \vec{H}_{+}^{\text{sca}} \right\rangle_{\partial V_{\text{sim sys}}} = P_{\text{sim sys}}^{\text{driving}} \quad (4\text{-}28)$$

where $P_{\text{sim sys}}^{\text{driving}}$ is just the frequency-domain version II of the DPO of material systems (for details see Section 4.2).

Above relationship (4-28) points out that: operators (4-19) and $(1/2) < \vec{J}_{\text{sim}}^{\text{SV}}, \vec{E}^{\text{inc}} >_{V_{\text{sim sys}}} + (1/2) < \vec{M}_{\text{sim}}^{\text{SV}}, \vec{H}^{\text{inc}} >_{V_{\text{sim sys}}}$ have the same physical meaning —— the frequency-domain version II of the DPO of material systems; the SIE-MatSca-CMT and VIE-MatSca-CMT for a single simply connected material body have the same physical destination —— constructing a series of fundamental modes having ability to orthogonalize frequency-domain DPO $P_{\text{sim sys}}^{\text{driving}}$ (i.e. constructing a series of steadily working modes not having net energy exchange in any integral period). To emphasize the physical picture of SIE-MatSca-CMT, this dissertation calls it as the surface formulation for constructing the DP-CMs of material systems.

Obviously, the process to establish surface DP-CM formulation in literature [34] and the process to derive above relationship (4-28) are only applicable to homogeneous isotropic material bodies. So whether or not relationship (4-28) is also suitable for inhomogeneous anisotropic material bodies? How to establish the surface DP-CM formulation of inhomogeneous anisotropic material bodies?

Our answer to the first question is YES, and we will, in the following Subsection 4.3.2, provide a rigorous mathematical proof for our above answer. During the process to answer the first question, we at the same time obtain a completely new surface DP-CM formulation, which is applicable to inhomogeneous anisotropic material bodies. Based on the new formulation obtained in Subsection 4.3.2, we provide in Subsections 4.3.3 and 4.3.4 the detailed operation process for constructing the DP-CMs of inhomogeneous anisotropic material bodies. Afterwards, we, in Subsection 4.3.5, expand any one working mode in terms of DP-CMs, and also provide the explicit expressions of the expansion coefficients.

### 4.3.2 Surface Formulation of the DPO Corresponding to Single Simply Connected Inhomogeneous Anisotropic Material Body

It is obvious that term $(1/2) < \vec{J}_{\text{sim}}^{\text{ES}}, \vec{E}^{\text{inc}} >_{\partial V_{\text{sim sys}}}$ can be alternatively written as follows:

$$
\begin{aligned}
(1/2)\left\langle \vec{J}_{\text{sim}}^{\text{ES}}, \vec{E}^{\text{inc}} \right\rangle_{\partial V_{\text{sim sys}}} &= (1/2)\oiint_{\partial V_{\text{sim sys}}} \left[ \vec{E}^{\text{inc}} \times \left( \vec{H}^{\text{tot}} \right)^* \right] \cdot \hat{n}_{\text{sim}}^{+} dS \\
&= (1/2)\iiint_{V_{\text{sim sys}}} \nabla \cdot \left[ \vec{E}^{\text{inc}} \times \left( \vec{H}^{\text{tot}} \right)^* \right] dV \\
&= (1/2)\left\langle \vec{H}^{\text{tot}}, \nabla \times \vec{E}^{\text{inc}} \right\rangle_{V_{\text{sim sys}}} - (1/2)\left\langle \nabla \times \vec{H}^{\text{tot}}, \vec{E}^{\text{inc}} \right\rangle_{V_{\text{sim sys}}} \\
&= (1/2)\left\langle \vec{H}^{\text{tot}}, -j\omega\mu_0 \vec{H}^{\text{inc}} \right\rangle_{V_{\text{sim sys}}} - (1/2)\left\langle j\omega \vec{\vec{\varepsilon}}_{\text{sim}}^{\text{c}} \cdot \vec{E}^{\text{tot}}, \vec{E}^{\text{inc}} \right\rangle_{V_{\text{sim sys}}} \\
&= (1/2)\left\langle \vec{H}^{\text{tot}}, -j\omega\mu_0 \vec{H}^{\text{inc}} \right\rangle_{V_{\text{sim sys}}} - (1/2)\left\langle j\omega \varepsilon_0 \vec{E}^{\text{tot}}, \vec{E}^{\text{inc}} \right\rangle_{V_{\text{sim sys}}} \\
&\quad - (1/2)\left\langle j\omega \Delta \vec{\vec{\varepsilon}}_{\text{sim}}^{\text{c}} \cdot \vec{E}^{\text{tot}}, \vec{E}^{\text{inc}} \right\rangle_{V_{\text{sim sys}}} \\
&= -\, j\, 2\omega \left[ (1/4)\left\langle \vec{H}^{\text{tot}}, \mu_0 \vec{H}^{\text{inc}} \right\rangle_{V_{\text{sim sys}}} - (1/4)\left\langle \varepsilon_0 \vec{E}^{\text{tot}}, \vec{E}^{\text{inc}} \right\rangle_{V_{\text{sim sys}}} \right] \\
&\quad - (1/2)\left\langle \vec{J}_{\text{sim}}^{\text{SV}}, \vec{E}^{\text{inc}} \right\rangle_{V_{\text{sim sys}}}
\end{aligned} \tag{4-29}
$$

In formulation (4-29), the first equality is based on the definition (C-41a) of equivalent surface electric current; the second equality is based on Gauss' divergence theorem[122]; the third equality is based on vectorial identity $\nabla \cdot (\vec{a} \times \vec{b}) = (\nabla \times \vec{a}) \cdot \vec{b} - \vec{a} \cdot (\nabla \times \vec{b})$ [122]; the fourth equality is based on the Maxwell's equations $\nabla \times \vec{E}^{\text{inc}} = -j\omega\mu_0 \vec{H}^{\text{inc}}$ and $\nabla \times \vec{H}^{\text{tot}} = j\omega \vec{\vec{\varepsilon}}_{\text{sim}}^{\text{c}} \cdot \vec{E}^{\text{tot}}$ on $V_{\text{sim sys}}$; the fifth equality is based on that $\vec{\vec{\varepsilon}}_{\text{sim}}^{\text{c}} = \vec{\vec{I}} \varepsilon_0 + \Delta \vec{\vec{\varepsilon}}_{\text{sim}}^{\text{c}}$; the sixth equality is based on that $\vec{J}_{\text{sim}}^{\text{SV}} = j\omega \Delta \vec{\vec{\varepsilon}}_{\text{sim}}^{\text{c}} \cdot \vec{E}^{\text{tot}}$. Obviously, all equalities mentioned above are valid for not only homogeneous isotropic material bodies but also inhomogeneous anisotropic material bodies.

Similarly to formulation (4-29), we can also derive the following relationship:

$$
\begin{aligned}
(1/2)\left\langle \vec{M}_{\text{sim}}^{\text{ES}}, \vec{H}^{\text{inc}} \right\rangle_{\partial V_{\text{sim sys}}} &= j\, 2\omega \left[ (1/4)\left\langle \vec{H}^{\text{tot}}, \mu_0 \vec{H}^{\text{inc}} \right\rangle_{V_{\text{sim sys}}} - (1/4)\left\langle \varepsilon_0 \vec{E}^{\text{tot}}, \vec{E}^{\text{inc}} \right\rangle_{V_{\text{sim sys}}} \right] \\
&\quad - (1/2)\left\langle \vec{M}_{\text{sim}}^{\text{SV}}, \vec{H}^{\text{inc}} \right\rangle_{V_{\text{sim sys}}}
\end{aligned} \tag{4-30}
$$

Obviously, formulation (4-29) and formulation (4-30) are dual to each other, i.e., doing replacements $(J,M) \rightarrow (M,J)$, $(E,H) \rightarrow (H,E)$, and $(\varepsilon,\mu) \rightarrow (\mu,\varepsilon)$ in any formulation will lead to the other formulation.

Summing formulation (4-29) and formulation (4-30), we can obtain the following relationship:

$$
\begin{aligned}
-(1/2)\left\langle \vec{J}_{\text{sim}}^{\text{ES}}, \vec{E}^{\text{inc}} \right\rangle_{\partial V_{\text{sim sys}}} - (1/2)\left\langle \vec{M}_{\text{sim}}^{\text{ES}}, \vec{H}^{\text{inc}} \right\rangle_{\partial V_{\text{sim sys}}} &= (1/2)\left\langle \vec{J}_{\text{sim}}^{\text{SV}}, \vec{E}^{\text{inc}} \right\rangle_{V_{\text{sim sys}}} + (1/2)\left\langle \vec{M}_{\text{sim}}^{\text{SV}}, \vec{H}^{\text{inc}} \right\rangle_{V_{\text{sim sys}}} \\
&= P_{\text{sim sys}}^{\text{driving}}
\end{aligned} \tag{4-31}
$$

If the $\vec{F}^{\text{inc}}$ on the LHS of the above first equality is replaced by $\vec{F}_{-}^{\text{tot}} - \vec{F}_{+}^{\text{sca}}$, then it can be found out that: the relationship (4-28) for homogeneous isotropic material bodies is also suitable for inhomogeneous anisotropic material bodies, in the aspect of mathematical form. This is just the reason why the result obtained in literature [34] and the result obtained in the following parts of this chapter are collectively referred to as the surface DP-CM formulation of material systems. In the following parts of this section, we will always select to use operator $P_{\text{sim sys}}^{\text{driving}} = -(1/2) < \vec{J}_{\text{sim}}^{\text{ES}}, \vec{E}^{\text{inc}} >_{\partial V_{\text{sim sys}}} -(1/2) < \vec{M}_{\text{sim}}^{\text{ES}}, \vec{H}^{\text{inc}} >_{\partial V_{\text{sim sys}}}$ rather than operator $P_{\text{sim sys}}^{\text{driving}} = -(1/2) < \vec{J}_{\text{sim}}^{\text{ES}}, \vec{E}_{-}^{\text{tot}} - \vec{E}_{+}^{\text{sca}} >_{\partial V_{\text{sim sys}}} -(1/2) < \vec{M}_{\text{sim}}^{\text{ES}}, \vec{H}_{-}^{\text{tot}} - \vec{H}_{+}^{\text{sca}} >_{\partial V_{\text{sim sys}}}$. Even in the face of the homogeneous isotropic material bodies, which SIE-MatSca-CMT [34] can be applicable to, we will also adopt the above habit, and this is a main difference[45] between this section and other literatures.

Based on GSEP (C-40) and that incident field is continuous on $\partial V_{\text{sim sys}}$, we can rewrite formulation (4-31) as follows[45]:

$$\begin{aligned} P_{\text{sim sys}}^{\text{driving}} &= (1/2)\left\langle -\vec{J}_{\text{sim}}^{\text{ES}}, \vec{E}^{\text{inc}} \right\rangle_{\partial V_{\text{sim sys}}} + (1/2)\left\langle -\vec{M}_{\text{sim}}^{\text{ES}}, \vec{H}^{\text{inc}} \right\rangle_{\partial V_{\text{sim sys}}} \\ &= -(1/2)\left\langle \vec{J}_{\text{sim}}^{\text{ES}}, \mathcal{E}_0\left(\vec{J}_{\text{sim}}^{\text{ES}}, \vec{M}_{\text{sim}}^{\text{ES}}\right)\right\rangle_{\partial V_{\text{sim sys}}^{-}} - (1/2)\left\langle \vec{M}_{\text{sim}}^{\text{ES}}, \mathcal{H}_0\left(\vec{J}_{\text{sim}}^{\text{ES}}, \vec{M}_{\text{sim}}^{\text{ES}}\right)\right\rangle_{\partial V_{\text{sim sys}}^{-}} \end{aligned} \tag{4-32}$$

In formulation (4-32), $\partial V_{\text{sim sys}}^{-}$ is the inner surface of $\partial V_{\text{sim sys}}$; operator $\mathcal{E}_0(\vec{J}, \vec{M})$ is defined as that $\mathcal{E}_0(\vec{J}, \vec{M}) = [\vec{G}_0^{JE} * \vec{J} + \vec{G}_0^{ME} * \vec{M}]_{\partial V_{\text{sim sys}}} = -j\omega\mu_0 \mathcal{L}_0(\vec{J}) - \mathcal{K}_0(\vec{M})$; operator $\mathcal{H}_0(\vec{J}, \vec{M})$ is defined as that $\mathcal{H}_0(\vec{J}, \vec{M}) = [\vec{G}_0^{JH} * \vec{J} + \vec{G}_0^{MH} * \vec{M}]_{\partial V_{\text{sim sys}}} = -j\omega\varepsilon_0 \mathcal{L}_0(\vec{M}) + \mathcal{K}_0(\vec{J})$. Here, the definitions of operators $\mathcal{L}_0(\vec{X})$ and $\mathcal{K}_0(\vec{X})$ have been given in the Chapter 3 of this dissertation, and they can also be found in literature [107], so they will not be explicitly provided here.

If $\{\vec{b}_\xi^{J_s}\}_{\xi=1}^{\Xi^{J_s}}$ and $\{\vec{b}_\xi^{M_s}\}_{\xi=1}^{\Xi^{M_s}}$ are respectively the complete and independent basis functions of the $\vec{J}_{\text{sim}}^{\text{ES}}$ and $\vec{M}_{\text{sim}}^{\text{ES}}$ on $\partial V_{\text{sim sys}}$, then any $\vec{J}_{\text{sim}}^{\text{ES}}$ and $\vec{M}_{\text{sim}}^{\text{ES}}$ can be expanded as follows[106-108,115]:

$$\vec{C}_{\text{sim}}^{\text{ES}}(\vec{r}) = \sum_{\xi=1}^{\Xi^{C_s}} a_\xi^{C_s} \vec{b}_\xi^{C_s}(\vec{r}) = \bar{\mathbf{B}}^{C_s} \cdot \bar{a}^{C_s} \quad , \quad \vec{r} \in \partial V_{\text{sim sys}} \tag{4-33}$$

where $C = J, M$. Inserting expansion formulation (4-33) into formulation (4-32), DPO $P_{\text{sim sys}}^{\text{driving}}$ is immediately discretized to the following matrix form[45]:

$$P_{\text{sim sys}}^{\text{driving}} = \begin{bmatrix} \bar{a}^{J_s} \\ \bar{a}^{M_s} \end{bmatrix}^H \cdot \Bigg\{ \overbrace{\underbrace{\begin{bmatrix} \bar{\bar{P}}_{0;\text{PVT}}^{J_sJ_s} & \bar{\bar{P}}_{0;\text{PVT}}^{J_sM_s} \\ \bar{\bar{P}}_{0;\text{PVT}}^{M_sJ_s} & \bar{\bar{P}}_{0;\text{PVT}}^{M_sM_s} \end{bmatrix}}_{\bar{\bar{P}}_{0;\text{PVT}}^{\text{sim sys}}} + \underbrace{\begin{bmatrix} 0 & \bar{\bar{P}}_{0;\text{SCT}}^{J_sM_s} \\ \bar{\bar{P}}_{0;\text{SCT}}^{M_sJ_s} & 0 \end{bmatrix}}_{\bar{\bar{P}}_{0;\text{SCT}}^{\text{sim sys}}}}^{\bar{\bar{P}}_{\text{sim sys}}^{\text{driving}}} \Bigg\} \cdot \begin{bmatrix} \bar{a}^{J_s} \\ \bar{a}^{M_s} \end{bmatrix} \tag{4-34}$$

where the elements of submatrices $\overline{\overline{P}}{}^{J_s J_s}_{0;\text{PVT}}$, $\overline{\overline{P}}{}^{J_s M_s}_{0;\text{PVT}}$, $\overline{\overline{P}}{}^{M_s J_s}_{0;\text{PVT}}$, $\overline{\overline{P}}{}^{M_s M_s}_{0;\text{PVT}}$, $\overline{\overline{P}}{}^{J_s M_s}_{0;\text{SCT}}$, and $\overline{\overline{P}}{}^{M_s J_s}_{0;\text{SCT}}$ are calculated as follows:

$$p^{J_s J_s}_{0;\text{PVT};\xi\zeta} = -(1/2)\left\langle \vec{b}^{J_s}_{\xi}, -j\omega\mu_0 \mathcal{L}_0\left(\vec{b}^{J_s}_{\zeta}\right)\right\rangle_{\partial V_{\text{sim sys}}} \tag{4-35a}$$

$$p^{J_s M_s}_{0;\text{PVT};\xi\zeta} = -(1/2)\left\langle \vec{b}^{J_s}_{\xi}, -\text{P.V.}\,\mathcal{K}_0\left(\vec{b}^{M_s}_{\zeta}\right)\right\rangle_{\partial V_{\text{sim sys}}} \tag{4-35b}$$

$$p^{M_s J_s}_{0;\text{PVT};\xi\zeta} = -(1/2)\left\langle \vec{b}^{M_s}_{\xi}, \text{P.V.}\,\mathcal{K}_0\left(\vec{b}^{J_s}_{\zeta}\right)\right\rangle_{\partial V_{\text{sim sys}}} \tag{4-35c}$$

$$p^{M_s M_s}_{0;\text{PVT};\xi\zeta} = -(1/2)\left\langle \vec{b}^{M_s}_{\xi}, -j\omega\varepsilon_0 \mathcal{L}_0\left(\vec{b}^{M_s}_{\zeta}\right)\right\rangle_{\partial V_{\text{sim sys}}} \tag{4-35d}$$

$$p^{J_s M_s}_{0;\text{SCT};\xi\zeta} = -(1/2)\left\langle \vec{b}^{J_s}_{\xi}, -\frac{1}{2}\vec{b}^{M_s}_{\zeta}\times\hat{n}^{-}_{\text{sim}}\right\rangle_{\partial V_{\text{sim sys}}} \tag{4-35e}$$

$$p^{M_s J_s}_{0;\text{SCT};\xi\zeta} = -(1/2)\left\langle \vec{b}^{M_s}_{\xi}, \frac{1}{2}\vec{b}^{J_s}_{\zeta}\times\hat{n}^{-}_{\text{sim}}\right\rangle_{\partial V_{\text{sim sys}}} \tag{4-35f}$$

In formulations (4-35b) and (4-35c), symbol "P.V." represents the principal value of the corresponding integral[107]. In formulations (4-35e) and (4-35f), $\hat{n}^{-}_{\text{sim}}$ is the normal direction of $\partial V_{\text{sim sys}}$, and points to the interior of $V_{\text{sim sys}}$, as shown in Figure 4-13, so it can be simply called as inner normal direction. In formulations (4-34) and (4-35), subscript "PVT" is the abbreviation of terminology "principal value term", and subscript "SCT" is the abbreviation of terminology "singular current term" ①.

In the above, we obtain three different forms of $P^{\text{driving}}_{\text{sim sys}}$ —— interaction form (4-31)②, current form (4-32)③, and matrix form (4-34)④. In interaction form (4-31), the variables in DPO $P^{\text{driving}}_{\text{sim sys}}$ contain both EM currents and EM fields; in current form (4-32) and matrix form (4-34), the variables in DPO $P^{\text{driving}}_{\text{sim sys}}$ contain EM currents only⑤.

In interaction form (4-31), incident fields $\{\vec{E}^{\text{inc}},\vec{H}^{\text{inc}}\}$ are not independent of equivalent currents $\{\vec{J}^{\text{ES}}_{\text{sim}},\vec{M}^{\text{ES}}_{\text{sim}}\}$. Before constructing DP-CMs by orthogonalizing DPO $P^{\text{driving}}_{\text{sim sys}}$, it is necessary to express the dependent variables as the functions of the independent variables, or it will lead to some spurious modes[35,42,44,45], and the variable set constituted by a series of independent and complete variables is called as basic

① Here, we don't do like literature [45] (in the literature, the principal value term and SCT contained in DPO are treated as a whole), but divide them into two terms explicitly. In the Section 6.3 of this dissertation, we will specially discuss the physical meaning of the SCT and the advantagement of the division.

② i.e., the surface formulation of $P^{\text{driving}}_{\text{sim sys}}$ is expressed as the interaction between EM field and EM current.

③ i.e., the surface formulation of $P^{\text{driving}}_{\text{sim sys}}$ is expressed in terms of the function of equivalent surface electric and magnetic currents, also i.e. the manifestation form of the $P^{\text{driving}}_{\text{sim sys}}$ in EM current space.

④ i.e., the surface formulation of $P^{\text{driving}}_{\text{sim sys}}$ is expressed as a matric quadratic form, also i.e. the manifestation form of the $P^{\text{driving}}_{\text{sim sys}}$ in expansion vector space.

⑤ Expansion vectors $\{\bar{a}^{J_s},\bar{a}^{M_s}\}$ are only another manifestation forms for EM currents $\{\vec{J}^{\text{ES}}_{\text{sim}},\vec{M}^{\text{ES}}_{\text{sim}}\}$.

variable (BV) set in this dissertation. The process to express the dependent variables as the functions of the BVs is called, in this dissertation, as “variable unification” or “to unify variables”. Actually, the process to transform interaction form (4-31) to current form (4-32) is essentially a process to unify variables, i.e., to express incident fields $\{\vec{E}^{\rm inc},\vec{H}^{\rm inc}\}$ as the functions of equivalent currents $\{\vec{J}_{\rm sim}^{\rm ES},\vec{M}_{\rm sim}^{\rm ES}\}$.

However, only transforming from interaction form (4-31) to current form (4-32) has not thoroughly finished variable unification, because: in current form (4-32), $\vec{J}_{\rm sim}^{\rm ES}$ and $\vec{M}_{\rm sim}^{\rm ES}$ are not independent of each other[35,42,44,45]. But, it is difficult to establish the dependent relationship between $\vec{J}_{\rm sim}^{\rm ES}$ and $\vec{M}_{\rm sim}^{\rm ES}$ in current space. Then, we firstly establish the one-to-one correspondence between the $\{\vec{J}_{\rm sim}^{\rm ES},\vec{M}_{\rm sim}^{\rm ES}\}$ in current space and the $\{\bar{a}^{J_{\rm s}},\bar{a}^{M_{\rm s}}\}$ in expansion vector space, and afterwards establish the transformation between $\bar{a}^{J_{\rm s}}$ and $\bar{a}^{M_{\rm s}}$ in expansion vector space[44,45]. Once the transformation between $\bar{a}^{J_{\rm s}}$ and $\bar{a}^{M_{\rm s}}$ is obtained, the matrix form (4-34) of the surface formulation of DPO $P_{\rm sim\,sys}^{\rm driving}$ include the BVs only.

Formulation (4-33) has established the one-to-one correspondence between $\{\vec{J}_{\rm sim}^{\rm ES},\vec{M}_{\rm sim}^{\rm ES}\}$ and $\{\bar{a}^{J_{\rm s}},\bar{a}^{M_{\rm s}}\}$, i.e., formulation (4-33) has achieved the transformation from EM current space to expansion vector space. In the following Subsection 4.3.3, we focus on establishing the transformation between $\bar{a}^{J_{\rm s}}$ and $\bar{a}^{M_{\rm s}}$ in expansion vector space, and then finish the thorough variable unification for DPO $P_{\rm sim\,sys}^{\rm driving}$. Afterwards, we will, in Subsection 4.3.4, construct DP-CMs by employing the matrix form of $P_{\rm sim\,sys}^{\rm driving}$ which only include BVs, and then we, in Subsection 4.3.5, expand any working mode in terms of the DP-CMs. During the process to derive the explicit expressions of expansion coefficients in Subsection 4.3.5, we will reuse the interaction form of $P_{\rm sim\,sys}^{\rm driving}$, because it is convenient to derive the explicit expressions by employing the interaction form.

In summary, three different forms of the surface formulation of $P_{\rm sim\,sys}^{\rm driving}$ have their own uses: the EM current form realizes the destination to express the EM fields in the surface formulation as the functions of EM currents, and then eliminates a part of dependent variables; the matrix form is convenient to completely eliminating all dependent variables in expansion vector space, and at the same time it enjoys unique advantage in the process of establishing and solving characteristic equation; the interaction form is suitable for the process of determining expansion coefficients, and of course a more important significance of the interaction form is its clear physical meaning —— power done by the EM fields acting on EM currents.

### 4.3.3 Variable Unification for the Equivalent Surface Sources on a Material Body

In the points 1) and 2) of this subsection, we will provide two variable unification schemes which have different mathematical forms. In the point 3) of this subsection, we will prove that the two schemes are equivalent to each other. In the point 4) of this subsection, we will provide detailed variable unification process based on the scheme given in point 2).

**1) Variable Unification Scheme Based on the Tangential Continuation of the Scattered Fields on $\partial V_{\text{sim sys}}$ (Scheme Developed in Literature [44])**

The Appendix A of this dissertation points out that there doesn't exist any surface scattered current distributing on $\partial V_{\text{sim sys}}$. Thus, the tangential components of scattered field $\vec{F}^{\text{sca}}$ are continuous on $\partial V_{\text{sim sys}}$. Based on the continuity and GSEPs (C-40) and (C-43), we have the following equations of $\vec{J}_{\text{sim}}^{\text{ES}}$ and $\vec{M}_{\text{sim}}^{\text{ES}}$ [44]:

$$\left[\Delta\mathcal{E}_{\text{sim}}\left(\vec{J}_{\text{sim}}^{\text{ES}},\vec{M}_{\text{sim}}^{\text{ES}}\right)\right]_{\vec{r}_{\text{sim}}\to\vec{r}}^{\tan} = -\left[\mathcal{E}_0\left(\vec{J}_{\text{sim}}^{\text{ES}},\vec{M}_{\text{sim}}^{\text{ES}}\right)\right]_{\vec{r}_0\to\vec{r}}^{\tan} \quad , \quad \vec{r}\in\partial V_{\text{sim sys}} \tag{4-36a}$$

$$\left[\Delta\mathcal{H}_{\text{sim}}\left(\vec{J}_{\text{sim}}^{\text{ES}},\vec{M}_{\text{sim}}^{\text{ES}}\right)\right]_{\vec{r}_{\text{sim}}\to\vec{r}}^{\tan} = -\left[\mathcal{H}_0\left(\vec{J}_{\text{sim}}^{\text{ES}},\vec{M}_{\text{sim}}^{\text{ES}}\right)\right]_{\vec{r}_0\to\vec{r}}^{\tan} \quad , \quad \vec{r}\in\partial V_{\text{sim sys}} \tag{4-36b}$$

In equation (4-36), operator $\Delta\mathcal{F}_{\text{sim}}(\vec{J},\vec{M})$ is defined as that $\Delta\mathcal{F}_{\text{sim}}(\vec{J},\vec{M}) = [\Delta\overset{\leftrightarrow}{G}{}_{\text{sim}}^{JF} * \vec{J} + \Delta\overset{\leftrightarrow}{G}{}_{\text{sim}}^{MF} * \vec{M}]_{\partial V_{\text{sim sys}}}$, where $(F,\mathcal{F}) = (E,\mathcal{E}),(H,\mathcal{H})$ and $\Delta\overset{\leftrightarrow}{G}{}_{\text{sim}}^{JF}$ and $\Delta\overset{\leftrightarrow}{G}{}_{\text{sim}}^{MF}$ are given in formulation (C-44); superscript "$\tan$" represents the tangential components of the corresponding fields; $\vec{r}_{\text{sim}}\in\text{int}V_{\text{sim sys}}$ and $\vec{r}_{\text{sim}}$ approaches $\vec{r}$ ; $\vec{r}_0\in\text{ext}V_{\text{sim sys}}$ and $\vec{r}_0$ approaches $\vec{r}$ . On boundary $\partial V_{\text{sim sys}}$, the tangential continuity of $\vec{E}^{\text{sca}}$ is equivalent to the tangential continuity of $\vec{H}^{\text{sca}}$ [123], so equation (4-36a) and equation (4-36b) are equivalent to each other. This implies that: we can unify variables by employing either equation (4-36a) or equation (4-36b).

**2) Variable Unification Scheme Based on the Definitions of the Equivalent Currents on $\partial V_{\text{sim sys}}$ (Scheme Developed in Literature [45])**

Inserting GSEP (C-42) into the definition (C-41) of $\vec{C}_{\text{sim}}^{\text{ES}}$, the following equations of $\vec{J}_{\text{sim}}^{\text{ES}}$ and $\vec{M}_{\text{sim}}^{\text{ES}}$ are immediately obtained[45]:

$$\left[\mathcal{E}_{\text{sim}}\left(\vec{J}_{\text{sim}}^{\text{ES}},\vec{M}_{\text{sim}}^{\text{ES}}\right)\right]_{\vec{r}_{\text{sim}}\to\vec{r}}^{\tan} = \hat{n}_{\text{sim}}^{-}\left(\vec{r}\right)\times\vec{M}_{\text{sim}}^{\text{ES}}\left(\vec{r}\right) \quad , \quad \vec{r}\in\partial V_{\text{sim sys}} \tag{4-37a}$$

$$\left[\mathcal{H}_{\text{sim}}\left(\vec{J}_{\text{sim}}^{\text{ES}},\vec{M}_{\text{sim}}^{\text{ES}}\right)\right]_{\vec{r}_{\text{sim}}\to\vec{r}}^{\tan} = \vec{J}_{\text{sim}}^{\text{ES}}\left(\vec{r}\right)\times\hat{n}_{\text{sim}}^{-}\left(\vec{r}\right) \quad , \quad \vec{r}\in\partial V_{\text{sim sys}} \tag{4-37b}$$

In equation (4-37), operator $\mathcal{E}_{\text{sim}}(\vec{J},\vec{M})$ is defined as that $\mathcal{E}_{\text{sim}}(\vec{J},\vec{M})=[\vec{G}_{\text{sim}}^{JE}*\vec{J}+\vec{G}_{\text{sim}}^{ME}*\vec{M}]_{\partial V_{\text{sim sys}}}$; operator $\mathcal{H}_{\text{sim}}(\vec{J},\vec{M})$ is defined as that $\mathcal{H}_{\text{sim}}(\vec{J},\vec{M})=[\vec{G}_{\text{sim}}^{JH}*\vec{J}+\vec{G}_{\text{sim}}^{MH}*\vec{M}]_{\partial V_{\text{sim sys}}}$. In the aspect of establishing the transformation between $\vec{J}_{\text{sim}}^{\text{ES}}$ and $\vec{M}_{\text{sim}}^{\text{ES}}$, equation (4-37a) is equivalent to equation (4-37b)[45] just like the equivalence between equation (4-36a) and equation (4-36b), and this conclusion will be proven in the following point 3). This implies that: we can unify variables by employing either equation (4-37a) or equation (4-37b).

**3) Equivalence Relationship Between the Above Two Variable Unification Schemes**

To prove the equivalence between scheme 1) and scheme 2), we equivalently rewrite scheme 1) as follows:

$$\left[\mathcal{E}_{\text{sim}}\left(\vec{J}_{\text{sim}}^{\text{ES}},\vec{M}_{\text{sim}}^{\text{ES}}\right)-\mathcal{E}_{0}\left(\vec{J}_{\text{sim}}^{\text{ES}},\vec{M}_{\text{sim}}^{\text{ES}}\right)\right]_{\vec{r}_{\text{sim}}\to\vec{r}}^{\tan}=-\left[\mathcal{E}_{0}\left(\vec{J}_{\text{sim}}^{\text{ES}},\vec{M}_{\text{sim}}^{\text{ES}}\right)\right]_{\vec{r}_{0}\to\vec{r}}^{\tan}\ ,\quad \vec{r}\in\partial V_{\text{sim sys}} \tag{4-38a}$$

$$\left[\mathcal{H}_{\text{sim}}\left(\vec{J}_{\text{sim}}^{\text{ES}},\vec{M}_{\text{sim}}^{\text{ES}}\right)-\mathcal{H}_{0}\left(\vec{J}_{\text{sim}}^{\text{ES}},\vec{M}_{\text{sim}}^{\text{ES}}\right)\right]_{\vec{r}_{\text{sim}}\to\vec{r}}^{\tan}=-\left[\mathcal{H}_{0}\left(\vec{J}_{\text{sim}}^{\text{ES}},\vec{M}_{\text{sim}}^{\text{ES}}\right)\right]_{\vec{r}_{0}\to\vec{r}}^{\tan}\ ,\quad \vec{r}\in\partial V_{\text{sim sys}} \tag{4-38b}$$

In fact, equations (4-38a) and (4-38b) can also be further rewritten as follows:

$$\begin{aligned}&\left[\mathcal{E}_{\text{sim}}\left(\vec{J}_{\text{sim}}^{\text{ES}},\vec{M}_{\text{sim}}^{\text{ES}}\right)\right]_{\vec{r}_{\text{sim}}\to\vec{r}}^{\tan}-\left[\text{P.V.}\mathcal{E}_{0}\left(\vec{J}_{\text{sim}}^{\text{ES}},\vec{M}_{\text{sim}}^{\text{ES}}\right)\right]^{\tan}-\hat{n}_{\text{sim}}^{-}\left(\vec{r}\right)\times\frac{1}{2}\vec{M}_{\text{sim}}^{\text{ES}}\left(\vec{r}\right)\\&=-\left[\text{P.V.}\mathcal{E}_{0}\left(\vec{J}_{\text{sim}}^{\text{ES}},\vec{M}_{\text{sim}}^{\text{ES}}\right)\right]^{\tan}-\frac{1}{2}\vec{M}_{\text{sim}}^{\text{ES}}\left(\vec{r}\right)\times\hat{n}_{\text{sim}}^{-}\left(\vec{r}\right)\ ,\quad \vec{r}\in\partial V_{\text{sim sys}}\end{aligned} \tag{4-39a}$$

$$\begin{aligned}&\left[\mathcal{H}_{\text{sim}}\left(\vec{J}_{\text{sim}}^{\text{ES}},\vec{M}_{\text{sim}}^{\text{ES}}\right)\right]_{\vec{r}_{\text{sim}}\to\vec{r}}^{\tan}-\left[\text{P.V.}\mathcal{H}_{0}\left(\vec{J}_{\text{sim}}^{\text{ES}},\vec{M}_{\text{sim}}^{\text{ES}}\right)\right]^{\tan}-\frac{1}{2}\vec{J}_{\text{sim}}^{\text{ES}}\left(\vec{r}\right)\times\hat{n}_{\text{sim}}^{-}\left(\vec{r}\right)\\&=-\left[\text{P.V.}\mathcal{H}_{0}\left(\vec{J}_{\text{sim}}^{\text{ES}},\vec{M}_{\text{sim}}^{\text{ES}}\right)\right]^{\tan}-\hat{n}_{\text{sim}}^{-}\left(\vec{r}\right)\times\frac{1}{2}\vec{J}_{\text{sim}}^{\text{ES}}\left(\vec{r}\right)\ ,\quad \vec{r}\in\partial V_{\text{sim sys}}\end{aligned} \tag{4-39b}$$

Obviously, equations (4-39a) and (4-39b) are just equations (4-37a) and (4-37b) respectively, and then the equivalence between schemes (4-36) and (4-37) is proven. In addition, it has been pointed out that equation (4-36a) is equivalent to equation (4-36b), so equation (4-37a) is also equivalent to equation (4-37b).

**4) Variable Unification**

Because of the equivalence between scheme 1) and scheme 2), we can unify variables based on either scheme 1) or scheme 2). Here, we select to use scheme 2), and provide four formally different transformations between $\vec{J}_{\text{sim}}^{\text{ES}}$ and $\vec{M}_{\text{sim}}^{\text{ES}}$.

**(4.1) Basing on the Definition of Equivalent Surface Magnetic Current and Electric-Current-Based Testing Method**

Inserting expansion formulation (4-33) into equation (4-37a), and testing equation (4-37a) with basis functions $\{\vec{b}_{\xi}^{J_{\rm s}}\}_{\xi=1}^{\Xi^{J_{\rm s}}}$, equation (4-37a) will be discretized into the following matrix equation[45]:

$$\overline{\overline{Z}}_{\rm sim}^{J_{\rm s}EJ_{\rm s}} \cdot \overline{a}^{J_{\rm s}} + \overline{\overline{Z}}_{\rm sim}^{J_{\rm s}EM_{\rm s}} \cdot \overline{a}^{M_{\rm s}} = \overline{\overline{C}}^{J_{\rm s}M_{\rm s}} \cdot \overline{a}^{M_{\rm s}} \tag{4-40}$$

where the elements of matrices $\overline{\overline{Z}}_{\rm sim}^{J_{\rm s}EJ_{\rm s}}$, $\overline{\overline{Z}}_{\rm sim}^{J_{\rm s}EM_{\rm s}}$, and $\overline{\overline{C}}^{J_{\rm s}M_{\rm s}}$ are calculated as follows:

$$z_{{\rm sim};\xi\zeta}^{J_{\rm s}EJ_{\rm s}} = \left\langle \vec{b}_{\xi}^{J_{\rm s}}, \mathcal{E}_{\rm sim}\left(\vec{b}_{\zeta}^{J_{\rm s}}, 0\right)\right\rangle_{\partial V_{\rm sim\ sys}^{-}} \tag{4-41a}$$

$$z_{{\rm sim};\xi\zeta}^{J_{\rm s}EM_{\rm s}} = \left\langle \vec{b}_{\xi}^{J_{\rm s}}, \mathcal{E}_{\rm sim}\left(0, \vec{b}_{\zeta}^{M_{\rm s}}\right)\right\rangle_{\partial V_{\rm sim\ sys}^{-}} \tag{4-41b}$$

$$c_{\xi\zeta}^{J_{\rm s}M_{\rm s}} = \left\langle \vec{b}_{\xi}^{J_{\rm s}}, \hat{n}_{\rm sim}^{-} \times \vec{b}_{\zeta}^{M_{\rm s}}\right\rangle_{\partial V_{\rm sim\ sys}} \tag{4-41c}$$

By solving the equation (4-40) of $\overline{a}^{J_{\rm s}}$ and $\overline{a}^{M_{\rm s}}$, the transformation from $\overline{a}^{M_{\rm s}}$ to $\overline{a}^{J_{\rm s}}$ is immediately obtained as follows[45]:

$$\overline{a}^{J_{\rm s}} = \underbrace{\left(\overline{\overline{Z}}_{\rm sim}^{J_{\rm s}EJ_{\rm s}}\right)^{-1} \cdot \left(\overline{\overline{C}}^{J_{\rm s}M_{\rm s}} - \overline{\overline{Z}}_{\rm sim}^{J_{\rm s}EM_{\rm s}}\right)}_{\overline{\overline{T}}_{\rm DESM}^{J_{\rm s}\leftarrow M_{\rm s}}} \cdot \overline{a}^{M_{\rm s}} \tag{4-42}$$

There exists one-to-one correspondence between EM currents and their expansion vectors, so above transformation from $\overline{a}^{M_{\rm s}}$ to $\overline{a}^{J_{\rm s}}$ is essentially the transformation from $\vec{M}_{\rm sim}^{\rm ES}$ to $\vec{J}_{\rm sim}^{\rm ES}$. Here, the subscript " DESM " in $\overline{\overline{T}}_{\rm DESM}^{J_{\rm s}\leftarrow M_{\rm s}}$ is to emphasize that: the transformation is derived from the definition of equivalent surface *M* (DESM).

**(4.2) Basing on the Definition of Equivalent Surface Magnetic Current and Magnetic-Current-Based Testing Method**

Inserting expansion formulation (4-33) into equation (4-37a), and testing equation (4-37a) with basis functions $\{\vec{b}_{\xi}^{M_{\rm s}}\}_{\xi=1}^{\Xi^{M_{\rm s}}}$, equation (4-37a) will be discretized into the following matrix equation[45]:

$$\overline{\overline{Z}}_{\rm sim}^{M_{\rm s}EJ_{\rm s}} \cdot \overline{a}^{J_{\rm s}} + \overline{\overline{Z}}_{\rm sim}^{M_{\rm s}EM_{\rm s}} \cdot \overline{a}^{M_{\rm s}} = \overline{\overline{C}}^{M_{\rm s}M_{\rm s}} \cdot \overline{a}^{M_{\rm s}} \tag{4-43}$$

where the elements of matrices $\overline{\overline{Z}}_{\rm sim}^{M_{\rm s}EJ_{\rm s}}$, $\overline{\overline{Z}}_{\rm sim}^{M_{\rm s}EM_{\rm s}}$, and $\overline{\overline{C}}^{M_{\rm s}M_{\rm s}}$ are calculated as follows:

$$z_{{\rm sim};\xi\zeta}^{M_{\rm s}EJ_{\rm s}} = \left\langle \vec{b}_{\xi}^{M_{\rm s}}, \mathcal{E}_{\rm sim}\left(\vec{b}_{\zeta}^{J_{\rm s}}, 0\right)\right\rangle_{\partial V_{\rm sim\ sys}^{-}} \tag{4-44a}$$

$$z_{{\rm sim};\xi\zeta}^{M_{\rm s}EM_{\rm s}} = \left\langle \vec{b}_{\xi}^{M_{\rm s}}, \mathcal{E}_{\rm sim}\left(0, \vec{b}_{\zeta}^{M_{\rm s}}\right)\right\rangle_{\partial V_{\rm sim\ sys}^{-}} \tag{4-44b}$$

$$c_{\xi\zeta}^{M_{\rm s}M_{\rm s}} = \left\langle \vec{b}_{\xi}^{M_{\rm s}}, \hat{n}_{\rm sim}^{-} \times \vec{b}_{\zeta}^{M_{\rm s}}\right\rangle_{\partial V_{\rm sim\ sys}} \tag{4-44c}$$

By solving the equation (4-43) of $\overline{a}^{J_{\rm s}}$ and $\overline{a}^{M_{\rm s}}$, the transformation from $\overline{a}^{J_{\rm s}}$ to $\overline{a}^{M_{\rm s}}$ is immediately obtained as follows[45]:

$$\bar{a}^{M_s} = \underbrace{\left(\bar{\bar{C}}^{M_s M_s} - \bar{\bar{Z}}_{\text{sim}}^{M_s EM_s}\right)^{-1} \cdot \bar{\bar{Z}}_{\text{sim}}^{M_s EJ_s}}_{\bar{\bar{T}}_{\text{DESM}}^{M_s \leftarrow J_s}} \cdot \bar{a}^{J_s} \tag{4-45}$$

There exists one-to-one correspondence between EM currents and their expansion vectors, so above transformation from $\bar{a}^{J_s}$ to $\bar{a}^{M_s}$ is essentially the transformation from $\vec{J}_{\text{sim}}^{\text{ES}}$ to $\vec{M}_{\text{sim}}^{\text{ES}}$.

**(4.3) Basing on the Definition of Equivalent Surface Electric Current and Electric-Current-Based Testing Method**

Inserting expansion formulation (4-33) into equation (4-37b), and testing equation (4-37b) with basis functions $\{\vec{b}_\xi^{J_s}\}_{\xi=1}^{\Xi^{J_s}}$, equation (4-37b) will be discretized into the following matrix equation[45]:

$$\bar{\bar{Z}}_{\text{sim}}^{J_s HJ_s} \cdot \bar{a}^{J_s} + \bar{\bar{Z}}_{\text{sim}}^{J_s HM_s} \cdot \bar{a}^{M_s} = \bar{\bar{C}}^{J_s J_s} \cdot \bar{a}^{J_s} \tag{4-46}$$

where the elements of matrices $\bar{\bar{Z}}_{\text{sim}}^{J_s HJ_s}$, $\bar{\bar{Z}}_{\text{sim}}^{J_s HM_s}$, and $\bar{\bar{C}}^{J_s J_s}$ are calculated as follows:

$$z_{\text{sim};\xi\zeta}^{J_s HJ_s} = \left\langle \vec{b}_\xi^{J_s}, \mathcal{H}_{\text{sim}}\left(\vec{b}_\zeta^{J_s}, 0\right)\right\rangle_{\partial V_{\text{sim sys}}^-} \tag{4-47a}$$

$$z_{\text{sim};\xi\zeta}^{J_s HM_s} = \left\langle \vec{b}_\xi^{J_s}, \mathcal{H}_{\text{sim}}\left(0, \vec{b}_\zeta^{M_s}\right)\right\rangle_{\partial V_{\text{sim sys}}^-} \tag{4-47b}$$

$$c_{\xi\zeta}^{J_s J_s} = \left\langle \vec{b}_\xi^{J_s}, \vec{b}_\zeta^{J_s} \times \hat{n}_{\text{sim}}^-\right\rangle_{\partial V_{\text{sim sys}}} \tag{4-47c}$$

By solving the equation (4-46) of $\bar{a}^{J_s}$ and $\bar{a}^{M_s}$, the transformation from $\bar{a}^{M_s}$ to $\bar{a}^{J_s}$ is immediately obtained as follows[45]:

$$\bar{a}^{J_s} = \underbrace{\left(\bar{\bar{C}}^{J_s J_s} - \bar{\bar{Z}}_{\text{sim}}^{J_s HJ_s}\right)^{-1} \cdot \bar{\bar{Z}}_{\text{sim}}^{J_s HM_s}}_{\bar{\bar{T}}_{\text{DESJ}}^{J_s \leftarrow M_s}} \cdot \bar{a}^{M_s} \tag{4-48}$$

Here, the subscript "DESJ" in $\bar{\bar{T}}_{\text{DESJ}}^{J_s \leftarrow M_s}$ is to emphasize that: the transformation is derived from the definition of equivalent surface $J$ (DESJ).

**(4.4) Basing on the Definition of Equivalent Surface Electric Current and Magnetic-Current-Based Testing Method**

Inserting expansion formulation (4-33) into equation (4-37b), and testing equation (4-37b) with basis functions $\{\vec{b}_\xi^{M_s}\}_{\xi=1}^{\Xi^{M_s}}$, equation (4-37b) will be discretized into the following matrix equation[45]:

$$\bar{\bar{Z}}_{\text{sim}}^{M_s HJ_s} \cdot \bar{a}^{J_s} + \bar{\bar{Z}}_{\text{sim}}^{M_s HM_s} \cdot \bar{a}^{M_s} = \bar{\bar{C}}^{M_s J_s} \cdot \bar{a}^{J_s} \tag{4-49}$$

where the elements of matrices $\bar{\bar{Z}}_{\text{sim}}^{M_s HJ_s}$, $\bar{\bar{Z}}_{\text{sim}}^{M_s HM_s}$, and $\bar{\bar{C}}^{M_s J_s}$ are calculated as follows:

$$z_{\text{sim};\xi\zeta}^{M_\text{s}HJ_\text{s}} = \left\langle \vec{b}_\xi^{M_\text{s}}, \mathcal{H}_\text{sim}\left(\vec{b}_\zeta^{J_\text{s}}, 0\right)\right\rangle_{\partial V_\text{sim sys}^-} \tag{4-50a}$$

$$z_{\text{sim};\xi\zeta}^{M_\text{s}HM_\text{s}} = \left\langle \vec{b}_\xi^{M_\text{s}}, \mathcal{H}_\text{sim}\left(0, \vec{b}_\zeta^{M_\text{s}}\right)\right\rangle_{\partial V_\text{sim sys}^-} \tag{4-50b}$$

$$c_{\xi\zeta}^{M_\text{s}J_\text{s}} = \left\langle \vec{b}_\xi^{M_\text{s}}, \vec{b}_\zeta^{J_\text{s}} \times \hat{n}_\text{sim}^-\right\rangle_{\partial V_\text{sim sys}} \tag{4-50c}$$

By solving the equation (4-49) of $\bar{a}^{J_\text{s}}$ and $\bar{a}^{M_\text{s}}$, the transformation from $\bar{a}^{J_\text{s}}$ to $\bar{a}^{M_\text{s}}$ is immediately obtained as follows[45]:

$$\bar{a}^{M_\text{s}} = \underbrace{\left(\overline{\overline{Z}}_\text{sim}^{M_\text{s}HM_\text{s}}\right)^{-1}\cdot\left(\overline{\overline{C}}^{M_\text{s}J_\text{s}} - \overline{\overline{Z}}_\text{sim}^{M_\text{s}HJ_\text{s}}\right)}_{\overline{\overline{T}}_\text{DESJ}^{M_\text{s}\leftarrow J_\text{s}}}\cdot\bar{a}^{J_\text{s}} \tag{4-51}$$

where subscript " DESJ " in $\overline{\overline{T}}_\text{DESJ}^{M_\text{s}\leftarrow J_\text{s}}$ has the same meaning as the one used in the $\overline{\overline{T}}_\text{DESJ}^{J_\text{s}\leftarrow M_\text{s}}$ in transformation (4-48).

So far, by discretizing the definitions of the equivalent surface sources into the corresponding matrix equations, we obtain four transformations between equivalent surface electric and magnetic currents —— transformations (4-42), (4-45), (4-48), and (4-51).

At the end of this section, we want to specifically emphasize that: there doesn't exist principle difference among transformations (4-42), (4-45), (4-48), and (4-51) (i.e., discretizing the definitions of $\vec{J}_\text{sim}^\text{ES}$ and $\vec{M}_\text{sim}^\text{ES}$ into matrix forms, and treating the matrix forms as the linear equations of $\bar{a}^{J_\text{s}}$ and $\bar{a}^{M_\text{s}}$, and then obtaining the transformations between $\bar{a}^{J_\text{s}}$ and $\bar{a}^{M_\text{s}}$ by solving the linear matrix equations), but there indeed exist some numerical differences among the transformations. Obviously, it is valuable to study the numerical differences. However, we will not discuss the topic immediately, and we will, firstly in the following Subsection 4.3.4, unify variables for operator $P_\text{sim sys}^\text{driving}$ and obtain DP-CMs. When whole WEP-ScaSys-CMT (i.e., WEP-MetSca-CMT & Vol-WEP-MatSca-CMT & Surf-WEP-MatSca-CMT & LS-WEP-ComSca-CMT) has been established, we will, in Section 6.2, do some numerical analysis for the above-mentioned four transformations, and the reasons doing like this are that: firstly, to prominent the principal line of WEP-ScaSys-CMT; secondly, some results given in Subsection 4.3.4 are necessary for doing the numerical analysis; thirdly, after establishing whole WEP-ScaSys-CMT formalism, we can obtain some relatively general conclusions instead of the special conclusions focusing on the simply connected material body shown in Figure 4-13.

### 4.3.4 DP-CMs and Their Orthogonality

Inserting the transformations derived in above Subsection 4.3.3 into the matrix form (4-34) of DPO, we immediately obtain the following matrix quadratic form (which only contains BVs)[45]:

$$P_{\text{sim sys}}^{\text{driving}} = \left(\bar{a}^{C_{\text{s}}}\right)^H \cdot \overline{\overline{P}}_{C_{\text{s}}}^{\text{driving}} \cdot \bar{a}^{C_{\text{s}}} \tag{4-52}$$

where $\overline{\overline{P}}_{C_{\text{s}}}^{\text{driving}}$ is given as follows:

$$\overline{\overline{P}}_{C_{\text{s}}}^{\text{driving}} = \begin{cases} \begin{bmatrix} \overline{\overline{I}}^{J_{\text{s}}} \\ \overline{\overline{T}}_{\text{DESS}}^{M_{\text{s}} \leftarrow J_{\text{s}}} \end{bmatrix}^H \cdot \overline{\overline{P}}_{\text{sim sys}}^{\text{driving}} \cdot \begin{bmatrix} \overline{\overline{I}}^{J_{\text{s}}} \\ \overline{\overline{T}}_{\text{DESS}}^{M_{\text{s}} \leftarrow J_{\text{s}}} \end{bmatrix} & , \left(\text{to select } \vec{J}_{\text{sim}}^{\text{ES}} \text{ as BV}\right) \\ \begin{bmatrix} \overline{\overline{T}}_{\text{DESS}}^{J_{\text{s}} \leftarrow M_{\text{s}}} \\ \overline{\overline{I}}^{M_{\text{s}}} \end{bmatrix}^H \cdot \overline{\overline{P}}_{\text{sim sys}}^{\text{driving}} \cdot \begin{bmatrix} \overline{\overline{T}}_{\text{DESS}}^{J_{\text{s}} \leftarrow M_{\text{s}}} \\ \overline{\overline{I}}^{M_{\text{s}}} \end{bmatrix} & , \left(\text{to select } \vec{M}_{\text{sim}}^{\text{ES}} \text{ as BV}\right) \end{cases} \tag{4-53}$$

where $\overline{\overline{I}}^{C_{\text{s}}}$ is a $\Xi^{C_{\text{s}}}$-order identity matrix; $\overline{\overline{T}}_{\text{DESS}}^{M_{\text{s}} \leftarrow J_{\text{s}}} = \overline{\overline{T}}_{\text{DESM}}^{M_{\text{s}} \leftarrow J_{\text{s}}}$ or $\overline{\overline{T}}_{\text{DESJ}}^{M_{\text{s}} \leftarrow J_{\text{s}}}$, and $\overline{\overline{T}}_{\text{DESS}}^{J_{\text{s}} \leftarrow M_{\text{s}}} = \overline{\overline{T}}_{\text{DESM}}^{J_{\text{s}} \leftarrow M_{\text{s}}}$ or $\overline{\overline{T}}_{\text{DESJ}}^{J_{\text{s}} \leftarrow M_{\text{s}}}$; subscript " DESS " is the abbreviation of terminology " definition of equivalent surface source".

Matrix $\overline{\overline{P}}_{C_{\text{s}}}^{\text{driving}}$ uniquely has the following Toeplitz's decomposition[116]:

$$\overline{\overline{P}}_{C_{\text{s}}}^{\text{driving}} = \overline{\overline{P}}_{C_{\text{s}};+}^{\text{driving}} + j\,\overline{\overline{P}}_{C_{\text{s}};-}^{\text{driving}} \tag{4-54}$$

where $\overline{\overline{P}}_{C_{\text{s}};+}^{\text{driving}} = [\overline{\overline{P}}_{C_{\text{s}}}^{\text{driving}} + (\overline{\overline{P}}_{C_{\text{s}}}^{\text{driving}})^H]/2$ and $\overline{\overline{P}}_{C_{\text{s}};-}^{\text{driving}} = [\overline{\overline{P}}_{C_{\text{s}}}^{\text{driving}} - (\overline{\overline{P}}_{C_{\text{s}}}^{\text{driving}})^H]/2j$. The DP-CMs of the simply connected material system shown in Figure 4-13 can be derived from solving the following generalized characteristic equation[45]:

$$\overline{\overline{P}}_{C_{\text{s}};-}^{\text{driving}} \cdot \bar{\alpha}_{C_{\text{s}};\xi}^{\text{driving}} = \lambda_{\text{sim sys};\xi}^{\text{driving}}\,\overline{\overline{P}}_{C_{\text{s}};+}^{\text{driving}} \cdot \bar{\alpha}_{C_{\text{s}};\xi}^{\text{driving}} \tag{4-55}$$

Inserting above characteristic vectors $\{\bar{\alpha}_{C_{\text{s}};\xi}^{\text{driving}}\}_{\xi=1}^{\Xi^{C_{\text{s}}}}$ into the transformations (4-42), (4-45), (4-48), and (4-51) given in Subsection 4.3.3, and then inserting characteristic vectors $\{\bar{\alpha}_{J_{\text{s}};\xi}^{\text{driving}}, \bar{\alpha}_{M_{\text{s}};\xi}^{\text{driving}}\}_{\xi=1}^{\Xi^{C_{\text{s}}}}$ into the expansion formulation (4-33) in Subsection 4.3.2, we can obtain characteristic equivalent surface currents $\{\vec{J}_{\text{sim};\xi}^{\text{ES}}, \vec{M}_{\text{sim};\xi}^{\text{ES}}\}_{\xi=1}^{\Xi^{C_{\text{s}}}}$. Inserting obtained $\{\vec{J}_{\text{sim};\xi}^{\text{ES}}, \vec{M}_{\text{sim};\xi}^{\text{ES}}\}_{\xi=1}^{\Xi^{C_{\text{s}}}}$ into the GSEP given in Appendix C5, we can obtain the corresponding various characteristic fields. Inserting obtained characteristic total fields into the volume equivalence principle given in Appendix A, we can obtain characteristic scattered volume currents $\{\vec{J}_{\text{sim};\xi}^{\text{SV}}, \vec{M}_{\text{sim};\xi}^{\text{SV}}\}_{\xi=1}^{\Xi^{C_{\text{s}}}}$.

Obviously, characteristic value $\lambda_{\text{sim sys};\xi}^{\text{driving}}$ and the modal power satisfy the following relationship:

$$\lambda_{\text{sim sys};\xi}^{\text{driving}} = \frac{\text{Im}\left\{P_{\text{sim sys};\xi}^{\text{driving}}\right\}}{\text{Re}\left\{P_{\text{sim sys};\xi}^{\text{driving}}\right\}} = \frac{\left(\bar{\alpha}_{C_s;\xi}^{\text{driving}}\right)^H \cdot \bar{\bar{P}}_{C_s;-}^{\text{driving}} \cdot \bar{\alpha}_{C_s;\xi}^{\text{driving}}}{\left(\bar{\alpha}_{C_s;\xi}^{\text{driving}}\right)^H \cdot \bar{\bar{P}}_{C_s;+}^{\text{driving}} \cdot \bar{\alpha}_{C_s;\xi}^{\text{driving}}}$$

$$= \frac{\text{Im}\left\{(1/2)\left\langle \vec{J}_{\text{sim};\xi}^{\text{SV}}, \vec{E}_{\xi}^{\text{inc}} \right\rangle_{V_{\text{sim sys}}} + (1/2)\left\langle \vec{M}_{\text{sim};\xi}^{\text{SV}}, \vec{H}_{\xi}^{\text{inc}} \right\rangle_{V_{\text{sim sys}}}\right\}}{\text{Re}\left\{(1/2)\left\langle \vec{J}_{\text{sim};\xi}^{\text{SV}}, \vec{E}_{\xi}^{\text{inc}} \right\rangle_{V_{\text{sim sys}}} + (1/2)\left\langle \vec{M}_{\text{sim};\xi}^{\text{SV}}, \vec{H}_{\xi}^{\text{inc}} \right\rangle_{V_{\text{sim sys}}}\right\}}$$

$$= \frac{\text{Im}\left\{-(1/2)\left\langle \vec{J}_{\text{sim};\xi}^{\text{ES}}, \vec{E}_{\xi}^{\text{inc}} \right\rangle_{\partial V_{\text{sim sys}}} - (1/2)\left\langle \vec{M}_{\text{sim};\xi}^{\text{ES}}, \vec{H}_{\xi}^{\text{inc}} \right\rangle_{\partial V_{\text{sim sys}}}\right\}}{\text{Re}\left\{-(1/2)\left\langle \vec{J}_{\text{sim};\xi}^{\text{ES}}, \vec{E}_{\xi}^{\text{inc}} \right\rangle_{\partial V_{\text{sim sys}}} - (1/2)\left\langle \vec{M}_{\text{sim};\xi}^{\text{ES}}, \vec{H}_{\xi}^{\text{inc}} \right\rangle_{\partial V_{\text{sim sys}}}\right\}} \tag{4-56}$$

where the third and fourth equalities are based on relationship (4-31).

Similarly to Section 3.2, it is easy to prove that the characteristic vectors derived from equation (4-55) satisfy the following orthogonality:

$$\text{Re}\left\{P_{\text{sim sys};\xi}^{\text{driving}}\right\}\delta_{\xi\zeta} = \left(\bar{\alpha}_{C_s;\xi}^{\text{driving}}\right)^H \cdot \bar{\bar{P}}_{C_s;+}^{\text{driving}} \cdot \bar{\alpha}_{C_s;\zeta}^{\text{driving}} \tag{4-57a}$$

$$\text{Im}\left\{P_{\text{sim sys};\xi}^{\text{driving}}\right\}\delta_{\xi\zeta} = \left(\bar{\alpha}_{C_s;\xi}^{\text{driving}}\right)^H \cdot \bar{\bar{P}}_{C_s;-}^{\text{driving}} \cdot \bar{\alpha}_{C_s;\zeta}^{\text{driving}} \tag{4-57b}$$

and

$$\underbrace{\left[\text{Re}\left\{P_{\text{sim sys};\xi}^{\text{driving}}\right\} + j\,\text{Im}\left\{P_{\text{sim sys};\xi}^{\text{driving}}\right\}\right]}_{P_{\text{sim sys};\xi}^{\text{driving}}}\delta_{\xi\zeta} = \left(\bar{\alpha}_{C_s;\xi}^{\text{driving}}\right)^H \cdot \underbrace{\left(\bar{\bar{P}}_{C_s;+}^{\text{driving}} + j\,\bar{\bar{P}}_{C_s;-}^{\text{driving}}\right)}_{\bar{\bar{P}}_{C_s}^{\text{driving}}} \cdot \bar{\alpha}_{C_s;\zeta}^{\text{driving}} \tag{4-58}$$

Based on orthogonality (4-58) and the physical meaning of the elements of matrix $\bar{\bar{P}}_{C_s}^{\text{driving}}$, we have the following orthogonality between characteristic fields and characteristic currents:

$$\begin{aligned} P_{\text{sim sys};\xi}^{\text{driving}}\delta_{\xi\zeta} &= (1/2)\left\langle \vec{J}_{\text{sim};\xi}^{\text{SV}}, \vec{E}_{\zeta}^{\text{inc}} \right\rangle_{V_{\text{sim sys}}} + (1/2)\left\langle \vec{M}_{\text{sim};\xi}^{\text{SV}}, \vec{H}_{\zeta}^{\text{inc}} \right\rangle_{V_{\text{sim sys}}} \\ &= -(1/2)\left\langle \vec{J}_{\text{sim};\xi}^{\text{ES}}, \vec{E}_{\zeta}^{\text{inc}} \right\rangle_{\partial V_{\text{sim sys}}} - (1/2)\left\langle \vec{M}_{\text{sim};\xi}^{\text{ES}}, \vec{H}_{\zeta}^{\text{inc}} \right\rangle_{\partial V_{\text{sim sys}}} \end{aligned} \tag{4-59}$$

where the second equality is based on relationship (4-31). Comparing formulations (4-59) and (4-15), it can be found out that: Vol-WEP-MatSca-CMT (this section) and Surf-WEP-MatSca-CMT (Section 4.2) have the same physical destination —— constructing a series of fundamental modes making frequency-domain DPO decoupled (i.e. constructing a series of steadily working modes not having net energy exchange in any integral period).

In the following Subsection 4.3.5, we will see that: above orthogonality (4-59) plays key role in the process to derive the explicit expressions of the modal expansion coefficients.

### 4.3.5 DP-CM-Based Modal Expansion

Based on the completeness of the DP-CMs, any one working mode can be expanded in terms of the DP-CMs as follows:

$$\vec{E}^{\text{inc}}(\vec{r}) = \sum_{\xi=1}^{\Xi^{C_s}} c_\xi \vec{E}_\xi^{\text{inc}}(\vec{r}) \quad , \quad \vec{r} \in V_{\text{sim sys}} \tag{4-60a}$$

$$\vec{H}^{\text{inc}}(\vec{r}) = \sum_{\xi=1}^{\Xi^{C_s}} c_\xi \vec{H}_\xi^{\text{inc}}(\vec{r}) \quad , \quad \vec{r} \in V_{\text{sim sys}} \tag{4-60b}$$

Testing modal expansion formulations (4-60a) and (4-60b) with basis functions $\{\vec{J}_{\text{sim};\xi}^{\text{SV}} / 2\}_{\xi=1}^{\Xi^{C_s}}$ and $\{\vec{M}_{\text{sim};\xi}^{\text{SV}} / 2\}_{\xi=1}^{\Xi^{C_s}}$ respectively, and summing the two formulations, we obtain the following equation:

$$\begin{aligned}
&(1/2)\langle \vec{J}_{\text{sim};\xi}^{\text{SV}}, \vec{E}^{\text{inc}} \rangle_{V_{\text{sim sys}}} + (1/2)\langle \vec{M}_{\text{sim};\xi}^{\text{SV}}, \vec{H}^{\text{inc}} \rangle_{V_{\text{sim sys}}} \\
=\ & (1/2)\Big\langle \vec{J}_{\text{sim};\xi}^{\text{SV}}, \sum_{\zeta=1}^{\Xi^{C_s}} c_\zeta \vec{E}_\zeta^{\text{inc}} \Big\rangle_{V_{\text{sim sys}}} + (1/2)\Big\langle \vec{M}_{\text{sim};\xi}^{\text{SV}}, \sum_{\zeta=1}^{\Xi^{C_s}} c_\zeta \vec{H}_\zeta^{\text{inc}} \Big\rangle_{V_{\text{sim sys}}} \\
=\ & \sum_{\zeta=1}^{\Xi^{C_s}} c_\zeta \Big[ (1/2)\langle \vec{J}_{\text{sim};\xi}^{\text{SV}}, \vec{E}_\zeta^{\text{inc}} \rangle_{V_{\text{sim sys}}} + (1/2)\langle \vec{M}_{\text{sim};\xi}^{\text{SV}}, \vec{H}_\zeta^{\text{inc}} \rangle_{V_{\text{sim sys}}} \Big] \ , \ \xi = 1, 2, \cdots, \Xi^{C_s}
\end{aligned} \tag{4-61}$$

where the second equality is based on the linear property of inner product. Applying orthogonality (4-59) to equation (4-61), we have that

$$(1/2)\langle \vec{J}_{\text{sim};\xi}^{\text{SV}}, \vec{E}^{\text{inc}} \rangle_{V_{\text{sim sys}}} + (1/2)\langle \vec{M}_{\text{sim};\xi}^{\text{SV}}, \vec{H}^{\text{inc}} \rangle_{V_{\text{sim sys}}} = c_\xi P_{\text{sim sys};\xi}^{\text{driving}} \tag{4-62}$$

In general, we have that $P_{\text{sim sys};\xi}^{\text{driving}} \neq 0$ for material systems, so

$$\begin{aligned}
c_\xi &= \frac{(1/2)\langle \vec{J}_{\text{sim};\xi}^{\text{SV}}, \vec{E}^{\text{inc}} \rangle_{V_{\text{sim sys}}} + (1/2)\langle \vec{M}_{\text{sim};\xi}^{\text{SV}}, \vec{H}^{\text{inc}} \rangle_{V_{\text{sim sys}}}}{P_{\text{sim sys};\xi}^{\text{driving}}} \\
&= \frac{-(1/2)\langle \vec{J}_{\text{sim};\xi}^{\text{ES}}, \vec{E}^{\text{inc}} \rangle_{\partial V_{\text{sim sys}}} - (1/2)\langle \vec{M}_{\text{sim};\xi}^{\text{ES}}, \vec{H}^{\text{inc}} \rangle_{\partial V_{\text{sim sys}}}}{P_{\text{sim sys};\xi}^{\text{driving}}}
\end{aligned} \tag{4-63}$$

where the second equality is based on relationship (4-31). Formulation (4-63) is just the explicit expression for determining the expansion coefficients used in modal expansion formulation (4-60).

To verify the validities of the pivotal conclusions given in this section, we provide some typical numerical examples in the following Subsection 4.3.6.

### 4.3.6 Numerical Examples Corresponding to Typical Structures

The Green's function theory for inhomogeneous anisotropic medium has not been mature, so all of the examples given in this subsection are homogeneous isotropic material systems. So, what values do the examples have? Our answers to this question are that:

**1)** The DPO used in this subsection is new $P_{\text{sim sys}}^{\text{driving}} = -(1/2) < \vec{J}_{\text{sim}}^{\text{ES}}, \vec{E}^{\text{inc}} >_{\partial V_{\text{sim sys}}} -(1/2) < \vec{M}_{\text{sim}}^{\text{ES}}, \vec{H}^{\text{inc}} >_{\partial V_{\text{sim sys}}}$ rather than traditional $P_{\text{sim sys}}^{\text{driving}} = -(1/2) < \vec{J}_{\text{sim}}^{\text{ES}}, \vec{E}_{-}^{\text{tot}} - \vec{E}_{+}^{\text{sca}} >_{\partial V_{\text{sim sys}}} -(1/2) < \vec{M}_{\text{sim}}^{\text{ES}}, \vec{H}_{-}^{\text{tot}} - \vec{H}_{+}^{\text{sca}} >_{\partial V_{\text{sim sys}}}$, and this has been emphasized in Subsection 4.3.2. One of the main destinations of this subsection is to confirm the validity of the new DPO.

**2)** Another destination of this subsection is to confirm the validities of the variable unification schemes proposed in Subsection 4.3.3.

In what follows, we will, respectively based on the theories established in literature [34] and this section, construct the DP-CMs of the lossless homogeneous isotropic material cylinder shown in Figure 4-14 (whose radius & height are 5.25mm & 4.60mm respectively and relative permeability & relative permittivity are 2 & 18 respectively), and do some necessary analysis and discussions for the obtained results.

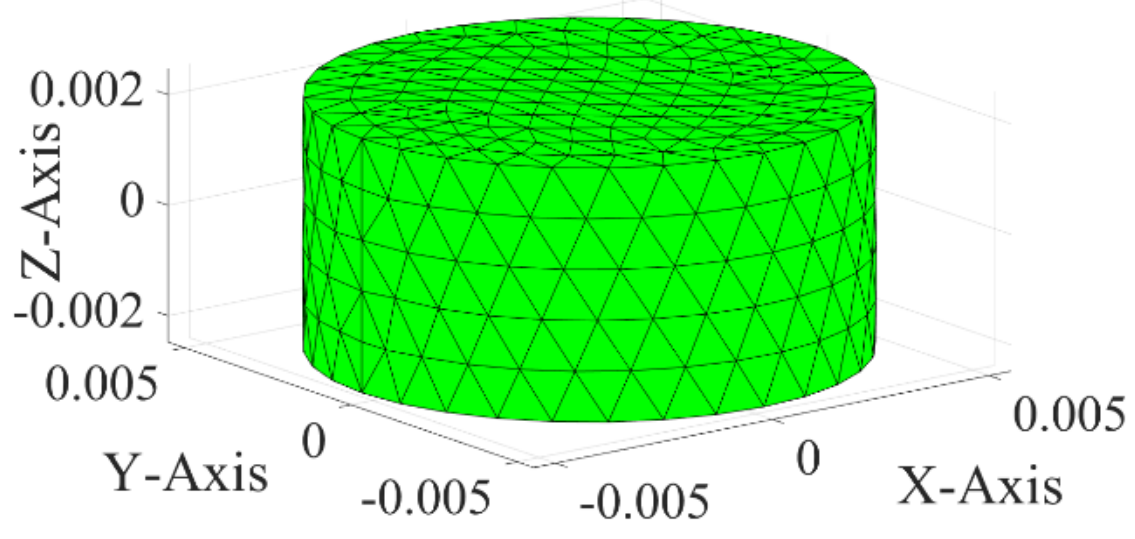

Figure 4-14 The topological structure and surface triangular meshes of a simply connected material cylinder whose radius and height are 5.25mm and 4.60mm

Focusing on the material system shown in Figure 4-14, the characteristic quantity curves corresponding to some typical DP-CMs derived from the formulation given in literature [34] are illustrated in Figure 4-15.

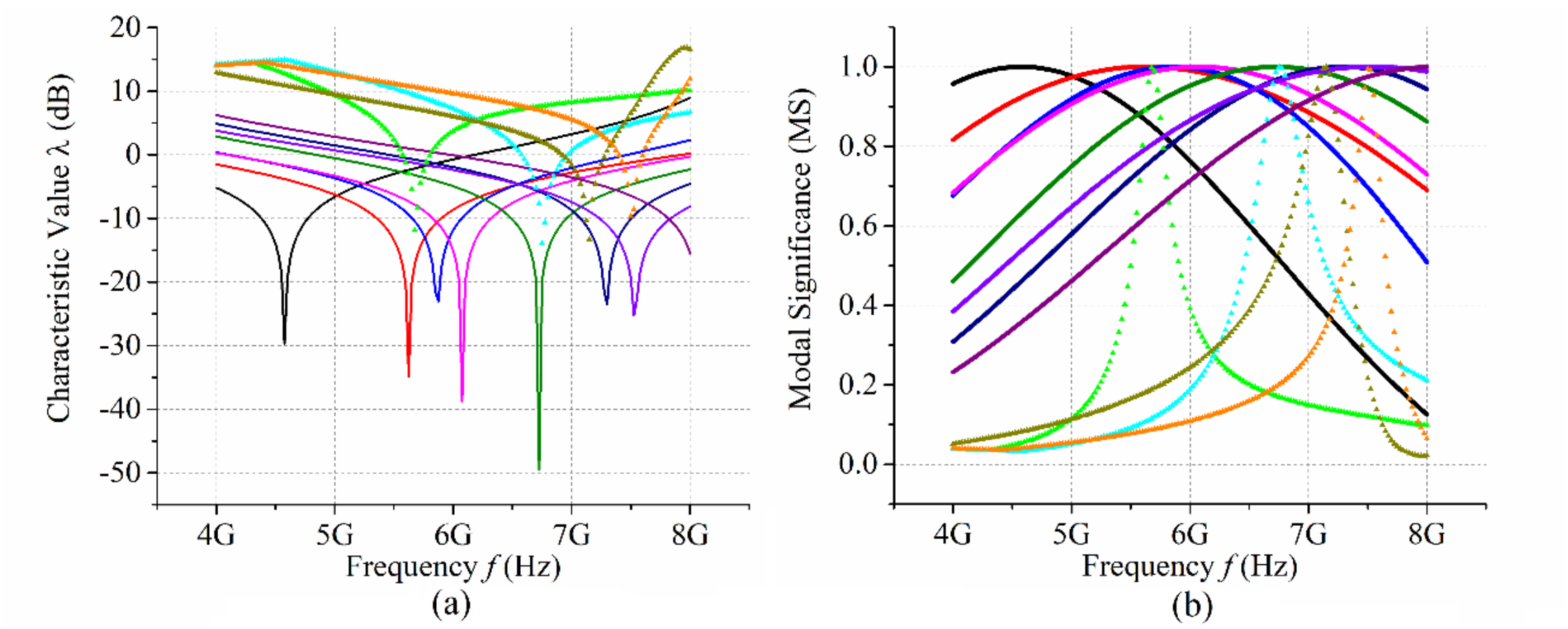

Figure 4-15 The characteristic quantity curves corresponding to several typical CMs (of the simply connected material system shown in Figure 4-14) derived from the theory given in literature [34]. (a) characteristic value dB curves; (b) MS curves

As pointed out in many literatures[35,42,44-49], the modes corresponding to the solid lines shown in Figure 4-15 are spurious modes. Literatures [35,42,44,45] and Subsections 4.3.2 and 4.3.3 have given the reason leading to the spurious modes —— being lack of proper variable unification pretreatment before constructing CMs by orthogonalizing SIE[34] operator (4-19). Based on SIE[34] operator (4-19) and employing variable unification scheme (4-36), literature [44] realized the effective suppression for the spurious modes, and can obtain the results shown in Figures 4-16 and 4-17, where the BVs selected in Figures 4-16 and 4-17 are equivalent surface electric and magnetic currents respectively, and the corresponding transformations from the BVs to the dependent variables are established basing on equations (4-36b) and (4-36a) respectively.

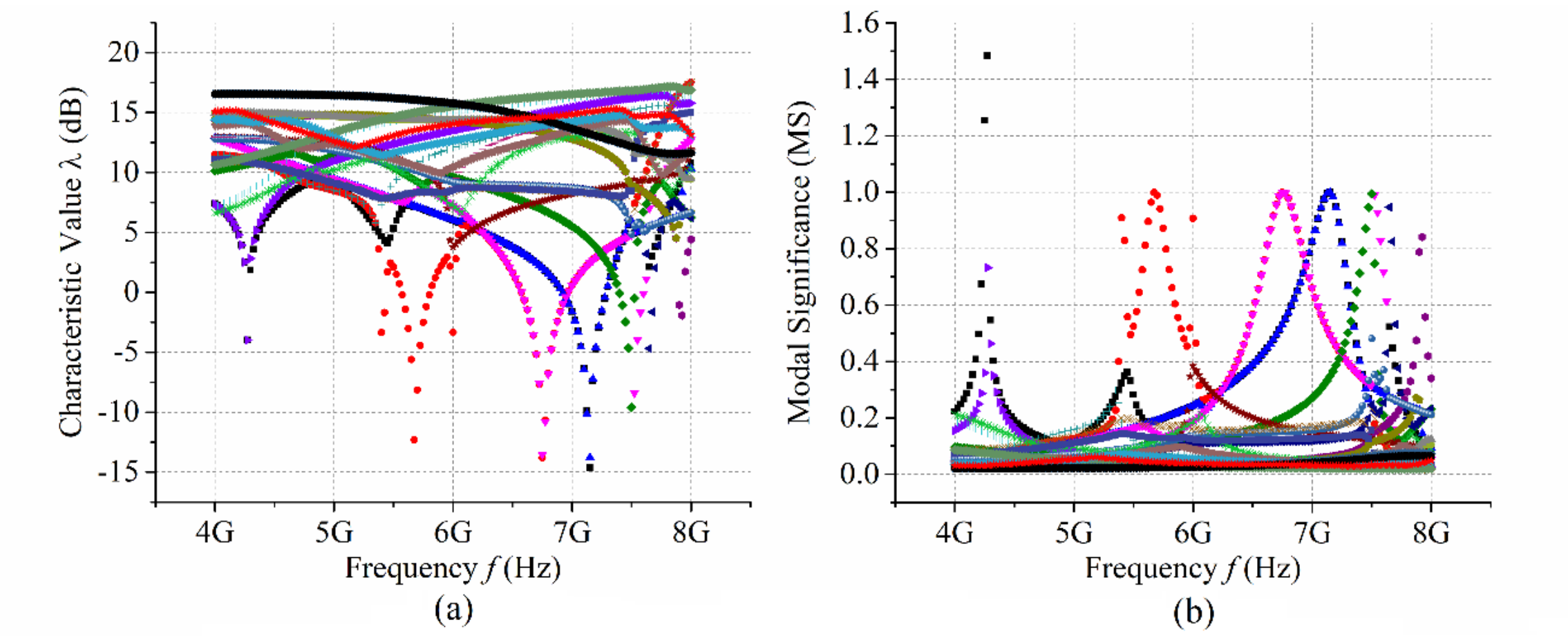

Figure 4-16 The characteristic quantity curves corresponding to several typical CMs (of the simply connected material system shown in Figure 4-14) derived from the SIE operator (4-19) with variable unification scheme (4-36b). (a) characteristic value dB curves; (b) MS curves

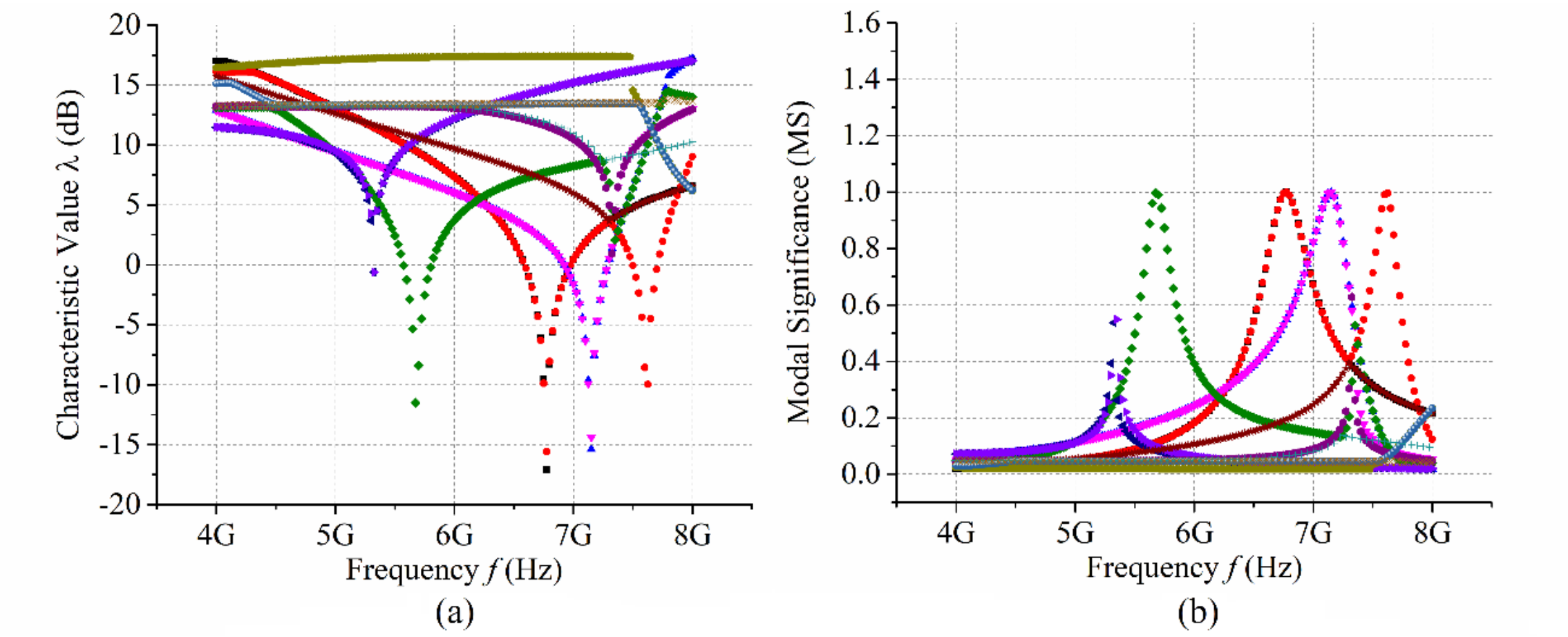

Figure 4-17 The characteristic quantity curves corresponding to several typical CMs (of the simply connected material system shown in Figure 4-14) derived from the SIE operator (4-19) with variable unification scheme (4-36a). (a) characteristic value dB curves; (b) MS curves

It is thus clear that variable unification scheme (4-36) indeed has ability to suppress the spurious modes, but the obtained results are not completely acceptable.

Focusing on the material system shown in Figure 4-14, the characteristic quantity curves corresponding to some typical DP-CMs derived from the formulation given in Subsections 4.3.2 and 4.3.3 are illustrated in Figures 4-18 and 4-19, where Figure 4-18 corresponds to variable unification scheme (4-51) and Figure 4-19 corresponds to variable unification scheme (4-42).

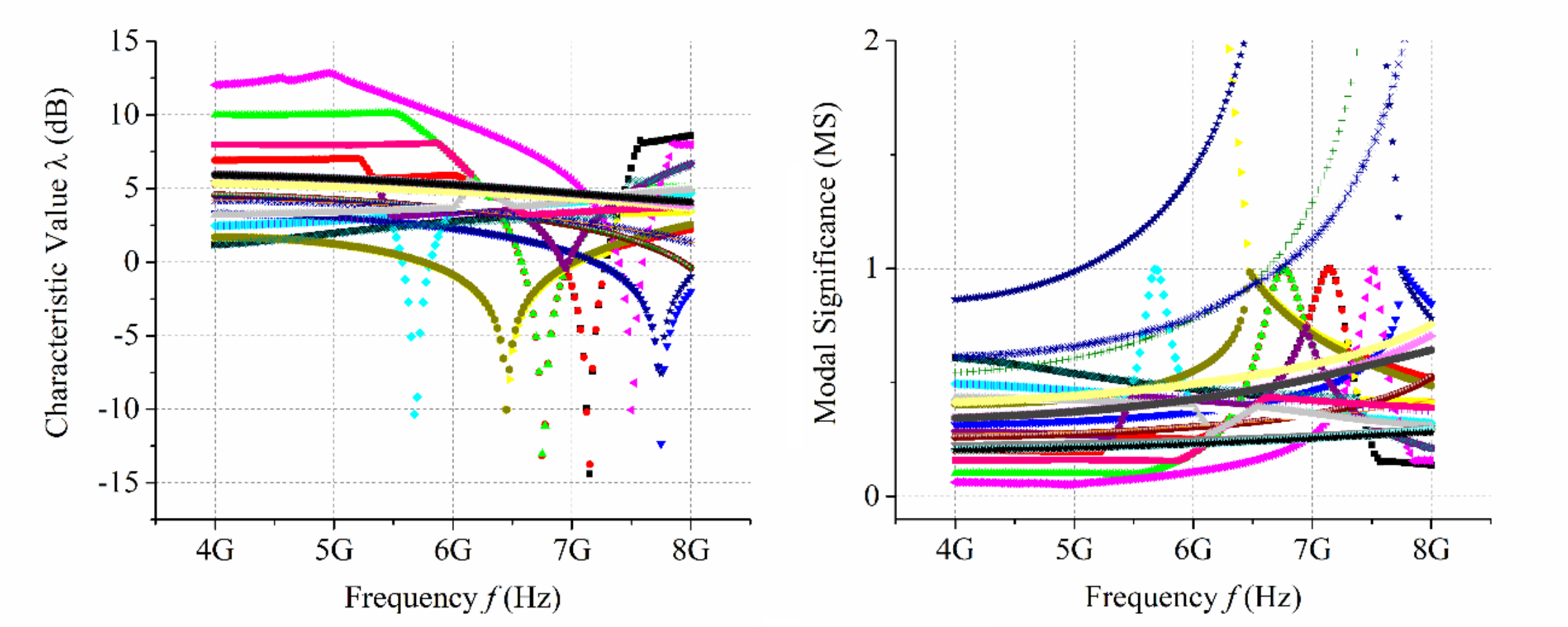

(a) (b)

Figure 4-18 The characteristic quantity curves corresponding to several typical DP-CMs (of the simply connected material system shown in Figure 4-14) derived from the DPO (4-32) with variable unification scheme (4-51). (a) characteristic value dB curves; (b) MS curves

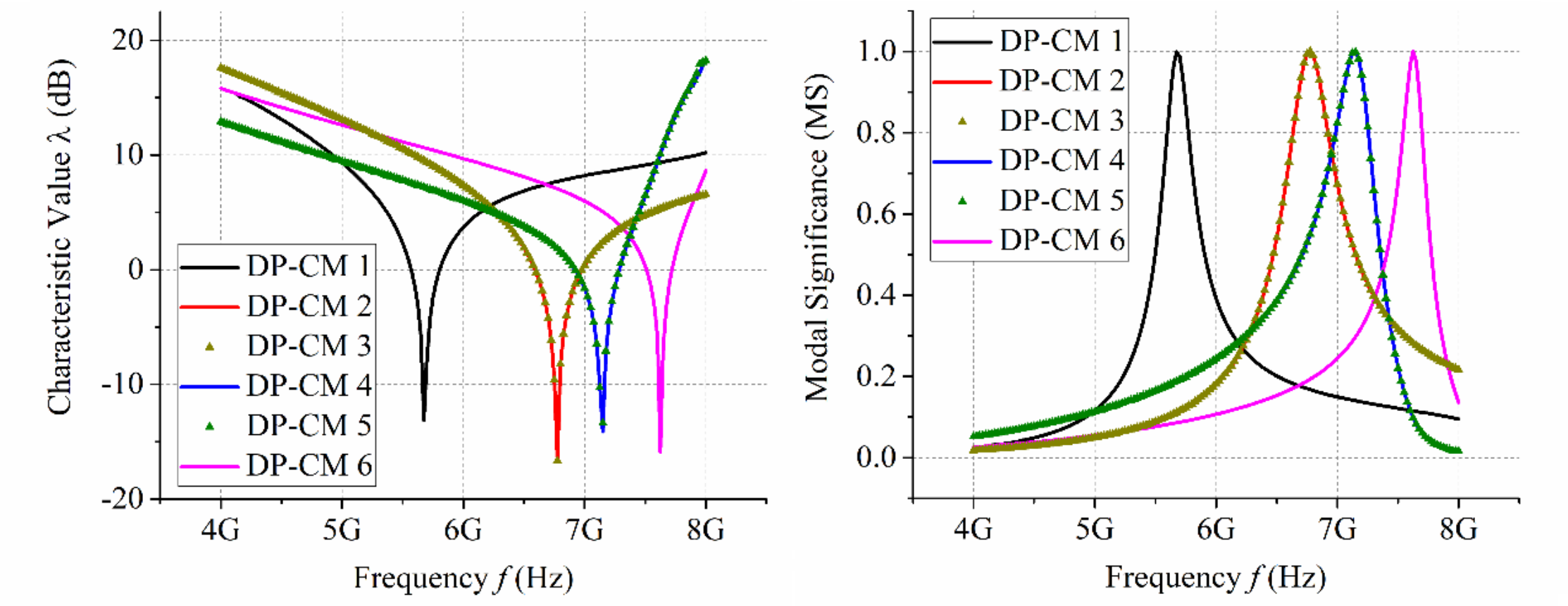

(a) (b)

Figure 4-19 The characteristic quantity curves corresponding to several typical DP-CMs (of the simply connected material system shown in Figure 4-14) derived from the DPO (4-32) with variable unification scheme (4-42). (a) characteristic value dB curves; (b) MS curves

By comparing all of the results mentioned above, it is easy to find out that: the modal set derived from the original SIE operator[34] (which is lack of proper variable unification) contains both all physical modes and some spurious modes; for the material system considered in this subsection (which is lossless homogeneous isotropic material cylinder and with material parameters $\vec{\mu}_{\text{sim}}^{\text{r}} = \vec{I}2$ and $\vec{\varepsilon}_{\text{sim}}^{\text{r}} = \vec{I}18$), the results derived from orthogonalizing the DPO (4-32) with variable unification scheme (4-42) are the most acceptable (as illustrated in Figure 4-19). Here, we want to emphasize that: the above-mentioned two conclusions are not generally correct; original SIE operator[34] sometimes cannot provide enough physical modes (for derails see the Section 6.2 of this dissertation); for some other material systems, the results derived from orthogonalizing the DPO (4-32) with variable unification scheme (4-42) will be worse than the results derived from orthogonalizing the DPO (4-32) with variable unification scheme (4-51) (for derails see the Section 6.2 of this dissertation).

Taking the calculated results shown in Figure 4-19, i.e. the DPO (4-32) with variable unification scheme (4-42), as examples, we provide the distributions of the modal equivalent surface currents, modal scattered volume currents, modal fields, and modal radiation patterns corresponding to some typical DP-CMs as below, to facilitate readers' references and comparisons.

It is easy, from Figure 4-19, to find out that DP-CM1 is "resonant" at 5.675GHz. For the "resonant" DP-CM, its equivalent surface magnetic current distribution and its equivalent surface electric current distribution are illustrated in Figure 4-20(a) and Figure 4-20(b) respectively, and its internal surface tangential total electric field distribution and its internal surface tangential total magnetic field distribution are illustrated in Figure 4-21(a) and Figure 4-21(b) respectively. Obviously, the distributions shown in Figure 4-20 and Figure 4-21 satisfy relationship (C-41) and relationship (4-37) ①. For the "resonant" DP-CM, its modal scattered volume electric current distribution and its modal scattered volume magnetic current distribution are illustrated in Figure 4-22 and Figure 4-23 respectively, and its modal incident electric field and its modal incident magnetic field distributing on the material cylinder are illustrated in Figure 4-24 and Figure 4-25 respectively. In addition, we also provide the modal radiation pattern of the "resonant" DP-CM as shown in Figure 4-26.

① Here, there exist some errors. When the meshes become more fine, and the order of the basis functions becomes higher, the errors will become smaller.

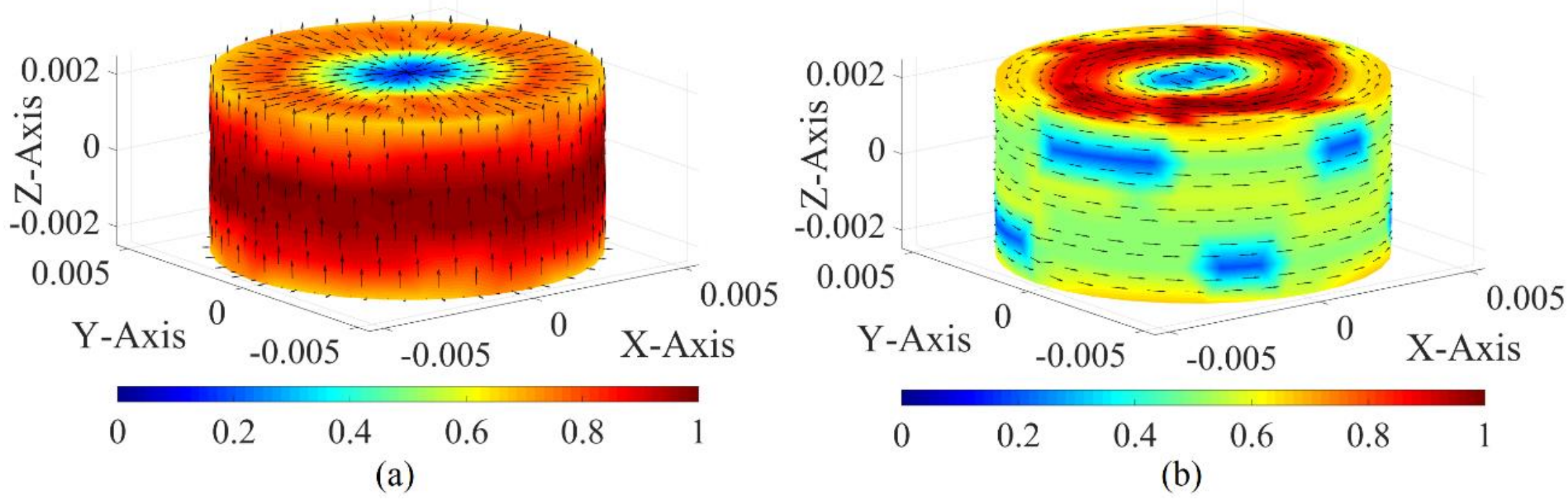

Figure 4-20 The modal equivalent source distributions of the DP-CM1 working at 5.675GHz and shown in Figure 4-19. (a) equivalent surface magnetic current; (b) equivalent surface electric current

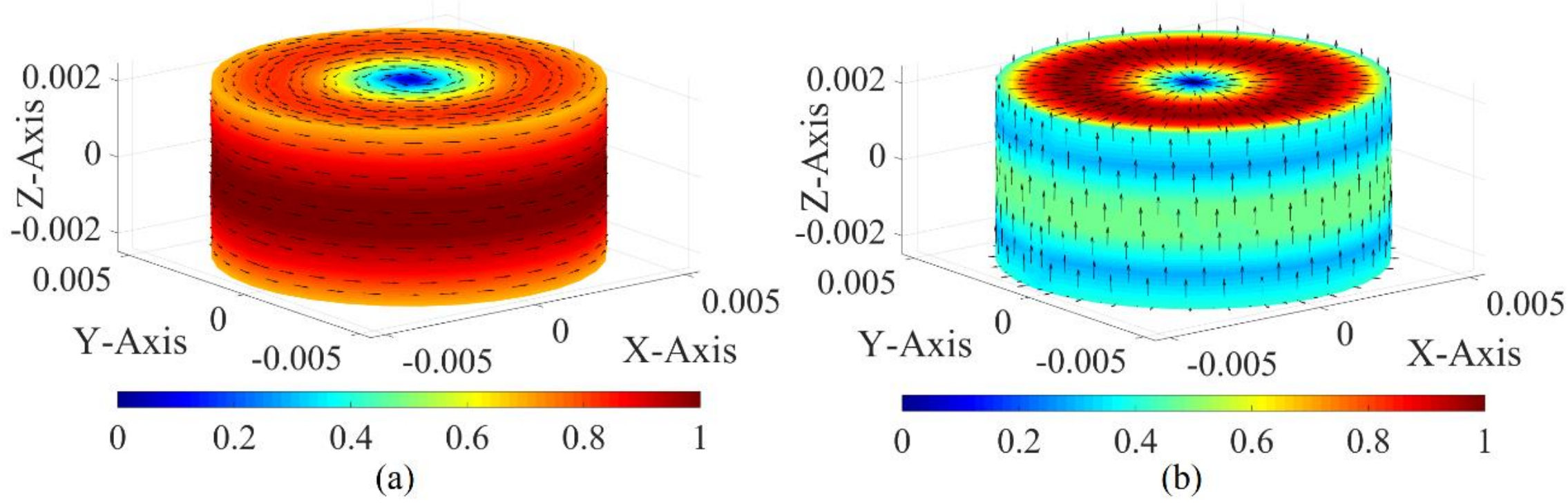

Figure 4-21 The inner surface tangential modal total field distributions of the DP-CM1 working at 5.675GHz and shown in Figure 4-19. (a) tangential total electric field; (b) tangential total magnetic field

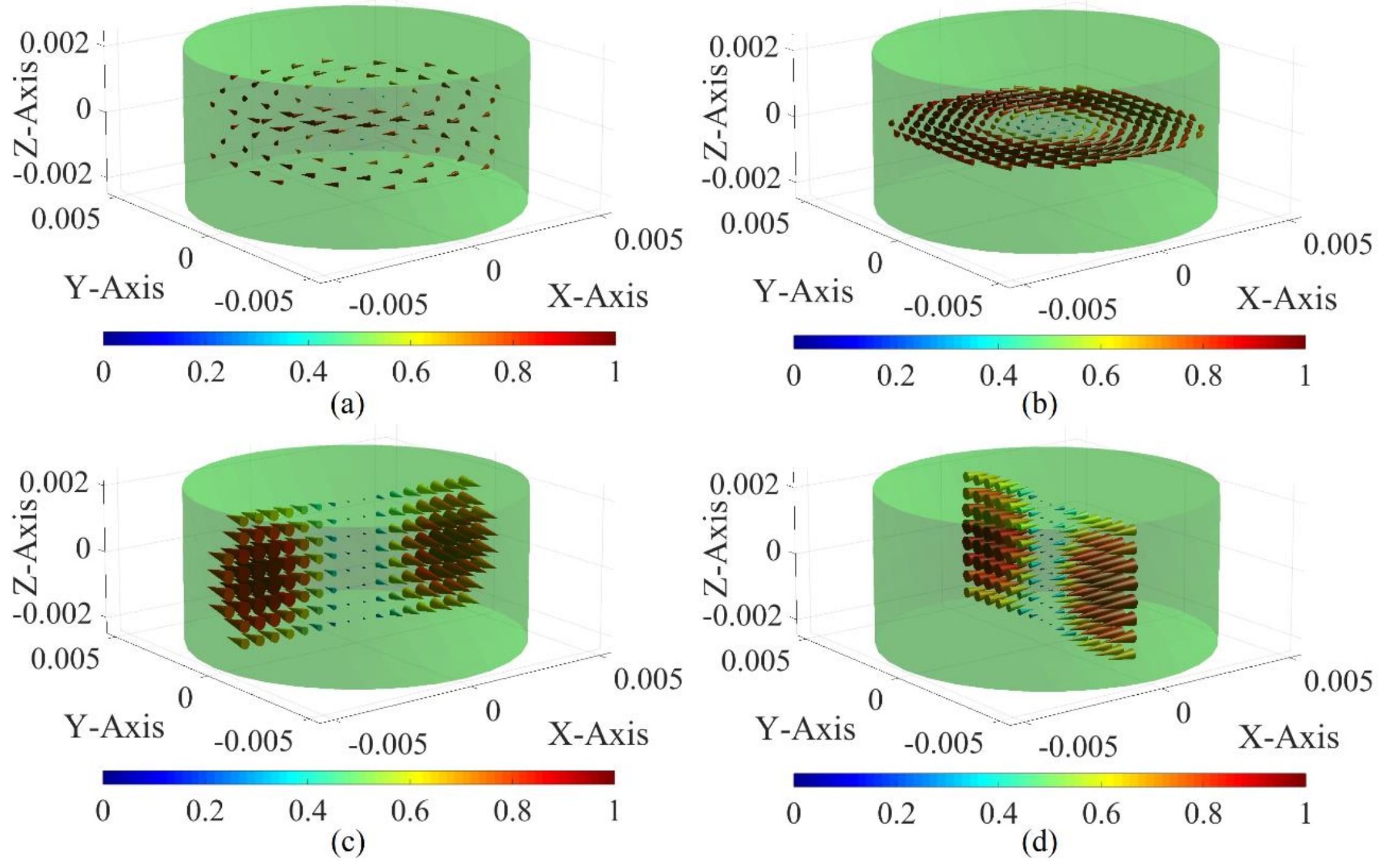

Figure 4-22 The modal scattered volume electric current distributions of the DP-CM1 working at 5.675GHz and shown in Figure 4-19. (a) the distribution on whole material system; (b) the distribution on xOy surface; (c) the distribution on xOz surface; (d) the distribution on yOz surface

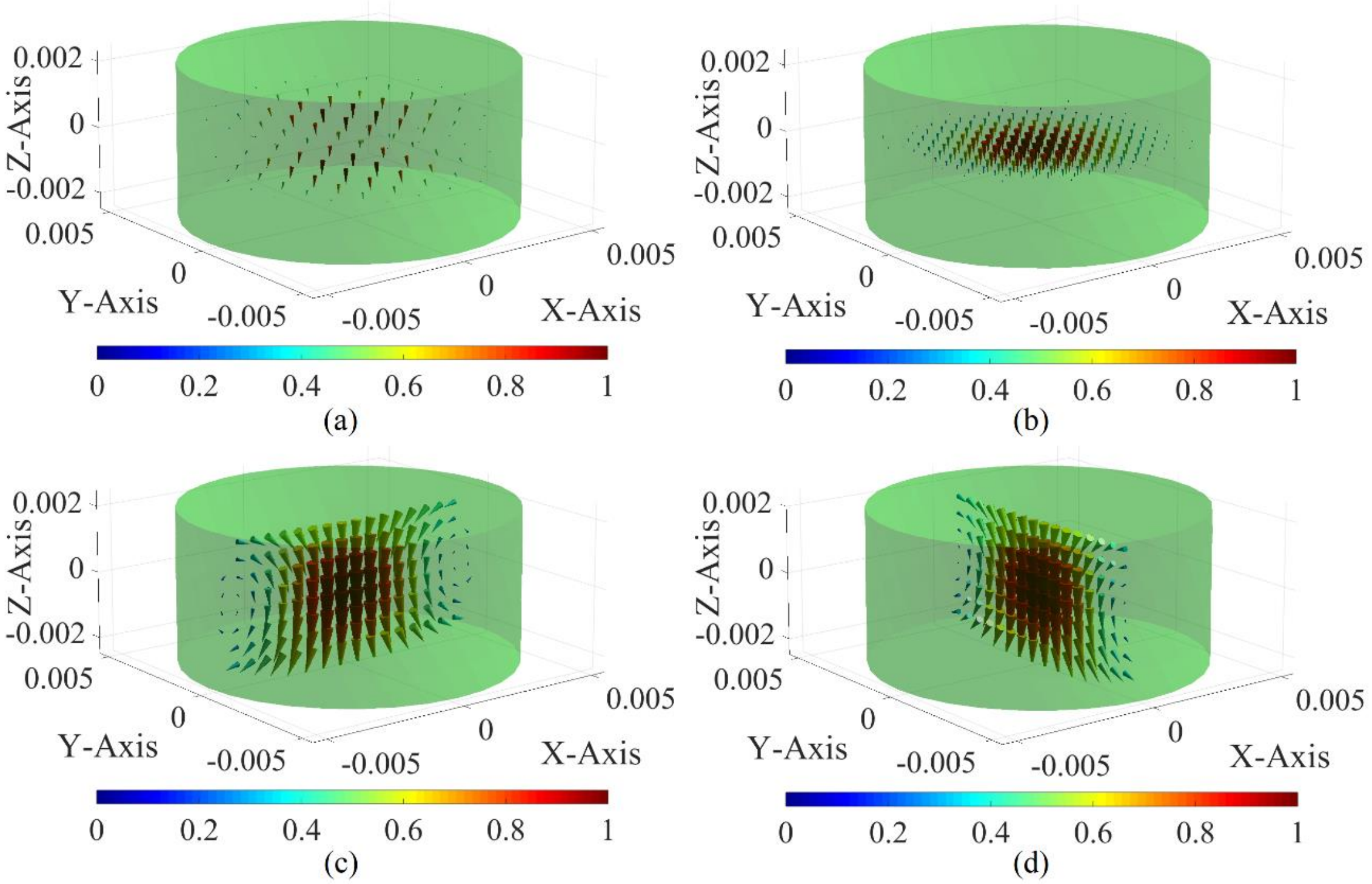

Figure 4-23 The modal scattered volume magnetic current distributions of the DP-CM1 working at 5.675GHz and shown in Figure 4-19. (a) the distribution on whole material system; (b) the distribution on xOy surface; (c) the distribution on xOz surface; (d) the distribution on yOz surface

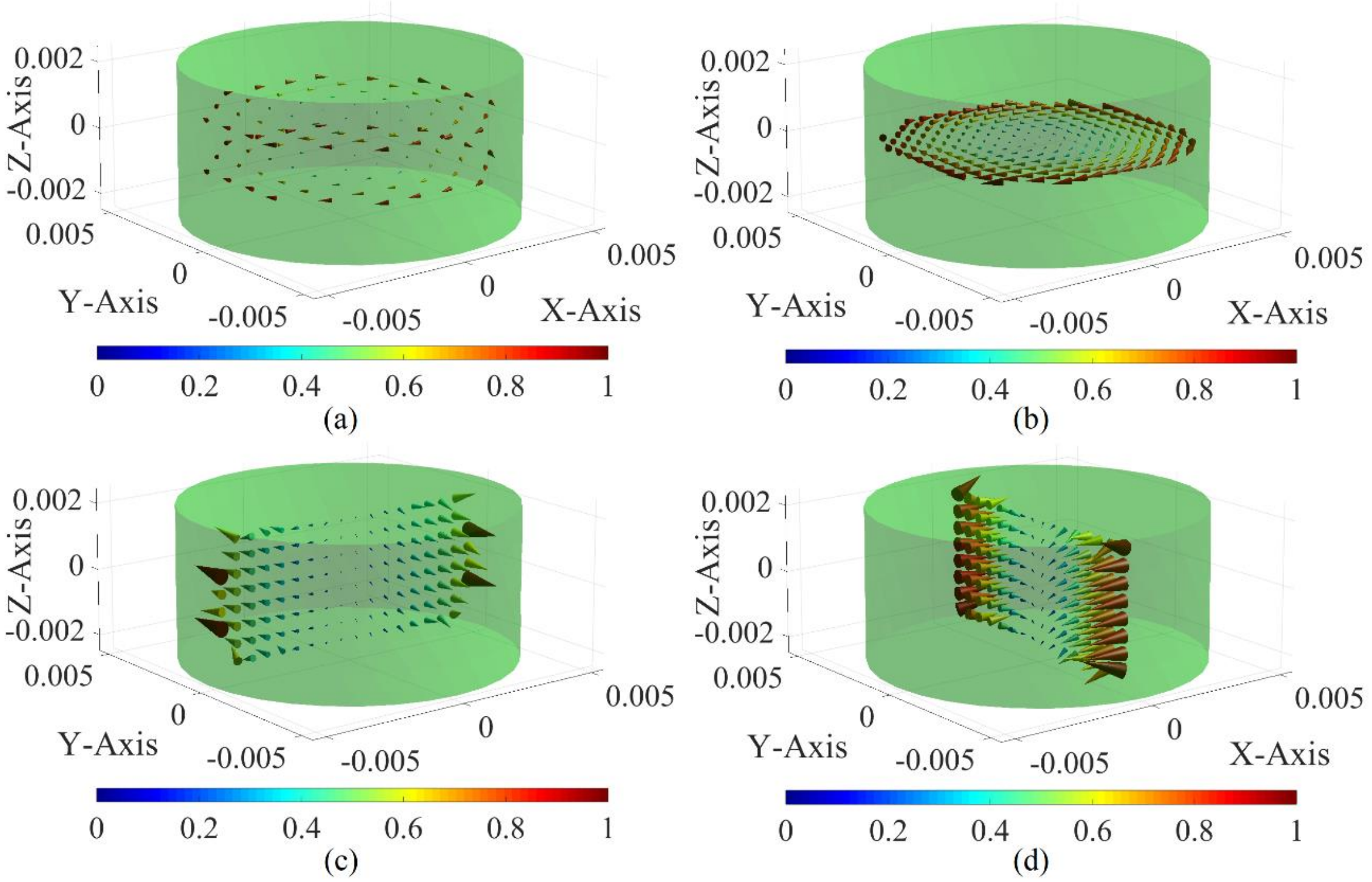

Figure 4-24 The modal incident electric field distributions of the DP-CM1 working at 5.675GHz and shown in Figure 4-19. (a) the distribution on whole material system; (b) the distribution on xOy surface; (c) the distribution on xOz surface; (d) the distribution on yOz surface

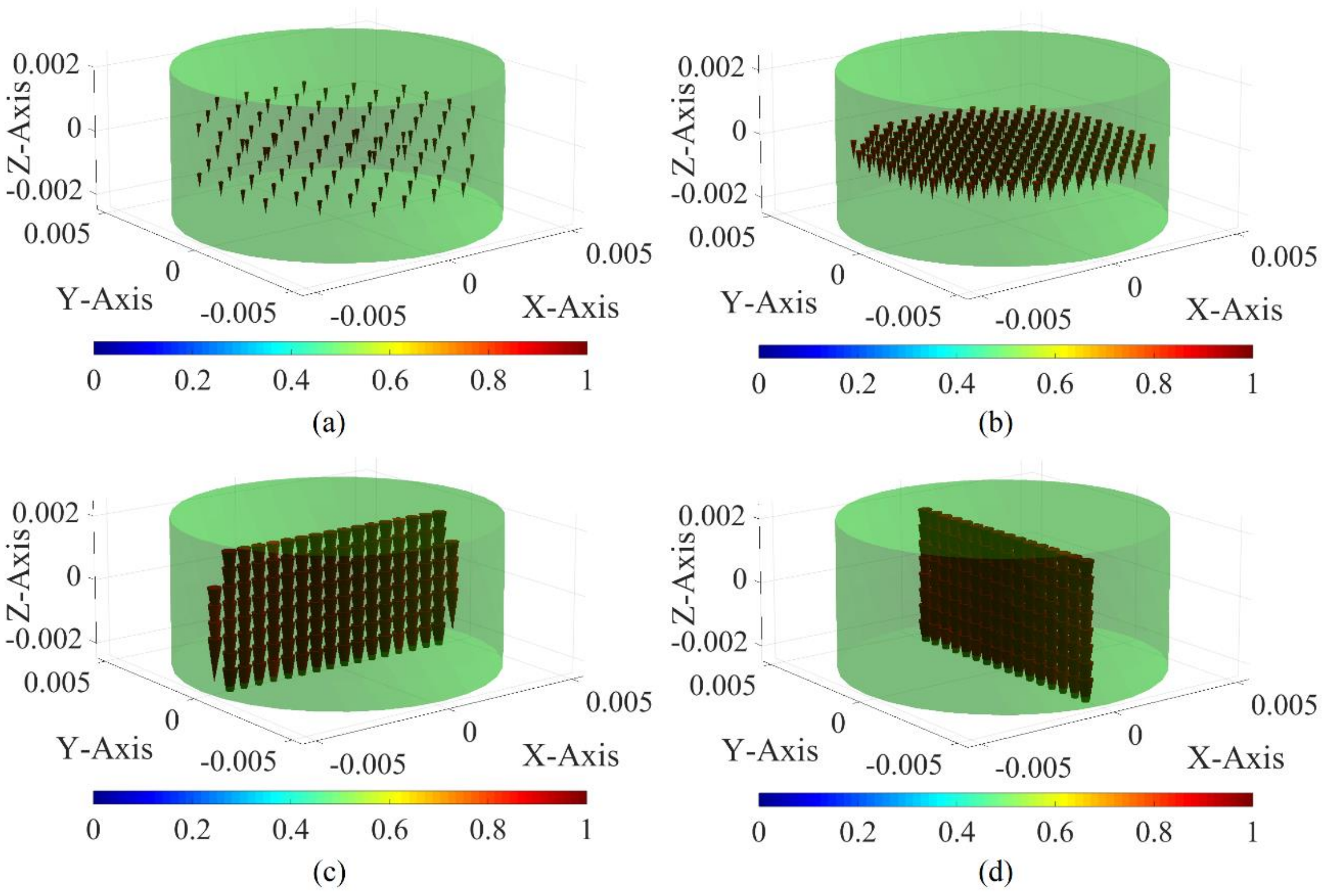

Figure 4-25 The modal incident magnetic field distributions of the DP-CM1 working at 5.675GHz and shown in Figure 4-19. (a) the distribution on whole material system; (b) the distribution on xOy surface; (c) the distribution on xOz surface; (d) the distribution on yOz surface

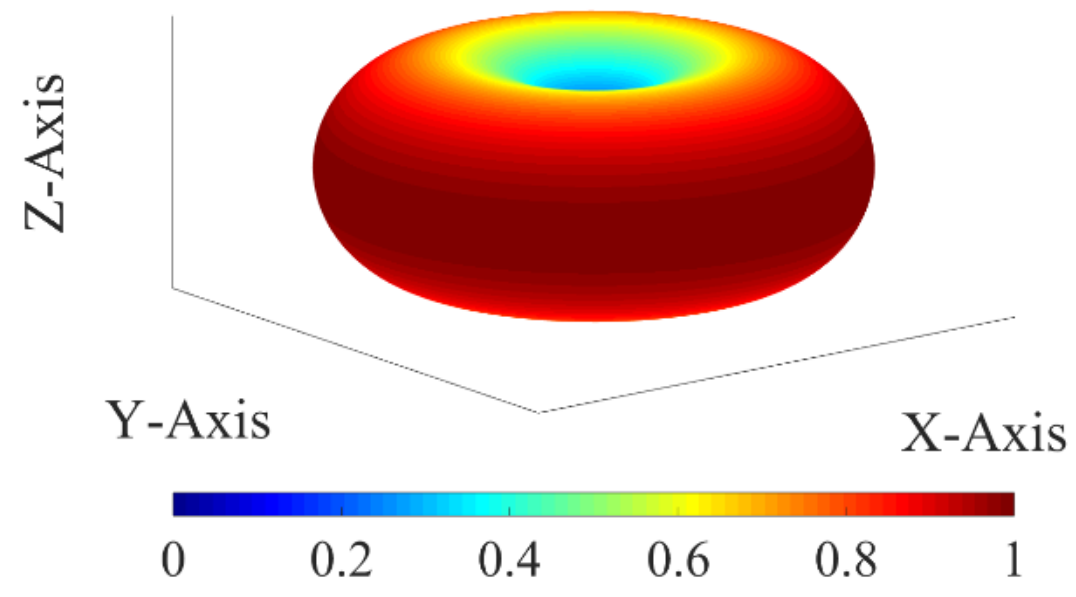

Figure 4-26 The radiation pattern of the DP-CM1 working at 5.675GHz and shown in Figure 4-19

## 4.4 The First Surface Formulation for Calculating the DP-CMs of the Systems Constructed by Double Simply Connected Material Bodies

In this section and subsequent Sections 4.5 and 4.6, we consider the two-body material system shown in Figure 4-27, and respectively provide three different surface formulations for constructing the DP-CMs of the two-body system. The surface formulation given in this section is based on expressing the tangential incident fields contained in DPO as the difference between the tangential total fields on the internal surfaces of material boundaries and the tangential scattered fields on the external surfaces of the material boundaries; the surface formulation given in Section 4.5 is based on expressing the tangential incident fields contained in DPO as the tangential incident fields on the internal surfaces of the material boundaries; the surface formulation given in Section 4.6 is derived from further simplifying the surface formulation given in Section 4.5. One of the main reasons that we provide three different surface DP-CM formulations in this section and Sections 4.5 and 4.6 is that: we want, from different perspectives, to verify the rationality to select WEP framework as the carrying framework of CMT.

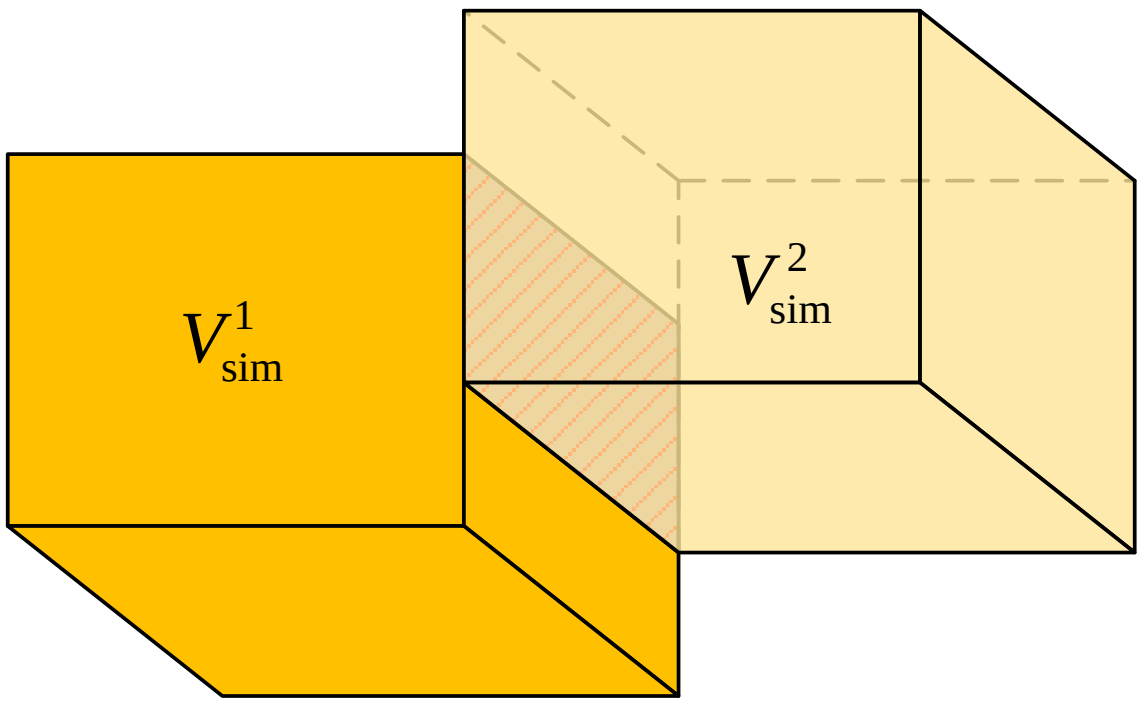

Figure 4-27 The topological structure of a two-body material system constituted by two simply connected material bodies which contact with each other

In Subsection 4.4.1, we provide the surface formulation of the DPO corresponding to the two-body system shown in Figure 4-27, and achieve the variable unification for the surface formulation; in Subsection 4.4.2, we construct the DP-CMs of the two-body system, and discuss the orthogonality among the DP-CMs; in Subsection 4.4.3, we expand the any working mode of the two-body system in terms of the DP-CMs, and derive the explicit expressions for the expansion coefficients.

## 4.4.1 The First Surface Formulation of the DPO Corresponding to Two-body Systems and the Related Variable Unification

Similarly to formulation (4-31), we can also prove that: for the any $i=1$ or $i=2$ in Figure 4-27, there exists the orthogonality that

$$-(1/2)\langle \vec{J}_{si}^{\text{ES}}, \vec{E}^{\text{inc}} \rangle_{\partial V_{\text{sim}}^{i}} -(1/2)\langle \vec{M}_{si}^{\text{ES}}, \vec{H}^{\text{inc}} \rangle_{\partial V_{\text{sim}}^{i}} =(1/2)\langle \vec{J}_{si}^{\text{SV}}, \vec{E}^{\text{inc}} \rangle_{V_{\text{sim}}^{i}} +(1/2)\langle \vec{M}_{si}^{\text{SV}}, \vec{H}^{\text{inc}} \rangle_{V_{\text{sim}}^{i}} = P_{si}^{\text{driving}} \tag{4-64}$$

Then, we have the interaction form of the surface formulation for the DPO corresponding to the two-body material system as follows:

$$\begin{aligned} P_{\text{ss sys}}^{\text{driving}} &= P_{\text{s1}}^{\text{driving}} + P_{\text{s2}}^{\text{driving}} \\ &= \sum_{i=1}^{2}\left[(1/2)\left\langle \vec{J}_{si}^{\text{SV}},\vec{E}^{\text{inc}}\right\rangle_{V_{\text{sim}}^{i}} + (1/2)\left\langle \vec{M}_{si}^{\text{SV}},\vec{H}^{\text{inc}}\right\rangle_{V_{\text{sim}}^{i}}\right] \\ &= -\sum_{i=1}^{2}\left[(1/2)\left\langle \vec{J}_{si}^{\text{ES}},\vec{E}^{\text{inc}}\right\rangle_{\partial V_{\text{sim}}^{i}} + (1/2)\left\langle \vec{M}_{si}^{\text{ES}},\vec{H}^{\text{inc}}\right\rangle_{\partial V_{\text{sim}}^{i}}\right] \end{aligned} \tag{4-65}$$

where $P_{\text{s1}}^{\text{driving}}$ and $P_{\text{s2}}^{\text{driving}}$ are the DPs done by incident fields $\{\vec{E}^{\text{inc}},\vec{H}^{\text{inc}}\}$ on $V_{\text{sim}}^{1}$ and $V_{\text{sim}}^{2}$ respectively.

In the following parts of this subsection, we accomplish, in four steps, the variable unification for the DPO $P_{\text{ss sys}}^{\text{driving}}$ of the two-body material system shown in Figure 4-27. **Step 1.** We transform the DPO from interaction form to current form, such that the DPO only includes various EM currents (at the same time, this step is to prepare for Step 3). **Step 2.** We, in EM current space, establish the transformations between the equivalent surface currents on $\partial V_{\text{sim}}^{1}$ and the equivalent surface currents on $\partial V_{\text{sim}}^{2}$. **Step 3.** We establish the one-to-one correspondence between the EM currents and their expansion vectors (i.e. establishing the transformation from EM current space to expansion vector space), and provide the manifestation form of the DPO in expansion vector space (i.e. the matrix form of the DPO). **Step 4.** We, in expansion vector space, establish the linear transformation from the BVs to the other dependent variables.

### Step 1. To Obtain the EM Current Form of DPO $P_{\text{ss sys}}^{\text{driving}}$

Based on superposition principle $\vec{F}^{\text{tot}} = \vec{F}^{\text{inc}} + \vec{F}^{\text{sca}}$, the tangential continuity of the fields $\vec{F}^{\text{tot}}$ and $\vec{F}^{\text{sca}}$ on material interfaces, and the convolution integral formulations given in Appendix C, the interaction form (4-65) of DPO $P_{\text{ss sys}}^{\text{driving}}$ can be transformed to the following current form:

$$\begin{aligned} &P_{\text{ss sys}}^{\text{driving}} \\ =& -\frac{1}{2}\left[\left\langle \vec{J}_{\text{s1}}^{\text{ES}},\mathcal{E}_0\left(\vec{J}_{\text{s1}}^{\text{ES}},\vec{M}_{\text{s1}}^{\text{ES}}\right)\right\rangle_{\partial V_{\text{sim}}^{1+}} + \left\langle \vec{J}_{\text{s1}}^{\text{ES}},\mathcal{E}_0\left(\vec{J}_{\text{s2}}^{\text{ES}},\vec{M}_{\text{s2}}^{\text{ES}}\right)\right\rangle_{\partial V_{\text{sim}}^{1-}} + \left\langle \vec{J}_{\text{s1}}^{\text{ES}},\mathcal{E}_{\text{sim}}^{1}\left(\vec{J}_{\text{s1}}^{\text{ES}},\vec{M}_{\text{s1}}^{\text{ES}}\right)\right\rangle_{\partial V_{\text{sim}}^{1-}}\right] \\ & -\frac{1}{2}\left[\left\langle \vec{J}_{\text{s2}}^{\text{ES}},\mathcal{E}_0\left(\vec{J}_{\text{s2}}^{\text{ES}},\vec{M}_{\text{s2}}^{\text{ES}}\right)\right\rangle_{\partial V_{\text{sim}}^{2+}} + \left\langle \vec{J}_{\text{s2}}^{\text{ES}},\mathcal{E}_0\left(\vec{J}_{\text{s1}}^{\text{ES}},\vec{M}_{\text{s1}}^{\text{ES}}\right)\right\rangle_{\partial V_{\text{sim}}^{2-}} + \left\langle \vec{J}_{\text{s2}}^{\text{ES}},\mathcal{E}_{\text{sim}}^{2}\left(\vec{J}_{\text{s2}}^{\text{ES}},\vec{M}_{\text{s2}}^{\text{ES}}\right)\right\rangle_{\partial V_{\text{sim}}^{2-}}\right] \\ & -\frac{1}{2}\left[\left\langle \vec{M}_{\text{s1}}^{\text{ES}},\mathcal{H}_0\left(\vec{J}_{\text{s1}}^{\text{ES}},\vec{M}_{\text{s1}}^{\text{ES}}\right)\right\rangle_{\partial V_{\text{sim}}^{1+}} + \left\langle \vec{M}_{\text{s1}}^{\text{ES}},\mathcal{H}_0\left(\vec{J}_{\text{s2}}^{\text{ES}},\vec{M}_{\text{s2}}^{\text{ES}}\right)\right\rangle_{\partial V_{\text{sim}}^{1-}} + \left\langle \vec{M}_{\text{s1}}^{\text{ES}},\mathcal{H}_{\text{sim}}^{1}\left(\vec{J}_{\text{s1}}^{\text{ES}},\vec{M}_{\text{s1}}^{\text{ES}}\right)\right\rangle_{\partial V_{\text{sim}}^{1-}}\right] \\ & -\frac{1}{2}\left[\left\langle \vec{M}_{\text{s2}}^{\text{ES}},\mathcal{H}_0\left(\vec{J}_{\text{s2}}^{\text{ES}},\vec{M}_{\text{s2}}^{\text{ES}}\right)\right\rangle_{\partial V_{\text{sim}}^{2+}} + \left\langle \vec{M}_{\text{s2}}^{\text{ES}},\mathcal{H}_0\left(\vec{J}_{\text{s1}}^{\text{ES}},\vec{M}_{\text{s1}}^{\text{ES}}\right)\right\rangle_{\partial V_{\text{sim}}^{2-}} + \left\langle \vec{M}_{\text{s2}}^{\text{ES}},\mathcal{H}_{\text{sim}}^{2}\left(\vec{J}_{\text{s2}}^{\text{ES}},\vec{M}_{\text{s2}}^{\text{ES}}\right)\right\rangle_{\partial V_{\text{sim}}^{2-}}\right] \end{aligned} \tag{4-66}$$

In formulation (4-66), integral domain $\partial V_{\text{sim}}^{1-}$ represents the inner surface of the boundary of $V_{\text{sim}}^{1}$ (as shown in Figure 4-28), and integral domain $\partial V_{\text{sim}}^{1+}$ represents the outer surface of the boundary of $V_{\text{sim}}^{1}$ (as shown in Figure 4-28), and the other integral domains can be explained similarly; the $\mathcal{E}_0(\vec{J}_{\text{s1}}^{\text{ES}},\vec{M}_{\text{s1}}^{\text{ES}})$ in term $<\vec{J}_{\text{s1}}^{\text{ES}},\mathcal{E}_0(\vec{J}_{\text{s1}}^{\text{ES}},\vec{M}_{\text{s1}}^{\text{ES}})>_{\partial V_{\text{sim}}^{1+}}$ corresponds to the opposite of the electric field which is generated by $V_{\text{sim}}^{1}$ and distributes on the outer boundary $\partial V_{\text{sim}}^{1+}$ of 1# material body (for details see Appendix C), and the $\mathcal{E}_0(\vec{J}_{\text{s2}}^{\text{ES}},\vec{M}_{\text{s2}}^{\text{ES}})$ in term $<\vec{J}_{\text{s1}}^{\text{ES}},\mathcal{E}_0(\vec{J}_{\text{s2}}^{\text{ES}},\vec{M}_{\text{s2}}^{\text{ES}})>_{\partial V_{\text{sim}}^{1-}}$ corresponds to the opposite of the electric field which is generated by $V_{\text{sim}}^{2}$ and distributes on the inner boundary $\partial V_{\text{sim}}^{1-}$ of 1# material body (for details see Appendix C), and the $\mathcal{E}_{\text{sim}}^{1}(\vec{J}_{\text{s1}}^{\text{ES}},\vec{M}_{\text{s1}}^{\text{ES}})$ in term $<\vec{J}_{\text{s1}}^{\text{ES}},\mathcal{E}_{\text{sim}}^{1}(\vec{J}_{\text{s1}}^{\text{ES}},\vec{M}_{\text{s1}}^{\text{ES}})>_{\partial V_{\text{sim}}^{1-}}$ corresponds to the total electric field distributing on the inner boundary $\partial V_{\text{sim}}^{1-}$ of 1# body (for details see Appendix C), and these above imply that term $-(1/2)[<\vec{J}_{\text{s1}}^{\text{ES}},\mathcal{E}_0(\vec{J}_{\text{s1}}^{\text{ES}},\vec{M}_{\text{s1}}^{\text{ES}})>_{\partial V_{\text{sim}}^{1+}}+<\vec{J}_{\text{s1}}^{\text{ES}},\mathcal{E}_0(\vec{J}_{\text{s2}}^{\text{ES}},\vec{M}_{\text{s2}}^{\text{ES}})>_{\partial V_{\text{sim}}^{1-}}+<\vec{J}_{\text{s1}}^{\text{ES}},\mathcal{E}_{\text{sim}}^{1}(\vec{J}_{\text{s1}}^{\text{ES}},\vec{M}_{\text{s1}}^{\text{ES}})>_{\partial V_{\text{sim}}^{1-}}]$ actually corresponds to the interaction between incident electric field $\vec{E}^{\text{inc}}$ and equivalent surface electric current $\vec{J}_{\text{s1}}^{\text{ES}}$; similarly, term $-(1/2)[<\vec{J}_{\text{s2}}^{\text{ES}},\mathcal{E}_0(\vec{J}_{\text{s2}}^{\text{ES}},\vec{M}_{\text{s2}}^{\text{ES}})>_{\partial V_{\text{sim}}^{2+}}+<\vec{J}_{\text{s2}}^{\text{ES}},\mathcal{E}_0(\vec{J}_{\text{s1}}^{\text{ES}},\vec{M}_{\text{s1}}^{\text{ES}})>_{\partial V_{\text{sim}}^{2-}}+<\vec{J}_{\text{s2}}^{\text{ES}},\mathcal{E}_{\text{sim}}^{2}(\vec{J}_{\text{s2}}^{\text{ES}},\vec{M}_{\text{s2}}^{\text{ES}})>_{\partial V_{\text{sim}}^{2-}}]$ corresponds to the interaction between incident electric field $\vec{E}^{\text{inc}}$ and equivalent surface electric current $\vec{J}_{\text{s2}}^{\text{ES}}$, and term $-(1/2)[<\vec{M}_{\text{s1}}^{\text{ES}},\mathcal{H}_0(\vec{J}_{\text{s1}}^{\text{ES}},\vec{M}_{\text{s1}}^{\text{ES}})>_{\partial V_{\text{sim}}^{1+}}+<\vec{M}_{\text{s1}}^{\text{ES}},\mathcal{H}_0(\vec{J}_{\text{s2}}^{\text{ES}},\vec{M}_{\text{s2}}^{\text{ES}})>_{\partial V_{\text{sim}}^{1-}}+<\vec{M}_{\text{s1}}^{\text{ES}},\mathcal{H}_{\text{sim}}^{1}(\vec{J}_{\text{s1}}^{\text{ES}},\vec{M}_{\text{s1}}^{\text{ES}})>_{\partial V_{\text{sim}}^{1-}}]$ corresponds to the interaction between incident magnetic field $\vec{H}^{\text{inc}}$ and equivalent surface magnetic current $\vec{M}_{\text{s1}}^{\text{ES}}$, and term $-(1/2)[<\vec{M}_{\text{s2}}^{\text{ES}},\mathcal{H}_0(\vec{J}_{\text{s2}}^{\text{ES}},\vec{M}_{\text{s2}}^{\text{ES}})>_{\partial V_{\text{sim}}^{2+}}+<\vec{M}_{\text{s2}}^{\text{ES}},\mathcal{H}_0(\vec{J}_{\text{s1}}^{\text{ES}},\vec{M}_{\text{s1}}^{\text{ES}})>_{\partial V_{\text{sim}}^{2-}}+<\vec{M}_{\text{s2}}^{\text{ES}},\mathcal{H}_{\text{sim}}^{2}(\vec{J}_{\text{s2}}^{\text{ES}},\vec{M}_{\text{s2}}^{\text{ES}})>_{\partial V_{\text{sim}}^{2-}}]$ corresponds to the interaction between incident magnetic field $\vec{H}^{\text{inc}}$ and equivalent surface magnetic current $\vec{M}_{\text{s2}}^{\text{ES}}$.

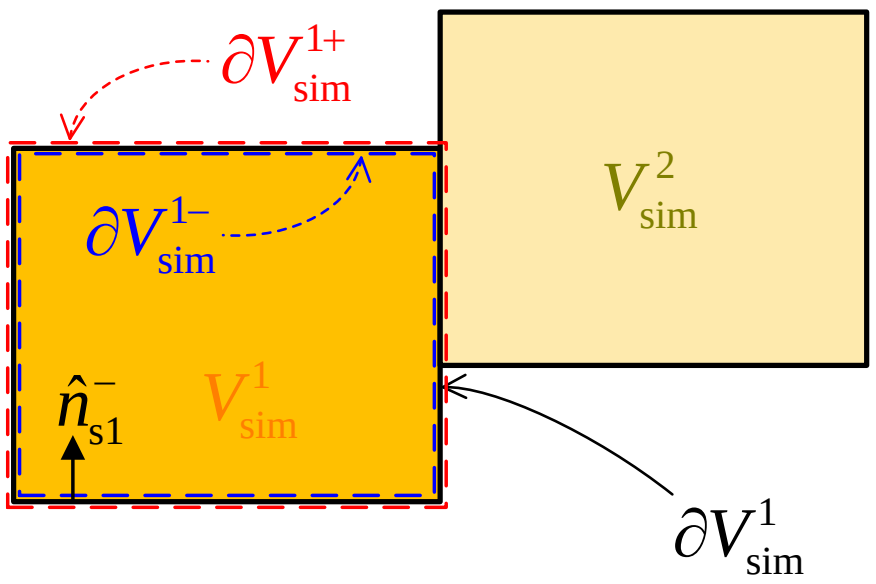

Figure 4-28 The inner surface $\partial V_{\text{sim}}^{1-}$ and outer surface $\partial V_{\text{sim}}^{1+}$ of simply connected material body $V_{\text{sim}}^{1}$. The inner and outer surfaces of simply connected material body $V_{\text{sim}}^{2}$ are similar, so we don't explicitly provide them here to concise this figure

**Step 2. To Unify Variables in EM Current Space**

To finish this step, we need to properly decompose $\partial V_{\text{sim}}^{1}$ & $\partial V_{\text{sim}}^{2}$ and $\vec{C}_{\text{s1}}^{\text{ES}}$ & $\vec{C}_{\text{s2}}^{\text{ES}}$ beforehand. The definitions for $\vec{C}_{\text{s1}}^{\text{ES}}$ and $\vec{C}_{\text{s2}}^{\text{ES}}$ are similar to formulation (C-41), and they will not be repeated here. Based on the relevant conclusions obtained in Appendix C, we can decompose $\partial V_{\text{sim}}^{1}$ and $\partial V_{\text{sim}}^{2}$ as that $\partial V_{\text{sim}}^{1} = \partial V_{\text{s10}} \bigcup \partial V_{\text{s12}}$ and $\partial V_{\text{sim}}^{2} = \partial V_{\text{s20}} \bigcup \partial V_{\text{s21}}$. The above various sub-boundaries are defined as that: $\partial V_{\text{s12}} = \partial V_{\text{sim}}^{1} \bigcap \partial V_{\text{sim}}^{2}$, and $\partial V_{\text{s10}} = \partial V_{\text{sim}}^{1} \setminus \partial V_{\text{s12}}$, and $\partial V_{\text{s21}} = \partial V_{\text{sim}}^{2} \bigcap \partial V_{\text{sim}}^{1}$, and $\partial V_{\text{s20}} = \partial V_{\text{sim}}^{2} \setminus \partial V_{\text{s21}}$. As shown in Figure 4-29, it is obvious that $\partial V_{\text{s12}} = \partial V_{\text{s21}}$. Based on the above boundary decomposition, we can correspondingly decompose equivalent surface currents $\vec{C}_{\text{s1}}^{\text{ES}}$ and $\vec{C}_{\text{s2}}^{\text{ES}}$ as follows:

$$\vec{C}_{\text{s1}}^{\text{ES}}(\vec{r}) = \vec{C}_{\text{s10}}^{\text{ES}}(\vec{r}) + \vec{C}_{\text{s12}}^{\text{ES}}(\vec{r}) \quad , \quad \vec{r} \in \partial V_{\text{sim}}^{1} \tag{4-67}$$

$$\vec{C}_{\text{s2}}^{\text{ES}}(\vec{r}) = \vec{C}_{\text{s20}}^{\text{ES}}(\vec{r}) + \vec{C}_{\text{s21}}^{\text{ES}}(\vec{r}) \quad , \quad \vec{r} \in \partial V_{\text{sim}}^{2} \tag{4-68}$$

where the various equivalent surface sub-currents are defined as follows:

$$\vec{C}_{\text{s10}}^{\text{ES}}(\vec{r}) = \begin{cases} \vec{C}_{\text{s1}}^{\text{ES}}(\vec{r}) & , \quad \vec{r} \in \partial V_{\text{s10}} \\ 0 & , \quad \vec{r} \in \partial V_{\text{s12}} \end{cases} \tag{4-69a}$$

$$\vec{C}_{\text{s12}}^{\text{ES}}(\vec{r}) = \begin{cases} 0 & , \quad \vec{r} \in \partial V_{\text{s10}} \\ \vec{C}_{\text{s1}}^{\text{ES}}(\vec{r}) & , \quad \vec{r} \in \partial V_{\text{s12}} \end{cases} \tag{4-69b}$$

$$\vec{C}_{\text{s20}}^{\text{ES}}(\vec{r}) = \begin{cases} \vec{C}_{\text{s2}}^{\text{ES}}(\vec{r}) & , \quad \vec{r} \in \partial V_{\text{s20}} \\ 0 & , \quad \vec{r} \in \partial V_{\text{s21}} \end{cases} \tag{4-70a}$$

$$\vec{C}_{\text{s21}}^{\text{ES}}(\vec{r}) = \begin{cases} 0 & , \quad \vec{r} \in \partial V_{\text{s20}} \\ \vec{C}_{\text{s2}}^{\text{ES}}(\vec{r}) & , \quad \vec{r} \in \partial V_{\text{s21}} \end{cases} \tag{4-70b}$$

According to the conclusions obtained in Appendix C, it immediately has that:

$$\vec{C}_{\text{s21}}^{\text{ES}}(\vec{r}) = -\vec{C}_{\text{s12}}^{\text{ES}}(\vec{r}) \tag{4-71}$$

The above relationship is just the transformation from $\vec{C}_{\text{s12}}^{\text{ES}}$ to $\vec{C}_{\text{s21}}^{\text{ES}}$ established in EM current space. So far, the variable unification in EM current space has been finished. But, we want to emphasize that: among the above sub-currents, there still are some dependent variables not been expressed as the functions of BVs. The work to thoroughly eliminate the remaining dependent variables will be finished in expansion vector space until Step 4.

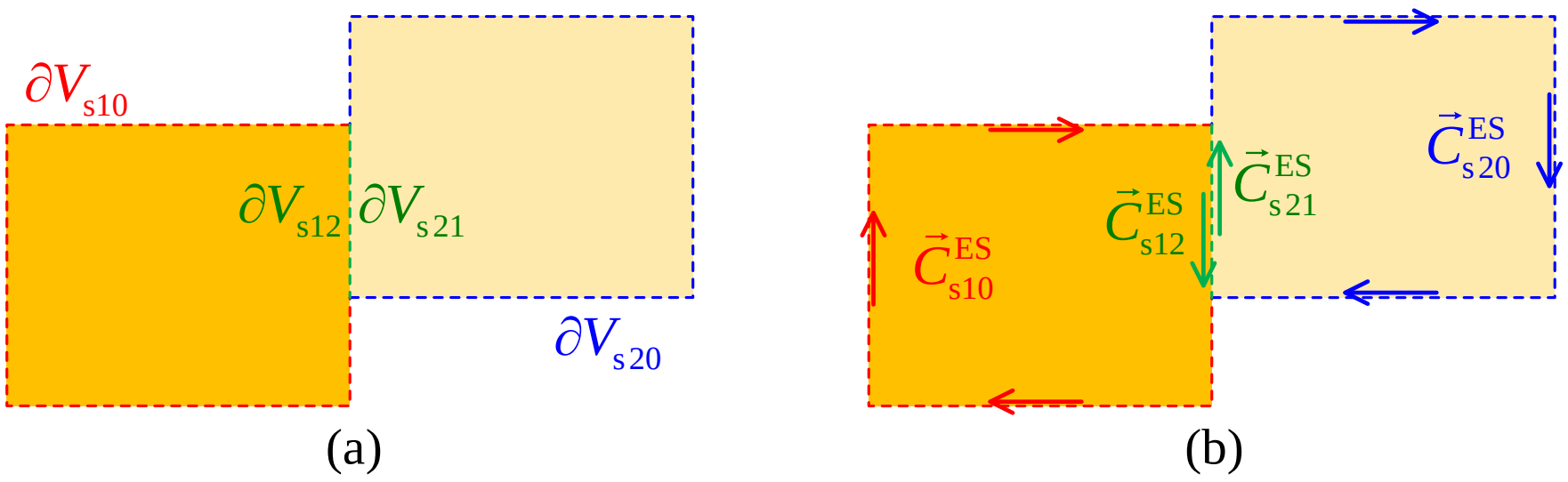

Figure 4-29 The decompositions related to the two-body material system shown in Figure 4-27. (a) boundary decomposition; (b) equivalent current decomposition

**Step 3. To Transform From EM Current Space to Expansion Vector Space**

As we know, it is relatively difficult to thoroughly establish the transformation from BVs to dependent variables①. This dissertation selects to establish the transformation in expansion vector space, so we firstly establish the one-to-one mapping between EM current space and expansion vector space as follows:

$$\vec{C}_{\text{s10}}^{\text{ES}}(\vec{r}) = \sum_{\xi=1}^{\Xi^{C_{\text{s10}}}} a_\xi^{C_{\text{s10}}} \vec{b}_\xi^{C_{\text{s10}}}(\vec{r}) = \bar{\boldsymbol{B}}^{C_{\text{s10}}} \cdot \bar{a}^{C_{\text{s10}}} \quad , \quad \vec{r} \in \partial V_{\text{s10}} \tag{4-72a}$$

$$\vec{C}_{\text{s12}}^{\text{ES}}(\vec{r}) = \sum_{\xi=1}^{\Xi^{C_{\text{s12}}}} a_\xi^{C_{\text{s12}}} \vec{b}_\xi^{C_{\text{s12}}}(\vec{r}) = \bar{\boldsymbol{B}}^{C_{\text{s12}}} \cdot \bar{a}^{C_{\text{s12}}} \quad , \quad \vec{r} \in \partial V_{\text{s12}} \tag{4-72b}$$

$$\begin{aligned}\vec{C}_{\text{s21}}^{\text{ES}}(\vec{r}) = \sum_{\xi=1}^{\Xi^{C_{\text{s12}}}} a_\xi^{C_{\text{s21}}} \vec{b}_\xi^{C_{\text{s12}}}(\vec{r}) &= \bar{\boldsymbol{B}}^{C_{\text{s12}}} \cdot \bar{a}^{C_{\text{s21}}} \\ &= -\bar{\boldsymbol{B}}^{C_{\text{s12}}} \cdot \bar{a}^{C_{\text{s12}}} \quad , \quad \vec{r} \in \partial V_{\text{s21}} = \partial V_{\text{s12}}\end{aligned} \tag{4-73a}$$

$$\vec{C}_{\text{s20}}^{\text{ES}}(\vec{r}) = \sum_{\xi=1}^{\Xi^{C_{\text{s20}}}} a_\xi^{C_{\text{s20}}} \vec{b}_\xi^{C_{\text{s20}}}(\vec{r}) = \bar{\boldsymbol{B}}^{C_{\text{s20}}} \cdot \bar{a}^{C_{\text{s20}}} \quad , \quad \vec{r} \in \partial V_{\text{s20}} \tag{4-73b}$$

where $\{\vec{b}_\xi^{C_{\text{s10}}}\}_{\xi=1}^{\Xi^{C_{\text{s10}}}}$, $\{\vec{b}_\xi^{C_{\text{s12}}}\}_{\xi=1}^{\Xi^{C_{\text{s12}}}}$, and $\{\vec{b}_\xi^{C_{\text{s20}}}\}_{\xi=1}^{\Xi^{C_{\text{s20}}}}$ are independent and complete basis functions. In formulation (4-72), $\bar{\boldsymbol{B}}^{C_{\text{s10}}} = [\vec{b}_1^{C_{\text{s10}}}, \cdots, \vec{b}_{\Xi^{C_{\text{s10}}}}^{C_{\text{s10}}}]$ and $\bar{a}^{C_{\text{s10}}} = [a_1^{C_{\text{s10}}}, \cdots, a_{\Xi^{C_{\text{s10}}}}^{C_{\text{s10}}}]^T$; $\bar{\boldsymbol{B}}^{C_{\text{s12}}} = [\vec{b}_1^{C_{\text{s12}}}, \cdots, \vec{b}_{\Xi^{C_{\text{s12}}}}^{C_{\text{s12}}}]$ and $\bar{a}^{C_{\text{s12}}} = [a_1^{C_{\text{s12}}}, \cdots, a_{\Xi^{C_{\text{s12}}}}^{C_{\text{s12}}}]^T$. In formulation (4-73), $\bar{a}^{C_{\text{s21}}} = [a_1^{C_{\text{s21}}}, \cdots, a_{\Xi^{C_{\text{s21}}}}^{C_{\text{s21}}}]^T$; $\bar{\boldsymbol{B}}^{C_{\text{s20}}} = [\vec{b}_1^{C_{\text{s20}}}, \cdots, \vec{b}_{\Xi^{C_{\text{s20}}}}^{C_{\text{s20}}}]$ and $\bar{a}^{C_{\text{s20}}} = [a_1^{C_{\text{s20}}}, \cdots, a_{\Xi^{C_{\text{s20}}}}^{C_{\text{s20}}}]^T$. Obviously, $\vec{C}_{\text{s10}}^{\text{ES}}$ and $\bar{a}^{C_{\text{s10}}}$ are one-to-one correspondence, and $\vec{C}_{\text{s12}}^{\text{ES}}$ and $\bar{a}^{C_{\text{s12}}}$ are one-to-one correspondence, and $\vec{C}_{\text{s21}}^{\text{ES}}$ and $\bar{a}^{C_{\text{s21}}}$ are one-to-one correspondence, and $\vec{C}_{\text{s20}}^{\text{ES}}$ and $\bar{a}^{C_{\text{s20}}}$ are one-to-one correspondence. Here, we want to emphasize that: the basis functions used in expansion formulation (4-73a) are completely the same as the ones used in expansion formulation (4-72b), and $\bar{a}^{C_{\text{s21}}} = -\bar{a}^{C_{\text{s12}}}$ (this is just the reason why the third equality holds), because of relationship (4-71). In fact, relationship $\bar{a}^{C_{\text{s21}}} = -\bar{a}^{C_{\text{s12}}}$ is just the manifestation form of relationship $\vec{C}_{\text{s21}}^{\text{ES}} = -\vec{C}_{\text{s12}}^{\text{ES}}$ in expansion vector space, and conversely relationship $\vec{C}_{\text{s21}}^{\text{ES}} = -\vec{C}_{\text{s12}}^{\text{ES}}$ is just the manifestation form of relationship $\bar{a}^{C_{\text{s21}}} = -\bar{a}^{C_{\text{s12}}}$ in EM current space.

To make the formulations to be appeared in the following parts of this section be more compact, the elements of $\{\vec{b}_\xi^{C_{\text{s10}}}\}_{\xi=1}^{\Xi^{C_{\text{s10}}}} \bigcup \{\vec{b}_\xi^{C_{\text{s12}}}\}_{\xi=1}^{\Xi^{C_{\text{s12}}}}$ and $\{\vec{b}_\xi^{C_{\text{s12}}}\}_{\xi=1}^{\Xi^{C_{\text{s12}}}} \bigcup \{\vec{b}_\xi^{C_{\text{s20}}}\}_{\xi=1}^{\Xi^{C_{\text{s20}}}}$ are renumbered and assembled as follows:

$$\overbrace{\begin{bmatrix} \vec{b}_1^{C_{\text{s1}}} & \vec{b}_2^{C_{\text{s1}}} & \cdots & \vec{b}_{\Xi^{C_{\text{s1}}}}^{C_{\text{s1}}} \end{bmatrix}}^{\bar{\boldsymbol{B}}^{C_{\text{s1}}}} = \overbrace{\begin{bmatrix} \vec{b}_1^{C_{\text{s10}}} & \vec{b}_2^{C_{\text{s10}}} & \cdots & \vec{b}_{\Xi^{C_{\text{s10}}}}^{C_{\text{s10}}} & \vec{b}_1^{C_{\text{s12}}} & \vec{b}_2^{C_{\text{s12}}} & \cdots & \vec{b}_{\Xi^{C_{\text{s12}}}}^{C_{\text{s12}}} \end{bmatrix}}^{\left[\bar{\boldsymbol{B}}^{C_{\text{s10}}} \quad \bar{\boldsymbol{B}}^{C_{\text{s12}}}\right]} \tag{4-74}$$

$$\underbrace{\begin{bmatrix} \vec{b}_1^{C_{\text{s2}}} & \vec{b}_2^{C_{\text{s2}}} & \cdots & \vec{b}_{\Xi^{C_{\text{s2}}}}^{C_{\text{s2}}} \end{bmatrix}}_{\bar{\boldsymbol{B}}^{C_{\text{s2}}}} = \underbrace{\begin{bmatrix} \vec{b}_1^{C_{\text{s12}}} & \vec{b}_2^{C_{\text{s12}}} & \cdots & \vec{b}_{\Xi^{C_{\text{s12}}}}^{C_{\text{s12}}} & \vec{b}_1^{C_{\text{s20}}} & \vec{b}_2^{C_{\text{s20}}} & \cdots & \vec{b}_{\Xi^{C_{\text{s20}}}}^{C_{\text{s20}}} \end{bmatrix}}_{\left[\bar{\boldsymbol{B}}^{C_{\text{s12}}} \quad \bar{\boldsymbol{B}}^{C_{\text{s20}}}\right]} \tag{4-75}$$

① In fact, it is also difficult to construct PD-CMs in EM current space, so this dissertation selects to construct DP-CMs in expansion vector space.

Obviously, $\Xi^{C_{s1}} = \Xi^{C_{s10}} + \Xi^{C_{s12}}$ and $\Xi^{C_{s2}} = \Xi^{C_{s12}} + \Xi^{C_{s20}}$, and

$$\vec{C}_{s1}^{ES}(\vec{r}) = \bar{\boldsymbol{B}}^{C_{s1}} \cdot \bar{a}^{C_{s1}} \quad , \quad \vec{r} \in \partial V_{sim}^{1} \tag{4-76}$$

$$\vec{C}_{s2}^{ES}(\vec{r}) = \bar{\boldsymbol{B}}^{C_{s2}} \cdot \bar{a}^{C_{s2}} \quad , \quad \vec{r} \in \partial V_{sim}^{2} \tag{4-77}$$

where

$$\bar{a}^{C_{s1}} = \begin{bmatrix} \bar{a}^{C_{s10}} \\ \bar{a}^{C_{s12}} \end{bmatrix} \tag{4-78}$$

$$\bar{a}^{C_{s2}} = \begin{bmatrix} \bar{a}^{C_{s21}} \\ \bar{a}^{C_{s20}} \end{bmatrix} = \begin{bmatrix} -\bar{a}^{C_{s12}} \\ \bar{a}^{C_{s20}} \end{bmatrix} \tag{4-79}$$

Inserting formulations (4-76) and (4-77) into formulation (4-66), the $P_{ss\,sys}^{driving}$ in formulation (4-66) is immediately discretized into the following matrix form:

$$P_{ss\,sys}^{driving} = \begin{bmatrix} \bar{a}^{J_{s1}} \\ \bar{a}^{J_{s2}} \\ \bar{a}^{M_{s1}} \\ \bar{a}^{M_{s2}} \end{bmatrix}^{H} \cdot \underbrace{\left( \bar{\bar{P}}_{1;0;PVT}^{ss\,sys} + \bar{\bar{P}}_{1;0;SCT}^{ss\,sys} + \bar{\bar{P}}_{1;m}^{ss\,sys} \right)}_{\bar{\bar{P}}_{ss\,sys;1}^{driving}} \cdot \begin{bmatrix} \bar{a}^{J_{s1}} \\ \bar{a}^{J_{s2}} \\ \bar{a}^{M_{s1}} \\ \bar{a}^{M_{s2}} \end{bmatrix} \tag{4-80}$$

In formulation (4-80), the subscripts "1" used in power matrices $\bar{\bar{P}}_{1;0;PVT}^{ss\,sys}$, $\bar{\bar{P}}_{1;0;SCT}^{ss\,sys}$, $\bar{\bar{P}}_{1;m}^{ss\,sys}$, and $\bar{\bar{P}}_{ss\,sys;1}^{driving}$ represent that the matrices are from discretizing the first surface formulation (4-66) of DPO $P_{ss\,sys}^{driving}$, but in subsequent Sections 4.5 and 4.6 the subscripts in the corresponding power matrices will be replaced by "2" and "3" respectively, because Sections 4.5 and 4.6 respectively select to use the second surface formulation (4-113) and the third surface formulation (4-117) of DPO $P_{ss\,sys}^{driving}$; the subscripts "0" used in power matrices $\bar{\bar{P}}_{1;0;PVT}^{ss\,sys}$ and $\bar{\bar{P}}_{1;0;SCT}^{ss\,sys}$ represent that the matrices correspond to the vacuum version of field operator $\mathcal{F}$ as illustrated in the subsequent formulation (4-82) for calculating the matrix elements; the subscript "m" used in power matrix $\bar{\bar{P}}_{1;m}^{ss\,sys}$ represents that the matrix corresponds to the matter version of field operator $\mathcal{F}$ as illustrated in the subsequent formulation (4-84) for calculating the matrix elements; the subscripts "PVT" and "SCT" used in power matrices $\bar{\bar{P}}_{1;0;PVT}^{ss\,sys}$ and $\bar{\bar{P}}_{1;0;SCT}^{ss\,sys}$ are the abbreviations of terminologies "principal value term" and "singular current term" respectively, and this point is similar to the formulation (4-34) of a single simply connected material body case. The power matrices in the matrix form (4-80) of the first surface formulation of DPO $P_{ss\,sys}^{driving}$ are as follows:

$$\bar{\bar{P}}_{1;0;PVT}^{ss\,sys} = \begin{bmatrix} \bar{\bar{P}}_{1;0;PVT}^{J_{s1}J_{s1}} & \bar{\bar{P}}_{1;0;PVT}^{J_{s1}J_{s2}} & \bar{\bar{P}}_{1;0;PVT}^{J_{s1}M_{s1}} & \bar{\bar{P}}_{1;0;PVT}^{J_{s1}M_{s2}} \\ \bar{\bar{P}}_{1;0;PVT}^{J_{s2}J_{s1}} & \bar{\bar{P}}_{1;0;PVT}^{J_{s2}J_{s2}} & \bar{\bar{P}}_{1;0;PVT}^{J_{s2}M_{s1}} & \bar{\bar{P}}_{1;0;PVT}^{J_{s2}M_{s2}} \\ \bar{\bar{P}}_{1;0;PVT}^{M_{s1}J_{s1}} & \bar{\bar{P}}_{1;0;PVT}^{M_{s1}J_{s2}} & \bar{\bar{P}}_{1;0;PVT}^{M_{s1}M_{s1}} & \bar{\bar{P}}_{1;0;PVT}^{M_{s1}M_{s2}} \\ \bar{\bar{P}}_{1;0;PVT}^{M_{s2}J_{s1}} & \bar{\bar{P}}_{1;0;PVT}^{M_{s2}J_{s2}} & \bar{\bar{P}}_{1;0;PVT}^{M_{s2}M_{s1}} & \bar{\bar{P}}_{1;0;PVT}^{M_{s2}M_{s2}} \end{bmatrix} \tag{4-81a}$$

$$\overline{\overline{P}}_{1;0;\text{SCT}}^{\text{ss sys}} = \begin{bmatrix} 0 & 0 & \overline{\overline{P}}_{1;0;\text{SCT}}^{J_{s1}M_{s1}} & \overline{\overline{P}}_{1;0;\text{SCT}}^{J_{s1}M_{s2}} \\ 0 & 0 & \overline{\overline{P}}_{1;0;\text{SCT}}^{J_{s2}M_{s1}} & \overline{\overline{P}}_{1;0;\text{SCT}}^{J_{s2}M_{s2}} \\ \overline{\overline{P}}_{1;0;\text{SCT}}^{M_{s1}J_{s1}} & \overline{\overline{P}}_{1;0;\text{SCT}}^{M_{s1}J_{s2}} & 0 & 0 \\ \overline{\overline{P}}_{1;0;\text{SCT}}^{M_{s2}J_{s1}} & \overline{\overline{P}}_{1;0;\text{SCT}}^{M_{s2}J_{s2}} & 0 & 0 \end{bmatrix} \tag{4-81b}$$

$$\overline{\overline{P}}_{1;\text{m}}^{\text{ss sys}} = \begin{bmatrix} \overline{\overline{P}}_{\text{m}}^{J_{s1}J_{s1}} & 0 & \overline{\overline{P}}_{\text{m}}^{J_{s1}M_{s1}} & 0 \\ 0 & \overline{\overline{P}}_{\text{m}}^{J_{s2}J_{s2}} & 0 & \overline{\overline{P}}_{\text{m}}^{J_{s2}M_{s2}} \\ \overline{\overline{P}}_{\text{m}}^{M_{s1}J_{s1}} & 0 & \overline{\overline{P}}_{\text{m}}^{M_{s1}M_{s1}} & 0 \\ 0 & \overline{\overline{P}}_{\text{m}}^{M_{s2}J_{s2}} & 0 & \overline{\overline{P}}_{\text{m}}^{M_{s2}M_{s2}} \end{bmatrix} \tag{4-81c}$$

In formulation (4-81a), the elements of the various sub-matrices are calculated as follows:

$$p_{1;0;\text{PVT};\xi\zeta}^{J_{s1}J_{s1}} = -(1/2)\left\langle \vec{b}_{\xi}^{J_{s1}}, -j\omega\mu_0\mathcal{L}_0\left(\vec{b}_{\zeta}^{J_{s1}}\right)\right\rangle_{\partial V_{\text{sim}}^1} \tag{4-82a}$$

$$p_{1;0;\text{PVT};\xi\zeta}^{J_{s1}J_{s2}} = -(1/2)\left\langle \vec{b}_{\xi}^{J_{s1}}, -j\omega\mu_0\mathcal{L}_0\left(\vec{b}_{\zeta}^{J_{s2}}\right)\right\rangle_{\partial V_{\text{sim}}^1} \tag{4-82b}$$

$$p_{1;0;\text{PVT};\xi\zeta}^{J_{s1}M_{s1}} = -(1/2)\left\langle \vec{b}_{\xi}^{J_{s1}}, -\text{P.V.}\,\mathcal{K}_0\left(\vec{b}_{\zeta}^{M_{s1}}\right)\right\rangle_{\partial V_{\text{sim}}^1} \tag{4-82c}$$

$$p_{1;0;\text{PVT};\xi\zeta}^{J_{s1}M_{s2}} = -(1/2)\left\langle \vec{b}_{\xi}^{J_{s1}}, -\text{P.V.}\,\mathcal{K}_0\left(\vec{b}_{\zeta}^{M_{s2}}\right)\right\rangle_{\partial V_{\text{sim}}^1} \tag{4-82d}$$

and

$$p_{1;0;\text{PVT};\xi\zeta}^{J_{s2}J_{s1}} = -(1/2)\left\langle \vec{b}_{\xi}^{J_{s2}}, -j\omega\mu_0\mathcal{L}_0\left(\vec{b}_{\zeta}^{J_{s1}}\right)\right\rangle_{\partial V_{\text{sim}}^2} \tag{4-82e}$$

$$p_{1;0;\text{PVT};\xi\zeta}^{J_{s2}J_{s2}} = -(1/2)\left\langle \vec{b}_{\xi}^{J_{s2}}, -j\omega\mu_0\mathcal{L}_0\left(\vec{b}_{\zeta}^{J_{s2}}\right)\right\rangle_{\partial V_{\text{sim}}^2} \tag{4-82f}$$

$$p_{1;0;\text{PVT};\xi\zeta}^{J_{s2}M_{s1}} = -(1/2)\left\langle \vec{b}_{\xi}^{J_{s2}}, -\text{P.V.}\,\mathcal{K}_0\left(\vec{b}_{\zeta}^{M_{s1}}\right)\right\rangle_{\partial V_{\text{sim}}^2} \tag{4-82g}$$

$$p_{1;0;\text{PVT};\xi\zeta}^{J_{s2}M_{s2}} = -(1/2)\left\langle \vec{b}_{\xi}^{J_{s2}}, -\text{P.V.}\,\mathcal{K}_0\left(\vec{b}_{\zeta}^{M_{s2}}\right)\right\rangle_{\partial V_{\text{sim}}^2} \tag{4-82h}$$

and

$$p_{1;0;\text{PVT};\xi\zeta}^{M_{s1}J_{s1}} = -(1/2)\left\langle \vec{b}_{\xi}^{M_{s1}}, \text{P.V.}\,\mathcal{K}_0\left(\vec{b}_{\zeta}^{J_{s1}}\right)\right\rangle_{\partial V_{\text{sim}}^1} \tag{4-82i}$$

$$p_{1;0;\text{PVT};\xi\zeta}^{M_{s1}J_{s2}} = -(1/2)\left\langle \vec{b}_{\xi}^{M_{s1}}, \text{P.V.}\,\mathcal{K}_0\left(\vec{b}_{\zeta}^{J_{s2}}\right)\right\rangle_{\partial V_{\text{sim}}^1} \tag{4-82j}$$

$$p_{1;0;\text{PVT};\xi\zeta}^{M_{s1}M_{s1}} = -(1/2)\left\langle \vec{b}_{\xi}^{M_{s1}}, -j\omega\varepsilon_0\mathcal{L}_0\left(\vec{b}_{\zeta}^{M_{s1}}\right)\right\rangle_{\partial V_{\text{sim}}^1} \tag{4-82k}$$

$$p_{1;0;\text{PVT};\xi\zeta}^{M_{s1}M_{s2}} = -(1/2)\left\langle \vec{b}_{\xi}^{M_{s1}}, -j\omega\varepsilon_0\mathcal{L}_0\left(\vec{b}_{\zeta}^{M_{s2}}\right)\right\rangle_{\partial V_{\text{sim}}^1} \tag{4-82l}$$

and

$$p_{1;0;\text{PVT};\xi\zeta}^{M_{s2}J_{s1}} = -(1/2)\left\langle \vec{b}_{\xi}^{M_{s2}}, \text{P.V.}\,\mathcal{K}_0\left(\vec{b}_{\zeta}^{J_{s1}}\right)\right\rangle_{\partial V_{\text{sim}}^2} \tag{4-82m}$$

$$p_{1;0;\text{PVT};\xi\zeta}^{M_{s2}J_{s2}} = -(1/2)\left\langle \vec{b}_{\xi}^{M_{s2}}, \text{P.V.}\,\mathcal{K}_0\left(\vec{b}_{\zeta}^{J_{s2}}\right)\right\rangle_{\partial V_{\text{sim}}^2} \tag{4-82n}$$

$$p_{1;0;\text{PVT};\xi\zeta}^{M_{s2}M_{s1}} = -(1/2)\left\langle \vec{b}_{\xi}^{M_{s2}}, -j\omega\varepsilon_0\mathcal{L}_0\left(\vec{b}_{\zeta}^{M_{s1}}\right)\right\rangle_{\partial V_{\text{sim}}^2} \tag{4-82o}$$

$$p_{1;0;\text{PVT};\xi\zeta}^{M_{s2}M_{s2}} = -(1/2)\left\langle \vec{b}_{\xi}^{M_{s2}}, -j\omega\varepsilon_0\mathcal{L}_0\left(\vec{b}_{\zeta}^{M_{s2}}\right)\right\rangle_{\partial V_{\text{sim}}^2} \tag{4-82p}$$

where "P.V." represents to get the principal values of the corresponding integrals, and this is just the common reason why this dissertation calls $\overline{\overline{P}}{}^{\text{ss sys}}_{1;0;\text{PVT}}$ as principal term and why subscript "PVT" is added to $\overline{\overline{P}}{}^{\text{ss sys}}_{1;0;\text{PVT}}$. In formulation (4-81b), the "0" are some zero matrices, which have proper line numbers and proper column numbers, and the elements of the nonzero sub-matrices are calculated as follows:

$$p^{J_{\text{s}1}M_{\text{s}1}}_{1;0;\text{SCT};\xi\zeta} = -(1/2)\left\langle \vec{b}^{J_{\text{s}1}}_{\xi}, \frac{1}{2}\vec{b}^{M_{\text{s}1}}_{\zeta} \times \hat{n}^{-}_{\text{s}1} \right\rangle_{\partial V^{1}_{\text{sim}}} \tag{4-83a}$$

$$p^{J_{\text{s}1}M_{\text{s}2}}_{1;0;\text{SCT};\xi\zeta} = -(1/2)\left\langle \vec{b}^{J_{\text{s}1}}_{\xi}, \hat{n}^{-}_{\text{s}1} \times \frac{1}{2}\vec{b}^{M_{\text{s}2}}_{\zeta} \right\rangle_{\partial V^{1}_{\text{sim}}} \tag{4-83b}$$

$$p^{J_{\text{s}2}M_{\text{s}1}}_{1;0;\text{SCT};\xi\zeta} = -(1/2)\left\langle \vec{b}^{J_{\text{s}2}}_{\xi}, \hat{n}^{-}_{\text{s}2} \times \frac{1}{2}\vec{b}^{M_{\text{s}1}}_{\zeta} \right\rangle_{\partial V^{2}_{\text{sim}}} \tag{4-83c}$$

$$p^{J_{\text{s}2}M_{\text{s}2}}_{1;0;\text{SCT};\xi\zeta} = -(1/2)\left\langle \vec{b}^{J_{\text{s}2}}_{\xi}, \frac{1}{2}\vec{b}^{M_{\text{s}2}}_{\zeta} \times \hat{n}^{-}_{\text{s}2} \right\rangle_{\partial V^{2}_{\text{sim}}} \tag{4-83d}$$

$$p^{M_{\text{s}1}J_{\text{s}1}}_{1;0;\text{SCT};\xi\zeta} = -(1/2)\left\langle \vec{b}^{M_{\text{s}1}}_{\xi}, \hat{n}^{-}_{\text{s}1} \times \frac{1}{2}\vec{b}^{J_{\text{s}1}}_{\zeta} \right\rangle_{\partial V^{1}_{\text{sim}}} \tag{4-83e}$$

$$p^{M_{\text{s}1}J_{\text{s}2}}_{1;0;\text{SCT};\xi\zeta} = -(1/2)\left\langle \vec{b}^{M_{\text{s}1}}_{\xi}, \frac{1}{2}\vec{b}^{J_{\text{s}2}}_{\zeta} \times \hat{n}^{-}_{\text{s}1} \right\rangle_{\partial V^{1}_{\text{sim}}} \tag{4-83f}$$

$$p^{M_{\text{s}2}J_{\text{s}1}}_{1;0;\text{SCT};\xi\zeta} = -(1/2)\left\langle \vec{b}^{M_{\text{s}2}}_{\xi}, \frac{1}{2}\vec{b}^{J_{\text{s}1}}_{\zeta} \times \hat{n}^{-}_{\text{s}2} \right\rangle_{\partial V^{2}_{\text{sim}}} \tag{4-83g}$$

$$p^{M_{\text{s}2}J_{\text{s}2}}_{1;0;\text{SCT};\xi\zeta} = -(1/2)\left\langle \vec{b}^{M_{\text{s}2}}_{\xi}, \hat{n}^{-}_{\text{s}2} \times \frac{1}{2}\vec{b}^{J_{\text{s}2}}_{\zeta} \right\rangle_{\partial V^{2}_{\text{sim}}} \tag{4-83h}$$

where $\hat{n}^{-}_{\text{s}1}$ is the normal vector of $\partial V^{1}_{\text{sim}}$ and points to the interior of $V^{1}_{\text{sim}}$, as shown in Figure 4-28; $\hat{n}^{-}_{\text{s}2}$ is the normal vector of $\partial V^{2}_{\text{sim}}$ and points to the interior of $V^{2}_{\text{sim}}$. In formulation (4-81c), the "0" are some zero matrices, which have proper line numbers and column numbers, and the elements of the nonzero sub-matrices are calculated as follows:

$$p^{J_{\text{s}1}J_{\text{s}1}}_{\text{m};\xi\zeta} = -(1/2)\left\langle \vec{b}^{J_{\text{s}1}}_{\xi}, \mathcal{E}^{1}_{\text{sim}}\left(\vec{b}^{J_{\text{s}1}}_{\zeta}, 0\right) \right\rangle_{\partial V^{1-}_{\text{sim}}} \tag{4-84a}$$

$$p^{J_{\text{s}1}M_{\text{s}1}}_{\text{m};\xi\zeta} = -(1/2)\left\langle \vec{b}^{J_{\text{s}1}}_{\xi}, \mathcal{E}^{1}_{\text{sim}}\left(0, \vec{b}^{M_{\text{s}1}}_{\zeta}\right) \right\rangle_{\partial V^{1-}_{\text{sim}}} \tag{4-84b}$$

$$p^{J_{\text{s}2}J_{\text{s}2}}_{\text{m};\xi\zeta} = -(1/2)\left\langle \vec{b}^{J_{\text{s}2}}_{\xi}, \mathcal{E}^{2}_{\text{sim}}\left(\vec{b}^{J_{\text{s}2}}_{\zeta}, 0\right) \right\rangle_{\partial V^{2-}_{\text{sim}}} \tag{4-84c}$$

$$p^{J_{\text{s}2}M_{\text{s}2}}_{\text{m};\xi\zeta} = -(1/2)\left\langle \vec{b}^{J_{\text{s}2}}_{\xi}, \mathcal{E}^{2}_{\text{sim}}\left(0, \vec{b}^{M_{\text{s}2}}_{\zeta}\right) \right\rangle_{\partial V^{2-}_{\text{sim}}} \tag{4-84d}$$

$$p^{M_{\text{s}1}J_{\text{s}1}}_{\text{m};\xi\zeta} = -(1/2)\left\langle \vec{b}^{M_{\text{s}1}}_{\xi}, \mathcal{H}^{1}_{\text{sim}}\left(\vec{b}^{J_{\text{s}1}}_{\zeta}, 0\right) \right\rangle_{\partial V^{1-}_{\text{sim}}} \tag{4-84e}$$

$$p^{M_{\text{s}1}M_{\text{s}1}}_{\text{m};\xi\zeta} = -(1/2)\left\langle \vec{b}^{M_{\text{s}1}}_{\xi}, \mathcal{H}^{1}_{\text{sim}}\left(0, \vec{b}^{M_{\text{s}1}}_{\zeta}\right) \right\rangle_{\partial V^{1-}_{\text{sim}}} \tag{4-84f}$$

$$p^{M_{\text{s}2}J_{\text{s}2}}_{\text{m};\xi\zeta} = -(1/2)\left\langle \vec{b}^{M_{\text{s}2}}_{\xi}, \mathcal{H}^{2}_{\text{sim}}\left(\vec{b}^{J_{\text{s}2}}_{\zeta}, 0\right) \right\rangle_{\partial V^{2-}_{\text{sim}}} \tag{4-84g}$$

$$p_{\text{m};\xi\zeta}^{M_{s2}M_{s2}} = -(1/2)\left\langle \vec{b}_{\xi}^{M_{s2}}, \mathcal{H}_{\text{sim}}^{2}\left(0, \vec{b}_{\zeta}^{M_{s2}}\right)\right\rangle_{\partial V_{\text{sim}}^{2-}} \tag{4-84h}$$

In what follows, we will thoroughly finish variable unification in expansion vector space, and then obtain the matrix form of DPO $P_{\text{ss sys}}^{\text{driving}}$ such that the matrix form only contains BVs.

**Step 4. To Unify Variables in Expansion Vector Space**

Here, we will thoroughly finish the variable unification for DPO $P_{\text{ss sys}}^{\text{driving}}$ in expansion vector space, based on "the definition of equivalent surface magnetic current and electric-current-based testing method" and "the tangential continuity of the total fields on material-material interfaces". The variable unification scheme based on "the definition of equivalent surface electric current and magnetic-current-based testing method" is completely similar, so it will not be repeated here.

Based on the definition for equivalent surface magnetic current and the GSEP for internal total electric field obtained in Appendix C, the following equations can be derived:

$$\left[\mathcal{E}_{\text{sim}}^{1}\left(\vec{J}_{s10}^{\text{ES}} + \vec{J}_{s12}^{\text{ES}}, \vec{M}_{s10}^{\text{ES}} + \vec{M}_{s12}^{\text{ES}}\right)\right]_{\vec{r}_{\text{sim}}^{1} \to \vec{r}}^{\tan} = \hat{n}_{s1}^{-}(\vec{r}) \times \vec{M}_{s10}^{\text{ES}}(\vec{r}) \quad , \quad \vec{r} \in \partial V_{s10} \tag{4-85}$$

$$\left[\mathcal{E}_{\text{sim}}^{2}\left(\vec{J}_{s20}^{\text{ES}} - \vec{J}_{s12}^{\text{ES}}, \vec{M}_{s20}^{\text{ES}} - \vec{M}_{s12}^{\text{ES}}\right)\right]_{\vec{r}_{\text{sim}}^{2} \to \vec{r}}^{\tan} = \hat{n}_{s2}^{-}(\vec{r}) \times \vec{M}_{s20}^{\text{ES}}(\vec{r}) \quad , \quad \vec{r} \in \partial V_{s20} \tag{4-86}$$

Based on the tangential continuity of the total fields on material-material interface $\partial V_{s12}$, the GSEP for internal total fields given in Appendix C, and current decompositions (4-67) and (4-68), we have the following equations:

$$\left[\mathcal{E}_{\text{sim}}^{1}\left(\vec{J}_{s10}^{\text{ES}} + \vec{J}_{s12}^{\text{ES}}, \vec{M}_{s10}^{\text{ES}} + \vec{M}_{s12}^{\text{ES}}\right)\right]_{\vec{r}_{\text{sim}}^{1} \to \vec{r}}^{\tan} = \left[\mathcal{E}_{\text{sim}}^{2}\left(\vec{J}_{s20}^{\text{ES}} - \vec{J}_{s12}^{\text{ES}}, \vec{M}_{s20}^{\text{ES}} - \vec{M}_{s12}^{\text{ES}}\right)\right]_{\vec{r}_{\text{sim}}^{2} \to \vec{r}}^{\tan} \quad , \quad \vec{r} \in \partial V_{s12} \tag{4-87}$$

$$\left[\mathcal{H}_{\text{sim}}^{1}\left(\vec{J}_{s10}^{\text{ES}} + \vec{J}_{s12}^{\text{ES}}, \vec{M}_{s10}^{\text{ES}} + \vec{M}_{s12}^{\text{ES}}\right)\right]_{\vec{r}_{\text{sim}}^{1} \to \vec{r}}^{\tan} = \left[\mathcal{H}_{\text{sim}}^{2}\left(\vec{J}_{s20}^{\text{ES}} - \vec{J}_{s12}^{\text{ES}}, \vec{M}_{s20}^{\text{ES}} - \vec{M}_{s12}^{\text{ES}}\right)\right]_{\vec{r}_{\text{sim}}^{2} \to \vec{r}}^{\tan} \quad , \quad \vec{r} \in \partial V_{s12} \tag{4-88}$$

In equations (4-85), (4-87), and (4-88), $\vec{r}_{\text{sim}}^{1} \in \text{int}V_{\text{sim}}^{1}$, and $\vec{r}_{\text{sim}}^{1}$ approaches $\vec{r}$. In equations (4-86)~(4-88), $\vec{r}_{\text{sim}}^{2} \in \text{int}V_{\text{sim}}^{2}$, and $\vec{r}_{\text{sim}}^{2}$ approaches $\vec{r}$. In addition, relationship (4-71) has been applied to equations (4-86)~(4-88).

Inserting expansion formulations (4-72) and (4-73) into equations (4-85)~(4-88), and testing equations (4-85), (4-86), (4-87), and (4-88) with basis functions $\{\vec{b}_{\xi}^{J_{s10}}\}_{\xi=1}^{\Xi^{J_{s10}}}$, $\{\vec{b}_{\xi}^{J_{s20}}\}_{\xi=1}^{\Xi^{J_{s20}}}$, $\{\vec{b}_{\xi}^{J_{s12}}\}_{\xi=1}^{\Xi^{J_{s12}}}$, and $\{\vec{b}_{\xi}^{M_{s12}}\}_{\xi=1}^{\Xi^{M_{s12}}}$ respectively, the equations will be discretized into the following matrix forms:

$$\begin{aligned}&\bar{\bar{Z}}_1^{J_{s10}EJ_{s10}}\cdot\bar{a}^{J_{s10}}+\bar{\bar{Z}}_1^{J_{s10}EJ_{s12}}\cdot\bar{a}^{J_{s12}}+\bar{\bar{Z}}_1^{J_{s10}EM_{s10}}\cdot\bar{a}^{M_{s10}}+\bar{\bar{Z}}_1^{J_{s10}EM_{s12}}\cdot\bar{a}^{M_{s12}}\\=\ &\bar{\bar{C}}^{J_{s10}M_{s10}}\cdot\bar{a}^{M_{s10}}\end{aligned}\tag{4-89}$$

$$\begin{aligned}&\bar{\bar{Z}}_2^{J_{s20}EJ_{s20}}\cdot\bar{a}^{J_{s20}}-\bar{\bar{Z}}_2^{J_{s20}EJ_{s12}}\cdot\bar{a}^{J_{s12}}+\bar{\bar{Z}}_2^{J_{s20}EM_{s20}}\cdot\bar{a}^{M_{s20}}-\bar{\bar{Z}}_2^{J_{s20}EM_{s12}}\cdot\bar{a}^{M_{s12}}\\=\ &\bar{\bar{C}}^{J_{s20}M_{s20}}\cdot\bar{a}^{M_{s20}}\end{aligned}\tag{4-90}$$

$$\begin{aligned}&\bar{\bar{Z}}_1^{J_{s12}EJ_{s10}}\cdot\bar{a}^{J_{s10}}+\bar{\bar{Z}}_1^{J_{s12}EJ_{s12}}\cdot\bar{a}^{J_{s12}}+\bar{\bar{Z}}_1^{J_{s12}EM_{s10}}\cdot\bar{a}^{M_{s10}}+\bar{\bar{Z}}_1^{J_{s12}EM_{s12}}\cdot\bar{a}^{M_{s12}}\\=\ &\bar{\bar{Z}}_2^{J_{s12}EJ_{s20}}\cdot\bar{a}^{J_{s20}}-\bar{\bar{Z}}_2^{J_{s12}EJ_{s12}}\cdot\bar{a}^{J_{s12}}+\bar{\bar{Z}}_2^{J_{s12}EM_{s20}}\cdot\bar{a}^{M_{s20}}-\bar{\bar{Z}}_2^{J_{s12}EM_{s12}}\cdot\bar{a}^{M_{s12}}\end{aligned}\tag{4-91}$$

$$\begin{aligned}&\bar{\bar{Z}}_1^{M_{s12}HJ_{s10}}\cdot\bar{a}^{J_{s10}}+\bar{\bar{Z}}_1^{M_{s12}HJ_{s12}}\cdot\bar{a}^{J_{s12}}+\bar{\bar{Z}}_1^{M_{s12}HM_{s10}}\cdot\bar{a}^{M_{s10}}+\bar{\bar{Z}}_1^{M_{s12}HM_{s12}}\cdot\bar{a}^{M_{s12}}\\=\ &\bar{\bar{Z}}_2^{M_{s12}HJ_{s20}}\cdot\bar{a}^{J_{s20}}-\bar{\bar{Z}}_2^{M_{s12}HJ_{s12}}\cdot\bar{a}^{J_{s12}}+\bar{\bar{Z}}_2^{M_{s12}HM_{s20}}\cdot\bar{a}^{M_{s20}}-\bar{\bar{Z}}_2^{M_{s12}HM_{s12}}\cdot\bar{a}^{M_{s12}}\end{aligned}\tag{4-92}$$

The elements of the matrices in equation (4-89) are calculated as follows:

$$z_{1;\xi\zeta}^{J_{s10}EJ_{s10}}=\left\langle\vec{b}_\xi^{J_{s10}},\mathcal{E}_{\text{sim}}^1\left(\vec{b}_\zeta^{J_{s10}},0\right)\right\rangle_{\partial V_{s10}^-}\tag{4-93a}$$

$$z_{1;\xi\zeta}^{J_{s10}EJ_{s12}}=\left\langle\vec{b}_\xi^{J_{s10}},\mathcal{E}_{\text{sim}}^1\left(\vec{b}_\zeta^{J_{s12}},0\right)\right\rangle_{\partial V_{s10}^-}\tag{4-93b}$$

$$z_{1;\xi\zeta}^{J_{s10}EM_{s10}}=\left\langle\vec{b}_\xi^{J_{s10}},\mathcal{E}_{\text{sim}}^1\left(0,\vec{b}_\zeta^{M_{s10}}\right)\right\rangle_{\partial V_{s10}^-}\tag{4-93c}$$

$$z_{1;\xi\zeta}^{J_{s10}EM_{s12}}=\left\langle\vec{b}_\xi^{J_{s10}},\mathcal{E}_{\text{sim}}^1\left(0,\vec{b}_\zeta^{M_{s12}}\right)\right\rangle_{\partial V_{s10}^-}\tag{4-93d}$$

$$c_{\xi\zeta}^{J_{s10}M_{s10}}=\left\langle\vec{b}_\xi^{J_{s10}},\hat{n}_{s1}^-\times\vec{b}_\zeta^{M_{s10}}\right\rangle_{\partial V_{s10}}\tag{4-93e}$$

The elements of the matrices in equation (4-90) are calculated as follows:

$$z_{2;\xi\zeta}^{J_{s20}EJ_{s20}}=\left\langle\vec{b}_\xi^{J_{s20}},\mathcal{E}_{\text{sim}}^2\left(\vec{b}_\zeta^{J_{s20}},0\right)\right\rangle_{\partial V_{s20}^-}\tag{4-94a}$$

$$z_{2;\xi\zeta}^{J_{s20}EJ_{s12}}=\left\langle\vec{b}_\xi^{J_{s20}},\mathcal{E}_{\text{sim}}^2\left(\vec{b}_\zeta^{J_{s12}},0\right)\right\rangle_{\partial V_{s20}^-}\tag{4-94b}$$

$$z_{2;\xi\zeta}^{J_{s20}EM_{s20}}=\left\langle\vec{b}_\xi^{J_{s20}},\mathcal{E}_{\text{sim}}^2\left(0,\vec{b}_\zeta^{M_{s20}}\right)\right\rangle_{\partial V_{s20}^-}\tag{4-94c}$$

$$z_{2;\xi\zeta}^{J_{s20}EM_{s12}}=\left\langle\vec{b}_\xi^{J_{s20}},\mathcal{E}_{\text{sim}}^2\left(0,\vec{b}_\zeta^{M_{s12}}\right)\right\rangle_{\partial V_{s20}^-}\tag{4-94d}$$

$$c_{\xi\zeta}^{J_{s20}M_{s20}}=\left\langle\vec{b}_\xi^{J_{s20}},\hat{n}_{s2}^-\times\vec{b}_\zeta^{M_{s20}}\right\rangle_{\partial V_{s20}}\tag{4-94e}$$

The elements of the matrices in equation (4-91) are calculated as follows:

$$z_{1;\xi\zeta}^{J_{s12}EJ_{s10}}=\left\langle\vec{b}_\xi^{J_{s12}},\mathcal{E}_{\text{sim}}^1\left(\vec{b}_\zeta^{J_{s10}},0\right)\right\rangle_{\partial V_{s12}^-}\tag{4-95a}$$

$$z_{1;\xi\zeta}^{J_{s12}EJ_{s12}}=\left\langle\vec{b}_\xi^{J_{s12}},\mathcal{E}_{\text{sim}}^1\left(\vec{b}_\zeta^{J_{s12}},0\right)\right\rangle_{\partial V_{s12}^-}\tag{4-95b}$$

$$z_{1;\xi\zeta}^{J_{s12}EM_{s10}}=\left\langle\vec{b}_\xi^{J_{s12}},\mathcal{E}_{\text{sim}}^1\left(0,\vec{b}_\zeta^{M_{s10}}\right)\right\rangle_{\partial V_{s12}^-}\tag{4-95c}$$

$$z_{1;\xi\zeta}^{J_{s12}EM_{s12}}=\left\langle\vec{b}_\xi^{J_{s12}},\mathcal{E}_{\text{sim}}^1\left(0,\vec{b}_\zeta^{M_{s12}}\right)\right\rangle_{\partial V_{s12}^-}\tag{4-95d}$$

$$z_{2;\xi\zeta}^{J_{s12}EJ_{s20}}=\left\langle\vec{b}_\xi^{J_{s12}},\mathcal{E}_{\text{sim}}^2\left(\vec{b}_\zeta^{J_{s20}},0\right)\right\rangle_{\partial V_{s21}^-}\tag{4-95e}$$

$$z_{2;\xi\zeta}^{J_{s12}EJ_{s12}}=\left\langle\vec{b}_\xi^{J_{s12}},\mathcal{E}_{\text{sim}}^2\left(\vec{b}_\zeta^{J_{s12}},0\right)\right\rangle_{\partial V_{s21}^-}\tag{4-95f}$$

$$z_{2;\xi\zeta}^{J_{s12}EM_{s20}}=\left\langle\vec{b}_\xi^{J_{s12}},\mathcal{E}_{\text{sim}}^2\left(0,\vec{b}_\zeta^{M_{s20}}\right)\right\rangle_{\partial V_{s21}^-}\tag{4-95g}$$

$$z_{2;\xi\zeta}^{J_{s12}EM_{s12}} = \left\langle \vec{b}_{\xi}^{J_{s12}}, \mathcal{E}_{\text{sim}}^{2}\left(0, \vec{b}_{\zeta}^{M_{s12}}\right)\right\rangle_{\partial V_{s21}^{-}} \tag{4-95h}$$

The elements of the matrices in equation (4-92) are calculated as follows:

$$z_{1;\xi\zeta}^{M_{s12}HJ_{s10}} = \left\langle \vec{b}_{\xi}^{M_{s12}}, \mathcal{H}_{\text{sim}}^{1}\left(\vec{b}_{\zeta}^{J_{s10}}, 0\right)\right\rangle_{\partial V_{s12}^{-}} \tag{4-96a}$$

$$z_{1;\xi\zeta}^{M_{s12}HJ_{s12}} = \left\langle \vec{b}_{\xi}^{M_{s12}}, \mathcal{H}_{\text{sim}}^{1}\left(\vec{b}_{\zeta}^{J_{s12}}, 0\right)\right\rangle_{\partial V_{s12}^{-}} \tag{4-96b}$$

$$z_{1;\xi\zeta}^{M_{s12}HM_{s10}} = \left\langle \vec{b}_{\xi}^{M_{s12}}, \mathcal{H}_{\text{sim}}^{1}\left(0, \vec{b}_{\zeta}^{M_{s10}}\right)\right\rangle_{\partial V_{s12}^{-}} \tag{4-96c}$$

$$z_{1;\xi\zeta}^{M_{s12}HM_{s12}} = \left\langle \vec{b}_{\xi}^{M_{s12}}, \mathcal{H}_{\text{sim}}^{1}\left(0, \vec{b}_{\zeta}^{M_{s12}}\right)\right\rangle_{\partial V_{s12}^{-}} \tag{4-96d}$$

$$z_{2;\xi\zeta}^{M_{s12}HJ_{s20}} = \left\langle \vec{b}_{\xi}^{M_{s12}}, \mathcal{H}_{\text{sim}}^{2}\left(\vec{b}_{\zeta}^{J_{s20}}, 0\right)\right\rangle_{\partial V_{s21}^{-}} \tag{4-96e}$$

$$z_{2;\xi\zeta}^{M_{s12}HJ_{s12}} = \left\langle \vec{b}_{\xi}^{M_{s12}}, \mathcal{H}_{\text{sim}}^{2}\left(\vec{b}_{\zeta}^{J_{s12}}, 0\right)\right\rangle_{\partial V_{s21}^{-}} \tag{4-96f}$$

$$z_{2;\xi\zeta}^{M_{s12}HM_{s20}} = \left\langle \vec{b}_{\xi}^{M_{s12}}, \mathcal{H}_{\text{sim}}^{2}\left(0, \vec{b}_{\zeta}^{M_{s20}}\right)\right\rangle_{\partial V_{s21}^{-}} \tag{4-96g}$$

$$z_{2;\xi\zeta}^{M_{s12}HM_{s12}} = \left\langle \vec{b}_{\xi}^{M_{s12}}, \mathcal{H}_{\text{sim}}^{2}\left(0, \vec{b}_{\zeta}^{M_{s12}}\right)\right\rangle_{\partial V_{s21}^{-}} \tag{4-96h}$$

Above integral domains $\partial V_{s10}^{-}$, $\partial V_{s20}^{-}$, $\partial V_{s12}^{-}$, and $\partial V_{s21}^{-}$ are respectively the $\partial V_{s10}$ on the side of $V_{\text{sim}}^{1}$, the $\partial V_{s20}$ on the side of $V_{\text{sim}}^{2}$, the $\partial V_{s12}$ on the side of $V_{\text{sim}}^{1}$, and the $\partial V_{s21}$ on the side of $V_{\text{sim}}^{2}$.

In simultaneous matrix equations (4-89)~(4-92), there are 6 different currents $\bar{a}^{J_{s10}}$, $\bar{a}^{J_{s12}}$, $\bar{a}^{J_{s20}}$, $\bar{a}^{M_{s10}}$, $\bar{a}^{M_{s12}}$, and $\bar{a}^{M_{s20}}$. We select $\bar{a}^{M_{s10}}$ and $\bar{a}^{M_{s20}}$ as BVs, and then the other 4 currents naturally become dependent variables①. To effectively establish the transformation from the BVs to the dependent variables, we properly re-arrange and assemble matrix equations (4-89)~(4-92) as the following single augmented matrix equation:

$$\begin{bmatrix} \bar{\bar{Z}}_{1}^{J_{s10}EJ_{s10}} & 0 & \bar{\bar{Z}}_{1}^{J_{s10}EJ_{s12}} & \bar{\bar{Z}}_{1}^{J_{s10}EM_{s12}} \\ 0 & -\bar{\bar{Z}}_{2}^{J_{s20}EJ_{s20}} & \bar{\bar{Z}}_{2}^{J_{s20}EJ_{s12}} & \bar{\bar{Z}}_{2}^{J_{s20}EM_{s12}} \\ \bar{\bar{Z}}_{1}^{J_{s12}EJ_{s10}} & -\bar{\bar{Z}}_{2}^{J_{s12}EJ_{s20}} & \bar{\bar{Z}}_{1}^{J_{s12}EJ_{s12}} + \bar{\bar{Z}}_{2}^{J_{s12}EJ_{s12}} & \bar{\bar{Z}}_{1}^{J_{s12}EM_{s12}} + \bar{\bar{Z}}_{2}^{J_{s12}EM_{s12}} \\ \bar{\bar{Z}}_{1}^{M_{s12}HJ_{s10}} & -\bar{\bar{Z}}_{2}^{M_{s12}HJ_{s20}} & \bar{\bar{Z}}_{1}^{M_{s12}HJ_{s12}} + \bar{\bar{Z}}_{2}^{M_{s12}HJ_{s12}} & \bar{\bar{Z}}_{1}^{M_{s12}HM_{s12}} + \bar{\bar{Z}}_{2}^{M_{s12}HM_{s12}} \end{bmatrix} \cdot \begin{bmatrix} \bar{a}^{J_{s10}} \\ \bar{a}^{J_{s20}} \\ \bar{a}^{J_{s12}} \\ \bar{a}^{M_{s12}} \end{bmatrix}$$

$$= \begin{bmatrix} \bar{\bar{C}}^{J_{s10}M_{s10}} - \bar{\bar{Z}}_{1}^{J_{s10}EM_{s10}} & 0 \\ 0 & \bar{\bar{Z}}_{2}^{J_{s20}EM_{s20}} - \bar{\bar{C}}^{J_{s20}M_{s20}} \\ -\bar{\bar{Z}}_{1}^{J_{s12}EM_{s10}} & \bar{\bar{Z}}_{2}^{J_{s12}EM_{s20}} \\ -\bar{\bar{Z}}_{1}^{M_{s12}HM_{s10}} & \bar{\bar{Z}}_{2}^{M_{s12}HM_{s20}} \end{bmatrix} \cdot \begin{bmatrix} \bar{a}^{M_{s10}} \\ \bar{a}^{M_{s20}} \end{bmatrix} \tag{4-97}$$

Obviously, the above equation implies the following linear transformation:

① 6 (variables) – 4 (complete and independent variables) = 2 (basic variables). But, it is not the case that any 2 EM currents are suitable for being basic variables. On how to select basic variables, we will carefully discuss it and completely answer to it in the Section 6.2 of this dissertation.

$$\begin{bmatrix} \overline{a}^{J_{s10}} \\ \overline{a}^{J_{s20}} \\ \overline{a}^{J_{s12}} \\ \overline{a}^{M_{s12}} \end{bmatrix} = \overline{\overline{T}}^{\{J_{s10},J_{s20},J_{s12},M_{s12}\}\leftarrow M_{s0}} \cdot \underbrace{\begin{bmatrix} \overline{a}^{M_{s10}} \\ \overline{a}^{M_{s20}} \end{bmatrix}}_{\overline{a}^{M_{s0}}} \tag{4-98}$$

where

$$\overline{\overline{T}}^{\{J_{s10},J_{s20},J_{s12},M_{s12}\}\leftarrow M_{s0}}$$

$$= \begin{bmatrix} \overline{\overline{Z}}_1^{J_{s10}EJ_{s10}} & 0 & \overline{\overline{Z}}_1^{J_{s10}EJ_{s12}} & \overline{\overline{Z}}_1^{J_{s10}EM_{s12}} \\ 0 & -\overline{\overline{Z}}_2^{J_{s20}EJ_{s20}} & \overline{\overline{Z}}_2^{J_{s20}EJ_{s12}} & \overline{\overline{Z}}_2^{J_{s20}EM_{s12}} \\ \overline{\overline{Z}}_1^{J_{s12}EJ_{s10}} & -\overline{\overline{Z}}_2^{J_{s12}EJ_{s20}} & \overline{\overline{Z}}_1^{J_{s12}EJ_{s12}} + \overline{\overline{Z}}_2^{J_{s12}EJ_{s12}} & \overline{\overline{Z}}_1^{J_{s12}EM_{s12}} + \overline{\overline{Z}}_2^{J_{s12}EM_{s12}} \\ \overline{\overline{Z}}_1^{M_{s12}HJ_{s10}} & -\overline{\overline{Z}}_2^{M_{s12}HJ_{s20}} & \overline{\overline{Z}}_1^{M_{s12}HJ_{s12}} + \overline{\overline{Z}}_2^{M_{s12}HJ_{s12}} & \overline{\overline{Z}}_1^{M_{s12}HM_{s12}} + \overline{\overline{Z}}_2^{M_{s12}HM_{s12}} \end{bmatrix}^{-1} \cdot$$

$$\begin{bmatrix} \overline{\overline{C}}^{J_{s10}M_{s10}} - \overline{\overline{Z}}_1^{J_{s10}EM_{s10}} & 0 \\ 0 & \overline{\overline{Z}}_2^{J_{s20}EM_{s20}} - \overline{\overline{C}}^{J_{s20}M_{s20}} \\ -\overline{\overline{Z}}_1^{J_{s12}EM_{s10}} & \overline{\overline{Z}}_2^{J_{s12}EM_{s20}} \\ -\overline{\overline{Z}}_1^{M_{s12}HM_{s10}} & \overline{\overline{Z}}_2^{M_{s12}HM_{s20}} \end{bmatrix} \tag{4-99}$$

If we partition above transformation matrix $\overline{\overline{T}}^{\{J_{s10},J_{s20},J_{s12},M_{s12}\}\leftarrow M_{s0}}$ as the partition way of the vector on the LHS of formulation (4-98), i.e.,

$$\overline{\overline{T}}^{\{J_{s10},J_{s20},J_{s12},M_{s12}\}\leftarrow M_{s0}} = \begin{bmatrix} \overline{\overline{T}}^{J_{s10}\leftarrow M_{s0}} \\ \overline{\overline{T}}^{J_{s20}\leftarrow M_{s0}} \\ \overline{\overline{T}}^{J_{s12}\leftarrow M_{s0}} \\ \overline{\overline{T}}^{M_{s12}\leftarrow M_{s0}} \end{bmatrix} \tag{4-100}$$

then we have the following a series of transformations:

$$\overline{a}^{J_{s10}} = \overline{\overline{T}}^{J_{s10}\leftarrow M_{s0}} \cdot \overline{a}^{M_{s0}} \tag{4-101a}$$

$$\overline{a}^{J_{s20}} = \overline{\overline{T}}^{J_{s20}\leftarrow M_{s0}} \cdot \overline{a}^{M_{s0}} \tag{4-101b}$$

$$\overline{a}^{J_{s12}} = \overline{\overline{T}}^{J_{s12}\leftarrow M_{s0}} \cdot \overline{a}^{M_{s0}} \tag{4-101c}$$

$$\overline{a}^{M_{s12}} = \overline{\overline{T}}^{M_{s12}\leftarrow M_{s0}} \cdot \overline{a}^{M_{s0}} \tag{4-101d}$$

If relationship (4-71) is utilized, then we further have the following transformations:

$$\overline{a}^{J_{s21}} = -\overline{\overline{T}}^{J_{s12}\leftarrow M_{s0}} \cdot \overline{a}^{M_{s0}} \tag{4-101e}$$

$$\overline{a}^{M_{s21}} = -\overline{\overline{T}}^{M_{s12}\leftarrow M_{s0}} \cdot \overline{a}^{M_{s0}} \tag{4-101f}$$

Inserting above transformations (4-101a)~(4-101f) into the original matrix form (4-80) of DPO $P_{\text{ss sys}}^{\text{driving}}$, it is immediately obtained that

$$P_{\text{ss sys}}^{\text{driving}} = \begin{bmatrix} \overline{a}^{J_{\text{s10}}} \\ \overline{a}^{J_{\text{s12}}} \\ \overline{a}^{J_{\text{s21}}} \\ \overline{a}^{J_{\text{s20}}} \\ \overline{a}^{M_{\text{s10}}} \\ \overline{a}^{M_{\text{s12}}} \\ \overline{a}^{M_{\text{s21}}} \\ \overline{a}^{M_{\text{s20}}} \end{bmatrix}^H \cdot \overline{\overline{P}}_{\text{ss sys;1}}^{\text{driving}} \cdot \begin{bmatrix} \overline{a}^{J_{\text{s10}}} \\ \overline{a}^{J_{\text{s12}}} \\ \overline{a}^{J_{\text{s21}}} \\ \overline{a}^{J_{\text{s20}}} \\ \overline{a}^{M_{\text{s10}}} \\ \overline{a}^{M_{\text{s12}}} \\ \overline{a}^{M_{\text{s21}}} \\ \overline{a}^{M_{\text{s20}}} \end{bmatrix} = \left(\overline{a}^{M_{\text{s0}}}\right)^H \cdot \underbrace{\begin{bmatrix} \overline{\overline{T}}^{J_{\text{s10}} \leftarrow M_{\text{s0}}} \\ \overline{\overline{T}}^{J_{\text{s12}} \leftarrow M_{\text{s0}}} \\ -\overline{\overline{T}}^{J_{\text{s12}} \leftarrow M_{\text{s0}}} \\ \overline{\overline{T}}^{J_{\text{s20}} \leftarrow M_{\text{s0}}} \\ \overline{\overline{\mathcal{I}}}^{M_{\text{s10}}} \\ \overline{\overline{T}}^{M_{\text{s12}} \leftarrow M_{\text{s0}}} \\ -\overline{\overline{T}}^{M_{\text{s12}} \leftarrow M_{\text{s0}}} \\ \overline{\overline{\mathcal{I}}}^{M_{\text{s20}}} \end{bmatrix}^H \cdot \overline{\overline{P}}_{\text{ss sys;1}}^{\text{driving}} \cdot \begin{bmatrix} \overline{\overline{T}}^{J_{\text{s10}} \leftarrow M_{\text{s0}}} \\ \overline{\overline{T}}^{J_{\text{s12}} \leftarrow M_{\text{s0}}} \\ -\overline{\overline{T}}^{J_{\text{s12}} \leftarrow M_{\text{s0}}} \\ \overline{\overline{T}}^{J_{\text{s20}} \leftarrow M_{\text{s0}}} \\ \overline{\overline{\mathcal{I}}}^{M_{\text{s10}}} \\ \overline{\overline{T}}^{M_{\text{s12}} \leftarrow M_{\text{s0}}} \\ -\overline{\overline{T}}^{M_{\text{s12}} \leftarrow M_{\text{s0}}} \\ \overline{\overline{\mathcal{I}}}^{M_{\text{s20}}} \end{bmatrix}}_{\overline{\overline{P}}_{1;M_{\text{s0}}}^{\text{driving}}} \cdot \overline{a}^{M_{\text{s0}}} \tag{4-102}$$

In formulation (4-102), the first equality is based on the partition ways of (4-78) and (4-79); $\overline{\overline{\mathcal{I}}}^{M_{\text{s10}}} = [\overline{\overline{I}}^{M_{\text{s10}}} \ \ 0]$ and $\overline{\overline{\mathcal{I}}}^{M_{\text{s20}}} = [0 \ \ \overline{\overline{I}}^{M_{\text{s20}}}]$, where $\overline{\overline{I}}^{M_{\text{s10}}}$ and $\overline{\overline{I}}^{M_{\text{s20}}}$ are the identity matrices whose orders are $\Xi^{M_{\text{s10}}}$ and $\Xi^{M_{\text{s20}}}$ respectively, and the 0s are some zero matrices having proper line numbers and proper column numbers. Formulation (4-102) is just the DPO's matrix form which contains only BVs. In the following Subsection 4.4.2, we will employ the matrix form to construct the DP-CMs of the two-body material system shown in Figure 4-27, and discuss the orthogonality among the DP-CMs.

## 4.4.2 DP-CMs and Their Orthogonality

Matrix $\overline{\overline{P}}_{1;M_{\text{s0}}}^{\text{driving}}$ uniquely has the following Toeplitz's decomposition[116]:

$$\overline{\overline{P}}_{1;M_{\text{s0}}}^{\text{driving}} = \overline{\overline{P}}_{1;M_{\text{s0}};+}^{\text{driving}} + j \, \overline{\overline{P}}_{1;M_{\text{s0}};-}^{\text{driving}} \tag{4-103}$$

where $\overline{\overline{P}}_{1;M_{\text{s0}};+}^{\text{driving}} = [\overline{\overline{P}}_{1;M_{\text{s0}}}^{\text{driving}} + (\overline{\overline{P}}_{1;M_{\text{s0}}}^{\text{driving}})^H]/2$ and $\overline{\overline{P}}_{1;M_{\text{s0}};-}^{\text{driving}} = [\overline{\overline{P}}_{1;M_{\text{s0}}}^{\text{driving}} - (\overline{\overline{P}}_{1;M_{\text{s0}}}^{\text{driving}})^H]/2j$ . The DP-CMs of two-body material systems can be derived from solving the following generalized characteristic equation:

$$\overline{\overline{P}}_{1;M_{\text{s0}};-}^{\text{driving}} \cdot \overline{\alpha}_{M_{\text{s0}};\xi}^{\text{driving}} = \lambda_{\text{ss sys};\xi}^{\text{driving}} \, \overline{\overline{P}}_{1;M_{\text{s0}};+}^{\text{driving}} \cdot \overline{\alpha}_{M_{\text{s0}};\xi}^{\text{driving}} \tag{4-104}$$

Inserting above characteristic vectors $\{\overline{\alpha}_{M_{\text{s0}};\xi}^{\text{driving}}\}_{\xi=1}^{\Xi^{M_{\text{s0}}}}$ into transformation formulation (4-101) and expansion formulations (4-72) and (4-73), characteristic equivalent surface sources $\{\vec{J}_{\text{s10};\xi}^{\text{ES}}, \vec{J}_{\text{s12};\xi}^{\text{ES}}, \vec{J}_{\text{s21};\xi}^{\text{ES}}, \vec{J}_{\text{s20};\xi}^{\text{ES}}; \vec{M}_{\text{s10};\xi}^{\text{ES}}, \vec{M}_{\text{s12};\xi}^{\text{ES}}, \vec{M}_{\text{s21};\xi}^{\text{ES}}, \vec{M}_{\text{s20};\xi}^{\text{ES}}\}_{\xi=1}^{\Xi^{M_{\text{s0}}}}$ can then be determined, where $\Xi^{M_{\text{s0}}} = \Xi^{M_{\text{s10}}} + \Xi^{M_{\text{s20}}}$ . Inserting the characteristic equivalent surface sources into the GSEP given in Appendix C, characteristic fields can be determined. Employing the characteristic total fields distributing on material systems and the volume equivalent principle given in Appendix A, characteristic scattered volume sources $\{\vec{J}_{\text{s1};\xi}^{\text{SV}}, \vec{M}_{\text{s1};\xi}^{\text{SV}}; \vec{J}_{\text{s2};\xi}^{\text{SV}}, \vec{M}_{\text{s2};\xi}^{\text{SV}}\}_{\xi=1}^{\Xi^{M_{\text{s0}}}}$ can be determined.

Obviously, characteristic values $\lambda_{\text{ss sys};\xi}^{\text{driving}}$ and the modal powers satisfy the following relationship:

$$\lambda_{\text{ss sys};\xi}^{\text{driving}} = \frac{\text{Im}\left\{P_{\text{ss sys};\xi}^{\text{driving}}\right\}}{\text{Re}\left\{P_{\text{ss sys};\xi}^{\text{driving}}\right\}} = \frac{\text{Im}\left\{\sum_{i=1}^{2}\left[\frac{1}{2}\left\langle \vec{J}_{\text{s}i;\xi}^{\text{SV}}, \vec{E}_{\xi}^{\text{inc}}\right\rangle_{V_{\text{sim}}^{i}} + \frac{1}{2}\left\langle \vec{M}_{\text{s}i;\xi}^{\text{SV}}, \vec{H}_{\xi}^{\text{inc}}\right\rangle_{V_{\text{sim}}^{i}}\right]\right\}}{\text{Re}\left\{\sum_{i=1}^{2}\left[\frac{1}{2}\left\langle \vec{J}_{\text{s}i;\xi}^{\text{SV}}, \vec{E}_{\xi}^{\text{inc}}\right\rangle_{V_{\text{sim}}^{i}} + \frac{1}{2}\left\langle \vec{M}_{\text{s}i;\xi}^{\text{SV}}, \vec{H}_{\xi}^{\text{inc}}\right\rangle_{V_{\text{sim}}^{i}}\right]\right\}}$$

$$= \frac{\text{Im}\left\{-\sum_{i=1}^{2}\left[\frac{1}{2}\left\langle \vec{J}_{\text{s}i;\xi}^{\text{ES}}, \vec{E}_{\xi}^{\text{inc}}\right\rangle_{\partial V_{\text{sim}}^{i}} + \frac{1}{2}\left\langle \vec{M}_{\text{s}i;\xi}^{\text{ES}}, \vec{H}_{\xi}^{\text{inc}}\right\rangle_{\partial V_{\text{sim}}^{i}}\right]\right\}}{\text{Re}\left\{-\sum_{i=1}^{2}\left[\frac{1}{2}\left\langle \vec{J}_{\text{s}i;\xi}^{\text{ES}}, \vec{E}_{\xi}^{\text{inc}}\right\rangle_{\partial V_{\text{sim}}^{i}} + \frac{1}{2}\left\langle \vec{M}_{\text{s}i;\xi}^{\text{ES}}, \vec{H}_{\xi}^{\text{inc}}\right\rangle_{\partial V_{\text{sim}}^{i}}\right]\right\}} \tag{4-105}$$

where $\xi = 1, 2, \cdots, \Xi^{M_{\text{s}0}}$, and the third equality is based on formulation (4-64).

In addition, the characteristic vectors satisfy the following orthogonality:

$$\text{Re}\left\{P_{\text{ss sys};\xi}^{\text{driving}}\right\}\delta_{\xi\zeta} = \left(\bar{\alpha}_{M_{\text{s}0};\xi}^{\text{driving}}\right)^{H} \cdot \bar{\bar{P}}_{1;M_{\text{s}0};+}^{\text{driving}} \cdot \bar{\alpha}_{M_{\text{s}0};\zeta}^{\text{driving}} \tag{4-106a}$$

$$\text{Im}\left\{P_{\text{ss sys};\xi}^{\text{driving}}\right\}\delta_{\xi\zeta} = \left(\bar{\alpha}_{M_{\text{s}0};\xi}^{\text{driving}}\right)^{H} \cdot \bar{\bar{P}}_{1;M_{\text{s}0};-}^{\text{driving}} \cdot \bar{\alpha}_{M_{\text{s}0};\zeta}^{\text{driving}} \tag{4-106b}$$

and

$$\underbrace{\left[\text{Re}\left\{P_{\text{ss sys};\xi}^{\text{driving}}\right\} + j\,\text{Im}\left\{P_{\text{ss sys};\xi}^{\text{driving}}\right\}\right]}_{P_{\text{ss sys};\xi}^{\text{driving}}}\delta_{\xi\zeta} = \left(\bar{\alpha}_{M_{\text{s}0};\xi}^{\text{driving}}\right)^{H} \cdot \underbrace{\left(\bar{\bar{P}}_{1;M_{\text{s}0};+}^{\text{driving}} + j\,\bar{\bar{P}}_{1;M_{\text{s}0};-}^{\text{driving}}\right)}_{\bar{\bar{P}}_{1;M_{\text{s}0}}^{\text{driving}}} \cdot \bar{\alpha}_{M_{\text{s}0};\zeta}^{\text{driving}} \tag{4-107}$$

Similarly to deriving formulation (4-59) from formulation (4-58), from above formulation (4-107) we can derive the following orthogonality among characteristic incident fields and characteristic scattered sources:

$$P_{\text{ss sys};\xi}^{\text{driving}}\delta_{\xi\zeta} = \sum_{i=1}^{2}\left[\left(1/2\right)\left\langle \vec{J}_{\text{s}i;\xi}^{\text{SV}}, \vec{E}_{\zeta}^{\text{inc}}\right\rangle_{V_{\text{sim}}^{i}} + \left(1/2\right)\left\langle \vec{M}_{\text{s}i;\xi}^{\text{SV}}, \vec{H}_{\zeta}^{\text{inc}}\right\rangle_{V_{\text{sim}}^{i}}\right]$$

$$= -\sum_{i=1}^{2}\left[\left(1/2\right)\left\langle \vec{J}_{\text{s}i;\xi}^{\text{ES}}, \vec{E}_{\zeta}^{\text{inc}}\right\rangle_{\partial V_{\text{sim}}^{i}} + \left(1/2\right)\left\langle \vec{M}_{\text{s}i;\xi}^{\text{ES}}, \vec{H}_{\zeta}^{\text{inc}}\right\rangle_{\partial V_{\text{sim}}^{i}}\right] \tag{4-108}$$

where the second equality is based on relationship (4-64).

### 4.4.3 DP-CM-Based Modal Expansion

Based on the completeness of the DP-CMs, the any working mode of the two-body material system shown in Figure 4-27 can be expanded in terms of the DP-CMs as follows:

$$\vec{E}^{\text{inc}}\left(\vec{r}\right) = \sum_{\xi=1}^{\Xi^{M_{\text{s}0}}} c_{\xi}\vec{E}_{\xi}^{\text{inc}}\left(\vec{r}\right) \quad , \quad \vec{r} \in V_{\text{sim}}^{1} \cup V_{\text{sim}}^{2} \tag{4-109a}$$

$$\vec{H}^{\text{inc}}\left(\vec{r}\right) = \sum_{\xi=1}^{\Xi^{M_{\text{s}0}}} c_{\xi}\vec{H}_{\xi}^{\text{inc}}\left(\vec{r}\right) \quad , \quad \vec{r} \in V_{\text{sim}}^{1} \cup V_{\text{sim}}^{2} \tag{4-109b}$$

Testing above modal expansion formulations (4-109a) and (4-109b) with functions $\{(\vec{J}_{\text{s}1;\xi}^{\text{SV}} + \vec{J}_{\text{s}2;\xi}^{\text{SV}})/2\}_{\xi=1}^{\Xi^{M_{\text{s}0}}}$ and $\{(\vec{M}_{\text{s}1;\xi}^{\text{SV}} + \vec{M}_{\text{s}2;\xi}^{\text{SV}})/2\}_{\xi=1}^{\Xi^{M_{\text{s}0}}}$ respectively, and summing the obtained results, we have the following equation:

$$
\begin{aligned}
&\sum\nolimits_{i=1}^{2}(1/2)\left\langle \vec{J}_{\text{s}i;\xi}^{\text{SV}},\vec{E}^{\text{inc}}\right\rangle_{V_{\text{sim}}^{i}}+\sum\nolimits_{i=1}^{2}(1/2)\left\langle \vec{M}_{\text{s}i;\xi}^{\text{SV}},\vec{H}^{\text{inc}}\right\rangle_{V_{\text{sim}}^{i}}\\
&=\sum\nolimits_{i=1}^{2}(1/2)\left\langle \vec{J}_{\text{s}i;\xi}^{\text{SV}},\sum\nolimits_{\zeta=1}^{\Xi^{M_{\text{s}0}}}c_{\zeta}\vec{E}_{\zeta}^{\text{inc}}\right\rangle_{V_{\text{sim}}^{i}}+\sum\nolimits_{i=1}^{2}(1/2)\left\langle \vec{M}_{\text{s}i;\xi}^{\text{SV}},\sum\nolimits_{\zeta=1}^{\Xi^{M_{\text{s}0}}}c_{\zeta}\vec{H}_{\zeta}^{\text{inc}}\right\rangle_{V_{\text{sim}}^{i}}\\
&=\sum\nolimits_{i=1}^{2}\left[(1/2)\left\langle \vec{J}_{\text{s}i;\xi}^{\text{SV}},\sum\nolimits_{\zeta=1}^{\Xi^{M_{\text{s}0}}}c_{\zeta}\vec{E}_{\zeta}^{\text{inc}}\right\rangle_{V_{\text{sim}}^{i}}+(1/2)\left\langle \vec{M}_{\text{s}i;\xi}^{\text{SV}},\sum\nolimits_{\zeta=1}^{\Xi^{M_{\text{s}0}}}c_{\zeta}\vec{H}_{\zeta}^{\text{inc}}\right\rangle_{V_{\text{sim}}^{i}}\right]\\
&=\sum\nolimits_{i=1}^{2}\sum\nolimits_{\zeta=1}^{\Xi^{M_{\text{s}0}}}c_{\zeta}\left[(1/2)\left\langle \vec{J}_{\text{s}i;\xi}^{\text{SV}},\vec{E}_{\zeta}^{\text{inc}}\right\rangle_{V_{\text{sim}}^{i}}+(1/2)\left\langle \vec{M}_{\text{s}i;\xi}^{\text{SV}},\vec{H}_{\zeta}^{\text{inc}}\right\rangle_{V_{\text{sim}}^{i}}\right]\\
&=\sum\nolimits_{\zeta=1}^{\Xi^{M_{\text{s}0}}}\left\{c_{\zeta}\sum\nolimits_{i=1}^{2}\left[(1/2)\left\langle \vec{J}_{\text{s}i;\xi}^{\text{SV}},\vec{E}_{\zeta}^{\text{inc}}\right\rangle_{V_{\text{sim}}^{i}}+(1/2)\left\langle \vec{M}_{\text{s}i;\xi}^{\text{SV}},\vec{H}_{\zeta}^{\text{inc}}\right\rangle_{V_{\text{sim}}^{i}}\right]\right\}
\end{aligned}
\tag{4-110}
$$

In above equation (4-110), $\xi=1,2,\cdots,\Xi^{M_{\text{s}0}}$; the second equality is evident; the third equality is based on the linear property of inner product; the fourth equality is based on the interchangeability[117] of the summations in equation (4-110). Applying orthogonality (4-108) to equation (4-110), we have that

$$
\begin{aligned}
&\sum\nolimits_{i=1}^{2}\left[(1/2)\left\langle \vec{J}_{\text{s}i;\xi}^{\text{SV}},\vec{E}^{\text{inc}}\right\rangle_{V_{\text{sim}}^{i}}+(1/2)\left\langle \vec{M}_{\text{s}i;\xi}^{\text{SV}},\vec{H}^{\text{inc}}\right\rangle_{V_{\text{sim}}^{i}}\right]\\
&=\sum\nolimits_{i=1}^{2}(1/2)\left\langle \vec{J}_{\text{s}i;\xi}^{\text{SV}},\vec{E}^{\text{inc}}\right\rangle_{V_{\text{sim}}^{i}}+\sum\nolimits_{i=1}^{2}(1/2)\left\langle \vec{M}_{\text{s}i;\xi}^{\text{SV}},\vec{H}^{\text{inc}}\right\rangle_{V_{\text{sim}}^{i}}\\
&=c_{\xi}P_{\text{ss sys};\xi}^{\text{driving}}
\end{aligned}
\tag{4-111}
$$

In general, $P_{\text{ss sys};\xi}^{\text{driving}}\neq 0$ for material systems, so

$$
\begin{aligned}
c_{\xi}&=\frac{1}{P_{\text{ss sys};\xi}^{\text{driving}}}\sum\nolimits_{i=1}^{2}\left[(1/2)\left\langle \vec{J}_{\text{s}i;\xi}^{\text{SV}},\vec{E}^{\text{inc}}\right\rangle_{V_{\text{sim}}^{i}}+(1/2)\left\langle \vec{M}_{\text{s}i;\xi}^{\text{SV}},\vec{H}^{\text{inc}}\right\rangle_{V_{\text{sim}}^{i}}\right]\\
&=-\frac{1}{P_{\text{ss sys};\xi}^{\text{driving}}}\sum\nolimits_{i=1}^{2}\left[(1/2)\left\langle \vec{J}_{\text{s}i;\xi}^{\text{ES}},\vec{E}^{\text{inc}}\right\rangle_{\partial V_{\text{sim}}^{i}}+(1/2)\left\langle \vec{M}_{\text{s}i;\xi}^{\text{ES}},\vec{H}^{\text{inc}}\right\rangle_{\partial V_{\text{sim}}^{i}}\right]
\end{aligned}
\tag{4-112}
$$

where the second equality is based on formulation (4-64). Formulation (4-112) is just the explicit expression for the expansion coefficients in modal expansion (4-109).

### 4.4.4 Numerical Examples Corresponding to Typical Structures

In this subsection, we provide some typical numerical examples to verify the validity of the theory established in this section in the aspect of constructing the DP-CMs of two-body material systems.

We consider the stacked material cylinders shown in Figure 4-30. The geometrical dimensions of the cylinders are the same —— radius and height are 5.25mm and 2.30mm respectively. We mark the upper one as 1#, and denote it as $V_{\text{sim}}^{1}$; we mark the lower one as 2#, and denote it as $V_{\text{sim}}^{2}$. The relative permeability, relative permittivity, and conductivity of 1# material cylinder $V_{\text{sim}}^{1}$ are 3, 12, and 0 respectively, and the relative permeability, relative permittivity, and conductivity of 2# material cylinder $V_{\text{sim}}^{2}$ are 6, 6, and 0 respectively.

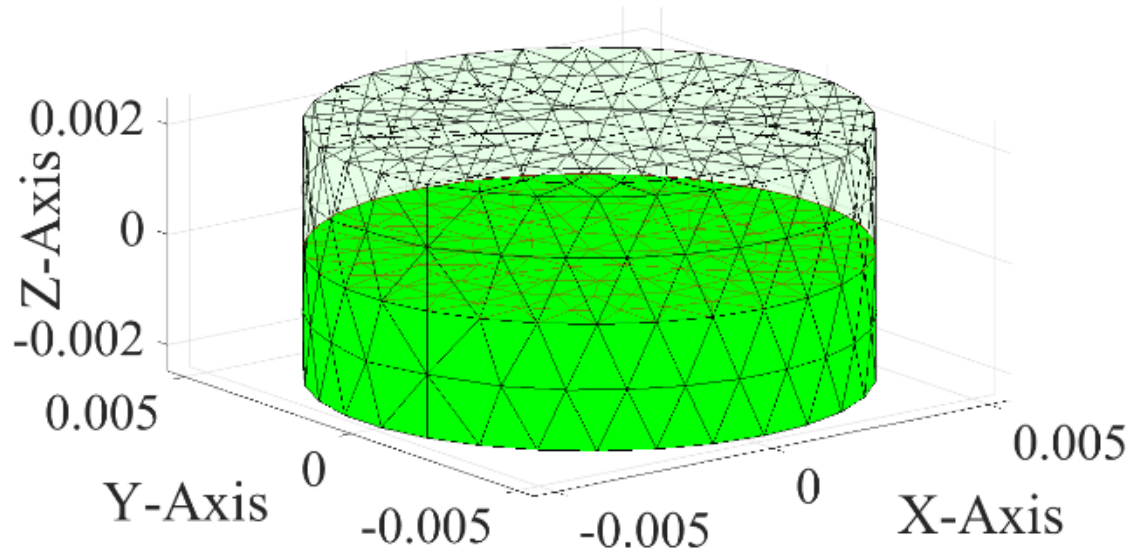

Figure 4-30 The topological structure and surface triangular meshes of a pair of stacked material cylinders

According to the boundary decomposition scheme proposed in Subsection 4.4.1, the boundary of $V_{\text{sim}}^{1}$ (which is denoted as $\partial V_{\text{sim}}^{1}$) and the boundary of $V_{\text{sim}}^{2}$ (which is denoted as $\partial V_{\text{sim}}^{2}$) can be decomposed into the sub-boundaries as illustrated in Figures 4-31 and 4-32.

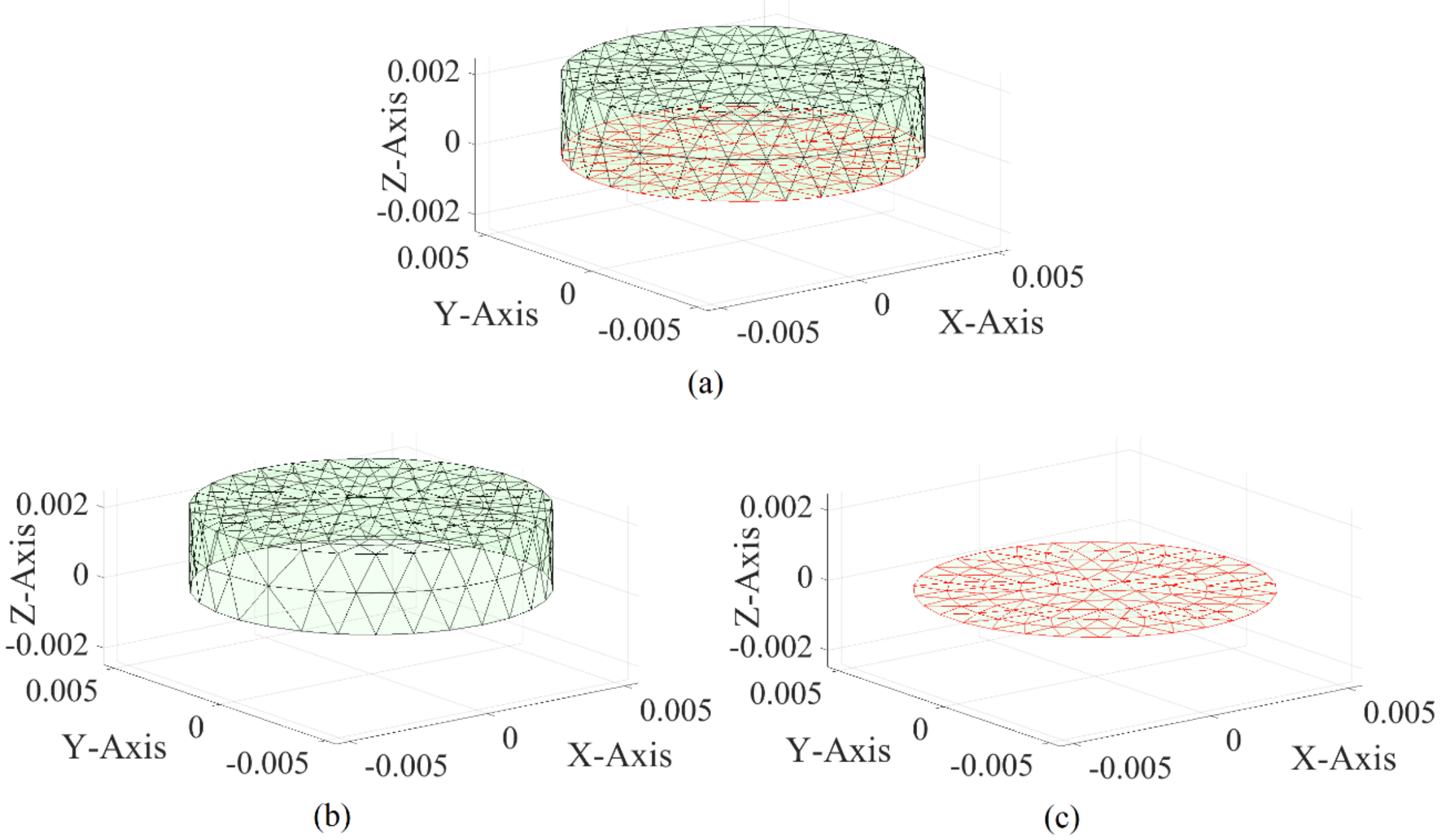

Figure 4-31 The boundary decomposition for the upper material cylinder $V_{\text{sim}}^{1}$ and the triangular meshes of the boundary and sub-boundaries. (a) boundary $\partial V_{\text{sim}}^{1}$ and its triangular mesh; (b) sub-boundary $\partial V_{\text{s10}}$ and its triangular mesh; (c) sub-boundary $\partial V_{\text{s12}}$ and its triangular mesh

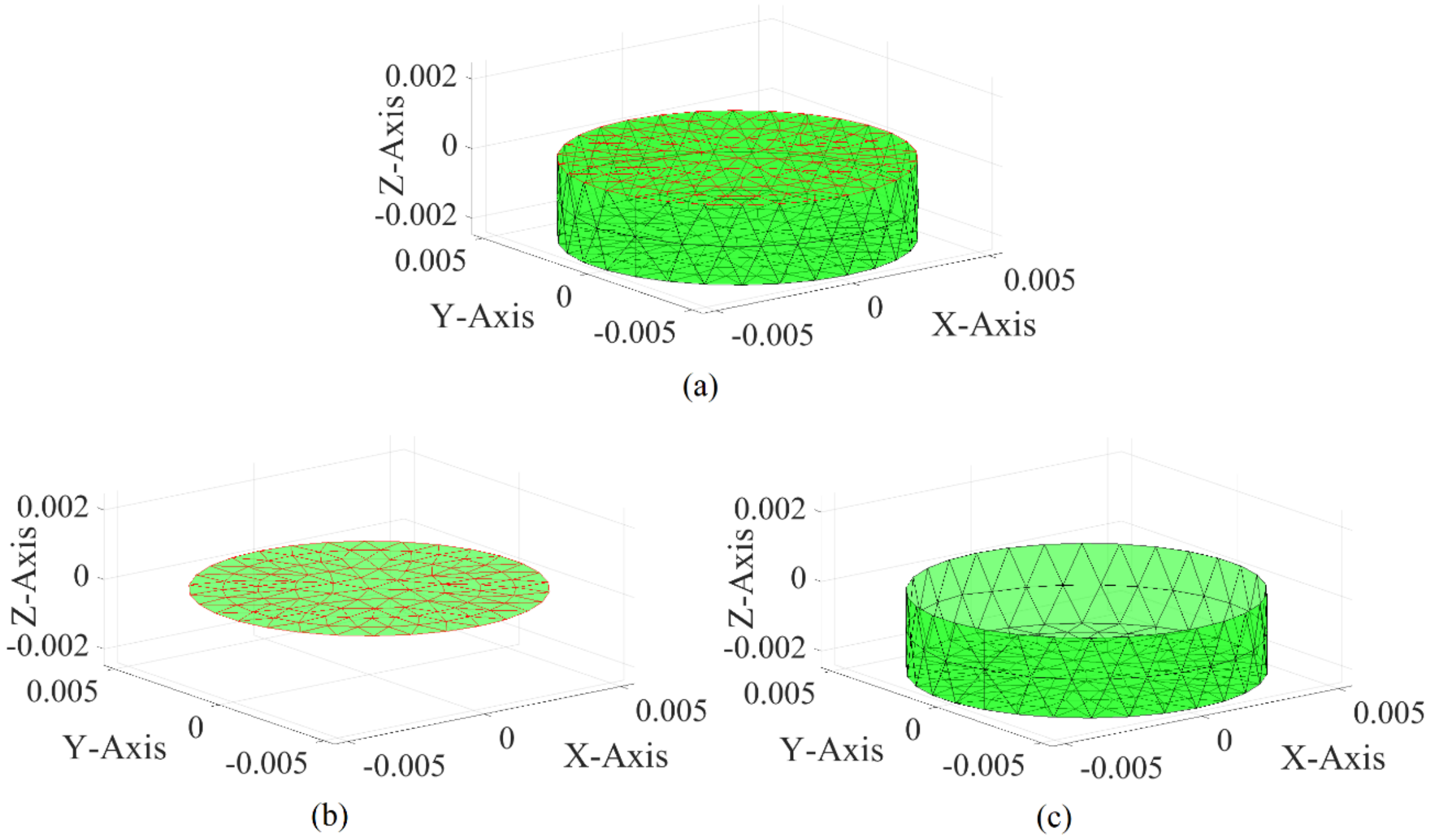

Figure 4-32 The boundary decomposition for the lower material cylinder $V_{\text{sim}}^{2}$ and the triangular meshes of the boundary and sub-boundaries. (a) boundary $\partial V_{\text{sim}}^{2}$ and its triangular mesh; (b) sub-boundary $\partial V_{\text{s}21}$ and its triangular mesh; (c) sub-boundary $\partial V_{\text{s}20}$ and its triangular mesh

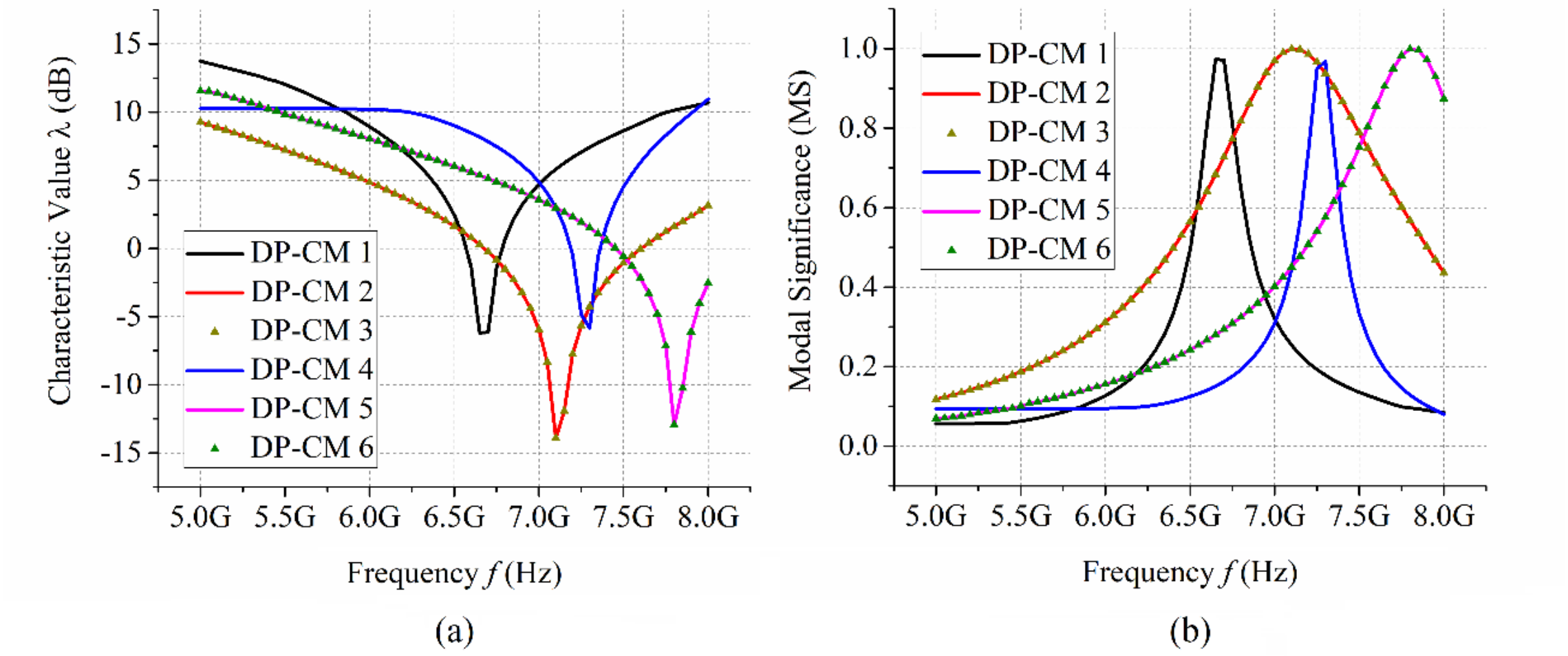

Figure 4-33 For the two-body system shown in Figure 4-30 ( $\mu_{\text{sim}}^{1\text{r}}=3$, $\varepsilon_{\text{sim}}^{1\text{r}}=12$ , $\sigma_{\text{sim}}^{1}=0$ ; $\mu_{\text{sim}}^{2\text{r}}=6$ , $\varepsilon_{\text{sim}}^{2\text{r}}=6$, $\sigma_{\text{sim}}^{2}=0$ ), the characteristic quantity curves corresponding to the first 6 typical DP-CMs derived from the Surf-WEP-MatSca-CMT established in this section. (a) characteristic value dB curves; (b) MS curves

For the stacked material cylinder system, some characteristic quantity curves corresponding to the first 6 typical DP-CMs calculated from the theory established in this section are illustrated in Figure 4-33. To verify the correctness of the results, we also construct the DP-CMs by employing the theory established in Section 4.2, and provide the corresponding characteristic quantity curves in Figure 4-34. By comparing Figures 4-33 and 4-34, it is easy to find out that the results derived from the two schemes basically coincide with each other, where the reason leading to a few errors is that the mesh is not dense enough①.

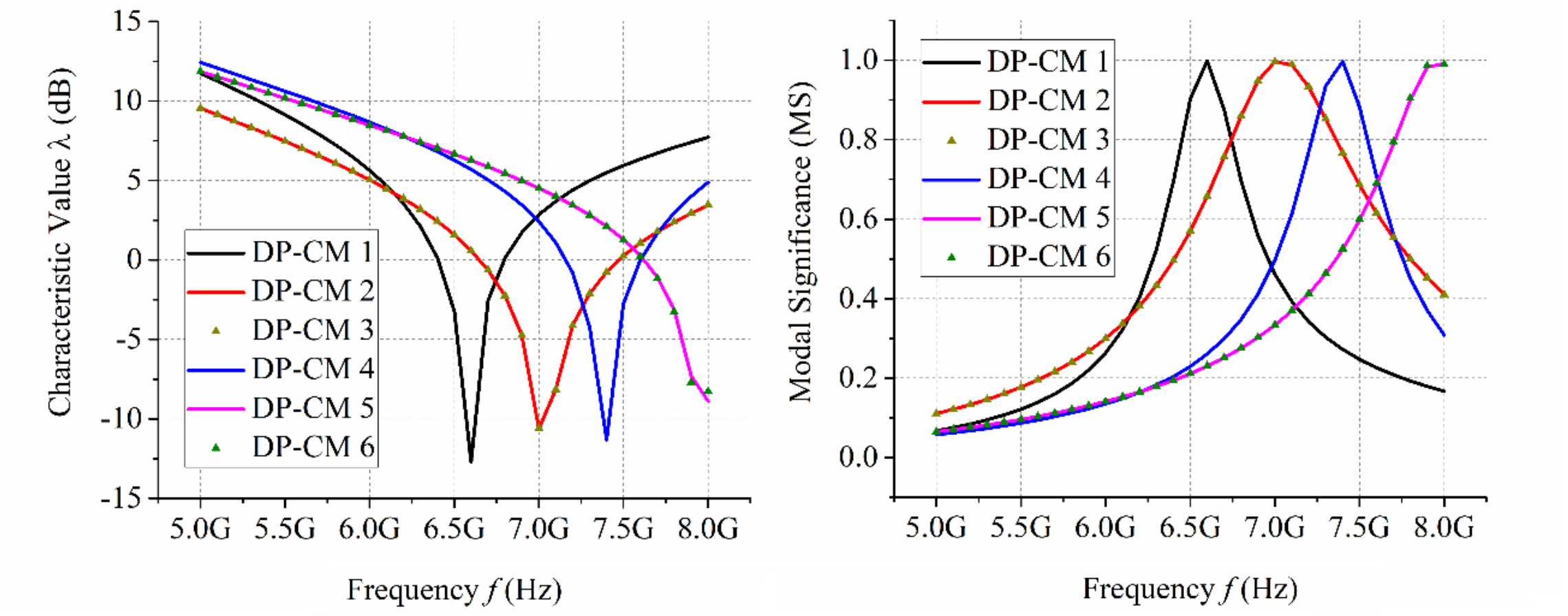

(a) (b)

Figure 4-34 For the two-body system shown in Figure 4-30 ( $\mu_{\text{sim}}^{1\text{r}} = 3$, $\varepsilon_{\text{sim}}^{1\text{r}} = 12$, $\sigma_{\text{sim}}^{1} = 0$; $\mu_{\text{sim}}^{2\text{r}} = 6$, $\varepsilon_{\text{sim}}^{2\text{r}} = 6$, $\sigma_{\text{sim}}^{2} = 0$ ), the characteristic quantity curves corresponding to the first 6 typical DP-CMs derived from the Vol-WEP-MatSca-CMT established in Section 4.2. (a) characteristic value dB curves; (b) MS curves

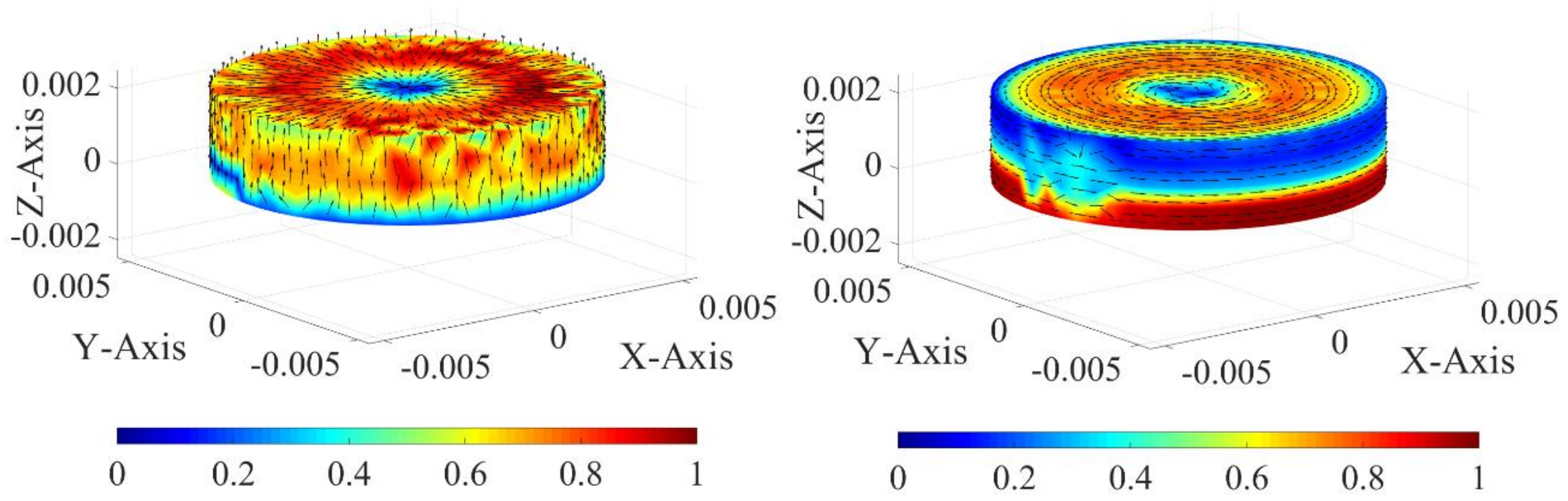

(a) (b)

Figure 4-35 The modal equivalent source distributions (on $\partial V_{\text{s10}}$ ) of the DP-CM1 working at 6.60GHz and shown in Figure 4-33. (a) equivalent surface magnetic current; (b) equivalent surface electric current

① All numerical examples given in this dissertation are calculated in Matlab computation environment and using personal computer (PC), and this dissertation never uses any large computational resource. Due to the constraint from computational resource, the meshes used in this dissertation are relatively sparse. When the meshes become more dense, the calculated results will become better.

From Figure 4-33, it is easy to find out that: DP-CM1 is "resonant" at 6.60GHz; DP-CM2 and DP-CM3 are "resonant" at 7.10GHz; DP-CM4 is "resonant" at 7.30GHz; DP-CM5 and DP-CM6 are "resonant" at 7.80GHz. In what follows, taking the first "resonant" DP-CM as a typical example, we draw its characteristic currents and characteristic fields as visual figures to facilitate references. For the "resonant" DP-CM1 working at 6.60GHz, its equivalent magnetic and electric currents distributing on $\partial V_{s10}$, $\partial V_{s12}$, $\partial V_{s21}$, and $\partial V_{s20}$ are illustrated in Figures 4-35, 4-36, 4-37, and 4-38 respectively.

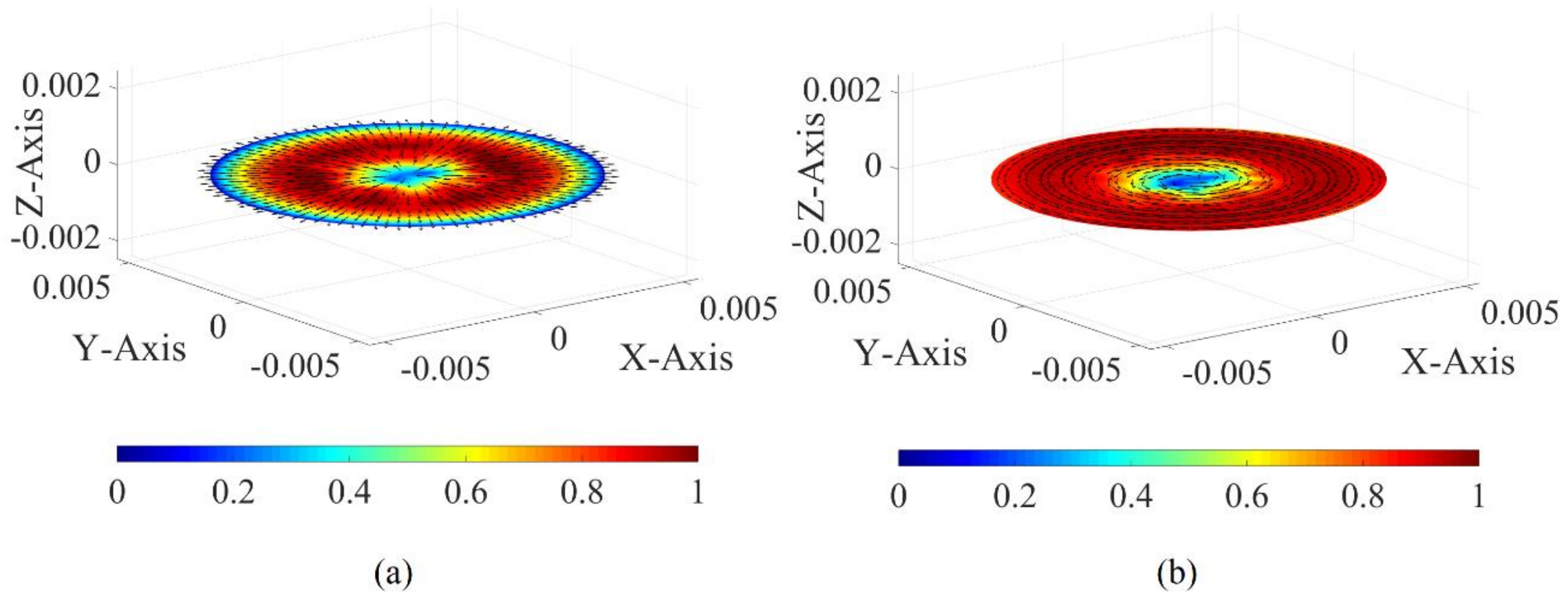

Figure 4-36 The modal equivalent source distributions (on $\partial V_{s12}$) of the DP-CM1 working at 6.60GHz and shown in Figure 4-33. (a) equivalent surface magnetic current; (b) equivalent surface electric current

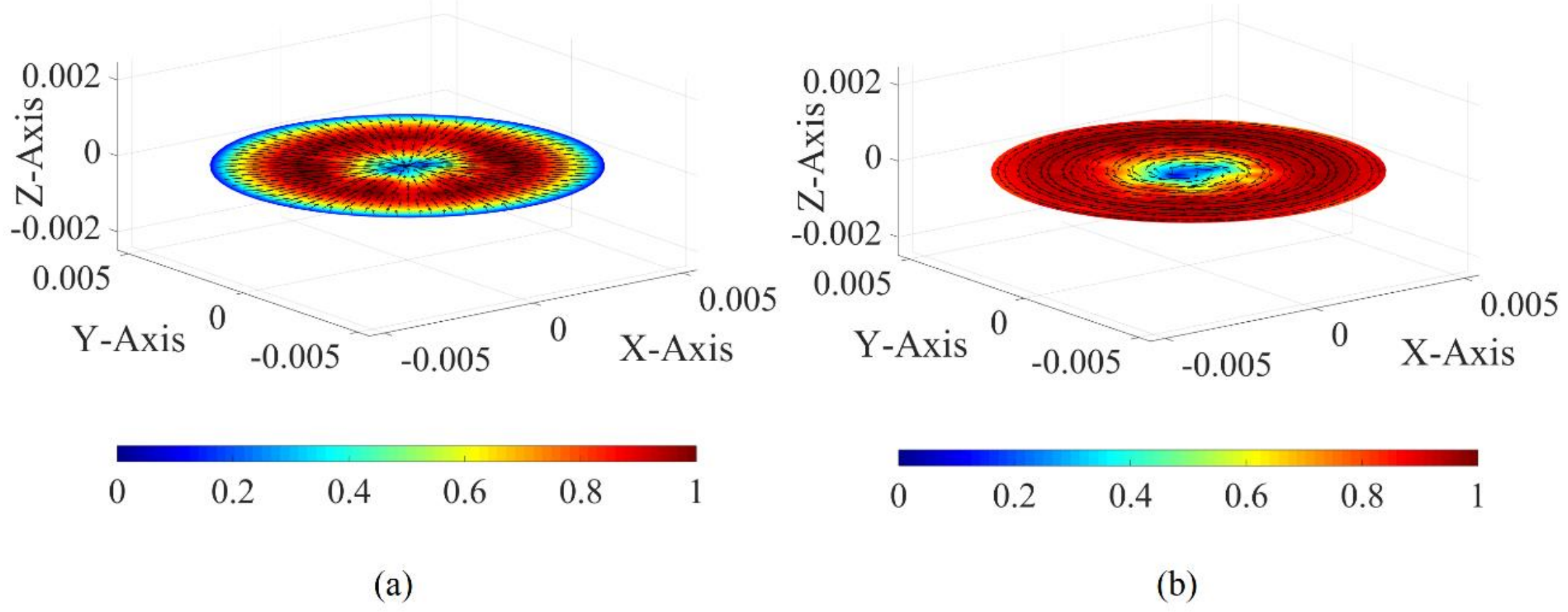

Figure 4-37 The modal equivalent source distributions (on $\partial V_{s21}$) of the DP-CM1 working at 6.60GHz and shown in Figure 4-33. (a) equivalent surface magnetic current; (b) equivalent surface electric current

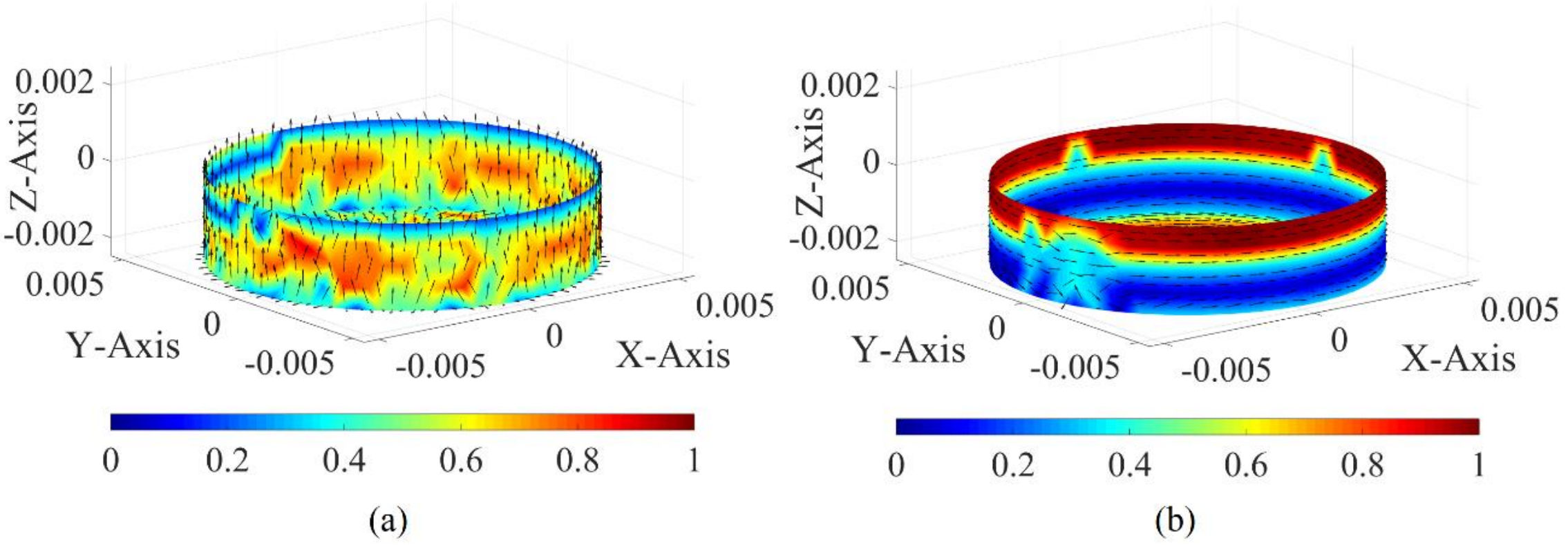

Figure 4-38 The modal equivalent source distributions (on $\partial V_{s20}$) of the DP-CM1 working at 6.60GHz and shown in Figure 4-33. (a) equivalent surface magnetic current; (b) equivalent surface electric current

For the "resonant" DP-CM1 working at 6.60GHz, its modal total fields, modal scattered fields, and modal incident fields distributing on $V_{\text{sim}}^{1}$ and $V_{\text{sim}}^{2}$ are illustrated in Figures 4-39 and 4-50 respectively.

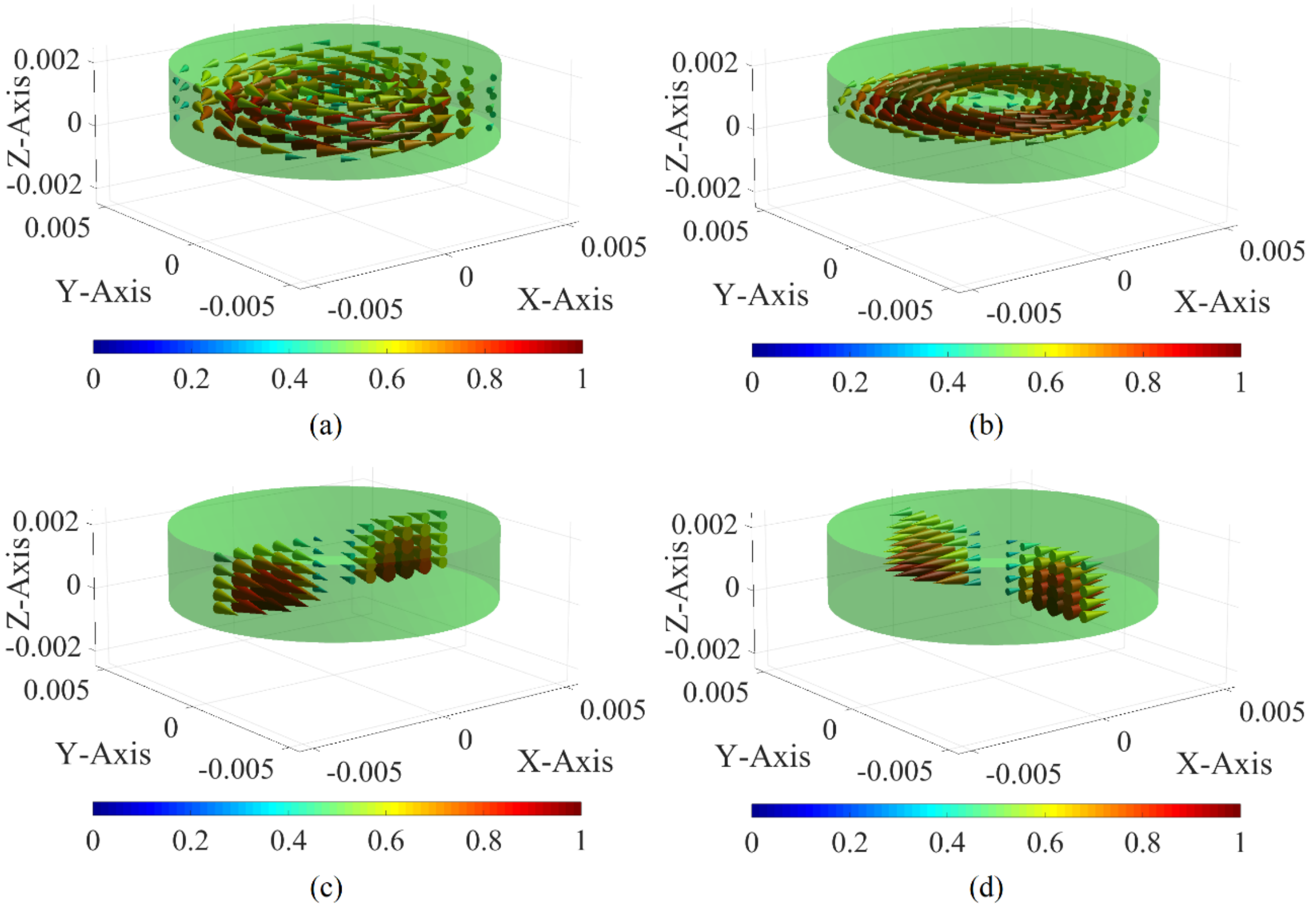

Figure 4-39 The modal total electric field distributions of the DP-CM1 working at 6.60GHz and shown in Figure 4-33. (a) the distribution on whole $V_{\text{sim}}^{1}$; (b) the distribution on $z = 1.15\text{mm}$ surface; (c) the distribution on xOz surface; (d) the distribution on yOz surface

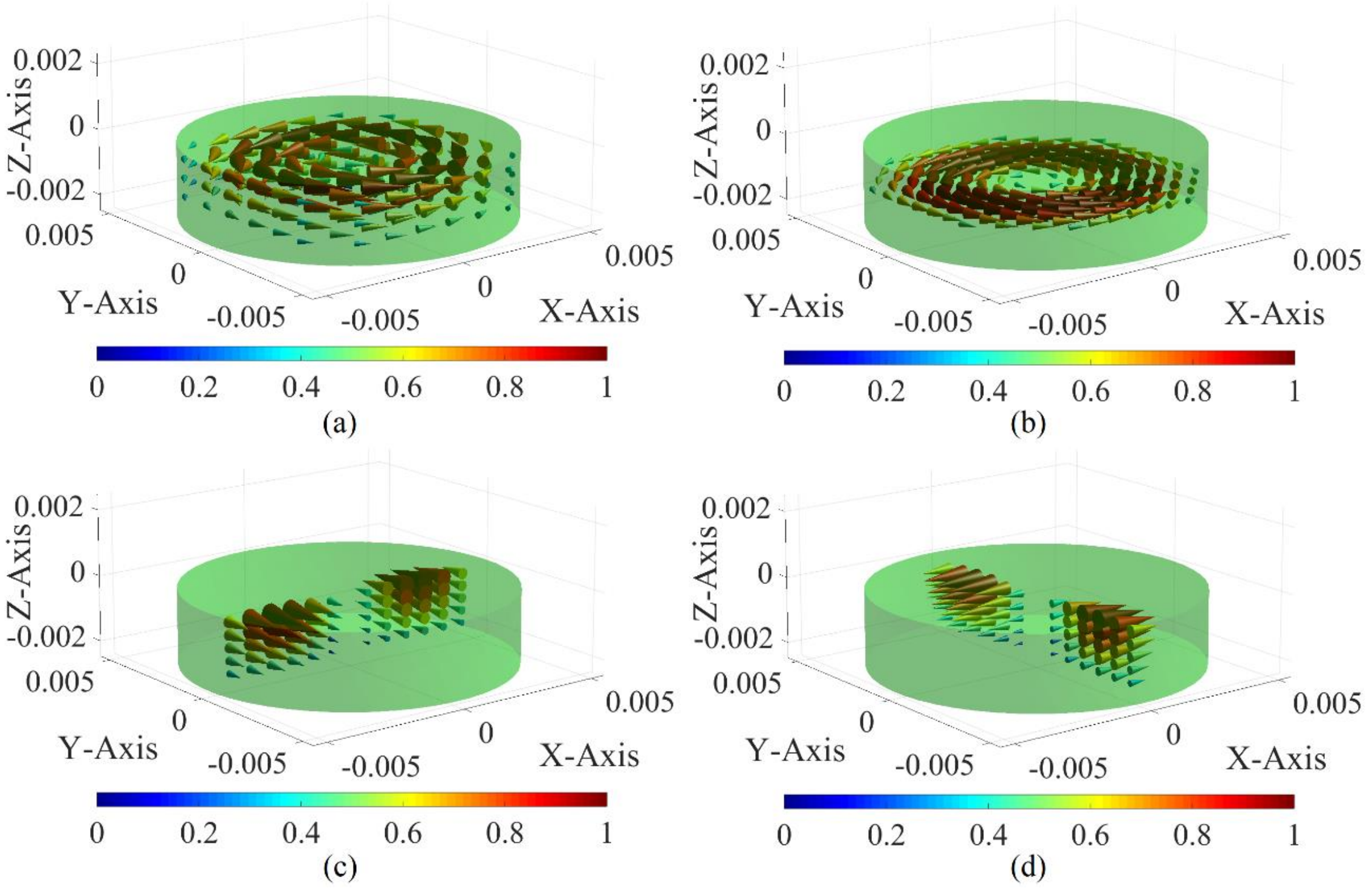

Figure 4-40 The modal total electric field distributions of the DP-CM1 working at 6.60GHz and shown in Figure 4-33. (a) the distribution on whole $V_{\rm sim}^{2}$; (b) the distribution on $z = -1.15\text{mm}$ surface; (c) the distribution on xOz surface; (d) the distribution on yOz surface

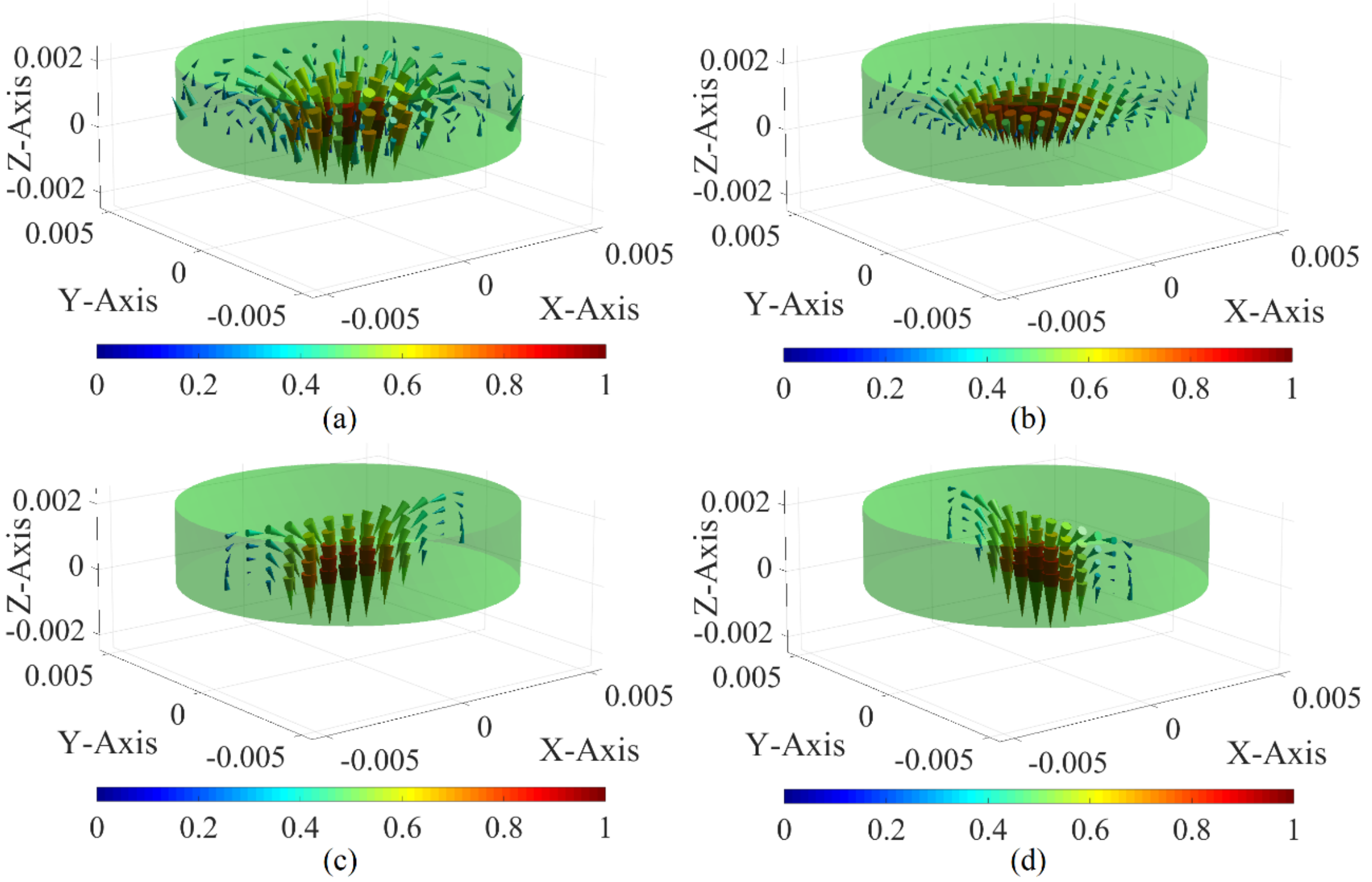

Figure 4-41 The modal total magnetic field distributions of the DP-CM1 working at 6.60GHz and shown in Figure 4-33. (a) the distribution on whole $V_{\rm sim}^{1}$; (b) the distribution on $z = 1.15\text{mm}$ surface; (c) the distribution on xOz surface; (d) the distribution on yOz surface

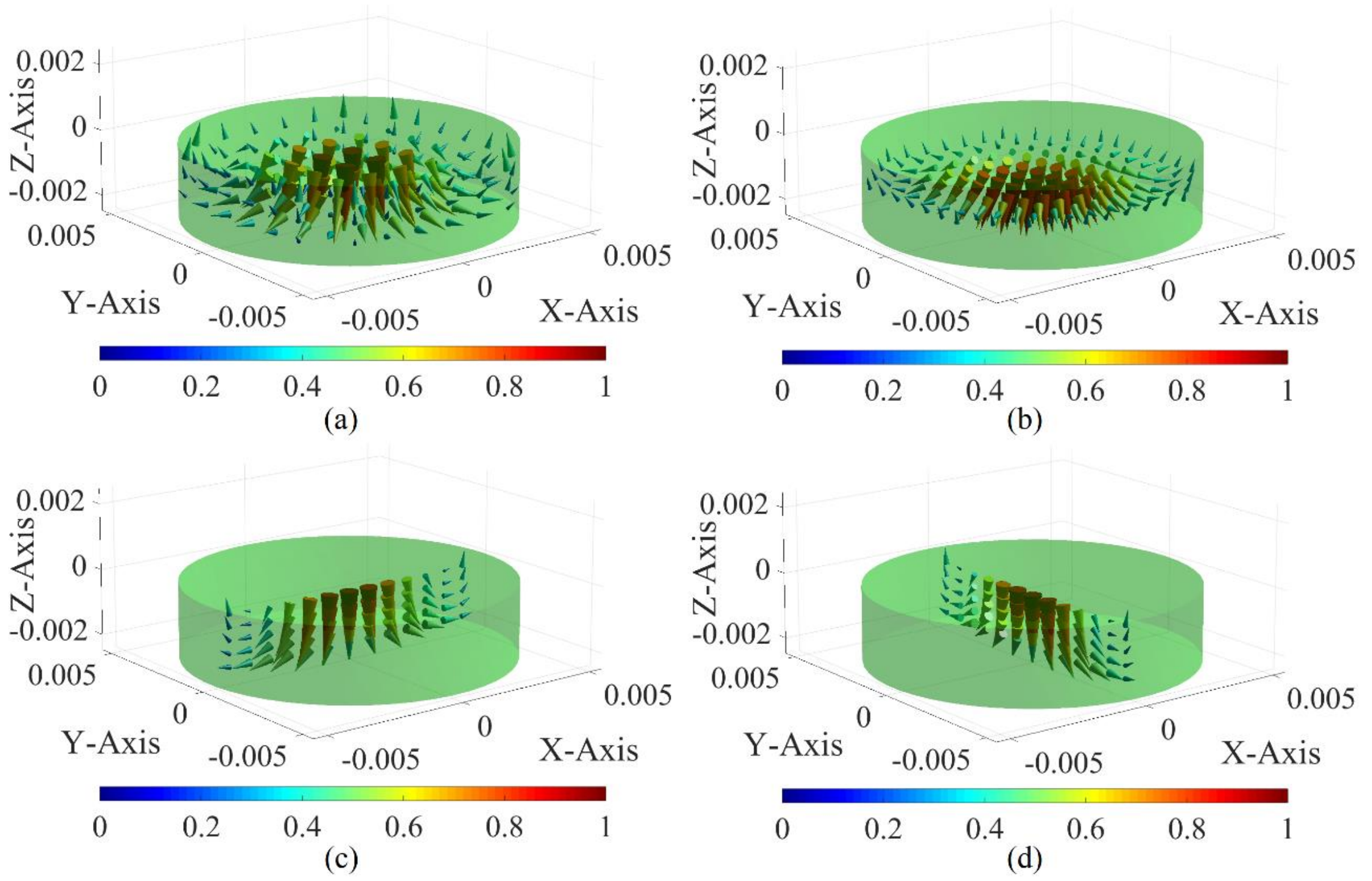

Figure 4-42 The modal total magnetic field distributions of the DP-CM1 working at 6.60GHz and shown in Figure 4-33. (a) the distribution on whole $V_{\text{sim}}^{2}$; (b) the distribution on $z = -1.15\text{mm}$ surface; (c) the distribution on xOz surface; (d) the distribution on yOz surface

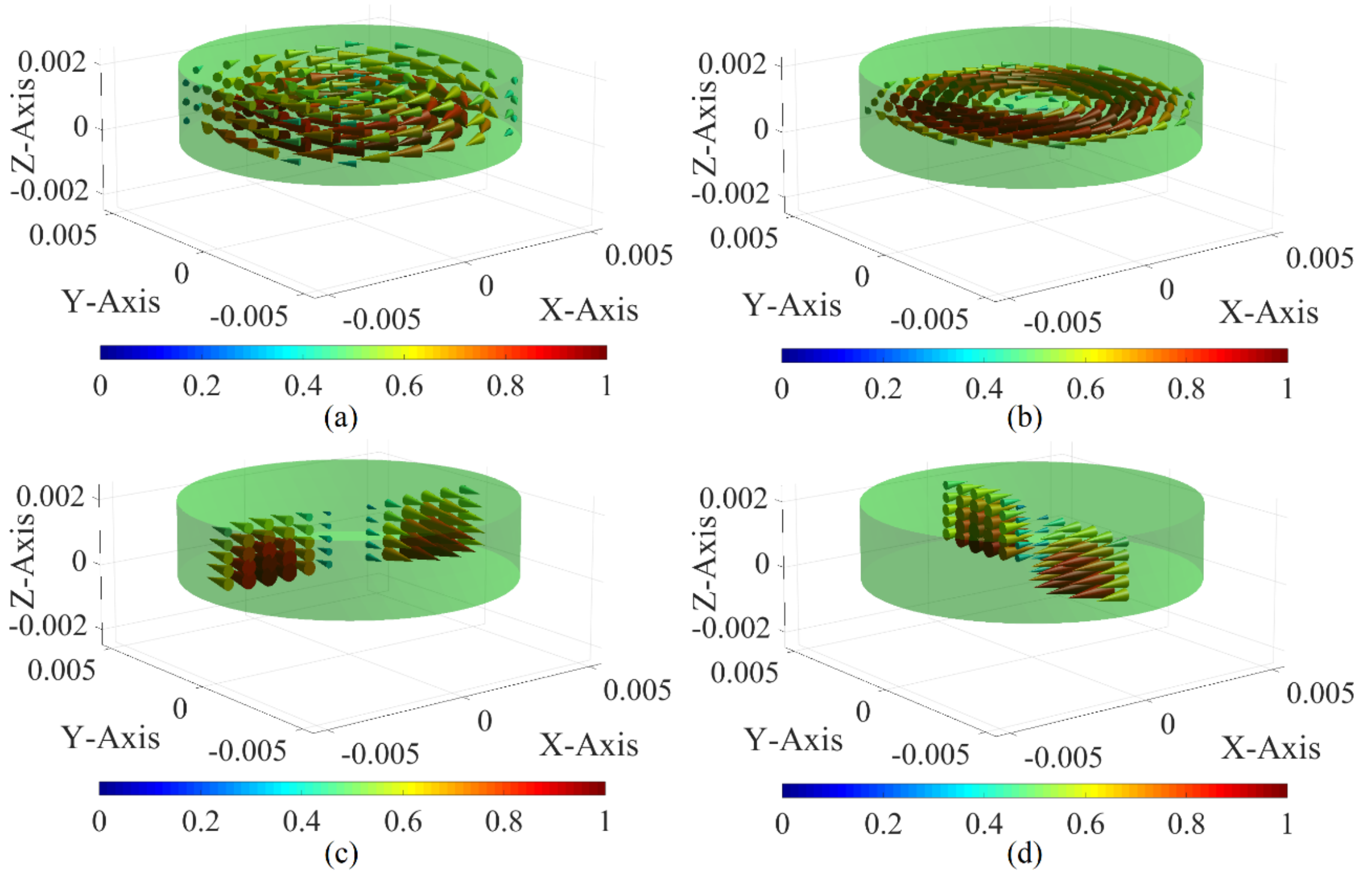

Figure 4-43 The modal scattered volume electric current distributions of the DP-CM1 working at 6.60GHz and shown in Figure 4-33. (a) the distribution on whole $V_{\text{sim}}^{1}$; (b) the distribution on $z = 1.15\text{mm}$ surface; (c) the distribution on xOz surface; (d) the distribution on yOz surface

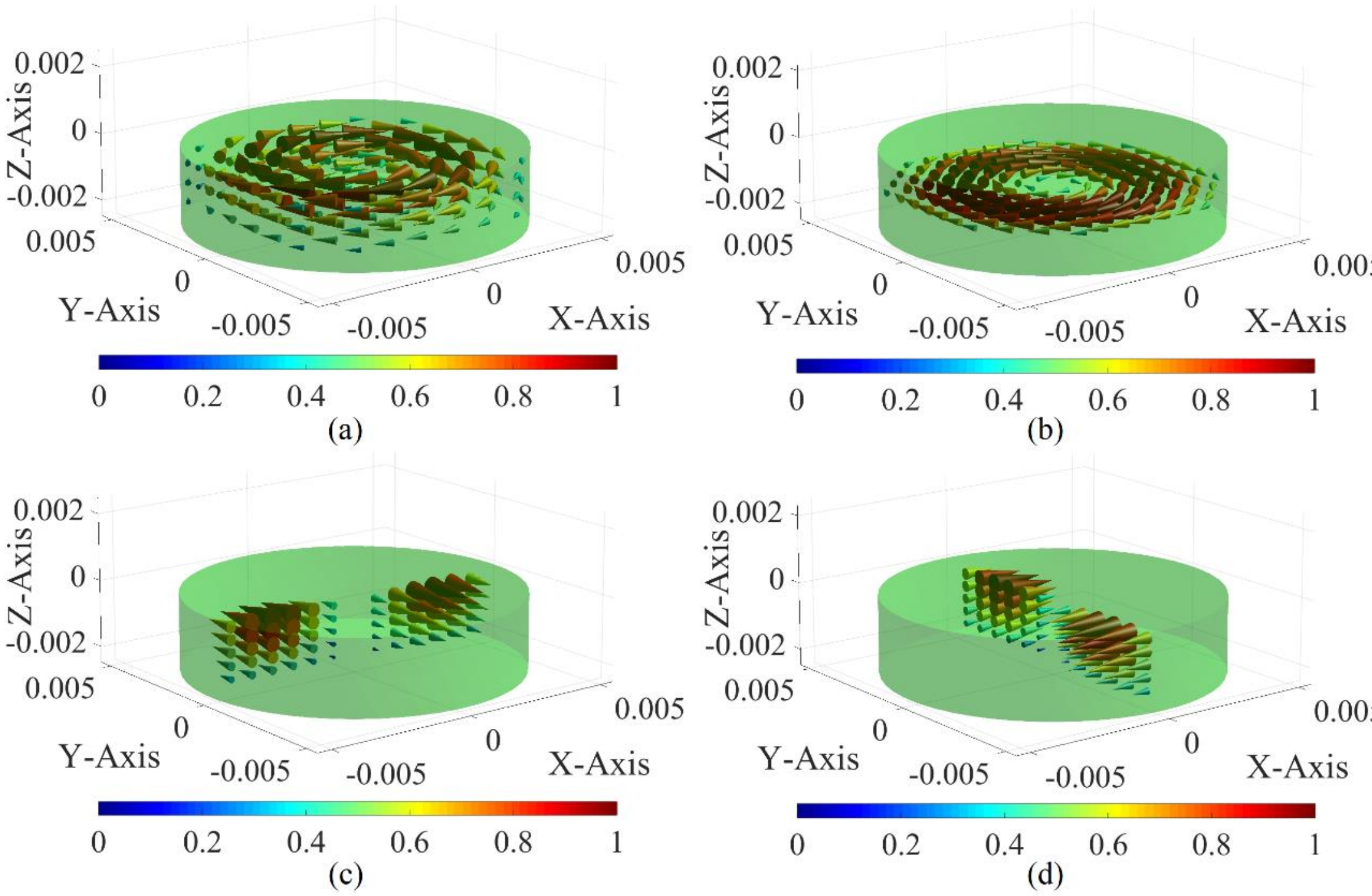

Figure 4-44 The modal scattered volume electric current distributions of the DP-CM1 working at 6.60GHz and shown in Figure 4-33. (a) the distribution on whole $V_{\rm sim}^{2}$; (b) the distribution on $z=-1.15\,\rm mm$ surface; (c) the distribution on xOz surface; (d) the distribution on yOz surface

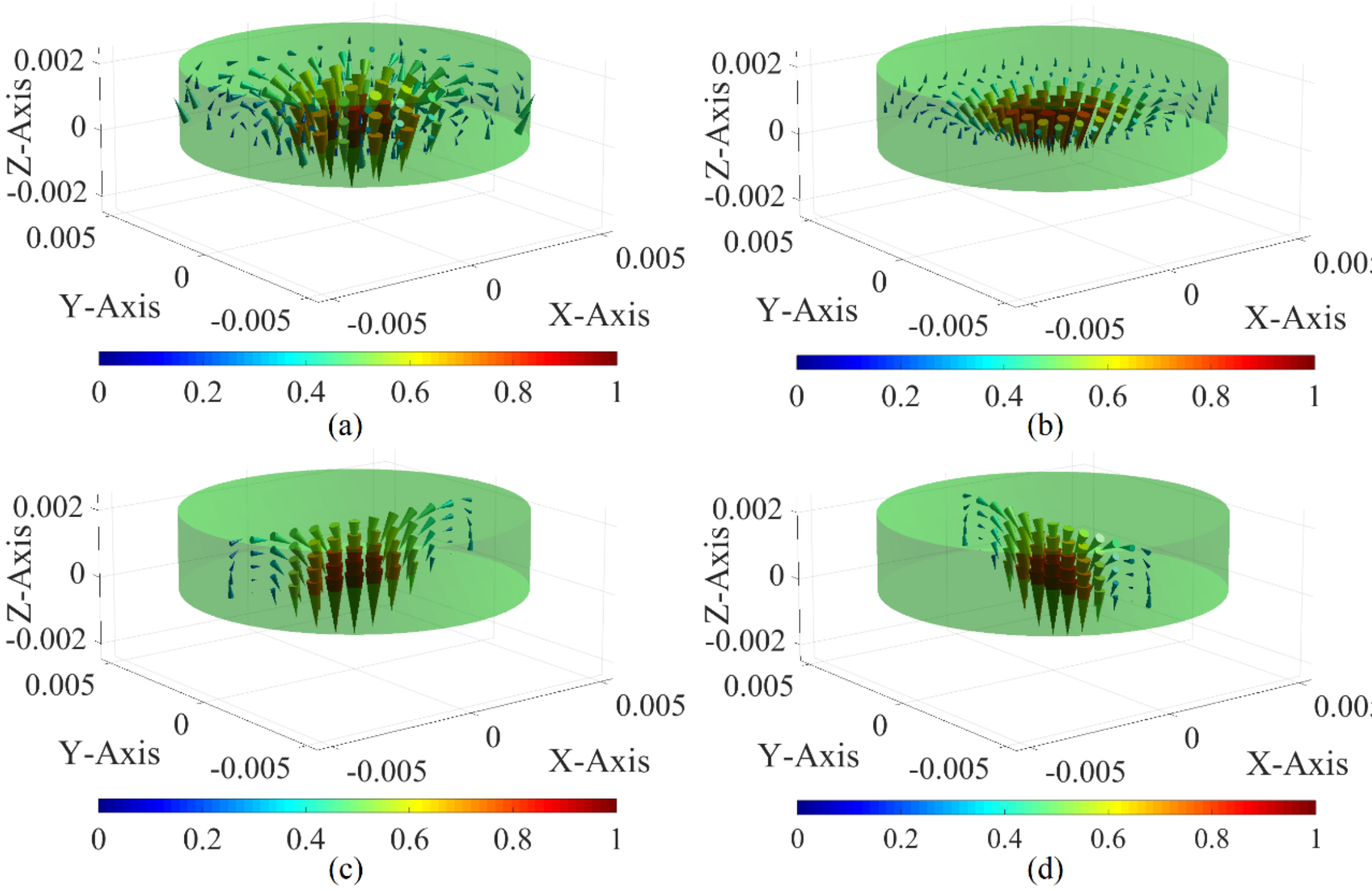

Figure 4-45 The modal scattered volume magnetic current distributions of the DP-CM1 working at 6.60GHz and shown in Figure 4-33. (a) the distribution on whole $V_{\rm sim}^{1}$; (b) the distribution on $z=1.15\,\rm mm$ surface; (c) the distribution on xOz surface; (d) the distribution on yOz surface

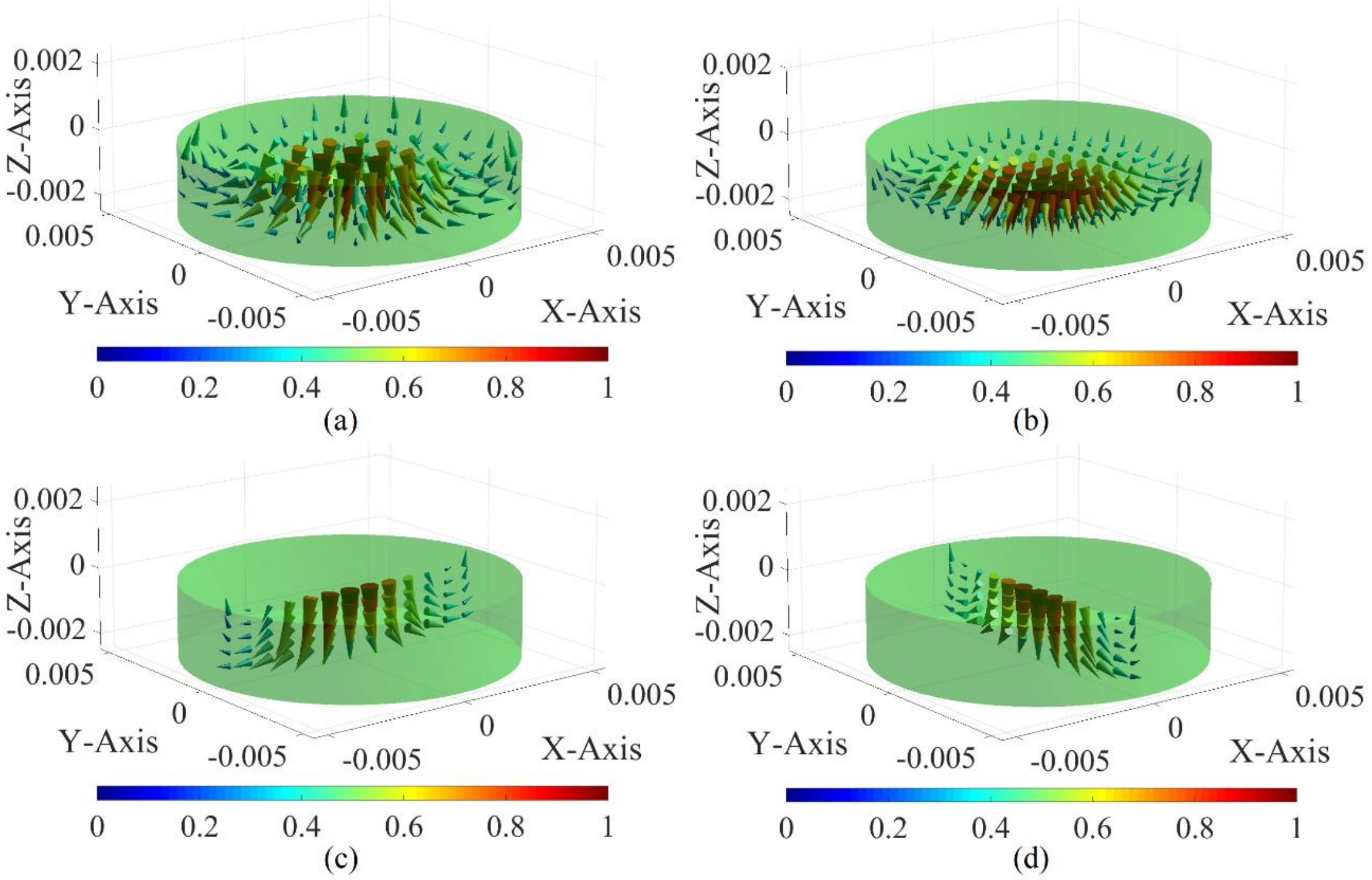

Figure 4-46 The modal scattered volume magnetic current distributions of the DP-CM1 working at 6.60GHz and shown in Figure 4-33. (a) the distribution on whole $V_{\text{sim}}^{2}$; (b) the distribution on $z = -1.15\,\text{mm}$ surface; (c) the distribution on xOz surface; (d) the distribution on yOz surface

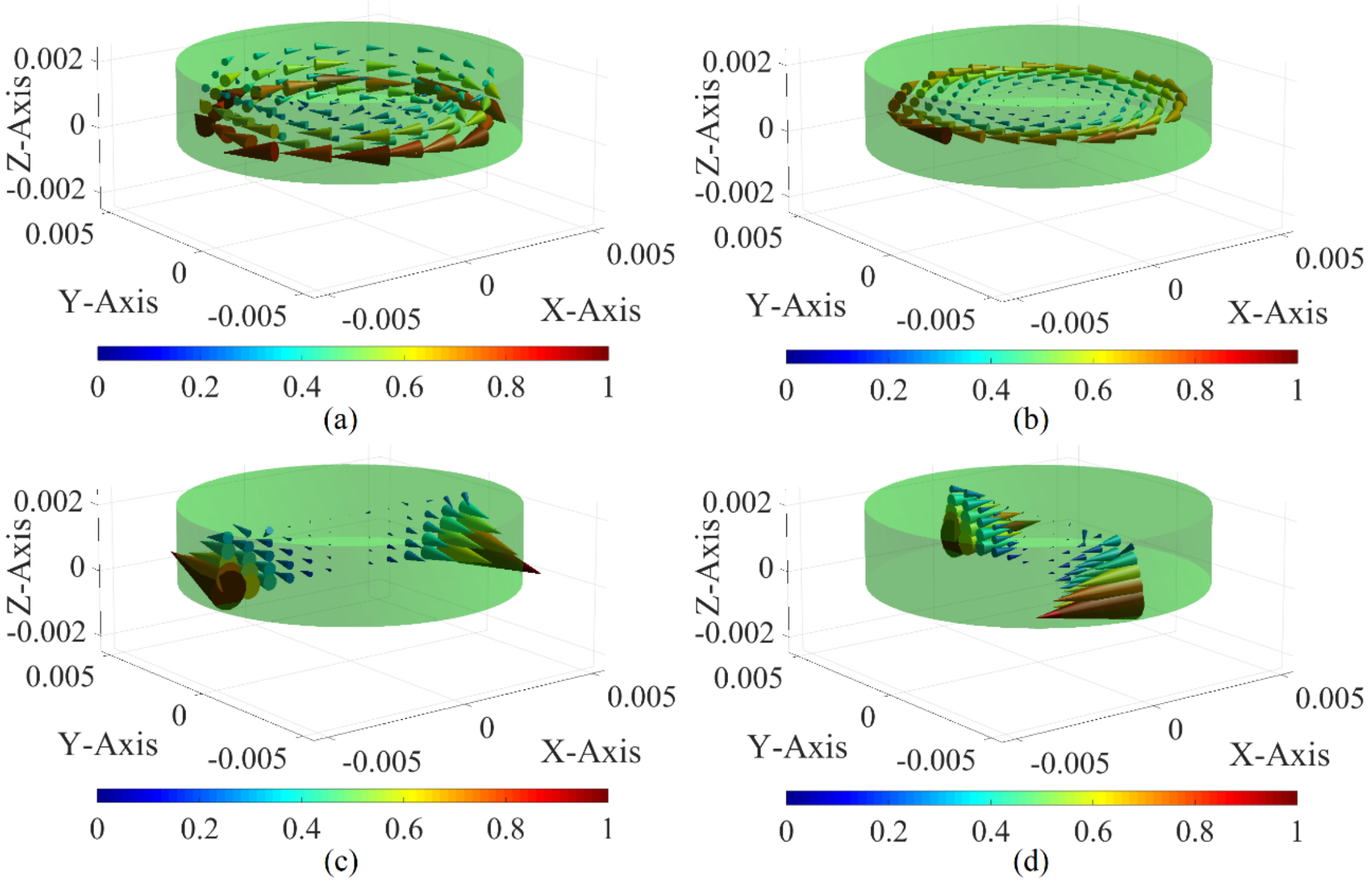

Figure 4-47 The modal incident electric field distributions of the DP-CM1 working at 6.60GHz and shown in Figure 4-33. (a) the distribution on whole $V_{\text{sim}}^{1}$; (b) the distribution on $z = 1.15\,\text{mm}$ surface; (c) the distribution on xOz surface; (d) the distribution on yOz surface

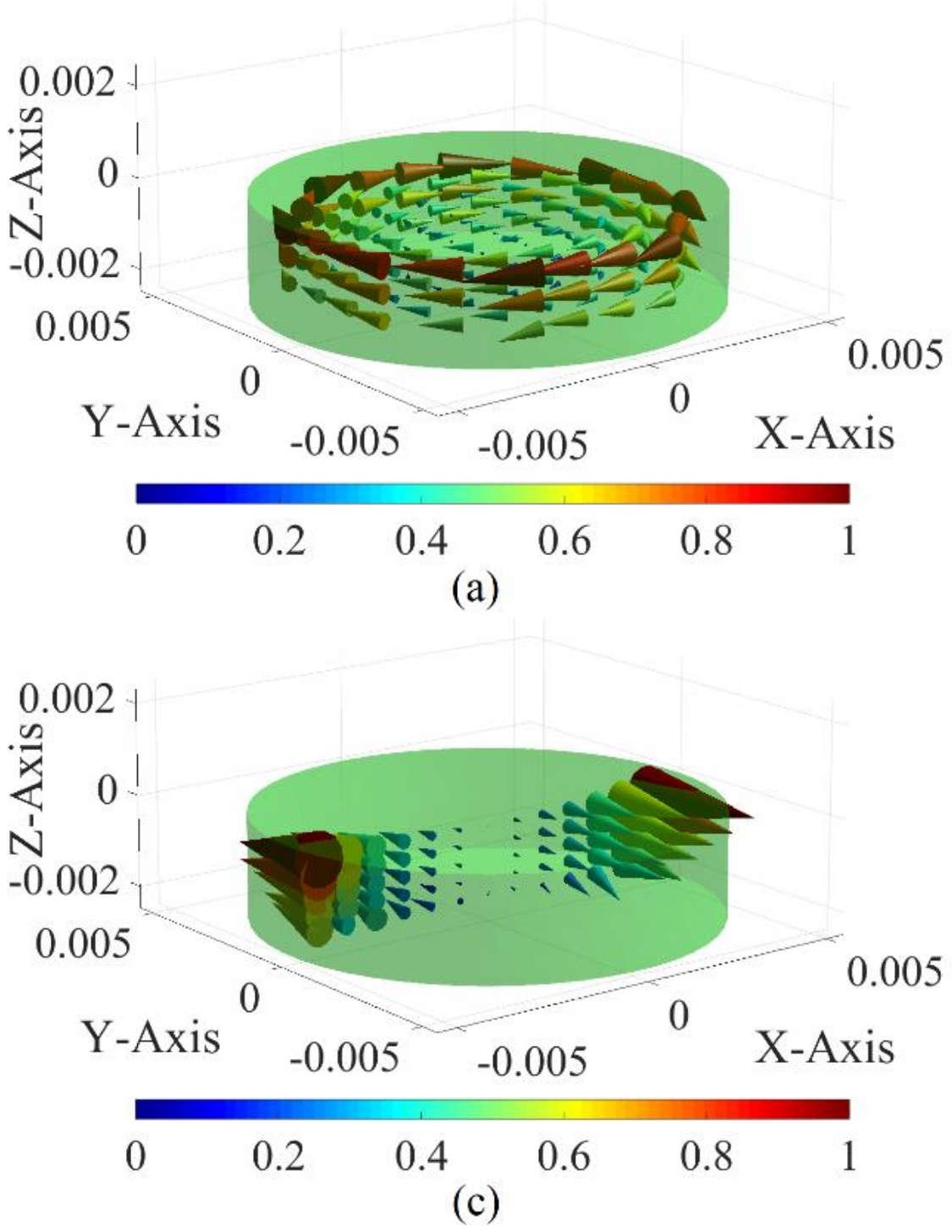

Figure 4-48 The modal incident electric field distributions of the DP-CM1 working at 6.60GHz and shown in Figure 4-33. (a) the distribution on whole $V_{\text{sim}}^{2}$; (b) the distribution on $z=-1.15\text{mm}$ surface; (c) the distribution on xOz surface; (d) the distribution on yOz surface

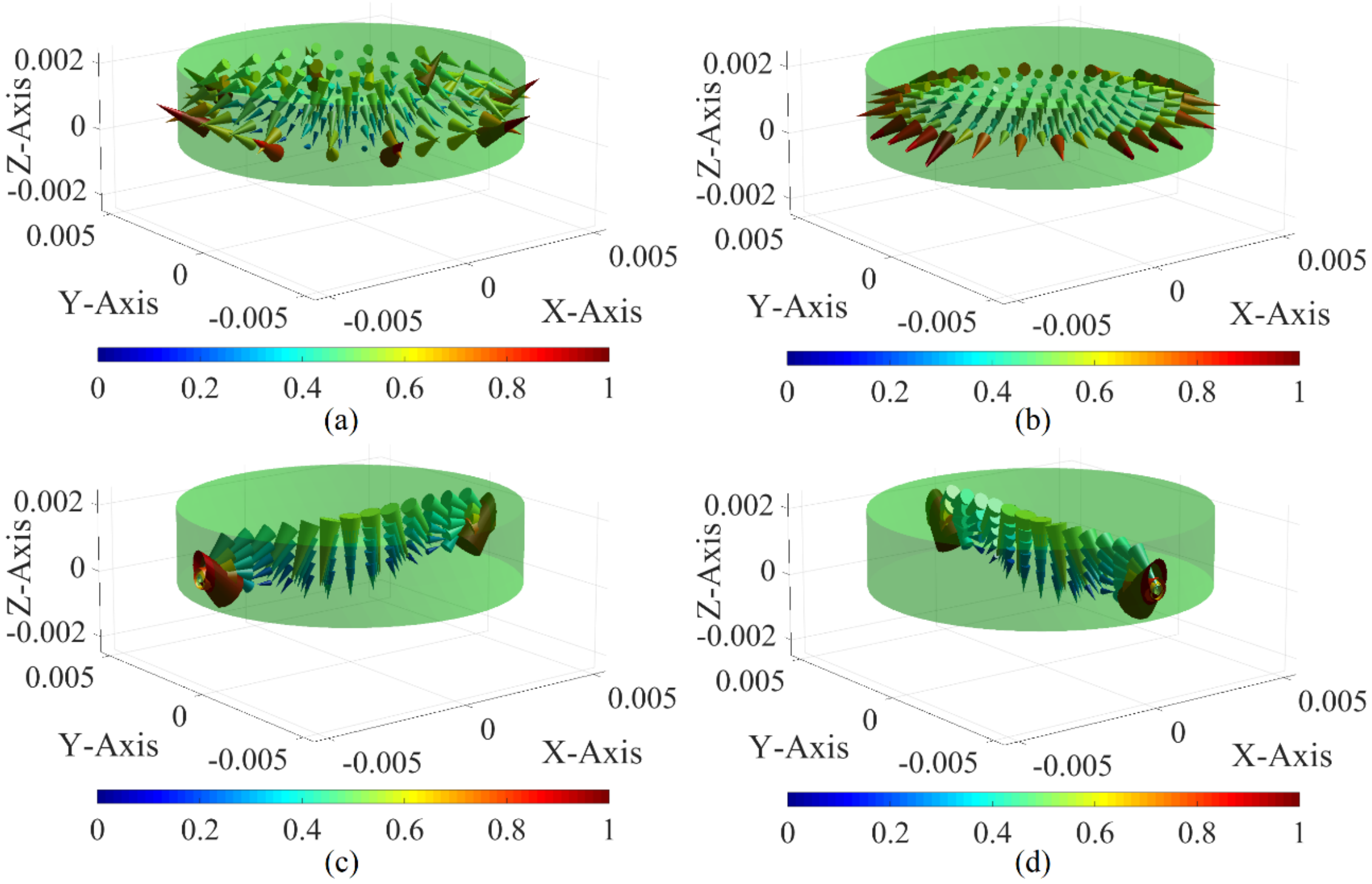

Figure 4-49 The modal incident magnetic field distributions of the DP-CM1 working at 6.60GHz and shown in Figure 4-33. (a) the distribution on whole $V_{\text{sim}}^{1}$; (b) the distribution on $z=1.15\text{mm}$ surface; (c) the distribution on xOz surface; (d) the distribution on yOz surface

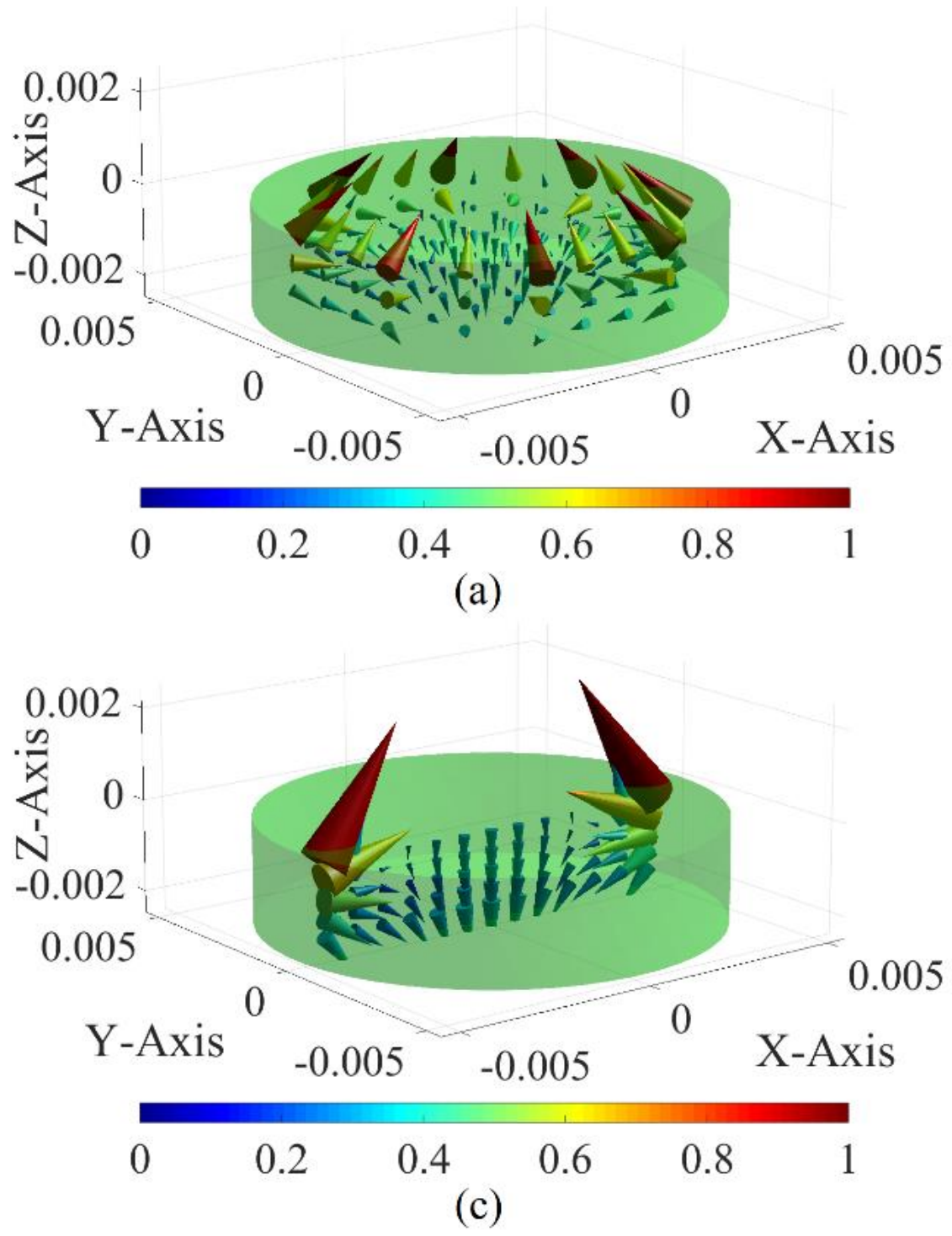

Figure 4-50 The modal incident magnetic field distributions of the DP-CM1 working at 6.60GHz and shown in Figure 4-33. (a) the distribution on whole $V_{\rm sim}^{2}$; (b) the distribution on $z = -1.15\text{mm}$ surface; (c) the distribution on xOz surface; (d) the distribution on yOz surface

For the “resonant” DP-CM1 working at 6.60GHz, its modal radiation pattern is shown in Figure 4-51.

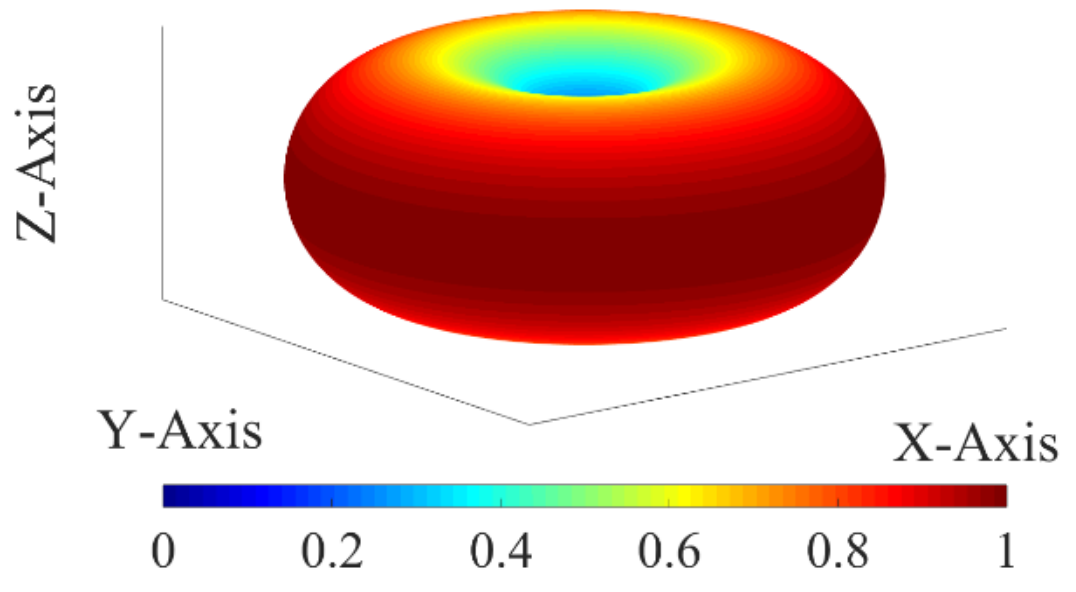

Figure 4-51 The radiation pattern of the DP-CM1 working at 6.60GHz and shown in Figure 4-33

## 4.5 The Second Surface Formulation for Calculating the DP-CMs of the Systems Constructed by Double Simply Connected Material Bodies

Similarly to transforming from formulation (4-28) to formulation (4-32), this section is committed to providing new surface formulations different from formulations (4-66)&(4-102) to the DPO $P_{\text{ss sys}}^{\text{driving}}$ corresponding to the two-body material system shown in Figure 4-27, and this section will construct the corresponding DP-CMs based on the new formulations, and this section also compare the results derived from the new formulations and the results derived from the formulations (4-66)&(4-102) given in Section 4.4 to verify the validity of the new formulations.

### 4.5.1 The Second Surface Formulation for Calculating the DP-CMs of Two-body Systems

Based on the results given in Appendix C, the DPO corresponding to the system shown in Figure 4-27 can be equivalently rewritten as follows:

$$\begin{aligned}
P_{\text{ss sys}}^{\text{driving}} &= -\sum_{i=1}^{2}\Big[(1/2)\big\langle \vec{J}_{si}^{\text{ES}},\vec{E}^{\text{inc}}\big\rangle_{\partial V_{\text{sim}}^{i}} + (1/2)\big\langle \vec{M}_{si}^{\text{ES}},\vec{H}^{\text{inc}}\big\rangle_{\partial V_{\text{sim}}^{i}}\Big] \\
&= -(1/2)\Big[\Big\langle \vec{J}_{\text{s1}}^{\text{ES}},\mathcal{E}_0\big(\vec{J}_{\text{s1}}^{\text{ES}},\vec{M}_{\text{s1}}^{\text{ES}}\big)\Big\rangle_{\partial V_{\text{sim}}^{1-}} + \Big\langle \vec{J}_{\text{s1}}^{\text{ES}},\mathcal{E}_0\big(\vec{J}_{\text{s2}}^{\text{ES}},\vec{M}_{\text{s2}}^{\text{ES}}\big)\Big\rangle_{\partial V_{\text{sim}}^{1-}}\Big] \\
&\quad -(1/2)\Big[\Big\langle \vec{J}_{\text{s2}}^{\text{ES}},\mathcal{E}_0\big(\vec{J}_{\text{s2}}^{\text{ES}},\vec{M}_{\text{s2}}^{\text{ES}}\big)\Big\rangle_{\partial V_{\text{sim}}^{2-}} + \Big\langle \vec{J}_{\text{s2}}^{\text{ES}},\mathcal{E}_0\big(\vec{J}_{\text{s1}}^{\text{ES}},\vec{M}_{\text{s1}}^{\text{ES}}\big)\Big\rangle_{\partial V_{\text{sim}}^{2-}}\Big] \\
&\quad -(1/2)\Big[\Big\langle \vec{M}_{\text{s1}}^{\text{ES}},\mathcal{H}_0\big(\vec{J}_{\text{s1}}^{\text{ES}},\vec{M}_{\text{s1}}^{\text{ES}}\big)\Big\rangle_{\partial V_{\text{sim}}^{1-}} + \Big\langle \vec{M}_{\text{s1}}^{\text{ES}},\mathcal{H}_0\big(\vec{J}_{\text{s2}}^{\text{ES}},\vec{M}_{\text{s2}}^{\text{ES}}\big)\Big\rangle_{\partial V_{\text{sim}}^{1-}}\Big] \\
&\quad -(1/2)\Big[\Big\langle \vec{M}_{\text{s2}}^{\text{ES}},\mathcal{H}_0\big(\vec{J}_{\text{s2}}^{\text{ES}},\vec{M}_{\text{s2}}^{\text{ES}}\big)\Big\rangle_{\partial V_{\text{sim}}^{2-}} + \Big\langle \vec{M}_{\text{s2}}^{\text{ES}},\mathcal{H}_0\big(\vec{J}_{\text{s1}}^{\text{ES}},\vec{M}_{\text{s1}}^{\text{ES}}\big)\Big\rangle_{\partial V_{\text{sim}}^{2-}}\Big]
\end{aligned} \tag{4-113}$$

where the second equality is based on Appendix C and the fact that: the resultant field used to excite $V_{\text{sim}}^{1}/V_{\text{sim}}^{2}$ consists of incident field $\vec{F}^{\text{inc}}$ and the scattered field generated by $V_{\text{sim}}^{2}/V_{\text{sim}}^{1}$, and the tangential components of $\vec{F}^{\text{inc}}$ are continuous on material boundaries.

Inserting expansion formulations (4-76) and (4-77) into above DPO (4-113), the DPO is immediately discretized into the following matrix form:

$$P_{\text{ss sys}}^{\text{driving}} = \begin{bmatrix} \bar{a}^{J_{\text{s1}}} \\ \bar{a}^{J_{\text{s2}}} \\ \bar{a}^{M_{\text{s1}}} \\ \bar{a}^{M_{\text{s2}}} \end{bmatrix}^{H} \cdot \underbrace{\left(\bar{\bar{P}}_{2;0;\text{PVT}}^{\text{ss sys}} + \bar{\bar{P}}_{2;0;\text{SCT}}^{\text{ss sys}}\right)}_{\bar{\bar{P}}_{\text{ss sys;2}}^{\text{driving}}} \cdot \begin{bmatrix} \bar{a}^{J_{\text{s1}}} \\ \bar{a}^{J_{\text{s2}}} \\ \bar{a}^{M_{\text{s1}}} \\ \bar{a}^{M_{\text{s2}}} \end{bmatrix} \tag{4-114}$$

where the explanations for the subscripts of the matrices in formulation (4-114) are similar to the explanations for the ones in formulation (4-80), and they will not be repeated here. The power matrices $\bar{\bar{P}}_{2;0;\text{PVT}}^{\text{ss sys}}$ and $\bar{\bar{P}}_{2;0;\text{SCT}}^{\text{ss sys}}$ in above formulation (4-114) are as follows:

$$\overline{\overline{P}}{}_{2;0;\text{PVT}}^{\text{ss sys}} = \overline{\overline{P}}{}_{1;0;\text{PVT}}^{\text{ss sys}} \tag{4-115a}$$

$$\overline{\overline{P}}{}_{2;0;\text{SCT}}^{\text{ss sys}} = \begin{bmatrix} 0 & 0 & -\overline{\overline{P}}{}_{1;0;\text{SCT}}^{J_{s1}M_{s1}} & \overline{\overline{P}}{}_{1;0;\text{SCT}}^{J_{s1}M_{s2}} \\ 0 & 0 & \overline{\overline{P}}{}_{1;0;\text{SCT}}^{J_{s2}M_{s1}} & -\overline{\overline{P}}{}_{1;0;\text{SCT}}^{J_{s2}M_{s2}} \\ -\overline{\overline{P}}{}_{1;0;\text{SCT}}^{M_{s1}J_{s1}} & \overline{\overline{P}}{}_{1;0;\text{SCT}}^{M_{s1}J_{s2}} & 0 & 0 \\ \overline{\overline{P}}{}_{1;0;\text{SCT}}^{M_{s2}J_{s1}} & -\overline{\overline{P}}{}_{1;0;\text{SCT}}^{M_{s2}J_{s2}} & 0 & 0 \end{bmatrix} \tag{4-115b}$$

where the power matrix $\overline{\overline{P}}{}_{1;0;\text{PVT}}^{\text{ss sys}}$ in formulation (4-115a) is just the one in formulation (4-81a), and the various sub-matrices in formulation (4-115b) (except the minus signs "$-$") are just the ones in formulation (4-81b).

Inserting transformation relationship (4-101) into matrix form (4-114), it is obtained that

$$P_{\text{ss sys}}^{\text{driving}} = \left(\overline{a}^{M_{s0}}\right)^H \cdot \underbrace{\begin{bmatrix} \overline{\overline{T}}{}^{J_{s10}\leftarrow M_{s0}} \\ \overline{\overline{T}}{}^{J_{s12}\leftarrow M_{s0}} \\ -\overline{\overline{T}}{}^{J_{s12}\leftarrow M_{s0}} \\ \overline{\overline{T}}{}^{J_{s20}\leftarrow M_{s0}} \\ \overline{\overline{\mathcal{I}}}{}^{M_{s10}} \\ \overline{\overline{T}}{}^{M_{s12}\leftarrow M_{s0}} \\ -\overline{\overline{T}}{}^{M_{s12}\leftarrow M_{s0}} \\ \overline{\overline{\mathcal{I}}}{}^{M_{s20}} \end{bmatrix}^H \cdot \overline{\overline{P}}{}_{\text{ss sys};2}^{\text{driving}} \cdot \begin{bmatrix} \overline{\overline{T}}{}^{J_{s10}\leftarrow M_{s0}} \\ \overline{\overline{T}}{}^{J_{s12}\leftarrow M_{s0}} \\ -\overline{\overline{T}}{}^{J_{s12}\leftarrow M_{s0}} \\ \overline{\overline{T}}{}^{J_{s20}\leftarrow M_{s0}} \\ \overline{\overline{\mathcal{I}}}{}^{M_{s10}} \\ \overline{\overline{T}}{}^{M_{s12}\leftarrow M_{s0}} \\ -\overline{\overline{T}}{}^{M_{s12}\leftarrow M_{s0}} \\ \overline{\overline{\mathcal{I}}}{}^{M_{s20}} \end{bmatrix}}_{\overline{\overline{P}}{}_{2;M_{s0}}^{\text{driving}}} \cdot \overline{a}^{M_{s0}} \tag{4-116}$$

This is just the matrix form of the DPO surface formulation with only BVs. The sub-matrices $\overline{\overline{\mathcal{I}}}{}^{M_{s10}}$ and $\overline{\overline{\mathcal{I}}}{}^{M_{s20}}$ in formulation (4-116) are the same as the ones in formulation (4-102). To emphasize the difference between above matrix form (4-116) and previous matrix form (4-102), we add subscript "2" to above power quadratic matrix $\overline{\overline{P}}{}_{2;M_{s0}}^{\text{driving}}$ instead of the subscript "1" added to the $\overline{\overline{P}}{}_{1;M_{s0}}^{\text{driving}}$ in formulation (4-102). Obviously, new formulations (4-113)&(4-114)&(4-116) have advantages over formulations (4-66)&(4-80)&(4-102) in the aspects of that: new formulation (4-113) not only is more concise than formulation (4-66), but also has a clearer physical picture; the matrix $\overline{\overline{P}}{}_{2;0;\text{PVT}}^{\text{ss sys}} + \overline{\overline{P}}{}_{2;0;\text{SCT}}^{\text{ss sys}} = \overline{\overline{P}}{}_{\text{ss sys};2}^{\text{driving}}$ in new formulation (4-114) is similar to the matrix $\overline{\overline{P}}{}_{1;0;\text{PVT}}^{\text{ss sys}} + \overline{\overline{P}}{}_{1;0;\text{SCT}}^{\text{ss sys}}$ in formulation (4-80) in the aspects of computational burden and memory usage, but new formulation (4-114) doesn't include the matrix $\overline{\overline{P}}{}_{1;\text{m}}^{\text{ss sys}}$ (which is related to material parameters and appears in formulation (4-80)), so new formulation (4-114) needs lesser computational resources than formulation (4-80).

The whole process to construct DP-CMs by orthogonalizing the above DPO is similar to the process given in previous Section 4.4, and the obtained DP-CMs also satisfy the same orthogonality and modal expansion as the ones given in previous Section 4.4, so they will not be repeated here.

## 4.5.2 Numerical Examples Corresponding to Typical Structures

In Subsection 4.4.4, we verified the validity of the surface formulation given in Section 4.4, by comparing the results calculated from the surface formulation given in Section 4.4 with the results calculated from the volume formulation given in Section 4.2. In this subsection, during the process to verify the validity of the formulation provided in this section, we will compare the results derived from the surface formulation developed in this section with the results derived from the volume formulation developed in Section 4.2; at the same time, we also compare the results derived from the surface formulation developed in this section with the results derived from the surface formulation developed in Section 4.4; in addition, by selecting a special two-body material system, we realize the comparison between the results derived from the surface formulation developed in this section and the results derived from the surface formulation developed in Section 4.3.

**1) Focusing on a Special Two-body Material System, Comparing the Results Derived From the Surface Formulation Developed in This Section with the Results Derived From the Surface Formulation Developed in Section 4.3**

Now, we still consider the two-body material system whose topological structure is shown in the Figure 4-30 given in Subsection 4.4.4. The lower material cylinder $V_{\text{sim}}^{2}$ is still the same as the one considered in Subsection 4.4.4 —— relative permeability $\vec{\vec{\mu}}_{\text{sim}}^{2\text{r}}$, relative permittivity $\vec{\vec{\varepsilon}}_{\text{sim}}^{2\text{r}}$, and conductivity $\vec{\vec{\sigma}}_{\text{sim}}^{2}$ are $\vec{\vec{I}}6$, $\vec{\vec{I}}6$, and $\vec{\vec{I}}0$ respectively. But, being different from the two-body material system considered in Subsection 4.4.4, this subsection sets the material parameters of the upper material cylinder as a series of different values, and calculates the DP-CMs corresponding to all cases by using the surface formulation developed in this section.

Figure 4-52(a) shows the characteristic value curves and MS curves of some typical DP-CMs corresponding to case $\{\vec{\vec{\mu}}_{\text{sim}}^{1\text{r}} = \vec{\vec{I}}4, \vec{\vec{\varepsilon}}_{\text{sim}}^{1\text{r}} = \vec{\vec{I}}4, \vec{\vec{\sigma}}_{\text{sim}}^{1} = \vec{\vec{I}}0\}$; Figure 4-52(b) shows

the characteristic value curves and MS curves of some typical DP-CMs corresponding to case $\{\overleftrightarrow{\mu}_{\text{sim}}^{1\text{r}} = \overleftrightarrow{I}3, \overleftrightarrow{\varepsilon}_{\text{sim}}^{1\text{r}} = \overleftrightarrow{I}3, \overleftrightarrow{\sigma}_{\text{sim}}^{1} = \overleftrightarrow{I}0\}$; Figure 4-52(c) shows the characteristic value curves and MS curves of some typical DP-CMs corresponding to case $\{\overleftrightarrow{\mu}_{\text{sim}}^{1\text{r}} = \overleftrightarrow{I}2, \overleftrightarrow{\varepsilon}_{\text{sim}}^{1\text{r}} = \overleftrightarrow{I}2, \overleftrightarrow{\sigma}_{\text{sim}}^{1} = \overleftrightarrow{I}0\}$; Figure 4-52(d) shows the characteristic value curves and MS curves of some typical DP-CMs corresponding to case $\{\overleftrightarrow{\mu}_{\text{sim}}^{1\text{r}} = \overleftrightarrow{I}1.5, \overleftrightarrow{\varepsilon}_{\text{sim}}^{1\text{r}} = \overleftrightarrow{I}1.5, \overleftrightarrow{\sigma}_{\text{sim}}^{1} = \overleftrightarrow{I}0\}$.

The one shown in Figure 4-53 is a two-body system whose upper material cylinder is an "air cylinder" (i.e., $\overleftrightarrow{\mu}_{\text{sim}}^{1\text{r}} = \overleftrightarrow{I}1$, $\overleftrightarrow{\varepsilon}_{\text{sim}}^{1\text{r}} = \overleftrightarrow{I}1$, and $\overleftrightarrow{\sigma}_{\text{sim}}^{1} = \overleftrightarrow{I}0$). In fact, the special two-body material system shown in Figure 4-53 is just a simply connected material cylinder $V_{\text{sim}}^{2}$, so the surface formulation developed in Section 4.3 is directly applicable to this case. Based on the surface formulation developed in Section 4.3, we calculate the DP-CMs of the material system shown in Figure 4-53, and provide the characteristic value curves and MS curves of some typical DP-CMs in Figure 4-54.

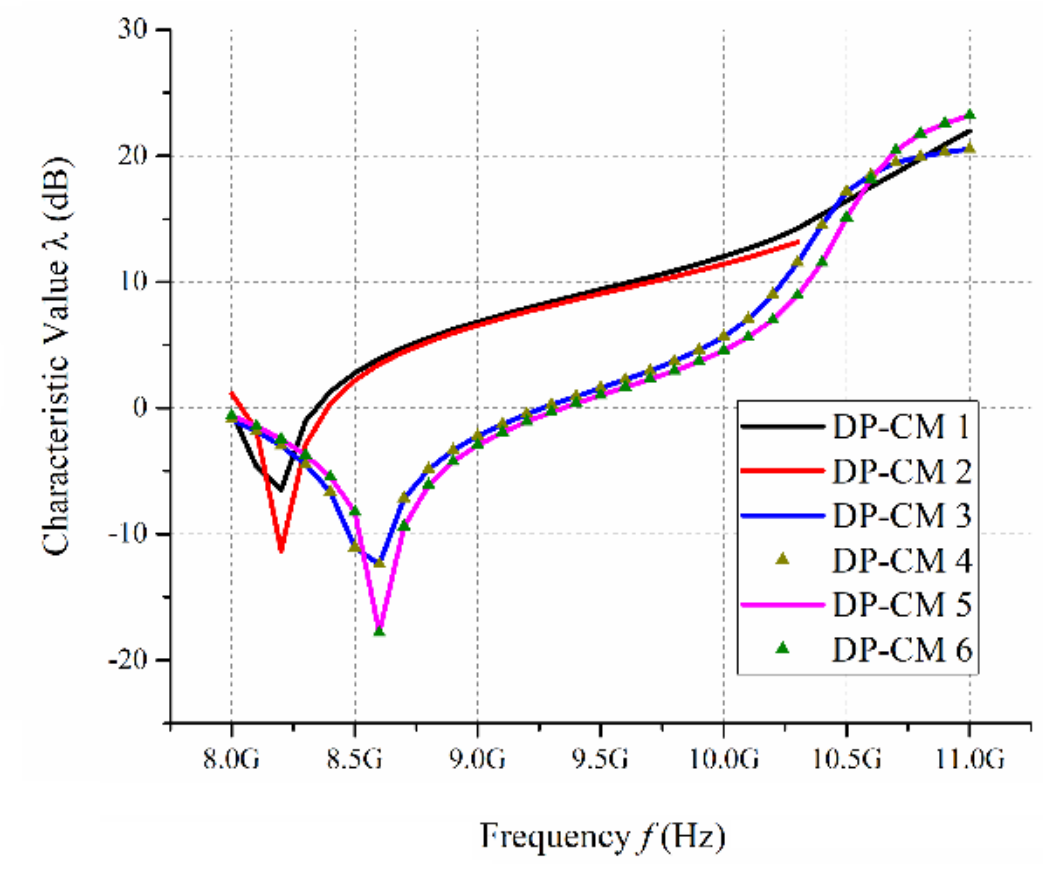

(a)

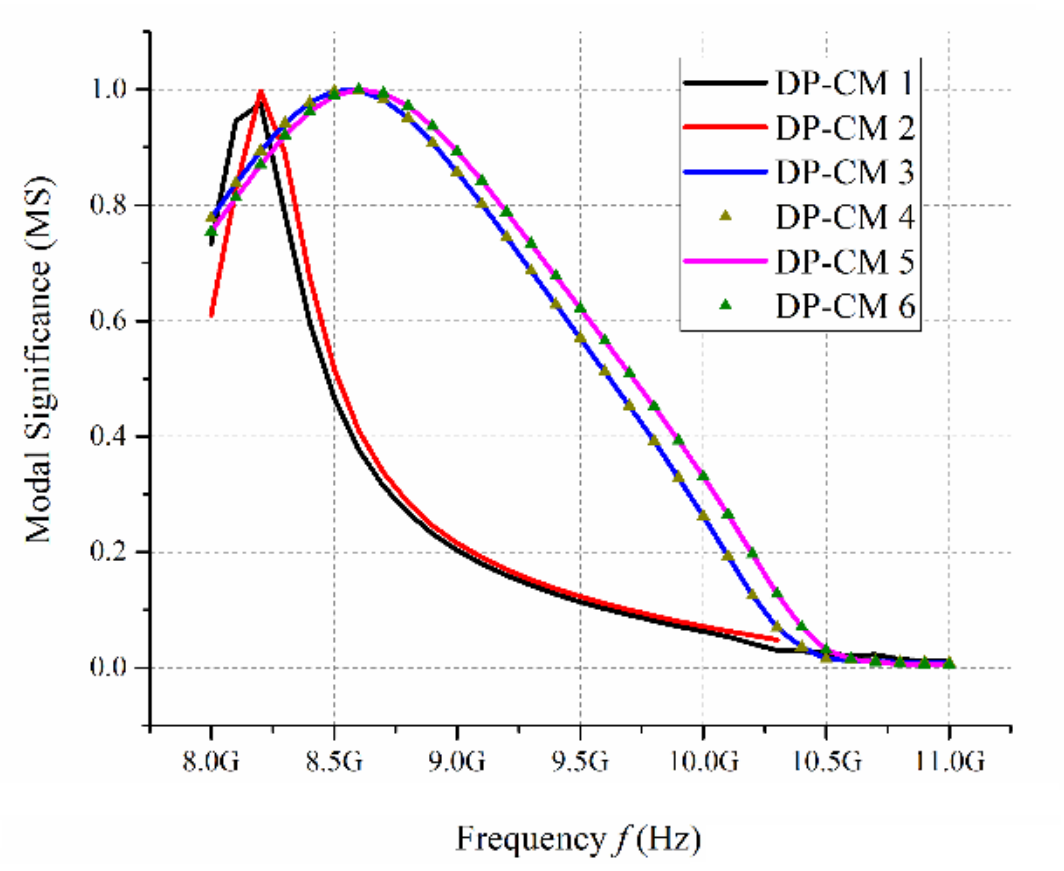

(b)

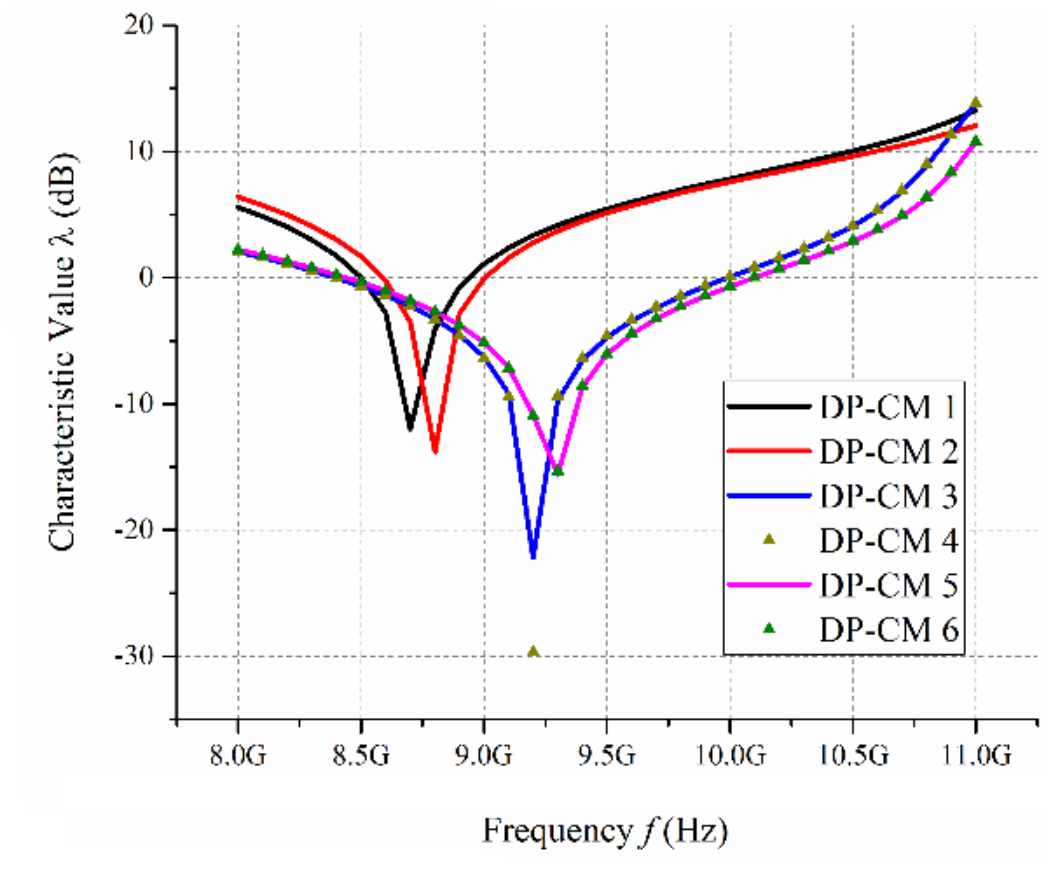

(c)

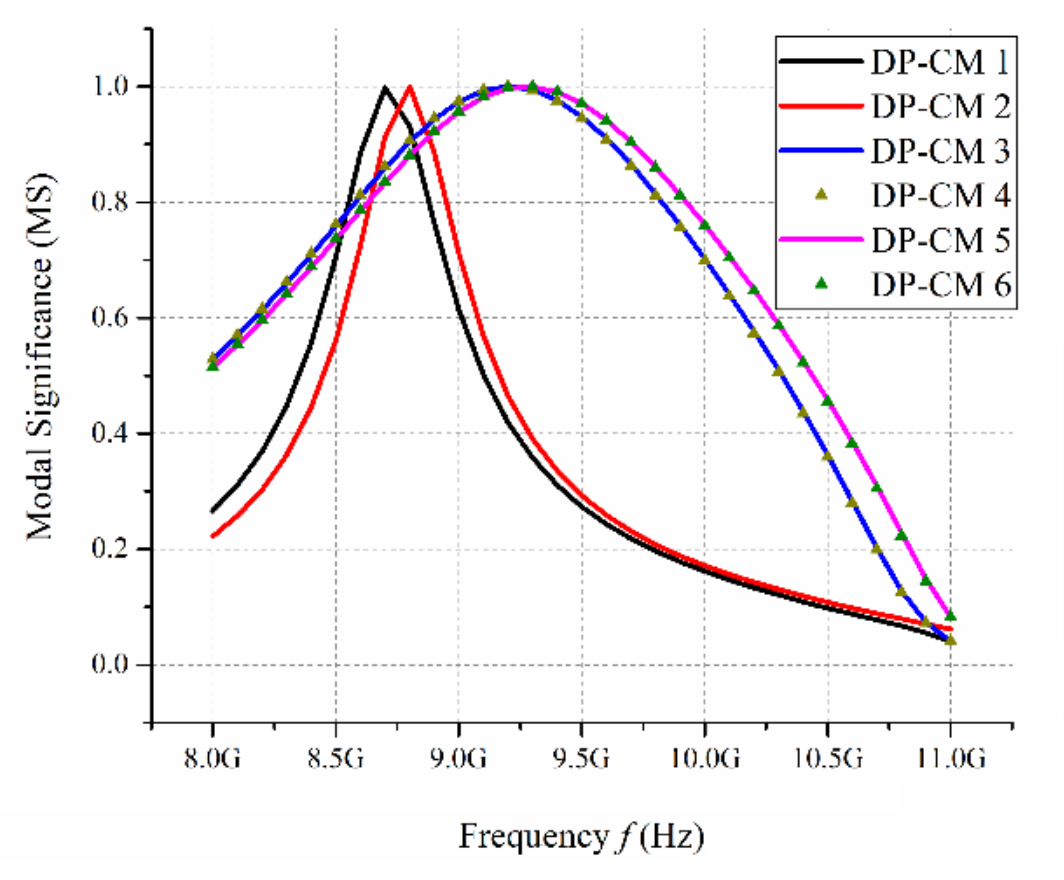

(d)

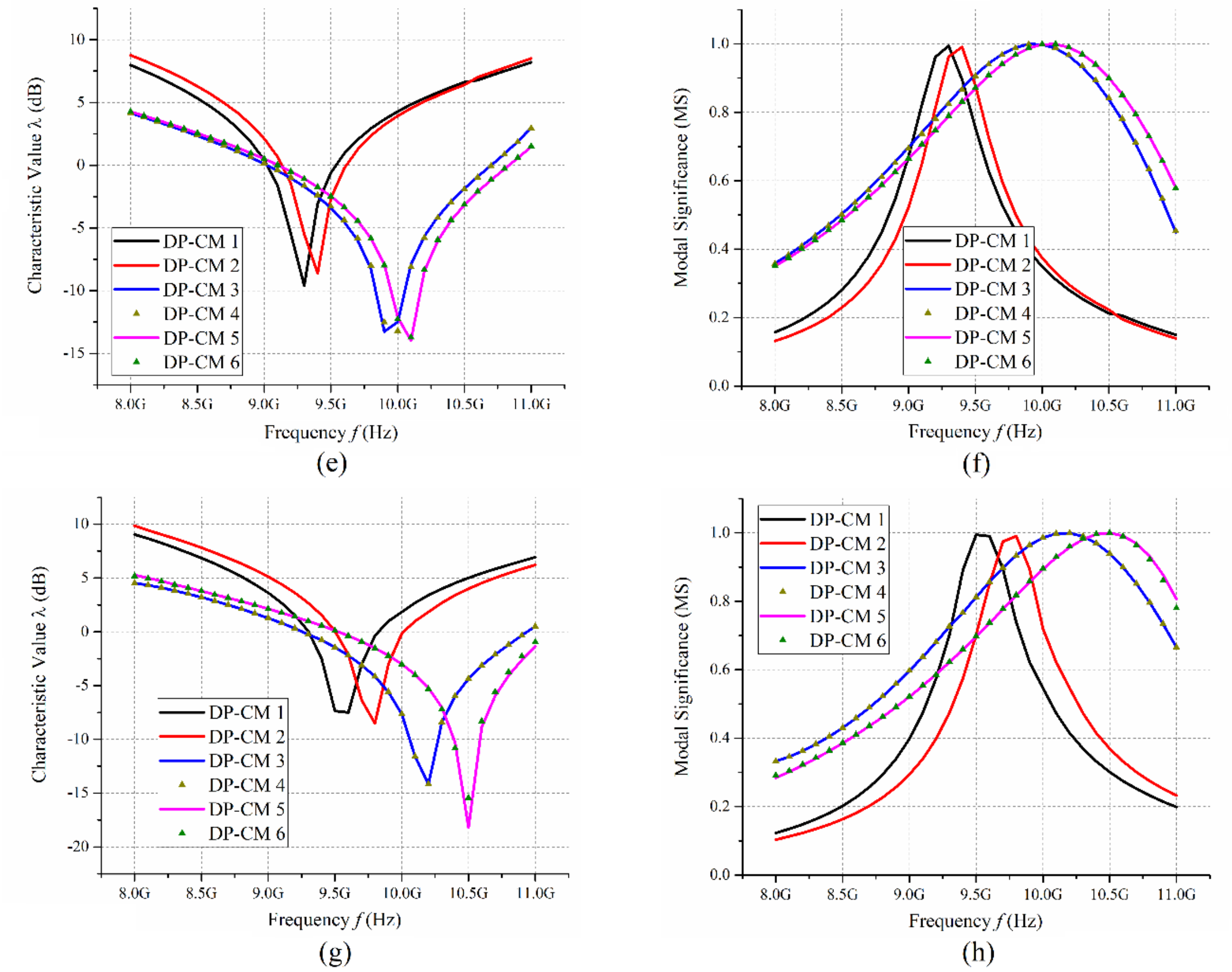

Figure 4-52 The characteristic quantity curves derived from the formulations given in this section. (a) the characteristic value curves corresponding to $\vec{\mu}_{\text{sim}}^{1\text{r}} = \vec{I}4 = \vec{\varepsilon}_{\text{sim}}^{1\text{r}}$; (b) the MS curves corresponding to $\vec{\mu}_{\text{sim}}^{1\text{r}} = \vec{I}4 = \vec{\varepsilon}_{\text{sim}}^{1\text{r}}$; (c) the characteristic value curves corresponding to $\vec{\mu}_{\text{sim}}^{1\text{r}} = \vec{I}3 = \vec{\varepsilon}_{\text{sim}}^{1\text{r}}$; (d) the MS curves corresponding to $\vec{\mu}_{\text{sim}}^{1\text{r}} = \vec{I}3 = \vec{\varepsilon}_{\text{sim}}^{1\text{r}}$; (e) the characteristic value curves corresponding to $\vec{\mu}_{\text{sim}}^{1\text{r}} = \vec{I}2 = \vec{\varepsilon}_{\text{sim}}^{1\text{r}}$; (f) the MS curves corresponding to $\vec{\mu}_{\text{sim}}^{1\text{r}} = \vec{I}2 = \vec{\varepsilon}_{\text{sim}}^{1\text{r}}$; (g) the characteristic value curves corresponding to $\vec{\mu}_{\text{sim}}^{1\text{r}} = \vec{I}1.5 = \vec{\varepsilon}_{\text{sim}}^{1\text{r}}$; (h) the MS curves corresponding to $\vec{\mu}_{\text{sim}}^{1\text{r}} = \vec{I}1.5 = \vec{\varepsilon}_{\text{sim}}^{1\text{r}}$

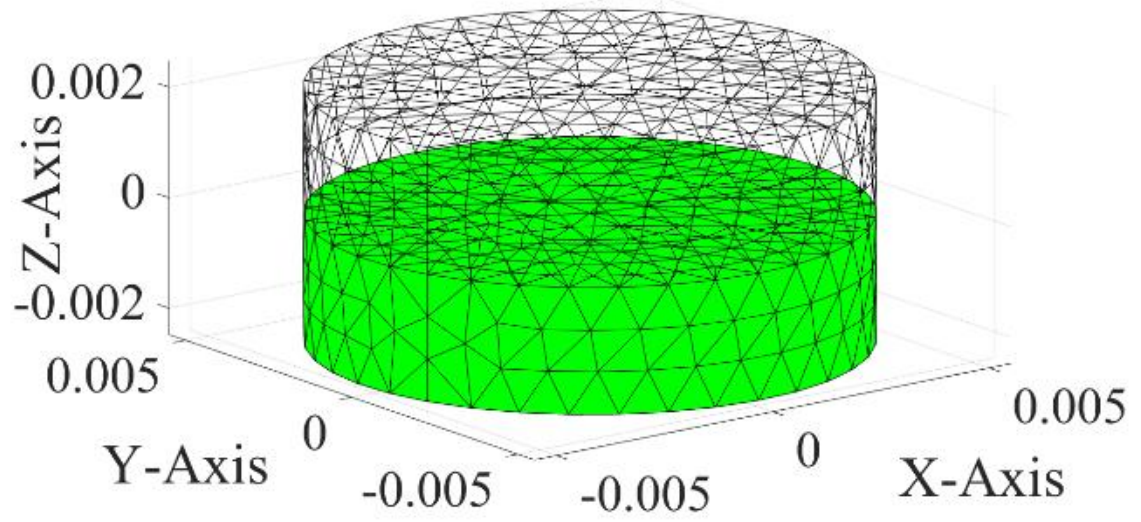

Figure 4-53 The topological structure and surface triangular meshes of a two-body material system constituted by an "air cylinder" and a material cylinder (the system shown in this figure is essentially a simply connected one-body material cylinder, i.e. the lower material cylinder)

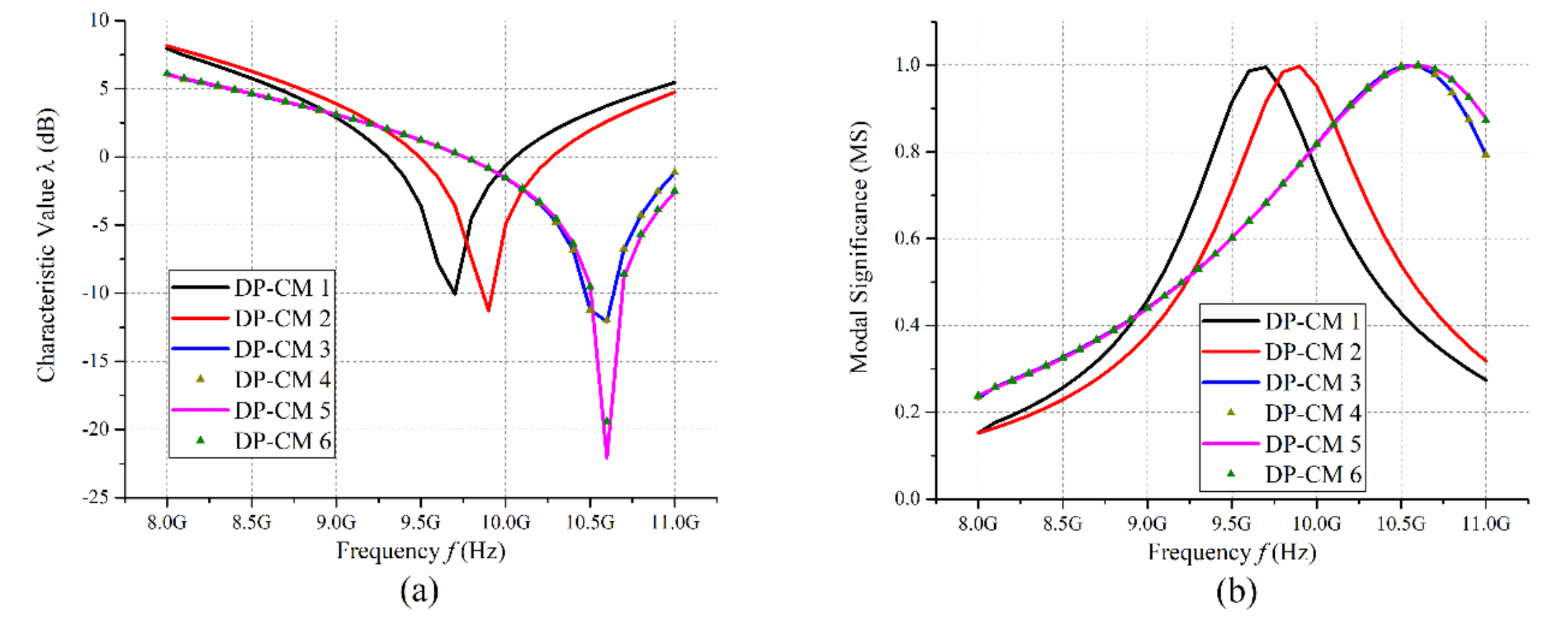

Figure 4-54 The characteristic quantity curves corresponding to several typical DP-CMs (of the simply connected one-body material cylinder shown in Figure 4-53) derived from the formulations given in Section 4.3. (a) characteristic value dB curves; (b) MS curves

Observing above Figures 4-52 and 4-54, it is easy to find out that: when the upper material cylinder approaches "air cylinder", the result calculated from the formulation developed in this section approaches the result calculated from the formulation developed in Section 4.3. This phenomenon is just a powerful proof for the validity of the surface formulation developed in this section.

**2) Focusing on the Two-body Material System Considered in Subsection 4.4.4, Comparing the Results Derived From the Surface Formulation Developed in This Section with the Results Derived From the Formulations Developed in Sections 4.2 and 4.4**

In Subsection 4.4.4, we, based on the surface formulation developed in Section 4.4, calculated the DP-CMs of the two-body material system whose topological structure was shown in Figure 4-30 and material parameters were $\{\vec{\vec{\mu}}_{\text{sim}}^{1\text{r}} = \vec{\vec{I}}\,3, \vec{\vec{\varepsilon}}_{\text{sim}}^{1\text{r}} = \vec{\vec{I}}\,12, \vec{\vec{\sigma}}_{\text{sim}}^{1} = \vec{\vec{I}}\,0\}$ & $\{\vec{\vec{\mu}}_{\text{sim}}^{2\text{r}} = \vec{\vec{I}}\,6, \vec{\vec{\varepsilon}}_{\text{sim}}^{2\text{r}} = \vec{\vec{I}}\,6, \vec{\vec{\sigma}}_{\text{sim}}^{2} = \vec{\vec{I}}\,0\}$, and also provided the corresponding characteristic value curves and MS curves in Figure 4-33. At that time, we also calculated the DP-CMs of the two-body material system based on the volume formulation developed in Section 4.2, and provided the corresponding characteristic value curves and MS curves in Figure 4-34. By comparing Figure 4-33 and Figure 4-34, we realized the verification for the validity of the surface formulation developed in Section 4.4. Now, we, based on the surface formulation developed in this section, calculated the DP-CMs of the two-body material system, and provide the corresponding characteristic value curves and MS curves in Figure 4-55.

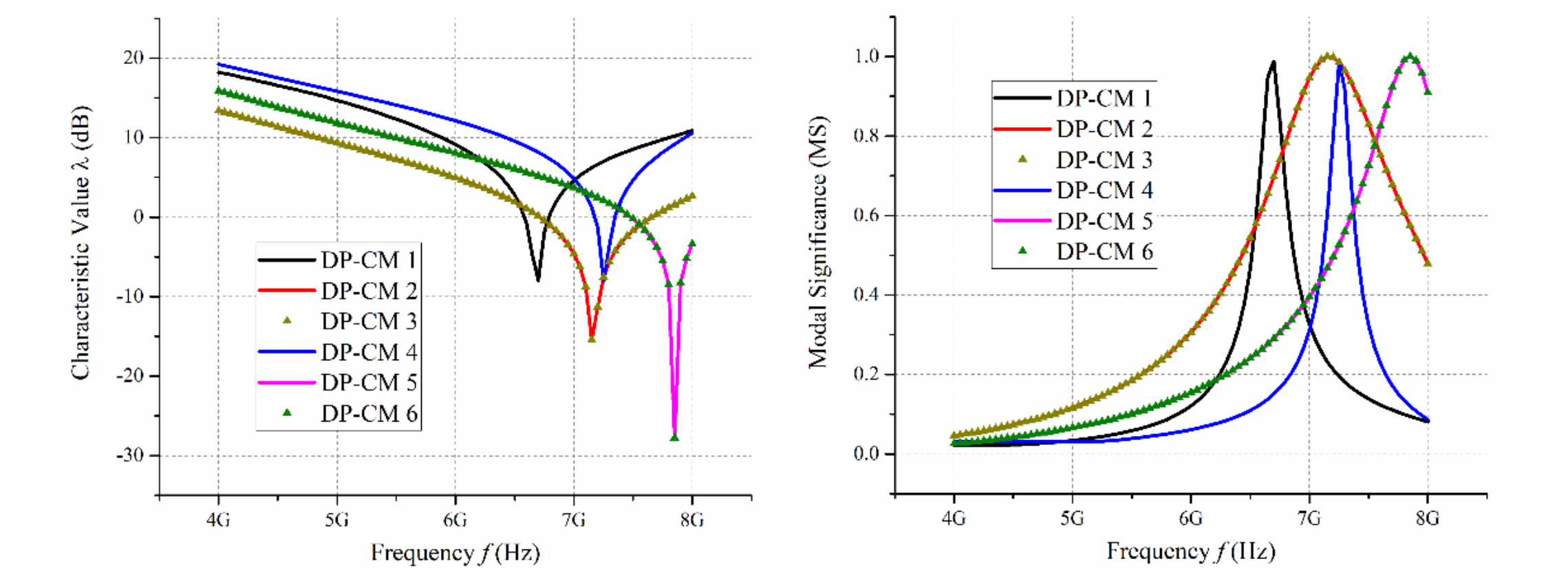

(a) (b)

Figure 4-55 For the two-body system shown in Figure 4-30 ( $\overset{\leftrightarrow}{\mu}{}^{1r}_{\text{sim}} = \overset{\leftrightarrow}{I}3$ , $\overset{\leftrightarrow}{\varepsilon}{}^{1r}_{\text{sim}} = \overset{\leftrightarrow}{I}12$ , $\overset{\leftrightarrow}{\sigma}{}^{1}_{\text{sim}} = \overset{\leftrightarrow}{I}0$ ; $\overset{\leftrightarrow}{\mu}{}^{2r}_{\text{sim}} = \overset{\leftrightarrow}{I}6 = \overset{\leftrightarrow}{\varepsilon}{}^{2r}_{\text{sim}}$ , $\overset{\leftrightarrow}{\sigma}{}^{2}_{\text{sim}} = \overset{\leftrightarrow}{I}0$ ), the characteristic quantity curves derived from the surface formulations given in this section. (a) characteristic value dB curves; (b) MS curves

Obviously, the results shown in above Figure 4-55 agree well with the results shown in Figures 4-33 and 4-34, and this implies the validity of the surface formulation developed in this section in the aspect of constructing the DP-CMs of the two-body material system shown in Figure 4-27.

## 4.6 The Third Surface Formulation for Calculating the DP-CMs of the Systems Constructed by Double Simply Connected Material Bodies

In Section 4.5, we, based on the GSEP obtained in Appendix C, did some identical transformations for the DPO (4-66) given in Section 4.4, and then derived a new surface formulation (4-113) of the DPO corresponding to the two-body material system shown in Figure 4-27. In the aspects of physical picture and conciseness, new formulation (4-113) is more advantageous than formulation (4-66), and then the (4-116)-based DP-CM calculation formulation requires less computational resources than the (4-102)-based DP-CM calculation formulation.

In this section, we will, based on current decompositions (4-67) & (4-68) and relationship (4-71), further simplify DPO (4-113), and obtain a more concise surface formulation of the DPO. After that, we provide the matrix form of the surface formulation, which only contains BVs. Based on the matrix form, we construct the corresponding DP-CMs. At the end of this section, we provide some typical numerical examples to verify the validy of the new formulation developed in this section in the aspect of constructing DP-CMs.

### 4.6.1 The Third Surface Formulation for Calculating the DP-CMs of Two-body Systems

For the two-body material system shown in Figure 4-27, its DPO surface formulation (4-113) can be further rewritten as follows:

$$
\begin{aligned}
& P_{\text{ss sys}}^{\text{driving}} \\
&= -(1/2)\Big[\big\langle \vec{J}_{\text{s1}}^{\text{ES}}, \mathcal{E}_0\big(\vec{J}_{\text{s1}}^{\text{ES}}, \vec{M}_{\text{s1}}^{\text{ES}}\big)\big\rangle_{\partial V_{\text{sim}}^{1-}} + \big\langle \vec{J}_{\text{s1}}^{\text{ES}}, \mathcal{E}_0\big(\vec{J}_{\text{s2}}^{\text{ES}}, \vec{M}_{\text{s2}}^{\text{ES}}\big)\big\rangle_{\partial V_{\text{sim}}^{1-}}\Big] \\
&\quad -(1/2)\Big[\big\langle \vec{J}_{\text{s2}}^{\text{ES}}, \mathcal{E}_0\big(\vec{J}_{\text{s2}}^{\text{ES}}, \vec{M}_{\text{s2}}^{\text{ES}}\big)\big\rangle_{\partial V_{\text{sim}}^{2-}} + \big\langle \vec{J}_{\text{s2}}^{\text{ES}}, \mathcal{E}_0\big(\vec{J}_{\text{s1}}^{\text{ES}}, \vec{M}_{\text{s1}}^{\text{ES}}\big)\big\rangle_{\partial V_{\text{sim}}^{2-}}\Big] \\
&\quad -(1/2)\Big[\big\langle \vec{M}_{\text{s1}}^{\text{ES}}, \mathcal{H}_0\big(\vec{J}_{\text{s1}}^{\text{ES}}, \vec{M}_{\text{s1}}^{\text{ES}}\big)\big\rangle_{\partial V_{\text{sim}}^{1-}} + \big\langle \vec{M}_{\text{s1}}^{\text{ES}}, \mathcal{H}_0\big(\vec{J}_{\text{s2}}^{\text{ES}}, \vec{M}_{\text{s2}}^{\text{ES}}\big)\big\rangle_{\partial V_{\text{sim}}^{1-}}\Big] \\
&\quad -(1/2)\Big[\big\langle \vec{M}_{\text{s2}}^{\text{ES}}, \mathcal{H}_0\big(\vec{J}_{\text{s2}}^{\text{ES}}, \vec{M}_{\text{s2}}^{\text{ES}}\big)\big\rangle_{\partial V_{\text{sim}}^{2-}} + \big\langle \vec{M}_{\text{s2}}^{\text{ES}}, \mathcal{H}_0\big(\vec{J}_{\text{s1}}^{\text{ES}}, \vec{M}_{\text{s1}}^{\text{ES}}\big)\big\rangle_{\partial V_{\text{sim}}^{2-}}\Big] \\
&= -\frac{1}{2}\big\langle \vec{J}_{\text{s1}}^{\text{ES}}, \mathcal{E}_0\big(\vec{J}_{\text{s1}}^{\text{ES}} + \vec{J}_{\text{s2}}^{\text{ES}}, \vec{M}_{\text{s1}}^{\text{ES}} + \vec{M}_{\text{s2}}^{\text{ES}}\big)\big\rangle_{\partial V_{\text{sim}}^{1-}} - \frac{1}{2}\Big[\big\langle \vec{J}_{\text{s2}}^{\text{ES}}, \mathcal{E}_0\big(\vec{J}_{\text{s1}}^{\text{ES}} + \vec{J}_{\text{s2}}^{\text{ES}}, \vec{M}_{\text{s1}}^{\text{ES}} + \vec{M}_{\text{s2}}^{\text{ES}}\big)\big\rangle_{\partial V_{\text{sim}}^{2-}}\Big] \\
&\quad -\frac{1}{2}\big\langle \vec{M}_{\text{s1}}^{\text{ES}}, \mathcal{H}_0\big(\vec{J}_{\text{s1}}^{\text{ES}} + \vec{J}_{\text{s2}}^{\text{ES}}, \vec{M}_{\text{s1}}^{\text{ES}} + \vec{M}_{\text{s2}}^{\text{ES}}\big)\big\rangle_{\partial V_{\text{sim}}^{1-}} - \frac{1}{2}\Big[\big\langle \vec{M}_{\text{s2}}^{\text{ES}}, \mathcal{H}_0\big(\vec{J}_{\text{s1}}^{\text{ES}} + \vec{J}_{\text{s2}}^{\text{ES}}, \vec{M}_{\text{s1}}^{\text{ES}} + \vec{M}_{\text{s2}}^{\text{ES}}\big)\big\rangle_{\partial V_{\text{sim}}^{2-}}\Big] \\
&= -\frac{1}{2}\big\langle \vec{J}_{\text{s1}}^{\text{ES}}, \mathcal{E}_0\big(\vec{J}_{\text{s10}}^{\text{ES}} + \vec{J}_{\text{s20}}^{\text{ES}}, \vec{M}_{\text{s10}}^{\text{ES}} + \vec{M}_{\text{s20}}^{\text{ES}}\big)\big\rangle_{\partial V_{\text{sim}}^{1-}} - \frac{1}{2}\Big[\big\langle \vec{J}_{\text{s2}}^{\text{ES}}, \mathcal{E}_0\big(\vec{J}_{\text{s10}}^{\text{ES}} + \vec{J}_{\text{s20}}^{\text{ES}}, \vec{M}_{\text{s10}}^{\text{ES}} + \vec{M}_{\text{s20}}^{\text{ES}}\big)\big\rangle_{\partial V_{\text{sim}}^{2-}}\Big] \\
&\quad -\frac{1}{2}\big\langle \vec{M}_{\text{s1}}^{\text{ES}}, \mathcal{H}_0\big(\vec{J}_{\text{s10}}^{\text{ES}} + \vec{J}_{\text{s20}}^{\text{ES}}, \vec{M}_{\text{s10}}^{\text{ES}} + \vec{M}_{\text{s20}}^{\text{ES}}\big)\big\rangle_{\partial V_{\text{sim}}^{1-}} - \frac{1}{2}\Big[\big\langle \vec{M}_{\text{s2}}^{\text{ES}}, \mathcal{H}_0\big(\vec{J}_{\text{s10}}^{\text{ES}} + \vec{J}_{\text{s20}}^{\text{ES}}, \vec{M}_{\text{s10}}^{\text{ES}} + \vec{M}_{\text{s20}}^{\text{ES}}\big)\big\rangle_{\partial V_{\text{sim}}^{2-}}\Big] \\
&= -(1/2)\big\langle \vec{J}_{\text{s10}}^{\text{ES}} + \vec{J}_{\text{s12}}^{\text{ES}}, \mathcal{E}_0\big(\vec{J}_{\text{s10}}^{\text{ES}} + \vec{J}_{\text{s20}}^{\text{ES}}, \vec{M}_{\text{s10}}^{\text{ES}} + \vec{M}_{\text{s20}}^{\text{ES}}\big)\big\rangle_{\partial V_{\text{s10}}^{-} \cup \partial V_{\text{s12}}} \\
&\quad -(1/2)\Big[\big\langle \vec{J}_{\text{s21}}^{\text{ES}} + \vec{J}_{\text{s20}}^{\text{ES}}, \mathcal{E}_0\big(\vec{J}_{\text{s10}}^{\text{ES}} + \vec{J}_{\text{s20}}^{\text{ES}}, \vec{M}_{\text{s10}}^{\text{ES}} + \vec{M}_{\text{s20}}^{\text{ES}}\big)\big\rangle_{\partial V_{\text{s21}} \cup \partial V_{\text{s20}}^{-}}\Big] \\
&\quad -(1/2)\big\langle \vec{M}_{\text{s10}}^{\text{ES}} + \vec{M}_{\text{s12}}^{\text{ES}}, \mathcal{H}_0\big(\vec{J}_{\text{s10}}^{\text{ES}} + \vec{J}_{\text{s20}}^{\text{ES}}, \vec{M}_{\text{s10}}^{\text{ES}} + \vec{M}_{\text{s20}}^{\text{ES}}\big)\big\rangle_{\partial V_{\text{s10}}^{-} \cup \partial V_{\text{s12}}} \\
&\quad -(1/2)\Big[\big\langle \vec{M}_{\text{s21}}^{\text{ES}} + \vec{M}_{\text{s20}}^{\text{ES}}, \mathcal{H}_0\big(\vec{J}_{\text{s10}}^{\text{ES}} + \vec{J}_{\text{s20}}^{\text{ES}}, \vec{M}_{\text{s10}}^{\text{ES}} + \vec{M}_{\text{s20}}^{\text{ES}}\big)\big\rangle_{\partial V_{\text{s21}} \cup \partial V_{\text{s20}}^{-}}\Big] \\
&= -(1/2)\big\langle \vec{J}_{\text{s10}}^{\text{ES}} + \vec{J}_{\text{s20}}^{\text{ES}}, \mathcal{E}_0\big(\vec{J}_{\text{s10}}^{\text{ES}} + \vec{J}_{\text{s20}}^{\text{ES}}, \vec{M}_{\text{s10}}^{\text{ES}} + \vec{M}_{\text{s20}}^{\text{ES}}\big)\big\rangle_{\partial V_{\text{s10}}^{-} \cup \partial V_{\text{s20}}^{-}} \\
&\quad -(1/2)\big\langle \vec{M}_{\text{s10}}^{\text{ES}} + \vec{M}_{\text{s20}}^{\text{ES}}, \mathcal{H}_0\big(\vec{J}_{\text{s10}}^{\text{ES}} + \vec{J}_{\text{s20}}^{\text{ES}}, \vec{M}_{\text{s10}}^{\text{ES}} + \vec{M}_{\text{s20}}^{\text{ES}}\big)\big\rangle_{\partial V_{\text{s10}}^{-} \cup \partial V_{\text{s20}}^{-}}
\end{aligned}
\tag{4-117}
$$

In formulation (4-117), the second equality is based on the linear property of inner product; the third equality is based on current decompositions (4-67)&(4-68) and relationship (4-71); the fourth equality is based on current decompositions (4-67)&(4-68) and the continuity of the field $\mathcal{F}_0(\vec{J}_{\text{s10}}^{\text{ES}} + \vec{J}_{\text{s20}}^{\text{ES}}, \vec{M}_{\text{s10}}^{\text{ES}} + \vec{M}_{\text{s20}}^{\text{ES}})$ on $\partial V_{\text{s12}} = \partial V_{\text{s21}}$; the fifth equality is based on relationship (4-71).

Inserting expansion formulations (4-72) and (4-73) into above DPO (4-117), the DPO is immediately decomposed into the following matrix form:

$$P_{\text{ss sys}}^{\text{driving}} = \begin{bmatrix} \bar{a}^{J_{\text{s}10}} \\ \bar{a}^{J_{\text{s}20}} \\ \bar{a}^{M_{\text{s}10}} \\ \bar{a}^{M_{\text{s}20}} \end{bmatrix}^H \cdot \underbrace{\left( \bar{\bar{P}}_{3;0;\text{PVT}}^{\text{ss sys}} + \bar{\bar{P}}_{3;0;\text{SCT}}^{\text{ss sys}} \right)}_{\bar{\bar{P}}_{\text{ss sys};3}^{\text{driving}}} \cdot \begin{bmatrix} \bar{a}^{J_{\text{s}10}} \\ \bar{a}^{J_{\text{s}20}} \\ \bar{a}^{M_{\text{s}10}} \\ \bar{a}^{M_{\text{s}20}} \end{bmatrix} \tag{4-118}$$

where

$$\bar{\bar{P}}_{3;0;\text{PVT}}^{\text{ss sys}} = \begin{bmatrix} \bar{\bar{P}}_{3;0;\text{PVT}}^{J_{\text{s}10}J_{\text{s}10}} & \bar{\bar{P}}_{3;0;\text{PVT}}^{J_{\text{s}10}J_{\text{s}20}} & \bar{\bar{P}}_{3;0;\text{PVT}}^{J_{\text{s}10}M_{\text{s}10}} & \bar{\bar{P}}_{3;0;\text{PVT}}^{J_{\text{s}10}M_{\text{s}20}} \\ \bar{\bar{P}}_{3;0;\text{PVT}}^{J_{\text{s}20}J_{\text{s}10}} & \bar{\bar{P}}_{3;0;\text{PVT}}^{J_{\text{s}20}J_{\text{s}20}} & \bar{\bar{P}}_{3;0;\text{PVT}}^{J_{\text{s}20}M_{\text{s}10}} & \bar{\bar{P}}_{3;0;\text{PVT}}^{J_{\text{s}20}M_{\text{s}20}} \\ \bar{\bar{P}}_{3;0;\text{PVT}}^{M_{\text{s}10}J_{\text{s}10}} & \bar{\bar{P}}_{3;0;\text{PVT}}^{M_{\text{s}10}J_{\text{s}20}} & \bar{\bar{P}}_{3;0;\text{PVT}}^{M_{\text{s}10}M_{\text{s}10}} & \bar{\bar{P}}_{3;0;\text{PVT}}^{M_{\text{s}10}M_{\text{s}20}} \\ \bar{\bar{P}}_{3;0;\text{PVT}}^{M_{\text{s}20}J_{\text{s}10}} & \bar{\bar{P}}_{3;0;\text{PVT}}^{M_{\text{s}20}J_{\text{s}20}} & \bar{\bar{P}}_{3;0;\text{PVT}}^{M_{\text{s}20}M_{\text{s}10}} & \bar{\bar{P}}_{3;0;\text{PVT}}^{M_{\text{s}20}M_{\text{s}20}} \end{bmatrix} \tag{4-119a}$$

$$\bar{\bar{P}}_{3;0;\text{SCT}}^{\text{ss sys}} = \begin{bmatrix} 0 & 0 & \bar{\bar{P}}_{3;0;\text{SCT}}^{J_{\text{s}10}M_{\text{s}10}} & 0 \\ 0 & 0 & 0 & \bar{\bar{P}}_{3;0;\text{SCT}}^{J_{\text{s}20}M_{\text{s}20}} \\ \bar{\bar{P}}_{3;0;\text{SCT}}^{M_{\text{s}10}J_{\text{s}10}} & 0 & 0 & 0 \\ 0 & \bar{\bar{P}}_{3;0;\text{SCT}}^{M_{\text{s}20}J_{\text{s}20}} & 0 & 0 \end{bmatrix} \tag{4-119b}$$

The elements of the sub-matrices in formulation (4-119a) are calculated as follows:

$$p_{3;0;\text{PVT};\xi\zeta}^{J_{\text{s}10}J_{\text{s}10}} = -(1/2)\left\langle \vec{b}_\xi^{J_{\text{s}10}}, -j\omega\mu_0\mathcal{L}_0\left(\vec{b}_\zeta^{J_{\text{s}10}}\right)\right\rangle_{\partial V_{\text{s}10}} \tag{4-120a}$$

$$p_{3;0;\text{PVT};\xi\zeta}^{J_{\text{s}10}J_{\text{s}20}} = -(1/2)\left\langle \vec{b}_\xi^{J_{\text{s}10}}, -j\omega\mu_0\mathcal{L}_0\left(\vec{b}_\zeta^{J_{\text{s}20}}\right)\right\rangle_{\partial V_{\text{s}10}} \tag{4-120b}$$

$$p_{3;0;\text{PVT};\xi\zeta}^{J_{\text{s}10}M_{\text{s}10}} = -(1/2)\left\langle \vec{b}_\xi^{J_{\text{s}10}}, -\text{P.V.}\,\mathcal{K}_0\left(\vec{b}_\zeta^{M_{\text{s}10}}\right)\right\rangle_{\partial V_{\text{s}10}} \tag{4-120c}$$

$$p_{3;0;\text{PVT};\xi\zeta}^{J_{\text{s}10}M_{\text{s}20}} = -(1/2)\left\langle \vec{b}_\xi^{J_{\text{s}10}}, -\text{P.V.}\,\mathcal{K}_0\left(\vec{b}_\zeta^{M_{\text{s}20}}\right)\right\rangle_{\partial V_{\text{s}10}} \tag{4-120d}$$

and

$$p_{3;0;\text{PVT};\xi\zeta}^{J_{\text{s}20}J_{\text{s}10}} = -(1/2)\left\langle \vec{b}_\xi^{J_{\text{s}20}}, -j\omega\mu_0\mathcal{L}_0\left(\vec{b}_\zeta^{J_{\text{s}10}}\right)\right\rangle_{\partial V_{\text{s}20}} \tag{4-120e}$$

$$p_{3;0;\text{PVT};\xi\zeta}^{J_{\text{s}20}J_{\text{s}20}} = -(1/2)\left\langle \vec{b}_\xi^{J_{\text{s}20}}, -j\omega\mu_0\mathcal{L}_0\left(\vec{b}_\zeta^{J_{\text{s}20}}\right)\right\rangle_{\partial V_{\text{s}20}} \tag{4-120f}$$

$$p_{3;0;\text{PVT};\xi\zeta}^{J_{\text{s}20}M_{\text{s}10}} = -(1/2)\left\langle \vec{b}_\xi^{J_{\text{s}20}}, -\text{P.V.}\,\mathcal{K}_0\left(\vec{b}_\zeta^{M_{\text{s}10}}\right)\right\rangle_{\partial V_{\text{s}20}} \tag{4-120g}$$

$$p_{3;0;\text{PVT};\xi\zeta}^{J_{\text{s}20}M_{\text{s}20}} = -(1/2)\left\langle \vec{b}_\xi^{J_{\text{s}20}}, -\text{P.V.}\,\mathcal{K}_0\left(\vec{b}_\zeta^{M_{\text{s}20}}\right)\right\rangle_{\partial V_{\text{s}20}} \tag{4-120h}$$

and

$$p_{3;0;\text{PVT};\xi\zeta}^{M_{\text{s}10}J_{\text{s}10}} = -(1/2)\left\langle \vec{b}_\xi^{M_{\text{s}10}}, \text{P.V.}\,\mathcal{K}_0\left(\vec{b}_\zeta^{J_{\text{s}10}}\right)\right\rangle_{\partial V_{\text{s}10}} \tag{4-120i}$$

$$p_{3;0;\text{PVT};\xi\zeta}^{M_{\text{s}10}J_{\text{s}20}} = -(1/2)\left\langle \vec{b}_\xi^{M_{\text{s}10}}, \text{P.V.}\,\mathcal{K}_0\left(\vec{b}_\zeta^{J_{\text{s}20}}\right)\right\rangle_{\partial V_{\text{s}10}} \tag{4-120j}$$

$$p_{3;0;\text{PVT};\xi\zeta}^{M_{s10}M_{s10}} = -(1/2)\left\langle \vec{b}_{\xi}^{M_{s10}}, -j\omega\varepsilon_0\mathcal{L}_0\left(\vec{b}_{\zeta}^{M_{s10}}\right)\right\rangle_{\partial V_{s10}} \tag{4-120k}$$

$$p_{3;0;\text{PVT};\xi\zeta}^{M_{s10}M_{s20}} = -(1/2)\left\langle \vec{b}_{\xi}^{M_{s10}}, -j\omega\varepsilon_0\mathcal{L}_0\left(\vec{b}_{\zeta}^{M_{s20}}\right)\right\rangle_{\partial V_{s10}} \tag{4-120l}$$

and

$$p_{3;0;\text{PVT};\xi\zeta}^{M_{s20}J_{s10}} = -(1/2)\left\langle \vec{b}_{\xi}^{M_{s20}}, \text{P.V.}\,\mathcal{K}_0\left(\vec{b}_{\zeta}^{J_{s10}}\right)\right\rangle_{\partial V_{s20}} \tag{4-120m}$$

$$p_{3;0;\text{PVT};\xi\zeta}^{M_{s20}J_{s20}} = -(1/2)\left\langle \vec{b}_{\xi}^{M_{s20}}, \text{P.V.}\,\mathcal{K}_0\left(\vec{b}_{\zeta}^{J_{s20}}\right)\right\rangle_{\partial V_{s20}} \tag{4-120n}$$

$$p_{3;0;\text{PVT};\xi\zeta}^{M_{s20}M_{s10}} = -(1/2)\left\langle \vec{b}_{\xi}^{M_{s20}}, -j\omega\varepsilon_0\mathcal{L}_0\left(\vec{b}_{\zeta}^{M_{s10}}\right)\right\rangle_{\partial V_{s20}} \tag{4-120o}$$

$$p_{3;0;\text{PVT};\xi\zeta}^{M_{s20}M_{s20}} = -(1/2)\left\langle \vec{b}_{\xi}^{M_{s20}}, -j\omega\varepsilon_0\mathcal{L}_0\left(\vec{b}_{\zeta}^{M_{s20}}\right)\right\rangle_{\partial V_{s20}} \tag{4-120p}$$

In formulation (4-119b), the 0s are some zero matrices with proper line numbers and proper column numbers, and the elements of the sub-matrices are calculated as follows:

$$p_{3;0;\text{SCT};\xi\zeta}^{J_{s10}M_{s10}} = -(1/2)\left\langle \vec{b}_{\xi}^{J_{s10}}, -\frac{1}{2}\vec{b}_{\zeta}^{M_{s10}} \times \hat{n}_{s1}^{-}\right\rangle_{\partial V_{s10}} \tag{4-121a}$$

$$p_{3;0;\text{SCT};\xi\zeta}^{J_{s20}M_{s20}} = -(1/2)\left\langle \vec{b}_{\xi}^{J_{s20}}, -\frac{1}{2}\vec{b}_{\zeta}^{M_{s20}} \times \hat{n}_{s2}^{-}\right\rangle_{\partial V_{s20}} \tag{4-121b}$$

$$p_{3;0;\text{SCT};\xi\zeta}^{M_{s10}J_{s10}} = -(1/2)\left\langle \vec{b}_{\xi}^{M_{s10}}, \frac{1}{2}\vec{b}_{\zeta}^{J_{s10}} \times \hat{n}_{s1}^{-}\right\rangle_{\partial V_{s10}} \tag{4-121c}$$

$$p_{3;0;\text{SCT};\xi\zeta}^{M_{s20}J_{s20}} = -(1/2)\left\langle \vec{b}_{\xi}^{M_{s20}}, \frac{1}{2}\vec{b}_{\zeta}^{J_{s20}} \times \hat{n}_{s2}^{-}\right\rangle_{\partial V_{s20}} \tag{4-121d}$$

Inserting transformation (4-101) into formulation (4-118), it can be obtained that

$$P_{\text{ss sys}}^{\text{driving}} = \left(\bar{a}^{M_{s0}}\right)^H \cdot \underbrace{\begin{bmatrix} \bar{\bar{T}}^{J_{s10}\leftarrow M_{s0}} \\ \bar{\bar{T}}^{J_{s20}\leftarrow M_{s0}} \\ \bar{\bar{\mathcal{I}}}^{M_{s10}} \\ \bar{\bar{\mathcal{I}}}^{M_{s20}} \end{bmatrix}^H \cdot \bar{\bar{P}}_{\text{ss sys;3}}^{\text{driving}} \cdot \begin{bmatrix} \bar{\bar{T}}^{J_{s10}\leftarrow M_{s0}} \\ \bar{\bar{T}}^{J_{s20}\leftarrow M_{s0}} \\ \bar{\bar{\mathcal{I}}}^{M_{s10}} \\ \bar{\bar{\mathcal{I}}}^{M_{s20}} \end{bmatrix}}_{\bar{\bar{P}}_{3;M_{s0}}^{\text{driving}}} \cdot \bar{a}^{M_{s0}} \tag{4-122}$$

This is just the matrix form of the surface formulation of DPO, which only contains BVs. In formulation (4-122), $\bar{\bar{\mathcal{I}}}^{M_{s10}}$ and $\bar{\bar{\mathcal{I}}}^{M_{s20}}$ are the same as the ones used in formulations (4-102) and (4-116). To emphasize the differences between above formulation (4-122) and previous formulations (4-102) and (4-116), we add subscript "3" to the $\bar{\bar{P}}_{3;M_{s0}}^{\text{driving}}$ in formulation (4-122) rather than the subscript "1" added to the $\bar{\bar{P}}_{1;M_{s0}}^{\text{driving}}$ in formulation (4-102) and the subscript "2" added to the $\bar{\bar{P}}_{2;M_{s0}}^{\text{driving}}$ in formulation (4-116). Obviously, new formulation (4-117) has a more concise manifestation form than formulation (4-113), and at the same time the order of the matrix $\overline{\overline{P}}{}^{\text{driving}}_{\text{ss sys;3}}$ in new formulation (4-122) is also smaller than the order of the matrix $\overline{\overline{P}}{}^{\text{driving}}_{\text{ss sys;2}}$ in new formulation (4-116), and then new formulation (4-117) realizes the further reduction for computational load.

The whole process to construct DP-CMs by orthogonalizing the above DPO is similar to the process given in previous Section 4.4, and the obtained DP-CMs also satisfy the same orthogonality and modal expansion as the ones given in previous Section 4.4, so they will not be repeated here.

### 4.6.2 A Special Case of the Result Obtained in Subsection 4.6.1

In this subsection, we consider a special case of the two-body system shown in Figure 4-27, and the special case is shown in Figure 4-56(b) where $V_{\text{sim}}^1$ and $V_{\text{sim}}^2$ have the same material parameters —— $\{\vec{\vec{\mu}}_{\text{sim}}^1 = \vec{\vec{\mu}}_{\text{sim}} = \vec{\vec{\mu}}_{\text{sim}}^2,\ \vec{\vec{\varepsilon}}_{\text{sim}}^1 = \vec{\vec{\varepsilon}}_{\text{sim}} = \vec{\vec{\varepsilon}}_{\text{sim}}^2,\ \vec{\vec{\sigma}}_{\text{sim}}^1 = \vec{\vec{\sigma}}_{\text{sim}} = \vec{\vec{\sigma}}_{\text{sim}}^2\}$.

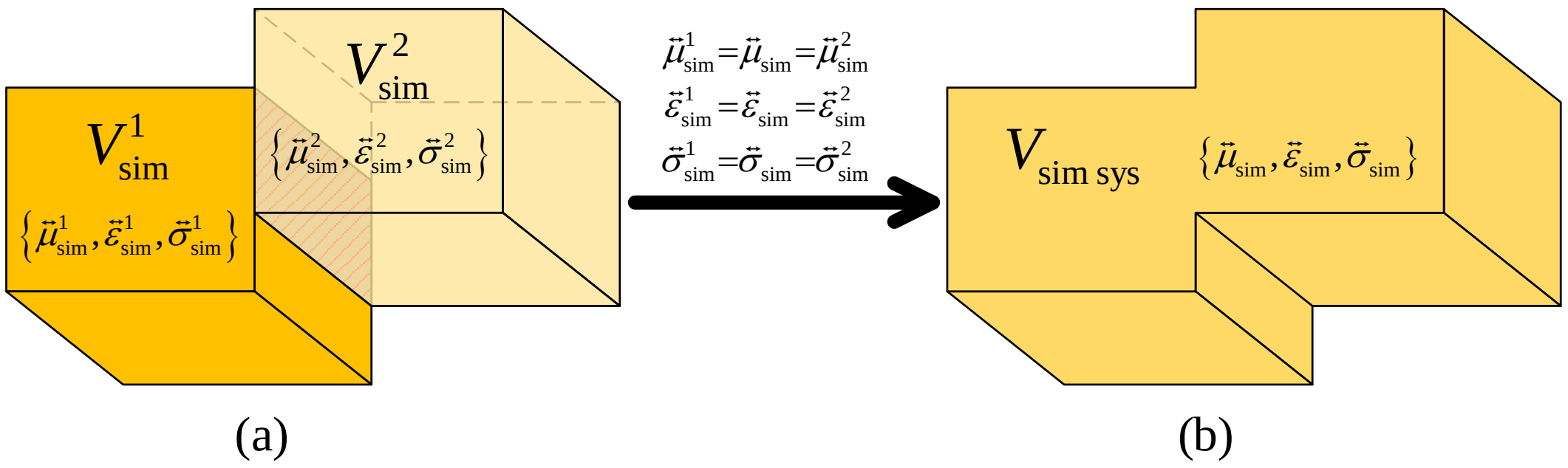

Figure 4-56 In the case that $\vec{\vec{\mu}}_{\text{sim}}^1 = \vec{\vec{\mu}}_{\text{sim}} = \vec{\vec{\mu}}_{\text{sim}}^2$ and $\vec{\vec{\varepsilon}}_{\text{sim}}^1 = \vec{\vec{\varepsilon}}_{\text{sim}} = \vec{\vec{\varepsilon}}_{\text{sim}}^2$ and $\vec{\vec{\sigma}}_{\text{sim}}^1 = \vec{\vec{\sigma}}_{\text{sim}} = \vec{\vec{\sigma}}_{\text{sim}}^2$, the two-body system becomes an one-body system. (a) two-body system; (b) one-body system

Obviously, the special case shown in Figure 4-56(b) is just the simply connected material body case discussed in Section 4.3. Then, if the results obtained in this section are correct, the DP-CMs of the system shown in Figure 4-56(b) derived from the method developed in this section should be the same as the ones derived from the method developed in Section 4.3. The destination of this subsection is to prove the above conclusion, and takes the conclusion as the verification for the validity of the surface formulation developed in this section. Our proof for the above conclusion will be done as the following three steps:

**Step 1.** In the case shown in Figure 4-56(b), we prove that DPO (4-117) degenerates into DPO (4-32);

**Step 2.** In the case shown in Figure 4-56(b), we prove that transformations (4-85)~(4-88) degenerate into transformation (4-37a);

**Step 3.** In the case shown in Figure 4-56(b), we, based on the results obtained in Steps 1 and 2, prove that the DPO $\overline{\overline{P}}_{3;M_{\text{s0}}}^{\text{driving}}$ in formulation (4-122) degenerates into the DPO $\overline{\overline{P}}_{M_{\text{s}}}^{\text{driving}}$ in formulation (4-52).

Because the two-body system $V_{\text{sim}}^1 \bigcup V_{\text{sim}}^2$ shown in Figure 4-56(b) is essentially equivalent to an one-body system, then we alternatively denote it as $V_{\text{sim sys}}$, i.e., $V_{\text{sim sys}} = V_{\text{sim}}^1 \bigcup V_{\text{sim}}^2$. Then, we can obtain the relationships that $\partial V_{\text{sim sys}} = \partial V_{\text{s10}} \bigcup \partial V_{\text{s20}}$ and $\partial V_{\text{sim sys}}^- = \partial V_{\text{s10}}^- \bigcup \partial V_{\text{s20}}^-$, where $\partial V_{\text{sim sys}}^-$, $\partial V_{\text{s10}}^-$, and $\partial V_{\text{s20}}^-$ are the inner surfaces of $\partial V_{\text{sim sys}}$, $\partial V_{\text{s10}}$, and $\partial V_{\text{s20}}$ respectively. Because $\partial V_{\text{sim sys}} = \partial V_{\text{s10}} \bigcup \partial V_{\text{s20}}$, then $\{\vec{J}_{\text{s10}}^{\text{ES}} + \vec{J}_{\text{s20}}^{\text{ES}}, \vec{M}_{\text{s10}}^{\text{ES}} + \vec{M}_{\text{s20}}^{\text{ES}}\}$ becomes the equivalent surface source defined on $\partial V_{\text{sim sys}}$. Thus, we alternatively denote $\{\vec{J}_{\text{s10}}^{\text{ES}} + \vec{J}_{\text{s20}}^{\text{ES}}, \vec{M}_{\text{s10}}^{\text{ES}} + \vec{M}_{\text{s20}}^{\text{ES}}\}$ as $\{\vec{J}_{\text{sim}}^{\text{ES}}, \vec{M}_{\text{sim}}^{\text{ES}}\}$, i.e., $\vec{J}_{\text{sim}}^{\text{ES}} = \vec{J}_{\text{s10}}^{\text{ES}} + \vec{J}_{\text{s20}}^{\text{ES}}$ and $\vec{M}_{\text{sim}}^{\text{ES}} = \vec{M}_{\text{s10}}^{\text{ES}} + \vec{M}_{\text{s20}}^{\text{ES}}$.

Inserting the above results into formulation (4-117), we can immediately rewrite the surface formulation of the DPO corresponding to the special two-body system shown in Figure 4-56(b) as follows:

$$P_{\text{ss sys}}^{\text{driving}} \xrightarrow{\text{Fig. 4-56(b)}} -\frac{1}{2}\left\langle \vec{J}_{\text{sim}}^{\text{ES}}, \mathcal{E}_0\left(\vec{J}_{\text{sim}}^{\text{ES}}, \vec{M}_{\text{sim}}^{\text{ES}}\right)\right\rangle_{\partial V_{\text{sim sys}}^-} - \frac{1}{2}\left\langle \vec{M}_{\text{sim}}^{\text{ES}}, \mathcal{H}_0\left(\vec{J}_{\text{sim}}^{\text{ES}}, \vec{M}_{\text{sim}}^{\text{ES}}\right)\right\rangle_{\partial V_{\text{sim sys}}^-} \tag{4-123}$$

Obviously, the above formulation is the same as formulation (4-32) completely.

For the system shown in Figure 4-56(b), the operators $\mathcal{F}_{\text{sim}}^1$ and $\mathcal{F}_{\text{sim}}^2$ used in formulations (4-85)~(4-86) will have the same functional forms, so we collectively denote them as $\mathcal{F}_{\text{sim}}$, i.e., $\mathcal{F}_{\text{sim}}^1 = \mathcal{F}_{\text{sim}} = \mathcal{F}_{\text{sim}}^2$; the inner normal directions $\hat{n}_{\text{s1}}^-$ and $\hat{n}_{\text{s2}}^-$ used in formulations (4-85) and (4-86) are the inner normal direction of $\partial V_{\text{sim sys}}$, so we collectively denote them as $\hat{n}_{\text{sim}}^-$, i.e., $\hat{n}_{\text{s1}}^- = \hat{n}_{\text{sim}}^- = \hat{n}_{\text{s2}}^-$; in formulations (4-85) and (4-86), both of the internal point $\vec{r}_{\text{sim}}^1$ of $V_{\text{sim}}^1$ and the internal point $\vec{r}_{\text{sim}}^2$ of $V_{\text{sim}}^2$ are the internal point of $V_{\text{sim sys}}$, so we collectively denote them as $\vec{r}_{\text{sim}}$, i.e., $\vec{r}_{\text{sim}}^1 = \vec{r}_{\text{sim}} = \vec{r}_{\text{sim}}^2$; equations (4-87) and (4-88) hold automatically, so there is no need to force them explicitly.

Based on the relationships mentioned above and relationships $\vec{C}_{\text{s10}}^{\text{ES}} + \vec{C}_{\text{s20}}^{\text{ES}} = \vec{C}_{\text{sim}}^{\text{ES}}$ and $\partial V_{\text{s10}} \bigcup \partial V_{\text{s20}} = \partial V_{\text{sim sys}}$, we can immediately simplify the summation of formulations (4-85) and (4-86) as follows:

$$\left[\mathcal{E}_{\text{sim}}\left(\vec{J}_{\text{sim}}^{\text{ES}}, \vec{M}_{\text{sim}}^{\text{ES}}\right)\right]_{\vec{r}_{\text{sim}} \to \vec{r}}^{\tan} = \hat{n}_{\text{sim}}^-(\vec{r}) \times \vec{M}_{\text{sim}}^{\text{ES}}(\vec{r}) \quad , \quad \vec{r} \in \partial V_{\text{sim sys}} \tag{4-124}$$

Obviously, the above equation of $\vec{J}_{\text{sim}}^{\text{ES}}$ and $\vec{M}_{\text{sim}}^{\text{ES}}$ is identical to equation (4-37a).

Based on formulations (4-123) and (4-124), it can be observed that the independent variable contained in the DPO of the special two-body system shown in Figure 4-56(b) is $\vec{J}_{\text{sim}}^{\text{ES}}$ or $\vec{M}_{\text{sim}}^{\text{ES}}$. In other words, if we discretize formulations (4-123) and (4-124) by using expansion formulation (4-33), we will obtain the power matrix being identical to formulation (4-34) and the transformation matrix being identical to formulation (4-42)/(4-45). Then, the $\overline{\overline{P}}_{M_{\text{s}0}}^{\text{driving}}$ corresponding to the special two-body system shown in Figure 4-56(b) will completely degenerate into the $\overline{\overline{P}}_{M_{\text{s}}}^{\text{driving}}$ in formulation (4-52). At this point, we have completed the proof for the conclusions given at the beginning of this section.

### 4.6.3 Numerical Examples Corresponding to Typical Structures

In Subsection 4.4.4, we realized the verification for the validity of the surface formulation developed in Section 4.4, by comparing the results derived from the surface formulation developed in Section 4.4 with the results derived from the volume formulation developed in Section 4.2. In Subsection 4.5.2, we used three different ways to verify the validity of the surface formulation given in Section 4.5: the first way is to compare the results derived from the surface formulation developed in Section 4.5 with the results derived from the volume formulation developed in Section 4.2; the second way is to compare the results derived from the surface formulation developed in Section 4.5 with the results derived from the surface formulation developed in Section 4.4; the third way is, by focusing on a special two-body material system, to compare the results derived from the surface formulation developed in Section 4.5 with the results derived from the surface formulation developed in Section 4.3.

Just like Subsection 4.5.2, we will, in this subsection, compare the results derived from the surface formulation developed in this section with the results derived from the surface formulation developed in Sections 4.2, 4.4, and 4.5, to verify the validity of the surface formulation established in this section. In addition, we will, by focusing on a special two-body system and employing the proof scheme developed in Subsection 4.6.2, compare the results derived from the surface formulation developed in this section with the results derived from the surface formulation developed in Section 4.3.

**1) Focusing on a Special Two-body Material System, Comparing the Results Derived From the Surface Formulation Developed in This Section with the Results Derived From the Surface Formulation Developed in Section 4.3**

Now, we still consider the two-body material system whose topological structure is shown in the Figure 4-30 given in Subsection 4.4.4. The lower material cylinder $V_{\rm sim}^{2}$ is still the same as the one considered in Subsection 4.4.4 —— relative permeability $\overleftrightarrow{\mu}_{\rm sim}^{\rm 2r}$, relative permittivity $\overleftrightarrow{\varepsilon}_{\rm sim}^{\rm 2r}$, and conductivity $\overleftrightarrow{\sigma}_{\rm sim}^{2}$ are $\overleftrightarrow{I}6$, $\overleftrightarrow{I}6$, and $\overleftrightarrow{I}0$ respectively. But, being different from the two-body material system considered in Subsection 4.4.4, this subsection sets the material parameters of the upper material cylinder as a series of different values, and calculates the DP-CMs corresponding to all cases by using the surface formulation developed in this section.

Figure 4-57(a) shows the characteristic value curves and MS curves of some typical DP-CMs corresponding to case $\{\overleftrightarrow{\mu}_{\rm sim}^{\rm 1r}=\overleftrightarrow{I}1,\overleftrightarrow{\varepsilon}_{\rm sim}^{\rm 1r}=\overleftrightarrow{I}36,\overleftrightarrow{\sigma}_{\rm sim}^{1}=\overleftrightarrow{I}0\}$; Figure 4-57(b) shows the characteristic value curves and MS curves of some typical DP-CMs corresponding to case $\{\overleftrightarrow{\mu}_{\rm sim}^{\rm 1r}=\overleftrightarrow{I}2,\overleftrightarrow{\varepsilon}_{\rm sim}^{\rm 1r}=\overleftrightarrow{I}18,\overleftrightarrow{\sigma}_{\rm sim}^{1}=\overleftrightarrow{I}0\}$; Figure 4-57(c) shows the characteristic value curves and MS curves of some typical DP-CMs corresponding to case $\{\overleftrightarrow{\mu}_{\rm sim}^{\rm 1r}=\overleftrightarrow{I}3,\overleftrightarrow{\varepsilon}_{\rm sim}^{\rm 1r}=\overleftrightarrow{I}12,\overleftrightarrow{\sigma}_{\rm sim}^{1}=\overleftrightarrow{I}0\}$; Figure 4-57(d) shows the characteristic value curves and MS curves of some typical DP-CMs corresponding to case $\{\overleftrightarrow{\mu}_{\rm sim}^{\rm 1r}=\overleftrightarrow{I}4,\overleftrightarrow{\varepsilon}_{\rm sim}^{\rm 1r}=\overleftrightarrow{I}9,\overleftrightarrow{\sigma}_{\rm sim}^{1}=\overleftrightarrow{I}0\}$; Figure 4-57(e) shows the characteristic value curves and MS curves of some typical DP-CMs corresponding to case $\{\overleftrightarrow{\mu}_{\rm sim}^{\rm 1r}=\overleftrightarrow{I}5,\overleftrightarrow{\varepsilon}_{\rm sim}^{\rm 1r}=\overleftrightarrow{I}7.2,\overleftrightarrow{\sigma}_{\rm sim}^{1}=\overleftrightarrow{I}0\}$; Figure 4-57(f) shows the characteristic value curves and MS curves of some typical DP-CMs corresponding to case $\{\overleftrightarrow{\mu}_{\rm sim}^{\rm 1r}=\overleftrightarrow{I}6,\overleftrightarrow{\varepsilon}_{\rm sim}^{\rm 1r}=\overleftrightarrow{I}6,\overleftrightarrow{\sigma}_{\rm sim}^{1}=\overleftrightarrow{I}0\}$. Obviously, the system shown in Figure 4-57(f) is essentially an one-body simply connected material system as shown in Figure 4-58(b). The formulations developed in Section 4.3 are applicable to the system shown in Figure 4-58(b). Based on the surface formulation developed in Section 4.3, we calculate the DP-CMs of the material system shown in Figure 4-58, and provide the characteristic value curves and MS curves of some typical DP-CMs in Figure 4-59.

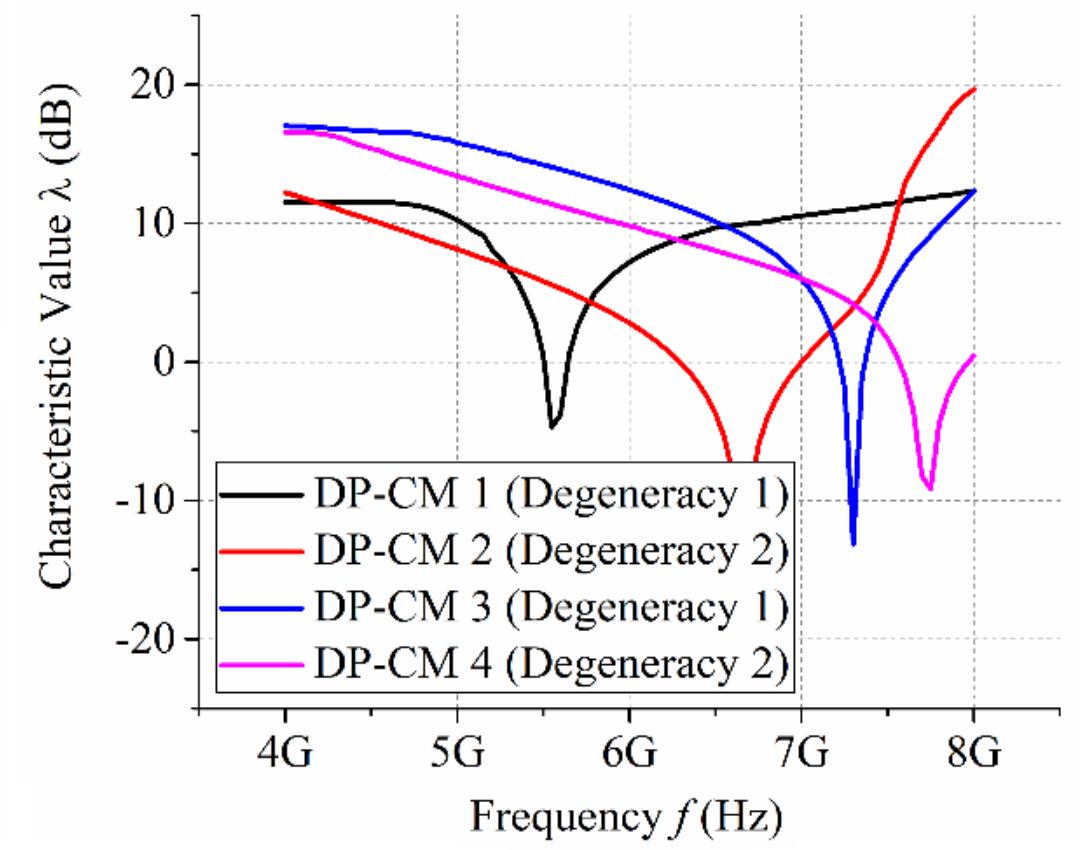

(a)

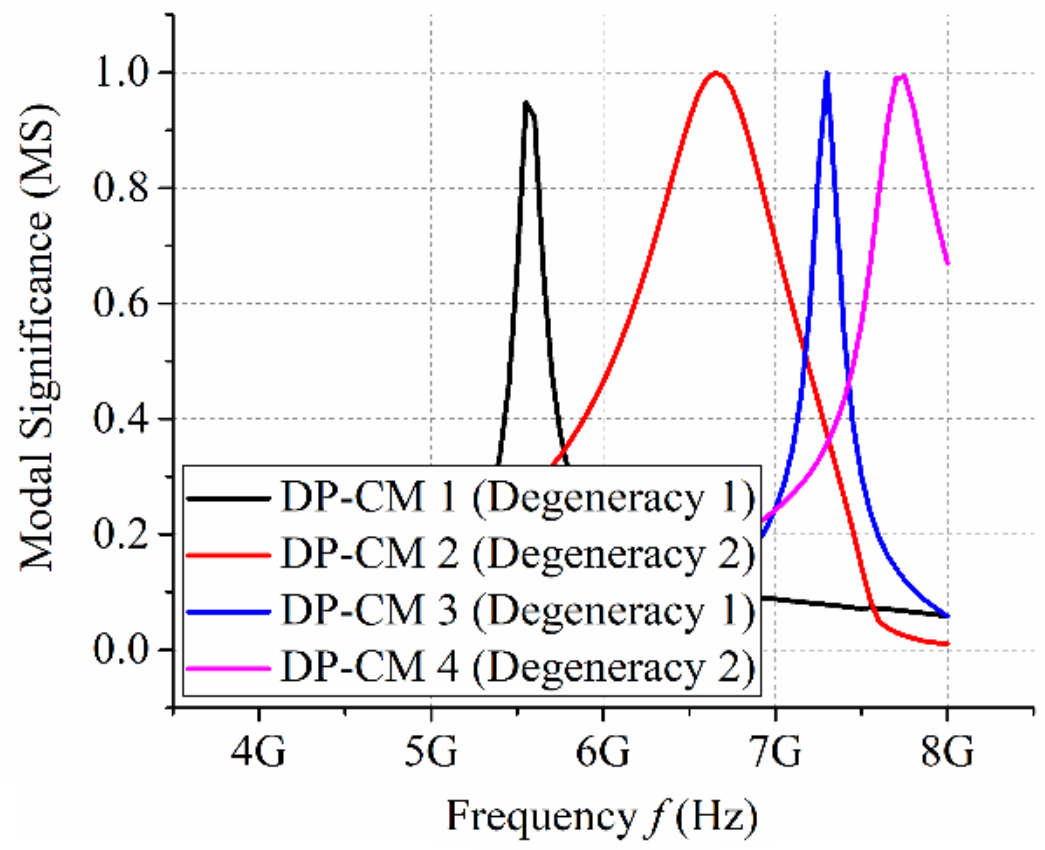

(b)

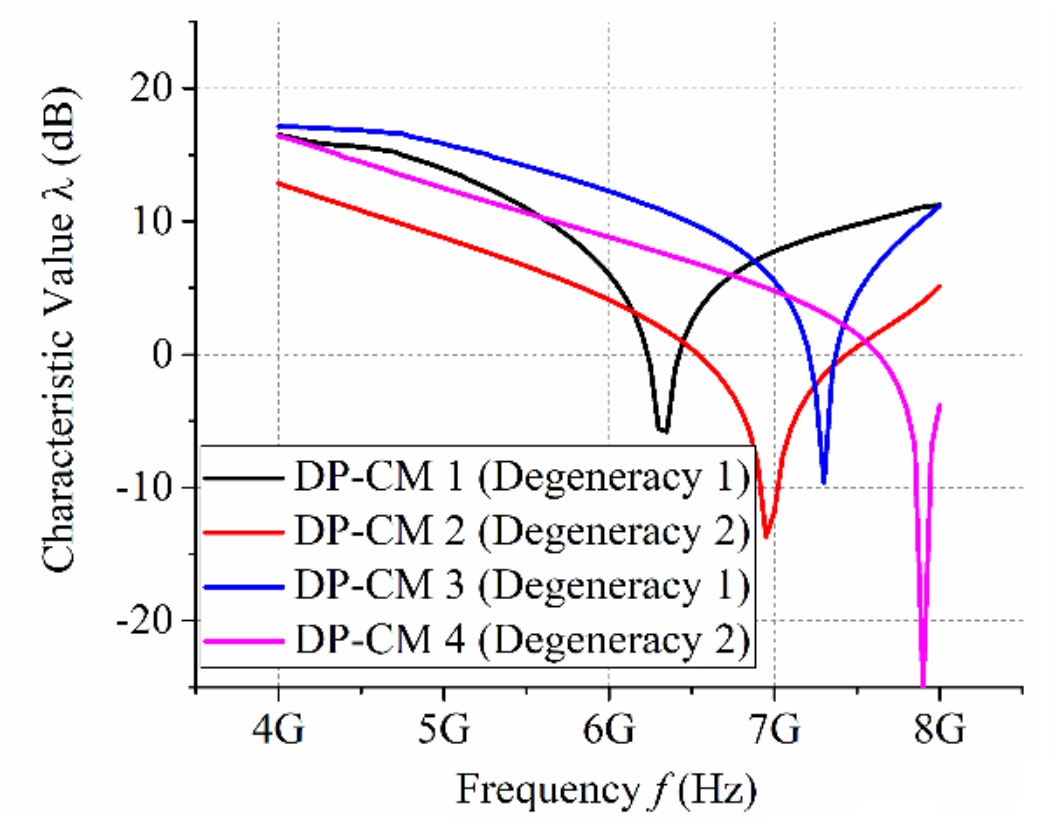
Characteristic Value λ (dB)
20
10
0
-10
-20
DP-CM 1 (Degeneracy 1)
DP-CM 2 (Degeneracy 2)
DP-CM 3 (Degeneracy 1)
DP-CM 4 (Degeneracy 2)
4G
5G
6G
7G
8G
Frequency *f* (Hz)

(c)

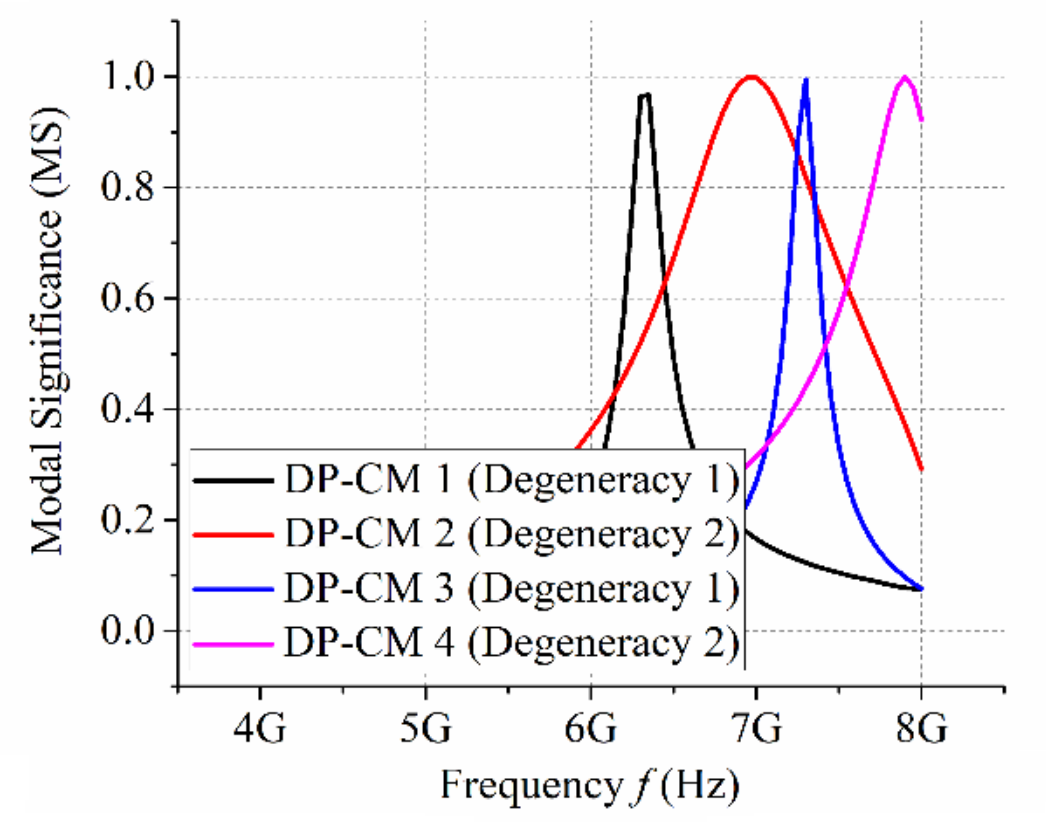
Modal Significance (MS)
1.0
0.8
0.6
0.4
0.2
0.0
DP-CM 1 (Degeneracy 1)
DP-CM 2 (Degeneracy 2)
DP-CM 3 (Degeneracy 1)
DP-CM 4 (Degeneracy 2)
4G
5G
6G
7G
8G
Frequency *f* (Hz)

(d)

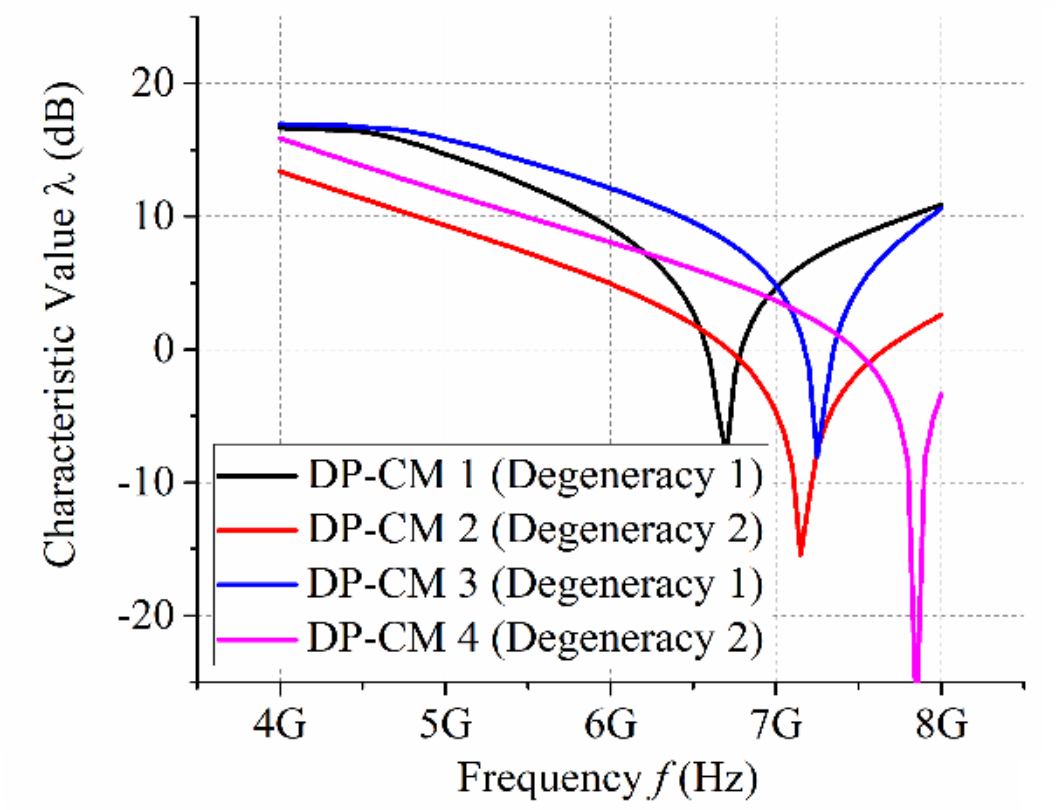
Characteristic Value λ (dB)
20
10
0
-10
-20
DP-CM 1 (Degeneracy 1)
DP-CM 2 (Degeneracy 2)
DP-CM 3 (Degeneracy 1)
DP-CM 4 (Degeneracy 2)
4G
5G
6G
7G
8G
Frequency *f* (Hz)

(e)

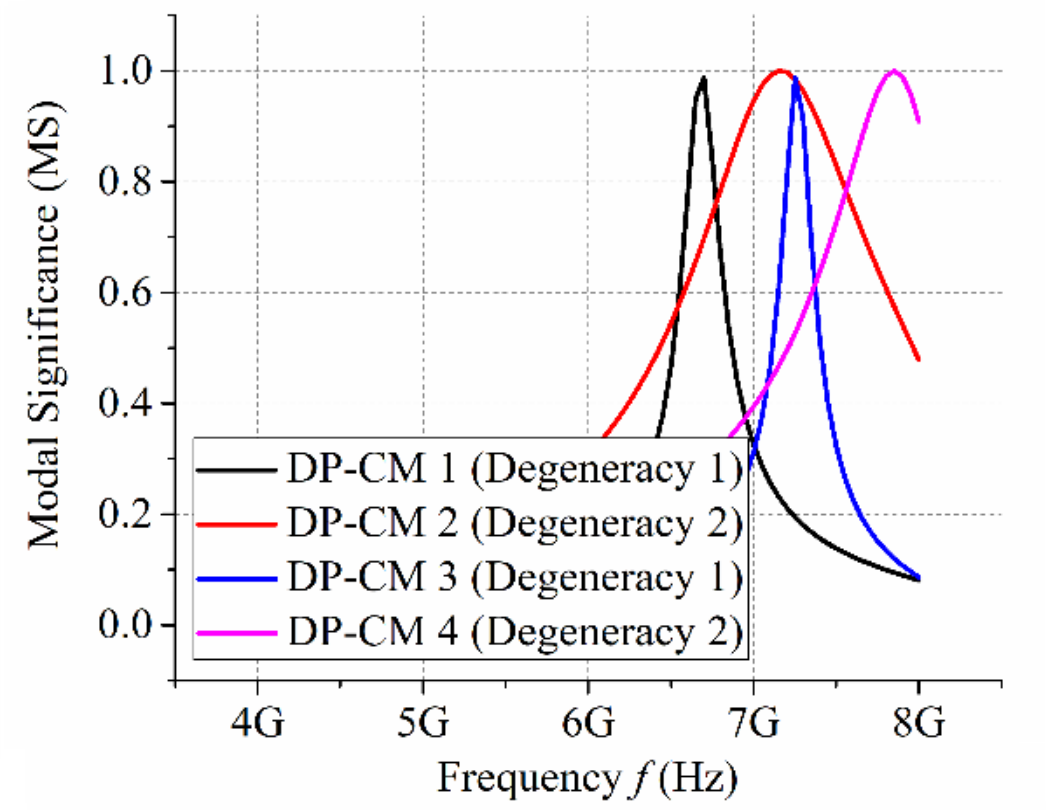
Modal Significance (MS)
1.0
0.8
0.6
0.4
0.2
0.0
DP-CM 1 (Degeneracy 1)
DP-CM 2 (Degeneracy 2)
DP-CM 3 (Degeneracy 1)
DP-CM 4 (Degeneracy 2)
4G
5G
6G
7G
8G
Frequency *f* (Hz)

(f)

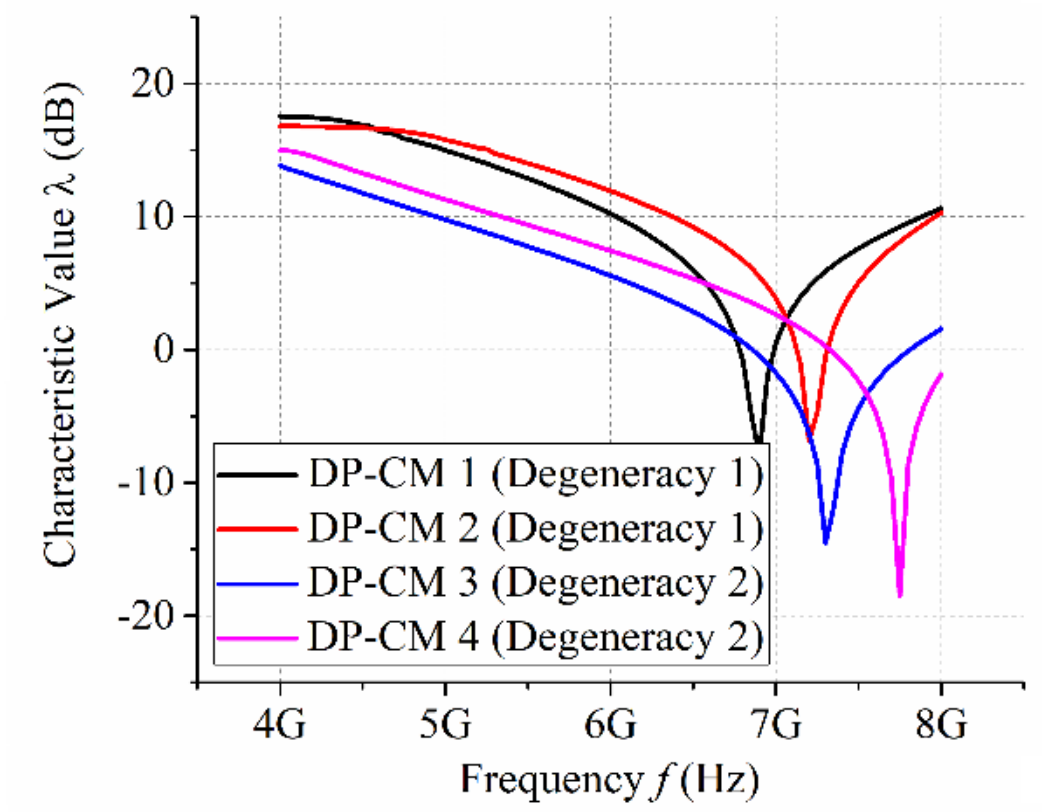
Characteristic Value λ (dB)
20
10
0
-10
-20
DP-CM 1 (Degeneracy 1)
DP-CM 2 (Degeneracy 1)
DP-CM 3 (Degeneracy 2)
DP-CM 4 (Degeneracy 2)
4G
5G
6G
7G
8G
Frequency *f* (Hz)

(g)

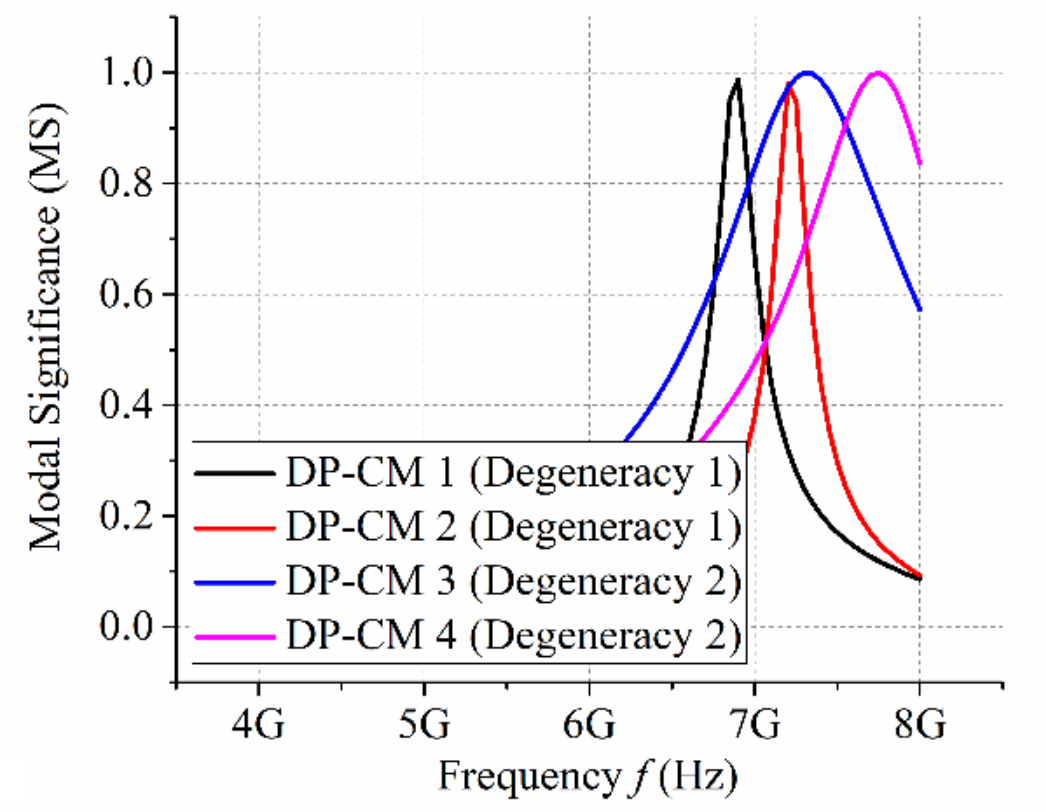
Modal Significance (MS)
1.0
0.8
0.6
0.4
0.2
0.0
DP-CM 1 (Degeneracy 1)
DP-CM 2 (Degeneracy 1)
DP-CM 3 (Degeneracy 2)
DP-CM 4 (Degeneracy 2)
4G
5G
6G
7G
8G
Frequency *f* (Hz)

(h)

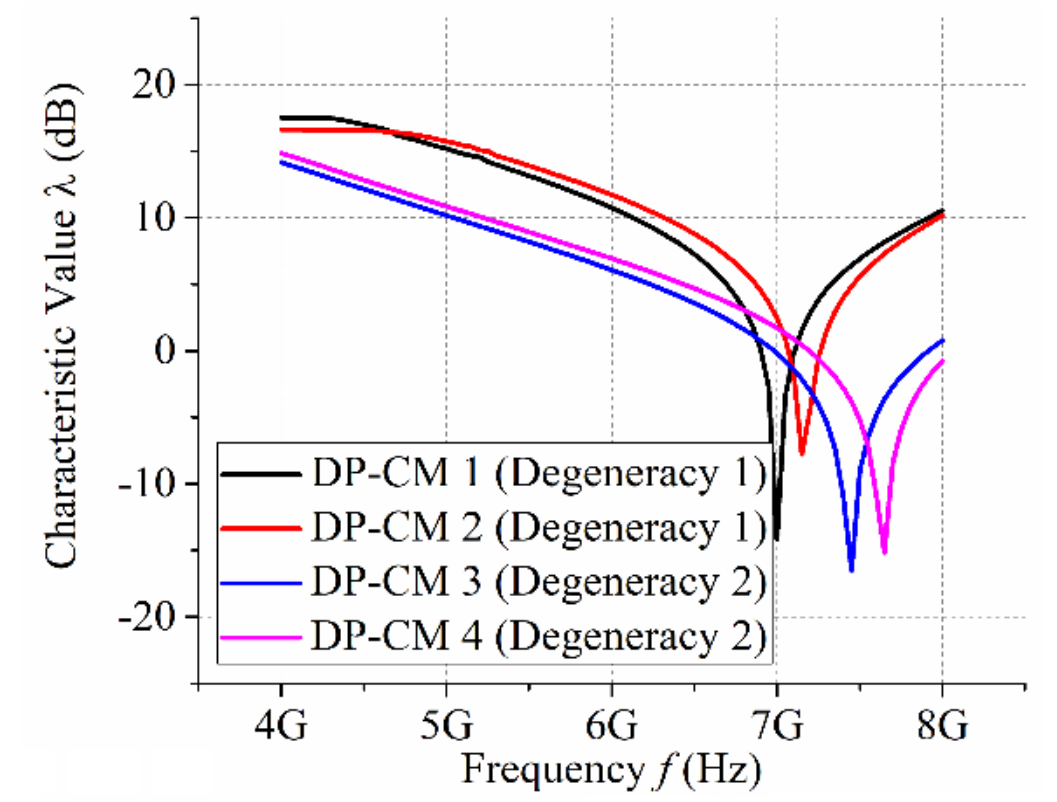
Characteristic Value λ (dB)
20
10
0
-10
-20
DP-CM 1 (Degeneracy 1)
DP-CM 2 (Degeneracy 1)
DP-CM 3 (Degeneracy 2)
DP-CM 4 (Degeneracy 2)
4G
5G
6G
7G
8G
Frequency *f* (Hz)

(i)

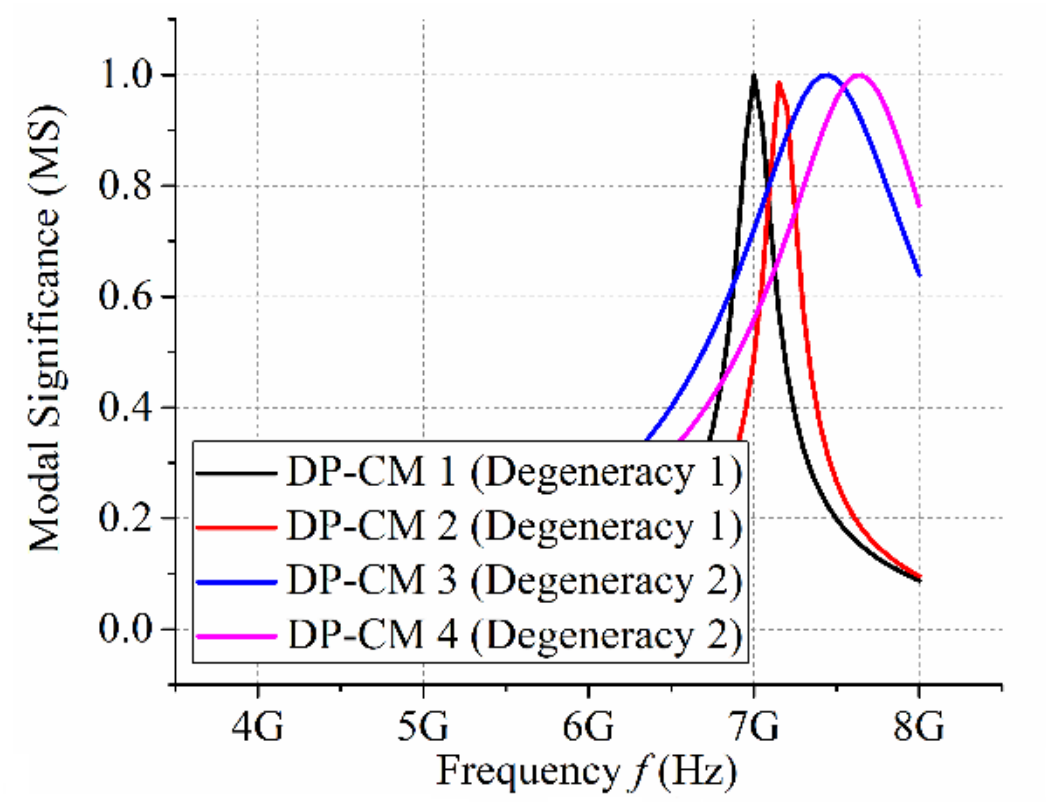
Modal Significance (MS)
1.0
0.8
0.6
0.4
0.2
0.0
DP-CM 1 (Degeneracy 1)
DP-CM 2 (Degeneracy 1)
DP-CM 3 (Degeneracy 2)
DP-CM 4 (Degeneracy 2)
4G
5G
6G
7G
8G
Frequency *f* (Hz)

(j)

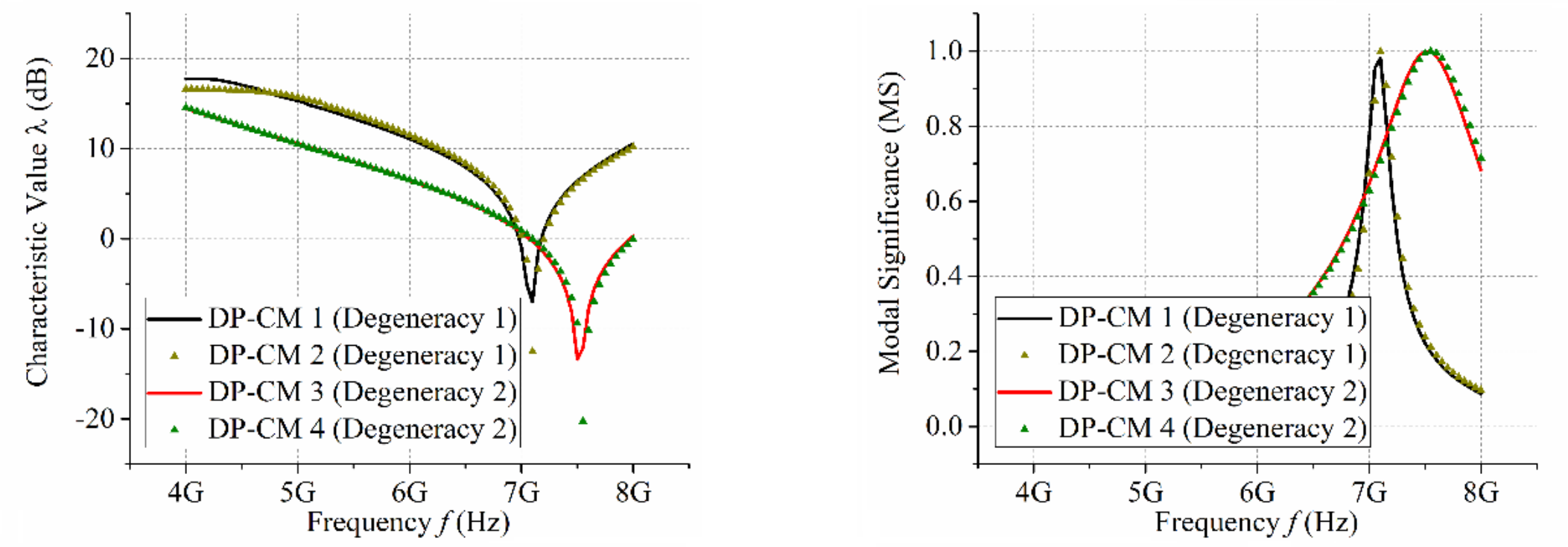

(k) (l)

Figure 4-57 The characteristic quantity curves derived from the formulations given in this section. (a) the characteristic value curves corresponding to $\overset{\leftrightarrow}{\mu}^{1\mathrm{r}}_{\mathrm{sim}}=1,\overset{\leftrightarrow}{\varepsilon}^{1\mathrm{r}}_{\mathrm{sim}}=36$; (b) the MS curves corresponding to $\overset{\leftrightarrow}{\mu}^{1\mathrm{r}}_{\mathrm{sim}}=1,\overset{\leftrightarrow}{\varepsilon}^{1\mathrm{r}}_{\mathrm{sim}}=36$; (c) the characteristic value curves corresponding to $\overset{\leftrightarrow}{\mu}^{1\mathrm{r}}_{\mathrm{sim}}=2,\overset{\leftrightarrow}{\varepsilon}^{1\mathrm{r}}_{\mathrm{sim}}=18$ ; (d) the MS curves corresponding to $\overset{\leftrightarrow}{\mu}^{1\mathrm{r}}_{\mathrm{sim}}=2,\overset{\leftrightarrow}{\varepsilon}^{1\mathrm{r}}_{\mathrm{sim}}=18$ ; (e) the characteristic value curves corresponding to $\overset{\leftrightarrow}{\mu}^{1\mathrm{r}}_{\mathrm{sim}}=3,\overset{\leftrightarrow}{\varepsilon}^{1\mathrm{r}}_{\mathrm{sim}}=12$ ; (f) the MS curves corresponding to $\overset{\leftrightarrow}{\mu}^{1\mathrm{r}}_{\mathrm{sim}}=3,\overset{\leftrightarrow}{\varepsilon}^{1\mathrm{r}}_{\mathrm{sim}}=12$ ; (g) the characteristic value curves corresponding to $\overset{\leftrightarrow}{\mu}^{1\mathrm{r}}_{\mathrm{sim}}=4,\overset{\leftrightarrow}{\varepsilon}^{1\mathrm{r}}_{\mathrm{sim}}=9$; (h) the MS curves corresponding to $\overset{\leftrightarrow}{\mu}^{1\mathrm{r}}_{\mathrm{sim}}=4,\overset{\leftrightarrow}{\varepsilon}^{1\mathrm{r}}_{\mathrm{sim}}=9$; (i) the characteristic value curves corresponding to $\overset{\leftrightarrow}{\mu}^{1\mathrm{r}}_{\mathrm{sim}}=5,\overset{\leftrightarrow}{\varepsilon}^{1\mathrm{r}}_{\mathrm{sim}}=7.2$; (j) the MS curves corresponding to $\overset{\leftrightarrow}{\mu}^{1\mathrm{r}}_{\mathrm{sim}}=5,\overset{\leftrightarrow}{\varepsilon}^{1\mathrm{r}}_{\mathrm{sim}}=7.2$; (k) the characteristic value curves corresponding to $\overset{\leftrightarrow}{\mu}^{1\mathrm{r}}_{\mathrm{sim}}=6,\overset{\leftrightarrow}{\varepsilon}^{1\mathrm{r}}_{\mathrm{sim}}=6$ ; (l) the MS curves corresponding to $\overset{\leftrightarrow}{\mu}^{1\mathrm{r}}_{\mathrm{sim}}=6,\overset{\leftrightarrow}{\varepsilon}^{1\mathrm{r}}_{\mathrm{sim}}=6$

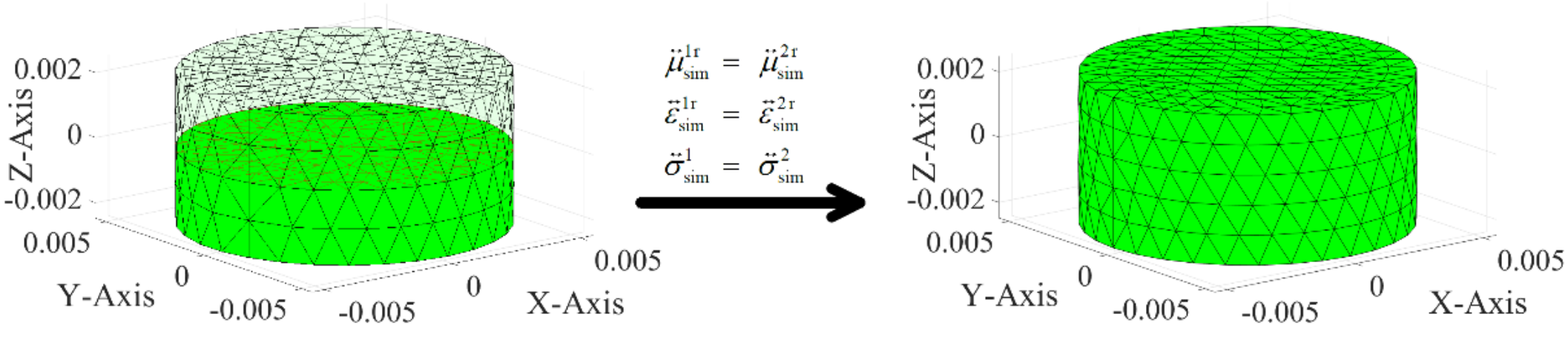

Figure 4-58 The transformation from two-body material system to one-body material system. (a) two-body material system; (b) one-body material system

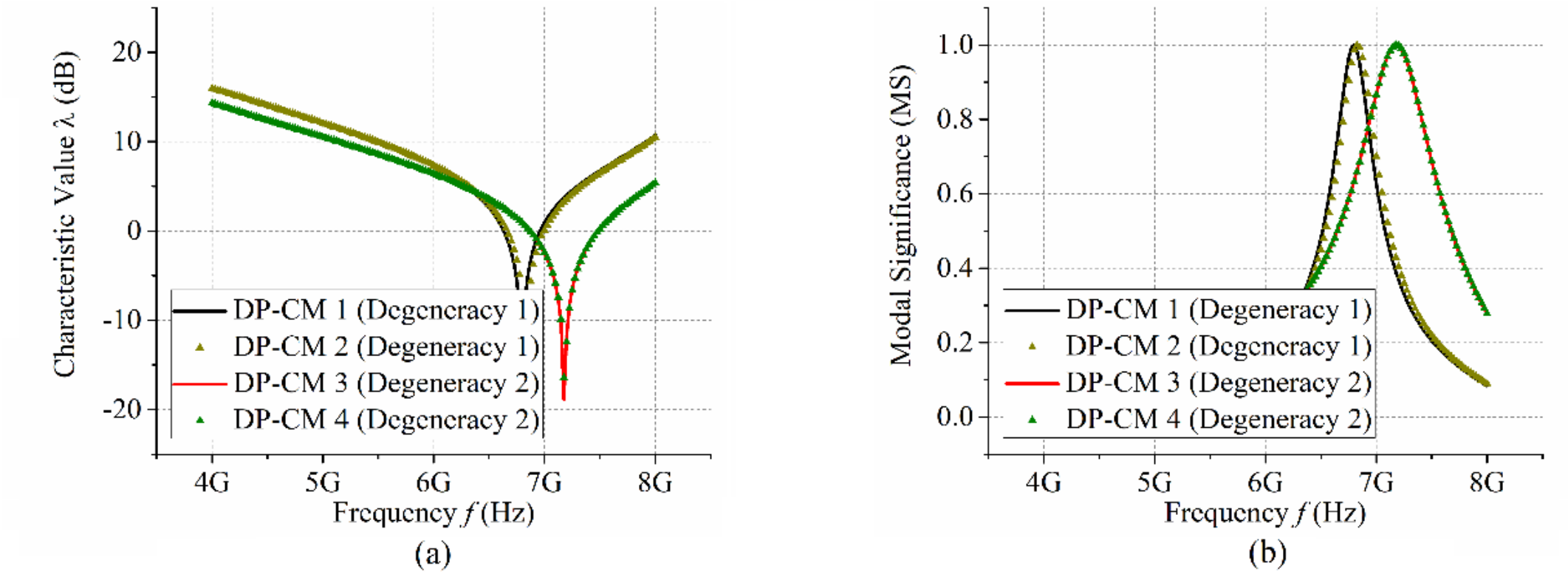

(a) (b)

Figure 4-59 The characteristic quantity curves corresponding to several typical DP-CMs (of the simply connected one-body material system shown in Figure 4-58(b)) derived from the formulations obtained in the Section 4.3 of this dissertation. (a) characteristic value dB curves; (b) MS curves

Observing above Figures 4-57 and 4-59, it is easy to find out that: when the upper material cylinder approaches the lower material cylinder, the result calculated from the formulation developed in this section approaches the result calculated from the formulation developed in Section 4.3. This phenomenon is just a powerful proof for the validity of the surface formulation developed in this section.

**2) Focusing on the Two-body Material System Considered in Subsection 4.4.4, Comparing the Results Derived From the Surface Formulation Developed in This Section with the Results Derived From the Formulations Developed in Sections 4.2, 4.4, and 4.5**

In Subsection 4.5.2, we, based on the surface formulation developed in Section 4.5, calculated the DP-CMs of the two-body material system whose topological structure was shown in Figure 4-30 and material parameters were $\{\vec{\mu}_{\text{sim}}^{1\text{r}} = \vec{I}3, \vec{\varepsilon}_{\text{sim}}^{1\text{r}} = \vec{I}12, \vec{\sigma}_{\text{sim}}^{1} = \vec{I}0\}$ & $\{\vec{\mu}_{\text{sim}}^{2\text{r}} = \vec{I}6, \vec{\varepsilon}_{\text{sim}}^{2\text{r}} = \vec{I}6, \vec{\sigma}_{\text{sim}}^{2} = \vec{I}0\}$, and also provided the corresponding characteristic value curves and MS curves in Figure 4-55, and compared the results with the results derived from the volume formulation developed in Section 4.2 (as shown in Figure 4-34) and the surface formulation developed in Section 4.4 (as shown in Figure 4-33), and then realized the verification for the validity of the surface formulation developed in Section 4.5. Now, we, based on the surface formulation developed in this section, calculate the DP-CMs of the two-body material system, and provide the corresponding characteristic value curves and MS curves in Figure 4-60.

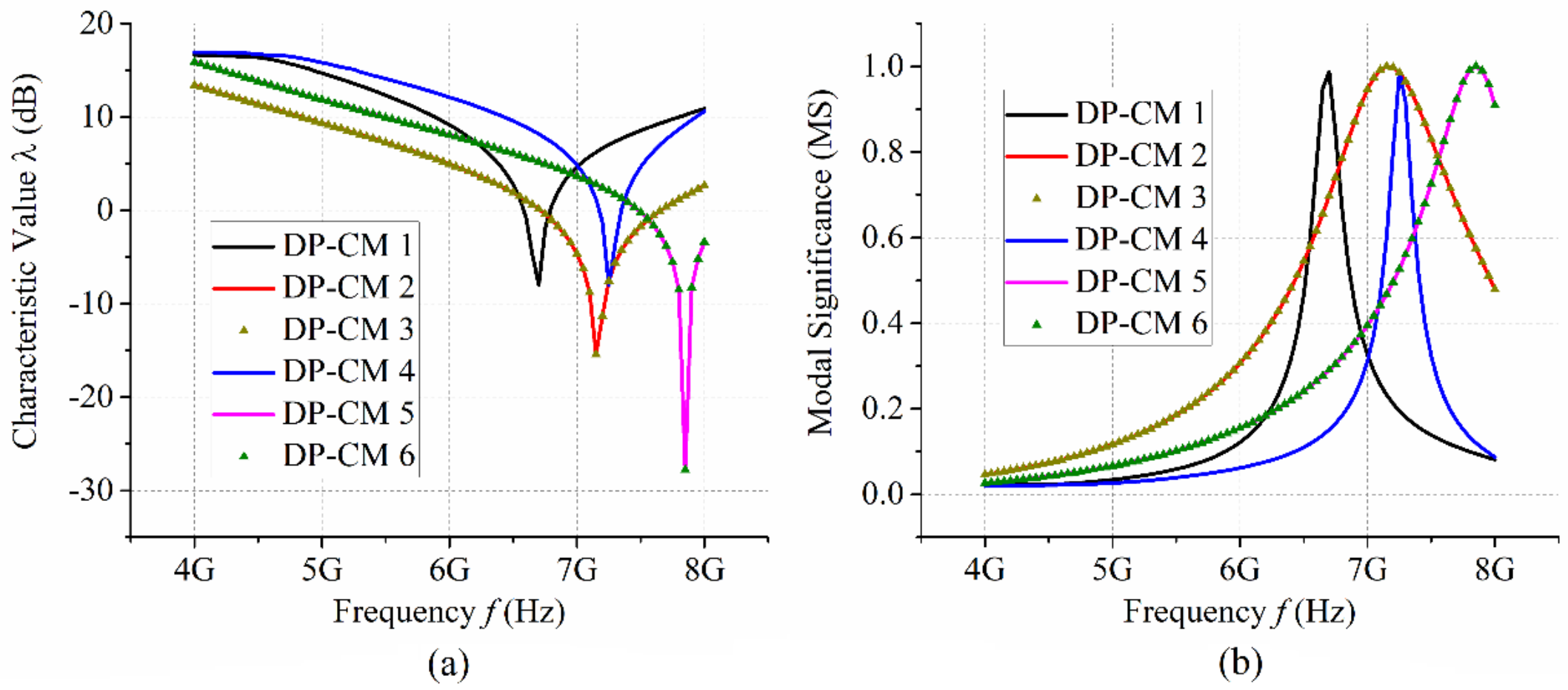

Figure 4-60 For the two-body system shown in Figure 4-30 ( $\vec{\mu}_{\text{sim}}^{1\text{r}} = \vec{I}3$ , $\vec{\varepsilon}_{\text{sim}}^{1\text{r}} = \vec{I}12$ , $\vec{\sigma}_{\text{sim}}^{1} = \vec{I}0$ ; $\vec{\mu}_{\text{sim}}^{2\text{r}} = \vec{I}6$ , $\vec{\varepsilon}_{\text{sim}}^{2\text{r}} = \vec{I}6$ , $\vec{\sigma}_{\text{sim}}^{2} = \vec{I}0$ ), the characteristic quantity curves derived from the surface formulations given in this section. (a) characteristic value dB curves; (b) MS curves

Obviously, the results shown in above Figure 4-60 agree well with the results shown in Figures 4-33, 4-34, and 4-55, and this implies the validity of the surface formulation developed in this section in the aspect of constructing the DP-CMs of the two-body material system shown in Figure 4-27.

## 4.7 Three Surface Formulations for Calculating the DP-CMs of the Systems Constructed by a Simply Connected Material Body and a Multiply Connected Material Body

For the completeness of this chapter, this section provides three different surface formulations to the two-body material systems constructed by a simply connected material body and a multiply connected material body. The various equivalent surface currents are defined as the ones used in Appendix C7, and they will not be repeated here.

The two-body material system considered in this section is shown in Figure 4-61, where $V_{\text{sim}}$ is a simply connected material body and $V_{\text{mul}}$ is a multiply connected material body. Whole two-body system is denoted as $V_{\text{sm sys}}$, i.e., $V_{\text{sm sys}} = V_{\text{sim}} \cup V_{\text{mul}}$.

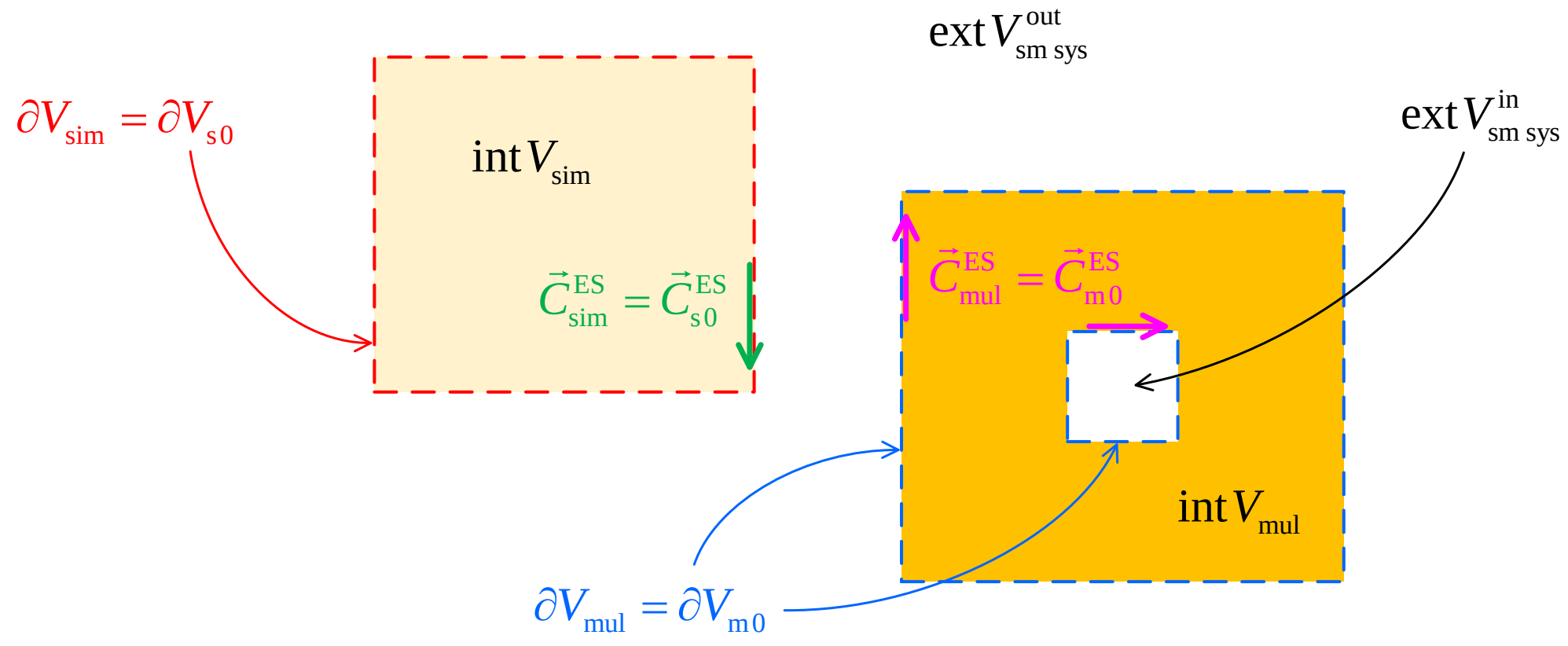

(a)

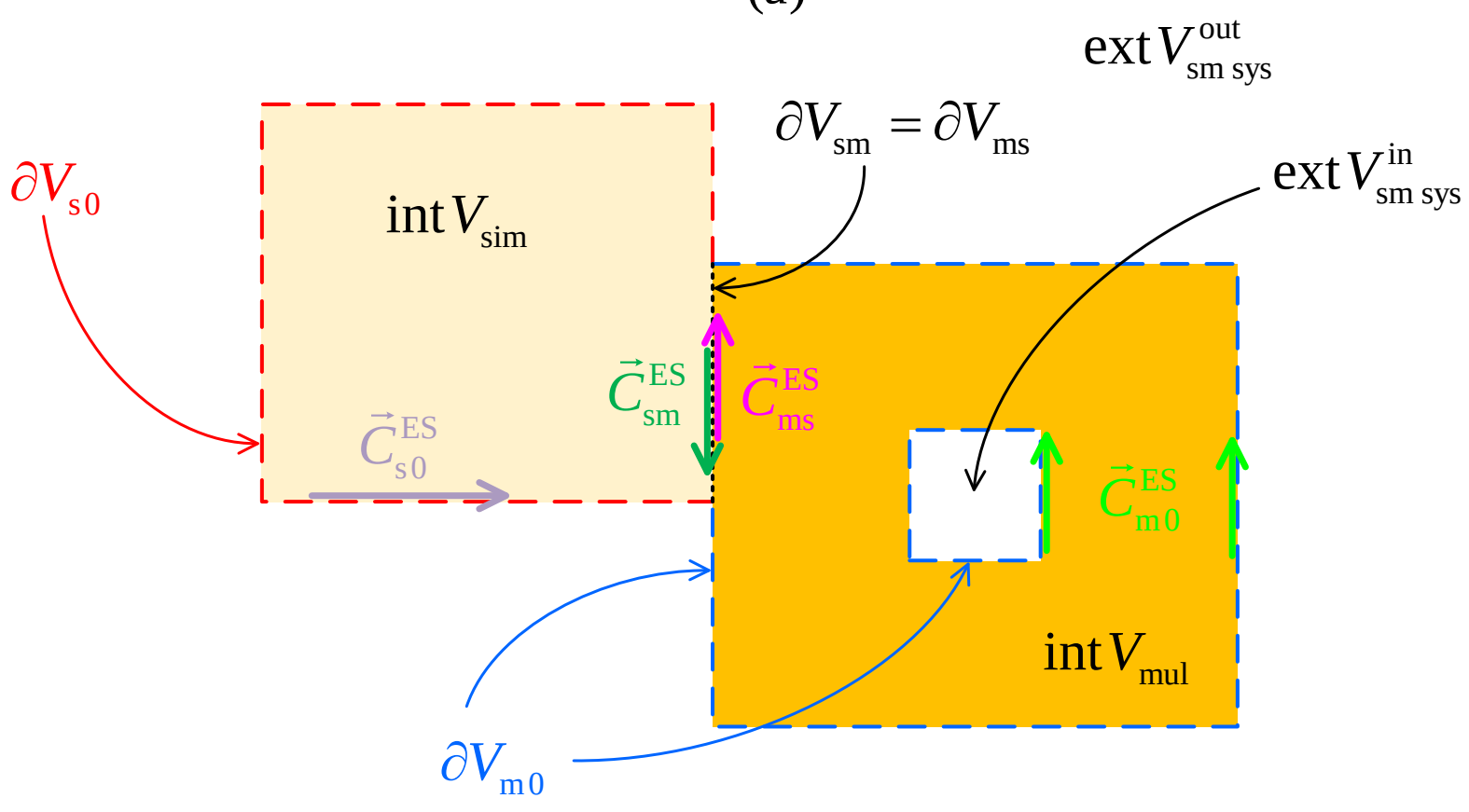

(b)

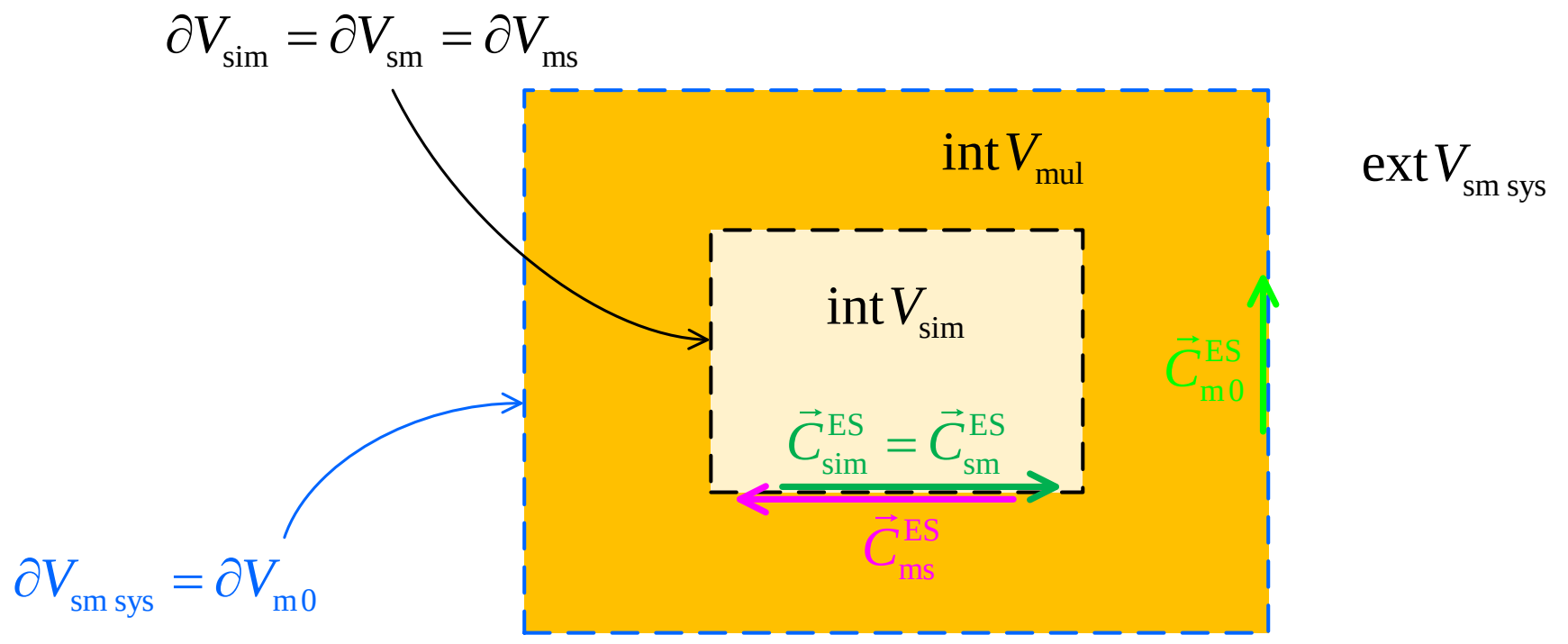

(c)

Figure 4-61 The topological structure of the two-body material system $V_{\text{sm sys}}$ constituted by a simply connected material body $V_{\text{sim}}$ and a multiply connected material body $V_{\text{mul}}$. (a) the case that $V_{\text{sim}}$ and $V_{\text{mul}}$ don't contact with each other; (b) the case that $V_{\text{sim}}$ and $V_{\text{mul}}$ contact with each other, but $V_{\text{sim}}$ is not submerged into $V_{\text{mul}}$; (c) the case that $V_{\text{sim}}$ and $V_{\text{mul}}$ contact with each other, and $V_{\text{sim}}$ is submerged into $V_{\text{mul}}$

## 4.7.1 Variable Unification

Similarly to the Section 4.4 of this dissertation, we expand the equivalent surface sub-currents distributing on various sub-boundaries in terms of the corresponding basis functions as follows:

$$\vec{C}_{\text{s}0}^{\text{ES}}(\vec{r}) = \sum_{\xi=1}^{\Xi^{C_{\text{s}0}}} a_{\xi}^{C_{\text{s}0}} \vec{b}_{\xi}^{C_{\text{s}0}}(\vec{r}) = \bar{\boldsymbol{B}}^{C_{\text{s}0}} \cdot \bar{a}^{C_{\text{s}0}} \quad , \quad \vec{r} \in \partial V_{\text{s}0} \tag{4-125a}$$

$$\vec{C}_{\text{sm}}^{\text{ES}}(\vec{r}) = \sum_{\xi=1}^{\Xi^{C_{\text{sm}}}} a_{\xi}^{C_{\text{sm}}} \vec{b}_{\xi}^{C_{\text{sm}}}(\vec{r}) = \bar{\boldsymbol{B}}^{C_{\text{sm}}} \cdot \bar{a}^{C_{\text{sm}}} \quad , \quad \vec{r} \in \partial V_{\text{sm}} \tag{4-125b}$$

$$\begin{aligned}\vec{C}_{\text{ms}}^{\text{ES}}(\vec{r}) = \sum_{\xi=1}^{\Xi^{C_{\text{sm}}}} a_{\xi}^{C_{\text{ms}}} \vec{b}_{\xi}^{C_{\text{sm}}}(\vec{r}) &= \bar{\boldsymbol{B}}^{C_{\text{sm}}} \cdot \bar{a}^{C_{\text{ms}}} \\ &= -\bar{\boldsymbol{B}}^{C_{\text{sm}}} \cdot \bar{a}^{C_{\text{sm}}} \quad , \quad \vec{r} \in \partial V_{\text{ms}} = \partial V_{\text{sm}}\end{aligned} \tag{4-126a}$$

$$\vec{C}_{\text{m}0}^{\text{ES}}(\vec{r}) = \sum_{\xi=1}^{\Xi^{C_{\text{m}0}}} a_{\xi}^{C_{\text{m}0}} \vec{b}_{\xi}^{C_{\text{m}0}}(\vec{r}) = \bar{\boldsymbol{B}}^{C_{\text{m}0}} \cdot \bar{a}^{C_{\text{m}0}} \quad , \quad \vec{r} \in \partial V_{\text{m}0} \tag{4-126b}$$

In the above formulations, $\bar{\boldsymbol{B}}^{C_{\text{s}0/\text{sm}/\text{m}0}} = [\vec{b}_1^{C_{\text{s}0/\text{sm}/\text{m}0}}, \vec{b}_2^{C_{\text{s}0/\text{sm}/\text{m}0}}, \cdots, \vec{b}_{\Xi^{C_{\text{s}0/\text{sm}/\text{m}0}}}^{C_{\text{s}0/\text{sm}/\text{m}0}}]$ and $\bar{a}^{C_{\text{s}0/\text{sm}/\text{ms}/\text{m}0}} = [a_1^{C_{\text{s}0/\text{sm}/\text{ms}/\text{m}0}}, a_2^{C_{\text{s}0/\text{sm}/\text{ms}/\text{m}0}}, \cdots, a_{\Xi^{C_{\text{s}0/\text{sm}/\text{sm}/\text{m}0}}}^{C_{\text{s}0/\text{sm}/\text{ms}/\text{m}0}}]^T$ ; $\bar{a}^{C_{\text{ms}}} = -\bar{a}^{C_{\text{sm}}}$ . In fact, above expansions (4-125) and (4-126) realize establishing the one-to-one mapping between equivalent currents and expansion vectors.

Similarly to the formulations (4-85)~(4-88) for the two-body system constructed by two simply connected sub-systems, the equivalent sub-currents distributing on the boundaries of the two sub-systems considered in this section satisfy the following integral equations:

$$\left[\mathcal{E}_{\text{sim}}\left(\vec{J}_{\text{s0}}^{\text{ES}}+\vec{J}_{\text{sm}}^{\text{ES}},\vec{M}_{\text{s0}}^{\text{ES}}+\vec{M}_{\text{sm}}^{\text{ES}}\right)\right]_{\vec{r}_{\text{sim}}\to\vec{r}}^{\tan} = \hat{n}_{\text{sim}}^{-}\left(\vec{r}\right)\times\vec{M}_{\text{s0}}^{\text{ES}}\left(\vec{r}\right) \quad , \quad \vec{r}\in\partial V_{\text{s0}} \tag{4-127}$$

$$\left[\mathcal{E}_{\text{mul}}\left(\vec{J}_{\text{m0}}^{\text{ES}}-\vec{J}_{\text{sm}}^{\text{ES}},\vec{M}_{\text{m0}}^{\text{ES}}-\vec{M}_{\text{sm}}^{\text{ES}}\right)\right]_{\vec{r}_{\text{mul}}\to\vec{r}}^{\tan} = \hat{n}_{\text{mul}}^{-}\left(\vec{r}\right)\times\vec{M}_{\text{m0}}^{\text{ES}}\left(\vec{r}\right) \quad , \quad \vec{r}\in\partial V_{\text{m0}} \tag{4-128}$$

$$\left[\mathcal{E}_{\text{sim}}\left(\vec{J}_{\text{s0}}^{\text{ES}}+\vec{J}_{\text{sm}}^{\text{ES}},\vec{M}_{\text{s0}}^{\text{ES}}+\vec{M}_{\text{sm}}^{\text{ES}}\right)\right]_{\vec{r}_{\text{sim}}\to\vec{r}}^{\tan} = \left[\mathcal{E}_{\text{mul}}\left(\vec{J}_{\text{m0}}^{\text{ES}}-\vec{J}_{\text{sm}}^{\text{ES}},\vec{M}_{\text{m0}}^{\text{ES}}-\vec{M}_{\text{sm}}^{\text{ES}}\right)\right]_{\vec{r}_{\text{mul}}\to\vec{r}}^{\tan} \quad , \quad \vec{r}\in\partial V_{\text{sm}} \tag{4-129}$$

$$\left[\mathcal{H}_{\text{sim}}\left(\vec{J}_{\text{s0}}^{\text{ES}}+\vec{J}_{\text{sm}}^{\text{ES}},\vec{M}_{\text{s0}}^{\text{ES}}+\vec{M}_{\text{sm}}^{\text{ES}}\right)\right]_{\vec{r}_{\text{sim}}\to\vec{r}}^{\tan} = \left[\mathcal{H}_{\text{mul}}\left(\vec{J}_{\text{m0}}^{\text{ES}}-\vec{J}_{\text{sm}}^{\text{ES}},\vec{M}_{\text{m0}}^{\text{ES}}-\vec{M}_{\text{sm}}^{\text{ES}}\right)\right]_{\vec{r}_{\text{mul}}\to\vec{r}}^{\tan} \quad , \quad \vec{r}\in\partial V_{\text{sm}} \tag{4-130}$$

In formulations (4-127), (4-129), and (4-130), $\vec{r}_{\text{sim}}\in\text{int}V_{\text{sim}}$, and $\vec{r}_{\text{sim}}$ approaches $\vec{r}$. In formulations (4-128)~(4-130), $\vec{r}_{\text{mul}}\in\text{int}V_{\text{mul}}$, and $\vec{r}_{\text{mul}}$ approaches $\vec{r}$. In addition, we have applied relationship (C-57) to formulations (4-128)~(4-130).

Inserting expansion formulations (4-125) and (4-126) into formulations (4-127)~(4-130), and testing formulations (4-127), (4-128), (4-129), and (4-130) with basis functions $\{\vec{b}_{\xi}^{J_{\text{s0}}}\}_{\xi=1}^{\Xi^{J_{\text{s0}}}}$, $\{\vec{b}_{\xi}^{J_{\text{m0}}}\}_{\xi=1}^{\Xi^{J_{\text{m0}}}}$, $\{\vec{b}_{\xi}^{J_{\text{sm}}}\}_{\xi=1}^{\Xi^{J_{\text{sm}}}}$, and $\{\vec{b}_{\xi}^{M_{\text{sm}}}\}_{\xi=1}^{\Xi^{M_{\text{sm}}}}$ respectively, formulations (4-127)~(4-130) can be discretized into the following matrix forms:

$$\begin{aligned}&\overline{\overline{Z}}_1^{J_{\text{s0}}EJ_{\text{s0}}}\cdot\overline{a}^{J_{\text{s0}}}+\overline{\overline{Z}}_1^{J_{\text{s0}}EJ_{\text{sm}}}\cdot\overline{a}^{J_{\text{sm}}}+\overline{\overline{Z}}_1^{J_{\text{s0}}EM_{\text{s0}}}\cdot\overline{a}^{M_{\text{s0}}}+\overline{\overline{Z}}_1^{J_{\text{s0}}EM_{\text{sm}}}\cdot\overline{a}^{M_{\text{sm}}}\\ =\ &\overline{\overline{C}}^{J_{\text{s0}}M_{\text{s0}}}\cdot\overline{a}^{M_{\text{s0}}}\end{aligned} \tag{4-131}$$

$$\begin{aligned}&\overline{\overline{Z}}_2^{J_{\text{m0}}EJ_{\text{m0}}}\cdot\overline{a}^{J_{\text{m0}}}-\overline{\overline{Z}}_2^{J_{\text{m0}}EJ_{\text{sm}}}\cdot\overline{a}^{J_{\text{sm}}}+\overline{\overline{Z}}_2^{J_{\text{m0}}EM_{\text{m0}}}\cdot\overline{a}^{M_{\text{m0}}}-\overline{\overline{Z}}_2^{J_{\text{m0}}EM_{\text{sm}}}\cdot\overline{a}^{M_{\text{sm}}}\\ =\ &\overline{\overline{C}}^{J_{\text{m0}}M_{\text{m0}}}\cdot\overline{a}^{M_{\text{m0}}}\end{aligned} \tag{4-132}$$

$$\begin{aligned}&\overline{\overline{Z}}_1^{J_{\text{sm}}EJ_{\text{s0}}}\cdot\overline{a}^{J_{\text{s0}}}+\overline{\overline{Z}}_1^{J_{\text{sm}}EJ_{\text{sm}}}\cdot\overline{a}^{J_{\text{sm}}}+\overline{\overline{Z}}_1^{J_{\text{sm}}EM_{\text{s0}}}\cdot\overline{a}^{M_{\text{s0}}}+\overline{\overline{Z}}_1^{J_{\text{sm}}EM_{\text{sm}}}\cdot\overline{a}^{M_{\text{sm}}}\\ =\ &\overline{\overline{Z}}_2^{J_{\text{sm}}EJ_{\text{m0}}}\cdot\overline{a}^{J_{\text{m0}}}-\overline{\overline{Z}}_2^{J_{\text{sm}}EJ_{\text{sm}}}\cdot\overline{a}^{J_{\text{sm}}}+\overline{\overline{Z}}_2^{J_{\text{sm}}EM_{\text{m0}}}\cdot\overline{a}^{M_{\text{m0}}}-\overline{\overline{Z}}_2^{J_{\text{sm}}EM_{\text{sm}}}\cdot\overline{a}^{M_{\text{sm}}}\end{aligned} \tag{4-133}$$

$$\begin{aligned}&\overline{\overline{Z}}_1^{M_{\text{sm}}HJ_{\text{s0}}}\cdot\overline{a}^{J_{\text{s0}}}+\overline{\overline{Z}}_1^{M_{\text{sm}}HJ_{\text{sm}}}\cdot\overline{a}^{J_{\text{sm}}}+\overline{\overline{Z}}_1^{M_{\text{sm}}HM_{\text{s0}}}\cdot\overline{a}^{M_{\text{s0}}}+\overline{\overline{Z}}_1^{M_{\text{sm}}HM_{\text{sm}}}\cdot\overline{a}^{M_{\text{sm}}}\\ =\ &\overline{\overline{Z}}_2^{M_{\text{sm}}HJ_{\text{m0}}}\cdot\overline{a}^{J_{\text{m0}}}-\overline{\overline{Z}}_2^{M_{\text{sm}}HJ_{\text{sm}}}\cdot\overline{a}^{J_{\text{sm}}}+\overline{\overline{Z}}_2^{M_{\text{sm}}HM_{\text{m0}}}\cdot\overline{a}^{M_{\text{m0}}}-\overline{\overline{Z}}_2^{M_{\text{sm}}HM_{\text{sm}}}\cdot\overline{a}^{M_{\text{sm}}}\end{aligned} \tag{4-134}$$

The elements of the matrices in above equations (4-131)~(4-134) can be similarly calculated as the ones in equations (4-89)~(4-92) (only need to properly replace the superscripts and subscripts), so the calculation formulations will not be repeated here. In equations (4-131)~(4-134), there exist 6 different equivalent EM currents $\overline{a}^{J_{\text{s0}}}$, $\overline{a}^{J_{\text{sm}}}$, $\overline{a}^{J_{\text{m0}}}$, $\overline{a}^{M_{\text{s0}}}$, $\overline{a}^{M_{\text{sm}}}$, and $\overline{a}^{M_{\text{m0}}}$. We select $\overline{a}^{M_{\text{s0}}}$ and $\overline{a}^{M_{\text{m0}}}$ as BVs, and then the other four currents become dependent variables naturally. To effectively establish the transformation from the BVs to the dependent variables, we properly re-arrange and assemble matrix equations (4-131)~(4-134) as the following single augmented matrix equation:

$$\begin{bmatrix} \overline{\overline{Z}}_{\text{sim}}^{J_{s0}EJ_{s0}} & 0 & \overline{\overline{Z}}_{\text{sim}}^{J_{s0}EJ_{sm}} & \overline{\overline{Z}}_{\text{sim}}^{J_{s0}EM_{sm}} \\ 0 & -\overline{\overline{Z}}_{\text{mul}}^{J_{m0}EJ_{m0}} & \overline{\overline{Z}}_{\text{mul}}^{J_{m0}EJ_{sm}} & \overline{\overline{Z}}_{\text{mul}}^{J_{m0}EM_{sm}} \\ \overline{\overline{Z}}_{\text{sim}}^{J_{sm}EJ_{s0}} & -\overline{\overline{Z}}_{\text{mul}}^{J_{sm}EJ_{m0}} & \overline{\overline{Z}}_{\text{sim}}^{J_{sm}EJ_{sm}}+\overline{\overline{Z}}_{\text{mul}}^{J_{sm}EJ_{sm}} & \overline{\overline{Z}}_{\text{sim}}^{J_{sm}EM_{sm}}+\overline{\overline{Z}}_{\text{mul}}^{J_{sm}EM_{sm}} \\ \overline{\overline{Z}}_{\text{sim}}^{M_{sm}HJ_{s0}} & -\overline{\overline{Z}}_{\text{mul}}^{M_{sm}HJ_{m0}} & \overline{\overline{Z}}_{\text{sim}}^{M_{sm}HJ_{sm}}+\overline{\overline{Z}}_{\text{mul}}^{M_{sm}HJ_{sm}} & \overline{\overline{Z}}_{\text{sim}}^{M_{sm}HM_{sm}}+\overline{\overline{Z}}_{\text{mul}}^{M_{sm}HM_{sm}} \end{bmatrix} \cdot \begin{bmatrix} \overline{a}^{J_{s0}} \\ \overline{a}^{J_{m0}} \\ \overline{a}^{J_{sm}} \\ \overline{a}^{M_{sm}} \end{bmatrix}$$

$$= \begin{bmatrix} \overline{\overline{C}}^{J_{s0}M_{s0}} - \overline{\overline{Z}}_{\text{sim}}^{J_{s0}EM_{s0}} & 0 \\ 0 & \overline{\overline{Z}}_{\text{mul}}^{J_{m0}EM_{m0}} - \overline{\overline{C}}^{J_{m0}M_{m0}} \\ -\overline{\overline{Z}}_{\text{sim}}^{J_{sm}EM_{s0}} & \overline{\overline{Z}}_{\text{mul}}^{J_{sm}EM_{m0}} \\ -\overline{\overline{Z}}_{\text{sim}}^{M_{sm}HM_{s0}} & \overline{\overline{Z}}_{\text{mul}}^{M_{sm}HM_{m0}} \end{bmatrix} \cdot \begin{bmatrix} \overline{a}^{M_{s0}} \\ \overline{a}^{M_{m0}} \end{bmatrix} \tag{4-135}$$

Obviously, the above equation implies the following transformation:

$$\begin{bmatrix} \overline{a}^{J_{s0}} \\ \overline{a}^{J_{m0}} \\ \overline{a}^{J_{sm}} \\ \overline{a}^{M_{sm}} \end{bmatrix} = \overline{\overline{T}}^{\{J_{s0},J_{m0},J_{sm},M_{sm}\}\leftarrow M_0} \cdot \underbrace{\begin{bmatrix} \overline{a}^{M_{s0}} \\ \overline{a}^{M_{m0}} \end{bmatrix}}_{\overline{a}^{M_0}} \tag{4-136}$$

where

$$\overline{\overline{T}}^{\{J_{s0},J_{m0},J_{sm},M_{sm}\}\leftarrow M_0} = \begin{bmatrix} \overline{\overline{Z}}_{\text{sim}}^{J_{s0}EJ_{s0}} & 0 & \overline{\overline{Z}}_{\text{sim}}^{J_{s0}EJ_{sm}} & \overline{\overline{Z}}_{\text{sim}}^{J_{s0}EM_{sm}} \\ 0 & -\overline{\overline{Z}}_{\text{mul}}^{J_{m0}EJ_{m0}} & \overline{\overline{Z}}_{\text{mul}}^{J_{m0}EJ_{sm}} & \overline{\overline{Z}}_{\text{mul}}^{J_{m0}EM_{sm}} \\ \overline{\overline{Z}}_{\text{sim}}^{J_{sm}EJ_{s0}} & -\overline{\overline{Z}}_{\text{mul}}^{J_{sm}EJ_{m0}} & \overline{\overline{Z}}_{\text{sim}}^{J_{sm}EJ_{sm}}+\overline{\overline{Z}}_{\text{mul}}^{J_{sm}EJ_{sm}} & \overline{\overline{Z}}_{\text{sim}}^{J_{sm}EM_{sm}}+\overline{\overline{Z}}_{\text{mul}}^{J_{sm}EM_{sm}} \\ \overline{\overline{Z}}_{\text{sim}}^{M_{sm}HJ_{s0}} & -\overline{\overline{Z}}_{\text{mul}}^{M_{sm}HJ_{m0}} & \overline{\overline{Z}}_{\text{sim}}^{M_{sm}HJ_{sm}}+\overline{\overline{Z}}_{\text{mul}}^{M_{sm}HJ_{sm}} & \overline{\overline{Z}}_{\text{sim}}^{M_{sm}HM_{sm}}+\overline{\overline{Z}}_{\text{mul}}^{M_{sm}HM_{sm}} \end{bmatrix}^{-1} \cdot \begin{bmatrix} \overline{\overline{C}}^{J_{s0}M_{s0}} - \overline{\overline{Z}}_{\text{sim}}^{J_{s0}EM_{s0}} & 0 \\ 0 & \overline{\overline{Z}}_{\text{mul}}^{J_{m0}EM_{m0}} - \overline{\overline{C}}^{J_{m0}M_{m0}} \\ -\overline{\overline{Z}}_{\text{sim}}^{J_{sm}EM_{s0}} & \overline{\overline{Z}}_{\text{mul}}^{J_{sm}EM_{m0}} \\ -\overline{\overline{Z}}_{\text{sim}}^{M_{sm}HM_{s0}} & \overline{\overline{Z}}_{\text{mul}}^{M_{sm}HM_{m0}} \end{bmatrix} \tag{4-137}$$

If we partition above transformation matrix $\overline{\overline{T}}^{\{J_{s0},J_{m0},J_{sm},M_{sm}\}\leftarrow M_0}$ as the partition way of the vector on the LHS of formulation (4-136), i.e.,

$$\overline{\overline{T}}^{\{J_{s0},J_{m0},J_{sm},M_{sm}\}\leftarrow M_0} = \begin{bmatrix} \overline{\overline{T}}^{J_{s0}\leftarrow M_0} \\ \overline{\overline{T}}^{J_{m0}\leftarrow M_0} \\ \overline{\overline{T}}^{J_{sm}\leftarrow M_0} \\ \overline{\overline{T}}^{M_{sm}\leftarrow M_0} \end{bmatrix} \tag{4-138}$$

then we have a series of transformations as follows:

$$\overline{a}^{J_{s0}} = \overline{\overline{T}}^{J_{s0}\leftarrow M_0} \cdot \overline{a}^{M_0} \tag{4-139a}$$

$$\overline{a}^{J_{m0}} = \overline{\overline{T}}^{J_{m0}\leftarrow M_0} \cdot \overline{a}^{M_0} \tag{4-139b}$$

$$\bar{a}^{J_{\text{sm}}} = \bar{\bar{T}}^{J_{\text{sm}} \leftarrow M_0} \cdot \bar{a}^{M_0} \tag{4-139c}$$

$$\bar{a}^{M_{\text{sm}}} = \bar{\bar{T}}^{M_{\text{sm}} \leftarrow M_0} \cdot \bar{a}^{M_0} \tag{4-139d}$$

If relationship (C-57) is utilized, then we further have the following transformations:

$$\bar{a}^{J_{\text{ms}}} = -\bar{\bar{T}}^{J_{\text{sm}} \leftarrow M_0} \cdot \bar{a}^{M_0} \tag{4-139e}$$

$$\bar{a}^{M_{\text{ms}}} = -\bar{\bar{T}}^{M_{\text{sm}} \leftarrow M_0} \cdot \bar{a}^{M_0} \tag{4-139f}$$

## 4.7.2 Three Surface Formulations of DPO

Similarly to the Subsections 4.4.1, 4.5.1, and 4.6.1 of this dissertation, this subsection will provide three different surface formulations to the system $V_{\text{sm sys}}$ shown in Figure 4-61.

### 1) The First Surface Formulation

Similarly to formulation (4-66), we provide the first surface formulation of the DPO corresponding to the system $V_{\text{sm sys}}$ shown in Figure 4-61 as follows:

$$\begin{aligned}
& P_{\text{sm sys}}^{\text{driving}} \\
= \; & P_{\text{sim}}^{\text{driving}} + P_{\text{mul}}^{\text{driving}} \\
= \; & \frac{1}{2}\left\langle \vec{J}_{\text{sim}}^{\text{SV}}, \vec{E}^{\text{inc}} \right\rangle_{V_{\text{sim}}} + \frac{1}{2}\left\langle \vec{M}_{\text{sim}}^{\text{SV}}, \vec{H}^{\text{inc}} \right\rangle_{V_{\text{sim}}} + \frac{1}{2}\left\langle \vec{J}_{\text{mul}}^{\text{SV}}, \vec{E}^{\text{inc}} \right\rangle_{V_{\text{mul}}} + \frac{1}{2}\left\langle \vec{M}_{\text{mul}}^{\text{SV}}, \vec{H}^{\text{inc}} \right\rangle_{V_{\text{mul}}} \\
= & -\frac{1}{2}\left\langle \vec{J}_{\text{sim}}^{\text{ES}}, \vec{E}^{\text{inc}} \right\rangle_{\partial V_{\text{sim}}} - \frac{1}{2}\left\langle \vec{M}_{\text{sim}}^{\text{ES}}, \vec{H}^{\text{inc}} \right\rangle_{\partial V_{\text{sim}}} - \frac{1}{2}\left\langle \vec{J}_{\text{mul}}^{\text{ES}}, \vec{E}^{\text{inc}} \right\rangle_{\partial V_{\text{mul}}} - \frac{1}{2}\left\langle \vec{M}_{\text{mul}}^{\text{ES}}, \vec{H}^{\text{inc}} \right\rangle_{\partial V_{\text{mul}}} \\
= & -\frac{1}{2}\left[ \left\langle \vec{J}_{\text{sim}}^{\text{ES}}, \mathcal{E}_0\left(\vec{J}_{\text{sim}}^{\text{ES}}, \vec{M}_{\text{sim}}^{\text{ES}}\right) \right\rangle_{\partial V_{\text{sim}}^{+}} + \left\langle \vec{J}_{\text{sim}}^{\text{ES}}, \mathcal{E}_0\left(\vec{J}_{\text{mul}}^{\text{ES}}, \vec{M}_{\text{mul}}^{\text{ES}}\right) \right\rangle_{\partial V_{\text{sim}}^{-}} + \left\langle \vec{J}_{\text{sim}}^{\text{ES}}, \mathcal{E}_{\text{sim}}\left(\vec{J}_{\text{sim}}^{\text{ES}}, \vec{M}_{\text{sim}}^{\text{ES}}\right) \right\rangle_{\partial V_{\text{sim}}^{-}} \right] \\
& -\frac{1}{2}\left[ \left\langle \vec{J}_{\text{mul}}^{\text{ES}}, \mathcal{E}_0\left(\vec{J}_{\text{mul}}^{\text{ES}}, \vec{M}_{\text{mul}}^{\text{ES}}\right) \right\rangle_{\partial V_{\text{mul}}^{+}} + \left\langle \vec{J}_{\text{mul}}^{\text{ES}}, \mathcal{E}_0\left(\vec{J}_{\text{sim}}^{\text{ES}}, \vec{M}_{\text{sim}}^{\text{ES}}\right) \right\rangle_{\partial V_{\text{mul}}^{-}} + \left\langle \vec{J}_{\text{mul}}^{\text{ES}}, \mathcal{E}_{\text{mul}}\left(\vec{J}_{\text{mul}}^{\text{ES}}, \vec{M}_{\text{mul}}^{\text{ES}}\right) \right\rangle_{\partial V_{\text{mul}}^{-}} \right] \\
& -\frac{1}{2}\left[ \left\langle \vec{M}_{\text{sim}}^{\text{ES}}, \mathcal{H}_0\left(\vec{J}_{\text{sim}}^{\text{ES}}, \vec{M}_{\text{sim}}^{\text{ES}}\right) \right\rangle_{\partial V_{\text{sim}}^{+}} + \left\langle \vec{M}_{\text{sim}}^{\text{ES}}, \mathcal{H}_0\left(\vec{J}_{\text{mul}}^{\text{ES}}, \vec{M}_{\text{mul}}^{\text{ES}}\right) \right\rangle_{\partial V_{\text{sim}}^{-}} + \left\langle \vec{M}_{\text{sim}}^{\text{ES}}, \mathcal{H}_{\text{sim}}\left(\vec{J}_{\text{sim}}^{\text{ES}}, \vec{M}_{\text{sim}}^{\text{ES}}\right) \right\rangle_{\partial V_{\text{sim}}^{-}} \right] \\
& -\frac{1}{2}\left[ \left\langle \vec{M}_{\text{mul}}^{\text{ES}}, \mathcal{H}_0\left(\vec{J}_{\text{mul}}^{\text{ES}}, \vec{M}_{\text{mul}}^{\text{ES}}\right) \right\rangle_{\partial V_{\text{mul}}^{+}} + \left\langle \vec{M}_{\text{mul}}^{\text{ES}}, \mathcal{H}_0\left(\vec{J}_{\text{sim}}^{\text{ES}}, \vec{M}_{\text{sim}}^{\text{ES}}\right) \right\rangle_{\partial V_{\text{mul}}^{-}} + \left\langle \vec{M}_{\text{mul}}^{\text{ES}}, \mathcal{H}_{\text{mul}}\left(\vec{J}_{\text{mul}}^{\text{ES}}, \vec{M}_{\text{mul}}^{\text{ES}}\right) \right\rangle_{\partial V_{\text{mul}}^{-}} \right]
\end{aligned} \tag{4-140}$$

Inserting expansion formulations (4-125)&(4-126) into the above formulation, we obtain that

$$P_{\text{sm sys}}^{\text{driving}} = \begin{bmatrix} \bar{a}^{J_{\text{s}0}} \\ \bar{a}^{J_{\text{sm}}} \\ \bar{a}^{J_{\text{ms}}} \\ \bar{a}^{J_{\text{m}0}} \\ \bar{a}^{M_{\text{s}0}} \\ \bar{a}^{M_{\text{sm}}} \\ \bar{a}^{M_{\text{ms}}} \\ \bar{a}^{M_{\text{m}0}} \end{bmatrix}^H \cdot \underbrace{\left( \bar{\bar{P}}_{1;0;\text{PVT}}^{\text{sm sys}} + \bar{\bar{P}}_{1;0;\text{SCT}}^{\text{sm sys}} + \bar{\bar{P}}_{1;\text{m}}^{\text{sm sys}} \right)}_{\bar{\bar{P}}_{\text{sm sys};1}^{\text{driving}}} \cdot \begin{bmatrix} \bar{a}^{J_{\text{s}0}} \\ \bar{a}^{J_{\text{sm}}} \\ \bar{a}^{J_{\text{ms}}} \\ \bar{a}^{J_{\text{m}0}} \\ \bar{a}^{M_{\text{s}0}} \\ \bar{a}^{M_{\text{sm}}} \\ \bar{a}^{M_{\text{ms}}} \\ \bar{a}^{M_{\text{m}0}} \end{bmatrix} \tag{4-141}$$

The elements of the power matrices $\overline{\overline{P}}{}_{1;0;\text{PVT}}^{\text{sm sys}}$, $\overline{\overline{P}}{}_{1;0;\text{SCT}}^{\text{sm sys}}$, and $\overline{\overline{P}}{}_{1;\text{m}}^{\text{sm sys}}$ in the above formulation can be calculated as the the elements of the power matrices $\overline{\overline{P}}{}_{1;0;\text{PVT}}^{\text{ss sys}}$, $\overline{\overline{P}}{}_{1;0;\text{SCT}}^{\text{ss sys}}$, and $\overline{\overline{P}}{}_{1;\text{m}}^{\text{ss sys}}$ used in formulation (4-80) (only need to properly replace the superscripts and subscripts), so the corresponding calculation formulations will not be explicitly provided here.

Inserting transformation (4-139) into formulation (4-141), it is immediately obtained that

$$P_{\text{sm sys}}^{\text{driving}} = \left(\bar{a}^{M_0}\right)^H \cdot \underbrace{\begin{bmatrix} \overline{\overline{T}}{}^{J_{\text{s}0} \leftarrow M_0} \\ \overline{\overline{T}}{}^{J_{\text{sm}} \leftarrow M_0} \\ -\overline{\overline{T}}{}^{J_{\text{sm}} \leftarrow M_0} \\ \overline{\overline{T}}{}^{J_{\text{m}0} \leftarrow M_0} \\ \overline{\overline{\mathcal{I}}}{}^{M_{\text{s}0}} \\ \overline{\overline{T}}{}^{M_{\text{sm}} \leftarrow M_0} \\ -\overline{\overline{T}}{}^{M_{\text{sm}} \leftarrow M_0} \\ \overline{\overline{\mathcal{I}}}{}^{M_{\text{m}0}} \end{bmatrix}^H \cdot \overline{\overline{P}}{}_{\text{sm sys;1}}^{\text{driving}} \cdot \begin{bmatrix} \overline{\overline{T}}{}^{J_{\text{s}0} \leftarrow M_0} \\ \overline{\overline{T}}{}^{J_{\text{sm}} \leftarrow M_0} \\ -\overline{\overline{T}}{}^{J_{\text{sm}} \leftarrow M_0} \\ \overline{\overline{T}}{}^{J_{\text{m}0} \leftarrow M_0} \\ \overline{\overline{\mathcal{I}}}{}^{M_{\text{s}0}} \\ \overline{\overline{T}}{}^{M_{\text{sm}} \leftarrow M_0} \\ -\overline{\overline{T}}{}^{M_{\text{sm}} \leftarrow M_0} \\ \overline{\overline{\mathcal{I}}}{}^{M_{\text{m}0}} \end{bmatrix}}_{\overline{\overline{P}}{}_{1;M_0}^{\text{driving}}} \cdot \bar{a}^{M_0} \tag{4-142}$$

In formulation (4-142), $\overline{\overline{\mathcal{I}}}{}^{M_{\text{s}0}} = [\overline{\overline{I}}{}^{M_{\text{s}0}} \;\; 0]$ and $\overline{\overline{\mathcal{I}}}{}^{M_{\text{m}0}} = [0 \;\; \overline{\overline{I}}{}^{M_{\text{m}0}}]$ (where $\overline{\overline{I}}{}^{M_{\text{s}0}}$ and $\overline{\overline{I}}{}^{M_{\text{m}0}}$ are $\Xi^{M_{\text{s}0}}$-order and $\Xi^{M_{\text{m}0}}$-order identity matrices respectively), and the 0s are some zero matrices with proper line numbers and column numbers. Formulation (4-142) is just the matrix form of the first DPO surface formulation which only contains BVs.

**2) The Second Surface Formulation**

Similarly to formulation (4-113), we provide the second surface formulation of the DPO corresponding to the system $V_{\text{sm sys}}$ shown in Figure 4-61 as follows:

$$\begin{aligned} P_{\text{sm sys}}^{\text{driving}} = & -(1/2)\left[\left\langle \vec{J}_{\text{sim}}^{\text{ES}}, \mathcal{E}_0\left(\vec{J}_{\text{sim}}^{\text{ES}}, \vec{M}_{\text{sim}}^{\text{ES}}\right)\right\rangle_{\partial V_{\text{sim}}^-} + \left\langle \vec{J}_{\text{sim}}^{\text{ES}}, \mathcal{E}_0\left(\vec{J}_{\text{mul}}^{\text{ES}}, \vec{M}_{\text{mul}}^{\text{ES}}\right)\right\rangle_{\partial V_{\text{sim}}^-}\right] \\ & -(1/2)\left[\left\langle \vec{J}_{\text{mul}}^{\text{ES}}, \mathcal{E}_0\left(\vec{J}_{\text{mul}}^{\text{ES}}, \vec{M}_{\text{mul}}^{\text{ES}}\right)\right\rangle_{\partial V_{\text{mul}}^-} + \left\langle \vec{J}_{\text{mul}}^{\text{ES}}, \mathcal{E}_0\left(\vec{J}_{\text{sim}}^{\text{ES}}, \vec{M}_{\text{sim}}^{\text{ES}}\right)\right\rangle_{\partial V_{\text{mul}}^-}\right] \\ & -(1/2)\left[\left\langle \vec{M}_{\text{sim}}^{\text{ES}}, \mathcal{H}_0\left(\vec{J}_{\text{sim}}^{\text{ES}}, \vec{M}_{\text{sim}}^{\text{ES}}\right)\right\rangle_{\partial V_{\text{sim}}^-} + \left\langle \vec{M}_{\text{sim}}^{\text{ES}}, \mathcal{H}_0\left(\vec{J}_{\text{mul}}^{\text{ES}}, \vec{M}_{\text{mul}}^{\text{ES}}\right)\right\rangle_{\partial V_{\text{sim}}^-}\right] \\ & -(1/2)\left[\left\langle \vec{M}_{\text{mul}}^{\text{ES}}, \mathcal{H}_0\left(\vec{J}_{\text{mul}}^{\text{ES}}, \vec{M}_{\text{mul}}^{\text{ES}}\right)\right\rangle_{\partial V_{\text{mul}}^-} + \left\langle \vec{M}_{\text{mul}}^{\text{ES}}, \mathcal{H}_0\left(\vec{J}_{\text{sim}}^{\text{ES}}, \vec{M}_{\text{sim}}^{\text{ES}}\right)\right\rangle_{\partial V_{\text{sim}}^-}\right] \end{aligned} \tag{4-143}$$

Inserting decomposition formulation (C-55) and expansion formulations (4-125)&(4-126) into the above formulation, we obtain that

$$P_{\text{sm sys}}^{\text{driving}} = \begin{bmatrix} \bar{a}^{J_{\text{s}0}} \\ \bar{a}^{J_{\text{sm}}} \\ \bar{a}^{J_{\text{ms}}} \\ \bar{a}^{J_{\text{m}0}} \\ \bar{a}^{M_{\text{s}0}} \\ \bar{a}^{M_{\text{sm}}} \\ \bar{a}^{M_{\text{ms}}} \\ \bar{a}^{M_{\text{m}0}} \end{bmatrix}^H \cdot \underbrace{\left( \bar{\bar{P}}_{2;0;\text{PVT}}^{\text{sm sys}} + \bar{\bar{P}}_{2;0;\text{SCT}}^{\text{sm sys}} \right)}_{\bar{\bar{P}}_{\text{sm sys};2}^{\text{driving}}} \cdot \begin{bmatrix} \bar{a}^{J_{\text{s}0}} \\ \bar{a}^{J_{\text{sm}}} \\ \bar{a}^{J_{\text{ms}}} \\ \bar{a}^{J_{\text{m}0}} \\ \bar{a}^{M_{\text{s}0}} \\ \bar{a}^{M_{\text{sm}}} \\ \bar{a}^{M_{\text{ms}}} \\ \bar{a}^{M_{\text{m}0}} \end{bmatrix} \tag{4-144}$$

The elements of the power matrices $\bar{\bar{P}}_{2;0;\text{PVT}}^{\text{sm sys}}$ and $\bar{\bar{P}}_{2;0;\text{SCT}}^{\text{sm sys}}$ in the above formulation can be calculated as the the elements of the power matrices $\bar{\bar{P}}_{2;0;\text{PVT}}^{\text{ss sys}}$ and $\bar{\bar{P}}_{2;0;\text{SCT}}^{\text{ss sys}}$ used in formulation (4-114) (only need to properly replace the superscripts and subscripts), so the corresponding calculation formulations will not be explicitly provided here.

Inserting transformation (4-139) into formulation (4-144), it is immediately obtained that

$$P_{\text{sm sys}}^{\text{driving}} = \left(\bar{a}^{M_0}\right)^H \cdot \underbrace{\begin{bmatrix} \bar{\bar{T}}^{J_{\text{s}0} \leftarrow M_0} \\ \bar{\bar{T}}^{J_{\text{sm}} \leftarrow M_0} \\ -\bar{\bar{T}}^{J_{\text{sm}} \leftarrow M_0} \\ \bar{\bar{T}}^{J_{\text{m}0} \leftarrow M_0} \\ \bar{\bar{\mathcal{I}}}^{M_{\text{s}0}} \\ \bar{\bar{T}}^{M_{\text{sm}} \leftarrow M_0} \\ -\bar{\bar{T}}^{M_{\text{sm}} \leftarrow M_0} \\ \bar{\bar{\mathcal{I}}}^{M_{\text{m}0}} \end{bmatrix}^H \cdot \bar{\bar{P}}_{\text{sm sys};2}^{\text{driving}} \cdot \begin{bmatrix} \bar{\bar{T}}^{J_{\text{s}0} \leftarrow M_0} \\ \bar{\bar{T}}^{J_{\text{sm}} \leftarrow M_0} \\ -\bar{\bar{T}}^{J_{\text{sm}} \leftarrow M_0} \\ \bar{\bar{T}}^{J_{\text{m}0} \leftarrow M_0} \\ \bar{\bar{\mathcal{I}}}^{M_{\text{s}0}} \\ \bar{\bar{T}}^{M_{\text{sm}} \leftarrow M_0} \\ -\bar{\bar{T}}^{M_{\text{sm}} \leftarrow M_0} \\ \bar{\bar{\mathcal{I}}}^{M_{\text{m}0}} \end{bmatrix}}_{\bar{\bar{P}}_{2;M_0}^{\text{driving}}} \cdot \bar{a}^{M_0} \tag{4-145}$$

This is just the matrix form of the second DPO surface formulation which only contains BVs.

### 3) The Third Surface Formulation

Similarly to formulation (4-117), we provide the third surface formulation of the DPO corresponding to the system $V_{\text{sm sys}}$ shown in Figure 4-61 as follows:

$$\begin{aligned} P_{\text{sm sys}}^{\text{driving}} = & -(1/2)\left\langle \vec{J}_{\text{s}0}^{\text{ES}} + \vec{J}_{\text{m}0}^{\text{ES}}, \mathcal{E}_0\left(\vec{J}_{\text{s}0}^{\text{ES}} + \vec{J}_{\text{m}0}^{\text{ES}}, \vec{M}_{\text{s}0}^{\text{ES}} + \vec{M}_{\text{m}0}^{\text{ES}}\right)\right\rangle_{\partial V_{\text{s}0}^- \cup \partial V_{\text{m}0}^-} \\ & -(1/2)\left\langle \vec{M}_{\text{s}0}^{\text{ES}} + \vec{M}_{\text{m}0}^{\text{ES}}, \mathcal{H}_0\left(\vec{J}_{\text{s}0}^{\text{ES}} + \vec{J}_{\text{m}0}^{\text{ES}}, \vec{M}_{\text{s}0}^{\text{ES}} + \vec{M}_{\text{m}0}^{\text{ES}}\right)\right\rangle_{\partial V_{\text{s}0}^- \cup \partial V_{\text{m}0}^-} \end{aligned} \tag{4-146}$$

Inserting expansion formulations (4-125)&(4-126) into the above formulation, we obtain

that

$$P_{\text{sm sys}}^{\text{driving}} = \begin{bmatrix} \overline{a}^{J_{\text{s}0}} \\ \overline{a}^{J_{\text{sm}}} \\ \overline{a}^{J_{\text{ms}}} \\ \overline{a}^{J_{\text{m}0}} \\ \overline{a}^{M_{\text{s}0}} \\ \overline{a}^{M_{\text{sm}}} \\ \overline{a}^{M_{\text{ms}}} \\ \overline{a}^{M_{\text{m}0}} \end{bmatrix}^H \cdot \underbrace{\left( \overline{\overline{P}}_{3;0;\text{PVT}}^{\text{sm sys}} + \overline{\overline{P}}_{3;0;\text{SCT}}^{\text{sm sys}} \right)}_{\overline{\overline{P}}_{\text{sm sys};3}^{\text{driving}}} \cdot \begin{bmatrix} \overline{a}^{J_{\text{s}0}} \\ \overline{a}^{J_{\text{sm}}} \\ \overline{a}^{J_{\text{ms}}} \\ \overline{a}^{J_{\text{m}0}} \\ \overline{a}^{M_{\text{s}0}} \\ \overline{a}^{M_{\text{sm}}} \\ \overline{a}^{M_{\text{ms}}} \\ \overline{a}^{M_{\text{m}0}} \end{bmatrix} \tag{4-147}$$

The elements of the power matrices $\overline{\overline{P}}_{3;0;\text{PVT}}^{\text{sm sys}}$ and $\overline{\overline{P}}_{3;0;\text{SCT}}^{\text{sm sys}}$ in the above formulation can be calculated as the elements of the power matrices $\overline{\overline{P}}_{1;0;\text{PVT}}^{\text{ss sys}}$, $\overline{\overline{P}}_{1;0;\text{SCT}}^{\text{ss sys}}$, and $\overline{\overline{P}}_{1;\text{m}}^{\text{ss sys}}$ used in formulation (4-118) (only need to properly replace the superscripts and subscripts), so the corresponding calculation formulations will not be explicitly provided here.

Inserting transformation (4-139) into formulation (4-147), it is immediately obtained that

$$P_{\text{sm sys}}^{\text{driving}} = \left(\overline{a}^{M_0}\right)^H \cdot \underbrace{\begin{bmatrix} \overline{\overline{T}}^{J_{\text{s}0} \leftarrow M_0} \\ \overline{\overline{T}}^{J_{\text{sm}} \leftarrow M_0} \\ -\overline{\overline{T}}^{J_{\text{sm}} \leftarrow M_0} \\ \overline{\overline{T}}^{J_{\text{m}0} \leftarrow M_0} \\ \overline{\overline{I}}^{M_{\text{s}0}} \\ \overline{\overline{T}}^{M_{\text{sm}} \leftarrow M_0} \\ -\overline{\overline{T}}^{M_{\text{sm}} \leftarrow M_0} \\ \overline{\overline{I}}^{M_{\text{m}0}} \end{bmatrix}^H \cdot \overline{\overline{P}}_{\text{sm sys};3}^{\text{driving}} \cdot \begin{bmatrix} \overline{\overline{T}}^{J_{\text{s}0} \leftarrow M_0} \\ \overline{\overline{T}}^{J_{\text{sm}} \leftarrow M_0} \\ -\overline{\overline{T}}^{J_{\text{sm}} \leftarrow M_0} \\ \overline{\overline{T}}^{J_{\text{m}0} \leftarrow M_0} \\ \overline{\overline{I}}^{M_{\text{s}0}} \\ \overline{\overline{T}}^{M_{\text{sm}} \leftarrow M_0} \\ -\overline{\overline{T}}^{M_{\text{sm}} \leftarrow M_0} \\ \overline{\overline{I}}^{M_{\text{m}0}} \end{bmatrix}}_{\overline{\overline{P}}_{3;M_0}^{\text{driving}}} \cdot \overline{a}^{M_0} \tag{4-148}$$

This is just the matrix form of the third DPO surface formulation which only contains BVs.

To simplify the symbolic system of the following parts of this section, we collectively denote $\overline{\overline{P}}_{1;M_0}^{\text{driving}}$, $\overline{\overline{P}}_{2;M_0}^{\text{driving}}$, and $\overline{\overline{P}}_{3;M_0}^{\text{driving}}$ as $\overline{\overline{P}}_{X;M_0}^{\text{driving}}$.

### 4.7.3 DP-CMs and Their Orthogonality

By doing the Toeplitz's decomposition for matrix $\overline{\overline{P}}_{X;M_0}^{\text{driving}}$, we obtain the characteristic equation for constructing DP-CMs as follows:

$$\overline{\overline{P}}_{X;M_0;-}^{\text{driving}} \cdot \overline{\alpha}_{M_0;\xi}^{\text{driving}} = \lambda_{\text{sm sys};\xi}^{\text{driving}} \, \overline{\overline{P}}_{X;M_0;+}^{\text{driving}} \cdot \overline{\alpha}_{M_0;\xi}^{\text{driving}} \tag{4-149}$$

where $\overline{\overline{P}}_{X;M_0;+}^{\text{driving}} = [\overline{\overline{P}}_{X;M_0}^{\text{driving}} + (\overline{\overline{P}}_{X;M_0}^{\text{driving}})^H]/2$ and $\overline{\overline{P}}_{X;M_0;-}^{\text{driving}} = [\overline{\overline{P}}_{X;M_0}^{\text{driving}} - (\overline{\overline{P}}_{X;M_0}^{\text{driving}})^H]/2j$. Inserting above characteristic vectors $\{\overline{\alpha}_{M_0;\xi}^{\text{driving}}\}_{\xi=1}^{\Xi^{M_0}}$ into the transformation (4-139) and the expansion formulations (4-125)&(4-126) given in Subsection 4.7.1, the corresponding characteristic equivalent surface currents $\{\vec{J}_{\text{sim};\xi}^{\text{ES}}, \vec{M}_{\text{sim};\xi}^{\text{ES}}, \vec{J}_{\text{mul};\xi}^{\text{ES}}, \vec{M}_{\text{mul};\xi}^{\text{ES}}\}_{\xi=1}^{\Xi^{M_0}}$ are immediately obtained, where $\Xi^{M_0} = \Xi^{M_{\text{s}0}} + \Xi^{M_{\text{m}0}}$. Inserting the characteristic equivalent surface currents into the GSEP given in Appendix C7, the corresponding characteristic fields can then be obtained. Utilizing the characteristic fields distributing on material system and the volume equivalence principle given in Appendix A, we can obtain the corresponding characteristic scattered volume currents $\{\vec{J}_{\text{sim};\xi}^{\text{SV}}, \vec{M}_{\text{sim};\xi}^{\text{SV}}, \vec{J}_{\text{mul};\xi}^{\text{SV}}, \vec{M}_{\text{mul};\xi}^{\text{SV}}\}_{\xi=1}^{\Xi^{M_0}}$. Obviously, characteristic values $\lambda_{\text{sm sys};\xi}^{\text{driving}}$ and the modal powers satisfy the following relationship:

$$\lambda_{\text{sm sys};\xi}^{\text{driving}} = \frac{\operatorname{Im}\left\{P_{\text{sm sys};\xi}^{\text{driving}}\right\}}{\operatorname{Re}\left\{P_{\text{sm sys};\xi}^{\text{driving}}\right\}} \tag{4-150}$$

where $\xi = 1, 2, \cdots, \Xi^{M_0}$.

In addition, the characteristic vectors satisfy the following orthogonality:

$$\operatorname{Re}\left\{P_{\text{sm sys};\xi}^{\text{driving}}\right\} \delta_{\xi\zeta} = \left(\overline{\alpha}_{M_0;\xi}^{\text{driving}}\right)^H \cdot \overline{\overline{P}}_{X;M_0;+}^{\text{driving}} \cdot \overline{\alpha}_{M_0;\zeta}^{\text{driving}} \tag{4-151a}$$

$$\operatorname{Im}\left\{P_{\text{sm sys};\xi}^{\text{driving}}\right\} \delta_{\xi\zeta} = \left(\overline{\alpha}_{M_0;\xi}^{\text{driving}}\right)^H \cdot \overline{\overline{P}}_{X;M_0;-}^{\text{driving}} \cdot \overline{\alpha}_{M_0;\zeta}^{\text{driving}} \tag{4-151b}$$

and

$$\underbrace{\left[\operatorname{Re}\left\{P_{\text{sm sys};\xi}^{\text{driving}}\right\} + j \operatorname{Im}\left\{P_{\text{sm sys};\xi}^{\text{driving}}\right\}\right]}_{P_{\text{sm sys};\xi}^{\text{driving}}} \delta_{\xi\zeta} = \left(\overline{\alpha}_{M_0;\xi}^{\text{driving}}\right)^H \cdot \underbrace{\left(\overline{\overline{P}}_{X;M_0;+}^{\text{driving}} + j\, \overline{\overline{P}}_{X;M_0;-}^{\text{driving}}\right)}_{\overline{\overline{P}}_{X;M_0}^{\text{driving}}} \cdot \overline{\alpha}_{M_0;\zeta}^{\text{driving}} \tag{4-152}$$

Similarly to the process to derive formulation (4-59) from formulation (4-58) and the process to derive formulation (4-108) from formulation (4-107), we can, from formulation (4-152), derive the orthogonality between the characteristic incident fields and the characteristic scattered currents as follows:

$$\begin{aligned} & P_{\text{sm sys};\xi}^{\text{driving}} \delta_{\xi\zeta} \\ = & \left[\frac{1}{2}\left\langle \vec{J}_{\text{sim};\xi}^{\text{SV}}, \vec{E}_{\zeta}^{\text{inc}} \right\rangle_{V_{\text{sim}}} + \frac{1}{2}\left\langle \vec{M}_{\text{sim};\xi}^{\text{SV}}, \vec{H}_{\zeta}^{\text{inc}} \right\rangle_{V_{\text{sim}}}\right] + \left[\frac{1}{2}\left\langle \vec{J}_{\text{mul};\xi}^{\text{SV}}, \vec{E}_{\zeta}^{\text{inc}} \right\rangle_{V_{\text{mul}}} + \frac{1}{2}\left\langle \vec{M}_{\text{mul};\xi}^{\text{SV}}, \vec{H}_{\zeta}^{\text{inc}} \right\rangle_{V_{\text{mul}}}\right] \\ = - & \left[\frac{1}{2}\left\langle \vec{J}_{\text{sim};\xi}^{\text{ES}}, \vec{E}_{\zeta}^{\text{inc}} \right\rangle_{\partial V_{\text{sim}}} + \frac{1}{2}\left\langle \vec{M}_{\text{sim};\xi}^{\text{ES}}, \vec{H}_{\zeta}^{\text{inc}} \right\rangle_{\partial V_{\text{sim}}}\right] - \left[\frac{1}{2}\left\langle \vec{J}_{\text{mul};\xi}^{\text{ES}}, \vec{E}_{\zeta}^{\text{inc}} \right\rangle_{\partial V_{\text{mul}}} + \frac{1}{2}\left\langle \vec{M}_{\text{mul};\xi}^{\text{ES}}, \vec{H}_{\zeta}^{\text{inc}} \right\rangle_{\partial V_{\text{mul}}}\right] \end{aligned} \tag{4-153}$$

## 4.7.4 DP-CM-Based Modal Expansion

Based on the completeness of the DP-CMs, any working mode can be expanded in terms of the DP-CMs as follows:

$$\vec{E}^{\text{inc}}\left(\vec{r}\right) = \sum_{\xi=1}^{\Xi^{M_{\text{sm sys}}}} c_\xi \vec{E}_\xi^{\text{inc}}\left(\vec{r}\right) \quad , \quad \vec{r} \in V_{\text{sim}} \cup V_{\text{mul}} \tag{4-154a}$$

$$\vec{H}^{\text{inc}}\left(\vec{r}\right) = \sum_{\xi=1}^{\Xi^{M_{\text{sm sys}}}} c_\xi \vec{H}_\xi^{\text{inc}}\left(\vec{r}\right) \quad , \quad \vec{r} \in V_{\text{sim}} \cup V_{\text{mul}} \tag{4-154b}$$

Testing above expansion formulations (4-154a) and (4-154b) with functions $\{(\vec{J}_{\text{sim};\xi}^{\text{SV}} + \vec{J}_{\text{mul};\xi}^{\text{SV}})/2\}_{\xi=1}^{\Xi^{M_0}}$ and $\{(\vec{M}_{\text{sim};\xi}^{\text{SV}} + \vec{M}_{\text{mul};\xi}^{\text{SV}})/2\}_{\xi=1}^{\Xi^{M_0}}$ respectively, and summing the obtained results, and employing orthogonality (4-153), we immediately obtain the explicit expressions for the expansion coefficients used in expansion formulation (4-154) as follows:

$$\begin{aligned} c_\xi &= \frac{\left[\left\langle \vec{J}_{\text{sim};\xi}^{\text{SV}}, \vec{E}^{\text{inc}} \right\rangle_{V_{\text{sim}}} + \left\langle \vec{M}_{\text{sim};\xi}^{\text{SV}}, \vec{H}^{\text{inc}} \right\rangle_{V_{\text{sim}}}\right] + \left[\left\langle \vec{J}_{\text{mul};\xi}^{\text{SV}}, \vec{E}^{\text{inc}} \right\rangle_{V_{\text{mul}}} + \left\langle \vec{M}_{\text{mul};\xi}^{\text{SV}}, \vec{H}^{\text{inc}} \right\rangle_{V_{\text{mul}}}\right]}{2P_{\text{sm sys};\xi}^{\text{driving}}} \\ &= \frac{-\left[\left\langle \vec{J}_{\text{sim};\xi}^{\text{ES}}, \vec{E}^{\text{inc}} \right\rangle_{\partial V_{\text{sim}}} + \left\langle \vec{M}_{\text{sim};\xi}^{\text{ES}}, \vec{H}^{\text{inc}} \right\rangle_{\partial V_{\text{sim}}}\right] - \left[\left\langle \vec{J}_{\text{mul};\xi}^{\text{ES}}, \vec{E}^{\text{inc}} \right\rangle_{\partial V_{\text{mul}}} + \left\langle \vec{M}_{\text{mul};\xi}^{\text{ES}}, \vec{H}^{\text{inc}} \right\rangle_{\partial V_{\text{mul}}}\right]}{2P_{\text{sm sys};\xi}^{\text{driving}}} \end{aligned} \tag{4-155}$$

where $\xi = 1, 2, \cdots, \Xi^{M_0}$ .

## 4.7.5 Numerical Examples Corresponding to Typical Structures

This subsection will provide some typical numerical examples to verify the validity of the formulations established in this section.

### 1) Two-Layered Material Sphere

Now, we consider the two-layered material sphere which is illustrated in Figure 4-62. In the figure, the internal radius and external radius of the outer material spherical shell are 2.50mm and 5.00mm respectively; the relative permeability, relative permittivity, and conductivity of the inner material sphere are 6, 6, and 0 respectively, and the relative permeability, relative permittivity, and conductivity of the outer material spherical shell are 2, 18, and 0 respectively. For the convenience of the following discussions, we denote the inner material sphere and the outer material spherical shell as $V_{\text{sim}}$ and $V_{\text{mul}}$ respectively.

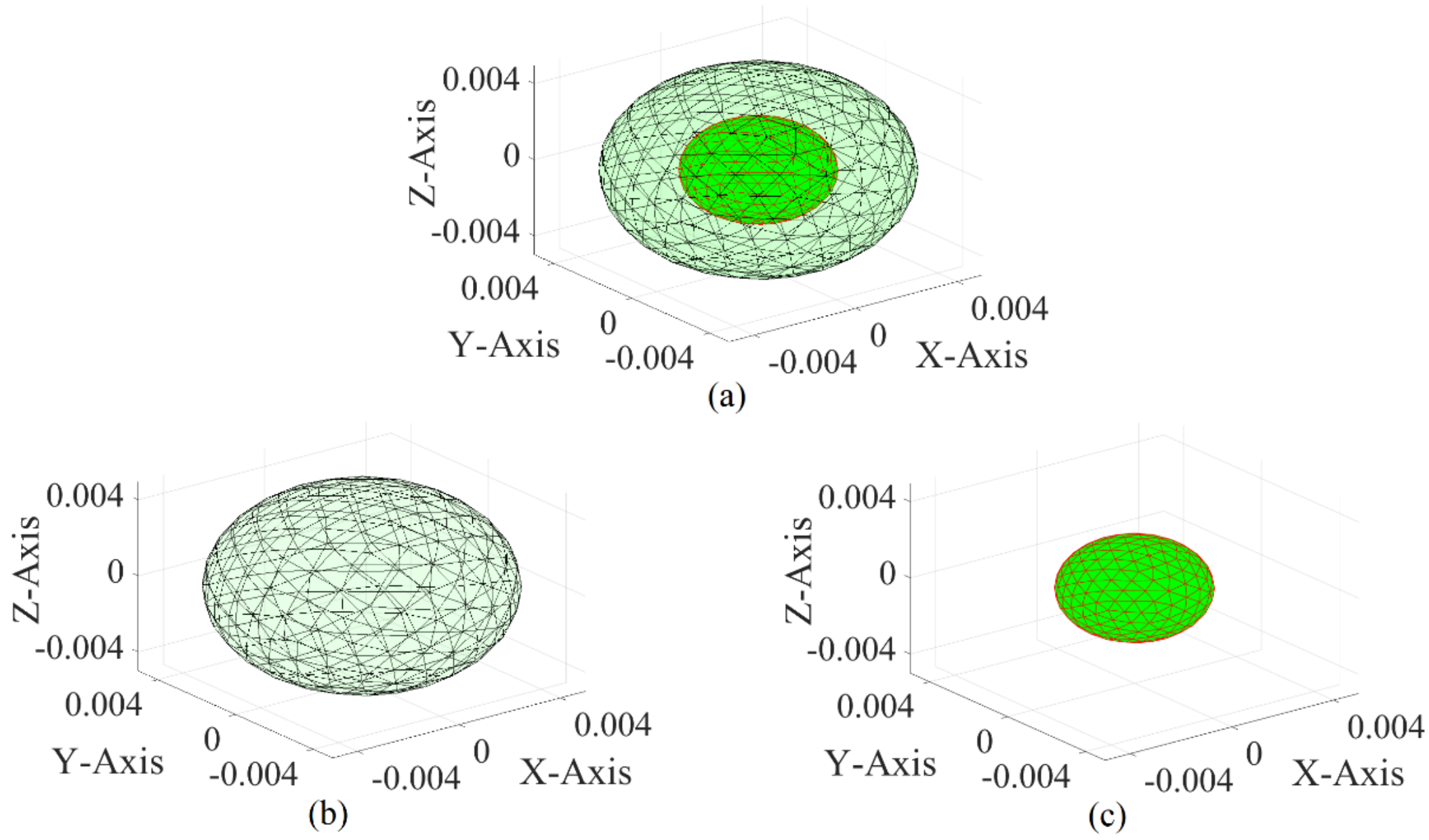

Figure 4-62 The topological structure and surface triangular meshes of a two-layer material sphere. (a) the surface triangular meshes of the whole two-layer material sphere; (b) the surface triangular meshes of the air-matter interface of the outer material spherical shell; (c) the surface triangular meshes of the matter-matter interface between the inner material sphere and the outer material spherical shell

Based on the three different surface formulations provided in this section, we construct the DP-CMs of the material system shown in Figure 4-62. The characteristic value (dB) curves and MS curves corresponding to the first several DP-CMs are shown in Figures 4-63, 4-64, and 4-65. Obviously, the results derived from the three different formulations agree well with each others.

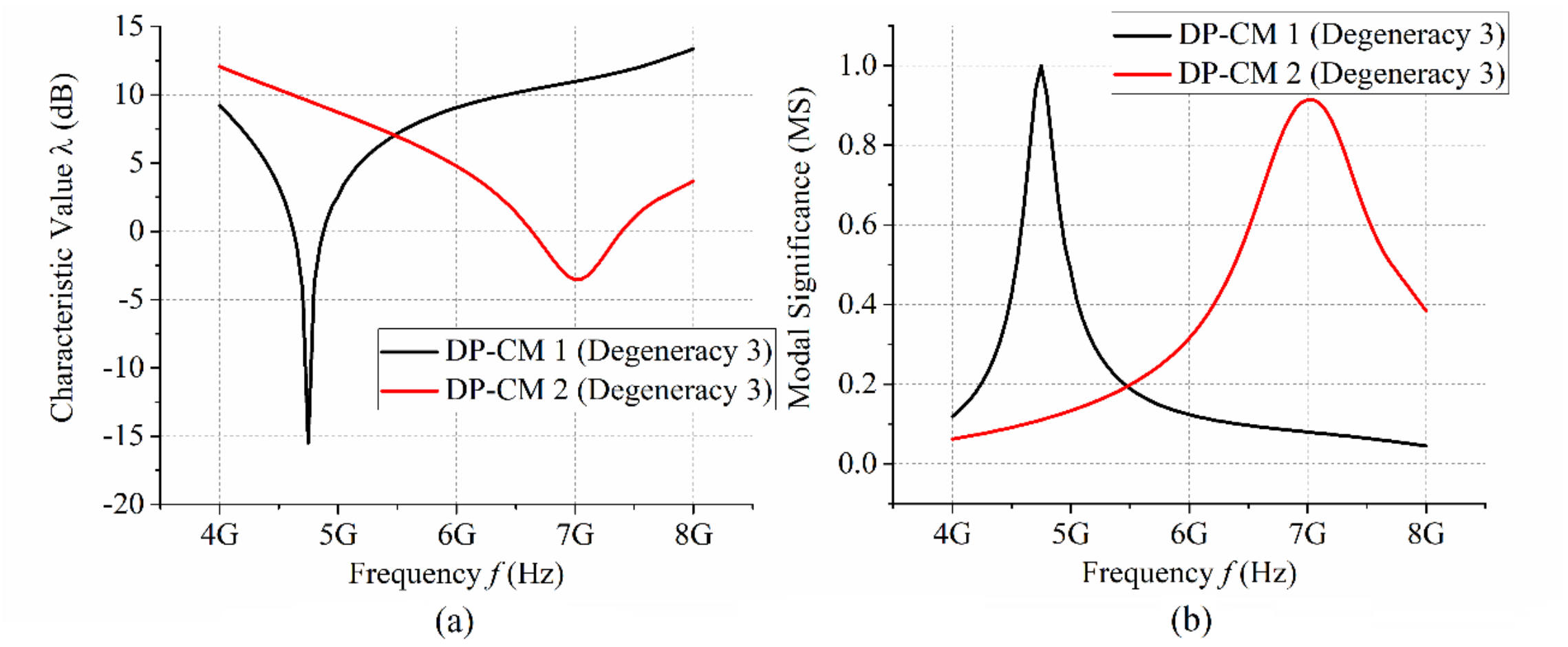

Figure 4-63 The characteristic quantity curves corresponding to several typical DP-CMs (of the two-layer material sphere shown in Figure 4-62) derived from the first formulation provided in this section. (a) characteristic value dB curves; (b) MS curves

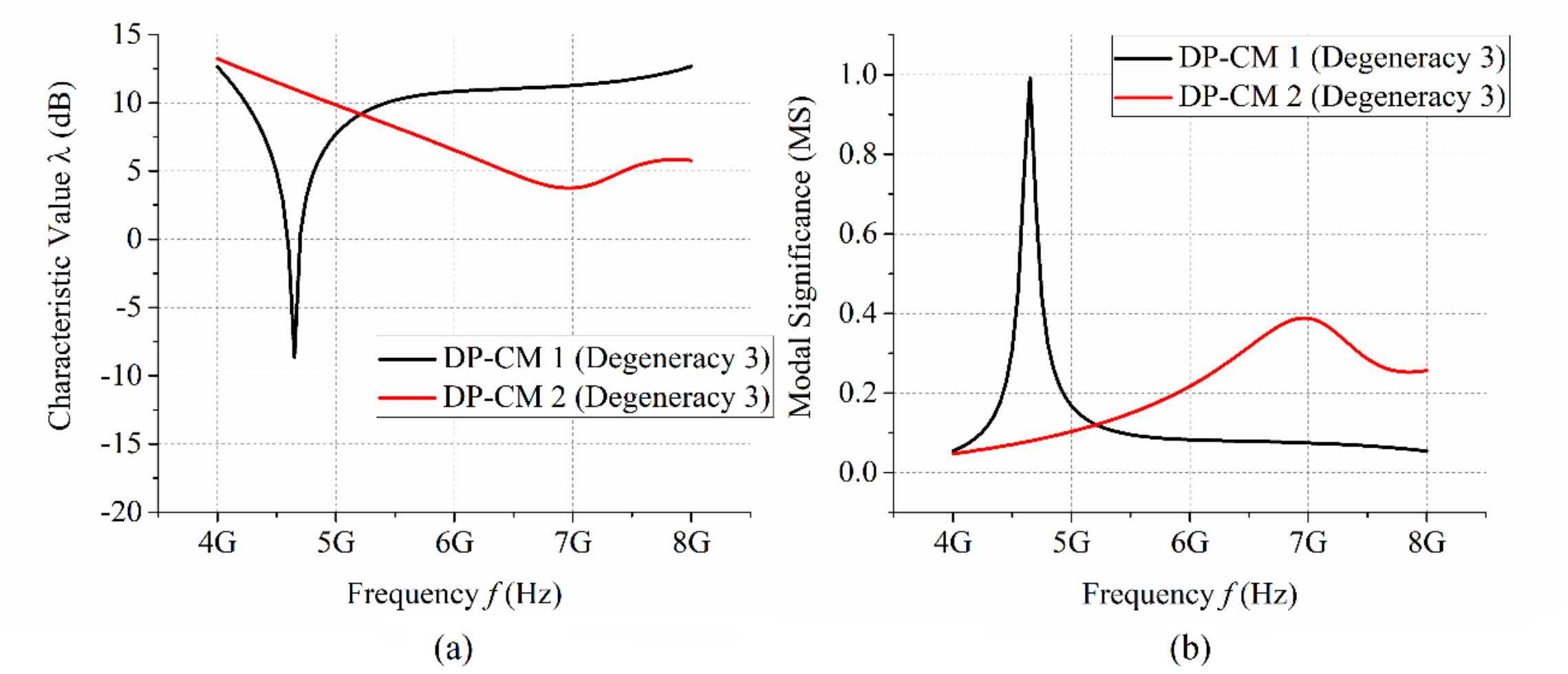

Figure 4-64 The characteristic quantity curves corresponding to several typical DP-CMs (of the two-layer material sphere shown in Figure 4-62) derived from the second formulation provided in this section. (a) characteristic value dB curves; (b) MS curves

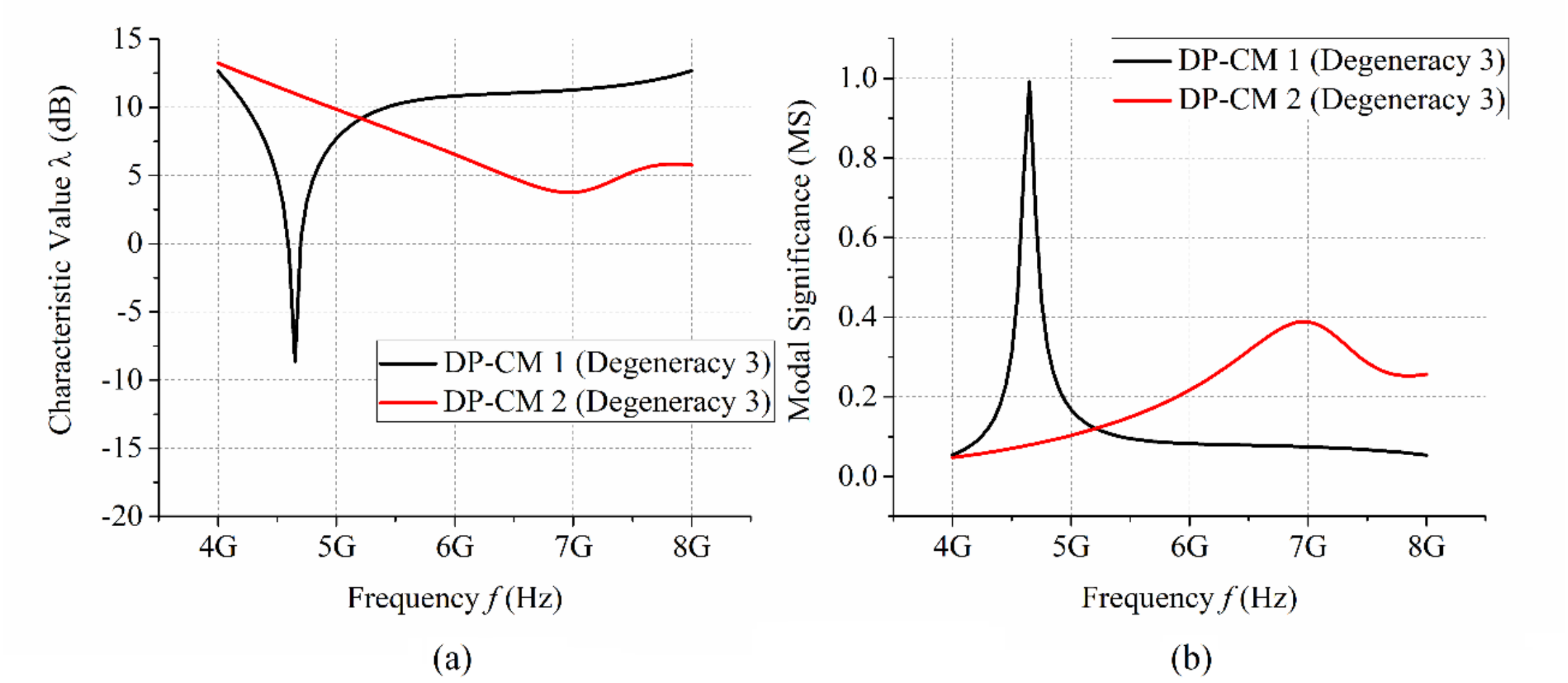

Figure 4-65 The characteristic quantity curves corresponding to several typical DP-CMs (of the two-layer material sphere shown in Figure 4-62) derived from the third formulation provided in this section. (a) characteristic value dB curves; (b) MS curves

Because the results shown in Figures 4-63, 4-64, and 4-65 agree with each others, then we take the modes shown in Figure 4-65 as typical examples, and provide the visual pictures of the characteristic physical quantities corresponding to the typical modes for readers' convenience of reference.

In Figure 4-65, DP-CM1 is "resonant" at 4.65GHz, and it has three different degenerate states. For the first degenerate state, its modal equivalent surface magnetic and electric currents distributing on $\partial V_{\text{mul}}^{\text{out}}$, $\partial V_{\text{mul}}^{\text{in}}$, and $\partial V_{\text{sim}}$ are shown in Figure 4-66;

the tangential components of its modal total fields distributing on $\partial V_{\text{mul}}^{\text{out}}$, $\partial V_{\text{mul}}^{\text{in}}$, and $\partial V_{\text{sim}}$ are shown in Figure 4-67; its modal total fields distributing on $V_{\text{mul}}^{\text{out}}$ and $V_{\text{mul}}^{\text{in}}$ are shown in Figure 4-68; its modal scattered volume sources distributing on $V_{\text{mul}}^{\text{out}}$ and $V_{\text{mul}}^{\text{in}}$ are shown in Figure 4-69; its modal incident fields distributing on $V_{\text{mul}}^{\text{out}}$ and $V_{\text{mul}}^{\text{in}}$ are shown in Figure 4-70; its modal radiation pattern is shown in Figure 4-71.

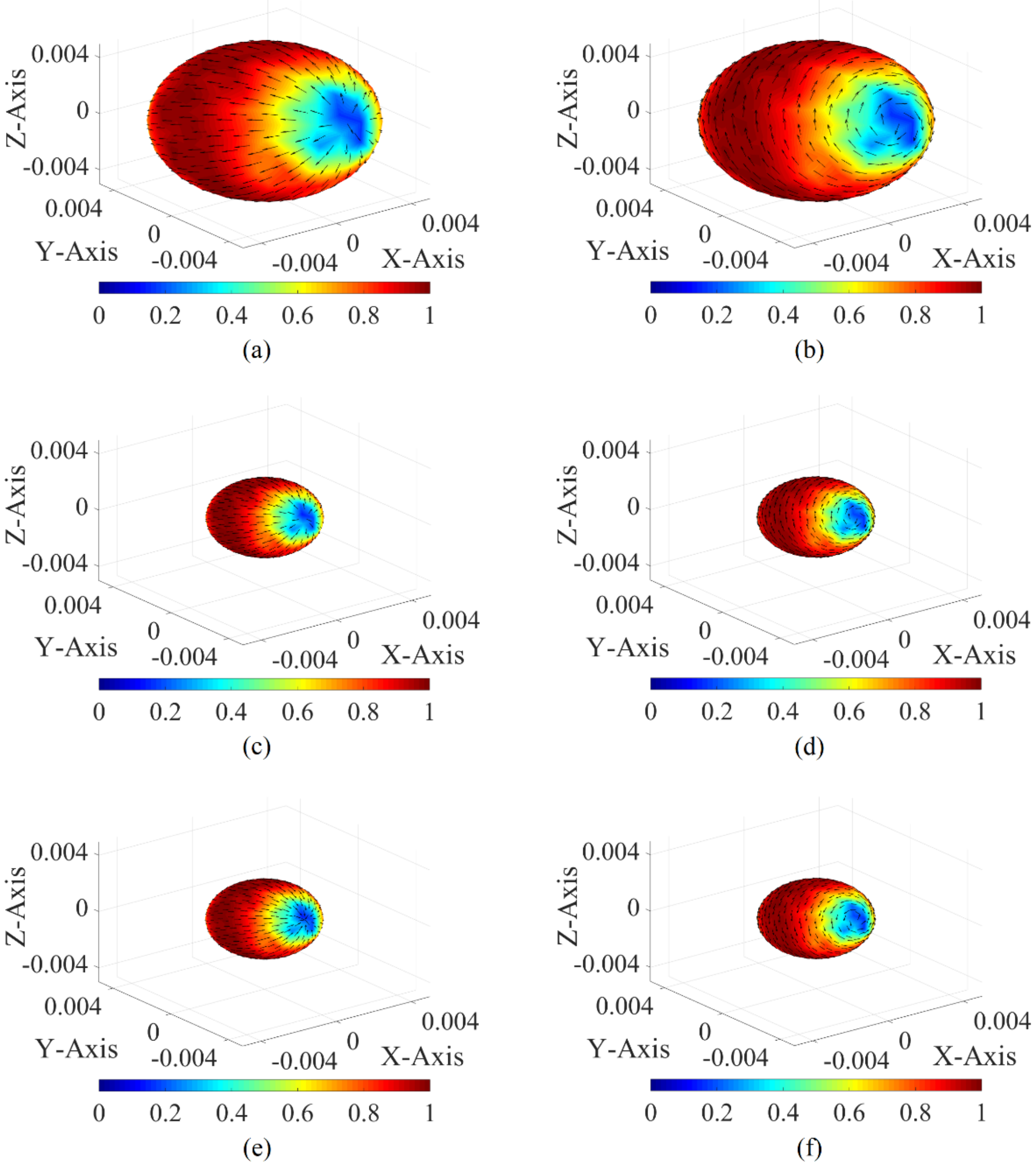

Figure 4-66 The modal equivalent source distributions of the first degenerate state of the DP-CM1 working at 4.65GHz and shown in Figure 4-65. (a) the equivalent surface magnetic current on $\partial V_{\text{mul}}^{\text{out}}$; (b) the equivalent surface electric current on $\partial V_{\text{mul}}^{\text{out}}$; (c) the equivalent surface magnetic current on $\partial V_{\text{mul}}^{\text{in}}$; (d) the equivalent surface electric current on $\partial V_{\text{mul}}^{\text{in}}$; (e) the equivalent surface magnetic current on $\partial V_{\text{sim}}$; (f) the equivalent surface electric current on $\partial V_{\text{sim}}$

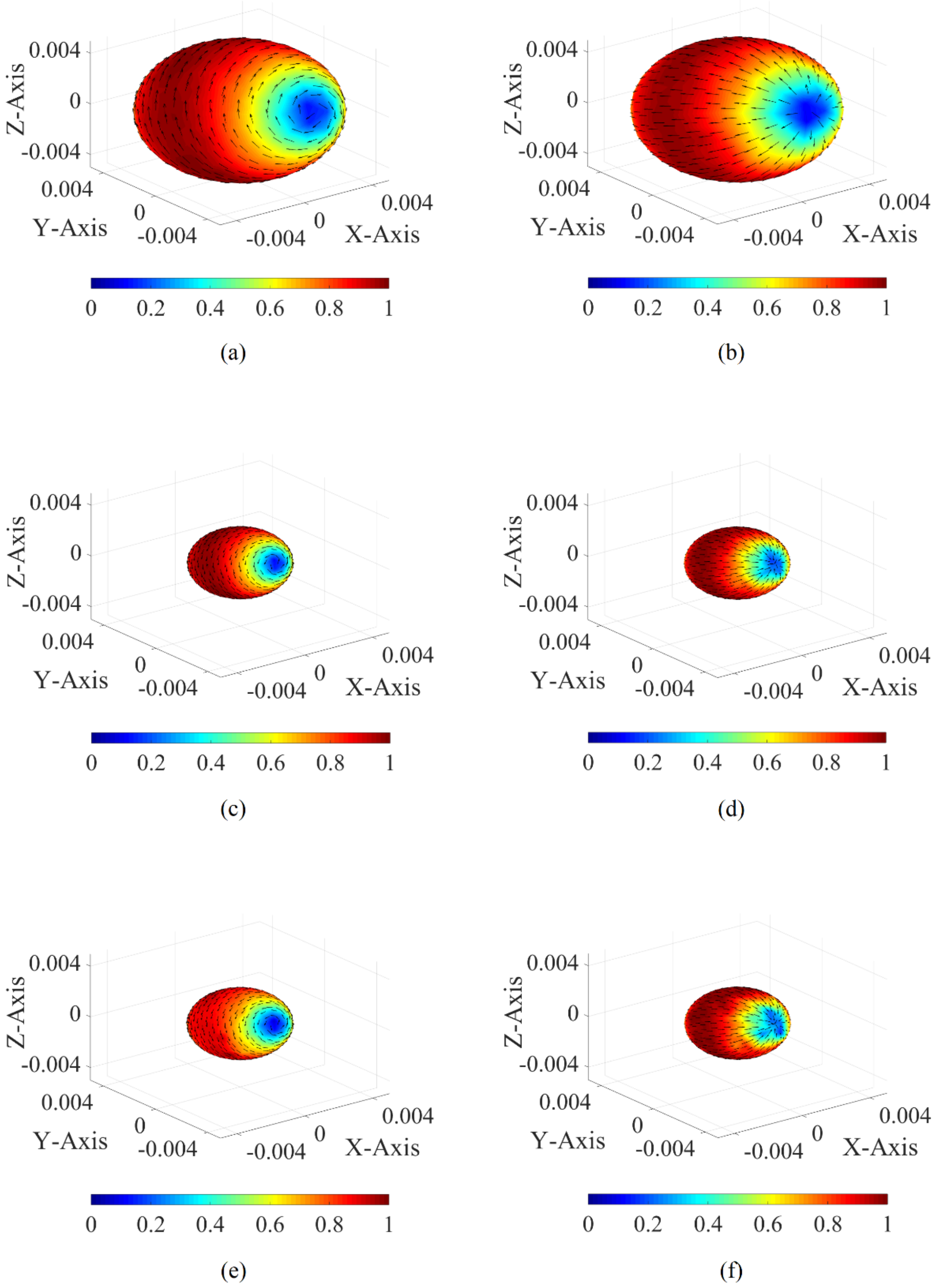

Figure 4-67 The modal tangential total field distributions of the first degenerate state of the DP-CM1 working at 4.65GHz and shown in Figure 4-65. (a) the tangential total electric field on $\partial V_{\text{mul}}^{\text{out}}$; (b) the tangential total magnetic field on $\partial V_{\text{mul}}^{\text{out}}$; (c) the tangential total electric field on $\partial V_{\text{mul}}^{\text{in}}$; (d) the tangential total magnetic field on $\partial V_{\text{mul}}^{\text{in}}$; (e) the tangential total electric field on $\partial V_{\text{sim}}$; (f) the tangential total magnetic field on $\partial V_{\text{sim}}$

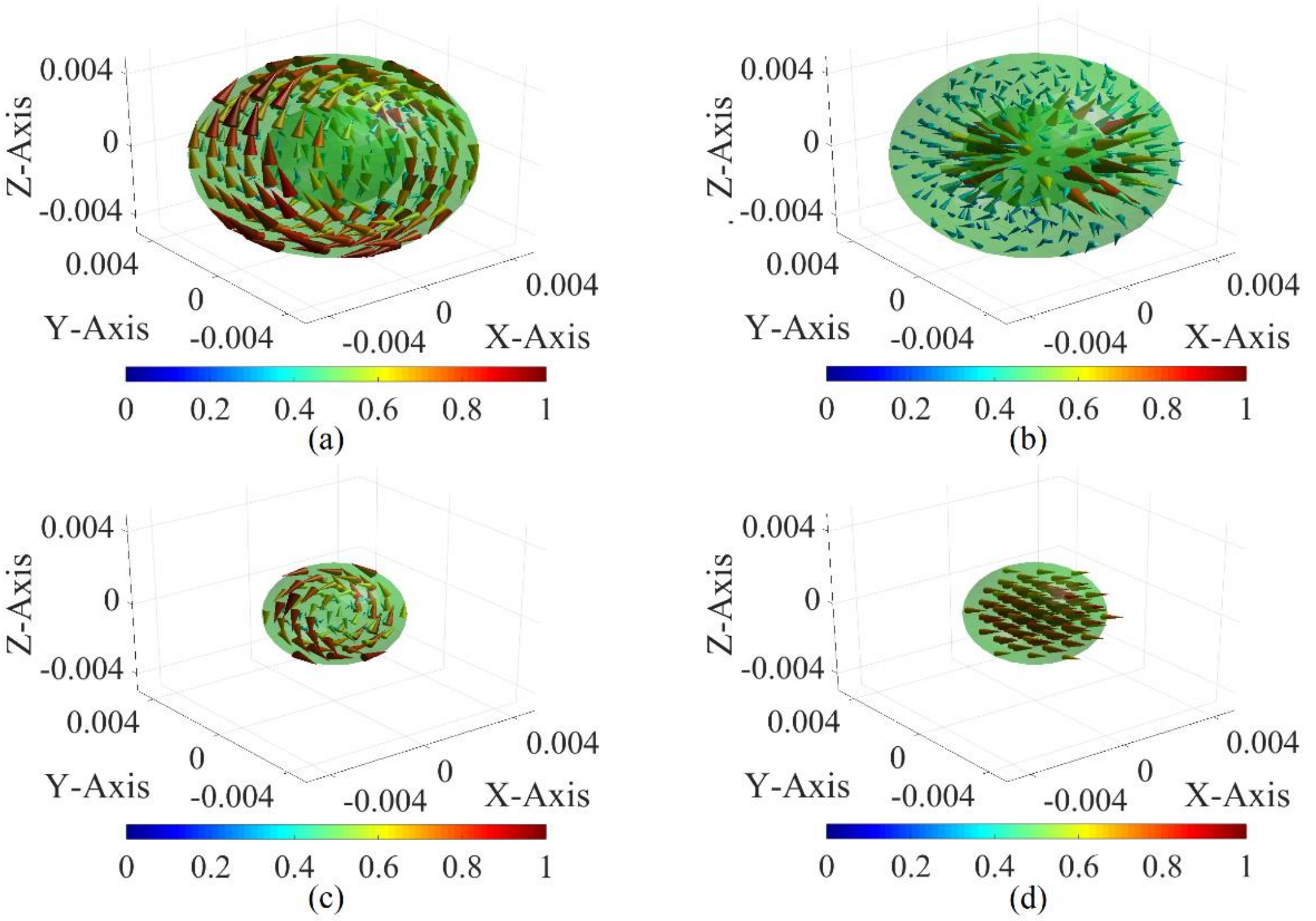

Figure 4-68 The modal total field distributions of the first degenerate state of the DP-CM1 working at 4.65GHz and shown in Figure 4-65. (a) the modal total electric field on $V_{\text{mul}}$; (b) the modal total magnetic field on $V_{\text{mul}}$; (c) the modal total electric field on $V_{\text{sim}}$; (d) the modal total magnetic field on $V_{\text{sim}}$

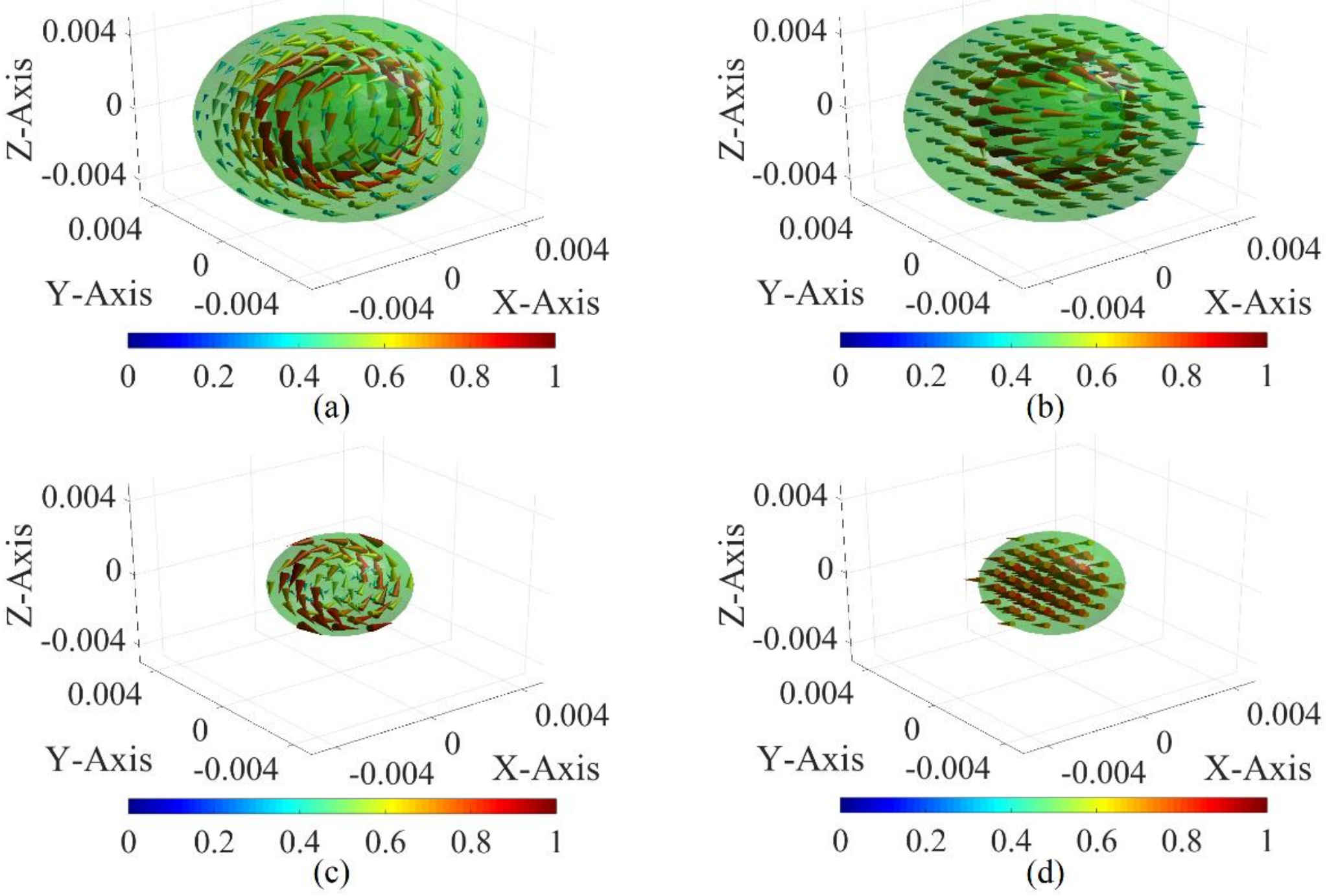

Figure 4-69 The modal scattered volume source distributions of the first degenerate state of the DP-CM1 working at 4.65GHz and shown in Figure 4-65. (a) the modal scattered volume electric current on $V_{\text{mul}}$; (b) the modal scattered volume magnetic current on $V_{\text{mul}}$; (c) the modal scattered volume electric current on $V_{\text{sim}}$; (d) the modal scattered volume magnetic current on $V_{\text{sim}}$

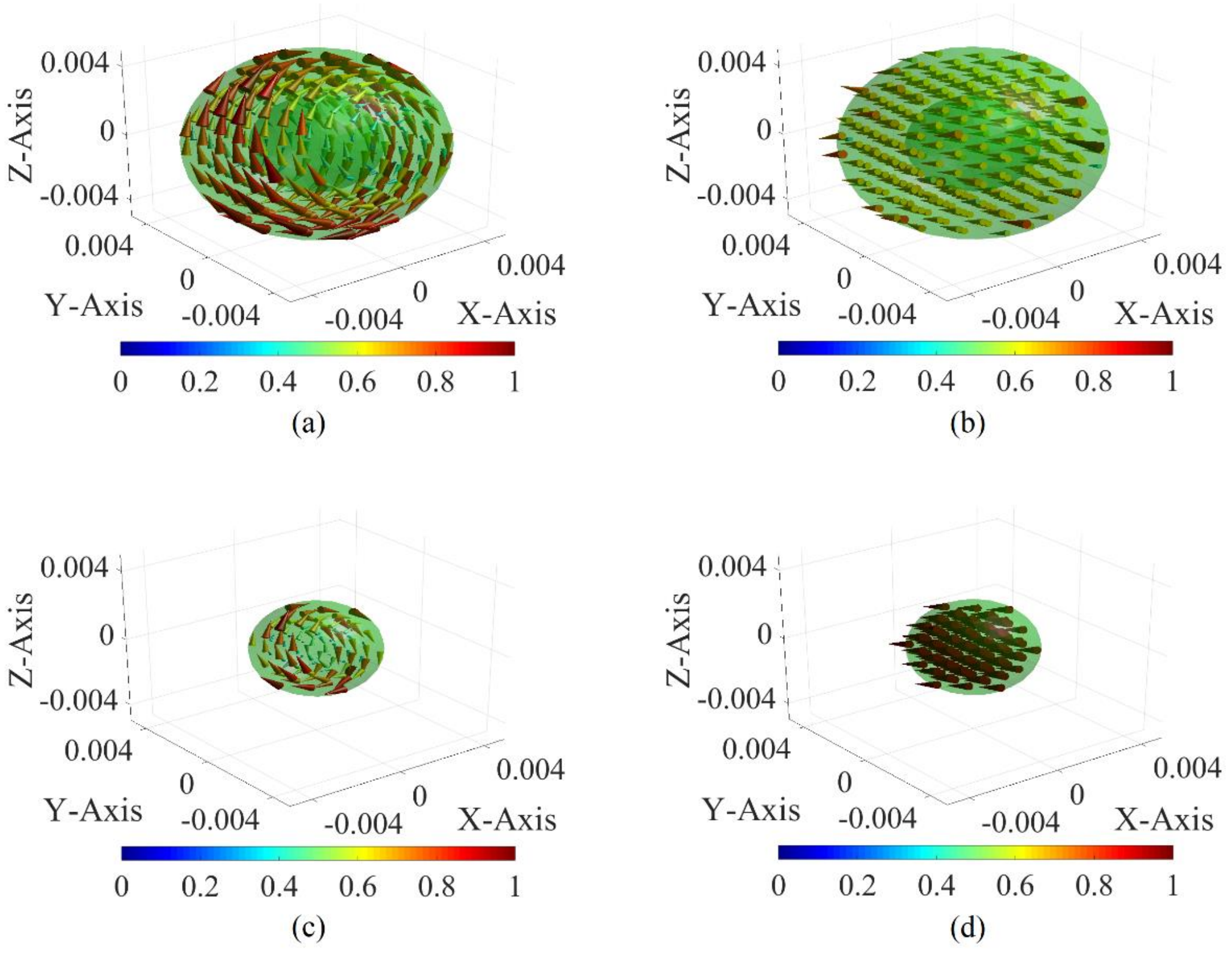

Figure 4-70 The modal incident field distributions of the first degenerate state of the DP-CM1 working at 4.65GHz and shown in Figure 4-65. (a) the modal incident electric field on $V_{\rm mul}$ ; (b) the modal incident magnetic field on $V_{\rm mul}$ ; (c) the modal incident electric field on $V_{\rm sim}$ ; (d) the modal incident magnetic field on $V_{\rm sim}$

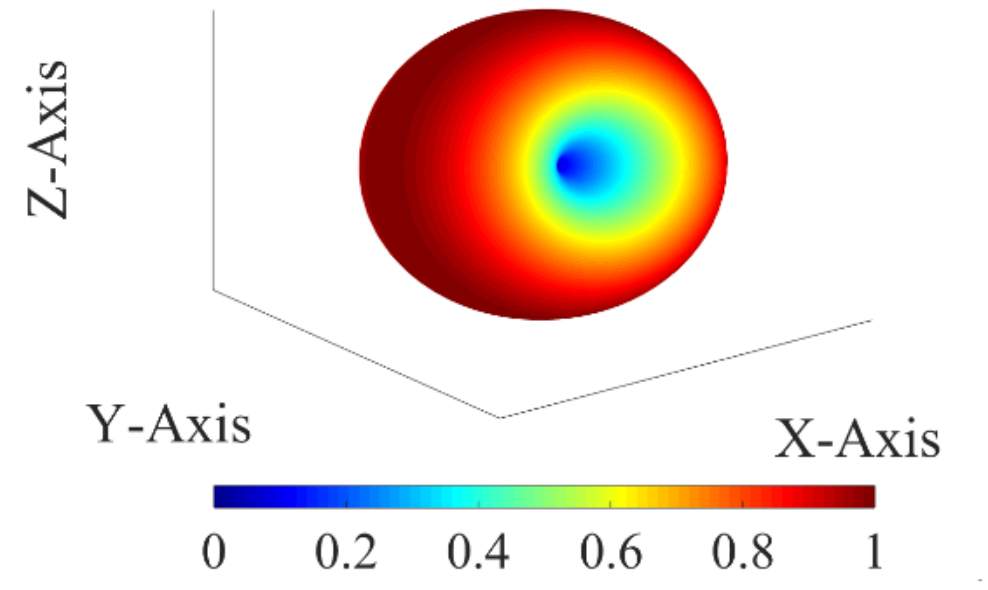

Figure 4-71 The radiation pattern of the first degenerate state of the DP-CM1 working at 4.65GHz and shown in Figure 4-65

In Figure 4-65, the "resonant" DP-CM1 working at 4.65GHz also has another two degenerate states, and their BV (the equivalent surface magnetic current on $\partial V_{\rm mul}^{\rm out}$ ) distributions and radiation patterns are illustrated in Figures 4-72 and 4-73 respectively.

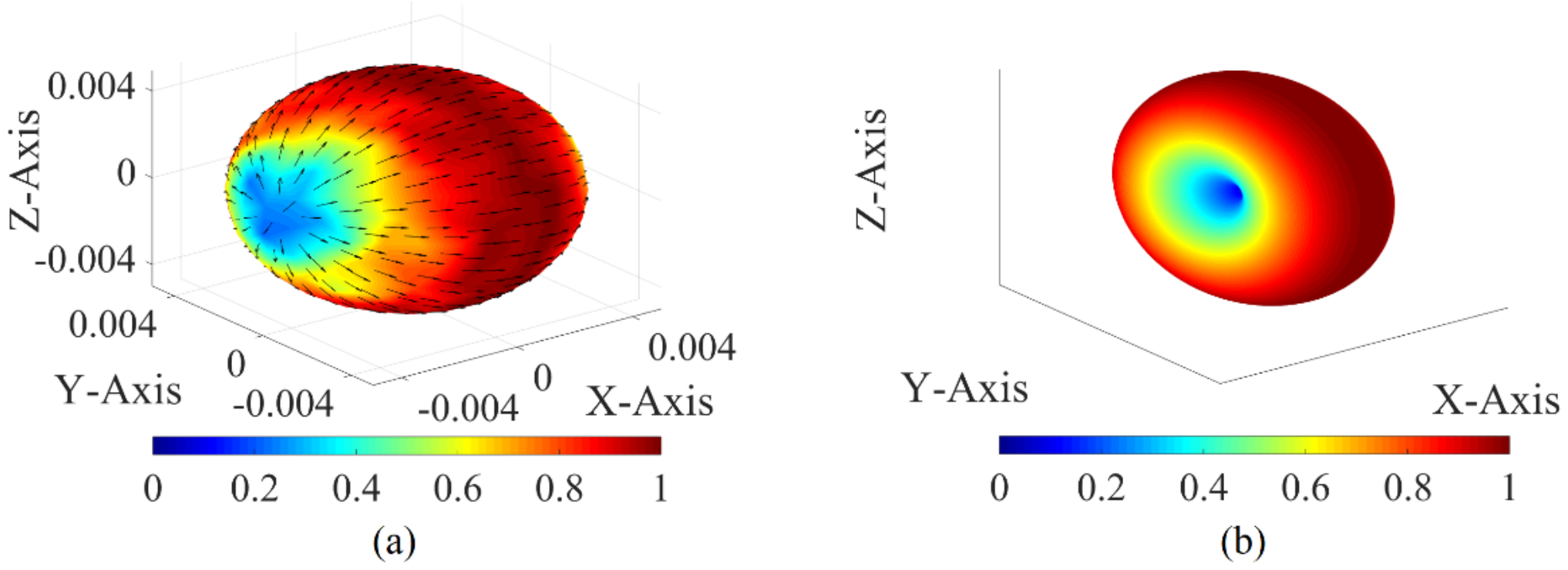

Figure 4-72 The basic variable and radiation pattern of the second degenerate state of the DP-CM1 working at 4.65GHz and shown in Figure 4-65. (a) the equivalent surface magnetic current (basic variable) on $\partial V_{\text{mul}}^{\text{out}}$; (b) the radiation pattern at infinity

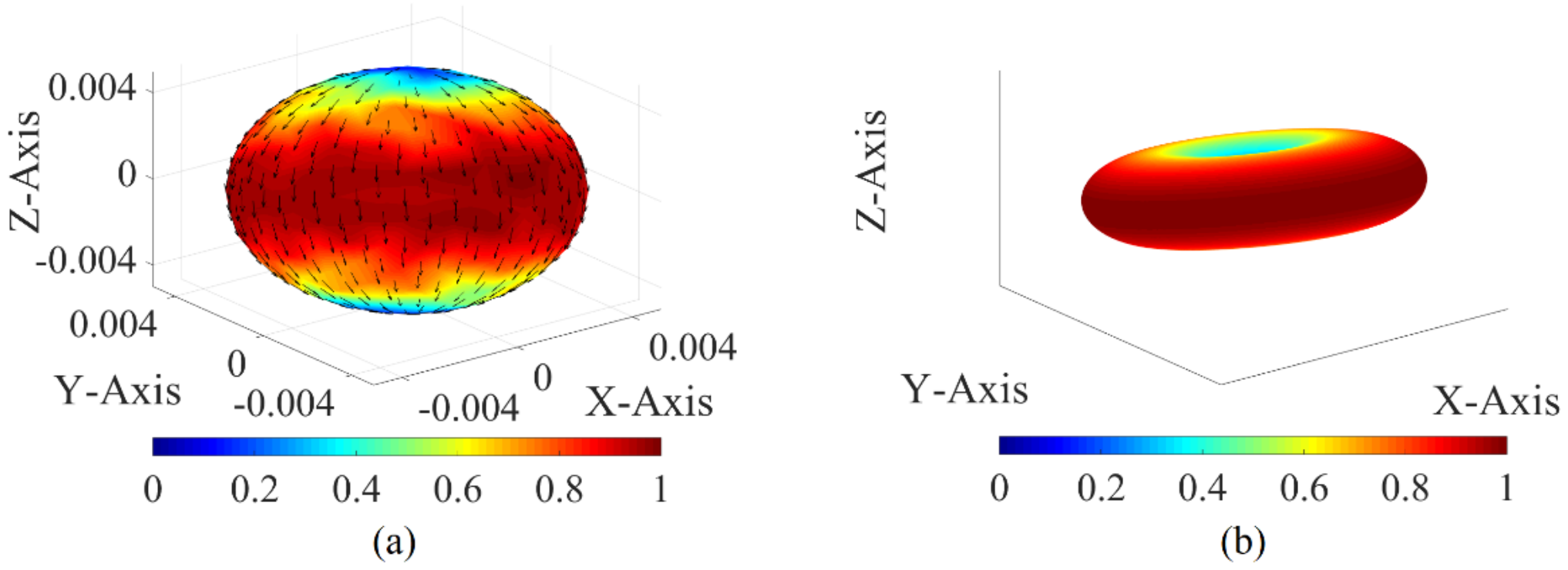

Figure 4-73 The basic variable and radiation pattern of the third degenerate state of the DP-CM1 working at 4.65GHz and shown in Figure 4-65. (a) the equivalent surface magnetic current (basic variable) on $\partial V_{\text{mul}}^{\text{out}}$; (b) the radiation pattern at infinity

**2) Multiply Connected Material Spherical Shell**

Now, we still consider the material system whose topological structure is the same as the one shown in Figure 4-62. In addition, the material system considered below is with the material parameters that the {relative permeability, relative permittivity, conductivity} of the internal material sphere and the external material spherical shell are {1, 1, 0} and {6, 6, 0} respectively. Obviously, this kind of material system is essentially a multiply connected material body.

Based on the third surface formulation established in this section, we construct the DP-CMs of the multiply connected material spherical shell, and illustrate the characteristic value (dB) curves and MS curves corresponding to two typical DP-CMs as Figure 4-74.

In Figure 4-74, the DP-CM1 working at 7.35GHz is "resonant", and it has three different degenerate states. Taking the first "resonant" degenerate state as a typical example, its modal equivalent surface magnetic and electric currents distributing on $\partial V_{\text{mul}}^{\text{out}}$ and $\partial V_{\text{mul}}^{\text{in}}$ are illustrated in Figure 4-75; the tangential components of its modal total fields distributing on $\partial V_{\text{mul}}^{\text{out}}$ and $\partial V_{\text{mul}}^{\text{in}}$ are illustrated in Figure 4-76; its modal total fields distributing on $V_{\text{mul}}^{\text{out}}$ are illustrated in Figure 4-77; its modal scattered volume sources distributing on $V_{\text{mul}}^{\text{out}}$ are illustrated in Figure 4-78; its modal incident fields distributing on $V_{\text{mul}}^{\text{out}}$ are illustrated in Figure 4-79; its modal radiation pattern is illustrated in Figure 4-80.

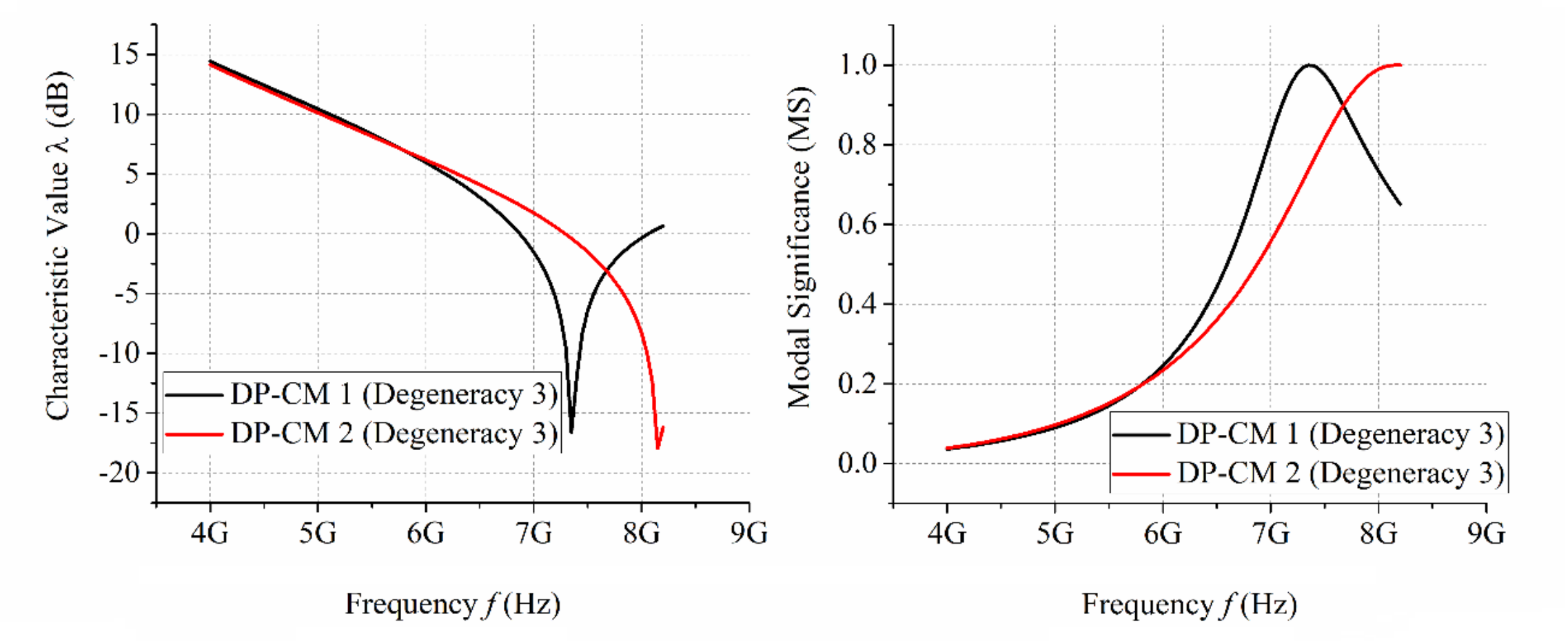

(a) (b)

Figure 4-74 The characteristic quantity curves corresponding to two typical DP-CMs (of the multiply connected material spherical shell shown in Figure 4-62) derived from the third formulation provided in this section. (a) characteristic value dB curves; (b) MS curves

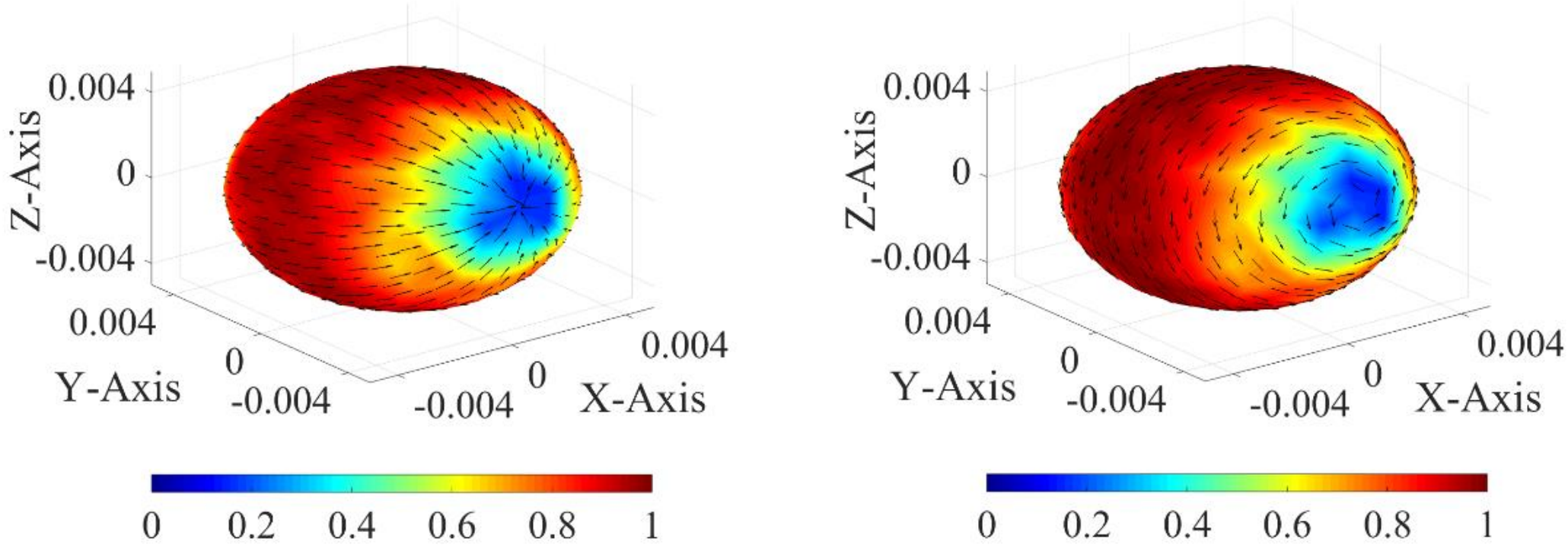

(a) (b)

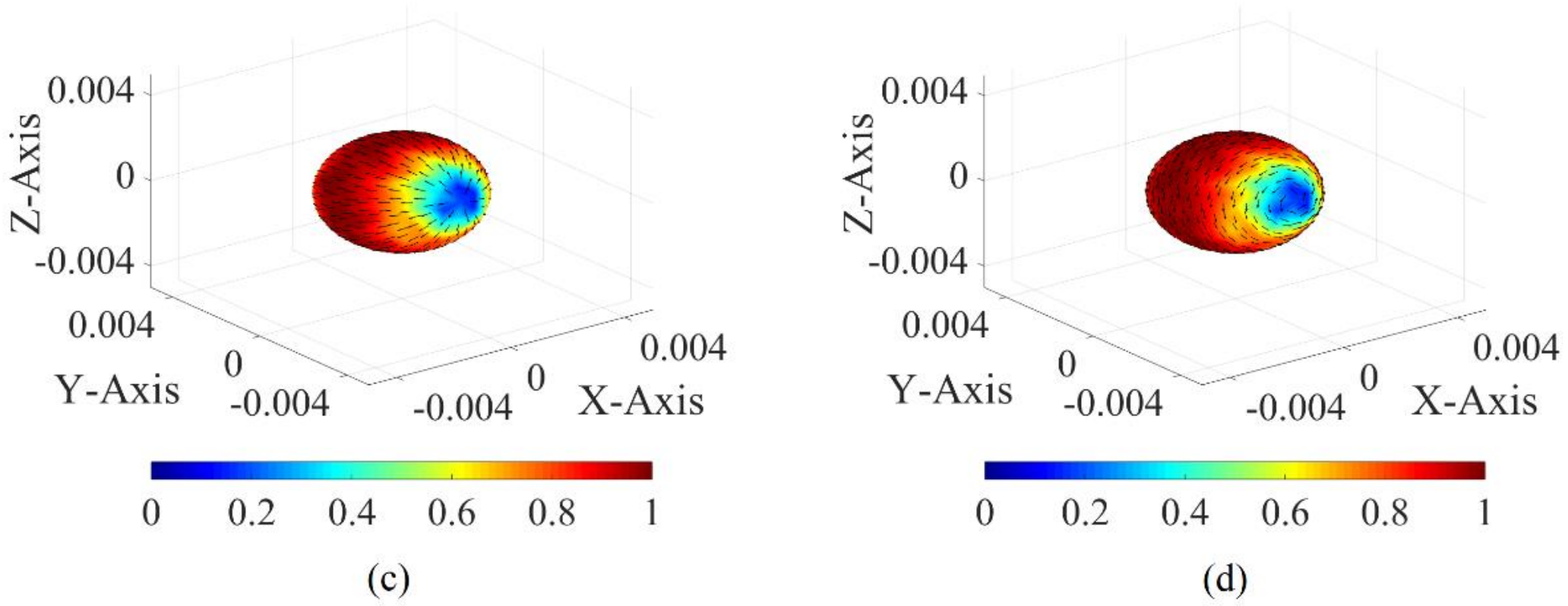

Figure 4-75 The modal equivalent source distributions of the first degenerate state of the DP-CM1 working at 7.35GHz and shown in Figure 4-74. (a) the equivalent surface magnetic current on $\partial V_{\text{mul}}^{\text{out}}$; (b) the equivalent surface electric current on $\partial V_{\text{mul}}^{\text{out}}$; (c) the equivalent surface magnetic current on $\partial V_{\text{mul}}^{\text{in}}$; (d) the equivalent surface electric current on $\partial V_{\text{mul}}^{\text{in}}$

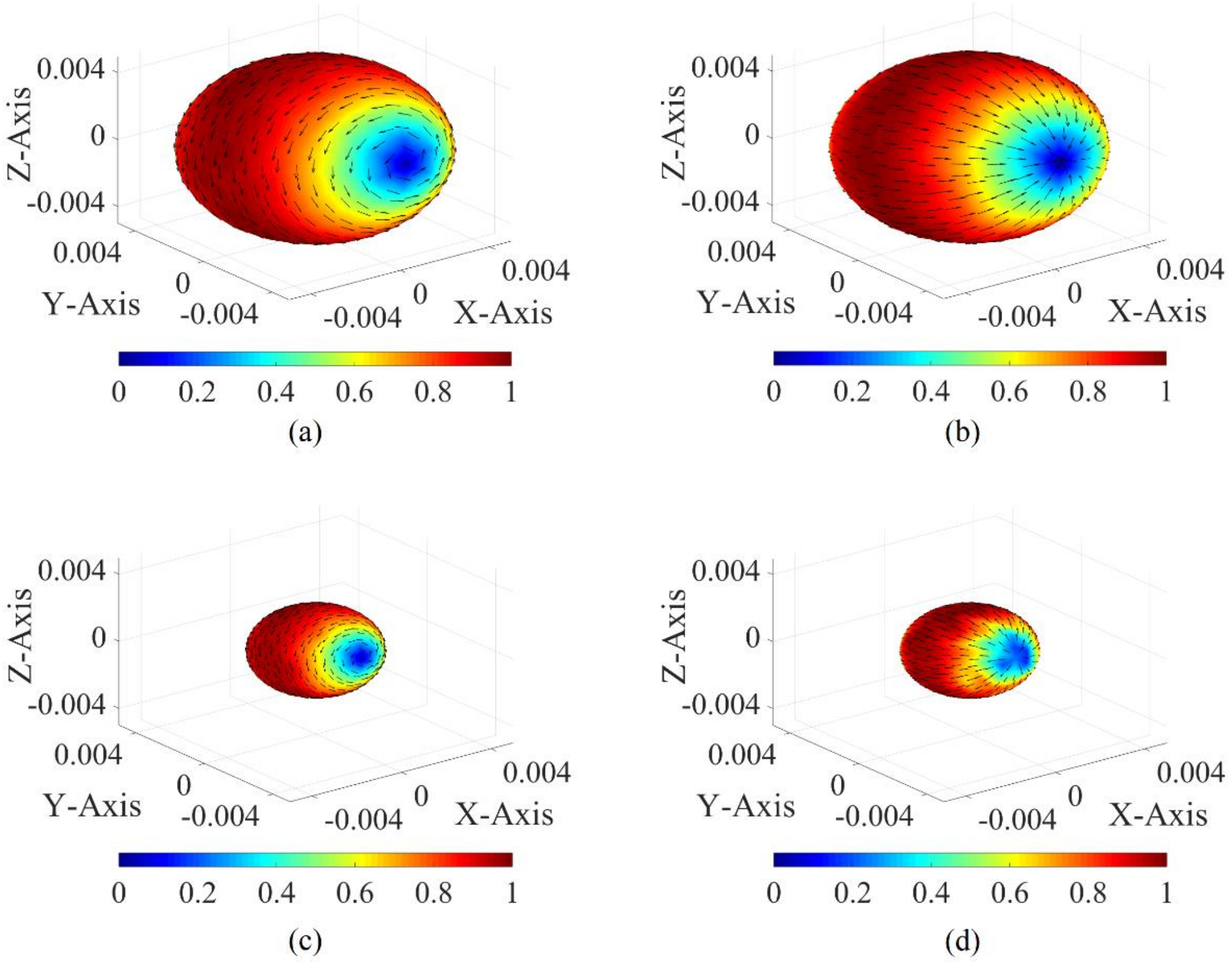

Figure 4-76 The modal tangential total field distributions of the first degenerate state of the DP-CM1 working at 7.35GHz and shown in Figure 4-74. (a) the tangential total electric field on $\partial V_{\text{mul}}^{\text{out}}$; (b) the tangential total magnetic field on $\partial V_{\text{mul}}^{\text{out}}$; (c) the tangential total electric field on $\partial V_{\text{mul}}^{\text{in}}$; (d) the tangential total magnetic field on $\partial V_{\text{mul}}^{\text{in}}$

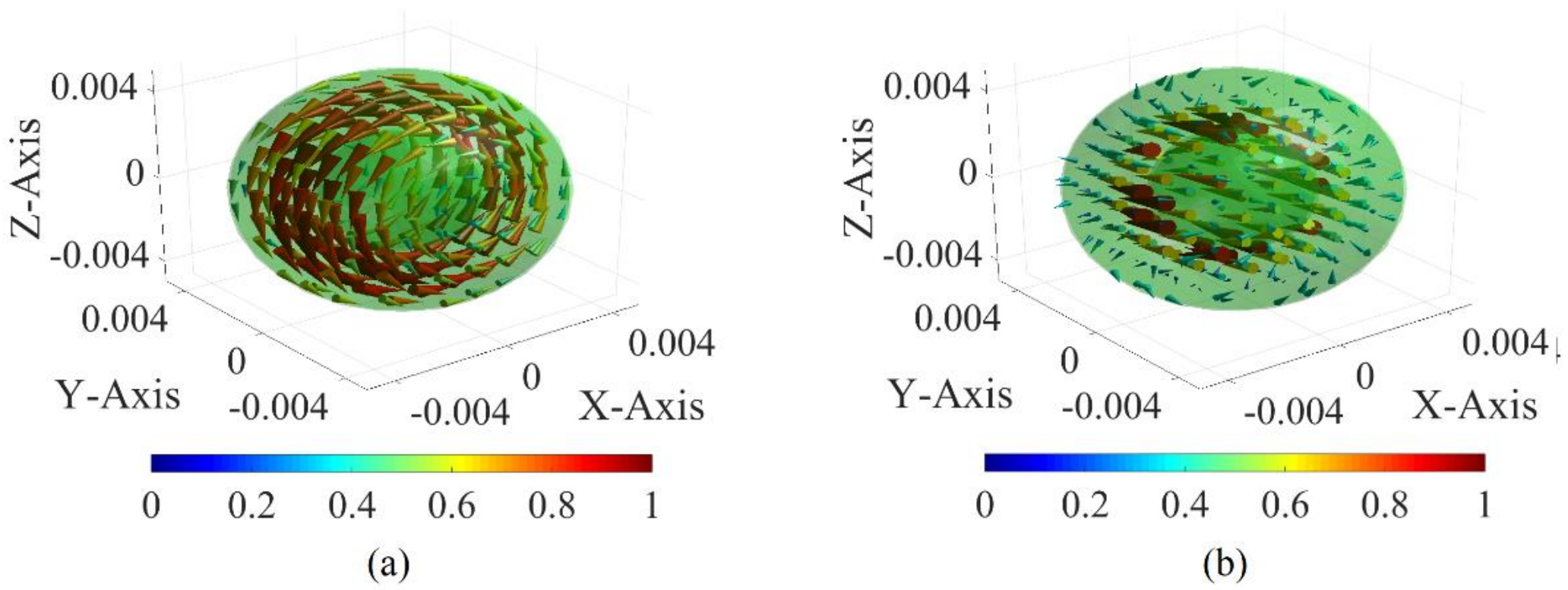

Figure 4-77 The modal total field distributions of the first degenerate state of the DP-CM1 working at 7.35GHz and shown in Figure 4-74. (a) the modal total electric field on $V_{\rm mul}$; (b) the modal total magnetic field on $V_{\rm mul}$

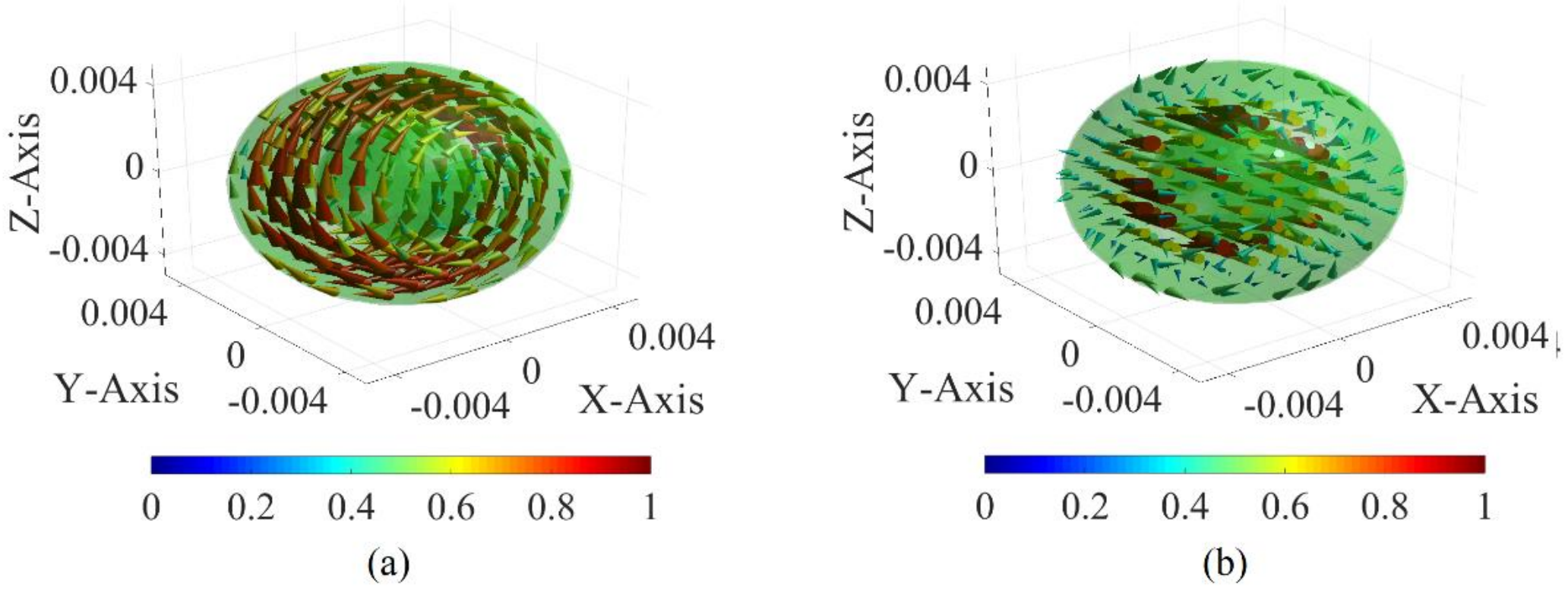

Figure 4-78 The modal scattered volume source distributions of the first degenerate state of the DP-CM1 working at 7.35GHz and shown in Figure 4-74. (a) the modal scattered volume electric current on $V_{\rm mul}$; (b) the modal scattered volume magnetic current on $V_{\rm mul}$

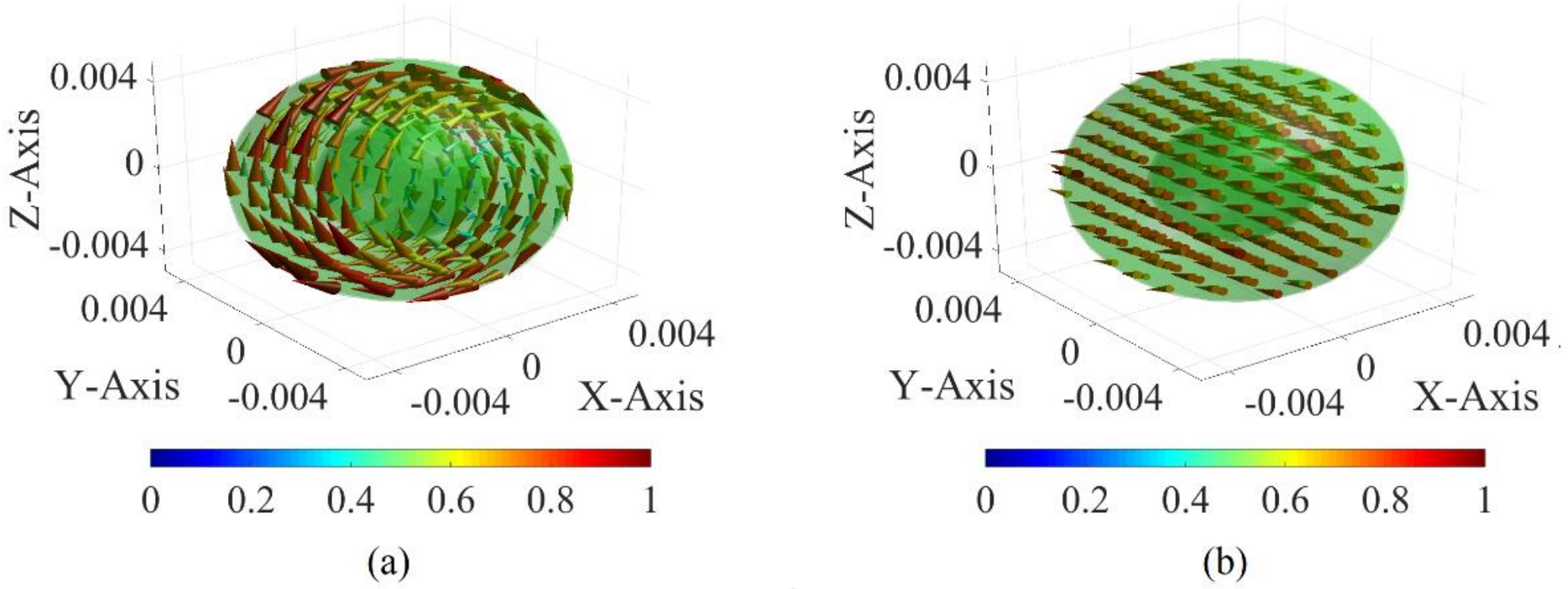

Figure 4-79 The modal incident field distributions of the first degenerate state of the DP-CM1 working at 7.35GHz and shown in Figure 4-74. (a) the modal incident electric field on $V_{\rm mul}$; (b) the modal incident magnetic field on $V_{\rm mul}$

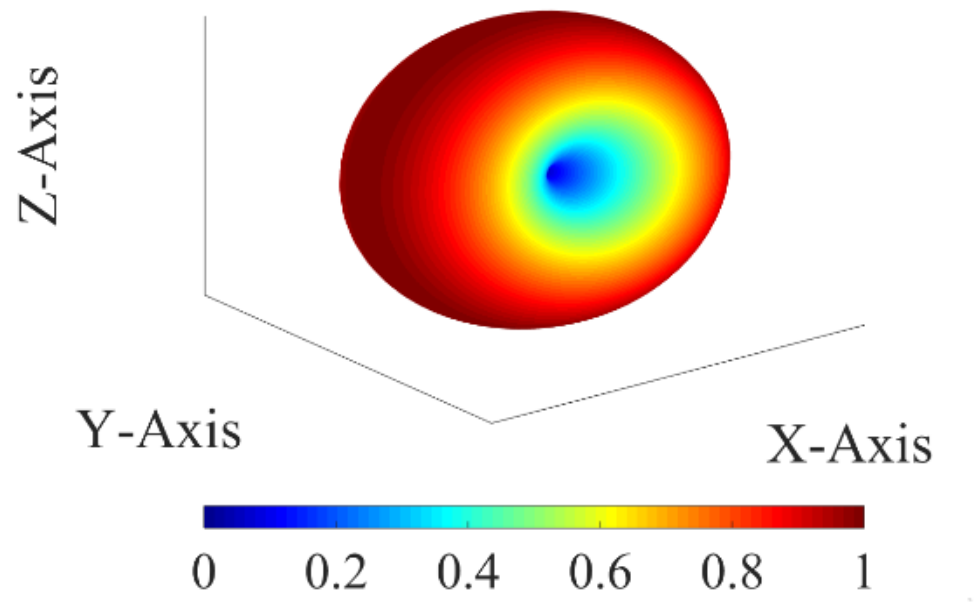

Figure 4-80 The radiation pattern of the first degenerate state of the DP-CM1 working at 7.35GHz and shown in Figure 4-74

In Figure 4-74, the "resonant" DP-CM1 working at 7.35GHz also has another two degenerate states, and their BV (the equivalent surface magnetic current on $\partial V_{\text{mul}}^{\text{out}}$) distributions and radiation patterns are illustrated in Figures 4-81 and 4-82 respectively.

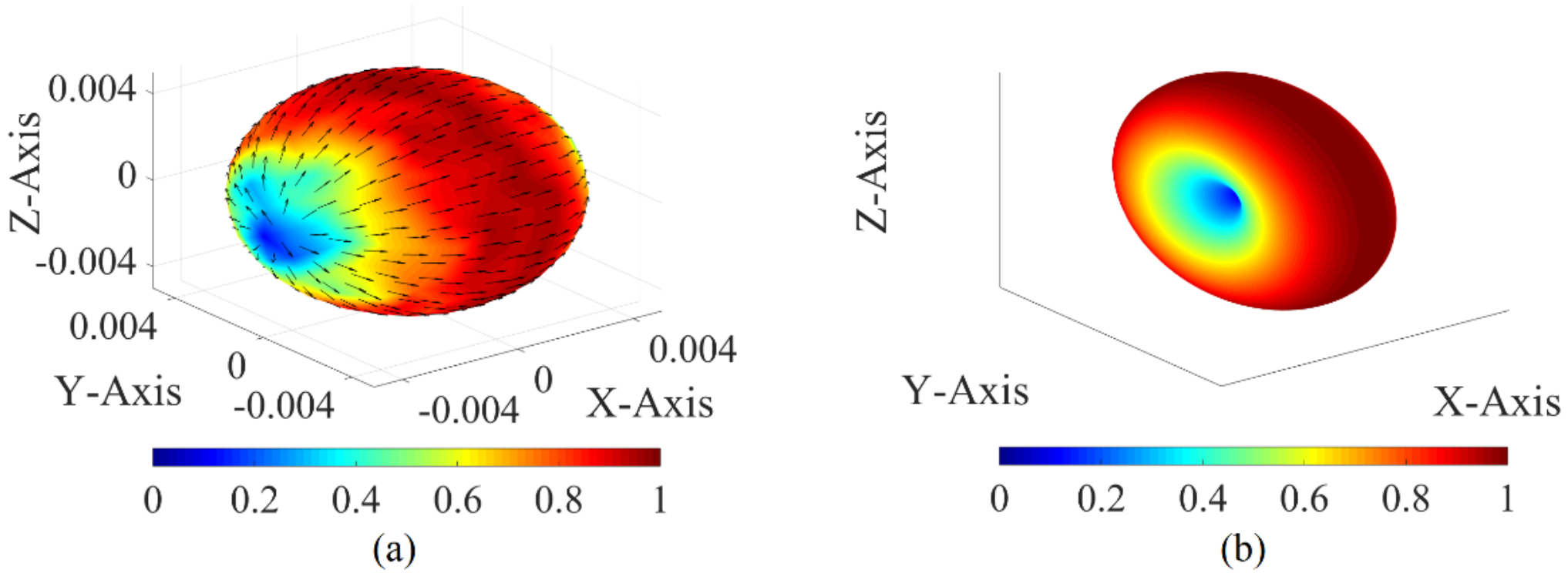

Figure 4-81 The basic variable and radiation pattern of the second degenerate state of the DP-CM1 working at 7.35GHz and shown in Figure 4-74. (a) the equivalent surface magnetic current (basic variable) on $\partial V_{\text{mul}}^{\text{out}}$; (b) the radiation pattern at infinity

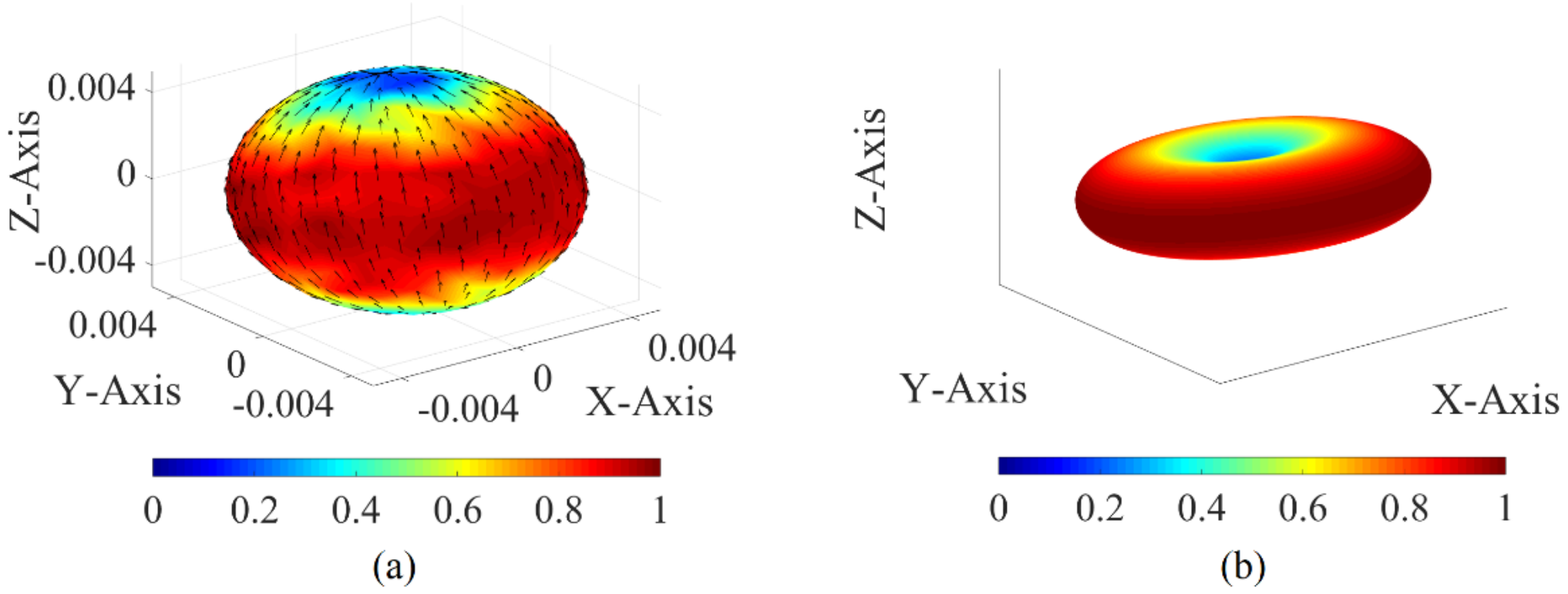

Figure 4-82 The basic variable and radiation pattern of the third degenerate state of the DP-CM1 working at 7.35GHz and shown in Figure 4-74. (a) the equivalent surface magnetic current (basic variable) on $\partial V_{\text{mul}}^{\text{out}}$; (b) the radiation pattern at infinity

By comparing Figures 4-75&4-80, 4-81, and 4-82, it is easy to find out that: the three degenerate states of “resonant” DP-CM1 working at 7.35GHz originate from the spacial rotation symmetry of the multiply connected material spherical shell considered in this subsection.

In the following Section 4.8, we will summarize this chapter from the aspects of fundamental principle, central philosophy, main method, and important conclusions.

## 4.8 Chapter Summary

In WEP framework, this chapter mainly studies the method to establish the CMT for material systems by orthogonalizing frequency-domain DPO.

Firstly, this chapter provides the mathematical expression for the WEP of material systems in electromagnetics, and then introduces the concept of DP for material systems —— the power done by the resultant fields acting on scattered sources. Originating from the concept of DP and employing the volume equivalence principle and GSEP of material systems, this chapter derives the scattered-volume-source-based volume formulation and equivalent-surface-source-based surface formulation of the DPO.

Afterwards, by orthogonalizing frequency-domain DPO, this chapter, for material systems, constructs a series of steadily working modes not having net energy exchange in any integral period (i.e. a series of orthogonal modes decoupling frequency-domain DPO) —— the DP-CMs of material systems. This chapter also proves, from the aspects of mathematical formulations and numerical examples, that the DP-CMs derived from orthogonalizing the volume formulation of DPO and the DP-CMs derived from orthogonalizing the surface formulation of DPO are the same.

At the same time, this chapter also proves that: traditional Harrington’s VIE-MatSca-CMT and SIE-MatSca-CMT and traditional Harrington’s EFIE-MetSca-CMT have the same physical picture —— constructing a series of steadily working modes not having net energy exchange in any integral period. Then, this chapter realizes the transformation for the carrying framework of whole Harrington’s CMT (including VIE-MatSca-CMT & SIE-MatSca-CMT & EFIE-MetSca-CMT) —— from IE framework to WEP framework.

The above transformation brings many benefits to CMT. **1)** Compared with the traditional Harrington’s VIE-MatSca-CMT and SIE-MatSca-CMT established in IE framework, the WEP-MatSca-CMT established in WEP framework provides a series of completely new surface formulations for constructing CMs, and the new formulations have wider applicable range and more concise manifestation form and clearer physical picture and less requirement for computational resources, and the new formulations have not been effectively established in IE framework. **2)** Compared with the traditional Harrington's VIE-MatSca-CMT and SIE-MatSca-CMT established in IE framework, the WEP-MatSca-CMT established in WEP framework is more helpful for revealing the generation mechanism of spurious modes and developing the suppression method for spurious modes. **3)** Compared with the traditional Harrington's VIE-MatSca-CMT and SIE-MatSca-CMT established in IE framework, the WEP-MatSca-CMT established in WEP framework has more general applicability in the aspect of external environment, i.e., WEP-MatSca-CMT is directly applicable to the material systems placed in complex external environment.

In the aspects of the topological structures and material parameters of material systems, WEP-MatSca-CMT also has a wider applicable range than Harrington's VIE-MatSca-CMT and SIE-MatSca-CMT. **A)** In the aspect of topological structures, WEP-MatSca-CMT is applicable to simply connected material bodies, multiply connected material bodies, and multi-body material systems. **B)** In the aspect of material parameters, WEP-MatSca-CMT is applicable to the systems formed by homogeneous isotropic matter, inhomogeneous anisotropic matter, and piecewise inhomogeneous anisotropic matter.

In addition, this chapter, using four different manners, demonstrates the validity of the new surface formulations in the aspect of constructing DP-CMs. **I)** This chapter compares the results derived from the new surface formulations with the results derived from volume formulation, and finds out that the results agree well with each other. **II)** This chapter, focusing on a certain material system, compares the results derived from three different surface formulations, and finds out that the results agree well with each others. **III)** This chapter particularly sets the material parameters of a complicated material system to some special values, such that the complicated material system degenerates into a simple material system. Focusing on the simple material system, this chapter compares the results derived from the new surface formulations for the complicated material system with the results derived from the new surface formulation for the simple material system, and finds out that the results agree well with each other. **IV)** Focusing on the material systems which can be analyzed by both of new formulations and traditional formulation, this chapter compares the results derived from the new surface formulations with the results derived from the traditional formulation, and finds out that the results agree well with each other.

Finally, we want to emphasize that: the traditional Harrington's VIE-MatSca-CMT and SIE-MatSca-CMT in IE framework can be effectively rebuilt in WEP framework, but the new formulations in WEP framework have not been able to be effectively built in IE framework, and the new formulations in WEP framework have many advantages over the traditional formulations in IE framework. At the same time, in WEP framework it is difficult to derive the inappropriate variants which had been derived in IE framework.

In summary, in the aspect of carrying CMT, WEP framework is more advantageous than IE framework; in the aspect of constructing CMs, orthogonalizing DPO method is more advantageous than orthogonalizing IMO method; in the aspect of presenting physical picture, WEP-based orthogonalizing DPO scheme is more advantageous than IE-based orthogonalizing IMO scheme; in the aspect of suppressing spurious modes, WEP-based orthogonalizing DPO scheme is more advantageous than IE-based orthogonalizing IMO scheme.

# Chapter 5 WEP-Based DP-CMs of Composite Scattering Systems

> To those who do not know mathematics it is difficult to get across a real feeling as to the beauty, the deepest beauty, of nature. … If you want to learn about nature, to appreciate nature, it is necessary to understand the language that she speaks in. … [152]
>
> —— Richard P. Feynman (Nobel Prize in Physics, 1965)

This chapter, in WEP framework, focuses on constructing the DP-CMs of metal-material composite systems (which don't have net energy exchange in any integral period) by orthogonalizing frequency-domain DPO, and doing some necessary analysis and discussions for the related topics.

## 5.1 Chapter Introduction

The composite scattering systems① constituted by metallic structures and material structures widely exist in EM engineering domain, such as printed microstrip antennas[9,10,17-19] and the stealth crafts with material coating[11,12] etc. To effectively construct the inherent working modes of an objective composite system is very valuable for extracting the inherent scattering characters of the objective composite system, and then has instructional significance to the related theoretical analysis and engineering design. Recently, by assembling Harrington's EFIE-OpeMetSca-CMT[31,32] and SIE-MatSca-CMT[34], some scholars[53,54], in IE framework, obtained the formulation for constructing the CMs of composite systems, and the result can be called as EFIE-SIE-based CMT for metal-material composite scattering systems (EFIE-SIE-ComSca-CMT).

Different from EFIE-SIE-ComSca-CMT[53,54] (which is established in IE framework and derives CMs from orthogonalizing IMO), this chapter, in WEP framework, is committed to constructing the DP-CMs of composite systems by orthogonalizing DPO, and the related theory is called as work-energy principle based CMT for metal-material composite scattering systems (WEP-ComSca-CMT). Besides possessing a clearer physical picture, WEP-ComSca-CMT has wider applicable ranges in the aspects of

① In what follows, we simply call it as composite system.

external EM environment, system topological structure, and system material parameter, and also provides a series of completely new formulations for constructing CMs. Comparing with traditional formulations, the new formulations have more concise manifestation forms, and require less computational resources, and the new formulations have not been able to be effectively established in traditional IE framework.

In this chapter, we, firstly in Sections 5.2~5.4, generalize the GHFP, GBET, and GFHF for the inhomogeneous anisotropic material systems placed in complex environments (which have been obtained in Appendix C) to the metal-material composite systems placed in complex environments, and then derive the line-surface equivalence principle (LSEP) for composite systems; we, afterwards in Section 5.5, derive the line-surface formulation of the DPO corresponding to composite systems by employing the LSEP and the WEP of composite systems, and then construct the DP-CMs of composite systems by orthogonalizing the DPO; we, finally in Section 5.6, provide the numerical examples corresponding to some typical composite systems to verify the validity of the formulations developed in this chapter.

## 5.2 Symbolic System and Structural Restrictions

In this section, we provide some necessary explanations for the topological structures of the composite systems considered in this chapter, and define the related symbolic system.

### 5.2.1 Symbolic System

In this chapter, we focus on the composite system $D_{\text{com sys}}$ shown in Figure 5-1, and the system is constructed by metallic line part $L_{\text{met}}$, metallic surface part $S_{\text{met}}$, metallic body part $V_{\text{met}}$, and material body part $V_{\text{mat}}$. In this chapter, we will utilize some commonly used concepts in point set topology, such as boundary, interior, and closure, etc. The rigorous mathematical definitions for these concepts can be found in literature [124], so the definitions will not be explicitly given here. In this chapter, the boundaries of sub-structures $L_{\text{met}}$, $S_{\text{met}}$, $V_{\text{met}}$, and $V_{\text{mat}}$ are denoted as $\partial L_{\text{met}}$, $\partial S_{\text{met}}$, $\partial V_{\text{met}}$, and $\partial V_{\text{mat}}$ respectively; the interior and closure of sub-structure $V_{\text{mat}}$ are denoted as $\text{int}\, V_{\text{mat}}$ and $\text{cl}\, V_{\text{mat}}$ respectively; the exterior of sub-structure $V_{\text{mat}}$ is denoted as $\text{ext}\, V_{\text{mat}}$, and $\text{ext}\, V_{\text{mat}}$ is defined as that $\text{ext}\, V_{\text{mat}} = \mathbb{R}^3 \setminus \text{cl}\, V_{\text{mat}}$, where $\mathbb{R}^3$ represents whole 3-D Euclidean space. Obviously, both of $\text{int}\, V_{\text{mat}}$ and $\text{ext}\, V_{\text{mat}}$ are open domains[124].

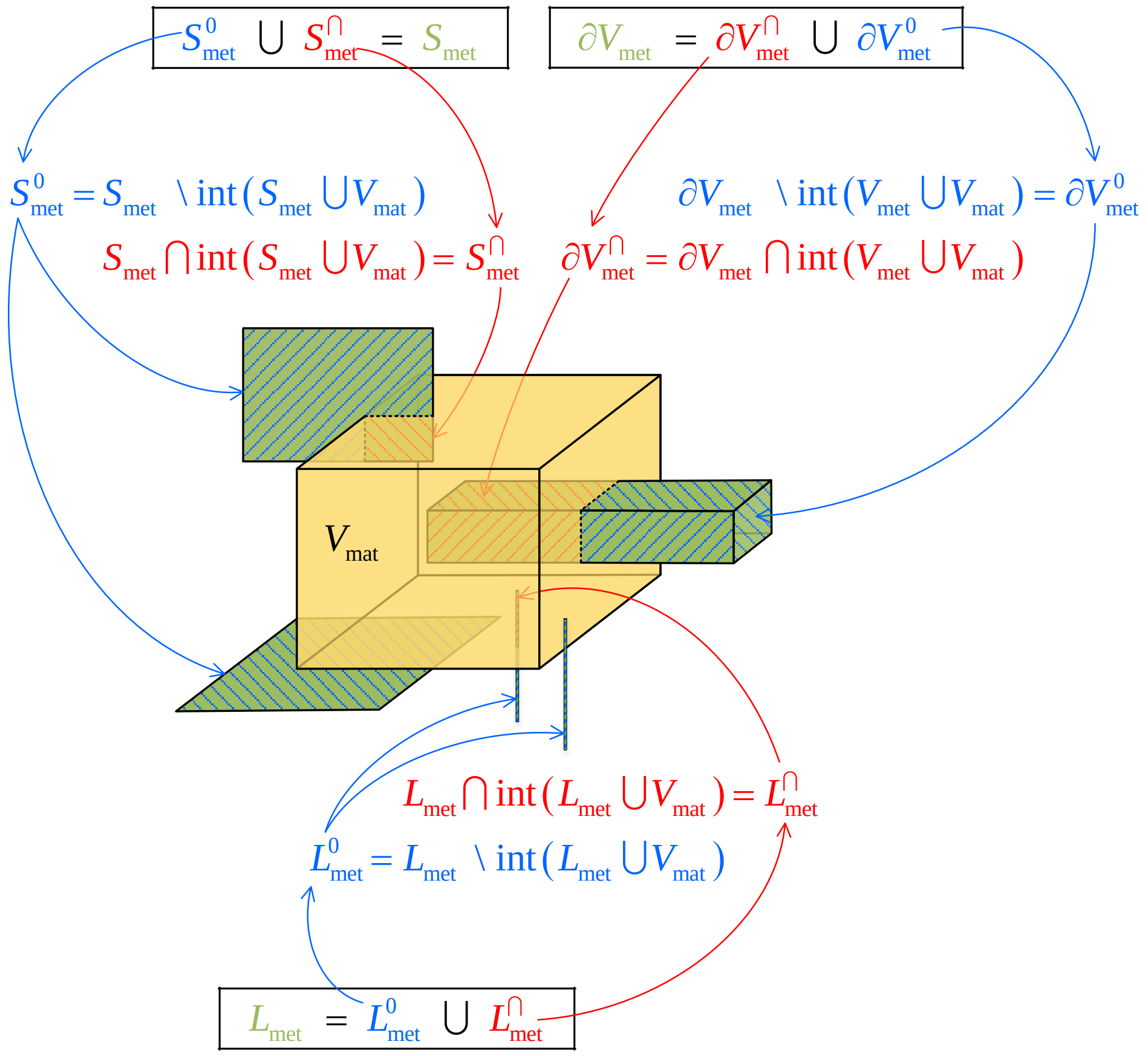

Figure 5-1 The topological structure of the metal-material composite system considered in this chapter and the boundary decomposition for the composite system

When a resultant field $\vec{F}^{\text{inc}}$ ① acts on $D_{\text{com sys}}$, scattered line electric current $\vec{J}^{\text{SL}}$, scattered surface line current $\vec{J}^{\text{SS}}_{\text{met surf}}$, and scattered surface electric current $\vec{J}^{\text{SS}}_{\text{met vol}}$ will be induced on $\partial L_{\text{met}}$, $\partial S_{\text{met}}$, and $\partial V_{\text{met}}$ respectively; conduction volume electric current $\vec{J}^{\text{CV}}_{\text{mat}}$, polarization volume electric current $\vec{J}^{\text{PV}}_{\text{mat}}$, and magnetization volume magnetic current $\vec{M}^{\text{MV}}_{\text{mat}}$ ② will be induced on $\text{int}V_{\text{mat}}$. Here, the superscript "SL" on $\vec{J}^{\text{SL}}$ is the abbreviation of "scattering line (electric current)", and the other superscripts can be explained similarly. To simplify the symbolic system of this chapter, we denote the summation of $\vec{J}^{\text{SS}}_{\text{met surf}}$ and $\vec{J}^{\text{SS}}_{\text{met vol}}$ as $\vec{J}^{\text{SS}}$, i.e., $\vec{J}^{\text{SS}} = \vec{J}^{\text{SS}}_{\text{met surf}} + \vec{J}^{\text{SS}}_{\text{met vol}}$, because they are both the surface electric current distributing on metallic boundary; we denote the summation of $\vec{J}^{\text{CV}}_{\text{mat}}$ and $\vec{J}^{\text{PV}}_{\text{mat}}$ as $\vec{J}^{\text{SV}}$, i.e., $\vec{J}^{\text{SV}} = \vec{J}^{\text{CV}}_{\text{mat}} + \vec{J}^{\text{PV}}_{\text{mat}}$, because they are both the volume electric current distributing on material body; similarly to the volume electric current distributing on material body, we denote $\vec{M}^{\text{MV}}_{\text{mat}}$ as $\vec{M}^{\text{SV}}$, i.e., $\vec{M}^{\text{SV}} = \vec{M}^{\text{MV}}_{\text{mat}}$.

① The resultant field is the summation of the field $\vec{F}_{\text{imp}}$ generated by external excitation source $\vec{J}_{\text{imp}}$ and the field $\vec{F}_{\text{env}}$ generated by external environment sources $\{\vec{J}_{\text{env}}, \vec{M}_{\text{env}}\}$, i.e., $\vec{F}^{\text{inc}} = \vec{F}_{\text{imp}} + \vec{F}_{\text{env}}$.

② For details see literature [110] and the conclusions given in the Appendix A of this dissertation.

Scattered electric currents $\{\vec{J}^{\rm SL},\vec{J}^{\rm SS}\}$ and scattered EM currents $\{\vec{J}^{\rm SV},\vec{M}^{\rm SV}\}$ will generate scattered field $\vec{F}^{\rm sca}$ together. We call the summation of $\vec{F}^{\rm sca}$ and $\vec{F}^{\rm inc}$ as total field, and denote the total field as $\vec{F}^{\rm tot}$, i.e., $\vec{F}^{\rm tot}=\vec{F}^{\rm inc}+\vec{F}^{\rm sca}$. For the convenience of the following discussions, we decompose $\vec{F}^{\rm sca}$ into the $\vec{F}^{\rm sca}_{\rm met}$ generated by metal-based scattered sources $\{\vec{J}^{\rm SL},\vec{J}^{\rm SS}\}$ and the $\vec{F}^{\rm sca}_{\rm mat}$ generated by material-based scattered sources $\{\vec{J}^{\rm SV},\vec{M}^{\rm SV}\}$, and it is obvious that $\vec{F}^{\rm sca}=\vec{F}^{\rm sca}_{\rm met}+\vec{F}^{\rm sca}_{\rm mat}$.

### 5.2.2 Structural Restrictions

From a purely mathematical point of view, we have relationships $L_{\rm met}\subseteq {\rm cl}\,L_{\rm met}$, $S_{\rm met}\subseteq {\rm cl}\,S_{\rm met}$, and $V_{\rm met}\subseteq {\rm cl}\,V_{\rm met}$ [124]. However, from a practical point of view, this chapter restricts $L_{\rm met}$, $S_{\rm met}$, and $V_{\rm met}$ as that:

$$\text{Restriction for } L_{\rm met} \;:\; L_{\rm met} = {\rm cl}\,L_{\rm met} \tag{5-1a}$$

$$\text{Restriction for } S_{\rm met} \;:\; S_{\rm met} = {\rm cl}\,S_{\rm met} \tag{5-1b}$$

$$\text{Restriction for } V_{\rm met} \;:\; V_{\rm met} = {\rm cl}\,V_{\rm met} \tag{5-1c}$$

The restrictions can be vividly understood as that: there doesn't exist any "point-type hole" on $L_{\rm met}$, i.e., $L_{\rm met}$ doesn't have any breakpoint; there doesn't exist any "point-type hole" and "line-type hole" on $S_{\rm met}$, i.e., $S_{\rm met}$ doesn't have any sand hole and crack; there doesn't exist any "point-type hole", "line-type hole", and "surface-type hole" on $V_{\rm met}$, i.e., $V_{\rm met}$ doesn't have any sand hole, pinhole, and crack. In addition, restrictions (5-1a) and (5-1b) also imply that: in $\mathbb{R}^3$, $L_{\rm met}=\partial L_{\rm met}$ and $S_{\rm met}=\partial S_{\rm met}$ [124]. Based on the above considerations, this chapter restricts the topological structure of $V_{\rm mat}$ as follows:

$$\text{Restriction for } V_{\rm mat} \;:\; {\rm cl}V_{\rm mat}\setminus V_{\rm mat} = \partial V_{\rm mat}\cap\left(L_{\rm met}\cup S_{\rm met}\cup\partial V_{\rm met}\right) \tag{5-1d}$$

Restriction (5-1d) can be vividly understood as that: on $V_{\rm mat}$, there doesn't exist any environment-filled "point-type hole", "line-type hole", and "surface-type hole", i.e., $V_{\rm mat}$ doesn't have any environment-filled sand hole, pinhole, and crack; all of the pinholes on $V_{\rm mat}$ originate from the submergence of $L_{\rm met}$; all of the cracks on $V_{\rm mat}$ originate from the submergence of $S_{\rm met}$. In summary, all of the pinholes and cracks on $V_{\rm mat}$ originate from the submergence of metallic structures.

From a practical point of view, this chapter also does some further restrictions for $L_{\rm met}$ and $S_{\rm met}$ as follows:

$$\text{Further restriction for } L_{\rm met} \;:\; L_{\rm met} = {\rm cl}\left(L_{\rm met}\setminus\left(S_{\rm met}\cup V_{\rm met}\right)\right) \tag{5-1e}$$

$$\text{Further restriction for } S_{\rm met} \;:\; S_{\rm met} = {\rm cl}\left(S_{\rm met}\setminus V_{\rm met}\right) \tag{5-1f}$$

Restriction (5-1e) is equivalent to that: the intersection between $L_{\text{met}}$ and $S_{\text{met}} \cup V_{\text{met}}$ can only be some discrete points, but cannot be any line. Restriction (5-1f) is equivalent to that: the intersection between $S_{\text{met}}$ and $V_{\text{met}}$ can only be some discrete points or lines, but cannot be any surface. Restrictions (5-1e) and (5-1f) imply that: the structures shown in Figure 5-2 are not considered in this chapter. In addition, this chapter also restricts $V_{\text{met}}$ to simply connected body, because there is no need to distinguish metallic bodies from their connectivity numbers.

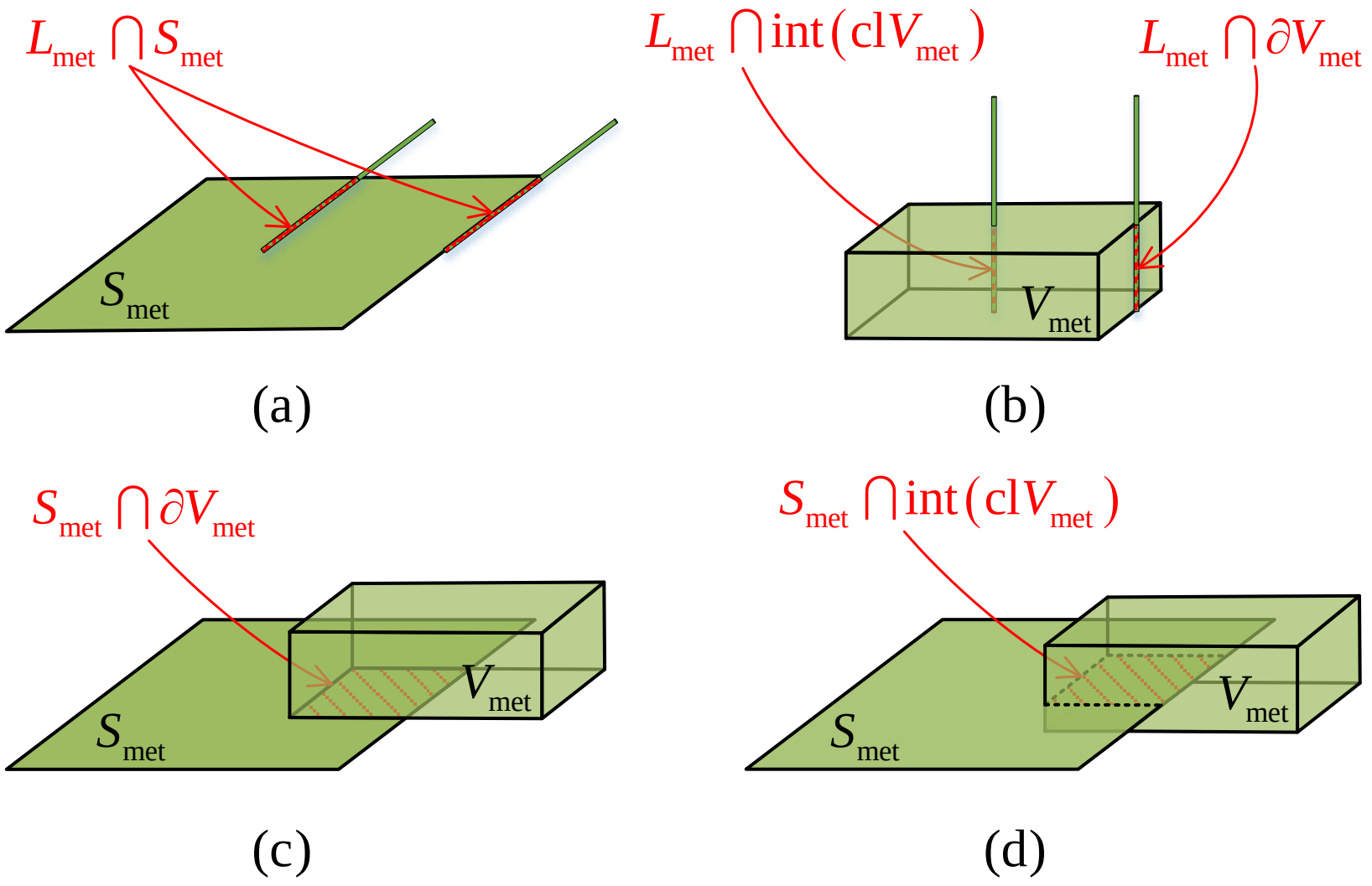

Figure 5-2 The topologies of some typical metallic structures which are not considered in this chapter. (a) a part of metallic line contacts with metallic surface; (b) a part of metallic line contacts with or is submerged into metallic body; (c) a part of metallic surface contacts with the boundary of metallic body; (d) a part of metallic surface is submerged into metallic body

Finally, this chapter also restricts $V_{\text{mat}}$ to simply connected inhomogeneous anisotropic material bodies, and the multiply connected cases and non-connected cases can be similarly discussed by employing the results given in the Appendix C and Chapter 4 of this dissertation. Thus, rigorously speaking, the $V_{\text{mat}}$ focused on by this chapter should be more completely denoted as $V_{\text{sim mat}}$. But, to simplify the symbolic system of this chapter, we omit the subscript "sim" on $V_{\text{sim mat}}$, and this will not lead to any confusion, and at the same time the formulations obtained in this chapter can be directly applied to the multiply connected $V_{\text{mat}}$ cases and the non-connected $V_{\text{mat}}$ cases. In addition, this chapter also restricts the material parameters $\vec{\sigma}_{\text{mat}}(\vec{r})$, $\vec{\varepsilon}_{\text{mat}}(\vec{r})$, and $\vec{\mu}_{\text{mat}}(\vec{r})$ of $V_{\text{mat}}$ to two-order symmetrical tensors, and the reason to do this restriction can be found in Appendix B1.

## 5.3 Boundary Decompositions and Current Decompositions for Composite Systems

In this chapter, the formulations for constructing the DP-CMs of composite systems only contain the scattered sources distributing on metallic boundaries and the equivalent sources distributing on material boundaries. This section firstly decomposes the various boundaries included in the composite system shown in Figure 5-1, and then decomposes the scattered and equivalent sources distributing on the boundaries by employing the obtained boundary decompositions.

### 5.3.1 Boundary Decompositions

In this Subsection, we respectively decompose the metallic boundaries and the material boundaries as below.

**1) Decompositions for Metallic Boundaries**

We decompose metallic boundaries $L_{\text{met}}$, $S_{\text{met}}$ ①, and $\partial V_{\text{met}}$ as follows:

$$L_{\text{met}} = L_{\text{met}}^{0} \cup L_{\text{met}}^{\cap} \tag{5-2}$$

$$S_{\text{met}} = S_{\text{met}}^{0} \cup S_{\text{met}}^{\cap} \tag{5-3}$$

$$\partial V_{\text{met}} = \partial V_{\text{met}}^{0} \cup \partial V_{\text{met}}^{\cap} \tag{5-4}$$

The sub-boundaries $L_{\text{met}}^{0}$ and $L_{\text{met}}^{\cap}$ in decomposition formulation (5-2) are defined as follows:

$$\begin{aligned} L_{\text{met}}^{0} &= L_{\text{met}} \setminus \text{int}\left(L_{\text{met}} \cup V_{\text{mat}}\right) \\ &= \partial L_{\text{met}} \setminus \text{int}\left(L_{\text{met}} \cup V_{\text{mat}}\right) \end{aligned} \tag{5-5a}$$

$$\begin{aligned} L_{\text{met}}^{\cap} &= L_{\text{met}} \cap \text{int}\left(L_{\text{met}} \cup V_{\text{mat}}\right) \\ &= \partial L_{\text{met}} \cap \text{int}\left(L_{\text{met}} \cup V_{\text{mat}}\right) \end{aligned} \tag{5-5b}$$

The sub-boundaries $S_{\text{met}}^{0}$ and $S_{\text{met}}^{\cap}$ in decomposition formulation (5-3) are defined as follows:

$$\begin{aligned} S_{\text{met}}^{0} &= S_{\text{met}} \setminus \text{int}\left(S_{\text{met}} \cup V_{\text{mat}}\right) \\ &= \partial S_{\text{met}} \setminus \text{int}\left(S_{\text{met}} \cup V_{\text{mat}}\right) \end{aligned} \tag{5-6a}$$

$$\begin{aligned} S_{\text{met}}^{\cap} &= S_{\text{met}} \cap \text{int}\left(S_{\text{met}} \cup V_{\text{mat}}\right) \\ &= \partial S_{\text{met}} \cap \text{int}\left(S_{\text{met}} \cup V_{\text{mat}}\right) \end{aligned} \tag{5-6b}$$

① Based on decompositions (5-1a) and (5-1b), it is known that: in Euclidean space $\mathbb{R}^3$, metallic line structure $L_{\text{met}}$ and metallic surface structure $S_{\text{met}}$ themselves are just their boundaries.

The sub-boundaries $\partial V_{\text{met}}^{0}$ and $\partial V_{\text{met}}^{\cap}$ in decomposition formulation (5-4) are defined as follows:

$$\partial V_{\text{met}}^{0} = \partial V_{\text{met}} \setminus \text{int}\left(V_{\text{met}} \cup V_{\text{mat}}\right) \quad (5\text{-}7a)$$

$$\partial V_{\text{met}}^{\cap} = \partial V_{\text{met}} \cap \text{int}\left(V_{\text{met}} \cup V_{\text{mat}}\right) \quad (5\text{-}7b)$$

The second equalities of above decomposition formulations (5-5a)~(5-6b) are based on the facts that: $L_{\text{met}} = \partial L_{\text{met}}$ and $S_{\text{met}} = \partial S_{\text{met}}$.

Above $L_{\text{met}}^{0}$ and $L_{\text{met}}^{\cap}$ can be vividly understood as the part exposed to environment and the part submerged into $V_{\text{mat}}$ respectively, and above $S_{\text{met}}^{0}$ and $S_{\text{met}}^{\cap}$ can be similarly understood, as shown in Figure 5-1. Above $\partial V_{\text{met}}^{0}$ and $\partial V_{\text{met}}^{\cap}$ can be vividly understood as the part being in contact with environment and the part being in contact with $V_{\text{mat}}$, as shown in Figure 5-1. In addition, the following decoupling relationships among various sub-boundaries are obvious:

$$L_{\text{met}}^{0} \cap L_{\text{met}}^{\cap} = \varnothing \quad (5\text{-}8)$$

$$S_{\text{met}}^{0} \cap S_{\text{met}}^{\cap} = \varnothing \quad (5\text{-}9)$$

$$\partial V_{\text{met}}^{0} \cap \partial V_{\text{met}}^{\cap} = \varnothing \quad (5\text{-}10)$$

So far, the decompositions for various metallic boundaries have been finished. In what follows, we will do some necessary decompositions for the material boundary.

**2) Decomposition for Material Boundary**

As pointed out in restriction (5-1), $V_{\text{mat}}$ doesn't have any environment-filled sand hole, pinhole, and crack, so we can decompose its boundary $\partial V_{\text{mat}}$ as the following four parts:

$$\text{Boundary Point Part} : \partial V_{\text{mat}}^{\text{point}} = \varnothing \quad (5\text{-}11a)$$

$$\text{Boundary line Part} : \partial V_{\text{mat}}^{\text{line}} = L_{\text{met}}^{\cap} \quad (5\text{-}11a)$$

$$\text{Boundary Open Surface Part} : \partial V_{\text{mat}}^{\text{open surf}} = S_{\text{met}}^{\cap} \quad (5\text{-}11a)$$

$$\text{Boundary Closed Surface Part} : \partial V_{\text{mat}}^{\text{closed surf}} = \partial V_{\text{mat}} \setminus \left(L_{\text{met}}^{\cap} \cup S_{\text{met}}^{\cap}\right) \quad (5\text{-}11a)$$

Obviously, the above four parts are pairwise disjoint, and

**A.** There doesn't exist discrete point on $\partial V_{\text{mat}}$, i.e., $V_{\text{mat}}$ doesn't have the sand hole occupied by metal, because we have restricted in (5-1) that there doesn't exist any discrete point in whole metallic sub-system;

**B.** All discrete line structures in $\partial V_{\text{mat}}$ originate from the submergence of metallic lines, as shown in Figure 5-1;

**C.** All open surface structures in $\partial V_{\rm mat}$ originate from the submergence of metallic surfaces, as shown in Figure 5-1;

**D.** The closed surface structure in $\partial V_{\rm mat}$ originates from the contact between $V_{\rm mat}$ and environment, the contact between $V_{\rm mat}$ and metallic lines (here, the metallic lines are not submerged into $V_{\rm mat}$), the contact between $V_{\rm mat}$ and metallic surfaces (here, the metallic surfaces are not submerged into $V_{\rm mat}$), and the contact between $V_{\rm mat}$ and metallic bodies, as shown in Figure 5-1. In fact, the closed surface structure in $\partial V_{\rm mat}$ can be further decomposed as follows:

$$\partial V_{\rm mat}^{\rm closed\ surf} = \partial V_{\rm mat}^{0} \bigcup \partial V_{\rm met}^{\cap} \tag{5-12}$$

where the definition for $\partial V_{\rm met}^{\cap}$ has been given in formulation (5-7b), and the definition for $\partial V_{\rm mat}^{0}$ is as follows:

$$\begin{aligned}\partial V_{\rm mat}^{0} &= \partial V_{\rm mat}^{\rm closed\ surf} \setminus \partial V_{\rm met}^{\cap} \\ &= \left(\partial V_{\rm mat} \setminus \left(L_{\rm met}^{\cap} \bigcup S_{\rm met}^{\cap}\right)\right) \setminus \partial V_{\rm met}^{\cap} \\ &= \partial V_{\rm mat} \setminus \left(L_{\rm met}^{\cap} \bigcup S_{\rm met}^{\cap} \bigcup \partial V_{\rm met}^{\cap}\right)\end{aligned} \tag{5-13}$$

To simplify the symbolic system of this chapter, we denote the union of $\partial V_{\rm mat}^{\rm open\ surf}$ and $\partial V_{\rm mat}^{\rm closed\ surf}$ as $\partial V_{\rm mat}^{\rm surf}$, i.e., the whole surface structure in material boundary is denoted as $\partial V_{\rm mat}^{\rm surf} = \partial V_{\rm mat}^{\rm open\ surf} \bigcup \partial V_{\rm mat}^{\rm closed\ surf}$. Then, we have the decomposition for whole material boundary as follows:

$$\partial V_{\rm mat} = \overbrace{\varnothing}^{\partial V_{\rm mat}^{\rm point}} \bigcup \overbrace{L_{\rm met}^{\cap}}^{\partial V_{\rm mat}^{\rm line}} \bigcup \overbrace{\underbrace{S_{\rm met}^{\cap}}_{\partial V_{\rm mat}^{\rm open\ surf}} \bigcup \underbrace{\partial V_{\rm met}^{\cap} \bigcup \partial V_{\rm mat}^{0}}_{\partial V_{\rm mat}^{\rm closed\ surf}}}^{\partial V_{\rm mat}^{\rm surf}} \tag{5-14}$$

So far, we have finished the decomposition for material boundary.

## 5.3.2 Current Decompositions

Based on the boundary decomposition method proposed in above Subsection 5.3.1, this subsection decomposes the currents distributing on the boundaries correspondingly. Based on the obtained current decompositions, we will derive the dependent relationships among the various electric sub-currents and the dependent relationships among the various magnetic sub-currents. The obtained relationships are indispensable for deriving the generalized Huygens-Fresnel principle (GHFP), generalized backward extinction theorem (GBET) and generalized Franz-Harrington formulation (GFHF) corresponding to metal-material composite scattering systems in subsequent Section 5.4.

**1) Decompositions for the Scattered Currents Distributing on Metallic Boundaries**

Based on formulations (5-2)~(5-4) and formulations (5-8)~(5-10), scattered line electric current $\vec{J}^{\text{SL}}$ and scattered surface electric current $\vec{J}^{\text{SS}}$ can be decomposed as follows:

$$\vec{J}^{\text{SL}}\left(\vec{r}\right)= \vec{J}_0^{\text{SL}}\left(\vec{r}\right)+\vec{J}_{\cap}^{\text{SL}}\left(\vec{r}\right) \quad , \quad \vec{r}\in L_{\text{met}} \tag{5-15}$$

$$\vec{J}^{\text{SS}}\left(\vec{r}\right)= \vec{J}_0^{\text{SS}}\left(\vec{r}\right)+\vec{J}_{\cap}^{\text{SS}}\left(\vec{r}\right) \quad , \quad \vec{r}\in S_{\text{met}}\bigcup\partial V_{\text{met}} \tag{5-16}$$

In decomposition formulation (5-15), sub-currents $\vec{J}_0^{\text{SL}}$ and $\vec{J}_{\cap}^{\text{SL}}$ are defined as follows:

$$\vec{J}_0^{\text{SL}}(\vec{r}) = \begin{cases} \vec{J}^{\text{SL}}(\vec{r}) & , \quad \vec{r}\in L_{\text{met}}^{0} \\ 0 & , \quad \vec{r}\in L_{\text{met}}^{\cap} \end{cases} \tag{5-17a}$$

$$\vec{J}_{\cap}^{\text{SL}}(\vec{r}) = \begin{cases} 0 & , \quad \vec{r}\in L_{\text{met}}^{0} \\ \vec{J}^{\text{SL}}(\vec{r}) & , \quad \vec{r}\in L_{\text{met}}^{\cap} \end{cases} \tag{5-17b}$$

In decomposition formulation (5-16), sub-currents $\vec{J}_0^{\text{SS}}$ and $\vec{J}_{\cap}^{\text{SS}}$ are defined as follows:

$$\vec{J}_0^{\text{SS}}(\vec{r}) = \begin{cases} \vec{J}^{\text{SS}}(\vec{r}) & , \quad \vec{r}\in S_{\text{met}}^{0}\bigcup\partial V_{\text{met}}^{0} \\ 0 & , \quad \vec{r}\in S_{\text{met}}^{\cap}\bigcup\partial V_{\text{met}}^{\cap} \end{cases} \tag{5-18a}$$

$$\vec{J}_{\cap}^{\text{SS}}(\vec{r}) = \begin{cases} 0 & , \quad \vec{r}\in S_{\text{met}}^{0}\bigcup\partial V_{\text{met}}^{0} \\ \vec{J}^{\text{SS}}(\vec{r}) & , \quad \vec{r}\in S_{\text{met}}^{\cap}\bigcup\partial V_{\text{met}}^{\cap} \end{cases} \tag{5-18b}$$

So far, the decompositions for the various scattered electric currents on metallic boundaries have been finished.

**2) Decompositions for the Equivalent Currents Distributing on Material Boundary**

In what follows, we will, based on the boundary decomposition (5-14) for material boundary, define the equivalent sub-currents distributing on various material sub-boundaries.

**(2.1) Decompositions for the Equivalent Surface Currents Distributing on $\partial V_{\text{mat}}^{\text{closed surf}}$ (Where $\partial V_{\text{mat}}^{\text{closed surf}} = \partial V_{\text{mat}}^{0}\bigcup\partial V_{\text{met}}^{\cap}$)**

Based on the boundary decompositions proposed in Subsection 5.3.1, the equivalent surface currents $\{\vec{J}_{\text{closed surf}}^{\text{ES}},\vec{M}_{\text{closed surf}}^{\text{ES}}\}$ on boundary closed surface part $\partial V_{\text{mat}}^{\text{closed surf}}$ can be decomposed as follows:

$$\vec{J}_{\text{closed surf}}^{\text{ES}}\left(\vec{r}\right) = \vec{J}_0^{\text{ES}}\left(\vec{r}\right)+\vec{J}_{\partial V_{\text{met}}^{\cap}}^{\text{ES}}\left(\vec{r}\right) \quad , \quad \vec{r}\in\partial V_{\text{mat}}^{\text{closed surf}} \tag{5-19a}$$

$$\vec{M}_{\text{closed surf}}^{\text{ES}}\left(\vec{r}\right)= \vec{M}_0^{\text{ES}}\left(\vec{r}\right)+\vec{M}_{\partial V_{\text{met}}^{\cap}}^{\text{ES}}\left(\vec{r}\right) \quad , \quad \vec{r}\in\partial V_{\text{mat}}^{\text{closed surf}} \tag{5-19b}$$

where equivalent sub-currents $\{\vec{J}_0^{\text{ES}}, \vec{M}_0^{\text{ES}}\}$ are defined as follows:

$$\vec{J}_0^{\text{ES}}(\vec{r}) = \hat{n}_{\text{mat}}^{-}(\vec{r}) \times \left[\vec{H}^{\text{tot}}(\vec{r}_{\text{mat}})\right]_{\vec{r}_{\text{mat}} \to \vec{r}} \quad , \quad \vec{r} \in \partial V_{\text{mat}}^{0} \tag{5-20a}$$

$$\vec{M}_0^{\text{ES}}(\vec{r}) = \left[\vec{E}^{\text{tot}}(\vec{r}_{\text{mat}})\right]_{\vec{r}_{\text{mat}} \to \vec{r}} \times \hat{n}_{\text{mat}}^{-}(\vec{r}) \quad , \quad \vec{r} \in \partial V_{\text{mat}}^{0} \tag{5-20b}$$

and equivalent sub-currents $\{\vec{J}_{\partial V_{\text{met}}^{\cap}}^{\text{ES}}, \vec{M}_{\partial V_{\text{met}}^{\cap}}^{\text{ES}}\}$ are defined as follows:

$$\vec{J}_{\partial V_{\text{met}}^{\cap}}^{\text{ES}}(\vec{r}) = \hat{n}_{\text{mat}}^{-}(\vec{r}) \times \left[\vec{H}^{\text{tot}}(\vec{r}_{\text{mat}})\right]_{\vec{r}_{\text{mat}} \to \vec{r}} \quad , \quad \vec{r} \in \partial V_{\text{met}}^{\cap} \tag{5-21a}$$

$$\vec{M}_{\partial V_{\text{met}}^{\cap}}^{\text{ES}}(\vec{r}) = \left[\vec{E}^{\text{tot}}(\vec{r}_{\text{mat}})\right]_{\vec{r}_{\text{mat}} \to \vec{r}} \times \hat{n}_{\text{mat}}^{-}(\vec{r}) \quad , \quad \vec{r} \in \partial V_{\text{met}}^{\cap} \tag{5-21b}$$

In definition formulations (5-20) and (5-21), $\vec{r}_{\text{mat}} \in \text{int}\, V_{\text{mat}}$, and $\vec{r}_{\text{mat}}$ approaches $\vec{r}$; $\hat{n}_{\text{mat}}^{-}$ is the normal vector of surface $\partial V_{\text{mat}}^{\text{closed surf}}$, and points to the interior of $V_{\text{mat}}$.

**(2.2) Decompositions for the Equivalent Surface Currents Distributing on $\partial V_{\text{mat}}^{\text{open surf}}$ (Where $\partial V_{\text{mat}}^{\text{open surf}} = S_{\text{met}}^{\cap}$)**

To effectively introduce the equivalent surface currents distributing on material boundary open surface $S_{\text{met}}^{\cap}$, firstly we consider the examples shown in Figures 5-3(a) and 5-3(c), i.e. a thick metallic slab $V_{\text{met}}^{\text{slab}}$ submerged into $V_{\text{mat}}$, and secondly we treat the $S_{\text{met}}^{\cap}$ shown in Figures 5-3(b) and 5-3(d) as the limitation of $V_{\text{met}}^{\text{slab}}$ when the thickness of $V_{\text{met}}^{\text{slab}}$ approaches zero.

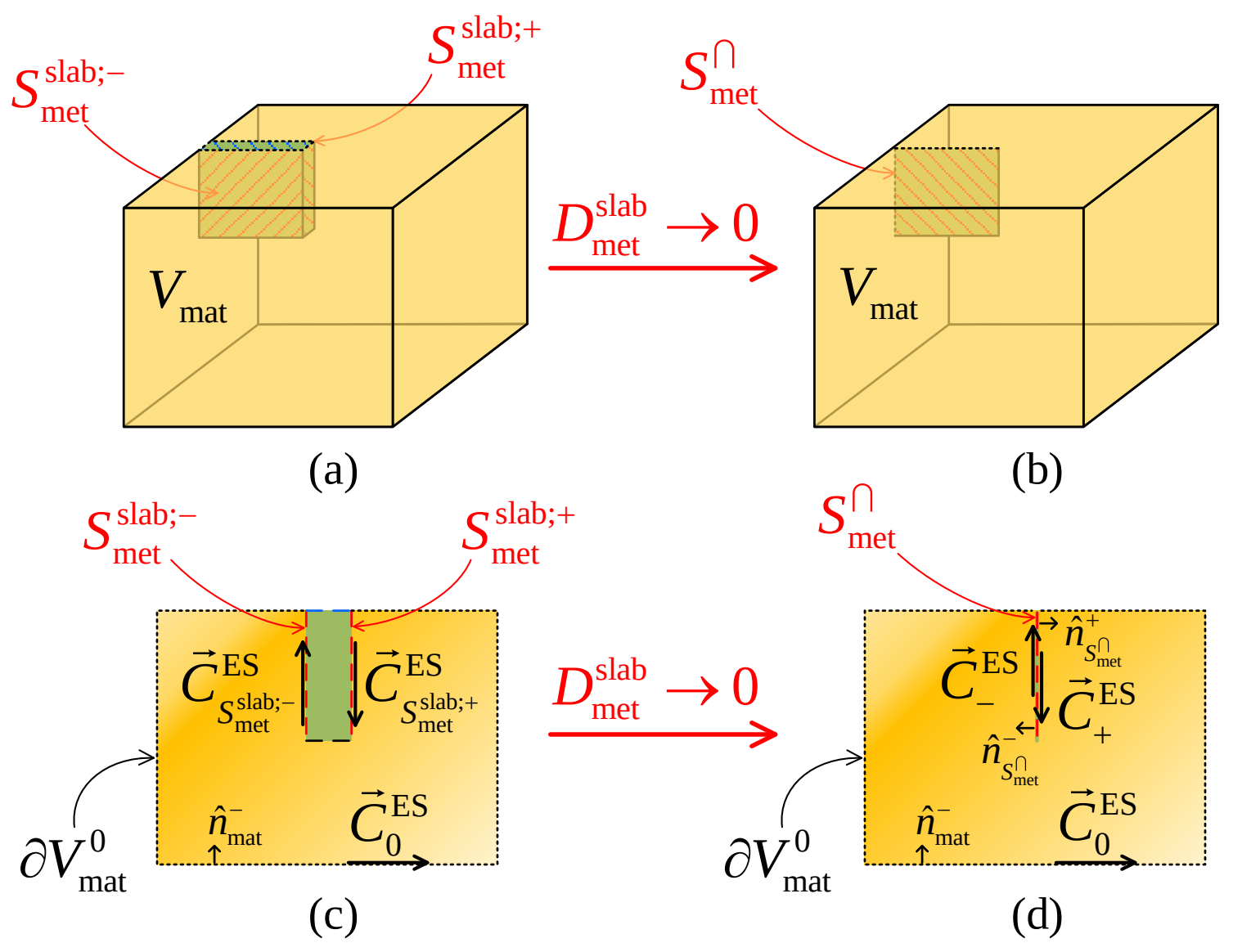

Figure 5-3 The transformation from a thick metallic slab to an infinitely thin metallic surface. (a) a thick metallic slab submerged into a material body; (b) the limitation case (that the thickness of the slab approaches zero) —— an infinitely thin metallic surface submerged into a material body; (c) the sectional view of Figure 5-3 (a) and the related equivalent surface currents; (d) the sectional view of Figure 5-3 (b) and the related equivalent surface currents

We denote the plus and minus sides of $V_{\text{met}}^{\text{slab}}$ as $S_{\text{met};+}^{\text{slab}}$ and $S_{\text{met};-}^{\text{slab}}$ respectively, and denote the scattered surface electric currents distributing on $S_{\text{met};+}^{\text{slab}}$ and $S_{\text{met};-}^{\text{slab}}$ as $\vec{J}_{S_{\text{met}}^{\text{slab};+}}^{\text{SS}}$ and $\vec{J}_{S_{\text{met}}^{\text{slab};-}}^{\text{SS}}$ respectively. Obviously, both of $S_{\text{met}}^{\text{slab};+}$ and $S_{\text{met}}^{\text{slab};-}$ are the parts of material boundary $\partial V_{\text{mat}}$, i.e., $S_{\text{met}}^{\text{slab};+}, S_{\text{met}}^{\text{slab};-} \subset \partial V_{\text{mat}}$ [124]. Thus, there will exist equivalent surface currents distributing on $S_{\text{met}}^{\text{slab};+}$ and $S_{\text{met}}^{\text{slab};-}$, and we correspondingly denote the currents as $\{\vec{J}_{S_{\text{met}}^{\text{slab};+}}^{\text{ES}}, \vec{M}_{S_{\text{met}}^{\text{slab};+}}^{\text{ES}}\}$ and $\{\vec{J}_{S_{\text{met}}^{\text{slab};-}}^{\text{ES}}, \vec{M}_{S_{\text{met}}^{\text{slab};-}}^{\text{ES}}\}$ respectively. If the thickness of $V_{\text{met}}^{\text{slab}}$ is denoted as $D_{\text{met}}^{\text{slab}}$, then the following limitations exist:

$$\lim_{D_{\text{met}}^{\text{slab}} \to 0} S_{\text{met}}^{\text{slab};\pm} = S_{\text{met}}^{\cap} \tag{5-22}$$

$$\lim_{D_{\text{met}}^{\text{slab}} \to 0} \left( \vec{J}_{S_{\text{met}}^{\text{slab};+}}^{\text{SS}} + \vec{J}_{S_{\text{met}}^{\text{slab};-}}^{\text{SS}} \right) = \vec{J}_{\cap}^{\text{SS}} \tag{5-23}$$

$$\lim_{D_{\text{met}}^{\text{slab}} \to 0} \vec{J}_{S_{\text{met}}^{\text{slab};\pm}}^{\text{ES}} = \vec{J}_{\pm}^{\text{ES}} \tag{5-24a}$$

$$\lim_{D_{\text{met}}^{\text{slab}} \to 0} \vec{M}_{S_{\text{met}}^{\text{slab};\pm}}^{\text{ES}} = \vec{M}_{\pm}^{\text{ES}} \tag{5-24b}$$

The limitation values $\vec{J}_{\pm}^{\text{ES}}$ and $\vec{M}_{\pm}^{\text{ES}}$ on the RHS of formulation (5-24) distribute on $S_{\text{met}}^{\cap}$, and they are defined as follows:

$$\vec{J}_{\pm}^{\text{ES}}(\vec{r}) = \hat{n}_{S_{\text{met}}^{\cap}}^{\pm}(\vec{r}) \times \left[ \vec{H}^{\text{tot}}(\vec{r}_{\text{mat}}^{\pm}) \right]_{\vec{r}_{\text{mat}}^{\pm} \to \vec{r}} \quad , \quad \vec{r} \in S_{\text{met}}^{\cap} \tag{5-25a}$$

$$\vec{M}_{\pm}^{\text{ES}}(\vec{r}) = \left[ \vec{E}^{\text{tot}}(\vec{r}_{\text{mat}}^{\pm}) \right]_{\vec{r}_{\text{mat}}^{\pm} \to \vec{r}} \times \hat{n}_{S_{\text{met}}^{\cap}}^{\pm}(\vec{r}) \quad , \quad \vec{r} \in S_{\text{met}}^{\cap} \tag{5-25b}$$

In definition (5-25), $\vec{r}_{\text{mat}}^{+}, \vec{r}_{\text{mat}}^{-} \in \operatorname{int} V_{\text{mat}}$, and $\vec{r}_{\text{mat}}^{+}$ and $\vec{r}_{\text{mat}}^{-}$ respectively approach $\vec{r}$ from the plus and minus sides of $S_{\text{met}}^{\cap}$; $\hat{n}_{S_{\text{met}}^{\cap}}^{+}$ and $\hat{n}_{S_{\text{met}}^{\cap}}^{-}$ are the normal vectors of $S_{\text{met}}^{\cap}$, and respectively point to the plus and minus sides of $S_{\text{met}}^{\cap}$.

Based on linear superposition principle, the summation of the fields generated by $\{\vec{J}_{+}^{\text{ES}}, \vec{M}_{+}^{\text{ES}}\}$ and $\{\vec{J}_{-}^{\text{ES}}, \vec{M}_{-}^{\text{ES}}\}$ equals the fields generated by $\{\vec{J}_{+}^{\text{ES}} + \vec{J}_{-}^{\text{ES}}, \vec{M}_{+}^{\text{ES}} + \vec{M}_{-}^{\text{ES}}\}$, so we treat $\{\vec{J}_{+}^{\text{ES}} + \vec{J}_{-}^{\text{ES}}, \vec{M}_{+}^{\text{ES}} + \vec{M}_{-}^{\text{ES}}\}$ as a whole instead of considering $\{\vec{J}_{+}^{\text{ES}}, \vec{M}_{+}^{\text{ES}}\}$ and $\{\vec{J}_{-}^{\text{ES}}, \vec{M}_{-}^{\text{ES}}\}$ respectively. In addition, both of $\{\vec{J}_{+}^{\text{ES}}, \vec{M}_{+}^{\text{ES}}\}$ and $\{\vec{J}_{-}^{\text{ES}}, \vec{M}_{-}^{\text{ES}}\}$ distribute on $S_{\text{met}}^{\cap}$, and relationship $\hat{n}_{S_{\text{met}}^{\cap}}^{-} = -\hat{n}_{S_{\text{met}}^{\cap}}^{+}$ holds on whole $S_{\text{met}}^{\cap}$, so we can define the equivalent surface currents distributing on $S_{\text{met}}^{\cap}$ as follows:

$$\begin{aligned} \vec{J}_{\text{open surf}}^{\text{ES}}(\vec{r}) &= \vec{J}_{+}^{\text{ES}}(\vec{r}) + \vec{J}_{-}^{\text{ES}}(\vec{r}) \\ &= \hat{n}_{S_{\text{met}}^{\cap}}^{\pm}(\vec{r}) \times \left[ \vec{H}^{\text{tot}}(\vec{r}_{\text{mat}}^{\pm}) - \vec{H}^{\text{tot}}(\vec{r}_{\text{mat}}^{\mp}) \right]_{\vec{r}_{\text{mat}}^{+}, \vec{r}_{\text{mat}}^{-} \to \vec{r}} \quad , \quad \vec{r} \in S_{\text{met}}^{\cap} \end{aligned} \tag{5-26a}$$

$$\begin{aligned} \vec{M}_{\text{open surf}}^{\text{ES}}(\vec{r}) &= \vec{M}_{+}^{\text{ES}}(\vec{r}) + \vec{M}_{-}^{\text{ES}}(\vec{r}) \\ &= \left[ \vec{E}^{\text{tot}}(\vec{r}_{\text{mat}}^{\pm}) - \vec{E}^{\text{tot}}(\vec{r}_{\text{mat}}^{\mp}) \right]_{\vec{r}_{\text{mat}}^{+}, \vec{r}_{\text{mat}}^{-} \to \vec{r}} \times \hat{n}_{S_{\text{met}}^{\cap}}^{\pm}(\vec{r}) \quad , \quad \vec{r} \in S_{\text{met}}^{\cap} \end{aligned} \tag{5-26b}$$

So far, the definitions for the equivalent surface currents distributing on $\partial V_{\text{mat}}^{\text{open surf}}$ have been finished.

**(2.3) Decompositions for the Equivalent Line Currents Distributing on $\partial V_{\text{mat}}^{\text{line}}$ (Where $\partial V_{\text{mat}}^{\text{line}} = L_{\text{met}}^{\cap}$)**

To effectively introduce the equivalent line currents $\{\vec{J}^{\text{EL}}, \vec{M}^{\text{EL}}\}$ distributing on material boundary line part $L_{\text{met}}^{\cap}$, firstly we consider the example shown in Figure 5-4(a), i.e. a metallic cylinder $V_{\text{met}}^{\text{cylinder}}$ submerged into $V_{\text{mat}}$, and secondly we treat the $L_{\text{met}}^{\cap}$ shown in Figure 5-4(b) as the limitation of $V_{\text{met}}^{\text{cylinder}}$ when the radius of $V_{\text{met}}^{\text{cylinder}}$ approaches zero.

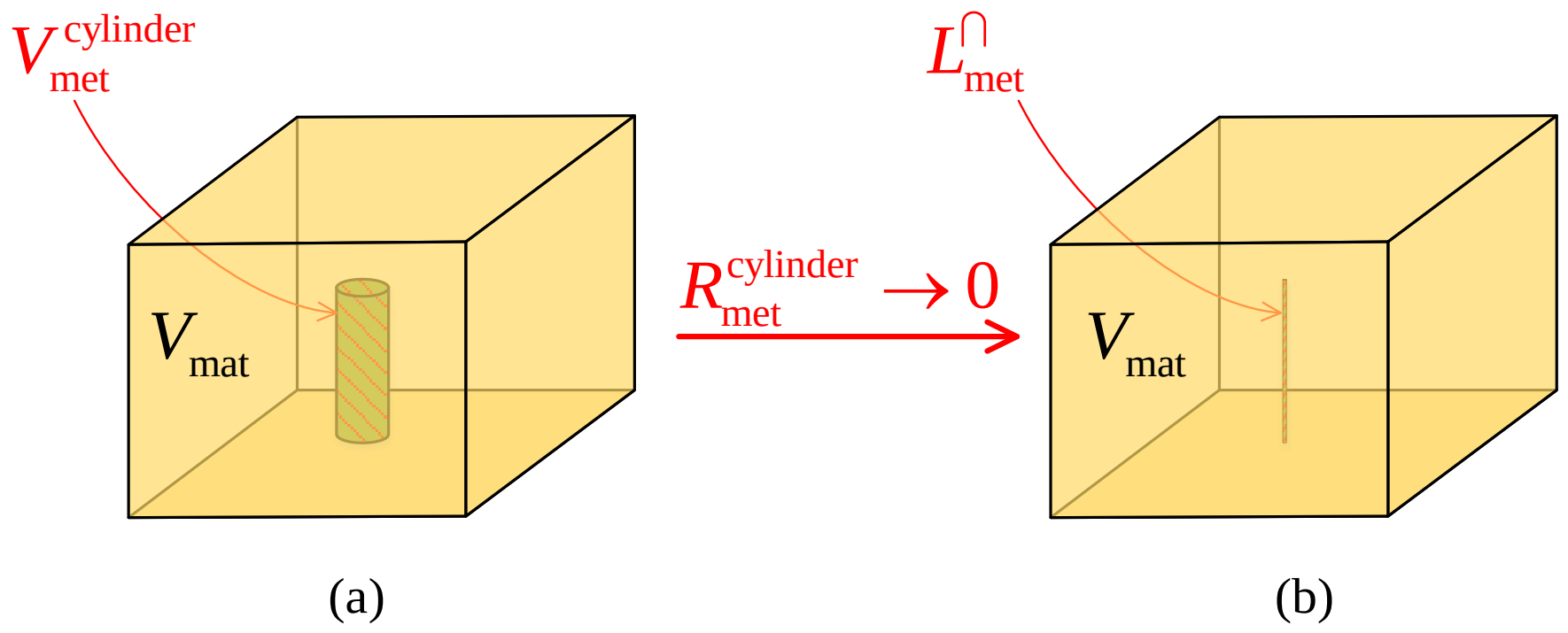

Figure 5-4 The transformation from a thick metallic cylinder to an infinitely thin metallic line. (a) a thick metallic cylinder submerged into a material body; (b) the limitation case (that the radius of the cylinder approaches zero) —— an infinitely thin metallic line submerged into a material body

We denote the boundary of $V_{\text{met}}^{\text{cylinder}}$ as $\partial V_{\text{met}}^{\text{cylinder}}$, and correspondingly denote the scattered surface electric current distributing on $\partial V_{\text{met}}^{\text{cylinder}}$ as $\vec{J}_{\partial V_{\text{met}}^{\text{cylinder}}}^{\text{SS}}$. Obviously, $\partial V_{\text{met}}^{\text{cylinder}}$ is a part of material boundary $\partial V_{\text{mat}}$, i.e., $\partial V_{\text{met}}^{\text{cylinder}} \subset \partial V_{\text{mat}}$ [124]. Thus, there will exist some equivalent surface currents distributing on $\partial V_{\text{met}}^{\text{cylinder}}$, and we correspondingly denote the currents as $\{\vec{J}_{\partial V_{\text{met}}^{\text{cylinder}}}^{\text{ES}}, \vec{M}_{\partial V_{\text{met}}^{\text{cylinder}}}^{\text{ES}}\}$. If the radius $R_{\text{met}}^{\text{cylinder}}$ of $V_{\text{met}}^{\text{cylinder}}$ approaches zero, the following limitations exist:

$$\partial V_{\text{met}}^{\text{cylinder}} \xrightarrow{R_{\text{met}}^{\text{cylinder}} \to 0} L_{\text{met}}^{\cap} \tag{5-27}$$

$$\vec{J}_{\partial V_{\text{met}}^{\text{cylinder}}}^{\text{SS}} \xrightarrow{R_{\text{met}}^{\text{cylinder}} \to 0} \vec{J}_{\cap}^{\text{SL}} \tag{5-28}$$

$$\vec{J}_{\partial V_{\text{met}}^{\text{cylinder}}}^{\text{ES}} \xrightarrow{R_{\text{met}}^{\text{cylinder}} \to 0} \vec{J}^{\text{EL}} \tag{5-29a}$$

$$\vec{M}_{\partial V_{\text{met}}^{\text{cylinder}}}^{\text{ES}} \xrightarrow{R_{\text{met}}^{\text{cylinder}} \to 0} \vec{M}^{\text{EL}} \tag{5-29b}$$

Based on these above, we can define the equivalent line currents $\{\vec{J}^{\text{EL}}, \vec{M}^{\text{EL}}\}$ distributing on $L_{\text{met}}^{\cap}$ as follows:

$$\vec{J}^{\,\text{EL}}(\vec{r}) = \hat{e}_l \lim_{\vec{r}'\to\vec{r}} \oint_{C(\vec{r}')} \vec{H}^{\text{tot}}(\vec{r}')\cdot d\vec{l}' \quad , \quad \vec{r} \in L_{\text{met}}^{\cap} \tag{5-30a}$$

$$\vec{M}^{\,\text{EL}}(\vec{r}) = -\hat{e}_l \lim_{\vec{r}'\to\vec{r}} \oint_{C(\vec{r}')} \vec{E}^{\text{tot}}(\vec{r}')\cdot d\vec{l}' \quad , \quad \vec{r} \in L_{\text{met}}^{\cap} \tag{5-30b}$$

In definition (5-30), integral path $C(\vec{r}')$ is a circle surrounding $L_{\text{met}}^{\cap}$, and the circle is constructed by the points $\vec{r}'$, which are in set $\text{int}V_{\text{mat}}$ and approach $\vec{r}$; unit vector $\hat{e}_l$ is the reference direction of equivalent line currents $\{\vec{J}^{\text{EL}},\vec{M}^{\text{EL}}\}$; $\hat{e}_l$ and the reference direction of integral path $C(\vec{r}')$ satisfy right-hand rule. So far, the definitions for the equivalent line currents on $\partial V_{\text{mat}}^{\text{line}}$ have been finished.

**(2.4) Summarization**

Whole material boundary $\partial V_{\text{mat}}$ can be decomposed into four parts as formulation (5-11), and can also be finely decomposed into five parts as formulation (5-14). Based on the boundary decompositions, the equivalent currents on various sub-boundaries can be defined as formulations (5-20), (5-21), (5-26), and (5-30). To simplify the symbolic system of the following parts of this chapter, we denote the summation of $\vec{C}_{\partial V_{\text{met}}^{\cap}}^{\text{ES}}$ and $\vec{C}_{\text{open surf}}^{\text{ES}}$ as $\vec{C}_{\cap}^{\text{ES}}$, i.e., $\vec{C}_{\cap}^{\text{ES}} = \vec{C}_{\partial V_{\text{met}}^{\cap}}^{\text{ES}} + \vec{C}_{\text{open surf}}^{\text{ES}}$, because $\vec{C}_{\partial V_{\text{met}}^{\cap}}^{\text{ES}}$ and $\vec{C}_{\text{open surf}}^{\text{ES}}$ distribute on the intersection of $\partial V_{\text{mat}}$ and $\partial V_{\text{met}}^{\cap} \bigcup S_{\text{met}}^{\cap}$; we denote the summation of $\vec{C}_{\text{closed surf}}^{\text{ES}}$ and $\vec{C}_{\text{open surf}}^{\text{ES}}$ as $\vec{C}^{\text{ES}}$, i.e., $\vec{C}^{\text{ES}} = \vec{C}_{\text{closed surf}}^{\text{ES}} + \vec{C}_{\text{open surf}}^{\text{ES}}$, because $\vec{C}_{\text{closed surf}}^{\text{ES}}$ and $\vec{C}_{\text{open surf}}^{\text{ES}}$ together constitute the whole of equivalent surface current. In summary, the various equivalent currents distributing on the material boundary of the composite system shown in Figure 5-1 are as follows:

$$\text{Equivalent Currents on } \partial V_{\text{mat}}: \left\{ \vec{C}^{\text{EL}},\ \overbrace{\overbrace{\vec{C}_0^{\text{ES}},\ \vec{C}_{\partial V_{\text{met}}^{\cap}}^{\text{ES}}}^{\vec{C}_{\text{closed surf}}^{\text{ES}}},\ \overbrace{\vec{C}_+^{\text{ES}},\ \vec{C}_-^{\text{ES}}}^{\vec{C}_{\text{open surf}}^{\text{ES}}}}^{\vec{C}^{\text{ES}}} \right\}, \quad \underbrace{\vec{C}_{\partial V_{\text{met}}^{\cap}}^{\text{ES}},\ \vec{C}_+^{\text{ES}},\ \vec{C}_-^{\text{ES}}}_{\vec{C}_{\cap}^{\text{ES}}} \tag{5-31}$$

where $C = J, M$.

**3) Relationships Among the Sub-currents Distributing on Various Sub-boundaries**

Based on the above discussions, we summarize all of the sub-currents distributing on the various boundaries of the composite system shown in Figure 5-1 as follows:

$$\text{Electric Currents}: \left\{ \overbrace{\overbrace{\vec{J}_0^{\text{SL}},\ \vec{J}_{\cap}^{\text{SL}}}^{\vec{J}^{\text{SL}}},\ \overbrace{\vec{J}_0^{\text{SS}},\ \vec{J}_{\cap}^{\text{SS}}}^{\vec{J}^{\text{SS}}}}^{\text{scattered } \vec{J} \text{ on metal boundary}},\ \overbrace{\vec{J}^{\text{EL}},\ \underbrace{\overbrace{\vec{J}_0^{\text{ES}},\ \vec{J}_{\partial V_{\text{met}}^{\cap}}^{\text{ES}}}^{\vec{J}_{\text{closed surf}}^{\text{ES}}},\ \overbrace{\vec{J}_+^{\text{ES}},\ \vec{J}_-^{\text{ES}}}^{\vec{J}_{\text{open surf}}^{\text{ES}}}}_{\vec{J}^{\text{ES}}}}^{\text{equivalent } \vec{J} \text{ on material boundary}} \right\}, \quad \underbrace{\vec{J}_{\partial V_{\text{met}}^{\cap}}^{\text{ES}},\ \vec{J}_+^{\text{ES}},\ \vec{J}_-^{\text{ES}}}_{\vec{J}_{\cap}^{\text{ES}}} \tag{5-32a}$$

$$\text{Magnetic Current}: \left\{ \vec{M}^{\text{EL}}, \overbrace{\underbrace{\vec{M}_0^{\text{ES}}, \vec{M}_{\partial V_{\text{met}}^{\cap}}^{\text{ES}}}_{\vec{M}_{\text{closed surf}}^{\text{ES}}}, \underbrace{\vec{M}_+^{\text{ES}}, \vec{M}_-^{\text{ES}}}_{\vec{M}_{\text{open surf}}^{\text{ES}}}}^{\vec{M}^{\text{ES}}} \right\} \tag{5-32b}$$

In what follows, we will establish some relatively elementary dependent relationships among the above sub-currents.

Based on the tangential boundary conditions of the $\vec{H}^{\text{tot}}$ and $\vec{E}^{\text{tot}}$ on $\partial V_{\text{met}}^{\cap}$, it is obvious that

$$\vec{J}_{\partial V_{\text{met}}^{\cap}}^{\text{ES}}(\vec{r}) = \vec{J}_{\partial V_{\text{met}}^{\cap}}^{\text{SS}}(\vec{r}) \quad , \quad \vec{r} \in \partial V_{\text{met}}^{\cap} \tag{5-33a}$$

$$\vec{M}_{\partial V_{\text{met}}^{\cap}}^{\text{ES}}(\vec{r}) = 0 \quad , \quad \vec{r} \in \partial V_{\text{met}}^{\cap} \tag{5-33b}$$

Similarly, formulations (5-23), (5-24), and (5-26) imply the following relationships:

$$\vec{J}_{S_{\text{met}}^{\text{slab};\pm}}^{\text{ES}}(\vec{r}) = \vec{J}_{S_{\text{met}}^{\text{slab};\pm}}^{\text{SS}}(\vec{r}) \quad , \quad \vec{r} \in S_{\text{met}}^{\text{slab};\pm} \tag{5-34a}$$

$$\vec{M}_{S_{\text{met}}^{\text{slab};\pm}}^{\text{ES}}(\vec{r}) = 0 \quad , \quad \vec{r} \in S_{\text{met}}^{\text{slab};\pm} \tag{5-34b}$$

and

$$\vec{J}_{\text{open surf}}^{\text{ES}}(\vec{r}) = \vec{J}_{\cap}^{\text{SS}}(\vec{r}) \quad , \quad \vec{r} \in S_{\text{met}}^{\cap} \tag{5-35a}$$

$$\vec{M}_{\text{open surf}}^{\text{ES}}(\vec{r}) = 0 \quad , \quad \vec{r} \in S_{\text{met}}^{\cap} \tag{5-35b}$$

In fact, above formulations (5-33) and (5-35) can be collectively written as follows:

$$\vec{J}_{\cap}^{\text{SS}}(\vec{r}) = \vec{J}_{\cap}^{\text{ES}}(\vec{r}) = \begin{cases} \vec{J}_{\text{open surf}}^{\text{ES}}(\vec{r}) & , \quad \vec{r} \in S_{\text{met}}^{\cap} \\ \vec{J}_{\partial V_{\text{met}}^{\cap}}^{\text{ES}}(\vec{r}) & , \quad \vec{r} \in \partial V_{\text{met}}^{\cap} \end{cases} \tag{5-36a}$$

$$0 = \vec{M}_{\cap}^{\text{ES}}(\vec{r}) = \begin{cases} \vec{M}_{\text{open surf}}^{\text{ES}}(\vec{r}) & , \quad \vec{r} \in S_{\text{met}}^{\cap} \\ \vec{M}_{\partial V_{\text{met}}^{\cap}}^{\text{ES}}(\vec{r}) & , \quad \vec{r} \in \partial V_{\text{met}}^{\cap} \end{cases} \tag{5-36b}$$

In addition, we emphasize here that: for any $\vec{r} \in \partial V_{\text{mat}}^{0} \cap S_{\text{met}}$, we have that $\vec{M}_0^{\text{ES}}(\vec{r}) = 0$, because of the homogeneous tangential boundary condition of the $\vec{E}^{\text{tot}}$ on $S_{\text{met}}$; for any $\vec{r} \in \partial V_{\text{mat}}^{0} \cap L_{\text{met}}$, we have that $[\hat{n}_{\text{mat}}^{-}(\vec{r}) \times \hat{e}_l(\vec{r})] \cdot \vec{M}_0^{\text{ES}}(\vec{r}) = 0$, because of the homogeneous tangential boundary condition of the $\vec{E}^{\text{tot}}$ on $L_{\text{met}}$.

In addition, formulations (5-28) and (5-29) imply the following relationships:

$$\vec{J}_{\partial V_{\text{met}}^{\text{cylinder}}}^{\text{ES}}(\vec{r}) = \vec{J}_{\partial V_{\text{met}}^{\text{cylinder}}}^{\text{SS}}(\vec{r}) \quad , \quad \vec{r} \in \partial V_{\text{met}}^{\text{cylinder}} \tag{5-37a}$$

$$\vec{M}_{\partial V_{\text{met}}^{\text{cylinder}}}^{\text{ES}}(\vec{r}) = 0 \quad , \quad \vec{r} \in \partial V_{\text{met}}^{\text{cylinder}} \tag{5-37b}$$

and

$$\vec{J}^{\text{EL}}(\vec{r}) = \vec{J}_{\cap}^{\text{SL}}(\vec{r}) \quad , \quad \vec{r} \in L_{\text{met}}^{\cap} \tag{5-38a}$$

$$\vec{M}^{\text{EL}}(\vec{r}) = 0 \quad , \quad \vec{r} \in L_{\text{met}}^{\cap} \tag{5-38b}$$

Above formulation (5-38a) is just the dependent relationship satisfied by the equivalent and scattered line electric currents distributing on the material boundary line part.

In summary, on the material boundary of the composite system shown in Figure 5-1, there also exists equivalent line electric current besides traditional equivalent surface electric and magnetic currents, so we call the corresponding equivalence principle as line-surface equivalence principle (LSEP) to be distinguished from the surface equivalence principle (SEP) of material systems. In Appendix C, we find out that the traditional Franz-Harrington formulation (FHF) for homogeneous isotropic material systems is just the mathematical expression for SEP, and obtain the GFHF and generalized SEP (GSEP) for inhomogeneous anisotropic material systems. In the following Section 5.4, we will further generalize the results obtained in Appendix C to metal-material composite systems, and obtain the corresponding LSEP and GFHF.

## 5.4 Generalized Equivalence Principle for Composite Systems

We denote whole metal-material composite system as $D_{\text{com sys}}$, and then

$$D_{\text{com sys}} = L_{\text{met}} \cup S_{\text{met}} \cup V_{\text{met}} \cup V_{\text{mat}} \tag{5-39}$$

Based on the theory related to point set topology[124], we have that

$$\partial D_{\text{com sys}} = L_{\text{met}}^{0} \cup S_{\text{met}}^{0} \cup \partial V_{\text{met}}^{0} \cup \partial V_{\text{mat}}^{0} \tag{5-40a}$$

$$\text{int}\, D_{\text{com sys}} = L_{\text{met}}^{\cap} \cup S_{\text{met}}^{\cap} \cup \partial V_{\text{met}}^{\cap} \cup \text{int}\, V_{\text{met}} \cup \text{int}\, V_{\text{mat}} \tag{5-40b}$$

$$\begin{aligned} \text{ext}\, D_{\text{com sys}} &= \mathbb{R}^3 \setminus \text{cl}\, D_{\text{com sys}} \\ &= \mathbb{R}^3 \setminus \quad D_{\text{com sys}} \\ &= \mathbb{R}^3 \setminus (L_{\text{met}} \cup S_{\text{met}} \cup V_{\text{met}} \cup V_{\text{mat}}) \end{aligned} \tag{5-40c}$$

In above formulation (5-40c), the second and third equalities are based on formulations (5-1) and (5-39) respectively.

### 5.4.1 Generalized Huygens-Fresnel Principle and Generalized Backward Extinction Theorem

For the composite system shown in Figure 5-1, the boundary of the system may be non-surface, because the boundary includes some line structures. This chapter calls the boundary as Huygens' boundary, and will derive the mathematical expressions of the GHFP and GBET corresponding to the Huygens' boundary.

**1) GHFP and GBET Corresponding to the Huygens' Boundary Selected as Material Boundary**

If Huygens' boundary is selected as whole material boundary $\partial V_{\text{mat}}$, then, based on the results obtained in Appendix C, we can mathematically express the GHFP of incident field as follows:

$$\begin{aligned}\left.\begin{array}{lll}\text{ext}\, D_{\text{com sys}} & : & 0 \\ \text{int}\, V_{\text{mat}} & : & \vec{F}^{\text{inc}} \\ \text{int}\, V_{\text{met}} & : & 0\end{array}\right\} = & \left[\overleftrightarrow{G}_0^{JF}(\vec{r},\vec{r}\,')*\hat{e}_l \lim_{\vec{r}\,''\to\vec{r}\,'}\oint_{C(\vec{r}\,'')}\vec{H}^{\text{inc}}(\vec{r}\,'')\cdot d\vec{l}\,''\right]_{L_{\text{met}}^{\cap}} \\ & -\left[\overleftrightarrow{G}_0^{MF}(\vec{r},\vec{r}\,')*\hat{e}_l \lim_{\vec{r}\,''\to\vec{r}\,'}\oint_{C(\vec{r}\,'')}\vec{E}^{\text{inc}}(\vec{r}\,'')\cdot d\vec{l}\,''\right]_{L_{\text{met}}^{\cap}} \\ & +\left[\overleftrightarrow{G}_0^{JF}*\left(\hat{n}_{S_{\text{met}}^{\cap+}}\times\vec{H}^{\text{inc}}\right)+\overleftrightarrow{G}_0^{MF}*\left(\vec{E}^{\text{inc}}\times\hat{n}_{S_{\text{met}}^{\cap+}}\right)\right]_{S_{\text{met}}^{\cap+}} \\ & +\left[\overleftrightarrow{G}_0^{JF}*\left(\hat{n}_{S_{\text{met}}^{\cap-}}\times\vec{H}^{\text{inc}}\right)+\overleftrightarrow{G}_0^{MF}*\left(\vec{E}^{\text{inc}}\times\hat{n}_{S_{\text{met}}^{\cap-}}\right)\right]_{S_{\text{met}}^{\cap-}} \\ & +\left[\overleftrightarrow{G}_0^{JF}*\left(\hat{n}_{V_{\text{mat}}^{-}}\times\vec{H}^{\text{inc}}\right)+\overleftrightarrow{G}_0^{MF}*\left(\vec{E}^{\text{inc}}\times\hat{n}_{V_{\text{mat}}^{-}}\right)\right]_{\partial V_{\text{met}}^{\cap+}} \\ & +\left[\overleftrightarrow{G}_0^{JF}*\left(\hat{n}_{V_{\text{mat}}^{-}}\times\vec{H}^{\text{inc}}\right)+\overleftrightarrow{G}_0^{MF}*\left(\vec{E}^{\text{inc}}\times\hat{n}_{V_{\text{mat}}^{-}}\right)\right]_{\partial V_{\text{mat}}^{0-}} \\ = & \left[\overleftrightarrow{G}_0^{JF}*\left(\hat{n}_{V_{\text{mat}}^{-}}\times\vec{H}^{\text{inc}}\right)+\overleftrightarrow{G}_0^{MF}*\left(\vec{E}^{\text{inc}}\times\hat{n}_{V_{\text{mat}}^{-}}\right)\right]_{\partial V_{\text{met}}^{\cap}} \\ & +\left[\overleftrightarrow{G}_0^{JF}*\left(\hat{n}_{V_{\text{mat}}^{-}}\times\vec{H}^{\text{inc}}\right)+\overleftrightarrow{G}_0^{MF}*\left(\vec{E}^{\text{inc}}\times\hat{n}_{V_{\text{mat}}^{-}}\right)\right]_{\partial V_{\text{mat}}^{0}}\end{aligned} \tag{5-41a}$$

where the second equality is based on the fact that the sources of incident field don't distribute on $L_{\text{met}}^{\cap}$, $S_{\text{met}}^{\cap}$, $\partial V_{\text{met}}^{\cap}$, and $\partial V_{\text{mat}}^{0}$. Based on the results obtained in the Appendix C of this dissertation, we mathematically express the GHFP of material-based scattered field as follows:

$$\begin{aligned}\left.\begin{array}{lll}\text{ext}\, D_{\text{com sys}} & : & \vec{F}_{\text{mat}}^{\text{sca}} \\ \text{int}\, V_{\text{mat}} & : & 0 \\ \text{int}\, V_{\text{met}} & : & \vec{F}_{\text{mat}}^{\text{sca}}\end{array}\right\} = & -\left[\overleftrightarrow{G}_0^{JF}(\vec{r},\vec{r}\,')*\hat{e}_l \lim_{\vec{r}\,''\to\vec{r}\,'}\oint_{C(\vec{r}\,'')}\vec{H}_{\text{mat}}^{\text{sca}}(\vec{r}\,'')\cdot d\vec{l}\,''\right]_{L_{\text{met}}^{\cap}} \\ & +\left[\overleftrightarrow{G}_0^{MF}(\vec{r},\vec{r}\,')*\hat{e}_l \lim_{\vec{r}\,''\to\vec{r}\,'}\oint_{C(\vec{r}\,'')}\vec{E}_{\text{mat}}^{\text{sca}}(\vec{r}\,'')\cdot d\vec{l}\,''\right]_{L_{\text{met}}^{\cap}} \\ & +\left[\overleftrightarrow{G}_0^{JF}*\left(\hat{n}_{S_{\text{met}}^{\cap-}}\times\vec{H}_{\text{mat}}^{\text{sca}}\right)+\overleftrightarrow{G}_0^{MF}*\left(\vec{E}_{\text{mat}}^{\text{sca}}\times\hat{n}_{S_{\text{met}}^{\cap-}}\right)\right]_{S_{\text{met}}^{\cap+}} \\ & +\left[\overleftrightarrow{G}_0^{JF}*\left(\hat{n}_{S_{\text{met}}^{\cap+}}\times\vec{H}_{\text{mat}}^{\text{sca}}\right)+\overleftrightarrow{G}_0^{MF}*\left(\vec{E}_{\text{mat}}^{\text{sca}}\times\hat{n}_{S_{\text{met}}^{\cap+}}\right)\right]_{S_{\text{met}}^{\cap-}} \\ & +\left[\overleftrightarrow{G}_0^{JF}*\left(\hat{n}_{V_{\text{mat}}^{+}}\times\vec{H}_{\text{mat}}^{\text{sca}}\right)+\overleftrightarrow{G}_0^{MF}*\left(\vec{E}_{\text{mat}}^{\text{sca}}\times\hat{n}_{V_{\text{mat}}^{+}}\right)\right]_{\partial V_{\text{met}}^{\cap-}} \\ & +\left[\overleftrightarrow{G}_0^{JF}*\left(\hat{n}_{V_{\text{mat}}^{+}}\times\vec{H}_{\text{mat}}^{\text{sca}}\right)+\overleftrightarrow{G}_0^{MF}*\left(\vec{E}_{\text{mat}}^{\text{sca}}\times\hat{n}_{V_{\text{mat}}^{+}}\right)\right]_{\partial V_{\text{mat}}^{0+}}\end{aligned}$$

$$
\begin{aligned}
= \quad & \Big[ \overset{\leftrightarrow}{G}{}_0^{JF} * \big( \hat{n}_{V_{\text{mat}}^{+}} \times \vec{H}_{\text{mat}}^{\text{sca}} \big) + \overset{\leftrightarrow}{G}{}_0^{MF} * \big( \vec{E}_{\text{mat}}^{\text{sca}} \times \hat{n}_{V_{\text{mat}}^{+}} \big) \Big]_{\partial V_{\text{met}}^{\cap}} \\
& + \Big[ \overset{\leftrightarrow}{G}{}_0^{JF} * \big( \hat{n}_{V_{\text{mat}}^{+}} \times \vec{H}_{\text{mat}}^{\text{sca}} \big) + \overset{\leftrightarrow}{G}{}_0^{MF} * \big( \vec{E}_{\text{mat}}^{\text{sca}} \times \hat{n}_{V_{\text{mat}}^{+}} \big) \Big]_{\partial V_{\text{mat}}^{0}} \qquad (5\text{-}42\text{a})
\end{aligned}
$$

where the second equality is based on the conclusion that the material-based scattered currents don't distribute on $L_{\text{met}}^{\cap}$, $S_{\text{met}}^{\cap}$, $\partial V_{\text{met}}^{\cap}$, and $\partial V_{\text{mat}}^{0}$ (for details see Appendix A). Based on the results obtained in Appendix C, we can mathematically express metal-based scattered field as follows:

$$
\left.
\begin{aligned}
\text{ext}\, D_{\text{com sys}} &: \quad 0 \\
\text{int}\, V_{\text{mat}} &: \quad \vec{F}_{\text{met}}^{\text{sca}} \\
\text{int}\, V_{\text{met}} &: \quad 0
\end{aligned}
\right\} =
\begin{aligned}[t]
& \Big[ \overset{\leftrightarrow}{G}{}_0^{JF}(\vec{r},\vec{r}\,') * \hat{e}_l \lim_{\vec{r}\,''\to\vec{r}\,'} \oint_{C(\vec{r}\,'')} \vec{H}_{\text{met}}^{\text{sca}}(\vec{r}\,'') \cdot d\vec{l}\,'' \Big]_{L_{\text{met}}^{\cap}} \\
& - \Big[ \overset{\leftrightarrow}{G}{}_0^{MF}(\vec{r},\vec{r}\,') * \hat{e}_l \lim_{\vec{r}\,''\to\vec{r}\,'} \oint_{C(\vec{r}\,'')} \vec{E}_{\text{met}}^{\text{sca}}(\vec{r}\,'') \cdot d\vec{l}\,'' \Big]_{L_{\text{met}}^{\cap}} \\
& + \Big[ \overset{\leftrightarrow}{G}{}_0^{JF} * \big( \hat{n}_{S_{\text{met}}^{\cap+}} \times \vec{H}_{\text{met}}^{\text{sca}} \big) + \overset{\leftrightarrow}{G}{}_0^{MF} * \big( \vec{E}_{\text{met}}^{\text{sca}} \times \hat{n}_{S_{\text{met}}^{\cap+}} \big) \Big]_{S_{\text{met}}^{\cap+}} \\
& + \Big[ \overset{\leftrightarrow}{G}{}_0^{JF} * \big( \hat{n}_{S_{\text{met}}^{\cap-}} \times \vec{H}_{\text{met}}^{\text{sca}} \big) + \overset{\leftrightarrow}{G}{}_0^{MF} * \big( \vec{E}_{\text{met}}^{\text{sca}} \times \hat{n}_{S_{\text{met}}^{\cap-}} \big) \Big]_{S_{\text{met}}^{\cap-}} \\
& + \Big[ \overset{\leftrightarrow}{G}{}_0^{JF} * \big( \hat{n}_{V_{\text{mat}}^{-}} \times \vec{H}_{\text{met}}^{\text{sca}} \big) + \overset{\leftrightarrow}{G}{}_0^{MF} * \big( \vec{E}_{\text{met}}^{\text{sca}} \times \hat{n}_{V_{\text{mat}}^{-}} \big) \Big]_{\partial V_{\text{met}}^{\cap+}} \\
& + \Big[ \overset{\leftrightarrow}{G}{}_0^{JF} * \big( \hat{n}_{V_{\text{mat}}^{-}} \times \vec{H}_{\text{met}}^{\text{sca}} \big) + \overset{\leftrightarrow}{G}{}_0^{MF} * \big( \vec{E}_{\text{met}}^{\text{sca}} \times \hat{n}_{V_{\text{mat}}^{-}} \big) \Big]_{\partial V_{\text{mat}}^{0-}} \\
= \quad & \Big[ \overset{\leftrightarrow}{G}{}_0^{JF} * \vec{J}_{\cap}^{\text{SL}} \Big]_{L_{\text{met}}^{\cap}} + \Big[ \overset{\leftrightarrow}{G}{}_0^{JF} * \vec{J}_{\cap}^{\text{SS}} \Big]_{S_{\text{met}}^{\cap}} \\
& + \Big[ \overset{\leftrightarrow}{G}{}_0^{JF} * \big( \hat{n}_{V_{\text{mat}}^{-}} \times \vec{H}_{\text{met}}^{\text{sca}} \big) + \overset{\leftrightarrow}{G}{}_0^{MF} * \big( \vec{E}_{\text{met}}^{\text{sca}} \times \hat{n}_{V_{\text{mat}}^{-}} \big) \Big]_{\partial V_{\text{met}}^{\cap+}} \\
& + \Big[ \overset{\leftrightarrow}{G}{}_0^{JF} * \big( \hat{n}_{V_{\text{mat}}^{-}} \times \vec{H}_{\text{met}}^{\text{sca}} \big) + \overset{\leftrightarrow}{G}{}_0^{MF} * \big( \vec{E}_{\text{met}}^{\text{sca}} \times \hat{n}_{V_{\text{mat}}^{-}} \big) \Big]_{\partial V_{\text{mat}}^{0-}} \qquad (5\text{-}42\text{b})
\end{aligned}
$$

where the second equality is based on the fact that there doesn't exist any scattered magnetic current distributing on $L_{\text{met}}^{\cap}$ and $S_{\text{met}}^{\cap}$.

In formulation (5-41a), integral domains $S_{\text{met}}^{\cap+}$ and $S_{\text{met}}^{\cap-}$ respectively represent the plus and minus sides of $S_{\text{met}}^{\cap}$; unit vectors $\hat{n}_{S_{\text{met}}^{\cap+}}$ and $\hat{n}_{S_{\text{met}}^{\cap-}}$ are the normal vectors of $S_{\text{met}}^{\cap}$, and respectively point to the plus and minus sides of $S_{\text{met}}^{\cap}$; integral domains $\partial V_{\text{met}}^{\cap+}$ and $\partial V_{\text{mat}}^{0-}$ respectively represent the outer metallic boundary corresponding to $\partial V_{\text{met}}^{\cap}$ and the inner material boundary corresponding to $\partial V_{\text{mat}}^{0}$; unit vector $\hat{n}_{V_{\text{mat}}^{-}}$ is the normal vector of $\partial V_{\text{mat}}$, and points to the interior of $V_{\text{mat}}$; the various Green's functions are vacuum Green's functions. The avrious symbols used in GHFP (5-42a) and GHFP (5-42b) can be explained similarly, and they will not be explicitly given here.

**2) GHFP and GBET Corresponding to the Huygens' Boundary Selected as Metallic Boundary**

If we select the Huygens' boundary as whole metallic boundary $L_{\text{met}} \bigcup S_{\text{met}} \bigcup \partial V_{\text{met}}$, then, based on the results given in Appendix C, we can mathematically express the GHFP of incident field as follows:

$$
\left.\begin{array}{lll}
\text{ext}\, D_{\text{com sys}} & : & 0 \\
\text{int}\, V_{\text{mat}} & : & 0 \\
\text{int}\, V_{\text{met}} & : & \vec{F}^{\text{inc}}
\end{array}\right\} = -\left[ \overset{\leftrightarrow}{G}{}_0^{JF}(\vec{r},\vec{r}\,') * \hat{e}_l \lim_{\vec{r}\,''\to\vec{r}\,'} \oint_{C(\vec{r}\,'')} \vec{H}^{\text{inc}}(\vec{r}\,'') \cdot d\vec{l}\,'' \right]_{L_{\text{met}}}
$$

$$
+\left[ \overset{\leftrightarrow}{G}{}_0^{MF}(\vec{r},\vec{r}\,') * \hat{e}_l \lim_{\vec{r}\,''\to\vec{r}\,'} \oint_{C(\vec{r}\,'')} \vec{E}^{\text{inc}}(\vec{r}\,'') \cdot d\vec{l}\,'' \right]_{L_{\text{met}}}
$$

$$
+\left[ \overset{\leftrightarrow}{G}{}_0^{JF} * \left( \hat{n}_{S_{\text{met}}^-} \times \vec{H}^{\text{inc}} \right) + \overset{\leftrightarrow}{G}{}_0^{MF} * \left( \vec{E}^{\text{inc}} \times \hat{n}_{S_{\text{met}}^-} \right) \right]_{S_{\text{met}}^+}
$$

$$
+\left[ \overset{\leftrightarrow}{G}{}_0^{JF} * \left( \hat{n}_{S_{\text{met}}^+} \times \vec{H}^{\text{inc}} \right) + \overset{\leftrightarrow}{G}{}_0^{MF} * \left( \vec{E}^{\text{inc}} \times \hat{n}_{S_{\text{met}}^+} \right) \right]_{S_{\text{met}}^-}
$$

$$
+\left[ \overset{\leftrightarrow}{G}{}_0^{JF} * \left( \hat{n}_{V_{\text{met}}^-} \times \vec{H}^{\text{inc}} \right) + \overset{\leftrightarrow}{G}{}_0^{MF} * \left( \vec{E}^{\text{inc}} \times \hat{n}_{V_{\text{met}}^-} \right) \right]_{\partial V_{\text{met}}^-}
$$

$$
= \left[ \overset{\leftrightarrow}{G}{}_0^{JF} * \left( \hat{n}_{V_{\text{met}}^-} \times \vec{H}^{\text{inc}} \right) + \overset{\leftrightarrow}{G}{}_0^{MF} * \left( \vec{E}^{\text{inc}} \times \hat{n}_{V_{\text{met}}^-} \right) \right]_{\partial V_{\text{met}}^-} \qquad (5\text{-}41\text{b})
$$

where the second equality is based on the fact that incident sources don't distribute on $L_{\text{met}}$, $S_{\text{met}}$, and $\partial V_{\text{met}}$. Based on the results given in Appendix C, we can mathematically express the GHFP of metal-based scattered field as follows:

$$
\left.\begin{array}{lll}
\text{ext}\, D_{\text{com sys}} & : & \vec{F}_{\text{met}}^{\text{sca}} \\
\text{int}\, V_{\text{mat}} & : & \vec{F}_{\text{met}}^{\text{sca}} \\
\text{int}\, V_{\text{met}} & : & 0
\end{array}\right\} = \left[ \overset{\leftrightarrow}{G}{}_0^{JF}(\vec{r},\vec{r}\,') * \hat{e}_l \lim_{\vec{r}\,''\to\vec{r}\,'} \oint_{C(\vec{r}\,'')} \vec{H}_{\text{met}}^{\text{sca}}(\vec{r}\,'') \cdot d\vec{l}\,'' \right]_{L_{\text{met}}}
$$

$$
-\left[ \overset{\leftrightarrow}{G}{}_0^{MF}(\vec{r},\vec{r}\,') * \hat{e}_l \lim_{\vec{r}\,''\to\vec{r}\,'} \oint_{C(\vec{r}\,'')} \vec{E}_{\text{met}}^{\text{sca}}(\vec{r}\,'') \cdot d\vec{l}\,'' \right]_{L_{\text{met}}}
$$

$$
+\left[ \overset{\leftrightarrow}{G}{}_0^{JF} * \left( \hat{n}_{S_{\text{met}}^+} \times \vec{H}_{\text{met}}^{\text{sca}} \right) + \overset{\leftrightarrow}{G}{}_0^{MF} * \left( \vec{E}_{\text{met}}^{\text{sca}} \times \hat{n}_{S_{\text{met}}^+} \right) \right]_{S_{\text{met}}^+}
$$

$$
+\left[ \overset{\leftrightarrow}{G}{}_0^{JF} * \left( \hat{n}_{S_{\text{met}}^-} \times \vec{H}_{\text{met}}^{\text{sca}} \right) + \overset{\leftrightarrow}{G}{}_0^{MF} * \left( \vec{E}_{\text{met}}^{\text{sca}} \times \hat{n}_{S_{\text{met}}^-} \right) \right]_{S_{\text{met}}^-}
$$

$$
+\left[ \overset{\leftrightarrow}{G}{}_0^{JF} * \left( \hat{n}_{V_{\text{met}}^+} \times \vec{H}_{\text{met}}^{\text{sca}} \right) + \overset{\leftrightarrow}{G}{}_0^{MF} * \left( \vec{E}_{\text{met}}^{\text{sca}} \times \hat{n}_{V_{\text{met}}^+} \right) \right]_{\partial V_{\text{met}}^+}
$$

$$
= \left[ \overset{\leftrightarrow}{G}{}_0^{JF} * \vec{J}^{\text{SL}} \right]_{L_{\text{met}}} + \left[ \overset{\leftrightarrow}{G}{}_0^{JF} * \vec{J}^{\text{SS}} \right]_{S_{\text{met}}}
$$

$$
+\left[ \overset{\leftrightarrow}{G}{}_0^{JF} * \left( \hat{n}_{V_{\text{met}}^+} \times \vec{H}_{\text{met}}^{\text{sca}} \right) + \overset{\leftrightarrow}{G}{}_0^{MF} * \left( \vec{E}_{\text{met}}^{\text{sca}} \times \hat{n}_{V_{\text{met}}^+} \right) \right]_{\partial V_{\text{met}}^+} \qquad (5\text{-}42\text{c})
$$

where the second equality is based on the fact that there doesn't exist any scattered magnetic current distributing on $L_{\text{met}}$ and $S_{\text{met}}$. Based on the results obtained in Appendix C, we can mathematically express the GHFP of material-based scattered field as follows:

$$\left.\begin{array}{lcc} \text{ext}\, D_{\text{com sys}} & : & 0 \\ \text{int}\, V_{\text{mat}} & : & 0 \\ \text{int}\, V_{\text{met}} & : & \vec{F}_{\text{mat}}^{\text{sca}} \end{array}\right\} = -\left[\overleftrightarrow{G}_0^{JF}(\vec{r},\vec{r}')*\hat{e}_l \lim_{\vec{r}''\to\vec{r}'}\oint_{C(\vec{r}'')}\vec{H}_{\text{mat}}^{\text{sca}}(\vec{r}'')\cdot d\vec{l}''\right]_{L_{\text{met}}}$$
$$+\left[\overleftrightarrow{G}_0^{MF}(\vec{r},\vec{r}')*\hat{e}_l \lim_{\vec{r}''\to\vec{r}'}\oint_{C(\vec{r}'')}\vec{E}_{\text{mat}}^{\text{sca}}(\vec{r}'')\cdot d\vec{l}''\right]_{L_{\text{met}}}$$
$$+\left[\overleftrightarrow{G}_0^{JF}*\left(\hat{n}_{S_{\text{met}}^-}\times\vec{H}_{\text{mat}}^{\text{sca}}\right)+\overleftrightarrow{G}_0^{MF}*\left(\vec{E}_{\text{mat}}^{\text{sca}}\times\hat{n}_{S_{\text{met}}^-}\right)\right]_{S_{\text{met}}^+}$$
$$+\left[\overleftrightarrow{G}_0^{JF}*\left(\hat{n}_{S_{\text{met}}^+}\times\vec{H}_{\text{mat}}^{\text{sca}}\right)+\overleftrightarrow{G}_0^{MF}*\left(\vec{E}_{\text{mat}}^{\text{sca}}\times\hat{n}_{S_{\text{met}}^+}\right)\right]_{S_{\text{met}}^-}$$
$$+\left[\overleftrightarrow{G}_0^{JF}*\left(\hat{n}_{V_{\text{met}}^-}\times\vec{H}_{\text{mat}}^{\text{sca}}\right)+\overleftrightarrow{G}_0^{MF}*\left(\vec{E}_{\text{mat}}^{\text{sca}}\times\hat{n}_{V_{\text{met}}^-}\right)\right]_{\partial V_{\text{met}}^-}$$
$$=\left[\overleftrightarrow{G}_0^{JF}*\left(\hat{n}_{V_{\text{met}}^-}\times\vec{H}_{\text{mat}}^{\text{sca}}\right)+\overleftrightarrow{G}_0^{MF}*\left(\vec{E}_{\text{mat}}^{\text{sca}}\times\hat{n}_{V_{\text{met}}^-}\right)\right]_{\partial V_{\text{met}}} \quad (5\text{-}42\text{d})$$

where the second equality is based on the conclusion that the material-based scattered currents don't distribute on $L_{\text{met}}$, $S_{\text{met}}$, and $\partial V_{\text{met}}$, and the rigorous mathematical proof for this conclusion can be found in Appendix A.

**3) GHFP and GBET Corresponding to the Huygens' Boundary Selected as Whole System Boundary —— Topological Additivity**

The summation of material-boundary-based GHFP (5-41a) and metallic-boundary-based GHFP (5-41b) gives the following system-boundary-based GHFP:

$$\left.\begin{array}{lcc} \text{ext}\, D_{\text{com sys}} & : & 0 \\ \text{int}\, V_{\text{mat}} & : & \vec{F}^{\text{inc}} \\ \text{int}\, V_{\text{met}} & : & \vec{F}^{\text{inc}} \end{array}\right\} = \left[\overleftrightarrow{G}_0^{JF}*\left(\hat{n}_{V_{\text{mat}}^-}\times\vec{H}^{\text{inc}}\right)+\overleftrightarrow{G}_0^{MF}*\left(\vec{E}^{\text{inc}}\times\hat{n}_{V_{\text{mat}}^-}\right)\right]_{\partial V_{\text{mat}}^0}$$
$$+\left[\overleftrightarrow{G}_0^{JF}*\left(\hat{n}_{V_{\text{met}}^-}\times\vec{H}^{\text{inc}}\right)+\overleftrightarrow{G}_0^{MF}*\left(\vec{E}^{\text{inc}}\times\hat{n}_{V_{\text{met}}^-}\right)\right]_{\partial V_{\text{met}}^0}$$
$$=\left[\overleftrightarrow{G}_0^{JF}*\left(\hat{n}_{D_{\text{com sys}}^-}\times\vec{H}^{\text{inc}}\right)+\overleftrightarrow{G}_0^{MF}*\left(\vec{E}^{\text{inc}}\times\hat{n}_{D_{\text{com sys}}^-}\right)\right]_{\partial V_{\text{mat}}^0\cup\partial V_{\text{met}}^0} \quad (5\text{-}43)$$

where the first equality is based on formulation (5-4), formulation (5-10), relationship $\hat{n}_{V_{\text{mat}}^-}=-\hat{n}_{V_{\text{met}}^-}$, and the continuity of the $\vec{F}^{\text{inc}}$ on $\partial V_{\text{met}}^{\cap}$; the integral domain $\partial V_{\text{mat}}^0\cup\partial V_{\text{met}}^0$ on the RHS of the second equality is just the surface part of whole system boundary $\partial D_{\text{com sys}}$; the unit vector $\hat{n}_{D_{\text{com sys}}^-}$ on the RHS of the second equality is the inner normal vector of closed surface $\partial V_{\text{mat}}^0\cup\partial V_{\text{met}}^0$.

Based on the GHFP (5-42a)&(5-42d) for material-based scattered field and the GHFP (5-42b)&(5-42c) for metal-based scattered field, we immediately have the following convolution integral formulation:

$$\left.\begin{array}{lll}\text{ext}\, D_{\text{com sys}} & : & \vec{F}^{\text{sca}} \\ \text{int}\, V_{\text{mat}} & : & 0 \\ \text{int}\, V_{\text{met}} & : & 0\end{array}\right\} = \left[\overleftrightarrow{G}_0^{JF} * \left(\hat{n}_{V_{\text{mat}}^+} \times \vec{H}^{\text{sca}}\right) + \overleftrightarrow{G}_0^{MF} * \left(\vec{E}^{\text{sca}} \times \hat{n}_{V_{\text{mat}}^+}\right)\right]_{\partial V_{\text{mat}}^{0-}}$$
$$+ \left[\overleftrightarrow{G}_0^{JF} * \left(\hat{n}_{V_{\text{met}}^+} \times \vec{H}^{\text{sca}}\right) + \overleftrightarrow{G}_0^{MF} * \left(\vec{E}^{\text{sca}} \times \hat{n}_{V_{\text{met}}^+}\right)\right]_{\partial V_{\text{met}}^{0+}}$$
$$+ \left[\overleftrightarrow{G}_0^{JF} * \vec{J}_0^{\text{SS}}\right]_{S_{\text{met}}^0} + \left[\overleftrightarrow{G}_0^{JF} * \vec{J}_0^{\text{SL}}\right]_{L_{\text{met}}^0}$$
$$= \left[\overleftrightarrow{G}_0^{JF} * \left(\hat{n}_{D_{\text{com sys}}^+} \times \vec{H}^{\text{sca}}\right) + \overleftrightarrow{G}_0^{MF} * \left(\vec{E}^{\text{sca}} \times \hat{n}_{D_{\text{com sys}}^+}\right)\right]_{\partial V_{\text{mat}}^{0-} \bigcup \partial V_{\text{met}}^{0+}}$$
$$+ \left[\overleftrightarrow{G}_0^{JF} * \vec{J}_0^{\text{SS}}\right]_{S_{\text{met}}^0} + \left[\overleftrightarrow{G}_0^{JF} * \vec{J}_0^{\text{SL}}\right]_{L_{\text{met}}^0} \tag{5-44}$$

where unit vector $\hat{n}_{D_{\text{com sys}}^+}$ is the inner normal vector of closed surface $\partial V_{\text{mat}}^0 \bigcup \partial V_{\text{met}}^0$.

Above formulations (5-43) and (5-44) are just the mathematical expressions of the GHFP for incident field and the GHFP for scattered field corresponding to the Huygens' boundary, which is selected as the boundary of whole composite system $D_{\text{com sys}}$. Obviously, both of GHFP (5-43) and GHFP (5-44) satisfy GBET and topological additivity.

## 5.4.2 Generalized Franz-Harrington Formulations

The difference between the GHFP (5-43) for incident field and the GHFP (5-44) for scattered field gives the following GFHF for incident and scattered fields:

$$\left.\begin{array}{lll}\text{ext}\, D_{\text{com sys}} & : & -\vec{F}^{\text{sca}} \\ \text{int}\, V_{\text{mat}} & : & \vec{F}^{\text{inc}} \\ \text{int}\, V_{\text{met}} & : & \vec{F}^{\text{inc}}\end{array}\right\} = \left[\overleftrightarrow{G}_0^{JF} * \vec{J}_0^{\text{ES}} + \overleftrightarrow{G}_0^{MF} * \vec{M}_0^{\text{ES}}\right]_{\partial V_{\text{mat}}^0} - \left[\overleftrightarrow{G}_0^{JF} * \vec{J}_0^{\text{SS}}\right]_{S_{\text{met}}^0 \bigcup \partial V_{\text{met}}^0}$$
$$- \left[\overleftrightarrow{G}_0^{JF} * \vec{J}_0^{\text{SL}}\right]_{L_{\text{met}}^0} \tag{5-45}$$

Based on the method to introduce the piecewise Green's functions for material systems in Appendix C, we can also introduce the piecewise Green's functions for composite systems as follows:

$$\tilde{\overleftrightarrow{G}}_{\text{com sys}}^{JF}\left(\vec{r},\vec{r}'\right) = \begin{cases}\overleftrightarrow{G}_{\text{mat}}^{JF}\left(\vec{r},\vec{r}'\right) & , \quad \left(\vec{r} \in \text{cl}V_{\text{mat}} \ , \vec{r}' \in \text{cl}V_{\text{mat}}\right) \\ 0 & , \quad \left(\vec{r} \in \text{ext}V_{\text{mat}} \ , \vec{r}' \in \text{cl}V_{\text{mat}}\right) \\ \overleftrightarrow{G}_0^{JF}\left(\vec{r},\vec{r}'\right) & , \quad \left( \qquad \vec{r}' \in \text{ext}V_{\text{mat}}\right)\end{cases} \tag{5-46a}$$

$$\tilde{\overleftrightarrow{G}}_{\text{com sys}}^{MF}\left(\vec{r},\vec{r}'\right) = \begin{cases}\overleftrightarrow{G}_{\text{mat}}^{MF}\left(\vec{r},\vec{r}'\right) & , \quad \left(\vec{r} \in \text{cl}V_{\text{mat}} \ , \vec{r}' \in \text{cl}V_{\text{mat}}\right) \\ 0 & , \quad \left(\vec{r} \in \text{ext}V_{\text{mat}} \ , \vec{r}' \in \text{cl}V_{\text{mat}}\right) \\ \overleftrightarrow{G}_0^{MF}\left(\vec{r},\vec{r}'\right) & , \quad \left( \qquad \vec{r}' \in \text{ext}V_{\text{mat}}\right)\end{cases} \tag{5-46b}$$

where $\overleftrightarrow{G}_{\text{mat}}^{JF}$ and $\overleftrightarrow{G}_{\text{mat}}^{MF}$ are the Green's functions corresponding to the material part in composite system. Based on above definition (5-46) and the results given in Appendix C, it is easy to obtain the following results:

$$\left.\begin{array}{lcl} \text{ext}\,D_{\text{com sys}} & : & 0 \\ \text{int}\,V_{\text{mat}} & : & \vec{F}^{\text{tot}} \\ \text{int}\,V_{\text{met}} & : & 0 \end{array}\right\} = \left[\tilde{\overset{\leftrightarrow}{G}}{}^{JF}_{\text{com sys}} * \vec{J}^{\text{EL}}\right]_{L^{\cap}_{\text{met}}} + \left[\tilde{\overset{\leftrightarrow}{G}}{}^{JF}_{\text{com sys}} * \vec{J}^{\text{ES}}_{\cap}\right]_{S^{\cap}_{\text{met}} \cup \partial V^{\cap}_{\text{met}}}$$
$$+\left[\tilde{\overset{\leftrightarrow}{G}}{}^{JF}_{\text{com sys}} * \vec{J}^{\text{ES}}_0 + \tilde{\overset{\leftrightarrow}{G}}{}^{MF}_{\text{com sys}} * \vec{M}^{\text{ES}}_0\right]_{\partial V^0_{\text{mat}}} \qquad (5\text{-}47\text{a})$$

Based on formulations (5-36a) and (5-38a) and the fact that the total field in the interior of metallic body is zero, above formulation (5-47a) can be equivalently rewritten as the following form:

$$\left.\begin{array}{lcl} \text{ext}\,D_{\text{com sys}} & : & 0 \\ \text{int}\,V_{\text{mat}} & : & \vec{F}^{\text{tot}} \\ \text{int}\,V_{\text{met}} & : & \vec{F}^{\text{tot}} \end{array}\right\} = \left[\tilde{\overset{\leftrightarrow}{G}}{}^{JF}_{\text{com sys}} * \vec{J}^{\text{SL}}_{\cap}\right]_{L^{\cap}_{\text{met}}} + \left[\tilde{\overset{\leftrightarrow}{G}}{}^{JF}_{\text{com sys}} * \vec{J}^{\text{SS}}_{\cap}\right]_{S^{\cap}_{\text{met}} \cup \partial V^{\cap}_{\text{met}}}$$
$$+\left[\tilde{\overset{\leftrightarrow}{G}}{}^{JF}_{\text{com sys}} * \vec{J}^{\text{ES}}_0 + \tilde{\overset{\leftrightarrow}{G}}{}^{MF}_{\text{com sys}} * \vec{M}^{\text{ES}}_0\right]_{\partial V^0_{\text{mat}}} \qquad (5\text{-}47\text{b})$$

Based on the method to introduce the piecewise delta Green's functions for material systems in Appendix C, we can also introduce the piecewise delta Green's functions for composite systems as follows:

$$\Delta\tilde{\overset{\leftrightarrow}{G}}{}^{JF}_{\text{com sys}}(\vec{r},\vec{r}\,') = \tilde{\overset{\leftrightarrow}{G}}{}^{JF}_{\text{com sys}}(\vec{r},\vec{r}\,') - \overset{\leftrightarrow}{G}{}^{JF}_0(\vec{r},\vec{r}\,')$$
$$= \begin{cases} \overset{\leftrightarrow}{G}{}^{JF}_{\text{mat}}(\vec{r},\vec{r}\,') - \overset{\leftrightarrow}{G}{}^{JF}_0(\vec{r},\vec{r}\,') & , \ (\vec{r} \in \text{cl}V_{\text{mat}} \ , \vec{r}\,' \in \text{cl}V_{\text{mat}}) \\ -\overset{\leftrightarrow}{G}{}^{JF}_0(\vec{r},\vec{r}\,') & , \ (\vec{r} \in \text{ext}V_{\text{mat}} \ , \vec{r}\,' \in \text{cl}V_{\text{mat}}) \\ 0 & , \ (\vec{r}\,' \in \text{ext}V_{\text{mat}}) \end{cases} \qquad (5\text{-}48\text{a})$$

$$\Delta\tilde{\overset{\leftrightarrow}{G}}{}^{MF}_{\text{com sys}}(\vec{r},\vec{r}\,') = \tilde{\overset{\leftrightarrow}{G}}{}^{MF}_{\text{com sys}}(\vec{r},\vec{r}\,') - \overset{\leftrightarrow}{G}{}^{MF}_0(\vec{r},\vec{r}\,')$$
$$= \begin{cases} \overset{\leftrightarrow}{G}{}^{MF}_{\text{mat}}(\vec{r},\vec{r}\,') - \overset{\leftrightarrow}{G}{}^{MF}_0(\vec{r},\vec{r}\,') & , \ (\vec{r} \in \text{cl}V_{\text{mat}} \ , \vec{r}\,' \in \text{cl}V_{\text{mat}}) \\ -\overset{\leftrightarrow}{G}{}^{MF}_0(\vec{r},\vec{r}\,') & , \ (\vec{r} \in \text{ext}V_{\text{mat}} \ , \vec{r}\,' \in \text{cl}V_{\text{mat}}) \\ 0 & , \ (\vec{r}\,' \in \text{ext}V_{\text{mat}}) \end{cases} \qquad (5\text{-}48\text{b})$$

Then, based on formulations (5-45) and (5-47b), we can obtain the GFHF corresponding to the scattered field $\vec{F}^{\text{sca}}$ generated by whole composite system as follows:

$$\left.\begin{array}{lcl} \text{ext}\,D_{\text{com sys}} & : & \vec{F}^{\text{sca}} \\ \text{int}\,V_{\text{mat}} & : & \vec{F}^{\text{sca}} \\ \text{int}\,V_{\text{met}} & : & \vec{F}^{\text{sca}} \end{array}\right\} = \left[\tilde{\overset{\leftrightarrow}{G}}{}^{JF}_{\text{com sys}} * \vec{J}^{\text{SL}}\right]_{L_{\text{met}}} + \left[\tilde{\overset{\leftrightarrow}{G}}{}^{JF}_{\text{com sys}} * \vec{J}^{\text{SS}}\right]_{S_{\text{met}} \cup \partial V_{\text{met}}}$$
$$+\left[\Delta\tilde{\overset{\leftrightarrow}{G}}{}^{JF}_{\text{com sys}} * \vec{J}^{\text{ES}}_0 + \Delta\tilde{\overset{\leftrightarrow}{G}}{}^{MF}_{\text{com sys}} * \vec{M}^{\text{ES}}_0\right]_{\partial V^0_{\text{mat}}} \qquad (5\text{-}49)$$

### 5.4.3 Topological Additivity

Obviously, the above GHFP, GBET, and GFHF for metal-material composite systems are consistent with the GHFP, GBET, and GFHF for material systems (which are given in Appendix C) in the aspect of manifestation form, and they satisfy the following topological additivity:

$$\begin{aligned}
&\text{Scattered field GHFP/GBET of metal-material composite system}\\
=\;&\text{Scattered field GHFP/GBET of metallic subsystem}\\
&+\text{Scattered field GHFP/GBET of material subsystem}\\
=\;&\sum\nolimits_{\xi}\text{Scattered field GHFP/GBET of metallic line } L_{\text{met}}^{\xi}\\
&+\sum\nolimits_{\zeta}\text{Scattered field GHFP/GBET of metallic surface } S_{\text{met}}^{\zeta}\\
&+\sum\nolimits_{\upsilon}\text{Scattered field GHFP/GBET of metallic body } V_{\text{met}}^{\upsilon}\\
&+\sum\nolimits_{\nu}\text{Scattered field GHFP/GBET of material body } V_{\text{mat}}^{\nu}
\end{aligned}\tag{5-50a}$$

$$\begin{aligned}
&\text{Incident field GHFP/GBET of metal-material composite system}\\
=\;&\text{Incident field GHFP/GBET of metallic subsystem}\\
&+\text{Incident field GHFP/GBET of material subsystem}\\
=\;&\sum\nolimits_{\xi}\text{Incident field GHFP/GBET of metallic line } L_{\text{met}}^{\xi}\\
&+\sum\nolimits_{\zeta}\text{Incident field GHFP/GBET of metallic surface } S_{\text{met}}^{\zeta}\\
&+\sum\nolimits_{\upsilon}\text{Incident field GHFP/GBET of metallic body } V_{\text{met}}^{\upsilon}\\
&+\sum\nolimits_{\nu}\text{Incident field GHFP/GBET of material body } V_{\text{mat}}^{\nu}
\end{aligned}\tag{5-50b}$$

$$\begin{aligned}
&\text{The GFHF of metal-material composite system}\\
=\;&\text{The GFHF of metallic subsystem}\\
&+\text{The GFHF of material subsystem}\\
=\;&\sum\nolimits_{\xi}\text{The GFHF of metallic line } L_{\text{met}}^{\xi}\\
&+\sum\nolimits_{\zeta}\text{The GFHF of metallic surface } S_{\text{met}}^{\zeta}\\
&+\sum\nolimits_{\upsilon}\text{The GFHF of metallic body } V_{\text{met}}^{\upsilon}\\
&+\sum\nolimits_{\nu}\text{The GFHF of material body } V_{\text{mat}}^{\nu}
\end{aligned}\tag{5-50c}$$

## 5.5 Line-Surface Formulation for Constructing the DP-CMs of Composite Systems

In 2017 and 2018, Dr. Guo and Prof. Chen[53,54] et al., by combining the method developed in literatures [31,32] with the method developed in literature [34], generalized the traditional Harrington's EFIE-MetSca-CMT[31,32] for metallic systems and the traditional Harrington's SIE-MatSca-CMT[34] for a single simply connected homogeneous isotropic material body to the metal-material composite system whose material part is a simply connected homogeneous isotropic material body, and this chapter will simply denotes the theory developed in literatures [53,54] as EFIE-SIE-ComSca-CMT.

In this section, we will, employing WEP, draw a clear physical picture for the EFIE-SIE-ComSca-CMT established in IE framework (Subsection 5.5.1), and then we will, based on the physical picture and the GFHF obtained in Section 5.4, generalize EFIE-SIE-ComSca-CMT to more general composite systems (Subsections 5.5.2~5.5.5). It needs to emphasize that: the above generalization is done in WEP framework rather than IE framework, so the corresponding formulations are completely new. In addition, the new formulations are more widely applicable, for example:

**A.** The metallic parts in composite systems can be metallic lines, or metallic surfaces, or metallic bodies, as shown in Figure 5-1.

**B.** The material parts in composite systems can only be material bodies, as shown in Figure 5-1. But, the material bodies can be either homogeneous isotropic or inhomogeneous anisotropic.

**C.** The composite systems can be placed in either vacuum environment or complex environment.

**D.** The metallic lines and material bodies can be separated from each other or in contact with each other, and can also be partially separated from each other and partially in contact with each other, as shown in Figure 5-1. In the case of contacting with each other, the metallic lines can be in contact with the outer surfaces of the material bodies, or be submerged into the interiors of the material bodies, as shown in Figure 5-1.

**E.** The metallic surfaces and material bodies can be separated from each other or in contact with each other, and can also be partially separated from each and partially in contact with each other, as shown in Figure 5-1. In the case of contacting with each other, the metallic surfaces can be in contact with the outer surfaces of the material bodies, or be submerged into the interiors of the material bodies, as shown in Figure 5-1.

**F.** The metallic bodies and the material bodies can be separated from each other or in contact with each other, and can also be partially separated from each other and partially in contact with each other, as shown in Figure 5-1.

### 5.5.1 Physical Picture of the Traditional Harrington's CMT for Composite Systems

In fact, the EFIE-SIE-ComSca-CMT[53,54] in IE framework is essentially the integration for the EFIE-MetSca-CMT[31] and SIE-MatSca-CMT[34] in IE framework, so the physical destination of EFIE-SIE-ComSca-CMT is to construct a series of fundamental modes which have ability to orthogonalize the following operator:

$$\begin{aligned}
&- (1/2)\langle \vec{J}^{\rm SS},\vec{E}^{\rm sca}\rangle_{\partial V_{\rm met}} -(1/2)\langle \vec{J}^{\rm ES},\vec{E}_-^{\rm tot}-\vec{E}_+^{\rm sca}\rangle_{\partial V_{\rm mat}} -(1/2)\langle \vec{M}^{\rm ES},\vec{H}_-^{\rm tot}-\vec{H}_+^{\rm sca}\rangle_{\partial V_{\rm mat}}\\
=&- (1/2)\langle \vec{J}^{\rm SS},\vec{E}^{\rm sca}\rangle_{\partial V_{\rm met}} -(1/2)\langle \vec{J}^{\rm ES},\vec{E}^{\rm inc}\rangle_{\partial V_{\rm mat}} -(1/2)\langle \vec{M}^{\rm ES},\vec{E}^{\rm inc}\rangle_{\partial V_{\rm mat}}\\
=&\ (1/2)\langle \vec{J}^{\rm SS},\vec{E}^{\rm inc}\rangle_{\partial V_{\rm met}} +(1/2)\langle \vec{J}^{\rm SV},\vec{E}^{\rm inc}\rangle_{V_{\rm mat}} +(1/2)\langle \vec{M}^{\rm SV},\vec{H}^{\rm inc}\rangle_{V_{\rm mat}}\\
=&\ (1/2)\langle \vec{J}^{\rm SS}\oplus\vec{J}^{\rm SV},\vec{E}^{\rm inc}\rangle_{\partial V_{\rm met}\cup V_{\rm mat}} +(1/2)\langle \vec{M}^{\rm SV},\vec{H}^{\rm inc}\rangle_{V_{\rm mat}}\\
=&\ P_{\rm com\ sys}^{\rm driving}
\end{aligned} \tag{5-51}$$

In formulation (5-51), the reason why the first line doesn't contain $\vec{J}^{\rm SL}$ and the $\vec{J}^{\rm SS}$ on $S_{\rm met}$ is that literatures [53,54] didn't consider composite structures $L_{\rm met}$-$V_{\rm mat}$ and $S_{\rm met}$-$V_{\rm mat}$; the first equality is based on superposition principle and the continuity of the incident field $\vec{F}^{\rm inc}$ on material boundary, where $\vec{F}_-^{\rm tot}$ and $\vec{F}_+^{\rm sca}$ are respectively the total field distributing on the interior of material body and the scattered field distributing on material body; the second equality is based on the tangential electric field boundary condition $-\vec{E}_{\rm tan}^{\rm sca}=\vec{E}_{\rm tan}^{\rm inc}$ on $\partial V_{\rm met}$ and relationship (4-31); the third equality is based on the linear property of inner product. The fourth equality in formulation (5-51) implies that: EFIE-SIE-ComSca-CMT[53,54] has the same physical destination as Harrington's EFIE-MetSca-CMT[31] and SIE-MatSca-CMT[34] —— constructing a series of modes having ability to orthogonalize frequency-domain DPO.

In the following parts of this section, we will, in WEP framework, establish a more widely applicable CMT for the composite system shown in Figure 5-1, according to the following order: firstly we, in Subsection 5.5.2, provide the variable unification scheme for the composite system; secondly we, in Subsection 5.5.3, derive the line-surface formulation of the DPO corresponding to the composite system, such that the formulation only contains BVs; thirdly we, in Subsection 5.5.4 and based on the results obtained in Subsection 5.5.3, construct the DP-CMs of the composite system and discuss their orthogonality; finally we, in Subsection 5.5.5, expand any working mode of the composite system in terms of the DP-CMs, and also provide the explicit expressions for the expansion coefficients.

### 5.5.2 Variable Unification

To effectively unify variables for the DPO corresponding to the composite system shown in Figure 5-1, this subsection firstly transforms the problem from EM current space to expansion vector space, and then accomplishes the variable unification in expansion vector space.

**1) To Transform From EM Current Space to Expansion Vector Space**

There exists the one-to-one mapping between EM current space and expansion vector space as follows:

$$\vec{J}^{\text{SL}}_{0/\cap}(\vec{r}) = \sum_{\xi=1}^{\Xi^{\vec{J}^{\text{SL}}_{0/\cap}}} a_{\xi}^{\vec{J}^{\text{SL}}_{0/\cap}} \vec{b}_{\xi}^{\vec{J}^{\text{SL}}_{0/\cap}}(\vec{r}) = \overline{\mathbf{B}}^{\vec{J}^{\text{SL}}_{0/\cap}} \cdot \bar{a}^{\vec{J}^{\text{SL}}_{0/\cap}} \quad , \quad \vec{r} \in L^{0/\cap}_{\text{met}} \tag{5-52a}$$

$$\vec{J}^{\text{SS}}_{0/\cap}(\vec{r}) = \sum_{\xi=1}^{\Xi^{\vec{J}^{\text{SS}}_{0/\cap}}} a_{\xi}^{\vec{J}^{\text{SS}}_{0/\cap}} \vec{b}_{\xi}^{\vec{J}^{\text{SS}}_{0/\cap}}(\vec{r}) = \overline{\mathbf{B}}^{\vec{J}^{\text{SS}}_{0/\cap}} \cdot \bar{a}^{\vec{J}^{\text{SS}}_{0/\cap}} \quad , \quad \vec{r} \in S^{0/\cap}_{\text{met}} \cup \partial V^{0/\cap}_{\text{met}} \tag{5-52b}$$

$$\vec{C}^{\text{ES}}_{0}(\vec{r}) = \sum_{\xi=1}^{\Xi^{\vec{C}^{\text{ES}}_{0}}} a_{\xi}^{\vec{C}^{\text{ES}}_{0}} \vec{b}_{\xi}^{\vec{C}^{\text{ES}}_{0}}(\vec{r}) = \overline{\mathbf{B}}^{\vec{C}^{\text{ES}}_{0}} \cdot \bar{a}^{\vec{C}^{\text{ES}}_{0}} \quad , \quad \vec{r} \in \partial V^{0}_{\text{mat}} \tag{5-52c}$$

In the above expansion formulations, $\overline{\mathbf{B}}^{\vec{J}^{\text{SL}}_{0/\cap}/\vec{J}^{\text{SS}}_{0/\cap}/\vec{C}^{\text{ES}}_{0}} = [\vec{b}_1^{\vec{J}^{\text{SL}}_{0/\cap}/\vec{J}^{\text{SS}}_{0/\cap}/\vec{C}^{\text{ES}}_{0}}, \vec{b}_2^{\vec{J}^{\text{SL}}_{0/\cap}/\vec{J}^{\text{SS}}_{0/\cap}/\vec{C}^{\text{ES}}_{0}}, \cdots, \vec{b}_{\Xi^{\vec{J}^{\text{SL}}_{0/\cap}/\vec{J}^{\text{SS}}_{0/\cap}/\vec{C}^{\text{ES}}_{0}}}^{\vec{J}^{\text{SL}}_{0/\cap}/\vec{J}^{\text{SS}}_{0/\cap}/\vec{C}^{\text{ES}}_{0}}]$ and $\bar{a}^{\vec{J}^{\text{SL}}_{0/\cap}/\vec{J}^{\text{SS}}_{0/\cap}/\vec{C}^{\text{ES}}_{0}} = [a_1^{\vec{J}^{\text{SL}}_{0/\cap}/\vec{J}^{\text{SS}}_{0/\cap}/\vec{C}^{\text{ES}}_{0}}, a_2^{\vec{J}^{\text{SL}}_{0/\cap}/\vec{J}^{\text{SS}}_{0/\cap}/\vec{C}^{\text{ES}}_{0}}, \cdots, a_{\Xi^{\vec{J}^{\text{SL}}_{0/\cap}/\vec{J}^{\text{SS}}_{0/\cap}/\vec{C}^{\text{ES}}_{0}}}^{\vec{J}^{\text{SL}}_{0/\cap}/\vec{J}^{\text{SS}}_{0/\cap}/\vec{C}^{\text{ES}}_{0}}]^T$, where superscript "$T$" represents the transposition operation for a matrix or vector.

**2) To Unify Variables in Expansion Vector Space**

Based on GFHF (5-47b) and the homogeneous tangential boundary condition of the total electric field $\vec{E}^{\text{tot}}$ on sub-boundaries $L^{\cap}_{\text{met}}$ and $S^{\cap}_{\text{met}} \cup \partial V^{\cap}_{\text{met}}$, the following integral equations of $\{\vec{J}^{\text{SL}}_{\cap}, \vec{J}^{\text{SS}}_{\cap}, \vec{J}^{\text{ES}}_{0}, \vec{M}^{\text{ES}}_{0}\}$ can be obtained:

$$\left[\mathcal{E}_{\text{mat}}\left(\vec{J}^{\text{SL}}_{\cap} \oplus \vec{J}^{\text{SS}}_{\cap} + \vec{J}^{\text{ES}}_{0}, \vec{M}^{\text{ES}}_{0}\right)\right]^{\tan} = 0 \quad , \quad \vec{r} \in L^{\cap}_{\text{met}} \tag{5-53}$$

$$\left[\mathcal{E}_{\text{mat}}\left(\vec{J}^{\text{SL}}_{\cap} \oplus \vec{J}^{\text{SS}}_{\cap} + \vec{J}^{\text{ES}}_{0}, \vec{M}^{\text{ES}}_{0}\right)\right]^{\tan} = 0 \quad , \quad \vec{r} \in S^{\cap}_{\text{met}} \cup \partial V^{\cap}_{\text{met}} \tag{5-54}$$

Based on GFHF (5-47b) and the definition of the $\vec{J}^{\text{ES}}_{0}$ on $\partial V^{0}_{\text{mat}}$, the following integral equation of $\{\vec{J}^{\text{SL}}_{\cap}, \vec{J}^{\text{SS}}_{\cap}, \vec{J}^{\text{ES}}_{0}, \vec{M}^{\text{ES}}_{0}\}$ can be obtained:

$$\left[\mathcal{H}_{\text{mat}}\left(\vec{J}^{\text{SL}}_{\cap} \oplus \vec{J}^{\text{SS}}_{\cap} + \vec{J}^{\text{ES}}_{0}, \vec{M}^{\text{ES}}_{0}\right)\right]^{\tan}_{\vec{r}_{\text{mat}} \to \vec{r}} = \vec{J}^{\text{ES}}_{0}(\vec{r}) \times \hat{n}^{-}_{\text{mat}}(\vec{r}) \quad , \quad \vec{r} \in \partial V^{0}_{\text{mat}} \tag{5-55}$$

where operator $\mathcal{F}_{\text{mat}}(\vec{J}, \vec{M})$ is defined as that $\mathcal{F}_{\text{mat}}(\vec{J}, \vec{M}) = [\overleftrightarrow{G}^{JF}_{\text{mat}} * \vec{J} + \overleftrightarrow{G}^{MF}_{\text{mat}} * \vec{M}]_{\partial V_{\text{mat}}}$. Inserting expansion formulation (5-52) into equations (5-53)~(5-55), and testing equations (5-53), (5-54), and (5-55) with basis functions $\{\vec{b}_{\xi}^{\vec{J}^{\text{SL}}_{\cap}}\}_{\xi=1}^{\Xi^{\vec{J}^{\text{SL}}_{\cap}}}$, $\{\vec{b}_{\xi}^{\vec{J}^{\text{SS}}_{\cap}}\}_{\xi=1}^{\Xi^{\vec{J}^{\text{SS}}_{\cap}}}$, and $\{\vec{b}_{\xi}^{\vec{M}^{\text{ES}}_{0}}\}_{\xi=1}^{\Xi^{\vec{M}^{\text{ES}}_{0}}}$ respectively, integral equations (5-53)~(5-55) will be discretized into the following matrix equations:

$$\overline{\overline{Z}}^{\vec{J}_{\cap}^{\rm SL}\vec{J}_{\cap}^{\rm SL}}\cdot\overline{a}^{\vec{J}_{\cap}^{\rm SL}}+\overline{\overline{Z}}^{\vec{J}_{\cap}^{\rm SL}\vec{J}_{\cap}^{\rm SS}}\cdot\overline{a}^{\vec{J}_{\cap}^{\rm SS}}+\overline{\overline{Z}}^{\vec{J}_{\cap}^{\rm SL}\vec{J}_{0}^{\rm ES}}\cdot\overline{a}^{\vec{J}_{0}^{\rm ES}}+\overline{\overline{Z}}^{\vec{J}_{\cap}^{\rm SL}\vec{M}_{0}^{\rm ES}}\cdot\overline{a}^{\vec{M}_{0}^{\rm ES}}=0 \tag{5-56}$$

$$\overline{\overline{Z}}^{\vec{J}_{\cap}^{\rm SS}\vec{J}_{\cap}^{\rm SL}}\cdot\overline{a}^{\vec{J}_{\cap}^{\rm SL}}+\overline{\overline{Z}}^{\vec{J}_{\cap}^{\rm SS}\vec{J}_{\cap}^{\rm SS}}\cdot\overline{a}^{\vec{J}_{\cap}^{\rm SS}}+\overline{\overline{Z}}^{\vec{J}_{\cap}^{\rm SS}\vec{J}_{0}^{\rm ES}}\cdot\overline{a}^{\vec{J}_{0}^{\rm ES}}+\overline{\overline{Z}}^{\vec{J}_{\cap}^{\rm SS}\vec{M}_{0}^{\rm ES}}\cdot\overline{a}^{\vec{M}_{0}^{\rm ES}}=0 \tag{5-57}$$

$$\overline{\overline{Z}}^{\vec{M}_{0}^{\rm ES}\vec{J}_{\cap}^{\rm SL}}\cdot\overline{a}^{\vec{J}_{\cap}^{\rm SL}}+\overline{\overline{Z}}^{\vec{M}_{0}^{\rm ES}\vec{J}_{\cap}^{\rm SS}}\cdot\overline{a}^{\vec{J}_{\cap}^{\rm SS}}+\overline{\overline{Z}}^{\vec{M}_{0}^{\rm ES}\vec{J}_{0}^{\rm ES}}\cdot\overline{a}^{\vec{J}_{0}^{\rm ES}}+\overline{\overline{Z}}^{\vec{M}_{0}^{\rm ES}\vec{M}_{0}^{\rm ES}}\cdot\overline{a}^{\vec{M}_{0}^{\rm ES}}=\overline{\overline{C}}^{\vec{M}_{0}^{\rm ES}\vec{J}_{0}^{\rm ES}}\cdot\overline{a}^{\vec{J}_{0}^{\rm ES}} \tag{5-58}$$

In equations (5-56)~(5-58), the elements of various matrices are calculated as follows:

$$z_{\xi\zeta}^{\vec{X}\vec{J}_{\cap}^{\rm SL}}=\left\langle\vec{b}_{\xi}^{\vec{X}},\mathcal{F}_{\rm mat}\left(\vec{b}_{\zeta}^{\vec{J}_{\cap}^{\rm SL}},0\right)\right\rangle_{\Omega} \tag{5-59a}$$

$$z_{\xi\zeta}^{\vec{X}\vec{J}_{\cap}^{\rm SS}}=\left\langle\vec{b}_{\xi}^{\vec{X}},\mathcal{F}_{\rm mat}\left(\vec{b}_{\zeta}^{\vec{J}_{\cap}^{\rm SS}},0\right)\right\rangle_{\Omega} \tag{5-59b}$$

$$z_{\xi\zeta}^{\vec{X}\vec{J}_{0}^{\rm ES}}=\left\langle\vec{b}_{\xi}^{\vec{X}},\mathcal{F}_{\rm mat}\left(\vec{b}_{\zeta}^{\vec{J}_{0}^{\rm ES}},0\right)\right\rangle_{\Omega} \tag{5-59c}$$

$$z_{\xi\zeta}^{\vec{X}\vec{M}_{0}^{\rm ES}}=\left\langle\vec{b}_{\xi}^{\vec{X}},\mathcal{F}_{\rm mat}\left(0,\vec{b}_{\zeta}^{\vec{M}_{0}^{\rm ES}}\right)\right\rangle_{\Omega} \tag{5-59d}$$

$$c_{\xi\zeta}^{\vec{M}_{0}^{\rm ES}\vec{J}_{0}^{\rm ES}}=\left\langle\vec{b}_{\xi}^{\vec{M}_{0}^{\rm ES}},\vec{b}_{\zeta}^{\vec{J}_{0}^{ES}}\times\hat{n}_{\rm mat}^{-}\right\rangle_{\partial V_{\rm mat}^{0}} \tag{5-59e}$$

where $(\vec{X},\mathcal{F},\Omega)=(\vec{J}_{\cap}^{\rm SL},\mathcal{E},L_{\rm met}^{\cap}),(\vec{J}_{\cap}^{\rm SS},\mathcal{E},S_{\rm met}^{\cap}\cup\partial V_{\rm met}^{\cap}),(\vec{M}_{0}^{\rm ES},\mathcal{H},\partial V_{\rm mat}^{0;-})$.

To effectively establish the transformation from BVs to dependent variables, we equivalently assemble equations (5-56)~(5-58) as follows:

$$\begin{bmatrix}\overline{\overline{Z}}^{\vec{J}_{\cap}^{\rm SL}\vec{J}_{\cap}^{\rm SL}} & \overline{\overline{Z}}^{\vec{J}_{\cap}^{\rm SL}\vec{J}_{\cap}^{\rm SS}} & \overline{\overline{Z}}^{\vec{J}_{\cap}^{\rm SL}\vec{M}_{0}^{\rm ES}}\\ \overline{\overline{Z}}^{\vec{J}_{\cap}^{\rm SS}\vec{J}_{\cap}^{\rm SL}} & \overline{\overline{Z}}^{\vec{J}_{\cap}^{\rm SS}\vec{J}_{\cap}^{\rm SS}} & \overline{\overline{Z}}^{\vec{J}_{\cap}^{\rm SS}\vec{M}_{0}^{\rm ES}}\\ \overline{\overline{Z}}^{\vec{M}_{0}^{\rm ES}\vec{J}_{\cap}^{\rm SL}} & \overline{\overline{Z}}^{\vec{M}_{0}^{\rm ES}\vec{J}_{\cap}^{\rm SS}} & \overline{\overline{Z}}^{\vec{M}_{0}^{\rm ES}\vec{M}_{0}^{\rm ES}}\end{bmatrix}\cdot\begin{bmatrix}\overline{a}^{\vec{J}_{\cap}^{\rm SL}}\\ \overline{a}^{\vec{J}_{\cap}^{\rm SS}}\\ \overline{a}^{\vec{M}_{0}^{\rm ES}}\end{bmatrix}=-\begin{bmatrix}\overline{\overline{Z}}^{\vec{J}_{\cap}^{\rm SL}\vec{J}_{0}^{\rm ES}}\\ \overline{\overline{Z}}^{\vec{J}_{\cap}^{\rm SS}\vec{J}_{0}^{\rm ES}}\\ \overline{\overline{Z}}^{\vec{M}_{0}^{\rm ES}\vec{J}_{0}^{\rm ES}}-\overline{\overline{C}}^{\vec{M}_{0}^{\rm ES}\vec{J}_{0}^{\rm ES}}\end{bmatrix}\cdot\overline{a}^{\vec{J}_{0}^{\rm ES}} \tag{5-60}$$

By solving the above single matrix equation, we obtain the following transformation:

$$\begin{bmatrix}\overline{a}^{\vec{J}_{\cap}^{\rm SL}}\\ \overline{a}^{\vec{J}_{\cap}^{\rm SS}}\\ \overline{a}^{\vec{M}_{0}^{\rm ES}}\end{bmatrix}=\underbrace{-\begin{bmatrix}\overline{\overline{Z}}^{\vec{J}_{\cap}^{\rm SL}\vec{J}_{\cap}^{\rm SL}} & \overline{\overline{Z}}^{\vec{J}_{\cap}^{\rm SL}\vec{J}_{\cap}^{\rm SS}} & \overline{\overline{Z}}^{\vec{J}_{\cap}^{\rm SL}\vec{M}_{0}^{\rm ES}}\\ \overline{\overline{Z}}^{\vec{J}_{\cap}^{\rm SS}\vec{J}_{\cap}^{\rm SL}} & \overline{\overline{Z}}^{\vec{J}_{\cap}^{\rm SS}\vec{J}_{\cap}^{\rm SS}} & \overline{\overline{Z}}^{\vec{J}_{\cap}^{\rm SS}\vec{M}_{0}^{\rm ES}}\\ \overline{\overline{Z}}^{\vec{M}_{0}^{\rm ES}\vec{J}_{\cap}^{\rm SL}} & \overline{\overline{Z}}^{\vec{M}_{0}^{\rm ES}\vec{J}_{\cap}^{\rm SS}} & \overline{\overline{Z}}^{\vec{M}_{0}^{\rm ES}\vec{M}_{0}^{\rm ES}}\end{bmatrix}^{-1}\cdot\begin{bmatrix}\overline{\overline{Z}}^{\vec{J}_{\cap}^{\rm SL}\vec{J}_{0}^{\rm ES}}\\ \overline{\overline{Z}}^{\vec{J}_{\cap}^{\rm SS}\vec{J}_{0}^{\rm ES}}\\ \overline{\overline{Z}}^{\vec{M}_{0}^{\rm ES}\vec{J}_{0}^{\rm ES}}-\overline{\overline{C}}^{\vec{M}_{0}^{\rm ES}\vec{J}_{0}^{\rm ES}}\end{bmatrix}}_{\overline{\overline{T}}^{\{\vec{J}_{\cap}^{\rm SL},\vec{J}_{\cap}^{\rm SS},\vec{M}_{0}^{\rm ES}\}\leftarrow\vec{J}_{0}^{\rm ES}}}\cdot\overline{a}^{\vec{J}_{0}^{\rm ES}} \tag{5-61}$$

If we partition transformation matrix $\overline{\overline{T}}^{\{\vec{J}_{\cap}^{\rm SL},\vec{J}_{\cap}^{\rm SS},\vec{M}_{0}^{\rm ES}\}\leftarrow\vec{J}_{0}^{\rm ES}}$ as follows:

$$\overline{\overline{T}}^{\{\vec{J}_{\cap}^{\rm SL},\vec{J}_{\cap}^{\rm SS},\vec{M}_{0}^{\rm ES}\}\leftarrow\vec{J}_{0}^{\rm ES}}=\begin{bmatrix}\overline{\overline{T}}^{\vec{J}_{\cap}^{\rm SL}\leftarrow\vec{J}_{0}^{\rm ES}}\\ \overline{\overline{T}}^{\vec{J}_{\cap}^{\rm SS}\leftarrow\vec{J}_{0}^{\rm ES}}\\ \overline{\overline{T}}^{\vec{M}_{0}^{\rm ES}\leftarrow\vec{J}_{0}^{\rm ES}}\end{bmatrix} \tag{5-62}$$

then we further have the following transformations:

$$\overline{a}^{\vec{J}_{\cap}^{\rm SL}}=\overline{\overline{T}}^{\vec{J}_{\cap}^{\rm SL}\leftarrow\vec{J}_{0}^{\rm ES}}\cdot\overline{a}^{\vec{J}_{0}^{\rm ES}} \tag{5-63a}$$

$$\overline{a}^{\vec{J}_{\cap}^{\rm SS}}=\overline{\overline{T}}^{\vec{J}_{\cap}^{\rm SS}\leftarrow\vec{J}_{0}^{\rm ES}}\cdot\overline{a}^{\vec{J}_{0}^{\rm ES}} \tag{5-63b}$$

$$\overline{a}^{\vec{M}_{0}^{\rm ES}}=\overline{\overline{T}}^{\vec{M}_{0}^{\rm ES}\leftarrow\vec{J}_{0}^{\rm ES}}\cdot\overline{a}^{\vec{J}_{0}^{\rm ES}} \tag{5-63c}$$

where the line numbers of $\overline{\overline{T}}^{\vec{J}_{\cap}^{\rm SL}\leftarrow\vec{J}_0^{\rm ES}}$, $\overline{\overline{T}}^{\vec{J}_{\cap}^{\rm SS}\leftarrow\vec{J}_0^{\rm ES}}$, and $\overline{\overline{T}}^{\vec{M}_0^{\rm ES}\leftarrow\vec{J}_0^{\rm ES}}$ are the same as the line numbers of $\bar{a}^{\vec{J}_{\cap}^{\rm SL}}$, $\bar{a}^{\vec{J}_{\cap}^{\rm SS}}$, and $\bar{a}^{\vec{M}_0^{\rm ES}}$ respectively.

### 5.5.3 Line-Surface Formulation of DPO

The driving power done by incident fields $\{\vec{E}^{\rm inc},\vec{H}^{\rm inc}\}$ on the scattered sources distributing on the composite system shown in Figure 5-1 is as follows:

$$\begin{aligned}P_{\rm com\,sys}^{\rm driving} &= (1/2)\left\langle \vec{J}^{\rm SL}\oplus\vec{J}^{\rm SS}\oplus\vec{J}^{\rm SV},\vec{E}^{\rm inc}\right\rangle_{L_{\rm met}\cup S_{\rm met}\cup\partial V_{\rm met}\cup V_{\rm mat}} + (1/2)\left\langle \vec{M}^{\rm SV},\vec{H}^{\rm inc}\right\rangle_{V_{\rm mat}}\\ &= (1/2)\left\langle \vec{J}_0^{\rm SL}\oplus\vec{J}_0^{\rm SS},\vec{E}^{\rm inc}\right\rangle_{L_{\rm met}^0\cup S_{\rm met}^0\cup\partial V_{\rm met}^0} + (1/2)\left\langle \vec{J}_{\cap}^{\rm SL}\oplus\vec{J}_{\cap}^{\rm SS},\vec{E}^{\rm inc}\right\rangle_{L_{\rm met}^{\cap}\cup S_{\rm met}^{\cap}\cup\partial V_{\rm met}^{\cap}}\\ &\quad + (1/2)\left\langle \vec{J}^{\rm SV},\vec{E}^{\rm inc}\right\rangle_{V_{\rm mat}} + (1/2)\left\langle \vec{M}^{\rm SV},\vec{H}^{\rm inc}\right\rangle_{V_{\rm mat}}\end{aligned} \tag{5-64}$$

where the second equality is based on current decompositions (5-15) and (5-16) and the linear property of inner product. Based on the first equality in the operator expression (4-31) given in Subsection 4.3.2, it can be known that

$$\begin{aligned}&(1/2)\left\langle \vec{J}^{\rm SV},\vec{E}^{\rm inc}\right\rangle_{V_{\rm mat}} + (1/2)\left\langle \vec{M}^{\rm SV},\vec{H}^{\rm inc}\right\rangle_{V_{\rm mat}}\\ &= -(1/2)\left\langle \vec{J}^{\rm EL}\oplus\left(\vec{J}_{\cap}^{\rm ES}+\vec{J}_0^{\rm ES}\right),\vec{E}^{\rm inc}\right\rangle_{\partial V_{\rm mat}} - (1/2)\left\langle \vec{M}_0^{\rm ES},\vec{H}^{\rm inc}\right\rangle_{\partial V_{\rm mat}}\\ &= -(1/2)\left\langle \vec{J}_{\cap}^{\rm SL}\oplus\left(\vec{J}_{\cap}^{\rm SS}+\vec{J}_0^{\rm ES}\right),\vec{E}^{\rm inc}\right\rangle_{L_{\rm met}^{\cap}\cup S_{\rm met}^{\cap}\cup\partial V_{\rm met}^{\cap}\cup\partial V_{\rm mat}^0} - (1/2)\left\langle \vec{M}_0^{\rm ES},\vec{H}^{\rm inc}\right\rangle_{\partial V_{\rm mat}^0}\end{aligned} \tag{5-65}$$

where the second equality is based on relationships (5-36) and (5-38). Inserting relationship (5-65) into operator expression (5-64), it can be obtained that

$$\begin{aligned}P_{\rm com\,sys}^{\rm driving} &= (1/2)\left\langle \vec{J}_0^{\rm SL}\oplus\vec{J}_0^{\rm SS},\vec{E}^{\rm inc}\right\rangle_{L_{\rm met}^0\cup S_{\rm met}^0\cup\partial V_{\rm met}^0} + (1/2)\left\langle \vec{J}_{\cap}^{\rm SL}\oplus\vec{J}_{\cap}^{\rm SS},\vec{E}^{\rm inc}\right\rangle_{L_{\rm met}^{\cap}\cup S_{\rm met}^{\cap}\cup\partial V_{\rm met}^{\cap}}\\ &\quad -(1/2)\left\langle \vec{J}_{\cap}^{\rm SL}\oplus\left(\vec{J}_{\cap}^{\rm SS}+\vec{J}_0^{\rm ES}\right),\vec{E}^{\rm inc}\right\rangle_{L_{\rm met}^{\cap}\cup S_{\rm met}^{\cap}\cup\partial V_{\rm met}^{\cap}\cup\partial V_{\rm mat}^0} - (1/2)\left\langle \vec{M}_0^{\rm ES},\vec{H}^{\rm inc}\right\rangle_{\partial V_{\rm mat}^0}\\ &= (1/2)\left\langle \vec{J}_0^{\rm SL}\oplus\vec{J}_0^{\rm SS},\vec{E}^{\rm inc}\right\rangle_{L_{\rm met}^0\cup S_{\rm met}^0\cup\partial V_{\rm met}^0} - (1/2)\left\langle \vec{J}_0^{\rm ES},\vec{E}^{\rm inc}\right\rangle_{\partial V_{\rm mat}^0} - (1/2)\left\langle \vec{M}_0^{\rm ES},\vec{H}^{\rm inc}\right\rangle_{\partial V_{\rm mat}^0}\\ &= (1/2)\left\langle \vec{J}_0^{\rm SL}\oplus\vec{J}_0^{\rm SS},-\vec{E}^{\rm sca}\right\rangle_{L_{\rm met}^0\cup S_{\rm met}^0\cup\partial V_{\rm met}^0} - (1/2)\left\langle \vec{J}_0^{\rm ES},\vec{E}^{\rm inc}\right\rangle_{\partial V_{\rm mat}^0} - (1/2)\left\langle \vec{M}_0^{\rm ES},\vec{H}^{\rm inc}\right\rangle_{\partial V_{\rm mat}^0}\\ &= (1/2)\left\langle \vec{J}_0^{\rm SL}\oplus\vec{J}_0^{\rm SS},-\vec{E}^{\rm sca}\right\rangle_{L_{\rm met}^0\cup S_{\rm met}^0\cup\partial V_{\rm met}^0}\\ &\quad -(1/2)\left\langle \vec{J}_0^{\rm ES},\vec{E}_-^{\rm tot}-\vec{E}_+^{\rm sca}\right\rangle_{\partial V_{\rm mat}^0} - (1/2)\left\langle \vec{M}_0^{\rm ES},\vec{H}_-^{\rm tot}-\vec{H}_+^{\rm sca}\right\rangle_{\partial V_{\rm mat}^0}\\ &= \frac{1}{2}\left\langle \vec{J}_0^{\rm SL}\oplus\vec{J}_0^{\rm SS},-\vec{E}^{\rm sca}\right\rangle_{L_{\rm met}^0\cup S_{\rm met}^0\cup\partial V_{\rm met}^0} - \frac{1}{2}\left\langle \vec{J}_0^{\rm ES},\vec{E}^{\rm inc}\right\rangle_{\partial V_{\rm mat}^{0;-}} - \frac{1}{2}\left\langle \vec{M}_0^{\rm ES},\vec{H}^{\rm inc}\right\rangle_{\partial V_{\rm mat}^{0;-}}\end{aligned} \tag{5-66}$$

In operator expression (5-66), the second equality is evident; the third equality is based on the tangential electric field boundary condition $\vec{E}_{\rm tan}^{\rm inc}=-\vec{E}_{\rm tan}^{\rm sca}$ on metallic boundary; the $\vec{F}_-^{\rm tot}$ and $\vec{F}_+^{\rm sca}$ in the fourth equality are respectively the $\vec{F}^{\rm tot}$ on material internal surface and the $\vec{F}^{\rm sca}$ on material external surface, and the equality is based on superposition principle and the tangential continuity of the $\vec{F}^{\rm tot}$ and $\vec{F}^{\rm sca}$ on $\partial V_{\rm mat}^0$;

the integral domain $\partial V_{\text{mat}}^{0;-}$ used in the fifth equality is the material side corresponding to $\partial V_{\text{mat}}^{0}$, and the equality is based on the continuity of the $\vec{F}^{\text{inc}}$ on $\partial V_{\text{mat}}^{0}$.

Inserting the GFHF (5-45) and GFHF (5-47b) obtained in Subsection 5.4.2 into the RHS of the fourth equality in operator expression (5-66), we obtain that

$$\begin{aligned} P_{\text{com sys}}^{\text{driving}} &= (1/2)\left\langle \vec{J}_0^{\text{SL}} \oplus \vec{J}_0^{\text{SS}}, -\vec{E}^{\text{sca}} \right\rangle_{L_{\text{met}}^0 \cup S_{\text{met}}^0 \cup \partial V_{\text{met}}^0} \\ &\quad - (1/2)\left\langle \vec{J}_0^{\text{ES}}, \vec{E}_-^{\text{tot}} - \vec{E}_+^{\text{sca}} \right\rangle_{\partial V_{\text{mat}}^0} - (1/2)\left\langle \vec{M}_0^{\text{ES}}, \vec{H}_-^{\text{tot}} - \vec{H}_+^{\text{sca}} \right\rangle_{\partial V_{\text{mat}}^0} \\ &= (1/2)\left\langle \vec{J}_0^{\text{SL}} \oplus \vec{J}_0^{\text{SS}} - \vec{J}_0^{\text{ES}}, \mathcal{E}_0\left(-\vec{J}_0^{\text{SL}} - \vec{J}_0^{\text{SS}} + \vec{J}_0^{\text{ES}}, \vec{M}_0^{\text{ES}}\right) \right\rangle_{L_{\text{met}}^0 \cup S_{\text{met}}^0 \cup \partial V_{\text{met}}^0 \cup \partial V_{\text{mat}}^{0;+}} \\ &\quad - (1/2)\left\langle \vec{M}_0^{\text{ES}}, \mathcal{H}_0\left(-\vec{J}_0^{\text{SL}} - \vec{J}_0^{\text{SS}} + \vec{J}_0^{\text{ES}}, \vec{M}_0^{\text{ES}}\right) \right\rangle_{\partial V_{\text{mat}}^{0;+}} \\ &\quad - (1/2)\left\langle \vec{J}_0^{\text{ES}}, \mathcal{E}_{\text{mat}}\left(\vec{J}_\cap^{\text{SL}} \oplus \vec{J}_\cap^{\text{SS}} + \vec{J}_0^{\text{ES}}, \vec{M}_0^{\text{ES}}\right) \right\rangle_{\partial V_{\text{mat}}^{0;-}} \\ &\quad - (1/2)\left\langle \vec{M}_0^{\text{ES}}, \mathcal{H}_{\text{mat}}\left(\vec{J}_\cap^{\text{SL}} \oplus \vec{J}_\cap^{\text{SS}} + \vec{J}_0^{\text{ES}}, \vec{M}_0^{\text{ES}}\right) \right\rangle_{\partial V_{\text{mat}}^{0;-}} \end{aligned} \tag{5-67}$$

Inserting expansion formulation (5-52) into operator expression (5-67), DPO $P_{\text{com sys}}^{\text{driving}}$ is immediately discretized into matrix form. Based on the matrix form and transformation (5-63), the matrix form of $P_{\text{com sys}}^{\text{driving}}$ which only contains BVs can be obtained, and then the DP-CMs can be derived from orthogonalizing the matrix form only containing BVs. The above scheme is very similar to the scheme used in Section 4.4 and the first scheme used in Section 4.7. We have no intention of providing the detailed process of the above scheme, and we will provide another scheme to the composite system shown in Figure 5-1, and the another scheme is very similar to the scheme used in Section 4.6 and the third scheme used in Section 4.7.

Inserting the GFHF (5-45) obtained in Section 5.4 into the RHS of the last equality in operator expression (5-66), we have that

$$\begin{aligned} P_{\text{com sys}}^{\text{driving}} &= (1/2)\left\langle \vec{J}_0^{\text{SL}} \oplus \vec{J}_0^{\text{SS}}, -\vec{E}^{\text{sca}} \right\rangle_{L_{\text{met}}^0 \cup S_{\text{met}}^0 \cup \partial V_{\text{met}}^0} - (1/2)\left\langle \vec{J}_0^{\text{ES}}, \vec{E}^{\text{inc}} \right\rangle_{\partial V_{\text{mat}}^{0;-}} - (1/2)\left\langle \vec{M}_0^{\text{ES}}, \vec{H}^{\text{inc}} \right\rangle_{\partial V_{\text{mat}}^{0;-}} \\ &= (1/2)\left\langle \vec{J}_0^{\text{SL}} \oplus \vec{J}_0^{\text{SS}}, \mathcal{E}_0\left(-\vec{J}_0^{\text{SL}} - \vec{J}_0^{\text{SS}} + \vec{J}_0^{\text{ES}}, \vec{M}_0^{\text{ES}}\right) \right\rangle_{L_{\text{met}}^0 \cup S_{\text{met}}^0 \cup \partial V_{\text{met}}^0} \\ &\quad - (1/2)\left\langle \vec{J}_0^{\text{ES}}, \mathcal{E}_0\left(-\vec{J}_0^{\text{SL}} - \vec{J}_0^{\text{SS}} + \vec{J}_0^{\text{ES}}, \vec{M}_0^{\text{ES}}\right) \right\rangle_{\partial V_{\text{mat}}^{0;-}} \\ &\quad - (1/2)\left\langle \vec{M}_0^{\text{ES}}, \mathcal{H}_0\left(-\vec{J}_0^{\text{SL}} - \vec{J}_0^{\text{SS}} + \vec{J}_0^{\text{ES}}, \vec{M}_0^{\text{ES}}\right) \right\rangle_{\partial V_{\text{mat}}^{0;-}} \\ &= (1/2)\left\langle \vec{J}_0^{\text{SL}} \oplus \vec{J}_0^{\text{SS}}, \mathcal{E}_0\left(-\vec{J}_0^{\text{SL}} - \vec{J}_0^{\text{SS}} + \vec{J}_0^{\text{ES}}, \vec{M}_0^{\text{ES}}\right) \right\rangle_{L_{\text{met}}^0 \cup S_{\text{met}}^0 \cup \partial V_{\text{met}}^0} \\ &\quad + (1/2)\left\langle \vec{J}_0^{\text{ES}}, \mathcal{E}_0\left(\vec{J}_0^{\text{SL}} + \vec{J}_0^{\text{SS}} - \vec{J}_0^{\text{ES}}, -\vec{M}_0^{\text{ES}}\right) \right\rangle_{\partial V_{\text{mat}}^{0;-}} \\ &\quad + (1/2)\left\langle \vec{M}_0^{\text{ES}}, \mathcal{H}_0\left(\vec{J}_0^{\text{SL}} + \vec{J}_0^{\text{SS}} - \vec{J}_0^{\text{ES}}, -\vec{M}_0^{\text{ES}}\right) \right\rangle_{\partial V_{\text{mat}}^{0;-}} \end{aligned} \tag{5-68}$$

The above third equality is evident. Inserting expansion formulation (5-52) into operator expression (5-68), operator $P_{\text{com sys}}^{\text{driving}}$ is immediately discretized into the following matrix form:

$$P_{\text{com sys}}^{\text{driving}} = \left(\bar{a}^{JM}\right)^H \cdot \underbrace{\left(\bar{\bar{P}}_{0;\text{PVT}}^{\text{driving}} + \bar{\bar{P}}_{0;\text{SCT}}^{\text{driving}}\right)}_{\bar{\bar{P}}_{\text{com sys}}^{\text{driving}}} \cdot \bar{a}^{JM} \tag{5-69}$$

where

$$\bar{\bar{P}}_{0;\text{PVT}}^{\text{driving}} = \begin{bmatrix} \bar{\bar{P}}_{0;\text{PVT}}^{\vec{J}_0^{\text{SL}}\vec{J}_0^{\text{SL}}} & \bar{\bar{P}}_{0;\text{PVT}}^{\vec{J}_0^{\text{SL}}\vec{J}_0^{\text{SS}}} & \bar{\bar{P}}_{0;\text{PVT}}^{\vec{J}_0^{\text{SL}}\vec{J}_0^{\text{ES}}} & \bar{\bar{P}}_{0;\text{PVT}}^{\vec{J}_0^{\text{SL}}\vec{M}_0^{\text{ES}}} \\ \bar{\bar{P}}_{0;\text{PVT}}^{\vec{J}_0^{\text{SS}}\vec{J}_0^{\text{SL}}} & \bar{\bar{P}}_{0;\text{PVT}}^{\vec{J}_0^{\text{SS}}\vec{J}_0^{\text{SS}}} & \bar{\bar{P}}_{0;\text{PVT}}^{\vec{J}_0^{\text{SS}}\vec{J}_0^{\text{ES}}} & \bar{\bar{P}}_{0;\text{PVT}}^{\vec{J}_0^{\text{SS}}\vec{M}_0^{\text{ES}}} \\ \bar{\bar{P}}_{0;\text{PVT}}^{\vec{J}_0^{\text{ES}}\vec{J}_0^{\text{SL}}} & \bar{\bar{P}}_{0;\text{PVT}}^{\vec{J}_0^{\text{ES}}\vec{J}_0^{\text{SS}}} & \bar{\bar{P}}_{0;\text{PVT}}^{\vec{J}_0^{\text{ES}}\vec{J}_0^{\text{ES}}} & \bar{\bar{P}}_{0;\text{PVT}}^{\vec{J}_0^{\text{ES}}\vec{M}_0^{\text{ES}}} \\ \bar{\bar{P}}_{0;\text{PVT}}^{\vec{M}_0^{\text{ES}}\vec{J}_0^{\text{SL}}} & \bar{\bar{P}}_{0;\text{PVT}}^{\vec{M}_0^{\text{ES}}\vec{J}_0^{\text{SS}}} & \bar{\bar{P}}_{0;\text{PVT}}^{\vec{M}_0^{\text{ES}}\vec{J}_0^{\text{ES}}} & \bar{\bar{P}}_{0;\text{PVT}}^{\vec{M}_0^{\text{ES}}\vec{M}_0^{\text{ES}}} \end{bmatrix} \tag{5-70a}$$

$$\bar{\bar{P}}_{0;\text{SCT}}^{\text{driving}} = \begin{bmatrix} 0 & 0 & 0 & 0 \\ 0 & 0 & 0 & 0 \\ 0 & 0 & 0 & \bar{\bar{P}}_{0;\text{SCT}}^{\vec{J}_0^{\text{ES}}\vec{M}_0^{\text{ES}}} \\ 0 & 0 & \bar{\bar{P}}_{0;\text{SCT}}^{\vec{M}_0^{\text{ES}}\vec{J}_0^{\text{ES}}} & 0 \end{bmatrix} \tag{5-70b}$$

$$\bar{a}^{JM} = \begin{bmatrix} \bar{a}^{\vec{J}_0^{\text{SL}}} \\ \bar{a}^{\vec{J}_0^{\text{SS}}} \\ \bar{a}^{\vec{J}_0^{\text{ES}}} \\ \bar{a}^{\vec{M}_0^{\text{ES}}} \end{bmatrix} \tag{5-70c}$$

The elements of the above various sub-matrices are calculated as follows:

$$p_{0;\text{PVT};\xi\zeta}^{\vec{J}_0^{\text{SL}}\vec{J}_0^{\text{SL}}} = -\left(1/2\right)\left\langle \vec{b}_\xi^{\vec{J}_0^{\text{SL}}}, -j\omega\mu_0\mathcal{L}_0\left(\vec{b}_\zeta^{\vec{J}_0^{\text{SL}}}\right)\right\rangle_{L_{\text{met}}^0} \tag{5-71a}$$

$$p_{0;\text{PVT};\xi\zeta}^{\vec{J}_0^{\text{SL}}\vec{J}_0^{\text{SS}}} = -\left(1/2\right)\left\langle \vec{b}_\xi^{\vec{J}_0^{\text{SL}}}, -j\omega\mu_0\mathcal{L}_0\left(\vec{b}_\zeta^{\vec{J}_0^{\text{SS}}}\right)\right\rangle_{L_{\text{met}}^0} \tag{5-71b}$$

$$p_{0;\text{PVT};\xi\zeta}^{\vec{J}_0^{\text{SL}}\vec{J}_0^{\text{ES}}} = \left(1/2\right)\left\langle \vec{b}_\xi^{\vec{J}_0^{\text{SL}}}, -j\omega\mu_0\mathcal{L}_0\left(\vec{b}_\zeta^{\vec{J}_0^{\text{ES}}}\right)\right\rangle_{L_{\text{met}}^0} \tag{5-71c}$$

$$p_{0;\text{PVT};\xi\zeta}^{\vec{J}_0^{\text{SL}}\vec{M}_0^{\text{ES}}} = \left(1/2\right)\left\langle \vec{b}_\xi^{\vec{J}_0^{\text{SL}}}, -\mathcal{K}_0\left(\vec{b}_\zeta^{\vec{M}_0^{\text{ES}}}\right)\right\rangle_{L_{\text{met}}^0} \tag{5-71d}$$

and

$$p_{0;\text{PVT};\xi\zeta}^{\vec{J}_0^{\text{SS}}\vec{J}_0^{\text{SL}}} = -\left(1/2\right)\left\langle \vec{b}_\xi^{\vec{J}_0^{\text{SS}}}, -j\omega\mu_0\mathcal{L}_0\left(\vec{b}_\zeta^{\vec{J}_0^{\text{SL}}}\right)\right\rangle_{S_{\text{met}}^0 \cup \partial V_{\text{met}}^0} \tag{5-72a}$$

$$p_{0;\text{PVT};\xi\zeta}^{\vec{J}_0^{\text{SS}}\vec{J}_0^{\text{SS}}} = -\left(1/2\right)\left\langle \vec{b}_\xi^{\vec{J}_0^{\text{SS}}}, -j\omega\mu_0\mathcal{L}_0\left(\vec{b}_\zeta^{\vec{J}_0^{\text{SS}}}\right)\right\rangle_{S_{\text{met}}^0 \cup \partial V_{\text{met}}^0} \tag{5-72b}$$

$$p_{0;\text{PVT};\xi\zeta}^{\vec{J}_0^{\text{SS}}\vec{J}_0^{\text{ES}}} = \left(1/2\right)\left\langle \vec{b}_\xi^{\vec{J}_0^{\text{SS}}}, -j\omega\mu_0\mathcal{L}_0\left(\vec{b}_\zeta^{\vec{J}_0^{\text{ES}}}\right)\right\rangle_{S_{\text{met}}^0 \cup \partial V_{\text{met}}^0} \tag{5-72c}$$

$$p_{0;\text{PVT};\xi\zeta}^{\vec{J}_0^{\text{SS}}\vec{M}_0^{\text{ES}}} = \left(1/2\right)\left\langle \vec{b}_\xi^{\vec{J}_0^{\text{SS}}}, -\mathcal{K}_0\left(\vec{b}_\zeta^{\vec{M}_0^{\text{ES}}}\right)\right\rangle_{S_{\text{met}}^0 \cup \partial V_{\text{met}}^0} \tag{5-72d}$$

and

$$p_{0;\text{PVT};\xi\zeta}^{\vec{J}_0^{\text{ES}}\vec{J}_0^{\text{SL}}} = \ (1/2)\left\langle \vec{b}_\xi^{\vec{J}_0^{\text{ES}}}, -j\omega\mu_0\mathcal{L}_0\left(\vec{b}_\zeta^{\vec{J}_0^{\text{SL}}}\right)\right\rangle_{\partial V_{\text{mat}}^0} \tag{5-73a}$$

$$p_{0;\text{PVT};\xi\zeta}^{\vec{J}_0^{\text{ES}}\vec{J}_0^{\text{SS}}} = \ (1/2)\left\langle \vec{b}_\xi^{\vec{J}_0^{\text{ES}}}, -j\omega\mu_0\mathcal{L}_0\left(\vec{b}_\zeta^{\vec{J}_0^{\text{SS}}}\right)\right\rangle_{\partial V_{\text{mat}}^0} \tag{5-73b}$$

$$p_{0;\text{PVT};\xi\zeta}^{\vec{J}_0^{\text{ES}}\vec{J}_0^{\text{ES}}} = -(1/2)\left\langle \vec{b}_\xi^{\vec{J}_0^{\text{ES}}}, -j\omega\mu_0\mathcal{L}_0\left(\vec{b}_\zeta^{\vec{J}_0^{\text{ES}}}\right)\right\rangle_{\partial V_{\text{mat}}^0} \tag{5-73c}$$

$$p_{0;\text{PVT};\xi\zeta}^{\vec{J}_0^{\text{ES}}\vec{M}_0^{\text{ES}}} = -(1/2)\left\langle \vec{b}_\xi^{\vec{J}_0^{\text{ES}}}, -\text{P.V.}\,\mathcal{K}_0\left(\vec{b}_\zeta^{\vec{M}_0^{\text{ES}}}\right)\right\rangle_{\partial V_{\text{mat}}^0} \tag{5-73d}$$

and

$$p_{0;\text{PVT};\xi\zeta}^{\vec{M}_0^{\text{ES}}\vec{J}_0^{\text{SL}}} = \ (1/2)\left\langle \vec{b}_\xi^{\vec{M}_0^{\text{ES}}}, \mathcal{K}_0\left(\vec{b}_\zeta^{\vec{J}_0^{\text{SL}}}\right)\right\rangle_{\partial V_{\text{mat}}^0} \tag{5-74a}$$

$$p_{0;\text{PVT};\xi\zeta}^{\vec{M}_0^{\text{ES}}\vec{J}_0^{\text{SS}}} = \ (1/2)\left\langle \vec{b}_\xi^{\vec{M}_0^{\text{ES}}}, \mathcal{K}_0\left(\vec{b}_\zeta^{\vec{J}_0^{\text{SS}}}\right)\right\rangle_{\partial V_{\text{mat}}^0} \tag{5-74b}$$

$$p_{0;\text{PVT};\xi\zeta}^{\vec{M}_0^{\text{ES}}\vec{J}_0^{\text{ES}}} = -(1/2)\left\langle \vec{b}_\xi^{\vec{M}_0^{\text{ES}}}, \text{P.V.}\,\mathcal{K}_0\left(\vec{b}_\zeta^{\vec{J}_0^{\text{ES}}}\right)\right\rangle_{\partial V_{\text{mat}}^0} \tag{5-74c}$$

$$p_{0;\text{PVT};\xi\zeta}^{\vec{M}_0^{\text{ES}}\vec{M}_0^{\text{ES}}} = -(1/2)\left\langle \vec{b}_\xi^{\vec{M}_0^{\text{ES}}}, -j\omega\varepsilon_0\mathcal{L}_0\left(\vec{b}_\zeta^{\vec{M}_0^{\text{ES}}}\right)\right\rangle_{\partial V_{\text{mat}}^0} \tag{5-74d}$$

and

$$p_{0;\text{SCT};\xi\zeta}^{\vec{J}_0^{\text{ES}}\vec{M}_0^{\text{ES}}} = -(1/2)\left\langle \vec{b}_\xi^{\vec{J}_0^{\text{ES}}}, \hat{n}_{\text{mat}}^{-} \times \frac{1}{2}\vec{b}_\zeta^{\vec{M}_0^{\text{ES}}}\right\rangle_{\partial V_{\text{mat}}^0} \tag{5-75a}$$

$$p_{0;\text{SCT};\xi\zeta}^{\vec{M}_0^{\text{ES}}\vec{J}_0^{\text{ES}}} = -(1/2)\left\langle \vec{b}_\xi^{\vec{M}_0^{\text{ES}}}, \frac{1}{2}\vec{b}_\zeta^{\vec{J}_0^{\text{ES}}} \times \hat{n}_{\text{mat}}^{-}\right\rangle_{\partial V_{\text{mat}}^0} \tag{5-75b}$$

Inserting transformation (5-63c) into matrix form (5-69), we obtain the matrix form of $P_{\text{com sys}}^{\text{driving}}$ only containing BV $\bar{a}^{J_0}$ as follows:

$$P_{\text{com sys}}^{\text{driving}} = \left(\bar{a}^{J_0}\right)^H \cdot \overline{\overline{P}}_{J_0}^{\text{driving}} \cdot \bar{a}^{J_0} \tag{5-76}$$

where

$$\overline{\overline{P}}_{J_0}^{\text{driving}} = \begin{bmatrix} \overline{\overline{I}}^{\vec{J}_0^{\text{SL}}} & 0 & 0 \\ 0 & \overline{\overline{I}}^{\vec{J}_0^{\text{SS}}} & 0 \\ 0 & 0 & \overline{\overline{I}}^{\vec{J}_0^{\text{ES}}} \\ 0 & 0 & \overline{\overline{T}}^{\vec{M}_0^{\text{ES}} \leftarrow \vec{J}_0^{\text{ES}}} \end{bmatrix}^H \cdot \overline{\overline{P}}_{\text{com sys}}^{\text{driving}} \cdot \begin{bmatrix} \overline{\overline{I}}^{\vec{J}_0^{\text{SL}}} & 0 & 0 \\ 0 & \overline{\overline{I}}^{\vec{J}_0^{\text{SS}}} & 0 \\ 0 & 0 & \overline{\overline{I}}^{\vec{J}_0^{\text{ES}}} \\ 0 & 0 & \overline{\overline{T}}^{\vec{M}_0^{\text{ES}} \leftarrow \vec{J}_0^{\text{ES}}} \end{bmatrix} \tag{5-77a}$$

$$\bar{a}^{J_0} = \begin{bmatrix} \bar{a}^{\vec{J}_0^{\text{SL}}} \\ \bar{a}^{\vec{J}_0^{\text{SS}}} \\ \bar{a}^{\vec{J}_0^{\text{ES}}} \end{bmatrix} \tag{5-77b}$$

In formulation (5-77a), the 0s are some zero matrices with proper line numbers and column numbers.

### 5.5.4 DP-CMs and Their Orthogonality

It is obvious that $\overline{\overline{P}}_{J_0}^{\text{driving}}$ is a square matrix, so there must be the following Toeplitz's decomposition[116]:

$$\overline{\overline{P}}_{J_0}^{\text{driving}} = \overline{\overline{P}}_{J_0;+}^{\text{driving}} + j\,\overline{\overline{P}}_{J_0;-}^{\text{driving}} \tag{5-78}$$

where

$$\overline{\overline{P}}_{J_0;+}^{\text{driving}} = \frac{1}{2}\left[\overline{\overline{P}}_{J_0}^{\text{driving}} + \left(\overline{\overline{P}}_{J_0}^{\text{driving}}\right)^H\right] \tag{5-79a}$$

$$\overline{\overline{P}}_{J_0;-}^{\text{driving}} = \frac{1}{2j}\left[\overline{\overline{P}}_{J_0}^{\text{driving}} - \left(\overline{\overline{P}}_{J_0}^{\text{driving}}\right)^H\right] \tag{5-79b}$$

The characteristic vectors corresponding to the DP-CMs of composite system $D_{\text{com sys}}$ can be derived from solving the following generalized characteristic equation:

$$\overline{\overline{P}}_{J_0;-}^{\text{driving}} \cdot \bar{\alpha}_{J_0;\xi}^{\text{driving}} = \lambda_{\text{com sys};\xi}^{\text{driving}}\, \overline{\overline{P}}_{J_0;+}^{\text{driving}} \cdot \bar{\alpha}_{J_0;\xi}^{\text{driving}} \tag{5-80}$$

and then

$$\begin{bmatrix} \bar{\alpha}_{\xi}^{\vec{J}_0^{\text{SL}}} \\ \bar{\alpha}_{\xi}^{\vec{J}_0^{\text{SS}}} \\ \bar{\alpha}_{\xi}^{\vec{J}_0^{\text{ES}}} \end{bmatrix} = \bar{\alpha}_{J_0;\xi}^{\text{driving}} \tag{5-81}$$

Based on transformation (5-63), $\bar{\alpha}_{\xi}^{\vec{J}_{\cap}^{\text{SL}}}$ , $\bar{\alpha}_{\xi}^{\vec{J}_{\cap}^{\text{SS}}}$, and $\bar{\alpha}_{\xi}^{\vec{M}_0^{\text{ES}}}$ can then be obtained. Inserting the above various characteristic sub-vectors into the corresponding expansion formulations, the various characteristic sub-currents can be determined. Inserting the characteristic sub-currents into GFHF (5-45), GFHF (5-47b), and GFHF (5-49), the various characteristic fields can be obtained. Inserting the characteristic total fields distributing on $D_{\text{com sys}}$ into the relationships $\vec{J}^{\text{SV}} = j\omega\Delta\ddot{\varepsilon}_{\text{mat}}^{\text{c}} \cdot \vec{E}^{\text{tot}}$ and $\vec{M}^{\text{SV}} = j\omega\Delta\ddot{\mu}_{\text{mat}} \cdot \vec{H}^{\text{tot}}$ given in Appendix A, the various characteristic scattered volume sources can be determined.

Obviously, characteristic values $\lambda_{\text{com sys};\xi}^{\text{driving}}$ and the modal powers satisfy the following relationship:

$$\begin{aligned}
\lambda_{\text{com sys};\xi}^{\text{driving}} &= \frac{\text{Im}\left\{P_{\text{com sys};\xi}^{\text{driving}}\right\}}{\text{Re}\left\{P_{\text{com sys};\xi}^{\text{driving}}\right\}} = \frac{\left(\bar{\alpha}_{J_0;\xi}^{\text{driving}}\right)^H \cdot \overline{\overline{P}}_{J_0;-}^{\text{driving}} \cdot \bar{\alpha}_{J_0;\xi}^{\text{driving}}}{\left(\bar{\alpha}_{J_0;\xi}^{\text{driving}}\right)^H \cdot \overline{\overline{P}}_{J_0;+}^{\text{driving}} \cdot \bar{\alpha}_{J_0;\xi}^{\text{driving}}} \\
&= \frac{\text{Im}\left\{(1/2)\left\langle \vec{J}_{\xi}^{\text{SL}} \oplus \vec{J}_{\xi}^{\text{SS}} \oplus \vec{J}_{\xi}^{\text{SV}}, \vec{E}_{\xi}^{\text{inc}} \right\rangle_{L_{\text{met}} \cup S_{\text{met}} \cup \partial V_{\text{met}} \cup V_{\text{mat}}} + (1/2)\left\langle \vec{M}_{\xi}^{\text{SV}}, \vec{H}_{\xi}^{\text{inc}} \right\rangle_{V_{\text{mat}}}\right\}}{\text{Re}\left\{(1/2)\left\langle \vec{J}_{\xi}^{\text{SL}} \oplus \vec{J}_{\xi}^{\text{SS}} \oplus \vec{J}_{\xi}^{\text{SV}}, \vec{E}_{\xi}^{\text{inc}} \right\rangle_{L_{\text{met}} \cup S_{\text{met}} \cup \partial V_{\text{met}} \cup V_{\text{mat}}} + (1/2)\left\langle \vec{M}_{\xi}^{\text{SV}}, \vec{H}_{\xi}^{\text{inc}} \right\rangle_{V_{\text{mat}}}\right\}} \\
&= \frac{\text{Im}\left\{\frac{1}{2}\left\langle \vec{J}_{0;\xi}^{\text{SL}} \oplus \vec{J}_{0;\xi}^{\text{SS}}, \vec{E}_{\xi}^{\text{inc}} \right\rangle_{L_{\text{met}}^0 \cup S_{\text{met}}^0 \cup \partial V_{\text{met}}^0} - \frac{1}{2}\left\langle \vec{J}_{0;\xi}^{\text{ES}}, \vec{E}_{\xi}^{\text{inc}} \right\rangle_{\partial V_{\text{mat}}^0} - \frac{1}{2}\left\langle \vec{M}_{0;\xi}^{\text{ES}}, \vec{H}_{\xi}^{\text{inc}} \right\rangle_{\partial V_{\text{mat}}^0}\right\}}{\text{Re}\left\{\frac{1}{2}\left\langle \vec{J}_{0;\xi}^{\text{SL}} \oplus \vec{J}_{0;\xi}^{\text{SS}}, \vec{E}_{\xi}^{\text{inc}} \right\rangle_{L_{\text{met}}^0 \cup S_{\text{met}}^0 \cup \partial V_{\text{met}}^0} - \frac{1}{2}\left\langle \vec{J}_{0;\xi}^{\text{ES}}, \vec{E}_{\xi}^{\text{inc}} \right\rangle_{\partial V_{\text{mat}}^0} - \frac{1}{2}\left\langle \vec{M}_{0;\xi}^{\text{ES}}, \vec{H}_{\xi}^{\text{inc}} \right\rangle_{\partial V_{\text{mat}}^0}\right\}}
\end{aligned} \tag{5-82}$$

where the fourth equality is based on the second equality in relationship (5-65). Comparing relationship (5-82) with relationships (4-56)&(4-105)&(4-150), it is easy to find out that: the characteristic values calculated from the line-surface formulation for constructing the DP-CMs of composite systems and the characteristic values calculated from the volume formulation (the result given in Section 4.2) and surface formulations (the results given in Sections 4.3~4.7) for constructing the DP-CMs of material systems have the same physical meaning —— the ratio of the imaginary part of modal driving power to the real part of modal driving power.

In addition, the characteristic vectors satisfy the following orthogonality:

$$\operatorname{Re}\left\{P_{\text{com sys};\xi}^{\text{driving}}\right\}\delta_{\xi\zeta} = \left(\bar{\alpha}_{J_0;\xi}^{\text{driving}}\right)^H \cdot \bar{\bar{P}}_{J_0;+}^{\text{driving}} \cdot \bar{\alpha}_{J_0;\zeta}^{\text{driving}} \tag{5-83a}$$

$$\operatorname{Im}\left\{P_{\text{com sys};\xi}^{\text{driving}}\right\}\delta_{\xi\zeta} = \left(\bar{\alpha}_{J_0;\xi}^{\text{driving}}\right)^H \cdot \bar{\bar{P}}_{J_0;-}^{\text{driving}} \cdot \bar{\alpha}_{J_0;\zeta}^{\text{driving}} \tag{5-83b}$$

and

$$\underbrace{\left[\operatorname{Re}\left\{P_{\text{com sys};\xi}^{\text{driving}}\right\} + j\operatorname{Im}\left\{P_{\text{com sys};\xi}^{\text{driving}}\right\}\right]}_{P_{\text{com sys};\xi}^{\text{driving}}}\delta_{\xi\zeta} = \left(\bar{\alpha}_{J_0;\xi}^{\text{driving}}\right)^H \cdot \underbrace{\left(\bar{\bar{P}}_{J_0;+}^{\text{driving}} + j\,\bar{\bar{P}}_{J_0;-}^{\text{driving}}\right)}_{\bar{\bar{P}}_{J_0}^{\text{driving}}} \cdot \bar{\alpha}_{J_0;\zeta}^{\text{driving}} \tag{5-84}$$

Based on the physical meaning of the elements in matrix $\bar{\bar{P}}_{J_0}^{\text{driving}}$ and employing orthogonality (5-84), the following orthogonality between the characteristic fields and the characteristic sources can be obtained immediately:

$$\begin{aligned} P_{\text{com sys};\xi}^{\text{driving}}\delta_{\xi\zeta} &= (1/2)\left\langle \vec{J}_\xi^{\text{SL}} \oplus \vec{J}_\xi^{\text{SS}} \oplus \vec{J}_\xi^{\text{SV}}, \vec{E}_\zeta^{\text{inc}} \right\rangle_{L_{\text{met}} \cup S_{\text{met}} \cup \partial V_{\text{met}} \cup V_{\text{mat}}} + (1/2)\left\langle \vec{M}_\xi^{\text{SV}}, \vec{H}_\zeta^{\text{inc}} \right\rangle_{V_{\text{mat}}} \\ &= \frac{1}{2}\left\langle \vec{J}_{0;\xi}^{\text{SL}} \oplus \vec{J}_{0;\xi}^{\text{SS}}, \vec{E}_\zeta^{\text{inc}} \right\rangle_{L_{\text{met}}^0 \cup S_{\text{met}}^0 \cup \partial V_{\text{met}}^0} - \frac{1}{2}\left\langle \vec{J}_{0;\xi}^{\text{ES}}, \vec{E}_\zeta^{\text{inc}} \right\rangle_{\partial V_{\text{mat}}^0} - \frac{1}{2}\left\langle \vec{M}_{0;\xi}^{\text{ES}}, \vec{H}_\zeta^{\text{inc}} \right\rangle_{\partial V_{\text{mat}}^0} \end{aligned} \tag{5-85}$$

where the second equality is based on the second equality in relationship (5-65). Comparing relationship (5-85) with relationships (4-15) and (4-59)&(4-108)&(4-153), it is easy to find out that: the line-surface formulation for composite systems (developed in this chapter) and the volume formulation (developed in Section 4.2) and surface formulations (developed in Sections 4.3~4.7) for material systems have the same physical destination —— constructing a series of steadily working modes not having net energy exchange in any integral period (i.e. constructing a series of orthogonal modes having ability to completely decouple frequency-domain DPO).

### 5.5.5 DP-CM-Based Modal Expansion

Based on the completeness of the DP-CMs, any working mode of the objective composite system can be expanded in terms of the DP-CMs as follows:

$$\vec{E}^{\text{inc}}(\vec{r}) = \sum_{\xi=1}^{\Xi^{J_0}} c_\xi \vec{E}_\xi^{\text{inc}}(\vec{r}) \quad , \quad \vec{r} \in D_{\text{com sys}} \tag{5-86a}$$

$$\vec{H}^{\text{inc}}(\vec{r}) = \sum_{\xi=1}^{\Xi^{J_0}} c_\xi \vec{H}_\xi^{\text{inc}}(\vec{r}) \quad , \quad \vec{r} \in D_{\text{com sys}} \tag{5-86b}$$

where $\Xi^{J_0} = \Xi^{\vec{J}_0^{\text{SL}}} + \Xi^{\vec{J}_0^{\text{SS}}} + \Xi^{\vec{J}_0^{\text{ES}}}$. Testing above modal expansions (5-86a) and (5-86b) with functions $\{(\vec{J}_\xi^{\text{SL}} \oplus \vec{J}_\xi^{\text{SS}} \oplus \vec{J}_\xi^{\text{SV}})/2\}_{\xi=1}^{\Xi^{J_0}}$ and $\{\vec{M}_\xi^{\text{SV}}/2\}_{\xi=1}^{\Xi^{J_0}}$ respectively, and summing the obtained two equations, it is immediately obtained that

$$\begin{aligned}
&(1/2)\left\langle \vec{J}_\xi^{\text{SL}} \oplus \vec{J}_\xi^{\text{SS}} \oplus \vec{J}_\xi^{\text{SV}}, \vec{E}^{\text{inc}} \right\rangle_{L_{\text{met}} \cup S_{\text{met}} \cup \partial V_{\text{met}} \cup V_{\text{mat}}} + (1/2)\left\langle \vec{M}_\xi^{\text{SV}}, \vec{H}^{\text{inc}} \right\rangle_{V_{\text{mat}}} \\
&= (1/2)\left\langle \vec{J}_\xi^{\text{SL}} \oplus \vec{J}_\xi^{\text{SS}} \oplus \vec{J}_\xi^{\text{SV}}, \sum_{\zeta=1}^{\Xi^{J_0}} c_\zeta \vec{E}_\zeta^{\text{inc}} \right\rangle_{L_{\text{met}} \cup S_{\text{met}} \cup \partial V_{\text{met}} \cup V_{\text{mat}}} + (1/2)\left\langle \vec{M}_\xi^{\text{SV}}, \sum_{\zeta=1}^{\Xi^{J_0}} c_\zeta \vec{H}_\zeta^{\text{inc}} \right\rangle_{V_{\text{mat}}} \\
&= \sum_{\zeta=1}^{\Xi^{J_0}} c_\zeta \left[ (1/2)\left\langle \vec{J}_\xi^{\text{SL}} \oplus \vec{J}_\xi^{\text{SS}} \oplus \vec{J}_\xi^{\text{SV}}, \vec{E}_\zeta^{\text{inc}} \right\rangle_{L_{\text{met}} \cup S_{\text{met}} \cup \partial V_{\text{met}} \cup V_{\text{mat}}} + (1/2)\left\langle \vec{M}_\xi^{\text{SV}}, \vec{H}_\zeta^{\text{inc}} \right\rangle_{V_{\text{mat}}} \right]
\end{aligned} \tag{5-87}$$

where the second equality is based on the linear property of inner product. Applying orthogonality (5-85) to the above equation, we immediately have that

$$(1/2)\left\langle \vec{J}_\xi^{\text{SL}} \oplus \vec{J}_\xi^{\text{SS}} \oplus \vec{J}_\xi^{\text{SV}}, \vec{E}^{\text{inc}} \right\rangle_{L_{\text{met}} \cup S_{\text{met}} \cup \partial V_{\text{met}} \cup V_{\text{mat}}} + (1/2)\left\langle \vec{M}_\xi^{\text{SV}}, \vec{H}^{\text{inc}} \right\rangle_{V_{\text{mat}}} = c_\xi P_{\text{com sys};\xi}^{\text{driving}} \tag{5-88}$$

Generally speaking, $P_{\text{com sys};\xi}^{\text{driving}} \neq 0$ for composite systems, so

$$\begin{aligned}
c_\xi &= \frac{(1/2)\left\langle \vec{J}_\xi^{\text{SL}} \oplus \vec{J}_\xi^{\text{SS}} \oplus \vec{J}_\xi^{\text{SV}}, \vec{E}^{\text{inc}} \right\rangle_{L_{\text{met}} \cup S_{\text{met}} \cup \partial V_{\text{met}} \cup V_{\text{mat}}} + (1/2)\left\langle \vec{M}_\xi^{\text{SV}}, \vec{H}^{\text{inc}} \right\rangle_{V_{\text{mat}}}}{P_{\text{com sys};\xi}^{\text{driving}}} \\
&= \frac{\frac{1}{2}\left\langle \vec{J}_{0;\xi}^{\text{SL}} \oplus \vec{J}_{0;\xi}^{\text{SS}}, \vec{E}^{\text{inc}} \right\rangle_{L_{\text{met}}^0 \cup S_{\text{met}}^0 \cup \partial V_{\text{met}}^0} - \frac{1}{2}\left\langle \vec{J}_{0;\xi}^{\text{ES}}, \vec{E}^{\text{inc}} \right\rangle_{\partial V_{\text{mat}}^0} - \frac{1}{2}\left\langle \vec{M}_{0;\xi}^{\text{ES}}, \vec{H}^{\text{inc}} \right\rangle_{\partial V_{\text{mat}}^0}}{P_{\text{com sys};\xi}^{\text{driving}}}
\end{aligned} \tag{5-89}$$

where $\xi = 1, 2, \cdots, \Xi^{J_0}$; the second equality is based on the second equality in relationship (5-65). Above formulation (5-89) is just the explicit expressions for the expansion coefficients used in modal expansion (5-86).

## 5.6 Numerical Examples Corresponding to Typical Structures

Here, we intend to use a whole section to construct the DP-CMs of some typical composite systems by employing the formulations developed in Sections 5.2~5.5, and take it as the verification for the validity of the formulations obtained in Sections 5.2~5.5.

### 5.6.1 Typical Structure I

In this subsection, we consider the composite system shown in Figure 5-5. The composite system is constructed by a metallic sphere, whose radius is 2.50mm, and a material spherical shell, whose inner and outer radiuses are 2.50mm and 5.00mm respectively and relative permeability, relative permittivity, and conductivity are 6, 6, and 0 respectively. From Figure 5-5, it is easy to find out that the topological structure of the composite system satisfies the topological restrictions given in the Section 5.2 of this dissertation, so we can, based on the boundary decomposition method introduced in the Section 5.3 of this dissertation, decompose the material boundary of the composite system as shown Figure 5-6. In addition, because whole metallic boundary is in contact with the material part of the composite system, then there is no need to further decompose the metallic boundary (for details see Section 5.3), and we provide the topological structure and triangular meshes of the metallic boundary in Figure 5-7. For the convenience of the following discussions and the consistency of the symbolic system used in previous sections, we denote the interface between the material spherical shell and environment as $\partial V_{\rm mat}^{0}$, and denote the interface between the material spherical shell and the metallic sphere as $\partial V_{\rm mat}^{\cap}$, and denote the boundary of the metallic sphere as $\partial V_{\rm met}$, and it is obvious that $\partial V_{\rm met} = \partial V_{\rm mat}^{\cap}$.

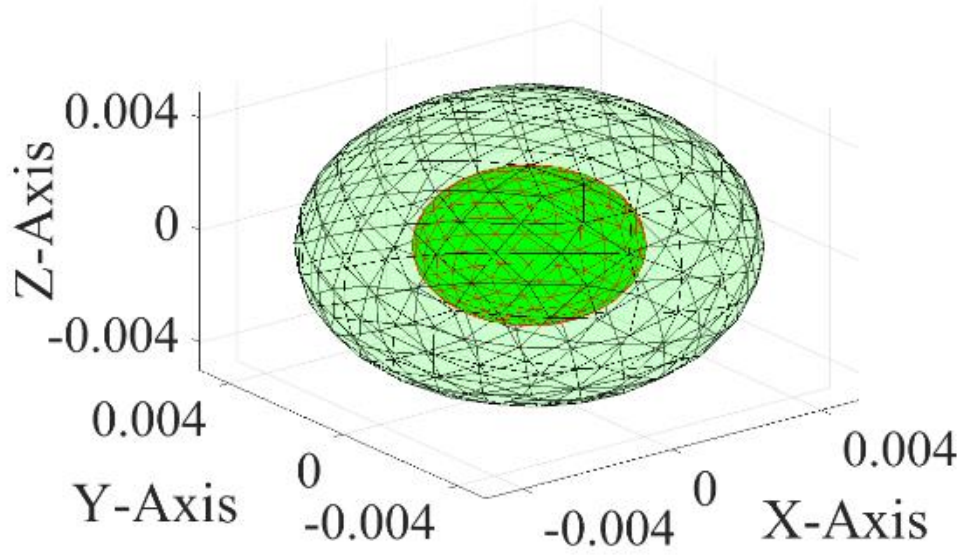

Figure 5-5 The topological structure and surface triangular meshes of the composite system constituted by “a metallic sphere whose radius is 2.50mm” and “a material spherical shell whose inner and outer radiuses are 2.50mm and 5.00mm respectively”

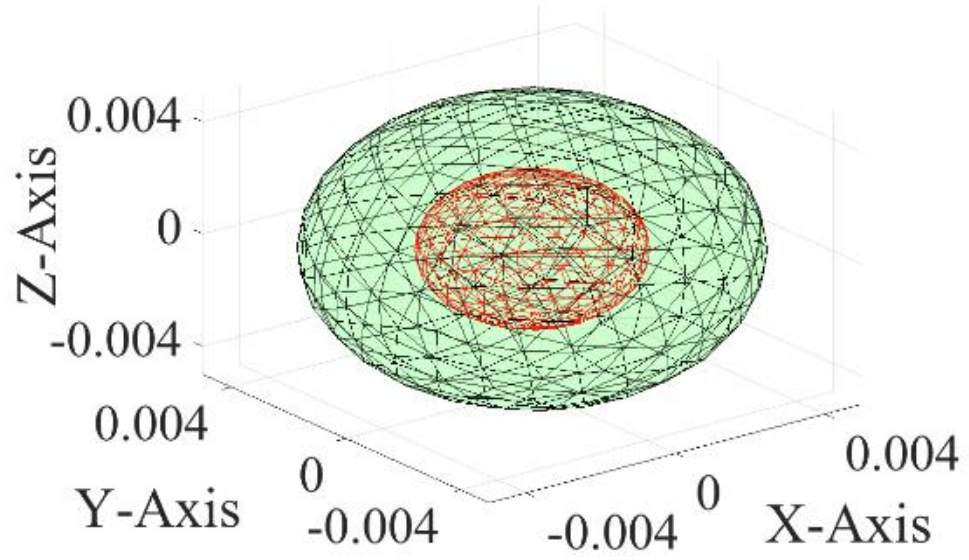

(a)

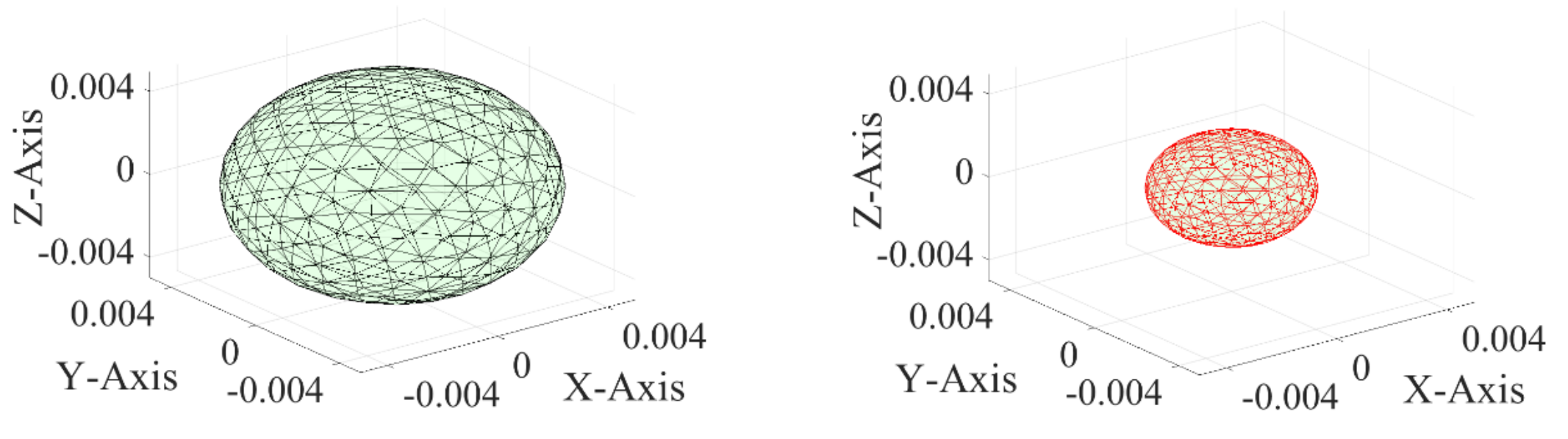

(b) (c)

Figure 5-6 The boundary decomposition for the material spherical shell shown in Figure 5-5. (a) the topological structure and surface triangular meshes of whole material spherical shell; (b) the material-environment boundary $\partial V_{\text{mat}}^{0}$ and its topological structure and surface triangular meshes; (c) the material-metal boundary $\partial V_{\text{mat}}^{\cap}$ and its topological structure and surface triangular meshes

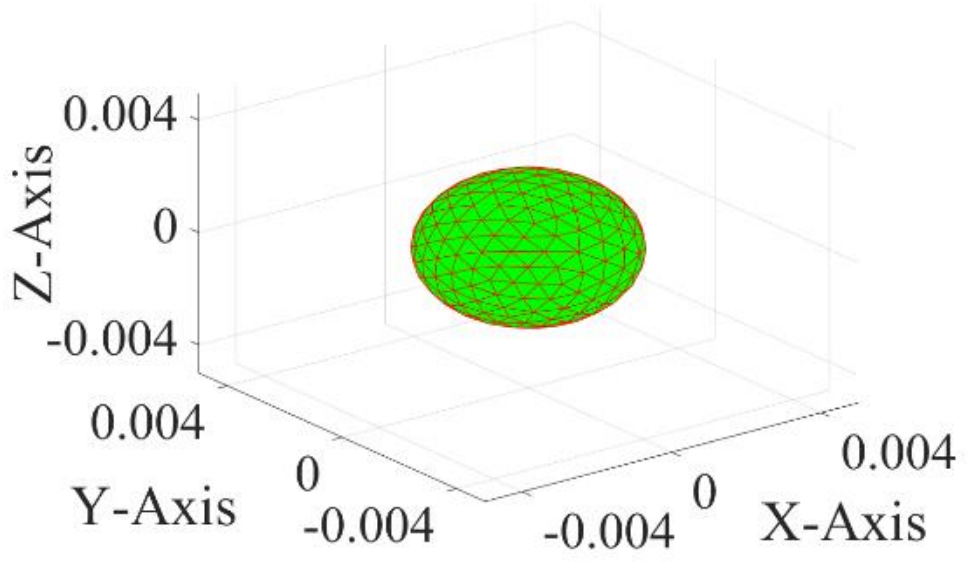

Figure 5-7 The boundary of the metallic sphere shown in Figure 5-5 is denoted as $\partial V_{\text{met}}$, and its topological structure and surface triangular meshes are shown in this figure, and it is obvious that $\partial V_{\text{met}} = \partial V_{\text{mat}}^{\cap}$

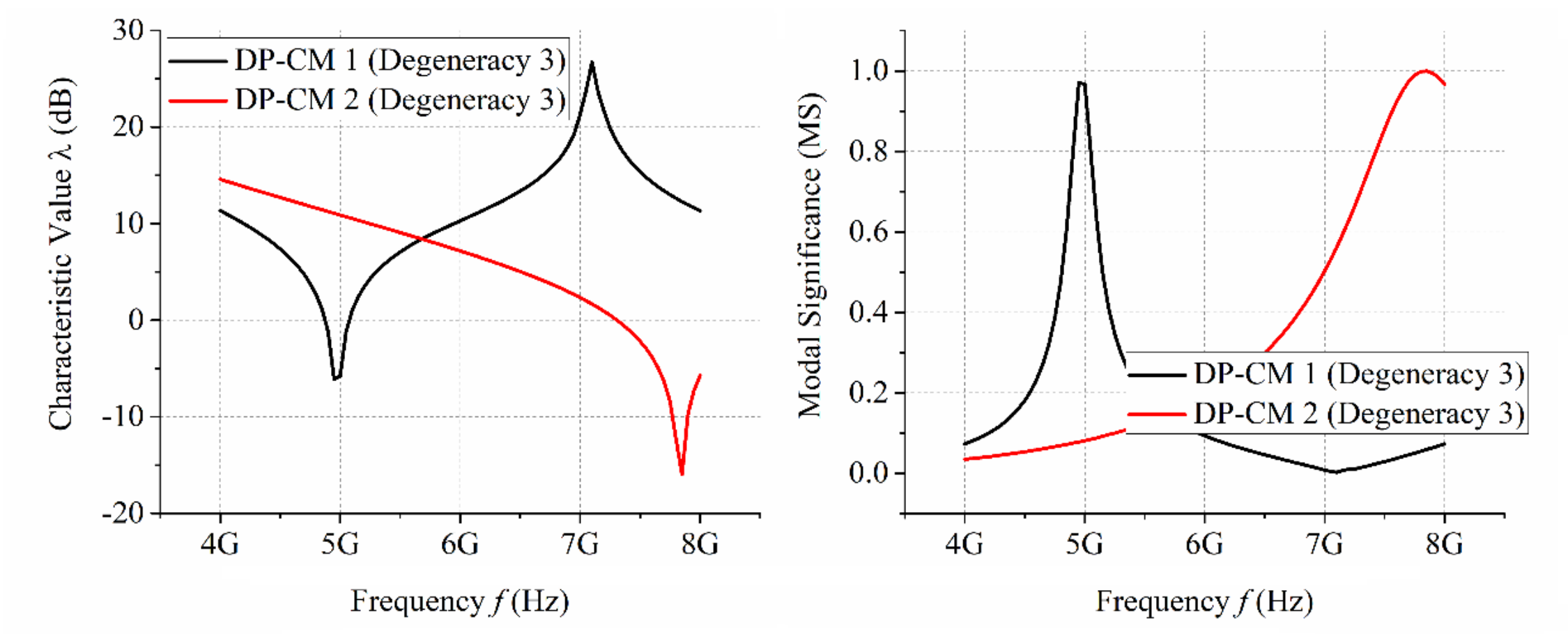

(a) (b)

Figure 5-8 The characteristic quantity curves corresponding to 2 typical DP-CMs (derived from the theory given in this chapter) of the composite system shown in Figure 5-5. (a) characteristic value dB curves; (b) MS curves

Focusing on the composite system shown in Figure 5-5, we obtain whole DP-CM set based on the formulation established in this chapter. The characteristic value (dB) and MS curves corresponding to some typical DP-CMs are illustrated in Figure 5-8. In addition, we also construct the CMs of the composite system shown in Figure 5-5 by employing the EFIE-SIE operator[53] with the variable unification scheme developed in this chapter, and provide the characteristic value (dB) curves and MS curves corresponding to some typical modes in Figure 5-9.

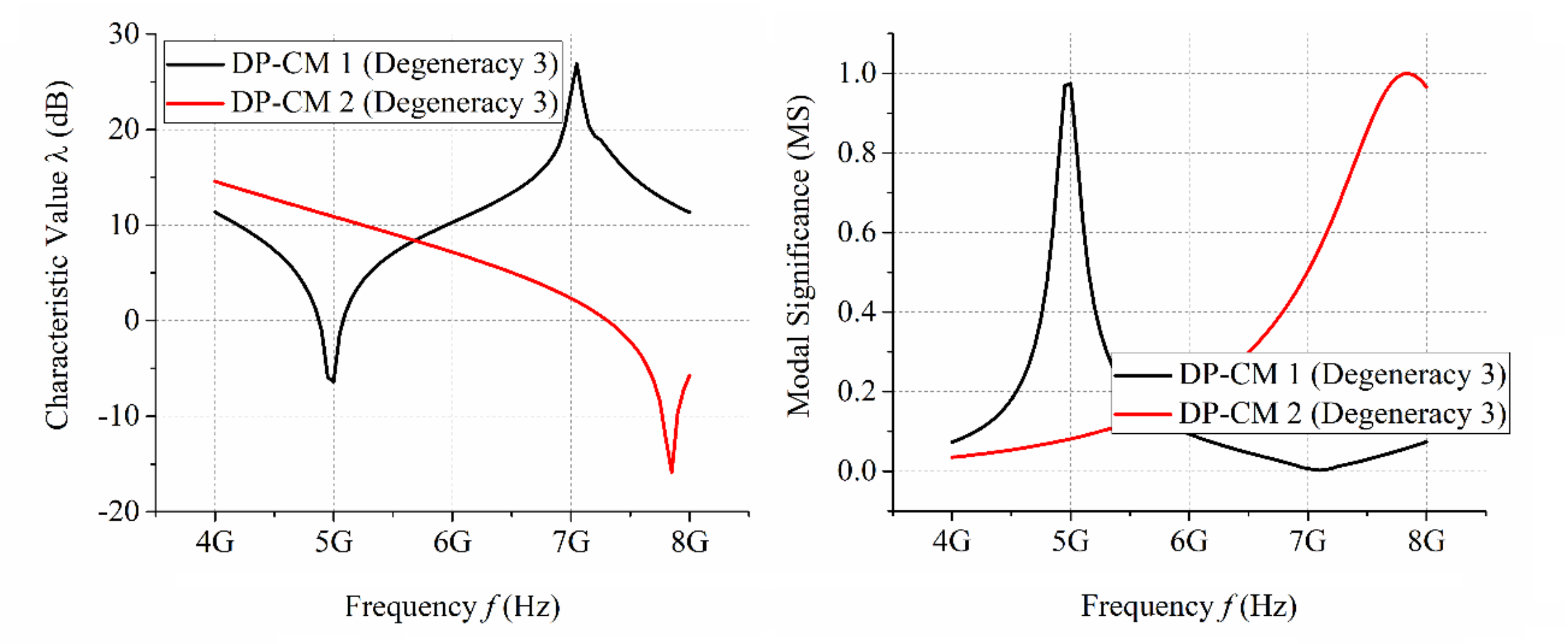

(a) (b)

Figure 5-9 The characteristic quantity curves corresponding to 2 typical DP-CMs (derived from the EFIE-SIE operator[53] with the variable unification scheme given in this chapter) of the composite system shown in Figure 5-5. (a) characteristic value dB curves; (b) MS curves

By comparing Figures 5-8 and 5-9, it is easy to find out that the results derived from the formulation developed in this chapter and the results derived from the EFIE-SIE operator[53] with the variable unification scheme developed in this chapter agree well with each other. Taking the DP-CMs shown in Figure 5-8 as typical examples, we provide their modal equivalent source distributions, modal scattered source distributions, modal field distributions, and modal radiation patterns as below.

From above Figure 5-8, it is easy to find out that DP-CM1 is “resonant” at 5.00GHz. For the first degenerate state of the “resonant” DP-CM1, its equivalent surface electric current and equivalent surface magnetic current are illustrated in Figure 5-10, and its tangential total magnetic field and tangential total electric field distributing on material inner boundaries are illustrated in Figure 5-11.

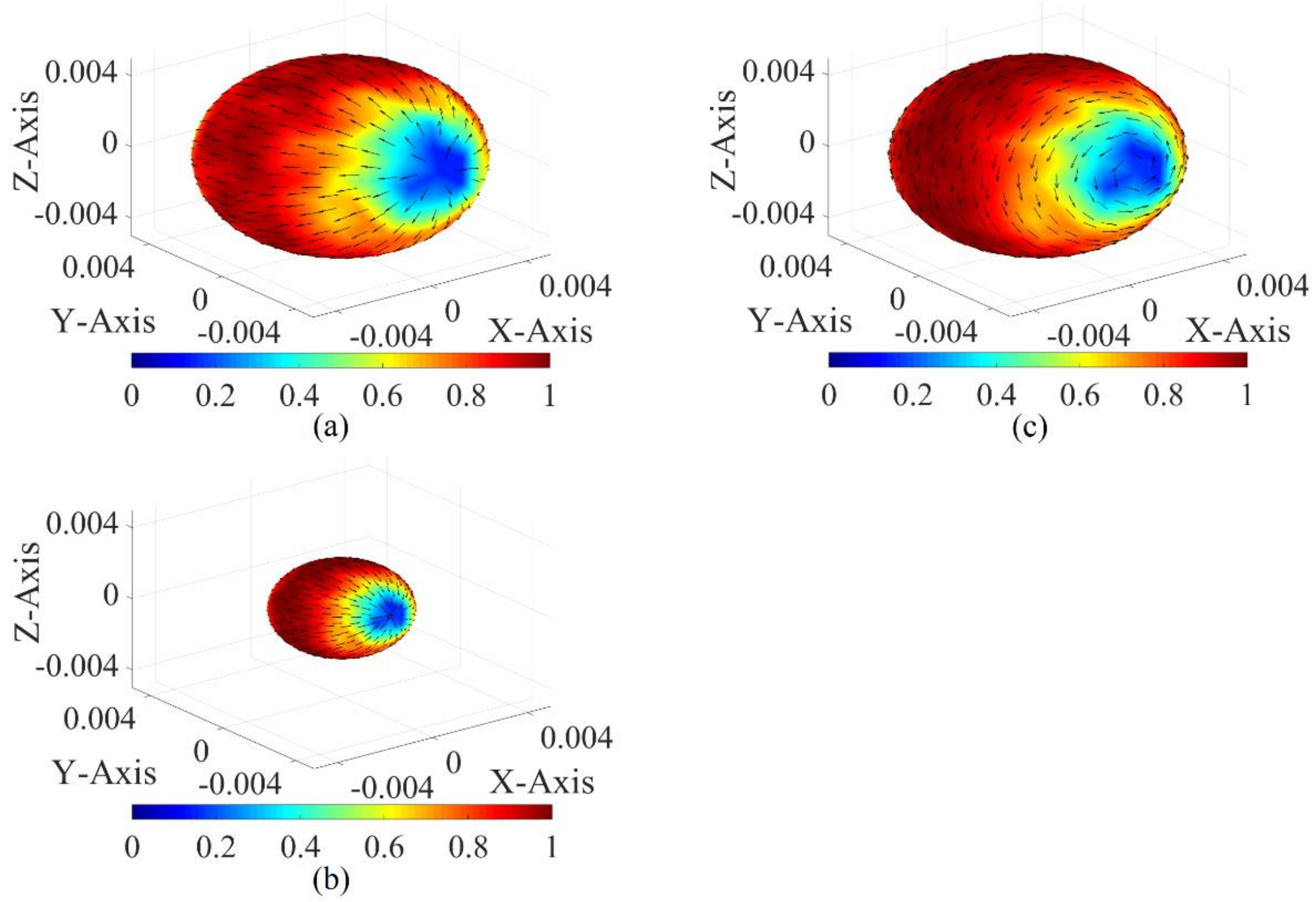

Figure 5-10 The modal equivalent surface source distributions of the first degenerate state of the DP-CM1 working at 5.00GHz and shown in Figure 5-8. (a) the equivalent surface electric current on $\partial V_{\text{mat}}^{0}$; (b) the equivalent surface electric current on $\partial V_{\text{mat}}^{\cap}$; (c) the equivalent surface magnetic current on $\partial V_{\text{mat}}^{0}$

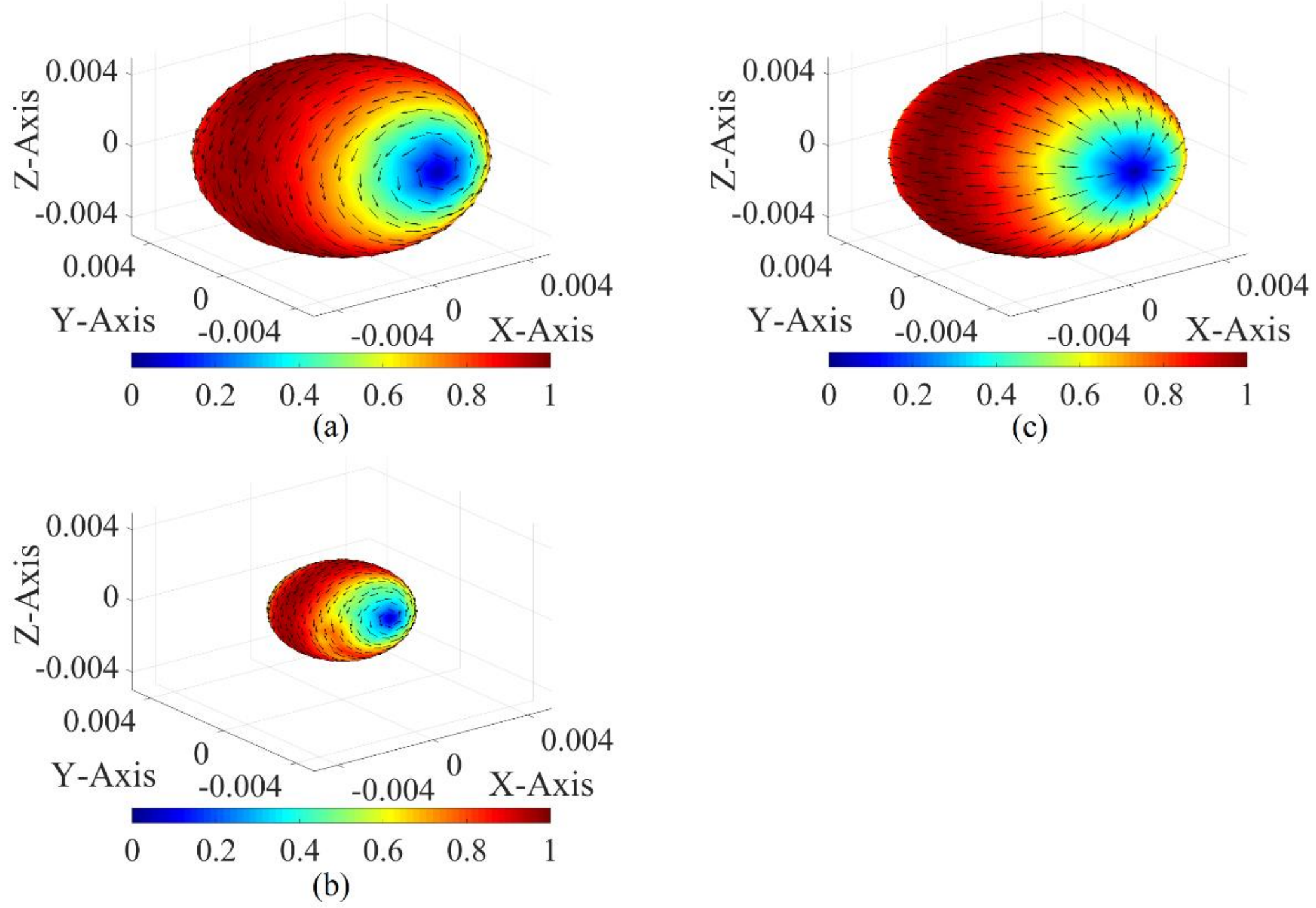

Figure 5-11 The modal tangential total field distributions of the first degenerate state of the DP-CM1 working at 5.00GHz and shown in Figure 5-8. (a) the tangential total magnetic field on the inner surface of $\partial V_{\text{mat}}^{0}$; (b) the tangential total magnetic field on the inner surface of $\partial V_{\text{mat}}^{\cap}$; (c) the tangential total electric field on the inner surface of $\partial V_{\text{mat}}^{0}$

Obviously, the distributions shown in Figure 5-10 and Figure 5-11 indeed satisfy relationships (5-20) and (5-21).

For the first degenerate state of the “resonant” DP-CM1 working at 5.00GHz, its modal total electric field and modal total magnetic field distributing on the material spherical shell are illustrated in the first line of Figure 5-12, and its modal scattered volume electric current and modal scattered volume magnetic current distributing on the material spherical shell are illustrated in the second line of Figure 5-12, and its modal incident electric field and modal incident magnetic field distributing on the material spherical shell are illustrated in the third line of Figure 5-12. In addition, we also provide the radiation pattern corresponding to the modal state in Figure 5-13.

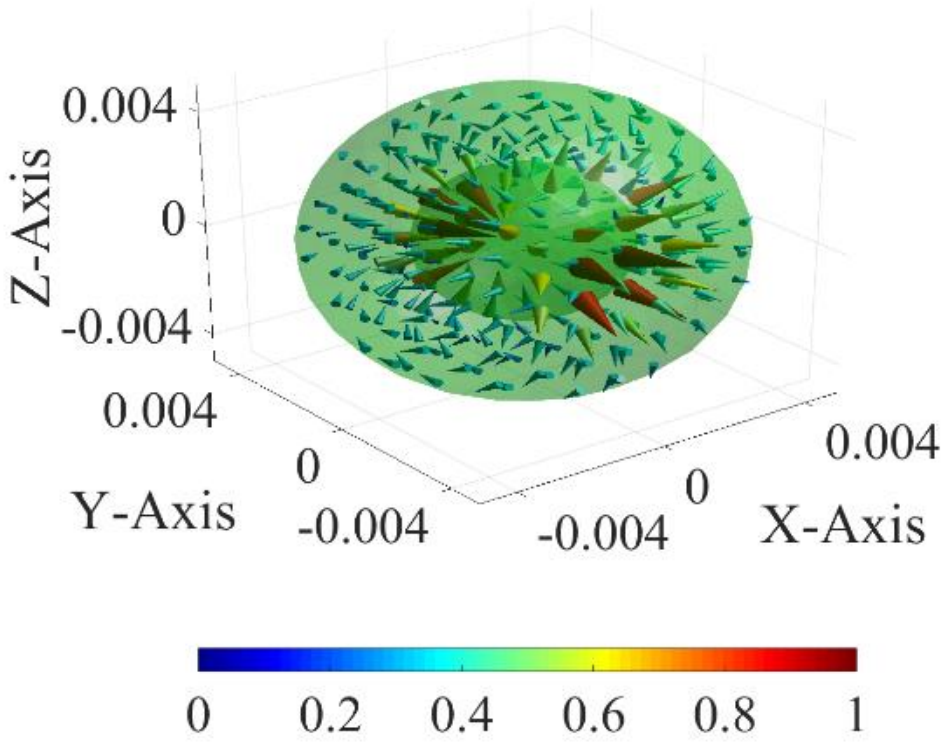

(a)

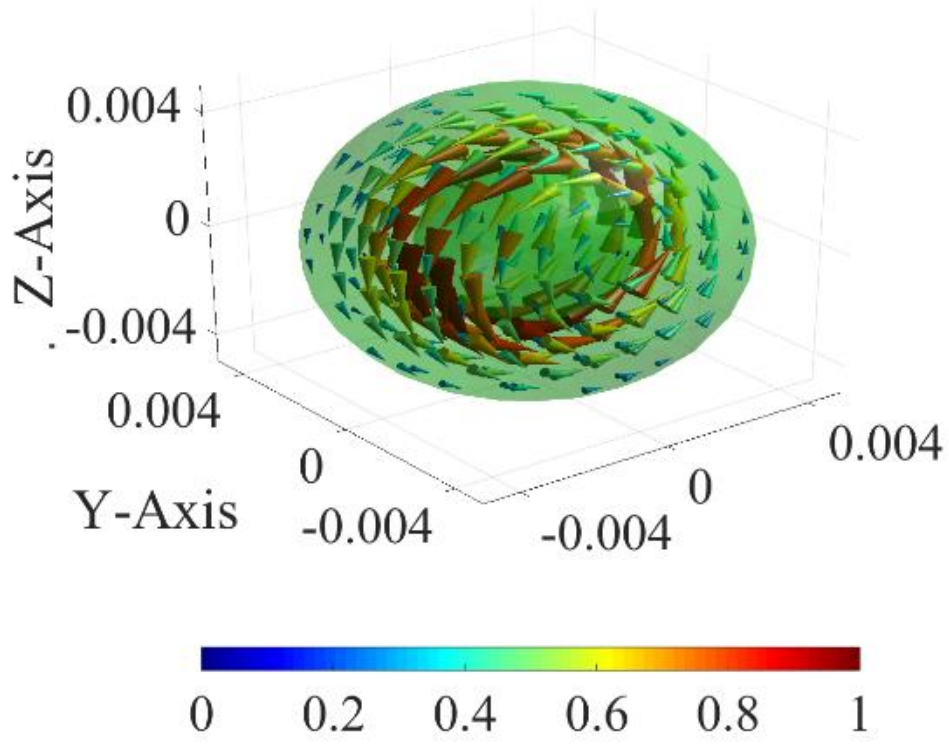

(d)

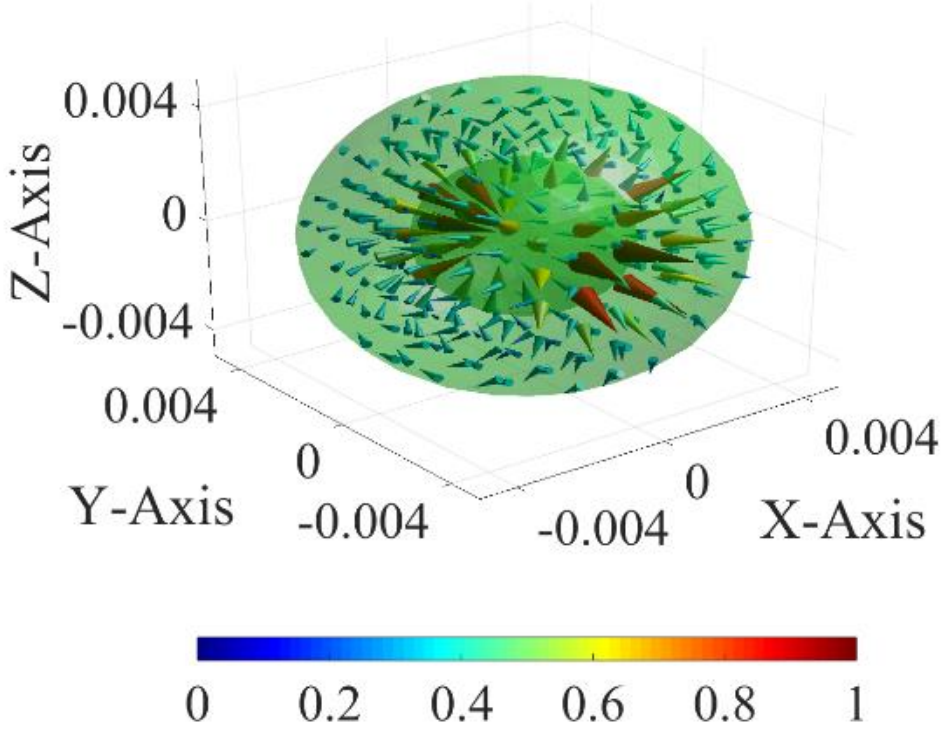

(b)

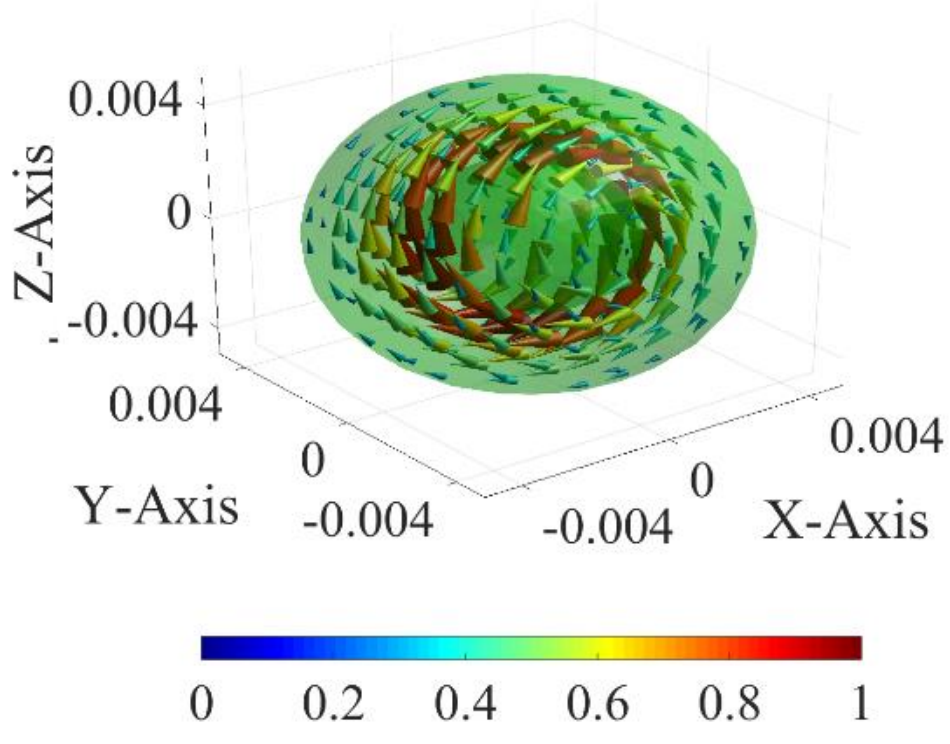

(e)

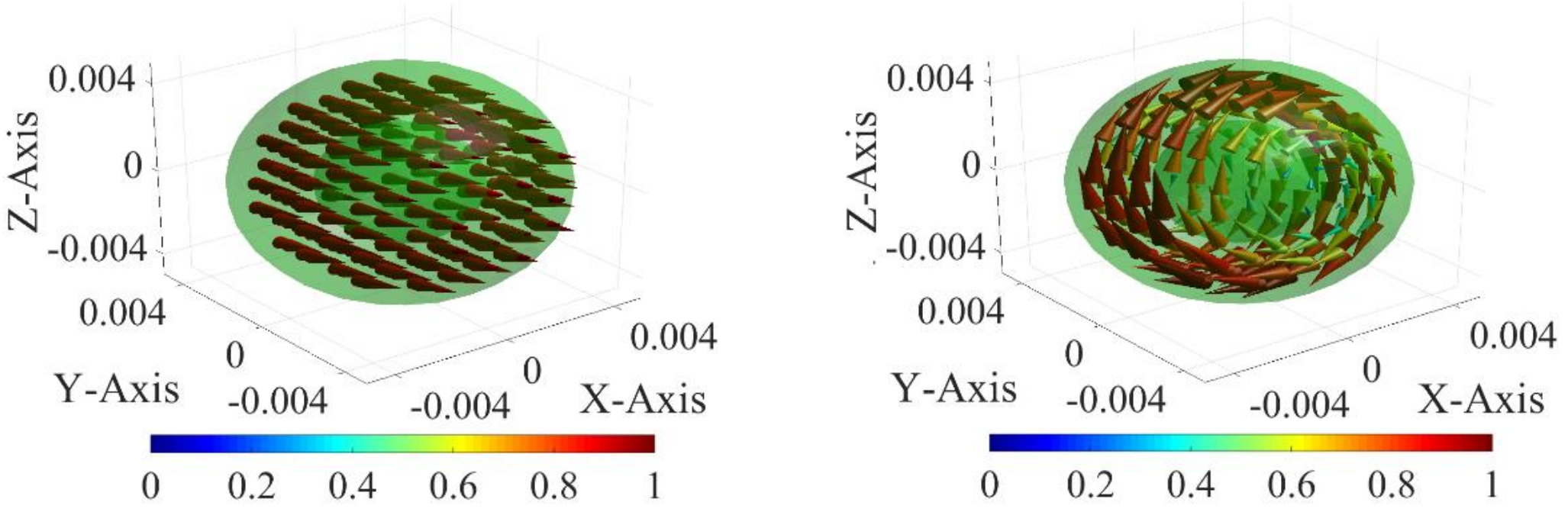

(c) (f)

Figure 5-12 The physical quantity distributions of the first degenerate state of the DP-CM1 working at 5.00GHz and shown in Figure 5-8. (a) total electric field; (b) scattered volume electric current; (c) incident electric field; (d) total magnetic field; (e) scattered volume magnetic current; (f) incident magnetic field

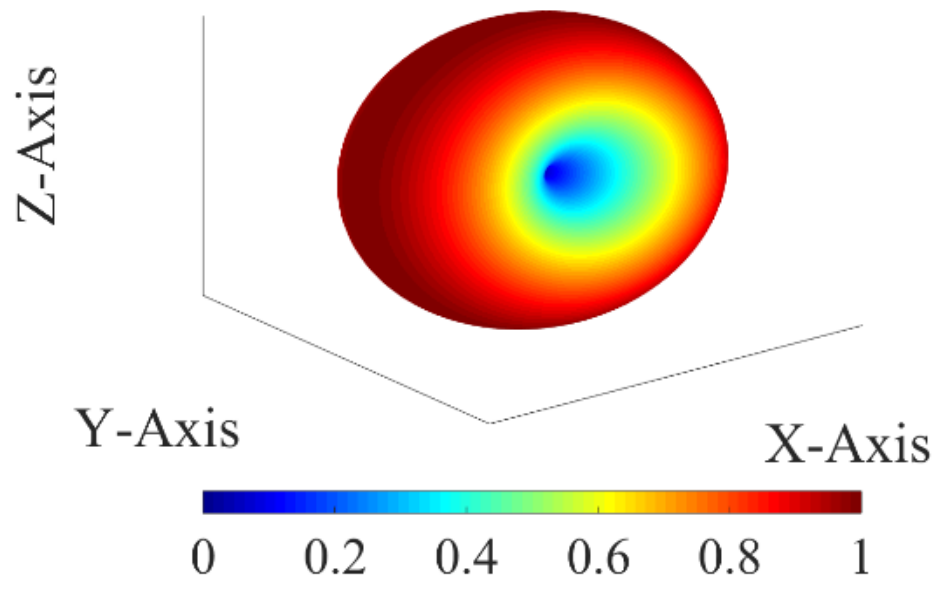

Figure 5-13 The radiation pattern of the first degenerate state of the DP-CM1 working at 5.00GHz and shown in Figure 5-8

For the second and third degenerate states of the “resonant” DP-CM1 working at 5.00GHz, their BV distributions (the modal equivalent surface electric currents distributing on $\partial V_{\text{mat}}^{0}$) and radiation patterns are illustrated in Figure 5-14 and Figure 5-15 respectively.

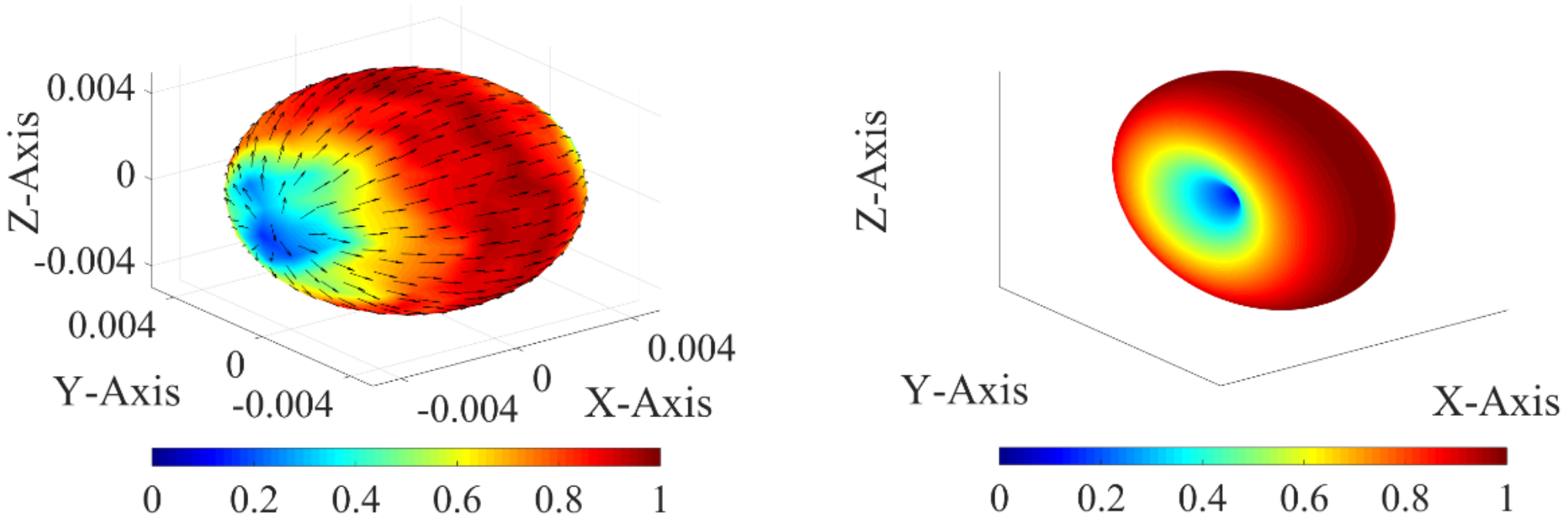

(a) (b)

Figure 5-14 The basic variable distribution and radiation pattern of the second degenerate state of the DP-CM1 working at 5.00GHz and shown in Figure 5-8. (a) basic variable (the equivalent surface electric current on $\partial V_{\text{mat}}^{0}$); (b) radiation pattern

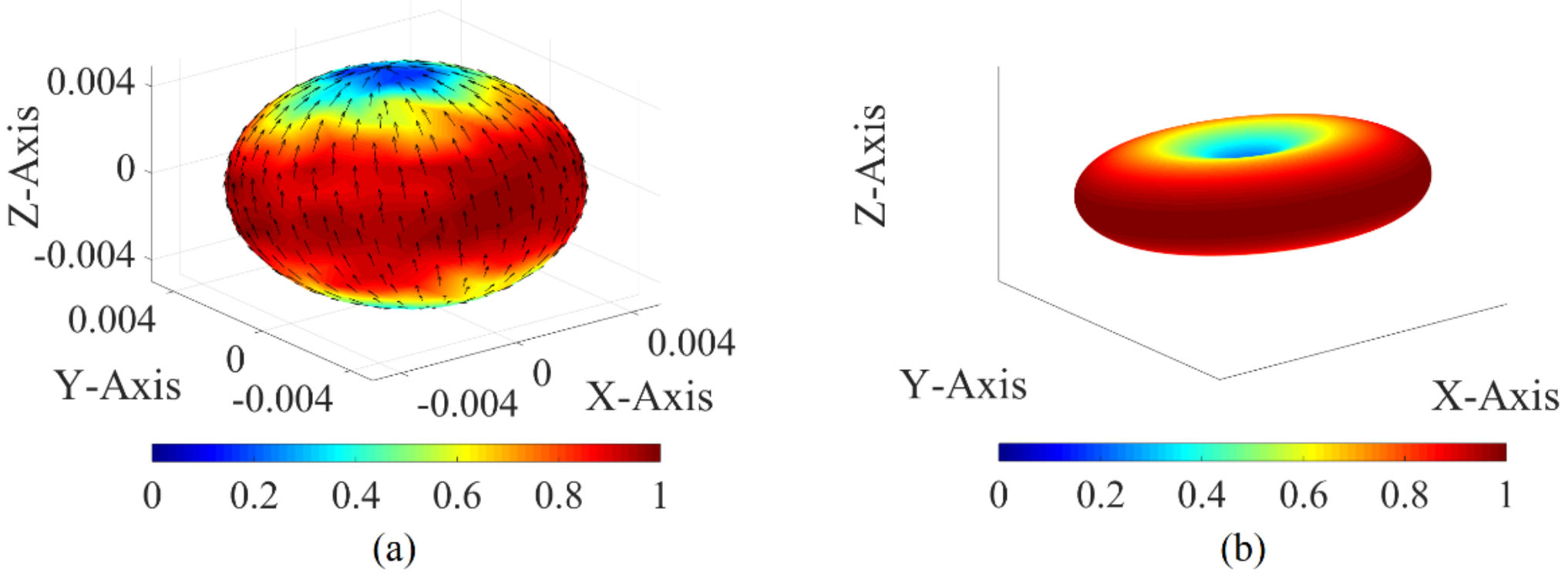

Figure 5-15 The basic variable distribution and radiation pattern of the third degenerate state of the DP-CM1 working at 5.00GHz and shown in Figure 5-8. (a) basic variable (the equivalent surface electric current on $\partial V_{\text{mat}}^{0}$); (b) radiation pattern

Obviously, the above three degenerate states of the DP-CM1 working at 5.00GHz originate from the spacially rotational symmetry of the composite system shown in Figure 5-5.

## 5.6.2 Typical Structure II

In this subsection, we consider the composite system shown in Figure 5-16, and the system is constructed by a metallic cylinder, whose radius and height are 5.25mm and 2.30mm respectively, and a material cylinder, whose radius and height are 5.25mm and 2.30mm respectively, and the {relative permeability, relative permittivity, conductivity} of the material cylinder are {6, 6, 0}, and the material cylinder is placed over the metallic cylinder.

From Figure 5-16, it is easy to find out that the topological structure of the composite system satisfies the topological restrictions given in the Section 5.2 of this dissertation, so we can, based on the boundary decomposition method introduced in the Section 5.3 of this dissertation, decompose the material and metallic boundaries of the composite system as shown Figures 5-17 and 5-18 respectively. For the convenience of the following discussions and the consistency of the symbolic system used in previous sections, we denote the interface between the material cylinder and environment as $\partial V_{\text{mat}}^{0}$, and denote the interface between the material cylinder and the metallic cylinder as $\partial V_{\text{mat}}^{\cap}$, and denote the interface between the metallic cylinder and environment as $\partial V_{\text{met}}^{0}$, and denote the interface between the metallic cylinder and the material cylinder as $\partial V_{\text{met}}^{\cap}$, and it is obvious that $\partial V_{\text{met}}^{\cap} = \partial V_{\text{mat}}^{\cap}$.

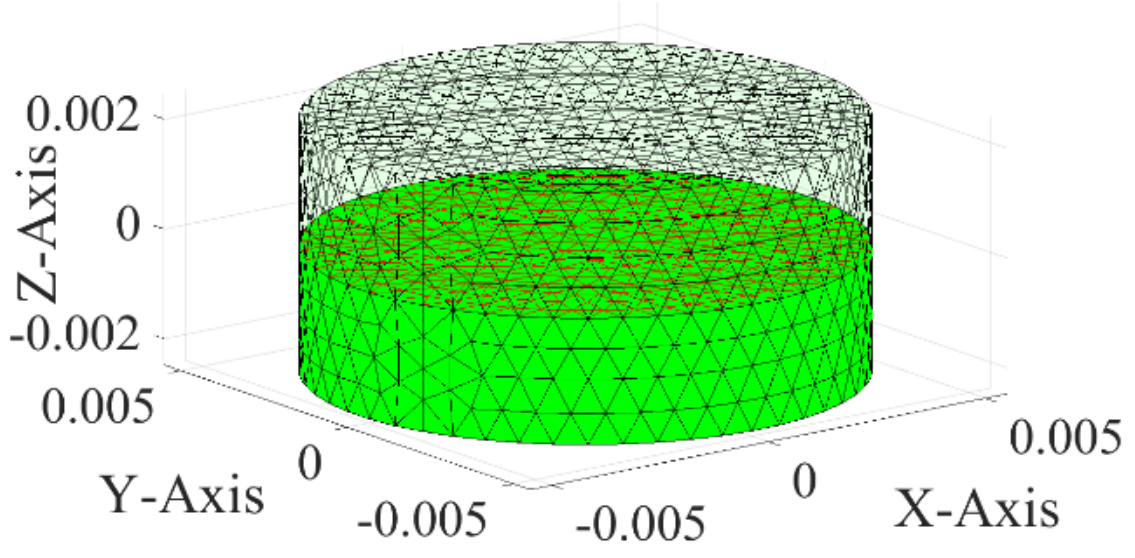

Figure 5-16 The topological structure and surface triangular meshes of the composite system constituted by “a metallic cylinder whose radius and height are 5.25mm and 2.30mm respectively” and “a material cylinder whose radius and height are 5.25mm and 2.30mm respectively”

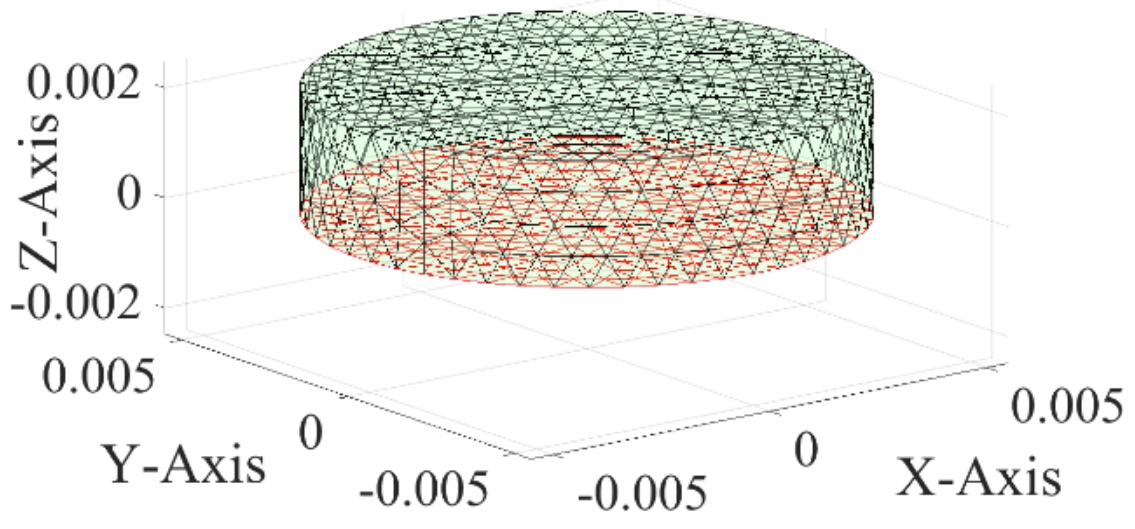

(a)

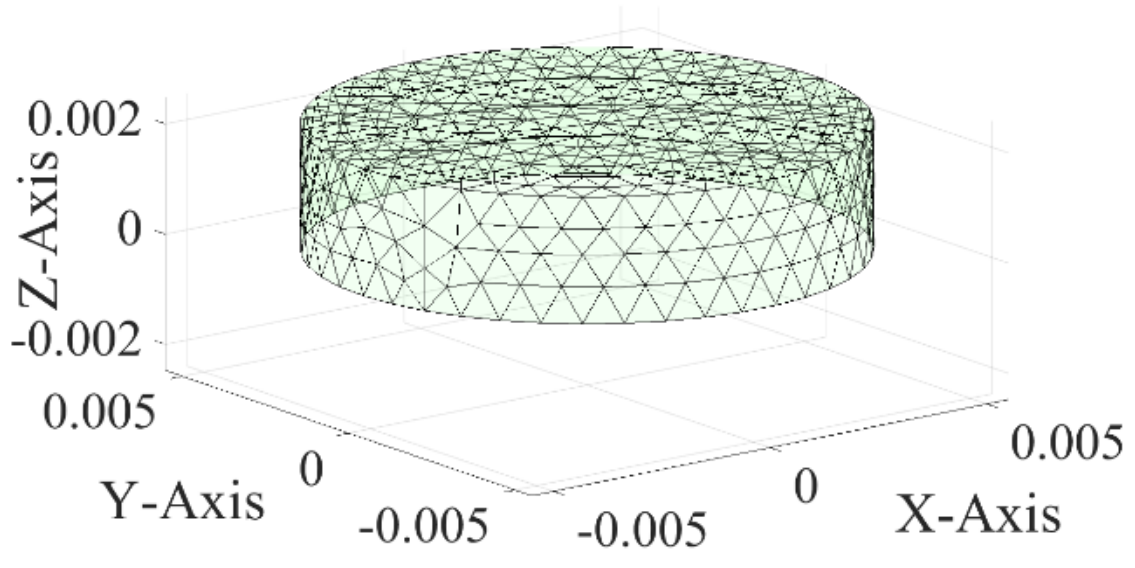

(b)

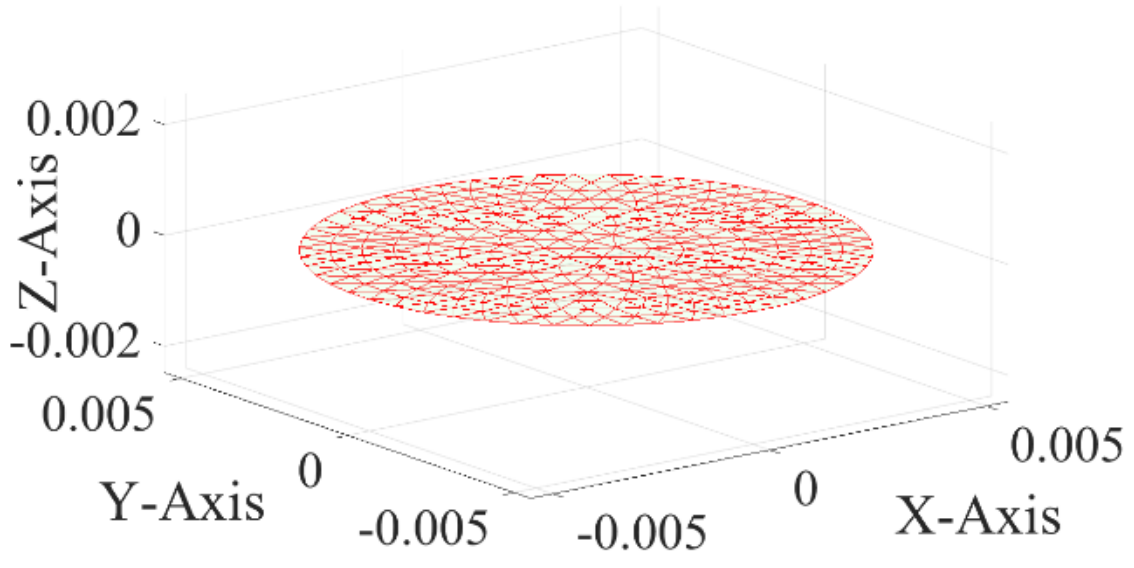

(c)

Figure 5-17 The boundary decomposition for the material cylinder shown in Figure 5-16. (a) the topological structure and surface triangular meshes of the whole boundary of the material cylinder; (b) the material-environment boundary $\partial V_{\text{mat}}^{0}$ and its topological structure and surface triangular meshes; (c) the material-metal boundary $\partial V_{\text{mat}}^{\cap}$ and its topological structure and surface triangular meshes

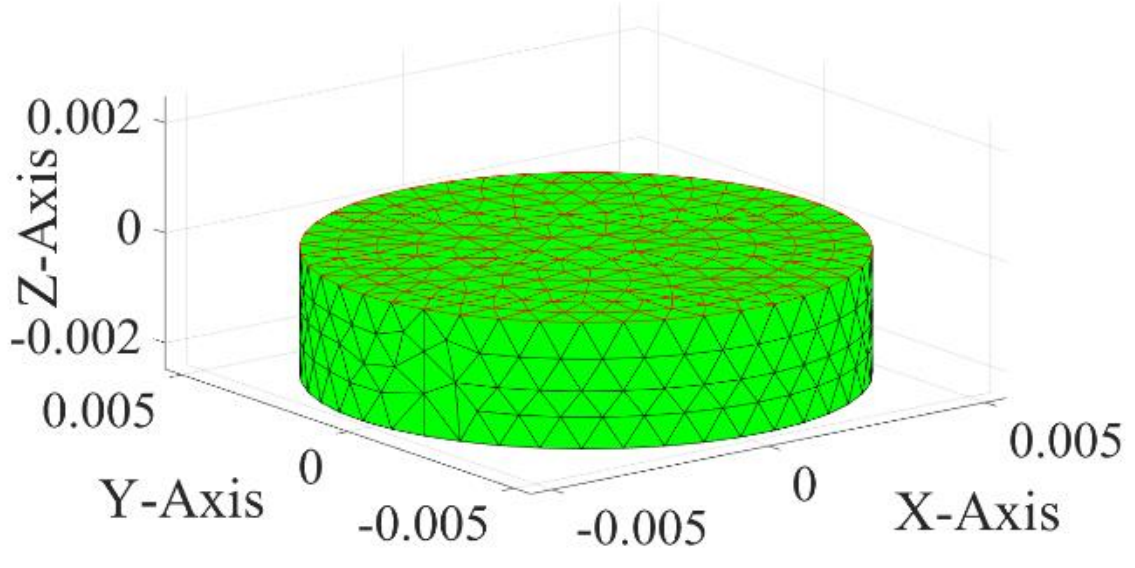

(a)

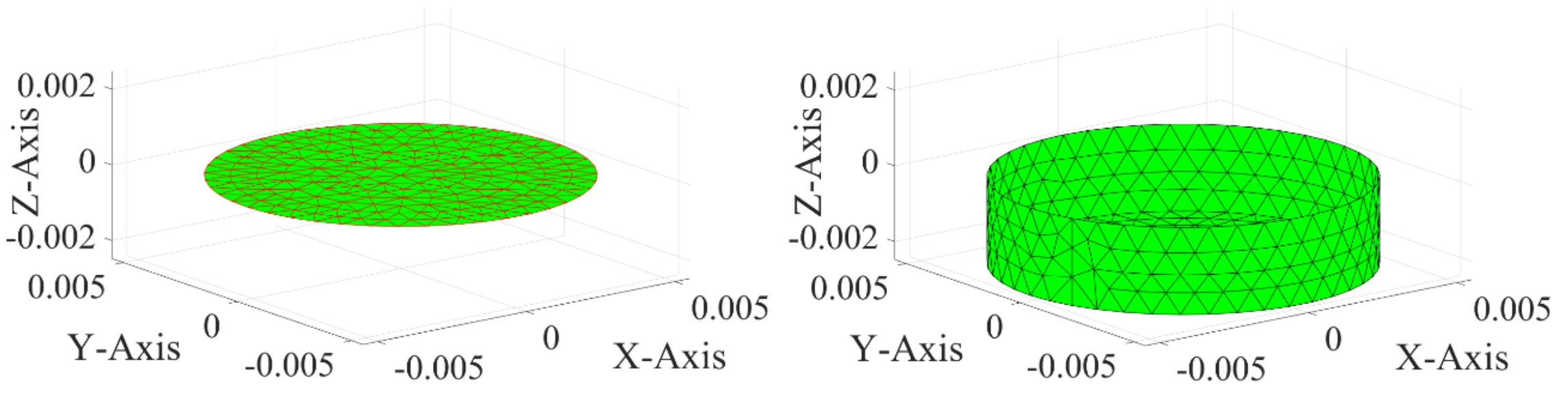

(b) (c)

Figure 5-18 The boundary decomposition for the metallic cylinder shown in Figure 5-16. (a) the topological structure and surface triangular meshes of the whole boundary of the metallic cylinder; (b) the metal-material boundary $\partial V_{\text{met}}^{\cap}$ and its topological structure and surface triangular meshes; (c) the metal-environment boundary $\partial V_{\text{met}}^{0}$ and its topological structure and surface triangular meshes

Focusing on the composite system shown in Figure 5-16, we obtain whole DP-CM set based on the formulation established in this chapter. The characteristic value (dB) curves and MS curves corresponding to some typical DP-CMs are illustrated in the following Figure 5-19.

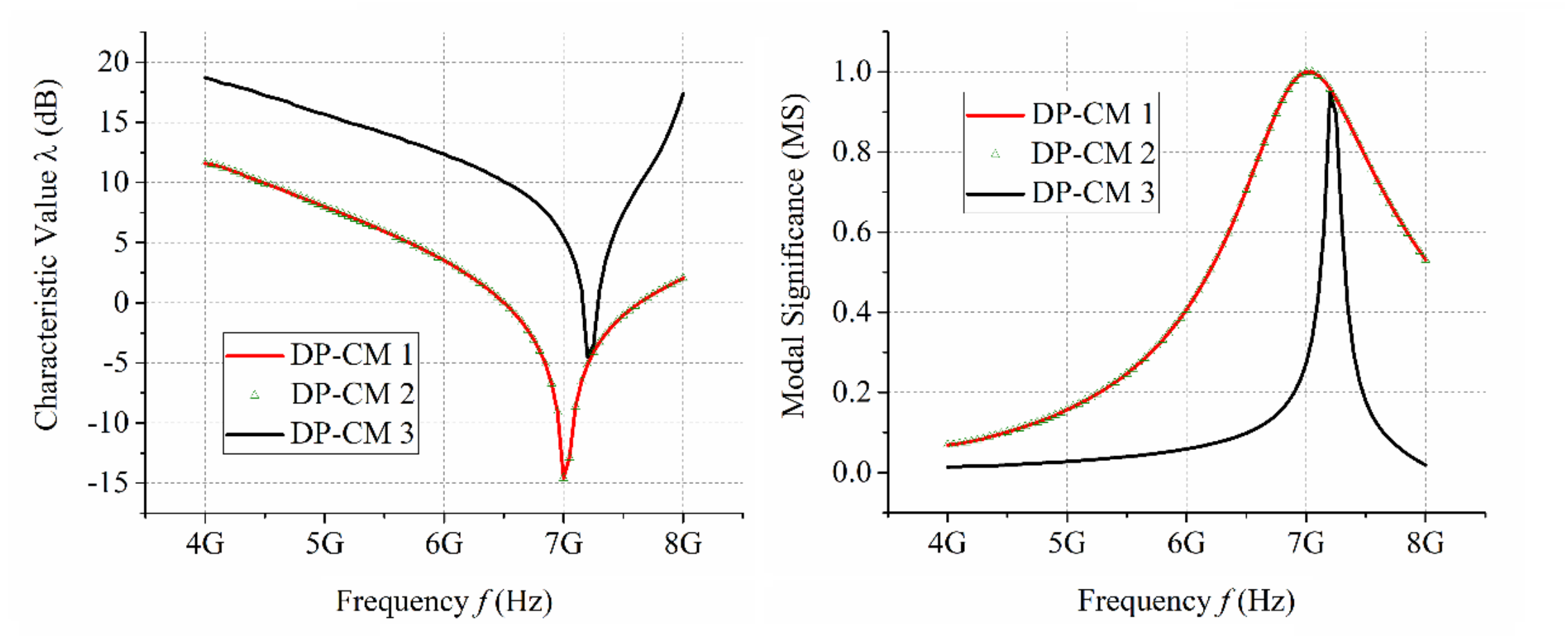

(a) (b)

Figure 5-19 The characteristic quantity curves corresponding to some typical DP-CMs (derived from the theory developed in this chapter) of the composite system shown in Figure 5-16. (a) characteristic value dB curves; (b) MS curves

In addition, we also construct the CMs of the composite system shown in Figure 5-16 by employing the EFIE-SIE operator[53] with the variable unification scheme developed in this chapter, and provide the characteristic value (dB) curves and MS curves corresponding to some typical modes in Figure 5-20.

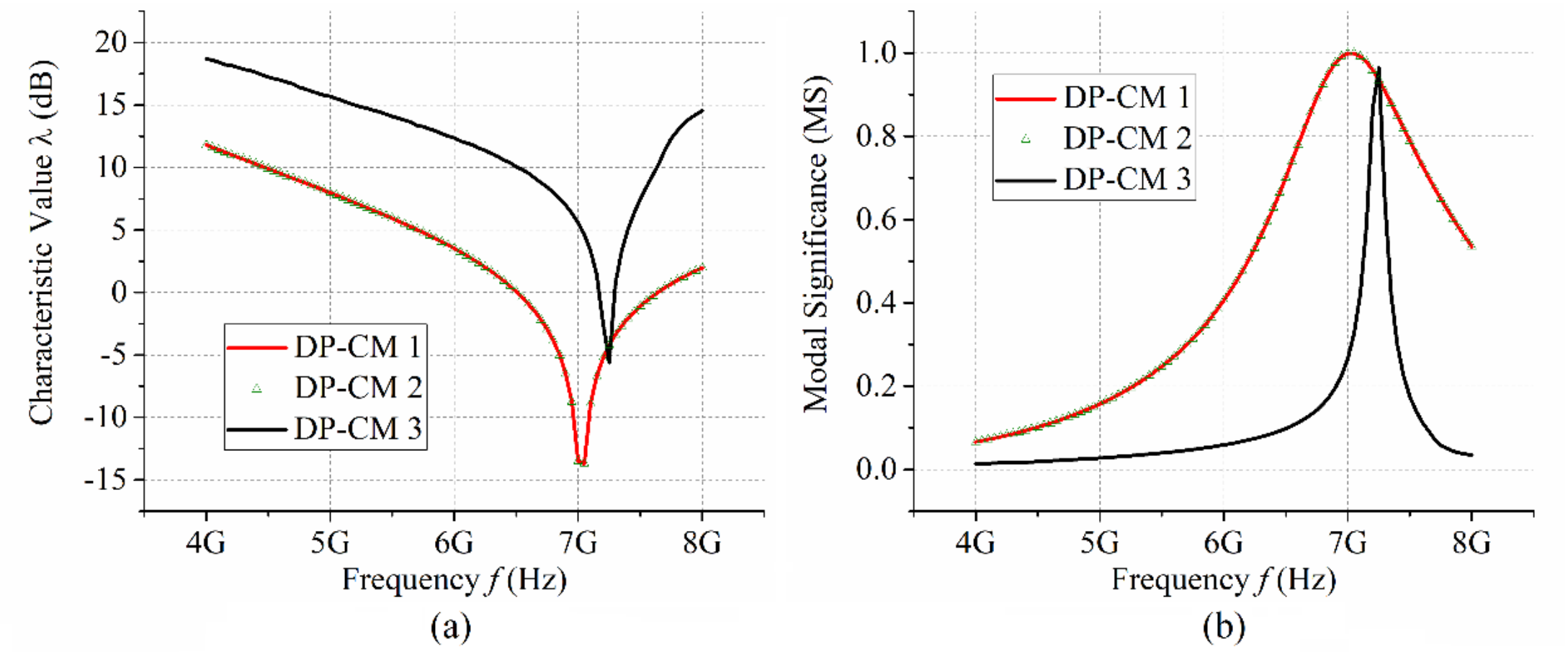

Figure 5-20 The characteristic quantity curves corresponding to some typical DP-CMs (derived from the EFIE-SIE operator[53] with the variable unification scheme given in this chapter) of the composite system shown in Figure 5-16. (a) characteristic value dB curves; (b) MS curves

By comparing Figure 5-19 and Figure 5-20, it is easy to find out that the results derived from the formulation developed in this chapter and the results derived from the EFIE-SIE operator[53] with the variable unification scheme developed in this chapter agree well with each other. Taking the DP-CMs shown in Figure 5-19 as typical examples, we provide their modal equivalent source distributions, modal scattered source distributions, modal field distributions, and modal radiation patterns as below.

From Figure 5-19, it is easy to find out that DP-CM3 is “resonant” at 7.20GHz. For the “resonant” DP-CM3, its modal radiation pattern is illustrated in Figure 5-21. For the “resonant” DP-CM3, its equivalent surface electric and magnetic currents distributing on material boundaries and its scattered surface electric current distributing on metallic boundary are illustrated in Figure 5-22. In addition, for the “resonant” DP-CM3, its modal total fields, modal scattered sources, and modal incident fields distributing on the material cylinder are illustrated in Figures 5-23, 5-24, and 5-25 respectively.

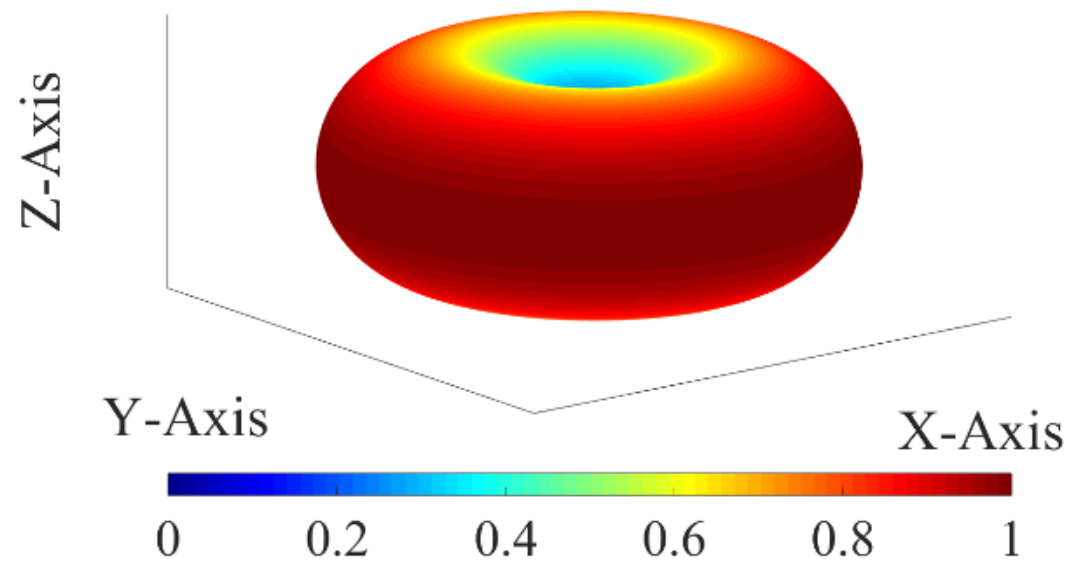

Figure 5-21 The modal radiation pattern of the DP-CM3 working at 7.20GHz and shown in Figure 5-19

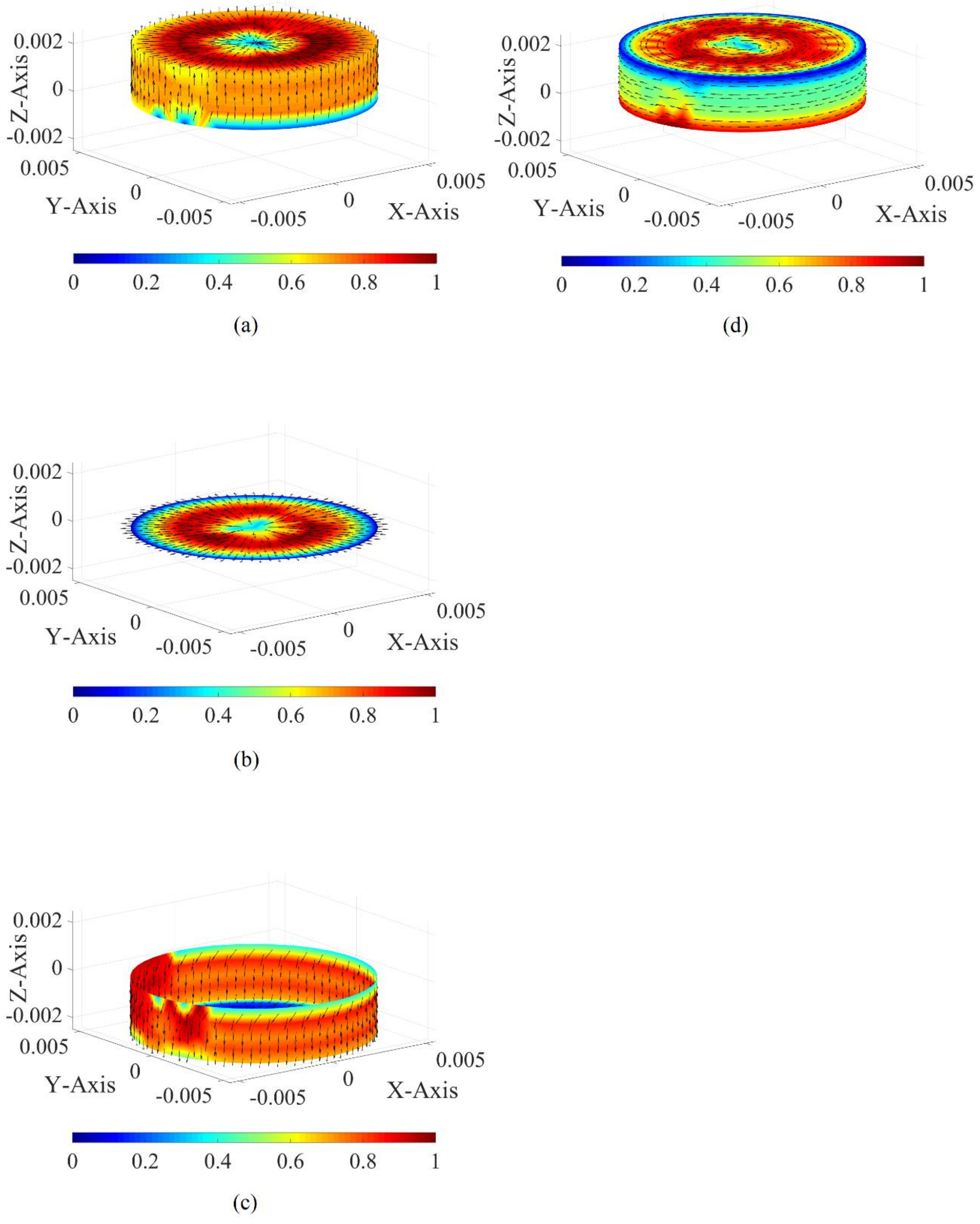

Figure 5-22 The modal equivalent surface source and scattered surface source distributions of the DP-CM3 working at 7.20GHz and shown in Figure 5-19. (a) the equivalent surface electric current on $\partial V_{\text{mat}}^{0}$; (b) the equivalent surface electric current on $\partial V_{\text{mat}}^{\cap}$ (i.e. the scattered surface electric current on $\partial V_{\text{met}}^{\cap}$); (c) the scattered surface electric current on $\partial V_{\text{met}}^{0}$; (d) the equivalent surface magnetic current on $\partial V_{\text{mat}}^{0}$

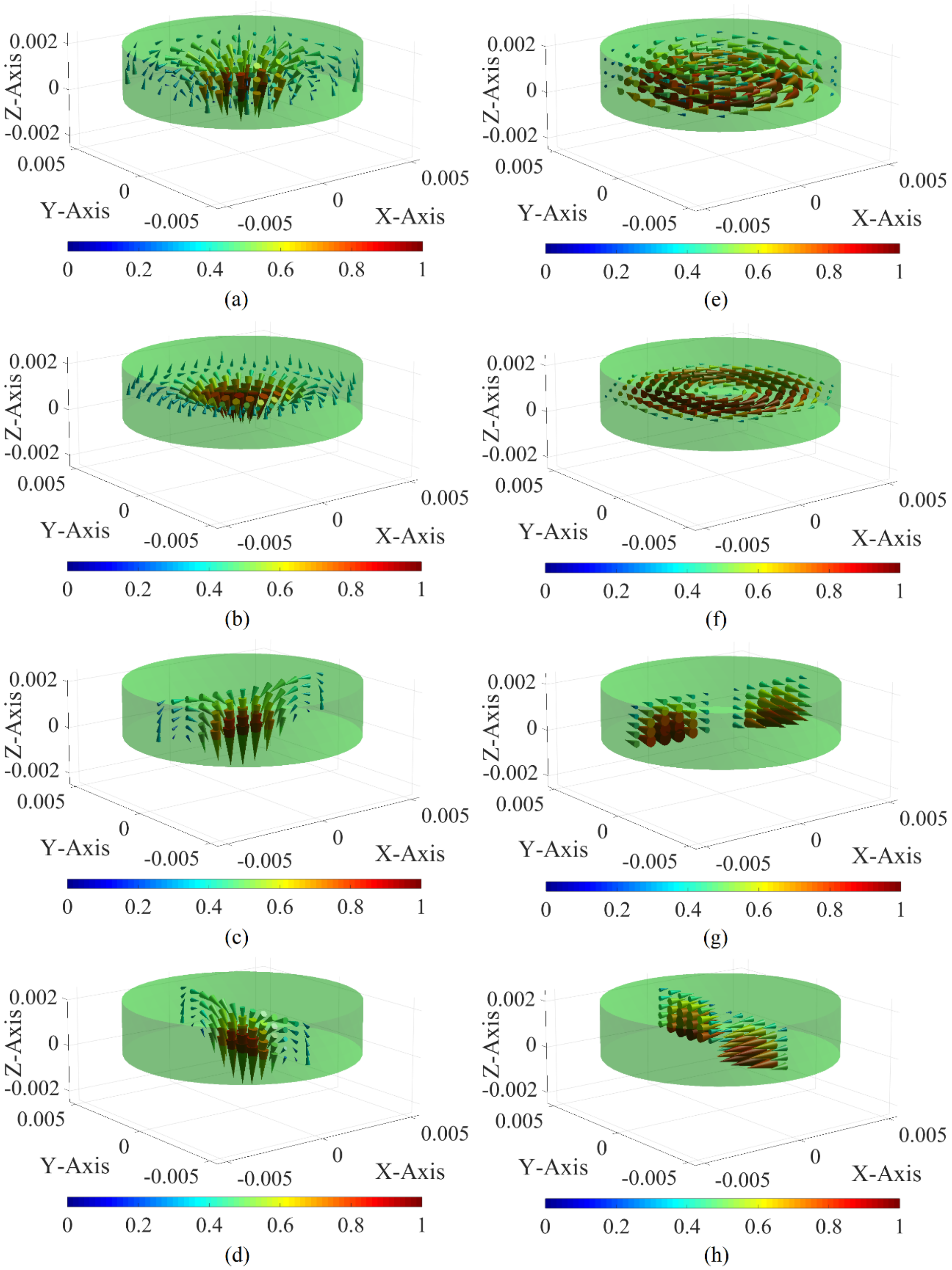

Figure 5-23 The modal total field distributions of the DP-CM3 working at 7.20GHz and shown in Figure 5-19. (a) the total electric field on $\partial V_{\text{mat}}^{0}$; (b) the total electric field on $z = 1.15\text{mm}$ surface; (c) the total electric field on xOz surface; (d) the total electric field on yOz surface; (e) the total magnetic field on $\partial V_{\text{mat}}^{0}$; (f) the total magnetic field on $z = 1.15\text{mm}$ surface; (g) the total magnetic field on xOz surface; (h) the total magnetic field on yOz surface

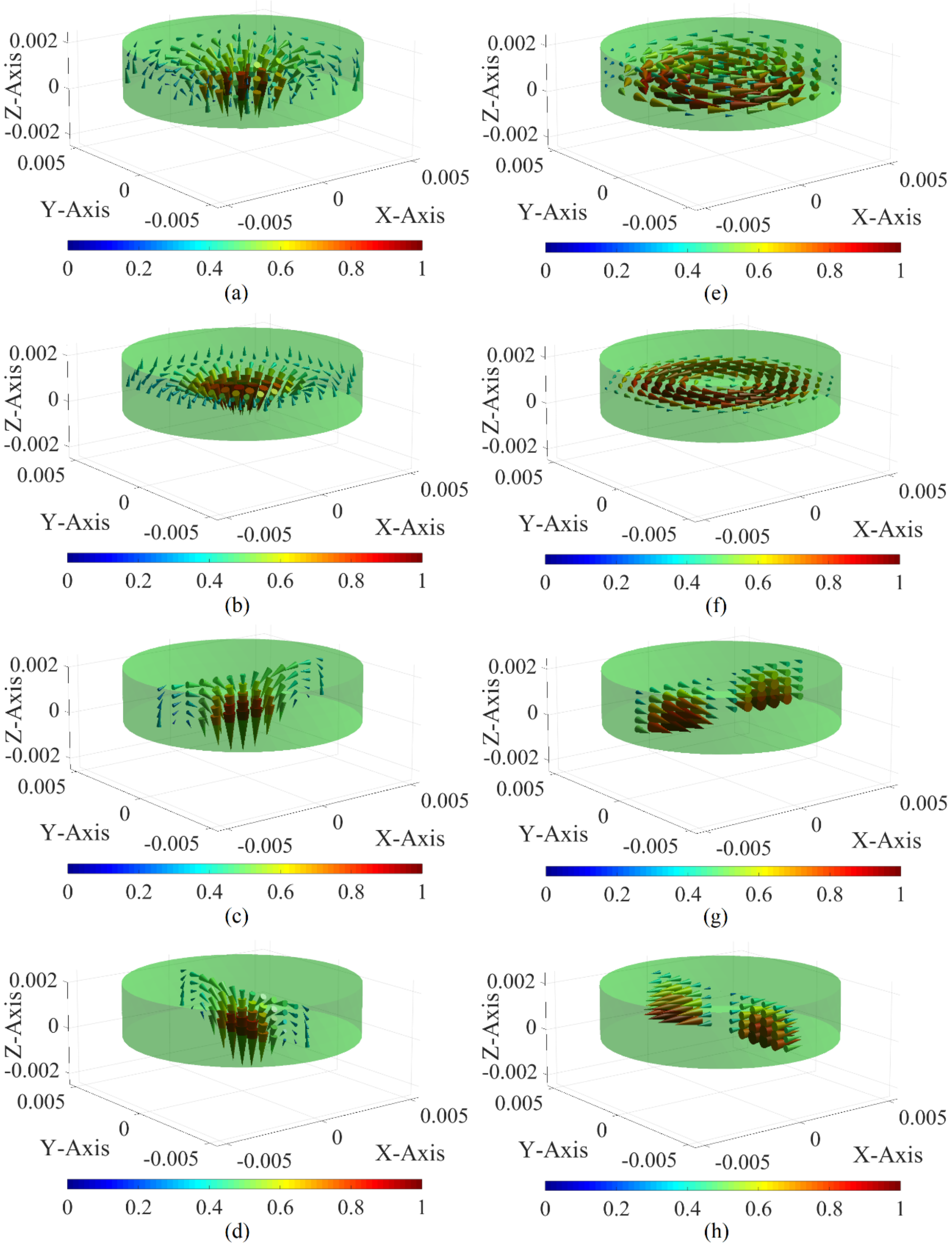

Figure 5-24 The modal scattered volume source distributions of the DP-CM3 working at 7.20GHz and shown in Figure 5-19. (a) the scattered electric current on $\partial V_{\rm mat}^{0}$; (b) the scattered electric current on $z = 1.15\text{mm}$ surface; (c) the scattered electric current on xOz surface; (d) the scattered electric current on yOz surface; (e) the scattered magnetic current on $\partial V_{\rm mat}^{0}$; (f) the scattered magnetic current on $z = 1.15\text{mm}$ surface; (g) the scattered magnetic current on xOz surface; (h) the scattered magnetic current on yOz surface

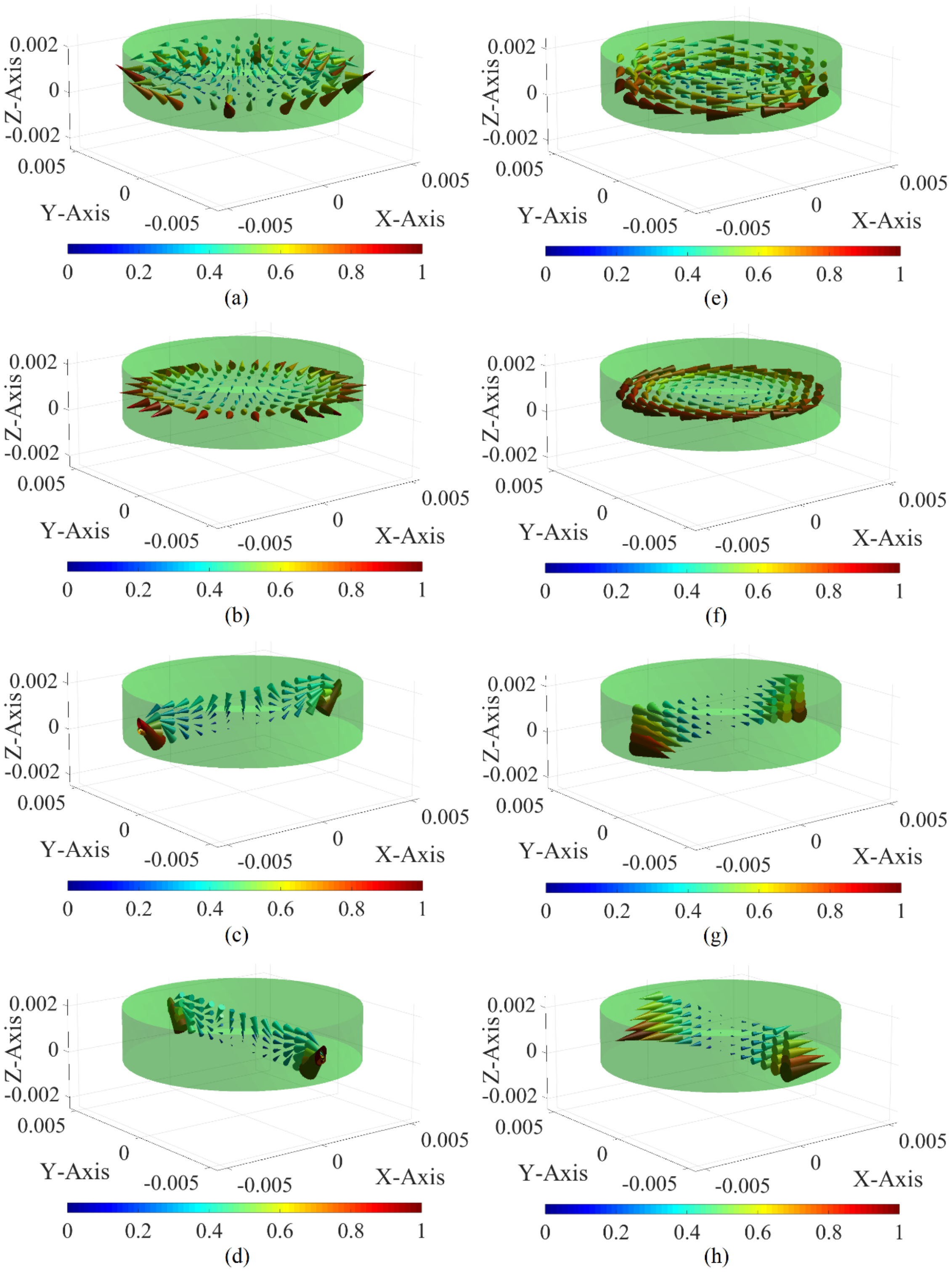

Figure 5-25 The modal incident field distributions of the DP-CM3 working at 7.20GHz and shown in Figure 5-19. (a) the incident electric field on $\partial V_{\text{mat}}^{0}$; (b) the incident electric field on z=1.15mm surface; (c) the incident electric field on xOz surface; (d) the incident electric field on yOz surface; (e) the incident magnetic field on $\partial V_{\text{mat}}^{0}$; (f) the incident magnetic field on z=1.15mm surface; (g) the incident magnetic field on xOz surface; (h) the incident magnetic field on yOz surface

### 5.6.3 Typical Structure III

In this subsection, we consider the composite system shown in Figure 5-26, and the system is constructed by an infinitely thin metallic dish, whose radius is 2.625mm, and a material cylinder, whose radius and height are 5.250mm and 4.600mm respectively, and the {relative permeability, relative permittivity, conductivity} of the material cylinder are {6, 6, 0}.

From Figure 5-26, it is easy to find out that the topological structure of the composite system satisfies the topological restrictions given in the Section 5.2 of this dissertation, so we can, based on the boundary decomposition method introduced in the Section 5.3 of this dissertation, decompose the material boundary of the composite system as shown Figure 5-27. In addition, because whole metallic boundary is in contact with the material part of the composite system, then there is no need to further decompose the metallic boundary (for details see Section 5.3), and we provide the topological structure and surface triangular meshes of the metallic boundary in Figure 5-28.

For the convenience of the following discussions and the consistency of the symbolic system used in previous sections, we denote the interface between the material cylinder and environment as $\partial V_{\text{mat}}^{0}$, and denote the interface between the material cylinder and the metallic dish as $\partial V_{\text{mat}}^{\cap}$, and denote the boundary of the metallic dish as $S_{\text{met}}$, and it is obvious that $S_{\text{met}} = \partial V_{\text{mat}}^{\cap}$.

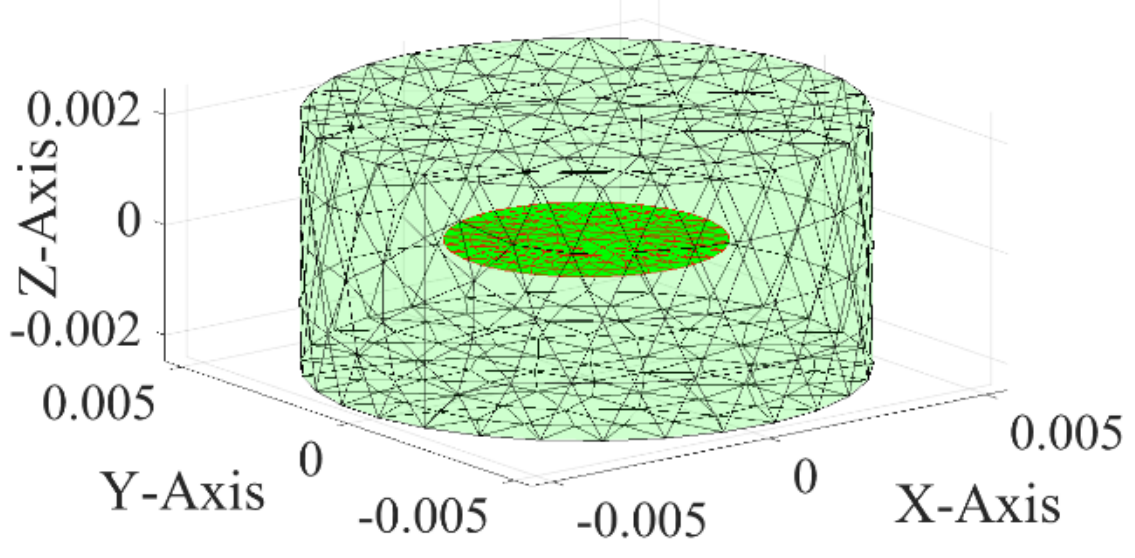

Figure 5-26 The topological structure and surface triangular meshes of the composite system constituted by "an infinitely thin metallic circular dish whose radius is 5.625mm" and "a material cylinder whose radius and height are 5.250mm and 4.600mm respectively"

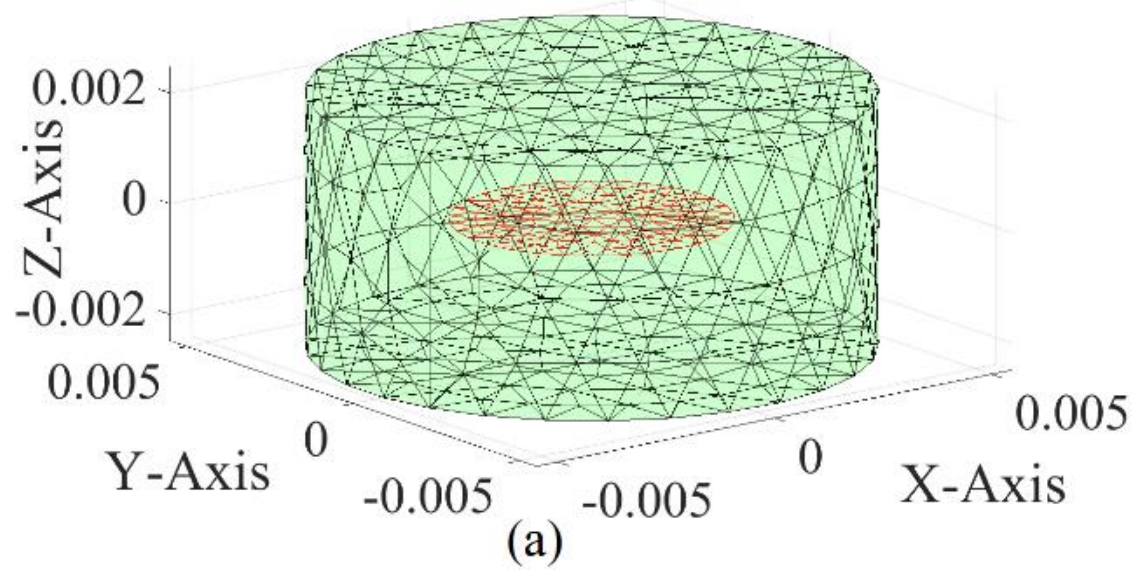

(a)

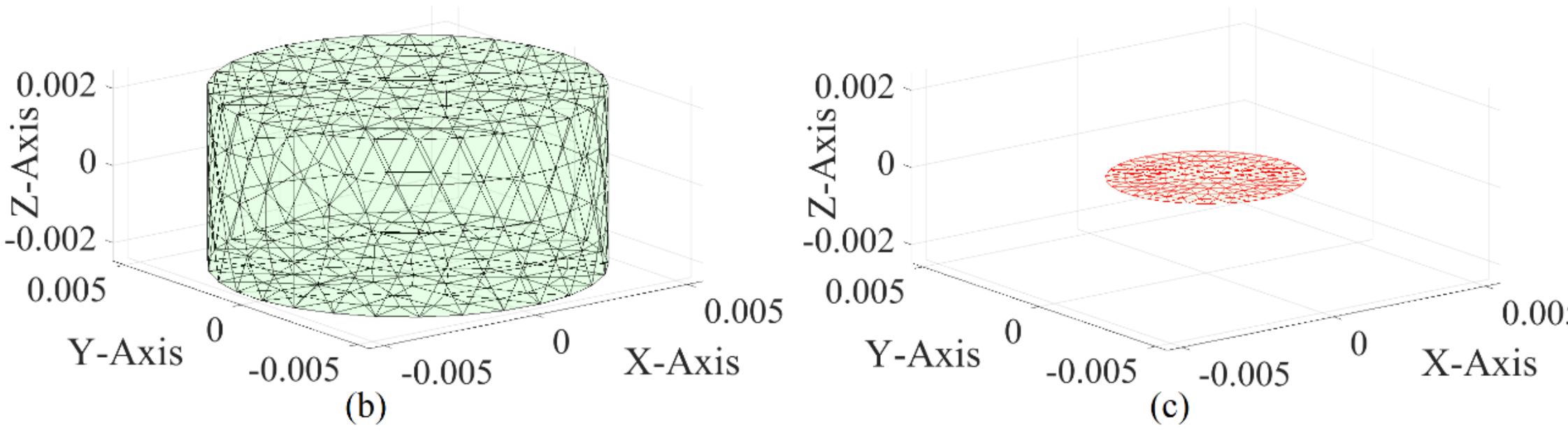

Figure 5-27 The boundary decomposition for the material cylinder shown in Figure 5-26. (a) The topological structure and surface triangular meshes of the whole boundary of material cylinder; (b) the material-environment boundary $\partial V_{\text{mat}}^{0}$ and its topological structure and surface triangular meshes; (c) the material-metal boundary $\partial V_{\text{mat}}^{\cap}$ and its topological structure and surface triangular meshes

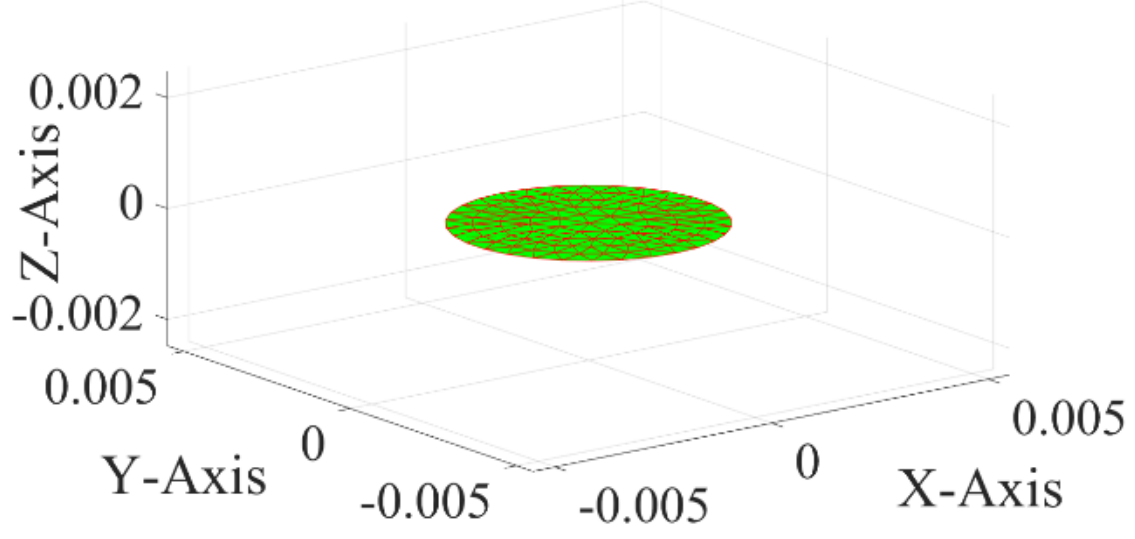

Figure 5-28 The topological structure and surface triangular meshes of the metallic dish $S_{\text{met}}$ shown in Figure 5-26. It is obvious that $S_{\text{met}} = \partial V_{\text{mat}}^{\cap}$

Focusing on the composite system shown in Figure 5-26, we obtain whole DP-CM set based on the formulation established in this chapter. The characteristic value (dB) curves and MS curves corresponding to some typical DP-CMs are shown in Figure 5-29.

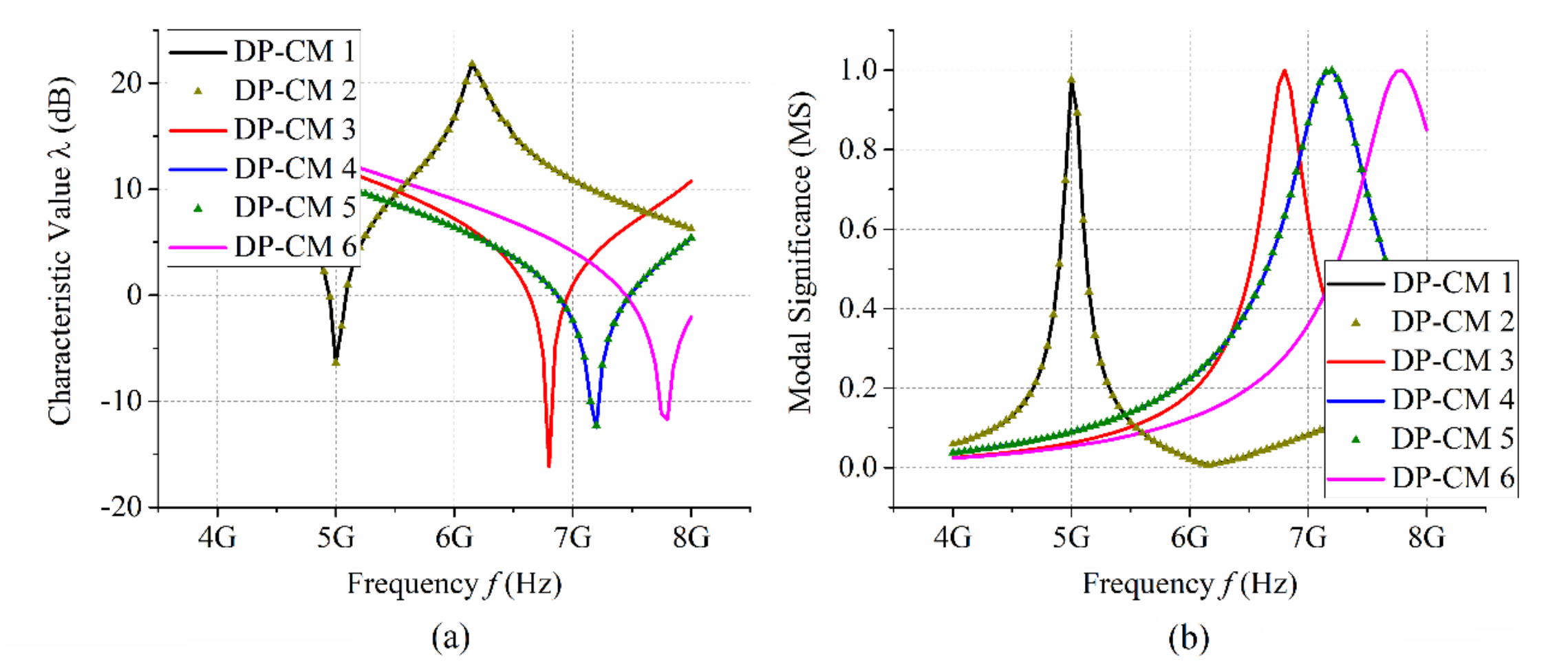

Figure 5-29 The characteristic quantity curves corresponding to some typical DP-CMs (derived from the theory developed in this chapter) of the composite system shown in Figure 5-26. (a) characteristic value dB curves; (b) MS curves

Taking the DP-CMs shown in Figure 5-29 as typical examples, we provide their modal equivalent source distributions, modal scattered source distributions, modal field distributions, and modal radiation patterns as below.

From Figure 5-29, it is easy to find out that DP-CM1 is “resonant” at 5.00GHz. For the “resonant” DP-CM1, its equivalent surface electric and magnetic currents are illustrated in Figure 5-30, and its tangential total magnetic and electric fields distributing on material inner boundaries are illustrated in Figure 5-31. Obviously, the distributions shown in Figures 5-30 and 5-31 indeed satisfy relationships (5-20) and (5-21). For the “resonant” DP-CM1 working at 5.00GHz, its modal scattered volume electric and magnetic currents distributing on the material spherical shell are illustrated in Figure 5-32. In addition, we also provide the radiation pattern corresponding to the modal state in Figure 5-33.

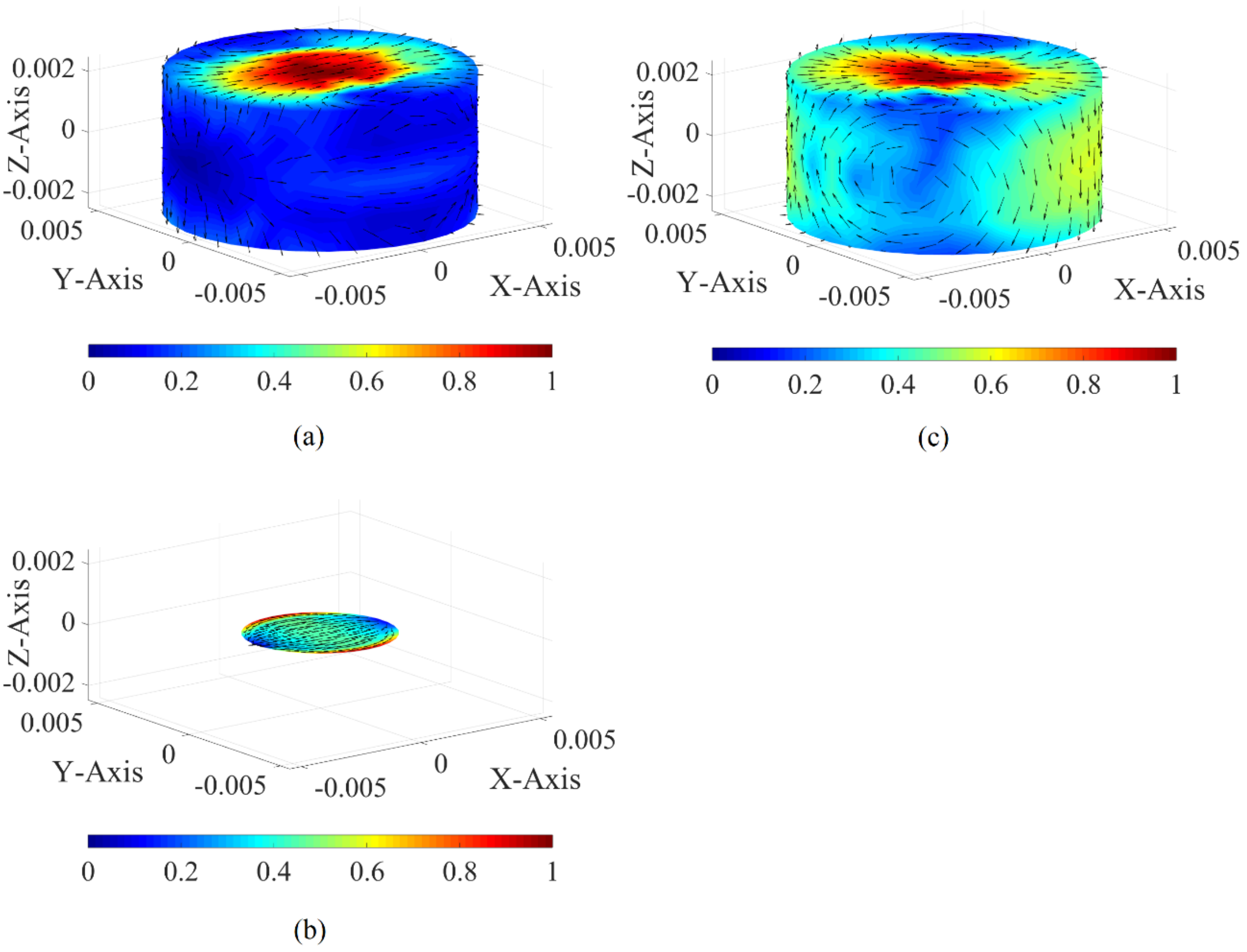

Figure 5-30 The modal equivalent surface source distributions of the DP-CM1 working at 5.00GHz and shown in Figure 5-29. (a) the equivalent surface electric current on $\partial V_{\text{mat}}^{0}$ ; (b) the equivalent surface electric current on $\partial V_{\text{mat}}^{\cap}$ ; (c) the equivalent surface magnetic current on $\partial V_{\text{mat}}^{0}$

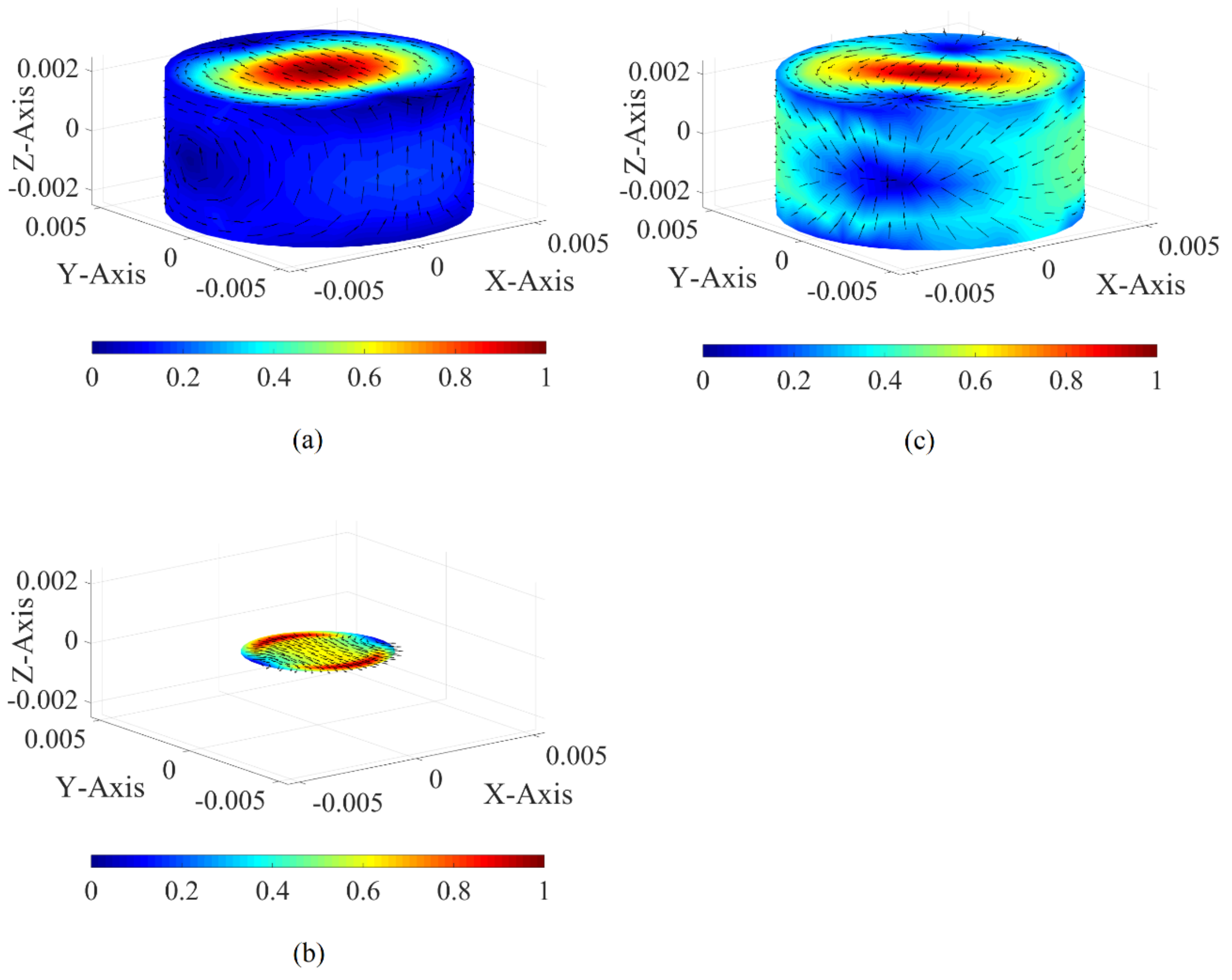

Figure 5-31 The modal tangential total field distributions of the DP-CM1 working at 5.00GHz and shown in Figure 5-29. (a) the tangential total magnetic field on the inner surface of $\partial V_{\text{mat}}^{0}$; (b) the difference between the tangential total magnetic fields on the upper and lower surfaces of $\partial V_{\text{mat}}^{\cap}$; (c) the tangential total electric field on the inner surface of $\partial V_{\text{mat}}^{0}$

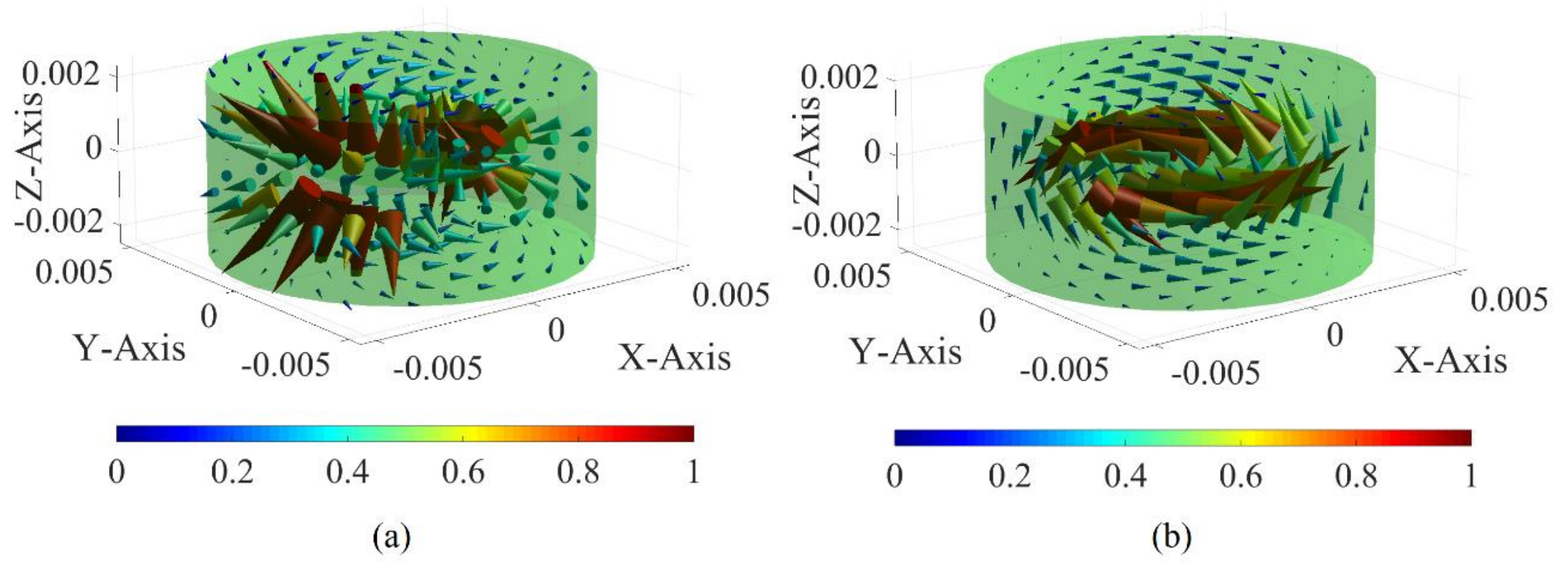

Figure 5-32 The modal scattered volume source distributions of the DP-CM1 working at 5.00GHz and shown in Figure 5-29. (a) the scattered electric current on $V_{\text{mat}}$; (b) the scattered magnetic current on $V_{\text{mat}}$

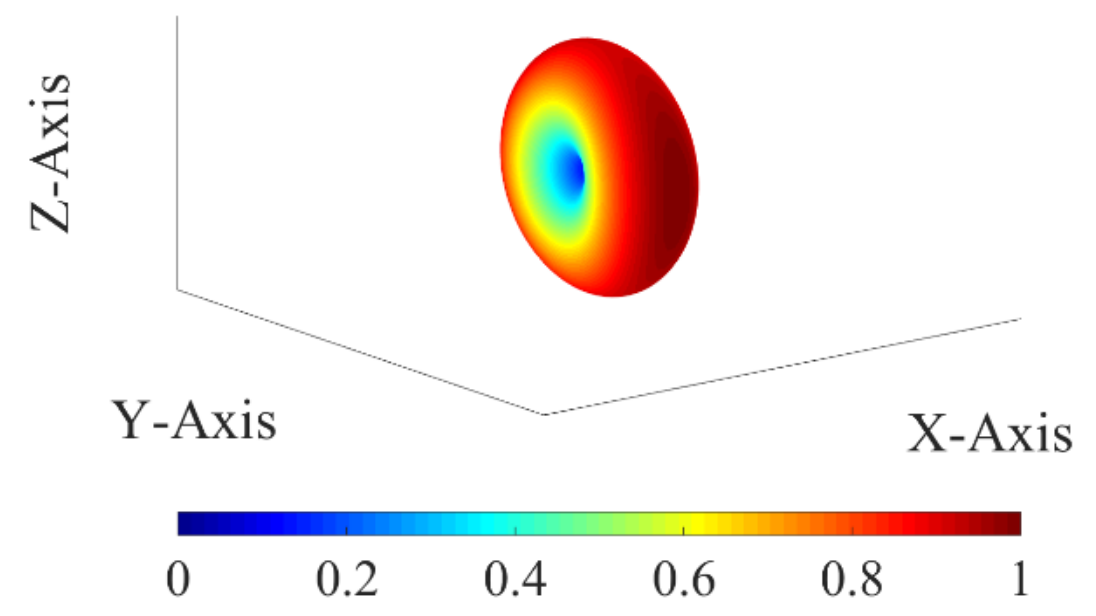

Figure 5-33 The modal radiation pattern of the DP-CM1 working at 5.00GHz and shown in Figure 5-29

For the "resonant" DP-CM2 working at 5.00GHz, its BV distribution (the modal equivalent surface electric current distributing on $\partial V_{\mathrm{mat}}^{0}$ ) and radiation pattern are illustrated in Figure 5-34.

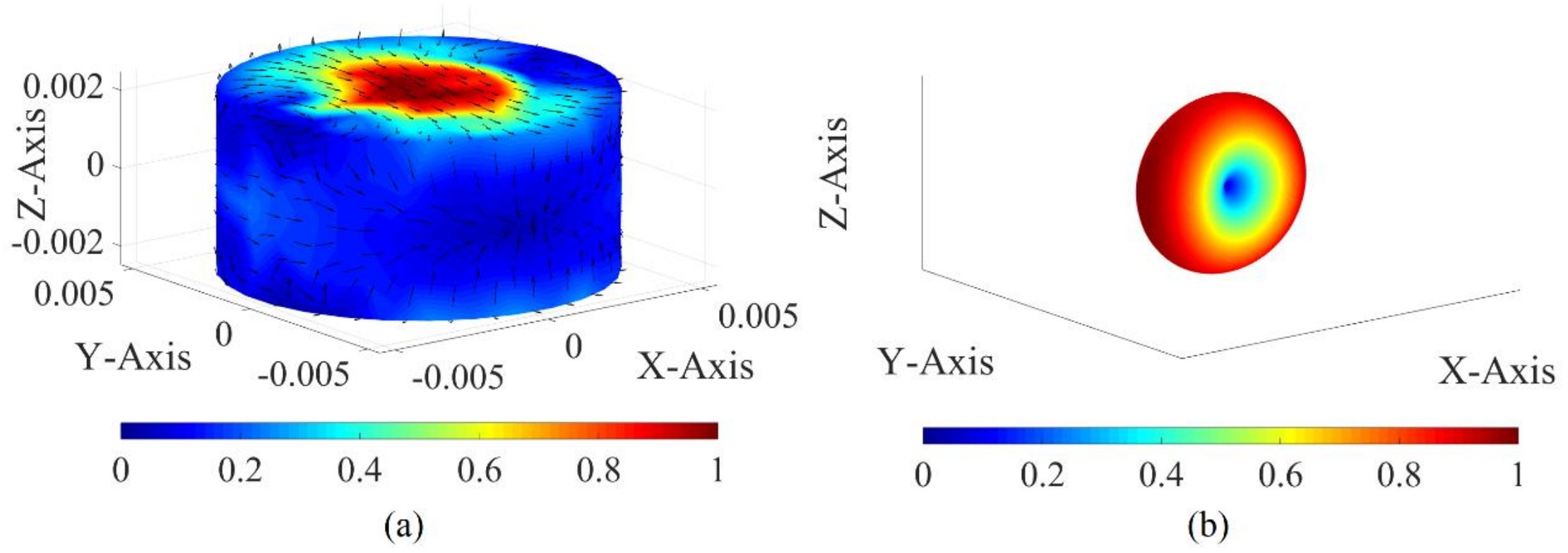

Figure 5-34 The modal basic variable and radiation pattern of the DP-CM2 working at 5.00GHz and shown in Figure 5-29. (a) basic variable (the equivalent surface electric current on $\partial V_{\mathrm{mat}}^{0}$ ); (b) radiation pattern

Obviously, the above two DP-CMs working at 5.00GHz constitute a pair of degenerate modes, and the degeneracy originates from the z-axial rotational symmetry of the composite system shown in Figure 5-26.

### 5.6.4 Typical Structure IV

In this subsection, we consider the composite system shown in Figure 5-35, and the system is constructed by a metallic dish, whose radius is 5.25mm, and a material cylinder, whose radius and height are 5.25mm and 2.30mm respectively, and the {relative permeability, relative permittivity, conductivity} of the material cylinder are {6, 6, 0}.

From Figure 5-35, it is easy to find out that the topological structure of the composite system satisfies the topological restrictions given in the Section 5.2 of this dissertation, so we can, based on the boundary decomposition method introduced in the Section 5.3 of this dissertation, decompose the material boundary of the composite system as shown Figure 5-36. In addition, because whole metallic boundary is in contact with the material part of the composite system, then there is no need to further decompose the metallic boundary (for details see Section 5.3), and we provide the topological structure and surface triangular meshes of the metallic boundary in Figure 5-37.

For the convenience of the following discussions and the consistency of the symbolic system used in previous sections, we denote the interface between the material cylinder and environment as $\partial V_{\mathrm{mat}}^{00}$, and denote the interface between the material cylinder and the metallic dish as $\partial V_{\mathrm{mat}}^{01}$, and denote the boundary of the metallic dish as $S_{\mathrm{met}}$, and it is obvious that $S_{\mathrm{met}} = \partial V_{\mathrm{mat}}^{01}$.

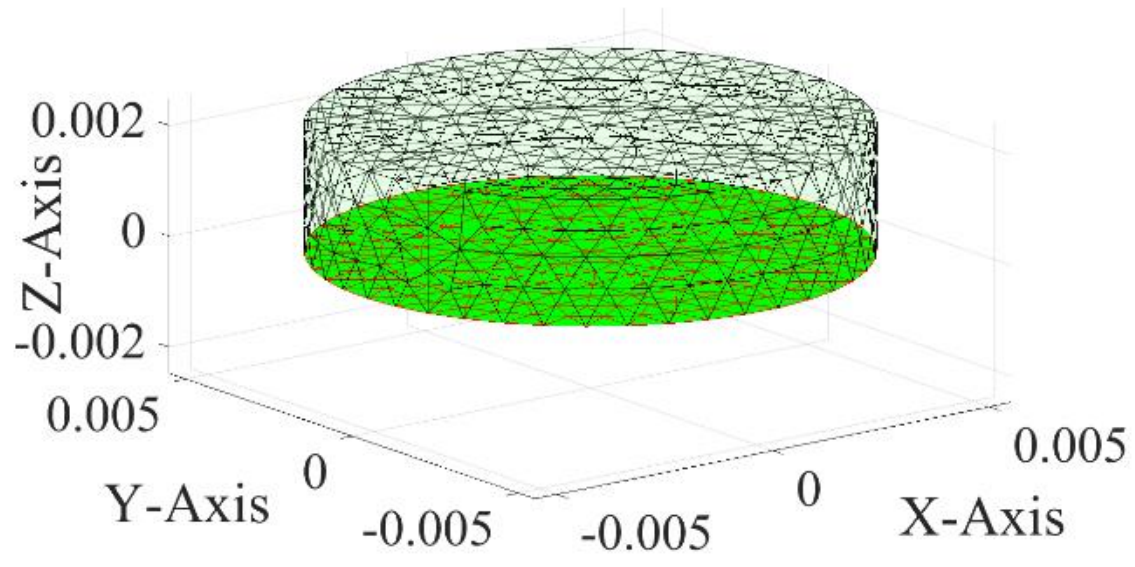

Figure 5-35 The topological structure and surface triangular meshes of the composite system constituted by "an infinitely thin metallic circular dish whose radius is 5.25mm" and "a material cylinder whose radius and height are 5.25mm and 2.30mm respectively"

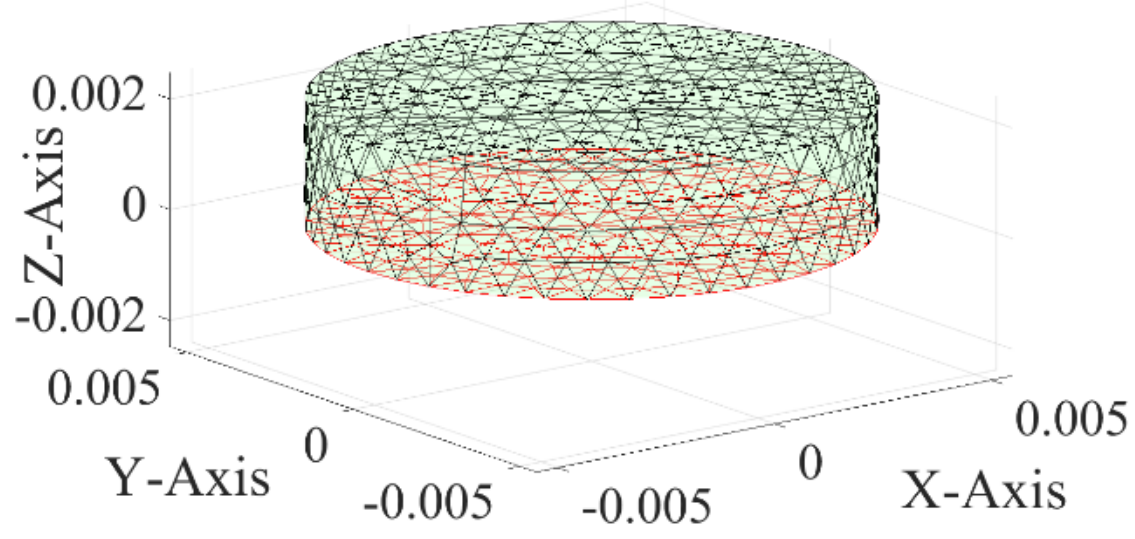

(a)

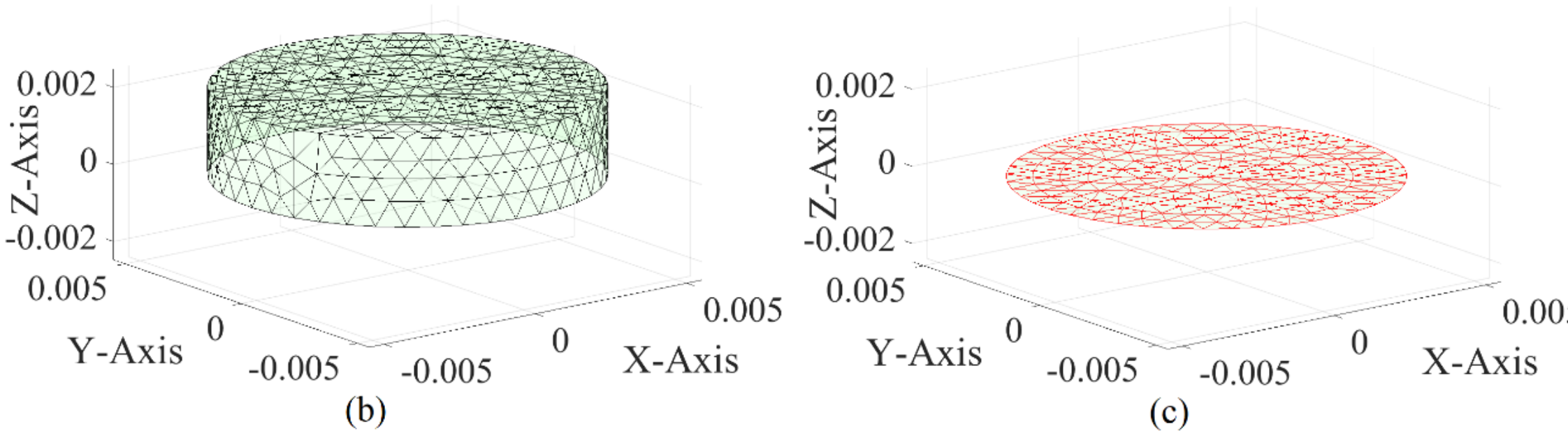

Figure 5-36 The boundary decomposition for the material cylinder shown in Figure 5-35. (a) the topological structure and surface triangular meshes of the whole boundary of the material cylinder; (b) the material-environment boundary $\partial V_{\text{mat}}^{00}$ and its topological structure and surface triangular meshes; (c) the material-metal boundary $\partial V_{\text{mat}}^{01}$ and its topological structure and surface triangular meshes

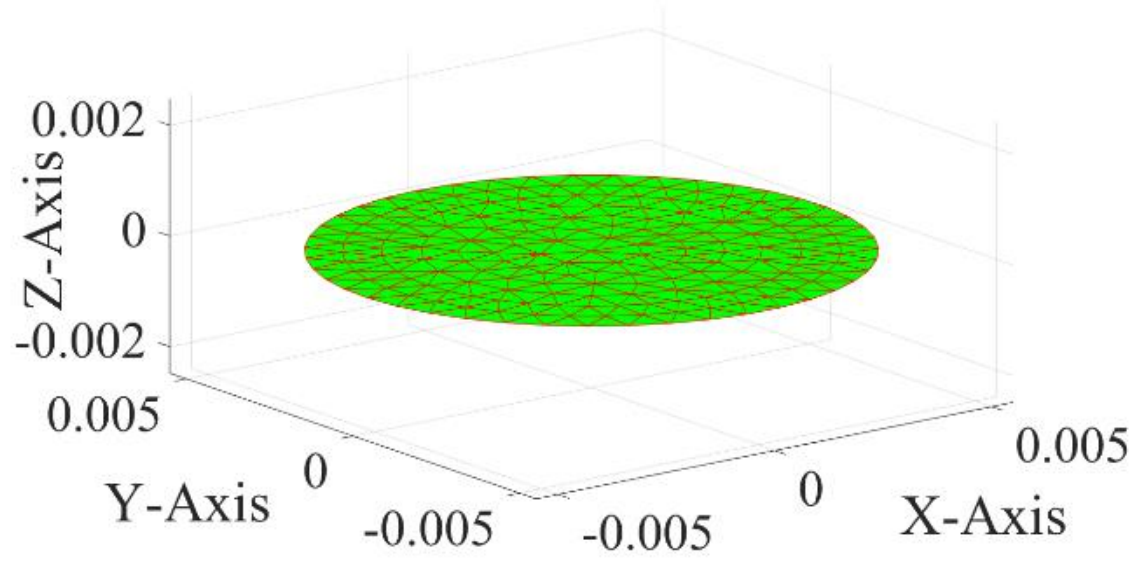

Figure 5-37 The topological structure and surface triangular meshes of the infinitely thin metallic dish $S_{\text{met}}$ shown in Figure 5-35. It is obvious that $S_{\text{met}} = \partial V_{\text{mat}}^{01}$

Focusing on the composite system shown in Figure 5-35, we obtain whole DP-CM set based on the formulation established in this chapter. The characteristic value (dB) curves and MS curves corresponding to some typical DP-CMs are shown in Figure 5-38.

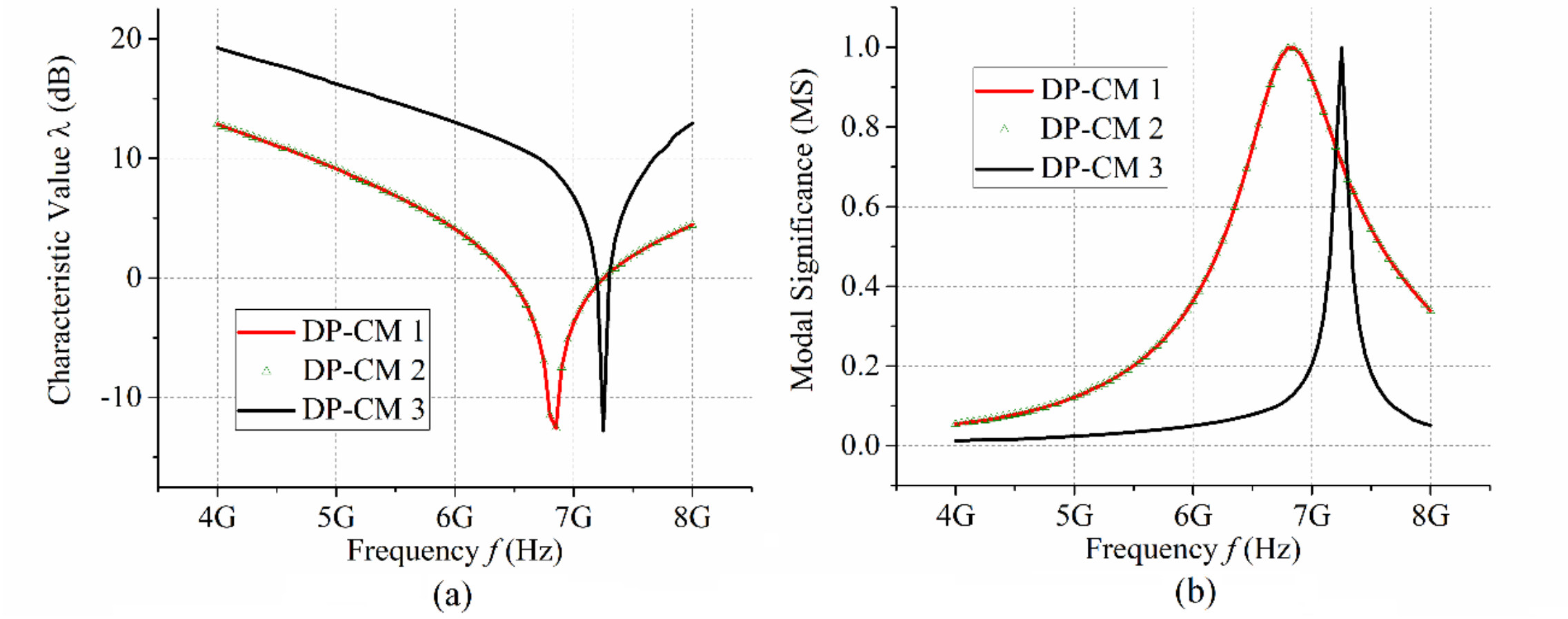

Figure 5-38 The characteristic quantity curves corresponding to some typical DP-CMs (derived from the theory developed in this chapter) of the composite system shown in Figure 5-35. (a) characteristic value dB curves; (b) MS curves

Taking the DP-CMs shown in Figure 5-38 as typical examples, we provide their modal equivalent source distributions, modal scattered source distributions, modal field distributions, and modal radiation patterns as below.

From Figure 5-38, it is easy to find out that DP-CM3 is “resonant” at 7.20GHz. For the “resonant” DP-CM3, its equivalent surface electric and magnetic currents on material boundaries and the scattered surface electric current on metallic boundary are illustrated in Figure 5-39, and its modal total fields, modal scattered volume sources, and modal incident fields distributing on the material cylinder are illustrated in Figures 5-40, 5-41, and 5-42 respectively. In addition, we also provide the radiation pattern corresponding to the modal state in Figure 5-43.

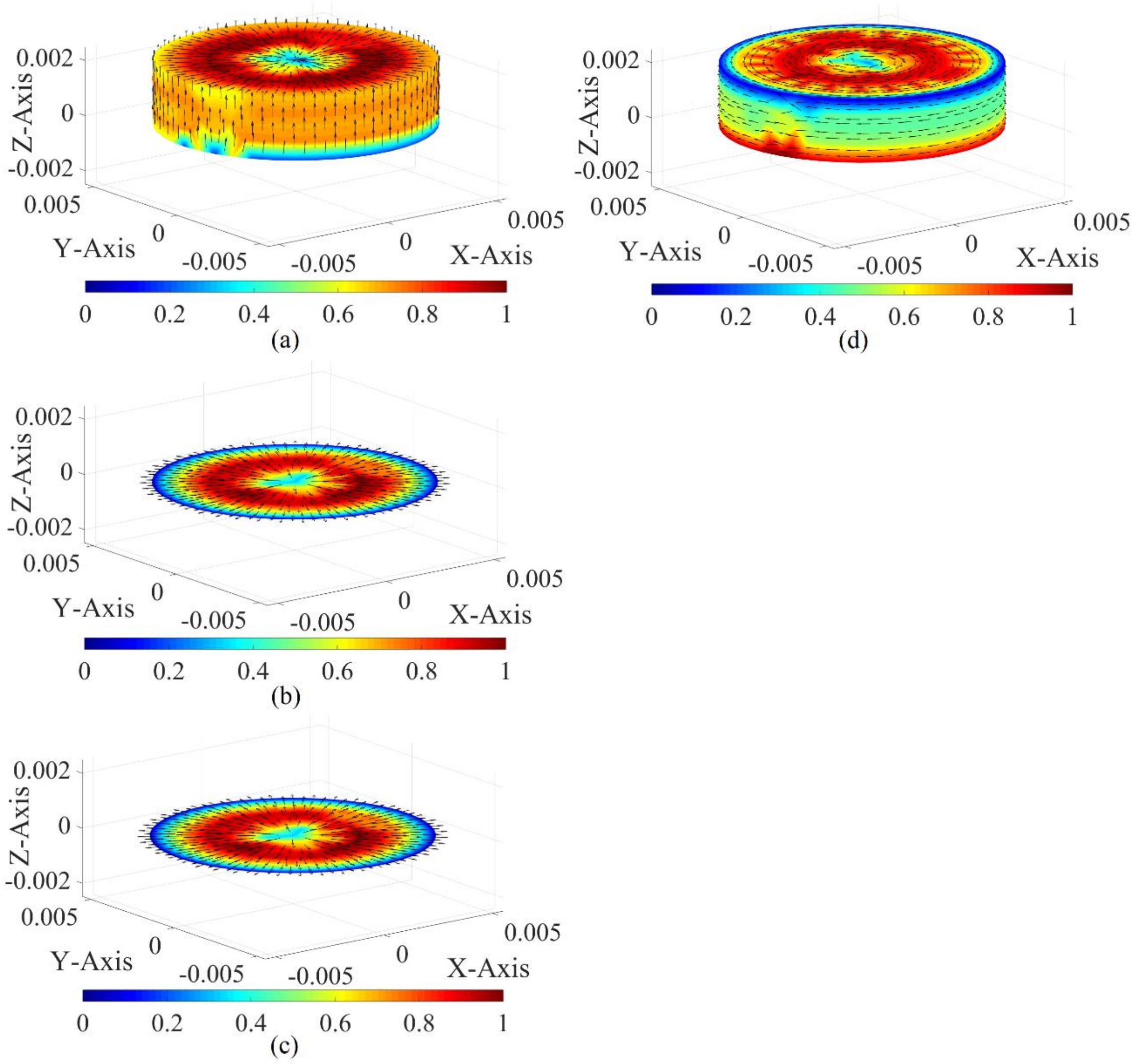

Figure 5-39 The modal equivalent surface source and scattered surface source distributions of the DP-CM3 working at 7.20GHz and shown in Figure 5-38. (a) the equivalent surface electric current on $\partial V_{\text{mat}}^{00}$; (b) the equivalent surface electric current on $\partial V_{\text{mat}}^{01}$; (c) the scattered surface electric current on $S_{\text{met}}$; (d) the equivalent surface magnetic current on $\partial V_{\text{mat}}^{00}$

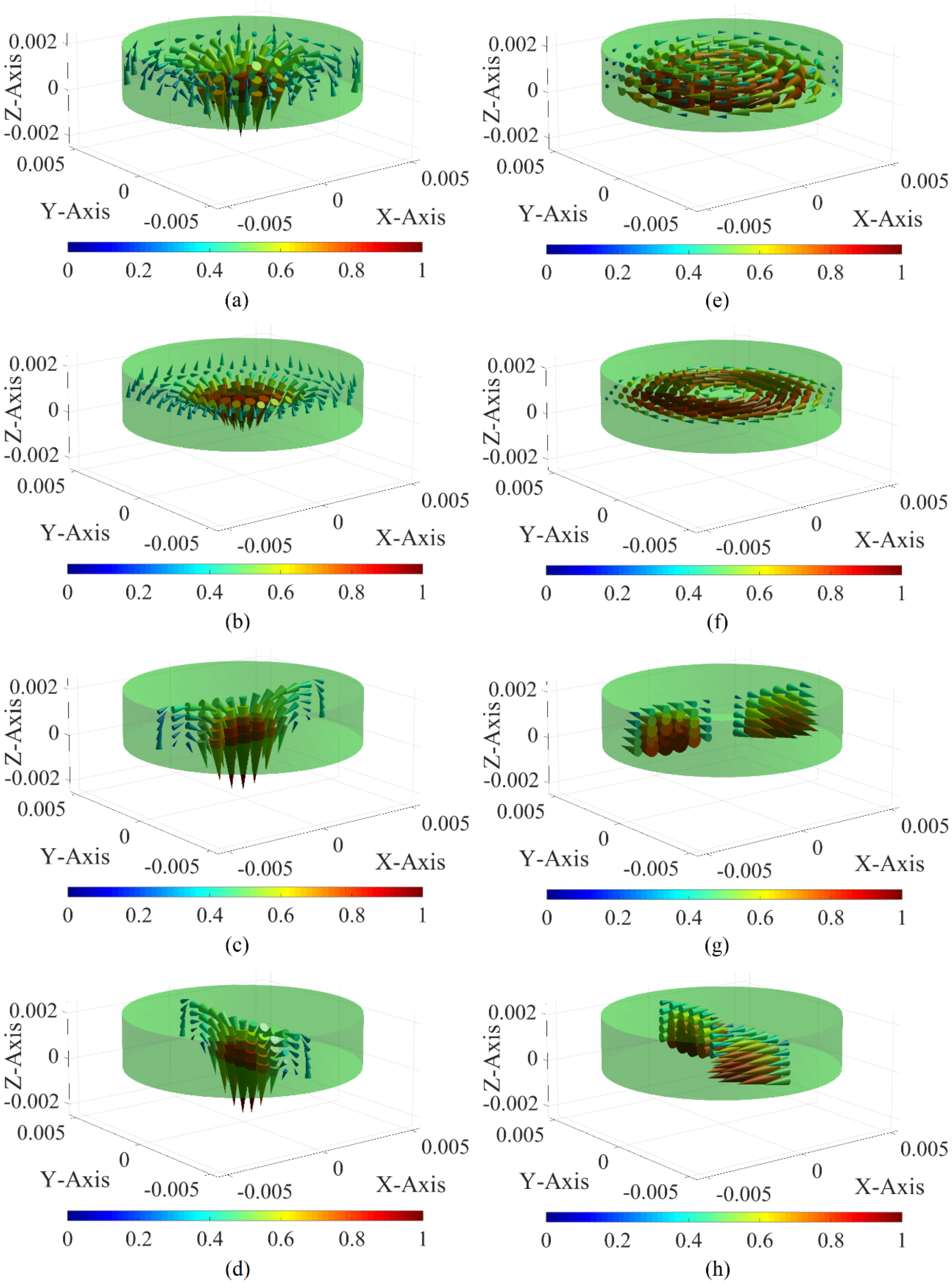

Figure 5-40 The modal total field distributions of the DP-CM3 working at 7.20GHz and shown in Figure 5-38. (a) the total electric field on $V_{\text{mat}}$; (b) the total electric field on $z = 1.15\text{mm}$ surface; (c) the total electric field on xOz surface; (d) the total electric field on yOz surface; (e) the total magnetic field on $V_{\text{mat}}$; (f) the total magnetic field on $z = 1.15\text{mm}$ surface; (g) the total magnetic field on xOz surface; (h) the total magnetic field on yOz surface

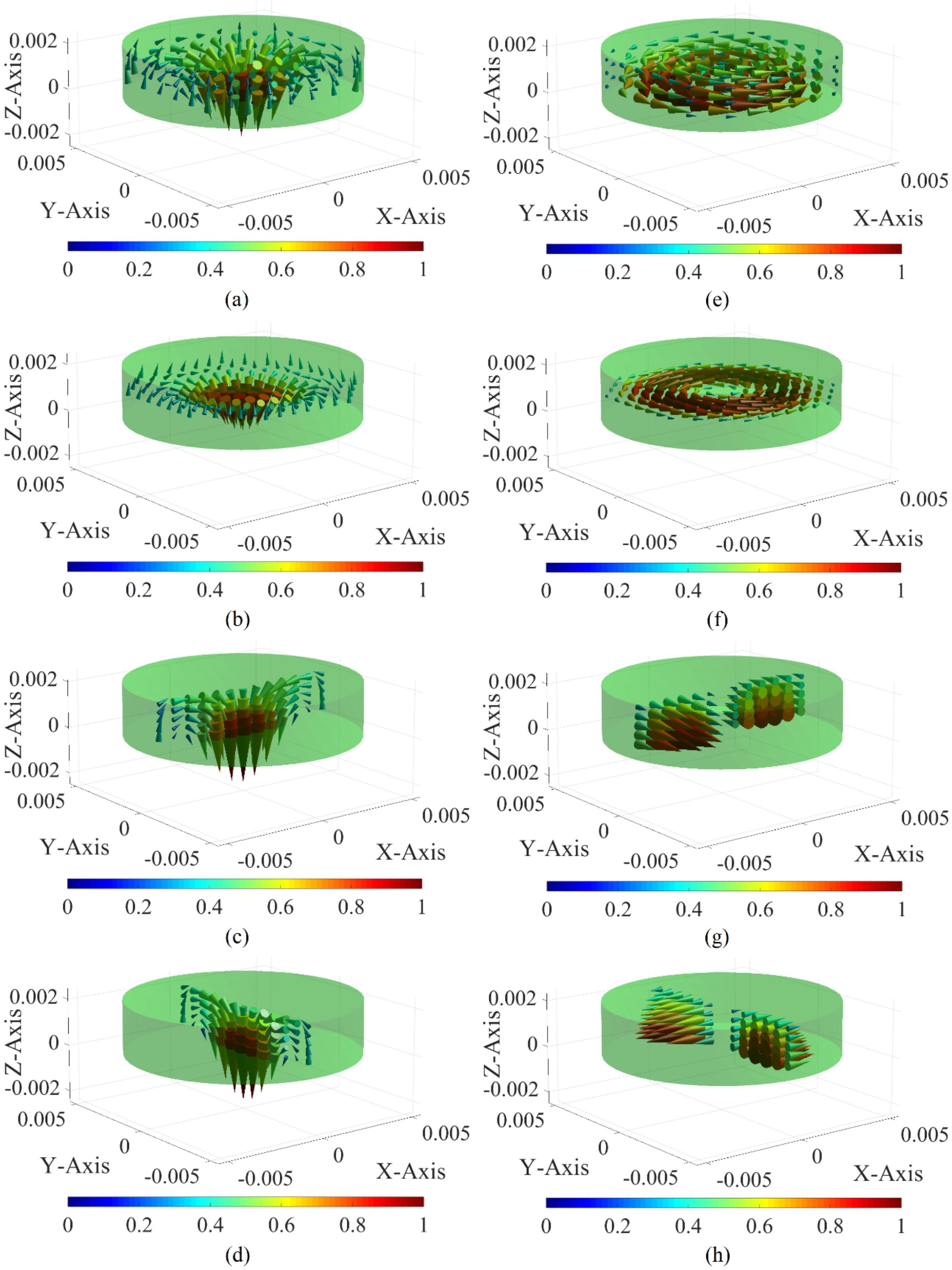

Figure 5-41 The modal scattered volume source distributions of the DP-CM3 working at 7.20GHz and shown in Figure 5-38. (a) the scattered volume electric current on $V_{\text{mat}}$ ; (b) the scattered volume electric current on $z = 1.15\text{mm}$ surface; (c) the scattered volume electric current on xOz surface; (d) the scattered volume electric current on yOz surface; (e) the scattered volume magnetic current on $V_{\text{mat}}$ ; (f) the scattered volume magnetic current on $z = 1.15\text{mm}$ surface; (g) the scattered volume magnetic current on xOz surface; (h) the scattered volume magnetic current on yOz surface

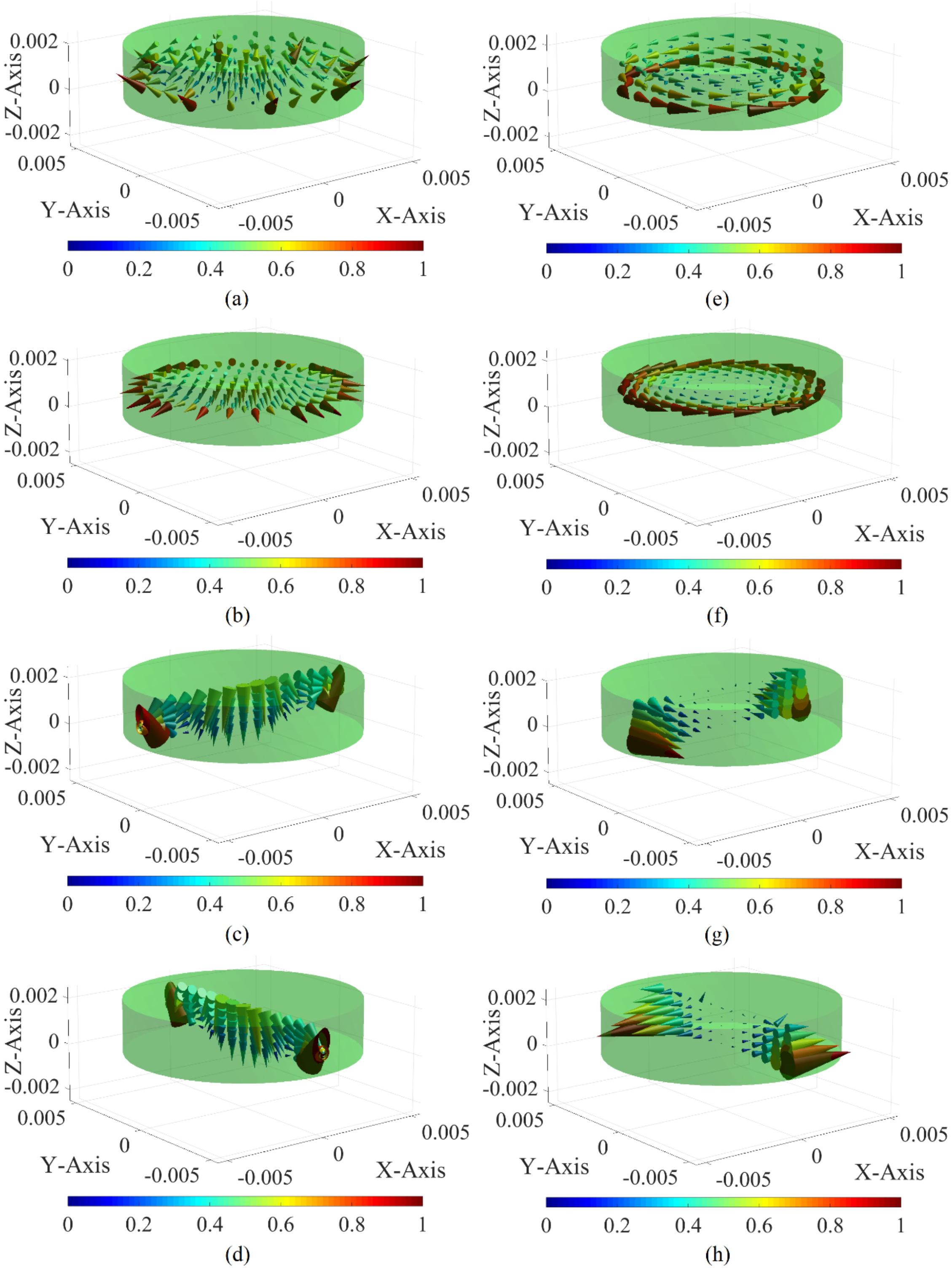

Figure 5-42 The modal incident field distributions of the DP-CM3 working at 7.20GHz and shown in Figure 5-38. (a) the incident electric field on $V_{mat}$ ; (b) the incident electric field on z=1.15mm surface; (c) the incident electric field on xOz surface; (d) the incident electric field on yOz surface; (e) the incident magnetic field on $V_{mat}$ ; (f) the incident magnetic field on z=1.15mm surface; (g) the incident magnetic field on xOz surface; (h) the incident magnetic field on yOz surface

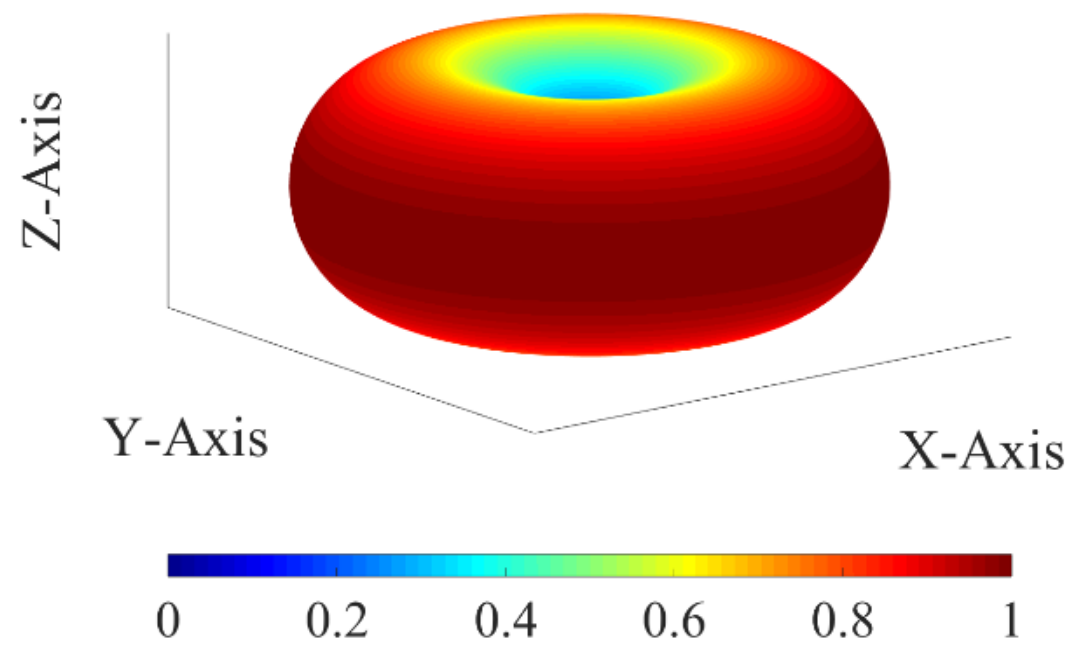

Figure 5-43 The modal radiation pattern of the DP-CM3 working at 7.20GHz and shown in Figure 5-38

In the following Section 5.7, we will summarize this chapter from the aspects of fundamental principle, central philosophy, main method, and important conclusions.

## 5.7 Chapter Summary

This chapter mainly studies how to establish the CMT for composite scattering systems by orthogonalizing frequency-domain DPO in WEP framework. In addition, the composite systems are very general in the aspects of external EM environment, system topological structure, and system material parameter: **1)** in the aspect of external EM environment, the composite systems can be placed in either a vacuum environment or a complex environment. **2)** in the aspect of system topological structure, the metallic parts in the composite systems can be metallic lines or metallic surfaces or metallic bodies or “line-surface-body composite structures”; the metallic and material parts in the composite systems can be contacted with each others or separated from each others, and the metallic parts can be completely or partially submerged into the material parts. **3)** in the aspect of system material parameter, the material parts in the composite systems can be either homogeneous isotropic or inhomogeneous anisotropic.

Firstly, we propose the methods to decompose the metallic boundaries and the material boundaries related to the composite systems. Then, we realize the decompositions for the scattered sources and the equivalent sources distributing on the various boundaries. Afterwards, based on the above decompositions and employing the generalized surface equivalence principle given in Appendix C, we establish the line-surface equivalence principle for composite systems (the line-surface equivalence principle involves both line sources and surface sources). After that, based on the WEP for composite systems and employing the line-surface equivalence principle, we derive the line-surface expression for the DPO of composite systems (the line-surface expression involves both line sources and surface sources). Finally, by orthogonalizing the DPO, we, for any objective composite system, construct a series of steadily working modes not having net energy exchange in any integral period —— the DP-CMs of the objective composite system. In addition, in the process to construct the DP-CMs, we also systematical establish a practical scheme for suppressing spurious modes.

The DP-CM calculation formulations obtained in this chapter are completely new. For some very simple composite systems, both the new formulations in WEP framework and the traditional formulations in IE framework (which need to use the variable unification scheme given in this chapter) are applicable, and in this special cases this chapter, by comparing the results calculated from the new and traditional formulations, verifies the effectiveness and correctness of the new formulations in the aspect of constructing the DP-CMs of composite systems. For some relatively complicated composite systems, the traditional formulations in IE framework have not been applicable to, but the new formulations provided by this chapter in WEP framework are applicable to.

# Chapter 6 Common Problems During Constructing DP-CMs

> We do not really deal with mathematical physics, but with physical mathematics; not with the mathematical formulation of physical facts, but with the physical motivation of mathematical methods. The oft-mentioned “prestabilized harmony” between what is mathematically interesting and what is physically important is met at each step and lends an esthetic – I should like to say metaphysical – attraction to our subject. [153]
>
> —— Arnold Sommerfeld (German physicist)

In the previous chapters, the fundamental formalism of the WEP-based CMT for scattering systems (WEP-ScaSys-CMT) is almost complete. But, there still exist some important technical topics which are valuable to study, and this chapter focuses on studying the topics.

## 6.1 Chapter Introduction

In 2014, literature [46] found out that the material CM set derived from the formulation developed in classical literature [34] contains many spurious modes. Since then, “what is the reason leading to the spurious modes and how to suppress the spurious modes” has been one of the research hotspots in CMT domain. To date, many scholars[35,42,44,45,47-49,53,54,56] have done some valuable studies for the related topics.

Many scattering systems considered in the Chapters 4 and 5 of this dissertation are much more complicated than the scattering systems focused on by literatures [35,42,44,45,47-49,53,54,56] in the aspects of topological structure and material parameters, so the suppression schemes proposed in literatures [35,42,44,45,47-49,53,54,56] cannot completely meet the requirements of the Chapters 4 and 5 of this dissertation. Because of this, we, in the Subsections 6.2.1~6.2.3 of this chapter, systematically study the causes and suppressions for the spurious modes of some typical complicated scattering systems discussed in the Chapters 4 and 5 of this dissertation, and summarize the systematical scheme for suppressing the spurious modes of general scattering systems in Subsection 6.2.4. After that, we, in Section 6.3, propose a compromise scheme for suppressing spurious modes based on the results obtained in Sections 4.2 and 6.2.

In addition, the physical meaning of the singular current terms (SCT) contained in the DPOs of material systems and composite systems is also the important topic to be discussed in this chapter.

## 6.2 Spurious Modes and Their Suppression

As demonstrated in some literatures[44,45] and the Chapters 4 and 5 of this dissertation, variable unification is indispensable for WEP-ScaSys-CMT, and "how to select basic variables (BVs) and how to establish the transformation from BVs to dependent variables (DVs)" is just the key for variable unification. This section is committed to discussing the related topics.

### 6.2.1 One-body Material System Case

In this subsection, we, taking the simply connected homogeneous isotropic material body considered in Section 4.3 as a typical example, do some quantitative and qualitative analysis for the variable unification schemes proposed in Section 4.3. The reason to select the homogeneous isotropic material body as example is that the Green's function in homogeneous isotropic medium has closed analytical expression, and this is conductive to doing quantitative discussions.

Based on formulations (4-34), (4-35), and (4-40)~(4-51), it is easy to find out that if basis functions $\{\vec{b}_\xi^{J_s}\}_{\xi=1}^{\Xi^{J_s}}$ and $\{\vec{b}_\xi^{M_s}\}_{\xi=1}^{\Xi^{M_s}}$ are selected as the same ones (i.e., $\vec{b}_\xi^{J_s} = \vec{b}_\xi^{M_s}$ and $\Xi^{J_s} = \Xi^{M_s}$), then we have the following conclusions:

**Conclusion 1.** $\bar{\bar{P}}_{0;\text{PVT}}^{J_s J_s} = \eta_0^2 \bar{\bar{P}}_{0;\text{PVT}}^{M_s M_s}$, so we denote $\bar{\bar{P}}_{0;\text{PVT}}^{J_s J_s} = \eta_0 \bar{\bar{A}}_0$ and $\bar{\bar{P}}_{0;\text{PVT}}^{M_s M_s} = \bar{\bar{A}}_0 / \eta_0$ in the following discussions;

**Conclusion 2.** $\bar{\bar{P}}_{0;\text{PVT}}^{J_s M_s} + \bar{\bar{P}}_{0;\text{SCT}}^{J_s M_s} = -(\bar{\bar{P}}_{0;\text{PVT}}^{M_s J_s} + \bar{\bar{P}}_{0;\text{SCT}}^{M_s J_s})$, so we denote $\bar{\bar{P}}_{0;\text{PVT}}^{M_s J_s} + \bar{\bar{P}}_{0;\text{SCT}}^{M_s J_s} = \bar{\bar{B}}_0$ and $\bar{\bar{P}}_{0;\text{PVT}}^{J_s M_s} + \bar{\bar{P}}_{0;\text{SCT}}^{J_s M_s} = -\bar{\bar{B}}_0$ in the following discussions;

**Conclusion 3.** $\bar{\bar{Z}}_{\text{sim}}^{J_s E J_s} = \bar{\bar{Z}}_{\text{sim}}^{M_s E J_s} = \eta_{\text{sim}}^2 \bar{\bar{Z}}_{\text{sim}}^{J_s H M_s} = \eta_{\text{sim}}^2 \bar{\bar{Z}}_{\text{sim}}^{M_s H M_s}$, so we denote $\bar{\bar{Z}}_{\text{sim}}^{J_s E J_s} = \bar{\bar{Z}}_{\text{sim}}^{M_s E J_s} = \eta_{\text{sim}} \bar{\bar{A}}_{\text{sim}}$ and $\bar{\bar{Z}}_{\text{sim}}^{J_s H M_s} = \bar{\bar{Z}}_{\text{sim}}^{M_s H M_s} = \bar{\bar{A}}_{\text{sim}} / \eta_{\text{sim}}$ in the following discussions, where $\eta_{\text{sim}} = \sqrt{\mu_{\text{sim}} / \varepsilon_{\text{sim}}}$.

**Conclusion 4.** $\bar{\bar{Z}}_{\text{sim}}^{J_s E M_s} = \bar{\bar{Z}}_{\text{sim}}^{M_s E M_s} = -\bar{\bar{Z}}_{\text{sim}}^{J_s H J_s} = -\bar{\bar{Z}}_{\text{sim}}^{M_s H J_s}$, so we denote $\bar{\bar{Z}}_{\text{sim}}^{J_s H J_s} = \bar{\bar{Z}}_{\text{sim}}^{M_s H J_s} = \bar{\bar{B}}_{\text{sim}}$ and $\bar{\bar{Z}}_{\text{sim}}^{J_s E M_s} = \bar{\bar{Z}}_{\text{sim}}^{M_s E M_s} = -\bar{\bar{B}}_{\text{sim}}$ in the following discussions;

**Conclusion 5.** $\overline{\overline{C}}^{J_s J_s} = \overline{\overline{C}}^{M_s J_s} = -\overline{\overline{C}}^{J_s M_s} = -\overline{\overline{C}}^{M_s M_s}$, so we denote $\overline{\overline{C}}^{J_s M_s} = \overline{\overline{C}}^{M_s M_s} = \overline{\overline{C}}$ and $\overline{\overline{C}}^{J_s J_s} = \overline{\overline{C}}^{M_s J_s} = -\overline{\overline{C}}$ in the following discussions.

Then, we have the following relationships:

$$\overbrace{\begin{bmatrix} \overline{\overline{I}}^{J_s} \\ \overline{\overline{T}}^{M_s \leftarrow J_s}_{\text{DESM}} \end{bmatrix}^H \cdot \overline{\overline{P}}^{\text{driving}}_{\text{sim sys}} \cdot \begin{bmatrix} \overline{\overline{I}}^{J_s} \\ \overline{\overline{T}}^{M_s \leftarrow J_s}_{\text{DESM}} \end{bmatrix}}^{\overline{\overline{P}}^{\text{driving}}_{J_s}} = \eta_0 \left[ \overline{\overline{A}}_0 + \eta^{\text{r}}_{\text{sim}} \overline{\overline{U}} - \eta^{\text{r}}_{\text{sim}} \overline{\overline{V}} - \left(\eta^{\text{r}}_{\text{sim}}\right)^2 \overline{\overline{W}} \right] \tag{6-1a}$$

$$\underbrace{\begin{bmatrix} \overline{\overline{T}}^{J_s \leftarrow M_s}_{\text{DESJ}} \\ \overline{\overline{I}}^{M_s} \end{bmatrix}^H \cdot \overline{\overline{P}}^{\text{driving}}_{\text{sim sys}} \cdot \begin{bmatrix} \overline{\overline{T}}^{J_s \leftarrow M_s}_{\text{DESJ}} \\ \overline{\overline{I}}^{M_s} \end{bmatrix}}_{\overline{\overline{P}}^{\text{driving}}_{M_s}} = -\frac{1}{\eta_0} \left[ \overline{\overline{A}}_0 + \frac{1}{\eta^{\text{r}}_{\text{sim}}} \overline{\overline{U}} - \frac{1}{\eta^{\text{r}}_{\text{sim}}} \overline{\overline{V}} + \left( \frac{1}{\eta^{\text{r}}_{\text{sim}}} \right)^2 \overline{\overline{W}} \right] \tag{6-1b}$$

and

$$\overbrace{\begin{bmatrix} \overline{\overline{I}}^{J_s} \\ \overline{\overline{T}}^{M_s \leftarrow J_s}_{\text{DESJ}} \end{bmatrix}^H \cdot \overline{\overline{P}}^{\text{driving}}_{\text{sim sys}} \cdot \begin{bmatrix} \overline{\overline{I}}^{J_s} \\ \overline{\overline{T}}^{M_s \leftarrow J_s}_{\text{DESJ}} \end{bmatrix}}^{\overline{\overline{P}}^{\text{driving}}_{J_s}} = \eta_0 \left[ \overline{\overline{A}}_0 + \eta^{\text{r}}_{\text{sim}} \overline{\overline{X}} - \eta^{\text{r}}_{\text{sim}} \overline{\overline{Y}} - \left(\eta^{\text{r}}_{\text{sim}}\right)^2 \overline{\overline{Z}} \right] \tag{6-2a}$$

$$\underbrace{\begin{bmatrix} \overline{\overline{T}}^{J_s \leftarrow M_s}_{\text{DESM}} \\ \overline{\overline{I}}^{M_s} \end{bmatrix}^H \cdot \overline{\overline{P}}^{\text{driving}}_{\text{sim sys}} \cdot \begin{bmatrix} \overline{\overline{T}}^{J_s \leftarrow M_s}_{\text{DESM}} \\ \overline{\overline{I}}^{M_s} \end{bmatrix}}_{\overline{\overline{P}}^{\text{driving}}_{M_s}} = -\frac{1}{\eta_0} \left[ \overline{\overline{A}}_0 + \frac{1}{\eta^{\text{r}}_{\text{sim}}} \overline{\overline{X}} - \frac{1}{\eta^{\text{r}}_{\text{sim}}} \overline{\overline{Y}} + \left( \frac{1}{\eta^{\text{r}}_{\text{sim}}} \right)^2 \overline{\overline{Z}} \right] \tag{6-2b}$$

where $\eta^{\text{r}}_{\text{sim}} = \sqrt{\mu^{\text{r}}_{\text{sim}} / \varepsilon^{\text{r}}_{\text{sim}}}$ is the relative wave impedance corresponding to the homogeneous isotropic matter, and

$$\overline{\overline{U}} = \overline{\overline{B}}_0 \cdot \left( \overline{\overline{C}} + \overline{\overline{B}}_{\text{sim}} \right)^{-1} \cdot \overline{\overline{A}}_{\text{sim}} \tag{6-3a}$$

$$\overline{\overline{V}} = \overline{\overline{A}}^H_{\text{sim}} \cdot \left( \overline{\overline{C}} + \overline{\overline{B}}_{\text{sim}} \right)^{-H} \cdot \overline{\overline{B}}_0 \tag{6-3b}$$

$$\overline{\overline{W}} = \overline{\overline{A}}^H_{\text{sim}} \cdot \left( \overline{\overline{C}} + \overline{\overline{B}}_{\text{sim}} \right)^{-H} \cdot \overline{\overline{A}}_0 \cdot \left( \overline{\overline{C}} + \overline{\overline{B}}_{\text{sim}} \right)^{-1} \cdot \overline{\overline{A}}_{\text{sim}} \tag{6-3c}$$

and

$$\overline{\overline{X}} = \overline{\overline{B}}_0 \cdot \overline{\overline{A}}^{-1}_{\text{sim}} \cdot \left( \overline{\overline{C}} + \overline{\overline{B}}_{\text{sim}} \right) \tag{6-4a}$$

$$\overline{\overline{Y}} = \left( \overline{\overline{C}} + \overline{\overline{B}}_{\text{sim}} \right)^H \cdot \overline{\overline{A}}^{-H}_{\text{sim}} \cdot \overline{\overline{B}}_0 \tag{6-4b}$$

$$\overline{\overline{Z}} = \left( \overline{\overline{C}} + \overline{\overline{B}}_{\text{sim}} \right)^H \cdot \overline{\overline{A}}^{-H}_{\text{sim}} \cdot \overline{\overline{A}}_0 \cdot \overline{\overline{A}}^{-1}_{\text{sim}} \cdot \left( \overline{\overline{C}} + \overline{\overline{B}}_{\text{sim}} \right) \tag{6-4c}$$

where symbol "$\overline{\overline{M}}^{-H}$" represents $(\overline{\overline{M}}^{-1})^H$. Because $(\overline{\overline{M}}^{-1})^H = (\overline{\overline{M}}^H)^{-1}$ [116], then symbol $\overline{\overline{M}}^{-H}$ also represents $(\overline{\overline{M}}^H)^{-1}$.

In what follows, we will provide some conclusions by employing the above relationships.

**Conclusion I.** Based on relationships (6-1a) and (6-1b), it is easy to find out that: in the aspect of the numerical performance of constructing DP-CMs, the $\overline{\overline{T}}{}_{\text{DESM}}^{M_{\text{s}} \leftarrow J_{\text{s}}}$-based power matrix $\overline{\overline{P}}{}_{J_{\text{s}}}^{\text{driving}}$ corresponding to material parameters $\{\mu_{\text{sim}}^{\text{r}} = a, \varepsilon_{\text{sim}}^{\text{r}} = b; \eta_{\text{sim}}^{\text{r}} = \sqrt{a/b}\}$ is the same as the $\overline{\overline{T}}{}_{\text{DESJ}}^{J_{\text{s}} \leftarrow M_{\text{s}}}$-based power matrix $\overline{\overline{P}}{}_{M_{\text{s}}}^{\text{driving}}$ corresponding to material parameters $\{\mu_{\text{sim}}^{\text{r}} = b, \varepsilon_{\text{sim}}^{\text{r}} = a; \eta_{\text{sim}}^{\text{cr}} = \sqrt{b/a}\}$. According to our observations for the calculated results of some typical material bodies, we obtain an experiential conclusion: when $\eta_{\text{sim}}^{\text{r}} < 1$, the $\overline{\overline{P}}{}_{J_{\text{s}}}^{\text{driving}}$ based on relationship (6-1a) is more advantageous than the $\overline{\overline{P}}{}_{M_{\text{s}}}^{\text{driving}}$ based on relationship (6-1b) in the aspect of the numerical performance of constructing DP-CMs; when $1/\eta_{\text{sim}}^{\text{r}} < 1$, the $\overline{\overline{P}}{}_{M_{\text{s}}}^{\text{driving}}$ based on relationship (6-1b) is more advantageous than the $\overline{\overline{P}}{}_{J_{\text{s}}}^{\text{driving}}$ based on relationship (6-1a) in the aspect of the numerical performance of constructing DP-CMs[44].

**Conclusion II.** Based on relationships (6-2a) and (6-2b), it is easy to find out that: in the aspect of the numerical performance of constructing DP-CMs, the $\overline{\overline{T}}{}_{\text{DESJ}}^{M_{\text{s}} \leftarrow J_{\text{s}}}$-based power matrix $\overline{\overline{P}}{}_{J_{\text{s}}}^{\text{driving}}$ corresponding to material parameters $\{\mu_{\text{sim}}^{\text{r}} = a, \varepsilon_{\text{sim}}^{\text{cr}} = b; \eta_{\text{sim}}^{\text{r}} = \sqrt{a/b}\}$ is the same as the $\overline{\overline{T}}{}_{\text{DESM}}^{J_{\text{s}} \leftarrow M_{\text{s}}}$-based power matrix $\overline{\overline{P}}{}_{M_{\text{s}}}^{\text{driving}}$ corresponding to material parameters $\{\mu_{\text{sim}}^{\text{r}} = b, \varepsilon_{\text{sim}}^{\text{r}} = a; \eta_{\text{sim}}^{\text{r}} = \sqrt{b/a}\}$. According to our observations for the calculated results of some typical material bodies, we obtain an experiential conclusion: when $\eta_{\text{sim}}^{\text{r}} > 1$, the $\overline{\overline{P}}{}_{J_{\text{s}}}^{\text{driving}}$ based on relationship (6-2a) is more advantageous than the $\overline{\overline{P}}{}_{M_{\text{s}}}^{\text{driving}}$ based on relationship (6-2b) in the aspect of the numerical performance of constructing DP-CMs; when $1/\eta_{\text{sim}}^{\text{r}} > 1$, the $\overline{\overline{P}}{}_{M_{\text{s}}}^{\text{driving}}$ based on relationship (6-2b) is more advantageous than the $\overline{\overline{P}}{}_{J_{\text{s}}}^{\text{driving}}$ based on relationship (6-2a) in the aspect of the numerical performance of constructing DP-CMs[44].

In what follows, we provide some typical examples to verify the above-mentioned experiential conclusions. The topological structure and surface triangular meshes of the material body considered here are illustrated in Figure 6-1 (i.e. Figure 4-14). For the convenience of the following discussions, we simply denote the material body as $V_{\text{sim sys}}$.

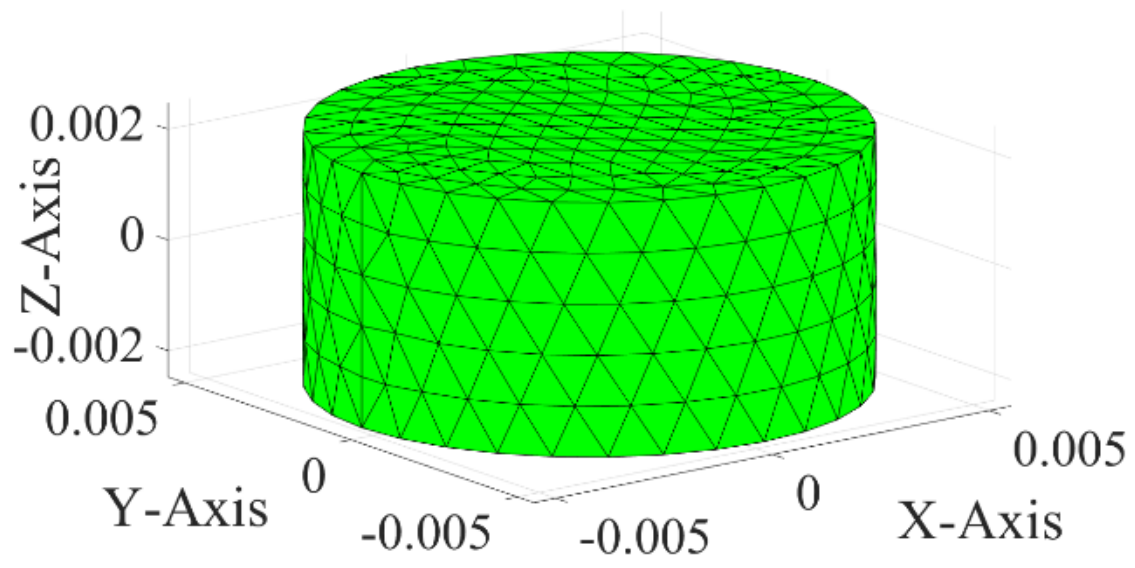

Figure 6-1 The topological structure and surface triangular meshes of a material cylinder whose radius and height are 5.25mm and 4.60mm respectively

When the material parameters of $V_{\text{sim sys}}$ are $\{\mu_{\text{sim}}^{\text{r}}=1,\varepsilon_{\text{sim}}^{\text{r}}=36,\sigma_{\text{sim}}=0\}$, the results calculated from the DPO (4-32) with variable unification schemes (4-51) and (4-42) are illustrated in Figure 6-2.

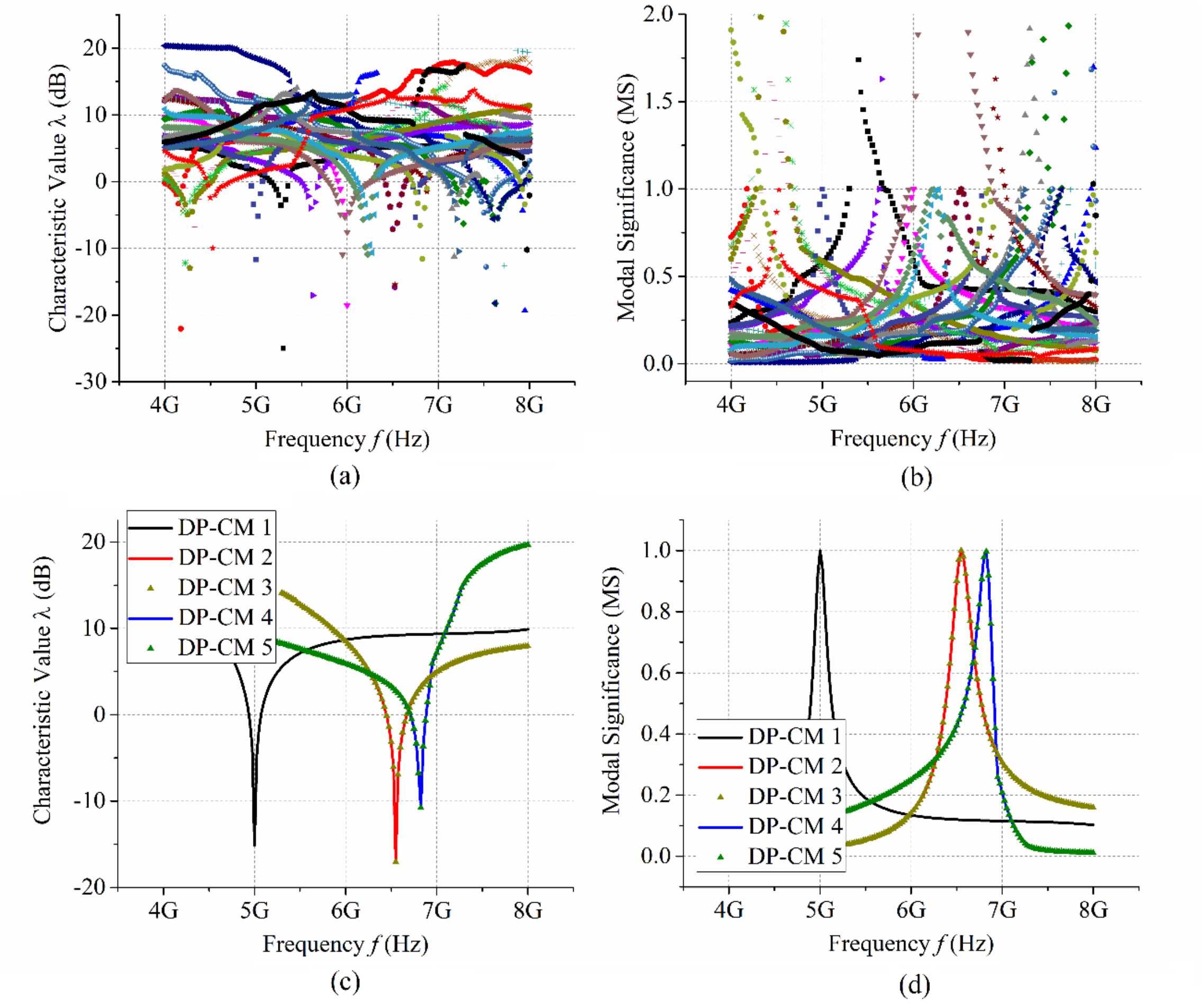

Figure 6-2 Some characteristic quantity curves of the material cylinder $V_{\text{sim sys}}$ with material parameters $\{\mu_{\text{sim}}^{\text{r}}=1,\varepsilon_{\text{sim}}^{\text{r}}=36,\sigma_{\text{sim}}=0\}$. (a) the (4-32)&(4-51)-based characteristic value dB curves; (b) the (4-32)&(4-51)-based MS curves; (c) the (4-32)&(4-42)-based characteristic value dB curves; (d) the (4-32)&(4-42)-based MS curves

When the material parameters of $V_{\text{sim sys}}$ are $\{\mu_{\text{sim}}^{\text{r}}=2,\varepsilon_{\text{sim}}^{\text{r}}=18,\sigma_{\text{sim}}=0\}$, the results calculated from the DPO (4-32) with variable unification schemes (4-51) and (4-42) are illustrated in Figure 6-3. When the material parameters of $V_{\text{sim sys}}$ are $\{\mu_{\text{sim}}^{\text{r}}=3,\varepsilon_{\text{sim}}^{\text{r}}=12,\sigma_{\text{sim}}=0\}$, the results calculated from the DPO (4-32) with variable unification schemes (4-51) and (4-42) are illustrated in Figure 6-4. When the material parameters of $V_{\text{sim sys}}$ are $\{\mu_{\text{sim}}^{\text{r}}=4,\varepsilon_{\text{sim}}^{\text{r}}=9,\sigma_{\text{sim}}=0\}$, the results calculated from the DPO (4-32) with variable unification schemes (4-51) and (4-42) are illustrated in Figure 6-5. When the material parameters of $V_{\text{sim sys}}$ are $\{\mu_{\text{sim}}^{\text{r}}=5,\varepsilon_{\text{sim}}^{\text{r}}=7.2,\sigma_{\text{sim}}=0\}$, the results calculated from the DPO (4-32) with variable unification schemes (4-51) and (4-42) are illustrated in Figure 6-6. When the material parameters of $V_{\text{sim sys}}$ are $\{\mu_{\text{sim}}^{\text{r}}=6,\varepsilon_{\text{sim}}^{\text{r}}=6,\sigma_{\text{sim}}=0\}$, the results calculated from the DPO (4-32) with variable unification schemes (4-51) and (4-42) are illustrated in Figure 6-7. Obviously, the above calculated results meet well with experiential Conclusions I and II. Thus, the Chapters 4 and 5 of this dissertation had properly selected variable unifications based on the conclusions.

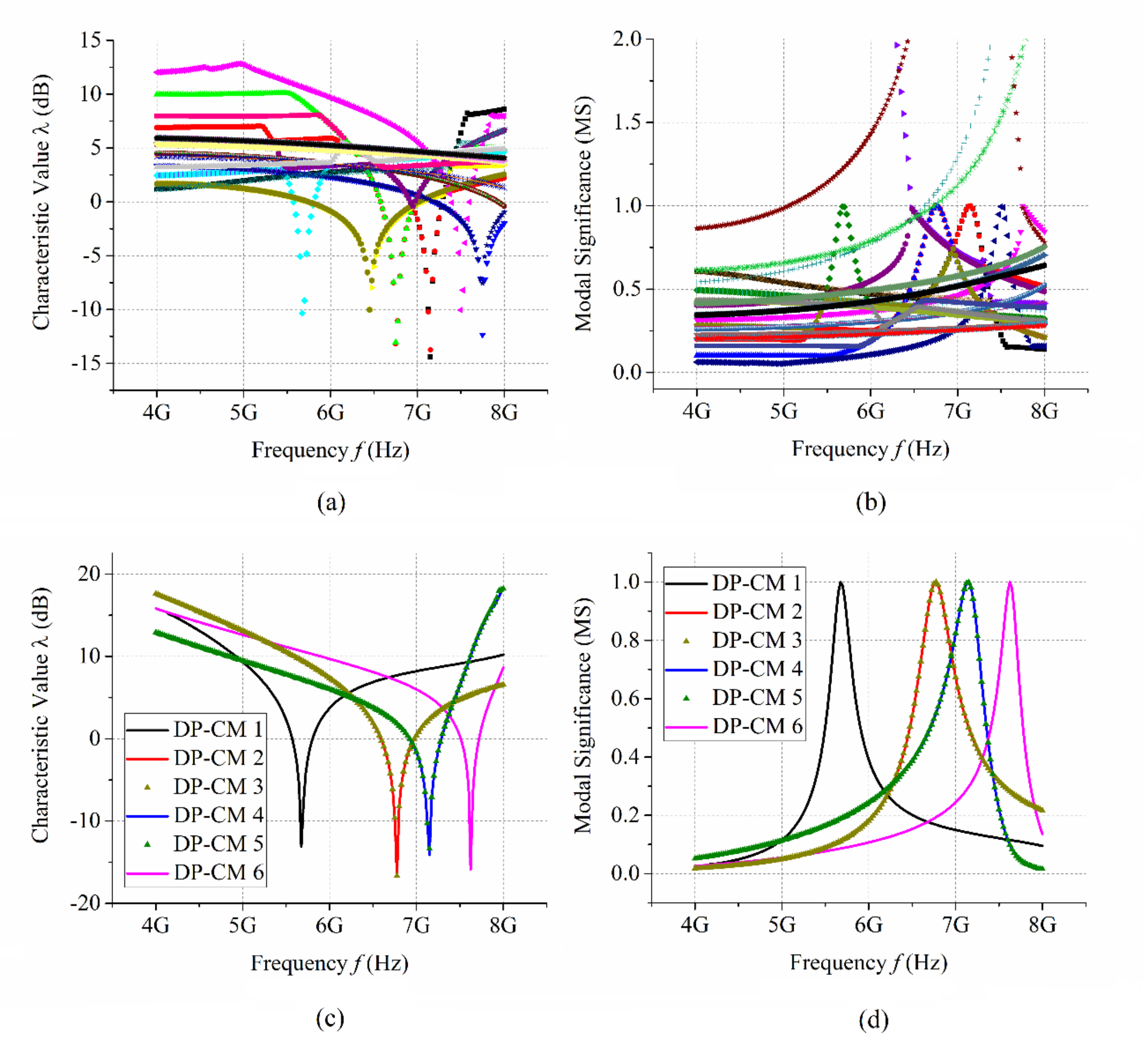

Figure 6-3 Some characteristic quantity curves of the material cylinder $V_{\text{sim sys}}$ with material parameters $\{\mu_{\text{sim}}^{\text{r}}=2,\varepsilon_{\text{sim}}^{\text{r}}=18,\sigma_{\text{sim}}=0\}$. (a) the (4-32)&(4-51)-based characteristic value dB curves; (b) the (4-32)&(4-51)-based MS curves; (c) the (4-32)&(4-42)-based characteristic value dB curves; (d) the (4-32)&(4-42)-based MS curves

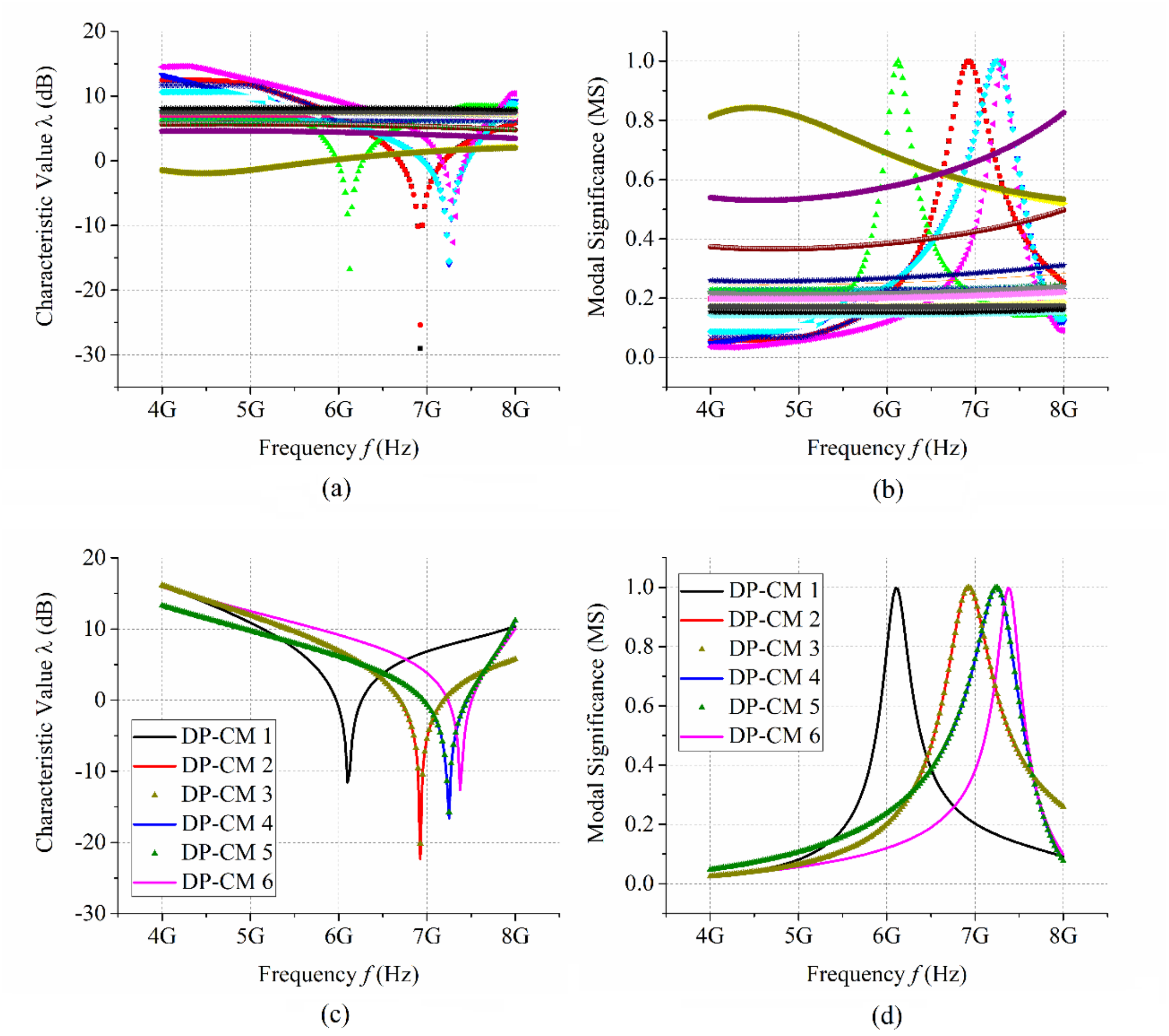

Figure 6-4 Some characteristic quantity curves of the material cylinder $V_{\text{sim sys}}$ with material parameters $\{\mu^{\text{r}}_{\text{sim}} = 3, \varepsilon^{\text{r}}_{\text{sim}} = 12, \sigma_{\text{sim}} = 0\}$. (a) the (4-32)&(4-51)-based characteristic value dB curves; (b) the (4-32)&(4-51)-based MS curves; (c) the (4-32)&(4-42)-based characteristic value dB curves; (d) the (4-32)&(4-42)-based MS curves

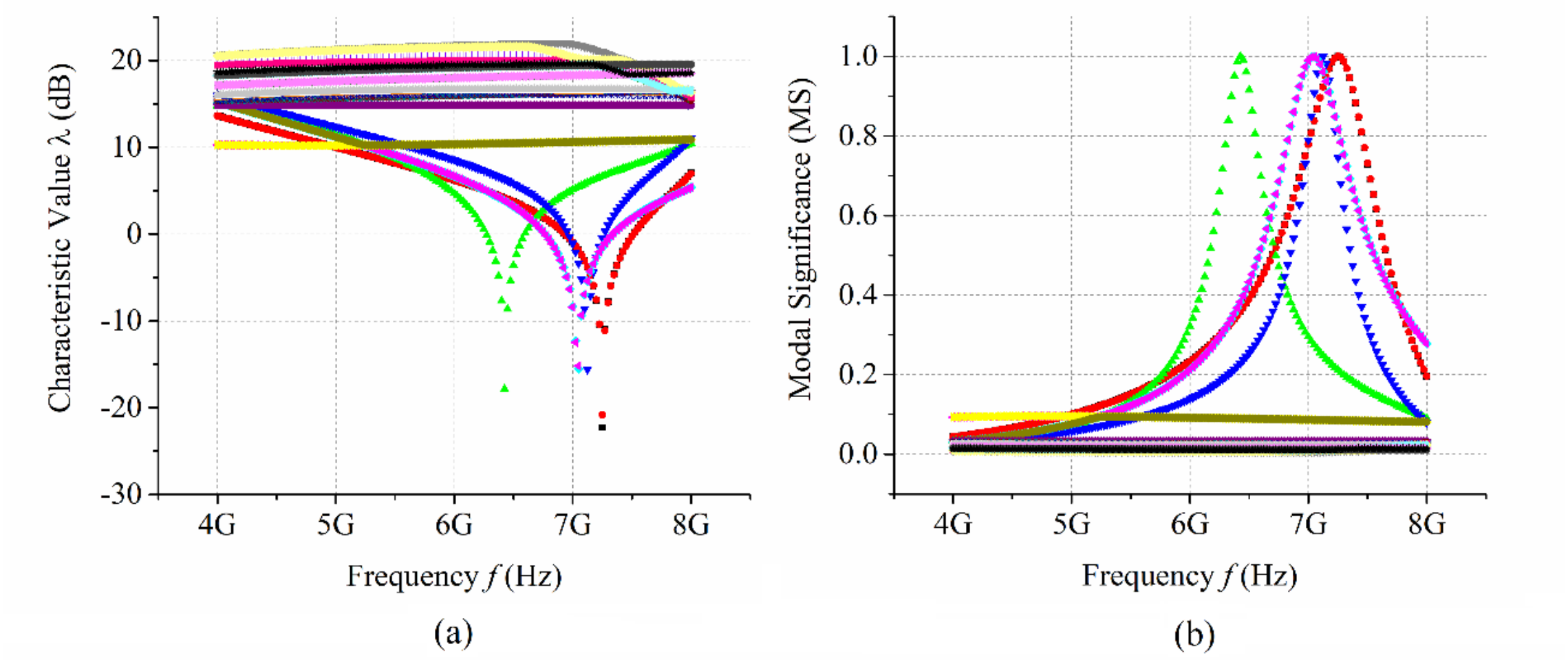

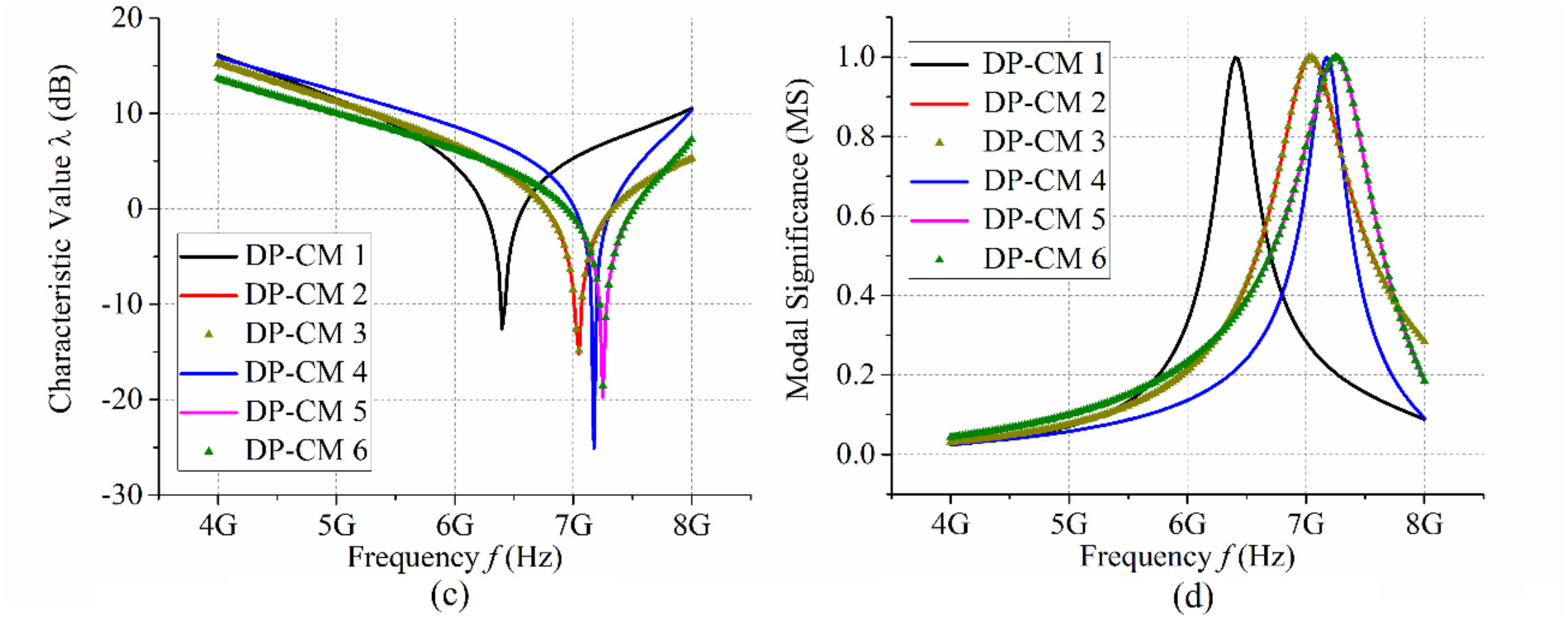

Figure 6-5 Some characteristic quantity curves of the material cylinder $V_{\text{sim sys}}$ with material parameters $\{\mu_{\text{sim}}^{\text{r}} = 4, \varepsilon_{\text{sim}}^{\text{r}} = 9, \sigma_{\text{sim}} = 0\}$. (a) the (4-32)&(4-51)-based characteristic value dB curves; (b) the (4-32)&(4-51)-based MS curves; (c) the (4-32)&(4-42)-based characteristic value dB curves; (d) the (4-32)&(4-42)-based MS curves

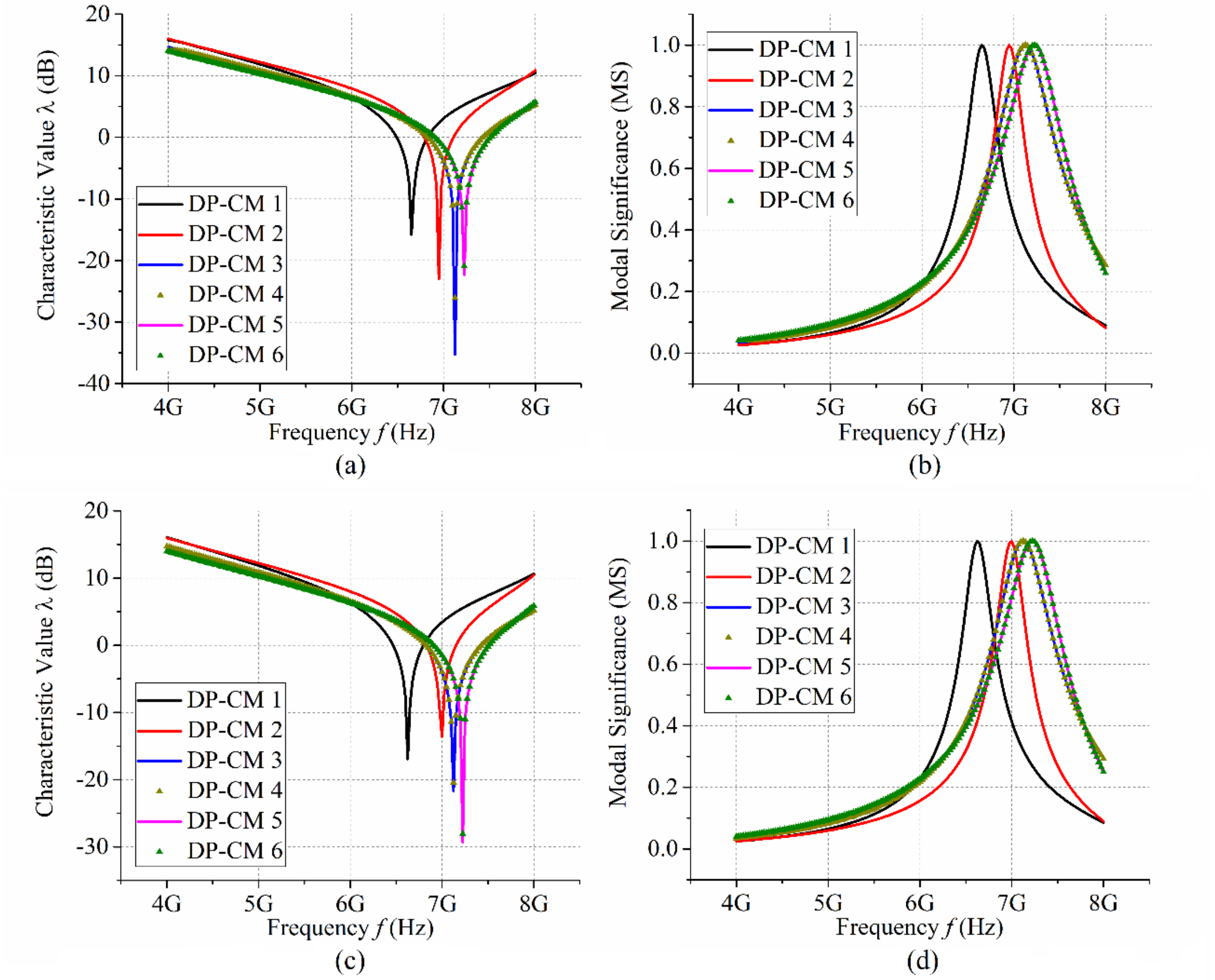

Figure 6-6 Some characteristic quantity curves of the material cylinder $V_{\text{sim sys}}$ with material parameters $\{\mu_{\text{sim}}^{\text{r}} = 5, \varepsilon_{\text{sim}}^{\text{r}} = 7.2, \sigma_{\text{sim}} = 0\}$. (a) the (4-32)&(4-51)-based characteristic value dB curves; (b) the (4-32)&(4-51)-based MS curves; (c) the (4-32)&(4-42)-based characteristic value dB curves; (d) the (4-32)&(4-42)-based MS curves

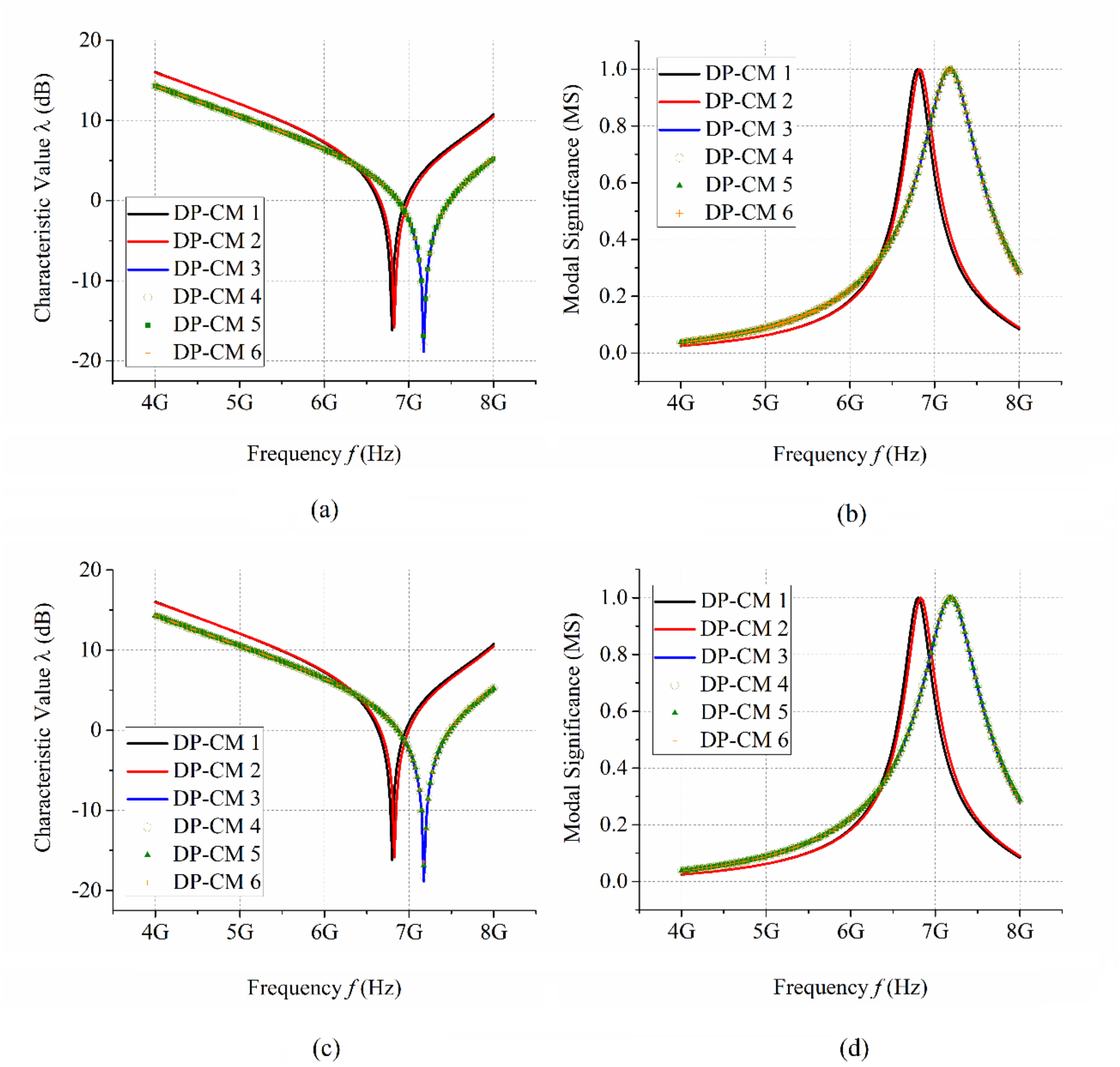

Figure 6-7 Some characteristic quantity curves of the material cylinder $V_{\text{sim sys}}$ with material parameters $\{\mu_{\text{sim}}^{\text{r}} = 6, \varepsilon_{\text{sim}}^{\text{r}} = 6, \sigma_{\text{sim}} = 0\}$. (a) the (4-32)&(4-51)-based characteristic value dB curves; (b) the (4-32)&(4-51)-based MS curves; (c) the (4-32)&(4-42)-based characteristic value dB curves; (d) the (4-32)&(4-42)-based MS curves

## 6.2.2 Two-body Material System Case

For the material system constituted by a single material body, the selection for BVs is relatively easy. For the two-body material system constituted by two material bodies considered in Sections 4.4~4.7, the scheme to select BVs is an important topic much worthy of discussion, and, so far as we know, there has not been systematical studies for the topic. In this section, taking the two-body system constituted by two simply connected material bodies considered in Sections 4.4~4.7 as an example, we give some further discussions related to the topic, to emphasize that: before constructing the DP-CMs of multi-body material systems, it is necessary to unify variables, and the variable unification must be done thoroughly. On the question “How to judge whether or not the variable unification has been done thoroughly?”, we will systematically answer it in Subsection 6.2.4 after finishing the discussions for the variable unification of metal-material composite systems (Subsection 6.2.3).

**1) Not Doing Any Variable Unification**

In Section 4.3, during the process of constructing the CMs of a single simply connected material body based on SIE operator[34], we found out that: if we don’t do variable unification preprocessing for the SIE operator, then the obtained CM set contains both physical modes (which we desire) and spurious modes (which we don’t desire); after doing proper variable unifications, the obtained CM set contains the physical modes only, and doesn’t contain any spurious mode.

In addition, we also emphasized that: before constructing DP-CMs by employing new DPO (4-32), it is necessary to unify variables, or the obtained DP-CM set doesn’t contain any desired physical mode. At the same time, we also provide the reason leading to the above phenomenon: in new operator (4-32), only the topological informations of the material body is involved, but the material parameter informations of the material body are not involved.

Then, we can obtain the following conclusions: during the process of constructing the DP-CMs of two-body material systems by employing DPOs (4-113) and (4-117), we must do some proper variable unifications, or we cannot obtain a desirable DP-CM set. Now the question is that: whether or not the modal set derived from the DPO without variable unification contains both the physical modes we desire and the spurious modes we don’t desire. Before answering this question, we firstly observe some calculated results.

Now, we consider the two-body material system shown in Figure 6-8 (i.e. the one shown in Figure 4-30). The two-body material system is constituted by two stacked material cylinders whose geometrical dimensions are the same as each other, and their radiuses and heights are both 5.25mm and 2.30mm respectively. The relative permeability, relative permittivity, and conductivity of the upper cylinder are 3, 12, and 0 respectively, and the relative permeability, relative permittivity, and conductivity of the lower cylinder are 6, 6, and 0 respectively.

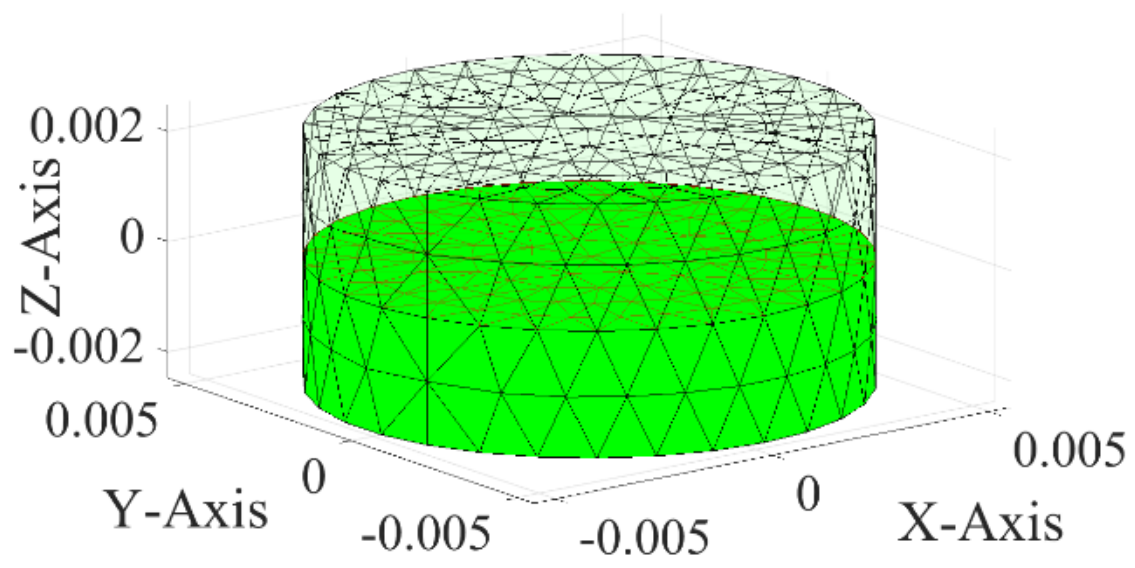

Figure 6-8 The topological structure and surface triangular meshes of the two-body material system constituted by two stacked material cylinders whose radiuses and heights are both 5.25mm and 2.30mm respectively. The upper cylinder is with material parameters $\{\mu_{\rm sim}^{\rm r}=3, \varepsilon_{\rm sim}^{\rm r}=12, \sigma_{\rm sim}=0\}$, and the lower cylinder is with material parameters $\{\mu_{\rm sim}^{\rm r}=6, \varepsilon_{\rm sim}^{\rm r}=6, \sigma_{\rm sim}=0\}$

Based on the surface formulation (4-66) given in the Section 4.4 of this dissertation, and not doing any variable unification (i.e. ignoring the dependent relationships among all of the EM currents contained in the surface formulation), the characteristic value curves and MS curves corresponding to the first 500 modes are illustrated in Figures 6-9(a) and 6-9(b).

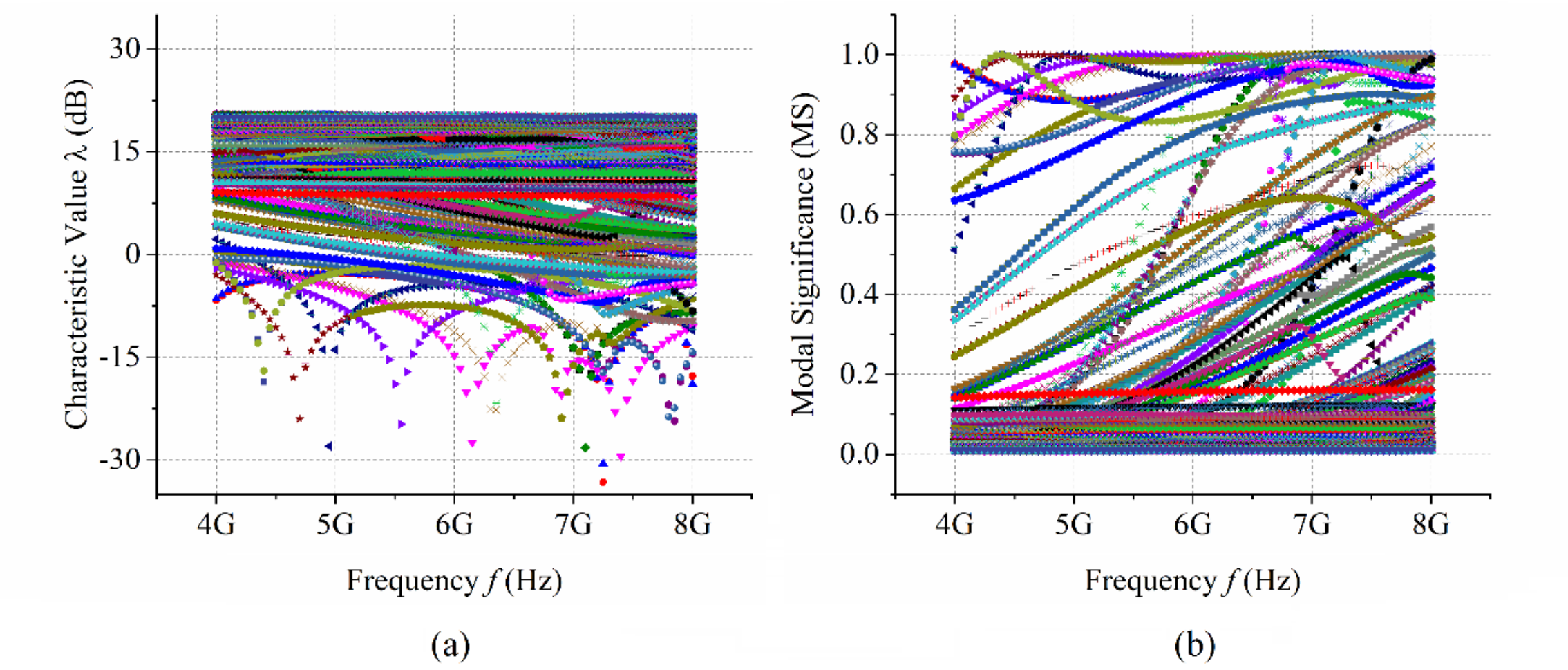

Figure 6-9 The characteristic quantity curves corresponding to the first 500 DP-CMs (derived from the surface formulation (4-66) without variable unification) of the two-body material system shown in Figure 6-8. (a) characteristic value dB curves; (b) MS curves

Comparing Figure 6-9 and Figure 4-33, it is easy to find out that: for the relatively complicated two-body material systems, the DPO (4-66) without variable unification preprocessing has no ability to provide enough physical modes which are desirable.

In summary, for the relatively complicated two-body material systems, before constructing CMs by employing DPOs (4-66), (4-113), and (4-117) we must properly unify variables, or we will not obtain enough physical modes which are desirable. So, what kind of variable unification scheme is “proper”? In the following parts of this section, we will provide some variable unification schemes having different unifiying degrees, and obtain some qualitative conclusions by observing the results outputted from the variable unification schemes.

**2) Only Doing Variable Unification in EM Current Space**

In this subsection, we, based on relationship (4-71), eliminate the $\vec{J}_{s21}^{ES}$ and $\vec{M}_{s21}^{ES}$ in all of the 8 sub-currents $\vec{J}_{s10}^{ES}$, $\vec{J}_{s12}^{ES}$, $\vec{J}_{s21}^{ES}$, $\vec{J}_{s20}^{ES}$, $\vec{M}_{s10}^{ES}$, $\vec{M}_{s12}^{ES}$, $\vec{M}_{s21}^{ES}$, and $\vec{M}_{s20}^{ES}$ contained in DPO (4-66), and then, based on the simplified DPO with 6 variables, construct DP-CMs, and at the same time observe the calculated results.

In Subsection 4.4.1, based on the tangential continuity conditions of the EM fields on material interfaces we obtain relationship (4-71), so there exists the following transformation:

$$\begin{bmatrix} \bar{a}^{J_{s10}} \\ \bar{a}^{J_{s12}} \\ \bar{a}^{J_{s21}} \\ \bar{a}^{J_{s20}} \\ \bar{a}^{M_{s10}} \\ \bar{a}^{M_{s12}} \\ \bar{a}^{M_{s21}} \\ \bar{a}^{M_{s20}} \end{bmatrix} = \underbrace{\begin{bmatrix} \bar{\bar{I}}^{J_{s10}} & 0 & 0 & 0 & 0 & 0 \\ 0 & \bar{\bar{I}}^{J_{s12}} & 0 & 0 & 0 & 0 \\ 0 & -\bar{\bar{I}}^{J_{s12}} & 0 & 0 & 0 & 0 \\ 0 & 0 & \bar{\bar{I}}^{J_{s20}} & 0 & 0 & 0 \\ 0 & 0 & 0 & \bar{\bar{I}}^{M_{s10}} & 0 & 0 \\ 0 & 0 & 0 & 0 & \bar{\bar{I}}^{M_{s12}} & 0 \\ 0 & 0 & 0 & 0 & -\bar{\bar{I}}^{M_{s12}} & 0 \\ 0 & 0 & 0 & 0 & 0 & \bar{\bar{I}}^{M_{s20}} \end{bmatrix}}_{\bar{\bar{T}}_1} \cdot \begin{bmatrix} \bar{a}^{\vec{J}_{10}^{ES}} \\ \bar{a}^{\vec{J}_{12}^{ES}} \\ \bar{a}^{\vec{J}_{20}^{ES}} \\ \bar{a}^{\vec{M}_{10}^{ES}} \\ \bar{a}^{\vec{M}_{12}^{ES}} \\ \bar{a}^{\vec{M}_{20}^{ES}} \end{bmatrix} \tag{6-5}$$

Inserting the above transformation into formulation (4-80), we can obtain the matrix form of the DPO $P_{\text{ss sys}}^{\text{driving}}$ which contains BVs $\{\bar{a}^{\vec{M}_{10}^{ES}}, \bar{a}^{\vec{M}_{20}^{ES}}\}$ and dependent variables $\{\bar{a}^{\vec{J}_{10}^{ES}}, \bar{a}^{\vec{J}_{12}^{ES}}, \bar{a}^{\vec{J}_{20}^{ES}}, \bar{a}^{\vec{M}_{12}^{ES}}\}$.

$$P_{\text{ss sys}}^{\text{driving}} = \begin{bmatrix} \bar{a}^{\vec{J}_{10}^{ES}} \\ \bar{a}^{\vec{J}_{12}^{ES}} \\ \bar{a}^{\vec{J}_{20}^{ES}} \\ \bar{a}^{\vec{M}_{10}^{ES}} \\ \bar{a}^{\vec{M}_{12}^{ES}} \\ \bar{a}^{\vec{M}_{20}^{ES}} \end{bmatrix}^H \cdot \left(\bar{\bar{T}}_1\right)^H \cdot \bar{\bar{P}}_{\text{ss sys;1}}^{\text{driving}} \cdot \bar{\bar{T}}_1 \cdot \begin{bmatrix} \bar{a}^{\vec{J}_{10}^{ES}} \\ \bar{a}^{\vec{J}_{12}^{ES}} \\ \bar{a}^{\vec{J}_{20}^{ES}} \\ \bar{a}^{\vec{M}_{10}^{ES}} \\ \bar{a}^{\vec{M}_{12}^{ES}} \\ \bar{a}^{\vec{M}_{20}^{ES}} \end{bmatrix} \tag{6-6}$$

Based on the matrix operator $(\overline{\overline{T}}_1)^H \cdot \overline{\overline{P}}_{\text{ss sys;1}}^{\text{driving}} \cdot \overline{\overline{T}}_1$ in formulation (6-6), we calculate the DP-CMs, and the characteristic value curves and MS curves corresponding to the first 500 modes are illustrated in Figure 6-10.

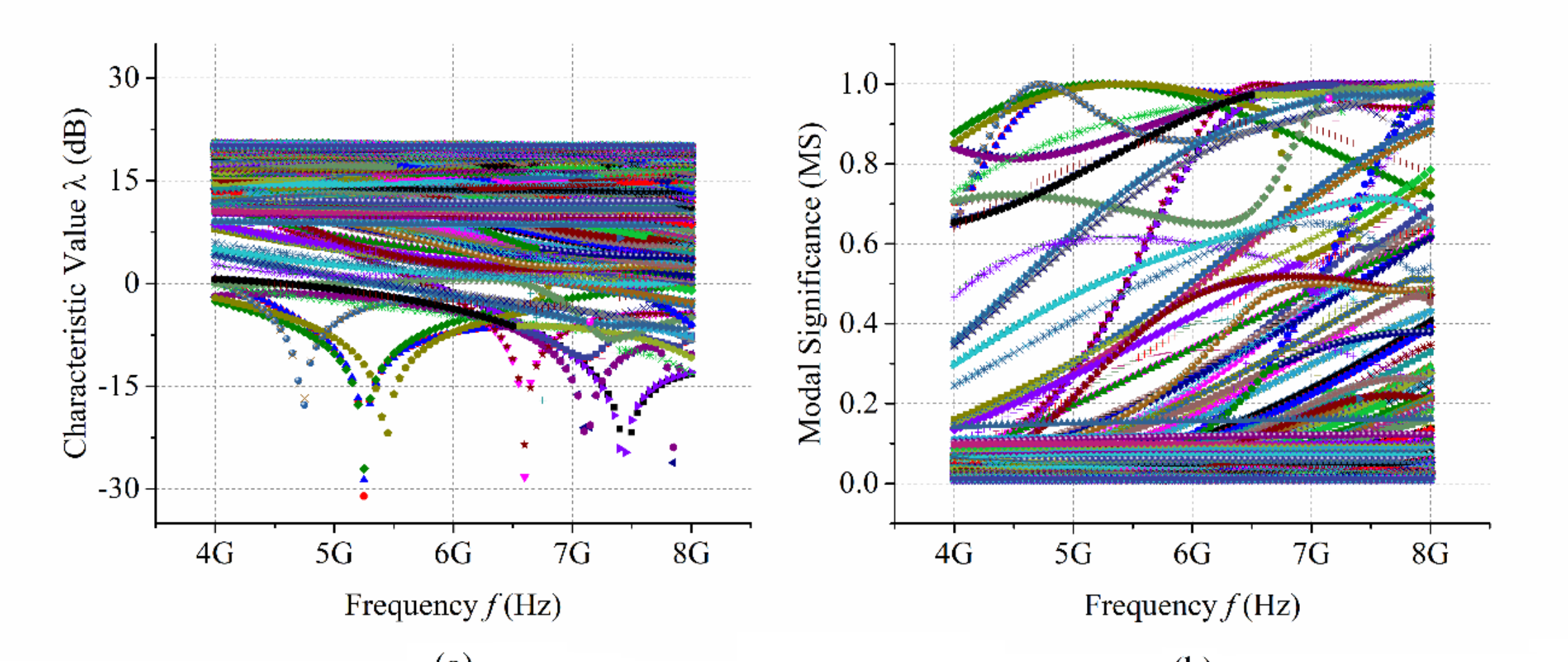

(a) (b)

Figure 6-10 The characteristic quantity curves corresponding to the first 500 DP-CMs (derived from the surface formulation (4-66) with the variable unification in EM current space) of the two-body material system shown in Figure 6-8. (a) characteristic value dB curves; (b) MS curves

Obviously, the incomplete variable unification leads to that we cannot obtain enough physical modes which are desirable.

**3) Doing Variable Unification in EM Current Space and Doing an Incomplete Variable Unification in Expansion Vector Space**

Now, we eliminate the $\vec{J}_{\text{s}10}^{\text{ES}}$, $\vec{J}_{\text{s}12}^{\text{ES}}$, $\vec{J}_{\text{s}21}^{\text{ES}}$, $\vec{J}_{\text{s}20}^{\text{ES}}$, and $\vec{M}_{\text{s}21}^{\text{ES}}$ in all of the 8 sub-currents $\vec{J}_{\text{s}10}^{\text{ES}}$, $\vec{J}_{\text{s}12}^{\text{ES}}$, $\vec{J}_{\text{s}21}^{\text{ES}}$, $\vec{J}_{\text{s}20}^{\text{ES}}$, $\vec{M}_{\text{s}10}^{\text{ES}}$, $\vec{M}_{\text{s}12}^{\text{ES}}$, $\vec{M}_{\text{s}21}^{\text{ES}}$, and $\vec{M}_{\text{s}20}^{\text{ES}}$ contained in DPO (4-66), and then, based on the simplified DPO with 3 variables ($\vec{M}_{\text{s}10}^{\text{ES}}$, $\vec{M}_{\text{s}12}^{\text{ES}}$, and $\vec{M}_{\text{s}20}^{\text{ES}}$), construct DP-CMs, and finally give some necessary discussions to the calculated results.

Based on the GFHF and the definitions for equivalent surface magnetic currents given in Appendix C, we have the following integral equations:

$$\left[\mathcal{E}_{\text{sim}}^{1}\left(\vec{J}_{\text{s}1}^{\text{ES}},\vec{M}_{\text{s}1}^{\text{ES}}\right)\right]_{\vec{r}_{\text{sim}}^{1}\to\vec{r}}^{\tan} = \hat{n}_{\text{s}1}^{-}\left(\vec{r}\right)\times\vec{M}_{\text{s}1}^{\text{ES}}\left(\vec{r}\right) \quad , \quad \vec{r}\in\partial V_{\text{sim}}^{1} \tag{6-7}$$

$$\left[\mathcal{E}_{\text{sim}}^{2}\left(\vec{J}_{\text{s}2}^{\text{ES}},\vec{M}_{\text{s}2}^{\text{ES}}\right)\right]_{\vec{r}_{\text{sim}}^{2}\to\vec{r}}^{\tan} = \hat{n}_{\text{s}2}^{-}\left(\vec{r}\right)\times\vec{M}_{\text{s}2}^{\text{ES}}\left(\vec{r}\right) \quad , \quad \vec{r}\in\partial V_{\text{sim}}^{2} \tag{6-8}$$

Inserting expansion formulations (4-76) and (4-77) into above integral equations (6-7) and (6-8), and testing integral equations (6-7) and (6-8) with basis functions $\{\vec{b}_{\xi}^{J_{\text{s}1}}\}_{\xi=1}^{\Xi^{J_{\text{s}1}}}$

and $\{\vec{b}_\xi^{J_{s2}}\}_{\xi=1}^{\Xi^{J_{s2}}}$ respectively, the integral equations will be discretized into the following matrix equations:

$$\bar{\bar{Z}}^{J_{s1}EJ_{s1}} \cdot \bar{a}^{J_{s1}} + \bar{\bar{Z}}^{J_{s1}EM_{s1}} \cdot \bar{a}^{M_{s1}} = \bar{\bar{C}}^{J_{s1}M_{s1}} \cdot \bar{a}^{M_{s1}} \tag{6-9}$$

$$\bar{\bar{Z}}^{J_{s2}EJ_{s2}} \cdot \bar{a}^{J_{s2}} + \bar{\bar{Z}}^{J_{s2}EM_{s2}} \cdot \bar{a}^{M_{s2}} = \bar{\bar{C}}^{J_{s2}M_{s2}} \cdot \bar{a}^{M_{s2}} \tag{6-10}$$

The elements of matrix equation (6-9) are calculated as follows:

$$z_{\xi\zeta}^{J_{s1}EJ_{s1}} = \left\langle \vec{b}_\xi^{J_{s1}}, \mathcal{E}_{\text{sim}}^1\left(\vec{b}_\zeta^{J_{s1}}, 0\right)\right\rangle_{\partial V_{\text{sim}}^{1-}} \tag{6-11a}$$

$$z_{\xi\zeta}^{J_{s1}EM_{s1}} = \left\langle \vec{b}_\xi^{J_{s1}}, \mathcal{E}_{\text{sim}}^1\left(0, \vec{b}_\zeta^{M_{s1}}\right)\right\rangle_{\partial V_{\text{sim}}^{1-}} \tag{6-11b}$$

$$c_{\xi\zeta}^{J_{s1}M_{s1}} = \left\langle \vec{b}_\xi^{J_{s1}}, \hat{n}_{s1}^- \times \vec{b}_\zeta^{M_{s1}}\right\rangle_{\partial V_{\text{sim}}^1} \tag{6-11c}$$

The elements of matrix equation (6-10) are calculated as follows:

$$z_{\xi\zeta}^{J_{s2}EJ_{s2}} = \left\langle \vec{b}_\xi^{J_{s2}}, \mathcal{E}_{\text{sim}}^2\left(\vec{b}_\zeta^{J_{s2}}, 0\right)\right\rangle_{\partial V_{\text{sim}}^{2-}} \tag{6-12a}$$

$$z_{\xi\zeta}^{J_{s2}EM_{s2}} = \left\langle \vec{b}_\xi^{J_{s2}}, \mathcal{E}_{\text{sim}}^2\left(0, \vec{b}_\zeta^{M_{s2}}\right)\right\rangle_{\partial V_{\text{sim}}^{2-}} \tag{6-12b}$$

$$c_{\xi\zeta}^{J_{s2}M_{s2}} = \left\langle \vec{b}_\xi^{J_{s2}}, \hat{n}_{s2}^- \times \vec{b}_\zeta^{M_{s2}}\right\rangle_{\partial V_{\text{sim}}^2} \tag{6-12c}$$

By solving matrix equations (6-9) and (6-10), the transformations from the equivalent surface magnetic currents to the equivalent surface electric currents are immediately obtained as follows:

$$\bar{a}^{J_{s1}} = \left(\bar{\bar{Z}}^{J_{s1}EJ_{s1}}\right)^{-1} \cdot \left[\bar{\bar{C}}^{J_{s1}M_{s1}} - \bar{\bar{Z}}^{J_{s1}EM_{s1}}\right] \cdot \bar{a}^{M_{s1}} \tag{6-13}$$

$$\bar{a}^{J_{s2}} = \left(\bar{\bar{Z}}^{J_{s2}EJ_{s2}}\right)^{-1} \cdot \left[\bar{\bar{C}}^{J_{s2}M_{s2}} - \bar{\bar{Z}}^{J_{s2}EM_{s2}}\right] \cdot \bar{a}^{M_{s2}} \tag{6-14}$$

and then we have the following transformation:

$$\begin{bmatrix} \bar{a}^{J_{s10}} \\ \bar{a}^{J_{s12}} \\ \bar{a}^{J_{s21}} \\ \bar{a}^{J_{s20}} \end{bmatrix} = \underbrace{\begin{bmatrix} \left(\bar{\bar{Z}}^{J_{s1}EJ_{s1}}\right)^{-1} \cdot \left(\bar{\bar{C}}^{J_{s1}M_{s1}} - \bar{\bar{Z}}^{J_{s1}EM_{s1}}\right) & 0 \\ 0 & \left(\bar{\bar{Z}}^{J_{s2}EJ_{s2}}\right)^{-1} \cdot \left(\bar{\bar{C}}^{J_{s2}M_{s2}} - \bar{\bar{Z}}^{J_{s2}EM_{s2}}\right) \end{bmatrix}}_{\bar{\bar{T}}_{\text{DESM}}^{J_{\text{ss sys}} \leftarrow M_{\text{ss sys}}}} \cdot \begin{bmatrix} \bar{a}^{M_{s10}} \\ \bar{a}^{M_{s12}} \\ \bar{a}^{M_{s21}} \\ \bar{a}^{M_{s20}} \end{bmatrix} \tag{6-15}$$

In addition, relationship $\bar{a}^{C_{s21}} = -\bar{a}^{C_{s12}}$ and the partition ways used in formulations (4-78) and (4-79) give the following transformation:

$$\begin{bmatrix} \bar{a}^{M_{s10}} \\ \bar{a}^{M_{s12}} \\ \bar{a}^{M_{s21}} \\ \bar{a}^{M_{s20}} \end{bmatrix} = \underbrace{\begin{bmatrix} \bar{\bar{I}}^{M_{s10}} & 0 & 0 \\ 0 & \bar{\bar{I}}^{M_{s12}} & 0 \\ 0 & -\bar{\bar{I}}^{M_{s12}} & 0 \\ 0 & 0 & \bar{\bar{I}}^{M_{s20}} \end{bmatrix}}_{\bar{\bar{T}}^{M_{\text{ss sys}} \leftarrow M_{\text{ss}}}} \cdot \begin{bmatrix} \bar{a}^{M_{s10}} \\ \bar{a}^{M_{s12}} \\ \bar{a}^{M_{s20}} \end{bmatrix} \tag{6-16}$$

where $\overline{\overline{I}}^{M_{\text{s10}}}$, $\overline{\overline{I}}^{M_{\text{s12}}}$, and $\overline{\overline{I}}^{M_{\text{s20}}}$ are $\Xi^{M_{\text{s10}}}$-order, $\Xi^{M_{\text{s12}}}$-order, and $\Xi^{M_{\text{s20}}}$-order identity matrices respectively. Then, we have the following transformation:

$$\begin{bmatrix} \overline{a}^{J_{\text{s10}}} \\ \overline{a}^{J_{\text{s12}}} \\ \overline{a}^{J_{\text{s21}}} \\ \overline{a}^{J_{\text{s20}}} \end{bmatrix} = \underbrace{\overline{\overline{T}}_{\text{DESM}}^{J_{\text{ss sys}} \leftarrow M_{\text{ss sys}}} \cdot \overline{\overline{T}}^{M_{\text{ss sys}} \leftarrow M_{\text{ss}}}}_{\overline{\overline{T}}_2} \cdot \begin{bmatrix} \overline{a}^{M_{\text{s10}}} \\ \overline{a}^{M_{\text{s12}}} \\ \overline{a}^{M_{\text{s20}}} \end{bmatrix} \tag{6-17}$$

Inserting transformation (6-17) into formulations (4-80), (4-114), and (4-118), we can obtain the following matrix form for the DPO $P_{\text{ss sys}}^{\text{driving}}$ containing BVs $\{\overline{a}^{M_{\text{s10}}}, \overline{a}^{M_{\text{s20}}}\}$ and dependent variable $\overline{a}^{M_{\text{s12}}}$:

$$P_{\text{ss sys}}^{\text{driving}} = \begin{bmatrix} \overline{a}^{M_{\text{s10}}} \\ \overline{a}^{M_{\text{s12}}} \\ \overline{a}^{M_{\text{s20}}} \end{bmatrix}^H \cdot \left(\overline{\overline{T}}_2\right)^H \cdot \overline{\overline{P}}_{\text{ss sys;1}}^{\text{driving}} \cdot \overline{\overline{T}}_2 \cdot \begin{bmatrix} \overline{a}^{M_{\text{s10}}} \\ \overline{a}^{M_{\text{s12}}} \\ \overline{a}^{M_{\text{s20}}} \end{bmatrix} \tag{6-18}$$

and

$$P_{\text{ss sys}}^{\text{driving}} = \begin{bmatrix} \overline{a}^{M_{\text{s10}}} \\ \overline{a}^{M_{\text{s12}}} \\ \overline{a}^{M_{\text{s20}}} \end{bmatrix}^H \cdot \left(\overline{\overline{T}}_2\right)^H \cdot \overline{\overline{P}}_{\text{ss sys;2}}^{\text{driving}} \cdot \overline{\overline{T}}_2 \cdot \begin{bmatrix} \overline{a}^{M_{\text{s10}}} \\ \overline{a}^{M_{\text{s12}}} \\ \overline{a}^{M_{\text{s20}}} \end{bmatrix} \tag{6-19}$$

and

$$P_{\text{ss sys}}^{\text{driving}} = \begin{bmatrix} \overline{a}^{M_{\text{s10}}} \\ \overline{a}^{M_{\text{s12}}} \\ \overline{a}^{M_{\text{s20}}} \end{bmatrix}^H \cdot \left(\overline{\overline{T}}_2\right)^H \cdot \overline{\overline{P}}_{\text{ss sys;3'}}^{\text{driving}} \cdot \overline{\overline{T}}_2 \cdot \begin{bmatrix} \overline{a}^{M_{\text{s10}}} \\ \overline{a}^{M_{\text{s12}}} \\ \overline{a}^{M_{\text{s20}}} \end{bmatrix} \tag{6-20}$$

In formulation (6-20),

$$\overline{\overline{P}}_{\text{ss sys;3'}}^{\text{driving}} = \begin{bmatrix} \overline{\overline{P}}_{3;0;\text{PVT}}^{J_{\text{s10}}J_{\text{s10}}} & 0 & 0 & \overline{\overline{P}}_{3;0;\text{PVT}}^{J_{\text{s10}}J_{\text{s20}}} & \overline{\overline{P}}_{3;0;\text{PVT}}^{J_{\text{s10}}M_{\text{s10}}} + \overline{\overline{P}}_{3;0;\text{SCT}}^{J_{\text{s10}}M_{\text{s10}}} & 0 & 0 & \overline{\overline{P}}_{3;0;\text{PVT}}^{J_{\text{s10}}M_{\text{s20}}} \\ 0 & 0 & 0 & 0 & 0 & 0 & 0 & 0 \\ 0 & 0 & 0 & 0 & 0 & 0 & 0 & 0 \\ \overline{\overline{P}}_{3;0;\text{PVT}}^{J_{\text{s20}}J_{\text{s10}}} & 0 & 0 & \overline{\overline{P}}_{3;0;\text{PVT}}^{J_{\text{s20}}J_{\text{s20}}} & \overline{\overline{P}}_{3;0;\text{PVT}}^{J_{\text{s20}}M_{\text{s10}}} & 0 & 0 & \overline{\overline{P}}_{3;0;\text{PVT}}^{J_{\text{s20}}M_{\text{s20}}} + \overline{\overline{P}}_{3;0;\text{SCT}}^{J_{\text{s20}}M_{\text{s20}}} \\ \overline{\overline{P}}_{3;0;\text{PVT}}^{M_{\text{s10}}J_{\text{s10}}} + \overline{\overline{P}}_{3;0;\text{SCT}}^{M_{\text{s10}}J_{\text{s10}}} & 0 & 0 & \overline{\overline{P}}_{3;0;\text{PVT}}^{M_{\text{s10}}J_{\text{s20}}} & \overline{\overline{P}}_{3;0;\text{PVT}}^{M_{\text{s10}}M_{\text{s10}}} & 0 & 0 & \overline{\overline{P}}_{3;0;\text{PVT}}^{M_{\text{s10}}M_{\text{s20}}} \\ 0 & 0 & 0 & 0 & 0 & 0 & 0 & 0 \\ 0 & 0 & 0 & 0 & 0 & 0 & 0 & 0 \\ \overline{\overline{P}}_{3;0;\text{PVT}}^{M_{\text{s20}}J_{\text{s10}}} & 0 & 0 & \overline{\overline{P}}_{3;0;\text{PVT}}^{M_{\text{s20}}J_{\text{s20}}} + \overline{\overline{P}}_{3;0;\text{SCT}}^{M_{\text{s20}}J_{\text{s20}}} & \overline{\overline{P}}_{3;0;\text{PVT}}^{M_{\text{s20}}M_{\text{s10}}} & 0 & 0 & \overline{\overline{P}}_{3;0;\text{PVT}}^{M_{\text{s20}}M_{\text{s20}}} \end{bmatrix} \tag{6-21}$$

Based on the matrix operator $(\overline{\overline{T}}_2)^H \cdot \overline{\overline{P}}_{\text{ss sys;1}}^{\text{driving}} \cdot \overline{\overline{T}}_2$ in formulation (6-18), we calculate the CMs, and the characteristic value curves and MS curves corresponding to the first 500 modes are illustrated in Figure 6-11.

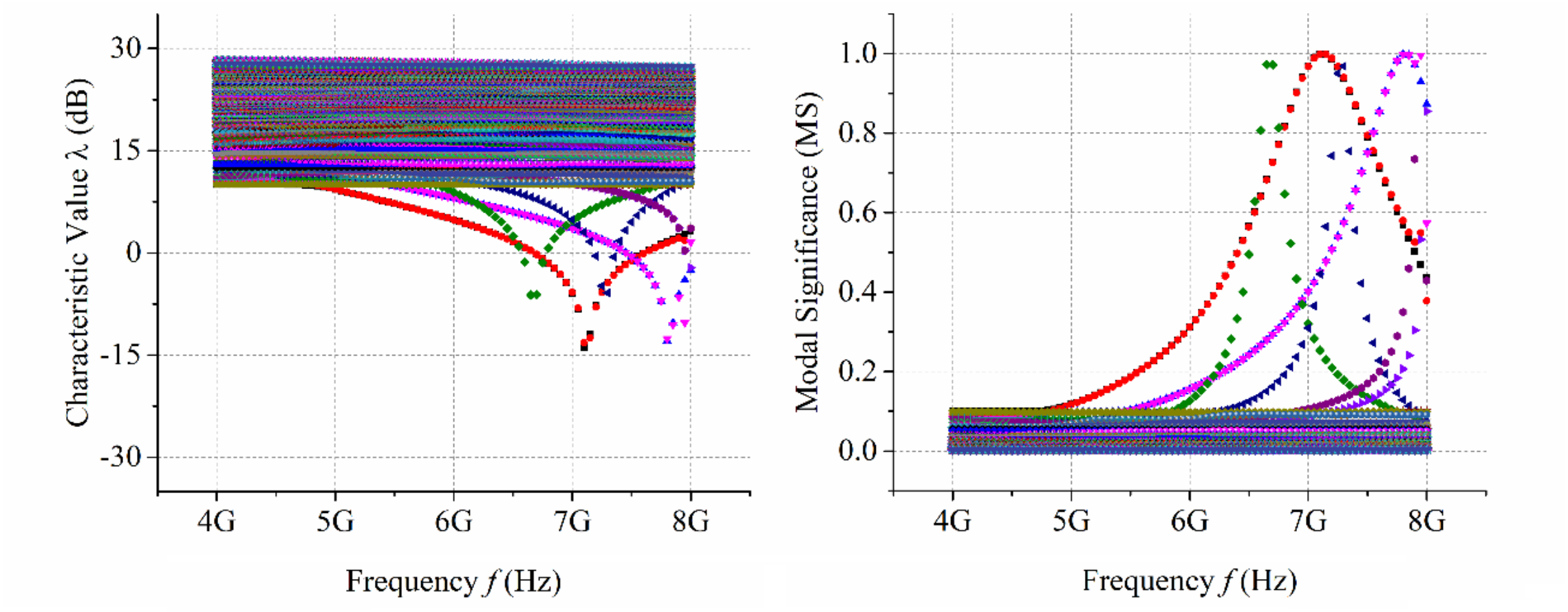

(a) (b)

Figure 6-11 The characteristic quantity curves corresponding to the first 500 DP-CMs (derived from surface formulation (6-18)) of the two-body material system shown in Figure 6-8. (a) characteristic value dB curves; (b) MS curves

Comparing Figure 6-11 and Figure 4-33, it is easy to find out that the figures are very similar to each other. So, whether or not this implies that transformation (6-17) gives a thorough variable unification scheme? Before answering this question, we firstly calculate DP-CMs by employing the matrix operator $(\overline{\overline{T}}_2)^H \cdot \overline{\overline{P}}_{\text{ss sys;2}}^{\text{driving}} \cdot \overline{\overline{T}}_2$ in formulation (6-19) and the matrix operator $(\overline{\overline{T}}_2)^H \cdot \overline{\overline{P}}_{\text{ss sys;3'}}^{\text{driving}} \cdot \overline{\overline{T}}_2$ in formulation (6-20), and provide the corresponding characteristic value curves and MS curves in Figure 6-12 and Figure 6-13 respectively. Afterwards, we answer the above question based on Figures 6-12 and 6-13.

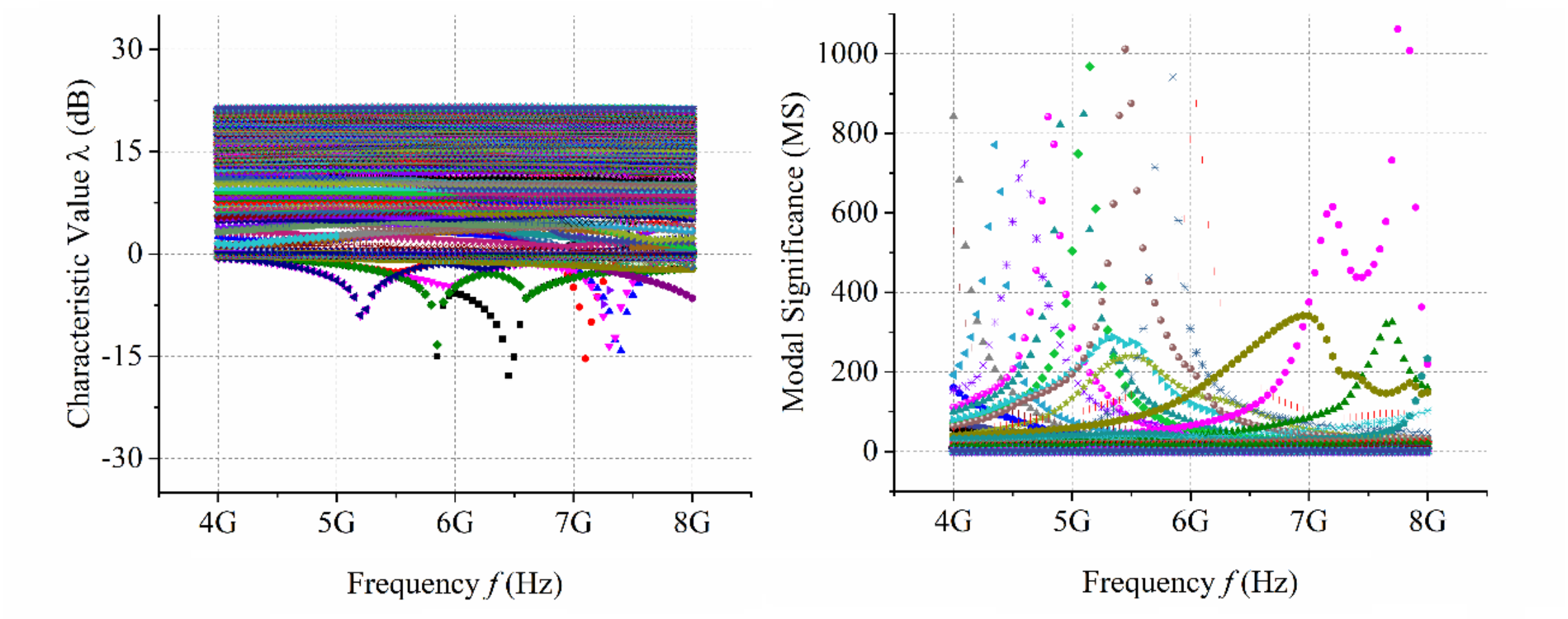

(a) (b)

Figure 6-12 The characteristic quantity curves corresponding to the first 500 DP-CMs (derived from surface formulation (6-19)) of the two-body material system shown in Figure 6-8. (a) characteristic value dB curves; (b) MS curves

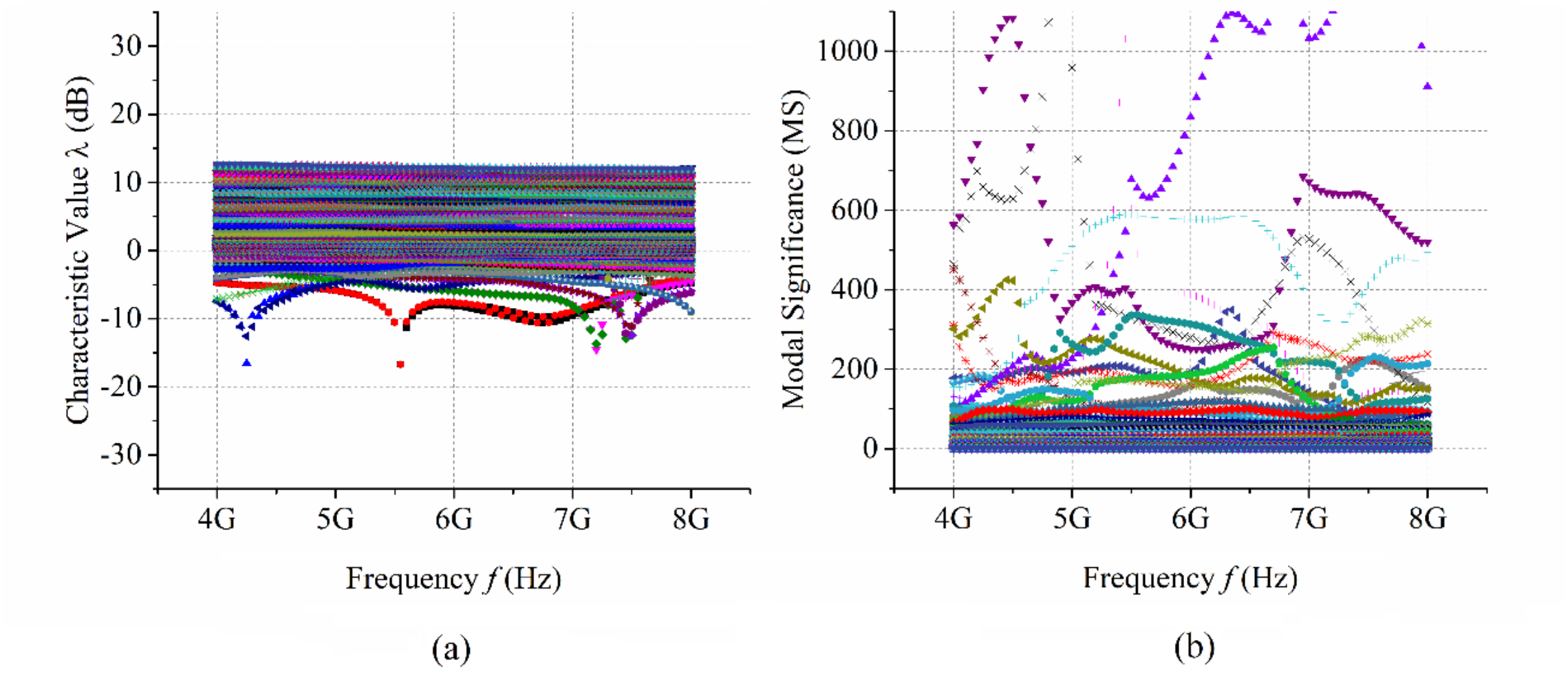

Figure 6-13 The characteristic quantity curves corresponding to the first 500 DP-CMs (derived from surface formulation (6-20)) of the two-body material system shown in Figure 6-8. (a) characteristic value dB curves; (b) MS curves

Obviously, Figures 6-12 and 6-13 don't include enough physical modes. This implies that the variable unification given in transformation (6-17) is not thorough.

### 4) Doing a Thorough Variable Unification

For the two-body material system shown in Figure 6-8 (i.e. the one shown in Figure 4-30), a thorough variabe unification scheme has been given in the transformation (4-101) in Section 4.4, and the scheme eliminates the 6 EM sub-currents $\vec{J}_{s10}^{ES}$, $\vec{J}_{s12}^{ES}$, $\vec{J}_{s21}^{ES}$, $\vec{J}_{s20}^{ES}$, $\vec{M}_{s12}^{ES}$, and $\vec{M}_{s21}^{ES}$ included in all EM sub-currents $\vec{J}_{s10}^{ES}$, $\vec{J}_{s12}^{ES}$, $\vec{J}_{s21}^{ES}$, $\vec{J}_{s20}^{ES}$, $\vec{M}_{s10}^{ES}$, $\vec{M}_{s12}^{ES}$, $\vec{M}_{s21}^{ES}$, and $\vec{M}_{s20}^{ES}$ such that the final DPO only contains two BVs $\vec{M}_{s10}^{ES}$ and $\vec{M}_{s20}^{ES}$. The corresponding calculated results agree well with the results calculated from volume formulation.

## 6.2.3 Metal-Material Composite System Case

Following Subsection 6.2.2, we, in this subsection, discuss the variable unification problem related to metal-material composite systems.

### 1) Not Doing Any Variable Unification

Now, taking the composite system shown in Figure 6-14 (i.e. the one shown in Figure 5-5) as an example, we study the variable unification problem related to metal-material composite systems. In Figure 6-14, the radius of the metallic sphere is 2.50mm; the inner radius and outer radius of the material spherical shell are 2.50mm and 5.00mm respectively; the relative permeability, relative permittivity, and conductivity of the material spherical shell are 6, 6, and 0 respectively, i.e., $\mu_{\rm mat}^{\rm r} = 6$, $\varepsilon_{\rm mat}^{\rm r} = 6$, and $\sigma_{\rm mat} = 0$. For the composite system, we, based on the operator (5-67) without any variable unification, calculate its DP-CMs, and the characteristic value curves and MS curves corresponding to the first 500 modes are illustrated in Figure 6-15.

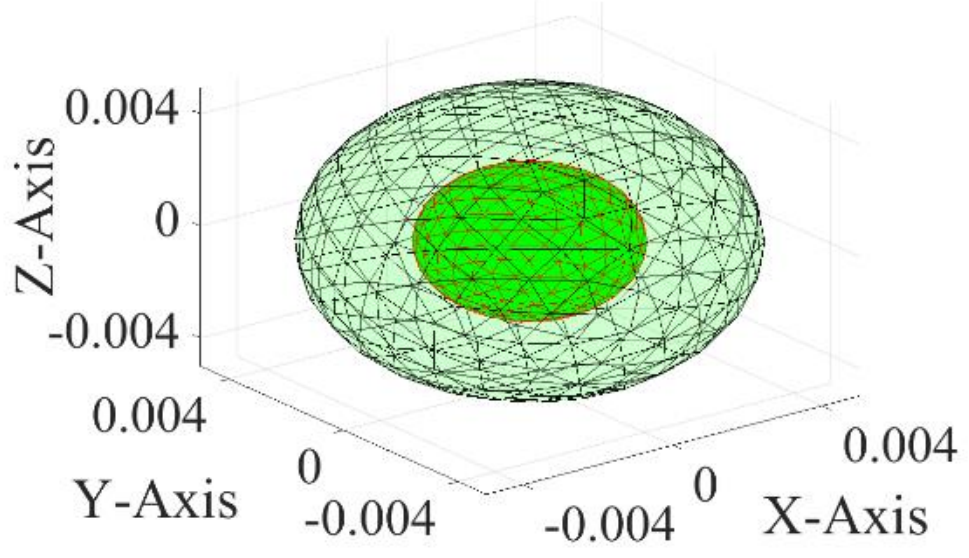

Figure 6-14 The topological structure and surface triangular meshes of the composite system constituted by "a metallic sphere whose radius is 2.50mm" and "a material spherical shell whose inner and outer radiuses are 2.50mm and 5.00mm respectively"

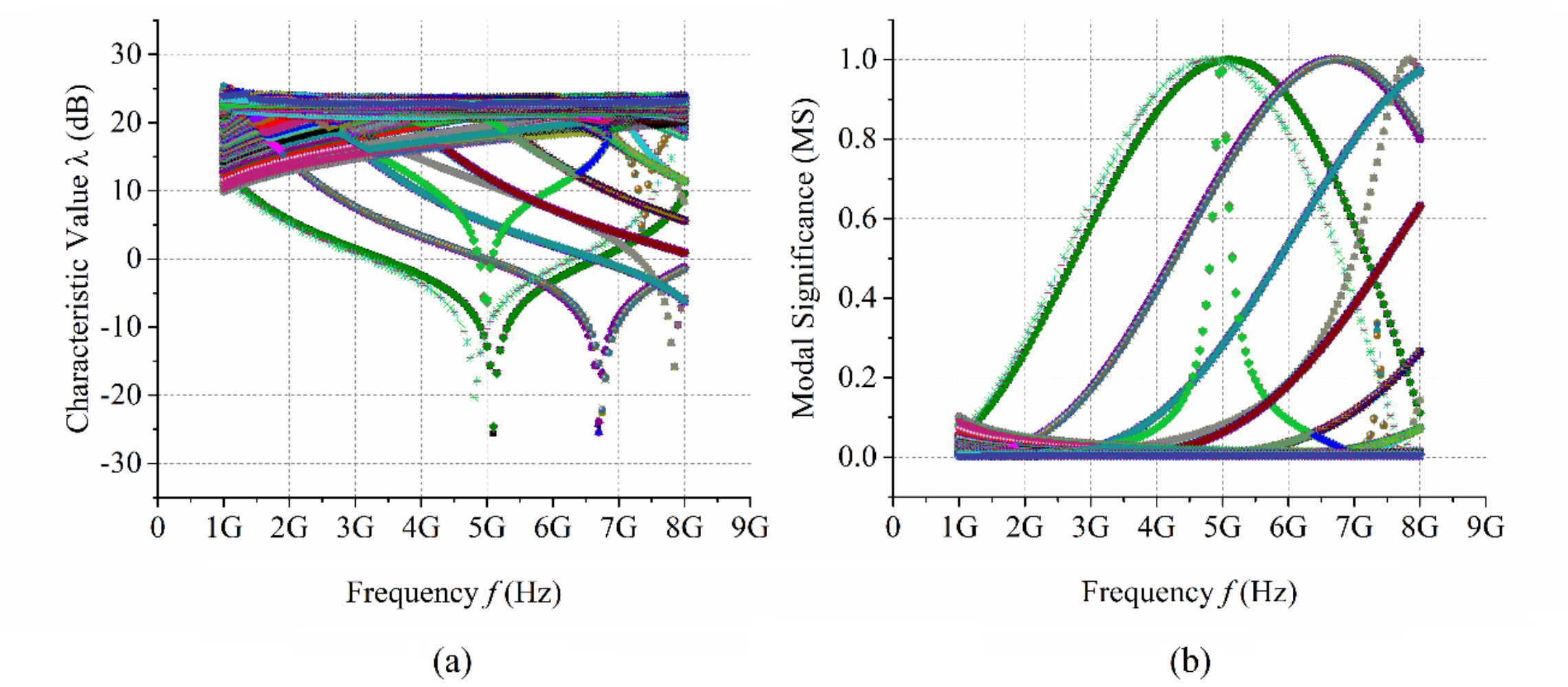

Figure 6-15 The characteristic quantity curves corresponding to the first 500 DP-CMs (derived from the line-surface formulation (5-67) without variable unification) of the composite system shown in Figure 6-14. (a) characteristic value dB curves; (b) MS curves

Comparing above Figure 6-15 with previous Figure 5-8, it is easy to find out that: for the composite system, the modal set derived from the DPO without variable unification includes many spurious modes.

**2) Only Doing Variable Unification in EM Current Space**

When the variables of operator (5-67) is only unified in EM current space, i.e., only relationships (5-33)~(5-38) are utilized, the calculated results are illustrated in the following Figure 6-16.

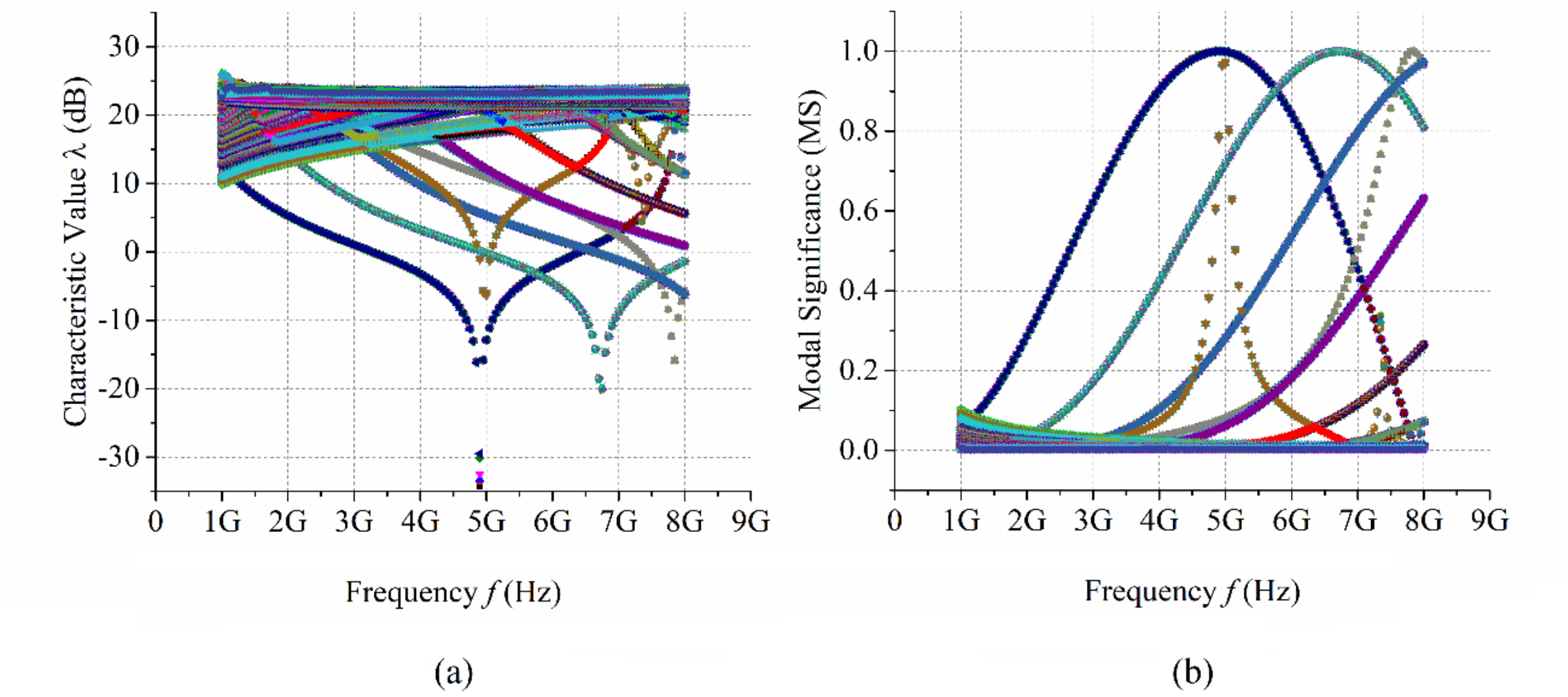

Figure 6-16 The characteristic quantity curves corresponding to the first 500 DP-CMs (derived from the line-surface formulation (5-67) with the variable unification in EM current space) of the composite system shown in Figure 6-14. (a) characteristic value dB curves; (b) MS curves

Comparing the above Figure 6-16 with the previous Figure 5-8, it is easy to find out that: for composite systems, the modal set derived from the DPO (5-67) whose variables are only unified in EM current space also includes many spurious modes.

**3) Doing Variable Unification in EM Current Space and Doing an Incomplete Variable Unification in Expansion Vector Space**

Now, we provide an incomplete variable unification scheme for the composite system. The formulations related to the incomplete scheme can be found in literature [55], and they will not be repeated here. Here, we provide some calculated results based on the incomplete scheme.

The characteristic value curves and MS curves corresponding to the DP-CMs derived from the operator (5-67) with the incomplete variable unification scheme given in literature [55] are illustrated in Figure 6-17.

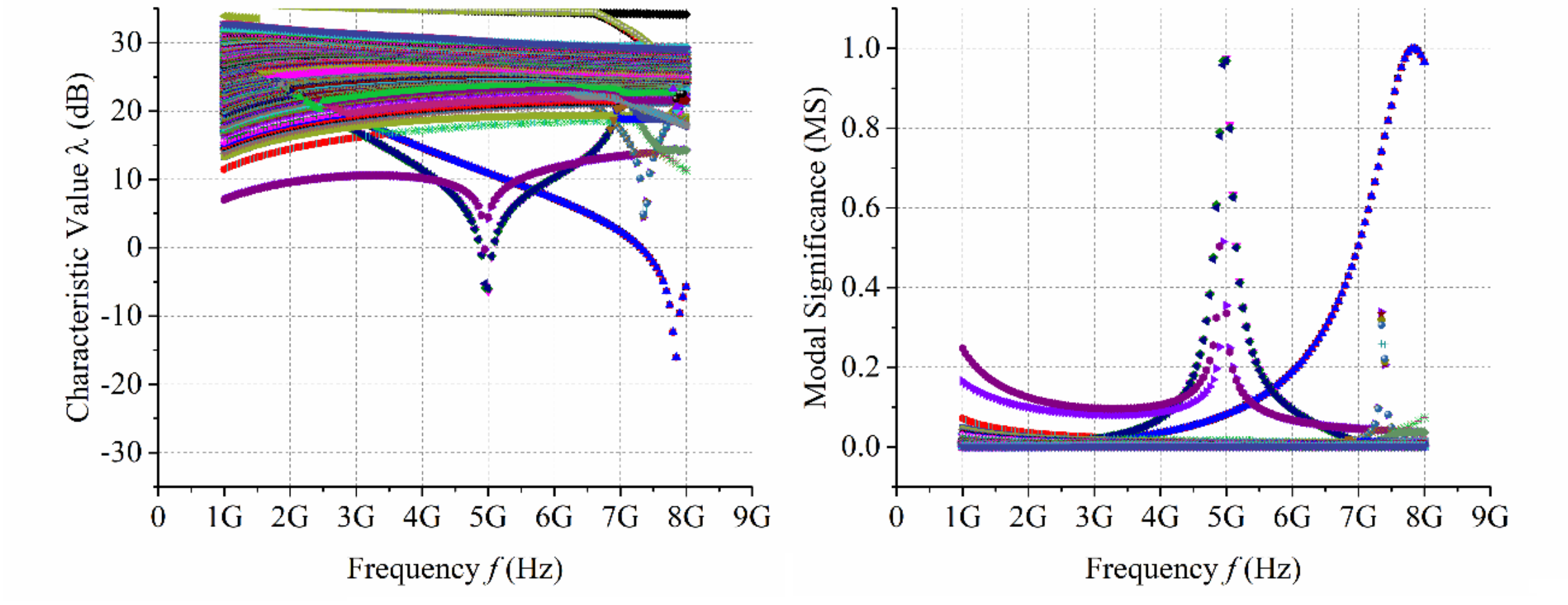

(a) (b)

Figure 6-17 The characteristic quantity curves corresponding to the first 500 DP-CMs (derived from line-surface formulation (5-67) with the incomplete variable unification scheme proposed in literature [55]) of the composite system shown in Figure 6-14. (a) characteristic value dB curves; (b) MS curves

Comparing above Figure 6-17 with previous Figure 5-8, it is easy to find out that the figures are very similar to each other. So, whether or not this implies that the scheme proposed in literature [55] is a thorough variable unification scheme? Before answering this question, we first construct DP-CMs by employing the DPO (5-68) with the variable unification scheme proposed in literature [55], and provide the corresponding characteristic value curves and MS curves in Figure 6-18. Afterwards, we answer the question based on the results shown in Figure 6-18.

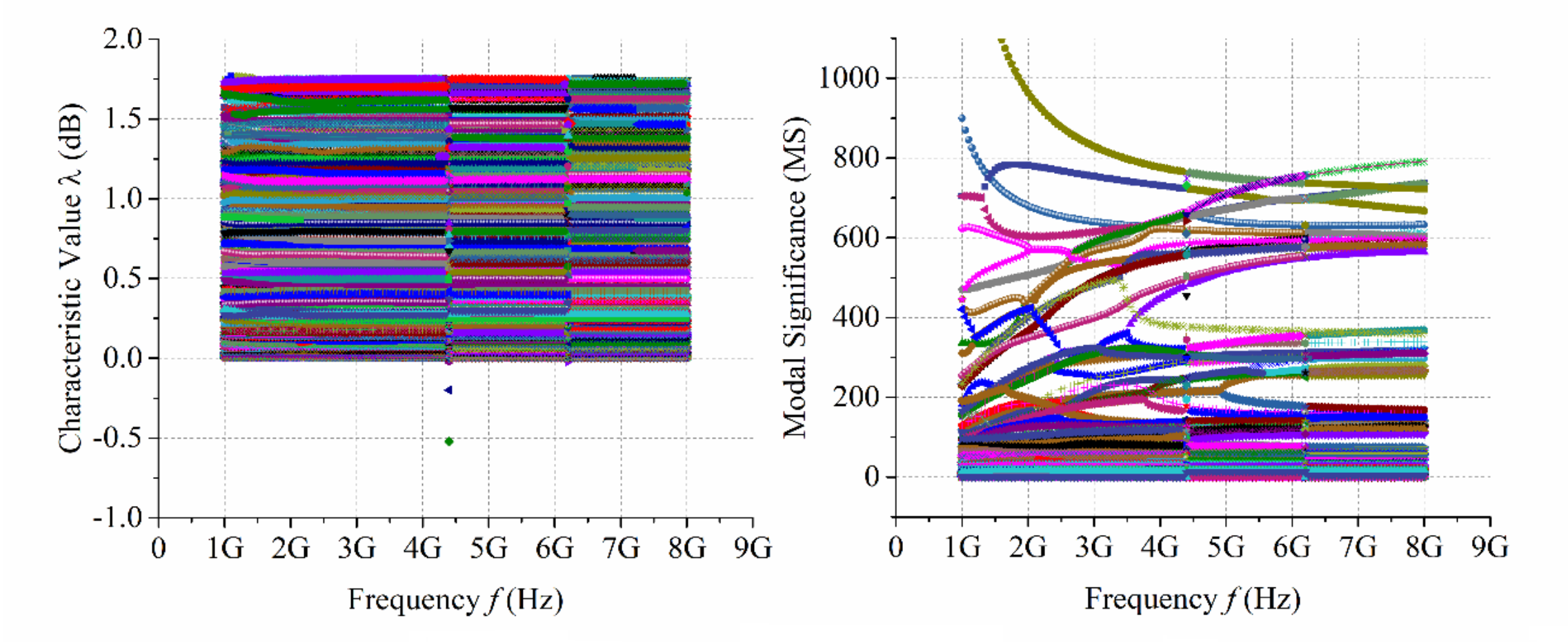

(a) (b)

Figure 6-18 The characteristic quantity curves corresponding to the first 500 DP-CMs (derived from the line-surface formulation (5-68) with the incomplete variable unification scheme proposed in literature [55]) of the composite system shown in Figure 6-14. (a) characteristic value dB curves; (b) MS curves

Obviously, Figure 6-18 includes many spurious modes. This implies that the variable unification scheme proposed in literature [55] is not a thorough scheme!

**4) Doing a Thorough Variable Unification**

For the composite system shown in Figure 6-14 (i.e. the one shown in Figure 5-5), a thorough variabe unification scheme has been given in the Subsection 5.5.2 of this dissertation, and the corresponding calculated results have also been given in the Subsection 5.6.1 of this dissertation and we will not repeat the results here.

### 6.2.4 Systematical Scheme for Suppressing Spurious Modes

For an objective scattering system, its DPO contains many variables. In the variables, some variables are independent, and some variables are dependent; some variables can uniquely determine the DPO numerically, and some variables cannot. In the variables, this dissertation calls the variables which are independent and can uniquely determine the DPO as basic variables (BVs), and calls the other variables as dependent variables (DVs).

Before constructing DP-CMs, it is necessary to obtain the DPO only containing BVs, i.e., it is necessary to eliminate all DVs in the DPO, or we will not be able to obtain enough physical modes or we will obtain some spurious modes. In this dissertation, the above process to eliminate the DVs is called as variable unification or to unify variables. Now, we summarize the process to unify variables and the process of the form variation of the DPO as below.

**1) The Variable Unification for the Objective Scattering System**

The generalized Franz-Harrington formulations (GFHFs)[①] for the objective scattering system are a series of formulations expressed as the convolution integrals of Green's functions and {the scattered line electric current $\vec{J}^{\text{SL}}$ on metallic boundary line, the scattered surface electric current $\vec{J}^{\text{SS}}$ on metallic boundary surface, the equivalent line electric current $\vec{J}^{\text{EL}}$ and equivalent line magnetic current $\vec{M}^{\text{EL}}$ on material boundary line, the equivalent surface electric current $\vec{J}^{\text{ES}}$ and equivalent surface magnetic current $\vec{M}^{\text{ES}}$ on material boundary surface}. The fundamental principles for unifying scattered sources $\{\vec{J}^{\text{SL}},\vec{J}^{\text{SS}}\}$ and equivalent sources $\{\vec{J}^{\text{EL}},\vec{M}^{\text{EL}},\vec{J}^{\text{ES}},\vec{M}^{\text{ES}}\}$

① The GFHF of a scattering system contains three parts: the GFHF for the incident field on the interior of the system, the GFHF for the scattered field on the exterior of the system, and the GFHF for the total field on the interior of the system. Because the first two parts have the same convolution integral expression (the only difference is the domains for carrying fields), so in what follows we collectively call them as partial field GFHF for the convenience of discussions. In what follows, we similarly call the third part as total field GFHF.

distributing on the scattering system are as follows: to determine BV set①; to provide enough independent equations for establishing the transformations from BVs to DVs; to establish the linear transformations from BVs to DVs by employing the independent equations. In addition, the above processes "to determine BV set", "to provide independent equations", and "to establish transformations" usually need to be done alternately. Specifically,

**(1.1)** Above $\{\vec{J}^{\rm SL},\vec{J}^{\rm SS},\vec{J}^{\rm EL},\vec{M}^{\rm EL},\vec{J}^{\rm ES},\vec{M}^{\rm ES}\}$ constitute the whole variable set used to describe the scattering problem. To effectively determine the BV set and establish the transformations from BVs to DVs, this dissertation does some proper decompositions for the boundaries of the scattering system and the sources on the boundaries, such that the above whole variable set is equivalently rewritten as {the scattered line electric current $\vec{J}_0^{\rm SL}$ on environment-metal boundary line $L_{\rm met}^0$, the scattered line electric current $\vec{J}_\cap^{\rm SL}$ on metal-material boundary line $L_{\rm met}^\cap$, the scattered surface electric current $\vec{J}_0^{\rm SS}$ on environment-metal boundary surface $S_{\rm met}^0 \bigcup \partial V_{\rm met}^0$, the scattered surface electric current $\vec{J}_\cap^{\rm SS}$ on metal-material boundary surface $S_{\rm met}^\cap \bigcup \partial V_{\rm met}^\cap$, the scattered surface electric current $\vec{J}_{0\cap}^{\rm SS}$ on environment-metal-material boundary surface $S_{\rm met}^{0\cap}$, and the equivalent line electric current $\vec{J}^{\rm EL}$ and equivalent line magnetic current $\vec{M}^{\rm EL}$ on metal-material boundary line $L_{\rm met}^\cap$, the equivalent surface electric current $\vec{J}_\cap^{\rm ES}$ and equivalent surface magnetic current $\vec{M}_\cap^{\rm ES}$ on metal-material boundary surface $S_{\rm met}^\cap \bigcup \partial V_{\rm met}^\cap$, the equivalent surface electric current $\vec{J}_{0\cap}^{\rm ES}$ and equivalent surface magnetic current $\vec{M}_{0\cap}^{\rm ES}$ on environment-metal-material boundary surface $S_{\rm met}^{0\cap}$, the equivalent surface electric current $\vec{J}_0^{\rm ES}$ and equivalent surface magnetic current $\vec{M}_0^{\rm ES}$ on environment-material boundary surface $\partial V_{\rm mat}^0$, the equivalent surface electric current $\vec{J}_{\rm mat+}^{\rm ES}$ and equivalent surface magnetic current $\vec{M}_{\rm mat+}^{\rm ES}$ on one side of material-material boundary surface $\partial V_{\rm mat}^{\rm mat}$, the equivalent surface electric current $\vec{J}_{\rm mat-}^{\rm ES}$ and equivalent surface magnetic current $\vec{M}_{\rm mat-}^{\rm ES}$ on the other side of material-material boundary surface $\partial V_{\rm mat}^{\rm mat}$}, i.e., $\{\vec{J}^{\rm SL},\vec{J}^{\rm SS},\vec{J}^{\rm EL},\vec{M}^{\rm EL},\vec{J}^{\rm ES},\vec{M}^{\rm ES}\}=\{\vec{J}_0^{\rm SL},\vec{J}_\cap^{\rm SL},\vec{J}_0^{\rm SS},\vec{J}_\cap^{\rm SS},\vec{J}_{0\cap}^{\rm SS},\vec{J}^{\rm EL},\vec{M}^{\rm EL},\vec{J}_\cap^{\rm ES},\vec{M}_\cap^{\rm ES},\vec{J}_{0\cap}^{\rm ES},\vec{M}_{0\cap}^{\rm ES},\vec{J}_0^{\rm ES},$ $\vec{M}_0^{\rm ES},\vec{J}_{\rm mat+}^{\rm ES},\vec{M}_{\rm mat+}^{\rm ES},\vec{J}_{\rm mat-}^{\rm ES},\vec{M}_{\rm mat-}^{\rm ES}\}$.

**(1.2)** The homogeneous tangential boundary condition $\hat{n}_{\rm met}^+ \times \vec{E}^{\rm tot} = 0$ satisfied by the total electric fields on the metal-material sub-boundaries requires that: some equivalent sources in the whole variable set are zero, i.e., $\vec{M}^{\rm EL}=0$, and $\vec{M}_\cap^{\rm ES}=0$, and $\vec{M}_{0\cap}^{\rm ES}=0$. The tangential boundary condition $\hat{n}_{\rm met}^+ \times \vec{H}^{\rm tot} = \vec{J}$ satisfied by the total

① When the BV set is determined, the DV set is also determined correspondingly.

magnetic fields on the metal-material sub-boundaries requires that: some scattered sources and some equivalent sources in the whole variable set satisfy some simple relationships, i.e., $\vec{J}_{\cap}^{\rm SL} = \vec{J}^{\rm EL}$ , and $\vec{J}_{\cap}^{\rm SS} = \vec{J}_{\cap}^{\rm ES}$ . The tangential continuation condition $\hat{n}_{\rm mat}^{+} \times (\vec{F}_{+}^{\rm tot} - \vec{F}_{-}^{\rm tot}) = 0$ satisfied by the total fields on the material-material sub-boundaries requires that: some equivalent sources and another equivalent sources in the whole variable set satisfy some simple relationships, i.e., $\vec{J}_{\rm mat+}^{\rm ES} = \vec{J}_{\rm mat}^{\rm ES} = -\vec{J}_{\rm mat-}^{\rm ES}$ , and $\vec{M}_{\rm mat+}^{\rm ES} = \vec{M}_{\rm mat}^{\rm ES} = -\vec{M}_{\rm mat-}^{\rm ES}$ . Applying the above relationships to the original GFHFs, the original GFHFs will be simplified into the convolution integrals of Green's functions and $\{\vec{J}_{0}^{\rm SL}, \vec{J}_{\cap}^{\rm SL}, \vec{J}_{0}^{\rm SS}, \vec{J}_{\cap}^{\rm SS}, \vec{J}_{0\cap}^{\rm SS}, \vec{J}_{0\cap}^{\rm ES}, \vec{J}_{0}^{\rm ES}, \vec{M}_{0}^{\rm ES}, \vec{J}_{\rm mat}^{\rm ES}, \vec{M}_{\rm mat}^{\rm ES}\}$ . In this dissertation, the latter convolution integrals are called as simplified GFHFs.

**(1.3)** Carefully comparing the simplified partial field GFHF and the simplified total field GFHF, it can be found out that: variables $\{\vec{J}_{\cap}^{\rm SL}, \vec{J}_{\cap}^{\rm SS}, \vec{J}_{\rm mat}^{\rm ES}, \vec{M}_{\rm mat}^{\rm ES}\}$ are not contained in the former, and only contained in the latter; the variables $\{\vec{J}_{0}^{\rm SL}, \vec{J}_{0}^{\rm SS}, \vec{J}_{0\cap}^{\rm SS}, \vec{J}_{0\cap}^{\rm ES}, \vec{J}_{0}^{\rm ES}, \vec{M}_{0}^{\rm ES}\}$ contained in the former are just the scattered sources and equivalent sources distributing on the outer boundary $\partial D_{\rm sca\,sys}$ of whole scattering system $D_{\rm sca\,sys}$ . In fact, the phenomenon implicitly implies an important conclusion —— the scattered sources and equivalent sources distributing on the outer boundary of the scattering system constitute a complete variable set for describing the scattering problem (but they are not necessarily independent!). The reason leading to the important conclusion is that: the simplified partial field GFHF expresses the incident fields on the scattering system as the convolution integral of vacuum Green's functions and the above complete variable set (i.e., the incident fields on the scattering system are uniquely determined by the above complete variable set), and the scattered sources distributing on the scattering system are uniquely determined by the incident fields on the scattering system.

**(1.4)** Whether or not all of the variables contained in above complete variable set $\{\vec{J}_{0}^{\rm SL}, \vec{J}_{0}^{\rm SS}, \vec{J}_{0\cap}^{\rm SS}, \vec{J}_{0\cap}^{\rm ES}, \vec{J}_{0}^{\rm ES}, \vec{M}_{0}^{\rm ES}\}$ are independent? This dissertation thinks that the answer to the question is NO. So, which variables contained in $\{\vec{J}_{0}^{\rm SL}, \vec{J}_{0}^{\rm SS}, \vec{J}_{0\cap}^{\rm SS}, \vec{J}_{0\cap}^{\rm ES}, \vec{J}_{0}^{\rm ES}, \vec{M}_{0}^{\rm ES}\}$ are both complete and independent? i.e., which variables contained in $\{\vec{J}_{0}^{\rm SL}, \vec{J}_{0}^{\rm SS}, \vec{J}_{0\cap}^{\rm SS}, \vec{J}_{0\cap}^{\rm ES}, \vec{J}_{0}^{\rm ES}, \vec{M}_{0}^{\rm ES}\}$ can constitute a BV set? This dissertation gives the answer to the question that: eliminating the anyone of $\vec{J}_{0}^{\rm ES}$ and $\vec{M}_{0}^{\rm ES}$ from the complete variable set, the remaining variables constitute a BV set. So far, we complete the determination for the BV set used to completely describe the scattering problem —— the BV set used to completely describe the scattering problem is $\{\vec{J}_{0}^{\rm SL}, \vec{J}_{0}^{\rm SS}, \vec{J}_{0\cap}^{\rm SS}, \vec{J}_{0\cap}^{\rm ES}, \vec{J}_{0}^{\rm ES}\}$ or $\{\vec{J}_0^{\text{SL}}, \vec{J}_0^{\text{SS}}, \vec{J}_{0\cap}^{\text{SS}}, \vec{J}_{0\cap}^{\text{ES}}, \vec{M}_0^{\text{ES}}\}$, and the DV set is correspondingly $\{\vec{J}_\cap^{\text{SL}}, \vec{J}_\cap^{\text{SS}}, \vec{M}_0^{\text{ES}}, \vec{J}_{\text{mat}}^{\text{ES}}, \vec{M}_{\text{mat}}^{\text{ES}}\}$ or $\{\vec{J}_\cap^{\text{SL}}, \vec{J}_\cap^{\text{SS}}, \vec{J}_0^{\text{ES}}, \vec{J}_{\text{mat}}^{\text{ES}}, \vec{M}_{\text{mat}}^{\text{ES}}\}$.

**(1.5)** To effectively establish the transformation from BV set to DV set, and employing the fact that all of the DVs are contained in the simplified total field GFHF, this dissertation establishes the integral equation of $\vec{J}_\cap^{\text{SL}}$ by using the boundary condition $\vec{E}_{\text{tan}}^{\text{tot}} = 0$ satisfied by the simplified total field GFHF on $L_{\text{met}}^\cap$, and establishes the integral equation of $\vec{J}_{\text{mat}}^{\text{ES}}$ by employing the boundary condition $\vec{E}_{\text{tan}}^{\text{tot}} = 0$ satisfied by the simplified total field GFHF on $S_{\text{met}}^\cap \bigcup \partial V_{\text{met}}^\cap$, and establishes the integral equation of $\vec{J}_{\text{mat}}^{\text{ES}}$ by employing the boundary condition $\vec{E}_{\text{tan}+}^{\text{tot}} = \vec{E}_{\text{tan}-}^{\text{tot}}$ satisfied by the simplified total field GFHF on $\partial V_{\text{mat}}^{\text{mat}}$, and establishes the integral equation of $\vec{M}_{\text{mat}}^{\text{ES}}$ by employing the boundary condition $\vec{H}_{\text{tan}+}^{\text{tot}} = \vec{H}_{\text{tan}-}^{\text{tot}}$ satisfied by the simplified total field GFHF on $\partial V_{\text{mat}}^{\text{mat}}$, and establishes the integral equation of $\vec{M}_0^{\text{ES}} / \vec{J}_0^{\text{ES}}$ by employing the definition $\vec{M}_0^{\text{ES}} = \vec{E}_-^{\text{tot}} \times \hat{n}_{\text{mat}}^- / \vec{J}_0^{\text{ES}} = \hat{n}_{\text{mat}}^- \times \vec{H}_-^{\text{tot}}$ for the $\vec{M}_0^{\text{ES}} / \vec{J}_0^{\text{ES}}$ on $\partial V_{\text{mat}}^0$.

**(1.6)** Based on the above integral equations, this dissertation successfully establishes the transformation from the BV set to the DV set, i.e., this dissertation successfully finished an important step in the whole process of constructing the DP-CMs of scattering systems —— variable unification. The reason to treat variable unification as an important step is that: only accomplishing the variable unification properly, we can obtain the DPO only containing BVs; only obtaining the DPO only containing BVs, we can effectively construct the DP-CMs of scattering systems.

**2) The Form Variation of the DPO Corresponding to the Objective Scattering System**

In WEP framework, this dissertation realizes the construction for the DP-CMs of scattering system by orthogonalizing the corresponding DPO, and this dissertation provides some different manifestation forms of the DPO. In all of the manifestation forms, some forms are suitable for constructing the DP-CMs, but some other forms are not suitable. If the form which can be directly utilized to construct the DP-CMs is called as the objective form of the DPO, then this dissertation transforms the original form of DPO to the objective form of the DPO by a series of operations as follows: to reduce the dimension of the DPO in field-current space, to transform the DPO from field-current space to EM current space, to unify variables in ME current space, to transform the DPO from EM current space to expansion vector space, and to unify variables in expansion vector space.

**(2.1)** Based on Maxwell's equations, this dissertation transforms the high-dimensional interaction form $P_{\text{sca sys}}^{\text{driving}} = (1/2) < \vec{J}^{\text{SL}} + \vec{J}^{\text{SS}} + \vec{J}^{\text{SV}}, \vec{E}^{\text{inc}} >_{D_{\text{sca sys}}} + (1/2) < \vec{M}^{\text{SV}}, \vec{H}^{\text{inc}} >_{D_{\text{sca sys}}}$ of the DPO of scattering system into the low-dimensional interaction form $P_{\text{sca sys}}^{\text{driving}} = (1/2) < \vec{J}^{\text{SL}} - \vec{J}^{\text{EL}} + \vec{J}^{\text{SS}} - \vec{J}^{\text{ES}}, \vec{E}^{\text{inc}} >_{D_{\text{sca sys}}} - (1/2) < \vec{M}^{\text{ES}}, \vec{H}^{\text{inc}} >_{D_{\text{sca sys}}}$. The reason to call the above two expressions as interaction forms is that they manifest themselves as the interactions between fields and sources. The reason to call the former as high-dimensional form and to call the latter as low-dimensional form is that the former contains both {line sources, surface sources} and {volume sources}, but the latter only contains {line sources, surface sources}. In this dissertation, the above process is particularly called as "to reduce the dimension of DPO".

**(2.2)** Inserting the previous partial field GFHF into the above low-dimensional interaction form, this dissertation obtains the DPO expression $P_{\text{sca sys}}^{\text{driving}} = (1/2) < \vec{J}^{\text{SL}} - \vec{J}^{\text{EL}} + \vec{J}^{\text{SS}} - \vec{J}^{\text{ES}}, \mathcal{E}_0(-\vec{J}_0^{\text{SL}} - \vec{J}_0^{\text{SS}} - \vec{J}_{0\cap}^{\text{SS}} + \vec{J}_{0\cap}^{\text{ES}} + \vec{J}_0^{\text{ES}}, \vec{M}_0^{\text{ES}}) >_{D_{\text{sca sys}}}$, which only contains low-dimensional EM currents $\{\vec{J}^{\text{SL}}, \vec{J}^{\text{EL}}, \vec{J}^{\text{SS}}, \vec{J}^{\text{ES}}, \vec{M}^{\text{ES}}\}$. Because the expression is the function of various EM currents, then this dissertation calls it as the original EM current form (or simply called as original current form) of DPO. In this dissertation, the above process is particularly called as "to transform DPO from field-current space to EM current space".

**(2.3)** In the process of establishing the GFHF for scattering system, we decompose $\{\vec{J}^{\text{SL}}, \vec{J}^{\text{EL}}, \vec{J}^{\text{SS}}, \vec{J}^{\text{ES}}, \vec{M}^{\text{ES}}\}$, and obtain the relationships among various sub-currents. Inserting the relationships into the above original EM current form, we obtain $P_{\text{sca sys}}^{\text{driving}} = (1/2) < \vec{J}_0^{\text{SL}} + \vec{J}_0^{\text{SS}} + \vec{J}_{0\cap}^{\text{SS}} - \vec{J}_{0\cap}^{\text{ES}} - \vec{J}_0^{\text{ES}}, \mathcal{E}_0(-\vec{J}_0^{\text{SL}} - \vec{J}_0^{\text{SS}} - \vec{J}_{0\cap}^{\text{SS}} + \vec{J}_{0\cap}^{\text{ES}} + \vec{J}_0^{\text{ES}}, \vec{M}_0^{\text{ES}}) >_{D_{\text{sca sys}}} + (1/2) < -\vec{M}_0^{\text{ES}}, \mathcal{H}_0(-\vec{J}_0^{\text{SL}} - \vec{J}_0^{\text{SS}} - \vec{J}_{0\cap}^{\text{SS}} + \vec{J}_{0\cap}^{\text{ES}} + \vec{J}_0^{\text{ES}}, \vec{M}_0^{\text{ES}}) >_{D_{\text{sca sys}}}$. This dissertation calls the obtained result as the simplified EM current form of DPO, to be distinguished from the original EM current form. In addition, this dissertation calls the above process to transform the original EM current form to the simplified EM current form as "to unify variables in EM current space".

**(2.4)** By expanding the EM currents contained in the simplified EM current form in terms of some proper basis functions, this dissertation discretizes the simplified EM current form into matrix form $P_{\text{sca sys}}^{\text{driving}} = \bar{a}^H \cdot \bar{\bar{P}}_{\text{sca sys}}^{\text{driving}} \cdot \bar{a}$, and calls $P_{\text{sca sys}}^{\text{driving}} = \bar{a}^H \cdot \bar{\bar{P}}_{\text{sca sys}}^{\text{driving}} \cdot \bar{a}$ as the original matrix form of DPO. In this dissertation, the above process is particularly called as "to transform DPO from EM current space to expansion vector space"

**(2.5)** Obviously, the $\{\vec{J}_0^{\text{SL}}, \vec{J}_0^{\text{SS}}, \vec{J}_{0\cap}^{\text{SS}}, \vec{J}_{0\cap}^{\text{ES}}, \vec{J}_0^{\text{ES}}, \vec{M}_0^{\text{ES}}\}$ used to express the simplified

EM current form includes whole BV set $\{\vec{J}_0^{\text{SL}},\vec{J}_0^{\text{SS}},\vec{J}_{0\cap}^{\text{SS}},\vec{J}_{0\cap}^{\text{ES}},\vec{J}_0^{\text{ES}}/\vec{M}_0^{\text{ES}}\}$, and at the same time includes DV set $\{\vec{M}_0^{\text{ES}}/\vec{J}_0^{\text{ES}}\}$. In EM current space, it is difficult to establish the transformation from $\{\vec{J}_0^{\text{SL}},\vec{J}_0^{\text{SS}},\vec{J}_{0\cap}^{\text{SS}},\vec{J}_{0\cap}^{\text{ES}},\vec{J}_0^{\text{ES}}/\vec{M}_0^{\text{ES}}\}$ to $\{\vec{M}_0^{\text{ES}}/\vec{J}_0^{\text{ES}}\}$, so this dissertation selects to establish the transformation in expansion vector space. By inserting the obtained transformation into the original matrix form, this dissertation obtains the matrix form $P_{\text{sca sys}}^{\text{driving}} = \bar{a}_{\text{BV}}^H \cdot \bar{\bar{P}}_{\text{BV}}^{\text{driving}} \cdot \bar{a}_{\text{BV}}$ which only contains BVs, and calls $P_{\text{sca sys}}^{\text{driving}} = \bar{a}_{\text{BV}}^H \cdot \bar{\bar{P}}_{\text{BV}}^{\text{driving}} \cdot \bar{a}_{\text{BV}}$ as the simplified matrix form of DPO. In this dissertation, the above process is particularly called as "to unify variables in expansion vector space".

In fact, the above simplified matrix form is just the objective form which can be directly utilized to construct DP-CMs.

### 3) The Flow Charts of Variable Unification and DPO Evolution

In summary, "the process to unify variables for scattering system" and "the evolution process of the manifestation form of DPO" can be more concisely illustrated in the following Figure 6-19.

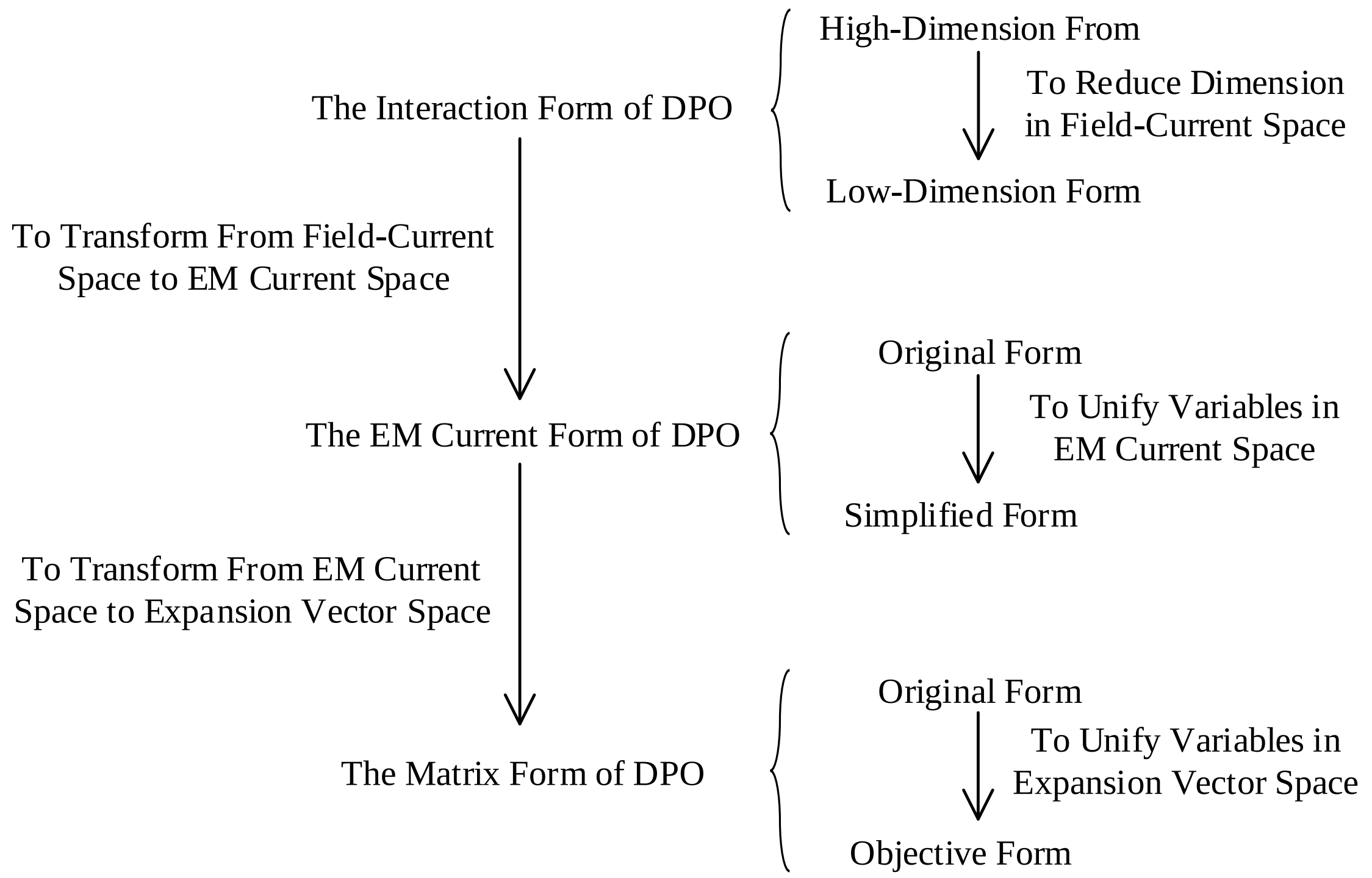

Figure 6-19 "The process to unify variables for scattering system" and "the evolution process of the manifestation form of DPO"

In addition, we also list the formulation indexes of the various manifestation forms of the DPOs corresponding to various scattering systems in Table 6-1.

Table 6-1 The various manifestation forms of the DPOs corresponding to various scattering systems

<table>
<tr><th rowspan="2">ScaSys / DPO</th><th rowspan="2">MetSca</th><th colspan="3">MatSca</th><th rowspan="2">ComSca</th></tr>
<tr><th>$\boldsymbol{V}_{\text{sim sys}}$</th><th>$\boldsymbol{V}_{\text{ss sys}}$</th><th>$\boldsymbol{V}_{\text{sm sys}}$</th></tr>
<tr><td>High-Dimension Interaction Form</td><td>The RHS of the 1st equality in (3-2)</td><td>The RHS of the 1st equality in (4-31)</td><td>The RHS of the 2nd equality in (4-65)</td><td>The RHS of the 2nd equality in (4-140)</td><td>Formulation (5-64)</td></tr>
<tr><td>↓</td><td colspan="5">To Unify Variables in Field-Current Space</td></tr>
<tr><td>Low-Dimension Interaction Form</td><td>Ditto</td><td>The RHS of the 1st equality in (4-32)</td><td>The RHS of the 3rd equality in (4-65)</td><td>The RHS of the 3rd equality in (4-140)</td><td>The RHS of 1st equality in (5-66)</td></tr>
<tr><td>↓</td><td colspan="5">To Transform From Field-Current Space to EM Current Space</td></tr>
<tr><td>Original EM Current Form</td><td>Formulation (3-3)</td><td>The RHS of the 2nd equality in (4-32)</td><td>The RHS of 1st equality in (4-117)</td><td>Formulation (4-143)</td><td>It is not explicitly provided in this dissertation</td></tr>
<tr><td>↓</td><td colspan="5">To Unify Variables in EM Current Space</td></tr>
<tr><td>Simplified EM Current Form</td><td>Ditto</td><td>Ditto</td><td>The RHS of the 5th equality in (4-117)</td><td>Formulation (4-146)</td><td>The RHS of the 3rd equality in (5-68)</td></tr>
<tr><td>↓</td><td colspan="5">To Transform From EM Current Space to Expansion Vector Space</td></tr>
<tr><td>Original Matrix Form</td><td>Formulation (3-5)</td><td>Formulation (4-34)</td><td>Formulation (4-118)</td><td>Formulation (4-147)</td><td>Formulation (5-69)</td></tr>
<tr><td>↓</td><td colspan="5">To Unify Variables in Expansion Vector Space</td></tr>
<tr><td>Objective Matrix Form</td><td>Ditto</td><td>Formulation (4-52)</td><td>Formulation (4-122)</td><td>Formulation (4-148)</td><td>Formulation (5-76)</td></tr>
</table>

## 6.3 A Compromise Scheme for Suppressing Spurious Modes

The Chapter 4, Chapter 5, and Section 6.2 of this dissertation have clearly revealed the reason why some CM calculation formulations output spurious modes —— the DPO contains some DVs, and have detailedly provided the systematical scheme to suppress the spurious modes (i.e. the systematical scheme to unify variables) —— to express all of the DVs contained in the DPO as the functions of basic variables such that the DPO doesn't contain any DV.

As illustrated by the typical numerical examples given in the Chapter 4, Chapter 5, and Section 6.2 of this dissertation, the above variable unification scheme indeed has ability to effectively suppress spurious modes. But, the scheme needs to calculate the inverses of some matrices during the process to unify variables, and it requires excessive computational resources.

In this section, we provide a compromise scheme for suppressing the spurious modes. In the premise of not calculating matrix inverses, the compromise scheme can effectively suppress the low-order spurious modes. But, it is possible that the compromise scheme will output some high-order spurious modes, and this is just the reason why this dissertation calls the scheme as "compromise scheme".

### 6.3.1 One-body Material System Case

In this subsection, we consider the scattering problem related to a simply connected material system shown in Figure 6-20, where the various symbols have the same meanings as the ones used in Section 4.3.

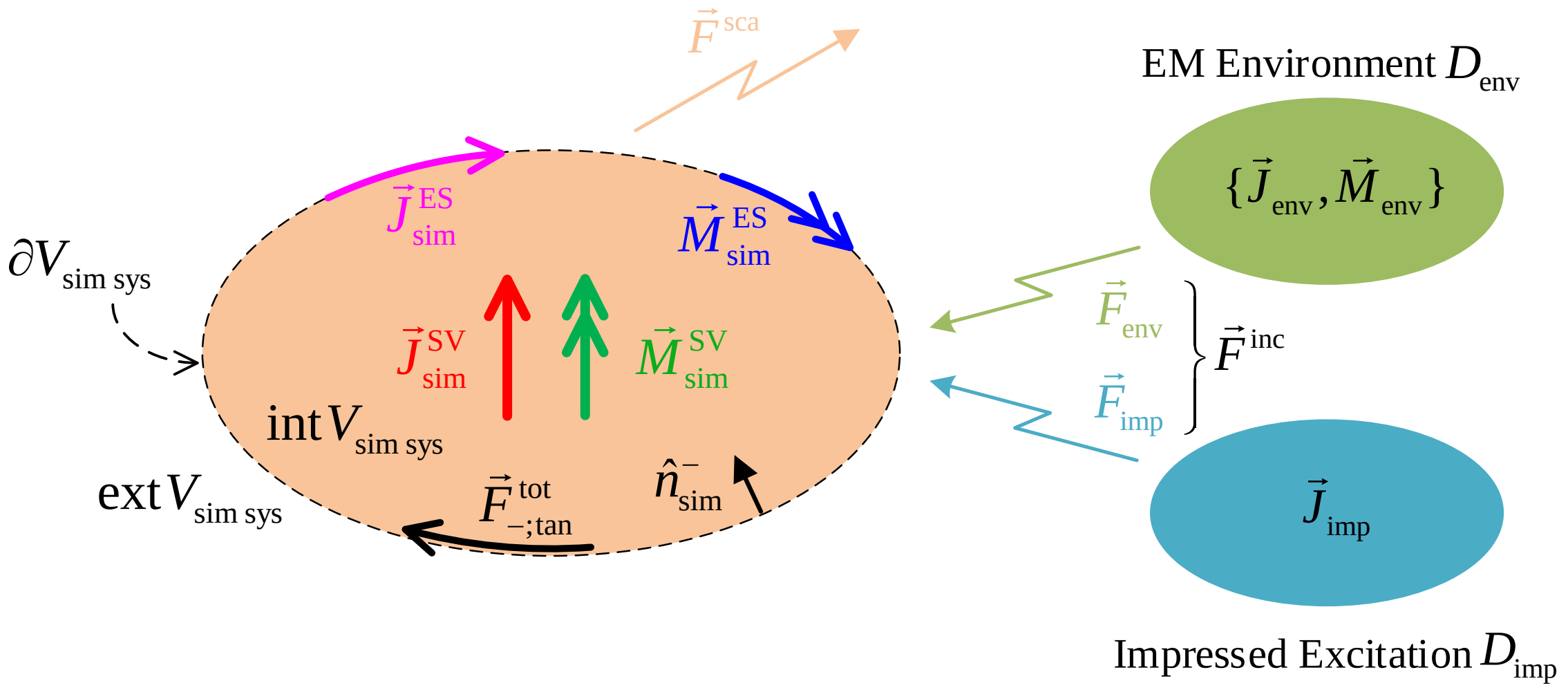

Figure 6-20 The scattering problem related to a simply connected material system

**1) The Source-Field Relationships Directly Related to Various Scattered Volume Sources and Equivalent Surface Sources**

For the simply connected material system $V_{\text{sim sys}}$ shown in Figure 6-20, all of the two groups of source-field relationships directly related to scattered volume sources $\{\vec{J}_{\text{sim}}^{\text{SV}},\vec{M}_{\text{sim}}^{\text{SV}}\}$ are as follows:

$$\text{int}V_{\text{sim sys}}\bigcup\text{ext}V_{\text{sim sys}}\ :\ \ \vec{F}^{\text{sca}}=\overleftrightarrow{G}_0^{JF}*\vec{J}_{\text{sim}}^{\text{SV}}+\overleftrightarrow{G}_0^{MF}*\vec{M}_{\text{sim}}^{\text{SV}}\tag{6-22}$$

and

$$\text{int}V_{\text{sim sys}}\ :\ \ \vec{E}^{\text{tot}}=\left(j\omega\Delta\overleftrightarrow{\varepsilon}_{\text{sim}}^{\text{c}}\right)^{-1}\cdot\vec{J}_{\text{sim}}^{\text{SV}}\tag{6-23a}$$

$$\text{int}V_{\text{sim sys}}\ :\ \ \vec{H}^{\text{tot}}=\left(j\omega\Delta\overleftrightarrow{\mu}_{\text{sim}}\right)^{-1}\cdot\vec{M}_{\text{sim}}^{\text{SV}}\tag{6-23b}$$

and their detailed derivations can be found in Appendix A. In source-field relationship (6-22), $F=E,H$; $\overleftrightarrow{G}_0^{JF}$ and $\overleftrightarrow{G}_0^{MF}$ are vacuum Green's functions.

For the simply connected material system $V_{\text{sim sys}}$ shown in Figure 6-20, all of the three groups of source-field relationships directly related to equivalent surface sources $\{\vec{J}_{\text{sim}}^{\text{ES}},\vec{M}_{\text{sim}}^{\text{ES}}\}$ are as follows:

$$\text{int}V_{\text{sim sys}}\ :\ \ \vec{F}^{\text{inc}}=\overleftrightarrow{G}_0^{JF}*\vec{J}_{\text{sim}}^{\text{ES}}+\overleftrightarrow{G}_0^{MF}*\vec{M}_{\text{sim}}^{\text{ES}}\tag{6-24}$$

$$\text{ext}V_{\text{sim sys}}\ :\ -\vec{F}^{\text{sca}}=\overleftrightarrow{G}_0^{JF}*\vec{J}_{\text{sim}}^{\text{ES}}+\overleftrightarrow{G}_0^{MF}*\vec{M}_{\text{sim}}^{\text{ES}}\tag{6-25}$$

$$\text{int}V_{\text{sim sys}}\ :\ \ \vec{F}^{\text{tot}}=\overleftrightarrow{G}_{\text{sim}}^{JF}*\vec{J}_{\text{sim}}^{\text{ES}}+\overleftrightarrow{G}_{\text{sim}}^{MF}*\vec{M}_{\text{sim}}^{\text{ES}}\tag{6-26}$$

and their detailed derivations can be found in Appendix C.

**2) Variable Unification Based Scheme for Suppressing Spurious Modes**

For the simply connected material body shown in Figure 6-20, the total fields $\{\vec{E}^{\text{tot}},\vec{H}^{\text{tot}}\}$ distributing on $\text{int}V_{\text{sim sys}}$ satisfy the following Maxwell's equations:

$$\left.\begin{aligned}\nabla\times\vec{H}^{\text{tot}}(\vec{r})&=\ \ j\omega\overleftrightarrow{\varepsilon}_{\text{sim}}^{\text{c}}(\vec{r})\cdot\vec{E}^{\text{tot}}(\vec{r})\\\nabla\times\vec{E}^{\text{tot}}(\vec{r})&=-j\omega\overleftrightarrow{\mu}_{\text{sim}}^{\text{c}}(\vec{r})\cdot\vec{H}^{\text{tot}}(\vec{r})\end{aligned}\right.\ ,\quad\vec{r}\in\text{int}V_{\text{sim sys}}\tag{6-27}$$

and their detailed derivations can be found in literature [121] and the Appendix A of this dissertation. Thus, source-field relationship (6-23) implies that scattered volume electric current $\vec{J}_{\text{sim}}^{\text{SV}}$ and scattered volume magnetic current $\vec{M}_{\text{sim}}^{\text{SV}}$ satisfy the following relationships:

$$\nabla\times\left\{\left[j\omega\Delta\overleftrightarrow{\mu}_{\text{sim}}(\vec{r})\right]^{-1}\cdot\vec{M}_{\text{sim}}^{\text{SV}}(\vec{r})\right\}=\ \ j\omega\overleftrightarrow{\varepsilon}_{\text{sim}}^{\text{c}}(\vec{r})\cdot\left[j\omega\Delta\overleftrightarrow{\varepsilon}_{\text{sim}}^{\text{c}}(\vec{r})\right]^{-1}\cdot\vec{J}_{\text{sim}}^{\text{SV}}(\vec{r})\tag{6-28a}$$

$$\nabla\times\left\{\left[j\omega\Delta\overleftrightarrow{\varepsilon}_{\text{sim}}^{\text{c}}(\vec{r})\right]^{-1}\cdot\vec{J}_{\text{sim}}^{\text{SV}}(\vec{r})\right\}=-j\omega\overleftrightarrow{\mu}_{\text{sim}}^{\text{c}}(\vec{r})\cdot\left[j\omega\Delta\overleftrightarrow{\mu}_{\text{sim}}(\vec{r})\right]^{-1}\cdot\vec{M}_{\text{sim}}^{\text{SV}}(\vec{r})\tag{6-28b}$$

where $\vec{r} \in \text{int}V_{\text{sim sys}}$.

For the simply connected material body shown in Figure 6-20, the tangential total fields $\{\vec{E}_{-;\tan}^{\text{tot}}, \vec{H}_{-;\tan}^{\text{tot}}\}$ distributing on the inner surface of $\partial V_{\text{sim sys}}$ and equivalent surface sources $\{\vec{J}_{\text{sim}}^{\text{ES}}, \vec{M}_{\text{sim}}^{\text{ES}}\}$ satisfy the following relationships:

$$\vec{J}_{\text{sim}}^{\text{ES}}(\vec{r}) = \hat{n}_{\text{sim}}^{-}(\vec{r}) \times \vec{H}_{-;\tan}^{\text{tot}}(\vec{r}) \quad , \quad \vec{r} \in \partial V_{\text{sim sys}} \tag{6-29a}$$

$$\vec{M}_{\text{sim}}^{\text{ES}}(\vec{r}) = \vec{E}_{-;\tan}^{\text{tot}}(\vec{r}) \times \hat{n}_{\text{sim}}^{-}(\vec{r}) \quad , \quad \vec{r} \in \partial V_{\text{sim sys}} \tag{6-29b}$$

and their detailed derivations can be found in Appendix C. Thus source-field relationship (6-26) implies that equivalent surface electric current $\vec{J}_{\text{sim}}^{\text{ES}}$ and equivalent surface magnetic current $\vec{M}_{\text{sim}}^{\text{ES}}$ satisfy the following relationships:

$$\vec{J}_{\text{sim}}^{\text{ES}}(\vec{r}) \times \hat{n}_{\text{sim}}^{-}(\vec{r}) = \left[ \overleftrightarrow{G}_{\text{sim}}^{JH} * \vec{J}_{\text{sim}}^{\text{ES}} + \overleftrightarrow{G}_{\text{sim}}^{MH} * \vec{M}_{\text{sim}}^{\text{ES}} \right]_{\vec{r}_{\text{sim}} \to \vec{r}}^{\tan} \quad , \quad \vec{r} \in \partial V_{\text{sim sys}} \tag{6-30a}$$

$$\hat{n}_{\text{sim}}^{-}(\vec{r}) \times \vec{M}_{\text{sim}}^{\text{ES}}(\vec{r}) = \left[ \overleftrightarrow{G}_{\text{sim}}^{JE} * \vec{J}_{\text{sim}}^{\text{ES}} + \overleftrightarrow{G}_{\text{sim}}^{ME} * \vec{M}_{\text{sim}}^{\text{ES}} \right]_{\vec{r}_{\text{sim}} \to \vec{r}}^{\tan} \quad , \quad \vec{r} \in \partial V_{\text{sim sys}} \tag{6-30b}$$

i.e. the relationship (4-37) given in this dissertation, where $\vec{r}_{\text{sim}} \in \text{int}V_{\text{sim sys}}$.

Above relationship (6-28) implies that $\vec{J}_{\text{sim}}^{\text{SV}}$ and $\vec{M}_{\text{sim}}^{\text{SV}}$ are not independent of each other. Above relationship (6-30) implies that $\vec{J}_{\text{sim}}^{\text{ES}}$ and $\vec{M}_{\text{sim}}^{\text{ES}}$ are not independent of each other. Thus, before constructing DP-CMs by orthogonalizing the DPO with variables $\{\vec{J}_{\text{sim}}^{\text{ES}}, \vec{M}_{\text{sim}}^{\text{ES}}\}$, it is necessary to establish the relationship between $\vec{J}_{\text{sim}}^{\text{ES}}$ and $\vec{M}_{\text{sim}}^{\text{ES}}$, or it will output many spurious modes (for details see the Sections 4.3 and 6.2 of this dissertation). But, before constructing DP-CMs by orthogonalizing the DPO with variables $\{\vec{J}_{\text{sim}}^{\text{SV}}, \vec{M}_{\text{sim}}^{\text{SV}}\}$, the Section 4.2 of this dissertation doesn't establish the relationship between $\vec{J}_{\text{sim}}^{\text{SV}}$ and $\vec{M}_{\text{sim}}^{\text{SV}}$, and the numerical examples given in Subsection 4.2.4 don't output low-order spurious modes.

### 3) Some Valuable Numerically Experimental Phenomena

Based on the above analysis and the numerical examples given in the Sections 4.2, 4.3, and 6.2 of this dissertation, it is easy to find out the following phenomena:

**Phenomenon a.** The volume formulation provided in the Section 4.2 of this dissertation doesn't establish the relationship between $\vec{J}_{\text{sim}}^{\text{SV}}$ and $\vec{M}_{\text{sim}}^{\text{SV}}$, but it simultaneously utilizes all of the two source-field relationships (6-22) and (6-23) directly related to $\vec{J}_{\text{sim}}^{\text{SV}}$ and $\vec{M}_{\text{sim}}^{\text{SV}}$, and then it successfully suppresses the low-order spurious modes;

**Phenomenon b1.** The surface formulation provided in literature [34] doesn't establish the relationship between $\vec{J}_{\text{sim}}^{\text{ES}}$ and $\vec{M}_{\text{sim}}^{\text{ES}}$, but it simultaneously utilizes two source-field relationships (6-25) and (6-26) directly related to $\vec{J}_{\text{sim}}^{\text{SV}}$ and $\vec{M}_{\text{sim}}^{\text{SV}}$, and then it successfully realizes that the outputted results contain enough low-order physical modes;

**Phenomenon b2.** The surface formulation provided in literature [34] properly establishes the relationship between $\vec{J}_{\text{sim}}^{\text{ES}}$ and $\vec{M}_{\text{sim}}^{\text{ES}}$, and then it can guarantee that the outputted results contain enough low-order physical modes and don't contain any low-order spurious mode;

**Phenomenon c1.** If the DPO surface formulation (4-32) provided in the Section 4.3 of this dissertation doesn't employ the relationship between $\vec{J}_{\text{sim}}^{\text{ES}}$ and $\vec{M}_{\text{sim}}^{\text{ES}}$, it will not be able to output enough low-order physical modes;

**Phenomenon c2.** If the DPO surface formulation (4-32) provided in the Section 4.3 of this dissertation employs a proper relationship between $\vec{J}_{\text{sim}}^{\text{ES}}$ and $\vec{M}_{\text{sim}}^{\text{ES}}$, it can guarantee that the outputted results contain enough low-order physical modes and don't contain any low-order spurious mode.

**4) A Compromise Scheme for Suppressing Spurious Modes**

Just based on the above observations, this dissertation proposes a compromise scheme for suppressing the spurious modes outputted by the surface CM calculation formulations of simply connected material bodies as follows —— to simultaneously utilize all of the three source-field relationships directly related to $\vec{J}_{\text{sim}}^{\text{ES}}$ and $\vec{M}_{\text{sim}}^{\text{ES}}$ and at the same time not to utilize the relationship between $\vec{J}_{\text{sim}}^{\text{ES}}$ and $\vec{M}_{\text{sim}}^{\text{ES}}$.

For the simply connected material body shown in Figure 6-20, its DPO has two different expressions as follows:

$$\begin{aligned} P_{\text{sim sys}}^{\text{driving}} &= (1/2)\left\langle -\vec{J}_{\text{sim}}^{\text{ES}}, \vec{E}_{-}^{\text{tot}} - \vec{E}_{+}^{\text{sca}} \right\rangle_{\partial V_{\text{sim sys}}} + (1/2)\left\langle -\vec{M}_{\text{sim}}^{\text{ES}}, \vec{H}_{-}^{\text{tot}} - \vec{H}_{+}^{\text{sca}} \right\rangle_{\partial V_{\text{sim sys}}} \\ &= -(1/2)\left\langle \vec{J}_{\text{sim}}^{\text{ES}}, \vec{E}^{\text{inc}} \right\rangle_{\partial V_{\text{sim sys}}} - (1/2)\left\langle \vec{M}_{\text{sim}}^{\text{ES}}, \vec{H}^{\text{inc}} \right\rangle_{\partial V_{\text{sim sys}}} \end{aligned} \quad (6\text{-}31)$$

i.e. the formulation (4-28) and the formulation (4-31) given in the Section 4.3 of this dissertation. In formulation (6-31), the RHS of the first equality involves source-field relationships (6-25) and (6-26), the RHS of the second equality involves source-field relationship (6-24). Inserting expansion formulation (4-33) into above DPO surface formulation (6-31), the surface formulation is immediately discretized into matrix form

$$P_{\text{sim sys}}^{\text{driving}} = \begin{bmatrix} \bar{a}^{J_s} \\ \bar{a}^{M_s} \end{bmatrix}^H \cdot \bar{\bar{P}}_{\text{sim sys;1}}^{\text{driving}} \cdot \begin{bmatrix} \bar{a}^{J_s} \\ \bar{a}^{M_s} \end{bmatrix} = \begin{bmatrix} \bar{a}^{J_s} \\ \bar{a}^{M_s} \end{bmatrix}^H \cdot \bar{\bar{P}}_{\text{sim sys;2}}^{\text{driving}} \cdot \begin{bmatrix} \bar{a}^{J_s} \\ \bar{a}^{M_s} \end{bmatrix} \quad (6\text{-}32)$$

Here, the method to calculate matrix $\overline{\overline{P}}{}^{\text{driving}}_{\text{sim sys;1}}$ has been given in literature [44], and matrix $\overline{\overline{P}}{}^{\text{driving}}_{\text{sim sys;2}}$ is just the matrix $\overline{\overline{P}}{}^{\text{driving}}_{\text{sim sys}}$ given in the formulation (4-34) of this dissertation. Obviously, in the process to obtain above matrix $\overline{\overline{P}}{}^{\text{driving}}_{\text{sim sys;1}}$, we need to utilize source-field relationships (6-25) and (6-26); in the process to obtain above matrix $\overline{\overline{P}}{}^{\text{driving}}_{\text{sim sys;2}}$, we need to utilize source-field relationship (6-24). In what follows, we try to establish a set of characteristic equations which simultaneously involve matrices $\overline{\overline{P}}{}^{\text{driving}}_{\text{sim sys;1}}$ and $\overline{\overline{P}}{}^{\text{driving}}_{\text{sim sys;2}}$, to achieve the destination —— the established characteristic value problem simultaneously involves all of the three source-field relationships directly related to $\vec{J}^{\text{ES}}_{\text{sim}}$ and $\vec{M}^{\text{ES}}_{\text{sim}}$.

The Sections 4.2 and 4.3 of this dissertation have proven that DP-CMs $\{\bar{\alpha}^{J_{\text{s}}}_{\xi},\bar{\alpha}^{M_{\text{s}}}_{\xi}\}$ have ability to orthogonalize the following four matrices simultaneously:

$$\begin{bmatrix}\bar{\alpha}^{J_{\text{s}}}_1 & \cdots & \bar{\alpha}^{J_{\text{s}}}_{\Xi^{\text{BV}}}\\ \bar{\alpha}^{M_{\text{s}}}_1 & \cdots & \bar{\alpha}^{M_{\text{s}}}_{\Xi^{\text{BV}}}\end{bmatrix}^H \cdot \overline{\overline{P}}{}^{\text{driving}}_{\text{sim sys;1;+}} \cdot \begin{bmatrix}\bar{\alpha}^{J_{\text{s}}}_1 & \cdots & \bar{\alpha}^{J_{\text{s}}}_{\Xi^{\text{BV}}}\\ \bar{\alpha}^{M_{\text{s}}}_1 & \cdots & \bar{\alpha}^{M_{\text{s}}}_{\Xi^{\text{BV}}}\end{bmatrix} = \begin{bmatrix}\text{Re}\left\{P^{\text{driving}}_{\text{sim sys;1}}\right\} & & \\ & \ddots & \\ & & \text{Re}\left\{P^{\text{driving}}_{\text{sim sys;}\Xi^{\text{BV}}}\right\}\end{bmatrix} \tag{6-33a}$$

$$\begin{aligned}\begin{bmatrix}\bar{\alpha}^{J_{\text{s}}}_1 & \cdots & \bar{\alpha}^{J_{\text{s}}}_{\Xi^{\text{BV}}}\\ \bar{\alpha}^{M_{\text{s}}}_1 & \cdots & \bar{\alpha}^{M_{\text{s}}}_{\Xi^{\text{BV}}}\end{bmatrix}^H \cdot \overline{\overline{P}}{}^{\text{driving}}_{\text{sim sys;1;}-} \cdot \begin{bmatrix}\bar{\alpha}^{J_{\text{s}}}_1 & \cdots & \bar{\alpha}^{J_{\text{s}}}_{\Xi^{\text{BV}}}\\ \bar{\alpha}^{M_{\text{s}}}_1 & \cdots & \bar{\alpha}^{M_{\text{s}}}_{\Xi^{\text{BV}}}\end{bmatrix} &= \begin{bmatrix}\text{Im}\left\{P^{\text{driving}}_{\text{sim sys;1}}\right\} & & \\ & \ddots & \\ & & \text{Im}\left\{P^{\text{driving}}_{\text{sim sys;}\Xi^{\text{BV}}}\right\}\end{bmatrix}\\ &= \begin{bmatrix}\lambda_\xi\,\text{Re}\left\{P^{\text{driving}}_{\text{sim sys;1}}\right\} & & \\ & \ddots & \\ & & \lambda_\xi\,\text{Re}\left\{P^{\text{driving}}_{\text{sim sys;}\Xi^{\text{BV}}}\right\}\end{bmatrix}\end{aligned} \tag{6-33b}$$

$$\begin{bmatrix}\bar{\alpha}^{J_{\text{s}}}_1 & \cdots & \bar{\alpha}^{J_{\text{s}}}_{\Xi^{\text{BV}}}\\ \bar{\alpha}^{M_{\text{s}}}_1 & \cdots & \bar{\alpha}^{M_{\text{s}}}_{\Xi^{\text{BV}}}\end{bmatrix}^H \cdot \overline{\overline{P}}{}^{\text{driving}}_{\text{sim sys;2;+}} \cdot \begin{bmatrix}\bar{\alpha}^{J_{\text{s}}}_1 & \cdots & \bar{\alpha}^{J_{\text{s}}}_{\Xi^{\text{BV}}}\\ \bar{\alpha}^{M_{\text{s}}}_1 & \cdots & \bar{\alpha}^{M_{\text{s}}}_{\Xi^{\text{BV}}}\end{bmatrix} = \begin{bmatrix}\text{Re}\left\{P^{\text{driving}}_{\text{sim sys;1}}\right\} & & \\ & \ddots & \\ & & \text{Re}\left\{P^{\text{driving}}_{\text{sim sys;}\Xi^{\text{BV}}}\right\}\end{bmatrix} \tag{6-34a}$$

$$\begin{aligned}\begin{bmatrix}\bar{\alpha}^{J_{\text{s}}}_1 & \cdots & \bar{\alpha}^{J_{\text{s}}}_{\Xi^{\text{BV}}}\\ \bar{\alpha}^{M_{\text{s}}}_1 & \cdots & \bar{\alpha}^{M_{\text{s}}}_{\Xi^{\text{BV}}}\end{bmatrix}^H \cdot \overline{\overline{P}}{}^{\text{driving}}_{\text{sim sys;2;}-} \cdot \begin{bmatrix}\bar{\alpha}^{J_{\text{s}}}_1 & \cdots & \bar{\alpha}^{J_{\text{s}}}_{\Xi^{\text{BV}}}\\ \bar{\alpha}^{M_{\text{s}}}_1 & \cdots & \bar{\alpha}^{M_{\text{s}}}_{\Xi^{\text{BV}}}\end{bmatrix} &= \begin{bmatrix}\text{Im}\left\{P^{\text{driving}}_{\text{sim sys;1}}\right\} & & \\ & \ddots & \\ & & \text{Im}\left\{P^{\text{driving}}_{\text{sim sys;}\Xi^{\text{BV}}}\right\}\end{bmatrix}\\ &= \begin{bmatrix}\lambda_\xi\,\text{Re}\left\{P^{\text{driving}}_{\text{sim sys;1}}\right\} & & \\ & \ddots & \\ & & \lambda_\xi\,\text{Re}\left\{P^{\text{driving}}_{\text{sim sys;}\Xi^{\text{BV}}}\right\}\end{bmatrix}\end{aligned} \tag{6-34b}$$

where $\lambda_\xi = \mathrm{Im}\{P^{\text{driving}}_{\text{sim sys};\xi}\} / \mathrm{Re}\{P^{\text{driving}}_{\text{sim sys};\xi}\}$ is a real number, and

$$\overline{\overline{P}}{}^{\text{driving}}_{\text{sim sys;1;+}} = \frac{1}{2}\left[\overline{\overline{P}}{}^{\text{driving}}_{\text{sim sys;1}} + \left(\overline{\overline{P}}{}^{\text{driving}}_{\text{sim sys;1}}\right)^H\right] \tag{6-35a}$$

$$\overline{\overline{P}}{}^{\text{driving}}_{\text{sim sys;1;}-} = \frac{1}{2j}\left[\overline{\overline{P}}{}^{\text{driving}}_{\text{sim sys;1}} - \left(\overline{\overline{P}}{}^{\text{driving}}_{\text{sim sys;1}}\right)^H\right] \tag{6-35b}$$

$$\overline{\overline{P}}{}^{\text{driving}}_{\text{sim sys;2;+}} = \frac{1}{2}\left[\overline{\overline{P}}{}^{\text{driving}}_{\text{sim sys;2}} + \left(\overline{\overline{P}}{}^{\text{driving}}_{\text{sim sys;2}}\right)^H\right] \tag{6-36a}$$

$$\overline{\overline{P}}{}^{\text{driving}}_{\text{sim sys;2;}-} = \frac{1}{2j}\left[\overline{\overline{P}}{}^{\text{driving}}_{\text{sim sys;2}} - \left(\overline{\overline{P}}{}^{\text{driving}}_{\text{sim sys;2}}\right)^H\right] \tag{6-36b}$$

Based on above orthogonalities (6-33) and (6-34), and inspired by literature [125], this dissertation proposes a series of new generalized characteristic equations for constructing the DP-CMs of the simply connected material body shown in Figure 6-20, as follows:

$$\left[\overbrace{\left(\overline{\overline{P}}{}^{\text{driving}}_{\text{sim sys};m;+} + j\,\overline{\overline{P}}{}^{\text{driving}}_{\text{sim sys};m;-}\right)}^{\overline{\overline{P}}{}^{\text{driving}}_{\text{sim sys};m}} + \vartheta\,\overline{\overline{P}}{}^{\text{driving}}_{\text{sim sys};n;-}\right]\cdot\bar{\alpha}_\xi = \left[1+\left(j+\vartheta\right)\lambda_\xi\right]\overline{\overline{P}}{}^{\text{driving}}_{\text{sim sys};n;+}\cdot\bar{\alpha}_\xi \tag{6-37a}$$

or

$$\left[\underbrace{\left(\overline{\overline{P}}{}^{\text{driving}}_{\text{sim sys};m;+} + j\,\overline{\overline{P}}{}^{\text{driving}}_{\text{sim sys};m;-}\right)}_{\overline{\overline{P}}{}^{\text{driving}}_{\text{sim sys};m}} + \vartheta\,\overline{\overline{P}}{}^{\text{driving}}_{\text{sim sys};n;+}\right]\cdot\bar{\alpha}_\xi = \left[\left(1+\vartheta\right)\frac{1}{\lambda_\xi}+j\right]\overline{\overline{P}}{}^{\text{driving}}_{\text{sim sys};n;-}\cdot\bar{\alpha}_\xi \tag{6-37b}$$

Here, $j$ is imaginary unit; $\vartheta$ is an adjustable constant scalar, but $\vartheta \neq -j,-1$; ordered numbers $(m,n)$ satisfy that $(m,n)=(1,2)$ or $(2,1)$ ; $\bar{\alpha}_\xi$ is the column vector constituted by $\bar{\alpha}_\xi^{J_s}$ and $\bar{\alpha}_\xi^{M_s}$ .

In what follows, we, under some special cases of $(m,n)$ and $\vartheta$, give the special manifestation forms of new generalized characteristic equation (6-37).

**Case 1.** When $(m,n)=(1,2)$ and $\vartheta=0$, new generalized characteristic equations (6-37a) and (6-37b) degenerate to the following forms:

$$\overline{\overline{P}}{}^{\text{driving}}_{\text{sim sys;1}}\cdot\bar{\alpha}_\xi = \left(1+j\lambda_\xi\right)\overline{\overline{P}}{}^{\text{driving}}_{\text{sim sys;2;+}}\cdot\bar{\alpha}_\xi \tag{6-38a}$$

$$\overline{\overline{P}}{}^{\text{driving}}_{\text{sim sys;1}}\cdot\bar{\alpha}_\xi = \left(\frac{1}{\lambda_\xi}+j\right)\overline{\overline{P}}{}^{\text{driving}}_{\text{sim sys;2;}-}\cdot\bar{\alpha}_\xi \tag{6-38b}$$

**Case 2.** When $(m,n)=(1,2)$ and $\vartheta=j$, new generalized characteristic equations (6-37a) and (6-37b) degenerate to the following forms:

$$\left(\overline{\overline{P}}{}^{\text{driving}}_{\text{sim sys;1}} + j\,\overline{\overline{P}}{}^{\text{driving}}_{\text{sim sys;2;}-}\right)\cdot\bar{\alpha}_\xi = \left(1+j\,2\lambda_\xi\right)\overline{\overline{P}}{}^{\text{driving}}_{\text{sim sys;2;+}}\cdot\bar{\alpha}_\xi \tag{6-39a}$$

$$\left(\overline{\overline{P}}{}^{\text{driving}}_{\text{sim sys;1}} + j\,\overline{\overline{P}}{}^{\text{driving}}_{\text{sim sys;2;+}}\right)\cdot\bar{\alpha}_\xi = \left[\left(1+j\right)\frac{1}{\lambda_\xi}+j\right]\overline{\overline{P}}{}^{\text{driving}}_{\text{sim sys;2;}-}\cdot\bar{\alpha}_\xi \tag{6-39b}$$

**Case 3.** When $(m,n)=(1,2)$ and $\vartheta=1$, new generalized characteristic equations (6-37a) and (6-37b) degenerate to the following forms:

$$\left(\overline{\overline{P}}{}^{\text{driving}}_{\text{sim sys;1}}+\overline{\overline{P}}{}^{\text{driving}}_{\text{sim sys;2;}-}\right)\cdot\bar{\alpha}_\xi=\left[1+\left(1+j\right)\lambda_\xi\right]\overline{\overline{P}}{}^{\text{driving}}_{\text{sim sys;2;}+}\cdot\bar{\alpha}_\xi \tag{6-40a}$$

$$\left(\overline{\overline{P}}{}^{\text{driving}}_{\text{sim sys;1}}+\overline{\overline{P}}{}^{\text{driving}}_{\text{sim sys;2;}+}\right)\cdot\bar{\alpha}_\xi=\left(\frac{2}{\lambda_\xi}+j\right)\overline{\overline{P}}{}^{\text{driving}}_{\text{sim sys;2;}-}\cdot\bar{\alpha}_\xi \tag{6-40b}$$

At last, it needs to be emphasized that: above new characteristic equation (6-37) is not completely the same as the one proposed in literature [125].

## 6.3.2 Two-body Material System Case

Similarly to the single simply connected material body focused on in Subsection 6.3.1, we propose the new generalized characteristic equation for the two-body material system shown in Figure 4-27 as follows:

$$\left[\overbrace{\left(\overline{\overline{P}}{}^{\text{driving}}_{\text{ss sys};m;+}+j\,\overline{\overline{P}}{}^{\text{driving}}_{\text{ss sys};m;-}\right)}^{\overline{\overline{P}}{}^{\text{driving}}_{\text{ss sys};m}}+\vartheta\,\overline{\overline{P}}{}^{\text{driving}}_{\text{ss sys};n;-}\right]\cdot\bar{\alpha}_\xi=\left[1+\left(j+\vartheta\right)\lambda_\xi\right]\overline{\overline{P}}{}^{\text{driving}}_{\text{ss sys};n;+}\cdot\bar{\alpha}_\xi \tag{6-41a}$$

or

$$\left[\underbrace{\left(\overline{\overline{P}}{}^{\text{driving}}_{\text{ss sys};m;+}+j\,\overline{\overline{P}}{}^{\text{driving}}_{\text{ss sys};m;-}\right)}_{\overline{\overline{P}}{}^{\text{driving}}_{\text{ss sys};m}}+\vartheta\,\overline{\overline{P}}{}^{\text{driving}}_{\text{ss sys};n;+}\right]\cdot\bar{\alpha}_\xi=\left[\left(1+\vartheta\right)\frac{1}{\lambda_\xi}+j\right]\overline{\overline{P}}{}^{\text{driving}}_{\text{ss sys};n;-}\cdot\bar{\alpha}_\xi \tag{6-41b}$$

Here, $\vartheta$ is an adjustable constant scalar, but $\vartheta\neq -j,-1$; ordered numbers $(m,n)$ satisfy that $(m,n)=(1,2)$ or $(2,1)$; the above matrices can be obtained from decomposing matrices $\overline{\overline{P}}{}^{\text{driving}}_{\text{ss sys;1}}$ and $\overline{\overline{P}}{}^{\text{driving}}_{\text{ss sys;2}}$ by employing decomposition methods (6-35) and (6-36), and the methods to calculate matrices $\overline{\overline{P}}{}^{\text{driving}}_{\text{ss sys;1}}$ and $\overline{\overline{P}}{}^{\text{driving}}_{\text{ss sys;2}}$ have been provided in Sections 4.4 and 4.5, and they will not be repeated here.

Similarly to the single simply connected material body focused on in Subsection 6.3.1, we propose the new generalized characteristic equation for the two-body material system shown in Figure 4-61 as follows:

$$\left[\overbrace{\left(\overline{\overline{P}}{}^{\text{driving}}_{\text{sm sys};m;+}+j\,\overline{\overline{P}}{}^{\text{driving}}_{\text{sm sys};m;-}\right)}^{\overline{\overline{P}}{}^{\text{driving}}_{\text{sm sys};m}}+\vartheta\,\overline{\overline{P}}{}^{\text{driving}}_{\text{sm sys};n;-}\right]\cdot\bar{\alpha}_\xi=\left[1+\left(j+\vartheta\right)\lambda_\xi\right]\overline{\overline{P}}{}^{\text{driving}}_{\text{sm sys};n;+}\cdot\bar{\alpha}_\xi \tag{6-42a}$$

or

$$\left[\underbrace{\left(\overline{\overline{P}}{}^{\text{driving}}_{\text{sm sys};m;+}+j\,\overline{\overline{P}}{}^{\text{driving}}_{\text{sm sys};m;-}\right)}_{\overline{\overline{P}}{}^{\text{driving}}_{\text{sm sys};m}}+\vartheta\,\overline{\overline{P}}{}^{\text{driving}}_{\text{sm sys};n;+}\right]\cdot\bar{\alpha}_\xi=\left[\left(1+\vartheta\right)\frac{1}{\lambda_\xi}+j\right]\overline{\overline{P}}{}^{\text{driving}}_{\text{sm sys};n;-}\cdot\bar{\alpha}_\xi \tag{6-42b}$$

Here, $\vartheta$ is an adjustable constant scalar, but $\vartheta \neq -j,-1$; ordered numbers $(m,n)$ satisfy that $(m,n)=(1,2)$ or $(2,1)$; the above matrices can be obtained from decomposing matrices $\overline{\overline{P}}{}^{\text{driving}}_{\text{sm sys;1}}$ and $\overline{\overline{P}}{}^{\text{driving}}_{\text{sm sys;2}}$ by employing decomposition methods (6-35) and (6-36), and the methods to calculate matrices $\overline{\overline{P}}{}^{\text{driving}}_{\text{sm sys;1}}$ and $\overline{\overline{P}}{}^{\text{driving}}_{\text{sm sys;2}}$ have been provided in Section 4.7, and they will not be repeated here.

### 6.3.3 Metal-Material Composite System Case

On the scheme to establish the new generalized characteristic equations for composite systems, it is completely similar to the material system cases discussed in Subsections 6.3.1 and 6.3.2, and it will not be repeated here.

## 6.4 Singular Current Term and Its Physical Meaning

In this section, we separately study the physical meanings of the singularity current trems (SCTs) contained in the DPOs corresponding to three typical scattering systems, and the typical scattering systems are an one-body material system, a two-body material system, and a metal-material composite system. After the studies, we provide the specific manifestation form of the SCT in the DPO corresponding to lossless scattering systems, and also discuss its influence to the numerical performance of the DP-CM formulation of lossless scattering systems.

### 6.4.1 One-body Material System Case

Firstly, we equivalently rewrite the surface formulation (4-32) of the DPO corresponding to the single simply connected material body discussed in the Subsection 4.3.2 of this dissertation as follows:

$$
\begin{aligned}
P^{\text{driving}}_{\text{sim sys}} &= -(1/2)\left\langle \vec{J}^{\text{ES}}_{\text{sim}}, \mathcal{E}_0\left(\vec{J}^{\text{ES}}_{\text{sim}}, \vec{M}^{\text{ES}}_{\text{sim}}\right)\right\rangle_{\partial V^-_{\text{sim sys}}} - (1/2)\left\langle \vec{M}^{\text{ES}}_{\text{sim}}, \mathcal{H}_0\left(\vec{J}^{\text{ES}}_{\text{sim}}, \vec{M}^{\text{ES}}_{\text{sim}}\right)\right\rangle_{\partial V^-_{\text{sim sys}}} \\
&= -(1/2)\left\langle \vec{J}^{\text{ES}}_{\text{sim}}, -j\omega\mu_0\mathcal{L}_0\left(\vec{J}^{\text{ES}}_{\text{sim}}\right) - \mathcal{K}_0\left(\vec{M}^{\text{ES}}_{\text{sim}}\right)\right\rangle_{\partial V^-_{\text{sim sys}}} \\
&\quad - (1/2)\left\langle \vec{M}^{\text{ES}}_{\text{sim}}, \mathcal{K}_0\left(\vec{J}^{\text{ES}}_{\text{sim}}\right) - j\omega\varepsilon_0\mathcal{L}_0\left(\vec{M}^{\text{ES}}_{\text{sim}}\right)\right\rangle_{\partial V^-_{\text{sim sys}}} \\
&= -\frac{1}{2}\left\langle \vec{J}^{\text{ES}}_{\text{sim}}, -j\omega\mu_0\mathcal{L}_0\left(\vec{J}^{\text{ES}}_{\text{sim}}\right)\right\rangle_{\partial V_{\text{sim sys}}} - \frac{1}{2}\left\langle \vec{J}^{\text{ES}}_{\text{sim}}, -\text{P.V.}\,\mathcal{K}_0\left(\vec{M}^{\text{ES}}_{\text{sim}}\right) - \frac{1}{2}\vec{M}^{\text{ES}}_{\text{sim}} \times \hat{n}^-_{\text{sim}}\right\rangle_{\partial V_{\text{sim sys}}} \\
&\quad - \frac{1}{2}\left\langle \vec{M}^{\text{ES}}_{\text{sim}}, \text{P.V.}\,\mathcal{K}_0\left(\vec{J}^{\text{ES}}_{\text{sim}}\right) + \frac{1}{2}\vec{J}^{\text{ES}}_{\text{sim}} \times \hat{n}^-_{\text{sim}}\right\rangle_{\partial V_{\text{sim sys}}} - \frac{1}{2}\left\langle \vec{M}^{\text{ES}}_{\text{sim}}, -j\omega\varepsilon_0\mathcal{L}_0\left(\vec{M}^{\text{ES}}_{\text{sim}}\right)\right\rangle_{\partial V_{\text{sim sys}}}
\end{aligned}
$$

$$= -(1/2)\left\langle \vec{J}_{\text{sim}}^{\text{ES}}, -j\omega\mu_0\mathcal{L}_0\left(\vec{J}_{\text{sim}}^{\text{ES}}\right)\right\rangle_{\partial V_{\text{sim sys}}} - (1/2)\left\langle \vec{J}_{\text{sim}}^{\text{ES}}, -\text{P.V.}\,\mathcal{K}_0\left(\vec{M}_{\text{sim}}^{\text{ES}}\right)\right\rangle_{\partial V_{\text{sim sys}}}$$
$$- (1/2)\left\langle \vec{M}_{\text{sim}}^{\text{ES}}, \text{P.V.}\,\mathcal{K}_0\left(\vec{J}_{\text{sim}}^{\text{ES}}\right)\right\rangle_{\partial V_{\text{sim sys}}} - (1/2)\left\langle \vec{M}_{\text{sim}}^{\text{ES}}, -j\omega\varepsilon_0\mathcal{L}_0\left(\vec{M}_{\text{sim}}^{\text{ES}}\right)\right\rangle_{\partial V_{\text{sim sys}}}$$
$$+ \text{Re}\left\{(1/2)\oiint_{\partial V_{\text{sim sys}}}\left[\vec{M}_{\text{sim}}^{\text{ES}} \times \left(\vec{J}_{\text{sim}}^{\text{ES}}\right)^*\right]\cdot \hat{n}_{\text{sim}}^{+} dS\right\} \tag{6-43}$$

In relationship (6-43), the second equality is based on operator relationships $\mathcal{E}_0(\vec{J},\vec{M}) = -j\omega\mu_0\mathcal{L}_0(\vec{J}) - \mathcal{K}_0(\vec{M})$ and $\mathcal{H}_0(\vec{J},\vec{M}) = \mathcal{K}_0(\vec{J}) - j\omega\varepsilon_0\mathcal{L}_0(\vec{M})$ ; the third equality is based on the well-known conclusion related to the operator $\mathcal{K}_0$ in computational electromagnetics (for details see literature [107]); the fourth equality is obvious. The last term in the RHS of the fourth equality originates from the singularity of the operator $\mathcal{K}_0$ in source domain, so this dissertation calls the term as singular current term (SCT). In what follows, we do some identical transformation for the SCT, to reveal its physical meaning.

Inserting definition (C-41) into the SCT, we obtain that

$$\text{Re}\left\{(1/2)\oiint_{\partial V_{\text{sim sys}}}\left[\vec{M}_{\text{sim}}^{\text{ES}} \times \left(\vec{J}_{\text{sim}}^{\text{ES}}\right)^*\right]\cdot \hat{n}_{\text{sim}}^{+} dS\right\}$$
$$= \text{Re}\left\{(1/2)\oiint_{\partial V_{\text{sim sys}}}\left[\left(\vec{E}_{-}^{\text{tot}} \times \hat{n}_{\text{sim}}^{-}\right)\times\left(\hat{n}_{\text{sim}}^{-} \times \vec{H}_{-}^{\text{tot}}\right)^*\right]\cdot \hat{n}_{\text{sim}}^{+} dS\right\}$$
$$= -\text{Re}\left\{(1/2)\oiint_{\partial V_{\text{sim sys}}}\left[\vec{E}^{\text{tot}} \times \left(\vec{H}^{\text{tot}}\right)^*\right]\cdot \hat{n}_{\text{sim}}^{+} dS\right\}$$
$$= -\text{Re}\left\{(1/2)\iiint_{V_{\text{sim sys}}}\nabla\cdot\left[\vec{E}^{\text{tot}} \times \left(\vec{H}^{\text{tot}}\right)^*\right] dV\right\}$$
$$= \text{Re}\left\{(1/2)\iiint_{V_{\text{sim sys}}}\vec{E}^{\text{tot}}\cdot\left(\nabla\times\vec{H}^{\text{tot}}\right)^* dV - (1/2)\iiint_{V_{\text{sim sys}}}\left(\nabla\times\vec{E}^{\text{tot}}\right)\cdot\left(\vec{H}^{\text{tot}}\right)^* dV\right\}$$
$$= \text{Re}\left\{\frac{1}{2}\iiint_{V_{\text{sim sys}}}\vec{E}^{\text{tot}}\cdot\left[\left(j\omega\vec{\vec{\varepsilon}}_{\text{sim}} + \vec{\vec{\sigma}}_{\text{sim}}\right)\cdot\vec{E}^{\text{tot}}\right]^* dV - \frac{1}{2}\iiint_{V_{\text{sim sys}}}\left(-j\omega\vec{\vec{\mu}}_{\text{sim}}\cdot\vec{H}^{\text{tot}}\right)\cdot\left(\vec{H}^{\text{tot}}\right)^* dV\right\}$$
$$= \text{Re}\left\{\frac{1}{2}\left\langle\vec{\vec{\sigma}}_{\text{sim}}\cdot\vec{E}^{\text{tot}},\vec{E}^{\text{tot}}\right\rangle_{V_{\text{sim sys}}} + j\,2\omega\left[\frac{1}{4}\left\langle\vec{H}^{\text{tot}},\vec{\vec{\mu}}_{\text{sim}}\cdot\vec{H}^{\text{tot}}\right\rangle_{V_{\text{sim sys}}} - \frac{1}{4}\left\langle\vec{\vec{\varepsilon}}_{\text{sim}}\cdot\vec{E}^{\text{tot}},\vec{E}^{\text{tot}}\right\rangle_{V_{\text{sim sys}}}\right]\right\}$$
$$= (1/2)\left\langle\vec{\vec{\sigma}}_{\text{sim}}\cdot\vec{E}^{\text{tot}},\vec{E}^{\text{tot}}\right\rangle_{V_{\text{sim sys}}} \tag{6-44}$$

In relationship (6-44), the second equality is obvious; the third equality is based on Gauss' divergence theorem; the fourth equality is based on vectorial differentiation rule $\nabla\cdot(\vec{a}\times\vec{b}) = (\nabla\times\vec{a})\cdot\vec{b} - \vec{a}\cdot(\nabla\times\vec{b})$; the fifth equality is based on Maxwell's equations $\nabla\times\vec{H}^{\text{tot}} = (j\omega\vec{\vec{\varepsilon}}_{\text{sim}} + \vec{\vec{\sigma}}_{\text{sim}})\cdot\vec{E}^{\text{tot}}$ and $\nabla\times\vec{E}^{\text{tot}} = -j\omega\vec{\vec{\mu}}_{\text{sim}}\cdot\vec{H}^{\text{tot}}$ ; the sixth equality is obvious; the seventh equality is based on the symmetries of material parameter tensors $\vec{\vec{\mu}}_{\text{sim}}$, $\vec{\vec{\varepsilon}}_{\text{sim}}$, and $\vec{\vec{\sigma}}_{\text{sim}}$.

Relationship (6-44) implies that: the SCT contained in the DPO (4-32) corresponding to a single simply connected material body is just the ohmic loss power (i.e., lossy power) of the material body. Thus, we immediately have that

$$\overbrace{\left(\bar{a}^{J_\mathrm{s}}\right)^H \cdot \begin{bmatrix} \bar{\bar{I}}^{J_\mathrm{s}} \\ \bar{\bar{T}}_{\mathrm{DESS}}^{M_\mathrm{s}\leftarrow J_\mathrm{s}} \end{bmatrix}^H \cdot \bar{\bar{P}}_{0;\mathrm{SCT}}^{\mathrm{sim\,sys}} \cdot \begin{bmatrix} \bar{\bar{I}}^{J_\mathrm{s}} \\ \bar{\bar{T}}_{\mathrm{DESS}}^{M_\mathrm{s}\leftarrow J_\mathrm{s}} \end{bmatrix} \cdot \bar{a}^{J_\mathrm{s}}}^{\bar{\bar{P}}_{0;\mathrm{SCT};J_\mathrm{s}}^{\mathrm{sim\,sys}}}$$
$$\|$$
$$(1/2)\left\langle \vec{\sigma}_\mathrm{sim} \cdot \vec{E}^\mathrm{tot}, \vec{E}^\mathrm{tot} \right\rangle_{V_\mathrm{sim\,sys}} \tag{6-45a}$$
$$\|$$
$$\underbrace{\left(\bar{a}^{M_\mathrm{s}}\right)^H \cdot \begin{bmatrix} \bar{\bar{T}}_{\mathrm{DESS}}^{J_\mathrm{s}\leftarrow M_\mathrm{s}} \\ \bar{\bar{I}}^{M_\mathrm{s}} \end{bmatrix}^H \cdot \bar{\bar{P}}_{0;\mathrm{SCT}}^{\mathrm{sim\,sys}} \cdot \begin{bmatrix} \bar{\bar{T}}_{\mathrm{DESS}}^{J_\mathrm{s}\leftarrow M_\mathrm{s}} \\ \bar{\bar{I}}^{M_\mathrm{s}} \end{bmatrix} \cdot \bar{a}^{M_\mathrm{s}}}_{\bar{\bar{P}}_{0;\mathrm{SCT};M_\mathrm{s}}^{\mathrm{sim\,sys}}}$$

where matrix $\bar{\bar{P}}_{0;\mathrm{SCT}}^{\mathrm{sim\,sys}}$ is just the one used in formulation (4-34), and matrices $\bar{\bar{T}}_{\mathrm{DESS}}^{M_\mathrm{s}\leftarrow J_\mathrm{s}}$ and $\bar{\bar{T}}_{\mathrm{DESS}}^{J_\mathrm{s}\leftarrow M_\mathrm{s}}$ are just the ones used in formulation (4-53). For the convenience of the following discussions, $\bar{\bar{P}}_{0;\mathrm{SCT};J_\mathrm{s}}^{\mathrm{sim\,sys}}$ and $\bar{\bar{P}}_{0;\mathrm{SCT};M_\mathrm{s}}^{\mathrm{sim\,sys}}$ are collectively denoted as $\bar{\bar{P}}_{0;\mathrm{SCT};C_\mathrm{s}}^{\mathrm{sim\,sys}}$, and then we have the following relationship:

$$(1/2)\left\langle \vec{\sigma}_\mathrm{sim} \cdot \vec{E}^\mathrm{tot}, \vec{E}^\mathrm{tot} \right\rangle_{V_\mathrm{sim\,sys}} = \left(\bar{a}^{C_\mathrm{s}}\right)^H \cdot \bar{\bar{P}}_{0;\mathrm{SCT};C_\mathrm{s}}^{\mathrm{sim\,sys}} \cdot \bar{a}^{C_\mathrm{s}} \tag{6-45b}$$

## 6.4.2 Two-body Material System Case

Similarly to the relationship (6-43) in the one-body material system case, we rewrite the surface formulation (4-117) of the DPO corresponding to the two-body material system constituted by two simply connected material bodies as follows:

$$\begin{aligned} P_\mathrm{ss\,sys}^\mathrm{driving} =& -(1/2)\left\langle \vec{J}_\mathrm{s10}^\mathrm{ES} + \vec{J}_\mathrm{s20}^\mathrm{ES}, \mathcal{E}_0\left(\vec{J}_\mathrm{s10}^\mathrm{ES} + \vec{J}_\mathrm{s20}^\mathrm{ES}, \vec{M}_\mathrm{s10}^\mathrm{ES} + \vec{M}_\mathrm{s20}^\mathrm{ES}\right)\right\rangle_{\partial V_\mathrm{s10}^- \cup \partial V_\mathrm{s20}^-} \\ & -(1/2)\left\langle \vec{M}_\mathrm{s10}^\mathrm{ES} + \vec{M}_\mathrm{s20}^\mathrm{ES}, \mathcal{H}_0\left(\vec{J}_\mathrm{s10}^\mathrm{ES} + \vec{J}_\mathrm{s20}^\mathrm{ES}, \vec{M}_\mathrm{s10}^\mathrm{ES} + \vec{M}_\mathrm{s20}^\mathrm{ES}\right)\right\rangle_{\partial V_\mathrm{s10}^- \cup \partial V_\mathrm{s20}^-} \\ =& -(1/2)\left\langle \vec{J}_\mathrm{s10}^\mathrm{ES} + \vec{J}_\mathrm{s20}^\mathrm{ES}, -j\omega\mu_0\mathcal{L}_0\left(\vec{J}_\mathrm{s10}^\mathrm{ES} + \vec{J}_\mathrm{s20}^\mathrm{ES}\right)\right\rangle_{\partial V_\mathrm{s10} \cup \partial V_\mathrm{s20}} \\ & -(1/2)\left\langle \vec{J}_\mathrm{s10}^\mathrm{ES} + \vec{J}_\mathrm{s20}^\mathrm{ES}, -\mathrm{P.V.}\,\mathcal{K}_0\left(\vec{M}_\mathrm{s10}^\mathrm{ES} + \vec{M}_\mathrm{s20}^\mathrm{ES}\right)\right\rangle_{\partial V_\mathrm{s10} \cup \partial V_\mathrm{s20}} \\ & -(1/2)\left\langle \vec{M}_\mathrm{s10}^\mathrm{ES} + \vec{M}_\mathrm{s20}^\mathrm{ES}, \mathrm{P.V.}\,\mathcal{K}_0\left(\vec{J}_\mathrm{s10}^\mathrm{ES} + \vec{J}_\mathrm{s20}^\mathrm{ES}\right)\right\rangle_{\partial V_\mathrm{s10} \cup \partial V_\mathrm{s20}} \\ & -(1/2)\left\langle \vec{M}_\mathrm{s10}^\mathrm{ES} + \vec{M}_\mathrm{s20}^\mathrm{ES}, -j\omega\varepsilon_0\mathcal{L}_0\left(\vec{M}_\mathrm{s10}^\mathrm{ES} + \vec{M}_\mathrm{s20}^\mathrm{ES}\right)\right\rangle_{\partial V_\mathrm{s10} \cup \partial V_\mathrm{s20}} \\ & + \mathrm{Re}\left\{\frac{1}{2}\oiint_{\partial V_\mathrm{s10}} \left[\vec{M}_\mathrm{s10}^\mathrm{ES} \times \left(\vec{J}_\mathrm{s10}^\mathrm{ES}\right)^*\right] \cdot \hat{n}_\mathrm{s10}^+ dS + \frac{1}{2}\oiint_{\partial V_\mathrm{s20}} \left[\vec{M}_\mathrm{s20}^\mathrm{ES} \times \left(\vec{J}_\mathrm{s20}^\mathrm{ES}\right)^*\right] \cdot \hat{n}_\mathrm{s20}^+ dS\right\} \end{aligned} \tag{6-46}$$

Similarly to the previous identical transformation for the SCT contained in relationship (6-44), we identically transform the SCT contained in above relationship (6-46) as follows:

$$
\begin{aligned}
&\operatorname{Re}\left\{(1/2)\iint_{\partial V_{s10}}\left[\vec{M}_{s10}^{ES}\times\left(\vec{J}_{s10}^{ES}\right)^*\right]\cdot\hat{n}_{s10}^{+}dS+(1/2)\iint_{\partial V_{s20}}\left[\vec{M}_{s20}^{ES}\times\left(\vec{J}_{s20}^{ES}\right)^*\right]\cdot\hat{n}_{s20}^{+}dS\right\}\\
=&-\operatorname{Re}\left\{(1/2)\iint_{\partial V_{s10}}\left[\vec{E}_{1-}^{tot}\times\left(\vec{H}_{1-}^{tot}\right)^*\right]\cdot\hat{n}_{s10}^{+}dS+(1/2)\iint_{\partial V_{s20}}\left[\vec{E}_{2-}^{tot}\times\left(\vec{H}_{2-}^{tot}\right)^*\right]\cdot\hat{n}_{s20}^{+}dS\right\}\\
=&-\operatorname{Re}\left\{(1/2)\iint_{\partial V_{s10}}\left[\vec{E}_{1-}^{tot}\times\left(\vec{H}_{1-}^{tot}\right)^*\right]\cdot\hat{n}_{s10}^{+}dS+(1/2)\iint_{\partial V_{s12}}\left[\vec{E}_{1-}^{tot}\times\left(\vec{H}_{1-}^{tot}\right)^*\right]\cdot\hat{n}_{s12}^{+}dS\right\}\\
&-\operatorname{Re}\left\{(1/2)\iint_{\partial V_{s20}}\left[\vec{E}_{2-}^{tot}\times\left(\vec{H}_{2-}^{tot}\right)^*\right]\cdot\hat{n}_{s20}^{+}dS+(1/2)\iint_{\partial V_{s21}}\left[\vec{E}_{2-}^{tot}\times\left(\vec{H}_{2-}^{tot}\right)^*\right]\cdot\hat{n}_{s21}^{+}dS\right\}\\
=&-\operatorname{Re}\left\{\frac{1}{2}\oiint_{\partial V_{sim}^{1}}\left[\vec{E}_{1-}^{tot}\times\left(\vec{H}_{1-}^{tot}\right)^*\right]\cdot\hat{n}_{sim}^{1+}dS\right\}-\operatorname{Re}\left\{\frac{1}{2}\oiint_{\partial V_{sim}^{2}}\left[\vec{E}_{2-}^{tot}\times\left(\vec{H}_{2-}^{tot}\right)^*\right]\cdot\hat{n}_{sim}^{2+}dS\right\}\\
=&\ (1/2)\left\langle\vec{\vec{\sigma}}_{sim}^{1}\cdot\vec{E}^{tot},\vec{E}^{tot}\right\rangle_{V_{sim}^{1}}+(1/2)\left\langle\vec{\vec{\sigma}}_{sim}^{2}\cdot\vec{E}^{tot},\vec{E}^{tot}\right\rangle_{V_{sim}^{2}}\\
=&\ (1/2)\left\langle\left(\vec{\vec{\sigma}}_{sim}^{1}+\vec{\vec{\sigma}}_{sim}^{2}\right)\cdot\vec{E}^{tot},\vec{E}^{tot}\right\rangle_{V_{ss\,sys}}
\end{aligned}
\tag{6-47}
$$

In relationship (6-47), the derivation for the first equality is similar to relationship (6-44); the second equality is based on the tangential continuation condition of the total field $\vec{F}^{tot}$ on $\partial V_{s12}=\partial V_{s21}$ and relationship $\hat{n}_{s12}^{+}=-\hat{n}_{s21}^{+}$; the third equality is based on the additivity of the integral domain of Riemann's integral[117]; the derivation for the fourth equality is similar to relationship (6-44); the fifth equality is based on the linear property of inner product and the fact that $\operatorname{int}V_{sim}^{1}\bigcap\operatorname{int}V_{sim}^{2}=\varnothing$. Relationship (6-47) implies that: the SCT contained in the DPO (4-117) corresponding to the two-body material system is just the lossy power of the material system. Thus, we immediately have that

$$
\underbrace{\left(\bar{a}^{M_{s0}}\right)^H\cdot\begin{bmatrix}\bar{\bar{T}}^{J_{s10}\leftarrow M_{s0}}\\\bar{\bar{T}}^{J_{s20}\leftarrow M_{s0}}\\\bar{\bar{\mathcal{I}}}^{M_{s10}}\\\bar{\bar{\mathcal{I}}}^{M_{s20}}\end{bmatrix}^H\cdot\bar{\bar{P}}_{3;0;SCT}^{ss\,sys}\cdot\begin{bmatrix}\bar{\bar{T}}^{J_{s10}\leftarrow M_{s0}}\\\bar{\bar{T}}^{J_{s20}\leftarrow M_{s0}}\\\bar{\bar{\mathcal{I}}}^{M_{s10}}\\\bar{\bar{\mathcal{I}}}^{M_{s20}}\end{bmatrix}}_{\bar{\bar{P}}_{3;0;SCT;M_{s0}}^{ss\,sys}}\cdot\bar{a}^{M_{s0}}=\frac{1}{2}\left\langle\left(\vec{\vec{\sigma}}_{sim}^{1}+\vec{\vec{\sigma}}_{sim}^{2}\right)\cdot\vec{E}^{tot},\vec{E}^{tot}\right\rangle_{V_{ss\,sys}}
\tag{6-48}
$$

where matrix $\bar{\bar{P}}_{3;0;SCT}^{ss\,sys}$ is just the one used in formulation (4-118), and matrices $\bar{\bar{T}}^{J_{s10}\leftarrow M_{s0}}$, $\bar{\bar{T}}^{J_{s20}\leftarrow M_{s0}}$, $\bar{\bar{\mathcal{I}}}^{M_{s10}}$, and $\bar{\bar{\mathcal{I}}}^{M_{s20}}$ are just the ones used in formulation (4-122).

For the material systems constituted by a simply connected material body and a multiply connected material body, the discussion for the physical meaning of the SCT contained in its DPO formulation (4-146) is similar, and here we only provide the final result as follows:

$$\left(\bar{a}^{M_0}\right)^H \cdot \underbrace{\begin{bmatrix} \overline{\overline{T}}^{J_{s0} \leftarrow M_0} \\ \overline{\overline{T}}^{J_{sm} \leftarrow M_0} \\ -\overline{\overline{T}}^{J_{sm} \leftarrow M_0} \\ \overline{\overline{T}}^{J_{m0} \leftarrow M_0} \\ \overline{\overline{\mathcal{I}}}^{M_{s0}} \\ \overline{\overline{T}}^{M_{sm} \leftarrow M_0} \\ -\overline{\overline{T}}^{M_{sm} \leftarrow M_0} \\ \overline{\overline{\mathcal{I}}}^{M_{m0}} \end{bmatrix}^H \cdot \overline{\overline{P}}_{3;0;\text{SCT}}^{\text{sm sys}} \cdot \begin{bmatrix} \overline{\overline{T}}^{J_{s0} \leftarrow M_0} \\ \overline{\overline{T}}^{J_{sm} \leftarrow M_0} \\ -\overline{\overline{T}}^{J_{sm} \leftarrow M_0} \\ \overline{\overline{T}}^{J_{m0} \leftarrow M_0} \\ \overline{\overline{\mathcal{I}}}^{M_{s0}} \\ \overline{\overline{T}}^{M_{sm} \leftarrow M_0} \\ -\overline{\overline{T}}^{M_{sm} \leftarrow M_0} \\ \overline{\overline{\mathcal{I}}}^{M_{m0}} \end{bmatrix}}_{\overline{\overline{P}}_{3;0;\text{SCT};M_0}^{\text{sm sys}}} \cdot \bar{a}^{M_0} = \frac{1}{2}\left\langle \left(\vec{\vec{\sigma}}_{\text{sim}} + \vec{\vec{\sigma}}_{\text{mul}}\right) \cdot \vec{E}^{\text{tot}}, \vec{E}^{\text{tot}} \right\rangle_{V_{\text{sm sys}}} \tag{6-49}$$

The derivation for the above formulation is not given here, and the various sub-matrices used in the formulation are jute the ones used in formulations (4-147) and (4-148).

### 6.4.3 Metal-Material Composite System Case

We rewrite the line-surface formulation (5-68) of the DPO corresponding to metal-material composite system as follows:

$$\begin{aligned} P_{\text{com sys}}^{\text{driving}} = \; & (1/2)\left\langle \vec{J}_0^{\text{SL}} \oplus \vec{J}_0^{\text{SS}}, \mathcal{E}_0\left(-\vec{J}_0^{\text{SL}} - \vec{J}_0^{\text{SS}} + \vec{J}_0^{\text{ES}}, \vec{M}_0^{\text{ES}}\right) \right\rangle_{L_{\text{met}}^0 \cup S_{\text{met}}^0 \cup \partial V_{\text{met}}^0} \\ & + (1/2)\left\langle \vec{J}_0^{\text{ES}}, \mathcal{E}_0\left(\vec{J}_0^{\text{SL}} + \vec{J}_0^{\text{SS}} - \vec{J}_0^{\text{ES}}, -\vec{M}_0^{\text{ES}}\right) \right\rangle_{\partial V_{\text{mat}}^{0;-}} \\ & + (1/2)\left\langle \vec{M}_0^{\text{ES}}, \mathcal{H}_0\left(\vec{J}_0^{\text{SL}} + \vec{J}_0^{\text{SS}} - \vec{J}_0^{\text{ES}}, -\vec{M}_0^{\text{ES}}\right) \right\rangle_{\partial V_{\text{mat}}^{0;-}} \\ = \; & (1/2)\left\langle \vec{J}_0^{\text{SL}} \oplus \vec{J}_0^{\text{SS}}, -j\omega\mu_0\mathcal{L}_0\left(-\vec{J}_0^{\text{SL}} - \vec{J}_0^{\text{SS}} + \vec{J}_0^{\text{ES}}\right) - \mathcal{K}_0\left(\vec{M}_0^{\text{ES}}\right) \right\rangle_{L_{\text{met}}^0 \cup S_{\text{met}}^0 \cup \partial V_{\text{met}}^0} \\ & + (1/2)\left\langle \vec{J}_0^{\text{ES}}, -j\omega\mu_0\mathcal{L}_0\left(\vec{J}_0^{\text{SL}} + \vec{J}_0^{\text{SS}} - \vec{J}_0^{\text{ES}}\right) + \text{P.V.}\,\mathcal{K}_0\left(\vec{M}_0^{\text{ES}}\right) \right\rangle_{\partial V_{\text{mat}}^0} \\ & + (1/2)\left\langle \vec{M}_0^{\text{ES}}, \text{P.V.}\,\mathcal{K}_0\left(\vec{J}_0^{\text{SL}} + \vec{J}_0^{\text{SS}}\right) - \text{P.V.}\,\mathcal{K}_0\left(\vec{J}_0^{\text{ES}}\right) + j\omega\varepsilon_0\mathcal{L}_0\left(\vec{M}_0^{\text{ES}}\right) \right\rangle_{\partial V_{\text{mat}}^0} \\ & + \text{Re}\left\{(1/2)\iint_{\partial V_{\text{mat}}^0}\left[\vec{M}_0^{\text{ES}} \times \left(\vec{J}_0^{\text{ES}}\right)^*\right] \cdot \hat{n}_{\text{mat}}^{0;+} dS\right\} \end{aligned} \tag{6-50}$$

where the derivation for the second equality is similar to relationships (6-43) and (6-46), and we don't provide the detailed derivation process here. We do the identical transformation for the SCT contained in relationship (6-50) as follows:

$$\begin{aligned} \text{Re}\left\{(1/2)\iint_{\partial V_{\text{mat}}^0}\left[\vec{M}_0^{\text{ES}} \times \left(\vec{J}_0^{\text{ES}}\right)^*\right] \cdot \hat{n}_{\text{mat}}^{0;+} dS\right\} & = -\text{Re}\left\{(1/2)\iint_{\partial V_{\text{mat}}^0}\left[\vec{E}_-^{\text{tot}} \times \left(\vec{H}_-^{\text{tot}}\right)^*\right] \cdot \hat{n}_{\text{mat}}^{0;+} dS\right\} \\ & = -\text{Re}\left\{(1/2)\oiint_{\partial V_{\text{mat}}}\left[\vec{E}_-^{\text{tot}} \times \left(\vec{H}_-^{\text{tot}}\right)^*\right] \cdot \hat{n}_{\text{mat}}^{+} dS\right\} \\ & = (1/2)\left\langle \vec{\vec{\sigma}}_{\text{mat}} \cdot \vec{E}^{\text{tot}}, \vec{E}^{\text{tot}} \right\rangle_{V_{\text{mat}}} \end{aligned} \tag{6-51}$$

In relationship (6-51), the derivation for the first equality is similar to relationships (6-44) and (6-47); the second equality is based on the homogeneous tangential electric field boundary condition on metal-material sub-boundary $\partial V_{\text{mat}} \setminus \partial V_{\text{mat}}^{0} = L_{\text{met}}^{\cap} \bigcup \partial V_{\text{met}}^{\cap} \bigcup \partial V_{\text{mat}}^{\text{open surf}}$ (for details see Section 5.3); the derivation for the third equality is similar to relationships (6-44) and (6-47).

Relationship (6-51) implies that: the SCT contained in the DPO (5-68) corresponding to the composite system is just the lossy power of the composite system. Thus, we immediately have that

$$\overbrace{\left(\bar{a}^{J_0}\right)^H \cdot \begin{bmatrix} \bar{\bar{I}}^{\vec{J}_0^{\text{SL}}} & 0 & 0 \\ 0 & \bar{\bar{I}}^{\vec{J}_0^{\text{SS}}} & 0 \\ 0 & 0 & \bar{\bar{I}}^{\vec{J}_0^{\text{ES}}} \\ 0 & 0 & \bar{\bar{T}}^{\vec{M}_0^{\text{ES}} \leftarrow \vec{J}_0^{\text{ES}}} \end{bmatrix}^H \cdot \bar{\bar{P}}_{0;\text{SCT}}^{\text{com sys}} \cdot \begin{bmatrix} \bar{\bar{I}}^{\vec{J}_0^{\text{SL}}} & 0 & 0 \\ 0 & \bar{\bar{I}}^{\vec{J}_0^{\text{SS}}} & 0 \\ 0 & 0 & \bar{\bar{I}}^{\vec{J}_0^{\text{ES}}} \\ 0 & 0 & \bar{\bar{T}}^{\vec{M}_0^{\text{ES}} \leftarrow \vec{J}_0^{\text{ES}}} \end{bmatrix} \cdot \bar{a}^{J_0}}^{\bar{\bar{P}}_{0;\text{SCT};J_0}^{\text{com sys}}}$$

$$= (1/2)\left\langle \vec{\vec{\sigma}}_{\text{mat}} \cdot \vec{E}^{\text{tot}}, \vec{E}^{\text{tot}} \right\rangle_{V_{\text{mat}}} \tag{6-52}$$

where matrix $\bar{\bar{P}}_{0;\text{SCT}}^{\text{com sys}}$ is just the one used in formulation (5-69), and matrices $\bar{\bar{I}}^{\vec{J}_0^{\text{SL}}}$, $\bar{\bar{I}}^{\vec{J}_0^{\text{SS}}}$, $\bar{\bar{I}}^{\vec{J}_0^{\text{ES}}}$, and $\bar{\bar{T}}^{\vec{M}_0^{\text{ES}} \leftarrow \vec{J}_0^{\text{ES}}}$ are just the ones used in formulation (5-77a).

### 6.4.4 Singular Current Term in the DPO of Lossless Scattering System

For the convenience of the following discussions, the $\{\bar{\bar{P}}_{0;\text{SCT};C_{\text{s}}}^{\text{sim sys}}, \bar{a}^{C_{\text{s}}}\}$ in the one-body material system case, the $\{\bar{\bar{P}}_{3;0;\text{SCT};M_{\text{s0}}}^{\text{ss sys}}, \bar{a}^{M_0}\} / \{\bar{\bar{P}}_{3;0;\text{SCT};M_0}^{\text{sm sys}}, \bar{a}^{M_0}\}$ in the two-body material system case, and the $\{\bar{\bar{P}}_{0;\text{SCT};J_0}^{\text{com sys}}, \bar{a}^{J_0}\}$ in the composite system case are collectively denoted as $\{\bar{\bar{P}}_{0;\text{SCT};\text{BV}}^{\text{sca sys}}, \bar{a}^{\text{BV}}\}$ (where superscript "BV" is the abbreviation of "basic variable"), and the $(1/2) < \vec{\vec{\sigma}}_{\text{sim}} \cdot \vec{E}^{\text{tot}}, \vec{E}^{\text{tot}} >_{V_{\text{sim sys}}}$ in one-body material system case, the $(1/2) < (\vec{\vec{\sigma}}_{\text{sim}}^{1} + \vec{\vec{\sigma}}_{\text{sim}}^{2}) \cdot \vec{E}^{\text{tot}}, \vec{E}^{\text{tot}} >_{V_{\text{ss sys}}}$ / $(1/2) < (\vec{\vec{\sigma}}_{\text{sim}} + \vec{\vec{\sigma}}_{\text{mul}}) \cdot \vec{E}^{\text{tot}}, \vec{E}^{\text{tot}} >_{V_{\text{sm sys}}}$ in two-body material system case and the $(1/2) < \vec{\vec{\sigma}}_{\text{mat}} \cdot \vec{E}^{\text{tot}}, \vec{E}^{\text{tot}} >_{V_{\text{mat}}}$ in composite system case are collectively denoted as $P^{\text{sca sys los}}$, and then we have that

$$\left(\bar{a}^{\text{BV}}\right)^H \cdot \bar{\bar{P}}_{0;\text{SCT};\text{BV}}^{\text{sca sys}} \cdot \bar{a}^{\text{BV}} = P^{\text{sca sys los}} \tag{6-53}$$

BV $\bar{a}^{\text{BV}}$ is arbitrary (because $\bar{a}^{\text{BV}}$ is independent) and $P^{\text{sca sys los}}$ is a real number, so matrix $\bar{\bar{P}}_{0;\text{SCT};\text{BV}}^{\text{sca sys}}$ must be Hermitian. In what follows, we study the SCT matrix $\bar{\bar{P}}_{0;\text{SCT};\text{BV}}^{\text{sca sys}}$ contained in lossless scattering system.

We denote the elements of $\bar{\bar{P}}_{0;\text{SCT};\text{BV}}^{\text{sca sys}}$ as $\theta_{\xi\zeta}$, and then

$$\overline{\overline{P}}_{0;\text{SCT;BV}}^{\text{sca sys}} = \begin{bmatrix} \theta_{11} & \theta_{12} & \theta_{13} & \cdots \\ \theta_{21} & \theta_{22} & \theta_{23} & \cdots \\ \theta_{31} & \theta_{32} & \theta_{33} & \cdots \\ \vdots & \vdots & \vdots & \ddots \end{bmatrix} \tag{6-54}$$

Because $\overline{\overline{P}}_{0;\text{SCT;BV}}^{\text{sca sys}}$ is Hermitian, then $\theta_{\zeta\xi}^{*} = \theta_{\xi\zeta}$, i.e.,

$$\overline{\overline{P}}_{0;\text{SCT;BV}}^{\text{sca sys}} = \begin{bmatrix} \theta_{11} & \theta_{12} & \theta_{13} & \cdots \\ \theta_{12}^{*} & \theta_{22} & \theta_{23} & \cdots \\ \theta_{13}^{*} & \theta_{23}^{*} & \theta_{33} & \cdots \\ \vdots & \vdots & \vdots & \ddots \end{bmatrix} \tag{6-55}$$

For lossless scattering systems, it always holds that $P^{\text{sca sys los}} = 0$, i.e., for any $\overline{a}^{\text{BV}} = [a_1^{\text{BV}} \quad a_2^{\text{BV}} \quad a_3^{\text{BV}} \quad \cdots]^T$ we have that

$$\left[\left(a_1^{\text{BV}}\right)^{*} \left(a_2^{\text{BV}}\right)^{*} \cdots \left(a_3^{\text{BV}}\right)^{*}\right] \cdot \begin{bmatrix} \theta_{11} & \theta_{12} & \theta_{13} & \cdots \\ \theta_{12}^{*} & \theta_{22} & \theta_{23} & \cdots \\ \theta_{13}^{*} & \theta_{23}^{*} & \theta_{33} & \cdots \\ \vdots & \vdots & \vdots & \ddots \end{bmatrix} \cdot \begin{bmatrix} a_1^{\text{BV}} \\ a_2^{\text{BV}} \\ a_3^{\text{BV}} \\ \vdots \end{bmatrix} = P^{\text{sca sys los}} \xrightarrow{\text{lossless}} 0 \tag{6-56}$$

In particular, we select $\overline{a}^{\text{BV}} = [1 \quad 0 \quad 0 \quad \cdots]^T$, and then

$$\theta_{11} = \begin{bmatrix} 1^{*} & 0 & 0 & \cdots \end{bmatrix} \cdot \begin{bmatrix} \theta_{11} & \theta_{12} & \theta_{13} & \cdots \\ \theta_{12}^{*} & \theta_{22} & \theta_{23} & \cdots \\ \theta_{13}^{*} & \theta_{23}^{*} & \theta_{33} & \cdots \\ \vdots & \vdots & \vdots & \ddots \end{bmatrix} \cdot \begin{bmatrix} 1 \\ 0 \\ 0 \\ \vdots \end{bmatrix} \xrightarrow{\text{lossless}} 0 \tag{6-57}$$

Similarly, it can be proven that: if the system is lossless, $\theta_{\xi\xi} = 0$ for any $\xi$, i.e., all of the diagonal elements of matrix $\overline{\overline{P}}_{0;\text{SCT;BV}}^{\text{sca sys}}$ are zero, and then

$$\overline{\overline{P}}_{0;\text{SCT;BV}}^{\text{sca sys}} \xrightarrow{\text{lossless}} \begin{bmatrix} 0 & \theta_{12} & \theta_{13} & \cdots \\ \theta_{12}^{*} & 0 & \theta_{23} & \cdots \\ \theta_{13}^{*} & \theta_{23}^{*} & 0 & \cdots \\ \vdots & \vdots & \vdots & \ddots \end{bmatrix} \tag{6-58}$$

If we select $\overline{a}^{\text{BV}} = [1 \quad 1 \quad 0 \quad \cdots]^T$, we have that

$$2\,\text{Re}\{\theta_{12}\} = \theta_{12} + \theta_{12}^{*} = \begin{bmatrix} 1^{*} & 1^{*} & 0 & \cdots \end{bmatrix} \cdot \begin{bmatrix} 0 & \theta_{12} & \theta_{13} & \cdots \\ \theta_{12}^{*} & 0 & \theta_{23} & \cdots \\ \theta_{13}^{*} & \theta_{23}^{*} & 0 & \cdots \\ \vdots & \vdots & \vdots & \ddots \end{bmatrix} \cdot \begin{bmatrix} 1 \\ 1 \\ 0 \\ \vdots \end{bmatrix} \xrightarrow{\text{lossless}} 0 \tag{6-59}$$

Similarly, we can prove that: if the system is lossless, $\text{Re}\{\theta_{\xi\zeta}\} = 0$ for any $\xi$ and $\zeta$, and then

$$\overline{\overline{P}}{}_{0;\text{SCT;BV}}^{\text{sca sys}} \xrightarrow{\text{lossless}} \begin{bmatrix} 0 & j\,\text{Im}\{\theta_{12}\} & j\,\text{Im}\{\theta_{13}\} & \cdots \\ -j\,\text{Im}\{\theta_{12}\} & 0 & j\,\text{Im}\{\theta_{23}\} & \cdots \\ -j\,\text{Im}\{\theta_{13}\} & -j\,\text{Im}\{\theta_{23}\} & 0 & \cdots \\ \vdots & \vdots & \vdots & \ddots \end{bmatrix} \tag{6-60}$$

If we select $\bar{a}^{\text{BV}} = [x \quad y \quad 0 \quad \cdots]^T$, we have that

$$\begin{aligned} & j\left(x^* y - x y^*\right)\text{Im}\{\theta_{12}\} \\ = \; & j\, x^* y\, \text{Im}\{\theta_{12}\} - j\, x y^*\, \text{Im}\{\theta_{12}\} \\ = \; & \begin{bmatrix} x^* & y^* & 0 & \cdots \end{bmatrix} \cdot \begin{bmatrix} 0 & j\,\text{Im}\{\theta_{12}\} & j\,\text{Im}\{\theta_{13}\} & \cdots \\ -j\,\text{Im}\{\theta_{12}\} & 0 & j\,\text{Im}\{\theta_{23}\} & \cdots \\ -j\,\text{Im}\{\theta_{13}\} & -j\,\text{Im}\{\theta_{23}\} & 0 & \cdots \\ \vdots & \vdots & \vdots & \ddots \end{bmatrix} \cdot \begin{bmatrix} x \\ y \\ 0 \\ \vdots \end{bmatrix} \\ \xrightarrow{\text{lossless}} \; & 0 \end{aligned} \tag{6-61}$$

The $x$ and $y$ in the above formulation are arbitrary (because $\bar{a}^{\text{BV}} = [x \quad y \quad 0 \quad \cdots]^T$ is independent), so we can make that $x^* y - x y^* = \text{Re}\{x^* y\} \neq 0$ by properly selecting $x$ and $y$, and this implies that: if the system is lossless, $\text{Im}\{\theta_{12}\} = 0$. Similarly, it can be proven that: when the system is lossless, $\text{Im}\{\theta_{\xi\zeta}\} = 0$ for any $\xi$ and $\zeta$, and then

$$\overline{\overline{P}}{}_{0;\text{SCT;BV}}^{\text{sca sys}} \xrightarrow{\text{lossless}} \begin{bmatrix} 0 & 0 & 0 & \cdots \\ 0 & 0 & 0 & \cdots \\ 0 & 0 & 0 & \cdots \\ \vdots & \vdots & \vdots & \ddots \end{bmatrix} = 0 \tag{6-62}$$

Above formulation (6-62) implies that: for lossless systems, SCT matrix $\overline{\overline{P}}{}_{0;\text{SCT;BV}}^{\text{sca sys}}$ is always zero.① Based on this conclusion, during the process to construct the DP-CMs of lossless systems in the Chapters 4 and 5 of this dissertation, we have directly set the SCTs to be zero. So, what will happen if we don't do like that? Let us see several examples.

**1) Lossless One-body Material System Case**

Now, we utilize two different schemes ("the scheme retaining SCT" and "the scheme deleting SCT") to construct the DP-CMs for the one-body material system shown in

---

① During the process to study the SCT contained in the DPO of lossless material systems, Dr. Xingyue Guo (ORCID: 0000-0002-9073-4213), who is with Peking University (PKU), participated in some discussions. Here, I would like to thank Dr. Guo for her valuable discussions.

Figure 6-1 (i.e., the one shown in Figure 4-14), and provide the characteristic value (dB) curves and MS curves derived from the two schemes in Figure 6-21 and Figure 6-22 respectively. In addition, we also provide the results derived from the volume formulation given in Section 4.2 in Figure 6-23, for comparison.

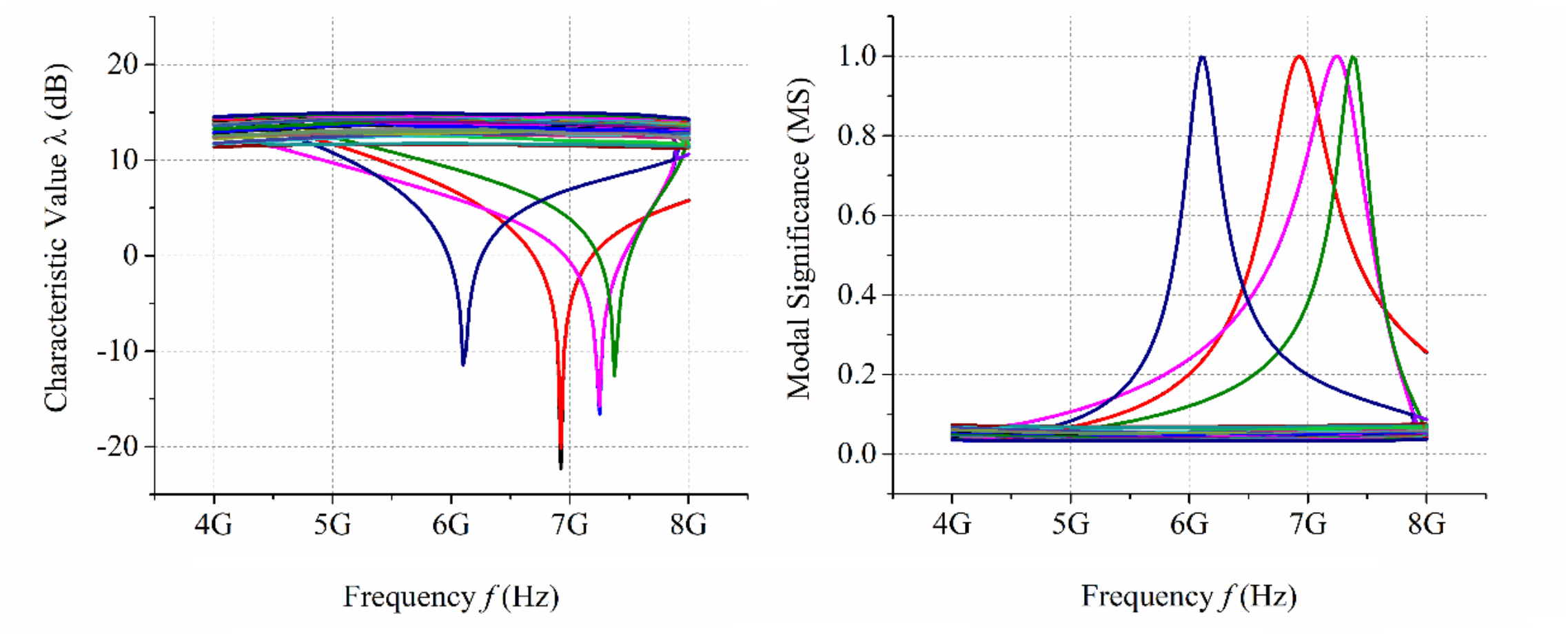

(a) (b)

Figure 6-21 The characteristic quantity curves corresponding to the first 50 DP-CMs (derived from the surface formulations (4-32)&(4-34)&(4-52) with SCT) of the one-body material system shown in Figure 6-1. (a) characteristic value dB curves; (b) MS curves

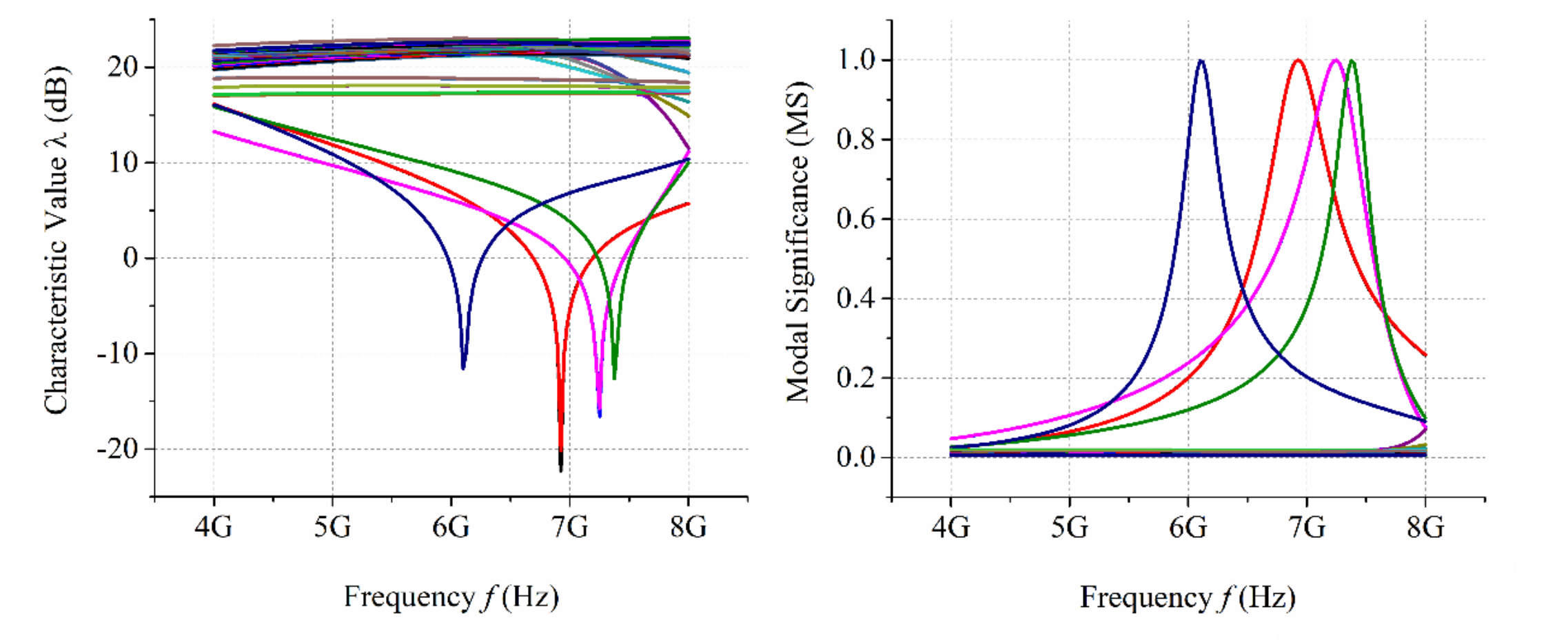

(a) (b)

Figure 6-22 The characteristic quantity curves corresponding to the first 50 DP-CMs (derived from the surface formulations (4-32)&(4-34)&(4-52) without SCT) of the one-body material system shown in Figure 6-1. (a) characteristic value dB curves; (b) MS curves

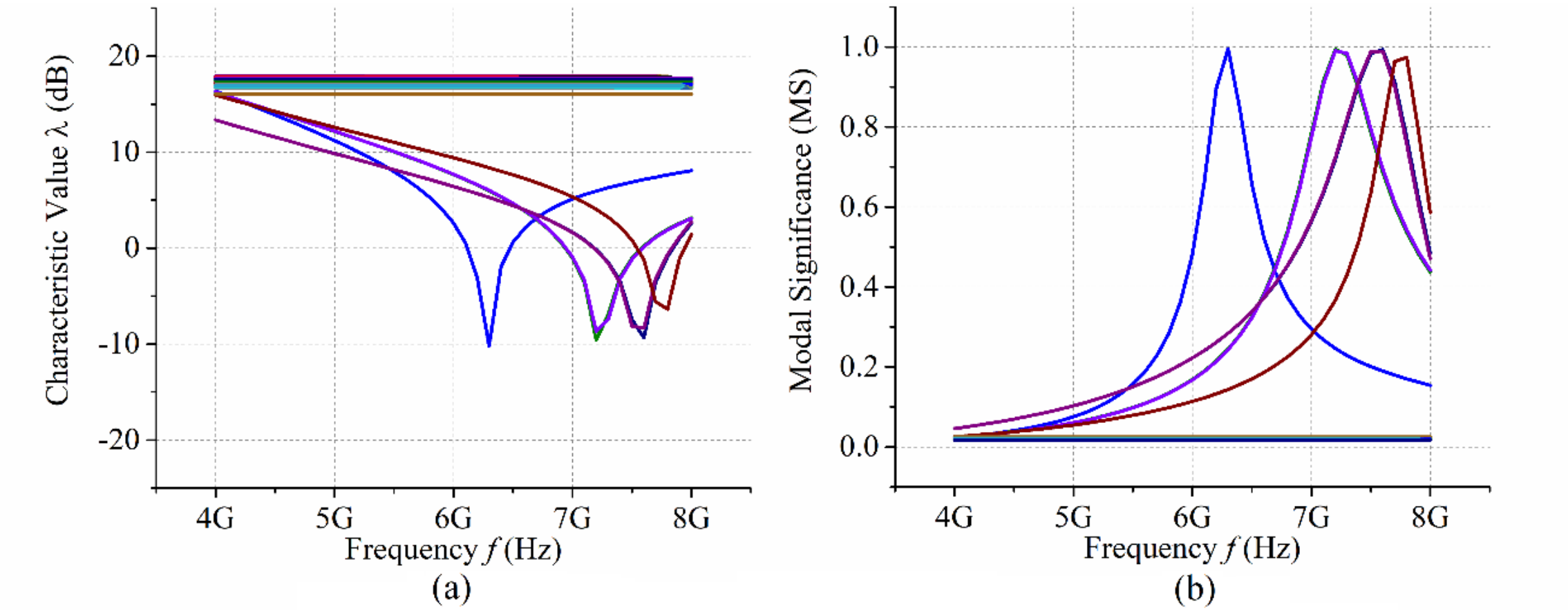

Figure 6-23 The characteristic quantity curves corresponding to the first 50 DP-CMs (derived from volume formulations (4-8)&(4-9)) of the one-body material system shown in Figure 6-1. (a) characteristic value dB curves; (b) MS curves

From the above figures, it is easy to find out that: compared with the results derived from "the scheme retaining SCT", the results derived from "the scheme deleting SCT" agree better with the results derived from the volume formulation.

### 2) Lossless Two-body Material System Case

Now, we utilize two different schemes ("the scheme retaining SCT" and "the scheme deleting SCT") to construct the DP-CMs for the two-body material system in Figure 6-8 (i.e., the one in Figure 4-30), and provide the characteristic value (dB) curves and MS curves derived from the two schemes in Figure 6-24 and Figure 6-25 respectively. The relative permeability, relative permittivity, and conductivity of the upper/lower material cylinder are 3/6, 12/6, and 0/0 respectively. In addition, we also provide the results derived from the volume formulation given in Section 4.2 in Figure 6-26, for comparison.

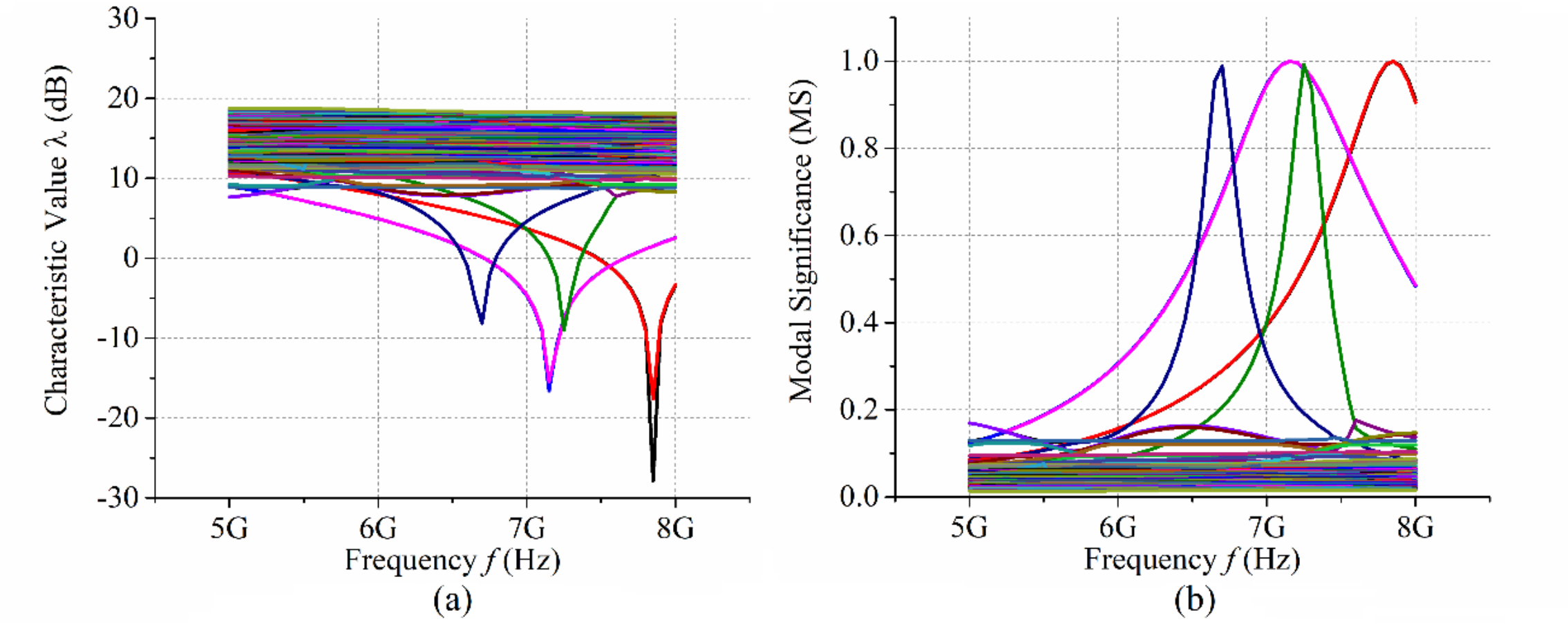

Figure 6-24 The characteristic quantity curves corresponding to the first 200 DP-CMs (derived from the surface formulations (4-117)&(4-122) with SCT) of the two-body material system shown in Figure 6-8. (a) characteristic value dB curves; (b) MS curves

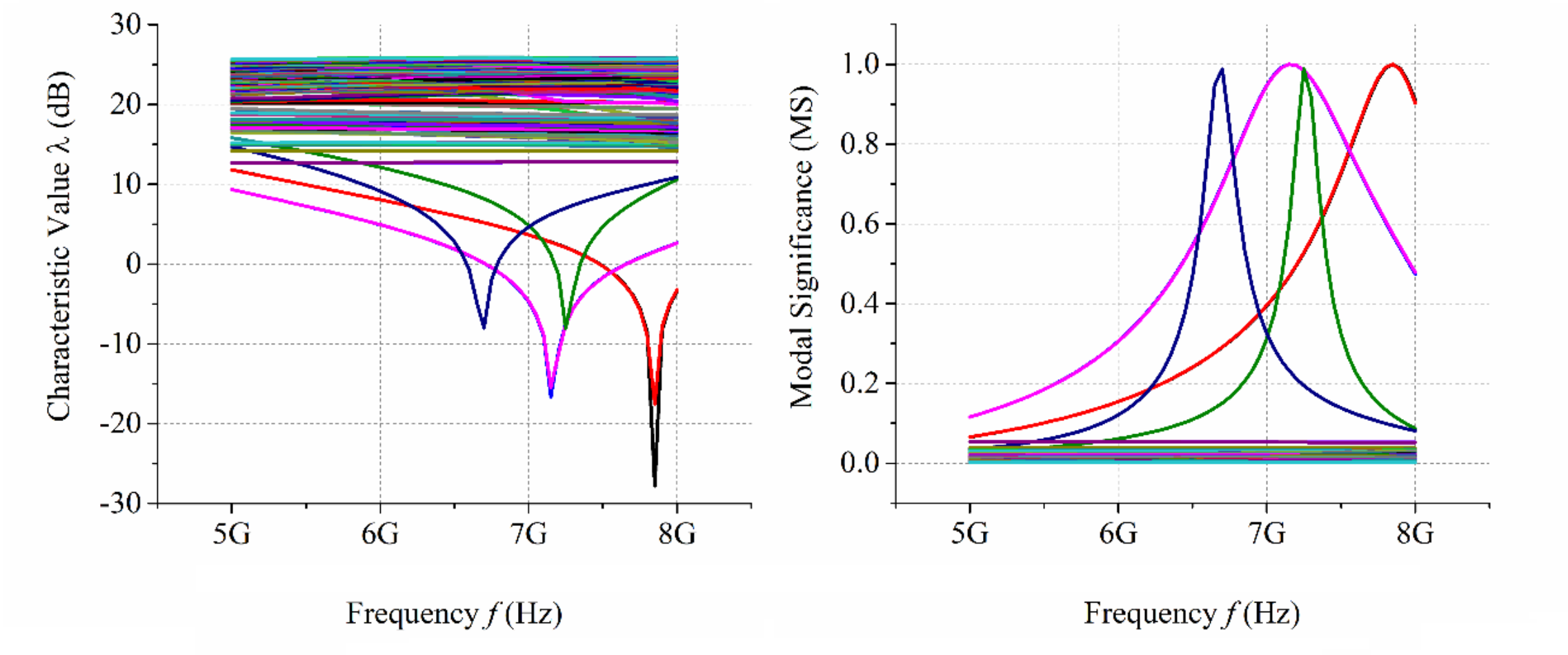

(a) (b)

Figure 6-25 The characteristic quantity curves corresponding to the first 200 DP-CMs (derived from the surface formulations (4-117)&(4-122) without SCT) of the two-body material system shown in Figure 6-8. (a) characteristic value dB curves; (b) MS curves

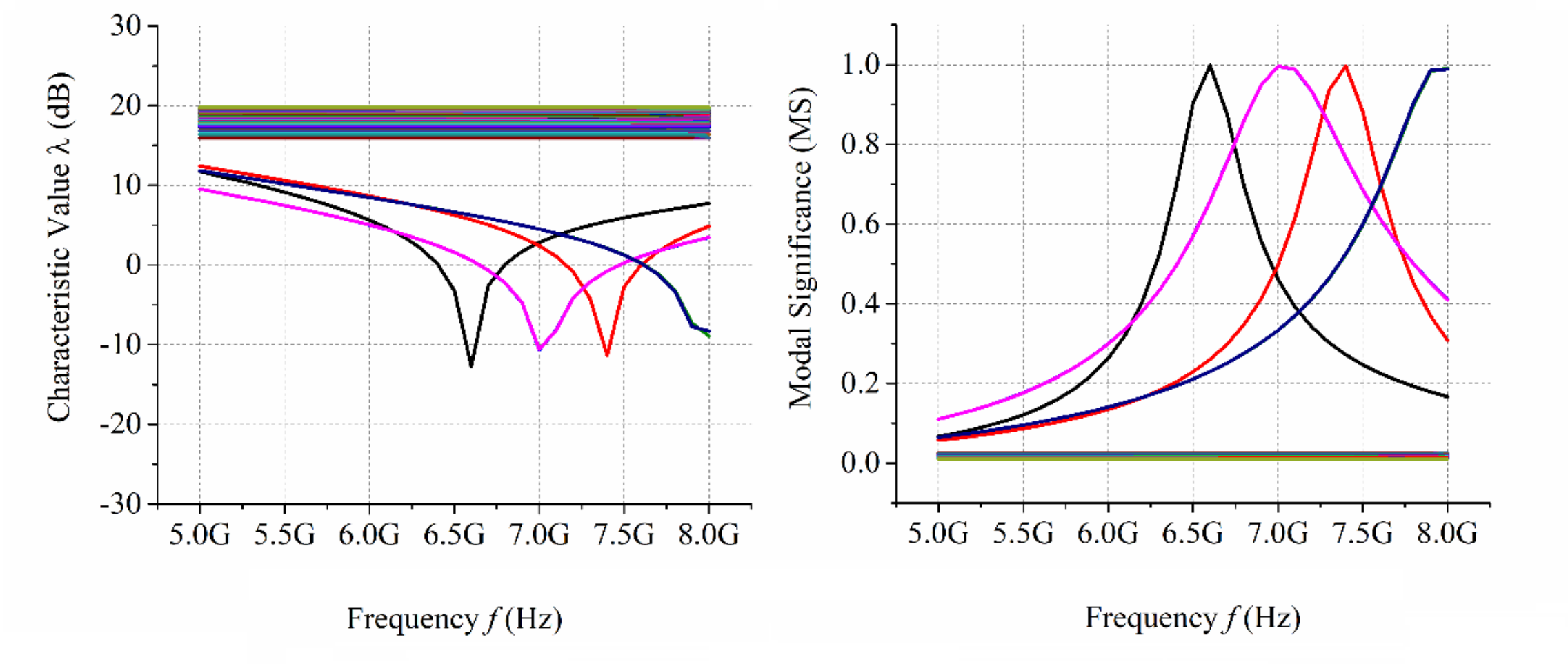

(a) (b)

Figure 6-26 The characteristic quantity curves corresponding to the first 200 DP-CMs (derived from the volume formulations (4-8)&(4-9)) of the two-body material system shown in Figure 6-8. (a) characteristic value dB curves; (b) MS curves

From above Figure 6-24, Figure 6-25, and Figure 6-26, it is easy to find out that: compared with the results derived from “the scheme retaining SCT”, the results derived from “the scheme deleting SCT” agree better with the results derived from the volume formulation.

### 3) Lossless Composite System Case

Now, we utilize two different schemes ("the scheme retaining SCT" and "the scheme deleting SCT") to construct the DP-CMs for the composite system shown in Figure 6-14 (i.e. the one shown in Figure 5-5), and provide the characteristic value (dB) curves and MS curves derived from the two schemes in Figure 6-27 and Figure 6-28 respectively.

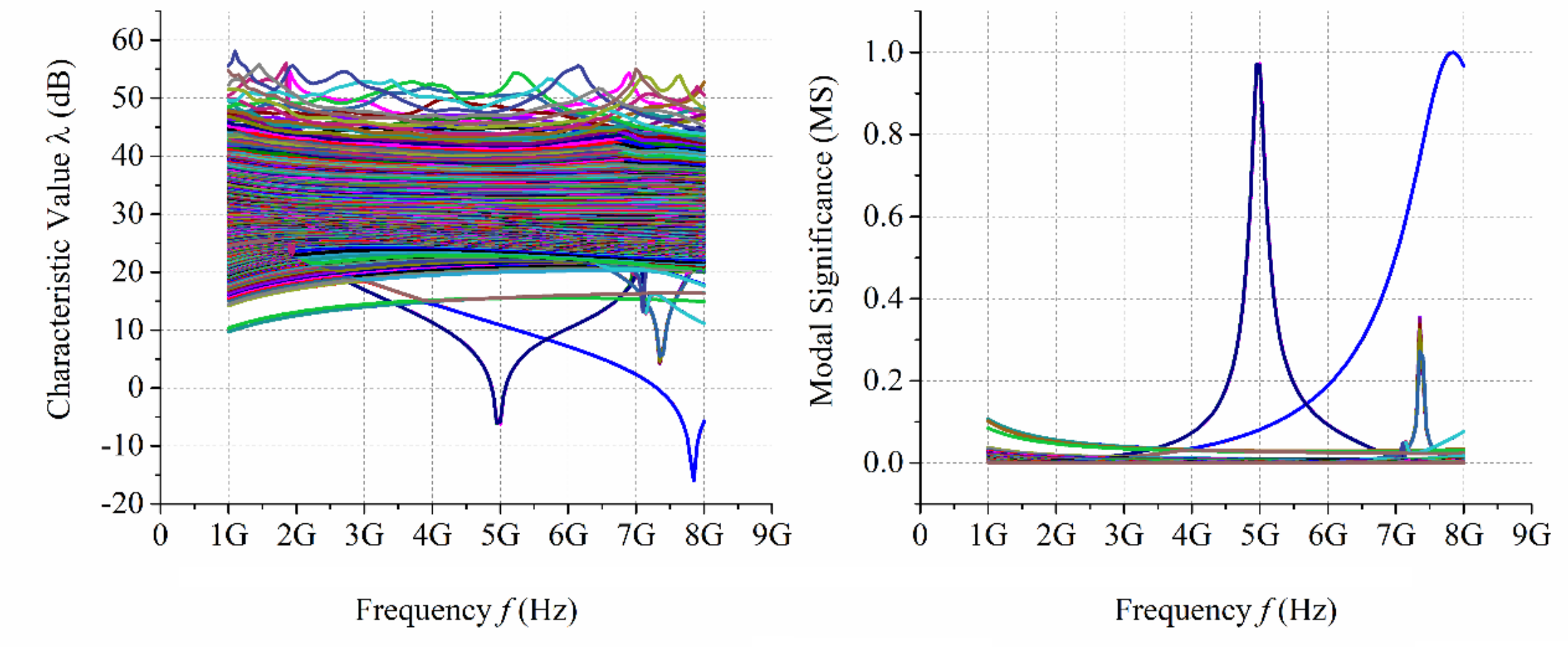

(a) (b)

Figure 6-27 The characteristic quantity curves corresponding to the first 700 DP-CMs (derived from the line-formulations (5-68)&(5-76) with SCT) of the composite system shown in Figure 6-14. (a) characteristic value dB curves; (b) MS curves

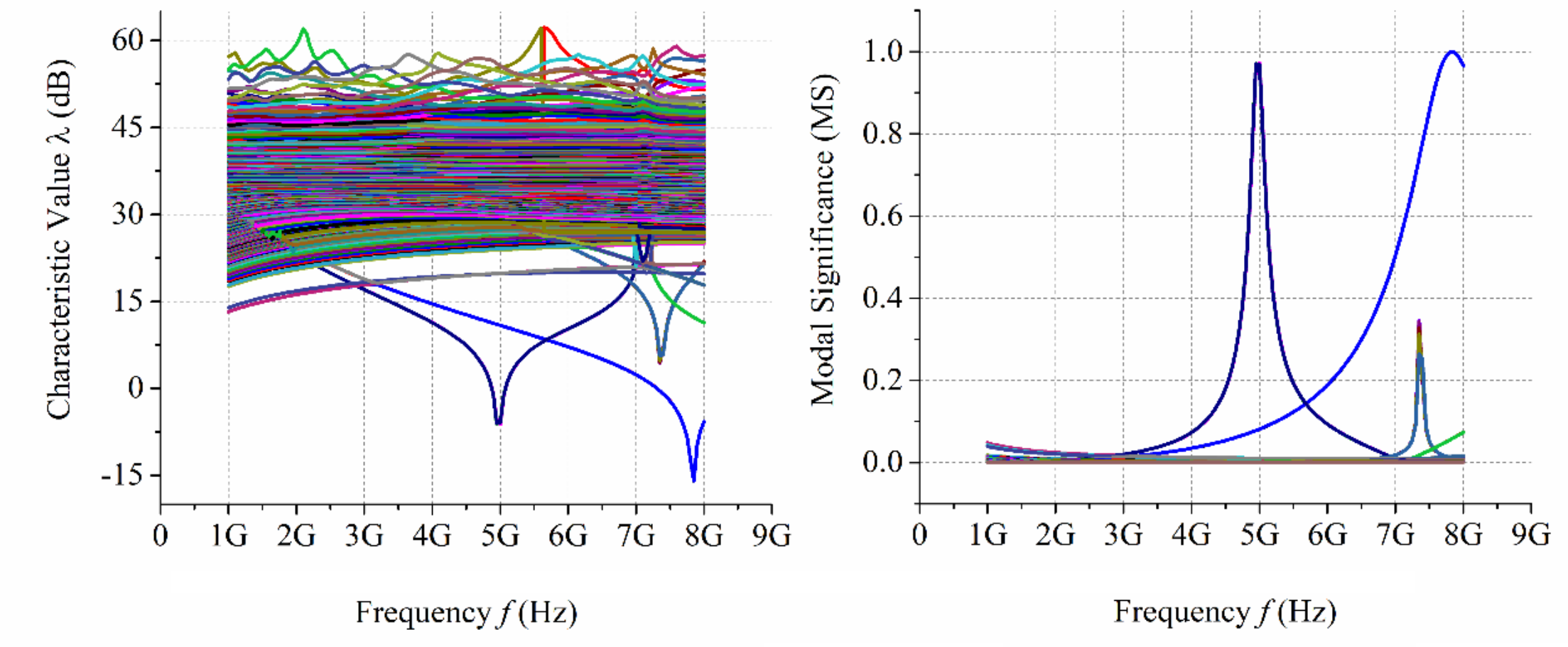

(a) (b)

Figure 6-28 The characteristic quantity curves corresponding to the first 700 DP-CMs (derived from the line-formulations (5-68)&(5-76) without SCT) of the composite system shown in Figure 6-14. (a) characteristic value dB curves; (b) MS curves

From the above figures, it is easy to find out that: compared with the results derived from "the scheme retaining SCT", the results derived from "the scheme deleting SCT" are more desirable, and this conclusion is consistent with the conclusions corresponding to the one-body and two-body material systems.

## 6.5 Chapter Summary

This chapter mainly studies two common topics related to the WEP-based CMTs for various scattering systems.

**1)** On the schemes to suppress the spurious modes outputted from the CM calculation formulations for various scattering systems. In fact, the topic can be divided into two sub-topics: on the selection for the independent and complete variable set contained in the DPO of various scattering systems (i.e., how to correctly select basic variables) and on the establishment for the linear transformation from the basic variables to the dependent variables (i.e., how to effectively unify variables). By doing some studies, it is found out that: the basic variables of scattering systems distribute on the outer boundaries of the scattering systems; the CM set derived from orthogonalizing the DPO with only basic variables doesn't contain any spurious mode; the CM set derived from orthogonalizing the DPO with dependent variables contains many spurious modes. This chapter summarizes a systematical scheme for suppressing the spurious modes of general scattering systems (as illustrated in Figure 6-19).

In addition, this chapter also provides a compromise scheme used to suppress spurious modes. In the premise of not calculating matrix inverses, the compromise scheme can effectively suppress the low-order spurious modes. But, it is possible that the compromise scheme can output some high-order spurious modes, and this is just the reason why this dissertation calls the scheme as "compromise scheme".

**2)** On the physical meaning of the singular EM current terms (SCTs) contained in the DPOs of various scattering systems. By some studies, it is found out that the SCTs equal to Ohmic loss powers of the corresponding scattering systems; for the lossless scattering systems, the SCTs contained in the matrix forms of the corresponding DPOs are manifested as 0 matrices; for the lossless scattering systems, the results derived from orthogonalizing the DPOs without SCTs are better than the results derived from orthogonalizing the DPOs with SCTs.

# Chapter 7 Conclusions

（数学的美体现在）主要是有新的东西。就是说，老的东西我真正有新的观念加进去，就改样了。要变化，要有新的东西，我对这有兴趣。……

—— 陈省身 （1983 年 Wolf 数学奖得主）

In this chapter, we summarize the works done by this dissertation firstly, and provide the prospects of future works finally.

## 7.1 Concluding Remarks

EM scattering systems widely exist in civilian field (such as wireless communication), commercial field (such as biomedical EM simulation), and military field (such as stealth and anti-stealth technology), etc. Characteristic mode theory (CMT) has ability to effectively construct a series of orthogonal characteristic modes (CMs) for any objective scattering system, and the obtained CMs only depend on the inherent physical properties of the objective scattering system. Thus, CMT has important instruction significance for extracting the inherent EM scattering characters of the objective scattering system, and has great application significance in the process of analyzing and designing the related engineering problems.

Taking a wide view of the whole development process of CMT (it has had a history of half a century, since Prof. Garbacz introduced the concept of CM in 1965 for the first time), it is easy to find out that: there was a transformation for the carrying framework of CMT —— from Prof. Garbacz's scattering matrix (SM) framework to Prof. Harrington's integral equation (IE) framework, and the framework transformation led to a transformation for the constructing method of CMs —— from Prof. Garbacz's orthogonalizing perturbation matrix operator (PMO) method to Prof. Harrington's orthogonalizing impedance matrix operator (IMO) method.

The above transformations greatly accelerate the development of CMT in the aspects of theoretical researches and engineering applications, because the orthogonalizing IMO method in IE framework is simpler and more efficient than the orthogonalizing PMO method in SM framework. But, Harrington's CMT (the CMT established by Prof. Harrington et al. in IE framework) still has some important problems which have not been completely solved (in what follows, they are simply called as old problems, and simply denoted as OPs), for example:

**Old Problem 1 (OP1).** The existing Harrington's CMT lacks a clear physical picture.

**Old Problem 2 (OP2).** The existing Harrington's CMT cannot construct complete CM sets for metallic systems.

**Old Problem 3 (OP3).** The existing Harrington's CMT cannot be applied to some relatively complicated material systems.

**Old Problem 4 (OP4).** The existing Harrington's CMT cannot be applied to some relatively complicated metal-material composite systems.

**Old Problem 5 (OP5).** The existing Harrington's CMT lacks the systematical scheme for suppressing the spurious modes outputted from the CM calculation formulations for general complicated scattering systems.

This dissertation is committed to further resolving the above-mentioned OP1~OP5 existing in Harrington's CMT. During solving OP1, OP3, and OP4, and after solving OP2 and OP5, this dissertation further extends a series of important problems which have not been widely concerned (in what follows they are simply called as new problems, and simply denoted as NPs), as follows:

**New Problem 1 (NP1).** During solving OP1, this dissertation, for the first time, puts forward the new problem on "the carrying framework for CMT".

**New Problem 2 (NP2).** After solving OP2, this dissertation, for the first time, puts forward the new problem on "finely classifying all working modes and orthogonally decomposing whole modal space".

**New Problem 3 (NP3).** During solving OP3, this dissertation puts forward the new problem on "deriving the mathematical expressions for the surface equivalence principle corresponding to complicated material systems".

**New Problem 4 (NP4).** During solving OP4, this dissertation, for the first time, puts forward the new problem on "establishing the mathematical expressions for the line-surface equivalence principle corresponding to complicated composite systems".

**New Problem 5 (NP5).** After solving OP5, this dissertation, for the first time, puts forward the new problem on "the physical meaning and the numerical performance of the singular EM current term contained in DPO".

Taking the above-mentioned 5 pairs of unsolved important problems {OP1, NP1}, {OP2, NP2}, …, and {OP5, NP5} as traction, this dissertation explores the new framework for carrying the CMT of scattering systems, and proposes the new method for constructing the CMs of scattering systems, and draws a clear physical picture for the Harrington's CMT of scattering systems, and extends the applicable range of the CMT for scattering systems, and derives a series of new formulations for calculating the CMs of various scattering systems, and improves some important steps in the process to construct the CMs of scattering systems, and reveals the essential differences between the Harrington's CMT of scattering systems and its some variants in IE framework, and enlarges whole theoretical formalism of CMT. Now, we summarize the main contributions and conclusions obtained in this dissertation (in what follows they are simply called as main contributions, and simply denoted as MCs) as follows:

**Main Contribution 1 (MC1). Contribution in the Aspect of Solving {OP1, NP1}**

In classical electromagnetics framework, this dissertation, for the first time, derives the counterpart of the work-energy principle (WEP) satisfied by the particles in classical mechanics —— the WEP satisfied by EM scattering systems: the work done by the resultant fields acting on the scattered sources are transverted into four parts, where a part is transferred from the scattering system to infinity, and a part is converted into Joule's heat, and a part is converted into the EM energies stored in the scattered fields, and a part is converted into the polarization and magnetization energies stored in the material part of the scattering system. Based on the WEP satisfied by scattering systems, this dissertation, for the first time, introduces the concept of the driving power (DP) for scattering systems —— power done by the resultant fields acting on scattered sources.

By orthogonalizing frequency-domain DP operator (DPO), this dissertation re-establishes the Harrington's CMT, which was established by Prof. Harrington et al. in IE framework in the 1970s. Actually, this is the second transformation for the carrying framework of CMT —— from IE framework to WEP framework, and also the second transformation for the constructing method of CMs —— from orthogonalizing IMO to orthogonalizing DPO, as illustrated in Figure 7-1. By transforming the carrying framework of CMT and the constructing method of CMs, this dissertation, for the first time, draws a clear physical picture for the Harrington's CMT of scattering systems —— constructing a series of steadily working modes not having net energy exchange in any integral period (i.e., constructing a series of orthogonal modes having ability to completely decouple the frequency-domain DPO of the scattering system), as illustrated in Figure 7-1.

From Figure 7-1, it is easy to find out that the Garbacz’s CMT in SM framework is a modal theory which is guided by a clear physical picture, so the physical features of Garbacz’s CMs have always been quite unequivocal for half a century (since 1965). But, the Harrington’s CMT in IE framework has been lack of a clear physical picture as guidance since its establishment (1971), so there have always been some doubts on the physical features satisfied by the Harrington’s CMs in IE framework, for example whether or not the far-fields of Harrington’s CMs must be orthogonal and whether or not the characteristic value calculated from Harrington’s characteristic equation must equal the ratio of modal imaginary power to modal real power; the Harrington’s CMT in IE framework has always had some imperfections, for example the traditional decomposition (i.e. to decompose the IMO into its real and imaginary parts) for EFIE-based IMO is not the most reasonable decomposition way; in IE framework, some derivatives of Harrington’s CMT were proposed, such as the MFIE-based CMT for metallic systems, the CFIE-based CMT for metallic systems, and the complex background Green’s function based CMT for metallic systems, etc.

Based on the physical picture that this dissertation draws for Harrington’s CMT, this dissertation, for the first time, explains the above-mentioned doubts on the far fields and characteristic values of Harrington’s CMs; this dissertation provides practicable schemes to improve some imperfections existing in Harrington’s CMT; this dissertation, for the first time, reveals the essential differences between the Harrington’s CMT for scattering systems and its some derivatives in IE framework, and then obtains the conclusion that “it should not confuse Harrington’s CMT with its some IE-based derivatives”.

In addition, orthogonalizing DPO method also clearly exhibits the active feature of Harrington’s CMT and the steadily working feature of Harrington’s CMs, i.e., Harrington’s CMT is an active modal theory, which is used to construct the steadily working modes of the scattering systems driven by external excitations, rather than a passive modal theory (so-called passive modal theory is source-free modal theory). This is just why some source-free derivatives of Harrington’s CMT cannot give satisfactory results, and then this dissertation obtains the conclusion that “it should not confuse Harrington’s CMT with its some source-free derivatives”.

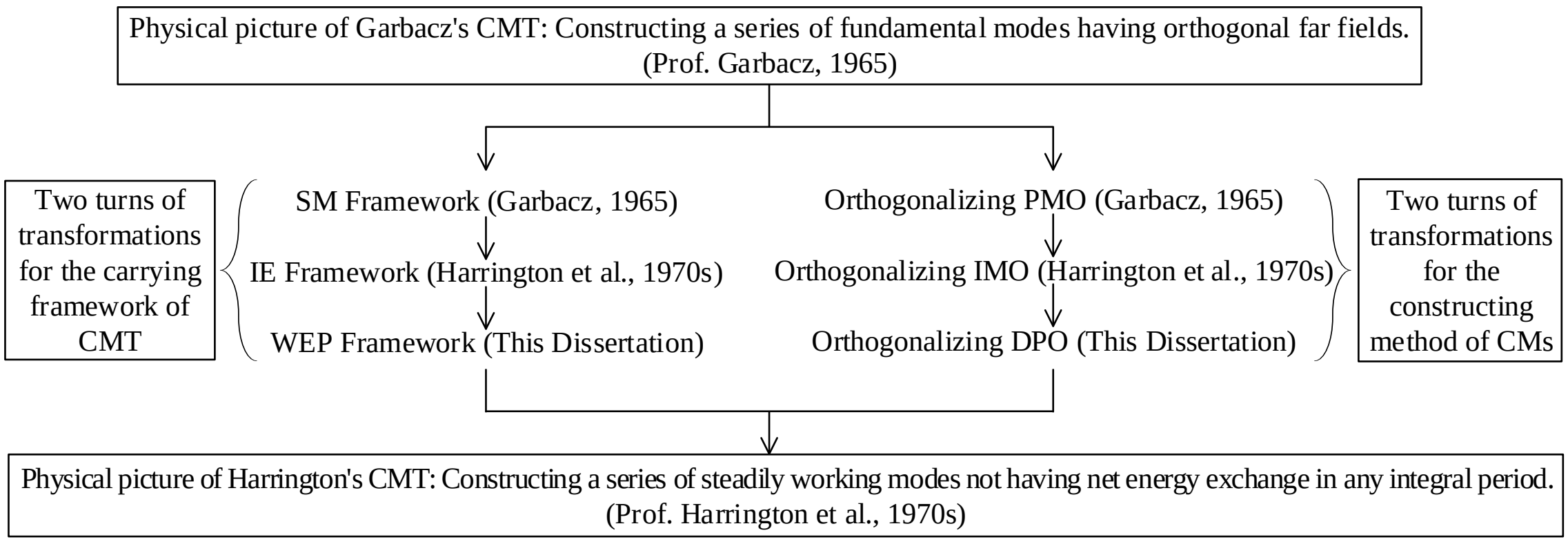

Figure 7-1 Two turns of transformations for the carrying framework of CMT, two turns of transformations for the constructing method of CMs, and the physical pictures of Garbacz's CMT and Harrington's CMT

In fact, the physical picture of Harrington's CMT and the second transformations for carrying framework and constructing method also benefit CMT in some other aspects (for details see the following MC2, MC3, MC4, and MC5). To emphasize that "the CMT discussed in this dissertation is the modal theory which is carried by WEP framework and which focuses on constructing the steadily working CMs not having net energy exchange in any integral period and which constructs the CMs by orthogonalizing frequency-domain DPO", we specially call the CMT and CMs as WEP-based CMT and driving power CMs (DP-CMs) respectively in the following MC2~MC5.

**Main Contribution 2 (MC2). Contribution in the Aspect of Solving {OP2, NP2}**

In new WEP framework, by orthogonalizing DPO, this dissertation, for any objective metallic system, constructs a series of steadily working modes not having net energy exchange in any integral period —— DP-CMs. This dissertation proves that the radiative DP-CMs are just the Harrington's CMs of open metallic systems, and also proves that the non-radiative DP-CMs are equivalent to the internally resonant eigen-modes of closed metallic systems, and then, for the first time, integrates the Harrington's CMT for open metallic systems and the eigen-mode theory for closed metallic systems as a whole modal theory —— WEP-based CMT for metallic scattering systems (WEP-MetSca-CMT). The above integration realizes the completeness for Harrington's CM set.

Employing the complete DP-CM set, this dissertation realizes the orthogonal decomposition for the modal space of any objective metallic system (i.e., modal space = pure capacitance space ⊕ nonradiation space ⊕ high-quality radiation space ⊕ pure inductance space) and the modal fine classification for all of the working modes of any objective metallic system (as illustrated in Figure 7-2).

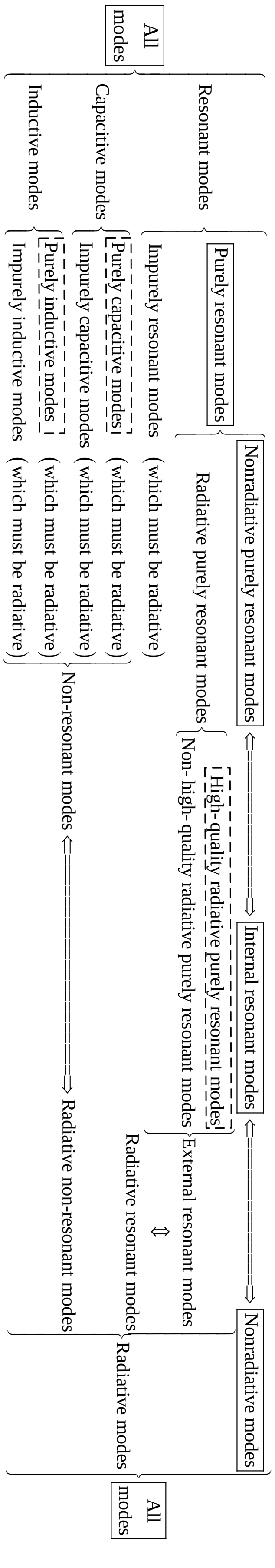

Figure 7-2 The fine classification for all of the steadily working modes in frequency domain

Then, this dissertation, for the first time, realizes the simplest orthogonal decompositions for various working modes (for example: a general working mode = purely capacitive component + nonradiative component + high-quality radiative component + purely inductive component), and also proves the uniqueness of the above decompositions. The above simplest orthogonal decompositions have important instruction significance for revealing the working mechanism of various modes.

**Main Contribution 3 (MC3). Contribution in the Aspect of Solving {OP3, NP3}**

From the aspects of external EM environment, system topological structure, and system material parameter, this dissertation generalizes the traditional surface equivalence principle (SEP), which is only suitable for simply connected homogeneous isotropic material bodies, to the generalized surface equivalence principle (GSEP), which is applicable to complicated material systems, and derives the corresponding mathematical expressions.

In new WEP framework, by orthogonalizing DPO, by employing the newly obtained GSEP, this dissertation establishes the volume and surface formulations of the WEP-based CMT for material systems (Vol-WEP-MatSca-CMT and Surf-WEP-MatSca-CMT). Vol-WEP-MatSca-CMT and Surf-WEP-MatSca-CMT can, for the objective material system, construct a series of steadily working modes not having net energy exchange in any integral period —— DP-CMs of the objective material system. From the aspects mathematical formulation and computational example, this dissertation verifies that the DP-CM set derived from Vol-WEP-MatSca-CMT and the DP-CM set derived from Surf-WEP-MatSca-CMT are equivalent to each other.

In addition, from the aspects of external EM environment, system topological structure, and system material parameter, WEP-MatSca-CMT (the collective name of Vol-WEP-MatSca-CMT and Surf-WEP-MatSca-CMT) has a very wide applicable range: in the aspect of EM environment, WEP-MatSca-CMT is independent of the EM environment surrounding the objective material system, and only depends on the inherent physical properties of the objective material system, so has great application significance in the process to extract the inherent EM scattering characters of the objective scattering system; in the aspect of topological structure, WEP-MatSca-CMT is not only suitable for a simply connected material body, but also valid for a multiply connected material body, even applicable to a multi-body material system, and the material bodies contained in the multi-body material system can be either simply connected or multiply connected (the topological structures of some typical multi-body material systems are illustrated in Figure 7-3); in the aspect of material parameter, WEP-MatSca-CMT is not only suitable for homogeneous isotropic material systems, but also valid for inhomogeneous anisotropic material systems, even applicable to piecewise inhomogeneous anisotropic material systems.

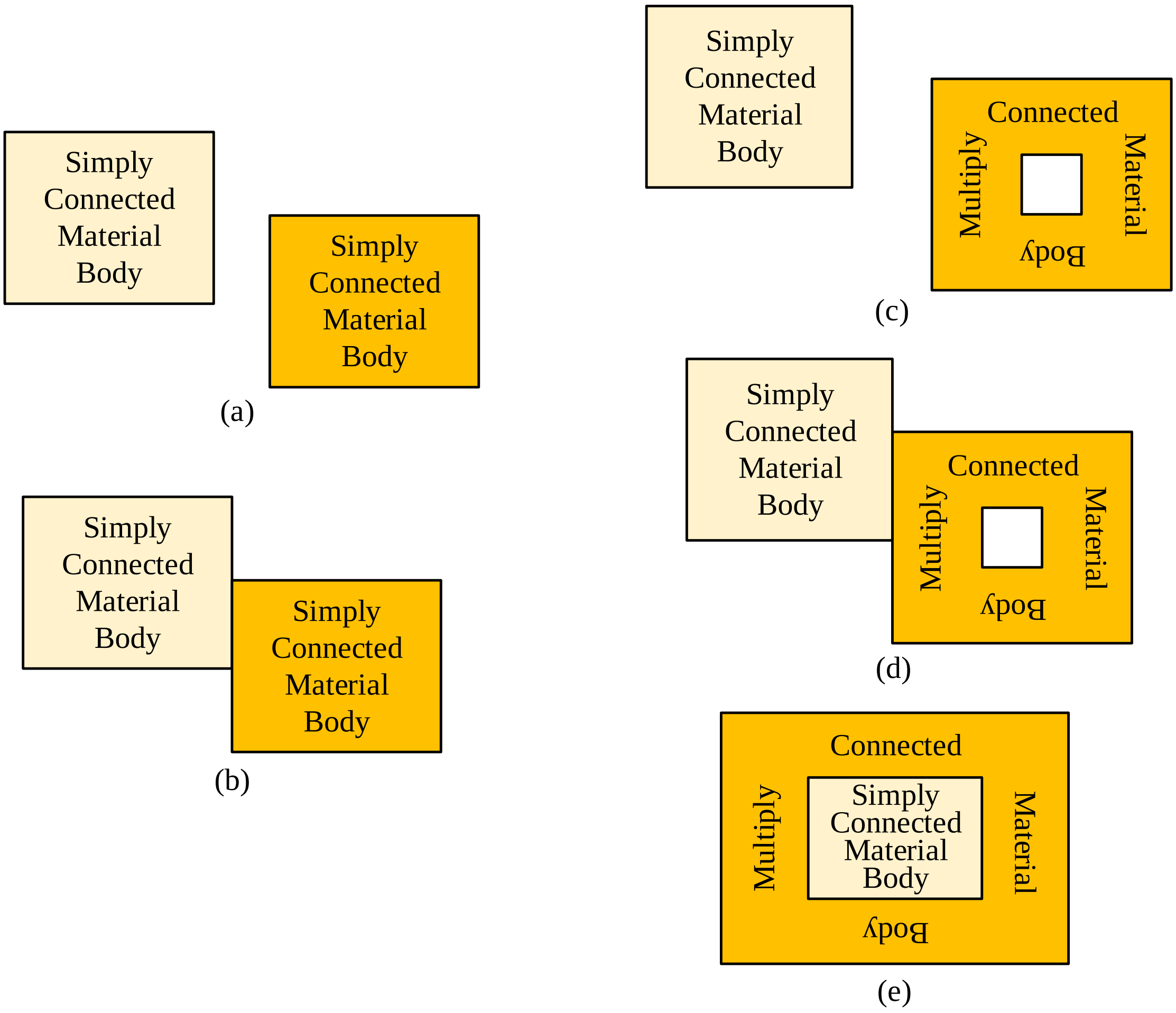

Figure 7-3 The topological structures of some typical two-body material systems to which WEP-MatSca-CMT is applicable. (a) the system constituted by two simply connected material bodies not contacting with each other; (b) the system constituted by two simply connected material bodies contacting with each other; (c) the system constituted by a simply connected material body and a multiply connected material body, where the bodies don't contact with each other; (d) the system constituted by a simply connected material body and a multiply connected material body, where the bodies contact with each other and the simply connected body is not surrounded by the multiply connected body; (e) the system constituted by a simply connected material body and a multiply connected material body, where the bodies contact with each other and the simply connected body is surrounded by the multiply connected body

It needs to be specially emphasized that: this dissertation establishes the Surf-WEP-MatSca-CMT for the inhomogeneous anisotropic material systems with various topological structures, but doesn't explicitly provide the closed analytical expressions corresponding to the Green's functions in inhomogeneous anisotropic medium. Thus, the Surf-WEP-MatSca-CMT for inhomogeneous anisotropic material systems still has values to be further studied. However, when all of the material bodies used to constitute whole objective material system are homogeneous isotropic or piecewise homogeneous isotropic, the above-mentioned Surf-WEP-MatSca-CMT for various system topological structures is applicable directly (as exhibited by the computational examples given in this dissertation), because the related problem needs to use the Green's functions in homogeneous isotropic medium only. In addition, the Vol-WEP-MatSca-CMT for the inhomogeneous anisotropic material systems with various topological structures needs to use the Green's functions in vacuum only, so it is applicable directly, as exhibited by the computational examples given in this dissertation.

**Main Contribution 4 (MC4). Contribution in the Aspect of Solving {OP4, NP4}**

Following the ideas used to obtain the GSEP for complicated material systems, this dissertation, for the first time, establishes the line-surface equivalence principle (LSEP) for complicated composite systems, and derives the corresponding mathematical expressions. In the aspects of external EM environment, system topological structure, and system material parameter, LSEP has a very wide applicable range.

In new WEP framework, by orthogonalizing DPO, and by employing the newly obtained LSEP, this dissertation establishes the line-surface formulation of the WEP-based CMT for composite systems (LS-WEP-ComSca-CMT). Utilizing LS-WEP-ComSca-CMT, this dissertation, for any objective composite system, constructs a series of steadily working modes not having net energy exchange in any integral period —— DP-CMs of the objective composite system.

In the aspects of external EM environment, system topological structure, and system material parameter, LS-WEP-ComSca-CMT has a very wide applicable range: in the aspect of EM environment, LS-WEP-ComSca-CMT is independent of the EM environment surrounding the objective composite system, just like WEP-MatSca-CMT; in the aspect of topological structure, the metallic part can be line structure, surface structure, volume structure, or "line-surface-volume composite structure", and the metallic part and the material part can be either contacted with each other or separated from each other, and the metallic part can be either completely or partially submerged into the material part (the topological structure of a typical composite system is illustrated in Figure 7-4); in the aspect of material parameter, the material part can be either homogeneous isotropic or inhomogeneous anisotropic.

It needs to be specially emphasized that: when the material part of the objective composite system is inhomogeneous anisotropic, the LS-WEP-ComSca-CMT established in this dissertation needs to use the Green's functions in inhomogeneous anisotropic medium; because there has not been the closed analytical expressions for the Green's functions in inhomogeneous anisotropic medium, then the LS-WEP-ComSca-CMT in this case has not been applicable directly, so the LS-WEP-ComSca-CMT in this case still has some researching spaces. But, when the material part of the objective composite system is homogeneous isotropic, the LS-WEP-ComSca-CMT established in this dissertation is applicable directly, as exhibited by the computational examples given in this dissertation.

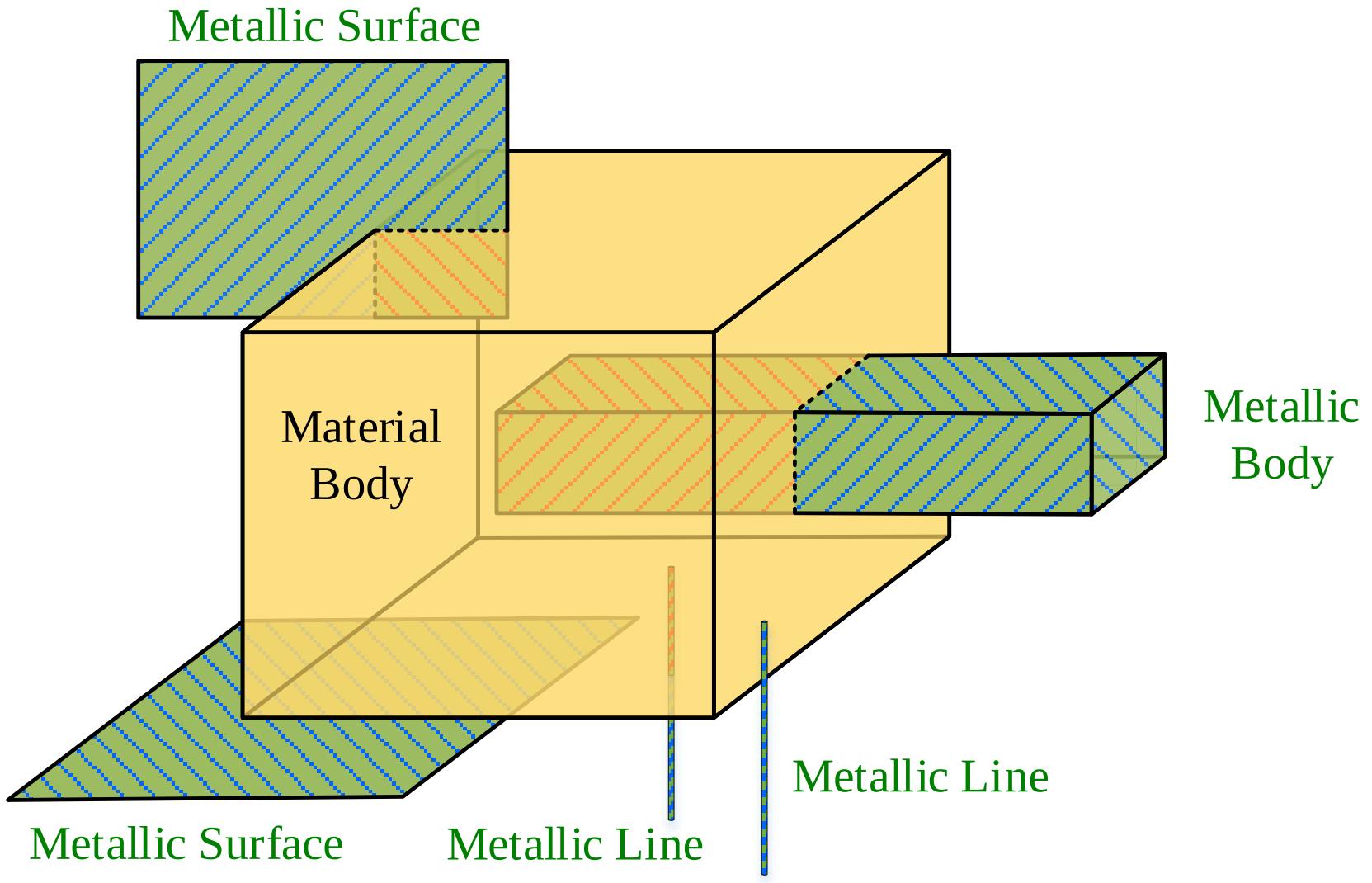

Figure 7-4 The typical topological structure of the composite systems to which LS-WEP-ComSca-CMT is applicable

**Main Contribution 5 (MC5). Contribution in the Aspect of Solving {OP5, NP5}**

In new WEP framework, focusing on orthogonalizing DPO, and employing the newly obtained GSEP and LSEP, this dissertation systematically studies the forming cause and suppressing method for the spurious modes which widely exist in CMT domain, and for the first time systematically summarizes the operating process for suppressing the spurious modes for arbitrary scattering systems, and the operating process is illustrated in Figure 7-5.

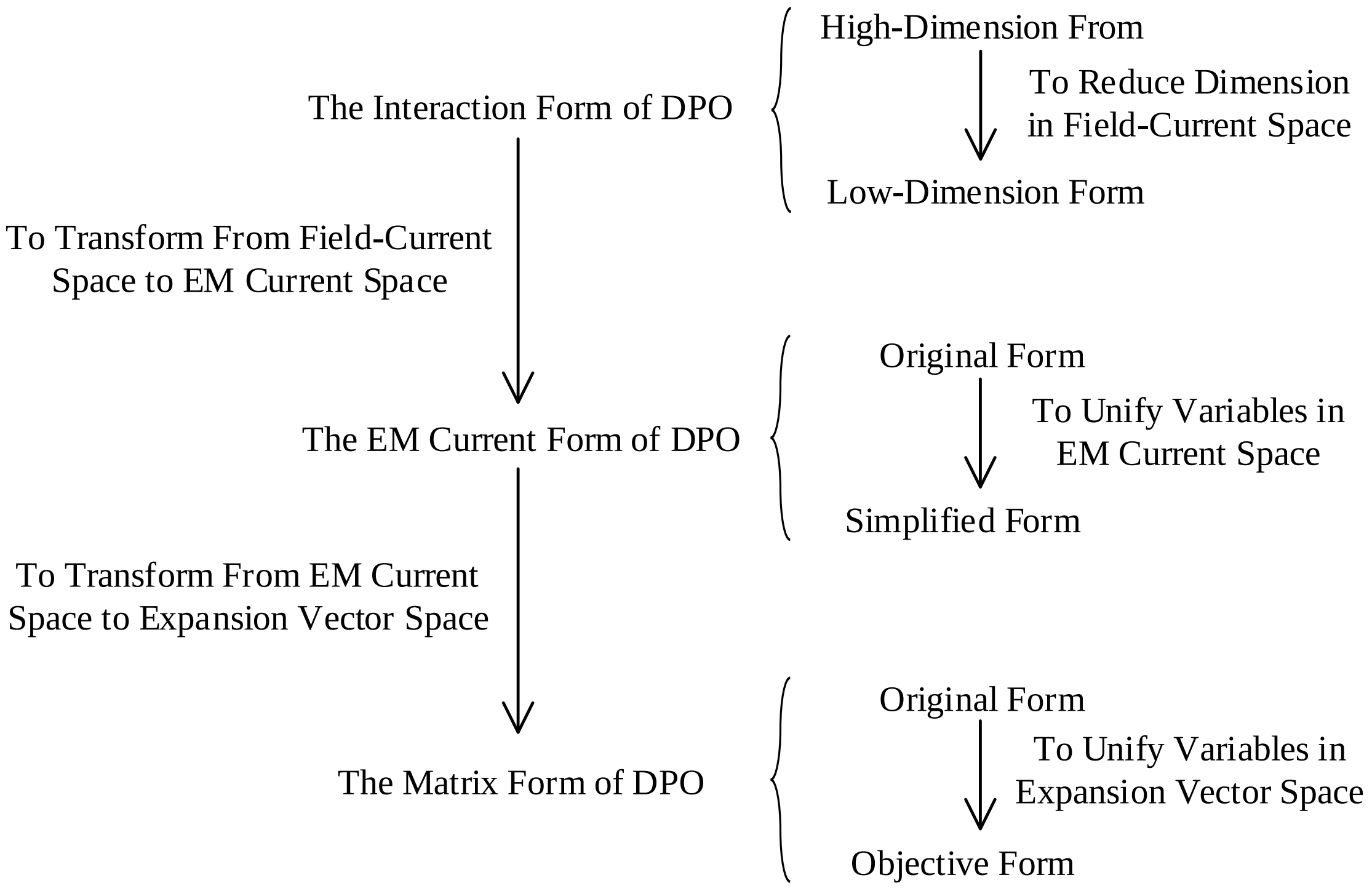

Figure 7-5 "The process to unify variables for scattering system" and "the evolution process of the manifestation form of DPO". For details see Subsection 6.2.4

This dissertation, for the first time, finds out that the physical meaning of the singular EM current term contained in the DPO of scattering systems is just the lossy power dissipated by scattering systems, and, for the first time, proves that the singular EM current term contained in the DPO of lossless scattering systems always appear as zero matrices, and points out that the above conclusion is valuable for improving the numerical performance of the Surf-WEP-MatSca-CMT and LS-WEP-ComSca-CMT for lossless systems.

In addition, in new WEP framework, by orthogonalizing DPO, this dissertation establishes some formulations for calculating the DP-CMs of various scattering systems, and the CM calculation formulations are completely new (except the one for metallic systems), but the new formulations have not been able to be effectively established in IE framework. However, the traditional Harrington's CM calculation formulations in IE framework can be effectively established in WEP framework. Moreover, the new formulations in WEP framework have wider applicable ranges, more concise manifestation forms, and clearer physical picture than the traditional formulations in IE framework. For the simple scattering systems to which both the new formulations and the traditional formulations are applicable, it is, by some numerical calculations, found out that: for the relatively simple scattering systems to which both the new and traditional formulations are applicable, it is verified by some numerical examples that the results derived from the new and traditional formulations agree well with each others; for the relatively complicated scattering systems to which the traditional formulations are not applicable, this dissertation, by specializing the parameters of the complicated systems, makes the complicated systems degenerate to the simple systems to which the traditional formulations are applicable, and constructs the CMs of the degenerated systems by employing both the new and traditional formulations, and finds out that the results derived from the different formulations agree well with each others. It is thus clear that: as the carrying framework for CMT, new WEP framework is more advantageous than traditional IE framework; as the constructing method for CMs, new orthogonalizing DPO method is more advantageous than traditional orthogonalizing IMO method.

Finally, the author summarizes this dissertation in one sentence as follows: *in work-energy principle (WEP) framework and by orthogonalizing frequency-domain driving power operator (DPO), this dissertation, for the objective scattering system, constructs a series of steadily working modes — driving power characteristic modes (DP-CMs) — which don't have net energy exchange in any integral period.*

## 7.2 Future Works

This dissertation analytically derives the convolution integral formulations of the surface equivalence principle for inhomogeneous anisotropic material systems and the line-surface equivalence principle for metal-material composite systems (the material parts of the composite systems can be inhomogeneous anisotropic), and also establishes the CM calculation formulations for the material systems and the composite systems by employing the convolution integrals.

But, the closed analytical expressions for the Green's functions in inhomogeneous anisotropic medium have not been obtained systematically, so the Surf-WEP-MatSca-CMT established in this dissertation has not been applicable directly when the objective material system contains inhomogeneous anisotropic matter (only due to the lack of the corresponding Green's functions); when the material part in the objective composite system is inhomogeneous anisotropic, the LS-WEP-ComSca-CMT established in this dissertation has not been applicable directly (only due to the lack of the corresponding Green's functions).

Obviously, "how to obtain the closed analytical expressions for the Green's functions in inhomogeneous anisotropic medium (at least in homogeneous anisotropic medium)" is an important topic which is worth to be further studied.

# Appendixes

… the most compact formulation appears to be the one based on the dyadic Green's function pertaining to the vector wave equation for $\vec{E}$ and $\vec{H}$ … [145]

—— Chen-To Tai

In this appendix, we provide the detailed derivations for some important formulations directly related to this dissertation.

## Appendix A Mathematical Expressions of the Volume Equivalence Principle for Inhomogeneous Anisotropic Material Systems

Based on conduction electric current model, polarization electric current model, and magnetization electric current model, literature [121] expressed the scattered fields in terms of the convolution integrals of vacuum Green's functions and {conduction volume electric current, polarization volume electric current, magnetization volume electric current, magnetization surface electric current}. Being different from literature [121], this Appendix A will express the scattered fields in terms of the convolution integrals of vacuum Green's functions and {conduction volume electric current, polarization volume electric current, magnetization volume magnetic current}, by employing conduction electric current model, polarization electric current model, and magnetization magnetic current model[110].

The common destination of literature [121] and this Appendix A is, in a mathematically rigorous way, to study the following important questions: when the conduction, polarization, and magnetization phenomena are described by employing some pre-selected models, which forms will the scattered sources distribute on material systems? Which relationships will the scattered sources and the EM fields satisfy? In what follows, the answers to the above questions are collectively referred to as the volume equivalence principle (VEP) or the induction principle for material systems.

### A1 Mathematical Expressions of the Frequency-Domain Volume Equivalence Principle for an Inhomogeneous Anisotropic Material Body Placed in Vacuum Environment

In this Appendix A1, we consider the scattering problem corresponding to the simply connected material body $V$ shown in Figure A-1, and rigorously establish the induced-source-based convolution integral formulations for EM fields.

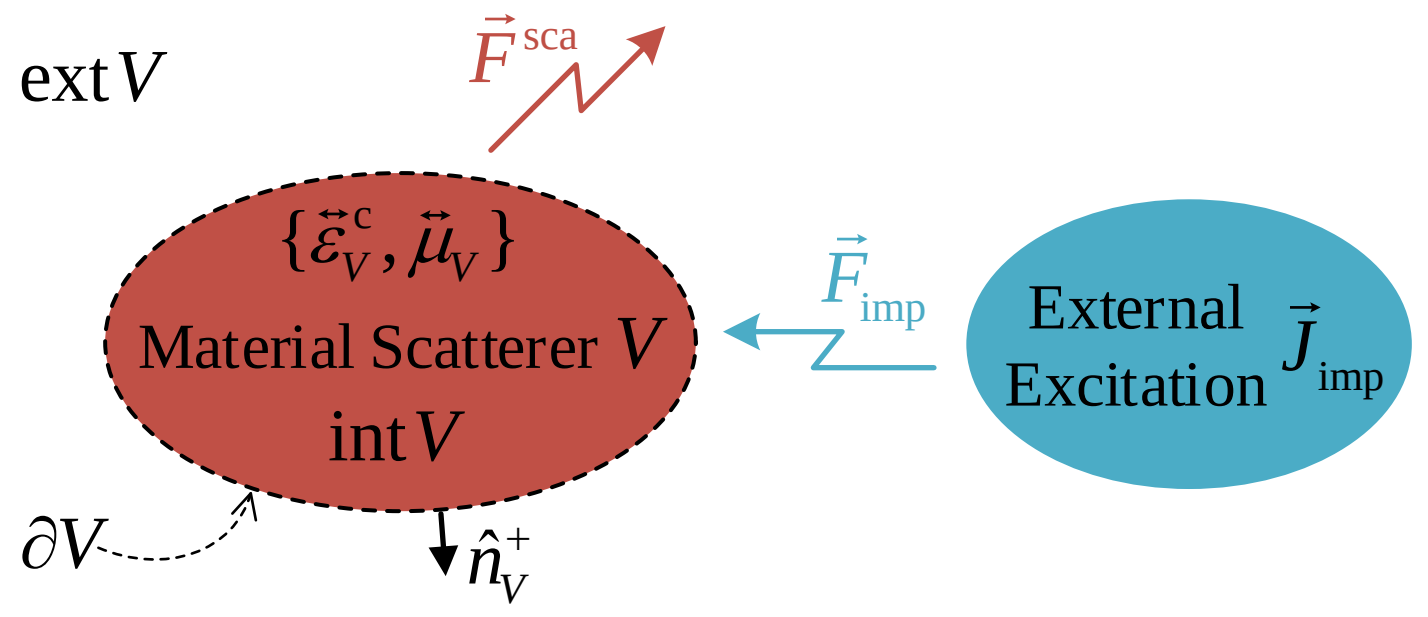

Figure A-1 The scattering problem corresponding to a material system placed in vacuum

In whole three-dimensional Euclidean space $\mathbb{R}^3$, the impressed fields $\{\vec{E}_{\text{imp}}, \vec{H}_{\text{imp}}\}$ ① generated by impressed source $\vec{J}_{\text{imp}}$ satisfy the following Maxwell's equations[121]:

$$\begin{aligned} \nabla \times \vec{H}_{\text{imp}}(\vec{r}) &= \ \ j\omega\varepsilon_0 \vec{E}_{\text{imp}}(\vec{r}) + \vec{J}_{\text{imp}}(\vec{r}) \\ \nabla \times \vec{E}_{\text{imp}}(\vec{r}) &= -j\omega\mu_0 \vec{H}_{\text{imp}}(\vec{r}) \end{aligned} \quad , \quad \vec{r} \in \mathbb{R}^3 \tag{A-1}$$

where $\vec{J}_{\text{imp}}$ doesn't distribute on $V$. We call the summation of impressed field and the field generated by induced sources as total field, and denote the summation as $\vec{F}^{\text{tot}}$, where $F = E, H$, and then the $\{\vec{E}^{\text{tot}}, \vec{H}^{\text{tot}}\}$ distributing on ext$V$ (which is the exterior of $V$) satisfy the following Maxwell's equations[121]:

$$\begin{aligned} \nabla \times \vec{H}^{\text{tot}}(\vec{r}) &= \ \ j\omega\varepsilon_0 \vec{E}^{\text{tot}}(\vec{r}) + \vec{J}_{\text{imp}}(\vec{r}) \\ \nabla \times \vec{E}^{\text{tot}}(\vec{r}) &= -j\omega\mu_0 \vec{H}^{\text{tot}}(\vec{r}) \end{aligned} \quad , \quad \vec{r} \in \text{ext}V \tag{A-2}$$

The $\{\vec{E}^{\text{tot}}, \vec{H}^{\text{tot}}\}$ distributing on int$V$ (which is the interior of $V$) satisfy the following Maxwell's equations[121]:

$$\begin{aligned} \nabla \times \vec{H}^{\text{tot}}(\vec{r}) &= \ \ j\omega\vec{\vec{\varepsilon}}_V^{\,c}(\vec{r}) \cdot \vec{E}^{\text{tot}}(\vec{r}) \\ \nabla \times \vec{E}^{\text{tot}}(\vec{r}) &= -j\omega\vec{\vec{\mu}}_V(\vec{r}) \cdot \vec{H}^{\text{tot}}(\vec{r}) \end{aligned} \quad , \quad \vec{r} \in \text{int}V \tag{A-3}$$

where $\vec{\vec{\varepsilon}}_V^{\,c} = \vec{\vec{\varepsilon}}_V + \vec{\vec{\sigma}}_V / j\omega$ and $\vec{\vec{\mu}}_V$ are respectively the complex permittivity and permeability of the material body, and $\vec{\vec{\varepsilon}}_V$ and $\vec{\vec{\sigma}}_V$ are respectively the permittivity and

① This dissertation always focuses on fields $\{\vec{E}, \vec{H}\}$ rather than $\{\vec{E}, \vec{B}\}$. As pointed out in literature [110] that: "… only by choosing $\vec{H}$ as the second macroscopic field vector it is possible to develop a macroscopic theory based on only two field vectors, which is both self-consistent and in agreement with experimental evidence. It turns out that this same choice also removes certain difficulties previously encountered in extending the theory to macroscopic bodies in relative motion. [110]".

conductivity of the material body, and these material parameters are two-order tensors.

In vacuum, the EM fields $\{\overleftrightarrow{G}_0^{JE}(\vec{r},\vec{r}\,'),\overleftrightarrow{G}_0^{JH}(\vec{r},\vec{r}\,')\}$ (i.e. the electric-type dyadic Green's functions) generated by electric-type unit dyadic point source $\overleftrightarrow{I}\delta(\vec{r}-\vec{r}\,')$ satisfy the following Maxwell's equations[120]:

$$\begin{aligned}\nabla\times\overleftrightarrow{G}_0^{JH}(\vec{r},\vec{r}\,') &= j\omega\varepsilon_0\overleftrightarrow{G}_0^{JE}(\vec{r},\vec{r}\,')+\overleftrightarrow{I}\delta(\vec{r}-\vec{r}\,')\\ \nabla\times\overleftrightarrow{G}_0^{JE}(\vec{r},\vec{r}\,') &= -j\omega\mu_0\overleftrightarrow{G}_0^{JH}(\vec{r},\vec{r}\,')\end{aligned}\quad,\quad(\vec{r},\vec{r}\,'\in\mathbb{R}^3)\qquad\text{(A-4)}$$

In the above equations, $\overleftrightarrow{I}=\hat{x}\hat{x}+\hat{y}\hat{y}+\hat{z}\hat{z}$ is two-order unit dyad; $\overleftrightarrow{G}_0^{JE}(\vec{r},\vec{r}\,')=\eta_0[(1/jk_0)\nabla\nabla-jk_0\overleftrightarrow{I}]G_0(\vec{r},\vec{r}\,')$ and $\overleftrightarrow{G}_0^{JH}(\vec{r},\vec{r}\,')=\nabla\times[\overleftrightarrow{I}G_0(\vec{r},\vec{r}\,')]$ [106]; $G_0(\vec{r},\vec{r}\,')$ is the scalar Green's function in vacuum. From equations (A-1)~(A-4), we can derive the following wave equations pertaining to incident fields, total fields, and dyadic Green's functions[121]:

$$\left(\nabla\times\nabla\times-k_0^2\right)\vec{E}_{\text{imp}}(\vec{r}) = -j\omega\mu_0\vec{J}_{\text{imp}}(\vec{r})\quad,\quad\vec{r}\in\mathbb{R}^3\qquad\text{(A-5)}$$

$$\left(\nabla\times\nabla\times-k_0^2\right)\vec{E}^{\text{tot}}(\vec{r}) = -j\omega\mu_0\vec{J}_{\text{imp}}(\vec{r})\quad,\quad\vec{r}\in\text{ext}V\qquad\text{(A-6)}$$

$$\left(\nabla\times\nabla\times-k_0^2\right)\vec{E}^{\text{tot}}(\vec{r}) = -j\omega\mu_0\left(j\omega\Delta\overleftrightarrow{\varepsilon}_V^{\text{c}}\cdot\vec{E}^{\text{tot}}\right)-\nabla\times\left[j\omega\Delta\overleftrightarrow{\mu}_V\cdot\vec{H}^{\text{tot}}\right]\quad,\quad\vec{r}\in\text{int}V\qquad\text{(A-7)}$$

$$\left(\nabla\times\nabla\times-k_0^2\right)\overleftrightarrow{G}_0^{JE}(\vec{r},\vec{r}\,') = -j\omega\mu_0\overleftrightarrow{I}\delta(\vec{r}-\vec{r}\,')\quad,\quad\vec{r}\in\mathbb{R}^3\qquad\text{(A-8)}$$

In wave equation (A-7), $\Delta\overleftrightarrow{\varepsilon}_V^{\text{c}}=\overleftrightarrow{\varepsilon}_V^{\text{c}}-\overleftrightarrow{I}\varepsilon_0$ and $\Delta\overleftrightarrow{\mu}_V=\overleftrightarrow{\mu}_V-\overleftrightarrow{I}\mu_0$.

In any open domain $\Omega$, there exists the following vector-dyad Green's second theorem[120,122]:

$$\iiint_\Omega\left[\vec{P}\cdot\left(\nabla\times\nabla\times\overleftrightarrow{Q}\right)-\left(\nabla\times\nabla\times\vec{P}\right)\cdot\overleftrightarrow{Q}\right]d\Omega=\oiint_{\partial\Omega}\hat{n}_\Omega^-\cdot\left[\vec{P}\times\left(\nabla\times\overleftrightarrow{Q}\right)+\left(\nabla\times\vec{P}\right)\times\overleftrightarrow{Q}\right]dS\qquad\text{(A-9)}$$

In Green's theorem (A-9), $\partial\Omega$ is the boundary of $\Omega$; $\hat{n}_\Omega^-$ is the normal vector of $\partial\Omega$ and points to the interior of $\Omega$. In Green's theorem (A-9), if we let $\vec{P}=\vec{E}_{\text{imp}}(\vec{r})$, and $\overleftrightarrow{Q}=\overleftrightarrow{G}_0^{JE}(\vec{r},\vec{r}\,')$, and $\Omega=\mathbb{R}^3$, and utilize the Maxwell's equations (A-1) & (A-4), wave equations (A-5) & (A-8), and Sommerfeld's radiation conditions $\lim_{|\vec{r}|\to+\infty}(\nabla\times\vec{F}_{\text{imp}}+jk_0\hat{n}_\infty^+\times\vec{F}_{\text{imp}})=0$ & $\lim_{|\vec{r}-\vec{r}\,'|\to+\infty}[\nabla\times\overleftrightarrow{G}_0^{JF}(\vec{r},\vec{r}\,')+jk_0\hat{n}_\infty^+\times\overleftrightarrow{G}_0^{JF}(\vec{r},\vec{r}\,')]=0$ [120] satisfied by $\vec{E}_{\text{imp}}(\vec{r})$ & $\overleftrightarrow{G}_0^{JE}(\vec{r},\vec{r}\,')$, we can obtain the convolution integral formulation pertaining to $\vec{E}_{\text{imp}}(\vec{r})$ as follows[121]:

$$\vec{E}_{\text{imp}}(\vec{r}) = \iiint_{\text{ext}V}\overleftrightarrow{G}_0^{JE}(\vec{r},\vec{r}\,')\cdot\vec{J}_{\text{imp}}(\vec{r}\,')dV'\quad,\quad\vec{r}\in\mathbb{R}^3\qquad\text{(A-10)}$$

Here, the reason to utilize $\text{ext}V$ as integral domain is that $\vec{J}_{\text{imp}}$ distributes on $\text{ext}V$ only.

For the convenience of the following discussions, we introduce a dyad $\overleftrightarrow{\Psi}(\vec{r}',\vec{r})$ as follows[121]:

$$\overleftrightarrow{\Psi}(\vec{r}',\vec{r}) = \vec{E}^{\text{tot}}(\vec{r}')\times\left[\nabla'\times\overleftrightarrow{G}_0^{JE}(\vec{r}',\vec{r})\right]+\left[\nabla'\times\vec{E}^{\text{tot}}(\vec{r}')\right]\times\overleftrightarrow{G}_0^{JE}(\vec{r}',\vec{r}) \tag{A-11}$$

where vectorial differential operator $\nabla'$ acts on variable $\vec{r}'$. Taking the $\vec{r}'$-based divergence of $\overleftrightarrow{\Psi}(\vec{r}',\vec{r})$, we have that

$$\begin{aligned}\nabla'\cdot\overleftrightarrow{\Psi}(\vec{r}',\vec{r}) = &\ \nabla'\cdot\left\{\vec{E}^{\text{tot}}(\vec{r}')\times\left[\nabla'\times\overleftrightarrow{G}_0^{JE}(\vec{r}',\vec{r})\right]\right\}+\nabla'\cdot\left\{\left[\nabla'\times\vec{E}^{\text{tot}}(\vec{r}')\right]\times\overleftrightarrow{G}_0^{JE}(\vec{r}',\vec{r})\right\}\\
= &\ \left[\nabla'\times\vec{E}^{\text{tot}}(\vec{r}')\right]\cdot\left[\nabla'\times\overleftrightarrow{G}_0^{JE}(\vec{r}',\vec{r})\right]-\vec{E}^{\text{tot}}(\vec{r}')\cdot\left[\nabla'\times\nabla'\times\overleftrightarrow{G}_0^{JE}(\vec{r}',\vec{r})\right]\\
&+\left[\nabla'\times\nabla'\times\vec{E}^{\text{tot}}(\vec{r}')\right]\cdot\overleftrightarrow{G}_0^{JE}(\vec{r}',\vec{r})-\left[\nabla'\times\vec{E}^{\text{tot}}(\vec{r}')\right]\cdot\left[\nabla'\times\overleftrightarrow{G}_0^{JE}(\vec{r}',\vec{r})\right]\\
= &\ \left[\nabla'\times\nabla'\times\vec{E}^{\text{tot}}(\vec{r}')\right]\cdot\overleftrightarrow{G}_0^{JE}(\vec{r}',\vec{r})-\vec{E}^{\text{tot}}(\vec{r}')\cdot\left[\nabla'\times\nabla'\times\overleftrightarrow{G}_0^{JE}(\vec{r}',\vec{r})\right]\\
= &\ \left[\nabla'\times\nabla'\times\vec{E}^{\text{tot}}(\vec{r}')-k_0^2\vec{E}^{\text{tot}}(\vec{r}')\right]\cdot\overleftrightarrow{G}_0^{JE}(\vec{r}',\vec{r})\\
&-\vec{E}^{\text{tot}}(\vec{r}')\cdot\left[\nabla'\times\nabla'\times\overleftrightarrow{G}_0^{JE}(\vec{r}',\vec{r})-k_0^2\overleftrightarrow{G}_0^{JE}(\vec{r}',\vec{r})\right]\end{aligned} \tag{A-12}$$

In formulation (A-12), the first equality is based on the linear property satisfied by divergence operation; the second equality is based on differential identity $\nabla\cdot(\vec{P}\times\overleftrightarrow{Q})=(\nabla\times\vec{P})\cdot\overleftrightarrow{Q}-\vec{P}\cdot(\nabla\times\overleftrightarrow{Q})$ [122]; the third and fourth equalities are obvious.

If $\vec{r}'\in\text{ext}V$, we can obtain the following formulation (A-13) by inserting wave equations (A-6) and (A-8) into the RHS of the fourth equality in relationship (A-12).

$$\begin{aligned}\nabla'\cdot\overleftrightarrow{\Psi}(\vec{r}',\vec{r}) &= -\,j\omega\mu_0\vec{J}_{\text{imp}}(\vec{r}')\cdot\overleftrightarrow{G}_0^{JE}(\vec{r}',\vec{r})-\vec{E}^{\text{tot}}(\vec{r}')\cdot\left[-j\omega\mu_0\overleftrightarrow{I}\delta(\vec{r}'-\vec{r})\right]\\
&= -\,j\omega\mu_0\overleftrightarrow{G}_0^{JE}(\vec{r},\vec{r}')\cdot\vec{J}_{\text{imp}}(\vec{r}')+j\omega\mu_0\vec{E}^{\text{tot}}(\vec{r}')\cdot\overleftrightarrow{I}\delta(\vec{r}'-\vec{r})\end{aligned} \tag{A-13}$$

The second equality in formulation (A-13) is based on the symmetry property $\overleftrightarrow{G}_0^{JE}(\vec{r}',\vec{r})=[\overleftrightarrow{G}_0^{JE}(\vec{r},\vec{r}')]^T$ [120] pertaining to Green's function $\overleftrightarrow{G}_0^{JE}(\vec{r}',\vec{r})$. If we integrate the variable $\vec{r}'$ in the LHS of formulation (A-13) on whole $\text{ext}V$, we have that

$$\begin{aligned}\iiint_{\text{ext}V}\nabla'\cdot\overleftrightarrow{\Psi}(\vec{r}',\vec{r})dV' &= -\oint\!\!\!\oint_{\partial V}\hat{n}_V^+\cdot\overleftrightarrow{\Psi}(\vec{r}',\vec{r})dS'+\oint\!\!\!\oint_{S_\infty}\hat{n}_\infty^+\cdot\overleftrightarrow{\Psi}(\vec{r}',\vec{r})dS'\\
&= -\oint\!\!\!\oint_{\partial V}\hat{n}_V^+\cdot\overleftrightarrow{\Psi}(\vec{r}',\vec{r})dS'\end{aligned} \tag{A-14}$$

In relationship (A-14), $\partial V$ is the boundary of material body $V$, and $\hat{n}_V^+$ is the outer normal vector of $\partial V$; the meanings of $S_\infty$ and $\hat{n}_\infty^+$ have been explained in Chapter 3, and it will not be repeated here; the first equality is based on Gauss' divergence theorem[122]; the second equality is based on the Sommerfeld's radiation conditions[120] satisfied by $\{\vec{E}^{\text{tot}}(\vec{r}'),\vec{H}^{\text{tot}}(\vec{r}')\}$ ① and $\{\overleftrightarrow{G}_0^{JE}(\vec{r}',\vec{r}),\overleftrightarrow{G}_0^{JH}(\vec{r}',\vec{r})\}$. If we integrate the variable

① Incident field $\vec{F}^{\text{inc}}$ and scattered field $\vec{F}^{\text{sca}}$ satisfy the Sommerfeld's radiation condition at infinity, and the

$\vec{r}'$ in the RHS of relationship (A-13) on whole $\mathrm{ext}V$, we have the following result:

$$\begin{aligned}\iiint_{\mathrm{ext}V}\nabla'\cdot\vec{\Psi}(\vec{r}',\vec{r})dV' &= -j\omega\mu_0\vec{E}_{\mathrm{imp}}(\vec{r})+j\omega\mu_0\iiint_{\mathrm{ext}V}\vec{E}^{\mathrm{tot}}(\vec{r}')\cdot\vec{I}\delta(\vec{r}'-\vec{r})dV' \\ &= \begin{cases}-j\omega\mu_0\vec{E}_{\mathrm{imp}}(\vec{r})+j\omega\mu_0\vec{E}^{\mathrm{tot}}(\vec{r}) &, \quad \vec{r}\in\mathrm{ext}V \\ -j\omega\mu_0\vec{E}_{\mathrm{imp}}(\vec{r}) &, \quad \vec{r}\in\mathrm{int}V\end{cases}\end{aligned} \tag{A-15}$$

In relationship (A-15), the first equality is based on convolution integral formulation (A-10); the second equality is based on the integral property of Dirac's delta function $\delta(\vec{r}'-\vec{r})$ [120,122]. Employing relationships (A-14) and (A-15), it is immediately obtained that

$$-\oiint_{\partial V}\hat{n}_V^+\cdot\vec{\Psi}(\vec{r}',\vec{r})dS' = \begin{cases}-j\omega\mu_0\vec{E}_{\mathrm{imp}}(\vec{r})+j\omega\mu_0\vec{E}^{\mathrm{tot}}(\vec{r}) &, \quad \vec{r}\in\mathrm{ext}V \\ -j\omega\mu_0\vec{E}_{\mathrm{imp}}(\vec{r}) &, \quad \vec{r}\in\mathrm{int}V\end{cases} \tag{A-16}$$

If $\vec{r}'\in\mathrm{int}V$, inserting wave equations (A-7) and (A-8) into the RHS of the fourth equality in relationship (A-12), we immediately obtain that

$$\begin{aligned}\nabla'\cdot\vec{\Psi}(\vec{r}',\vec{r}) =& -j\omega\mu_0\left[j\omega\Delta\vec{\varepsilon}_V^{\,\mathrm{c}}\cdot\vec{E}^{\mathrm{tot}}(\vec{r}')\right]\cdot\vec{G}_0^{JE}(\vec{r}',\vec{r})-\left\{\nabla'\times\left[j\omega\Delta\vec{\mu}_V\cdot\vec{H}^{\mathrm{tot}}(\vec{r}')\right]\right\}\cdot\vec{G}_0^{JE}(\vec{r}',\vec{r}) \\ &-\vec{E}^{\mathrm{tot}}(\vec{r}')\cdot\left[-j\omega\mu_0\vec{I}\delta(\vec{r}'-\vec{r})\right] \\ =& -j\omega\mu_0\vec{G}_0^{JE}(\vec{r},\vec{r}')\cdot\left[j\omega\Delta\vec{\varepsilon}_V^{\,\mathrm{c}}\cdot\vec{E}^{\mathrm{tot}}(\vec{r}')\right]-\left\{\nabla'\times\left[j\omega\Delta\vec{\mu}_V\cdot\vec{H}^{\mathrm{tot}}(\vec{r}')\right]\right\}\cdot\vec{G}_0^{JE}(\vec{r}',\vec{r}) \\ &+j\omega\mu_0\vec{E}^{\mathrm{tot}}(\vec{r}')\cdot\vec{I}\delta(\vec{r}'-\vec{r})\end{aligned} \tag{A-17}$$

In above formulation (A-17), the second equality is based on symmetry $\vec{G}_0^{JE}(\vec{r}',\vec{r})=[\vec{G}_0^{JE}(\vec{r},\vec{r}')]^T$ [120]. Integrating the LHS of formulation (A-17) on whole $\mathrm{int}V$, and employing Gauss' divergence theorem[122], we have that

$$\iiint_{\mathrm{int}V}\nabla'\cdot\vec{\Psi}(\vec{r}',\vec{r})dV' = \oiint_{\partial V}\hat{n}_V^+\cdot\vec{\Psi}(\vec{r}',\vec{r})dS' \tag{A-18}$$

In what follows, we do some identical transformations for the RHS of relationship (A-17) in two different ways, and then obtain two different convolution integral formulations for scattered fields, where the second formulation is just the one given in literature [121].

**1) To Describe Polarization Phenomenon and Magnetization Phenomenon by Employing Polarization Electric Current Model and Magnetization Magnetic Current Model Respectively[110]**

Integrating the RHS of formulation (A-17) on whole $\mathrm{int}V$, we have that

---

condition is linear, so total field $\vec{F}^{\mathrm{tot}}$ also satisfies the same condition due to linear superposition principle $\vec{F}^{\mathrm{tot}}=\vec{F}^{\mathrm{inc}}+\vec{F}^{\mathrm{sca}}$.

$$
\begin{aligned}
&\iiint_{\mathrm{int}V} \nabla' \cdot \vec{\vec{\Psi}}(\vec{r}',\vec{r})dV' \\
&= -j\omega\mu_0 \iiint_{\mathrm{int}V} \vec{\vec{G}}_0^{JE}(\vec{r},\vec{r}')\cdot\left[j\omega\Delta\vec{\vec{\varepsilon}}_V^{\,\mathrm{c}}\cdot\vec{E}^{\mathrm{tot}}(\vec{r}')\right]dV' - \iiint_{\mathrm{int}V}\nabla'\cdot\left\{\left[j\omega\Delta\vec{\vec{\mu}}_V\cdot\vec{H}^{\mathrm{tot}}(\vec{r}')\right]\times\vec{\vec{G}}_0^{JE}(\vec{r}',\vec{r})\right\}dV' \\
&\quad - \iiint_{\mathrm{int}V}\left[j\omega\Delta\vec{\vec{\mu}}_V\cdot\vec{H}^{\mathrm{tot}}(\vec{r}')\right]\cdot\left[\nabla'\times\vec{\vec{G}}_0^{JE}(\vec{r}',\vec{r})\right]dV' + j\omega\mu_0\iiint_{\mathrm{int}V}\vec{E}^{\mathrm{tot}}(\vec{r}')\cdot\vec{\vec{I}}\delta(\vec{r}'-\vec{r})dV' \\
&= -j\omega\mu_0 \iiint_{\mathrm{int}V} \vec{\vec{G}}_0^{JE}(\vec{r},\vec{r}')\cdot\left[j\omega\Delta\vec{\vec{\varepsilon}}_V^{\,\mathrm{c}}\cdot\vec{E}^{\mathrm{tot}}(\vec{r}')\right]dV' - \oint\!\!\!\oint_{\partial V}\hat{n}_V^+\cdot\left\{\left[j\omega\Delta\vec{\vec{\mu}}_V\cdot\vec{H}^{\mathrm{tot}}(\vec{r}')\right]\times\vec{\vec{G}}_0^{JE}(\vec{r}',\vec{r})\right\}dS' \\
&\quad - \iiint_{\mathrm{int}V}\left[j\omega\Delta\vec{\vec{\mu}}_V\cdot\vec{H}^{\mathrm{tot}}(\vec{r}')\right]\cdot\left[\nabla'\times\vec{\vec{G}}_0^{JE}(\vec{r}',\vec{r})\right]dV' + j\omega\mu_0\iiint_{\mathrm{int}V}\vec{E}^{\mathrm{tot}}(\vec{r}')\cdot\vec{\vec{I}}\delta(\vec{r}'-\vec{r})dV' \\
&= -j\omega\mu_0 \iiint_{\mathrm{int}V} \vec{\vec{G}}_0^{JE}(\vec{r},\vec{r}')\cdot\left[j\omega\Delta\vec{\vec{\varepsilon}}_V^{\,\mathrm{c}}\cdot\vec{E}^{\mathrm{tot}}(\vec{r}')\right]dV' - \oint\!\!\!\oint_{\partial V}\hat{n}_V^+\cdot\left\{\left[j\omega\Delta\vec{\vec{\mu}}_V\cdot\vec{H}^{\mathrm{tot}}(\vec{r}')\right]\times\vec{\vec{G}}_0^{JE}(\vec{r}',\vec{r})\right\}dS' \\
&\quad - \iiint_{\mathrm{int}V}\left[j\omega\Delta\vec{\vec{\mu}}_V\cdot\vec{H}^{\mathrm{tot}}(\vec{r}')\right]\cdot\left[-j\omega\mu_0\vec{\vec{G}}_0^{JH}(\vec{r}',\vec{r})\right]dV' + j\omega\mu_0\iiint_{\mathrm{int}V}\vec{E}^{\mathrm{tot}}(\vec{r}')\cdot\vec{\vec{I}}\delta(\vec{r}'-\vec{r})dV' \\
&= -j\omega\mu_0 \iiint_{\mathrm{int}V} \vec{\vec{G}}_0^{JE}(\vec{r},\vec{r}')\cdot\left[j\omega\Delta\vec{\vec{\varepsilon}}_V^{\,\mathrm{c}}\cdot\vec{E}^{\mathrm{tot}}(\vec{r}')\right]dV' - \oint\!\!\!\oint_{\partial V}\hat{n}_V^+\cdot\left\{\left[j\omega\Delta\vec{\vec{\mu}}_V\cdot\vec{H}^{\mathrm{tot}}(\vec{r}')\right]\times\vec{\vec{G}}_0^{JE}(\vec{r}',\vec{r})\right\}dS' \\
&\quad - j\omega\mu_0\iiint_{\mathrm{int}V}\vec{\vec{G}}_0^{ME}(\vec{r},\vec{r}')\cdot\left[j\omega\Delta\vec{\vec{\mu}}_V\cdot\vec{H}^{\mathrm{tot}}(\vec{r}')\right]dV' + j\omega\mu_0\iiint_{\mathrm{int}V}\vec{E}^{\mathrm{tot}}(\vec{r}')\cdot\vec{\vec{I}}\delta(\vec{r}'-\vec{r})dV' \\
&= -j\omega\mu_0 \iiint_{\mathrm{int}V} \vec{\vec{G}}_0^{JE}(\vec{r},\vec{r}')\cdot\left[j\omega\Delta\vec{\vec{\varepsilon}}_V^{\,\mathrm{c}}\cdot\vec{E}^{\mathrm{tot}}(\vec{r}')\right]dV' - \oint\!\!\!\oint_{\partial V}\hat{n}_V^+\cdot\left\{\left[j\omega\Delta\vec{\vec{\mu}}_V\cdot\vec{H}^{\mathrm{tot}}(\vec{r}')\right]\times\vec{\vec{G}}_0^{JE}(\vec{r}',\vec{r})\right\}dS' \\
&\quad - j\omega\mu_0\iiint_{\mathrm{int}V}\vec{\vec{G}}_0^{ME}(\vec{r},\vec{r}')\cdot\left[j\omega\Delta\vec{\vec{\mu}}_V\cdot\vec{H}^{\mathrm{tot}}(\vec{r}')\right]dV' + \begin{cases} 0 & , \quad \vec{r}\in\mathrm{ext}V \\ j\omega\mu_0\vec{E}^{\mathrm{tot}}(\vec{r}) & , \quad \vec{r}\in\mathrm{int}V \end{cases} \qquad \text{(A-19)}
\end{aligned}
$$

In formulation (A-19), the first equality is based on differential identity $\nabla\cdot(\vec{P}\times\vec{Q}) = (\nabla\times\vec{P})\cdot\vec{Q} - \vec{P}\cdot(\nabla\times\vec{Q})$ [122]; the second equality is based on Gauss' divergence theorem[122]; the third equality is based on formulation (A-4); the fourth equality is based on symmetry $\vec{\vec{G}}_0^{ME}(\vec{r},\vec{r}') = -[\vec{\vec{G}}_0^{JH}(\vec{r}',\vec{r})]^T$ [120]; the fifth equality is based on the integral property of Dirac's delta function $\delta(\vec{r}'-\vec{r})$ [120,122]. Employing formulations (A-18) and (A-19), it is immediately obtained that

$$
\begin{aligned}
\oint\!\!\!\oint_{\partial V}\hat{n}_V^+\cdot\vec{\vec{\Psi}}(\vec{r}',\vec{r})dS' = &-j\omega\mu_0\iiint_{\mathrm{int}V}\vec{\vec{G}}_0^{JE}(\vec{r},\vec{r}')\cdot\left[j\omega\Delta\vec{\vec{\varepsilon}}_V^{\,\mathrm{c}}\cdot\vec{E}^{\mathrm{tot}}(\vec{r}')\right]dV' \\
&-j\omega\mu_0\iiint_{\mathrm{int}V}\vec{\vec{G}}_0^{ME}(\vec{r},\vec{r}')\cdot\left[j\omega\Delta\vec{\vec{\mu}}_V\cdot\vec{H}^{\mathrm{tot}}(\vec{r}')\right]dV' \\
&-\oint\!\!\!\oint_{\partial V}\hat{n}_V^+\cdot\left\{\left[j\omega\Delta\vec{\vec{\mu}}_V\cdot\vec{H}^{\mathrm{tot}}(\vec{r}')\right]\times\vec{\vec{G}}_0^{JE}(\vec{r}',\vec{r})\right\}dS' \\
&+\begin{cases} 0 & , \quad \vec{r}\in\mathrm{ext}V \\ j\omega\mu_0\vec{E}^{\mathrm{tot}}(\vec{r}) & , \quad \vec{r}\in\mathrm{int}V \end{cases} \qquad \text{(A-20)}
\end{aligned}
$$

The summation of relationships (A-16) and (A-20) gives that

$$
\begin{aligned}
\oint\!\!\!\oint_{\partial V}\hat{n}_V^+\cdot\left[\vec{\vec{\Psi}}(\vec{r}_V^-,\vec{r})-\vec{\vec{\Psi}}(\vec{r}_V^+,\vec{r})\right]dS' = &-j\omega\mu_0\iiint_{\mathrm{int}V}\vec{\vec{G}}_0^{JE}(\vec{r},\vec{r}')\cdot\left[j\omega\Delta\vec{\vec{\varepsilon}}_V^{\,\mathrm{c}}\cdot\vec{E}^{\mathrm{tot}}(\vec{r}')\right]dV' \\
&-j\omega\mu_0\iiint_{\mathrm{int}V}\vec{\vec{G}}_0^{ME}(\vec{r},\vec{r}')\cdot\left[j\omega\Delta\vec{\vec{\mu}}_V\cdot\vec{H}^{\mathrm{tot}}(\vec{r}')\right]dV' \\
&-\oint\!\!\!\oint_{\partial V}\hat{n}_V^+\cdot\left\{\left[j\omega\Delta\vec{\vec{\mu}}_V\cdot\vec{H}^{\mathrm{tot}}(\vec{r}_V^-)\right]\times\vec{\vec{G}}_0^{JE}(\vec{r}_V^-,\vec{r})\right\}dS' \\
&+\begin{cases} -j\omega\mu_0\vec{E}_{\mathrm{imp}}(\vec{r})+j\omega\mu_0\vec{E}^{\mathrm{tot}}(\vec{r}) & , \quad \vec{r}\in\mathrm{ext}V \\ -j\omega\mu_0\vec{E}_{\mathrm{imp}}(\vec{r})+j\omega\mu_0\vec{E}^{\mathrm{tot}}(\vec{r}) & , \quad \vec{r}\in\mathrm{int}V \end{cases}
\end{aligned}
$$

$$
\begin{aligned}
&= - j\omega\mu_0 \iiint_{\text{int}V} \vec{G}_0^{JE}(\vec{r},\vec{r}\,')\cdot\left[j\omega\Delta\vec{\varepsilon}_V^{\,\text{c}}\cdot\vec{E}^{\text{tot}}(\vec{r}\,')\right]dV' \\
&\quad - j\omega\mu_0 \iiint_{\text{int}V} \vec{G}_0^{ME}(\vec{r},\vec{r}\,')\cdot\left[j\omega\Delta\vec{\mu}_V\cdot\vec{H}^{\text{tot}}(\vec{r}\,')\right]dV' \\
&\quad - \oiint_{\partial V} \hat{n}_V^+\cdot\left\{\left[j\omega\Delta\vec{\mu}_V\cdot\vec{H}^{\text{tot}}\left(\vec{r}_V^{\,-}\right)\right]\times\vec{G}_0^{JE}(\vec{r}\,',\vec{r})\right\}dS' \\
&\quad + \begin{cases} -j\omega\mu_0\vec{E}_{\text{imp}}(\vec{r}) + j\omega\mu_0\vec{E}^{\text{tot}}(\vec{r}) \;,\quad \vec{r}\in\text{ext}V \\ -j\omega\mu_0\vec{E}_{\text{imp}}(\vec{r}) + j\omega\mu_0\vec{E}^{\text{tot}}(\vec{r}) \;,\quad \vec{r}\in\text{int}V \end{cases}
\end{aligned}
\tag{A-21}
$$

In formulation (A-21), $\vec{r}_V^{\,-}\in\text{int}V$, and $\vec{r}_V^{\,+}\in\text{ext}V$, and $\vec{r}_V^{\,-}$ and $\vec{r}_V^{\,+}$ approach the point $\vec{r}\,'$ on $\partial V$; the second equality is based on the continuity of the Green's function $\vec{G}_0^{JE}(\vec{r}\,',\vec{r})$ on $\partial V$, because its source $\vec{I}\delta(\vec{r}\,'-\vec{r})$ doesn't distribute on $\partial V$. Now, we do some necessary identical transformations for the integral $\oiint_{\partial V}\hat{n}_V^+\cdot[\vec{\Psi}(\vec{r}_V^{\,-},\vec{r})-\vec{\Psi}(\vec{r}_V^{\,+},\vec{r})]dS'$ on the LHS of formulation (A-21), as follows:

$$
\begin{aligned}
&\oiint_{\partial V}\hat{n}_V^+\cdot\left[\vec{\Psi}\left(\vec{r}_V^{\,-},\vec{r}\right)-\vec{\Psi}\left(\vec{r}_V^{\,+},\vec{r}\right)\right]dS' \\
=\;& \oiint_{\partial V}\hat{n}_V^+\cdot\left\{\vec{E}^{\text{tot}}\left(\vec{r}_V^{\,-}\right)\times\left[\nabla_V^-\times\vec{G}_0^{JE}\left(\vec{r}_V^{\,-},\vec{r}\right)\right]+\left[\nabla_V^-\times\vec{E}^{\text{tot}}\left(\vec{r}_V^{\,-}\right)\right]\times\vec{G}_0^{JE}\left(\vec{r}_V^{\,-},\vec{r}\right)\right\}dS' \\
&-\oiint_{\partial V}\hat{n}_V^+\cdot\left\{\vec{E}^{\text{tot}}\left(\vec{r}_V^{\,+}\right)\times\left[\nabla_V^+\times\vec{G}_0^{JE}\left(\vec{r}_V^{\,+},\vec{r}\right)\right]+\left[\nabla_V^+\times\vec{E}^{\text{tot}}\left(\vec{r}_V^{\,+}\right)\right]\times\vec{G}_0^{JE}\left(\vec{r}_V^{\,+},\vec{r}\right)\right\}dS' \\
=\;& \oiint_{\partial V}\hat{n}_V^+\cdot\left\{\vec{E}^{\text{tot}}\left(\vec{r}_V^{\,-}\right)\times\left[\nabla'\times\vec{G}_0^{JE}(\vec{r}\,',\vec{r})\right]+\left[\nabla_V^-\times\vec{E}^{\text{tot}}\left(\vec{r}_V^{\,-}\right)\right]\times\vec{G}_0^{JE}(\vec{r}\,',\vec{r})\right\}dS' \\
&-\oiint_{\partial V}\hat{n}_V^+\cdot\left\{\vec{E}^{\text{tot}}\left(\vec{r}_V^{\,+}\right)\times\left[\nabla'\times\vec{G}_0^{JE}(\vec{r}\,',\vec{r})\right]+\left[\nabla_V^+\times\vec{E}^{\text{tot}}\left(\vec{r}_V^{\,+}\right)\right]\times\vec{G}_0^{JE}(\vec{r}\,',\vec{r})\right\}dS' \\
=\;& \oiint_{\partial V}\hat{n}_V^+\cdot\left\{\left[\vec{E}^{\text{tot}}\left(\vec{r}_V^{\,-}\right)-\vec{E}^{\text{tot}}\left(\vec{r}_V^{\,+}\right)\right]\times\left[\nabla'\times\vec{G}_0^{JE}(\vec{r}\,',\vec{r})\right]\right\}dS' \\
&+\oiint_{\partial V}\hat{n}_V^+\cdot\left[\left\{\left[\nabla_V^-\times\vec{E}^{\text{tot}}\left(\vec{r}_V^{\,-}\right)-\nabla_V^+\times\vec{E}^{\text{tot}}\left(\vec{r}_V^{\,+}\right)\right]\right\}\times\vec{G}_0^{JE}(\vec{r}\,',\vec{r})\right]dS' \\
=\;& - j\omega\mu_0\oiint_{\partial V}\hat{n}_V^+\cdot\left\{\left[\vec{E}^{\text{tot}}\left(\vec{r}_V^{\,-}\right)-\vec{E}^{\text{tot}}\left(\vec{r}_V^{\,+}\right)\right]\times\vec{G}_0^{JH}(\vec{r}\,',\vec{r})\right\}dS' \\
&- j\omega\oiint_{\partial V}\hat{n}_V^+\cdot\left\{\left[\vec{\mu}_V\cdot\vec{H}^{\text{tot}}\left(\vec{r}_V^{\,-}\right)-\mu_0\vec{H}^{\text{tot}}\left(\vec{r}_V^{\,+}\right)\right]\times\vec{G}_0^{JE}(\vec{r}\,',\vec{r})\right\}dS' \\
=\;& - j\omega\mu_0\oiint_{\partial V}\hat{n}_V^+\cdot\left\{\left[\vec{E}^{\text{tot}}\left(\vec{r}_V^{\,-}\right)-\vec{E}^{\text{tot}}\left(\vec{r}_V^{\,+}\right)\right]\times\vec{G}_0^{JH}(\vec{r}\,',\vec{r})\right\}dS' \\
&- j\omega\mu_0\oiint_{\partial V}\hat{n}_V^+\cdot\left\{\left[\vec{H}^{\text{tot}}\left(\vec{r}_V^{\,-}\right)-\vec{H}^{\text{tot}}\left(\vec{r}_V^{\,+}\right)\right]\times\vec{G}_0^{JE}(\vec{r}\,',\vec{r})\right\}dS' \\
&- j\omega\oiint_{\partial V}\hat{n}_V^+\cdot\left\{\left[\vec{\mu}_V\cdot\vec{H}^{\text{tot}}\left(\vec{r}_V^{\,-}\right)-\mu_0\vec{H}^{\text{tot}}\left(\vec{r}_V^{\,-}\right)\right]\times\vec{G}_0^{JE}(\vec{r}\,',\vec{r})\right\}dS' \\
=\;& - j\omega\mu_0\oiint_{\partial V}\hat{n}_V^+\cdot\left\{\left[\vec{E}^{\text{tot}}\left(\vec{r}_V^{\,-}\right)-\vec{E}^{\text{tot}}\left(\vec{r}_V^{\,+}\right)\right]\times\vec{G}_0^{JH}(\vec{r}\,',\vec{r})\right\}dS' \\
&- j\omega\mu_0\oiint_{\partial V}\hat{n}_V^+\cdot\left\{\left[\vec{H}^{\text{tot}}\left(\vec{r}_V^{\,-}\right)-\vec{H}^{\text{tot}}\left(\vec{r}_V^{\,+}\right)\right]\times\vec{G}_0^{JE}(\vec{r}\,',\vec{r})\right\}dS' \\
&- \oiint_{\partial V}\hat{n}_V^+\cdot\left\{\left[j\omega\Delta\vec{\mu}_V\cdot\vec{H}^{\text{tot}}\left(\vec{r}_V^{\,-}\right)\right]\times\vec{G}_0^{JE}(\vec{r}\,',\vec{r})\right\}dS' \\
=\;& - j\omega\mu_0\oiint_{\partial V}\left\{\hat{n}_V^+\times\left[\vec{E}^{\text{tot}}\left(\vec{r}_V^{\,-}\right)-\vec{E}^{\text{tot}}\left(\vec{r}_V^{\,+}\right)\right]\right\}\cdot\vec{G}_0^{JH}(\vec{r}\,',\vec{r})\,dS' \\
&- j\omega\mu_0\oiint_{\partial V}\left\{\hat{n}_V^+\times\left[\vec{H}^{\text{tot}}\left(\vec{r}_V^{\,-}\right)-\vec{H}^{\text{tot}}\left(\vec{r}_V^{\,+}\right)\right]\right\}\cdot\vec{G}_0^{JE}(\vec{r}\,',\vec{r})\,dS' \\
&- \oiint_{\partial V}\hat{n}_V^+\cdot\left\{\left[j\omega\Delta\vec{\mu}_V\cdot\vec{H}^{\text{tot}}\left(\vec{r}_V^{\,-}\right)\right]\times\vec{G}_0^{JE}(\vec{r}\,',\vec{r})\right\}dS' \\
=\;& - \oiint_{\partial V}\hat{n}_V^+\cdot\left\{\left[j\omega\Delta\vec{\mu}_V\cdot\vec{H}^{\text{tot}}\left(\vec{r}_V^{\,-}\right)\right]\times\vec{G}_0^{JE}(\vec{r}\,',\vec{r})\right\}dS'
\end{aligned}
\tag{A-22}
$$

In relationship (A-22), the first equality is based on definition (A-11); the second equality is based on the continuity of the $\vec{G}_0^{JE}(\vec{r}',\vec{r})$ on $\partial V$; the third equality is based on the distributive law of the multiplication among vectors[122]; the fourth equality is based on equations (A-2)~(A-4) and the fact that $\vec{J}_{\text{imp}}$ doesn't distribute on $\partial V$; the fifth equality is just an identical transformation; the sixth equality is based on $\Delta\vec{\mu}_V = \vec{\mu}_V - \vec{I}\mu_0$; the seventh equality is based on identity $\vec{a}\cdot(\vec{b}\times\vec{c}) = (\vec{a}\times\vec{b})\cdot\vec{c}$ [122]; the eighth equality is based on Ampere's law[110-113] and the fact that $\vec{J}_{\text{imp}}$ doesn't distribute on $\partial V$ and the fact that there doesn't exist $\vec{M}_{\text{imp}}$. By comparing formulation (A-21) with formulation (A-22), it is immediately obtained that

$$
\begin{aligned}
\vec{E}^{\text{tot}}(\vec{r}) = & \ \vec{E}_{\text{imp}}(\vec{r}) + \iiint_{\text{int}V} \vec{G}_0^{JE}(\vec{r},\vec{r}')\cdot\left[j\omega\Delta\vec{\varepsilon}_V^{\text{c}}\cdot\vec{E}^{\text{tot}}(\vec{r}')\right]dV' \\
& + \iiint_{\text{int}V} \vec{G}_0^{ME}(\vec{r},\vec{r}')\cdot\left[j\omega\Delta\vec{\mu}_V\cdot\vec{H}^{\text{tot}}(\vec{r}')\right]dV' \\
= & \ \vec{E}_{\text{imp}}(\vec{r}) + \iiint_{\text{int}V} \vec{G}_0^{JE}(\vec{r},\vec{r}')\cdot\left[\vec{\sigma}_V\cdot\vec{E}^{\text{tot}}(\vec{r}')\right]dV' \\
& + \iiint_{\text{int}V} \vec{G}_0^{JE}(\vec{r},\vec{r}')\cdot\left[j\omega\Delta\vec{\varepsilon}_V\cdot\vec{E}^{\text{tot}}(\vec{r}')\right]dV' \\
& + \iiint_{\text{int}V} \vec{G}_0^{ME}(\vec{r},\vec{r}')\cdot\left[j\omega\Delta\vec{\mu}_V\cdot\vec{H}^{\text{tot}}(\vec{r}')\right]dV' \\
= & \ \vec{E}_{\text{imp}}(\vec{r}) + \iiint_{\text{int}V} \vec{G}_0^{JE}(\vec{r},\vec{r}')\cdot\vec{J}_V^{\text{CV}}(\vec{r}')dV' \\
& + \iiint_{\text{int}V} \vec{G}_0^{JE}(\vec{r},\vec{r}')\cdot\vec{J}_V^{\text{PV}}(\vec{r}')dV' \\
& + \iiint_{\text{int}V} \vec{G}_0^{ME}(\vec{r},\vec{r}')\cdot\vec{M}_V^{\text{MV}}(\vec{r}')dV' \quad , \quad \vec{r}\in\text{ext}V\bigcup\text{int}V \qquad (\text{A-23})
\end{aligned}
$$

In above convolution integral (A-23), the second equality is based on $j\omega\Delta\vec{\varepsilon}_{\text{mat}}^{\text{c}} = \vec{\sigma}_V + j\omega\Delta\vec{\varepsilon}_V$; the third equality is based on $\vec{J}_V^{\text{CV}} = \vec{\sigma}_V\cdot\vec{E}^{\text{tot}}$, and $\vec{J}_V^{\text{PV}} = j\omega\Delta\vec{\varepsilon}_V\cdot\vec{E}^{\text{tot}}$, and $\vec{M}_V^{\text{MV}} = j\omega\Delta\vec{\mu}_V\cdot\vec{H}^{\text{tot}}$, where $\vec{J}_V^{\text{CV}}$, $\vec{J}_V^{\text{PV}}$, and $\vec{M}_V^{\text{MV}}$ are conduction volume electric current, polarization volume electric current, and magnetization volume magnetic current respectively, and the superscript "CV" used in $\vec{J}_V^{\text{CV}}$ is the abbreviation of "conduction volume (electric current)", and the subscript "$V$" used in $\vec{J}_V^{\text{CV}}$ is to emphasize that $\vec{J}_V^{\text{CV}}$ distributes on material body $V$, and the superscripts and subscripts used in the other currents can be explained similarly.

**2) To Describe Polarization Phenomenon and Magnetization Phenomenon by Employing Polarization Electric Current Model and Magnetization Electric Current Model Respectively[121]**

Integrating the RHS of relationship (A-17) on whole $\text{int}V$, we have that

$$
\begin{aligned}
& \iiint_{\text{int}V} \nabla'\cdot\vec{\Psi}(\vec{r}',\vec{r})dV' \\
= & -j\omega\mu_0 \iiint_{\text{int}V} \vec{G}_0^{JE}(\vec{r},\vec{r}')\cdot\left[j\omega\Delta\vec{\varepsilon}_V^{\text{c}}\cdot\vec{E}^{\text{tot}}(\vec{r}')\right]dV' - \iiint_{\text{int}V} \vec{G}_0^{JE}(\vec{r},\vec{r}')\cdot\left\{\nabla'\times\left[j\omega\Delta\vec{\mu}_V\cdot\vec{H}^{\text{tot}}(\vec{r}')\right]\right\}dV' \\
& + j\omega\mu_0 \iiint_{\text{int}V} \vec{E}^{\text{tot}}(\vec{r}')\cdot\vec{I}\delta(\vec{r}'-\vec{r})dV'
\end{aligned}
$$

$$= - j\omega\mu_0 \iiint_{\text{int}V} \vec{G}_0^{JE}(\vec{r},\vec{r}\,') \cdot \left[ j\omega\Delta\vec{\varepsilon}_V^{\text{c}} \cdot \vec{E}^{\text{tot}}(\vec{r}\,') \right] dV' - \iiint_{\text{int}V} \vec{G}_0^{JE}(\vec{r},\vec{r}\,') \cdot \left\{ \nabla' \times \left[ j\omega\Delta\vec{\mu}_V \cdot \vec{H}^{\text{tot}}(\vec{r}\,') \right] \right\} dV'$$
$$+ \begin{cases} 0 & , \quad \vec{r} \in \text{ext}V \\ j\omega\mu_0 \vec{E}^{\text{tot}}(\vec{r}) & , \quad \vec{r} \in \text{int}V \end{cases} \tag{A-24}$$

In relationship (A-24), the first equality is based on symmetry $\vec{G}_0^{JE}(\vec{r}\,',\vec{r}) = [\vec{G}_0^{JE}(\vec{r},\vec{r}\,')]^T$ [120]; the second equality is based on the integral property of Dirac's delta function $\delta(\vec{r}\,' - \vec{r})$ [120,122]. Employing relationships (A-18) and (A-24), we have that

$$\begin{aligned} \oiint_{\partial V} \hat{n}_V^+ \cdot \vec{\Psi}(\vec{r}\,',\vec{r}) dS' = & - j\omega\mu_0 \iiint_{\text{int}V} \vec{G}_0^{JE}(\vec{r},\vec{r}\,') \cdot \left[ j\omega\Delta\vec{\varepsilon}_V^{\text{c}} \cdot \vec{E}^{\text{tot}}(\vec{r}\,') \right] dV' \\ & - j\omega\mu_0 \iiint_{\text{int}V} \vec{G}_0^{JE}(\vec{r},\vec{r}\,') \cdot \left\{ \nabla' \times \left[ \Delta\vec{\mu}_V^{\text{r}} \cdot \vec{H}^{\text{tot}}(\vec{r}\,') \right] \right\} dV' \\ & + \begin{cases} 0 & , \quad \vec{r} \in \text{ext}V \\ j\omega\mu_0 \vec{E}^{\text{tot}}(\vec{r}) & , \quad \vec{r} \in \text{int}V \end{cases} \end{aligned} \tag{A-25}$$

where $\Delta\vec{\mu}_V^{\text{r}} = \Delta\vec{\mu}_V / \mu_0$. The summation of relationship (A-16) and relationship (A-25) gives that

$$\begin{aligned} \oiint_{\partial V} \hat{n}_V^+ \cdot \left[ \vec{\Psi}(\vec{r}_V^-,\vec{r}) - \vec{\Psi}(\vec{r}_V^+,\vec{r}) \right] dS' = & - j\omega\mu_0 \iiint_{\text{int}V} \vec{G}_0^{JE}(\vec{r},\vec{r}\,') \cdot \left[ j\omega\Delta\vec{\varepsilon}_V^{\text{c}} \cdot \vec{E}^{\text{tot}}(\vec{r}\,') \right] dV' \\ & - j\omega\mu_0 \iiint_{\text{int}V} \vec{G}_0^{JE}(\vec{r},\vec{r}\,') \cdot \left\{ \nabla' \times \left[ \Delta\vec{\mu}_V^{\text{r}} \cdot \vec{H}^{\text{tot}}(\vec{r}\,') \right] \right\} dV' \\ & + \begin{cases} - j\omega\mu_0 \vec{E}_{\text{imp}}(\vec{r}) + j\omega\mu_0 \vec{E}^{\text{tot}}(\vec{r}) & , \quad \vec{r} \in \text{ext}V \\ - j\omega\mu_0 \vec{E}_{\text{imp}}(\vec{r}) + j\omega\mu_0 \vec{E}^{\text{tot}}(\vec{r}) & , \quad \vec{r} \in \text{int}V \end{cases} \end{aligned} \tag{A-26}$$

By comparing relationship (A-22) with relationship (A-26), it is obtained that

$$\begin{aligned} \vec{E}^{\text{tot}}(\vec{r}) = & \vec{E}_{\text{imp}}(\vec{r}) + \iiint_{\text{int}V} \vec{G}_0^{JE}(\vec{r},\vec{r}\,') \cdot \vec{J}_V^{\text{CV}}(\vec{r}\,') dV' \\ & + \iiint_{\text{int}V} \vec{G}_0^{JE}(\vec{r},\vec{r}\,') \cdot \vec{J}_V^{\text{PV}}(\vec{r}\,') dV' \\ & + \iiint_{\text{int}V} \vec{G}_0^{JE}(\vec{r},\vec{r}\,') \cdot \left\{ \nabla' \times \left[ \Delta\vec{\mu}_V^{\text{r}} \cdot \vec{H}^{\text{tot}}(\vec{r}\,') \right] \right\} dV' \\ & - \oiint_{\partial V} \hat{n}_V^+ \cdot \left\{ \left[ \Delta\vec{\mu}_V^{\text{r}} \cdot \vec{H}^{\text{tot}}(\vec{r}_V^-) \right] \times \vec{G}_0^{JE}(\vec{r}\,',\vec{r}) \right\} dS' \\ = & \vec{E}_{\text{imp}}(\vec{r}) + \iiint_{\text{int}V} \vec{G}_0^{JE}(\vec{r},\vec{r}\,') \cdot \vec{J}_V^{\text{CV}}(\vec{r}\,') dV' \\ & + \iiint_{\text{int}V} \vec{G}_0^{JE}(\vec{r},\vec{r}\,') \cdot \vec{J}_V^{\text{PV}}(\vec{r}\,') dV' \\ & + \iiint_{\text{int}V} \vec{G}_0^{JE}(\vec{r},\vec{r}\,') \cdot \left\{ \nabla' \times \left[ \Delta\vec{\mu}_V^{\text{r}} \cdot \vec{H}^{\text{tot}}(\vec{r}\,') \right] \right\} dV' \\ & + \oiint_{\partial V} \vec{G}_0^{JE}(\vec{r},\vec{r}\,') \cdot \left\{ \hat{n}_V^+ \times \left[ 0 - \Delta\vec{\mu}_V^{\text{r}} \cdot \vec{H}^{\text{tot}}(\vec{r}_V^-) \right] \right\} dS' \\ = & \vec{E}_{\text{imp}}(\vec{r}) + \iiint_{\text{int}V} \vec{G}_0^{JE}(\vec{r},\vec{r}\,') \cdot \vec{J}_V^{\text{CV}}(\vec{r}\,') dV' \\ & + \iiint_{\text{int}V} \vec{G}_0^{JE}(\vec{r},\vec{r}\,') \cdot \vec{J}_V^{\text{PV}}(\vec{r}\,') dV' \\ & + \iiint_{\text{int}V} \vec{G}_0^{JE}(\vec{r},\vec{r}\,') \cdot \left[ \nabla' \times \vec{M}^{\text{mag}}(\vec{r}\,') \right] dV' \\ & + \oiint_{\partial V} \vec{G}_0^{JE}(\vec{r},\vec{r}\,') \cdot \left\{ \hat{n}_V^+ \times \left[ 0 - \vec{M}^{\text{mag}}(\vec{r}\,') \right] \right\} dS' \end{aligned}$$

$$\begin{aligned} = & \vec{E}_{\text{imp}}(\vec{r}) + \iiint_{\text{int}V} \vec{G}_0^{JE}(\vec{r},\vec{r}') \cdot \vec{J}_V^{\text{CV}}(\vec{r}')dV' \\ & + \iiint_{\text{int}V} \vec{G}_0^{JE}(\vec{r},\vec{r}') \cdot \vec{J}_V^{\text{PV}}(\vec{r}')dV' \\ & + \iiint_{\text{int}V} \vec{G}_0^{JE}(\vec{r},\vec{r}') \cdot \vec{J}_V^{\text{MV}}(\vec{r}')dV' \\ & + \oiint_{\partial V} \vec{G}_0^{JE}(\vec{r},\vec{r}') \cdot \vec{J}_{\partial V}^{\text{MS}}(\vec{r}')dS' \quad , \quad \vec{r} \in \text{ext}V \bigcup \text{int}V \end{aligned} \quad \text{(A-27)}$$

In above convolution integral formulation (A-27), the second equality is based on symmetry $\vec{G}_0^{JE}(\vec{r}',\vec{r}) = [\vec{G}_0^{JE}(\vec{r},\vec{r}')]^T$ [120] and identity $\vec{a} \cdot (\vec{b} \times \vec{c}) = (\vec{a} \times \vec{b}) \cdot \vec{c}$ [122]; the third equality is based on $\vec{M}^{\text{mag}} = \Delta\vec{\mu}_V^{\text{r}} \cdot \vec{H}^{\text{tot}}$ [126], where $\vec{M}^{\text{mag}}$ is so-called magnetization intensity; the fourth equality is based on $\vec{J}_V^{\text{MV}}(\vec{r}') = \nabla' \times \vec{M}^{\text{mag}}(\vec{r}')$ and $\vec{J}_{\partial V}^{\text{MS}} = \hat{n}_V^+ \times (0 - \vec{M}^{\text{mag}})$ [126], where $\vec{J}_V^{\text{MV}}$ and $\vec{J}_{\partial V}^{\text{MS}}$ are magnetization volume electric current and magnetization surface electric current respectively, and the subscripts used in magnetization electric currents $\vec{J}_V^{\text{MV}}$ and $\vec{J}_{\partial V}^{\text{MS}}$ are to emphasize that the currents distribute on $V$ and $\partial V$ respectively.

**3) To Compare the Two Different Models**

Convolution integral formulation (A-23) implies that: when polarization electric current model and magnetization magnetic current model are utilized to describe polarization phenomenon and magnetization phenomenon respectively, there exist induced volume sources only, but doesn't exist any induced surface source.

Convolution integral formulation (A-27) implies that: when polarization electric current model and magnetization electric current model are utilized to describe polarization phenomenon and magnetization phenomenon respectively, there exist both induced volume sources and induced surface source[121].

In some aspects, the above two models are equivalent to each other, such as determining the scattered fields in the exterior of source domain. But, the models are not equivalent in all of aspects, as stated by Prof. Lan Jen Chu et al. that[110]

> "… the critical consequences that lead to the choice of the magnetic-charge model as opposed to the amperian-current model concern the flow, storage, and dissipation of energy … [110]"

In this dissertation, we will always select to use the polarization electric current and magnetization magnetic current models corresponding to convolution integral formulation (A-23), and this selection has many advantages as pointed out by Prof. Lan Jen Chu et al. that[110]

> "… some textbooks present a macroscopic theory of electromagnetism based on the amperian-current model of magnetized matter, and avoid the above interpretation by introducing in the theory what appears to the authors of this book as a logical inconsistency. [110]"

## A2 Mathematical Expressions of the Frequency-Domain Volume Equivalence Principle for an Inhomogeneous Anisotropic Material Body Placed in Complex Environment

In what follows, we consider the EM scattering problem illustrated in Figure A-2. In the figure, $D_{\text{mat sys}}$ is a material system, and $D_{\text{env}}$ represents EM environment, and $D_{\text{imp}}$ is an external excitation. Both $D_{\text{mat sys}}$ and $D_{\text{env}}$ can be inhomogeneous anisotropic, and the conductivity, permittivity, and permeability of $D_{\text{mat sys}}$ are denoted as $\ddot{\sigma}_{\text{mat}}(\vec{r})$, $\ddot{\varepsilon}_{\text{mat}}(\vec{r})$, and $\ddot{\mu}_{\text{mat}}(\vec{r})$ respectively, and the conductivity, permittivity, and permeability of $D_{\text{env}}$ are denoted as $\ddot{\sigma}_{\text{env}}(\vec{r})$, $\ddot{\varepsilon}_{\text{env}}(\vec{r})$, and $\ddot{\mu}_{\text{env}}(\vec{r})$ respectively. In this dissertation, all of the parameters are restricted to being symmetrical two-order tensors, and the physical explanation for doing this restriction will be given in subsequent Appendix B.

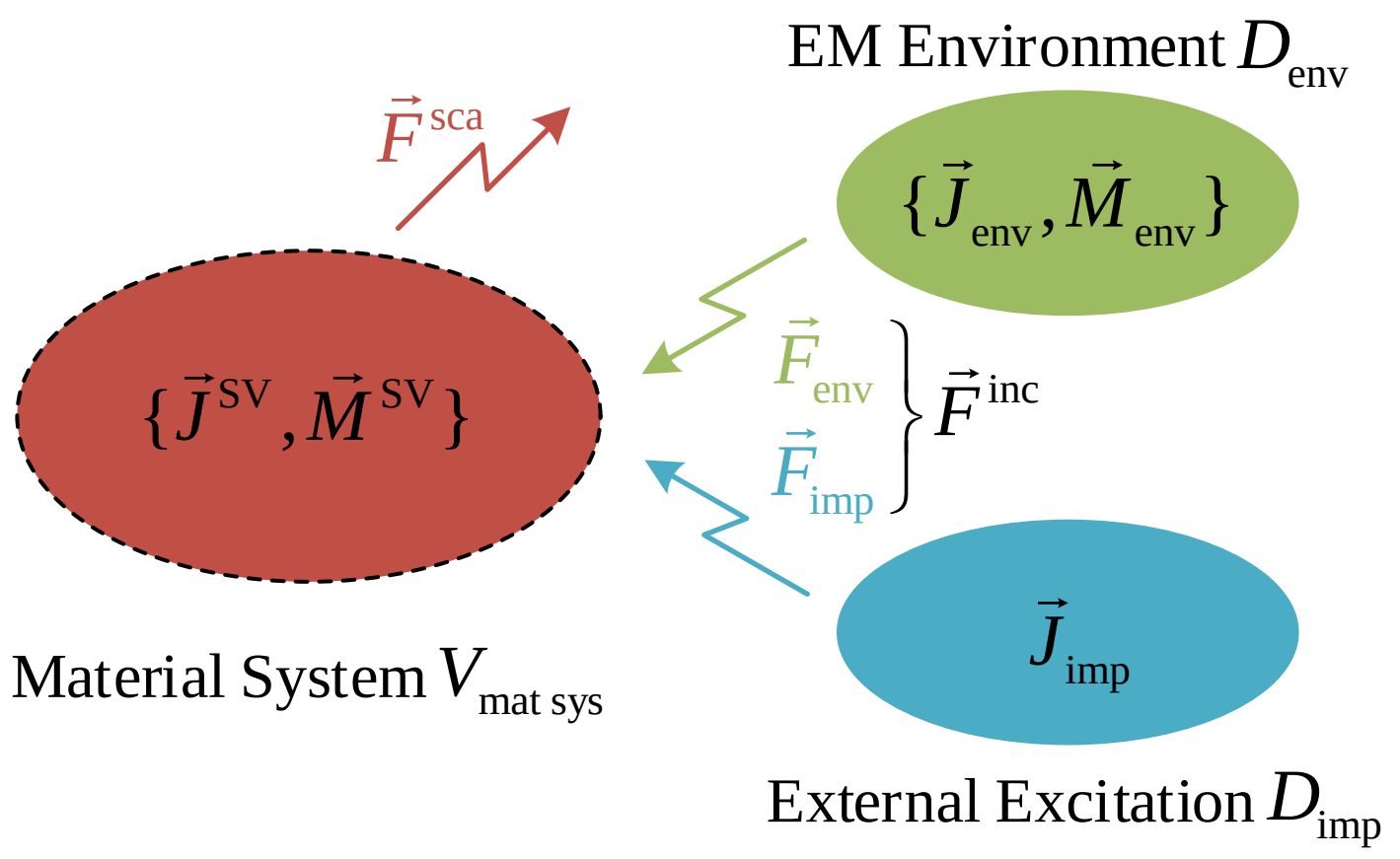

Figure A-2 The scattering problem corresponding to a material system placed in complicated environment

Based on the conclusions obtained in Appendix A1, it is easy to know that: under the action of the excitation field $\vec{F}_{\text{imp}}$, which is generated by the excitation source $\vec{J}_{\text{imp}}$ on $D_{\text{imp}}$, conduction phenomenon, polarization phenomenon, and magnetization phenomenon will appear on $D_{\text{mat sys}}$ and $D_{\text{env}}$; the conduction phenomenon will lead to some conduction volume electric currents $\vec{J}_{\text{mat}}^{\text{CV}}$ and $\vec{J}_{\text{env}}^{\text{CV}}$ on $D_{\text{mat sys}}$ and $D_{\text{env}}$ respectively; if the polarization phenomenon and the magnetization phenomenon are described by polarization electric current model and magnetization magnetic current model respectively, some polarization volume electric currents $\vec{J}_{\text{mat}}^{\text{PV}}$ and some magnetization volume magnetic currents $\vec{M}_{\text{mat}}^{\text{MV}}$ will appear on $D_{\text{mat sys}}$, and some polarization volume electric currents $\vec{J}_{\text{env}}^{\text{PV}}$ and some magnetization volume magnetic currents $\vec{M}_{\text{env}}^{\text{MV}}$ will appear on $D_{\text{env}}$. To simplify the symbolic system of this Appendix A2, we denote the summation of $\vec{J}_{\text{mat}}^{\text{CV}}$ and $\vec{J}_{\text{mat}}^{\text{PV}}$ as $\vec{J}^{\text{SV}}$, i.e., $\vec{J}^{\text{SV}} = \vec{J}_{\text{mat}}^{\text{CV}} + \vec{J}_{\text{mat}}^{\text{PV}}$; we simply denote $\vec{M}_{\text{mat}}^{\text{MV}}$ as $\vec{M}^{\text{SV}}$ similarly, i.e., $\vec{M}^{\text{SV}} = \vec{M}_{\text{mat}}^{\text{MV}}$; we denote the summation of $\vec{J}_{\text{env}}^{\text{CV}}$ and $\vec{J}_{\text{env}}^{\text{PV}}$ as $\vec{J}_{\text{env}}$, i.e., $\vec{J}_{\text{env}} = \vec{J}_{\text{env}}^{\text{CV}} + \vec{J}_{\text{env}}^{\text{PV}}$; we simply denote $\vec{M}_{\text{env}}^{\text{MV}}$ as $\vec{M}_{\text{env}}$ similarly, i.e., $\vec{M}_{\text{env}} = \vec{M}_{\text{env}}^{\text{MV}}$. Here, $\vec{J}^{\text{SV}}$ and $\vec{M}^{\text{SV}}$ are called as scattered volume electric current and scattered volume magnetic current respectively, and their common superscript "SV" is just the abbreviation of term "scattered volume (current)".

Just as Chapter 3, the EM fields generated by $\{\vec{J}^{\text{SV}}, \vec{M}^{\text{SV}}\}$ and $\{\vec{J}_{\text{env}}, \vec{M}_{\text{env}}\}$ are denoted as $\vec{F}^{\text{sca}}$ and $\vec{F}_{\text{env}}$ respectively; the summation of $\vec{F}_{\text{imp}}$ and $\vec{F}_{\text{env}}$ is denoted as $\vec{F}^{\text{inc}}$, i.e., $\vec{F}^{\text{inc}} = \vec{F}_{\text{imp}} + \vec{F}_{\text{env}}$ ①; the summation of $\vec{F}^{\text{inc}}$ and $\vec{F}^{\text{sca}}$ is denoted as $\vec{F}^{\text{tot}}$, i.e., $\vec{F}^{\text{tot}} = \vec{F}^{\text{inc}} + \vec{F}^{\text{sca}}$. Here, $\vec{F}^{\text{inc}}$ is just the resultant field used to drive the working of $D_{\text{mat sys}}$, and $\vec{F}^{\text{tot}}$ is just the total field related to whole scattering problem.

Based on volume equivalence principle, scattered field $\vec{F}^{\text{sca}}$ and environment field $\vec{F}_{\text{env}}$ satisfy the following Maxwell's equations:

$$\begin{aligned} \nabla \times \vec{H}^{\text{sca}}(\vec{r}) &= \ j\omega\varepsilon_0 \vec{E}^{\text{sca}}(\vec{r}) + \overbrace{j\omega\Delta\vec{\vec{\varepsilon}}_{\text{mat}}^{\text{c}}(\vec{r}) \cdot \vec{E}^{\text{tot}}(\vec{r})}^{\vec{J}^{\text{SV}}(\vec{r})} \\ \nabla \times \vec{E}^{\text{sca}}(\vec{r}) &= -j\omega\mu_0 \vec{H}^{\text{sca}}(\vec{r}) - \underbrace{j\omega\Delta\vec{\vec{\mu}}_{\text{mat}}(\vec{r}) \cdot \vec{H}^{\text{tot}}(\vec{r})}_{\vec{M}^{\text{SV}}(\vec{r})} \end{aligned} \quad , \quad \vec{r} \in \mathbb{R}^3 \tag{A-28}$$

and

$$\begin{aligned} \nabla \times \vec{H}_{\text{env}}(\vec{r}) &= \ j\omega\varepsilon_0 \vec{E}_{\text{env}}(\vec{r}) + \overbrace{j\omega\Delta\vec{\vec{\varepsilon}}_{\text{env}}^{\text{c}}(\vec{r}) \cdot \vec{E}^{\text{tot}}(\vec{r})}^{\vec{J}_{\text{env}}(\vec{r})} \\ \nabla \times \vec{E}_{\text{env}}(\vec{r}) &= -j\omega\mu_0 \vec{H}_{\text{env}}(\vec{r}) - \underbrace{j\omega\Delta\vec{\vec{\mu}}_{\text{env}}(\vec{r}) \cdot \vec{H}^{\text{tot}}(\vec{r})}_{\vec{M}_{\text{env}}(\vec{r})} \end{aligned} \quad , \quad \vec{r} \in \mathbb{R}^3 \tag{A-29}$$

In Maxwell's equations (A-28), $\Delta\vec{\vec{\varepsilon}}_{\text{mat}}^{\text{c}} = \vec{\vec{\varepsilon}}_{\text{mat}}^{\text{c}} - \vec{\vec{I}}\varepsilon_0$ and $\Delta\vec{\vec{\mu}}_{\text{mat}} = \vec{\vec{\mu}}_{\text{mat}} - \vec{\vec{I}}\mu_0$. In Maxwell's equations (A-29), $\Delta\vec{\vec{\varepsilon}}_{\text{env}}^{\text{c}} = \vec{\vec{\varepsilon}}_{\text{env}}^{\text{c}} - \vec{\vec{I}}\varepsilon_0$ and $\Delta\vec{\vec{\mu}}_{\text{env}} = \vec{\vec{\mu}}_{\text{env}} - \vec{\vec{I}}\mu_0$. Based on Maxwell's equations (A-1) and (A-29) and relationship $\vec{F}^{\text{inc}} = \vec{F}_{\text{imp}} + \vec{F}_{\text{env}}$, we immediately have the Maxwell's equations pertaining to $\vec{F}^{\text{inc}}$ as follows:

$$\begin{aligned} \nabla \times \vec{H}^{\text{inc}}(\vec{r}) &= \ j\omega\varepsilon_0 \vec{E}^{\text{inc}}(\vec{r}) + \overbrace{\vec{J}_{\text{env}}(\vec{r}) + \vec{J}_{\text{imp}}(\vec{r})}^{\vec{J}^{\text{inc}}(\vec{r})} \\ \nabla \times \vec{E}^{\text{inc}}(\vec{r}) &= -j\omega\mu_0 \vec{H}^{\text{inc}}(\vec{r}) - \underbrace{\vec{M}_{\text{env}}(\vec{r})}_{\vec{M}^{\text{inc}}(\vec{r})} \end{aligned} \quad , \quad \vec{r} \in \mathbb{R}^3 \tag{A-30}$$

① The reason to do like this has been given in Chapter 3, and it will not be repeated here.

where $\vec{J}^{\text{inc}} = \vec{J}_{\text{env}} + \vec{J}_{\text{imp}}$ and $\vec{M}^{\text{inc}} = \vec{M}_{\text{env}}$.

The above result is just the frequency-domain volume equivalence principle corresponding to the material systems placed in EM environment. For the convenience of discussing the time-domain WEP pertaining to material systems in Subsection 4.2, we provide the time-domain version of the above frequency-domain volume equivalence principle in the following Appendix A3.

## A3 Mathematical Expressions of the Time-Domain Volume Equivalence Principle for an Inhomogeneous Anisotropic Material Body Placed in Complex Environment

The time-domain versions of the various scattered sources on material systems are as follows:

$$\vec{J}_{\text{mat}}^{\text{CV}}(t) = \overset{\leftrightarrow}{\sigma}_{\text{mat}} \cdot \vec{E}^{\text{tot}}(t) \tag{A-31a}$$

$$\vec{J}_{\text{mat}}^{\text{PV}}(t) = \frac{\partial}{\partial t}\left[\Delta\overset{\leftrightarrow}{\varepsilon}_{\text{mat}} \cdot \vec{E}^{\text{tot}}(t)\right] \tag{A-31b}$$

$$\vec{M}_{\text{mat}}^{\text{MV}}(t) = \frac{\partial}{\partial t}\left[\Delta\overset{\leftrightarrow}{\mu}_{\text{mat}} \cdot \vec{H}^{\text{tot}}(t)\right] \tag{A-31c}$$

and $\vec{J}^{\text{SV}}(t) = \vec{J}_{\text{mat}}^{\text{CV}}(t) + \vec{J}_{\text{mat}}^{\text{PV}}(t)$ and $\vec{M}^{\text{SV}}(t) = \vec{M}_{\text{mat}}^{\text{MV}}(t)$. The time-domain versions of the induced sources on environment are as follows:

$$\vec{J}_{\text{env}}^{\text{CV}}(t) = \overset{\leftrightarrow}{\sigma}_{\text{env}} \cdot \vec{E}^{\text{tot}}(t) \tag{A-32a}$$

$$\vec{J}_{\text{env}}^{\text{PV}}(t) = \frac{\partial}{\partial t}\left[\Delta\overset{\leftrightarrow}{\varepsilon}_{\text{env}} \cdot \vec{E}^{\text{tot}}(t)\right] \tag{A-32b}$$

$$\vec{M}_{\text{env}}^{\text{MV}}(t) = \frac{\partial}{\partial t}\left[\Delta\overset{\leftrightarrow}{\mu}_{\text{env}} \cdot \vec{H}^{\text{tot}}(t)\right] \tag{A-32c}$$

and $\vec{J}_{\text{env}}(t) = \vec{J}_{\text{env}}^{\text{CV}}(t) + \vec{J}_{\text{env}}^{\text{PV}}(t)$ and $\vec{M}_{\text{env}}(t) = \vec{M}_{\text{env}}^{\text{MV}}(t)$. The time-domain versions of the incident sources generating incident field are as follows:

$$\vec{J}^{\text{inc}}(t) = \vec{J}_{\text{env}}(t) + \vec{J}_{\text{imp}}(t) \tag{A-33a}$$

$$\vec{M}^{\text{inc}}(t) = \vec{M}_{\text{env}}(t) \tag{A-33b}$$

In Section 4.2, we will establish the volume formulation for constructing the DP-CMs of material systems based on the above volume equivalence principle.

# Appendix B Symmetrical Material Parameter Tensors and Several Algebraic Identities Satisfied by Them

In this Appendix B, we firstly provide the physical explanation for the symmetry satisfied by material parameter tensors, and then provide several identical relationships satisfied by the symmetrical tensors.

## B1 The Symmetry Satisfied by Material Parameter Tensors and Its Physical Explanation

In this dissertation, material parameters $\vec{\vec{\sigma}}_{\text{mat}}(\vec{r})$ , $\vec{\vec{\varepsilon}}_{\text{mat}}(\vec{r})$ , and $\vec{\vec{\mu}}_{\text{mat}}(\vec{r})$ are restricted to symmetrical two-order tensors. Now taking $\vec{\vec{\sigma}}_{\text{mat}}(\vec{r})$ as an example, we provide the physical explanation for the above restriction, and the physical explanations for the symmetries satisfied by $\vec{\vec{\varepsilon}}_{\text{mat}}(\vec{r})$ and $\vec{\vec{\mu}}_{\text{mat}}(\vec{r})$ are completely similar. For any point $\vec{r}_0$ on $V_{\text{mat sys}}$ and any Maxwellian field $\vec{E}^{\text{tot}}(\vec{r}_0)$, the dissipation power density at $\vec{r}_0$ must be a real number, i.e., $[\vec{J}^{\text{CV}}(\vec{r}_0)]^* \cdot \vec{E}^{\text{tot}}(\vec{r}_0) \in \mathbb{R}$, so $[\vec{E}^{\text{tot}}(\vec{r}_0)]^* \cdot \vec{\vec{\sigma}}_{\text{mat}}(\vec{r}_0) \cdot \vec{E}^{\text{tot}}(\vec{r}_0) = [\vec{E}^{\text{tot}}(\vec{r}_0)]^* \cdot \vec{J}^{\text{CV}}(\vec{r}_0) \in \mathbb{R}$. If we expand $\vec{E}^{\text{tot}}(\vec{r}_0)$ in terms of its Cartesian coordinates, i.e., $\vec{E}^{\text{tot}}(\vec{r}_0) = \hat{x}E_x^{\text{tot}}(\vec{r}_0) + \hat{y}E_y^{\text{tot}}(\vec{r}_0) + \hat{z}E_z^{\text{tot}}(\vec{r}_0)$, then we have that

$$\begin{bmatrix} E_x^{\text{tot}}(\vec{r}_0) \\ E_y^{\text{tot}}(\vec{r}_0) \\ E_z^{\text{tot}}(\vec{r}_0) \end{bmatrix}^H \cdot \begin{bmatrix} \sigma_{\text{mat}}^{xx}(\vec{r}_0) & \sigma_{\text{mat}}^{xy}(\vec{r}_0) & \sigma_{\text{mat}}^{xz}(\vec{r}_0) \\ \sigma_{\text{mat}}^{yx}(\vec{r}_0) & \sigma_{\text{mat}}^{yy}(\vec{r}_0) & \sigma_{\text{mat}}^{yz}(\vec{r}_0) \\ \sigma_{\text{mat}}^{zx}(\vec{r}_0) & \sigma_{\text{mat}}^{zy}(\vec{r}_0) & \sigma_{\text{mat}}^{zz}(\vec{r}_0) \end{bmatrix} \cdot \begin{bmatrix} E_x^{\text{tot}}(\vec{r}_0) \\ E_y^{\text{tot}}(\vec{r}_0) \\ E_z^{\text{tot}}(\vec{r}_0) \end{bmatrix} = \left[\vec{E}^{\text{tot}}(\vec{r}_0)\right]^* \cdot \vec{\vec{\sigma}}_{\text{mat}}(\vec{r}_0) \cdot \vec{E}^{\text{tot}}(\vec{r}_0) \in \mathbb{R} \tag{B-1}$$

Because $\vec{E}^{\text{tot}}(\vec{r}_0)$ is arbitrary and $\vec{\vec{\sigma}}_{\text{mat}}(\vec{r}_0)$ is real, then $\vec{\vec{\sigma}}_{\text{mat}}(\vec{r}_0)$ must be symmetrical[116]. Because the $\vec{r}_0$ mentioned above is an arbitrary point on $V_{\text{mat sys}}$, then $\vec{\vec{\sigma}}_{\text{mat}}(\vec{r})$ is symmetrical for any $\vec{r} \in V_{\text{mat sys}}$.

## B2 Several Algebraic Identities Satisfied by Symmetrical Dyads

Now, there are vector $\vec{a}$ and dyads $\vec{\vec{A}}$ and $\vec{\vec{B}}$ as follows:

$$\vec{a} = \hat{x}a_x + \hat{y}a_y + \hat{z}a_z \tag{B-2}$$

$$\vec{\vec{A}} = \hat{x}\hat{x}A_{xx} + \hat{x}\hat{y}A_{xy} + \hat{x}\hat{z}A_{xz} + \hat{y}\hat{x}A_{yx} + \hat{y}\hat{y}A_{yy} + \hat{y}\hat{z}A_{yz} + \hat{z}\hat{x}A_{zx} + \hat{z}\hat{y}A_{zy} + \hat{z}\hat{z}A_{zz} \tag{B-3}$$

$$\vec{\vec{B}} = \hat{x}\hat{x}B_{xx} + \hat{x}\hat{y}B_{xy} + \hat{x}\hat{z}B_{xz} + \hat{y}\hat{x}B_{yx} + \hat{y}\hat{y}B_{yy} + \hat{y}\hat{z}B_{yz} + \hat{z}\hat{x}B_{zx} + \hat{z}\hat{y}B_{zy} + \hat{z}\hat{z}B_{zz} \tag{B-4}$$

As everyone knows, $\vec{a}$ , $\vec{\vec{A}}$ , and $\vec{\vec{B}}$ satisfy the following associative law for multiplication[120,122]:

$$\vec{a} \cdot \left(\vec{\vec{A}} \cdot \vec{\vec{B}}\right) = \left(\vec{a} \cdot \vec{\vec{A}}\right) \cdot \vec{\vec{B}} \tag{B-5}$$

and its proof can be found in classical literatures [120,122].

If $\vec{\vec{A}}$ is a symmetrical dyad, i.e., $\vec{\vec{A}}^T = \vec{\vec{A}}$, then $\vec{a}$ and $\vec{\vec{A}}$ satisfy the following commutative law for multiplication:

$$\vec{a}\cdot\overleftrightarrow{A} = \overleftrightarrow{A}\cdot\vec{a} \tag{B-6}$$

Now, we prove above relationship (B-6) as below. Obviously,

$$\begin{aligned}\vec{a}\cdot\overleftrightarrow{A} &= \left(\hat{x}a_x+\hat{y}a_y+\hat{z}a_z\right)\\ &\quad\cdot\left(\hat{x}\hat{x}A_{xx}+\hat{x}\hat{y}A_{xy}+\hat{x}\hat{z}A_{xz}+\hat{y}\hat{x}A_{yx}+\hat{y}\hat{y}A_{yy}+\hat{y}\hat{z}A_{yz}+\hat{z}\hat{x}A_{zx}+\hat{z}\hat{y}A_{zy}+\hat{z}\hat{z}A_{zz}\right)\\ &= \left(\hat{x}a_x+\hat{y}a_y+\hat{z}a_z\right)\\ &\quad\cdot\left[\hat{x}\left(\hat{x}A_{xx}+\hat{y}A_{xy}+\hat{z}A_{xz}\right)+\hat{y}\left(\hat{x}A_{yx}+\hat{y}A_{yy}+\hat{z}A_{yz}\right)+\hat{z}\left(\hat{x}A_{zx}+\hat{y}A_{zy}+\hat{z}A_{zz}\right)\right]\\ &= a_x\left(\hat{x}A_{xx}+\hat{y}A_{xy}+\hat{z}A_{xz}\right)+a_y\left(\hat{x}A_{yx}+\hat{y}A_{yy}+\hat{z}A_{yz}\right)+a_z\left(\hat{x}A_{zx}+\hat{y}A_{zy}+\hat{z}A_{zz}\right)\end{aligned} \tag{B-7}$$

and

$$\begin{aligned}\overleftrightarrow{A}\cdot\vec{a} &= \left(\hat{x}\hat{x}A_{xx}+\hat{x}\hat{y}A_{xy}+\hat{x}\hat{z}A_{xz}+\hat{y}\hat{x}A_{yx}+\hat{y}\hat{y}A_{yy}+\hat{y}\hat{z}A_{yz}+\hat{z}\hat{x}A_{zx}+\hat{z}\hat{y}A_{zy}+\hat{z}\hat{z}A_{zz}\right)\\ &\quad\cdot\left(\hat{x}a_x+\hat{y}a_y+\hat{z}a_z\right)\\ &= \left[\left(\hat{x}A_{xx}+\hat{y}A_{yx}+\hat{z}A_{zx}\right)\hat{x}+\left(\hat{x}A_{xy}+\hat{y}A_{yy}+\hat{z}A_{zy}\right)\hat{y}+\left(\hat{x}A_{xz}+\hat{y}A_{yz}+\hat{z}A_{zz}\right)\hat{z}\right]\\ &\quad\cdot\left(\hat{x}a_x+\hat{y}a_y+\hat{z}a_z\right)\\ &= \left[\left(\hat{x}A_{xx}+\hat{y}A_{yx}+\hat{z}A_{zx}\right)a_x+\left(\hat{x}A_{xy}+\hat{y}A_{yy}+\hat{z}A_{zy}\right)a_y+\left(\hat{x}A_{xz}+\hat{y}A_{yz}+\hat{z}A_{zz}\right)a_z\right]\end{aligned} \tag{B-8}$$

Because $\overleftrightarrow{A}^T=\overleftrightarrow{A}$, then we have

$$A_{mn} = A_{nm} \quad , \quad (m,n=x,y,z) \tag{B-9}$$

Comparing relationship (B-7) with relationship (B-8) and employing relationship (B-9), it is easy to know that relationship (B-6) holds.

If $\overleftrightarrow{A}$ is a symmetrical dyad, i.e., $\overleftrightarrow{A}^T=\overleftrightarrow{A}$, then $\vec{a}$, $\overleftrightarrow{A}$, and $\overleftrightarrow{B}$ satisfy the following identical relationship:

$$\vec{a}\cdot\left(\overleftrightarrow{A}\cdot\overleftrightarrow{B}\right) = \left(\overleftrightarrow{A}\cdot\vec{a}\right)\cdot\overleftrightarrow{B} \tag{B-10}$$

and it can be proven by inserting relationship (B-6) into the RHS of relationship (B-5).

## Appendix C Mathematical Expressions of the Surface Equivalence Principle for Inhomogeneous Anisotropic Material Systems

Huygens-Fresnel principle (HFP), backward extinction theorem (BET), and Franz-Harrington formulation (FHF) are three most commonly used manifestations of surface equivalence principle (SEP)①, and at the same time they are also the important components of EM theory. In this Appendix C, we will generalize SEP from the following aspects:

① In what follows, Huygens-Fresnel principle (HFP), backward extinction theorem (BET), and Franz-Harrington formulation (FHF) are collectively referred to as surface equivalence principle (SEP).

**1.** The traditional SEP, which is only applicable to homogeneous isotropic EM environment①, will be generalized to inhomogeneous anisotropic EM environment.

**2.** The traditional SEP, which is only applicable to homogeneous isotropic material systems, will be generalized to inhomogeneous anisotropic material systems②.

**3.** In traditional HFP and BET, Huygens' surface is a single simply closed surface. This Appendix C will generalize it to the case that whole Huygens' surface is constituted by multiple simply closed surfaces③.

**4.** For a material body, traditional FHF has ability to express the external scattered field and the internal total field of material body as the functions of the equivalent surface sources on material boundary. This Appendix C generalizes the traditional FHF, and the generalized version can also express the internal incident field and internal scattered field as the functions of the equivalent surface sources on material boundary.④

In addition, this Appendix C will also discuss the differences and relations among HFP, BET, and FHF.

## C1 Research Background and Content Arrangement of Appendix C

SEP provides a method to express the interesting EM fields in an interesting region in terms of the functions of equivalent surface sources (instead of real sources). The earliest studies on SEP can be dated back to C. Huygens. In 1969, Huygens published a seminal book on the propagation of light, *Traite de la Lumiere*[127], and at the same time introduced the following famous Huygens' construction:

> "… every point of a wave-front may be considered as a centre of a secondary disturbance which gives rise to spherical wavelets, and the wave-front at any later instant may be regarded as the envelope of these wavelets. [128]"

Based on above Huygens' construction, Huygens provided an effective geometrography to research the propagation, reflection, and transmission phenomena of light. In fact, above Huygens' construction is a qualitative method rather than a quantitative method.

① In what follows, we simply call "EM environment" as "environment".

② In the Chapter 5 of this dissertation, we will further generalize the SEP for material scattering systems to the line-surface equivalence principle (LSEP) for metal-material composite scattering systems.

③ In the Chapter 5 of this dissertation, we will further generalize it to the line-surface composite cases. At that time, the surfaces can be either some closed surfaces or some open surfaces.

④ Here, the total field represents the summation of incident field and scattered field. In what follows, we will simply call "the scattered field on the exterior of the material body", "the total field on the interior of the material body", "the incident field on the interior of the material body", and "the scattered field on the interior of the material body" as "external scattered field", "internal total field", "internal incident field", and "internal scattered field" respectively.

The earliest quantitative researches on various optical phenomena started with T. Young and A. Fresnel. Around 1800, Young[129] did his famous double-slit interference experiment, and deeply studied the diffraction phenomenon of light. To mathematically explain the new phenomena observed by himself, Young introduced, in addition to the geometric-optical principle of propagation of locally-plane waves in the direction of rays, the phenomenological conception of transverse transmission of the oscillation amplitudes directly along the wave-fronts[130]. However, his method cannot generally explain all of the phenomena related to the diffraction of light. Until 1819, Fresnel[131], by introducing the interferences among secondary wavelets into Huygens' construction, phenomenologically explained some commonly observed phenomena related to the diffraction of light. Thus, Huygens' construction and Fresnel's interferences among secondary wavelets are usually collectively referred to as Huygens-Fresnel principle[128]. However, Fresnel's original theory cannot provide a completely correct description to the propagation of light in vacuum, because the theory generates the backward waves which propagate towards wave source①, and then the theory conflicts with the law of causality. To suppress the backward waves, Fresnel introduced the oblique factor into his theory, so his theory is essentially a phenomenological theory just like Young's theory.

The efforts for establishing HFP on a rigorous mathematical foundation started with H. Helmholtz and G. Kirchhoff. In 1859, Helmholtz[132] established the rigorous mathematical foundation for stead-state optics (or called as monochromatic optics), and his results are usually called as scalar Green's second theorem. In 1882, Kirchhoff[133,134] established the rigorous mathematical foundation for time-dependent optics, and his results are usually called as Fresnel-Kirchhoff diffraction integral formulation. Later on, Lord Rayleigh[135] and A. Sommerfeld[136], by employing half-space scalar Green's function, further generalized Fresnel-Kirchhoff diffraction integral formulation, and their generalized results are usually called as Rayleigh-Sommerfeld diffraction integral formulation. Just based on the above famous works, the wave nature of light, which was hidden in Sir Isaac Newton's corpuscular theory for a long time, was revealed gradually. A very comprehensive review for the history related to the above famous works can be found in literatures [128,131,137].

In fact, all of the above-mentioned formulations are the scalar formulations for describing optical phenomena, so they are usually called as scalar diffraction formulations.

① So-called backward wave is the wave which propagates towards the sources generating it.

However, since J. Maxwell[138] established his famous Maxwell's equations, it has been realised that light is essentially EM wave and EM wave is a vectorial wave instead of scalar wave, and it naturally guides the successors to update the scalar diffraction formulations established by the predecessors to the corresponding vectorial versions.

The earliest scholars focusing on deriving vectorial diffraction formulations from Maxwell's equations are A. Love (1901)[139] and H. MacDonald (1911)[140]. Love introduced the concept of equivalent surface source to act as Huygens' second source for the first time, and his work is now known as Love's equivalence principle. In 1936, S. Schelkunoff[141] further extended Love's results, and his results are now called as Schelkunoff's equivalence principle. In addition, Schelkunoff also discussed the relationships between his results and the results given by J. Larmor[142] in 1903. By employing so-called vector Green's second theorem, J. Stratton and L. Chu[143], in 1938, provided a formulation to express EM field in terms of both the normal and tangential components of the field on a closed surface, and their formulation is now called as Stratton-Chu formulation. In 1948, W. Franz[144] established so-called Franz's formulation①. Different from Stratton-Chu formulation, Franz's formulation can express EM field in terms of only tangential surface field. Later on, Prof. C.-T. Tai[145] proved that: Stratton-Chu formulation and Franz's formulation are equivalent to each other, and Stratton-Chu formulation is essentially identical to Larmor-Tedone formulation. A relatively comprehensive summarization for the above vectorial diffraction formulations can be found in literatures [26,27,115].

In addition, based on his studies on the Cauchy's problem for partial differential equations, mathematician J. Hadamard[146] gave HFP a mathematically rigorous and somewhat philosophical description, and revealed that the crucial building blocks of HFP are the following three: **i)** the concept of action through medium②, **ii)** the law of causality, and **iii)** the linear superposition principle③. The concept of action through medium requires that the mathematical expression of HFP needs to employ the field propagator, which is usually the field generated by unit point source —— Green's function, and the frequency-domain Green's function must include a phase term which depends on spacial position coordinate; the law of causality requires that the propagator should propagate

① Sometimes, it is alternatively called as Kottler-Franz formulation, because F. Kottler also obtained a similar result in 1923.

② This view point originates from M. Faraday and Maxwell.

③ In what follows, we simply call it as superposition principle.

away from source rather than being towards source, i.e., the phase term included in Green's function should gradually lag along radial direction and Green's function should satisfy Sommerfeld's radiation condition[147]; linear superposition principle requires that the mathematical expression of HFP will appear as an integral. Hence, all of the above scalar and vectorial diffraction formulations appear as the convolution integrals between Green's functions and Huygens's second sources or equivalent surface sources.

Obviously, the predecessors' establishment for HFP and its mathematical expression experienced an evolution from qualitative to quantitative and from scalar to vectorial. In classical electromagnetics framework, Franz's formulation reaches the peak of the evolution, as stated by Prof. C.-T. Tai in literature [145] that:

> "It seems obvious that the Franz formula is conceptually simpler since it requires only the tangential components of the field on the closed surface, while the Stratton-Chu formula requires the normal components as well. Most important of all, when the field has an edge singularity on the surface of integration the Larmor-Tedone formula or Stratton-Chu formula must be modified as shown by Kottler in order to make the resultant field Maxwellian. [145]"

Because of this, Franz's formulation has had many important applications in theoretical, computational, and engineering electromagnetics domains, for example, the famous PMCHWT-based (A. Poggio & E. Miller[148], Y. Chang & R. Harrington[34], W. Wu[149], Tsai) scattering integral equation and CM calculation formulation are just established by employing Franz's formulation.

The formulation utilized by Dr. Y. Chang and Prof. Harrington in literature [34] originates from Franz's formulation, but the former is more advantageous than the latter in the following aspects: in the aspect of mathematical manifestation, the former is more concise than the latter; in the aspect of expressing EM field, the former expresses external scattering field and internal total field in terms of an identical set of equivalent surface sources. Based on these features, the former is more popular than the latter in computational electromagnetics and EM engineering. Because the former originates from the latter and Dr. Chang and Prof. Harrington gave the former in literature [34] by employing Prof. Harrington's classical literature *Time-Harmonic Electromagnetic Fields*[114], then this dissertation calls the former as Franz-Harrington formulation (FHF).

Although FHF has had many successful applications in engineering and computational electromagnetics domains, it also has some limitations. This Appendix C will do some further improvements and generalizations for FHF from the following aspects.

**1. Limitation in the Aspect of Propagation Medium and The Corresponding Improvements and Generalizations Done by This Appendix C**

Traditional HFP, BET, and FHF are only suitable for the homogeneous isotropic material systems placed in homogeneous isotropic environment. It is still an unsolved problem how to establish the HFP, BET, and FHF which are applicable to the inhomogeneous anisotropic material systems placed in inhomogeneous anisotropic environment.

In Appendixes C2 and C3, we, based on generalized vector-dyad Green's second theorem, express the fields in an inhomogeneous anisotropic open domain① in terms of the convolution integrals of dyadic Green's functions and the tangential field components on the boundary of the domain. At that time, we will utilize the vector-dyad version of Green's second theorem rather than the vector-vector version, and the reason can be found from Prof. C.-T. Tai's the following statement:

> "… the most compact formulation appears to be the one based on the dyadic Green's function pertaining to the vector wave equation for $\vec{E}$ and $\vec{H}$ … [145]"

At that time, we will utilize the generalized version of vector-dyad Green's second theorem rather than the traditional version, and the reason is that: the traditional version is suitable for neither inhomogeneous domain nor anisotropic domain. Based on the results obtained in Appendixes C2 and C3, we generalize traditional HFP, BET, and FHF from the following aspects:

**(1.1)** Based on the results obtained in Appendixes C2 and C3, we, in Appendix C4, derive the convolution integrals for the various EM fields related to a simply connected② inhomogeneous anisotropic material body.

**(1.2)** Based on the results obtained in Appendixes C4 and A, Appendixes C5 and C6 generalize traditional FHF to the inhomogeneous anisotropic material body placed in an inhomogeneous anisotropic environment.

**(1.3)** Based on the results obtained in Appendixes C5 and C6, Appendix C7 further generalizes the FHF obtained in Appendix C6 to the piecewisely inhomogeneous

---

① Open domain is a commonly used concept in point set topology[124], and it represents the domain which doesn't include its boundary.

② The concept of "simply connected" will be carefully explained in the following point 2), and its rigorous mathematical definition can be found in literature [124].

anisotropic material system placed in an inhomogeneous anisotropic environment. Here, modifier "piecewisely" means that: the material parameters are discontinuous on two sides of the interface between two different material bodies, and a typical example is illustrated in Figure C-1(d).

**2. Limitation in the Aspect of Topological Structure and The Corresponding Improvements and Generalizations Done by This Appendix C**

In traditional HFP and BET, the Huygens' surface must be a single closed surface. Traditional FHF is only suitable for the material system $V_{\text{mat sys}}$ which is constituted by a single simply connected material body $V_{\text{sim}}$, i.e., it is only suitable for the $V_{\text{mat sys}} = V_{\text{sim}}$ case. In this Appendix C, we generalize the traditional HFP, BET, and FHF from the following aspects:

**(2.1)** In Appendixes C6 and C7, Huygens' surface is generalized to the multiple closed surfaces case.

**(2.2)** In Appendix C6, we consider the case that material system $V_{\text{mat sys}}$ is constituted by a single multiply connected material body $V_{\text{mul}}$, i.e. the case that $V_{\text{mat sys}} = V_{\text{mul}}$, and derive the HFP, BET, and FHF which are applicable to $V_{\text{mul}}$.

**(2.3)** The material systems focused on by Appendixes C5 and C6 are connected, and the results obtained in Appendixes C5 and C6 will be further generalized to non-connected material systems.

The above modifiers "connected, non-connected, simply connected, and multiply connected" are the commonly used terms in point set topology, and their rigorous definitions can be found in literature [124]. The "non-connected" can be vividly understood as that there exist some different parts of system, such that the parts don't contact with each other, as shown in Figure C-1(c); if the system is not non-connected, it is connected, as shown in Figure C-1(a); the "simply connected" can be vividly understood as that there doesn't exist any hole on material body, as shown in Figure C-1(a); the "multiply connected" can be vividly understood as that there exist some holes on material body, as shown in Figure C-1(b).

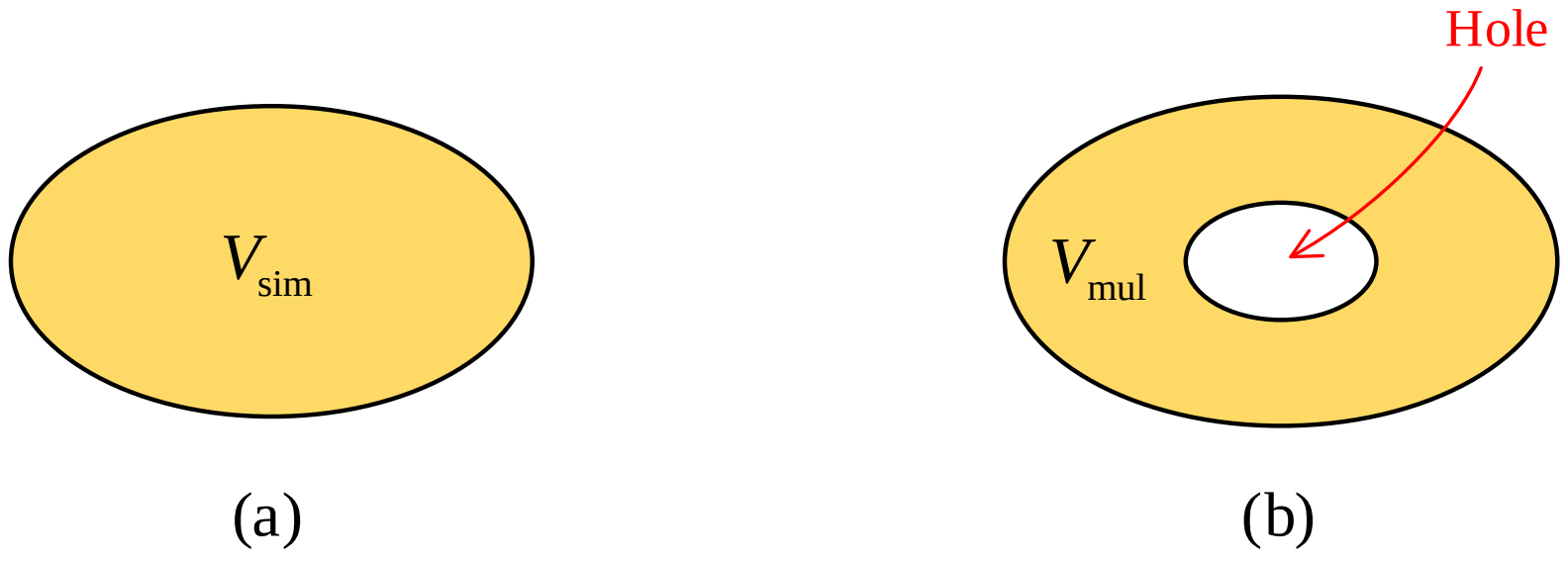

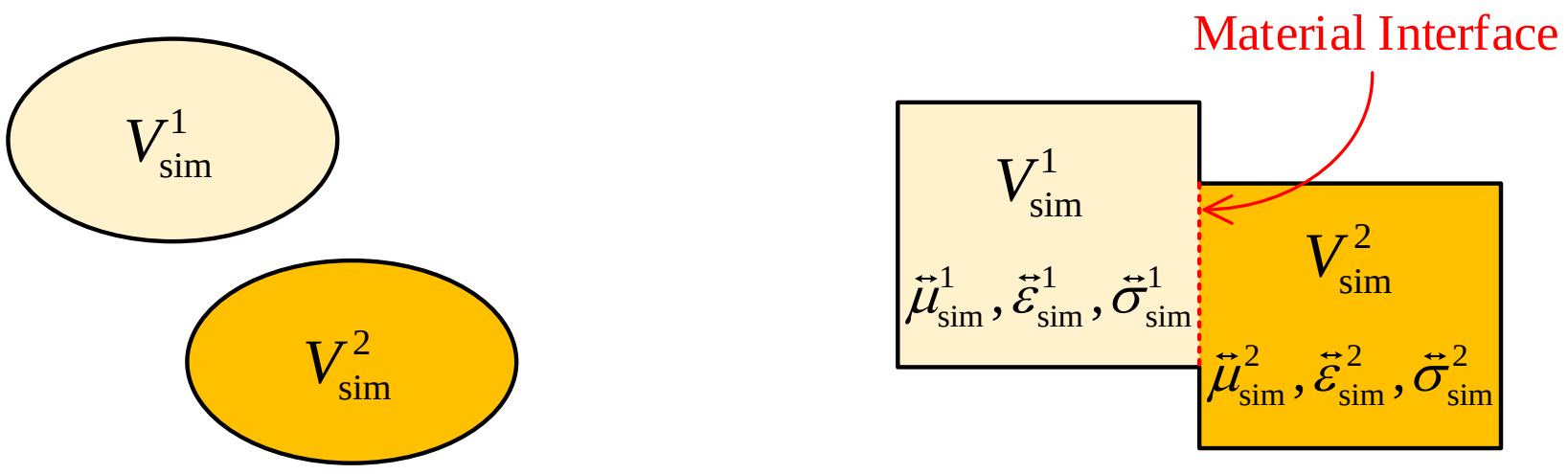

(c) (d)

Figure C-1 The topological structures of several typical material systems. (a) a simply connected material body; (b) a 2-connected material body ("2-connected" is the most special "multiply connected"); (c) a non-connected material system constituted by two non-contacted simply connected material bodies; (d) a piecewisely inhomogeneous anisotropic material system constituted by two simply connected material bodies, where the two bodies are contacted with each other and have different material parameters

### 3. Limitation in the Aspect of Formulating EM Fields and the Corresponding Improvements and Generalizations Done by This Appendix C

When a simply connected material body $V_{\text{sim}}$ is considered, whole three-dimensional Euclidean space $\mathbb{R}^3$ is divided into two parts (the interior of $V_{\text{sim}}$ and the exterior of $V_{\text{sim}}$) by material boundary $\partial V_{\text{sim}}$, as illustrated in Figure C-2. Traditional FHF has only ability to express external scattered field $\vec{F}_+^{\text{sca}}$ and internal total field $\vec{F}_-^{\text{tot}}$ ($\vec{F}_-^{\text{tot}}$ is the summation of internal incident field $\vec{F}_-^{\text{inc}}$ and internal scattered field $\vec{F}_-^{\text{sca}}$) in terms of the convolution integrals of Green's functions and the equivalent surface currents on $\partial V_{\text{sim}}$. In Appendixes C4~C7, the traditional FHF for $\vec{F}_+^{\text{sca}}$ and $\vec{F}_-^{\text{tot}}$ will be further generalized, such that the generalized version has also ability to formulate $\vec{F}_-^{\text{inc}}$ and $\vec{F}_-^{\text{sca}}$ by employing the identical set of equivalent surface currents.

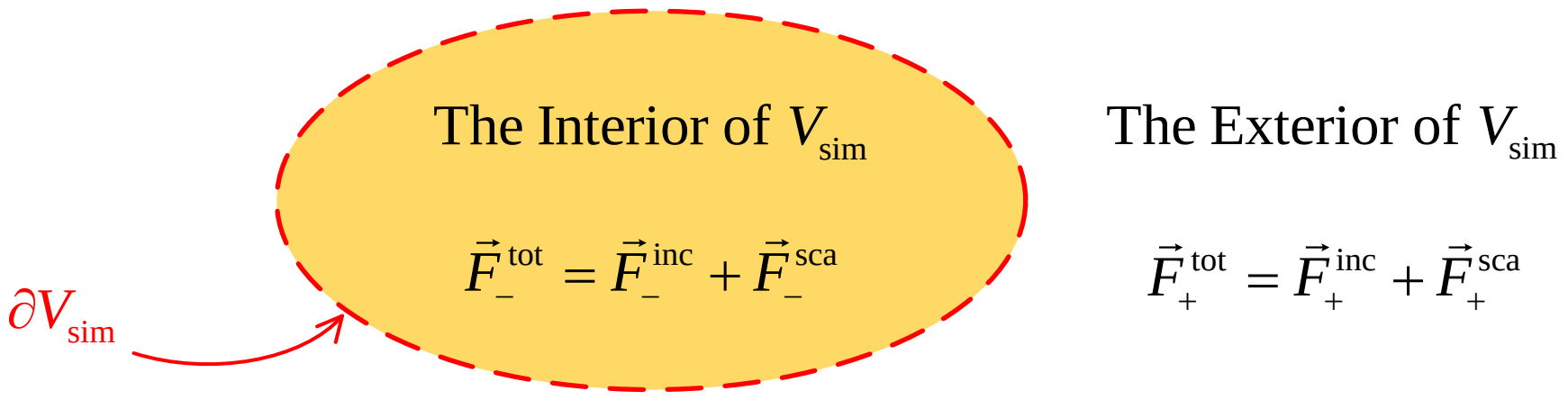

Figure C-2 The various domains related to the scattering problem of a simply connected materia body and the various EM fields distributing on the domains

In addition, this Appendix C also does some studies in the following aspects.

**A. On the Mathematical Expression of Huygens-Fresnel Principle**

The FHF for expressing $\vec{F}_+^{\text{sca}}$ is usually viewed as the mathematical expression for the HFP corresponding to scattered field, but it will generate backward waves in the interior of material body, so it conflicts with the law of causality. In addition, the FHF for expressing $\vec{F}_{-}^{\text{tot}}$ is sometimes classified into the BET family, but it cannot guarantee the null field in whole exterior of material body, though it indeed guarantees null tangential field on the external surface of material boundary. To clarify the above confusions, Appendix C5 discusses the differences and relationships among HFP, BET, and FHF.

**B. On the Topological Additivity of Surface Equivalence Principle**

The material systems $V_{\text{mat sys}}$ considered in Appendixes C5~C7 have different topological structures, i.e., the system considered in Appendix C5 satisfies that $V_{\text{mat sys}} = V_{\text{sim}}$; the system considered in Appendix C6 satisfies that $V_{\text{mat sys}} = V_{\text{mul}}$; the system considered in Appendix C7 satisfies that $V_{\text{mat sys}} = V_{\text{sim}} \bigcup V_{\text{mul}}$. In Appendix C7, we will prove that: both BET and the mathematical expression of HFP satisfy so-called topological additivity, i.e., the HFP/BET for whole system equals the summation of the HFP/BET for all of the sub-systems. Obviously, this property is consistent with linear superposition principle. In addition, Appendix C5 will prove that: the FHF used to formulate $\vec{F}_{-}^{\text{inc}}$ and $\vec{F}_{+}^{\text{sca}}$ is actually the summation of the HFP pertaining to incident field and the HFP pertaining to scattered field. Then, it is immediately known that: the FHF used to formulate $\vec{F}_{-}^{\text{inc}}$ and $\vec{F}_{+}^{\text{sca}}$ must satisfy topological additivity. However, the FHFs used to formulate $\vec{F}_{-}^{\text{tot}}$ and $\vec{F}_{-}^{\text{sca}}$ don't satisfy the above topological additivity. To resolve the problem, we will introduce the concept of piecewise Green's function in Appendix C7.

In the end of this Appendix C1, we provide some brief explanations for the symbolic system used in whole Appendix C.

**I.** The conductivity, permittivity, and permeability of $V_{\text{sim/mul}}$ are denoted as $\vec{\vec{\sigma}}_{\text{sim/mul}}(\vec{r})$, $\vec{\vec{\varepsilon}}_{\text{sim/mul}}(\vec{r})$, and $\vec{\vec{\mu}}_{\text{sim/mul}}(\vec{r})$. The conductivity, permittivity, and permeability of $D_{\text{env}}$ are denoted as $\vec{\vec{\sigma}}_{\text{env}}(\vec{r})$, $\vec{\vec{\varepsilon}}_{\text{env}}(\vec{r})$, and $\vec{\vec{\mu}}_{\text{env}}(\vec{r})$. This Appendix C focuses on inhomogeneous anisotropic matter, so spacial variable $\vec{r}$ is explicitly added to the above material parameters, and all of the parameters are two-order tensors as marked by superscript "$\leftrightarrow$". This Appendix C will separately consider the material parameter related to ohmic loss and the material parameter related to polarization, and ignore the magnetic loss of matter, so the permittivities and permeabilities are real. In addition, this Appendix C also restricts the above material parameters to symmetrical tensors, and the reason to do this restriction has been carefully explained in Appendix B.

**II.** This Appendix C will frequently utilize some concepts related to point set topology, such as open set $\Omega$ ①, boundary $\partial\Omega$, closure $\text{cl}\,\Omega$, interior $\text{int}\,\Omega$, and exterior $\text{ext}\,\Omega$, etc. The rigorous mathematical definitions for $\Omega$, $\partial\Omega$, $\text{cl}\,\Omega$, and $\text{int}\,\Omega$ can be found in general monographs on point set topology[124], so they will not be repeated here. Here, we specially provide the definition for $\text{ext}\,\Omega$ as that $\text{ext}\,\Omega = \mathbb{R}^3 \setminus \text{cl}\,\Omega$. In addition, it is obvious that both $\text{int}\,\Omega$ and $\text{ext}\,\Omega$ are open sets[124].

**III.** When the field $\vec{F}_{\text{imp}}$ generated by impressed source $\vec{J}_{\text{imp}}$ exists, some currents $\{\vec{J}^{\text{SV}},\vec{M}^{\text{SV}}\}$ and $\{\vec{J}_{\text{env}},\vec{M}_{\text{env}}\}$ ② will be induced on system $V_{\text{mat sys}}$ and environment $D_{\text{env}}$. In whole three-dimensional Euclidean space $\mathbb{R}^3$, induced currents $\{\vec{J}^{\text{SV}},\vec{M}^{\text{SV}}\}$ and $\{\vec{J}_{\text{env}},\vec{M}_{\text{env}}\}$ will generate fields $\vec{F}^{\text{sca}}$ and $\vec{F}_{\text{env}}$ respectively. Due to the same reason as Chapter 3 and Section 4.2, this Appendix C treats $\vec{F}_{\text{imp}}$ and $\vec{F}_{\text{env}}$ as a whole, and the whole is called as incident field (or resultant field), and correspondingly denoted as $\vec{F}^{\text{inc}}$, i.e., $\vec{F}^{\text{inc}} = \vec{F}_{\text{imp}} + \vec{F}_{\text{env}}$. The source used to generate $\vec{F}^{\text{inc}}$ is correspondingly denoted as $\{\vec{J}^{\text{inc}},\vec{M}^{\text{inc}}\}$, and it is obvious that $\vec{J}^{\text{inc}} = \vec{J}_{\text{imp}} + \vec{J}_{\text{env}}$ and $\vec{M}^{\text{inc}} = \vec{M}_{\text{env}}$ ③.

**IV.** When $\vec{F}^{\text{inc}}$ is incident on $V_{\text{mat sys}}$ (where $V_{\text{mat sys}} = V_{\text{sim}} \cup V_{\text{mul}}$), some scattered sources $\{\vec{J}^{\text{SV}}_{\text{sim}},\vec{M}^{\text{SV}}_{\text{sim}}\}$ and $\{\vec{J}^{\text{SV}}_{\text{mul}},\vec{M}^{\text{SV}}_{\text{mul}}\}$ will be induced on $V_{\text{sim}}$ and $V_{\text{mul}}$ respectively, and sources $\{\vec{J}^{\text{SV}}_{\text{sim}},\vec{M}^{\text{SV}}_{\text{sim}}\}$ and $\{\vec{J}^{\text{SV}}_{\text{mul}},\vec{M}^{\text{SV}}_{\text{mul}}\}$ will generate scattered fields $\vec{F}^{\text{sca}}_{\text{sim}}$ and $\vec{F}^{\text{sca}}_{\text{mul}}$ respectively④. The summation of $\vec{F}^{\text{sca}}_{\text{sim}}$ and $\vec{F}^{\text{sca}}_{\text{mul}}$ is just the scattered field $\vec{F}^{\text{sca}}$ generated by whole system, i.e., $\vec{F}^{\text{sca}} = \vec{F}^{\text{sca}}_{\text{sim}} + \vec{F}^{\text{sca}}_{\text{mul}}$. In addition, the summation of $\vec{F}^{\text{inc}}$ and $\vec{F}^{\text{sca}}$ is just the total field $\vec{F}^{\text{tot}}$ corresponding to the whole scattering problem, i.e., $\vec{F}^{\text{tot}} = \vec{F}^{\text{inc}} + \vec{F}^{\text{sca}}$. The above-mentioned various EM fields are illustrated in Figure C-3.

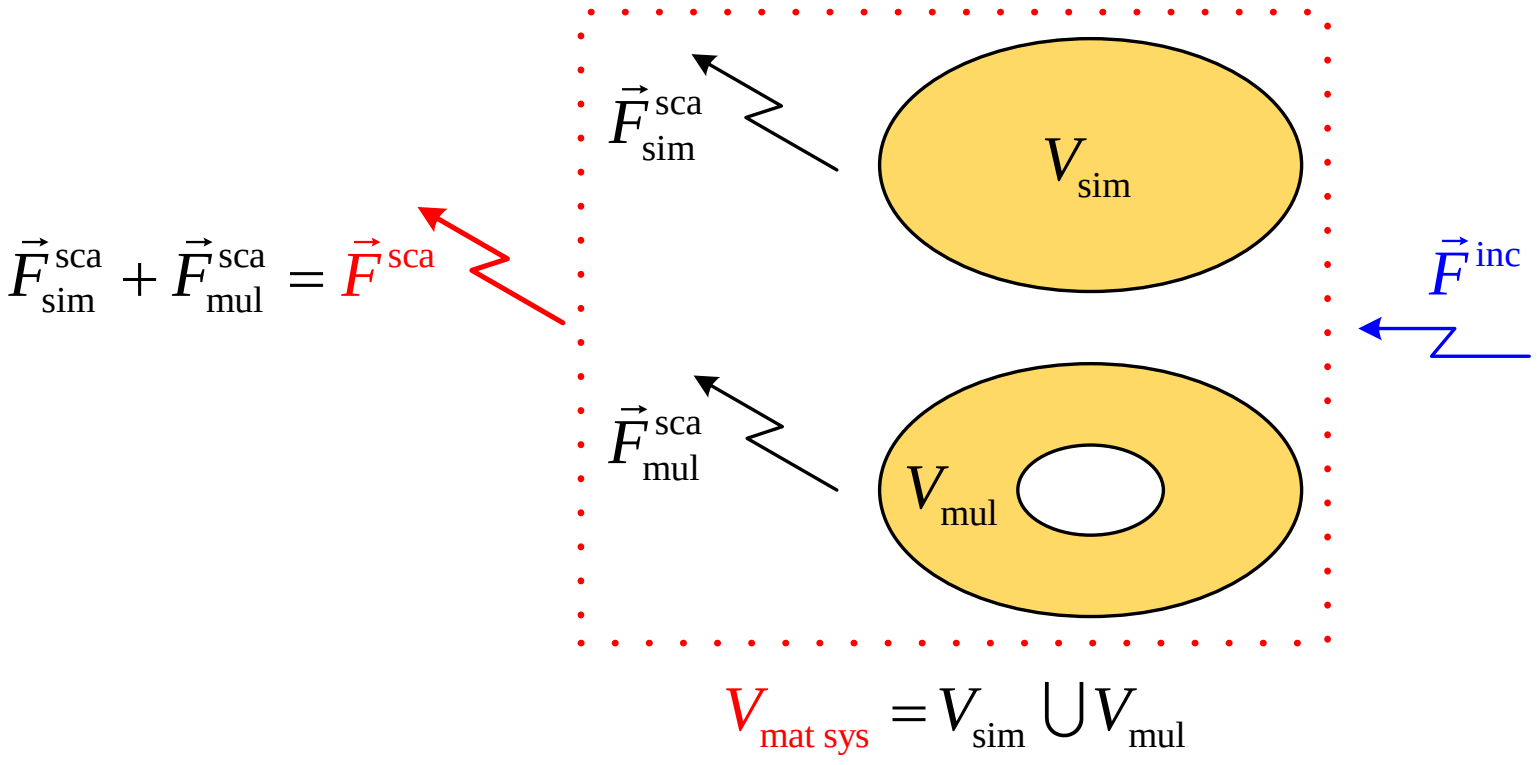

Figure C-3 The scattering problem related to a material system $V_{\text{mat sys}}$ which is constituted by a simply connected material body $V_{\text{sim}}$ and a multiply connected material body $V_{\text{mul}}$ and excited by an incident field $\vec{F}^{\text{inc}}$

① Here, so-called open set is not the counterpart of the closed structure usually mentioned in engineering electromagnetics, but the point set which doesn't include its boundary.
② For details see Appendix A.
③ For details see Appendix A.
④ For details see Appendix A.

## C2 Non-standard Convolution Integral Formulations for the EM Fields in an Inhomogeneous Anisotropic Open Domain

In this Appendix C2, we express the EM fields in an inhomogeneous anisotropic open domain in terms of the convolution integrals of the EM dyadic Green's functions in the domain and the tangential EM fields on the boundary of the domain.

In the open domain with material parameters $\{\vec{\vec{\varepsilon}}_\Omega, \vec{\vec{\mu}}_\Omega\}$, EM fields $\{\vec{E}_\Omega, \vec{H}_\Omega\}$ and EM currents $\{\vec{J}_\Omega, \vec{M}_\Omega\}$ satisfy the following Maxwell's equations①:

$$\begin{aligned} \nabla \times \vec{H}_\Omega(\vec{r}) &= \vec{J}_\Omega(\vec{r}) + j\omega\vec{\vec{\varepsilon}}_\Omega(\vec{r}) \cdot \vec{E}_\Omega(\vec{r}) \\ \nabla \times \vec{E}_\Omega(\vec{r}) &= -\vec{M}_\Omega(\vec{r}) - j\omega\vec{\vec{\mu}}_\Omega(\vec{r}) \cdot \vec{H}_\Omega(\vec{r}) \end{aligned} \quad , \quad \vec{r} \in \Omega \tag{C-1}$$

Here, the subscript "$\Omega$" is to emphasize that: the physical quantities with the subscript distribute on $\Omega$. On domain $\Omega$, we define the dyadic Green's functions corresponding to electric-typy unit dyadic point source as follows②:

$$\begin{aligned} \nabla \times \vec{\vec{G}}_\Omega^{JH}(\vec{r}, \vec{r}\,') &= \vec{\vec{I}}\delta(\vec{r} - \vec{r}\,') + j\omega\vec{\vec{\varepsilon}}_\Omega(\vec{r}) \cdot \vec{\vec{G}}_\Omega^{JE}(\vec{r}, \vec{r}\,') \\ \nabla \times \vec{\vec{G}}_\Omega^{JE}(\vec{r}, \vec{r}\,') &= -j\omega\vec{\vec{\mu}}_\Omega(\vec{r}) \cdot \vec{\vec{G}}_\Omega^{JH}(\vec{r}, \vec{r}\,') \end{aligned} \quad , \quad (\vec{r}, \vec{r}\,' \in \Omega) \tag{C-2a}$$

and the dyadic Green's functions corresponding to magnetic-typy unit dyadic point source as follows③:

$$\begin{aligned} \nabla \times \vec{\vec{G}}_\Omega^{MH}(\vec{r}, \vec{r}\,') &= j\omega\vec{\vec{\varepsilon}}_\Omega(\vec{r}) \cdot \vec{\vec{G}}_\Omega^{ME}(\vec{r}, \vec{r}\,') \\ \nabla \times \vec{\vec{G}}_\Omega^{ME}(\vec{r}, \vec{r}\,') &= -\vec{\vec{I}}\delta(\vec{r} - \vec{r}\,') - j\omega\vec{\vec{\mu}}_\Omega(\vec{r}) \cdot \vec{\vec{G}}_\Omega^{MH}(\vec{r}, \vec{r}\,') \end{aligned} \quad , \quad (\vec{r}, \vec{r}\,' \in \Omega) \tag{C-2b}$$

In equations (C-2a) and (C-2b), $\vec{\vec{I}}$ is unit dyad, i.e., $\vec{\vec{I}} = \hat{x}\hat{x} + \hat{y}\hat{y} + \hat{z}\hat{z}$ ; $\delta(\vec{r} - \vec{r}\,')$ is Dirac's delta function.

If $\vec{\vec{\vartheta}}$ is a symmetrical two-order complex tensor, then its inverse $\vec{\vec{\vartheta}}^{-1}$ must be also symmetrical, because: relationship $(\vec{\vec{\vartheta}}^{-1})^T \cdot \vec{\vec{\vartheta}}^T = (\vec{\vec{\vartheta}} \cdot \vec{\vec{\vartheta}}^{-1})^T = \vec{\vec{I}}^T = \vec{\vec{I}}$ implies relationship $(\vec{\vec{\vartheta}}^{-1})^T = (\vec{\vec{\vartheta}}^T)^{-1} = \vec{\vec{\vartheta}}^{-1}$ [116]. Based on vectorial differential identity $\nabla \cdot (\vec{a} \times \vec{b}) = (\nabla \times \vec{a}) \cdot \vec{b} - \vec{a} \cdot (\nabla \times \vec{b})$ [122] and the symmetry of $\vec{\vec{\vartheta}}^{-1}$ and the identity (B-10) given in Appendix D, we can obtain the following relationship:

$$\begin{aligned} &\nabla \cdot \left\{ \vec{P} \times \left[ \vec{\vec{\vartheta}}^{-1} \cdot \left( \nabla \times \vec{Q} \right) \right] + \left[ \vec{\vec{\vartheta}}^{-1} \cdot \left( \nabla \times \vec{P} \right) \right] \times \vec{Q} \right\} \\ &= \left\{ \nabla \times \left[ \vec{\vec{\vartheta}}^{-1} \cdot \left( \nabla \times \vec{P} \right) \right] \right\} \cdot \vec{Q} - \vec{P} \cdot \left\{ \nabla \times \left[ \vec{\vec{\vartheta}}^{-1} \cdot \left( \nabla \times \vec{Q} \right) \right] \right\} \end{aligned} \tag{C-3}$$

---

① Here, $\{\vec{J}_\Omega, \vec{M}_\Omega\}$ are not necessarily the sources used to generate $\{\vec{E}_\Omega, \vec{H}_\Omega\}$, and it is only necessary that $\{\vec{J}_\Omega, \vec{M}_\Omega\}$ satisfy Maxwell's equations (C-1). But, we want to emphasize that: if $\{\vec{J}_\Omega, \vec{M}_\Omega\}$ are indeed the sources used to generate $\{\vec{E}_\Omega, \vec{H}_\Omega\}$, then they must satisfy Maxwell's equations (C-1).

② Here, $\vec{\vec{I}}\delta(\vec{r} - \vec{r}\,')$ is not necessarily the source used to generate $\{\vec{\vec{G}}_\Omega^{JE}(\vec{r}, \vec{r}\,'), \vec{\vec{G}}_\Omega^{JH}(\vec{r}, \vec{r}\,')\}$, and it is only necessary that $\vec{\vec{I}}\delta(\vec{r} - \vec{r}\,')$ satisfies Maxwell's equations (C-2a).

③ Here, $\vec{\vec{I}}\delta(\vec{r} - \vec{r}\,')$ is not necessarily the source used to generate $\{\vec{\vec{G}}_\Omega^{ME}(\vec{r}, \vec{r}\,'), \vec{\vec{G}}_\Omega^{MH}(\vec{r}, \vec{r}\,')\}$, and it is only necessary that $\vec{\vec{I}}\delta(\vec{r} - \vec{r}\,')$ satisfies Maxwell's equations (C-2b).

Applying Gauss' divergence theorem[122] to above relationship (C-3), we can obtain the following generalized vector-dyad Green's second theorem:

$$\begin{aligned}&\iiint_{\Omega}\Big(\vec{P}\cdot\Big\{\nabla\times\Big[\vec{\vec{\vartheta}}^{-1}\cdot\big(\nabla\times\vec{\vec{Q}}\big)\Big]\Big\}-\Big\{\nabla\times\Big[\vec{\vec{\vartheta}}^{-1}\cdot\big(\nabla\times\vec{P}\big)\Big]\Big\}\cdot\vec{\vec{Q}}\Big)dV\\&=\oiint_{\partial\Omega}\hat{n}_{\Omega}^{-}\cdot\Big\{\vec{P}\times\Big[\vec{\vec{\vartheta}}^{-1}\cdot\big(\nabla\times\vec{\vec{Q}}\big)\Big]+\Big[\vec{\vec{\vartheta}}^{-1}\cdot\big(\nabla\times\vec{P}\big)\Big]\times\vec{\vec{Q}}\Big\}dS\end{aligned}\qquad\text{(C-4)}$$

Here, closed surface $\partial\Omega$ is the boundary of $\Omega$; $\hat{n}_{\Omega}^{-}$ is the unit normal vector of closed surface $\partial\Omega$, and it points to the interior of $\Omega$.

Inserting $\vec{P}=\vec{E}_{\Omega}(\vec{r})$ and $\vec{\vec{Q}}=\vec{\vec{G}}_{\Omega}^{JE}(\vec{r},\vec{r}\,')$ and $\vec{\vec{\vartheta}}=\vec{\vec{\mu}}_{\Omega}(\vec{r})$ into Green's second theorem (C-4), and restricting that there is not any magnetization surface magnetic current distributing on $\partial\Omega$ ①, the following relationship can be obtained:

$$\begin{aligned}&-j\omega\iiint_{\Omega}\vec{E}_{\Omega}(\vec{r})\cdot\vec{\vec{I}}\delta(\vec{r}-\vec{r}\,')d\Omega\\&+j\omega\iiint_{\Omega}\vec{J}_{\Omega}(\vec{r})\cdot\vec{\vec{G}}_{\Omega}^{JE}(\vec{r},\vec{r}\,')d\Omega\\&+\omega^{2}\iiint_{\Omega}\vec{E}_{\Omega}(\vec{r})\cdot\Big[\vec{\vec{\varepsilon}}_{\Omega}(\vec{r})\cdot\vec{\vec{G}}_{\Omega}^{JE}(\vec{r},\vec{r}\,')\Big]d\Omega\\&-\omega^{2}\iiint_{\Omega}\Big[\vec{\vec{\varepsilon}}_{\Omega}(\vec{r})\cdot\vec{E}_{\Omega}(\vec{r})\Big]\cdot\vec{\vec{G}}_{\Omega}^{JE}(\vec{r},\vec{r}\,')d\Omega\\&+\iiint_{\Omega}\Big\{\nabla\times\Big[\vec{\vec{\mu}}_{\Omega}^{-1}(\vec{r})\cdot\vec{M}_{\Omega}(\vec{r})\Big]\Big\}\cdot\vec{\vec{G}}_{\Omega}^{JE}(\vec{r},\vec{r}\,')d\Omega\\&=-j\omega\oiint_{\partial\Omega}\hat{n}_{\rightarrow\Omega}\cdot\Big[\Big\{\vec{\vec{\mu}}_{\Omega}^{-1}(\vec{r})\cdot\Big[\vec{\vec{\mu}}_{\Omega}(\vec{r})\cdot\vec{H}_{\Omega}(\vec{r})\Big]\Big\}\times\vec{\vec{G}}_{\Omega}^{JE}(\vec{r},\vec{r}\,')\Big]dS\\&\quad-j\omega\oiint_{\partial\Omega}\hat{n}_{\rightarrow\Omega}\cdot\Big[\vec{E}_{\Omega}(\vec{r})\times\Big\{\vec{\vec{\mu}}_{\Omega}^{-1}(\vec{r})\cdot\Big[\vec{\vec{\mu}}_{\Omega}(\vec{r})\cdot\vec{\vec{G}}_{\Omega}^{JH}(\vec{r},\vec{r}\,')\Big]\Big\}\Big]dS\\&\quad-\oiint_{\partial\Omega}\hat{n}_{\rightarrow\Omega}\cdot\Big\{\Big[\vec{\vec{\mu}}_{\Omega}^{-1}(\vec{r})\cdot\vec{M}_{\Omega}(\vec{r})\Big]\times\vec{\vec{G}}_{\Omega}^{JE}(\vec{r},\vec{r}\,')\Big\}dS\end{aligned}\qquad\text{(C-5)}$$

The above integral on $\partial\Omega$ is defined as that $\oiint_{\partial\Omega}\vec{A}(\vec{r},\vec{r}\,')dS=\lim_{\Gamma\rightarrow\partial\Omega}\oiint_{\Gamma}\vec{A}(\vec{r},\vec{r}\,')dS$, where $\Gamma$ is a closed surface contained in $\Omega$ and $\Gamma$ approaches $\partial\Omega$. The last term in the LHS of relationship (C-5) can be equivalently rewritten as follows:

$$\begin{aligned}&\iiint_{\Omega}\Big\{\nabla\times\Big[\vec{\vec{\mu}}_{\Omega}^{-1}(\vec{r})\cdot\vec{M}_{\Omega}(\vec{r})\Big]\Big\}\cdot\vec{\vec{G}}_{\Omega}^{JE}(\vec{r},\vec{r}\,')d\Omega\\&=-\oiint_{\partial\Omega}\hat{n}_{\rightarrow\Omega}\cdot\Big\{\Big[\vec{\vec{\mu}}_{\Omega}^{-1}(\vec{r})\cdot\vec{M}_{\Omega}(\vec{r})\Big]\times\vec{\vec{G}}_{\Omega}^{JE}(\vec{r},\vec{r}\,')\Big\}dS\\&\quad+\iiint_{\Omega}\Big[\vec{\vec{\mu}}_{\Omega}^{-1}(\vec{r})\cdot\vec{M}_{\Omega}(\vec{r})\Big]\cdot\Big[\nabla\times\vec{\vec{G}}_{\Omega}^{JE}(\vec{r},\vec{r}\,')\Big]d\Omega\\&=-\oiint_{\partial\Omega}\hat{n}_{\rightarrow\Omega}\cdot\Big\{\Big[\vec{\vec{\mu}}_{\Omega}^{-1}(\vec{r})\cdot\vec{M}_{\Omega}(\vec{r})\Big]\times\vec{\vec{G}}_{\Omega}^{JE}(\vec{r},\vec{r}\,')\Big\}dS\\&\quad-j\omega\iiint_{\Omega}\Big[\vec{\vec{\mu}}_{\Omega}^{-1}(\vec{r})\cdot\vec{M}_{\Omega}(\vec{r})\Big]\cdot\Big[\vec{\vec{\mu}}_{\Omega}(\vec{r})\cdot\vec{\vec{G}}_{\Omega}^{JH}(\vec{r},\vec{r}\,')\Big]d\Omega\end{aligned}\qquad\text{(C-6)}$$

① The rationality of this restriction has been proven in Appendix A.

In formulation (C-6), the first equality is based on differential identity $\nabla\cdot(\vec{a}\times\vec{b})=(\nabla\times\vec{a})\cdot\vec{b}-\vec{a}\cdot(\nabla\times\vec{b})$ [122] and Gauss' divergence theorem[122]; the second equality is based on equations (C-2a). Because $\overleftrightarrow{\varepsilon}_{\Omega}$ is symmetrical and vector-dyad operations satisfy the associative property (B-5)[122] and the identity (B-10) given in Appendix D, then we have that

$$\begin{aligned}\vec{E}_{\Omega}(\vec{r})\cdot\left[\overleftrightarrow{\varepsilon}_{\Omega}(\vec{r})\cdot\overleftrightarrow{G}_{\Omega}^{JE}(\vec{r},\vec{r}')\right] &= \left[\vec{E}_{\Omega}(\vec{r})\cdot\overleftrightarrow{\varepsilon}_{\Omega}(\vec{r})\right]\cdot\overleftrightarrow{G}_{\Omega}^{JE}(\vec{r},\vec{r}')\\ &= \left[\overleftrightarrow{\varepsilon}_{\Omega}(\vec{r})\cdot\vec{E}_{\Omega}(\vec{r})\right]\cdot\overleftrightarrow{G}_{\Omega}^{JE}(\vec{r},\vec{r}') \qquad \text{(C-7a)}\end{aligned}$$

Similarly to above relationship (C-7a), we, based on the symmetry of $\overleftrightarrow{\mu}_{\Omega}$ and $\overleftrightarrow{\mu}_{\Omega}^{-1}$, can also obtain the following relationships:

$$\left[\overleftrightarrow{\mu}_{\Omega}^{-1}(\vec{r})\cdot\vec{M}_{\Omega}(\vec{r})\right]\cdot\left[\overleftrightarrow{\mu}_{\Omega}(\vec{r})\cdot\overleftrightarrow{G}_{\Omega}^{JH}(\vec{r},\vec{r}')\right] = \vec{M}_{\Omega}(\vec{r})\cdot\overleftrightarrow{G}_{\Omega}^{JH}(\vec{r},\vec{r}') \qquad \text{(C-7b)}$$

$$\left\{\overleftrightarrow{\mu}_{\Omega}^{-1}(\vec{r})\cdot\left[\overleftrightarrow{\mu}_{\Omega}(\vec{r})\cdot\vec{H}_{\Omega}(\vec{r})\right]\right\}\times\overleftrightarrow{G}_{\Omega}^{JE}(\vec{r},\vec{r}') = \vec{H}_{\Omega}(\vec{r})\times\overleftrightarrow{G}_{\Omega}^{JE}(\vec{r},\vec{r}') \qquad \text{(C-7c)}$$

$$\vec{E}_{\Omega}(\vec{r})\times\left\{\overleftrightarrow{\mu}_{\Omega}^{-1}(\vec{r})\cdot\left[\overleftrightarrow{\mu}_{\Omega}(\vec{r})\cdot\overleftrightarrow{G}_{\Omega}^{JH}(\vec{r},\vec{r}')\right]\right\} = \vec{E}_{\Omega}(\vec{r})\times\overleftrightarrow{G}_{\Omega}^{JH}(\vec{r},\vec{r}') \qquad \text{(C-7d)}$$

Inserting relationships (C-6) and (C-7) into relationship (C-5), and utilizing relationship $\vec{E}_{\Omega}(\vec{r}')=\iiint_{\Omega}\vec{E}_{\Omega}(\vec{r})\cdot\overleftrightarrow{I}\delta(\vec{r}-\vec{r}')d\Omega$ (because $\vec{r},\vec{r}'\in\Omega$), we can obtain the following convolution integral formulation for electric field:

$$\begin{aligned}\vec{E}_{\Omega}(\vec{r}) = &\iiint_{\Omega}\vec{J}_{\Omega}(\vec{r}')\cdot\overleftrightarrow{G}_{\Omega}^{JE}(\vec{r}',\vec{r})dV'\\ &-\iiint_{\Omega}\vec{M}_{\Omega}(\vec{r}')\cdot\overleftrightarrow{G}_{\Omega}^{JH}(\vec{r}',\vec{r})dV'\\ &+\oiint_{\partial\Omega}\left[\hat{n}_{\Omega}^{-}\times\vec{H}_{\Omega}(\vec{r}')\right]\cdot\overleftrightarrow{G}_{\Omega}^{JE}(\vec{r}',\vec{r})dS'\\ &-\oiint_{\partial\Omega}\left[\vec{E}_{\Omega}(\vec{r}')\times\hat{n}_{\Omega}^{-}\right]\cdot\overleftrightarrow{G}_{\Omega}^{JH}(\vec{r}',\vec{r})dS' \quad , \quad \vec{r}\in\Omega \qquad \text{(C-8)}\end{aligned}$$

Here, the boundary integral represents the limitation during that integrated variable $\vec{r}'$ approaches boundary $\partial\Omega$. If we introduce the following boundary electric and magnetic currents:

$$\vec{J}_{\partial\Omega}(\vec{r}) = \hat{n}_{\Omega}^{-}(\vec{r})\times\left[\vec{H}_{\Omega}(\vec{r}_{\Omega})\right]_{\vec{r}_{\Omega}\to\vec{r}} \quad , \quad \vec{r}\in\partial\Omega \qquad \text{(C-9a)}$$

$$\vec{M}_{\partial\Omega}(\vec{r}) = \left[\vec{E}_{\Omega}(\vec{r}_{\Omega})\right]_{\vec{r}_{\Omega}\to\vec{r}}\times\hat{n}_{\Omega}^{-}(\vec{r}) \quad , \quad \vec{r}\in\partial\Omega \qquad \text{(C-9b)}$$

(where $\vec{r}_{\Omega}\in\Omega$ and $\vec{r}_{\Omega}$ approaches $\vec{r}$), then convolution integral formulation (C-8) can be rewritten as the following form:

$$\begin{aligned}\vec{E}_{\Omega}(\vec{r}) = &\iiint_{\Omega}\vec{J}_{\Omega}(\vec{r}')\cdot\overleftrightarrow{G}_{\Omega}^{JE}(\vec{r}',\vec{r})dV'\\ &-\iiint_{\Omega}\vec{M}_{\Omega}(\vec{r}')\cdot\overleftrightarrow{G}_{\Omega}^{JH}(\vec{r}',\vec{r})dV'\\ &+\oiint_{\partial\Omega}\vec{J}_{\partial\Omega}(\vec{r}')\cdot\overleftrightarrow{G}_{\Omega}^{JE}(\vec{r}',\vec{r})dS'\\ &-\oiint_{\partial\Omega}\vec{M}_{\partial\Omega}(\vec{r}')\cdot\overleftrightarrow{G}_{\Omega}^{JH}(\vec{r}',\vec{r})dS' \quad , \quad \vec{r}\in\Omega \qquad \text{(C-10)}\end{aligned}$$

Similarly, inserting $\vec{P}=\vec{H}_{\Omega}(\vec{r})$ and $\vec{Q}=\vec{G}_{\Omega}^{MH}(\vec{r},\vec{r}')$ and $\vec{\vartheta}=\vec{\varepsilon}_{\Omega}(\vec{r})$ into Green's second theorem (C-4), and doing some necessary mathematical operations, we can obtain the following convolution integral formulation for magnetic field:

$$\begin{aligned}\vec{H}_{\Omega}(\vec{r}) = &-\iiint_{\Omega}\vec{J}_{\Omega}(\vec{r}')\cdot\vec{G}_{\Omega}^{ME}(\vec{r}',\vec{r})dV'\\ &+\iiint_{\Omega}\vec{M}_{\Omega}(\vec{r}')\cdot\vec{G}_{\Omega}^{MH}(\vec{r}',\vec{r})dV'\\ &-\oiint_{\partial\Omega}\vec{J}_{\partial\Omega}(\vec{r}')\cdot\vec{G}_{\Omega}^{ME}(\vec{r}',\vec{r})dS'\\ &+\oiint_{\partial\Omega}\vec{M}_{\partial\Omega}(\vec{r}')\cdot\vec{G}_{\Omega}^{MH}(\vec{r}',\vec{r})dS' \quad , \quad \vec{r}\in\Omega \end{aligned} \tag{C-11}$$

At this point, we have obtained the convolution integral formulations for the EM fields distributing on an inhomogeneous anisotropic open domain.

In above convolution integral formulations (C-10) and (C-11), the dyadic Green's functions appear on the right sides of the sources. In engineering electromagnetics society, the Green's functions used in the convolution integral formulations for the EM fields distributing on a homogeneous isotropic open domain usually appear on the left sides of the sources. To effectively distinguish the above two different convolution integral forms, we call the case that the Green's functions appear on the left of the sources as standard convolution integral formulation (or the standard form of convolution integral formulation), and call the case that the Green's functions appear on the right of the sources as non-standard convolution integral formulation (or the non-standard form of convolution integral formulation). Then, it is natural to ask that: whether or not the convolution integral formulations for the EM field distributing on an inhomogeneous anisotropic open domain also have their standard forms similarly to the EM fields distributing on a homogeneous isotropic open domain? If the standard forms exist, how to transform above non-standard forms (C-10) and (C-11) to standard forms? In the following Appendix C3, we will focus on answering to these important questions?

## C3 Standard Convolution Integral Formulations for the EM Fields in an Inhomogeneous Anisotropic Open Domain

Based on equations (C-2b), magnetic-type unit vectorial point source $\hat{\xi}\delta(\vec{r}-\vec{r}'')$ and vectorial fields $\{\vec{G}_{\Omega;\bullet\xi}^{ME}(\vec{r},\vec{r}''),\vec{G}_{\Omega;\bullet\xi}^{MH}(\vec{r},\vec{r}'')\}$ satisfy the following Maxwell's equations:

$$\left.\begin{aligned}\nabla\times\vec{G}_{\Omega;\bullet\xi}^{MH}(\vec{r},\vec{r}'') &= j\omega\vec{\varepsilon}_{\Omega}(\vec{r})\cdot\vec{G}_{\Omega;\bullet\xi}^{ME}(\vec{r},\vec{r}'')\\ \nabla\times\vec{G}_{\Omega;\bullet\xi}^{ME}(\vec{r},\vec{r}'') &= -\hat{\xi}\delta(\vec{r}-\vec{r}'')-j\omega\vec{\mu}_{\Omega}(\vec{r})\cdot\vec{G}_{\Omega;\bullet\xi}^{MH}(\vec{r},\vec{r}'')\end{aligned}\right\} \quad , \quad (\vec{r},\vec{r}''\in\Omega) \tag{C-12}$$

In above equations (C-12), $\xi = x, y, z$ and $\vec{G}^{MF}_{\Omega;\bullet\xi}(\vec{r},\vec{r}'') = \hat{x}G^{MF}_{\Omega;x\xi}(\vec{r},\vec{r}'') + \hat{y}G^{MF}_{\Omega;y\xi}(\vec{r},\vec{r}'')$ $+\hat{z}G^{MF}_{\Omega;z\xi}(\vec{r},\vec{r}'')$, where $F = E, H$.

If we let the electric and magnetic currents in convolution integral formulation (C-10) satisfy that $\{\vec{J}_\Omega(\vec{r}), \vec{J}_{\partial\Omega}(\vec{r})\} = \{0\}$ and $\{\vec{M}_\Omega(\vec{r}), \vec{M}_{\partial\Omega}(\vec{r})\} = \{\hat{\xi}\delta(\vec{r} - \vec{r}'')\}$, then convolution integral formulation (C-10) will be simplified into the following form:

$$\begin{aligned}\vec{G}^{ME}_{\Omega;\bullet\xi}(\vec{r},\vec{r}'') &= -\int_{\Omega\bigcup\partial\Omega}\left[\hat{\xi}\delta(\vec{r}'-\vec{r}'')\right]\cdot\overleftrightarrow{G}^{JH}_{\Omega}(\vec{r}',\vec{r})\,d\Pi' \\ &= -\int_{\Omega\bigcup\partial\Omega}\delta(\vec{r}'-\vec{r}'')\vec{G}^{JH}_{\Omega;\xi\bullet}(\vec{r}',\vec{r})\,d\Pi' \\ &= -\vec{G}^{JH}_{\Omega;\xi\bullet}(\vec{r}'',\vec{r}) \qquad , \qquad (\vec{r},\vec{r}''\in\Omega\bigcup\partial\Omega)\end{aligned} \tag{C-13}$$

In above relationship (C-13), the second equality is based on the operational rule among vector and dyad[122], and $\vec{G}^{JH}_{\Omega;\xi\bullet}(\vec{r}',\vec{r}) = \hat{x}G^{JH}_{\Omega;\xi x}(\vec{r}',\vec{r}) + \hat{y}G^{JH}_{\Omega;\xi y}(\vec{r}',\vec{r}) + \hat{z}G^{JH}_{\Omega;\xi z}(\vec{r}',\vec{r})$; the third equality is based on the fact that $\vec{r},\vec{r}'' \in \Omega\bigcup\partial\Omega$ and the integral rule satisfied by Dirac's delta function[122]; the case $\vec{r},\vec{r}'' \in \partial\Omega$ in definition domain represents that position vectors $\vec{r}$ and $\vec{r}''$ can approach $\partial\Omega$. In fact, relationship (C-13) can be equivalently written as the following symmetry relationships:

$$\overleftrightarrow{G}^{ME}_{\Omega}(\vec{r},\vec{r}') = -\left[\overleftrightarrow{G}^{JH}_{\Omega}(\vec{r}',\vec{r})\right]^T \quad , \quad (\vec{r},\vec{r}'\in\Omega\bigcup\partial\Omega) \tag{C-14a}$$

$$\overleftrightarrow{G}^{JH}_{\Omega}(\vec{r},\vec{r}') = -\left[\overleftrightarrow{G}^{ME}_{\Omega}(\vec{r}',\vec{r})\right]^T \quad , \quad (\vec{r},\vec{r}'\in\Omega\bigcup\partial\Omega) \tag{C-14b}$$

In symmetry relationship (C-14), superscript "$T$" represents the transposition operation for dyad, and the rigorous definition for the transposition operation can be found in literatures [120,122]. Relationship (C-14) is just the symmetry relationship satisfied by dyadic Green's functions $\overleftrightarrow{G}^{ME}_{\Omega}(\vec{r},\vec{r}')$ and $\overleftrightarrow{G}^{JH}_{\Omega}(\vec{r},\vec{r}')$.

Similarly to generalized vector-dyad Green's second theorem (C-4), there also exists the following generalized vector-vector Green's second theorem:

$$\begin{aligned}&\iiint_\Omega\left(\vec{P}\cdot\left\{\nabla\times\left[\overleftrightarrow{\vartheta}^{-1}\cdot\left(\nabla\times\vec{Q}\right)\right]\right\}-\left\{\nabla\times\left[\overleftrightarrow{\vartheta}^{-1}\cdot\left(\nabla\times\vec{P}\right)\right]\right\}\cdot\vec{Q}\right)dV \\ &= \oiint_{\partial\Omega}\hat{n}^-_\Omega\cdot\left\{\vec{P}\times\left[\overleftrightarrow{\vartheta}^{-1}\cdot\left(\nabla\times\vec{Q}\right)\right]+\left[\overleftrightarrow{\vartheta}^{-1}\cdot\left(\nabla\times\vec{P}\right)\right]\times\vec{Q}\right\}dS\end{aligned} \tag{C-15}$$

If EM fields $\{\vec{E}^{\vec{J}^a_\Omega}_\Omega, \vec{H}^{\vec{J}^a_\Omega}_\Omega\}/\{\vec{E}^{\vec{J}^b_\Omega}_\Omega, \vec{H}^{\vec{J}^b_\Omega}_\Omega\}$ and the corresponding electric current $\vec{J}^a_\Omega / \vec{J}^b_\Omega$ satisfy the following Maxwell's equations:

$$\begin{aligned}\nabla\times\vec{H}^{\vec{J}^{a/b}_\Omega}_\Omega(\vec{r}) &= \vec{J}^{a/b}_\Omega(\vec{r}) + j\omega\overleftrightarrow{\varepsilon}_\Omega(\vec{r})\cdot\vec{E}^{\vec{J}^{a/b}_\Omega}_\Omega(\vec{r}) \\ \nabla\times\vec{E}^{\vec{J}^{a/b}_\Omega}_\Omega(\vec{r}) &= -j\omega\overleftrightarrow{\mu}_\Omega(\vec{r})\cdot\vec{H}^{\vec{J}^{a/b}_\Omega}_\Omega(\vec{r})\end{aligned} \quad , \quad \vec{r}\in\Omega \tag{C-16}$$

then the following relationship can be obtained by inserting $\vec{P} = \vec{E}^{\vec{J}^a_\Omega}_\Omega(\vec{r})$ and $\vec{Q} = \vec{E}_\Omega^{\vec{J}_\Omega^b}(\vec{r})$ and $\vec{\vartheta} = \vec{\mu}_\Omega(\vec{r})$ into formulation (C-15):

$$\begin{aligned}&\iiint_\Omega \vec{E}_\Omega^{\vec{J}_\Omega^a}(\vec{r}) \cdot \vec{J}_\Omega^b(\vec{r}) dV + \oiint_{\partial\Omega} \vec{E}_\Omega^{\vec{J}_\Omega^a}(\vec{r}) \cdot \left[ \hat{n}_\Omega^- \times \vec{H}_\Omega^{\vec{J}_\Omega^b}(\vec{r}) \right] dS \\ = &\iiint_\Omega \vec{J}_\Omega^a(\vec{r}) \cdot \vec{E}_\Omega^{\vec{J}_\Omega^b}(\vec{r}) dV + \oiint_{\partial\Omega} \left[ \hat{n}_\Omega^- \times \vec{H}_\Omega^{\vec{J}_\Omega^a}(\vec{r}) \right] \cdot \vec{E}_\Omega^{\vec{J}_\Omega^b}(\vec{r}) dS \qquad \text{(C-17)}\end{aligned}$$

Similarly to definition (C-9a), we introduce the following boundary electric currents:

$$\vec{J}_{\partial\Omega}^a(\vec{r}) = \hat{n}_\Omega^-(\vec{r}) \times \left[ \vec{H}_\Omega^{\vec{J}_\Omega^a}(\vec{r}_\Omega) \right]_{\vec{r}_\Omega \to \vec{r}} \quad , \quad \vec{r} \in \partial\Omega \qquad \text{(C-18a)}$$

$$\vec{J}_{\partial\Omega}^b(\vec{r}) = \hat{n}_\Omega^-(\vec{r}) \times \left[ \vec{H}_\Omega^{\vec{J}_\Omega^b}(\vec{r}_\Omega) \right]_{\vec{r}_\Omega \to \vec{r}} \quad , \quad \vec{r} \in \partial\Omega \qquad \text{(C-18b)}$$

and then above relationship (C-17) can be simplified into the following reciprocity theorem form:

$$\int_{\Omega\cup\partial\Omega} \vec{E}_\Omega^{\vec{J}_\Omega^a}(\vec{r}) \cdot \left[ \vec{J}_\Omega^b(\vec{r}) \oplus \vec{J}_{\partial\Omega}^b(\vec{r}) \right] d\Pi = \int_{\Omega\cup\partial\Omega} \left[ \vec{J}_\Omega^a(\vec{r}) \oplus \vec{J}_{\partial\Omega}^a(\vec{r}) \right] \cdot \vec{E}_\Omega^{\vec{J}_\Omega^b}(\vec{r}) d\Pi \quad \text{(C-19)}$$

Here, the reason to utilize symbol "$\oplus$" instead of symbol "$+$" is that $\vec{J}_\Omega^a / \vec{J}_\Omega^b$ and $\vec{J}_{\partial\Omega}^a / \vec{J}_{\partial\Omega}^b$ have different dimensions.

If we let $\{\vec{J}_\Omega^a(\vec{r}), \vec{J}_{\partial\Omega}^a(\vec{r})\} = \{\hat{\xi}\delta(\vec{r} - \vec{r}_a)\}$ and $\{\vec{J}_\Omega^b(\vec{r}), \vec{J}_{\partial\Omega}^b(\vec{r})\} = \{\hat{\zeta}\delta(\vec{r} - \vec{r}_b)\}$, equations (C-2a) and (C-16) imply that: $\vec{E}_\Omega^{\vec{J}_\Omega^a}(\vec{r}) = \vec{G}_{\Omega;\bullet\xi}^{JE}(\vec{r}, \vec{r}_a) = \hat{x}G_{\Omega;x\xi}^{JE}(\vec{r}, \vec{r}_a) + \hat{y}G_{\Omega;y\xi}^{JE}(\vec{r}, \vec{r}_a) + \hat{z}G_{\Omega;z\xi}^{JE}(\vec{r}, \vec{r}_a)$ and $\vec{E}_\Omega^{\vec{J}_\Omega^b}(\vec{r}) = \vec{G}_{\Omega;\bullet\zeta}^{JE}(\vec{r}, \vec{r}_b) = \hat{x}G_{\Omega;x\zeta}^{JE}(\vec{r}, \vec{r}_b) + \hat{y}G_{\Omega;y\zeta}^{JE}(\vec{r}, \vec{r}_b) + \hat{z}G_{\Omega;z\zeta}^{JE}(\vec{r}, \vec{r}_b)$, where $\xi, \zeta = x, y, z$. Inserting the above electric currents and electric fields into relationship (C-19), we immediately obtain the following relationship:

$$\begin{aligned}G_{\Omega;\zeta\xi}^{JE}(\vec{r}_b, \vec{r}_a) &= \int_{\Omega\cup\partial\Omega} \vec{G}_{\Omega;\bullet\xi}^{JE}(\vec{r}, \vec{r}_a) \cdot \left[ \hat{\zeta}\delta(\vec{r} - \vec{r}_b) \right] d\Pi \\ &= \int_{\Omega\cup\partial\Omega} \left[ \hat{\xi}\delta(\vec{r} - \vec{r}_a) \right] \cdot \vec{G}_{\Omega;\bullet\zeta}^{JE}(\vec{r}, \vec{r}_b) d\Pi \\ &= G_{\Omega;\xi\zeta}^{JE}(\vec{r}_a, \vec{r}_b) \quad , \quad (\vec{r}_a, \vec{r}_b \in \Omega\cup\partial\Omega) \qquad \text{(C-20)}\end{aligned}$$

Here, the case $\vec{r}_a, \vec{r}_b \in \partial\Omega$ in definition domain represents that the position vectors $\vec{r}_a$ and $\vec{r}_b$ in open domain $\Omega$ can approach $\partial\Omega$. In fact, relationship (C-20) can be equivalently rewritten as the following symmetry relationship for $\overleftrightarrow{G}_\Omega^{JE}(\vec{r}, \vec{r}')$:

$$\overleftrightarrow{G}_\Omega^{JE}(\vec{r}, \vec{r}') = \left[ \overleftrightarrow{G}_\Omega^{JE}(\vec{r}', \vec{r}) \right]^T \quad , \quad (\vec{r}, \vec{r}' \in \Omega\cup\partial\Omega) \qquad \text{(C-21)}$$

where $\vec{r}, \vec{r}' \in \partial\Omega$ means that the $\vec{r}$ and $\vec{r}'$ in open domain $\Omega$ can approach $\partial\Omega$.

Similarly to symmetry relationship (C-21), we can also obtain the following symmetry relationship on $\overleftrightarrow{G}_\Omega^{MH}(\vec{r}, \vec{r}')$:

$$\overleftrightarrow{G}_\Omega^{MH}(\vec{r}, \vec{r}') = \left[ \overleftrightarrow{G}_\Omega^{MH}(\vec{r}', \vec{r}) \right]^T \quad , \quad (\vec{r}, \vec{r}' \in \Omega\cup\partial\Omega) \qquad \text{(C-22)}$$

Above relationships (C-21) and (C-22) are just the symmetries satisfied by Green's functions $\overset{\leftrightarrow}{G}{}_{\Omega}^{JE}(\vec{r},\vec{r}')$ and $\overset{\leftrightarrow}{G}{}_{\Omega}^{MH}(\vec{r},\vec{r}')$ respectively.

Based on the symmetry relationships (C-14), (C-21), and (C-22) obtained above, convolution integral formulations (C-10) and (C-11) can be equivalently rewritten as the following standard forms:

$$\begin{aligned}
\vec{E}_{\Omega}(\vec{r}) = \; & \iiint_{\Omega} \overset{\leftrightarrow}{G}{}_{\Omega}^{JE}(\vec{r},\vec{r}') \cdot \vec{J}_{\Omega}(\vec{r}') dV' \\
& + \iiint_{\Omega} \overset{\leftrightarrow}{G}{}_{\Omega}^{ME}(\vec{r},\vec{r}') \cdot \vec{M}_{\Omega}(\vec{r}') dV' \\
& + \oiint_{\partial\Omega} \overset{\leftrightarrow}{G}{}_{\Omega}^{JE}(\vec{r},\vec{r}') \cdot \left[ \hat{n}_{\Omega}^{-} \times \vec{H}_{\Omega}(\vec{r}') \right] dS' \\
& + \oiint_{\partial\Omega} \overset{\leftrightarrow}{G}{}_{\Omega}^{ME}(\vec{r},\vec{r}') \cdot \left[ \vec{E}_{\Omega}(\vec{r}') \times \hat{n}_{\Omega}^{-} \right] dS' \quad , \quad \vec{r} \in \Omega
\end{aligned} \tag{C-23}$$

$$\begin{aligned}
\vec{H}_{\Omega}(\vec{r}) = \; & \iiint_{\Omega} \overset{\leftrightarrow}{G}{}_{\Omega}^{JH}(\vec{r},\vec{r}') \cdot \vec{J}_{\Omega}(\vec{r}') dV' \\
& + \iiint_{\Omega} \overset{\leftrightarrow}{G}{}_{\Omega}^{MH}(\vec{r},\vec{r}') \cdot \vec{M}_{\Omega}(\vec{r}') dV' \\
& + \oiint_{\partial\Omega} \overset{\leftrightarrow}{G}{}_{\Omega}^{JH}(\vec{r},\vec{r}') \cdot \left[ \hat{n}_{\Omega}^{-} \times \vec{H}_{\Omega}(\vec{r}') \right] dS' \\
& + \oiint_{\partial\Omega} \overset{\leftrightarrow}{G}{}_{\Omega}^{MH}(\vec{r},\vec{r}') \cdot \left[ \vec{E}_{\Omega}(\vec{r}') \times \hat{n}_{\Omega}^{-} \right] dS' \quad , \quad \vec{r} \in \Omega
\end{aligned} \tag{C-24}$$

Obviously, above convolution integral formulations (C-23) and (C-24) can be uniformly written as the following form:

$$\begin{aligned}
\vec{F}_{\Omega}(\vec{r}) = \; & \iiint_{\Omega} \overset{\leftrightarrow}{G}{}_{\Omega}^{JF}(\vec{r},\vec{r}') \cdot \vec{J}_{\Omega}(\vec{r}') d\Omega' \\
& + \iiint_{\Omega} \overset{\leftrightarrow}{G}{}_{\Omega}^{MF}(\vec{r},\vec{r}') \cdot \vec{M}_{\Omega}(\vec{r}') d\Omega' \\
& + \oiint_{\partial\Omega} \overset{\leftrightarrow}{G}{}_{\Omega}^{JF}(\vec{r},\vec{r}') \cdot \left[ \hat{n}_{\Omega}^{-} \times \vec{H}_{\Omega}(\vec{r}') \right] dS' \\
& + \oiint_{\partial\Omega} \overset{\leftrightarrow}{G}{}_{\Omega}^{MF}(\vec{r},\vec{r}') \cdot \left[ \vec{E}_{\Omega}(\vec{r}') \times \hat{n}_{\Omega}^{-} \right] dS' \quad , \quad \vec{r} \in \Omega
\end{aligned} \tag{C-25}$$

where $F = E, H$. If we select the simplified representation method used in literature [106] to express the convolution integrals, convolution integral formulation (C-25) can be further simplified into the following more compact form:

$$\begin{aligned}
\Omega \; : \; \vec{F}_{\Omega} = \; & \left[ \overset{\leftrightarrow}{G}{}_{\Omega}^{JF} * \vec{J}_{\Omega} \right]_{\Omega} + \left[ \overset{\leftrightarrow}{G}{}_{\Omega}^{JF} * \left( \hat{n}_{\Omega}^{-} \times \vec{H}_{\Omega} \right) \right]_{\partial\Omega} \\
& + \left[ \overset{\leftrightarrow}{G}{}_{\Omega}^{MF} * \vec{M}_{\Omega} \right]_{\Omega} + \left[ \overset{\leftrightarrow}{G}{}_{\Omega}^{MF} * \left( \vec{E}_{\Omega} \times \hat{n}_{\Omega}^{-} \right) \right]_{\partial\Omega} \\
= \; & \left[ \overset{\leftrightarrow}{G}{}_{\Omega}^{JF} * \vec{J}_{\Omega} + \overset{\leftrightarrow}{G}{}_{\Omega}^{MF} * \vec{M}_{\Omega} \right]_{\Omega} \\
& + \left[ \overset{\leftrightarrow}{G}{}_{\Omega}^{JF} * \left( \hat{n}_{\Omega}^{-} \times \vec{H}_{\Omega} \right) + \overset{\leftrightarrow}{G}{}_{\Omega}^{MF} * \left( \vec{E}_{\Omega} \times \hat{n}_{\Omega}^{-} \right) \right]_{\partial\Omega}
\end{aligned} \tag{C-26}$$

where $F = E, H$. Obviously, all of the Green's functions used in the convolution integral formulations (C-23)~(C-26) for the EM fields distributing on an inhomogeneous anisotropic open domain appear on the right sides of the sources, and convolution integral formulations (C-23)~(C-26) have the similar manifestation forms to the convolution integral formulations for the EM fields distributing on a homogeneous isotropic open domain, i.e., formulations (C-23)~(C-26) are just the standard forms of the convolution integral formulations for the EM fields distributing on an inhomogeneous anisotropic open domain. In the following parts of this Appendix C and the main body of this dissertation, we will always select to use the standard forms, so we will not repeat the term “standard form”.

## C4 Convolution Integral Formulations for the EM Fields Corresponding to a Simply Connected Inhomogeneous Anisotropic Material Body

In this Appendix C4, we consider a simply connected inhomogeneous anisotropic material body $V_{\text{sim sys}}$ placed in an inhomogeneous anisotropic environment, where the subscript “$\text{sim sys}$” in $V_{\text{sim sys}}$ is to emphasize that whole scattering system is constituted by a simply connected material body only. The boundary $\partial V_{\text{sim sys}}$ of the material body divides whole three-dimensional Euclidean space $\mathbb{R}^3$ into two parts[124]: $\text{int}\,V_{\text{sim sys}}$ (the interior of the material body) and $\text{ext}\,V_{\text{sim sys}}$ (the exterior of the material body), as illustrated in Figure C-4. Obviously, both $\text{int}\,V_{\text{sim sys}}$ and $\text{ext}\,V_{\text{sim sys}}$ are open domains[124]. Under the action of incident field $\vec{F}^{\text{inc}}$ (where $\vec{F}^{\text{inc}} = \vec{F}_{\text{imp}} + \vec{F}_{\text{env}}$), some polarization volume electric currents and magnetization volume magnetic currents will be induced on $V_{\text{sim sys}}$, but there will not be any polarization or magnetization surface current on boundary $\partial V_{\text{sim sys}}$ (based on the conclusions obtained in Appendix A). We denote the above polarization volume electric current and magnetization volume magnetic current as $\vec{J}_{\text{sim}}^{\text{PV}}$ and $\vec{M}_{\text{sim}}^{\text{MV}}$ respectively. In addition, when $V_{\text{sim sys}}$ is lossy, there will also exist conduction volume electric current $\vec{J}_{\text{sim}}^{\text{CV}}$ on $V_{\text{sim sys}}$ [121], and there doesn’t exist any conduction surface electric current distributing on boundary $\partial V_{\text{sim sys}}$. To simplify the symbolic system of the following discussions, we treat $\vec{J}_{\text{sim}}^{\text{PV}}$ and $\vec{J}_{\text{sim}}^{\text{CV}}$ as a whole, and call the whole as scattered volume electric current, and denote the whole as $\vec{J}_{\text{sim}}^{\text{SV}}$, i.e., $\vec{J}_{\text{sim}}^{\text{SV}} = \vec{J}_{\text{sim}}^{\text{PV}} + \vec{J}_{\text{sim}}^{\text{CV}}$; similarly, we denote $\vec{M}_{\text{sim}}^{\text{MV}}$ as $\vec{M}_{\text{sim}}^{\text{SV}}$, i.e., $\vec{M}_{\text{sim}}^{\text{SV}} = \vec{M}_{\text{sim}}^{\text{MV}}$. The superscript “$\text{SV}$” used in $\vec{J}_{\text{sim}}^{\text{SV}}$ and $\vec{M}_{\text{sim}}^{\text{SV}}$ is just the abbreviation of term “scattered volume (current)”. Obviously, the sources $\{\vec{J}^{\text{inc}}, \vec{M}^{\text{inc}}\}$ used to generate field $\vec{F}^{\text{inc}}$ distribute on $\text{ext}\,V_{\text{sim sys}}$, and are zero on $\text{int}\,V_{\text{sim sys}}$ and $\partial V_{\text{sim sys}}$.

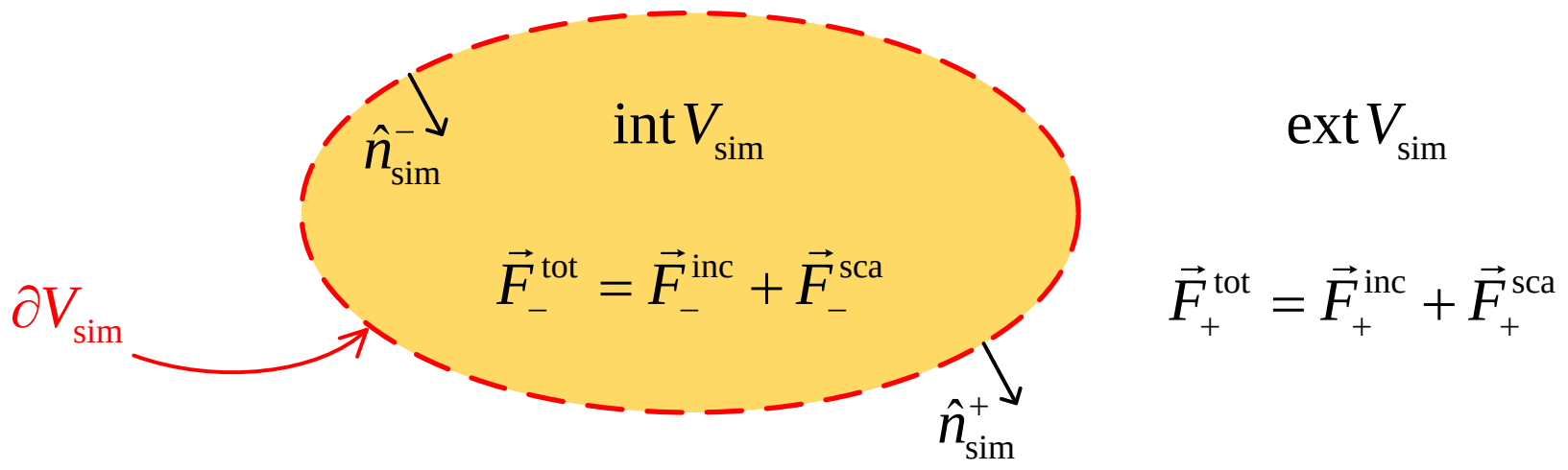

Figure C-4 The topological structure of a simply connected inhomogeneous anisotropic material system

Based on the conclusions obtained in Appendix A, it is known that incident field $\vec{F}^{\text{inc}}$ satisfies the following Maxwell's equations:

$$\begin{aligned} \nabla\times\vec{H}^{\text{inc}}(\vec{r}) &= \vec{J}^{\text{inc}}(\vec{r}) + j\omega\varepsilon_0\vec{E}^{\text{inc}}(\vec{r}) \\ \nabla\times\vec{E}^{\text{inc}}(\vec{r}) &= -\vec{M}^{\text{inc}}(\vec{r}) - j\omega\mu_0\vec{H}^{\text{inc}}(\vec{r}) \end{aligned} \quad , \quad \vec{r}\in\mathbb{R}^3 \tag{C-27}$$

In equations (C-27), $\vec{J}^{\text{inc}} = \vec{J}_{\text{env}} + \vec{J}_{\text{imp}}$ and $\vec{M}^{\text{inc}} = \vec{M}_{\text{env}}$ ; $\vec{J}_{\text{env}} = j\omega\Delta\overset{\leftrightarrow}{\varepsilon}_{\text{env}}^{\text{c}}\cdot\vec{E}^{\text{tot}}$ and $\vec{M}_{\text{env}} = j\omega\Delta\overset{\leftrightarrow}{\mu}_{\text{env}}\cdot\vec{H}^{\text{tot}}$ ; $\Delta\overset{\leftrightarrow}{\varepsilon}_{\text{env}}^{\text{c}} = \overset{\leftrightarrow}{\varepsilon}_{\text{env}}^{\text{c}} - \overset{\leftrightarrow}{I}\varepsilon_0$ and $\Delta\overset{\leftrightarrow}{\mu}_{\text{env}} = \overset{\leftrightarrow}{\mu}_{\text{env}} - \overset{\leftrightarrow}{I}\mu_0$ ; $\overset{\leftrightarrow}{\varepsilon}_{\text{env}}^{\text{c}} = \overset{\leftrightarrow}{\varepsilon}_{\text{env}} + \overset{\leftrightarrow}{\sigma}_{\text{env}}/j\omega$. Based on the conclusions obtained in Appendix A, it is known that scattered field $\vec{F}^{\text{sca}}$ satisfies the following Maxwell's equations:

$$\begin{aligned} \nabla\times\vec{H}^{\text{sca}}(\vec{r}) &= \vec{J}^{\text{SV}}(\vec{r}) + j\omega\varepsilon_0\vec{E}^{\text{sca}}(\vec{r}) \\ \nabla\times\vec{E}^{\text{sca}}(\vec{r}) &= -\vec{M}^{\text{SV}}(\vec{r}) - j\omega\mu_0\vec{H}^{\text{sca}}(\vec{r}) \end{aligned} \quad , \quad \vec{r}\in\mathbb{R}^3 \tag{C-28}$$

In equations (C-28), $\vec{J}_{\text{sim}}^{\text{SV}} = j\omega\Delta\overset{\leftrightarrow}{\varepsilon}_{\text{sim}}^{\text{c}}\cdot\vec{E}^{\text{tot}}$ and $\vec{M}_{\text{sim}}^{\text{SV}} = j\omega\Delta\overset{\leftrightarrow}{\mu}_{\text{sim}}\cdot\vec{H}^{\text{tot}}$; $\Delta\overset{\leftrightarrow}{\varepsilon}_{\text{sim}}^{\text{c}} = \overset{\leftrightarrow}{\varepsilon}_{\text{sim}}^{\text{c}} - \overset{\leftrightarrow}{I}\varepsilon_0$ and $\Delta\overset{\leftrightarrow}{\mu}_{\text{sim}} = \overset{\leftrightarrow}{\mu}_{\text{sim}} - \overset{\leftrightarrow}{I}\mu_0$ ; $\overset{\leftrightarrow}{\varepsilon}_{\text{sim}}^{\text{c}} = \overset{\leftrightarrow}{\varepsilon}_{\text{sim}} + \overset{\leftrightarrow}{\sigma}_{\text{sim}}/j\omega$. In open domain $\text{int}\,V_{\text{sim sys}}$, total field $\vec{F}^{\text{tot}}$ satisfies the following Maxwell's equations[121]:

$$\begin{aligned} \nabla\times\vec{H}^{\text{tot}}(\vec{r}) &= j\omega\overset{\leftrightarrow}{\varepsilon}_{\text{sim}}^{\text{c}}(\vec{r})\cdot\vec{E}^{\text{tot}}(\vec{r}) \\ \nabla\times\vec{E}^{\text{tot}}(\vec{r}) &= -j\omega\overset{\leftrightarrow}{\mu}_{\text{sim}}(\vec{r})\cdot\vec{H}^{\text{tot}}(\vec{r}) \end{aligned} \quad , \quad \vec{r}\in\text{int}\,V_{\text{sim sys}} \tag{C-29}$$

In what follows, we will, based on the above Maxwell's equations and the convolution integral formulation (C-26) obtained in Appendix C3, derive some important convolution integral formulations for the EM fields related to the simply connected material system shown in Figure C-4.

Inserting equations (C-27) into convolution integral formulation (C-26), and letting $\Omega=\mathbb{R}^3$, and utilizing the Sommerfeld's radiation conditions satisfied by $\vec{F}^{\text{inc}}$ and Green's functions, we can obtain the following convolution integral formulation for the $\vec{F}^{\text{inc}}$ on whole $\mathbb{R}^3$:

$$\mathbb{R}^3 \quad : \quad \vec{F}^{\text{inc}} = \left[\overset{\leftrightarrow}{G}_0^{JF} * \vec{J}^{\text{inc}} + \overset{\leftrightarrow}{G}_0^{MF} * \vec{M}^{\text{inc}}\right]_{\text{ext}\,V_{\text{sim sys}}} \tag{C-30}$$

In convolution integral formulation (C-30), $F = E, H$ ; subscript "0" represents that the corresponding Green's functions are the vacuum versions, and $\overleftrightarrow{G}_0^{JE}(\vec{r},\vec{r}') = (1/j\omega\varepsilon_0)(\nabla\nabla + k_0^2\overleftrightarrow{I})G_0(\vec{r},\vec{r}')$ and $\overleftrightarrow{G}_0^{JH}(\vec{r},\vec{r}') = \nabla\times\overleftrightarrow{I}G_0(\vec{r},\vec{r}')$ and $\overleftrightarrow{G}_0^{ME}(\vec{r},\vec{r}') = -\overleftrightarrow{G}_0^{JH}(\vec{r},\vec{r}')$ and $\overleftrightarrow{G}_0^{MH}(\vec{r},\vec{r}') = (\varepsilon_0/\mu_0)\overleftrightarrow{G}_0^{JE}(\vec{r},\vec{r}')$ [106,120], where $G_0(\vec{r},\vec{r}')$ is just the vacuum Green's function given in Chapter 3. Inserting equations (C-27) into convolution integral formulation (C-26), and letting $\Omega = \text{int}V_{\text{sim sys}}$, we can obtain the convolution integral formulation for the $\vec{F}^{\text{inc}}$ on $\text{int}V_{\text{sim sys}}$ as follows:

$$\begin{aligned}
\text{int}V_{\text{sim sys}} \quad : \quad \vec{F}^{\text{inc}} &= \left[\overleftrightarrow{G}_0^{JF} * \left(\hat{n}_{\text{sim}}^{-} \times \vec{H}_{-}^{\text{inc}}\right) + \overleftrightarrow{G}_0^{MF} * \left(\vec{E}_{-}^{\text{inc}} \times \hat{n}_{\text{sim}}^{-}\right)\right]_{\partial V_{\text{sim sys}}} \\
&= \left[\overleftrightarrow{G}_0^{JF} * \left(\hat{n}_{\text{sim}}^{-} \times \vec{H}^{\text{inc}}\right) + \overleftrightarrow{G}_0^{MF} * \left(\vec{E}^{\text{inc}} \times \hat{n}_{\text{sim}}^{-}\right)\right]_{\partial V_{\text{sim sys}}} \qquad \text{(C-31)}
\end{aligned}$$

In formulation (C-31), $F = E, H$ ; $\hat{n}_{\text{sim}}^{-}$ is the normal vector of $\partial V_{\text{sim sys}}$, and points to the interior of $V_{\text{sim sys}}$ ①, as illustrated in Figure C-4; the subscript "$-$" used in the fields in the RHS of the first equality is to emphasize that all of the fields distribute on the inner surface of $\partial V_{\text{sim sys}}$; the second equality is based on the continuity of the $\vec{F}^{\text{inc}}$ on $\partial V_{\text{sim sys}}$, because its sources $\{\vec{J}^{\text{inc}}, \vec{M}^{\text{inc}}\}$ don't distribute on $\partial V_{\text{sim sys}}$. Inserting formulation (C-27) into formulation (C-26), and letting $\Omega = \text{ext}V_{\text{sim sys}}$, and utilizing the Sommerfeld's radiation conditions satisfied by $\vec{F}^{\text{inc}}$ and various Green's functions, we can obtain the following convolution integral formulation for the $\vec{F}^{\text{inc}}$ on $\text{ext}V_{\text{sim sys}}$:

$$\begin{aligned}
\text{ext}V_{\text{sim sys}} \quad : \quad \vec{F}^{\text{inc}} = &\left[\overleftrightarrow{G}_0^{JF} * \vec{J}^{\text{inc}} + \overleftrightarrow{G}_0^{MF} * \vec{M}^{\text{inc}}\right]_{\text{ext}V_{\text{sim sys}}} \\
&+ \left[\overleftrightarrow{G}_0^{JF} * \left(\hat{n}_{\text{sim}}^{+} \times \vec{H}^{\text{inc}}\right) + \overleftrightarrow{G}_0^{MF} * \left(\vec{E}^{\text{inc}} \times \hat{n}_{\text{sim}}^{+}\right)\right]_{\partial V_{\text{sim sys}}} \qquad \text{(C-32)}
\end{aligned}$$

In convolution integral formulation (C-32), $F = E, H$ ; $\hat{n}_{\text{sim}}^{+}$ is the normal vector of $\partial V_{\text{sim sys}}$, and points to the exterior of $V_{\text{sim sys}}$ ②, as illustrated in Figure C-4. By comparing convolution integral formulation (C-30) with convolution integral formulation (C-32), it is immediately known that:

$$\begin{aligned}
\text{ext}V_{\text{sim sys}} \quad : \quad 0 &= \left[\overleftrightarrow{G}_0^{JF} * \left(\hat{n}_{\text{sim}}^{+} \times \vec{H}^{\text{inc}}\right) + \overleftrightarrow{G}_0^{MF} * \left(\vec{E}^{\text{inc}} \times \hat{n}_{\text{sim}}^{+}\right)\right]_{\partial V_{\text{sim sys}}} \\
&= -\left[\overleftrightarrow{G}_0^{JF} * \left(\hat{n}_{\text{sim}}^{-} \times \vec{H}^{\text{inc}}\right) + \overleftrightarrow{G}_0^{MF} * \left(\vec{E}^{\text{inc}} \times \hat{n}_{\text{sim}}^{-}\right)\right]_{\partial V_{\text{sim sys}}} \qquad \text{(C-33)}
\end{aligned}$$

where the second equality is based on the fact that relationship $\hat{n}_{\text{sim}}^{-} = -\hat{n}_{\text{sim}}^{+}$ holds on whole $\partial V_{\text{sim sys}}$.

① In what follows, we simply call $\hat{n}_{\text{sim}}^{-}$ as the inner normal direction of $\partial V_{\text{sim sys}}$.
② In what follows, we simply call $\hat{n}_{\text{sim}}^{+}$ as the outer normal direction of $\partial V_{\text{sim sys}}$.

Similarly to the above process to derive convolution integral formulations (C-30)~(C-33) from convolution integral formulation (C-26) and equations (C-27), we can, from convolution integral formulation (C-26) and equations (C-28), derive the convolution integral formulations for the scattered fields on various domains as follows:

$$\mathbb{R}^3 \quad : \quad \vec{F}^{\text{sca}} = \left[ \overleftrightarrow{G}_0^{JF} * \vec{J}_{\text{sim}}^{\text{SV}} + \overleftrightarrow{G}_0^{MF} * \vec{M}_{\text{sim}}^{\text{SV}} \right]_{\text{int}\,V_{\text{sim sys}}} \tag{C-34}$$

$$\text{ext}\,V_{\text{sim sys}} \quad : \quad \vec{F}^{\text{sca}} = \left[ \overleftrightarrow{G}_0^{JF} * \left( \hat{n}_{\text{sim}}^{+} \times \vec{H}^{\text{sca}} \right) + \overleftrightarrow{G}_0^{MF} * \left( \vec{E}^{\text{sca}} \times \hat{n}_{\text{sim}}^{+} \right) \right]_{\partial V_{\text{sim sys}}} \tag{C-35}$$

$$\text{int}\,V_{\text{sim sys}} \quad : \quad 0 = \left[ \overleftrightarrow{G}_0^{JF} * \left( \hat{n}_{\text{sim}}^{+} \times \vec{H}^{\text{sca}} \right) + \overleftrightarrow{G}_0^{MF} * \left( \vec{E}^{\text{sca}} \times \hat{n}_{\text{sim}}^{+} \right) \right]_{\partial V_{\text{sim sys}}} \tag{C-36}$$

where $F = E, H$. In the process to derive convolution integral formulations (C-35) and (C-36), we have utilized the tangential continuation condition of the $\vec{F}^{\text{sca}}$ on $\partial V_{\text{sim sys}}$, and the foundation to guarantee the tangential continuation condition is based on an important conclusion obtained in Appendix A: if polarization electric current model and magnetization magnetic current model are utilized to describe polarization phenomenon and magnetization phenomenon respectively, then there doesn't exist any scattered surface current on $\partial V_{\text{sim sys}}$.

In addition, the following convolution integral formulation for the total field $\vec{F}^{\text{tot}}$ on $\text{int}\,V_{\text{sim sys}}$ can be derived from formulations (C-26) and (C-29):

$$\text{int}\,V_{\text{sim sys}} \quad : \quad \vec{F}^{\text{tot}} = \left[ \overleftrightarrow{G}_{\text{sim}}^{JF} * \left( \hat{n}_{\text{sim}}^{-} \times \vec{H}^{\text{tot}} \right) + \overleftrightarrow{G}_{\text{sim}}^{MF} * \left( \vec{E}^{\text{tot}} \times \hat{n}_{\text{sim}}^{-} \right) \right]_{\partial V_{\text{sim sys}}} \tag{C-37}$$

In convolution integral formulation (C-37), $F = E, H$; the subscript "sim" used in the Green's functions is to emphasize that the Green's functions correspond to simply connected inhomogeneous anisotropic open domain $\text{int}\,V_{\text{sim sys}}$.

## C5 Mathematical Expressions of the Generalized Surface Equivalence Principle for a Simply Connected Inhomogeneous Anisotropic Material Body

In this Appendix C5, we consider the material system illustrated in Figure C-4, and the system is constituted by a single simply connected inhomogeneous anisotropic material body. Just like Appendix C4, we denote the material system as $V_{\text{sim sys}}$. Thus, $\vec{F}_{\text{mul}}^{\text{sca}} = 0$, and $\vec{F}^{\text{sca}} = \vec{F}_{\text{sim}}^{\text{sca}}$, and $\vec{F}^{\text{tot}} = \vec{F}^{\text{inc}} + \vec{F}^{\text{sca}} = \vec{F}^{\text{inc}} + \vec{F}_{\text{sim}}^{\text{sca}}$. Based on the results obtained in Appendix C4, this Appendix C5 generalizes traditional HFP, BET, and FHF, which are only suitable for a simply connected homogeneous isotropic material body placed in a homogeneous isotropic environment, to the generalized ones, which are applicable to the inhomogeneous anisotropic case.

**1) Generalized Huygens-Fresnel Principle and Generalized Backward Extinction Theorem**

In fact, the convolution integral formulations (C-31) and (C-33) obtained in Appendix C4 can be uniformly written as the following form:

$$\left.\begin{array}{lcc} \mathrm{ext}V_{\mathrm{sim\ sys}} & : & 0 \\ \mathrm{int}V_{\mathrm{sim\ sys}} & : & \vec{F}^{\mathrm{inc}} \end{array}\right\} = \left[\overleftrightarrow{G}_0^{JF} * \left(\hat{n}_{\mathrm{sim}}^{-} \times \vec{H}^{\mathrm{inc}}\right)\right]_{\partial V_{\mathrm{sim\ sys}}} + \left[\overleftrightarrow{G}_0^{MF} * \left(\vec{E}^{\mathrm{inc}} \times \hat{n}_{\mathrm{sim}}^{-}\right)\right]_{\partial V_{\mathrm{sim\ sys}}}$$

$$= \left[\overleftrightarrow{G}_0^{JF} * \left(\hat{n}_{\mathrm{sim}}^{-} \times \vec{H}^{\mathrm{inc}}\right) + \overleftrightarrow{G}_0^{MF} * \left(\vec{E}^{\mathrm{inc}} \times \hat{n}_{\mathrm{sim}}^{-}\right)\right]_{\partial V_{\mathrm{sim\ sys}}} \qquad \text{(C-38)}$$

where $F = E, H$. Similarly, the convolution integral formulations (C-35) and (C-36) obtained in Appendix C4 can be uniformly written as the following form:

$$\left.\begin{array}{lcc} \mathrm{ext}V_{\mathrm{sim\ sys}} & : & \vec{F}^{\mathrm{sca}} \\ \mathrm{int}V_{\mathrm{sim\ sys}} & : & 0 \end{array}\right\} = \left[\overleftrightarrow{G}_0^{JF} * \left(\hat{n}_{\mathrm{sim}}^{+} \times \vec{H}^{\mathrm{sca}}\right) + \overleftrightarrow{G}_0^{MF} * \left(\vec{E}^{\mathrm{sca}} \times \hat{n}_{\mathrm{sim}}^{+}\right)\right]_{\partial V_{\mathrm{sim\ sys}}} \qquad \text{(C-39)}$$

where $F = E, H$.

Above convolution integral formulations (C-38) and (C-39) point out that: second sources $\{\hat{n}_{\mathrm{sim}}^{-} \times \vec{H}^{\mathrm{inc}}, \vec{E}^{\mathrm{inc}} \times \hat{n}_{\mathrm{sim}}^{-}\}$ and $\{\hat{n}_{\mathrm{sim}}^{+} \times \vec{H}^{\mathrm{sca}}, \vec{E}^{\mathrm{sca}} \times \hat{n}_{\mathrm{sim}}^{+}\}$ will establish null fields in whole $\mathrm{ext}V_{\mathrm{sim\,sys}}$ and $\mathrm{int}V_{\mathrm{sim\,sys}}$ respectively, i.e., they will not generate any backward wave. Thus, this dissertation calls formulations (C-38) and (C-39) as generalized backward extinction theorem (GBET). In fact, formulations (C-38) and (C-39) are just the mathematical expressions for the generalized Huygens-Fresnel principles (GHFPs) used to describe the propagation characters of incident field $\vec{F}^{\mathrm{inc}}$ and scattered field $\vec{F}^{\mathrm{sca}}$. Here, the reason to use adjective "generalized" is that: the generalized versions are not only suitable for the simply connected homogeneous isotropic material bodies placed in homogeneous isotropic environment just like their traditional versions, but also applicable to the simply connected inhomogeneous anisotropic material bodies placed in inhomogeneous anisotropic environment.

**2) Generalized Franz-Harrington Formulation and Generalized Weak Backward Extinction Theorem**

Because $\vec{F}^{\mathrm{tot}} = \vec{F}^{\mathrm{inc}} + \vec{F}^{\mathrm{sca}}$, then the difference between convolution integral formulation (C-38) and convolution integral formulation (C-39) gives the following convolution integral formulation:

$$\left.\begin{aligned} \text{ext}V_{\text{sim sys}} &: -\vec{F}^{\text{sca}} \\ \text{int}V_{\text{sim sys}} &: \vec{F}^{\text{inc}} \end{aligned}\right\} = \left[\overleftrightarrow{G}_0^{JF} * \vec{J}_{\text{sim}}^{\text{ES}} + \overleftrightarrow{G}_0^{MF} * \vec{M}_{\text{sim}}^{\text{ES}}\right]_{\partial V_{\text{sim sys}}} \tag{C-40}$$

where $F = E, H$, and $\vec{J}_{\text{sim}}^{\text{ES}}$ and $\vec{M}_{\text{sim}}^{\text{ES}}$ are respectively defined as follows:

$$\vec{J}_{\text{sim}}^{\text{ES}}(\vec{r}) = \hat{n}_{\text{sim}}^{-}(\vec{r}) \times \left[\vec{H}^{\text{tot}}(\vec{r}_{\text{sim}})\right]_{\vec{r}_{\text{sim}} \to \vec{r}} \quad , \quad \vec{r} \in \partial V_{\text{sim}} \tag{C-41a}$$

$$\vec{M}_{\text{sim}}^{\text{ES}}(\vec{r}) = \left[\vec{E}^{\text{tot}}(\vec{r}_{\text{sim}})\right]_{\vec{r}_{\text{sim}} \to \vec{r}} \times \hat{n}_{\text{sim}}^{-}(\vec{r}) \quad , \quad \vec{r} \in \partial V_{\text{sim}} \tag{C-41b}$$

In definition (C-41), $\vec{r}_{\text{sim}} \in \text{int}V_{\text{sim sys}}$ and $\vec{r}_{\text{sim}}$ approaches $\vec{r}$; superscript "ES" is the abbreviation of "equivalent surface (current)". In fact, formulation (C-41) can also be utilized to simplify convolution integral formulation (C-37), and the simplified version of convolution integral formulation (C-37) is as follows:

$$\text{int}V_{\text{sim sys}} \quad : \quad \vec{F}^{\text{tot}} = \left[\overleftrightarrow{G}_{\text{sim}}^{JF} * \vec{J}_{\text{sim}}^{\text{ES}} + \overleftrightarrow{G}_{\text{sim}}^{MF} * \vec{M}_{\text{sim}}^{\text{ES}}\right]_{\partial V_{\text{sim sys}}} \tag{C-42}$$

where $F = E, H$. Based on convolution integral formulations (C-40) and (C-42) and superposition principle $\vec{F}^{\text{tot}} = \vec{F}^{\text{inc}} + \vec{F}^{\text{sca}}$, we can also obtain the convolution integral formulation for the $\vec{F}^{\text{sca}}$ on $\text{int}V_{\text{sim sys}}$ as follows:

$$\text{int}V_{\text{sim sys}} \quad : \quad \vec{F}^{\text{sca}} = \left[\Delta\overleftrightarrow{G}_{\text{sim}}^{JF} * \vec{J}_{\text{sim}}^{\text{ES}} + \Delta\overleftrightarrow{G}_{\text{sim}}^{MF} * \vec{M}_{\text{sim}}^{\text{ES}}\right]_{\partial V_{\text{sim sys}}} \tag{C-43}$$

where $F = E, H$, and $\Delta\overleftrightarrow{G}_{\text{sim}}^{JF}(\vec{r}, \vec{r}')$ and $\Delta\overleftrightarrow{G}_{\text{sim}}^{MF}(\vec{r}, \vec{r}')$ are defined as follows:

$$\Delta\overleftrightarrow{G}_{\text{sim}}^{JF}(\vec{r}, \vec{r}') = \overleftrightarrow{G}_{\text{sim}}^{JF}(\vec{r}, \vec{r}') - \overleftrightarrow{G}_0^{JF}(\vec{r}, \vec{r}') \quad , \quad \left(\vec{r}, \vec{r}' \in \text{int}V_{\text{sim sys}}\right) \tag{C-44a}$$

$$\Delta\overleftrightarrow{G}_{\text{sim}}^{MF}(\vec{r}, \vec{r}') = \overleftrightarrow{G}_{\text{sim}}^{MF}(\vec{r}, \vec{r}') - \overleftrightarrow{G}_0^{MF}(\vec{r}, \vec{r}') \quad , \quad \left(\vec{r}, \vec{r}' \in \text{int}V_{\text{sim sys}}\right) \tag{C-44b}$$

In convolution integral formulations (C-40), (C-42), and (C-43), internal incident field $\vec{F}_{-}^{\text{inc}}$, internal scattered field $\vec{F}_{-}^{\text{sca}}$, internal total field $\vec{F}_{-}^{\text{tot}}$, and external scattered field $\vec{F}_{+}^{\text{sca}}$ are expressed in terms of the convolution integrals of various Green's functions and an identical set of equivalent surface sources $\{\vec{J}_{\text{sim}}^{\text{ES}}, \vec{M}_{\text{sim}}^{\text{ES}}\}$. In this dissertation, convolution integral formulations (C-40), (C-42), and (C-43) are collectively referred to as the generalized Franz-Harrington formulation (GFHF) for the simply connected inhomogeneous anisotropic material bodies placed in inhomogeneous anisotropic environment. The reason to utilize the adjective "generalized" is that: the traditional Franz-Harrington formulation is only suitable for the simply connected homogeneous isotropic material bodies placed in homogeneous isotropic environment, and has only ability to formulate the internal total field $\vec{F}_{-}^{\text{tot}}$ and "external scattered field $\vec{F}_{+}^{\text{sca}}$" ①. In addition, the above GFHF for simply connected material bodies will be further

① Here, the reason to specially utilize a quotation mark is to emphasize that: when we utilize traditional FHF to generalized to multiply connected material bodies and non-connected material systems in the following Appendixes C6 and C7.

Sometimes, formulation is also viewed as a special case of backward extinction theorem. However, we want to emphasize that: formulation (C-42) indeed gives null tangential field on the outer surface of $\partial V_{\text{sim sys}}$, but it cannot guarantee null field in whole $\text{ext}\, V_{\text{sim sys}}$. In addition, the GHFP, GBET, and GFHF obtained above are collectively referred to as generalized surface equivalence principle (GSEP).

## C6 Mathematical Expressions of the Generalized Surface Equivalence Principle for a Multiply Connected Inhomogeneous Anisotropic Material Body

In Appendix C5, we generalize the traditional GHFP, GBET, and GFHF for the simply connected homogeneous isotropic material bodies placed in homogeneous isotropic environment to the ones for the simply connected inhomogeneous anisotropic material bodies $V_{\text{sim}}$ placed in inhomogeneous anisotropic environment. In this Appendix C6, we further generalize the results obtained in Appendix C5 to the multiply connected inhomogeneous anisotropic material bodies $V_{\text{mul}}$ placed in inhomogeneous anisotropic environment, and we restrict the $V_{\text{mul}}$ to being 2-connected (as illustrated in Figure C-5)①. Thus, we, in this Appendix C6, have that: $V_{\text{sim}} = \varnothing$, and $V_{\text{mat sys}} = V_{\text{mul}}$, and $\vec{F}^{\text{sca}} = \vec{F}^{\text{sca}}_{\text{mul}}$, and $\vec{F}^{\text{tot}} = \vec{F}^{\text{inc}} + \vec{F}^{\text{sca}} = \vec{F}^{\text{inc}} + \vec{F}^{\text{sca}}_{\text{mul}}$. To emphasize that $V_{\text{mat sys}} = V_{\text{mul}}$, this Appendix C6 denotes the material system as $V_{\text{mul sys}}$ specially.

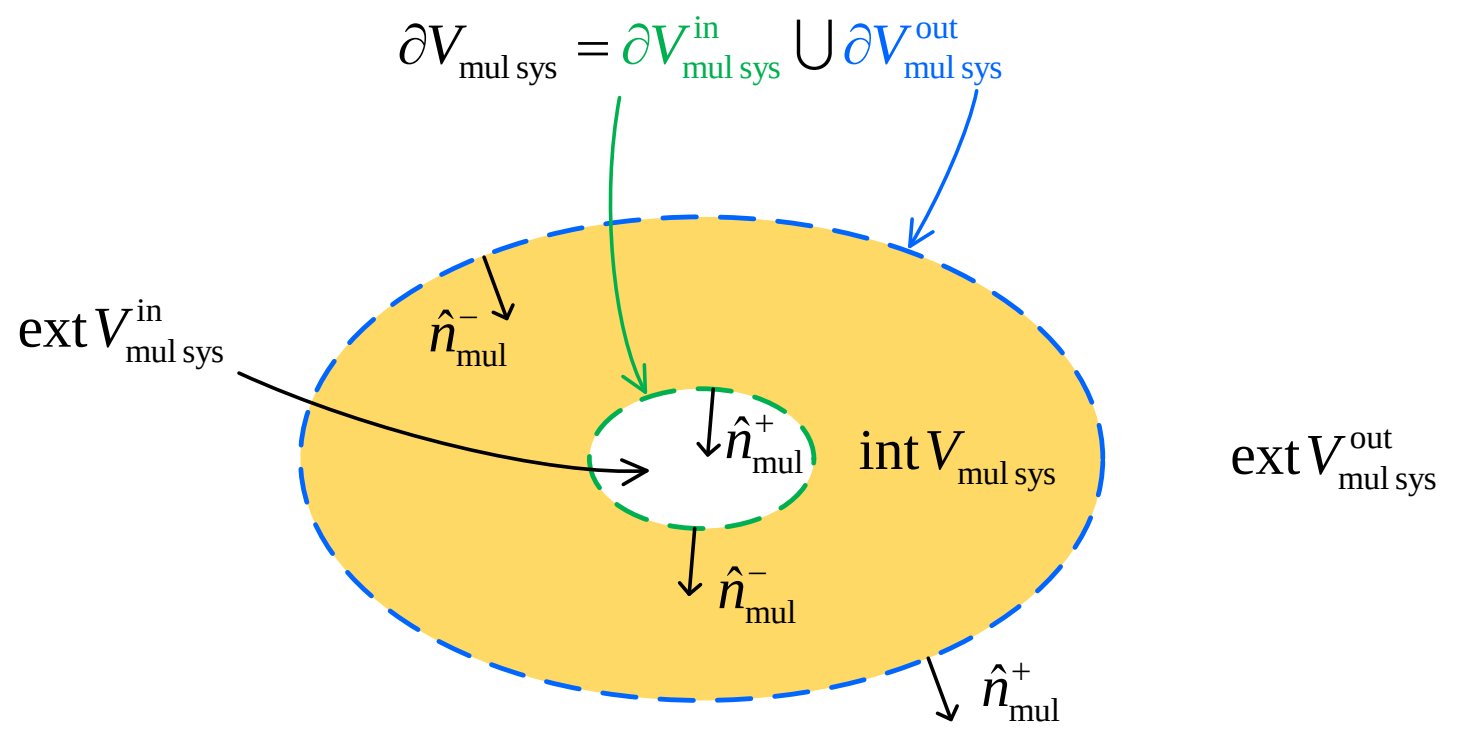

Figure C-5 The topological structure of a 2-connected (the simplest multiply connected[124]) material system

express external scattered field, if the Green's functions are selected as the non-vacuum version, then the obtained result may not be the correct external scattered field. The reason leading to this conclusion can be found in the subscript "0" of the Green's functions used in convolution integral formulation (C-40).

① The method to discuss an arbitrary $l(l > 2)$ -connected material system is completely similar to the method to discuss the 2-connected case, so it will not be repeated here.

Whole boundary $\partial V_{\text{mul sys}}$ can be divided into two parts $\partial V_{\text{mul sys}}^{\text{in}}$ and $\partial V_{\text{mul sys}}^{\text{out}}$, i.e., $\partial V_{\text{mul sys}} = \partial V_{\text{mul sys}}^{\text{in}} \bigcup \partial V_{\text{mul sys}}^{\text{out}}$. Whole $\text{ext} V_{\text{mul sys}}$ can also be divided into two parts $\text{ext} V_{\text{mul sys}}^{\text{in}}$ and $\text{ext} V_{\text{mul sys}}^{\text{out}}$, i.e., $\text{ext} V_{\text{mul sys}} = \text{ext} V_{\text{mul sys}}^{\text{in}} \bigcup \text{ext} V_{\text{mul sys}}^{\text{out}}$ [124]. Above $\partial V_{\text{mul sys}}^{\text{in}}$, $\partial V_{\text{mul sys}}^{\text{out}}$, $\text{ext} V_{\text{mul sys}}^{\text{in}}$, and $\text{ext} V_{\text{mul sys}}^{\text{out}}$ are illustrated in Figure C-5. In this Appendix C6, we restrict that $\{\vec{J}^{\text{inc}}, \vec{M}^{\text{inc}}\}$ distribute on $\text{ext} V_{\text{mul sys}}^{\text{out}}$. The case that $\{\vec{J}^{\text{inc}}, \vec{M}^{\text{inc}}\}$ distribute on $\text{ext} V_{\text{mul sys}}^{\text{in}}$ can be considered similarly, and the results corresponding to the case are similar to the results obtained in this Appendix C6 in the aspect of mathematical manifestation form. The case that $\{\vec{J}^{\text{inc}}, \vec{M}^{\text{inc}}\}$ simultaneously distribute on $\text{ext} V_{\text{mul sys}}^{\text{out}}$ and $\text{ext} V_{\text{mul sys}}^{\text{in}}$ can be viewed as the superposition of the two cases mentioned above, and its results are also similar to the results obtained in this Appendix C6 in the aspect of mathematical manifestation form.

**1) Generalized Huygens-Fresnel Principle and Generalized Backward Extinction Theorem**

Similarly to the method and process to derive the convolution integral formulations (C-38) and (C-39) for simply connected case, we can derive the following convolution integral formulations for incident field $\vec{F}^{\text{inc}}$ and scattered field $\vec{F}^{\text{sca}}$ by selecting the Huygens' surface as $\partial V_{\text{mul sys}}^{\text{in}}$ ①:

$$\left.\begin{array}{lcc} \text{ext} V_{\text{mul sys}}^{\text{out}} & : & 0 \\ \text{int} V_{\text{mul sys}} & : & 0 \\ \text{ext} V_{\text{mul sys}}^{\text{in}} & : & -\vec{F}^{\text{inc}} \end{array}\right\} = \left[ \overset{\leftrightarrow}{G}_0^{JF} * \left( \hat{n}_{\text{mul}}^{-} \times \vec{H}^{\text{inc}} \right) + \overset{\leftrightarrow}{G}_0^{MF} * \left( \vec{E}^{\text{inc}} \times \hat{n}_{\text{mul}}^{-} \right) \right]_{\partial V_{\text{mul sys}}^{\text{in}}} \tag{C-45a}$$

$$\left.\begin{array}{lcc} \text{ext} V_{\text{mul sys}}^{\text{out}} & : & 0 \\ \text{int} V_{\text{mul sys}} & : & 0 \\ \text{ext} V_{\text{mul sys}}^{\text{in}} & : & \vec{F}^{\text{sca}} \end{array}\right\} = \left[ \overset{\leftrightarrow}{G}_0^{JF} * \left( \hat{n}_{\text{mul}}^{+} \times \vec{H}^{\text{sca}} \right) + \overset{\leftrightarrow}{G}_0^{MF} * \left( \vec{E}^{\text{sca}} \times \hat{n}_{\text{mul}}^{+} \right) \right]_{\partial V_{\text{mul sys}}^{\text{in}}} \tag{C-46a}$$

or the following convolution integral formulations for incident field $\vec{F}^{\text{inc}}$ and scattered field $\vec{F}^{\text{sca}}$ by selecting the Huygens' surface as $\partial V_{\text{mul sys}}^{\text{out}}$ ②:

$$\left.\begin{array}{lcc} \text{ext} V_{\text{mul sys}}^{\text{out}} & : & 0 \\ \text{int} V_{\text{mul sys}} & : & \vec{F}^{\text{inc}} \\ \text{ext} V_{\text{mul sys}}^{\text{in}} & : & \vec{F}^{\text{inc}} \end{array}\right\} = \left[ \overset{\leftrightarrow}{G}_0^{JF} * \left( \hat{n}_{\text{mul}}^{-} \times \vec{H}^{\text{inc}} \right) + \overset{\leftrightarrow}{G}_0^{MF} * \left( \vec{E}^{\text{inc}} \times \hat{n}_{\text{mul}}^{-} \right) \right]_{\partial V_{\text{mul sys}}^{\text{out}}} \tag{C-45b}$$

---

① Because $\partial V_{\text{mul sys}}^{\text{in}} \bigcup S_\infty$ completely surrounds incident sources $\{\vec{J}^{\text{inc}}, \vec{M}^{\text{inc}}\}$ and scattered sources $\{\vec{J}_{\text{mul}}^{\text{SV}}, \vec{M}_{\text{mul}}^{\text{SV}}\}$, and incident field $\vec{F}^{\text{inc}}$ and scattered field $\vec{F}^{\text{sca}}$ satisfy Sommerfeld's radiation condition on $S_\infty$.

② Because $\partial V_{\text{mul sys}}^{\text{out}} \bigcup S_\infty$ completely surrounds incident sources $\{\vec{J}^{\text{inc}}, \vec{M}^{\text{inc}}\}$, and $\partial V_{\text{mul sys}}^{\text{out}}$ completely surrounds scattered sources $\{\vec{J}_{\text{mul}}^{\text{SV}}, \vec{M}_{\text{mul}}^{\text{SV}}\}$, and incident field $\vec{F}^{\text{inc}}$ satisfies Sommerfeld's radiation condition on $S_\infty$.

$$\left.\begin{array}{lcc} \text{ext}V_{\text{mul sys}}^{\text{out}} & : & \vec{F}^{\text{sca}} \\ \text{int}V_{\text{mul sys}} & : & 0 \\ \text{ext}V_{\text{mul sys}}^{\text{in}} & : & 0 \end{array}\right\} = \left[\overleftrightarrow{G}_0^{JF} * \left(\hat{n}_{\text{mul}}^{+} \times \vec{H}^{\text{sca}}\right) + \overleftrightarrow{G}_0^{MF} * \left(\vec{E}^{\text{sca}} \times \hat{n}_{\text{mul}}^{+}\right)\right]_{\partial V_{\text{mul sys}}^{\text{out}}} \quad \text{(C-46b)}$$

The unit normal vectors $\hat{n}_{\text{mul}}^{+}$ and $\hat{n}_{\text{mul}}^{-}$ used in convolution integral formulations (C-45) and (C-46) are illustrated in Figure C-5. The summation of convolution integral formulations (C-45a) and (C-45b) and the summation of convolution integral formulations (C-46a) and (C-46b) respectively give the following convolution integral formulations for $\vec{F}^{\text{inc}}$ and $\vec{F}^{\text{sca}}$:

$$\left.\begin{array}{lcc} \text{ext}V_{\text{mul sys}} & : & 0 \\ \text{int}V_{\text{mul sys}} & : & \vec{F}^{\text{inc}} \end{array}\right\} = \left[\overleftrightarrow{G}_0^{JF} * \left(\hat{n}_{\text{mul}}^{-} \times \vec{H}^{\text{inc}}\right) + \overleftrightarrow{G}_0^{MF} * \left(\vec{E}^{\text{inc}} \times \hat{n}_{\text{mul}}^{-}\right)\right]_{\partial V_{\text{mul sys}}} \quad \text{(C-47)}$$

$$\left.\begin{array}{lcc} \text{ext}V_{\text{mul sys}} & : & \vec{F}^{\text{sca}} \\ \text{int}V_{\text{mul sys}} & : & 0 \end{array}\right\} = \left[\overleftrightarrow{G}_0^{JF} * \left(\hat{n}_{\text{mul}}^{+} \times \vec{H}^{\text{sca}}\right) + \overleftrightarrow{G}_0^{MF} * \left(\vec{E}^{\text{sca}} \times \hat{n}_{\text{mul}}^{+}\right)\right]_{\partial V_{\text{mul sys}}} \quad \text{(C-48)}$$

In the process to derive convolution integral formulations (C-47) and (C-48) from convolution integral formulations (C-45) and (C-46), we have utilized relationships $\partial V_{\text{mul sys}} = \partial V_{\text{mul sys}}^{\text{in}} \bigcup \partial V_{\text{mul sys}}^{\text{out}}$ and $\text{ext}V_{\text{mul}} = \text{ext}V_{\text{mul sys}}^{\text{in}} \bigcup \text{ext}V_{\text{mul sys}}^{\text{out}}$ to simplify the symbolic system of this dissertation.

As the counterparts of convolution integral formulations (C-38) and (C-39), we respectively call above generalized backward extinction theorems (C-47) and (C-48) as the generalized Huygens-Fresnel principles for the $\vec{F}^{\text{inc}}$ and $\vec{F}^{\text{sca}}$ related to multiply connected material bodies. Here, the reason to use adjective “generalized” is that: the Huygens’ surface used in traditional Huygens-Fresnel principle and traditional backward extinction theorem is a single closed surface, but the Huygens’ surface used in the generalized version obtained above is constituted by two closed surfaces. In fact, we will, in Section 5.4, further generalize the concept of Huygens’ surface to the concept of Huygens’ boundary, and at that time the Huygens’ boundary contains not only some closed surfaces but also some open surfaces and even some line structures.

**2) Generalized Franz-Harrington Formulation and Generalized Weak Backward Extinction Theorem**

Similarly to the process to derive convolution integral formulation (C-40) from convolution integral formulations (C-38) and (C-39), it can be derived from convolution integral formulations (C-47) and (C-48) that

$$\left.\begin{array}{lll}\text{ext}\,V_{\text{mul sys}} & : & -\vec{F}^{\text{sca}} \\ \text{int}\,V_{\text{mul sys}} & : & \vec{F}^{\text{inc}}\end{array}\right\} = \left[\overleftrightarrow{G}_0^{JF} * \vec{J}_{\text{mul}}^{\text{ES}} + \overleftrightarrow{G}_0^{MF} * \vec{M}_{\text{mul}}^{\text{ES}}\right]_{\partial V_{\text{mul sys}}} \tag{C-49}$$

Here, $\vec{J}_{\text{mul}}^{\text{ES}}$ and $\vec{M}_{\text{mul}}^{\text{ES}}$ can be defined according to the manner used in formulation (C-41), and the only difference is that it needs to replace the subscript "sim" in formulation (C-41) with subscript "mul".

Similarly to the process to generalize the convolution integral formulation (C-40) in simply connected case to the convolution integral formulation (C-49) in multiply connected case, the convolution integral formulation (C-42) in simply connected case can also be generalized to multiply connected case as follows:

$$\text{int}\,V_{\text{mul sys}} \quad : \quad \vec{F}^{\text{tot}} = \left[\overleftrightarrow{G}_{\text{mul}}^{JF} * \vec{J}_{\text{mul}}^{\text{ES}} + \overleftrightarrow{G}_{\text{mul}}^{MF} * \vec{M}_{\text{mul}}^{\text{ES}}\right]_{\partial V_{\text{mul sys}}} \tag{C-50}$$

where $\overleftrightarrow{G}_{\text{mul}}^{JF}(\vec{r},\vec{r}\,')$ and $\overleftrightarrow{G}_{\text{mul}}^{MF}(\vec{r},\vec{r}\,')$ are the dyadic Green's functions corresponding to multiply connected open domain $\text{int}\,V_{\text{mul sys}}$ , and then the convolution integral formulation (C-43) in simply connected case can be generalized to multiply connected case as follows:

$$\text{int}\,V_{\text{mul sys}} \quad : \quad \vec{F}^{\text{sca}} = \left[\Delta\overleftrightarrow{G}_{\text{mul}}^{JF} * \vec{J}_{\text{mul}}^{\text{ES}} + \Delta\overleftrightarrow{G}_{\text{mul}}^{MF} * \vec{M}_{\text{mul}}^{\text{ES}}\right]_{\partial V_{\text{mul sys}}} \tag{C-51}$$

Here, the definitions for $\Delta\overleftrightarrow{G}_{\text{mul}}^{JF}(\vec{r},\vec{r}\,')$ and $\Delta\overleftrightarrow{G}_{\text{mul}}^{MF}(\vec{r},\vec{r}\,')$ are similar to the ones defined in formulation (C-44) except replacing the subscript "sim" used in formulation (C-44) with subscript "mul".

## C7 Mathematical Expressions of the Generalized Surface Equivalence Principle for a Multi-body Inhomogeneous Anisotropic Material System

Previous Appendix C5 and Appendix C6 focus on the system which is constituted by a single material body, but this Appendix C7 considers the system which is constituted by two material bodies, and the system constituted by $l(l>2)$ material bodies can be discussed by employing the method used in this Appendix C7.

We exhibit some typical two-body material systems $V_{\text{mat sys}}$ in Figure C-6, and all of the $V_{\text{mat sys}}$ contain a simply connected material body $V_{\text{sim}}$ and a multiply connected material body $V_{\text{mul}}$, i.e., $V_{\text{mat sys}} = V_{\text{sim}} \cup V_{\text{mul}}$. Thus, this Appendix C7 denotes the material systems as $V_{\text{sm sys}}$ correspondingly. The systems constituted by two simply connected material bodies or two multiply connected material bodies can be similarly discussed by using the method used in this Appendix C7. In the following parts of this Appendix C7, we decompose the boundaries of material bodies $V_{\text{sim}}$ and $V_{\text{mul}}$ firstly, and then decompose the equivalent surface currents on the boundaries correspondingly, and generalize the results obtained in Appendixes C5 and C6 to the two-body system $V_{\text{sm sys}}$ finally.

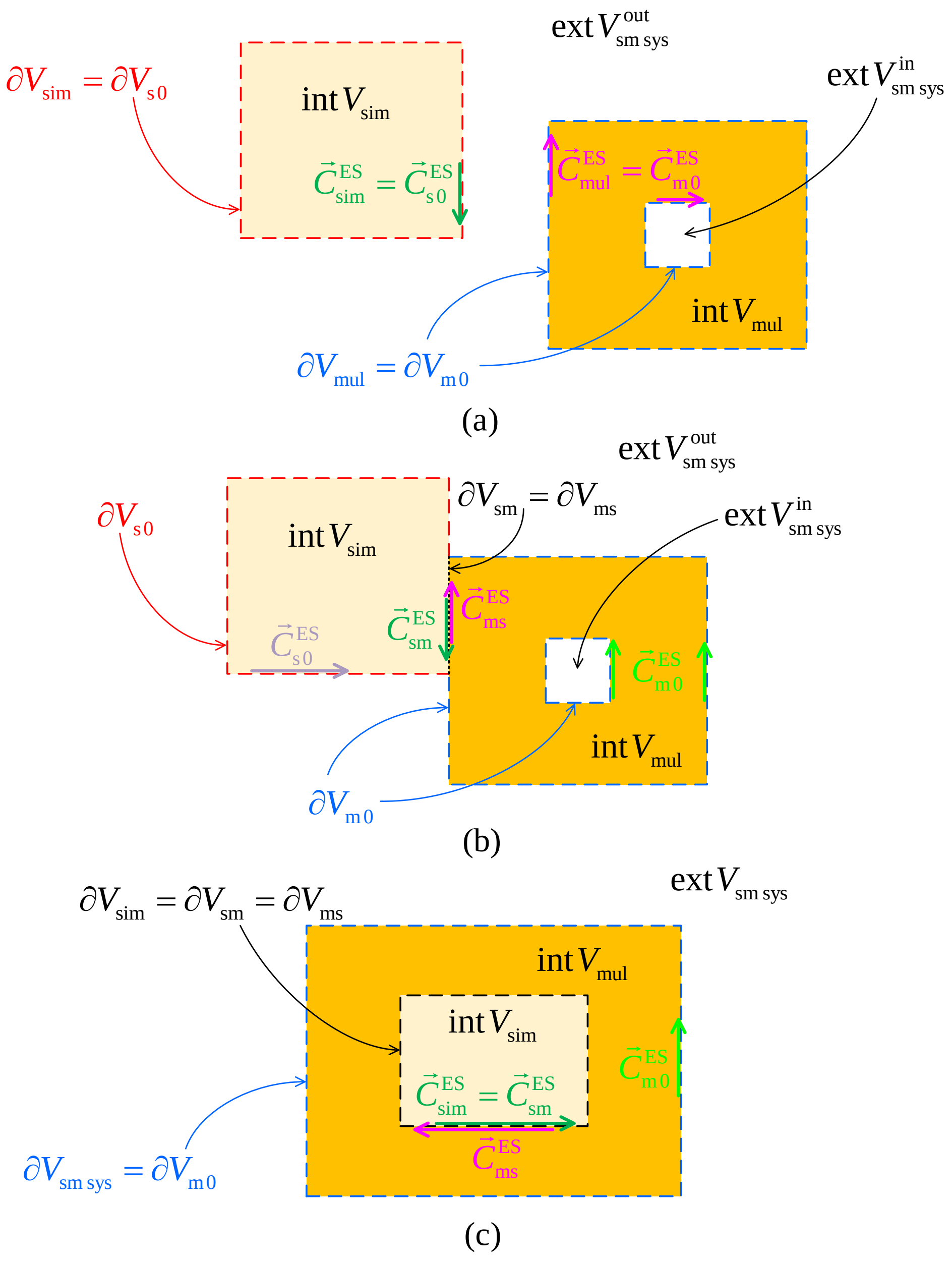

Figure C-6 The topological structure of a 2-body material system constituted by a simply connected material body $V_{\text{sim}}$ and a multiply connected material body $V_{\text{mul}}$ and the related equivalent surface sources. (a) the case that $V_{\text{sim}}$ and $V_{\text{mul}}$ don't contact with each other; (b) the case that $V_{\text{sim}}$ and $V_{\text{mul}}$ contact with each other, where $V_{\text{sim}}$ is not surrounded by $V_{\text{mul}}$; (c) the case that $V_{\text{sim}}$ and $V_{\text{mul}}$ contact with each other, where $V_{\text{sim}}$ is completely submerged into $V_{\text{mul}}$

**1) Boundary Decompositions**

If the boundaries of $V_{\text{sim}}$ and $V_{\text{mul}}$ are denoted as $\partial V_{\text{sim}}$ and $\partial V_{\text{mul}}$ respectively, then $\partial V_{\text{sim}}$ and $\partial V_{\text{mul}}$ can be decomposed as follows:

$$\partial V_{\text{sim/mul}} = \partial V_{\text{s0/m0}} \bigcup \partial V_{\text{sm/ms}} \tag{C-52}$$

where $\partial V_{\text{s0/m0}}$ and $\partial V_{\text{sm/ms}}$ are respectively defined as follows:

$$\partial V_{\text{s0/m0}} = \text{cl}\left(\partial V_{\text{sim/mul}} \setminus \partial V_{\text{mul/sim}}\right) \tag{C-53a}$$

$$\partial V_{\text{sm/ms}} = \partial V_{\text{sim/mul}} \setminus \partial V_{\text{s0/m0}} \tag{C-53b}$$

and the various sub-boundaries are illustrated in Figure C-6. Obviously, $\partial V_{\text{sm}} = \partial V_{\text{ms}}$, and $\partial V_{\text{s0/m0}}$ doesn't intersect with both $\partial V_{\text{sm}}$ and $\partial V_{\text{ms}}$, i.e.,

$$\partial V_{\text{s0/m0}} \bigcap \partial V_{\text{sm}} = \varnothing = \partial V_{\text{ms}} \bigcap \partial V_{\text{m0/s0}} \tag{C-54}$$

Relationship (C-54) implies that (C-52) is a completely decoupled decomposition for $\partial V_{\text{sim}}$ and $\partial V_{\text{mul}}$.

**2) Equivalence Surface Current Decompositions**

Based on decomposition (C-52) and relationship (C-54), we can decompose the equivalent surface currents distributing on $\partial V_{\text{sim}}$ and $\partial V_{\text{mul}}$ as follows:

$$\vec{C}_{\text{sim/mul}}^{\text{ES}}\left(\vec{r}\right) = \vec{C}_{\text{s0/m0}}^{\text{ES}}\left(\vec{r}\right) + \vec{C}_{\text{sm/ms}}^{\text{ES}}\left(\vec{r}\right) \quad , \quad \vec{r} \in \partial V_{\text{sim/mul}} \tag{C-55}$$

where $C = J, M$, and

$$\vec{C}_{\text{s0/m0}}^{\text{ES}}\left(\vec{r}\right) = \begin{cases} \vec{C}_{\text{sim/mul}}^{\text{ES}}\left(\vec{r}\right) & , \quad \vec{r} \in \partial V_{\text{s0/m0}} \\ 0 & , \quad \vec{r} \in \partial V_{\text{sm/ms}} \end{cases} \tag{C-56a}$$

$$\vec{C}_{\text{sm/ms}}^{\text{ES}}\left(\vec{r}\right) = \begin{cases} 0 & , \quad \vec{r} \in \partial V_{\text{s0/m0}} \\ \vec{C}_{\text{sim/mul}}^{\text{ES}}\left(\vec{r}\right) & , \quad \vec{r} \in \partial V_{\text{sm/ms}} \end{cases} \tag{C-56b}$$

Here, $\vec{C}_{\text{sim/mul}}^{\text{ES}}$ can be defined similarly to the way used in formulation (C-41) except replacing the $\partial V_{\text{sim}}$ in formulation (C-41) with $\partial V_{\text{sim/mul}}$. Because there doesn't exist any scattered surface current distributing on $\partial V_{\text{sim}}$ and $\partial V_{\text{mul}}$ (for details see Appendix A), then the tangential components of $\vec{F}^{\text{tot}}$ are continuous on $\partial V_{\text{sm/ms}}$, so there exists the following relationship:

$$\vec{C}_{\text{sm}}^{\text{ES}}\left(\vec{r}\right) = -\vec{C}_{\text{ms}}^{\text{ES}}\left(\vec{r}\right) \quad , \quad \vec{r} \in \partial V_{\text{sm}} = \partial V_{\text{ms}} \tag{C-57}$$

**3) Generalized Huygens-Fresnel Principle, Generalized Backward Extinction Theorem, and Generalized Franz-Harrington Formulation —— General Case**

Now, we further generalize the results obtained in Appendixes C5 and C6 to a general two-body material system, and in 4) we provide the special manifestation forms of the generalized results corresponding to some special two-body material systems.

**(3.1) Generalized Huygens-Fresnel Principle and Generalized Backward Extinction Theorem**

Based on convolution integral formulation (C-38) and convolution integral formulation (C-39), we have

$$\left.\begin{array}{lcl} \text{ext}V_{\text{sm sys}} & : & 0 \\ \text{int}V_{\text{sim}} & : & \vec{F}^{\text{inc}} \\ \text{int}V_{\text{mul}} & : & 0 \end{array}\right\} = \left[\overleftrightarrow{G}_0^{JF} * \left(\hat{n}_{\text{sim}}^{-} \times \vec{H}^{\text{inc}}\right) + \overleftrightarrow{G}_0^{MF} * \left(\vec{E}^{\text{inc}} \times \hat{n}_{\text{sim}}^{-}\right)\right]_{\partial V_{\text{sim}}} \quad \text{(C-58a)}$$

$$\left.\begin{array}{lcl} \text{ext}V_{\text{sm sys}} & : & \vec{F}_{\text{sim}}^{\text{sca}} \\ \text{int}V_{\text{sim}} & : & 0 \\ \text{int}V_{\text{mul}} & : & \vec{F}_{\text{sim}}^{\text{sca}} \end{array}\right\} = \left[\overleftrightarrow{G}_0^{JF} * \left(\hat{n}_{\text{sim}}^{+} \times \vec{H}_{\text{sim}}^{\text{sca}}\right) + \overleftrightarrow{G}_0^{MF} * \left(\vec{E}_{\text{sim}}^{\text{sca}} \times \hat{n}_{\text{sim}}^{+}\right)\right]_{\partial V_{\text{sim}}} \quad \text{(C-59a)}$$

In convolution integral formulation (C-58a) and convolution integral formulation (C-59a), $\text{ext}V_{\text{sm sys}}$ is the exterior of whole system $V_{\text{sm sys}}$, as illustrated in Figure C-6; $\hat{n}_{\text{sim}}^{-}$ and $\hat{n}_{\text{sim}}^{+}$ are the normal vectors of $\partial V_{\text{sim}}$, and they point to the interior and exterior of $V_{\text{sim}}$ respectively. Based on convolution integral formulation (C-47) and convolution integral formulation (C-48), we have

$$\left.\begin{array}{lcl} \text{ext}V_{\text{sm sys}} & : & 0 \\ \text{int}V_{\text{sim}} & : & 0 \\ \text{int}V_{\text{mul}} & : & \vec{F}^{\text{inc}} \end{array}\right\} = \left[\overleftrightarrow{G}_0^{JF} * \left(\hat{n}_{\text{mul}}^{-} \times \vec{H}^{\text{inc}}\right) + \overleftrightarrow{G}_0^{MF} * \left(\vec{E}^{\text{inc}} \times \hat{n}_{\text{mul}}^{-}\right)\right]_{\partial V_{\text{mul}}} \quad \text{(C-58b)}$$

$$\left.\begin{array}{lcl} \text{ext}V_{\text{sm sys}} & : & \vec{F}_{\text{mul}}^{\text{sca}} \\ \text{int}V_{\text{sim}} & : & \vec{F}_{\text{mul}}^{\text{sca}} \\ \text{int}V_{\text{mul}} & : & 0 \end{array}\right\} = \left[\overleftrightarrow{G}_0^{JF} * \left(\hat{n}_{\text{mul}}^{+} \times \vec{H}_{\text{mul}}^{\text{sca}}\right) + \overleftrightarrow{G}_0^{MF} * \left(\vec{E}_{\text{mul}}^{\text{sca}} \times \hat{n}_{\text{mul}}^{+}\right)\right]_{\partial V_{\text{mul}}} \quad \text{(C-59b)}$$

In convolution integral formulation (C-58b) and convolution integral formulation (C-59b), $\hat{n}_{\text{mul}}^{-}$ and $\hat{n}_{\text{mul}}^{+}$ are the normal vectors of $\partial V_{\text{mul}}$, and they point to the interior and exterior of $V_{\text{mul}}$ respectively.

In addition, similarly to deriving convolution integral formulation (C-38) and convolution integral formulation (C-47), we can derive the following convolution integral formulations for $\vec{F}_{\text{mul}}^{\text{sca}}$ and $\vec{F}_{\text{sim}}^{\text{sca}}$:

$$\left.\begin{array}{lcl} \text{ext}V_{\text{sm sys}} & : & 0 \\ \text{int}V_{\text{sim}} & : & -\vec{F}_{\text{mul}}^{\text{sca}} \\ \text{int}V_{\text{mul}} & : & 0 \end{array}\right\} = \left[\overleftrightarrow{G}_0^{JF} * \left(\hat{n}_{\text{sim}}^{+} \times \vec{H}_{\text{mul}}^{\text{sca}}\right) + \overleftrightarrow{G}_0^{MF} * \left(\vec{E}_{\text{mul}}^{\text{sca}} \times \hat{n}_{\text{sim}}^{+}\right)\right]_{\partial V_{\text{sim}}} \quad \text{(C-59c)}$$

$$\left.\begin{array}{lcc} \text{ext}V_{\text{sm sys}} & : & 0 \\ \text{int}V_{\text{sim}} & : & 0 \\ \text{int}V_{\text{mul}} & : & -\vec{F}^{\text{sca}}_{\text{sim}} \end{array}\right\} = \left[\overleftrightarrow{G}^{JF}_0 * \left(\hat{n}^+_{\text{mul}} \times \vec{H}^{\text{sca}}_{\text{sim}}\right) + \overleftrightarrow{G}^{MF}_0 * \left(\vec{E}^{\text{sca}}_{\text{sim}} \times \hat{n}^+_{\text{mul}}\right)\right]_{\partial V_{\text{mul}}} \quad \text{(C-59d)}$$

Summing convolution integral formulation (C-58a) and convolution integral formulation (C-58b), it is obtained that

$$\left.\begin{array}{lcc} \text{ext}V_{\text{sm sys}} & : & 0 \\ \text{int}V_{\text{sim}} & : & \vec{F}^{\text{inc}} \\ \text{int}V_{\text{mul}} & : & \vec{F}^{\text{inc}} \end{array}\right\} = \begin{array}{l} \left[\overleftrightarrow{G}^{JF}_0 * \left(\hat{n}^-_{\text{sim}} \times \vec{H}^{\text{inc}}\right) + \overleftrightarrow{G}^{MF}_0 * \left(\vec{E}^{\text{inc}} \times \hat{n}^-_{\text{sim}}\right)\right]_{\partial V_{\text{sim}}} \\ + \left[\overleftrightarrow{G}^{JF}_0 * \left(\hat{n}^-_{\text{mul}} \times \vec{H}^{\text{inc}}\right) + \overleftrightarrow{G}^{MF}_0 * \left(\vec{E}^{\text{inc}} \times \hat{n}^-_{\text{mul}}\right)\right]_{\partial V_{\text{mul}}} \end{array} \quad \text{(C-60)}$$

Summing convolution integral formulations (C-59a)~(C-59d), it is obtained that

$$\left.\begin{array}{lcc} \text{ext}V_{\text{sm sys}} & : & \vec{F}^{\text{sca}} \\ \text{int}V_{\text{sim}} & : & 0 \\ \text{int}V_{\text{mul}} & : & 0 \end{array}\right\} = \begin{array}{l} \left[\overleftrightarrow{G}^{JF}_0 * \left(\hat{n}^+_{\text{sim}} \times \vec{H}^{\text{sca}}\right) + \overleftrightarrow{G}^{MF}_0 * \left(\vec{E}^{\text{sca}} \times \hat{n}^+_{\text{sim}}\right)\right]_{\partial V_{\text{sim}}} \\ + \left[\overleftrightarrow{G}^{JF}_0 * \left(\hat{n}^+_{\text{mul}} \times \vec{H}^{\text{sca}}\right) + \overleftrightarrow{G}^{MF}_0 * \left(\vec{E}^{\text{sca}} \times \hat{n}^+_{\text{mul}}\right)\right]_{\partial V_{\text{mul}}} \end{array} \quad \text{(C-61)}$$

We call above convolution integral formulation (C-60) and convolution integral formulation (C-61) as the topological additivity satisfied by GHFP and GBET, i.e., the GHFP/GBET for whole system equals the summation of the GHFPs/GBETs for all of the sub-systems. The topological additivity can be mathematically expressed as follows:

$$\begin{aligned} &\text{Incident field GHFP/GBET of whole system} \\ =\ &\textstyle\sum_{\xi} \text{Incident field GHFP/GBET of sub-system } V_{\xi} \end{aligned} \quad \text{(C-62)}$$

$$\begin{aligned} &\text{Scattered field GHFP/GBET of whole system} \\ =\ &\textstyle\sum_{\xi} \text{Scattered field GHFP/GBET of sub-system } V_{\xi} \end{aligned} \quad \text{(C-63)}$$

Obviously, the topological additivity is consistent with superposition principle.

**(3.2) Generalized Franz-Harrington Formulation**

Similarly to deriving convolution integral formulation (C-40) from convolution integral formulations (C-38) and (C-39) and deriving convolution integral formulation (C-49) from convolution integral formulations (C-47) and (C-48), we can derive the following convolution integral formulation from convolution integral formulations (C-60) and (C-61):

$$\left.\begin{array}{lcc} \text{ext}V_{\text{sm sys}} & : & -\vec{F}^{\text{sca}} \\ \text{int}V_{\text{sim}} & : & \vec{F}^{\text{inc}} \\ \text{int}V_{\text{mul}} & : & \vec{F}^{\text{inc}} \end{array}\right\} \begin{array}{l} = \left[\overleftrightarrow{G}^{JF}_0 * \vec{J}^{\text{ES}}_{\text{sim}} + \overleftrightarrow{G}^{MF}_0 * \vec{M}^{\text{ES}}_{\text{sim}}\right]_{\partial V_{\text{sim}}} + \left[\overleftrightarrow{G}^{JF}_0 * \vec{J}^{\text{ES}}_{\text{mul}} + \overleftrightarrow{G}^{MF}_0 * \vec{M}^{\text{ES}}_{\text{mul}}\right]_{\partial V_{\text{mul}}} \\ = \left[\overleftrightarrow{G}^{JF}_0 * \vec{J}^{\text{ES}}_{\text{s}0} + \overleftrightarrow{G}^{MF}_0 * \vec{M}^{\text{ES}}_{\text{s}0}\right]_{\partial V_{\text{s}0}} + \left[\overleftrightarrow{G}^{JF}_0 * \vec{J}^{\text{ES}}_{\text{m}0} + \overleftrightarrow{G}^{MF}_0 * \vec{M}^{\text{ES}}_{\text{m}0}\right]_{\partial V_{\text{m}0}} \end{array} \quad \text{(C-64)}$$

where the second equality is based on boundary decomposition (C-52), current decomposition (C-55), and relationship (C-57). Obviously, just like the topological additivity (C-62) and (C-63) satisfied by the GHFP and GBET of the two-body system, the GFHFs for the internal incident field and external scattered field related to the two-body system satisfy the following topological additivity:

$$\begin{aligned}&\text{Internal incident field GFHF of whole system}\\ =&\sum\nolimits_{\xi}\text{Internal incident field GFHF of sub-system } V_{\xi}\end{aligned}\tag{C-65a}$$

$$\begin{aligned}&\text{External scattered field GFHF of whole system}\\ =&\sum\nolimits_{\xi}\text{External scattered field GFHF of sub-system } V_{\xi}\end{aligned}\tag{C-65b}$$

because the GFHF for internal incident field and external scattered field is just the summation of the GHFP for internal incident field and the GHFP for external scattered field.

However, the GFHFs for the internal total field and internal scattered field related to the two-body system don't satisfy topological additivity, because they don't satisfy GBET. To resolve this problem, we introduce the following piecewise Green's functions:

$$\tilde{\overleftrightarrow{G}}{}^{JF}_{\text{sim/mul}}(\vec{r},\vec{r}\,') = \begin{cases}\overleftrightarrow{G}{}^{JF}_{\text{sim/mul}}(\vec{r},\vec{r}\,') & , \left(\vec{r}\in \text{cl}V_{\text{sim/mul}}, \vec{r}\,'\in \text{cl}V_{\text{sim/mul}}\right)\\ 0 & , \left(\vec{r}\in \text{ext}V_{\text{sim/mul}}, \vec{r}\,'\in \text{cl}V_{\text{sim/mul}}\right)\end{cases}\tag{C-66a}$$

$$\tilde{\overleftrightarrow{G}}{}^{MF}_{\text{sim/mul}}(\vec{r},\vec{r}\,') = \begin{cases}\overleftrightarrow{G}{}^{MF}_{\text{sim/mul}}(\vec{r},\vec{r}\,') & , \left(\vec{r}\in \text{cl}V_{\text{sim/mul}}, \vec{r}\,'\in \text{cl}V_{\text{sim/mul}}\right)\\ 0 & , \left(\vec{r}\in \text{ext}V_{\text{sim/mul}}, \vec{r}\,'\in \text{cl}V_{\text{sim/mul}}\right)\end{cases}\tag{C-66b}$$

Based on definitions (C-66a) and (C-66b), we can rewrite the convolution integral formulation (C-42) in simply connected case and the convolution integral formulation (C-50) in multiply connected case as follows:

$$\left.\begin{matrix}\text{ext}V_{\text{sim}} & : & 0\\ \text{int}V_{\text{sim}} & : & \vec{F}^{\text{tot}}\end{matrix}\right\} = \left[\tilde{\overleftrightarrow{G}}{}^{JF}_{\text{sim}} * \vec{J}^{\text{ES}}_{\text{sim}} + \tilde{\overleftrightarrow{G}}{}^{MF}_{\text{sim}} * \vec{M}^{\text{ES}}_{\text{sim}}\right]_{\partial V_{\text{sim}}}\tag{C-67}$$

$$\left.\begin{matrix}\text{ext}V_{\text{mul}} & : & 0\\ \text{int}V_{\text{mul}} & : & \vec{F}^{\text{tot}}\end{matrix}\right\} = \left[\tilde{\overleftrightarrow{G}}{}^{JF}_{\text{mul}} * \vec{J}^{\text{ES}}_{\text{mul}} + \tilde{\overleftrightarrow{G}}{}^{MF}_{\text{mul}} * \vec{M}^{\text{ES}}_{\text{mul}}\right]_{\partial V_{\text{mul}}}\tag{C-68}$$

and then we have the following convolution integral formulation of two-body version:

$$\left.\begin{matrix}\text{ext}V_{\text{sm sys}} & : & 0\\ \text{int}V_{\text{sim}} & : & \vec{F}^{\text{tot}}\\ \text{int}V_{\text{mul}} & : & \vec{F}^{\text{tot}}\end{matrix}\right\} = \left[\tilde{\overleftrightarrow{G}}{}^{JF}_{\text{sim}} * \vec{J}^{\text{ES}}_{\text{sim}} + \tilde{\overleftrightarrow{G}}{}^{MF}_{\text{sim}} * \vec{M}^{\text{ES}}_{\text{sim}}\right]_{\partial V_{\text{sim}}} + \left[\tilde{\overleftrightarrow{G}}{}^{JF}_{\text{mul}} * \vec{J}^{\text{ES}}_{\text{mul}} + \tilde{\overleftrightarrow{G}}{}^{MF}_{\text{mul}} * \vec{M}^{\text{ES}}_{\text{mul}}\right]_{\partial V_{\text{mul}}}\tag{C-69}$$

Similarly, if we introduce the following piecewise delta Green's functions:

$$\begin{aligned}\Delta\tilde{\overset{\leftrightarrow}{G}}{}^{JF}_{\text{sim/mul}}(\vec{r},\vec{r}') &= \tilde{\overset{\leftrightarrow}{G}}{}^{JF}_{\text{sim/mul}}(\vec{r},\vec{r}') - \overset{\leftrightarrow}{G}{}^{JF}_{0}(\vec{r},\vec{r}') \\ &= \begin{cases} \overset{\leftrightarrow}{G}{}^{JF}_{\text{sim/mul}}(\vec{r},\vec{r}') - \overset{\leftrightarrow}{G}{}^{JF}_{0}(\vec{r},\vec{r}') &, \left(\vec{r}\in \text{cl}V_{\text{sim/mul}} \,, \vec{r}'\in \text{cl}V_{\text{sim/mul}}\right) \\ -\overset{\leftrightarrow}{G}{}^{JF}_{0}(\vec{r},\vec{r}') &, \left(\vec{r}\in \text{ext}V_{\text{sim/mul}} \,, \vec{r}'\in \text{cl}V_{\text{sim/mul}}\right) \end{cases}\end{aligned} \quad (\text{C-70a})$$

$$\begin{aligned}\Delta\tilde{\overset{\leftrightarrow}{G}}{}^{MF}_{\text{sim/mul}}(\vec{r},\vec{r}') &= \tilde{\overset{\leftrightarrow}{G}}{}^{MF}_{\text{sim/mul}}(\vec{r},\vec{r}') - \overset{\leftrightarrow}{G}{}^{MF}_{0}(\vec{r},\vec{r}') \\ &= \begin{cases} \overset{\leftrightarrow}{G}{}^{MF}_{\text{sim/mul}}(\vec{r},\vec{r}') - \overset{\leftrightarrow}{G}{}^{MF}_{0}(\vec{r},\vec{r}') &, \left(\vec{r}\in \text{cl}V_{\text{sim/mul}} \,, \vec{r}'\in \text{cl}V_{\text{sim/mul}}\right) \\ -\overset{\leftrightarrow}{G}{}^{MF}_{0}(\vec{r},\vec{r}') &, \left(\vec{r}\in \text{ext}V_{\text{sim/mul}} \,, \vec{r}'\in \text{cl}V_{\text{sim/mul}}\right) \end{cases}\end{aligned} \quad (\text{C-70b})$$

then the convolution integral formulation (C-43) of simply connected version and the convolution integral formulation (C-51) of multiply connected version can be rewritten as follows:

$$\left.\begin{aligned} \text{ext}V_{\text{sim}} &: \vec{F}^{\text{sca}}_{\text{sim}} \\ \text{int}V_{\text{sim}} &: \vec{F}^{\text{sca}}_{\text{sim}} \end{aligned}\right\} = \left[\Delta\tilde{\overset{\leftrightarrow}{G}}{}^{JF}_{\text{sim}} * \vec{J}^{\text{ES}}_{\text{sim}} + \Delta\tilde{\overset{\leftrightarrow}{G}}{}^{MF}_{\text{sim}} * \vec{M}^{\text{ES}}_{\text{sim}}\right]_{\partial V_{\text{sim}}} \quad (\text{C-71})$$

$$\left.\begin{aligned} \text{ext}V_{\text{mul}} &: \vec{F}^{\text{sca}}_{\text{mul}} \\ \text{int}V_{\text{mul}} &: \vec{F}^{\text{sca}}_{\text{mul}} \end{aligned}\right\} = \left[\Delta\tilde{\overset{\leftrightarrow}{G}}{}^{JF}_{\text{mul}} * \vec{J}^{\text{ES}}_{\text{mul}} + \Delta\tilde{\overset{\leftrightarrow}{G}}{}^{MF}_{\text{mul}} * \vec{M}^{\text{ES}}_{\text{mul}}\right]_{\partial V_{\text{mul}}} \quad (\text{C-72})$$

and then we have the convolution integral formulation of two-body version as follows:

$$\left.\begin{aligned} \text{ext}V_{\text{sm sys}} &: \vec{F}^{\text{sca}} \\ \text{int}V_{\text{sim}} &: \vec{F}^{\text{sca}} \\ \text{int}V_{\text{mul}} &: \vec{F}^{\text{sca}} \end{aligned}\right\} = \left[\Delta\tilde{\overset{\leftrightarrow}{G}}{}^{JF}_{\text{sim}} * \vec{J}^{\text{ES}}_{\text{sim}} + \Delta\tilde{\overset{\leftrightarrow}{G}}{}^{MF}_{\text{sim}} * \vec{M}^{\text{ES}}_{\text{sim}}\right]_{\partial V_{\text{sim}}} + \left[\Delta\tilde{\overset{\leftrightarrow}{G}}{}^{JF}_{\text{mul}} * \vec{J}^{\text{ES}}_{\text{mul}} + \Delta\tilde{\overset{\leftrightarrow}{G}}{}^{MF}_{\text{mul}} * \vec{M}^{\text{ES}}_{\text{mul}}\right]_{\partial V_{\text{mul}}} \quad (\text{C-73})$$

Obviously, the above GFHFs employing piecewise Green's functions and piecewise delta Green's functions satisfy the following topological additivity:

$$\begin{aligned}&\text{Internal total field GFHF of whole system} \\ &= \sum\nolimits_{\xi} \text{Internal total field GFHF of sub-system } V_{\xi}\end{aligned} \quad (\text{C-74})$$

$$\begin{aligned}&\text{Internal scattered field GFHF of whole system} \\ &= \sum\nolimits_{\xi} \text{Internal scattered field GFHF of sub-system } V_{\xi}\end{aligned} \quad (\text{C-75})$$

In fact, topological additivity (C-65), (C-74), and (C-75) can also be uniformly written as the following form:

$$\text{GFHF of whole system} = \sum\nolimits_{\xi} \text{GFHF of sub-system } V_{\xi} \quad (\text{C-76})$$

**4) Generalized Huygens-Fresnel Principle, Generalized Backward Extinction Theorem, and Generalized Franz-Harrington Formulation —— Special Cases**

In this point 4), we will, focusing on some special cases, give the special manifestation forms of the above general results.

**(4.1) The Two Material Bodies Used to Construct the Two-body System Are Not Contact With Each Other**

In this case, there exist relationships: $\partial V_{\text{sm sys}} = \partial V_{\text{sim}} \bigcup \partial V_{\text{mul}}$, $\text{ext} V_{\text{sm sys}} = \text{ext} V_{\text{sm sys}}^{\text{in}} \bigcup \text{ext} V_{\text{sm sys}}^{\text{out}}$, and $\text{int} V_{\text{sm sys}} = \text{int} V_{\text{sim}} \bigcup \text{int} V_{\text{mul}}$, where $\partial V_{\text{sim}}$, $\partial V_{\text{mul}}$, $\text{ext} V_{\text{sm sys}}^{\text{in}}$, $\text{ext} V_{\text{sm sys}}^{\text{out}}$, $\text{int} V_{\text{sim}}$, and $\text{int} V_{\text{mul}}$ are illustrated in Figure C-6(a). Thus, the GHFP (C-60) for incident field and the GHFP (C-61) for scattered field can be concretized as follows:

$$\left.\begin{array}{lcl} \text{ext} V_{\text{sm sys}} & : & 0 \\ \text{int} V_{\text{sm sys}} & : & \vec{F}^{\text{inc}} \end{array}\right\} = \left[ \overset{\leftrightarrow}{G}_0^{JF} * \left( \hat{n}_{\text{sys}}^{-} \times \vec{H}^{\text{inc}} \right) + \overset{\leftrightarrow}{G}_0^{MF} * \left( \vec{E}^{\text{inc}} \times \hat{n}_{\text{sys}}^{-} \right) \right]_{\partial V_{\text{sm sys}}} \tag{C-77}$$

$$\left.\begin{array}{lcl} \text{ext} V_{\text{sm sys}} & : & \vec{F}^{\text{sca}} \\ \text{int} V_{\text{sm sys}} & : & 0 \end{array}\right\} = \left[ \overset{\leftrightarrow}{G}_0^{JF} * \left( \hat{n}_{\text{sys}}^{+} \times \vec{H}^{\text{sca}} \right) + \overset{\leftrightarrow}{G}_0^{MF} * \left( \vec{E}^{\text{sca}} \times \hat{n}_{\text{sys}}^{+} \right) \right]_{\partial V_{\text{sm sys}}} \tag{C-78}$$

and GFHF (C-64) can be concretized as follows:

$$\left.\begin{array}{lcl} \text{ext} V_{\text{sm sys}} & : & -\vec{F}^{\text{sca}} \\ \text{int} V_{\text{sm sys}} & : & \vec{F}^{\text{inc}} \end{array}\right\} = \left[ \overset{\leftrightarrow}{G}_0^{JF} * \vec{J}_{\text{sys}}^{\text{ES}} + \overset{\leftrightarrow}{G}_0^{MF} * \vec{M}_{\text{sys}}^{\text{ES}} \right]_{\partial V_{\text{sm sys}}} \tag{C-79}$$

and GFHF (C-69) and GFHF (C-73) can be respectively concretized as follows:

$$\left.\begin{array}{lcl} \text{ext} V_{\text{sm sys}} & : & 0 \\ \text{int} V_{\text{sm sys}} & : & \vec{F}^{\text{tot}} \end{array}\right\} = \left[ \overset{\approx}{G}_{\text{sys}}^{JF} * \vec{J}_{\text{sys}}^{\text{ES}} + \overset{\approx}{G}_{\text{sys}}^{MF} * \vec{M}_{\text{sys}}^{\text{ES}} \right]_{\partial V_{\text{sm sys}}} \tag{C-80}$$

$$\left.\begin{array}{lcl} \text{ext} V_{\text{sm sys}} & : & \vec{F}^{\text{sca}} \\ \text{int} V_{\text{sm sys}} & : & \vec{F}^{\text{sca}} \end{array}\right\} = \left[ \Delta\overset{\approx}{G}_{\text{sys}}^{JF} * \vec{J}_{\text{sys}}^{\text{ES}} + \Delta\overset{\approx}{G}_{\text{sys}}^{MF} * \vec{M}_{\text{sys}}^{\text{ES}} \right]_{\partial V_{\text{sm sys}}} \tag{C-81}$$

The Green's functions and equivalent surface sources used in convolution integral formulations (C-80) and (C-81) are respectively as follows:

$$\overset{\approx}{G}_{\text{sys}}^{JF/MF} \left( \vec{r}, \vec{r}' \right) = \left\{ \begin{array}{lcl} \overset{\approx}{G}_{\text{sim}}^{JF/MF} \left( \vec{r}, \vec{r}' \right) & , & \vec{r}' \in \text{cl} V_{\text{sim}} \\ \overset{\approx}{G}_{\text{mul}}^{JF/MF} \left( \vec{r}, \vec{r}' \right) & , & \vec{r}' \in \text{cl} V_{\text{mul}} \end{array} \right. \tag{C-82}$$

$$\Delta\overset{\approx}{G}_{\text{sys}}^{JF/MF} \left( \vec{r}, \vec{r}' \right) = \left\{ \begin{array}{lcl} \Delta\overset{\approx}{G}_{\text{sim}}^{JF/MF} \left( \vec{r}, \vec{r}' \right) & , & \vec{r}' \in \text{cl} V_{\text{sim}} \\ \Delta\overset{\approx}{G}_{\text{mul}}^{JF/MF} \left( \vec{r}, \vec{r}' \right) & , & \vec{r}' \in \text{cl} V_{\text{mul}} \end{array} \right. \tag{C-83}$$

and

$$\vec{C}_{\text{sys}}^{\text{ES}} \left( \vec{r} \right) = \left\{ \begin{array}{lcl} \vec{C}_{\text{sim}}^{\text{ES}} \left( \vec{r} \right) & , & \vec{r} \in \partial V_{\text{sim}} \\ \vec{C}_{\text{mul}}^{\text{ES}} \left( \vec{r} \right) & , & \vec{r} \in \partial V_{\text{mul}} \end{array} \right. \tag{C-84}$$

In definitions (C-82) and (C-83), $F = E, H$; in definition (C-84), $C = J, M$.

**(4.2) The Two Material Bodies Used to Construct the Two-body System Are Contact With Each Other, But Anyone Is Not Completely Coated by the Other**

In this case, there exist relationships: $\partial V_{\text{sm sys}} = \partial V_{\text{s}0} \bigcup \partial V_{\text{m}0}$ , and $\text{ext}\, V_{\text{sm sys}} = \text{ext}\, V_{\text{sm sys}}^{\text{in}} \bigcup \text{ext}\, V_{\text{sm sys}}^{\text{out}}$ , and $\text{int}\, V_{\text{sm sys}} = \text{int}\, V_{\text{sim}} \bigcup \text{int}\, V_{\text{mul}} \bigcup \partial V_{\text{sm}} = \text{int}\, V_{\text{sim}} \bigcup \text{int}\, V_{\text{mul}} \bigcup \partial V_{\text{ms}}$ , where $\partial V_{\text{s}0}$ , $\partial V_{\text{m}0}$ , $\partial V_{\text{sm}}$ , $\partial V_{\text{ms}}$ , $\text{ext}\, V_{\text{sm sys}}^{\text{in}}$ , $\text{ext}\, V_{\text{sm sys}}^{\text{out}}$ , $\text{int}\, V_{\text{sim}}$ , and $\text{int}\, V_{\text{mul}}$ are illustrated in Figure C-6(b).

Because $\hat{n}_{\text{sim}}^{\pm} = -\hat{n}_{\text{mul}}^{\pm}$ on whole $\partial V_{\text{sm}}$ , then GHFP (C-60) and GHFP (C-61) will become GHFP (C-77) and GHFP (C-78) formally, and GFHF (C-64) will become GFHF (C-79) formally.

**(4.3) The Two Material Bodies Used to Construct the Two-body System Are Contact with Each Other, and One Is Completely Coated by the Other**

In this case, there exist relationships: $\partial V_{\text{sm}} = \partial V_{\text{sim}}$ , and $\partial V_{\text{s}0} = \varnothing$ , and $\partial V_{\text{sm sys}} = \partial V_{\text{m}0}$ , and $\text{int}\, V_{\text{sm sys}} = \text{int}\, V_{\text{mul}} \bigcup \text{int}\, V_{\text{sim}} \bigcup \partial V_{\text{sm/ms}} = \text{int}\, V_{\text{mul}} \bigcup \text{int}\, V_{\text{sim}} \bigcup \partial V_{\text{sim}} = \text{int}\, V_{\text{mul}} \bigcup \text{cl}\, V_{\text{sim}}$ , where boundaries $\partial V_{\text{sim}}$ , $\partial V_{\text{sm}}$ , $\partial V_{\text{ms}}$ , $\partial V_{\text{sm sys}}$ , and $\partial V_{\text{m}0}$ and open domains $\text{int}\, V_{\text{sim}}$ and $\text{int}\, V_{\text{mul}}$ are respectively illustrated in Figure C-6(c).

Because $\hat{n}_{\text{sim}}^{\pm} = -\hat{n}_{\text{mul}}^{\pm}$ on whole $\partial V_{\text{sm}}$ (i.e. $\partial V_{\text{ms}}$ ), then GHFP (C-60) and GHFP (C-61) will become GHFP (C-77) and GHFP (C-78) formally, and GFHF (C-64) will become GFHF (C-79) formally.

# Acknowledgements

This doctoral dissertation is dedicated to my family. The dissertation cannot be finished without their constant understanding, support, and encouragement.

# Research Results Obtained During the Study for Doctor Degree